\def\sphinxdocclass{report}
\documentclass[letterpaper,10pt,english]{sphinxmanual}
\ifdefined\pdfpxdimen
   \let\sphinxpxdimen\pdfpxdimen\else\newdimen\sphinxpxdimen
\fi \sphinxpxdimen=.75bp\relax

\PassOptionsToPackage{warn}{textcomp}
\usepackage[utf8]{inputenc}
\ifdefined\DeclareUnicodeCharacter
  \ifdefined\DeclareUnicodeCharacterAsOptional
    \def\sphinxDUC#1{\DeclareUnicodeCharacter{"#1}}
  \else
    \let\sphinxDUC\DeclareUnicodeCharacter
  \fi
  \sphinxDUC{00A0}{\nobreakspace}
  \sphinxDUC{2500}{\sphinxunichar{2500}}
  \sphinxDUC{2502}{\sphinxunichar{2502}}
  \sphinxDUC{2514}{\sphinxunichar{2514}}
  \sphinxDUC{251C}{\sphinxunichar{251C}}
  \sphinxDUC{2572}{\textbackslash}
\fi
\usepackage{cmap}
\usepackage[T1]{fontenc}
\usepackage{amsmath,amssymb,amstext}
\usepackage{babel}

\usepackage{times}
\expandafter\ifx\csname T@LGR\endcsname\relax
\else
  \substitutefont{LGR}{\rmdefault}{cmr}
  \substitutefont{LGR}{\sfdefault}{cmss}
  \substitutefont{LGR}{\ttdefault}{cmtt}
\fi
\expandafter\ifx\csname T@X2\endcsname\relax
  \expandafter\ifx\csname T@T2A\endcsname\relax
  \else
    \substitutefont{T2A}{\rmdefault}{cmr}
    \substitutefont{T2A}{\sfdefault}{cmss}
    \substitutefont{T2A}{\ttdefault}{cmtt}
  \fi
\else
  \substitutefont{X2}{\rmdefault}{cmr}
  \substitutefont{X2}{\sfdefault}{cmss}
  \substitutefont{X2}{\ttdefault}{cmtt}
\fi

\usepackage[Bjarne]{fncychap}
\usepackage{sphinx}

\fvset{fontsize=\small}
\usepackage{geometry}

\usepackage{hyperref}
\usepackage{hypcap}
\usepackage{sphinxmessages}
\title{symjax}
\date{May 21, 2020}
\release{0.0.1}
\author{Randall Balestriero}
\newcommand{\sphinxlogo}{\vbox{}}
\renewcommand{\releasename}{Release}
\makeindex
\begin{document}

\pagestyle{empty}
\sphinxmaketitle
\pagestyle{plain}
\sphinxtableofcontents
\pagestyle{normal}
\phantomsection\label{\detokenize{index::doc}}

\begin{itemize}
\item {} 
\sphinxhref{https://github.com/google/jax}{JAX} = \sphinxhref{https://www.tensorflow.org/xla}{XLA} + \sphinxhref{https://github.com/hips/autograd}{Autograd}

\item {} 
\sphinxhref{https://github.com/RandallBalestriero/SymJAX}{SymJAX} =  \sphinxhref{https://github.com/google/jax}{JAX} + symbolic programming + deep Learning

\end{itemize}

\sphinxhref{https://www.tensorflow.org/xla}{XLA} is a compiler that optimizes a computational graph by fusing multiple kernels into one preventing intermediate computation, reducing memory operations and increasing performances.

\sphinxhref{https://github.com/google/jax}{JAX} is a python interface that provides a \sphinxhref{https://numpy.org}{Numpy}\sphinxhyphen{}like software on top of XLA and providing just\sphinxhyphen{}in\sphinxhyphen{}time compilation a well as advanced automatic differenciation.

\sphinxhref{https://github.com/RandallBalestriero/SymJAX}{SymJAX} is a symbolic programming version of \sphinxhref{https://github.com/google/jax}{JAX} simplifying graph input, output and updates and providing additional functionalities for general machine learning and deep learning applications. From an user perspective \sphinxhref{https://github.com/RandallBalestriero/SymJAX}{SymJAX} apparents to \sphinxhref{https://github.com/Theano/Theano}{Theano} with fast graph optimization/compilation and broad hardware support, along with \sphinxhref{https://github.com/Lasagne/Lasagne}{Lasagne}\sphinxhyphen{}like deep learning functionalities

This is an under development research project, not an official product, expect bugs and sharp edges; please help by trying it out, reporting bugs.

\chapter{Contents}
\label{\detokenize{index:contents}}

\section{Installation}
\label{\detokenize{user/installation:installation}}\label{\detokenize{user/installation::doc}}
This installation is restricted to GPU support only.

\subsection{Installation with pip}
\label{\detokenize{user/installation:installation-with-pip}}\begin{enumerate}
\sphinxsetlistlabels{\arabic}{enumi}{enumii}{}{.}%
\item {} 
Install all GPU divers and compilers (\sphinxcode{\sphinxupquote{cuda}}, \sphinxcode{\sphinxupquote{cudnn}}, and GPU drivers).

\item {} 
Install \sphinxcode{\sphinxupquote{jax}} following \sphinxhref{https://github.com/google/jax\#installation}{Jax Installation}. Here is a minimal instruction to install the GPU version
\begin{quote}

\begin{sphinxVerbatim}[commandchars=\\\{\}]
\PYGZdl{} \PYG{n+nv}{PYTHON\PYGZus{}VERSION}\PYG{o}{=}cp37  \PYG{c+c1}{\PYGZsh{} alternatives: cp35, cp36, cp37, cp38}
\PYGZdl{} \PYG{n+nv}{CUDA\PYGZus{}VERSION}\PYG{o}{=}cuda92  \PYG{c+c1}{\PYGZsh{} alternatives: cuda92, cuda100, cuda101, cuda102}
\PYGZdl{} \PYG{n+nv}{PLATFORM}\PYG{o}{=}linux\PYGZus{}x86\PYGZus{}64  \PYG{c+c1}{\PYGZsh{} alternatives: linux\PYGZus{}x86\PYGZus{}64}
\PYGZdl{} \PYG{n+nv}{BASE\PYGZus{}URL}\PYG{o}{=}\PYG{l+s+s1}{\PYGZsq{}https://storage.googleapis.com/jax\PYGZhy{}releases\PYGZsq{}}
\PYGZdl{} pip install \PYGZhy{}\PYGZhy{}upgrade \PYG{n+nv}{\PYGZdl{}BASE\PYGZus{}URL}/\PYG{n+nv}{\PYGZdl{}CUDA\PYGZus{}VERSION} jaxlib\PYGZhy{}0.1.38\PYGZhy{}\PYG{n+nv}{\PYGZdl{}PYTHON\PYGZus{}VERSION}\PYGZhy{}none\PYGZhy{}\PYG{n+nv}{\PYGZdl{}PLATFORM}.whl
\end{sphinxVerbatim}
\end{quote}

\item {} 
Install SymJAX with
\begin{quote}

\begin{sphinxVerbatim}[commandchars=\\\{\}]
\PYGZdl{} pip install symjax
\end{sphinxVerbatim}
\end{quote}

\end{enumerate}

\subsection{Manual installation}
\label{\detokenize{user/installation:manual-installation}}\begin{enumerate}
\sphinxsetlistlabels{\arabic}{enumi}{enumii}{}{.}%
\item {} 
Clone this repository with
\begin{quote}

\begin{sphinxVerbatim}[commandchars=\\\{\}]
\PYGZdl{} git clone https://github.com/RandallBalestriero/SymJAX
\end{sphinxVerbatim}
\end{quote}

\item {} 
Install.
\begin{quote}

\begin{sphinxVerbatim}[commandchars=\\\{\}]
\PYGZdl{} \PYG{n+nb}{cd} SymJAX
\PYGZdl{} pip install \PYGZhy{}r requirements.txt
\PYGZdl{} pip install .
\end{sphinxVerbatim}
\end{quote}

\end{enumerate}

\section{Examples}
\label{\detokenize{user/examples:examples}}\label{\detokenize{user/examples::doc}}

\subsection{Multivariate Gaussian}
\label{\detokenize{user/examples:multivariate-gaussian}}
\begin{sphinxVerbatim}[commandchars=\\\{\}]
\PYG{k+kn}{import} \PYG{n+nn}{sys}
\PYG{n}{sys}\PYG{o}{.}\PYG{n}{path}\PYG{o}{.}\PYG{n}{insert}\PYG{p}{(}\PYG{l+m+mi}{0}\PYG{p}{,} \PYG{l+s+s2}{\PYGZdq{}}\PYG{l+s+s2}{../}\PYG{l+s+s2}{\PYGZdq{}}\PYG{p}{)}

\PYG{k+kn}{import} \PYG{n+nn}{symjax} \PYG{k}{as} \PYG{n+nn}{sj}
\PYG{k+kn}{import} \PYG{n+nn}{symjax}\PYG{n+nn}{.}\PYG{n+nn}{tensor} \PYG{k}{as} \PYG{n+nn}{T}
\PYG{k+kn}{import} \PYG{n+nn}{matplotlib}\PYG{n+nn}{.}\PYG{n+nn}{pyplot} \PYG{k}{as} \PYG{n+nn}{plt}
\PYG{k+kn}{import} \PYG{n+nn}{matplotlib}
\PYG{n}{matplotlib}\PYG{o}{.}\PYG{n}{use}\PYG{p}{(}\PYG{l+s+s1}{\PYGZsq{}}\PYG{l+s+s1}{Agg}\PYG{l+s+s1}{\PYGZsq{}}\PYG{p}{)}

\PYG{c+c1}{\PYGZsh{}\PYGZsh{}\PYGZsh{}\PYGZsh{}\PYGZsh{}\PYGZsh{} 2D GAUSSIAN EXAMPLE}

\PYG{n}{t} \PYG{o}{=} \PYG{n}{T}\PYG{o}{.}\PYG{n}{linspace}\PYG{p}{(}\PYG{o}{\PYGZhy{}}\PYG{l+m+mi}{5}\PYG{p}{,} \PYG{l+m+mi}{5}\PYG{p}{,} \PYG{l+m+mi}{5}\PYG{p}{)}
\PYG{n}{x}\PYG{p}{,} \PYG{n}{y} \PYG{o}{=} \PYG{n}{T}\PYG{o}{.}\PYG{n}{meshgrid}\PYG{p}{(}\PYG{n}{t}\PYG{p}{,} \PYG{n}{t}\PYG{p}{)}
\PYG{n}{X} \PYG{o}{=} \PYG{n}{T}\PYG{o}{.}\PYG{n}{stack}\PYG{p}{(}\PYG{p}{[}\PYG{n}{x}\PYG{o}{.}\PYG{n}{flatten}\PYG{p}{(}\PYG{p}{)}\PYG{p}{,} \PYG{n}{y}\PYG{o}{.}\PYG{n}{flatten}\PYG{p}{(}\PYG{p}{)}\PYG{p}{]}\PYG{p}{,} \PYG{l+m+mi}{1}\PYG{p}{)}
\PYG{n}{p} \PYG{o}{=} \PYG{n}{T}\PYG{o}{.}\PYG{n}{pdfs}\PYG{o}{.}\PYG{n}{multivariate\PYGZus{}normal}\PYG{o}{.}\PYG{n}{pdf}\PYG{p}{(}\PYG{n}{X}\PYG{p}{,} \PYG{n}{T}\PYG{o}{.}\PYG{n}{zeros}\PYG{p}{(}\PYG{l+m+mi}{2}\PYG{p}{)}\PYG{p}{,} \PYG{n}{T}\PYG{o}{.}\PYG{n}{eye}\PYG{p}{(}\PYG{l+m+mi}{2}\PYG{p}{)}\PYG{p}{)}
\PYG{n}{p} \PYG{o}{=} \PYG{n}{p}\PYG{o}{.}\PYG{n}{reshape}\PYG{p}{(}\PYG{p}{(}\PYG{l+m+mi}{5}\PYG{p}{,} \PYG{l+m+mi}{5}\PYG{p}{)}\PYG{p}{)}\PYG{o}{.}\PYG{n}{round}\PYG{p}{(}\PYG{l+m+mi}{2}\PYG{p}{)}

\PYG{n+nb}{print}\PYG{p}{(}\PYG{n}{p}\PYG{p}{)}
\PYG{c+c1}{\PYGZsh{} Tensor(Op=round\PYGZus{}, shape=(5, 5), dtype=float32)}

\PYG{c+c1}{\PYGZsh{} lazy evaluation (not compiled nor optimized)}
\PYG{n+nb}{print}\PYG{p}{(}\PYG{n}{p}\PYG{o}{.}\PYG{n}{get}\PYG{p}{(}\PYG{p}{)}\PYG{p}{)}
\PYG{c+c1}{\PYGZsh{} [[0.   0.   0.   0.   0.  ]}
\PYG{c+c1}{\PYGZsh{}  [0.   0.   0.01 0.   0.  ]}
\PYG{c+c1}{\PYGZsh{}  [0.   0.01 0.16 0.01 0.  ]}
\PYG{c+c1}{\PYGZsh{}  [0.   0.   0.01 0.   0.  ]}
\PYG{c+c1}{\PYGZsh{}  [0.   0.   0.   0.   0.  ]]}

\PYG{c+c1}{\PYGZsh{} create the function which internall compiles and optimizes}
\PYG{c+c1}{\PYGZsh{} the function does not take any arguments and only outputs the}
\PYG{c+c1}{\PYGZsh{} computed tensor p}
\PYG{n}{f} \PYG{o}{=} \PYG{n}{sj}\PYG{o}{.}\PYG{n}{function}\PYG{p}{(}\PYG{n}{outputs}\PYG{o}{=}\PYG{n}{p}\PYG{p}{)}
\PYG{n+nb}{print}\PYG{p}{(}\PYG{n}{f}\PYG{p}{(}\PYG{p}{)}\PYG{p}{)}
\PYG{c+c1}{\PYGZsh{} [[0.   0.   0.   0.   0.  ]}
\PYG{c+c1}{\PYGZsh{}  [0.   0.   0.01 0.   0.  ]}
\PYG{c+c1}{\PYGZsh{}  [0.   0.01 0.16 0.01 0.  ]}
\PYG{c+c1}{\PYGZsh{}  [0.   0.   0.01 0.   0.  ]}
\PYG{c+c1}{\PYGZsh{}  [0.   0.   0.   0.   0.  ]]}

\end{sphinxVerbatim}

\subsection{Stochastic Gradient Descent}
\label{\detokenize{user/examples:stochastic-gradient-descent}}
\begin{sphinxVerbatim}[commandchars=\\\{\}]
\PYG{k+kn}{import} \PYG{n+nn}{sys}
\PYG{n}{sys}\PYG{o}{.}\PYG{n}{path}\PYG{o}{.}\PYG{n}{insert}\PYG{p}{(}\PYG{l+m+mi}{0}\PYG{p}{,} \PYG{l+s+s2}{\PYGZdq{}}\PYG{l+s+s2}{../}\PYG{l+s+s2}{\PYGZdq{}}\PYG{p}{)}
\PYG{k+kn}{import} \PYG{n+nn}{symjax}
\PYG{k+kn}{import} \PYG{n+nn}{symjax}\PYG{n+nn}{.}\PYG{n+nn}{tensor} \PYG{k}{as} \PYG{n+nn}{T}

\PYG{c+c1}{\PYGZsh{} create our variable to be optimized}
\PYG{n}{mu} \PYG{o}{=} \PYG{n}{T}\PYG{o}{.}\PYG{n}{Variable}\PYG{p}{(}\PYG{n}{T}\PYG{o}{.}\PYG{n}{random}\PYG{o}{.}\PYG{n}{normal}\PYG{p}{(}\PYG{p}{(}\PYG{p}{)}\PYG{p}{,} \PYG{n}{seed}\PYG{o}{=}\PYG{l+m+mi}{1}\PYG{p}{)}\PYG{p}{)}

\PYG{c+c1}{\PYGZsh{} create our cost}
\PYG{n}{cost} \PYG{o}{=} \PYG{n}{T}\PYG{o}{.}\PYG{n}{exp}\PYG{p}{(}\PYG{o}{\PYGZhy{}}\PYG{p}{(}\PYG{n}{mu}\PYG{o}{\PYGZhy{}}\PYG{l+m+mi}{1}\PYG{p}{)}\PYG{o}{*}\PYG{o}{*}\PYG{l+m+mi}{2}\PYG{p}{)}

\PYG{c+c1}{\PYGZsh{} get the gradient, notice that it is itself a tensor that can then}
\PYG{c+c1}{\PYGZsh{} be manipulated as well}
\PYG{n}{g} \PYG{o}{=} \PYG{n}{symjax}\PYG{o}{.}\PYG{n}{gradients}\PYG{p}{(}\PYG{n}{cost}\PYG{p}{,} \PYG{n}{mu}\PYG{p}{)}
\PYG{n+nb}{print}\PYG{p}{(}\PYG{n}{g}\PYG{p}{)}

\PYG{c+c1}{\PYGZsh{} (Tensor: shape=(), dtype=float32)}

\PYG{c+c1}{\PYGZsh{} create the compield function that will compute the cost and apply}
\PYG{c+c1}{\PYGZsh{} the update onto the variable}
\PYG{n}{f} \PYG{o}{=} \PYG{n}{symjax}\PYG{o}{.}\PYG{n}{function}\PYG{p}{(}\PYG{n}{outputs}\PYG{o}{=}\PYG{n}{cost}\PYG{p}{,} \PYG{n}{updates}\PYG{o}{=}\PYG{p}{\PYGZob{}}\PYG{n}{mu}\PYG{p}{:}\PYG{n}{mu}\PYG{o}{\PYGZhy{}}\PYG{l+m+mf}{0.2}\PYG{o}{*}\PYG{n}{g}\PYG{p}{\PYGZcb{}}\PYG{p}{)}

\PYG{k}{for} \PYG{n}{i} \PYG{o+ow}{in} \PYG{n+nb}{range}\PYG{p}{(}\PYG{l+m+mi}{10}\PYG{p}{)}\PYG{p}{:}
    \PYG{n+nb}{print}\PYG{p}{(}\PYG{n}{f}\PYG{p}{(}\PYG{p}{)}\PYG{p}{)}

\PYG{c+c1}{\PYGZsh{} 0.008471076}
\PYG{c+c1}{\PYGZsh{} 0.008201109}
\PYG{c+c1}{\PYGZsh{} 0.007946267}
\PYG{c+c1}{\PYGZsh{} 0.007705368}
\PYG{c+c1}{\PYGZsh{} 0.0074773384}
\PYG{c+c1}{\PYGZsh{} 0.007261208}
\PYG{c+c1}{\PYGZsh{} 0.0070561105}
\PYG{c+c1}{\PYGZsh{} 0.006861261}
\PYG{c+c1}{\PYGZsh{} 0.006675923}
\PYG{c+c1}{\PYGZsh{} 0.006499458}

\end{sphinxVerbatim}

\subsection{CIFAR\sphinxhyphen{}10 Classification}
\label{\detokenize{user/examples:cifar-10-classification}}
\begin{sphinxVerbatim}[commandchars=\\\{\}]
\PYG{k+kn}{import} \PYG{n+nn}{sys}
\PYG{n}{sys}\PYG{o}{.}\PYG{n}{path}\PYG{o}{.}\PYG{n}{insert}\PYG{p}{(}\PYG{l+m+mi}{0}\PYG{p}{,} \PYG{l+s+s2}{\PYGZdq{}}\PYG{l+s+s2}{../}\PYG{l+s+s2}{\PYGZdq{}}\PYG{p}{)}
\PYG{k+kn}{import} \PYG{n+nn}{symjax}\PYG{n+nn}{.}\PYG{n+nn}{tensor} \PYG{k}{as} \PYG{n+nn}{T}                                                   
\PYG{k+kn}{import} \PYG{n+nn}{symjax} \PYG{k}{as} \PYG{n+nn}{sj}
\PYG{k+kn}{import} \PYG{n+nn}{numpy} \PYG{k}{as} \PYG{n+nn}{np}

\PYG{c+c1}{\PYGZsh{} load the dataset}
\PYG{n}{images\PYGZus{}train}\PYG{p}{,} \PYG{n}{labels\PYGZus{}train}\PYG{p}{,} \PYG{n}{images\PYGZus{}test}\PYG{p}{,} \PYG{n}{labels\PYGZus{}test} \PYG{o}{=} \PYG{n}{sj}\PYG{o}{.}\PYG{n}{datasets}\PYG{o}{.}\PYG{n}{cifar10}\PYG{o}{.}\PYG{n}{load}\PYG{p}{(}\PYG{p}{)}

\PYG{c+c1}{\PYGZsh{} some renormalization}
\PYG{n}{images\PYGZus{}train} \PYG{o}{/}\PYG{o}{=} \PYG{n}{images\PYGZus{}train}\PYG{o}{.}\PYG{n}{max}\PYG{p}{(}\PYG{p}{(}\PYG{l+m+mi}{1}\PYG{p}{,} \PYG{l+m+mi}{2}\PYG{p}{,} \PYG{l+m+mi}{3}\PYG{p}{)}\PYG{p}{,} \PYG{n}{keepdims}\PYG{o}{=}\PYG{k+kc}{True}\PYG{p}{)}
\PYG{n}{images\PYGZus{}test} \PYG{o}{/}\PYG{o}{=} \PYG{n}{images\PYGZus{}test}\PYG{o}{.}\PYG{n}{max}\PYG{p}{(}\PYG{p}{(}\PYG{l+m+mi}{1}\PYG{p}{,} \PYG{l+m+mi}{2}\PYG{p}{,} \PYG{l+m+mi}{3}\PYG{p}{)}\PYG{p}{,} \PYG{n}{keepdims}\PYG{o}{=}\PYG{k+kc}{True}\PYG{p}{)}

\PYG{c+c1}{\PYGZsh{} create the network}
\PYG{n}{BATCH\PYGZus{}SIZE} \PYG{o}{=} \PYG{l+m+mi}{32}
\PYG{n}{inputs} \PYG{o}{=} \PYG{n}{T}\PYG{o}{.}\PYG{n}{Placeholder}\PYG{p}{(}\PYG{p}{(}\PYG{n}{BATCH\PYGZus{}SIZE}\PYG{p}{,}\PYG{p}{)} \PYG{o}{+} \PYG{n}{images\PYGZus{}train}\PYG{o}{.}\PYG{n}{shape}\PYG{p}{[}\PYG{l+m+mi}{1}\PYG{p}{:}\PYG{p}{]}\PYG{p}{,} \PYG{l+s+s1}{\PYGZsq{}}\PYG{l+s+s1}{float32}\PYG{l+s+s1}{\PYGZsq{}}\PYG{p}{)}
\PYG{n}{outputs} \PYG{o}{=} \PYG{n}{T}\PYG{o}{.}\PYG{n}{Placeholder}\PYG{p}{(}\PYG{p}{(}\PYG{n}{BATCH\PYGZus{}SIZE}\PYG{p}{,}\PYG{p}{)}\PYG{p}{,} \PYG{l+s+s1}{\PYGZsq{}}\PYG{l+s+s1}{int32}\PYG{l+s+s1}{\PYGZsq{}}\PYG{p}{)}
\PYG{n}{deterministic} \PYG{o}{=} \PYG{n}{T}\PYG{o}{.}\PYG{n}{Placeholder}\PYG{p}{(}\PYG{p}{(}\PYG{l+m+mi}{1}\PYG{p}{,}\PYG{p}{)}\PYG{p}{,} \PYG{l+s+s1}{\PYGZsq{}}\PYG{l+s+s1}{bool}\PYG{l+s+s1}{\PYGZsq{}}\PYG{p}{)}

\PYG{n}{layer} \PYG{o}{=} \PYG{p}{[}\PYG{n}{sj}\PYG{o}{.}\PYG{n}{layers}\PYG{o}{.}\PYG{n}{RandomCrop}\PYG{p}{(}\PYG{n}{inputs}\PYG{p}{,} \PYG{n}{crop\PYGZus{}shape}\PYG{o}{=}\PYG{p}{(}\PYG{l+m+mi}{3}\PYG{p}{,} \PYG{l+m+mi}{32}\PYG{p}{,} \PYG{l+m+mi}{32}\PYG{p}{)}\PYG{p}{,}
                \PYG{n}{padding}\PYG{o}{=}\PYG{p}{[}\PYG{p}{(}\PYG{l+m+mi}{0}\PYG{p}{,} \PYG{l+m+mi}{0}\PYG{p}{)}\PYG{p}{,} \PYG{p}{(}\PYG{l+m+mi}{4}\PYG{p}{,} \PYG{l+m+mi}{4}\PYG{p}{)}\PYG{p}{,} \PYG{p}{(}\PYG{l+m+mi}{4}\PYG{p}{,} \PYG{l+m+mi}{4}\PYG{p}{)}\PYG{p}{]}\PYG{p}{,}
                \PYG{n}{deterministic}\PYG{o}{=}\PYG{n}{deterministic}\PYG{p}{)}\PYG{p}{]}
\PYG{n}{layer}\PYG{o}{.}\PYG{n}{append}\PYG{p}{(}\PYG{n}{sj}\PYG{o}{.}\PYG{n}{layers}\PYG{o}{.}\PYG{n}{Conv2D}\PYG{p}{(}\PYG{n}{layer}\PYG{p}{[}\PYG{o}{\PYGZhy{}}\PYG{l+m+mi}{1}\PYG{p}{]}\PYG{p}{,} \PYG{p}{(}\PYG{l+m+mi}{32}\PYG{p}{,} \PYG{l+m+mi}{3}\PYG{p}{,} \PYG{l+m+mi}{3}\PYG{p}{,} \PYG{l+m+mi}{3}\PYG{p}{)}\PYG{p}{)}\PYG{p}{)}
\PYG{n}{layer}\PYG{o}{.}\PYG{n}{append}\PYG{p}{(}\PYG{n}{sj}\PYG{o}{.}\PYG{n}{layers}\PYG{o}{.}\PYG{n}{BatchNormalization}\PYG{p}{(}\PYG{n}{layer}\PYG{p}{[}\PYG{o}{\PYGZhy{}}\PYG{l+m+mi}{1}\PYG{p}{]}\PYG{p}{,} \PYG{p}{[}\PYG{l+m+mi}{0}\PYG{p}{,} \PYG{l+m+mi}{2}\PYG{p}{,} \PYG{l+m+mi}{3}\PYG{p}{]}\PYG{p}{,}
                                \PYG{n}{deterministic}\PYG{p}{)}\PYG{p}{)}
\PYG{n}{layer}\PYG{o}{.}\PYG{n}{append}\PYG{p}{(}\PYG{n}{sj}\PYG{o}{.}\PYG{n}{layers}\PYG{o}{.}\PYG{n}{Activation}\PYG{p}{(}\PYG{n}{layer}\PYG{p}{[}\PYG{o}{\PYGZhy{}}\PYG{l+m+mi}{1}\PYG{p}{]}\PYG{p}{,} \PYG{n}{T}\PYG{o}{.}\PYG{n}{relu}\PYG{p}{)}\PYG{p}{)}
\PYG{n}{layer}\PYG{o}{.}\PYG{n}{append}\PYG{p}{(}\PYG{n}{sj}\PYG{o}{.}\PYG{n}{layers}\PYG{o}{.}\PYG{n}{Pool2D}\PYG{p}{(}\PYG{n}{layer}\PYG{p}{[}\PYG{o}{\PYGZhy{}}\PYG{l+m+mi}{1}\PYG{p}{]}\PYG{p}{,} \PYG{p}{(}\PYG{l+m+mi}{2}\PYG{p}{,} \PYG{l+m+mi}{2}\PYG{p}{)}\PYG{p}{)}\PYG{p}{)}                       

\PYG{n}{layer}\PYG{o}{.}\PYG{n}{append}\PYG{p}{(}\PYG{n}{sj}\PYG{o}{.}\PYG{n}{layers}\PYG{o}{.}\PYG{n}{Conv2D}\PYG{p}{(}\PYG{n}{layer}\PYG{p}{[}\PYG{o}{\PYGZhy{}}\PYG{l+m+mi}{1}\PYG{p}{]}\PYG{p}{,} \PYG{p}{(}\PYG{l+m+mi}{64}\PYG{p}{,} \PYG{l+m+mi}{32}\PYG{p}{,} \PYG{l+m+mi}{3}\PYG{p}{,} \PYG{l+m+mi}{3}\PYG{p}{)}\PYG{p}{)}\PYG{p}{)}
\PYG{n}{layer}\PYG{o}{.}\PYG{n}{append}\PYG{p}{(}\PYG{n}{sj}\PYG{o}{.}\PYG{n}{layers}\PYG{o}{.}\PYG{n}{BatchNormalization}\PYG{p}{(}\PYG{n}{layer}\PYG{p}{[}\PYG{o}{\PYGZhy{}}\PYG{l+m+mi}{1}\PYG{p}{]}\PYG{p}{,} \PYG{p}{[}\PYG{l+m+mi}{0}\PYG{p}{,} \PYG{l+m+mi}{2}\PYG{p}{,} \PYG{l+m+mi}{3}\PYG{p}{]}\PYG{p}{,}
                                \PYG{n}{deterministic}\PYG{p}{)}\PYG{p}{)}
\PYG{n}{layer}\PYG{o}{.}\PYG{n}{append}\PYG{p}{(}\PYG{n}{sj}\PYG{o}{.}\PYG{n}{layers}\PYG{o}{.}\PYG{n}{Activation}\PYG{p}{(}\PYG{n}{layer}\PYG{p}{[}\PYG{o}{\PYGZhy{}}\PYG{l+m+mi}{1}\PYG{p}{]}\PYG{p}{,} \PYG{n}{T}\PYG{o}{.}\PYG{n}{relu}\PYG{p}{)}\PYG{p}{)}
\PYG{n}{layer}\PYG{o}{.}\PYG{n}{append}\PYG{p}{(}\PYG{n}{sj}\PYG{o}{.}\PYG{n}{layers}\PYG{o}{.}\PYG{n}{Pool2D}\PYG{p}{(}\PYG{n}{layer}\PYG{p}{[}\PYG{o}{\PYGZhy{}}\PYG{l+m+mi}{1}\PYG{p}{]}\PYG{p}{,} \PYG{p}{(}\PYG{l+m+mi}{2}\PYG{p}{,} \PYG{l+m+mi}{2}\PYG{p}{)}\PYG{p}{)}\PYG{p}{)}

\PYG{n}{layer}\PYG{o}{.}\PYG{n}{append}\PYG{p}{(}\PYG{n}{sj}\PYG{o}{.}\PYG{n}{layers}\PYG{o}{.}\PYG{n}{Dense}\PYG{p}{(}\PYG{n}{layer}\PYG{p}{[}\PYG{o}{\PYGZhy{}}\PYG{l+m+mi}{1}\PYG{p}{]}\PYG{p}{,} \PYG{l+m+mi}{128}\PYG{p}{)}\PYG{p}{)} 
\PYG{n}{layer}\PYG{o}{.}\PYG{n}{append}\PYG{p}{(}\PYG{n}{sj}\PYG{o}{.}\PYG{n}{layers}\PYG{o}{.}\PYG{n}{BatchNormalization}\PYG{p}{(}\PYG{n}{layer}\PYG{p}{[}\PYG{o}{\PYGZhy{}}\PYG{l+m+mi}{1}\PYG{p}{]}\PYG{p}{,} \PYG{p}{[}\PYG{l+m+mi}{0}\PYG{p}{]}\PYG{p}{,}
                                        \PYG{n}{deterministic}\PYG{p}{)}\PYG{p}{)}
\PYG{n}{layer}\PYG{o}{.}\PYG{n}{append}\PYG{p}{(}\PYG{n}{sj}\PYG{o}{.}\PYG{n}{layers}\PYG{o}{.}\PYG{n}{Activation}\PYG{p}{(}\PYG{n}{layer}\PYG{p}{[}\PYG{o}{\PYGZhy{}}\PYG{l+m+mi}{1}\PYG{p}{]}\PYG{p}{,} \PYG{n}{T}\PYG{o}{.}\PYG{n}{relu}\PYG{p}{)}\PYG{p}{)}
\PYG{n}{layer}\PYG{o}{.}\PYG{n}{append}\PYG{p}{(}\PYG{n}{sj}\PYG{o}{.}\PYG{n}{layers}\PYG{o}{.}\PYG{n}{Dense}\PYG{p}{(}\PYG{n}{layer}\PYG{p}{[}\PYG{o}{\PYGZhy{}}\PYG{l+m+mi}{1}\PYG{p}{]}\PYG{p}{,} \PYG{l+m+mi}{10}\PYG{p}{)}\PYG{p}{)}

\PYG{c+c1}{\PYGZsh{} each layer is itself a tensor which represents its output and thus}
\PYG{c+c1}{\PYGZsh{} any tensor operation can be used on the layer instance, for example}
\PYG{k}{for} \PYG{n}{l} \PYG{o+ow}{in} \PYG{n}{layer}\PYG{p}{:}
    \PYG{n+nb}{print}\PYG{p}{(}\PYG{n}{l}\PYG{o}{.}\PYG{n}{shape}\PYG{p}{)}

\PYG{c+c1}{\PYGZsh{} (32, 3, 32, 32)}
\PYG{c+c1}{\PYGZsh{} (32, 32, 30, 30)}
\PYG{c+c1}{\PYGZsh{} (32, 32, 30, 30)}
\PYG{c+c1}{\PYGZsh{} (32, 32, 30, 30)}
\PYG{c+c1}{\PYGZsh{} (32, 32, 15, 15)}
\PYG{c+c1}{\PYGZsh{} (32, 64, 13, 13)}
\PYG{c+c1}{\PYGZsh{} (32, 64, 13, 13)}
\PYG{c+c1}{\PYGZsh{} (32, 64, 13, 13)}
\PYG{c+c1}{\PYGZsh{} (32, 64, 6, 6)}
\PYG{c+c1}{\PYGZsh{} (32, 128)}
\PYG{c+c1}{\PYGZsh{} (32, 128)}
\PYG{c+c1}{\PYGZsh{} (32, 128)}
\PYG{c+c1}{\PYGZsh{} (32, 10)}

\PYG{n}{loss} \PYG{o}{=} \PYG{n}{sj}\PYG{o}{.}\PYG{n}{losses}\PYG{o}{.}\PYG{n}{sparse\PYGZus{}crossentropy\PYGZus{}logits}\PYG{p}{(}\PYG{n}{outputs}\PYG{p}{,} \PYG{n}{layer}\PYG{p}{[}\PYG{o}{\PYGZhy{}}\PYG{l+m+mi}{1}\PYG{p}{]}\PYG{p}{)}\PYG{o}{.}\PYG{n}{mean}\PYG{p}{(}\PYG{p}{)}
\PYG{n}{accuracy} \PYG{o}{=} \PYG{n}{sj}\PYG{o}{.}\PYG{n}{losses}\PYG{o}{.}\PYG{n}{accuracy}\PYG{p}{(}\PYG{n}{outputs}\PYG{p}{,} \PYG{n}{layer}\PYG{p}{[}\PYG{o}{\PYGZhy{}}\PYG{l+m+mi}{1}\PYG{p}{]}\PYG{p}{)}

\PYG{n}{params} \PYG{o}{=} \PYG{n+nb}{sum}\PYG{p}{(}\PYG{p}{[}\PYG{n}{lay}\PYG{o}{.}\PYG{n}{variables}\PYG{p}{(}\PYG{p}{)} \PYG{k}{for} \PYG{n}{lay} \PYG{o+ow}{in} \PYG{n}{layer}\PYG{p}{]}\PYG{p}{,} \PYG{p}{[}\PYG{p}{]}\PYG{p}{)}

\PYG{n}{lr}\PYG{o}{=}\PYG{n}{sj}\PYG{o}{.}\PYG{n}{schedules}\PYG{o}{.}\PYG{n}{PiecewiseConstant}\PYG{p}{(}\PYG{l+m+mf}{0.005}\PYG{p}{,} \PYG{p}{\PYGZob{}}\PYG{l+m+mi}{50}\PYG{p}{:} \PYG{l+m+mf}{0.001}\PYG{p}{,} \PYG{l+m+mi}{75}\PYG{p}{:} \PYG{l+m+mf}{0.0005}\PYG{p}{\PYGZcb{}}\PYG{p}{)}
\PYG{n}{opt} \PYG{o}{=} \PYG{n}{sj}\PYG{o}{.}\PYG{n}{optimizers}\PYG{o}{.}\PYG{n}{Adam}\PYG{p}{(}\PYG{n}{loss}\PYG{p}{,} \PYG{n}{params}\PYG{p}{,} \PYG{n}{lr}\PYG{p}{)}

\PYG{k}{for} \PYG{n}{l} \PYG{o+ow}{in} \PYG{n}{layer}\PYG{p}{:}
    \PYG{n}{opt}\PYG{o}{.}\PYG{n}{updates}\PYG{o}{.}\PYG{n}{update}\PYG{p}{(}\PYG{n}{l}\PYG{o}{.}\PYG{n}{updates}\PYG{p}{)}

\PYG{n}{test} \PYG{o}{=} \PYG{n}{sj}\PYG{o}{.}\PYG{n}{function}\PYG{p}{(}\PYG{n}{inputs}\PYG{p}{,} \PYG{n}{outputs}\PYG{p}{,} \PYG{n}{deterministic}\PYG{p}{,}
                            \PYG{n}{outputs}\PYG{o}{=}\PYG{p}{[}\PYG{n}{loss}\PYG{p}{,} \PYG{n}{accuracy}\PYG{p}{]}\PYG{p}{)}

\PYG{n}{train} \PYG{o}{=} \PYG{n}{sj}\PYG{o}{.}\PYG{n}{function}\PYG{p}{(}\PYG{n}{inputs}\PYG{p}{,} \PYG{n}{outputs}\PYG{p}{,} \PYG{n}{deterministic}\PYG{p}{,}
                    \PYG{n}{outputs}\PYG{o}{=}\PYG{p}{[}\PYG{n}{loss}\PYG{p}{,} \PYG{n}{accuracy}\PYG{p}{]}\PYG{p}{,} \PYG{n}{updates}\PYG{o}{=}\PYG{n}{opt}\PYG{o}{.}\PYG{n}{updates}\PYG{p}{)}

\PYG{k}{for} \PYG{n}{epoch} \PYG{o+ow}{in} \PYG{n+nb}{range}\PYG{p}{(}\PYG{l+m+mi}{100}\PYG{p}{)}\PYG{p}{:}
    \PYG{n}{L} \PYG{o}{=} \PYG{n+nb}{list}\PYG{p}{(}\PYG{p}{)}
    \PYG{k}{for} \PYG{n}{x}\PYG{p}{,} \PYG{n}{y} \PYG{o+ow}{in} \PYG{n}{sj}\PYG{o}{.}\PYG{n}{utils}\PYG{o}{.}\PYG{n}{batchify}\PYG{p}{(}\PYG{n}{images\PYGZus{}test}\PYG{p}{,} \PYG{n}{labels\PYGZus{}test}\PYG{p}{,} \PYG{n}{batch\PYGZus{}size}\PYG{o}{=}\PYG{n}{BATCH\PYGZus{}SIZE}\PYG{p}{,}
                                      \PYG{n}{option}\PYG{o}{=}\PYG{l+s+s1}{\PYGZsq{}}\PYG{l+s+s1}{continuous}\PYG{l+s+s1}{\PYGZsq{}}\PYG{p}{)}\PYG{p}{:}
        \PYG{n}{L}\PYG{o}{.}\PYG{n}{append}\PYG{p}{(}\PYG{n}{test}\PYG{p}{(}\PYG{n}{x}\PYG{p}{,} \PYG{n}{y}\PYG{p}{,} \PYG{l+m+mi}{1}\PYG{p}{)}\PYG{p}{)}
    \PYG{n+nb}{print}\PYG{p}{(}\PYG{l+s+s1}{\PYGZsq{}}\PYG{l+s+s1}{Test Loss and Accu:}\PYG{l+s+s1}{\PYGZsq{}}\PYG{p}{,} \PYG{n}{np}\PYG{o}{.}\PYG{n}{mean}\PYG{p}{(}\PYG{n}{L}\PYG{p}{,} \PYG{l+m+mi}{0}\PYG{p}{)}\PYG{p}{)}
    \PYG{n}{L} \PYG{o}{=} \PYG{n+nb}{list}\PYG{p}{(}\PYG{p}{)}
    \PYG{k}{for} \PYG{n}{x}\PYG{p}{,} \PYG{n}{y} \PYG{o+ow}{in} \PYG{n}{sj}\PYG{o}{.}\PYG{n}{utils}\PYG{o}{.}\PYG{n}{batchify}\PYG{p}{(}\PYG{n}{images\PYGZus{}train}\PYG{p}{,} \PYG{n}{labels\PYGZus{}train}\PYG{p}{,}
                            \PYG{n}{batch\PYGZus{}size}\PYG{o}{=}\PYG{n}{BATCH\PYGZus{}SIZE}\PYG{p}{,} \PYG{n}{option}\PYG{o}{=}\PYG{l+s+s1}{\PYGZsq{}}\PYG{l+s+s1}{random\PYGZus{}see\PYGZus{}all}\PYG{l+s+s1}{\PYGZsq{}}\PYG{p}{)}\PYG{p}{:}
        \PYG{n}{L}\PYG{o}{.}\PYG{n}{append}\PYG{p}{(}\PYG{n}{train}\PYG{p}{(}\PYG{n}{x}\PYG{p}{,} \PYG{n}{y}\PYG{p}{,} \PYG{l+m+mi}{0}\PYG{p}{)}\PYG{p}{)}
    \PYG{n+nb}{print}\PYG{p}{(}\PYG{l+s+s1}{\PYGZsq{}}\PYG{l+s+s1}{Train Loss and Accu}\PYG{l+s+s1}{\PYGZsq{}}\PYG{p}{,} \PYG{n}{np}\PYG{o}{.}\PYG{n}{mean}\PYG{p}{(}\PYG{n}{L}\PYG{p}{,} \PYG{l+m+mi}{0}\PYG{p}{)}\PYG{p}{)}
    \PYG{n}{lr}\PYG{o}{.}\PYG{n}{update}\PYG{p}{(}\PYG{p}{)}

\PYG{c+c1}{\PYGZsh{} Test Loss and Accu: [2.6886015  0.09194712]}
\PYG{c+c1}{\PYGZsh{} Train Loss and Accu [1.3671544  0.51288414]}
\PYG{c+c1}{\PYGZsh{} Test Loss and Accu: [1.7053369  0.43449518]}
\PYG{c+c1}{\PYGZsh{} Train Loss and Accu [1.1127299 0.6065541]}
\PYG{c+c1}{\PYGZsh{} Test Loss and Accu: [1.1878427  0.59094554]}
\PYG{c+c1}{\PYGZsh{} Train Loss and Accu [1.0067393 0.6460667]}
\PYG{c+c1}{\PYGZsh{} Test Loss and Accu: [1.1366144 0.6133814]}
\PYG{c+c1}{\PYGZsh{} Train Loss and Accu [0.9416873  0.66995436]}
\PYG{c+c1}{\PYGZsh{} Test Loss and Accu: [0.95114607 0.6744792 ]}
\PYG{c+c1}{\PYGZsh{} Train Loss and Accu [0.891217   0.68737996]}
\PYG{c+c1}{\PYGZsh{} Test Loss and Accu: [1.272816  0.5885417]}
\PYG{c+c1}{\PYGZsh{} Train Loss and Accu [0.84912854 0.7034651 ]}
\PYG{c+c1}{\PYGZsh{} Test Loss and Accu: [0.81524473 0.7214543 ]}
\PYG{c+c1}{\PYGZsh{} .....}

\end{sphinxVerbatim}

\chapter{API}
\label{\detokenize{index:api}}

\section{General}
\label{\detokenize{index:general}}

\subsection{\sphinxstyleliteralintitle{\sphinxupquote{symjax}}}
\label{\detokenize{modules/symjax:symjax}}\label{\detokenize{modules/symjax::doc}}\index{function (class in symjax)@\spxentry{function}\spxextra{class in symjax}}

\begin{fulllineitems}
\phantomsection\label{\detokenize{modules/symjax:symjax.function}}\pysiglinewithargsret{\sphinxbfcode{\sphinxupquote{class }}\sphinxbfcode{\sphinxupquote{function}}}{\emph{\DUrole{o}{*}\DUrole{n}{classargs}}, \emph{\DUrole{n}{outputs}\DUrole{o}{=}\DUrole{default_value}{{[}{]}}}, \emph{\DUrole{n}{updates}\DUrole{o}{=}\DUrole{default_value}{None}}, \emph{\DUrole{n}{device}\DUrole{o}{=}\DUrole{default_value}{None}}, \emph{\DUrole{n}{backend}\DUrole{o}{=}\DUrole{default_value}{None}}, \emph{\DUrole{n}{default\_value}\DUrole{o}{=}\DUrole{default_value}{None}}}{}
Generate a user function that compiles a computational graph.

Based on given inputs, outputs and update policy of variables. This
function internally jit compile the underlying jax computational
graph for performances and thus should be favored to the get
method of tensors.
\begin{quote}\begin{description}
\item[{Parameters}] \leavevmode\begin{itemize}
\item {} 
\sphinxstyleliteralstrong{\sphinxupquote{classargs}} (\sphinxstyleliteralemphasis{\sphinxupquote{trailing tuple}}) \textendash{} the inputs to the function to be compiled. The tuple should
contain all the placeholders that are roots of any output
given of the function and update values

\item {} 
\sphinxstyleliteralstrong{\sphinxupquote{outputs}} (\sphinxstyleliteralemphasis{\sphinxupquote{List}}\sphinxstyleliteralemphasis{\sphinxupquote{ (}}\sphinxstyleliteralemphasis{\sphinxupquote{optional}}\sphinxstyleliteralemphasis{\sphinxupquote{)}}) \textendash{} the outputs of the function, if a single element, it can be
given as a standalone and not a list

\item {} 
\sphinxstyleliteralstrong{\sphinxupquote{updates}} (\sphinxstyleliteralemphasis{\sphinxupquote{Dict}}\sphinxstyleliteralemphasis{\sphinxupquote{ (}}\sphinxstyleliteralemphasis{\sphinxupquote{optional}}\sphinxstyleliteralemphasis{\sphinxupquote{)}}) \textendash{} the dictionnary of updates as per \{var:new\_value\} for any
variable of the graph

\item {} 
\sphinxstyleliteralstrong{\sphinxupquote{device}} \textendash{} 
??

\item {} 
\sphinxstyleliteralstrong{\sphinxupquote{backend}} (\sphinxstyleliteralemphasis{\sphinxupquote{\textquotesingle{}cpu\textquotesingle{}}}\sphinxstyleliteralemphasis{\sphinxupquote{ or }}\sphinxstyleliteralemphasis{\sphinxupquote{\textquotesingle{}gpu\textquotesingle{}}}) \textendash{} the backend to use to run the function on

\item {} 
\sphinxstyleliteralstrong{\sphinxupquote{default\_value}} (\sphinxstyleliteralemphasis{\sphinxupquote{not implemented}}) \textendash{} not implemented

\end{itemize}

\item[{Returns}] \leavevmode
the user frontend function that takes the specified inputs,
returns the specified outputs and perform internally the
updates

\item[{Return type}] \leavevmode
callable

\end{description}\end{quote}
\subsubsection*{Examples}

\begin{sphinxVerbatim}[commandchars=\\\{\}]
\PYG{g+gp}{\PYGZgt{}\PYGZgt{}\PYGZgt{} }\PYG{k+kn}{import} \PYG{n+nn}{jaxonn}
\PYG{g+gp}{\PYGZgt{}\PYGZgt{}\PYGZgt{} }\PYG{k+kn}{import} \PYG{n+nn}{jaxonn}\PYG{n+nn}{.}\PYG{n+nn}{tensor} \PYG{k}{as} \PYG{n+nn}{T}
\PYG{g+gp}{\PYGZgt{}\PYGZgt{}\PYGZgt{} }\PYG{n}{x} \PYG{o}{=} \PYG{n}{T}\PYG{o}{.}\PYG{n}{ones}\PYG{p}{(}\PYG{p}{(}\PYG{l+m+mi}{4}\PYG{p}{,} \PYG{l+m+mi}{4}\PYG{p}{)}\PYG{p}{)}
\PYG{g+gp}{\PYGZgt{}\PYGZgt{}\PYGZgt{} }\PYG{n}{xs} \PYG{o}{=} \PYG{n}{x}\PYG{o}{.}\PYG{n}{sum}\PYG{p}{(}\PYG{p}{)} \PYG{o}{+} \PYG{l+m+mi}{1}
\PYG{g+gp}{\PYGZgt{}\PYGZgt{}\PYGZgt{} }\PYG{n}{f} \PYG{o}{=} \PYG{n}{jaxonn}\PYG{o}{.}\PYG{n}{function}\PYG{p}{(}\PYG{n}{outputs}\PYG{o}{=}\PYG{n}{xs}\PYG{p}{)}
\PYG{g+gp}{\PYGZgt{}\PYGZgt{}\PYGZgt{} }\PYG{n+nb}{print}\PYG{p}{(}\PYG{n}{f}\PYG{p}{(}\PYG{p}{)}\PYG{p}{)} \PYG{c+c1}{\PYGZsh{} returns 17}
\end{sphinxVerbatim}

\begin{sphinxVerbatim}[commandchars=\\\{\}]
\PYG{g+gp}{\PYGZgt{}\PYGZgt{}\PYGZgt{} }\PYG{n}{w} \PYG{o}{=} \PYG{n}{T}\PYG{o}{.}\PYG{n}{Variable}\PYG{p}{(}\PYG{l+m+mf}{0.}\PYG{p}{,} \PYG{n}{name}\PYG{o}{=}\PYG{l+s+s1}{\PYGZsq{}}\PYG{l+s+s1}{w}\PYG{l+s+s1}{\PYGZsq{}}\PYG{p}{)}
\PYG{g+gp}{\PYGZgt{}\PYGZgt{}\PYGZgt{} }\PYG{n}{increment} \PYG{o}{=} \PYG{n}{jaxonn}\PYG{o}{.}\PYG{n}{function}\PYG{p}{(}\PYG{n}{updates}\PYG{o}{=}\PYG{p}{\PYGZob{}}\PYG{n}{w}\PYG{p}{:} \PYG{n}{w} \PYG{o}{+} \PYG{l+m+mi}{1}\PYG{p}{\PYGZcb{}}\PYG{p}{)}
\PYG{g+gp}{\PYGZgt{}\PYGZgt{}\PYGZgt{} }\PYG{k}{for} \PYG{n}{i} \PYG{o+ow}{in} \PYG{n+nb}{range}\PYG{p}{(}\PYG{l+m+mi}{10}\PYG{p}{)}\PYG{p}{:}
\PYG{g+gp}{\PYGZgt{}\PYGZgt{}\PYGZgt{} }    \PYG{n}{increment}\PYG{p}{(}\PYG{p}{)}
\PYG{g+gp}{\PYGZgt{}\PYGZgt{}\PYGZgt{} }\PYG{n+nb}{print}\PYG{p}{(}\PYG{n}{w}\PYG{o}{.}\PYG{n}{value}\PYG{p}{)} \PYG{c+c1}{\PYGZsh{} returns 10}
\end{sphinxVerbatim}

\end{fulllineitems}

\phantomsection\label{\detokenize{modules/symjax:module-symjax}}\index{module@\spxentry{module}!symjax@\spxentry{symjax}}\index{symjax@\spxentry{symjax}!module@\spxentry{module}}\index{gradients() (in module symjax)@\spxentry{gradients()}\spxextra{in module symjax}}

\begin{fulllineitems}
\phantomsection\label{\detokenize{modules/symjax:symjax.gradients}}\pysiglinewithargsret{\sphinxbfcode{\sphinxupquote{gradients}}}{\emph{\DUrole{n}{scalar}}, \emph{\DUrole{n}{variables}}}{}
Compute the gradients of a scalar w.r.t to a given list of variables.
\begin{quote}\begin{description}
\item[{Parameters}] \leavevmode\begin{itemize}
\item {} 
\sphinxstyleliteralstrong{\sphinxupquote{scalar}} (\sphinxcode{\sphinxupquote{symjax.tensor.base.Tensor}}) \textendash{} the variable to differentiate

\item {} 
\sphinxstyleliteralstrong{\sphinxupquote{variables}} (\sphinxstyleliteralemphasis{\sphinxupquote{List}}\sphinxstyleliteralemphasis{\sphinxupquote{ or }}\sphinxstyleliteralemphasis{\sphinxupquote{Tuple}}) \textendash{} the variables used to compute the derivative.

\end{itemize}

\item[{Returns}] \leavevmode
\sphinxstylestrong{gradients} \textendash{} the sequency of gradients ordered as given in the input variables

\item[{Return type}] \leavevmode
Tuple

\end{description}\end{quote}

\end{fulllineitems}

\index{jacobians() (in module symjax)@\spxentry{jacobians()}\spxextra{in module symjax}}

\begin{fulllineitems}
\phantomsection\label{\detokenize{modules/symjax:symjax.jacobians}}\pysiglinewithargsret{\sphinxbfcode{\sphinxupquote{jacobians}}}{\emph{\DUrole{n}{tensor}}, \emph{\DUrole{n}{variables}}, \emph{\DUrole{n}{mode}\DUrole{o}{=}\DUrole{default_value}{\textquotesingle{}forward\textquotesingle{}}}}{}
Compute the jacobians of a tensor w.r.t to a given list of variables.

The tensor needs not to be a vector, but will be treated as such. For
example if tensor.shape is (10, 3, 3) and a variable shape if (10, 10)
the resulting jacobian has shape (10, 3, 3, 10, 10). It is possible
to specify the mode forward or backward. For tall jacobians, forward
is faster and vice\sphinxhyphen{}versa.
\begin{quote}\begin{description}
\item[{Parameters}] \leavevmode\begin{itemize}
\item {} 
\sphinxstyleliteralstrong{\sphinxupquote{vector}} (\sphinxstyleliteralemphasis{\sphinxupquote{Tensor}}) \textendash{} the variable to differentiate

\item {} 
\sphinxstyleliteralstrong{\sphinxupquote{variables}} (\sphinxstyleliteralemphasis{\sphinxupquote{List}}\sphinxstyleliteralemphasis{\sphinxupquote{ or }}\sphinxstyleliteralemphasis{\sphinxupquote{Tuple}}) \textendash{} the variables used to compute the derivative.

\end{itemize}

\item[{Returns}] \leavevmode
\sphinxstylestrong{jacobians} \textendash{} the sequency of gradients ordered as given in the input variables

\item[{Return type}] \leavevmode
Tuple

\end{description}\end{quote}

\end{fulllineitems}

\subsection{\sphinxstyleliteralintitle{\sphinxupquote{symjax.tensor}}}
\label{\detokenize{modules/tensor:module-symjax.tensor}}\label{\detokenize{modules/tensor:symjax-tensor}}\label{\detokenize{modules/tensor::doc}}\index{module@\spxentry{module}!symjax.tensor@\spxentry{symjax.tensor}}\index{symjax.tensor@\spxentry{symjax.tensor}!module@\spxentry{module}}
Implements the NumPy API, using the primitives in \sphinxcode{\sphinxupquote{jax.lax}}.
As SymJAX follows the JAX restrictions, not all NumPy functins are present.
\begin{itemize}
\item {} 
Notably, since JAX arrays are immutable, NumPy APIs that mutate arrays
in\sphinxhyphen{}place cannot be implemented in JAX. However, often JAX is able to provide a
alternative API that is purely functional. For example, instead of in\sphinxhyphen{}place
array updates (\sphinxcode{\sphinxupquote{x{[}i{]} = y}}), JAX provides an alternative pure indexed
update function \sphinxcode{\sphinxupquote{jax.ops.index\_update()}}.

\item {} 
NumPy is very aggressive at promoting values to \sphinxcode{\sphinxupquote{float64}} type. JAX
sometimes is less aggressive about type promotion.

\end{itemize}

Finally, since SymJAX uses jit\sphinxhyphen{}compilation, any function that returns
data\sphinxhyphen{}dependent output shapes are incompatible and thus not implemented.
In fact, The XLA compiler requires that shapes of arrays be known at
compile time. While it would be possible to provide. Thus an implementation of an API such as \sphinxcode{\sphinxupquote{numpy.nonzero()}}, we would be unable
to JIT\sphinxhyphen{}compile it because the shape of its output depends on the contents
of the input data.

Not every function in NumPy is implemented; contributions are welcome!

\subsubsection{Control Flow}
\label{\detokenize{modules/tensor:module-symjax.tensor.control_flow}}\label{\detokenize{modules/tensor:control-flow}}\index{module@\spxentry{module}!symjax.tensor.control\_flow@\spxentry{symjax.tensor.control\_flow}}\index{symjax.tensor.control\_flow@\spxentry{symjax.tensor.control\_flow}!module@\spxentry{module}}\index{cond() (in module symjax.tensor.control\_flow)@\spxentry{cond()}\spxextra{in module symjax.tensor.control\_flow}}

\begin{fulllineitems}
\phantomsection\label{\detokenize{modules/tensor:symjax.tensor.control_flow.cond}}\pysiglinewithargsret{\sphinxbfcode{\sphinxupquote{cond}}}{\emph{\DUrole{n}{predicate}}, \emph{\DUrole{n}{true\_predicate}}, \emph{\DUrole{n}{true\_fun}}, \emph{\DUrole{n}{false\_predicate}}, \emph{\DUrole{n}{false\_fun}}}{}~\begin{description}
\item[{predicate should be a boolean tensor with shape ()}] \leavevmode
true\_input is the input passed to true\_fn that will give the output
if the predicate evaluates to True, and conversely for False…

\end{description}

LAX\sphinxhyphen{}backend implementation of \sphinxcode{\sphinxupquote{\_cond()}}.
ADDITIONOriginal docstring below.

\end{fulllineitems}

\index{scan() (in module symjax.tensor.control\_flow)@\spxentry{scan()}\spxextra{in module symjax.tensor.control\_flow}}

\begin{fulllineitems}
\phantomsection\label{\detokenize{modules/tensor:symjax.tensor.control_flow.scan}}\pysiglinewithargsret{\sphinxbfcode{\sphinxupquote{scan}}}{\emph{\DUrole{n}{fn}}, \emph{\DUrole{n}{init}}, \emph{\DUrole{n}{xs}}, \emph{\DUrole{n}{constants}\DUrole{o}{=}\DUrole{default_value}{()}}, \emph{\DUrole{n}{length}\DUrole{o}{=}\DUrole{default_value}{None}}}{}
Scan a function over leading array axes while carrying along state.

The type signature in brief is

\begin{sphinxVerbatim}[commandchars=\\\{\}]
\PYG{n+nf}{scan} \PYG{o+ow}{::} \PYG{p}{(}\PYG{n}{c} \PYG{o+ow}{\PYGZhy{}\PYGZgt{}} \PYG{n}{a} \PYG{o+ow}{\PYGZhy{}\PYGZgt{}} \PYG{p}{(}\PYG{n}{c}\PYG{p}{,} \PYG{n}{b}\PYG{p}{)}\PYG{p}{)} \PYG{o+ow}{\PYGZhy{}\PYGZgt{}} \PYG{n}{c} \PYG{o+ow}{\PYGZhy{}\PYGZgt{}} \PYG{p}{[}\PYG{n}{a}\PYG{p}{]} \PYG{o+ow}{\PYGZhy{}\PYGZgt{}} \PYG{p}{(}\PYG{n}{c}\PYG{p}{,} \PYG{p}{[}\PYG{n}{b}\PYG{p}{]}\PYG{p}{)}
\end{sphinxVerbatim}

where we use {[}t{]} here to denote the type t with an additional leading axis.
That is, if t is an array type then {[}t{]} represents the type with an additional
leading axis, and if t is a pytree (container) type with array leaves then {[}t{]}
represents the type with the same pytree structure and corresponding leaves
each with an additional leading axis.

When \sphinxcode{\sphinxupquote{a}} is an array type or None, and \sphinxcode{\sphinxupquote{b}} is an array type, the semantics
of \sphinxcode{\sphinxupquote{scan}} are given roughly by this Python implementation:

\begin{sphinxVerbatim}[commandchars=\\\{\}]
\PYG{k}{def} \PYG{n+nf}{scan}\PYG{p}{(}\PYG{n}{f}\PYG{p}{,} \PYG{n}{init}\PYG{p}{,} \PYG{n}{xs}\PYG{p}{,} \PYG{n}{length}\PYG{o}{=}\PYG{k+kc}{None}\PYG{p}{)}\PYG{p}{:}
  \PYG{k}{if} \PYG{n}{xs} \PYG{o+ow}{is} \PYG{k+kc}{None}\PYG{p}{:}
    \PYG{n}{xs} \PYG{o}{=} \PYG{p}{[}\PYG{k+kc}{None}\PYG{p}{]} \PYG{o}{*} \PYG{n}{length}
  \PYG{n}{carry} \PYG{o}{=} \PYG{n}{init}
  \PYG{n}{ys} \PYG{o}{=} \PYG{p}{[}\PYG{p}{]}
  \PYG{k}{for} \PYG{n}{x} \PYG{o+ow}{in} \PYG{n}{xs}\PYG{p}{:}
    \PYG{n}{carry}\PYG{p}{,} \PYG{n}{y} \PYG{o}{=} \PYG{n}{f}\PYG{p}{(}\PYG{n}{carry}\PYG{p}{,} \PYG{n}{x}\PYG{p}{)}
    \PYG{n}{ys}\PYG{o}{.}\PYG{n}{append}\PYG{p}{(}\PYG{n}{y}\PYG{p}{)}
  \PYG{k}{return} \PYG{n}{carry}\PYG{p}{,} \PYG{n}{np}\PYG{o}{.}\PYG{n}{stack}\PYG{p}{(}\PYG{n}{ys}\PYG{p}{)}
\end{sphinxVerbatim}

Unlike that Python version, both \sphinxcode{\sphinxupquote{a}} and \sphinxcode{\sphinxupquote{b}} may be arbitrary pytree
types, and so multiple arrays can be scanned over at once and produce multiple
output arrays. (None is actually an empty pytree.)

Also unlike that Python version, \sphinxcode{\sphinxupquote{scan}} is a JAX primitive and is lowered to
a single XLA While HLO. That makes it useful for reducing compilation times
for jit\sphinxhyphen{}compiled functions, since native Python loop constructs in an \sphinxcode{\sphinxupquote{@jit}}
function are unrolled, leading to large XLA computations.

Finally, the loop\sphinxhyphen{}carried value \sphinxcode{\sphinxupquote{carry}} must hold a fixed shape and dtype
across all iterations (and not just be consistent up to NumPy rank/shape
broadcasting and dtype promotion rules, for example). In other words, the type
\sphinxcode{\sphinxupquote{c}} in the type signature above represents an array with a fixed shape and
dtype (or a nested tuple/list/dict container data structure with a fixed
structure and arrays with fixed shape and dtype at the leaves).
\begin{quote}\begin{description}
\item[{Parameters}] \leavevmode\begin{itemize}
\item {} 
\sphinxstyleliteralstrong{\sphinxupquote{f}} \textendash{} a Python function to be scanned of type \sphinxcode{\sphinxupquote{c \sphinxhyphen{}\textgreater{} a \sphinxhyphen{}\textgreater{} (c, b)}}, meaning
that \sphinxcode{\sphinxupquote{f}} accepts two arguments where the first is a value of the loop
carry and the second is a slice of \sphinxcode{\sphinxupquote{xs}} along its leading axis, and that
\sphinxcode{\sphinxupquote{f}} returns a pair where the first element represents a new value for
the loop carry and the second represents a slice of the output.

\item {} 
\sphinxstyleliteralstrong{\sphinxupquote{init}} \textendash{} an initial loop carry value of type \sphinxcode{\sphinxupquote{c}}, which can be a scalar,
array, or any pytree (nested Python tuple/list/dict) thereof, representing
the initial loop carry value. This value must have the same structure as
the first element of the pair returned by \sphinxcode{\sphinxupquote{f}}.

\item {} 
\sphinxstyleliteralstrong{\sphinxupquote{xs}} \textendash{} the value of type \sphinxcode{\sphinxupquote{{[}a{]}}} over which to scan along the leading axis,
where \sphinxcode{\sphinxupquote{{[}a{]}}} can be an array or any pytree (nested Python
tuple/list/dict) thereof with consistent leading axis sizes.

\item {} 
\sphinxstyleliteralstrong{\sphinxupquote{length}} \textendash{} optional integer specifying the number of loop iterations, which
must agree with the sizes of leading axes of the arrays in \sphinxcode{\sphinxupquote{xs}} (but can
be used to perform scans where no input \sphinxcode{\sphinxupquote{xs}} are needed).

\end{itemize}

\item[{Returns}] \leavevmode
A pair of type \sphinxcode{\sphinxupquote{(c, {[}b{]})}} where the first element represents the final
loop carry value and the second element represents the stacked outputs of
the second output of \sphinxcode{\sphinxupquote{f}} when scanned over the leading axis of the inputs.

\end{description}\end{quote}

\end{fulllineitems}

\subsubsection{Index Operations}
\label{\detokenize{modules/tensor:module-symjax.tensor.index_ops}}\label{\detokenize{modules/tensor:index-operations}}\index{module@\spxentry{module}!symjax.tensor.index\_ops@\spxentry{symjax.tensor.index\_ops}}\index{symjax.tensor.index\_ops@\spxentry{symjax.tensor.index\_ops}!module@\spxentry{module}}\index{index\_update() (in module symjax.tensor.index\_ops)@\spxentry{index\_update()}\spxextra{in module symjax.tensor.index\_ops}}

\begin{fulllineitems}
\phantomsection\label{\detokenize{modules/tensor:symjax.tensor.index_ops.index_update}}\pysiglinewithargsret{\sphinxbfcode{\sphinxupquote{index\_update}}}{\emph{\DUrole{n}{x}}, \emph{\DUrole{n}{idx}}, \emph{\DUrole{n}{y}}}{}
Pure equivalent of \sphinxcode{\sphinxupquote{x{[}idx{]} = y}}.

LAX\sphinxhyphen{}backend implementation of {\hyperref[\detokenize{modules/tensor:symjax.tensor.index_ops.index_update}]{\sphinxcrossref{\sphinxcode{\sphinxupquote{index\_update()}}}}}.
ADDITIONOriginal docstring below.
\begin{quote}
\begin{quote}

Returns the value of \sphinxtitleref{x} that would result from the
\end{quote}
\begin{description}
\item[{NumPy\sphinxhyphen{}style \sphinxcode{\sphinxupquote{indexed assignment}}::}] \leavevmode
x{[}idx{]} = y

\end{description}

Note the \sphinxtitleref{index\_update} operator is pure; \sphinxtitleref{x} itself is
not modified, instead the new value that \sphinxtitleref{x} would have taken is returned.

Unlike NumPy’s \sphinxcode{\sphinxupquote{x{[}idx{]} = y}}, if multiple indices refer to the same
location it is undefined which update is chosen; JAX may choose the order of
updates arbitrarily and nondeterministically (e.g., due to concurrent
updates on some hardware platforms).
\begin{description}
\item[{Args:}] \leavevmode
x: an array with the values to be updated.
idx: a Numpy\sphinxhyphen{}style index, consisting of \sphinxtitleref{None}, integers, \sphinxtitleref{slice} objects,
\begin{quote}

ellipses, ndarrays with integer dtypes, or a tuple of the above. A
convenient syntactic sugar for forming indices is via the
\sphinxcode{\sphinxupquote{jax.ops.index}} object.
\end{quote}
\begin{description}
\item[{y: the array of updates. \sphinxtitleref{y} must be broadcastable to the shape of the}] \leavevmode
array that would be returned by \sphinxtitleref{x{[}idx{]}}.

\end{description}

\item[{Returns:}] \leavevmode
An array.

\end{description}

\begin{sphinxVerbatim}[commandchars=\\\{\}]
\PYG{g+gp}{\PYGZgt{}\PYGZgt{}\PYGZgt{} }\PYG{n}{x} \PYG{o}{=} \PYG{n}{jax}\PYG{o}{.}\PYG{n}{numpy}\PYG{o}{.}\PYG{n}{ones}\PYG{p}{(}\PYG{p}{(}\PYG{l+m+mi}{5}\PYG{p}{,} \PYG{l+m+mi}{6}\PYG{p}{)}\PYG{p}{)}
\PYG{g+gp}{\PYGZgt{}\PYGZgt{}\PYGZgt{} }\PYG{n}{jax}\PYG{o}{.}\PYG{n}{ops}\PYG{o}{.}\PYG{n}{index\PYGZus{}update}\PYG{p}{(}\PYG{n}{x}\PYG{p}{,} \PYG{n}{jax}\PYG{o}{.}\PYG{n}{ops}\PYG{o}{.}\PYG{n}{index}\PYG{p}{[}\PYG{p}{:}\PYG{p}{:}\PYG{l+m+mi}{2}\PYG{p}{,} \PYG{l+m+mi}{3}\PYG{p}{:}\PYG{p}{]}\PYG{p}{,} \PYG{l+m+mf}{6.}\PYG{p}{)}
\PYG{g+go}{array([[1., 1., 1., 6., 6., 6.],}
\PYG{g+go}{       [1., 1., 1., 1., 1., 1.],}
\PYG{g+go}{       [1., 1., 1., 6., 6., 6.],}
\PYG{g+go}{       [1., 1., 1., 1., 1., 1.],}
\PYG{g+go}{       [1., 1., 1., 6., 6., 6.]], dtype=float32)}
\end{sphinxVerbatim}
\end{quote}

\end{fulllineitems}

\index{index\_add() (in module symjax.tensor.index\_ops)@\spxentry{index\_add()}\spxextra{in module symjax.tensor.index\_ops}}

\begin{fulllineitems}
\phantomsection\label{\detokenize{modules/tensor:symjax.tensor.index_ops.index_add}}\pysiglinewithargsret{\sphinxbfcode{\sphinxupquote{index\_add}}}{\emph{\DUrole{n}{x}}, \emph{\DUrole{n}{idx}}, \emph{\DUrole{n}{y}}}{}
Pure equivalent of \sphinxcode{\sphinxupquote{x{[}idx{]} += y}}.

LAX\sphinxhyphen{}backend implementation of {\hyperref[\detokenize{modules/tensor:symjax.tensor.index_ops.index_add}]{\sphinxcrossref{\sphinxcode{\sphinxupquote{index\_add()}}}}}.
ADDITIONOriginal docstring below.
\begin{quote}
\begin{quote}

Returns the value of \sphinxtitleref{x} that would result from the
\end{quote}
\begin{description}
\item[{NumPy\sphinxhyphen{}style \sphinxcode{\sphinxupquote{indexed assignment}}::}] \leavevmode
x{[}idx{]} += y

\end{description}

Note the \sphinxtitleref{index\_add} operator is pure; \sphinxtitleref{x} itself is
not modified, instead the new value that \sphinxtitleref{x} would have taken is returned.

Unlike the NumPy code \sphinxcode{\sphinxupquote{x{[}idx{]} += y}}, if multiple indices refer to the
same location the updates will be summed. (NumPy would only apply the last
update, rather than summing the updates.) The order in which conflicting
updates are applied is implementation\sphinxhyphen{}defined and may be nondeterministic
(e.g., due to concurrency on some hardware platforms).
\begin{description}
\item[{Args:}] \leavevmode
x: an array with the values to be updated.
idx: a Numpy\sphinxhyphen{}style index, consisting of \sphinxtitleref{None}, integers, \sphinxtitleref{slice} objects,
\begin{quote}

ellipses, ndarrays with integer dtypes, or a tuple of the above. A
convenient syntactic sugar for forming indices is via the
\sphinxcode{\sphinxupquote{jax.ops.index}} object.
\end{quote}
\begin{description}
\item[{y: the array of updates. \sphinxtitleref{y} must be broadcastable to the shape of the}] \leavevmode
array that would be returned by \sphinxtitleref{x{[}idx{]}}.

\end{description}

\item[{Returns:}] \leavevmode
An array.

\end{description}

\begin{sphinxVerbatim}[commandchars=\\\{\}]
\PYG{g+gp}{\PYGZgt{}\PYGZgt{}\PYGZgt{} }\PYG{n}{x} \PYG{o}{=} \PYG{n}{jax}\PYG{o}{.}\PYG{n}{numpy}\PYG{o}{.}\PYG{n}{ones}\PYG{p}{(}\PYG{p}{(}\PYG{l+m+mi}{5}\PYG{p}{,} \PYG{l+m+mi}{6}\PYG{p}{)}\PYG{p}{)}
\PYG{g+gp}{\PYGZgt{}\PYGZgt{}\PYGZgt{} }\PYG{n}{jax}\PYG{o}{.}\PYG{n}{ops}\PYG{o}{.}\PYG{n}{index\PYGZus{}add}\PYG{p}{(}\PYG{n}{x}\PYG{p}{,} \PYG{n}{jax}\PYG{o}{.}\PYG{n}{ops}\PYG{o}{.}\PYG{n}{index}\PYG{p}{[}\PYG{l+m+mi}{2}\PYG{p}{:}\PYG{l+m+mi}{4}\PYG{p}{,} \PYG{l+m+mi}{3}\PYG{p}{:}\PYG{p}{]}\PYG{p}{,} \PYG{l+m+mf}{6.}\PYG{p}{)}
\PYG{g+go}{array([[1., 1., 1., 1., 1., 1.],}
\PYG{g+go}{       [1., 1., 1., 1., 1., 1.],}
\PYG{g+go}{       [1., 1., 1., 7., 7., 7.],}
\PYG{g+go}{       [1., 1., 1., 7., 7., 7.],}
\PYG{g+go}{       [1., 1., 1., 1., 1., 1.]], dtype=float32)}
\end{sphinxVerbatim}
\end{quote}

\end{fulllineitems}

\index{index\_max() (in module symjax.tensor.index\_ops)@\spxentry{index\_max()}\spxextra{in module symjax.tensor.index\_ops}}

\begin{fulllineitems}
\phantomsection\label{\detokenize{modules/tensor:symjax.tensor.index_ops.index_max}}\pysiglinewithargsret{\sphinxbfcode{\sphinxupquote{index\_max}}}{\emph{\DUrole{n}{x}}, \emph{\DUrole{n}{idx}}, \emph{\DUrole{n}{y}}}{}
Pure equivalent of \sphinxcode{\sphinxupquote{x{[}idx{]} = maximum(x{[}idx{]}, y)}}.

LAX\sphinxhyphen{}backend implementation of {\hyperref[\detokenize{modules/tensor:symjax.tensor.index_ops.index_max}]{\sphinxcrossref{\sphinxcode{\sphinxupquote{index\_max()}}}}}.
ADDITIONOriginal docstring below.
\begin{quote}
\begin{quote}

Returns the value of \sphinxtitleref{x} that would result from the
\end{quote}
\begin{description}
\item[{NumPy\sphinxhyphen{}style \sphinxcode{\sphinxupquote{indexed assignment}}::}] \leavevmode
x{[}idx{]} = maximum(x{[}idx{]}, y)

\end{description}

Note the \sphinxtitleref{index\_max} operator is pure; \sphinxtitleref{x} itself is
not modified, instead the new value that \sphinxtitleref{x} would have taken is returned.

Unlike the NumPy code \sphinxcode{\sphinxupquote{x{[}idx{]} = maximum(x{[}idx{]}, y)}}, if multiple indices
refer to the same location the final value will be the overall max. (NumPy
would only look at the last update, rather than all of the updates.)
\begin{description}
\item[{Args:}] \leavevmode
x: an array with the values to be updated.
idx: a Numpy\sphinxhyphen{}style index, consisting of \sphinxtitleref{None}, integers, \sphinxtitleref{slice} objects,
\begin{quote}

ellipses, ndarrays with integer dtypes, or a tuple of the above. A
convenient syntactic sugar for forming indices is via the
\sphinxcode{\sphinxupquote{jax.ops.index}} object.
\end{quote}
\begin{description}
\item[{y: the array of updates. \sphinxtitleref{y} must be broadcastable to the shape of the}] \leavevmode
array that would be returned by \sphinxtitleref{x{[}idx{]}}.

\end{description}

\item[{Returns:}] \leavevmode
An array.

\end{description}

\begin{sphinxVerbatim}[commandchars=\\\{\}]
\PYG{g+gp}{\PYGZgt{}\PYGZgt{}\PYGZgt{} }\PYG{n}{x} \PYG{o}{=} \PYG{n}{jax}\PYG{o}{.}\PYG{n}{numpy}\PYG{o}{.}\PYG{n}{ones}\PYG{p}{(}\PYG{p}{(}\PYG{l+m+mi}{5}\PYG{p}{,} \PYG{l+m+mi}{6}\PYG{p}{)}\PYG{p}{)}
\PYG{g+gp}{\PYGZgt{}\PYGZgt{}\PYGZgt{} }\PYG{n}{jax}\PYG{o}{.}\PYG{n}{ops}\PYG{o}{.}\PYG{n}{index\PYGZus{}max}\PYG{p}{(}\PYG{n}{x}\PYG{p}{,} \PYG{n}{jax}\PYG{o}{.}\PYG{n}{ops}\PYG{o}{.}\PYG{n}{index}\PYG{p}{[}\PYG{l+m+mi}{2}\PYG{p}{:}\PYG{l+m+mi}{4}\PYG{p}{,} \PYG{l+m+mi}{3}\PYG{p}{:}\PYG{p}{]}\PYG{p}{,} \PYG{l+m+mf}{6.}\PYG{p}{)}
\PYG{g+go}{array([[1., 1., 1., 1., 1., 1.],}
\PYG{g+go}{       [1., 1., 1., 1., 1., 1.],}
\PYG{g+go}{       [1., 1., 1., 6., 6., 6.],}
\PYG{g+go}{       [1., 1., 1., 6., 6., 6.],}
\PYG{g+go}{       [1., 1., 1., 1., 1., 1.]], dtype=float32)}
\end{sphinxVerbatim}
\end{quote}

\end{fulllineitems}

\index{index\_min() (in module symjax.tensor.index\_ops)@\spxentry{index\_min()}\spxextra{in module symjax.tensor.index\_ops}}

\begin{fulllineitems}
\phantomsection\label{\detokenize{modules/tensor:symjax.tensor.index_ops.index_min}}\pysiglinewithargsret{\sphinxbfcode{\sphinxupquote{index\_min}}}{\emph{\DUrole{n}{x}}, \emph{\DUrole{n}{idx}}, \emph{\DUrole{n}{y}}}{}
Pure equivalent of \sphinxcode{\sphinxupquote{x{[}idx{]} = minimum(x{[}idx{]}, y)}}.

LAX\sphinxhyphen{}backend implementation of {\hyperref[\detokenize{modules/tensor:symjax.tensor.index_ops.index_min}]{\sphinxcrossref{\sphinxcode{\sphinxupquote{index\_min()}}}}}.
ADDITIONOriginal docstring below.
\begin{quote}
\begin{quote}

Returns the value of \sphinxtitleref{x} that would result from the
\end{quote}
\begin{description}
\item[{NumPy\sphinxhyphen{}style \sphinxcode{\sphinxupquote{indexed assignment}}::}] \leavevmode
x{[}idx{]} = minimum(x{[}idx{]}, y)

\end{description}

Note the \sphinxtitleref{index\_min} operator is pure; \sphinxtitleref{x} itself is
not modified, instead the new value that \sphinxtitleref{x} would have taken is returned.

Unlike the NumPy code \sphinxcode{\sphinxupquote{x{[}idx{]} = minimum(x{[}idx{]}, y)}}, if multiple indices
refer to the same location the final value will be the overall min. (NumPy
would only look at the last update, rather than all of the updates.)
\begin{description}
\item[{Args:}] \leavevmode
x: an array with the values to be updated.
idx: a Numpy\sphinxhyphen{}style index, consisting of \sphinxtitleref{None}, integers, \sphinxtitleref{slice} objects,
\begin{quote}

ellipses, ndarrays with integer dtypes, or a tuple of the above. A
convenient syntactic sugar for forming indices is via the
\sphinxcode{\sphinxupquote{jax.ops.index}} object.
\end{quote}
\begin{description}
\item[{y: the array of updates. \sphinxtitleref{y} must be broadcastable to the shape of the}] \leavevmode
array that would be returned by \sphinxtitleref{x{[}idx{]}}.

\end{description}

\item[{Returns:}] \leavevmode
An array.

\end{description}

\begin{sphinxVerbatim}[commandchars=\\\{\}]
\PYG{g+gp}{\PYGZgt{}\PYGZgt{}\PYGZgt{} }\PYG{n}{x} \PYG{o}{=} \PYG{n}{jax}\PYG{o}{.}\PYG{n}{numpy}\PYG{o}{.}\PYG{n}{ones}\PYG{p}{(}\PYG{p}{(}\PYG{l+m+mi}{5}\PYG{p}{,} \PYG{l+m+mi}{6}\PYG{p}{)}\PYG{p}{)}
\PYG{g+gp}{\PYGZgt{}\PYGZgt{}\PYGZgt{} }\PYG{n}{jax}\PYG{o}{.}\PYG{n}{ops}\PYG{o}{.}\PYG{n}{index\PYGZus{}minimum}\PYG{p}{(}\PYG{n}{x}\PYG{p}{,} \PYG{n}{jax}\PYG{o}{.}\PYG{n}{ops}\PYG{o}{.}\PYG{n}{index}\PYG{p}{[}\PYG{l+m+mi}{2}\PYG{p}{:}\PYG{l+m+mi}{4}\PYG{p}{,} \PYG{l+m+mi}{3}\PYG{p}{:}\PYG{p}{]}\PYG{p}{,} \PYG{l+m+mf}{0.}\PYG{p}{)}
\PYG{g+go}{array([[1., 1., 1., 1., 1., 1.],}
\PYG{g+go}{       [1., 1., 1., 1., 1., 1.],}
\PYG{g+go}{       [1., 1., 1., 0., 0., 0.],}
\PYG{g+go}{       [1., 1., 1., 0., 0., 0.],}
\PYG{g+go}{       [1., 1., 1., 1., 1., 1.]], dtype=float32)}
\end{sphinxVerbatim}
\end{quote}

\end{fulllineitems}

\subsubsection{Numpy Like}
\label{\detokenize{modules/tensor:module-symjax.tensor.ops_math}}\label{\detokenize{modules/tensor:numpy-like}}\index{module@\spxentry{module}!symjax.tensor.ops\_math@\spxentry{symjax.tensor.ops\_math}}\index{symjax.tensor.ops\_math@\spxentry{symjax.tensor.ops\_math}!module@\spxentry{module}}\index{hat\_1D() (in module symjax.tensor.ops\_math)@\spxentry{hat\_1D()}\spxextra{in module symjax.tensor.ops\_math}}

\begin{fulllineitems}
\phantomsection\label{\detokenize{modules/tensor:symjax.tensor.ops_math.hat_1D}}\pysiglinewithargsret{\sphinxbfcode{\sphinxupquote{hat\_1D}}}{\emph{\DUrole{n}{x}}, \emph{\DUrole{n}{t\_left}}, \emph{\DUrole{n}{t\_center}}, \emph{\DUrole{n}{t\_right}}}{}
hat basis function in 1\sphinxhyphen{}D

Hat function, continuous piecewise linear
\begin{quote}\begin{description}
\item[{Parameters}] \leavevmode\begin{itemize}
\item {} 
\sphinxstyleliteralstrong{\sphinxupquote{x}} (\sphinxstyleliteralemphasis{\sphinxupquote{array\sphinxhyphen{}like}}) \textendash{} the sampled input space

\item {} 
\sphinxstyleliteralstrong{\sphinxupquote{t\_left}} (\sphinxstyleliteralemphasis{\sphinxupquote{scalar}}) \textendash{} the position of the left knot

\item {} 
\sphinxstyleliteralstrong{\sphinxupquote{t\_center}} (\sphinxstyleliteralemphasis{\sphinxupquote{scalar}}) \textendash{} the position of the center knot

\item {} 
\sphinxstyleliteralstrong{\sphinxupquote{t\_right}} (\sphinxstyleliteralemphasis{\sphinxupquote{scalar}}) \textendash{} the position of the right knot

\end{itemize}

\item[{Returns}] \leavevmode
\sphinxstylestrong{output} \textendash{} same shape as x with applied hat function

\item[{Return type}] \leavevmode
array

\end{description}\end{quote}

\end{fulllineitems}

\index{one\_hot() (in module symjax.tensor.ops\_math)@\spxentry{one\_hot()}\spxextra{in module symjax.tensor.ops\_math}}

\begin{fulllineitems}
\phantomsection\label{\detokenize{modules/tensor:symjax.tensor.ops_math.one_hot}}\pysiglinewithargsret{\sphinxbfcode{\sphinxupquote{one\_hot}}}{\emph{\DUrole{n}{i}}, \emph{\DUrole{n}{N}}, \emph{\DUrole{n}{dtype}\DUrole{o}{=}\DUrole{default_value}{\textquotesingle{}float32\textquotesingle{}}}}{}
Create a one\sphinxhyphen{}hot encoding of x of size k.

\end{fulllineitems}

\subsubsection{Other}
\label{\detokenize{modules/tensor:other}}\index{abs() (in module symjax.tensor)@\spxentry{abs()}\spxextra{in module symjax.tensor}}

\begin{fulllineitems}
\phantomsection\label{\detokenize{modules/tensor:symjax.tensor.abs}}\pysiglinewithargsret{\sphinxbfcode{\sphinxupquote{abs}}}{\emph{\DUrole{n}{x}}}{}
Calculate the absolute value element\sphinxhyphen{}wise.

LAX\sphinxhyphen{}backend implementation of {\hyperref[\detokenize{modules/tensor:symjax.tensor.absolute}]{\sphinxcrossref{\sphinxcode{\sphinxupquote{absolute()}}}}}.
ADDITIONOriginal docstring below.

LAX\sphinxhyphen{}backend implementation of {\hyperref[\detokenize{modules/tensor:symjax.tensor.absolute}]{\sphinxcrossref{\sphinxcode{\sphinxupquote{absolute()}}}}}.
Original docstring below.

absolute(x, /, out=None, {\color{red}\bfseries{}*}, where=True, casting=’same\_kind’, order=’K’, dtype=None, subok=True{[}, signature, extobj{]})

\sphinxcode{\sphinxupquote{np.abs}} is a shorthand for this function.
\begin{quote}\begin{description}
\item[{Returns}] \leavevmode
\sphinxstylestrong{absolute} \textendash{} An ndarray containing the absolute value of
each element in \sphinxtitleref{x}.  For complex input, \sphinxcode{\sphinxupquote{a + ib}}, the
absolute value is \(\sqrt{ a^2 + b^2 }\).
This is a scalar if \sphinxtitleref{x} is a scalar.

\item[{Return type}] \leavevmode
ndarray

\end{description}\end{quote}
\subsubsection*{Examples}

\begin{sphinxVerbatim}[commandchars=\\\{\}]
\PYG{g+gp}{\PYGZgt{}\PYGZgt{}\PYGZgt{} }\PYG{n}{x} \PYG{o}{=} \PYG{n}{np}\PYG{o}{.}\PYG{n}{array}\PYG{p}{(}\PYG{p}{[}\PYG{o}{\PYGZhy{}}\PYG{l+m+mf}{1.2}\PYG{p}{,} \PYG{l+m+mf}{1.2}\PYG{p}{]}\PYG{p}{)}
\PYG{g+gp}{\PYGZgt{}\PYGZgt{}\PYGZgt{} }\PYG{n}{np}\PYG{o}{.}\PYG{n}{absolute}\PYG{p}{(}\PYG{n}{x}\PYG{p}{)}
\PYG{g+go}{array([ 1.2,  1.2])}
\PYG{g+gp}{\PYGZgt{}\PYGZgt{}\PYGZgt{} }\PYG{n}{np}\PYG{o}{.}\PYG{n}{absolute}\PYG{p}{(}\PYG{l+m+mf}{1.2} \PYG{o}{+} \PYG{l+m+mi}{1}\PYG{n}{j}\PYG{p}{)}
\PYG{g+go}{1.5620499351813308}
\end{sphinxVerbatim}

Plot the function over \sphinxcode{\sphinxupquote{{[}\sphinxhyphen{}10, 10{]}}}:

\begin{sphinxVerbatim}[commandchars=\\\{\}]
\PYG{g+gp}{\PYGZgt{}\PYGZgt{}\PYGZgt{} }\PYG{k+kn}{import} \PYG{n+nn}{matplotlib}\PYG{n+nn}{.}\PYG{n+nn}{pyplot} \PYG{k}{as} \PYG{n+nn}{plt}
\end{sphinxVerbatim}

\begin{sphinxVerbatim}[commandchars=\\\{\}]
\PYG{g+gp}{\PYGZgt{}\PYGZgt{}\PYGZgt{} }\PYG{n}{x} \PYG{o}{=} \PYG{n}{np}\PYG{o}{.}\PYG{n}{linspace}\PYG{p}{(}\PYG{n}{start}\PYG{o}{=}\PYG{o}{\PYGZhy{}}\PYG{l+m+mi}{10}\PYG{p}{,} \PYG{n}{stop}\PYG{o}{=}\PYG{l+m+mi}{10}\PYG{p}{,} \PYG{n}{num}\PYG{o}{=}\PYG{l+m+mi}{101}\PYG{p}{)}
\PYG{g+gp}{\PYGZgt{}\PYGZgt{}\PYGZgt{} }\PYG{n}{plt}\PYG{o}{.}\PYG{n}{plot}\PYG{p}{(}\PYG{n}{x}\PYG{p}{,} \PYG{n}{np}\PYG{o}{.}\PYG{n}{absolute}\PYG{p}{(}\PYG{n}{x}\PYG{p}{)}\PYG{p}{)}
\PYG{g+gp}{\PYGZgt{}\PYGZgt{}\PYGZgt{} }\PYG{n}{plt}\PYG{o}{.}\PYG{n}{show}\PYG{p}{(}\PYG{p}{)}
\end{sphinxVerbatim}

Plot the function over the complex plane:

\begin{sphinxVerbatim}[commandchars=\\\{\}]
\PYG{g+gp}{\PYGZgt{}\PYGZgt{}\PYGZgt{} }\PYG{n}{xx} \PYG{o}{=} \PYG{n}{x} \PYG{o}{+} \PYG{l+m+mi}{1}\PYG{n}{j} \PYG{o}{*} \PYG{n}{x}\PYG{p}{[}\PYG{p}{:}\PYG{p}{,} \PYG{n}{np}\PYG{o}{.}\PYG{n}{newaxis}\PYG{p}{]}
\PYG{g+gp}{\PYGZgt{}\PYGZgt{}\PYGZgt{} }\PYG{n}{plt}\PYG{o}{.}\PYG{n}{imshow}\PYG{p}{(}\PYG{n}{np}\PYG{o}{.}\PYG{n}{abs}\PYG{p}{(}\PYG{n}{xx}\PYG{p}{)}\PYG{p}{,} \PYG{n}{extent}\PYG{o}{=}\PYG{p}{[}\PYG{o}{\PYGZhy{}}\PYG{l+m+mi}{10}\PYG{p}{,} \PYG{l+m+mi}{10}\PYG{p}{,} \PYG{o}{\PYGZhy{}}\PYG{l+m+mi}{10}\PYG{p}{,} \PYG{l+m+mi}{10}\PYG{p}{]}\PYG{p}{,} \PYG{n}{cmap}\PYG{o}{=}\PYG{l+s+s1}{\PYGZsq{}}\PYG{l+s+s1}{gray}\PYG{l+s+s1}{\PYGZsq{}}\PYG{p}{)}
\PYG{g+gp}{\PYGZgt{}\PYGZgt{}\PYGZgt{} }\PYG{n}{plt}\PYG{o}{.}\PYG{n}{show}\PYG{p}{(}\PYG{p}{)}
\end{sphinxVerbatim}

\end{fulllineitems}

\index{absolute() (in module symjax.tensor)@\spxentry{absolute()}\spxextra{in module symjax.tensor}}

\begin{fulllineitems}
\phantomsection\label{\detokenize{modules/tensor:symjax.tensor.absolute}}\pysiglinewithargsret{\sphinxbfcode{\sphinxupquote{absolute}}}{\emph{\DUrole{n}{x}}}{}
Calculate the absolute value element\sphinxhyphen{}wise.

LAX\sphinxhyphen{}backend implementation of {\hyperref[\detokenize{modules/tensor:symjax.tensor.absolute}]{\sphinxcrossref{\sphinxcode{\sphinxupquote{absolute()}}}}}.
ADDITIONOriginal docstring below.

LAX\sphinxhyphen{}backend implementation of {\hyperref[\detokenize{modules/tensor:symjax.tensor.absolute}]{\sphinxcrossref{\sphinxcode{\sphinxupquote{absolute()}}}}}.
Original docstring below.

absolute(x, /, out=None, {\color{red}\bfseries{}*}, where=True, casting=’same\_kind’, order=’K’, dtype=None, subok=True{[}, signature, extobj{]})

\sphinxcode{\sphinxupquote{np.abs}} is a shorthand for this function.
\begin{quote}\begin{description}
\item[{Returns}] \leavevmode
\sphinxstylestrong{absolute} \textendash{} An ndarray containing the absolute value of
each element in \sphinxtitleref{x}.  For complex input, \sphinxcode{\sphinxupquote{a + ib}}, the
absolute value is \(\sqrt{ a^2 + b^2 }\).
This is a scalar if \sphinxtitleref{x} is a scalar.

\item[{Return type}] \leavevmode
ndarray

\end{description}\end{quote}
\subsubsection*{Examples}

\begin{sphinxVerbatim}[commandchars=\\\{\}]
\PYG{g+gp}{\PYGZgt{}\PYGZgt{}\PYGZgt{} }\PYG{n}{x} \PYG{o}{=} \PYG{n}{np}\PYG{o}{.}\PYG{n}{array}\PYG{p}{(}\PYG{p}{[}\PYG{o}{\PYGZhy{}}\PYG{l+m+mf}{1.2}\PYG{p}{,} \PYG{l+m+mf}{1.2}\PYG{p}{]}\PYG{p}{)}
\PYG{g+gp}{\PYGZgt{}\PYGZgt{}\PYGZgt{} }\PYG{n}{np}\PYG{o}{.}\PYG{n}{absolute}\PYG{p}{(}\PYG{n}{x}\PYG{p}{)}
\PYG{g+go}{array([ 1.2,  1.2])}
\PYG{g+gp}{\PYGZgt{}\PYGZgt{}\PYGZgt{} }\PYG{n}{np}\PYG{o}{.}\PYG{n}{absolute}\PYG{p}{(}\PYG{l+m+mf}{1.2} \PYG{o}{+} \PYG{l+m+mi}{1}\PYG{n}{j}\PYG{p}{)}
\PYG{g+go}{1.5620499351813308}
\end{sphinxVerbatim}

Plot the function over \sphinxcode{\sphinxupquote{{[}\sphinxhyphen{}10, 10{]}}}:

\begin{sphinxVerbatim}[commandchars=\\\{\}]
\PYG{g+gp}{\PYGZgt{}\PYGZgt{}\PYGZgt{} }\PYG{k+kn}{import} \PYG{n+nn}{matplotlib}\PYG{n+nn}{.}\PYG{n+nn}{pyplot} \PYG{k}{as} \PYG{n+nn}{plt}
\end{sphinxVerbatim}

\begin{sphinxVerbatim}[commandchars=\\\{\}]
\PYG{g+gp}{\PYGZgt{}\PYGZgt{}\PYGZgt{} }\PYG{n}{x} \PYG{o}{=} \PYG{n}{np}\PYG{o}{.}\PYG{n}{linspace}\PYG{p}{(}\PYG{n}{start}\PYG{o}{=}\PYG{o}{\PYGZhy{}}\PYG{l+m+mi}{10}\PYG{p}{,} \PYG{n}{stop}\PYG{o}{=}\PYG{l+m+mi}{10}\PYG{p}{,} \PYG{n}{num}\PYG{o}{=}\PYG{l+m+mi}{101}\PYG{p}{)}
\PYG{g+gp}{\PYGZgt{}\PYGZgt{}\PYGZgt{} }\PYG{n}{plt}\PYG{o}{.}\PYG{n}{plot}\PYG{p}{(}\PYG{n}{x}\PYG{p}{,} \PYG{n}{np}\PYG{o}{.}\PYG{n}{absolute}\PYG{p}{(}\PYG{n}{x}\PYG{p}{)}\PYG{p}{)}
\PYG{g+gp}{\PYGZgt{}\PYGZgt{}\PYGZgt{} }\PYG{n}{plt}\PYG{o}{.}\PYG{n}{show}\PYG{p}{(}\PYG{p}{)}
\end{sphinxVerbatim}

Plot the function over the complex plane:

\begin{sphinxVerbatim}[commandchars=\\\{\}]
\PYG{g+gp}{\PYGZgt{}\PYGZgt{}\PYGZgt{} }\PYG{n}{xx} \PYG{o}{=} \PYG{n}{x} \PYG{o}{+} \PYG{l+m+mi}{1}\PYG{n}{j} \PYG{o}{*} \PYG{n}{x}\PYG{p}{[}\PYG{p}{:}\PYG{p}{,} \PYG{n}{np}\PYG{o}{.}\PYG{n}{newaxis}\PYG{p}{]}
\PYG{g+gp}{\PYGZgt{}\PYGZgt{}\PYGZgt{} }\PYG{n}{plt}\PYG{o}{.}\PYG{n}{imshow}\PYG{p}{(}\PYG{n}{np}\PYG{o}{.}\PYG{n}{abs}\PYG{p}{(}\PYG{n}{xx}\PYG{p}{)}\PYG{p}{,} \PYG{n}{extent}\PYG{o}{=}\PYG{p}{[}\PYG{o}{\PYGZhy{}}\PYG{l+m+mi}{10}\PYG{p}{,} \PYG{l+m+mi}{10}\PYG{p}{,} \PYG{o}{\PYGZhy{}}\PYG{l+m+mi}{10}\PYG{p}{,} \PYG{l+m+mi}{10}\PYG{p}{]}\PYG{p}{,} \PYG{n}{cmap}\PYG{o}{=}\PYG{l+s+s1}{\PYGZsq{}}\PYG{l+s+s1}{gray}\PYG{l+s+s1}{\PYGZsq{}}\PYG{p}{)}
\PYG{g+gp}{\PYGZgt{}\PYGZgt{}\PYGZgt{} }\PYG{n}{plt}\PYG{o}{.}\PYG{n}{show}\PYG{p}{(}\PYG{p}{)}
\end{sphinxVerbatim}

\end{fulllineitems}

\index{add() (in module symjax.tensor)@\spxentry{add()}\spxextra{in module symjax.tensor}}

\begin{fulllineitems}
\phantomsection\label{\detokenize{modules/tensor:symjax.tensor.add}}\pysiglinewithargsret{\sphinxbfcode{\sphinxupquote{add}}}{\emph{\DUrole{n}{x1}}, \emph{\DUrole{n}{x2}}}{}
Add arguments element\sphinxhyphen{}wise.

LAX\sphinxhyphen{}backend implementation of {\hyperref[\detokenize{modules/tensor:symjax.tensor.add}]{\sphinxcrossref{\sphinxcode{\sphinxupquote{add()}}}}}.
ADDITIONOriginal docstring below.

LAX\sphinxhyphen{}backend implementation of {\hyperref[\detokenize{modules/tensor:symjax.tensor.add}]{\sphinxcrossref{\sphinxcode{\sphinxupquote{add()}}}}}.
Original docstring below.

add(x1, x2, /, out=None, {\color{red}\bfseries{}*}, where=True, casting=’same\_kind’, order=’K’, dtype=None, subok=True{[}, signature, extobj{]})
\begin{quote}\begin{description}
\item[{Returns}] \leavevmode
\sphinxstylestrong{add} \textendash{} The sum of \sphinxtitleref{x1} and \sphinxtitleref{x2}, element\sphinxhyphen{}wise.
This is a scalar if both \sphinxtitleref{x1} and \sphinxtitleref{x2} are scalars.

\item[{Return type}] \leavevmode
ndarray or scalar

\end{description}\end{quote}
\subsubsection*{Notes}

Equivalent to \sphinxtitleref{x1} + \sphinxtitleref{x2} in terms of array broadcasting.
\subsubsection*{Examples}

\begin{sphinxVerbatim}[commandchars=\\\{\}]
\PYG{g+gp}{\PYGZgt{}\PYGZgt{}\PYGZgt{} }\PYG{n}{np}\PYG{o}{.}\PYG{n}{add}\PYG{p}{(}\PYG{l+m+mf}{1.0}\PYG{p}{,} \PYG{l+m+mf}{4.0}\PYG{p}{)}
\PYG{g+go}{5.0}
\PYG{g+gp}{\PYGZgt{}\PYGZgt{}\PYGZgt{} }\PYG{n}{x1} \PYG{o}{=} \PYG{n}{np}\PYG{o}{.}\PYG{n}{arange}\PYG{p}{(}\PYG{l+m+mf}{9.0}\PYG{p}{)}\PYG{o}{.}\PYG{n}{reshape}\PYG{p}{(}\PYG{p}{(}\PYG{l+m+mi}{3}\PYG{p}{,} \PYG{l+m+mi}{3}\PYG{p}{)}\PYG{p}{)}
\PYG{g+gp}{\PYGZgt{}\PYGZgt{}\PYGZgt{} }\PYG{n}{x2} \PYG{o}{=} \PYG{n}{np}\PYG{o}{.}\PYG{n}{arange}\PYG{p}{(}\PYG{l+m+mf}{3.0}\PYG{p}{)}
\PYG{g+gp}{\PYGZgt{}\PYGZgt{}\PYGZgt{} }\PYG{n}{np}\PYG{o}{.}\PYG{n}{add}\PYG{p}{(}\PYG{n}{x1}\PYG{p}{,} \PYG{n}{x2}\PYG{p}{)}
\PYG{g+go}{array([[  0.,   2.,   4.],}
\PYG{g+go}{       [  3.,   5.,   7.],}
\PYG{g+go}{       [  6.,   8.,  10.]])}
\end{sphinxVerbatim}

\end{fulllineitems}

\index{all() (in module symjax.tensor)@\spxentry{all()}\spxextra{in module symjax.tensor}}

\begin{fulllineitems}
\phantomsection\label{\detokenize{modules/tensor:symjax.tensor.all}}\pysiglinewithargsret{\sphinxbfcode{\sphinxupquote{all}}}{\emph{\DUrole{n}{a}}, \emph{\DUrole{n}{axis}\DUrole{o}{=}\DUrole{default_value}{None}}, \emph{\DUrole{n}{dtype}\DUrole{o}{=}\DUrole{default_value}{None}}, \emph{\DUrole{n}{out}\DUrole{o}{=}\DUrole{default_value}{None}}, \emph{\DUrole{n}{keepdims}\DUrole{o}{=}\DUrole{default_value}{False}}}{}
Test whether all array elements along a given axis evaluate to True.

LAX\sphinxhyphen{}backend implementation of {\hyperref[\detokenize{modules/tensor:symjax.tensor.all}]{\sphinxcrossref{\sphinxcode{\sphinxupquote{all()}}}}}.
ADDITIONOriginal docstring below.

LAX\sphinxhyphen{}backend implementation of {\hyperref[\detokenize{modules/tensor:symjax.tensor.all}]{\sphinxcrossref{\sphinxcode{\sphinxupquote{all()}}}}}.
Original docstring below.
\begin{quote}\begin{description}
\item[{Returns}] \leavevmode
\sphinxstylestrong{all} \textendash{} A new boolean or array is returned unless \sphinxtitleref{out} is specified,
in which case a reference to \sphinxtitleref{out} is returned.

\item[{Return type}] \leavevmode
ndarray, bool

\end{description}\end{quote}

\sphinxstrong{See also:}

\begin{description}
\item[{\sphinxcode{\sphinxupquote{ndarray.all()}}}] \leavevmode
equivalent method

\item[{{\hyperref[\detokenize{modules/tensor:symjax.tensor.any}]{\sphinxcrossref{\sphinxcode{\sphinxupquote{any()}}}}}}] \leavevmode
Test whether any element along a given axis evaluates to True.

\end{description}

\subsubsection*{Notes}

Not a Number (NaN), positive infinity and negative infinity
evaluate to \sphinxtitleref{True} because these are not equal to zero.
\subsubsection*{Examples}

\begin{sphinxVerbatim}[commandchars=\\\{\}]
\PYG{g+gp}{\PYGZgt{}\PYGZgt{}\PYGZgt{} }\PYG{n}{np}\PYG{o}{.}\PYG{n}{all}\PYG{p}{(}\PYG{p}{[}\PYG{p}{[}\PYG{k+kc}{True}\PYG{p}{,}\PYG{k+kc}{False}\PYG{p}{]}\PYG{p}{,}\PYG{p}{[}\PYG{k+kc}{True}\PYG{p}{,}\PYG{k+kc}{True}\PYG{p}{]}\PYG{p}{]}\PYG{p}{)}
\PYG{g+go}{False}
\end{sphinxVerbatim}

\begin{sphinxVerbatim}[commandchars=\\\{\}]
\PYG{g+gp}{\PYGZgt{}\PYGZgt{}\PYGZgt{} }\PYG{n}{np}\PYG{o}{.}\PYG{n}{all}\PYG{p}{(}\PYG{p}{[}\PYG{p}{[}\PYG{k+kc}{True}\PYG{p}{,}\PYG{k+kc}{False}\PYG{p}{]}\PYG{p}{,}\PYG{p}{[}\PYG{k+kc}{True}\PYG{p}{,}\PYG{k+kc}{True}\PYG{p}{]}\PYG{p}{]}\PYG{p}{,} \PYG{n}{axis}\PYG{o}{=}\PYG{l+m+mi}{0}\PYG{p}{)}
\PYG{g+go}{array([ True, False])}
\end{sphinxVerbatim}

\begin{sphinxVerbatim}[commandchars=\\\{\}]
\PYG{g+gp}{\PYGZgt{}\PYGZgt{}\PYGZgt{} }\PYG{n}{np}\PYG{o}{.}\PYG{n}{all}\PYG{p}{(}\PYG{p}{[}\PYG{o}{\PYGZhy{}}\PYG{l+m+mi}{1}\PYG{p}{,} \PYG{l+m+mi}{4}\PYG{p}{,} \PYG{l+m+mi}{5}\PYG{p}{]}\PYG{p}{)}
\PYG{g+go}{True}
\end{sphinxVerbatim}

\begin{sphinxVerbatim}[commandchars=\\\{\}]
\PYG{g+gp}{\PYGZgt{}\PYGZgt{}\PYGZgt{} }\PYG{n}{np}\PYG{o}{.}\PYG{n}{all}\PYG{p}{(}\PYG{p}{[}\PYG{l+m+mf}{1.0}\PYG{p}{,} \PYG{n}{np}\PYG{o}{.}\PYG{n}{nan}\PYG{p}{]}\PYG{p}{)}
\PYG{g+go}{True}
\end{sphinxVerbatim}

\begin{sphinxVerbatim}[commandchars=\\\{\}]
\PYG{g+gp}{\PYGZgt{}\PYGZgt{}\PYGZgt{} }\PYG{n}{o}\PYG{o}{=}\PYG{n}{np}\PYG{o}{.}\PYG{n}{array}\PYG{p}{(}\PYG{k+kc}{False}\PYG{p}{)}
\PYG{g+gp}{\PYGZgt{}\PYGZgt{}\PYGZgt{} }\PYG{n}{z}\PYG{o}{=}\PYG{n}{np}\PYG{o}{.}\PYG{n}{all}\PYG{p}{(}\PYG{p}{[}\PYG{o}{\PYGZhy{}}\PYG{l+m+mi}{1}\PYG{p}{,} \PYG{l+m+mi}{4}\PYG{p}{,} \PYG{l+m+mi}{5}\PYG{p}{]}\PYG{p}{,} \PYG{n}{out}\PYG{o}{=}\PYG{n}{o}\PYG{p}{)}
\PYG{g+gp}{\PYGZgt{}\PYGZgt{}\PYGZgt{} }\PYG{n+nb}{id}\PYG{p}{(}\PYG{n}{z}\PYG{p}{)}\PYG{p}{,} \PYG{n+nb}{id}\PYG{p}{(}\PYG{n}{o}\PYG{p}{)}\PYG{p}{,} \PYG{n}{z}
\PYG{g+go}{(28293632, 28293632, array(True)) \PYGZsh{} may vary}
\end{sphinxVerbatim}

\end{fulllineitems}

\index{allclose() (in module symjax.tensor)@\spxentry{allclose()}\spxextra{in module symjax.tensor}}

\begin{fulllineitems}
\phantomsection\label{\detokenize{modules/tensor:symjax.tensor.allclose}}\pysiglinewithargsret{\sphinxbfcode{\sphinxupquote{allclose}}}{\emph{\DUrole{n}{a}}, \emph{\DUrole{n}{b}}, \emph{\DUrole{n}{rtol}\DUrole{o}{=}\DUrole{default_value}{1e\sphinxhyphen{}05}}, \emph{\DUrole{n}{atol}\DUrole{o}{=}\DUrole{default_value}{1e\sphinxhyphen{}08}}}{}
Returns True if two arrays are element\sphinxhyphen{}wise equal within a tolerance.

LAX\sphinxhyphen{}backend implementation of {\hyperref[\detokenize{modules/tensor:symjax.tensor.allclose}]{\sphinxcrossref{\sphinxcode{\sphinxupquote{allclose()}}}}}.
ADDITIONOriginal docstring below.

LAX\sphinxhyphen{}backend implementation of {\hyperref[\detokenize{modules/tensor:symjax.tensor.allclose}]{\sphinxcrossref{\sphinxcode{\sphinxupquote{allclose()}}}}}.
Original docstring below.

The tolerance values are positive, typically very small numbers.  The
relative difference (\sphinxtitleref{rtol} * abs(\sphinxtitleref{b})) and the absolute difference
\sphinxtitleref{atol} are added together to compare against the absolute difference
between \sphinxtitleref{a} and \sphinxtitleref{b}.

NaNs are treated as equal if they are in the same place and if
\sphinxcode{\sphinxupquote{equal\_nan=True}}.  Infs are treated as equal if they are in the same
place and of the same sign in both arrays.
\begin{quote}\begin{description}
\item[{Returns}] \leavevmode
\sphinxstylestrong{allclose} \textendash{} Returns True if the two arrays are equal within the given
tolerance; False otherwise.

\item[{Return type}] \leavevmode
bool

\end{description}\end{quote}

\sphinxstrong{See also:}

{\hyperref[\detokenize{modules/tensor:symjax.tensor.isclose}]{\sphinxcrossref{\sphinxcode{\sphinxupquote{isclose()}}}}}, {\hyperref[\detokenize{modules/tensor:symjax.tensor.all}]{\sphinxcrossref{\sphinxcode{\sphinxupquote{all()}}}}}, {\hyperref[\detokenize{modules/tensor:symjax.tensor.any}]{\sphinxcrossref{\sphinxcode{\sphinxupquote{any()}}}}}, {\hyperref[\detokenize{modules/tensor:symjax.tensor.equal}]{\sphinxcrossref{\sphinxcode{\sphinxupquote{equal()}}}}}

\subsubsection*{Notes}

If the following equation is element\sphinxhyphen{}wise True, then allclose returns
True.
\begin{quote}

absolute(\sphinxtitleref{a} \sphinxhyphen{} \sphinxtitleref{b}) \textless{}= (\sphinxtitleref{atol} + \sphinxtitleref{rtol} * absolute(\sphinxtitleref{b}))
\end{quote}

The above equation is not symmetric in \sphinxtitleref{a} and \sphinxtitleref{b}, so that
\sphinxcode{\sphinxupquote{allclose(a, b)}} might be different from \sphinxcode{\sphinxupquote{allclose(b, a)}} in
some rare cases.

The comparison of \sphinxtitleref{a} and \sphinxtitleref{b} uses standard broadcasting, which
means that \sphinxtitleref{a} and \sphinxtitleref{b} need not have the same shape in order for
\sphinxcode{\sphinxupquote{allclose(a, b)}} to evaluate to True.  The same is true for
\sphinxtitleref{equal} but not \sphinxtitleref{array\_equal}.
\subsubsection*{Examples}

\begin{sphinxVerbatim}[commandchars=\\\{\}]
\PYG{g+gp}{\PYGZgt{}\PYGZgt{}\PYGZgt{} }\PYG{n}{np}\PYG{o}{.}\PYG{n}{allclose}\PYG{p}{(}\PYG{p}{[}\PYG{l+m+mf}{1e10}\PYG{p}{,}\PYG{l+m+mf}{1e\PYGZhy{}7}\PYG{p}{]}\PYG{p}{,} \PYG{p}{[}\PYG{l+m+mf}{1.00001e10}\PYG{p}{,}\PYG{l+m+mf}{1e\PYGZhy{}8}\PYG{p}{]}\PYG{p}{)}
\PYG{g+go}{False}
\PYG{g+gp}{\PYGZgt{}\PYGZgt{}\PYGZgt{} }\PYG{n}{np}\PYG{o}{.}\PYG{n}{allclose}\PYG{p}{(}\PYG{p}{[}\PYG{l+m+mf}{1e10}\PYG{p}{,}\PYG{l+m+mf}{1e\PYGZhy{}8}\PYG{p}{]}\PYG{p}{,} \PYG{p}{[}\PYG{l+m+mf}{1.00001e10}\PYG{p}{,}\PYG{l+m+mf}{1e\PYGZhy{}9}\PYG{p}{]}\PYG{p}{)}
\PYG{g+go}{True}
\PYG{g+gp}{\PYGZgt{}\PYGZgt{}\PYGZgt{} }\PYG{n}{np}\PYG{o}{.}\PYG{n}{allclose}\PYG{p}{(}\PYG{p}{[}\PYG{l+m+mf}{1e10}\PYG{p}{,}\PYG{l+m+mf}{1e\PYGZhy{}8}\PYG{p}{]}\PYG{p}{,} \PYG{p}{[}\PYG{l+m+mf}{1.0001e10}\PYG{p}{,}\PYG{l+m+mf}{1e\PYGZhy{}9}\PYG{p}{]}\PYG{p}{)}
\PYG{g+go}{False}
\PYG{g+gp}{\PYGZgt{}\PYGZgt{}\PYGZgt{} }\PYG{n}{np}\PYG{o}{.}\PYG{n}{allclose}\PYG{p}{(}\PYG{p}{[}\PYG{l+m+mf}{1.0}\PYG{p}{,} \PYG{n}{np}\PYG{o}{.}\PYG{n}{nan}\PYG{p}{]}\PYG{p}{,} \PYG{p}{[}\PYG{l+m+mf}{1.0}\PYG{p}{,} \PYG{n}{np}\PYG{o}{.}\PYG{n}{nan}\PYG{p}{]}\PYG{p}{)}
\PYG{g+go}{False}
\PYG{g+gp}{\PYGZgt{}\PYGZgt{}\PYGZgt{} }\PYG{n}{np}\PYG{o}{.}\PYG{n}{allclose}\PYG{p}{(}\PYG{p}{[}\PYG{l+m+mf}{1.0}\PYG{p}{,} \PYG{n}{np}\PYG{o}{.}\PYG{n}{nan}\PYG{p}{]}\PYG{p}{,} \PYG{p}{[}\PYG{l+m+mf}{1.0}\PYG{p}{,} \PYG{n}{np}\PYG{o}{.}\PYG{n}{nan}\PYG{p}{]}\PYG{p}{,} \PYG{n}{equal\PYGZus{}nan}\PYG{o}{=}\PYG{k+kc}{True}\PYG{p}{)}
\PYG{g+go}{True}
\end{sphinxVerbatim}

\end{fulllineitems}

\index{alltrue() (in module symjax.tensor)@\spxentry{alltrue()}\spxextra{in module symjax.tensor}}

\begin{fulllineitems}
\phantomsection\label{\detokenize{modules/tensor:symjax.tensor.alltrue}}\pysiglinewithargsret{\sphinxbfcode{\sphinxupquote{alltrue}}}{\emph{\DUrole{n}{a}}, \emph{\DUrole{n}{axis}\DUrole{o}{=}\DUrole{default_value}{None}}, \emph{\DUrole{n}{dtype}\DUrole{o}{=}\DUrole{default_value}{None}}, \emph{\DUrole{n}{out}\DUrole{o}{=}\DUrole{default_value}{None}}, \emph{\DUrole{n}{keepdims}\DUrole{o}{=}\DUrole{default_value}{False}}}{}
Test whether all array elements along a given axis evaluate to True.

LAX\sphinxhyphen{}backend implementation of {\hyperref[\detokenize{modules/tensor:symjax.tensor.all}]{\sphinxcrossref{\sphinxcode{\sphinxupquote{all()}}}}}.
ADDITIONOriginal docstring below.

LAX\sphinxhyphen{}backend implementation of {\hyperref[\detokenize{modules/tensor:symjax.tensor.all}]{\sphinxcrossref{\sphinxcode{\sphinxupquote{all()}}}}}.
Original docstring below.
\begin{quote}\begin{description}
\item[{Returns}] \leavevmode
\sphinxstylestrong{all} \textendash{} A new boolean or array is returned unless \sphinxtitleref{out} is specified,
in which case a reference to \sphinxtitleref{out} is returned.

\item[{Return type}] \leavevmode
ndarray, bool

\end{description}\end{quote}

\sphinxstrong{See also:}

\begin{description}
\item[{\sphinxcode{\sphinxupquote{ndarray.all()}}}] \leavevmode
equivalent method

\item[{{\hyperref[\detokenize{modules/tensor:symjax.tensor.any}]{\sphinxcrossref{\sphinxcode{\sphinxupquote{any()}}}}}}] \leavevmode
Test whether any element along a given axis evaluates to True.

\end{description}

\subsubsection*{Notes}

Not a Number (NaN), positive infinity and negative infinity
evaluate to \sphinxtitleref{True} because these are not equal to zero.
\subsubsection*{Examples}

\begin{sphinxVerbatim}[commandchars=\\\{\}]
\PYG{g+gp}{\PYGZgt{}\PYGZgt{}\PYGZgt{} }\PYG{n}{np}\PYG{o}{.}\PYG{n}{all}\PYG{p}{(}\PYG{p}{[}\PYG{p}{[}\PYG{k+kc}{True}\PYG{p}{,}\PYG{k+kc}{False}\PYG{p}{]}\PYG{p}{,}\PYG{p}{[}\PYG{k+kc}{True}\PYG{p}{,}\PYG{k+kc}{True}\PYG{p}{]}\PYG{p}{]}\PYG{p}{)}
\PYG{g+go}{False}
\end{sphinxVerbatim}

\begin{sphinxVerbatim}[commandchars=\\\{\}]
\PYG{g+gp}{\PYGZgt{}\PYGZgt{}\PYGZgt{} }\PYG{n}{np}\PYG{o}{.}\PYG{n}{all}\PYG{p}{(}\PYG{p}{[}\PYG{p}{[}\PYG{k+kc}{True}\PYG{p}{,}\PYG{k+kc}{False}\PYG{p}{]}\PYG{p}{,}\PYG{p}{[}\PYG{k+kc}{True}\PYG{p}{,}\PYG{k+kc}{True}\PYG{p}{]}\PYG{p}{]}\PYG{p}{,} \PYG{n}{axis}\PYG{o}{=}\PYG{l+m+mi}{0}\PYG{p}{)}
\PYG{g+go}{array([ True, False])}
\end{sphinxVerbatim}

\begin{sphinxVerbatim}[commandchars=\\\{\}]
\PYG{g+gp}{\PYGZgt{}\PYGZgt{}\PYGZgt{} }\PYG{n}{np}\PYG{o}{.}\PYG{n}{all}\PYG{p}{(}\PYG{p}{[}\PYG{o}{\PYGZhy{}}\PYG{l+m+mi}{1}\PYG{p}{,} \PYG{l+m+mi}{4}\PYG{p}{,} \PYG{l+m+mi}{5}\PYG{p}{]}\PYG{p}{)}
\PYG{g+go}{True}
\end{sphinxVerbatim}

\begin{sphinxVerbatim}[commandchars=\\\{\}]
\PYG{g+gp}{\PYGZgt{}\PYGZgt{}\PYGZgt{} }\PYG{n}{np}\PYG{o}{.}\PYG{n}{all}\PYG{p}{(}\PYG{p}{[}\PYG{l+m+mf}{1.0}\PYG{p}{,} \PYG{n}{np}\PYG{o}{.}\PYG{n}{nan}\PYG{p}{]}\PYG{p}{)}
\PYG{g+go}{True}
\end{sphinxVerbatim}

\begin{sphinxVerbatim}[commandchars=\\\{\}]
\PYG{g+gp}{\PYGZgt{}\PYGZgt{}\PYGZgt{} }\PYG{n}{o}\PYG{o}{=}\PYG{n}{np}\PYG{o}{.}\PYG{n}{array}\PYG{p}{(}\PYG{k+kc}{False}\PYG{p}{)}
\PYG{g+gp}{\PYGZgt{}\PYGZgt{}\PYGZgt{} }\PYG{n}{z}\PYG{o}{=}\PYG{n}{np}\PYG{o}{.}\PYG{n}{all}\PYG{p}{(}\PYG{p}{[}\PYG{o}{\PYGZhy{}}\PYG{l+m+mi}{1}\PYG{p}{,} \PYG{l+m+mi}{4}\PYG{p}{,} \PYG{l+m+mi}{5}\PYG{p}{]}\PYG{p}{,} \PYG{n}{out}\PYG{o}{=}\PYG{n}{o}\PYG{p}{)}
\PYG{g+gp}{\PYGZgt{}\PYGZgt{}\PYGZgt{} }\PYG{n+nb}{id}\PYG{p}{(}\PYG{n}{z}\PYG{p}{)}\PYG{p}{,} \PYG{n+nb}{id}\PYG{p}{(}\PYG{n}{o}\PYG{p}{)}\PYG{p}{,} \PYG{n}{z}
\PYG{g+go}{(28293632, 28293632, array(True)) \PYGZsh{} may vary}
\end{sphinxVerbatim}

\end{fulllineitems}

\index{amax() (in module symjax.tensor)@\spxentry{amax()}\spxextra{in module symjax.tensor}}

\begin{fulllineitems}
\phantomsection\label{\detokenize{modules/tensor:symjax.tensor.amax}}\pysiglinewithargsret{\sphinxbfcode{\sphinxupquote{amax}}}{\emph{\DUrole{n}{a}}, \emph{\DUrole{n}{axis}\DUrole{o}{=}\DUrole{default_value}{None}}, \emph{\DUrole{n}{dtype}\DUrole{o}{=}\DUrole{default_value}{None}}, \emph{\DUrole{n}{out}\DUrole{o}{=}\DUrole{default_value}{None}}, \emph{\DUrole{n}{keepdims}\DUrole{o}{=}\DUrole{default_value}{False}}}{}
Return the maximum of an array or maximum along an axis.

LAX\sphinxhyphen{}backend implementation of {\hyperref[\detokenize{modules/tensor:symjax.tensor.amax}]{\sphinxcrossref{\sphinxcode{\sphinxupquote{amax()}}}}}.
ADDITIONOriginal docstring below.

LAX\sphinxhyphen{}backend implementation of {\hyperref[\detokenize{modules/tensor:symjax.tensor.amax}]{\sphinxcrossref{\sphinxcode{\sphinxupquote{amax()}}}}}.
Original docstring below.
\begin{quote}\begin{description}
\item[{Returns}] \leavevmode
\sphinxstylestrong{amax} \textendash{} Maximum of \sphinxtitleref{a}. If \sphinxtitleref{axis} is None, the result is a scalar value.
If \sphinxtitleref{axis} is given, the result is an array of dimension
\sphinxcode{\sphinxupquote{a.ndim \sphinxhyphen{} 1}}.

\item[{Return type}] \leavevmode
ndarray or scalar

\end{description}\end{quote}

\sphinxstrong{See also:}

\begin{description}
\item[{{\hyperref[\detokenize{modules/tensor:symjax.tensor.amin}]{\sphinxcrossref{\sphinxcode{\sphinxupquote{amin()}}}}}}] \leavevmode
The minimum value of an array along a given axis, propagating any NaNs.

\item[{{\hyperref[\detokenize{modules/tensor:symjax.tensor.nanmax}]{\sphinxcrossref{\sphinxcode{\sphinxupquote{nanmax()}}}}}}] \leavevmode
The maximum value of an array along a given axis, ignoring any NaNs.

\item[{{\hyperref[\detokenize{modules/tensor:symjax.tensor.maximum}]{\sphinxcrossref{\sphinxcode{\sphinxupquote{maximum()}}}}}}] \leavevmode
Element\sphinxhyphen{}wise maximum of two arrays, propagating any NaNs.

\item[{\sphinxcode{\sphinxupquote{fmax()}}}] \leavevmode
Element\sphinxhyphen{}wise maximum of two arrays, ignoring any NaNs.

\item[{{\hyperref[\detokenize{modules/tensor:symjax.tensor.argmax}]{\sphinxcrossref{\sphinxcode{\sphinxupquote{argmax()}}}}}}] \leavevmode
Return the indices of the maximum values.

\end{description}

{\hyperref[\detokenize{modules/tensor:symjax.tensor.nanmin}]{\sphinxcrossref{\sphinxcode{\sphinxupquote{nanmin()}}}}}, {\hyperref[\detokenize{modules/tensor:symjax.tensor.minimum}]{\sphinxcrossref{\sphinxcode{\sphinxupquote{minimum()}}}}}, \sphinxcode{\sphinxupquote{fmin()}}

\subsubsection*{Notes}

NaN values are propagated, that is if at least one item is NaN, the
corresponding max value will be NaN as well. To ignore NaN values
(MATLAB behavior), please use nanmax.

Don’t use \sphinxtitleref{amax} for element\sphinxhyphen{}wise comparison of 2 arrays; when
\sphinxcode{\sphinxupquote{a.shape{[}0{]}}} is 2, \sphinxcode{\sphinxupquote{maximum(a{[}0{]}, a{[}1{]})}} is faster than
\sphinxcode{\sphinxupquote{amax(a, axis=0)}}.
\subsubsection*{Examples}

\begin{sphinxVerbatim}[commandchars=\\\{\}]
\PYG{g+gp}{\PYGZgt{}\PYGZgt{}\PYGZgt{} }\PYG{n}{a} \PYG{o}{=} \PYG{n}{np}\PYG{o}{.}\PYG{n}{arange}\PYG{p}{(}\PYG{l+m+mi}{4}\PYG{p}{)}\PYG{o}{.}\PYG{n}{reshape}\PYG{p}{(}\PYG{p}{(}\PYG{l+m+mi}{2}\PYG{p}{,}\PYG{l+m+mi}{2}\PYG{p}{)}\PYG{p}{)}
\PYG{g+gp}{\PYGZgt{}\PYGZgt{}\PYGZgt{} }\PYG{n}{a}
\PYG{g+go}{array([[0, 1],}
\PYG{g+go}{       [2, 3]])}
\PYG{g+gp}{\PYGZgt{}\PYGZgt{}\PYGZgt{} }\PYG{n}{np}\PYG{o}{.}\PYG{n}{amax}\PYG{p}{(}\PYG{n}{a}\PYG{p}{)}           \PYG{c+c1}{\PYGZsh{} Maximum of the flattened array}
\PYG{g+go}{3}
\PYG{g+gp}{\PYGZgt{}\PYGZgt{}\PYGZgt{} }\PYG{n}{np}\PYG{o}{.}\PYG{n}{amax}\PYG{p}{(}\PYG{n}{a}\PYG{p}{,} \PYG{n}{axis}\PYG{o}{=}\PYG{l+m+mi}{0}\PYG{p}{)}   \PYG{c+c1}{\PYGZsh{} Maxima along the first axis}
\PYG{g+go}{array([2, 3])}
\PYG{g+gp}{\PYGZgt{}\PYGZgt{}\PYGZgt{} }\PYG{n}{np}\PYG{o}{.}\PYG{n}{amax}\PYG{p}{(}\PYG{n}{a}\PYG{p}{,} \PYG{n}{axis}\PYG{o}{=}\PYG{l+m+mi}{1}\PYG{p}{)}   \PYG{c+c1}{\PYGZsh{} Maxima along the second axis}
\PYG{g+go}{array([1, 3])}
\PYG{g+gp}{\PYGZgt{}\PYGZgt{}\PYGZgt{} }\PYG{n}{np}\PYG{o}{.}\PYG{n}{amax}\PYG{p}{(}\PYG{n}{a}\PYG{p}{,} \PYG{n}{where}\PYG{o}{=}\PYG{p}{[}\PYG{k+kc}{False}\PYG{p}{,} \PYG{k+kc}{True}\PYG{p}{]}\PYG{p}{,} \PYG{n}{initial}\PYG{o}{=}\PYG{o}{\PYGZhy{}}\PYG{l+m+mi}{1}\PYG{p}{,} \PYG{n}{axis}\PYG{o}{=}\PYG{l+m+mi}{0}\PYG{p}{)}
\PYG{g+go}{array([\PYGZhy{}1,  3])}
\PYG{g+gp}{\PYGZgt{}\PYGZgt{}\PYGZgt{} }\PYG{n}{b} \PYG{o}{=} \PYG{n}{np}\PYG{o}{.}\PYG{n}{arange}\PYG{p}{(}\PYG{l+m+mi}{5}\PYG{p}{,} \PYG{n}{dtype}\PYG{o}{=}\PYG{n+nb}{float}\PYG{p}{)}
\PYG{g+gp}{\PYGZgt{}\PYGZgt{}\PYGZgt{} }\PYG{n}{b}\PYG{p}{[}\PYG{l+m+mi}{2}\PYG{p}{]} \PYG{o}{=} \PYG{n}{np}\PYG{o}{.}\PYG{n}{NaN}
\PYG{g+gp}{\PYGZgt{}\PYGZgt{}\PYGZgt{} }\PYG{n}{np}\PYG{o}{.}\PYG{n}{amax}\PYG{p}{(}\PYG{n}{b}\PYG{p}{)}
\PYG{g+go}{nan}
\PYG{g+gp}{\PYGZgt{}\PYGZgt{}\PYGZgt{} }\PYG{n}{np}\PYG{o}{.}\PYG{n}{amax}\PYG{p}{(}\PYG{n}{b}\PYG{p}{,} \PYG{n}{where}\PYG{o}{=}\PYG{o}{\PYGZti{}}\PYG{n}{np}\PYG{o}{.}\PYG{n}{isnan}\PYG{p}{(}\PYG{n}{b}\PYG{p}{)}\PYG{p}{,} \PYG{n}{initial}\PYG{o}{=}\PYG{o}{\PYGZhy{}}\PYG{l+m+mi}{1}\PYG{p}{)}
\PYG{g+go}{4.0}
\PYG{g+gp}{\PYGZgt{}\PYGZgt{}\PYGZgt{} }\PYG{n}{np}\PYG{o}{.}\PYG{n}{nanmax}\PYG{p}{(}\PYG{n}{b}\PYG{p}{)}
\PYG{g+go}{4.0}
\end{sphinxVerbatim}

You can use an initial value to compute the maximum of an empty slice, or
to initialize it to a different value:

\begin{sphinxVerbatim}[commandchars=\\\{\}]
\PYG{g+gp}{\PYGZgt{}\PYGZgt{}\PYGZgt{} }\PYG{n}{np}\PYG{o}{.}\PYG{n}{max}\PYG{p}{(}\PYG{p}{[}\PYG{p}{[}\PYG{o}{\PYGZhy{}}\PYG{l+m+mi}{50}\PYG{p}{]}\PYG{p}{,} \PYG{p}{[}\PYG{l+m+mi}{10}\PYG{p}{]}\PYG{p}{]}\PYG{p}{,} \PYG{n}{axis}\PYG{o}{=}\PYG{o}{\PYGZhy{}}\PYG{l+m+mi}{1}\PYG{p}{,} \PYG{n}{initial}\PYG{o}{=}\PYG{l+m+mi}{0}\PYG{p}{)}
\PYG{g+go}{array([ 0, 10])}
\end{sphinxVerbatim}

Notice that the initial value is used as one of the elements for which the
maximum is determined, unlike for the default argument Python’s max
function, which is only used for empty iterables.

\begin{sphinxVerbatim}[commandchars=\\\{\}]
\PYG{g+gp}{\PYGZgt{}\PYGZgt{}\PYGZgt{} }\PYG{n}{np}\PYG{o}{.}\PYG{n}{max}\PYG{p}{(}\PYG{p}{[}\PYG{l+m+mi}{5}\PYG{p}{]}\PYG{p}{,} \PYG{n}{initial}\PYG{o}{=}\PYG{l+m+mi}{6}\PYG{p}{)}
\PYG{g+go}{6}
\PYG{g+gp}{\PYGZgt{}\PYGZgt{}\PYGZgt{} }\PYG{n+nb}{max}\PYG{p}{(}\PYG{p}{[}\PYG{l+m+mi}{5}\PYG{p}{]}\PYG{p}{,} \PYG{n}{default}\PYG{o}{=}\PYG{l+m+mi}{6}\PYG{p}{)}
\PYG{g+go}{5}
\end{sphinxVerbatim}

\end{fulllineitems}

\index{amin() (in module symjax.tensor)@\spxentry{amin()}\spxextra{in module symjax.tensor}}

\begin{fulllineitems}
\phantomsection\label{\detokenize{modules/tensor:symjax.tensor.amin}}\pysiglinewithargsret{\sphinxbfcode{\sphinxupquote{amin}}}{\emph{\DUrole{n}{a}}, \emph{\DUrole{n}{axis}\DUrole{o}{=}\DUrole{default_value}{None}}, \emph{\DUrole{n}{dtype}\DUrole{o}{=}\DUrole{default_value}{None}}, \emph{\DUrole{n}{out}\DUrole{o}{=}\DUrole{default_value}{None}}, \emph{\DUrole{n}{keepdims}\DUrole{o}{=}\DUrole{default_value}{False}}}{}
Return the minimum of an array or minimum along an axis.

LAX\sphinxhyphen{}backend implementation of {\hyperref[\detokenize{modules/tensor:symjax.tensor.amin}]{\sphinxcrossref{\sphinxcode{\sphinxupquote{amin()}}}}}.
ADDITIONOriginal docstring below.

LAX\sphinxhyphen{}backend implementation of {\hyperref[\detokenize{modules/tensor:symjax.tensor.amin}]{\sphinxcrossref{\sphinxcode{\sphinxupquote{amin()}}}}}.
Original docstring below.
\begin{quote}\begin{description}
\item[{Returns}] \leavevmode
\sphinxstylestrong{amin} \textendash{} Minimum of \sphinxtitleref{a}. If \sphinxtitleref{axis} is None, the result is a scalar value.
If \sphinxtitleref{axis} is given, the result is an array of dimension
\sphinxcode{\sphinxupquote{a.ndim \sphinxhyphen{} 1}}.

\item[{Return type}] \leavevmode
ndarray or scalar

\end{description}\end{quote}

\sphinxstrong{See also:}

\begin{description}
\item[{{\hyperref[\detokenize{modules/tensor:symjax.tensor.amax}]{\sphinxcrossref{\sphinxcode{\sphinxupquote{amax()}}}}}}] \leavevmode
The maximum value of an array along a given axis, propagating any NaNs.

\item[{{\hyperref[\detokenize{modules/tensor:symjax.tensor.nanmin}]{\sphinxcrossref{\sphinxcode{\sphinxupquote{nanmin()}}}}}}] \leavevmode
The minimum value of an array along a given axis, ignoring any NaNs.

\item[{{\hyperref[\detokenize{modules/tensor:symjax.tensor.minimum}]{\sphinxcrossref{\sphinxcode{\sphinxupquote{minimum()}}}}}}] \leavevmode
Element\sphinxhyphen{}wise minimum of two arrays, propagating any NaNs.

\item[{\sphinxcode{\sphinxupquote{fmin()}}}] \leavevmode
Element\sphinxhyphen{}wise minimum of two arrays, ignoring any NaNs.

\item[{{\hyperref[\detokenize{modules/tensor:symjax.tensor.argmin}]{\sphinxcrossref{\sphinxcode{\sphinxupquote{argmin()}}}}}}] \leavevmode
Return the indices of the minimum values.

\end{description}

{\hyperref[\detokenize{modules/tensor:symjax.tensor.nanmax}]{\sphinxcrossref{\sphinxcode{\sphinxupquote{nanmax()}}}}}, {\hyperref[\detokenize{modules/tensor:symjax.tensor.maximum}]{\sphinxcrossref{\sphinxcode{\sphinxupquote{maximum()}}}}}, \sphinxcode{\sphinxupquote{fmax()}}

\subsubsection*{Notes}

NaN values are propagated, that is if at least one item is NaN, the
corresponding min value will be NaN as well. To ignore NaN values
(MATLAB behavior), please use nanmin.

Don’t use \sphinxtitleref{amin} for element\sphinxhyphen{}wise comparison of 2 arrays; when
\sphinxcode{\sphinxupquote{a.shape{[}0{]}}} is 2, \sphinxcode{\sphinxupquote{minimum(a{[}0{]}, a{[}1{]})}} is faster than
\sphinxcode{\sphinxupquote{amin(a, axis=0)}}.
\subsubsection*{Examples}

\begin{sphinxVerbatim}[commandchars=\\\{\}]
\PYG{g+gp}{\PYGZgt{}\PYGZgt{}\PYGZgt{} }\PYG{n}{a} \PYG{o}{=} \PYG{n}{np}\PYG{o}{.}\PYG{n}{arange}\PYG{p}{(}\PYG{l+m+mi}{4}\PYG{p}{)}\PYG{o}{.}\PYG{n}{reshape}\PYG{p}{(}\PYG{p}{(}\PYG{l+m+mi}{2}\PYG{p}{,}\PYG{l+m+mi}{2}\PYG{p}{)}\PYG{p}{)}
\PYG{g+gp}{\PYGZgt{}\PYGZgt{}\PYGZgt{} }\PYG{n}{a}
\PYG{g+go}{array([[0, 1],}
\PYG{g+go}{       [2, 3]])}
\PYG{g+gp}{\PYGZgt{}\PYGZgt{}\PYGZgt{} }\PYG{n}{np}\PYG{o}{.}\PYG{n}{amin}\PYG{p}{(}\PYG{n}{a}\PYG{p}{)}           \PYG{c+c1}{\PYGZsh{} Minimum of the flattened array}
\PYG{g+go}{0}
\PYG{g+gp}{\PYGZgt{}\PYGZgt{}\PYGZgt{} }\PYG{n}{np}\PYG{o}{.}\PYG{n}{amin}\PYG{p}{(}\PYG{n}{a}\PYG{p}{,} \PYG{n}{axis}\PYG{o}{=}\PYG{l+m+mi}{0}\PYG{p}{)}   \PYG{c+c1}{\PYGZsh{} Minima along the first axis}
\PYG{g+go}{array([0, 1])}
\PYG{g+gp}{\PYGZgt{}\PYGZgt{}\PYGZgt{} }\PYG{n}{np}\PYG{o}{.}\PYG{n}{amin}\PYG{p}{(}\PYG{n}{a}\PYG{p}{,} \PYG{n}{axis}\PYG{o}{=}\PYG{l+m+mi}{1}\PYG{p}{)}   \PYG{c+c1}{\PYGZsh{} Minima along the second axis}
\PYG{g+go}{array([0, 2])}
\PYG{g+gp}{\PYGZgt{}\PYGZgt{}\PYGZgt{} }\PYG{n}{np}\PYG{o}{.}\PYG{n}{amin}\PYG{p}{(}\PYG{n}{a}\PYG{p}{,} \PYG{n}{where}\PYG{o}{=}\PYG{p}{[}\PYG{k+kc}{False}\PYG{p}{,} \PYG{k+kc}{True}\PYG{p}{]}\PYG{p}{,} \PYG{n}{initial}\PYG{o}{=}\PYG{l+m+mi}{10}\PYG{p}{,} \PYG{n}{axis}\PYG{o}{=}\PYG{l+m+mi}{0}\PYG{p}{)}
\PYG{g+go}{array([10,  1])}
\end{sphinxVerbatim}

\begin{sphinxVerbatim}[commandchars=\\\{\}]
\PYG{g+gp}{\PYGZgt{}\PYGZgt{}\PYGZgt{} }\PYG{n}{b} \PYG{o}{=} \PYG{n}{np}\PYG{o}{.}\PYG{n}{arange}\PYG{p}{(}\PYG{l+m+mi}{5}\PYG{p}{,} \PYG{n}{dtype}\PYG{o}{=}\PYG{n+nb}{float}\PYG{p}{)}
\PYG{g+gp}{\PYGZgt{}\PYGZgt{}\PYGZgt{} }\PYG{n}{b}\PYG{p}{[}\PYG{l+m+mi}{2}\PYG{p}{]} \PYG{o}{=} \PYG{n}{np}\PYG{o}{.}\PYG{n}{NaN}
\PYG{g+gp}{\PYGZgt{}\PYGZgt{}\PYGZgt{} }\PYG{n}{np}\PYG{o}{.}\PYG{n}{amin}\PYG{p}{(}\PYG{n}{b}\PYG{p}{)}
\PYG{g+go}{nan}
\PYG{g+gp}{\PYGZgt{}\PYGZgt{}\PYGZgt{} }\PYG{n}{np}\PYG{o}{.}\PYG{n}{amin}\PYG{p}{(}\PYG{n}{b}\PYG{p}{,} \PYG{n}{where}\PYG{o}{=}\PYG{o}{\PYGZti{}}\PYG{n}{np}\PYG{o}{.}\PYG{n}{isnan}\PYG{p}{(}\PYG{n}{b}\PYG{p}{)}\PYG{p}{,} \PYG{n}{initial}\PYG{o}{=}\PYG{l+m+mi}{10}\PYG{p}{)}
\PYG{g+go}{0.0}
\PYG{g+gp}{\PYGZgt{}\PYGZgt{}\PYGZgt{} }\PYG{n}{np}\PYG{o}{.}\PYG{n}{nanmin}\PYG{p}{(}\PYG{n}{b}\PYG{p}{)}
\PYG{g+go}{0.0}
\end{sphinxVerbatim}

\begin{sphinxVerbatim}[commandchars=\\\{\}]
\PYG{g+gp}{\PYGZgt{}\PYGZgt{}\PYGZgt{} }\PYG{n}{np}\PYG{o}{.}\PYG{n}{min}\PYG{p}{(}\PYG{p}{[}\PYG{p}{[}\PYG{o}{\PYGZhy{}}\PYG{l+m+mi}{50}\PYG{p}{]}\PYG{p}{,} \PYG{p}{[}\PYG{l+m+mi}{10}\PYG{p}{]}\PYG{p}{]}\PYG{p}{,} \PYG{n}{axis}\PYG{o}{=}\PYG{o}{\PYGZhy{}}\PYG{l+m+mi}{1}\PYG{p}{,} \PYG{n}{initial}\PYG{o}{=}\PYG{l+m+mi}{0}\PYG{p}{)}
\PYG{g+go}{array([\PYGZhy{}50,   0])}
\end{sphinxVerbatim}

Notice that the initial value is used as one of the elements for which the
minimum is determined, unlike for the default argument Python’s max
function, which is only used for empty iterables.

Notice that this isn’t the same as Python’s \sphinxcode{\sphinxupquote{default}} argument.

\begin{sphinxVerbatim}[commandchars=\\\{\}]
\PYG{g+gp}{\PYGZgt{}\PYGZgt{}\PYGZgt{} }\PYG{n}{np}\PYG{o}{.}\PYG{n}{min}\PYG{p}{(}\PYG{p}{[}\PYG{l+m+mi}{6}\PYG{p}{]}\PYG{p}{,} \PYG{n}{initial}\PYG{o}{=}\PYG{l+m+mi}{5}\PYG{p}{)}
\PYG{g+go}{5}
\PYG{g+gp}{\PYGZgt{}\PYGZgt{}\PYGZgt{} }\PYG{n+nb}{min}\PYG{p}{(}\PYG{p}{[}\PYG{l+m+mi}{6}\PYG{p}{]}\PYG{p}{,} \PYG{n}{default}\PYG{o}{=}\PYG{l+m+mi}{5}\PYG{p}{)}
\PYG{g+go}{6}
\end{sphinxVerbatim}

\end{fulllineitems}

\index{angle() (in module symjax.tensor)@\spxentry{angle()}\spxextra{in module symjax.tensor}}

\begin{fulllineitems}
\phantomsection\label{\detokenize{modules/tensor:symjax.tensor.angle}}\pysiglinewithargsret{\sphinxbfcode{\sphinxupquote{angle}}}{\emph{\DUrole{n}{z}}}{}
Return the angle of the complex argument.

LAX\sphinxhyphen{}backend implementation of {\hyperref[\detokenize{modules/tensor:symjax.tensor.angle}]{\sphinxcrossref{\sphinxcode{\sphinxupquote{angle()}}}}}.
ADDITIONOriginal docstring below.

LAX\sphinxhyphen{}backend implementation of {\hyperref[\detokenize{modules/tensor:symjax.tensor.angle}]{\sphinxcrossref{\sphinxcode{\sphinxupquote{angle()}}}}}.
Original docstring below.
\begin{quote}\begin{description}
\item[{Returns}] \leavevmode

\sphinxstylestrong{angle} \textendash{} The counterclockwise angle from the positive real axis on the complex
plane in the range \sphinxcode{\sphinxupquote{(\sphinxhyphen{}pi, pi{]}}}, with dtype as numpy.float64.
\begin{description}
\item[{..versionchanged:: 1.16.0}] \leavevmode
This function works on subclasses of ndarray like \sphinxtitleref{ma.array}.

\end{description}

\item[{Return type}] \leavevmode
ndarray or scalar

\end{description}\end{quote}

\sphinxstrong{See also:}

{\hyperref[\detokenize{modules/tensor:symjax.tensor.arctan2}]{\sphinxcrossref{\sphinxcode{\sphinxupquote{arctan2()}}}}}, {\hyperref[\detokenize{modules/tensor:symjax.tensor.absolute}]{\sphinxcrossref{\sphinxcode{\sphinxupquote{absolute()}}}}}

\subsubsection*{Examples}

\begin{sphinxVerbatim}[commandchars=\\\{\}]
\PYG{g+gp}{\PYGZgt{}\PYGZgt{}\PYGZgt{} }\PYG{n}{np}\PYG{o}{.}\PYG{n}{angle}\PYG{p}{(}\PYG{p}{[}\PYG{l+m+mf}{1.0}\PYG{p}{,} \PYG{l+m+mf}{1.0}\PYG{n}{j}\PYG{p}{,} \PYG{l+m+mi}{1}\PYG{o}{+}\PYG{l+m+mi}{1}\PYG{n}{j}\PYG{p}{]}\PYG{p}{)}               \PYG{c+c1}{\PYGZsh{} in radians}
\PYG{g+go}{array([ 0.        ,  1.57079633,  0.78539816]) \PYGZsh{} may vary}
\PYG{g+gp}{\PYGZgt{}\PYGZgt{}\PYGZgt{} }\PYG{n}{np}\PYG{o}{.}\PYG{n}{angle}\PYG{p}{(}\PYG{l+m+mi}{1}\PYG{o}{+}\PYG{l+m+mi}{1}\PYG{n}{j}\PYG{p}{,} \PYG{n}{deg}\PYG{o}{=}\PYG{k+kc}{True}\PYG{p}{)}                  \PYG{c+c1}{\PYGZsh{} in degrees}
\PYG{g+go}{45.0}
\end{sphinxVerbatim}

\end{fulllineitems}

\index{any() (in module symjax.tensor)@\spxentry{any()}\spxextra{in module symjax.tensor}}

\begin{fulllineitems}
\phantomsection\label{\detokenize{modules/tensor:symjax.tensor.any}}\pysiglinewithargsret{\sphinxbfcode{\sphinxupquote{any}}}{\emph{\DUrole{n}{a}}, \emph{\DUrole{n}{axis}\DUrole{o}{=}\DUrole{default_value}{None}}, \emph{\DUrole{n}{dtype}\DUrole{o}{=}\DUrole{default_value}{None}}, \emph{\DUrole{n}{out}\DUrole{o}{=}\DUrole{default_value}{None}}, \emph{\DUrole{n}{keepdims}\DUrole{o}{=}\DUrole{default_value}{False}}}{}
Test whether any array element along a given axis evaluates to True.

LAX\sphinxhyphen{}backend implementation of {\hyperref[\detokenize{modules/tensor:symjax.tensor.any}]{\sphinxcrossref{\sphinxcode{\sphinxupquote{any()}}}}}.
ADDITIONOriginal docstring below.

LAX\sphinxhyphen{}backend implementation of {\hyperref[\detokenize{modules/tensor:symjax.tensor.any}]{\sphinxcrossref{\sphinxcode{\sphinxupquote{any()}}}}}.
Original docstring below.

Returns single boolean unless \sphinxtitleref{axis} is not \sphinxcode{\sphinxupquote{None}}
\begin{quote}\begin{description}
\item[{Returns}] \leavevmode
\sphinxstylestrong{any} \textendash{} A new boolean or \sphinxtitleref{ndarray} is returned unless \sphinxtitleref{out} is specified,
in which case a reference to \sphinxtitleref{out} is returned.

\item[{Return type}] \leavevmode
bool or ndarray

\end{description}\end{quote}

\sphinxstrong{See also:}

\begin{description}
\item[{\sphinxcode{\sphinxupquote{ndarray.any()}}}] \leavevmode
equivalent method

\item[{{\hyperref[\detokenize{modules/tensor:symjax.tensor.all}]{\sphinxcrossref{\sphinxcode{\sphinxupquote{all()}}}}}}] \leavevmode
Test whether all elements along a given axis evaluate to True.

\end{description}

\subsubsection*{Notes}

Not a Number (NaN), positive infinity and negative infinity evaluate
to \sphinxtitleref{True} because these are not equal to zero.
\subsubsection*{Examples}

\begin{sphinxVerbatim}[commandchars=\\\{\}]
\PYG{g+gp}{\PYGZgt{}\PYGZgt{}\PYGZgt{} }\PYG{n}{np}\PYG{o}{.}\PYG{n}{any}\PYG{p}{(}\PYG{p}{[}\PYG{p}{[}\PYG{k+kc}{True}\PYG{p}{,} \PYG{k+kc}{False}\PYG{p}{]}\PYG{p}{,} \PYG{p}{[}\PYG{k+kc}{True}\PYG{p}{,} \PYG{k+kc}{True}\PYG{p}{]}\PYG{p}{]}\PYG{p}{)}
\PYG{g+go}{True}
\end{sphinxVerbatim}

\begin{sphinxVerbatim}[commandchars=\\\{\}]
\PYG{g+gp}{\PYGZgt{}\PYGZgt{}\PYGZgt{} }\PYG{n}{np}\PYG{o}{.}\PYG{n}{any}\PYG{p}{(}\PYG{p}{[}\PYG{p}{[}\PYG{k+kc}{True}\PYG{p}{,} \PYG{k+kc}{False}\PYG{p}{]}\PYG{p}{,} \PYG{p}{[}\PYG{k+kc}{False}\PYG{p}{,} \PYG{k+kc}{False}\PYG{p}{]}\PYG{p}{]}\PYG{p}{,} \PYG{n}{axis}\PYG{o}{=}\PYG{l+m+mi}{0}\PYG{p}{)}
\PYG{g+go}{array([ True, False])}
\end{sphinxVerbatim}

\begin{sphinxVerbatim}[commandchars=\\\{\}]
\PYG{g+gp}{\PYGZgt{}\PYGZgt{}\PYGZgt{} }\PYG{n}{np}\PYG{o}{.}\PYG{n}{any}\PYG{p}{(}\PYG{p}{[}\PYG{o}{\PYGZhy{}}\PYG{l+m+mi}{1}\PYG{p}{,} \PYG{l+m+mi}{0}\PYG{p}{,} \PYG{l+m+mi}{5}\PYG{p}{]}\PYG{p}{)}
\PYG{g+go}{True}
\end{sphinxVerbatim}

\begin{sphinxVerbatim}[commandchars=\\\{\}]
\PYG{g+gp}{\PYGZgt{}\PYGZgt{}\PYGZgt{} }\PYG{n}{np}\PYG{o}{.}\PYG{n}{any}\PYG{p}{(}\PYG{n}{np}\PYG{o}{.}\PYG{n}{nan}\PYG{p}{)}
\PYG{g+go}{True}
\end{sphinxVerbatim}

\begin{sphinxVerbatim}[commandchars=\\\{\}]
\PYG{g+gp}{\PYGZgt{}\PYGZgt{}\PYGZgt{} }\PYG{n}{o}\PYG{o}{=}\PYG{n}{np}\PYG{o}{.}\PYG{n}{array}\PYG{p}{(}\PYG{k+kc}{False}\PYG{p}{)}
\PYG{g+gp}{\PYGZgt{}\PYGZgt{}\PYGZgt{} }\PYG{n}{z}\PYG{o}{=}\PYG{n}{np}\PYG{o}{.}\PYG{n}{any}\PYG{p}{(}\PYG{p}{[}\PYG{o}{\PYGZhy{}}\PYG{l+m+mi}{1}\PYG{p}{,} \PYG{l+m+mi}{4}\PYG{p}{,} \PYG{l+m+mi}{5}\PYG{p}{]}\PYG{p}{,} \PYG{n}{out}\PYG{o}{=}\PYG{n}{o}\PYG{p}{)}
\PYG{g+gp}{\PYGZgt{}\PYGZgt{}\PYGZgt{} }\PYG{n}{z}\PYG{p}{,} \PYG{n}{o}
\PYG{g+go}{(array(True), array(True))}
\PYG{g+gp}{\PYGZgt{}\PYGZgt{}\PYGZgt{} }\PYG{c+c1}{\PYGZsh{} Check now that z is a reference to o}
\PYG{g+gp}{\PYGZgt{}\PYGZgt{}\PYGZgt{} }\PYG{n}{z} \PYG{o+ow}{is} \PYG{n}{o}
\PYG{g+go}{True}
\PYG{g+gp}{\PYGZgt{}\PYGZgt{}\PYGZgt{} }\PYG{n+nb}{id}\PYG{p}{(}\PYG{n}{z}\PYG{p}{)}\PYG{p}{,} \PYG{n+nb}{id}\PYG{p}{(}\PYG{n}{o}\PYG{p}{)} \PYG{c+c1}{\PYGZsh{} identity of z and o              }
\PYG{g+go}{(191614240, 191614240)}
\end{sphinxVerbatim}

\end{fulllineitems}

\index{append() (in module symjax.tensor)@\spxentry{append()}\spxextra{in module symjax.tensor}}

\begin{fulllineitems}
\phantomsection\label{\detokenize{modules/tensor:symjax.tensor.append}}\pysiglinewithargsret{\sphinxbfcode{\sphinxupquote{append}}}{\emph{\DUrole{n}{arr}}, \emph{\DUrole{n}{values}}, \emph{\DUrole{n}{axis}\DUrole{o}{=}\DUrole{default_value}{None}}}{}
Append values to the end of an array.

LAX\sphinxhyphen{}backend implementation of {\hyperref[\detokenize{modules/tensor:symjax.tensor.append}]{\sphinxcrossref{\sphinxcode{\sphinxupquote{append()}}}}}.
ADDITIONOriginal docstring below.

LAX\sphinxhyphen{}backend implementation of {\hyperref[\detokenize{modules/tensor:symjax.tensor.append}]{\sphinxcrossref{\sphinxcode{\sphinxupquote{append()}}}}}.
Original docstring below.
\begin{quote}\begin{description}
\item[{Returns}] \leavevmode
\sphinxstylestrong{append} \textendash{} A copy of \sphinxtitleref{arr} with \sphinxtitleref{values} appended to \sphinxtitleref{axis}.  Note that
\sphinxtitleref{append} does not occur in\sphinxhyphen{}place: a new array is allocated and
filled.  If \sphinxtitleref{axis} is None, \sphinxtitleref{out} is a flattened array.

\item[{Return type}] \leavevmode
ndarray

\end{description}\end{quote}

\sphinxstrong{See also:}

\begin{description}
\item[{\sphinxcode{\sphinxupquote{insert()}}}] \leavevmode
Insert elements into an array.

\item[{\sphinxcode{\sphinxupquote{delete()}}}] \leavevmode
Delete elements from an array.

\end{description}

\subsubsection*{Examples}

\begin{sphinxVerbatim}[commandchars=\\\{\}]
\PYG{g+gp}{\PYGZgt{}\PYGZgt{}\PYGZgt{} }\PYG{n}{np}\PYG{o}{.}\PYG{n}{append}\PYG{p}{(}\PYG{p}{[}\PYG{l+m+mi}{1}\PYG{p}{,} \PYG{l+m+mi}{2}\PYG{p}{,} \PYG{l+m+mi}{3}\PYG{p}{]}\PYG{p}{,} \PYG{p}{[}\PYG{p}{[}\PYG{l+m+mi}{4}\PYG{p}{,} \PYG{l+m+mi}{5}\PYG{p}{,} \PYG{l+m+mi}{6}\PYG{p}{]}\PYG{p}{,} \PYG{p}{[}\PYG{l+m+mi}{7}\PYG{p}{,} \PYG{l+m+mi}{8}\PYG{p}{,} \PYG{l+m+mi}{9}\PYG{p}{]}\PYG{p}{]}\PYG{p}{)}
\PYG{g+go}{array([1, 2, 3, ..., 7, 8, 9])}
\end{sphinxVerbatim}

When \sphinxtitleref{axis} is specified, \sphinxtitleref{values} must have the correct shape.

\begin{sphinxVerbatim}[commandchars=\\\{\}]
\PYG{g+gp}{\PYGZgt{}\PYGZgt{}\PYGZgt{} }\PYG{n}{np}\PYG{o}{.}\PYG{n}{append}\PYG{p}{(}\PYG{p}{[}\PYG{p}{[}\PYG{l+m+mi}{1}\PYG{p}{,} \PYG{l+m+mi}{2}\PYG{p}{,} \PYG{l+m+mi}{3}\PYG{p}{]}\PYG{p}{,} \PYG{p}{[}\PYG{l+m+mi}{4}\PYG{p}{,} \PYG{l+m+mi}{5}\PYG{p}{,} \PYG{l+m+mi}{6}\PYG{p}{]}\PYG{p}{]}\PYG{p}{,} \PYG{p}{[}\PYG{p}{[}\PYG{l+m+mi}{7}\PYG{p}{,} \PYG{l+m+mi}{8}\PYG{p}{,} \PYG{l+m+mi}{9}\PYG{p}{]}\PYG{p}{]}\PYG{p}{,} \PYG{n}{axis}\PYG{o}{=}\PYG{l+m+mi}{0}\PYG{p}{)}
\PYG{g+go}{array([[1, 2, 3],}
\PYG{g+go}{       [4, 5, 6],}
\PYG{g+go}{       [7, 8, 9]])}
\PYG{g+gp}{\PYGZgt{}\PYGZgt{}\PYGZgt{} }\PYG{n}{np}\PYG{o}{.}\PYG{n}{append}\PYG{p}{(}\PYG{p}{[}\PYG{p}{[}\PYG{l+m+mi}{1}\PYG{p}{,} \PYG{l+m+mi}{2}\PYG{p}{,} \PYG{l+m+mi}{3}\PYG{p}{]}\PYG{p}{,} \PYG{p}{[}\PYG{l+m+mi}{4}\PYG{p}{,} \PYG{l+m+mi}{5}\PYG{p}{,} \PYG{l+m+mi}{6}\PYG{p}{]}\PYG{p}{]}\PYG{p}{,} \PYG{p}{[}\PYG{l+m+mi}{7}\PYG{p}{,} \PYG{l+m+mi}{8}\PYG{p}{,} \PYG{l+m+mi}{9}\PYG{p}{]}\PYG{p}{,} \PYG{n}{axis}\PYG{o}{=}\PYG{l+m+mi}{0}\PYG{p}{)}
\PYG{g+gt}{Traceback (most recent call last):}
    \PYG{o}{.}\PYG{o}{.}\PYG{o}{.}
\PYG{g+gr}{ValueError}: \PYG{n}{all the input arrays must have same number of dimensions}
\end{sphinxVerbatim}

\end{fulllineitems}

\index{arange() (in module symjax.tensor)@\spxentry{arange()}\spxextra{in module symjax.tensor}}

\begin{fulllineitems}
\phantomsection\label{\detokenize{modules/tensor:symjax.tensor.arange}}\pysiglinewithargsret{\sphinxbfcode{\sphinxupquote{arange}}}{\emph{\DUrole{n}{start}}, \emph{\DUrole{n}{stop}\DUrole{o}{=}\DUrole{default_value}{None}}, \emph{\DUrole{n}{step}\DUrole{o}{=}\DUrole{default_value}{None}}, \emph{\DUrole{n}{dtype}\DUrole{o}{=}\DUrole{default_value}{None}}}{}
Return evenly spaced values within a given interval.

LAX\sphinxhyphen{}backend implementation of {\hyperref[\detokenize{modules/tensor:symjax.tensor.arange}]{\sphinxcrossref{\sphinxcode{\sphinxupquote{arange()}}}}}.
ADDITIONOriginal docstring below.

LAX\sphinxhyphen{}backend implementation of {\hyperref[\detokenize{modules/tensor:symjax.tensor.arange}]{\sphinxcrossref{\sphinxcode{\sphinxupquote{arange()}}}}}.
Original docstring below.

arange({[}start,{]} stop{[}, step,{]}, dtype=None)
\begin{quote}
\begin{quote}

Values are generated within the half\sphinxhyphen{}open interval \sphinxcode{\sphinxupquote{{[}start, stop)}}
(in other words, the interval including \sphinxtitleref{start} but excluding \sphinxtitleref{stop}).
For integer arguments the function is equivalent to the Python built\sphinxhyphen{}in
\sphinxtitleref{range} function, but returns an ndarray rather than a list.

When using a non\sphinxhyphen{}integer step, such as 0.1, the results will often not
be consistent.  It is better to use \sphinxtitleref{numpy.linspace} for these cases.
\end{quote}
\begin{description}
\item[{Returns}] \leavevmode\begin{description}
\item[{arange}] \leavevmode{[}ndarray{]}
Array of evenly spaced values.

For floating point arguments, the length of the result is
\sphinxcode{\sphinxupquote{ceil((stop \sphinxhyphen{} start)/step)}}.  Because of floating point overflow,
this rule may result in the last element of \sphinxtitleref{out} being greater
than \sphinxtitleref{stop}.

\end{description}

numpy.linspace : Evenly spaced numbers with careful handling of endpoints.
numpy.ogrid: Arrays of evenly spaced numbers in N\sphinxhyphen{}dimensions.
numpy.mgrid: Grid\sphinxhyphen{}shaped arrays of evenly spaced numbers in N\sphinxhyphen{}dimensions.

\begin{sphinxVerbatim}[commandchars=\\\{\}]
\PYG{g+gp}{\PYGZgt{}\PYGZgt{}\PYGZgt{} }\PYG{n}{np}\PYG{o}{.}\PYG{n}{arange}\PYG{p}{(}\PYG{l+m+mi}{3}\PYG{p}{)}
\PYG{g+go}{array([0, 1, 2])}
\PYG{g+gp}{\PYGZgt{}\PYGZgt{}\PYGZgt{} }\PYG{n}{np}\PYG{o}{.}\PYG{n}{arange}\PYG{p}{(}\PYG{l+m+mf}{3.0}\PYG{p}{)}
\PYG{g+go}{array([ 0.,  1.,  2.])}
\PYG{g+gp}{\PYGZgt{}\PYGZgt{}\PYGZgt{} }\PYG{n}{np}\PYG{o}{.}\PYG{n}{arange}\PYG{p}{(}\PYG{l+m+mi}{3}\PYG{p}{,}\PYG{l+m+mi}{7}\PYG{p}{)}
\PYG{g+go}{array([3, 4, 5, 6])}
\PYG{g+gp}{\PYGZgt{}\PYGZgt{}\PYGZgt{} }\PYG{n}{np}\PYG{o}{.}\PYG{n}{arange}\PYG{p}{(}\PYG{l+m+mi}{3}\PYG{p}{,}\PYG{l+m+mi}{7}\PYG{p}{,}\PYG{l+m+mi}{2}\PYG{p}{)}
\PYG{g+go}{array([3, 5])}
\end{sphinxVerbatim}

\end{description}
\end{quote}

\end{fulllineitems}

\index{arccos() (in module symjax.tensor)@\spxentry{arccos()}\spxextra{in module symjax.tensor}}

\begin{fulllineitems}
\phantomsection\label{\detokenize{modules/tensor:symjax.tensor.arccos}}\pysiglinewithargsret{\sphinxbfcode{\sphinxupquote{arccos}}}{\emph{\DUrole{n}{x}}}{}
Trigonometric inverse cosine, element\sphinxhyphen{}wise.

LAX\sphinxhyphen{}backend implementation of {\hyperref[\detokenize{modules/tensor:symjax.tensor.arccos}]{\sphinxcrossref{\sphinxcode{\sphinxupquote{arccos()}}}}}.
ADDITIONOriginal docstring below.

LAX\sphinxhyphen{}backend implementation of {\hyperref[\detokenize{modules/tensor:symjax.tensor.arccos}]{\sphinxcrossref{\sphinxcode{\sphinxupquote{arccos()}}}}}.
Original docstring below.

arccos(x, /, out=None, {\color{red}\bfseries{}*}, where=True, casting=’same\_kind’, order=’K’, dtype=None, subok=True{[}, signature, extobj{]})

The inverse of \sphinxtitleref{cos} so that, if \sphinxcode{\sphinxupquote{y = cos(x)}}, then \sphinxcode{\sphinxupquote{x = arccos(y)}}.
\begin{quote}\begin{description}
\item[{Returns}] \leavevmode
\sphinxstylestrong{angle} \textendash{} The angle of the ray intersecting the unit circle at the given
\sphinxtitleref{x}\sphinxhyphen{}coordinate in radians {[}0, pi{]}.
This is a scalar if \sphinxtitleref{x} is a scalar.

\item[{Return type}] \leavevmode
ndarray

\end{description}\end{quote}

\sphinxstrong{See also:}

{\hyperref[\detokenize{modules/tensor:symjax.tensor.cos}]{\sphinxcrossref{\sphinxcode{\sphinxupquote{cos()}}}}}, {\hyperref[\detokenize{modules/tensor:symjax.tensor.arctan}]{\sphinxcrossref{\sphinxcode{\sphinxupquote{arctan()}}}}}, {\hyperref[\detokenize{modules/tensor:symjax.tensor.arcsin}]{\sphinxcrossref{\sphinxcode{\sphinxupquote{arcsin()}}}}}, \sphinxcode{\sphinxupquote{emath.arccos()}}

\subsubsection*{Notes}

\sphinxtitleref{arccos} is a multivalued function: for each \sphinxtitleref{x} there are infinitely
many numbers \sphinxtitleref{z} such that \sphinxtitleref{cos(z) = x}. The convention is to return
the angle \sphinxtitleref{z} whose real part lies in \sphinxtitleref{{[}0, pi{]}}.

For real\sphinxhyphen{}valued input data types, \sphinxtitleref{arccos} always returns real output.
For each value that cannot be expressed as a real number or infinity,
it yields \sphinxcode{\sphinxupquote{nan}} and sets the \sphinxtitleref{invalid} floating point error flag.

For complex\sphinxhyphen{}valued input, \sphinxtitleref{arccos} is a complex analytic function that
has branch cuts \sphinxtitleref{{[}\sphinxhyphen{}inf, \sphinxhyphen{}1{]}} and \sphinxtitleref{{[}1, inf{]}} and is continuous from
above on the former and from below on the latter.

The inverse \sphinxtitleref{cos} is also known as \sphinxtitleref{acos} or cos\textasciicircum{}\sphinxhyphen{}1.
\subsubsection*{References}

M. Abramowitz and I.A. Stegun, “Handbook of Mathematical Functions”,
10th printing, 1964, pp. 79. \sphinxurl{http://www.math.sfu.ca/~cbm/aands/}
\subsubsection*{Examples}

We expect the arccos of 1 to be 0, and of \sphinxhyphen{}1 to be pi:

\begin{sphinxVerbatim}[commandchars=\\\{\}]
\PYG{g+gp}{\PYGZgt{}\PYGZgt{}\PYGZgt{} }\PYG{n}{np}\PYG{o}{.}\PYG{n}{arccos}\PYG{p}{(}\PYG{p}{[}\PYG{l+m+mi}{1}\PYG{p}{,} \PYG{o}{\PYGZhy{}}\PYG{l+m+mi}{1}\PYG{p}{]}\PYG{p}{)}
\PYG{g+go}{array([ 0.        ,  3.14159265])}
\end{sphinxVerbatim}

Plot arccos:

\begin{sphinxVerbatim}[commandchars=\\\{\}]
\PYG{g+gp}{\PYGZgt{}\PYGZgt{}\PYGZgt{} }\PYG{k+kn}{import} \PYG{n+nn}{matplotlib}\PYG{n+nn}{.}\PYG{n+nn}{pyplot} \PYG{k}{as} \PYG{n+nn}{plt}
\PYG{g+gp}{\PYGZgt{}\PYGZgt{}\PYGZgt{} }\PYG{n}{x} \PYG{o}{=} \PYG{n}{np}\PYG{o}{.}\PYG{n}{linspace}\PYG{p}{(}\PYG{o}{\PYGZhy{}}\PYG{l+m+mi}{1}\PYG{p}{,} \PYG{l+m+mi}{1}\PYG{p}{,} \PYG{n}{num}\PYG{o}{=}\PYG{l+m+mi}{100}\PYG{p}{)}
\PYG{g+gp}{\PYGZgt{}\PYGZgt{}\PYGZgt{} }\PYG{n}{plt}\PYG{o}{.}\PYG{n}{plot}\PYG{p}{(}\PYG{n}{x}\PYG{p}{,} \PYG{n}{np}\PYG{o}{.}\PYG{n}{arccos}\PYG{p}{(}\PYG{n}{x}\PYG{p}{)}\PYG{p}{)}
\PYG{g+gp}{\PYGZgt{}\PYGZgt{}\PYGZgt{} }\PYG{n}{plt}\PYG{o}{.}\PYG{n}{axis}\PYG{p}{(}\PYG{l+s+s1}{\PYGZsq{}}\PYG{l+s+s1}{tight}\PYG{l+s+s1}{\PYGZsq{}}\PYG{p}{)}
\PYG{g+gp}{\PYGZgt{}\PYGZgt{}\PYGZgt{} }\PYG{n}{plt}\PYG{o}{.}\PYG{n}{show}\PYG{p}{(}\PYG{p}{)}
\end{sphinxVerbatim}

\end{fulllineitems}

\index{arccosh() (in module symjax.tensor)@\spxentry{arccosh()}\spxextra{in module symjax.tensor}}

\begin{fulllineitems}
\phantomsection\label{\detokenize{modules/tensor:symjax.tensor.arccosh}}\pysiglinewithargsret{\sphinxbfcode{\sphinxupquote{arccosh}}}{\emph{\DUrole{n}{x}}}{}
Inverse hyperbolic cosine, element\sphinxhyphen{}wise.

LAX\sphinxhyphen{}backend implementation of {\hyperref[\detokenize{modules/tensor:symjax.tensor.arccosh}]{\sphinxcrossref{\sphinxcode{\sphinxupquote{arccosh()}}}}}.
ADDITIONOriginal docstring below.

LAX\sphinxhyphen{}backend implementation of {\hyperref[\detokenize{modules/tensor:symjax.tensor.arccosh}]{\sphinxcrossref{\sphinxcode{\sphinxupquote{arccosh()}}}}}.
Original docstring below.

arccosh(x, /, out=None, {\color{red}\bfseries{}*}, where=True, casting=’same\_kind’, order=’K’, dtype=None, subok=True{[}, signature, extobj{]})
\begin{quote}\begin{description}
\item[{Returns}] \leavevmode
\sphinxstylestrong{arccosh} \textendash{} Array of the same shape as \sphinxtitleref{x}.
This is a scalar if \sphinxtitleref{x} is a scalar.

\item[{Return type}] \leavevmode
ndarray

\end{description}\end{quote}

\sphinxstrong{See also:}

{\hyperref[\detokenize{modules/tensor:symjax.tensor.cosh}]{\sphinxcrossref{\sphinxcode{\sphinxupquote{cosh()}}}}}, {\hyperref[\detokenize{modules/tensor:symjax.tensor.arcsinh}]{\sphinxcrossref{\sphinxcode{\sphinxupquote{arcsinh()}}}}}, {\hyperref[\detokenize{modules/tensor:symjax.tensor.sinh}]{\sphinxcrossref{\sphinxcode{\sphinxupquote{sinh()}}}}}, {\hyperref[\detokenize{modules/tensor:symjax.tensor.arctanh}]{\sphinxcrossref{\sphinxcode{\sphinxupquote{arctanh()}}}}}, {\hyperref[\detokenize{modules/tensor:symjax.tensor.tanh}]{\sphinxcrossref{\sphinxcode{\sphinxupquote{tanh()}}}}}

\subsubsection*{Notes}

\sphinxtitleref{arccosh} is a multivalued function: for each \sphinxtitleref{x} there are infinitely
many numbers \sphinxtitleref{z} such that \sphinxtitleref{cosh(z) = x}. The convention is to return the
\sphinxtitleref{z} whose imaginary part lies in \sphinxtitleref{{[}\sphinxhyphen{}pi, pi{]}} and the real part in
\sphinxcode{\sphinxupquote{{[}0, inf{]}}}.

For real\sphinxhyphen{}valued input data types, \sphinxtitleref{arccosh} always returns real output.
For each value that cannot be expressed as a real number or infinity, it
yields \sphinxcode{\sphinxupquote{nan}} and sets the \sphinxtitleref{invalid} floating point error flag.

For complex\sphinxhyphen{}valued input, \sphinxtitleref{arccosh} is a complex analytical function that
has a branch cut \sphinxtitleref{{[}\sphinxhyphen{}inf, 1{]}} and is continuous from above on it.
\subsubsection*{References}
\subsubsection*{Examples}

\begin{sphinxVerbatim}[commandchars=\\\{\}]
\PYG{g+gp}{\PYGZgt{}\PYGZgt{}\PYGZgt{} }\PYG{n}{np}\PYG{o}{.}\PYG{n}{arccosh}\PYG{p}{(}\PYG{p}{[}\PYG{n}{np}\PYG{o}{.}\PYG{n}{e}\PYG{p}{,} \PYG{l+m+mf}{10.0}\PYG{p}{]}\PYG{p}{)}
\PYG{g+go}{array([ 1.65745445,  2.99322285])}
\PYG{g+gp}{\PYGZgt{}\PYGZgt{}\PYGZgt{} }\PYG{n}{np}\PYG{o}{.}\PYG{n}{arccosh}\PYG{p}{(}\PYG{l+m+mi}{1}\PYG{p}{)}
\PYG{g+go}{0.0}
\end{sphinxVerbatim}

\end{fulllineitems}

\index{arcsin() (in module symjax.tensor)@\spxentry{arcsin()}\spxextra{in module symjax.tensor}}

\begin{fulllineitems}
\phantomsection\label{\detokenize{modules/tensor:symjax.tensor.arcsin}}\pysiglinewithargsret{\sphinxbfcode{\sphinxupquote{arcsin}}}{\emph{\DUrole{n}{x}}}{}
Inverse sine, element\sphinxhyphen{}wise.

LAX\sphinxhyphen{}backend implementation of {\hyperref[\detokenize{modules/tensor:symjax.tensor.arcsin}]{\sphinxcrossref{\sphinxcode{\sphinxupquote{arcsin()}}}}}.
ADDITIONOriginal docstring below.

LAX\sphinxhyphen{}backend implementation of {\hyperref[\detokenize{modules/tensor:symjax.tensor.arcsin}]{\sphinxcrossref{\sphinxcode{\sphinxupquote{arcsin()}}}}}.
Original docstring below.

arcsin(x, /, out=None, {\color{red}\bfseries{}*}, where=True, casting=’same\_kind’, order=’K’, dtype=None, subok=True{[}, signature, extobj{]})
\begin{quote}\begin{description}
\item[{Returns}] \leavevmode
\sphinxstylestrong{angle} \textendash{} The inverse sine of each element in \sphinxtitleref{x}, in radians and in the
closed interval \sphinxcode{\sphinxupquote{{[}\sphinxhyphen{}pi/2, pi/2{]}}}.
This is a scalar if \sphinxtitleref{x} is a scalar.

\item[{Return type}] \leavevmode
ndarray

\end{description}\end{quote}

\sphinxstrong{See also:}

{\hyperref[\detokenize{modules/tensor:symjax.tensor.sin}]{\sphinxcrossref{\sphinxcode{\sphinxupquote{sin()}}}}}, {\hyperref[\detokenize{modules/tensor:symjax.tensor.cos}]{\sphinxcrossref{\sphinxcode{\sphinxupquote{cos()}}}}}, {\hyperref[\detokenize{modules/tensor:symjax.tensor.arccos}]{\sphinxcrossref{\sphinxcode{\sphinxupquote{arccos()}}}}}, {\hyperref[\detokenize{modules/tensor:symjax.tensor.tan}]{\sphinxcrossref{\sphinxcode{\sphinxupquote{tan()}}}}}, {\hyperref[\detokenize{modules/tensor:symjax.tensor.arctan}]{\sphinxcrossref{\sphinxcode{\sphinxupquote{arctan()}}}}}, {\hyperref[\detokenize{modules/tensor:symjax.tensor.arctan2}]{\sphinxcrossref{\sphinxcode{\sphinxupquote{arctan2()}}}}}, \sphinxcode{\sphinxupquote{emath.arcsin()}}

\subsubsection*{Notes}

\sphinxtitleref{arcsin} is a multivalued function: for each \sphinxtitleref{x} there are infinitely
many numbers \sphinxtitleref{z} such that \(sin(z) = x\).  The convention is to
return the angle \sphinxtitleref{z} whose real part lies in {[}\sphinxhyphen{}pi/2, pi/2{]}.

For real\sphinxhyphen{}valued input data types, \sphinxstyleemphasis{arcsin} always returns real output.
For each value that cannot be expressed as a real number or infinity,
it yields \sphinxcode{\sphinxupquote{nan}} and sets the \sphinxtitleref{invalid} floating point error flag.

For complex\sphinxhyphen{}valued input, \sphinxtitleref{arcsin} is a complex analytic function that
has, by convention, the branch cuts {[}\sphinxhyphen{}inf, \sphinxhyphen{}1{]} and {[}1, inf{]}  and is
continuous from above on the former and from below on the latter.

The inverse sine is also known as \sphinxtitleref{asin} or sin\textasciicircum{}\{\sphinxhyphen{}1\}.
\subsubsection*{References}

Abramowitz, M. and Stegun, I. A., \sphinxstyleemphasis{Handbook of Mathematical Functions},
10th printing, New York: Dover, 1964, pp. 79ff.
\sphinxurl{http://www.math.sfu.ca/~cbm/aands/}
\subsubsection*{Examples}

\begin{sphinxVerbatim}[commandchars=\\\{\}]
\PYG{g+gp}{\PYGZgt{}\PYGZgt{}\PYGZgt{} }\PYG{n}{np}\PYG{o}{.}\PYG{n}{arcsin}\PYG{p}{(}\PYG{l+m+mi}{1}\PYG{p}{)}     \PYG{c+c1}{\PYGZsh{} pi/2}
\PYG{g+go}{1.5707963267948966}
\PYG{g+gp}{\PYGZgt{}\PYGZgt{}\PYGZgt{} }\PYG{n}{np}\PYG{o}{.}\PYG{n}{arcsin}\PYG{p}{(}\PYG{o}{\PYGZhy{}}\PYG{l+m+mi}{1}\PYG{p}{)}    \PYG{c+c1}{\PYGZsh{} \PYGZhy{}pi/2}
\PYG{g+go}{\PYGZhy{}1.5707963267948966}
\PYG{g+gp}{\PYGZgt{}\PYGZgt{}\PYGZgt{} }\PYG{n}{np}\PYG{o}{.}\PYG{n}{arcsin}\PYG{p}{(}\PYG{l+m+mi}{0}\PYG{p}{)}
\PYG{g+go}{0.0}
\end{sphinxVerbatim}

\end{fulllineitems}

\index{arcsinh() (in module symjax.tensor)@\spxentry{arcsinh()}\spxextra{in module symjax.tensor}}

\begin{fulllineitems}
\phantomsection\label{\detokenize{modules/tensor:symjax.tensor.arcsinh}}\pysiglinewithargsret{\sphinxbfcode{\sphinxupquote{arcsinh}}}{\emph{\DUrole{n}{x}}}{}
Inverse hyperbolic sine element\sphinxhyphen{}wise.

LAX\sphinxhyphen{}backend implementation of {\hyperref[\detokenize{modules/tensor:symjax.tensor.arcsinh}]{\sphinxcrossref{\sphinxcode{\sphinxupquote{arcsinh()}}}}}.
ADDITIONOriginal docstring below.

LAX\sphinxhyphen{}backend implementation of {\hyperref[\detokenize{modules/tensor:symjax.tensor.arcsinh}]{\sphinxcrossref{\sphinxcode{\sphinxupquote{arcsinh()}}}}}.
Original docstring below.

arcsinh(x, /, out=None, {\color{red}\bfseries{}*}, where=True, casting=’same\_kind’, order=’K’, dtype=None, subok=True{[}, signature, extobj{]})
\begin{quote}\begin{description}
\item[{Returns}] \leavevmode
\sphinxstylestrong{out} \textendash{} Array of the same shape as \sphinxtitleref{x}.
This is a scalar if \sphinxtitleref{x} is a scalar.

\item[{Return type}] \leavevmode
ndarray or scalar

\end{description}\end{quote}
\subsubsection*{Notes}

\sphinxtitleref{arcsinh} is a multivalued function: for each \sphinxtitleref{x} there are infinitely
many numbers \sphinxtitleref{z} such that \sphinxtitleref{sinh(z) = x}. The convention is to return the
\sphinxtitleref{z} whose imaginary part lies in \sphinxtitleref{{[}\sphinxhyphen{}pi/2, pi/2{]}}.

For real\sphinxhyphen{}valued input data types, \sphinxtitleref{arcsinh} always returns real output.
For each value that cannot be expressed as a real number or infinity, it
returns \sphinxcode{\sphinxupquote{nan}} and sets the \sphinxtitleref{invalid} floating point error flag.

For complex\sphinxhyphen{}valued input, \sphinxtitleref{arccos} is a complex analytical function that
has branch cuts \sphinxtitleref{{[}1j, infj{]}} and \sphinxtitleref{{[}\sphinxhyphen{}1j, \sphinxhyphen{}infj{]}} and is continuous from
the right on the former and from the left on the latter.

The inverse hyperbolic sine is also known as \sphinxtitleref{asinh} or \sphinxcode{\sphinxupquote{sinh\textasciicircum{}\sphinxhyphen{}1}}.
\subsubsection*{References}
\subsubsection*{Examples}

\begin{sphinxVerbatim}[commandchars=\\\{\}]
\PYG{g+gp}{\PYGZgt{}\PYGZgt{}\PYGZgt{} }\PYG{n}{np}\PYG{o}{.}\PYG{n}{arcsinh}\PYG{p}{(}\PYG{n}{np}\PYG{o}{.}\PYG{n}{array}\PYG{p}{(}\PYG{p}{[}\PYG{n}{np}\PYG{o}{.}\PYG{n}{e}\PYG{p}{,} \PYG{l+m+mf}{10.0}\PYG{p}{]}\PYG{p}{)}\PYG{p}{)}
\PYG{g+go}{array([ 1.72538256,  2.99822295])}
\end{sphinxVerbatim}

\end{fulllineitems}

\index{arctan() (in module symjax.tensor)@\spxentry{arctan()}\spxextra{in module symjax.tensor}}

\begin{fulllineitems}
\phantomsection\label{\detokenize{modules/tensor:symjax.tensor.arctan}}\pysiglinewithargsret{\sphinxbfcode{\sphinxupquote{arctan}}}{\emph{\DUrole{n}{x}}}{}
Trigonometric inverse tangent, element\sphinxhyphen{}wise.

LAX\sphinxhyphen{}backend implementation of {\hyperref[\detokenize{modules/tensor:symjax.tensor.arctan}]{\sphinxcrossref{\sphinxcode{\sphinxupquote{arctan()}}}}}.
ADDITIONOriginal docstring below.

LAX\sphinxhyphen{}backend implementation of {\hyperref[\detokenize{modules/tensor:symjax.tensor.arctan}]{\sphinxcrossref{\sphinxcode{\sphinxupquote{arctan()}}}}}.
Original docstring below.

arctan(x, /, out=None, {\color{red}\bfseries{}*}, where=True, casting=’same\_kind’, order=’K’, dtype=None, subok=True{[}, signature, extobj{]})

The inverse of tan, so that if \sphinxcode{\sphinxupquote{y = tan(x)}} then \sphinxcode{\sphinxupquote{x = arctan(y)}}.
\begin{quote}\begin{description}
\item[{Returns}] \leavevmode
\sphinxstylestrong{out} \textendash{} Out has the same shape as \sphinxtitleref{x}.  Its real part is in
\sphinxcode{\sphinxupquote{{[}\sphinxhyphen{}pi/2, pi/2{]}}} (\sphinxcode{\sphinxupquote{arctan(+/\sphinxhyphen{}inf)}} returns \sphinxcode{\sphinxupquote{+/\sphinxhyphen{}pi/2}}).
This is a scalar if \sphinxtitleref{x} is a scalar.

\item[{Return type}] \leavevmode
ndarray or scalar

\end{description}\end{quote}

\sphinxstrong{See also:}

\begin{description}
\item[{{\hyperref[\detokenize{modules/tensor:symjax.tensor.arctan2}]{\sphinxcrossref{\sphinxcode{\sphinxupquote{arctan2()}}}}}}] \leavevmode
The “four quadrant” arctan of the angle formed by (\sphinxtitleref{x}, \sphinxtitleref{y}) and the positive \sphinxtitleref{x}\sphinxhyphen{}axis.

\item[{{\hyperref[\detokenize{modules/tensor:symjax.tensor.angle}]{\sphinxcrossref{\sphinxcode{\sphinxupquote{angle()}}}}}}] \leavevmode
Argument of complex values.

\end{description}

\subsubsection*{Notes}

\sphinxtitleref{arctan} is a multi\sphinxhyphen{}valued function: for each \sphinxtitleref{x} there are infinitely
many numbers \sphinxtitleref{z} such that tan(\sphinxtitleref{z}) = \sphinxtitleref{x}.  The convention is to return
the angle \sphinxtitleref{z} whose real part lies in {[}\sphinxhyphen{}pi/2, pi/2{]}.

For real\sphinxhyphen{}valued input data types, \sphinxtitleref{arctan} always returns real output.
For each value that cannot be expressed as a real number or infinity,
it yields \sphinxcode{\sphinxupquote{nan}} and sets the \sphinxtitleref{invalid} floating point error flag.

For complex\sphinxhyphen{}valued input, \sphinxtitleref{arctan} is a complex analytic function that
has {[}\sphinxtitleref{1j, infj}{]} and {[}\sphinxtitleref{\sphinxhyphen{}1j, \sphinxhyphen{}infj}{]} as branch cuts, and is continuous
from the left on the former and from the right on the latter.

The inverse tangent is also known as \sphinxtitleref{atan} or tan\textasciicircum{}\{\sphinxhyphen{}1\}.
\subsubsection*{References}

Abramowitz, M. and Stegun, I. A., \sphinxstyleemphasis{Handbook of Mathematical Functions},
10th printing, New York: Dover, 1964, pp. 79.
\sphinxurl{http://www.math.sfu.ca/~cbm/aands/}
\subsubsection*{Examples}

We expect the arctan of 0 to be 0, and of 1 to be pi/4:

\begin{sphinxVerbatim}[commandchars=\\\{\}]
\PYG{g+gp}{\PYGZgt{}\PYGZgt{}\PYGZgt{} }\PYG{n}{np}\PYG{o}{.}\PYG{n}{arctan}\PYG{p}{(}\PYG{p}{[}\PYG{l+m+mi}{0}\PYG{p}{,} \PYG{l+m+mi}{1}\PYG{p}{]}\PYG{p}{)}
\PYG{g+go}{array([ 0.        ,  0.78539816])}
\end{sphinxVerbatim}

\begin{sphinxVerbatim}[commandchars=\\\{\}]
\PYG{g+gp}{\PYGZgt{}\PYGZgt{}\PYGZgt{} }\PYG{n}{np}\PYG{o}{.}\PYG{n}{pi}\PYG{o}{/}\PYG{l+m+mi}{4}
\PYG{g+go}{0.78539816339744828}
\end{sphinxVerbatim}

Plot arctan:

\begin{sphinxVerbatim}[commandchars=\\\{\}]
\PYG{g+gp}{\PYGZgt{}\PYGZgt{}\PYGZgt{} }\PYG{k+kn}{import} \PYG{n+nn}{matplotlib}\PYG{n+nn}{.}\PYG{n+nn}{pyplot} \PYG{k}{as} \PYG{n+nn}{plt}
\PYG{g+gp}{\PYGZgt{}\PYGZgt{}\PYGZgt{} }\PYG{n}{x} \PYG{o}{=} \PYG{n}{np}\PYG{o}{.}\PYG{n}{linspace}\PYG{p}{(}\PYG{o}{\PYGZhy{}}\PYG{l+m+mi}{10}\PYG{p}{,} \PYG{l+m+mi}{10}\PYG{p}{)}
\PYG{g+gp}{\PYGZgt{}\PYGZgt{}\PYGZgt{} }\PYG{n}{plt}\PYG{o}{.}\PYG{n}{plot}\PYG{p}{(}\PYG{n}{x}\PYG{p}{,} \PYG{n}{np}\PYG{o}{.}\PYG{n}{arctan}\PYG{p}{(}\PYG{n}{x}\PYG{p}{)}\PYG{p}{)}
\PYG{g+gp}{\PYGZgt{}\PYGZgt{}\PYGZgt{} }\PYG{n}{plt}\PYG{o}{.}\PYG{n}{axis}\PYG{p}{(}\PYG{l+s+s1}{\PYGZsq{}}\PYG{l+s+s1}{tight}\PYG{l+s+s1}{\PYGZsq{}}\PYG{p}{)}
\PYG{g+gp}{\PYGZgt{}\PYGZgt{}\PYGZgt{} }\PYG{n}{plt}\PYG{o}{.}\PYG{n}{show}\PYG{p}{(}\PYG{p}{)}
\end{sphinxVerbatim}

\end{fulllineitems}

\index{arctan2() (in module symjax.tensor)@\spxentry{arctan2()}\spxextra{in module symjax.tensor}}

\begin{fulllineitems}
\phantomsection\label{\detokenize{modules/tensor:symjax.tensor.arctan2}}\pysiglinewithargsret{\sphinxbfcode{\sphinxupquote{arctan2}}}{\emph{\DUrole{n}{x1}}, \emph{\DUrole{n}{x2}}}{}
Element\sphinxhyphen{}wise arc tangent of \sphinxcode{\sphinxupquote{x1/x2}} choosing the quadrant correctly.

LAX\sphinxhyphen{}backend implementation of {\hyperref[\detokenize{modules/tensor:symjax.tensor.arctan2}]{\sphinxcrossref{\sphinxcode{\sphinxupquote{arctan2()}}}}}.
ADDITIONOriginal docstring below.

LAX\sphinxhyphen{}backend implementation of {\hyperref[\detokenize{modules/tensor:symjax.tensor.arctan2}]{\sphinxcrossref{\sphinxcode{\sphinxupquote{arctan2()}}}}}.
Original docstring below.

arctan2(x1, x2, /, out=None, {\color{red}\bfseries{}*}, where=True, casting=’same\_kind’, order=’K’, dtype=None, subok=True{[}, signature, extobj{]})

The quadrant (i.e., branch) is chosen so that \sphinxcode{\sphinxupquote{arctan2(x1, x2)}} is
the signed angle in radians between the ray ending at the origin and
passing through the point (1,0), and the ray ending at the origin and
passing through the point (\sphinxtitleref{x2}, \sphinxtitleref{x1}).  (Note the role reversal: the
“\sphinxtitleref{y}\sphinxhyphen{}coordinate” is the first function parameter, the “\sphinxtitleref{x}\sphinxhyphen{}coordinate”
is the second.)  By IEEE convention, this function is defined for
\sphinxtitleref{x2} = +/\sphinxhyphen{}0 and for either or both of \sphinxtitleref{x1} and \sphinxtitleref{x2} = +/\sphinxhyphen{}inf (see
Notes for specific values).

This function is not defined for complex\sphinxhyphen{}valued arguments; for the
so\sphinxhyphen{}called argument of complex values, use \sphinxtitleref{angle}.
\begin{quote}\begin{description}
\item[{Returns}] \leavevmode
\sphinxstylestrong{angle} \textendash{} Array of angles in radians, in the range \sphinxcode{\sphinxupquote{{[}\sphinxhyphen{}pi, pi{]}}}.
This is a scalar if both \sphinxtitleref{x1} and \sphinxtitleref{x2} are scalars.

\item[{Return type}] \leavevmode
ndarray

\end{description}\end{quote}

\sphinxstrong{See also:}

{\hyperref[\detokenize{modules/tensor:symjax.tensor.arctan}]{\sphinxcrossref{\sphinxcode{\sphinxupquote{arctan()}}}}}, {\hyperref[\detokenize{modules/tensor:symjax.tensor.tan}]{\sphinxcrossref{\sphinxcode{\sphinxupquote{tan()}}}}}, {\hyperref[\detokenize{modules/tensor:symjax.tensor.angle}]{\sphinxcrossref{\sphinxcode{\sphinxupquote{angle()}}}}}

\subsubsection*{Notes}

\sphinxstyleemphasis{arctan2} is identical to the \sphinxtitleref{atan2} function of the underlying
C library.  The following special values are defined in the C
standard: {\color{red}\bfseries{}{[}1{]}\_}

\begin{savenotes}\sphinxattablestart
\centering
\begin{tabulary}{\linewidth}[t]{|T|T|T|}
\hline
\sphinxstyletheadfamily 
\sphinxtitleref{x1}
&\sphinxstyletheadfamily 
\sphinxtitleref{x2}
&\sphinxstyletheadfamily 
\sphinxtitleref{arctan2(x1,x2)}
\\
\hline
+/\sphinxhyphen{} 0
&
+0
&
+/\sphinxhyphen{} 0
\\
\hline
+/\sphinxhyphen{} 0
&
\sphinxhyphen{}0
&
+/\sphinxhyphen{} pi
\\
\hline
\textgreater{} 0
&
+/\sphinxhyphen{}inf
&
+0 / +pi
\\
\hline
\textless{} 0
&
+/\sphinxhyphen{}inf
&
\sphinxhyphen{}0 / \sphinxhyphen{}pi
\\
\hline
+/\sphinxhyphen{}inf
&
+inf
&
+/\sphinxhyphen{} (pi/4)
\\
\hline
+/\sphinxhyphen{}inf
&
\sphinxhyphen{}inf
&
+/\sphinxhyphen{} (3*pi/4)
\\
\hline
\end{tabulary}
\par
\sphinxattableend\end{savenotes}

Note that +0 and \sphinxhyphen{}0 are distinct floating point numbers, as are +inf
and \sphinxhyphen{}inf.
\subsubsection*{References}
\subsubsection*{Examples}

Consider four points in different quadrants:

\begin{sphinxVerbatim}[commandchars=\\\{\}]
\PYG{g+gp}{\PYGZgt{}\PYGZgt{}\PYGZgt{} }\PYG{n}{x} \PYG{o}{=} \PYG{n}{np}\PYG{o}{.}\PYG{n}{array}\PYG{p}{(}\PYG{p}{[}\PYG{o}{\PYGZhy{}}\PYG{l+m+mi}{1}\PYG{p}{,} \PYG{o}{+}\PYG{l+m+mi}{1}\PYG{p}{,} \PYG{o}{+}\PYG{l+m+mi}{1}\PYG{p}{,} \PYG{o}{\PYGZhy{}}\PYG{l+m+mi}{1}\PYG{p}{]}\PYG{p}{)}
\PYG{g+gp}{\PYGZgt{}\PYGZgt{}\PYGZgt{} }\PYG{n}{y} \PYG{o}{=} \PYG{n}{np}\PYG{o}{.}\PYG{n}{array}\PYG{p}{(}\PYG{p}{[}\PYG{o}{\PYGZhy{}}\PYG{l+m+mi}{1}\PYG{p}{,} \PYG{o}{\PYGZhy{}}\PYG{l+m+mi}{1}\PYG{p}{,} \PYG{o}{+}\PYG{l+m+mi}{1}\PYG{p}{,} \PYG{o}{+}\PYG{l+m+mi}{1}\PYG{p}{]}\PYG{p}{)}
\PYG{g+gp}{\PYGZgt{}\PYGZgt{}\PYGZgt{} }\PYG{n}{np}\PYG{o}{.}\PYG{n}{arctan2}\PYG{p}{(}\PYG{n}{y}\PYG{p}{,} \PYG{n}{x}\PYG{p}{)} \PYG{o}{*} \PYG{l+m+mi}{180} \PYG{o}{/} \PYG{n}{np}\PYG{o}{.}\PYG{n}{pi}
\PYG{g+go}{array([\PYGZhy{}135.,  \PYGZhy{}45.,   45.,  135.])}
\end{sphinxVerbatim}

Note the order of the parameters. \sphinxtitleref{arctan2} is defined also when \sphinxtitleref{x2} = 0
and at several other special points, obtaining values in
the range \sphinxcode{\sphinxupquote{{[}\sphinxhyphen{}pi, pi{]}}}:

\begin{sphinxVerbatim}[commandchars=\\\{\}]
\PYG{g+gp}{\PYGZgt{}\PYGZgt{}\PYGZgt{} }\PYG{n}{np}\PYG{o}{.}\PYG{n}{arctan2}\PYG{p}{(}\PYG{p}{[}\PYG{l+m+mf}{1.}\PYG{p}{,} \PYG{o}{\PYGZhy{}}\PYG{l+m+mf}{1.}\PYG{p}{]}\PYG{p}{,} \PYG{p}{[}\PYG{l+m+mf}{0.}\PYG{p}{,} \PYG{l+m+mf}{0.}\PYG{p}{]}\PYG{p}{)}
\PYG{g+go}{array([ 1.57079633, \PYGZhy{}1.57079633])}
\PYG{g+gp}{\PYGZgt{}\PYGZgt{}\PYGZgt{} }\PYG{n}{np}\PYG{o}{.}\PYG{n}{arctan2}\PYG{p}{(}\PYG{p}{[}\PYG{l+m+mf}{0.}\PYG{p}{,} \PYG{l+m+mf}{0.}\PYG{p}{,} \PYG{n}{np}\PYG{o}{.}\PYG{n}{inf}\PYG{p}{]}\PYG{p}{,} \PYG{p}{[}\PYG{o}{+}\PYG{l+m+mf}{0.}\PYG{p}{,} \PYG{o}{\PYGZhy{}}\PYG{l+m+mf}{0.}\PYG{p}{,} \PYG{n}{np}\PYG{o}{.}\PYG{n}{inf}\PYG{p}{]}\PYG{p}{)}
\PYG{g+go}{array([ 0.        ,  3.14159265,  0.78539816])}
\end{sphinxVerbatim}

\end{fulllineitems}

\index{arctanh() (in module symjax.tensor)@\spxentry{arctanh()}\spxextra{in module symjax.tensor}}

\begin{fulllineitems}
\phantomsection\label{\detokenize{modules/tensor:symjax.tensor.arctanh}}\pysiglinewithargsret{\sphinxbfcode{\sphinxupquote{arctanh}}}{\emph{\DUrole{n}{x}}}{}
Inverse hyperbolic tangent element\sphinxhyphen{}wise.

LAX\sphinxhyphen{}backend implementation of {\hyperref[\detokenize{modules/tensor:symjax.tensor.arctanh}]{\sphinxcrossref{\sphinxcode{\sphinxupquote{arctanh()}}}}}.
ADDITIONOriginal docstring below.

LAX\sphinxhyphen{}backend implementation of {\hyperref[\detokenize{modules/tensor:symjax.tensor.arctanh}]{\sphinxcrossref{\sphinxcode{\sphinxupquote{arctanh()}}}}}.
Original docstring below.

arctanh(x, /, out=None, {\color{red}\bfseries{}*}, where=True, casting=’same\_kind’, order=’K’, dtype=None, subok=True{[}, signature, extobj{]})
\begin{quote}\begin{description}
\item[{Returns}] \leavevmode
\sphinxstylestrong{out} \textendash{} Array of the same shape as \sphinxtitleref{x}.
This is a scalar if \sphinxtitleref{x} is a scalar.

\item[{Return type}] \leavevmode
ndarray or scalar

\end{description}\end{quote}

\sphinxstrong{See also:}

\sphinxcode{\sphinxupquote{emath.arctanh()}}

\subsubsection*{Notes}

\sphinxtitleref{arctanh} is a multivalued function: for each \sphinxtitleref{x} there are infinitely
many numbers \sphinxtitleref{z} such that \sphinxtitleref{tanh(z) = x}. The convention is to return
the \sphinxtitleref{z} whose imaginary part lies in \sphinxtitleref{{[}\sphinxhyphen{}pi/2, pi/2{]}}.

For real\sphinxhyphen{}valued input data types, \sphinxtitleref{arctanh} always returns real output.
For each value that cannot be expressed as a real number or infinity,
it yields \sphinxcode{\sphinxupquote{nan}} and sets the \sphinxtitleref{invalid} floating point error flag.

For complex\sphinxhyphen{}valued input, \sphinxtitleref{arctanh} is a complex analytical function
that has branch cuts \sphinxtitleref{{[}\sphinxhyphen{}1, \sphinxhyphen{}inf{]}} and \sphinxtitleref{{[}1, inf{]}} and is continuous from
above on the former and from below on the latter.

The inverse hyperbolic tangent is also known as \sphinxtitleref{atanh} or \sphinxcode{\sphinxupquote{tanh\textasciicircum{}\sphinxhyphen{}1}}.
\subsubsection*{References}
\subsubsection*{Examples}

\begin{sphinxVerbatim}[commandchars=\\\{\}]
\PYG{g+gp}{\PYGZgt{}\PYGZgt{}\PYGZgt{} }\PYG{n}{np}\PYG{o}{.}\PYG{n}{arctanh}\PYG{p}{(}\PYG{p}{[}\PYG{l+m+mi}{0}\PYG{p}{,} \PYG{o}{\PYGZhy{}}\PYG{l+m+mf}{0.5}\PYG{p}{]}\PYG{p}{)}
\PYG{g+go}{array([ 0.        , \PYGZhy{}0.54930614])}
\end{sphinxVerbatim}

\end{fulllineitems}

\index{argmax() (in module symjax.tensor)@\spxentry{argmax()}\spxextra{in module symjax.tensor}}

\begin{fulllineitems}
\phantomsection\label{\detokenize{modules/tensor:symjax.tensor.argmax}}\pysiglinewithargsret{\sphinxbfcode{\sphinxupquote{argmax}}}{\emph{\DUrole{n}{a}}, \emph{\DUrole{n}{axis}\DUrole{o}{=}\DUrole{default_value}{None}}}{}
Returns the indices of the maximum values along an axis.

LAX\sphinxhyphen{}backend implementation of {\hyperref[\detokenize{modules/tensor:symjax.tensor.argmax}]{\sphinxcrossref{\sphinxcode{\sphinxupquote{argmax()}}}}}.
ADDITIONOriginal docstring below.

LAX\sphinxhyphen{}backend implementation of {\hyperref[\detokenize{modules/tensor:symjax.tensor.argmax}]{\sphinxcrossref{\sphinxcode{\sphinxupquote{argmax()}}}}}.
Original docstring below.
\begin{quote}\begin{description}
\item[{Returns}] \leavevmode
\sphinxstylestrong{index\_array} \textendash{} Array of indices into the array. It has the same shape as \sphinxtitleref{a.shape}
with the dimension along \sphinxtitleref{axis} removed.

\item[{Return type}] \leavevmode
ndarray of ints

\end{description}\end{quote}

\sphinxstrong{See also:}

\sphinxcode{\sphinxupquote{ndarray.argmax()}}, {\hyperref[\detokenize{modules/tensor:symjax.tensor.argmin}]{\sphinxcrossref{\sphinxcode{\sphinxupquote{argmin()}}}}}
\begin{description}
\item[{{\hyperref[\detokenize{modules/tensor:symjax.tensor.amax}]{\sphinxcrossref{\sphinxcode{\sphinxupquote{amax()}}}}}}] \leavevmode
The maximum value along a given axis.

\item[{\sphinxcode{\sphinxupquote{unravel\_index()}}}] \leavevmode
Convert a flat index into an index tuple.

\item[{{\hyperref[\detokenize{modules/tensor:symjax.tensor.take_along_axis}]{\sphinxcrossref{\sphinxcode{\sphinxupquote{take\_along\_axis()}}}}}}] \leavevmode
Apply \sphinxcode{\sphinxupquote{np.expand\_dims(index\_array, axis)}} from argmax to an array as if by calling max.

\end{description}

\subsubsection*{Notes}

In case of multiple occurrences of the maximum values, the indices
corresponding to the first occurrence are returned.
\subsubsection*{Examples}

\begin{sphinxVerbatim}[commandchars=\\\{\}]
\PYG{g+gp}{\PYGZgt{}\PYGZgt{}\PYGZgt{} }\PYG{n}{a} \PYG{o}{=} \PYG{n}{np}\PYG{o}{.}\PYG{n}{arange}\PYG{p}{(}\PYG{l+m+mi}{6}\PYG{p}{)}\PYG{o}{.}\PYG{n}{reshape}\PYG{p}{(}\PYG{l+m+mi}{2}\PYG{p}{,}\PYG{l+m+mi}{3}\PYG{p}{)} \PYG{o}{+} \PYG{l+m+mi}{10}
\PYG{g+gp}{\PYGZgt{}\PYGZgt{}\PYGZgt{} }\PYG{n}{a}
\PYG{g+go}{array([[10, 11, 12],}
\PYG{g+go}{       [13, 14, 15]])}
\PYG{g+gp}{\PYGZgt{}\PYGZgt{}\PYGZgt{} }\PYG{n}{np}\PYG{o}{.}\PYG{n}{argmax}\PYG{p}{(}\PYG{n}{a}\PYG{p}{)}
\PYG{g+go}{5}
\PYG{g+gp}{\PYGZgt{}\PYGZgt{}\PYGZgt{} }\PYG{n}{np}\PYG{o}{.}\PYG{n}{argmax}\PYG{p}{(}\PYG{n}{a}\PYG{p}{,} \PYG{n}{axis}\PYG{o}{=}\PYG{l+m+mi}{0}\PYG{p}{)}
\PYG{g+go}{array([1, 1, 1])}
\PYG{g+gp}{\PYGZgt{}\PYGZgt{}\PYGZgt{} }\PYG{n}{np}\PYG{o}{.}\PYG{n}{argmax}\PYG{p}{(}\PYG{n}{a}\PYG{p}{,} \PYG{n}{axis}\PYG{o}{=}\PYG{l+m+mi}{1}\PYG{p}{)}
\PYG{g+go}{array([2, 2])}
\end{sphinxVerbatim}

Indexes of the maximal elements of a N\sphinxhyphen{}dimensional array:

\begin{sphinxVerbatim}[commandchars=\\\{\}]
\PYG{g+gp}{\PYGZgt{}\PYGZgt{}\PYGZgt{} }\PYG{n}{ind} \PYG{o}{=} \PYG{n}{np}\PYG{o}{.}\PYG{n}{unravel\PYGZus{}index}\PYG{p}{(}\PYG{n}{np}\PYG{o}{.}\PYG{n}{argmax}\PYG{p}{(}\PYG{n}{a}\PYG{p}{,} \PYG{n}{axis}\PYG{o}{=}\PYG{k+kc}{None}\PYG{p}{)}\PYG{p}{,} \PYG{n}{a}\PYG{o}{.}\PYG{n}{shape}\PYG{p}{)}
\PYG{g+gp}{\PYGZgt{}\PYGZgt{}\PYGZgt{} }\PYG{n}{ind}
\PYG{g+go}{(1, 2)}
\PYG{g+gp}{\PYGZgt{}\PYGZgt{}\PYGZgt{} }\PYG{n}{a}\PYG{p}{[}\PYG{n}{ind}\PYG{p}{]}
\PYG{g+go}{15}
\end{sphinxVerbatim}

\begin{sphinxVerbatim}[commandchars=\\\{\}]
\PYG{g+gp}{\PYGZgt{}\PYGZgt{}\PYGZgt{} }\PYG{n}{b} \PYG{o}{=} \PYG{n}{np}\PYG{o}{.}\PYG{n}{arange}\PYG{p}{(}\PYG{l+m+mi}{6}\PYG{p}{)}
\PYG{g+gp}{\PYGZgt{}\PYGZgt{}\PYGZgt{} }\PYG{n}{b}\PYG{p}{[}\PYG{l+m+mi}{1}\PYG{p}{]} \PYG{o}{=} \PYG{l+m+mi}{5}
\PYG{g+gp}{\PYGZgt{}\PYGZgt{}\PYGZgt{} }\PYG{n}{b}
\PYG{g+go}{array([0, 5, 2, 3, 4, 5])}
\PYG{g+gp}{\PYGZgt{}\PYGZgt{}\PYGZgt{} }\PYG{n}{np}\PYG{o}{.}\PYG{n}{argmax}\PYG{p}{(}\PYG{n}{b}\PYG{p}{)}  \PYG{c+c1}{\PYGZsh{} Only the first occurrence is returned.}
\PYG{g+go}{1}
\end{sphinxVerbatim}

\begin{sphinxVerbatim}[commandchars=\\\{\}]
\PYG{g+gp}{\PYGZgt{}\PYGZgt{}\PYGZgt{} }\PYG{n}{x} \PYG{o}{=} \PYG{n}{np}\PYG{o}{.}\PYG{n}{array}\PYG{p}{(}\PYG{p}{[}\PYG{p}{[}\PYG{l+m+mi}{4}\PYG{p}{,}\PYG{l+m+mi}{2}\PYG{p}{,}\PYG{l+m+mi}{3}\PYG{p}{]}\PYG{p}{,} \PYG{p}{[}\PYG{l+m+mi}{1}\PYG{p}{,}\PYG{l+m+mi}{0}\PYG{p}{,}\PYG{l+m+mi}{3}\PYG{p}{]}\PYG{p}{]}\PYG{p}{)}
\PYG{g+gp}{\PYGZgt{}\PYGZgt{}\PYGZgt{} }\PYG{n}{index\PYGZus{}array} \PYG{o}{=} \PYG{n}{np}\PYG{o}{.}\PYG{n}{argmax}\PYG{p}{(}\PYG{n}{x}\PYG{p}{,} \PYG{n}{axis}\PYG{o}{=}\PYG{o}{\PYGZhy{}}\PYG{l+m+mi}{1}\PYG{p}{)}
\PYG{g+gp}{\PYGZgt{}\PYGZgt{}\PYGZgt{} }\PYG{c+c1}{\PYGZsh{} Same as np.max(x, axis=\PYGZhy{}1, keepdims=True)}
\PYG{g+gp}{\PYGZgt{}\PYGZgt{}\PYGZgt{} }\PYG{n}{np}\PYG{o}{.}\PYG{n}{take\PYGZus{}along\PYGZus{}axis}\PYG{p}{(}\PYG{n}{x}\PYG{p}{,} \PYG{n}{np}\PYG{o}{.}\PYG{n}{expand\PYGZus{}dims}\PYG{p}{(}\PYG{n}{index\PYGZus{}array}\PYG{p}{,} \PYG{n}{axis}\PYG{o}{=}\PYG{o}{\PYGZhy{}}\PYG{l+m+mi}{1}\PYG{p}{)}\PYG{p}{,} \PYG{n}{axis}\PYG{o}{=}\PYG{o}{\PYGZhy{}}\PYG{l+m+mi}{1}\PYG{p}{)}
\PYG{g+go}{array([[4],}
\PYG{g+go}{       [3]])}
\PYG{g+gp}{\PYGZgt{}\PYGZgt{}\PYGZgt{} }\PYG{c+c1}{\PYGZsh{} Same as np.max(x, axis=\PYGZhy{}1)}
\PYG{g+gp}{\PYGZgt{}\PYGZgt{}\PYGZgt{} }\PYG{n}{np}\PYG{o}{.}\PYG{n}{take\PYGZus{}along\PYGZus{}axis}\PYG{p}{(}\PYG{n}{x}\PYG{p}{,} \PYG{n}{np}\PYG{o}{.}\PYG{n}{expand\PYGZus{}dims}\PYG{p}{(}\PYG{n}{index\PYGZus{}array}\PYG{p}{,} \PYG{n}{axis}\PYG{o}{=}\PYG{o}{\PYGZhy{}}\PYG{l+m+mi}{1}\PYG{p}{)}\PYG{p}{,} \PYG{n}{axis}\PYG{o}{=}\PYG{o}{\PYGZhy{}}\PYG{l+m+mi}{1}\PYG{p}{)}\PYG{o}{.}\PYG{n}{squeeze}\PYG{p}{(}\PYG{n}{axis}\PYG{o}{=}\PYG{o}{\PYGZhy{}}\PYG{l+m+mi}{1}\PYG{p}{)}
\PYG{g+go}{array([4, 3])}
\end{sphinxVerbatim}

\end{fulllineitems}

\index{argmin() (in module symjax.tensor)@\spxentry{argmin()}\spxextra{in module symjax.tensor}}

\begin{fulllineitems}
\phantomsection\label{\detokenize{modules/tensor:symjax.tensor.argmin}}\pysiglinewithargsret{\sphinxbfcode{\sphinxupquote{argmin}}}{\emph{\DUrole{n}{a}}, \emph{\DUrole{n}{axis}\DUrole{o}{=}\DUrole{default_value}{None}}}{}
Returns the indices of the minimum values along an axis.

LAX\sphinxhyphen{}backend implementation of {\hyperref[\detokenize{modules/tensor:symjax.tensor.argmin}]{\sphinxcrossref{\sphinxcode{\sphinxupquote{argmin()}}}}}.
ADDITIONOriginal docstring below.

LAX\sphinxhyphen{}backend implementation of {\hyperref[\detokenize{modules/tensor:symjax.tensor.argmin}]{\sphinxcrossref{\sphinxcode{\sphinxupquote{argmin()}}}}}.
Original docstring below.
\begin{quote}\begin{description}
\item[{Returns}] \leavevmode
\sphinxstylestrong{index\_array} \textendash{} Array of indices into the array. It has the same shape as \sphinxtitleref{a.shape}
with the dimension along \sphinxtitleref{axis} removed.

\item[{Return type}] \leavevmode
ndarray of ints

\end{description}\end{quote}

\sphinxstrong{See also:}

\sphinxcode{\sphinxupquote{ndarray.argmin()}}, {\hyperref[\detokenize{modules/tensor:symjax.tensor.argmax}]{\sphinxcrossref{\sphinxcode{\sphinxupquote{argmax()}}}}}
\begin{description}
\item[{{\hyperref[\detokenize{modules/tensor:symjax.tensor.amin}]{\sphinxcrossref{\sphinxcode{\sphinxupquote{amin()}}}}}}] \leavevmode
The minimum value along a given axis.

\item[{\sphinxcode{\sphinxupquote{unravel\_index()}}}] \leavevmode
Convert a flat index into an index tuple.

\item[{{\hyperref[\detokenize{modules/tensor:symjax.tensor.take_along_axis}]{\sphinxcrossref{\sphinxcode{\sphinxupquote{take\_along\_axis()}}}}}}] \leavevmode
Apply \sphinxcode{\sphinxupquote{np.expand\_dims(index\_array, axis)}} from argmin to an array as if by calling min.

\end{description}

\subsubsection*{Notes}

In case of multiple occurrences of the minimum values, the indices
corresponding to the first occurrence are returned.
\subsubsection*{Examples}

\begin{sphinxVerbatim}[commandchars=\\\{\}]
\PYG{g+gp}{\PYGZgt{}\PYGZgt{}\PYGZgt{} }\PYG{n}{a} \PYG{o}{=} \PYG{n}{np}\PYG{o}{.}\PYG{n}{arange}\PYG{p}{(}\PYG{l+m+mi}{6}\PYG{p}{)}\PYG{o}{.}\PYG{n}{reshape}\PYG{p}{(}\PYG{l+m+mi}{2}\PYG{p}{,}\PYG{l+m+mi}{3}\PYG{p}{)} \PYG{o}{+} \PYG{l+m+mi}{10}
\PYG{g+gp}{\PYGZgt{}\PYGZgt{}\PYGZgt{} }\PYG{n}{a}
\PYG{g+go}{array([[10, 11, 12],}
\PYG{g+go}{       [13, 14, 15]])}
\PYG{g+gp}{\PYGZgt{}\PYGZgt{}\PYGZgt{} }\PYG{n}{np}\PYG{o}{.}\PYG{n}{argmin}\PYG{p}{(}\PYG{n}{a}\PYG{p}{)}
\PYG{g+go}{0}
\PYG{g+gp}{\PYGZgt{}\PYGZgt{}\PYGZgt{} }\PYG{n}{np}\PYG{o}{.}\PYG{n}{argmin}\PYG{p}{(}\PYG{n}{a}\PYG{p}{,} \PYG{n}{axis}\PYG{o}{=}\PYG{l+m+mi}{0}\PYG{p}{)}
\PYG{g+go}{array([0, 0, 0])}
\PYG{g+gp}{\PYGZgt{}\PYGZgt{}\PYGZgt{} }\PYG{n}{np}\PYG{o}{.}\PYG{n}{argmin}\PYG{p}{(}\PYG{n}{a}\PYG{p}{,} \PYG{n}{axis}\PYG{o}{=}\PYG{l+m+mi}{1}\PYG{p}{)}
\PYG{g+go}{array([0, 0])}
\end{sphinxVerbatim}

Indices of the minimum elements of a N\sphinxhyphen{}dimensional array:

\begin{sphinxVerbatim}[commandchars=\\\{\}]
\PYG{g+gp}{\PYGZgt{}\PYGZgt{}\PYGZgt{} }\PYG{n}{ind} \PYG{o}{=} \PYG{n}{np}\PYG{o}{.}\PYG{n}{unravel\PYGZus{}index}\PYG{p}{(}\PYG{n}{np}\PYG{o}{.}\PYG{n}{argmin}\PYG{p}{(}\PYG{n}{a}\PYG{p}{,} \PYG{n}{axis}\PYG{o}{=}\PYG{k+kc}{None}\PYG{p}{)}\PYG{p}{,} \PYG{n}{a}\PYG{o}{.}\PYG{n}{shape}\PYG{p}{)}
\PYG{g+gp}{\PYGZgt{}\PYGZgt{}\PYGZgt{} }\PYG{n}{ind}
\PYG{g+go}{(0, 0)}
\PYG{g+gp}{\PYGZgt{}\PYGZgt{}\PYGZgt{} }\PYG{n}{a}\PYG{p}{[}\PYG{n}{ind}\PYG{p}{]}
\PYG{g+go}{10}
\end{sphinxVerbatim}

\begin{sphinxVerbatim}[commandchars=\\\{\}]
\PYG{g+gp}{\PYGZgt{}\PYGZgt{}\PYGZgt{} }\PYG{n}{b} \PYG{o}{=} \PYG{n}{np}\PYG{o}{.}\PYG{n}{arange}\PYG{p}{(}\PYG{l+m+mi}{6}\PYG{p}{)} \PYG{o}{+} \PYG{l+m+mi}{10}
\PYG{g+gp}{\PYGZgt{}\PYGZgt{}\PYGZgt{} }\PYG{n}{b}\PYG{p}{[}\PYG{l+m+mi}{4}\PYG{p}{]} \PYG{o}{=} \PYG{l+m+mi}{10}
\PYG{g+gp}{\PYGZgt{}\PYGZgt{}\PYGZgt{} }\PYG{n}{b}
\PYG{g+go}{array([10, 11, 12, 13, 10, 15])}
\PYG{g+gp}{\PYGZgt{}\PYGZgt{}\PYGZgt{} }\PYG{n}{np}\PYG{o}{.}\PYG{n}{argmin}\PYG{p}{(}\PYG{n}{b}\PYG{p}{)}  \PYG{c+c1}{\PYGZsh{} Only the first occurrence is returned.}
\PYG{g+go}{0}
\end{sphinxVerbatim}

\begin{sphinxVerbatim}[commandchars=\\\{\}]
\PYG{g+gp}{\PYGZgt{}\PYGZgt{}\PYGZgt{} }\PYG{n}{x} \PYG{o}{=} \PYG{n}{np}\PYG{o}{.}\PYG{n}{array}\PYG{p}{(}\PYG{p}{[}\PYG{p}{[}\PYG{l+m+mi}{4}\PYG{p}{,}\PYG{l+m+mi}{2}\PYG{p}{,}\PYG{l+m+mi}{3}\PYG{p}{]}\PYG{p}{,} \PYG{p}{[}\PYG{l+m+mi}{1}\PYG{p}{,}\PYG{l+m+mi}{0}\PYG{p}{,}\PYG{l+m+mi}{3}\PYG{p}{]}\PYG{p}{]}\PYG{p}{)}
\PYG{g+gp}{\PYGZgt{}\PYGZgt{}\PYGZgt{} }\PYG{n}{index\PYGZus{}array} \PYG{o}{=} \PYG{n}{np}\PYG{o}{.}\PYG{n}{argmin}\PYG{p}{(}\PYG{n}{x}\PYG{p}{,} \PYG{n}{axis}\PYG{o}{=}\PYG{o}{\PYGZhy{}}\PYG{l+m+mi}{1}\PYG{p}{)}
\PYG{g+gp}{\PYGZgt{}\PYGZgt{}\PYGZgt{} }\PYG{c+c1}{\PYGZsh{} Same as np.min(x, axis=\PYGZhy{}1, keepdims=True)}
\PYG{g+gp}{\PYGZgt{}\PYGZgt{}\PYGZgt{} }\PYG{n}{np}\PYG{o}{.}\PYG{n}{take\PYGZus{}along\PYGZus{}axis}\PYG{p}{(}\PYG{n}{x}\PYG{p}{,} \PYG{n}{np}\PYG{o}{.}\PYG{n}{expand\PYGZus{}dims}\PYG{p}{(}\PYG{n}{index\PYGZus{}array}\PYG{p}{,} \PYG{n}{axis}\PYG{o}{=}\PYG{o}{\PYGZhy{}}\PYG{l+m+mi}{1}\PYG{p}{)}\PYG{p}{,} \PYG{n}{axis}\PYG{o}{=}\PYG{o}{\PYGZhy{}}\PYG{l+m+mi}{1}\PYG{p}{)}
\PYG{g+go}{array([[2],}
\PYG{g+go}{       [0]])}
\PYG{g+gp}{\PYGZgt{}\PYGZgt{}\PYGZgt{} }\PYG{c+c1}{\PYGZsh{} Same as np.max(x, axis=\PYGZhy{}1)}
\PYG{g+gp}{\PYGZgt{}\PYGZgt{}\PYGZgt{} }\PYG{n}{np}\PYG{o}{.}\PYG{n}{take\PYGZus{}along\PYGZus{}axis}\PYG{p}{(}\PYG{n}{x}\PYG{p}{,} \PYG{n}{np}\PYG{o}{.}\PYG{n}{expand\PYGZus{}dims}\PYG{p}{(}\PYG{n}{index\PYGZus{}array}\PYG{p}{,} \PYG{n}{axis}\PYG{o}{=}\PYG{o}{\PYGZhy{}}\PYG{l+m+mi}{1}\PYG{p}{)}\PYG{p}{,} \PYG{n}{axis}\PYG{o}{=}\PYG{o}{\PYGZhy{}}\PYG{l+m+mi}{1}\PYG{p}{)}\PYG{o}{.}\PYG{n}{squeeze}\PYG{p}{(}\PYG{n}{axis}\PYG{o}{=}\PYG{o}{\PYGZhy{}}\PYG{l+m+mi}{1}\PYG{p}{)}
\PYG{g+go}{array([2, 0])}
\end{sphinxVerbatim}

\end{fulllineitems}

\index{argsort() (in module symjax.tensor)@\spxentry{argsort()}\spxextra{in module symjax.tensor}}

\begin{fulllineitems}
\phantomsection\label{\detokenize{modules/tensor:symjax.tensor.argsort}}\pysiglinewithargsret{\sphinxbfcode{\sphinxupquote{argsort}}}{\emph{\DUrole{n}{a}}, \emph{\DUrole{n}{axis}\DUrole{o}{=}\DUrole{default_value}{\sphinxhyphen{} 1}}, \emph{\DUrole{n}{kind}\DUrole{o}{=}\DUrole{default_value}{\textquotesingle{}quicksort\textquotesingle{}}}, \emph{\DUrole{n}{order}\DUrole{o}{=}\DUrole{default_value}{None}}}{}
Returns the indices that would sort an array.

LAX\sphinxhyphen{}backend implementation of {\hyperref[\detokenize{modules/tensor:symjax.tensor.argsort}]{\sphinxcrossref{\sphinxcode{\sphinxupquote{argsort()}}}}}.
ADDITIONOriginal docstring below.

LAX\sphinxhyphen{}backend implementation of {\hyperref[\detokenize{modules/tensor:symjax.tensor.argsort}]{\sphinxcrossref{\sphinxcode{\sphinxupquote{argsort()}}}}}.
Original docstring below.

Perform an indirect sort along the given axis using the algorithm specified
by the \sphinxtitleref{kind} keyword. It returns an array of indices of the same shape as
\sphinxtitleref{a} that index data along the given axis in sorted order.
\begin{quote}\begin{description}
\item[{Returns}] \leavevmode
\sphinxstylestrong{index\_array} \textendash{} Array of indices that sort \sphinxtitleref{a} along the specified \sphinxtitleref{axis}.
If \sphinxtitleref{a} is one\sphinxhyphen{}dimensional, \sphinxcode{\sphinxupquote{a{[}index\_array{]}}} yields a sorted \sphinxtitleref{a}.
More generally, \sphinxcode{\sphinxupquote{np.take\_along\_axis(a, index\_array, axis=axis)}}
always yields the sorted \sphinxtitleref{a}, irrespective of dimensionality.

\item[{Return type}] \leavevmode
ndarray, int

\end{description}\end{quote}

\sphinxstrong{See also:}

\begin{description}
\item[{{\hyperref[\detokenize{modules/tensor:symjax.tensor.sort}]{\sphinxcrossref{\sphinxcode{\sphinxupquote{sort()}}}}}}] \leavevmode
Describes sorting algorithms used.

\item[{\sphinxcode{\sphinxupquote{lexsort()}}}] \leavevmode
Indirect stable sort with multiple keys.

\item[{\sphinxcode{\sphinxupquote{ndarray.sort()}}}] \leavevmode
Inplace sort.

\item[{\sphinxcode{\sphinxupquote{argpartition()}}}] \leavevmode
Indirect partial sort.

\item[{{\hyperref[\detokenize{modules/tensor:symjax.tensor.take_along_axis}]{\sphinxcrossref{\sphinxcode{\sphinxupquote{take\_along\_axis()}}}}}}] \leavevmode
Apply \sphinxcode{\sphinxupquote{index\_array}} from argsort to an array as if by calling sort.

\end{description}

\subsubsection*{Notes}

See \sphinxtitleref{sort} for notes on the different sorting algorithms.

As of NumPy 1.4.0 \sphinxtitleref{argsort} works with real/complex arrays containing
nan values. The enhanced sort order is documented in \sphinxtitleref{sort}.
\subsubsection*{Examples}

One dimensional array:

\begin{sphinxVerbatim}[commandchars=\\\{\}]
\PYG{g+gp}{\PYGZgt{}\PYGZgt{}\PYGZgt{} }\PYG{n}{x} \PYG{o}{=} \PYG{n}{np}\PYG{o}{.}\PYG{n}{array}\PYG{p}{(}\PYG{p}{[}\PYG{l+m+mi}{3}\PYG{p}{,} \PYG{l+m+mi}{1}\PYG{p}{,} \PYG{l+m+mi}{2}\PYG{p}{]}\PYG{p}{)}
\PYG{g+gp}{\PYGZgt{}\PYGZgt{}\PYGZgt{} }\PYG{n}{np}\PYG{o}{.}\PYG{n}{argsort}\PYG{p}{(}\PYG{n}{x}\PYG{p}{)}
\PYG{g+go}{array([1, 2, 0])}
\end{sphinxVerbatim}

Two\sphinxhyphen{}dimensional array:

\begin{sphinxVerbatim}[commandchars=\\\{\}]
\PYG{g+gp}{\PYGZgt{}\PYGZgt{}\PYGZgt{} }\PYG{n}{x} \PYG{o}{=} \PYG{n}{np}\PYG{o}{.}\PYG{n}{array}\PYG{p}{(}\PYG{p}{[}\PYG{p}{[}\PYG{l+m+mi}{0}\PYG{p}{,} \PYG{l+m+mi}{3}\PYG{p}{]}\PYG{p}{,} \PYG{p}{[}\PYG{l+m+mi}{2}\PYG{p}{,} \PYG{l+m+mi}{2}\PYG{p}{]}\PYG{p}{]}\PYG{p}{)}
\PYG{g+gp}{\PYGZgt{}\PYGZgt{}\PYGZgt{} }\PYG{n}{x}
\PYG{g+go}{array([[0, 3],}
\PYG{g+go}{       [2, 2]])}
\end{sphinxVerbatim}

\begin{sphinxVerbatim}[commandchars=\\\{\}]
\PYG{g+gp}{\PYGZgt{}\PYGZgt{}\PYGZgt{} }\PYG{n}{ind} \PYG{o}{=} \PYG{n}{np}\PYG{o}{.}\PYG{n}{argsort}\PYG{p}{(}\PYG{n}{x}\PYG{p}{,} \PYG{n}{axis}\PYG{o}{=}\PYG{l+m+mi}{0}\PYG{p}{)}  \PYG{c+c1}{\PYGZsh{} sorts along first axis (down)}
\PYG{g+gp}{\PYGZgt{}\PYGZgt{}\PYGZgt{} }\PYG{n}{ind}
\PYG{g+go}{array([[0, 1],}
\PYG{g+go}{       [1, 0]])}
\PYG{g+gp}{\PYGZgt{}\PYGZgt{}\PYGZgt{} }\PYG{n}{np}\PYG{o}{.}\PYG{n}{take\PYGZus{}along\PYGZus{}axis}\PYG{p}{(}\PYG{n}{x}\PYG{p}{,} \PYG{n}{ind}\PYG{p}{,} \PYG{n}{axis}\PYG{o}{=}\PYG{l+m+mi}{0}\PYG{p}{)}  \PYG{c+c1}{\PYGZsh{} same as np.sort(x, axis=0)}
\PYG{g+go}{array([[0, 2],}
\PYG{g+go}{       [2, 3]])}
\end{sphinxVerbatim}

\begin{sphinxVerbatim}[commandchars=\\\{\}]
\PYG{g+gp}{\PYGZgt{}\PYGZgt{}\PYGZgt{} }\PYG{n}{ind} \PYG{o}{=} \PYG{n}{np}\PYG{o}{.}\PYG{n}{argsort}\PYG{p}{(}\PYG{n}{x}\PYG{p}{,} \PYG{n}{axis}\PYG{o}{=}\PYG{l+m+mi}{1}\PYG{p}{)}  \PYG{c+c1}{\PYGZsh{} sorts along last axis (across)}
\PYG{g+gp}{\PYGZgt{}\PYGZgt{}\PYGZgt{} }\PYG{n}{ind}
\PYG{g+go}{array([[0, 1],}
\PYG{g+go}{       [0, 1]])}
\PYG{g+gp}{\PYGZgt{}\PYGZgt{}\PYGZgt{} }\PYG{n}{np}\PYG{o}{.}\PYG{n}{take\PYGZus{}along\PYGZus{}axis}\PYG{p}{(}\PYG{n}{x}\PYG{p}{,} \PYG{n}{ind}\PYG{p}{,} \PYG{n}{axis}\PYG{o}{=}\PYG{l+m+mi}{1}\PYG{p}{)}  \PYG{c+c1}{\PYGZsh{} same as np.sort(x, axis=1)}
\PYG{g+go}{array([[0, 3],}
\PYG{g+go}{       [2, 2]])}
\end{sphinxVerbatim}

Indices of the sorted elements of a N\sphinxhyphen{}dimensional array:

\begin{sphinxVerbatim}[commandchars=\\\{\}]
\PYG{g+gp}{\PYGZgt{}\PYGZgt{}\PYGZgt{} }\PYG{n}{ind} \PYG{o}{=} \PYG{n}{np}\PYG{o}{.}\PYG{n}{unravel\PYGZus{}index}\PYG{p}{(}\PYG{n}{np}\PYG{o}{.}\PYG{n}{argsort}\PYG{p}{(}\PYG{n}{x}\PYG{p}{,} \PYG{n}{axis}\PYG{o}{=}\PYG{k+kc}{None}\PYG{p}{)}\PYG{p}{,} \PYG{n}{x}\PYG{o}{.}\PYG{n}{shape}\PYG{p}{)}
\PYG{g+gp}{\PYGZgt{}\PYGZgt{}\PYGZgt{} }\PYG{n}{ind}
\PYG{g+go}{(array([0, 1, 1, 0]), array([0, 0, 1, 1]))}
\PYG{g+gp}{\PYGZgt{}\PYGZgt{}\PYGZgt{} }\PYG{n}{x}\PYG{p}{[}\PYG{n}{ind}\PYG{p}{]}  \PYG{c+c1}{\PYGZsh{} same as np.sort(x, axis=None)}
\PYG{g+go}{array([0, 2, 2, 3])}
\end{sphinxVerbatim}

Sorting with keys:

\begin{sphinxVerbatim}[commandchars=\\\{\}]
\PYG{g+gp}{\PYGZgt{}\PYGZgt{}\PYGZgt{} }\PYG{n}{x} \PYG{o}{=} \PYG{n}{np}\PYG{o}{.}\PYG{n}{array}\PYG{p}{(}\PYG{p}{[}\PYG{p}{(}\PYG{l+m+mi}{1}\PYG{p}{,} \PYG{l+m+mi}{0}\PYG{p}{)}\PYG{p}{,} \PYG{p}{(}\PYG{l+m+mi}{0}\PYG{p}{,} \PYG{l+m+mi}{1}\PYG{p}{)}\PYG{p}{]}\PYG{p}{,} \PYG{n}{dtype}\PYG{o}{=}\PYG{p}{[}\PYG{p}{(}\PYG{l+s+s1}{\PYGZsq{}}\PYG{l+s+s1}{x}\PYG{l+s+s1}{\PYGZsq{}}\PYG{p}{,} \PYG{l+s+s1}{\PYGZsq{}}\PYG{l+s+s1}{\PYGZlt{}i4}\PYG{l+s+s1}{\PYGZsq{}}\PYG{p}{)}\PYG{p}{,} \PYG{p}{(}\PYG{l+s+s1}{\PYGZsq{}}\PYG{l+s+s1}{y}\PYG{l+s+s1}{\PYGZsq{}}\PYG{p}{,} \PYG{l+s+s1}{\PYGZsq{}}\PYG{l+s+s1}{\PYGZlt{}i4}\PYG{l+s+s1}{\PYGZsq{}}\PYG{p}{)}\PYG{p}{]}\PYG{p}{)}
\PYG{g+gp}{\PYGZgt{}\PYGZgt{}\PYGZgt{} }\PYG{n}{x}
\PYG{g+go}{array([(1, 0), (0, 1)],}
\PYG{g+go}{      dtype=[(\PYGZsq{}x\PYGZsq{}, \PYGZsq{}\PYGZlt{}i4\PYGZsq{}), (\PYGZsq{}y\PYGZsq{}, \PYGZsq{}\PYGZlt{}i4\PYGZsq{})])}
\end{sphinxVerbatim}

\begin{sphinxVerbatim}[commandchars=\\\{\}]
\PYG{g+gp}{\PYGZgt{}\PYGZgt{}\PYGZgt{} }\PYG{n}{np}\PYG{o}{.}\PYG{n}{argsort}\PYG{p}{(}\PYG{n}{x}\PYG{p}{,} \PYG{n}{order}\PYG{o}{=}\PYG{p}{(}\PYG{l+s+s1}{\PYGZsq{}}\PYG{l+s+s1}{x}\PYG{l+s+s1}{\PYGZsq{}}\PYG{p}{,}\PYG{l+s+s1}{\PYGZsq{}}\PYG{l+s+s1}{y}\PYG{l+s+s1}{\PYGZsq{}}\PYG{p}{)}\PYG{p}{)}
\PYG{g+go}{array([1, 0])}
\end{sphinxVerbatim}

\begin{sphinxVerbatim}[commandchars=\\\{\}]
\PYG{g+gp}{\PYGZgt{}\PYGZgt{}\PYGZgt{} }\PYG{n}{np}\PYG{o}{.}\PYG{n}{argsort}\PYG{p}{(}\PYG{n}{x}\PYG{p}{,} \PYG{n}{order}\PYG{o}{=}\PYG{p}{(}\PYG{l+s+s1}{\PYGZsq{}}\PYG{l+s+s1}{y}\PYG{l+s+s1}{\PYGZsq{}}\PYG{p}{,}\PYG{l+s+s1}{\PYGZsq{}}\PYG{l+s+s1}{x}\PYG{l+s+s1}{\PYGZsq{}}\PYG{p}{)}\PYG{p}{)}
\PYG{g+go}{array([0, 1])}
\end{sphinxVerbatim}

\end{fulllineitems}

\index{around() (in module symjax.tensor)@\spxentry{around()}\spxextra{in module symjax.tensor}}

\begin{fulllineitems}
\phantomsection\label{\detokenize{modules/tensor:symjax.tensor.around}}\pysiglinewithargsret{\sphinxbfcode{\sphinxupquote{around}}}{\emph{\DUrole{n}{a}}, \emph{\DUrole{n}{decimals}\DUrole{o}{=}\DUrole{default_value}{0}}}{}
Round an array to the given number of decimals.

LAX\sphinxhyphen{}backend implementation of \sphinxcode{\sphinxupquote{round\_()}}.
ADDITIONOriginal docstring below.

LA

\end{fulllineitems}

\index{asarray() (in module symjax.tensor)@\spxentry{asarray()}\spxextra{in module symjax.tensor}}

\begin{fulllineitems}
\phantomsection\label{\detokenize{modules/tensor:symjax.tensor.asarray}}\pysiglinewithargsret{\sphinxbfcode{\sphinxupquote{asarray}}}{\emph{\DUrole{n}{a}}, \emph{\DUrole{n}{dtype}\DUrole{o}{=}\DUrole{default_value}{None}}, \emph{\DUrole{n}{order}\DUrole{o}{=}\DUrole{default_value}{None}}}{}
Convert the input to an array.

LAX\sphinxhyphen{}backend implementation of {\hyperref[\detokenize{modules/tensor:symjax.tensor.asarray}]{\sphinxcrossref{\sphinxcode{\sphinxupquote{asarray()}}}}}.
ADDITIONOriginal docstring below.

LAX\sphinxhyphen{}backend implementation of {\hyperref[\detokenize{modules/tensor:symjax.tensor.asarray}]{\sphinxcrossref{\sphinxcode{\sphinxupquote{asarray()}}}}}.
Original docstring below.
\begin{quote}\begin{description}
\item[{Parameters}] \leavevmode
\sphinxstyleliteralstrong{\sphinxupquote{dtype}} (\sphinxstyleliteralemphasis{\sphinxupquote{data\sphinxhyphen{}type}}\sphinxstyleliteralemphasis{\sphinxupquote{, }}\sphinxstyleliteralemphasis{\sphinxupquote{optional}}) \textendash{} By default, the data\sphinxhyphen{}type is inferred from the input data.

\item[{Returns}] \leavevmode
\sphinxstylestrong{out} \textendash{} Array interpretation of \sphinxtitleref{a}.  No copy is performed if the input
is already an ndarray with matching dtype and order.  If \sphinxtitleref{a} is a
subclass of ndarray, a base class ndarray is returned.

\item[{Return type}] \leavevmode
ndarray

\end{description}\end{quote}

\sphinxstrong{See also:}

\begin{description}
\item[{\sphinxcode{\sphinxupquote{asanyarray()}}}] \leavevmode
Similar function which passes through subclasses.

\item[{\sphinxcode{\sphinxupquote{ascontiguousarray()}}}] \leavevmode
Convert input to a contiguous array.

\item[{\sphinxcode{\sphinxupquote{asfarray()}}}] \leavevmode
Convert input to a floating point ndarray.

\item[{\sphinxcode{\sphinxupquote{asfortranarray()}}}] \leavevmode
Convert input to an ndarray with column\sphinxhyphen{}major memory order.

\item[{\sphinxcode{\sphinxupquote{asarray\_chkfinite()}}}] \leavevmode
Similar function which checks input for NaNs and Infs.

\item[{\sphinxcode{\sphinxupquote{fromiter()}}}] \leavevmode
Create an array from an iterator.

\item[{\sphinxcode{\sphinxupquote{fromfunction()}}}] \leavevmode
Construct an array by executing a function on grid positions.

\end{description}

\subsubsection*{Examples}

Convert a list into an array:

\begin{sphinxVerbatim}[commandchars=\\\{\}]
\PYG{g+gp}{\PYGZgt{}\PYGZgt{}\PYGZgt{} }\PYG{n}{a} \PYG{o}{=} \PYG{p}{[}\PYG{l+m+mi}{1}\PYG{p}{,} \PYG{l+m+mi}{2}\PYG{p}{]}
\PYG{g+gp}{\PYGZgt{}\PYGZgt{}\PYGZgt{} }\PYG{n}{np}\PYG{o}{.}\PYG{n}{asarray}\PYG{p}{(}\PYG{n}{a}\PYG{p}{)}
\PYG{g+go}{array([1, 2])}
\end{sphinxVerbatim}

Existing arrays are not copied:

\begin{sphinxVerbatim}[commandchars=\\\{\}]
\PYG{g+gp}{\PYGZgt{}\PYGZgt{}\PYGZgt{} }\PYG{n}{a} \PYG{o}{=} \PYG{n}{np}\PYG{o}{.}\PYG{n}{array}\PYG{p}{(}\PYG{p}{[}\PYG{l+m+mi}{1}\PYG{p}{,} \PYG{l+m+mi}{2}\PYG{p}{]}\PYG{p}{)}
\PYG{g+gp}{\PYGZgt{}\PYGZgt{}\PYGZgt{} }\PYG{n}{np}\PYG{o}{.}\PYG{n}{asarray}\PYG{p}{(}\PYG{n}{a}\PYG{p}{)} \PYG{o+ow}{is} \PYG{n}{a}
\PYG{g+go}{True}
\end{sphinxVerbatim}

If \sphinxtitleref{dtype} is set, array is copied only if dtype does not match:

\begin{sphinxVerbatim}[commandchars=\\\{\}]
\PYG{g+gp}{\PYGZgt{}\PYGZgt{}\PYGZgt{} }\PYG{n}{a} \PYG{o}{=} \PYG{n}{np}\PYG{o}{.}\PYG{n}{array}\PYG{p}{(}\PYG{p}{[}\PYG{l+m+mi}{1}\PYG{p}{,} \PYG{l+m+mi}{2}\PYG{p}{]}\PYG{p}{,} \PYG{n}{dtype}\PYG{o}{=}\PYG{n}{np}\PYG{o}{.}\PYG{n}{float32}\PYG{p}{)}
\PYG{g+gp}{\PYGZgt{}\PYGZgt{}\PYGZgt{} }\PYG{n}{np}\PYG{o}{.}\PYG{n}{asarray}\PYG{p}{(}\PYG{n}{a}\PYG{p}{,} \PYG{n}{dtype}\PYG{o}{=}\PYG{n}{np}\PYG{o}{.}\PYG{n}{float32}\PYG{p}{)} \PYG{o+ow}{is} \PYG{n}{a}
\PYG{g+go}{True}
\PYG{g+gp}{\PYGZgt{}\PYGZgt{}\PYGZgt{} }\PYG{n}{np}\PYG{o}{.}\PYG{n}{asarray}\PYG{p}{(}\PYG{n}{a}\PYG{p}{,} \PYG{n}{dtype}\PYG{o}{=}\PYG{n}{np}\PYG{o}{.}\PYG{n}{float64}\PYG{p}{)} \PYG{o+ow}{is} \PYG{n}{a}
\PYG{g+go}{False}
\end{sphinxVerbatim}

Contrary to \sphinxtitleref{asanyarray}, ndarray subclasses are not passed through:

\begin{sphinxVerbatim}[commandchars=\\\{\}]
\PYG{g+gp}{\PYGZgt{}\PYGZgt{}\PYGZgt{} }\PYG{n+nb}{issubclass}\PYG{p}{(}\PYG{n}{np}\PYG{o}{.}\PYG{n}{recarray}\PYG{p}{,} \PYG{n}{np}\PYG{o}{.}\PYG{n}{ndarray}\PYG{p}{)}
\PYG{g+go}{True}
\PYG{g+gp}{\PYGZgt{}\PYGZgt{}\PYGZgt{} }\PYG{n}{a} \PYG{o}{=} \PYG{n}{np}\PYG{o}{.}\PYG{n}{array}\PYG{p}{(}\PYG{p}{[}\PYG{p}{(}\PYG{l+m+mf}{1.0}\PYG{p}{,} \PYG{l+m+mi}{2}\PYG{p}{)}\PYG{p}{,} \PYG{p}{(}\PYG{l+m+mf}{3.0}\PYG{p}{,} \PYG{l+m+mi}{4}\PYG{p}{)}\PYG{p}{]}\PYG{p}{,} \PYG{n}{dtype}\PYG{o}{=}\PYG{l+s+s1}{\PYGZsq{}}\PYG{l+s+s1}{f4,i4}\PYG{l+s+s1}{\PYGZsq{}}\PYG{p}{)}\PYG{o}{.}\PYG{n}{view}\PYG{p}{(}\PYG{n}{np}\PYG{o}{.}\PYG{n}{recarray}\PYG{p}{)}
\PYG{g+gp}{\PYGZgt{}\PYGZgt{}\PYGZgt{} }\PYG{n}{np}\PYG{o}{.}\PYG{n}{asarray}\PYG{p}{(}\PYG{n}{a}\PYG{p}{)} \PYG{o+ow}{is} \PYG{n}{a}
\PYG{g+go}{False}
\PYG{g+gp}{\PYGZgt{}\PYGZgt{}\PYGZgt{} }\PYG{n}{np}\PYG{o}{.}\PYG{n}{asanyarray}\PYG{p}{(}\PYG{n}{a}\PYG{p}{)} \PYG{o+ow}{is} \PYG{n}{a}
\PYG{g+go}{True}
\end{sphinxVerbatim}

\end{fulllineitems}

\index{atleast\_1d() (in module symjax.tensor)@\spxentry{atleast\_1d()}\spxextra{in module symjax.tensor}}

\begin{fulllineitems}
\phantomsection\label{\detokenize{modules/tensor:symjax.tensor.atleast_1d}}\pysiglinewithargsret{\sphinxbfcode{\sphinxupquote{atleast\_1d}}}{\emph{\DUrole{o}{*}\DUrole{n}{arys}}}{}
Convert inputs to arrays with at least one dimension.

LAX\sphinxhyphen{}backend implementation of {\hyperref[\detokenize{modules/tensor:symjax.tensor.atleast_1d}]{\sphinxcrossref{\sphinxcode{\sphinxupquote{atleast\_1d()}}}}}.
ADDITIONOriginal docstring below.

LAX\sphinxhyphen{}backend implementation of {\hyperref[\detokenize{modules/tensor:symjax.tensor.atleast_1d}]{\sphinxcrossref{\sphinxcode{\sphinxupquote{atleast\_1d()}}}}}.
Original docstring below.
\begin{quote}
\begin{quote}

Scalar inputs are converted to 1\sphinxhyphen{}dimensional arrays, whilst
higher\sphinxhyphen{}dimensional inputs are preserved.
\end{quote}
\begin{description}
\item[{Returns}] \leavevmode\begin{description}
\item[{ret}] \leavevmode{[}ndarray{]}
An array, or list of arrays, each with \sphinxcode{\sphinxupquote{a.ndim \textgreater{}= 1}}.
Copies are made only if necessary.

\end{description}

atleast\_2d, atleast\_3d

\begin{sphinxVerbatim}[commandchars=\\\{\}]
\PYG{g+gp}{\PYGZgt{}\PYGZgt{}\PYGZgt{} }\PYG{n}{np}\PYG{o}{.}\PYG{n}{atleast\PYGZus{}1d}\PYG{p}{(}\PYG{l+m+mf}{1.0}\PYG{p}{)}
\PYG{g+go}{array([1.])}
\end{sphinxVerbatim}

\begin{sphinxVerbatim}[commandchars=\\\{\}]
\PYG{g+gp}{\PYGZgt{}\PYGZgt{}\PYGZgt{} }\PYG{n}{x} \PYG{o}{=} \PYG{n}{np}\PYG{o}{.}\PYG{n}{arange}\PYG{p}{(}\PYG{l+m+mf}{9.0}\PYG{p}{)}\PYG{o}{.}\PYG{n}{reshape}\PYG{p}{(}\PYG{l+m+mi}{3}\PYG{p}{,}\PYG{l+m+mi}{3}\PYG{p}{)}
\PYG{g+gp}{\PYGZgt{}\PYGZgt{}\PYGZgt{} }\PYG{n}{np}\PYG{o}{.}\PYG{n}{atleast\PYGZus{}1d}\PYG{p}{(}\PYG{n}{x}\PYG{p}{)}
\PYG{g+go}{array([[0., 1., 2.],}
\PYG{g+go}{       [3., 4., 5.],}
\PYG{g+go}{       [6., 7., 8.]])}
\PYG{g+gp}{\PYGZgt{}\PYGZgt{}\PYGZgt{} }\PYG{n}{np}\PYG{o}{.}\PYG{n}{atleast\PYGZus{}1d}\PYG{p}{(}\PYG{n}{x}\PYG{p}{)} \PYG{o+ow}{is} \PYG{n}{x}
\PYG{g+go}{True}
\end{sphinxVerbatim}

\begin{sphinxVerbatim}[commandchars=\\\{\}]
\PYG{g+gp}{\PYGZgt{}\PYGZgt{}\PYGZgt{} }\PYG{n}{np}\PYG{o}{.}\PYG{n}{atleast\PYGZus{}1d}\PYG{p}{(}\PYG{l+m+mi}{1}\PYG{p}{,} \PYG{p}{[}\PYG{l+m+mi}{3}\PYG{p}{,} \PYG{l+m+mi}{4}\PYG{p}{]}\PYG{p}{)}
\PYG{g+go}{[array([1]), array([3, 4])]}
\end{sphinxVerbatim}

\end{description}
\end{quote}

\end{fulllineitems}

\index{atleast\_2d() (in module symjax.tensor)@\spxentry{atleast\_2d()}\spxextra{in module symjax.tensor}}

\begin{fulllineitems}
\phantomsection\label{\detokenize{modules/tensor:symjax.tensor.atleast_2d}}\pysiglinewithargsret{\sphinxbfcode{\sphinxupquote{atleast\_2d}}}{\emph{\DUrole{o}{*}\DUrole{n}{arys}}}{}
View inputs as arrays with at least two dimensions.

LAX\sphinxhyphen{}backend implementation of {\hyperref[\detokenize{modules/tensor:symjax.tensor.atleast_2d}]{\sphinxcrossref{\sphinxcode{\sphinxupquote{atleast\_2d()}}}}}.
ADDITIONOriginal docstring below.

LAX\sphinxhyphen{}backend implementation of {\hyperref[\detokenize{modules/tensor:symjax.tensor.atleast_2d}]{\sphinxcrossref{\sphinxcode{\sphinxupquote{atleast\_2d()}}}}}.
Original docstring below.
\begin{quote}
\begin{quote}
\end{quote}
\begin{description}
\item[{Returns}] \leavevmode\begin{description}
\item[{res, res2, …}] \leavevmode{[}ndarray{]}
An array, or list of arrays, each with \sphinxcode{\sphinxupquote{a.ndim \textgreater{}= 2}}.
Copies are avoided where possible, and views with two or more
dimensions are returned.

\end{description}

atleast\_1d, atleast\_3d

\begin{sphinxVerbatim}[commandchars=\\\{\}]
\PYG{g+gp}{\PYGZgt{}\PYGZgt{}\PYGZgt{} }\PYG{n}{np}\PYG{o}{.}\PYG{n}{atleast\PYGZus{}2d}\PYG{p}{(}\PYG{l+m+mf}{3.0}\PYG{p}{)}
\PYG{g+go}{array([[3.]])}
\end{sphinxVerbatim}

\begin{sphinxVerbatim}[commandchars=\\\{\}]
\PYG{g+gp}{\PYGZgt{}\PYGZgt{}\PYGZgt{} }\PYG{n}{x} \PYG{o}{=} \PYG{n}{np}\PYG{o}{.}\PYG{n}{arange}\PYG{p}{(}\PYG{l+m+mf}{3.0}\PYG{p}{)}
\PYG{g+gp}{\PYGZgt{}\PYGZgt{}\PYGZgt{} }\PYG{n}{np}\PYG{o}{.}\PYG{n}{atleast\PYGZus{}2d}\PYG{p}{(}\PYG{n}{x}\PYG{p}{)}
\PYG{g+go}{array([[0., 1., 2.]])}
\PYG{g+gp}{\PYGZgt{}\PYGZgt{}\PYGZgt{} }\PYG{n}{np}\PYG{o}{.}\PYG{n}{atleast\PYGZus{}2d}\PYG{p}{(}\PYG{n}{x}\PYG{p}{)}\PYG{o}{.}\PYG{n}{base} \PYG{o+ow}{is} \PYG{n}{x}
\PYG{g+go}{True}
\end{sphinxVerbatim}

\begin{sphinxVerbatim}[commandchars=\\\{\}]
\PYG{g+gp}{\PYGZgt{}\PYGZgt{}\PYGZgt{} }\PYG{n}{np}\PYG{o}{.}\PYG{n}{atleast\PYGZus{}2d}\PYG{p}{(}\PYG{l+m+mi}{1}\PYG{p}{,} \PYG{p}{[}\PYG{l+m+mi}{1}\PYG{p}{,} \PYG{l+m+mi}{2}\PYG{p}{]}\PYG{p}{,} \PYG{p}{[}\PYG{p}{[}\PYG{l+m+mi}{1}\PYG{p}{,} \PYG{l+m+mi}{2}\PYG{p}{]}\PYG{p}{]}\PYG{p}{)}
\PYG{g+go}{[array([[1]]), array([[1, 2]]), array([[1, 2]])]}
\end{sphinxVerbatim}

\end{description}
\end{quote}

\end{fulllineitems}

\index{atleast\_3d() (in module symjax.tensor)@\spxentry{atleast\_3d()}\spxextra{in module symjax.tensor}}

\begin{fulllineitems}
\phantomsection\label{\detokenize{modules/tensor:symjax.tensor.atleast_3d}}\pysiglinewithargsret{\sphinxbfcode{\sphinxupquote{atleast\_3d}}}{\emph{\DUrole{o}{*}\DUrole{n}{arys}}}{}
View inputs as arrays with at least three dimensions.

LAX\sphinxhyphen{}backend implementation of {\hyperref[\detokenize{modules/tensor:symjax.tensor.atleast_3d}]{\sphinxcrossref{\sphinxcode{\sphinxupquote{atleast\_3d()}}}}}.
ADDITIONOriginal docstring below.

LAX\sphinxhyphen{}backend implementation of {\hyperref[\detokenize{modules/tensor:symjax.tensor.atleast_3d}]{\sphinxcrossref{\sphinxcode{\sphinxupquote{atleast\_3d()}}}}}.
Original docstring below.
\begin{quote}
\begin{quote}
\end{quote}
\begin{description}
\item[{Returns}] \leavevmode\begin{description}
\item[{res1, res2, …}] \leavevmode{[}ndarray{]}
An array, or list of arrays, each with \sphinxcode{\sphinxupquote{a.ndim \textgreater{}= 3}}.  Copies are
avoided where possible, and views with three or more dimensions are
returned.  For example, a 1\sphinxhyphen{}D array of shape \sphinxcode{\sphinxupquote{(N,)}} becomes a view
of shape \sphinxcode{\sphinxupquote{(1, N, 1)}}, and a 2\sphinxhyphen{}D array of shape \sphinxcode{\sphinxupquote{(M, N)}} becomes a
view of shape \sphinxcode{\sphinxupquote{(M, N, 1)}}.

\end{description}

atleast\_1d, atleast\_2d

\begin{sphinxVerbatim}[commandchars=\\\{\}]
\PYG{g+gp}{\PYGZgt{}\PYGZgt{}\PYGZgt{} }\PYG{n}{np}\PYG{o}{.}\PYG{n}{atleast\PYGZus{}3d}\PYG{p}{(}\PYG{l+m+mf}{3.0}\PYG{p}{)}
\PYG{g+go}{array([[[3.]]])}
\end{sphinxVerbatim}

\begin{sphinxVerbatim}[commandchars=\\\{\}]
\PYG{g+gp}{\PYGZgt{}\PYGZgt{}\PYGZgt{} }\PYG{n}{x} \PYG{o}{=} \PYG{n}{np}\PYG{o}{.}\PYG{n}{arange}\PYG{p}{(}\PYG{l+m+mf}{3.0}\PYG{p}{)}
\PYG{g+gp}{\PYGZgt{}\PYGZgt{}\PYGZgt{} }\PYG{n}{np}\PYG{o}{.}\PYG{n}{atleast\PYGZus{}3d}\PYG{p}{(}\PYG{n}{x}\PYG{p}{)}\PYG{o}{.}\PYG{n}{shape}
\PYG{g+go}{(1, 3, 1)}
\end{sphinxVerbatim}

\begin{sphinxVerbatim}[commandchars=\\\{\}]
\PYG{g+gp}{\PYGZgt{}\PYGZgt{}\PYGZgt{} }\PYG{n}{x} \PYG{o}{=} \PYG{n}{np}\PYG{o}{.}\PYG{n}{arange}\PYG{p}{(}\PYG{l+m+mf}{12.0}\PYG{p}{)}\PYG{o}{.}\PYG{n}{reshape}\PYG{p}{(}\PYG{l+m+mi}{4}\PYG{p}{,}\PYG{l+m+mi}{3}\PYG{p}{)}
\PYG{g+gp}{\PYGZgt{}\PYGZgt{}\PYGZgt{} }\PYG{n}{np}\PYG{o}{.}\PYG{n}{atleast\PYGZus{}3d}\PYG{p}{(}\PYG{n}{x}\PYG{p}{)}\PYG{o}{.}\PYG{n}{shape}
\PYG{g+go}{(4, 3, 1)}
\PYG{g+gp}{\PYGZgt{}\PYGZgt{}\PYGZgt{} }\PYG{n}{np}\PYG{o}{.}\PYG{n}{atleast\PYGZus{}3d}\PYG{p}{(}\PYG{n}{x}\PYG{p}{)}\PYG{o}{.}\PYG{n}{base} \PYG{o+ow}{is} \PYG{n}{x}\PYG{o}{.}\PYG{n}{base}  \PYG{c+c1}{\PYGZsh{} x is a reshape, so not base itself}
\PYG{g+go}{True}
\end{sphinxVerbatim}

\begin{sphinxVerbatim}[commandchars=\\\{\}]
\PYG{g+gp}{\PYGZgt{}\PYGZgt{}\PYGZgt{} }\PYG{k}{for} \PYG{n}{arr} \PYG{o+ow}{in} \PYG{n}{np}\PYG{o}{.}\PYG{n}{atleast\PYGZus{}3d}\PYG{p}{(}\PYG{p}{[}\PYG{l+m+mi}{1}\PYG{p}{,} \PYG{l+m+mi}{2}\PYG{p}{]}\PYG{p}{,} \PYG{p}{[}\PYG{p}{[}\PYG{l+m+mi}{1}\PYG{p}{,} \PYG{l+m+mi}{2}\PYG{p}{]}\PYG{p}{]}\PYG{p}{,} \PYG{p}{[}\PYG{p}{[}\PYG{p}{[}\PYG{l+m+mi}{1}\PYG{p}{,} \PYG{l+m+mi}{2}\PYG{p}{]}\PYG{p}{]}\PYG{p}{]}\PYG{p}{)}\PYG{p}{:}
\PYG{g+gp}{... }    \PYG{n+nb}{print}\PYG{p}{(}\PYG{n}{arr}\PYG{p}{,} \PYG{n}{arr}\PYG{o}{.}\PYG{n}{shape}\PYG{p}{)} 
\PYG{g+gp}{...}
\PYG{g+go}{[[[1]}
\PYG{g+go}{  [2]]] (1, 2, 1)}
\PYG{g+go}{[[[1]}
\PYG{g+go}{  [2]]] (1, 2, 1)}
\PYG{g+go}{[[[1 2]]] (1, 1, 2)}
\end{sphinxVerbatim}

\end{description}
\end{quote}

\end{fulllineitems}

\index{bitwise\_and() (in module symjax.tensor)@\spxentry{bitwise\_and()}\spxextra{in module symjax.tensor}}

\begin{fulllineitems}
\phantomsection\label{\detokenize{modules/tensor:symjax.tensor.bitwise_and}}\pysiglinewithargsret{\sphinxbfcode{\sphinxupquote{bitwise\_and}}}{\emph{\DUrole{n}{x1}}, \emph{\DUrole{n}{x2}}}{}
Compute the bit\sphinxhyphen{}wise AND of two arrays element\sphinxhyphen{}wise.

LAX\sphinxhyphen{}backend implementation of {\hyperref[\detokenize{modules/tensor:symjax.tensor.bitwise_and}]{\sphinxcrossref{\sphinxcode{\sphinxupquote{bitwise\_and()}}}}}.
ADDITIONOriginal docstring below.

LAX\sphinxhyphen{}backend implementation of {\hyperref[\detokenize{modules/tensor:symjax.tensor.bitwise_and}]{\sphinxcrossref{\sphinxcode{\sphinxupquote{bitwise\_and()}}}}}.
Original docstring below.

bitwise\_and(x1, x2, /, out=None, {\color{red}\bfseries{}*}, where=True, casting=’same\_kind’, order=’K’, dtype=None, subok=True{[}, signature, extobj{]})

Computes the bit\sphinxhyphen{}wise AND of the underlying binary representation of
the integers in the input arrays. This ufunc implements the C/Python
operator \sphinxcode{\sphinxupquote{\&}}.
\begin{quote}\begin{description}
\item[{Returns}] \leavevmode
\sphinxstylestrong{out} \textendash{} Result.
This is a scalar if both \sphinxtitleref{x1} and \sphinxtitleref{x2} are scalars.

\item[{Return type}] \leavevmode
ndarray or scalar

\end{description}\end{quote}

\sphinxstrong{See also:}

{\hyperref[\detokenize{modules/tensor:symjax.tensor.logical_and}]{\sphinxcrossref{\sphinxcode{\sphinxupquote{logical\_and()}}}}}, {\hyperref[\detokenize{modules/tensor:symjax.tensor.bitwise_or}]{\sphinxcrossref{\sphinxcode{\sphinxupquote{bitwise\_or()}}}}}, {\hyperref[\detokenize{modules/tensor:symjax.tensor.bitwise_xor}]{\sphinxcrossref{\sphinxcode{\sphinxupquote{bitwise\_xor()}}}}}
\begin{description}
\item[{\sphinxcode{\sphinxupquote{binary\_repr()}}}] \leavevmode
Return the binary representation of the input number as a string.

\end{description}

\subsubsection*{Examples}

The number 13 is represented by \sphinxcode{\sphinxupquote{00001101}}.  Likewise, 17 is
represented by \sphinxcode{\sphinxupquote{00010001}}.  The bit\sphinxhyphen{}wise AND of 13 and 17 is
therefore \sphinxcode{\sphinxupquote{000000001}}, or 1:

\begin{sphinxVerbatim}[commandchars=\\\{\}]
\PYG{g+gp}{\PYGZgt{}\PYGZgt{}\PYGZgt{} }\PYG{n}{np}\PYG{o}{.}\PYG{n}{bitwise\PYGZus{}and}\PYG{p}{(}\PYG{l+m+mi}{13}\PYG{p}{,} \PYG{l+m+mi}{17}\PYG{p}{)}
\PYG{g+go}{1}
\end{sphinxVerbatim}

\begin{sphinxVerbatim}[commandchars=\\\{\}]
\PYG{g+gp}{\PYGZgt{}\PYGZgt{}\PYGZgt{} }\PYG{n}{np}\PYG{o}{.}\PYG{n}{bitwise\PYGZus{}and}\PYG{p}{(}\PYG{l+m+mi}{14}\PYG{p}{,} \PYG{l+m+mi}{13}\PYG{p}{)}
\PYG{g+go}{12}
\PYG{g+gp}{\PYGZgt{}\PYGZgt{}\PYGZgt{} }\PYG{n}{np}\PYG{o}{.}\PYG{n}{binary\PYGZus{}repr}\PYG{p}{(}\PYG{l+m+mi}{12}\PYG{p}{)}
\PYG{g+go}{\PYGZsq{}1100\PYGZsq{}}
\PYG{g+gp}{\PYGZgt{}\PYGZgt{}\PYGZgt{} }\PYG{n}{np}\PYG{o}{.}\PYG{n}{bitwise\PYGZus{}and}\PYG{p}{(}\PYG{p}{[}\PYG{l+m+mi}{14}\PYG{p}{,}\PYG{l+m+mi}{3}\PYG{p}{]}\PYG{p}{,} \PYG{l+m+mi}{13}\PYG{p}{)}
\PYG{g+go}{array([12,  1])}
\end{sphinxVerbatim}

\begin{sphinxVerbatim}[commandchars=\\\{\}]
\PYG{g+gp}{\PYGZgt{}\PYGZgt{}\PYGZgt{} }\PYG{n}{np}\PYG{o}{.}\PYG{n}{bitwise\PYGZus{}and}\PYG{p}{(}\PYG{p}{[}\PYG{l+m+mi}{11}\PYG{p}{,}\PYG{l+m+mi}{7}\PYG{p}{]}\PYG{p}{,} \PYG{p}{[}\PYG{l+m+mi}{4}\PYG{p}{,}\PYG{l+m+mi}{25}\PYG{p}{]}\PYG{p}{)}
\PYG{g+go}{array([0, 1])}
\PYG{g+gp}{\PYGZgt{}\PYGZgt{}\PYGZgt{} }\PYG{n}{np}\PYG{o}{.}\PYG{n}{bitwise\PYGZus{}and}\PYG{p}{(}\PYG{n}{np}\PYG{o}{.}\PYG{n}{array}\PYG{p}{(}\PYG{p}{[}\PYG{l+m+mi}{2}\PYG{p}{,}\PYG{l+m+mi}{5}\PYG{p}{,}\PYG{l+m+mi}{255}\PYG{p}{]}\PYG{p}{)}\PYG{p}{,} \PYG{n}{np}\PYG{o}{.}\PYG{n}{array}\PYG{p}{(}\PYG{p}{[}\PYG{l+m+mi}{3}\PYG{p}{,}\PYG{l+m+mi}{14}\PYG{p}{,}\PYG{l+m+mi}{16}\PYG{p}{]}\PYG{p}{)}\PYG{p}{)}
\PYG{g+go}{array([ 2,  4, 16])}
\PYG{g+gp}{\PYGZgt{}\PYGZgt{}\PYGZgt{} }\PYG{n}{np}\PYG{o}{.}\PYG{n}{bitwise\PYGZus{}and}\PYG{p}{(}\PYG{p}{[}\PYG{k+kc}{True}\PYG{p}{,} \PYG{k+kc}{True}\PYG{p}{]}\PYG{p}{,} \PYG{p}{[}\PYG{k+kc}{False}\PYG{p}{,} \PYG{k+kc}{True}\PYG{p}{]}\PYG{p}{)}
\PYG{g+go}{array([False,  True])}
\end{sphinxVerbatim}

\end{fulllineitems}

\index{bitwise\_not() (in module symjax.tensor)@\spxentry{bitwise\_not()}\spxextra{in module symjax.tensor}}

\begin{fulllineitems}
\phantomsection\label{\detokenize{modules/tensor:symjax.tensor.bitwise_not}}\pysiglinewithargsret{\sphinxbfcode{\sphinxupquote{bitwise\_not}}}{\emph{\DUrole{n}{x}}}{}
Compute bit\sphinxhyphen{}wise inversion, or bit\sphinxhyphen{}wise NOT, element\sphinxhyphen{}wise.

LAX\sphinxhyphen{}backend implementation of \sphinxcode{\sphinxupquote{invert()}}.
ADDITIONOriginal docstring below.

LAX\sphinxhyphen{}backend implementation of \sphinxcode{\sphinxupquote{invert()}}.
Original docstring below.

invert(x, /, out=None, {\color{red}\bfseries{}*}, where=True, casting=’same\_kind’, order=’K’, dtype=None, subok=True{[}, signature, extobj{]})

Computes the bit\sphinxhyphen{}wise NOT of the underlying binary representation of
the integers in the input arrays. This ufunc implements the C/Python
operator \sphinxcode{\sphinxupquote{\textasciitilde{}}}.

For signed integer inputs, the two’s complement is returned.  In a
two’s\sphinxhyphen{}complement system negative numbers are represented by the two’s
complement of the absolute value. This is the most common method of
representing signed integers on computers {\color{red}\bfseries{}{[}1{]}\_}. A N\sphinxhyphen{}bit
two’s\sphinxhyphen{}complement system can represent every integer in the range
\(-2^{N-1}\) to \(+2^{N-1}-1\).
\begin{quote}\begin{description}
\item[{Returns}] \leavevmode
\sphinxstylestrong{out} \textendash{} Result.
This is a scalar if \sphinxtitleref{x} is a scalar.

\item[{Return type}] \leavevmode
ndarray or scalar

\end{description}\end{quote}

\sphinxstrong{See also:}

{\hyperref[\detokenize{modules/tensor:symjax.tensor.bitwise_and}]{\sphinxcrossref{\sphinxcode{\sphinxupquote{bitwise\_and()}}}}}, {\hyperref[\detokenize{modules/tensor:symjax.tensor.bitwise_or}]{\sphinxcrossref{\sphinxcode{\sphinxupquote{bitwise\_or()}}}}}, {\hyperref[\detokenize{modules/tensor:symjax.tensor.bitwise_xor}]{\sphinxcrossref{\sphinxcode{\sphinxupquote{bitwise\_xor()}}}}}, {\hyperref[\detokenize{modules/tensor:symjax.tensor.logical_not}]{\sphinxcrossref{\sphinxcode{\sphinxupquote{logical\_not()}}}}}
\begin{description}
\item[{\sphinxcode{\sphinxupquote{binary\_repr()}}}] \leavevmode
Return the binary representation of the input number as a string.

\end{description}

\subsubsection*{Notes}

\sphinxtitleref{bitwise\_not} is an alias for \sphinxtitleref{invert}:

\begin{sphinxVerbatim}[commandchars=\\\{\}]
\PYG{g+gp}{\PYGZgt{}\PYGZgt{}\PYGZgt{} }\PYG{n}{np}\PYG{o}{.}\PYG{n}{bitwise\PYGZus{}not} \PYG{o+ow}{is} \PYG{n}{np}\PYG{o}{.}\PYG{n}{invert}
\PYG{g+go}{True}
\end{sphinxVerbatim}
\subsubsection*{References}
\subsubsection*{Examples}

We’ve seen that 13 is represented by \sphinxcode{\sphinxupquote{00001101}}.
The invert or bit\sphinxhyphen{}wise NOT of 13 is then:

\begin{sphinxVerbatim}[commandchars=\\\{\}]
\PYG{g+gp}{\PYGZgt{}\PYGZgt{}\PYGZgt{} }\PYG{n}{x} \PYG{o}{=} \PYG{n}{np}\PYG{o}{.}\PYG{n}{invert}\PYG{p}{(}\PYG{n}{np}\PYG{o}{.}\PYG{n}{array}\PYG{p}{(}\PYG{l+m+mi}{13}\PYG{p}{,} \PYG{n}{dtype}\PYG{o}{=}\PYG{n}{np}\PYG{o}{.}\PYG{n}{uint8}\PYG{p}{)}\PYG{p}{)}
\PYG{g+gp}{\PYGZgt{}\PYGZgt{}\PYGZgt{} }\PYG{n}{x}
\PYG{g+go}{242}
\PYG{g+gp}{\PYGZgt{}\PYGZgt{}\PYGZgt{} }\PYG{n}{np}\PYG{o}{.}\PYG{n}{binary\PYGZus{}repr}\PYG{p}{(}\PYG{n}{x}\PYG{p}{,} \PYG{n}{width}\PYG{o}{=}\PYG{l+m+mi}{8}\PYG{p}{)}
\PYG{g+go}{\PYGZsq{}11110010\PYGZsq{}}
\end{sphinxVerbatim}

The result depends on the bit\sphinxhyphen{}width:

\begin{sphinxVerbatim}[commandchars=\\\{\}]
\PYG{g+gp}{\PYGZgt{}\PYGZgt{}\PYGZgt{} }\PYG{n}{x} \PYG{o}{=} \PYG{n}{np}\PYG{o}{.}\PYG{n}{invert}\PYG{p}{(}\PYG{n}{np}\PYG{o}{.}\PYG{n}{array}\PYG{p}{(}\PYG{l+m+mi}{13}\PYG{p}{,} \PYG{n}{dtype}\PYG{o}{=}\PYG{n}{np}\PYG{o}{.}\PYG{n}{uint16}\PYG{p}{)}\PYG{p}{)}
\PYG{g+gp}{\PYGZgt{}\PYGZgt{}\PYGZgt{} }\PYG{n}{x}
\PYG{g+go}{65522}
\PYG{g+gp}{\PYGZgt{}\PYGZgt{}\PYGZgt{} }\PYG{n}{np}\PYG{o}{.}\PYG{n}{binary\PYGZus{}repr}\PYG{p}{(}\PYG{n}{x}\PYG{p}{,} \PYG{n}{width}\PYG{o}{=}\PYG{l+m+mi}{16}\PYG{p}{)}
\PYG{g+go}{\PYGZsq{}1111111111110010\PYGZsq{}}
\end{sphinxVerbatim}

When using signed integer types the result is the two’s complement of
the result for the unsigned type:

\begin{sphinxVerbatim}[commandchars=\\\{\}]
\PYG{g+gp}{\PYGZgt{}\PYGZgt{}\PYGZgt{} }\PYG{n}{np}\PYG{o}{.}\PYG{n}{invert}\PYG{p}{(}\PYG{n}{np}\PYG{o}{.}\PYG{n}{array}\PYG{p}{(}\PYG{p}{[}\PYG{l+m+mi}{13}\PYG{p}{]}\PYG{p}{,} \PYG{n}{dtype}\PYG{o}{=}\PYG{n}{np}\PYG{o}{.}\PYG{n}{int8}\PYG{p}{)}\PYG{p}{)}
\PYG{g+go}{array([\PYGZhy{}14], dtype=int8)}
\PYG{g+gp}{\PYGZgt{}\PYGZgt{}\PYGZgt{} }\PYG{n}{np}\PYG{o}{.}\PYG{n}{binary\PYGZus{}repr}\PYG{p}{(}\PYG{o}{\PYGZhy{}}\PYG{l+m+mi}{14}\PYG{p}{,} \PYG{n}{width}\PYG{o}{=}\PYG{l+m+mi}{8}\PYG{p}{)}
\PYG{g+go}{\PYGZsq{}11110010\PYGZsq{}}
\end{sphinxVerbatim}

Booleans are accepted as well:

\begin{sphinxVerbatim}[commandchars=\\\{\}]
\PYG{g+gp}{\PYGZgt{}\PYGZgt{}\PYGZgt{} }\PYG{n}{np}\PYG{o}{.}\PYG{n}{invert}\PYG{p}{(}\PYG{n}{np}\PYG{o}{.}\PYG{n}{array}\PYG{p}{(}\PYG{p}{[}\PYG{k+kc}{True}\PYG{p}{,} \PYG{k+kc}{False}\PYG{p}{]}\PYG{p}{)}\PYG{p}{)}
\PYG{g+go}{array([False,  True])}
\end{sphinxVerbatim}

\end{fulllineitems}

\index{bitwise\_or() (in module symjax.tensor)@\spxentry{bitwise\_or()}\spxextra{in module symjax.tensor}}

\begin{fulllineitems}
\phantomsection\label{\detokenize{modules/tensor:symjax.tensor.bitwise_or}}\pysiglinewithargsret{\sphinxbfcode{\sphinxupquote{bitwise\_or}}}{\emph{\DUrole{n}{x1}}, \emph{\DUrole{n}{x2}}}{}
Compute the bit\sphinxhyphen{}wise OR of two arrays element\sphinxhyphen{}wise.

LAX\sphinxhyphen{}backend implementation of {\hyperref[\detokenize{modules/tensor:symjax.tensor.bitwise_or}]{\sphinxcrossref{\sphinxcode{\sphinxupquote{bitwise\_or()}}}}}.
ADDITIONOriginal docstring below.

LAX\sphinxhyphen{}backend implementation of {\hyperref[\detokenize{modules/tensor:symjax.tensor.bitwise_or}]{\sphinxcrossref{\sphinxcode{\sphinxupquote{bitwise\_or()}}}}}.
Original docstring below.

bitwise\_or(x1, x2, /, out=None, {\color{red}\bfseries{}*}, where=True, casting=’same\_kind’, order=’K’, dtype=None, subok=True{[}, signature, extobj{]})

Computes the bit\sphinxhyphen{}wise OR of the underlying binary representation of
the integers in the input arrays. This ufunc implements the C/Python
operator \sphinxcode{\sphinxupquote{|}}.
\begin{quote}\begin{description}
\item[{Returns}] \leavevmode
\sphinxstylestrong{out} \textendash{} Result.
This is a scalar if both \sphinxtitleref{x1} and \sphinxtitleref{x2} are scalars.

\item[{Return type}] \leavevmode
ndarray or scalar

\end{description}\end{quote}

\sphinxstrong{See also:}

{\hyperref[\detokenize{modules/tensor:symjax.tensor.logical_or}]{\sphinxcrossref{\sphinxcode{\sphinxupquote{logical\_or()}}}}}, {\hyperref[\detokenize{modules/tensor:symjax.tensor.bitwise_and}]{\sphinxcrossref{\sphinxcode{\sphinxupquote{bitwise\_and()}}}}}, {\hyperref[\detokenize{modules/tensor:symjax.tensor.bitwise_xor}]{\sphinxcrossref{\sphinxcode{\sphinxupquote{bitwise\_xor()}}}}}
\begin{description}
\item[{\sphinxcode{\sphinxupquote{binary\_repr()}}}] \leavevmode
Return the binary representation of the input number as a string.

\end{description}

\subsubsection*{Examples}

The number 13 has the binaray representation \sphinxcode{\sphinxupquote{00001101}}. Likewise,
16 is represented by \sphinxcode{\sphinxupquote{00010000}}.  The bit\sphinxhyphen{}wise OR of 13 and 16 is
then \sphinxcode{\sphinxupquote{000111011}}, or 29:

\begin{sphinxVerbatim}[commandchars=\\\{\}]
\PYG{g+gp}{\PYGZgt{}\PYGZgt{}\PYGZgt{} }\PYG{n}{np}\PYG{o}{.}\PYG{n}{bitwise\PYGZus{}or}\PYG{p}{(}\PYG{l+m+mi}{13}\PYG{p}{,} \PYG{l+m+mi}{16}\PYG{p}{)}
\PYG{g+go}{29}
\PYG{g+gp}{\PYGZgt{}\PYGZgt{}\PYGZgt{} }\PYG{n}{np}\PYG{o}{.}\PYG{n}{binary\PYGZus{}repr}\PYG{p}{(}\PYG{l+m+mi}{29}\PYG{p}{)}
\PYG{g+go}{\PYGZsq{}11101\PYGZsq{}}
\end{sphinxVerbatim}

\begin{sphinxVerbatim}[commandchars=\\\{\}]
\PYG{g+gp}{\PYGZgt{}\PYGZgt{}\PYGZgt{} }\PYG{n}{np}\PYG{o}{.}\PYG{n}{bitwise\PYGZus{}or}\PYG{p}{(}\PYG{l+m+mi}{32}\PYG{p}{,} \PYG{l+m+mi}{2}\PYG{p}{)}
\PYG{g+go}{34}
\PYG{g+gp}{\PYGZgt{}\PYGZgt{}\PYGZgt{} }\PYG{n}{np}\PYG{o}{.}\PYG{n}{bitwise\PYGZus{}or}\PYG{p}{(}\PYG{p}{[}\PYG{l+m+mi}{33}\PYG{p}{,} \PYG{l+m+mi}{4}\PYG{p}{]}\PYG{p}{,} \PYG{l+m+mi}{1}\PYG{p}{)}
\PYG{g+go}{array([33,  5])}
\PYG{g+gp}{\PYGZgt{}\PYGZgt{}\PYGZgt{} }\PYG{n}{np}\PYG{o}{.}\PYG{n}{bitwise\PYGZus{}or}\PYG{p}{(}\PYG{p}{[}\PYG{l+m+mi}{33}\PYG{p}{,} \PYG{l+m+mi}{4}\PYG{p}{]}\PYG{p}{,} \PYG{p}{[}\PYG{l+m+mi}{1}\PYG{p}{,} \PYG{l+m+mi}{2}\PYG{p}{]}\PYG{p}{)}
\PYG{g+go}{array([33,  6])}
\end{sphinxVerbatim}

\begin{sphinxVerbatim}[commandchars=\\\{\}]
\PYG{g+gp}{\PYGZgt{}\PYGZgt{}\PYGZgt{} }\PYG{n}{np}\PYG{o}{.}\PYG{n}{bitwise\PYGZus{}or}\PYG{p}{(}\PYG{n}{np}\PYG{o}{.}\PYG{n}{array}\PYG{p}{(}\PYG{p}{[}\PYG{l+m+mi}{2}\PYG{p}{,} \PYG{l+m+mi}{5}\PYG{p}{,} \PYG{l+m+mi}{255}\PYG{p}{]}\PYG{p}{)}\PYG{p}{,} \PYG{n}{np}\PYG{o}{.}\PYG{n}{array}\PYG{p}{(}\PYG{p}{[}\PYG{l+m+mi}{4}\PYG{p}{,} \PYG{l+m+mi}{4}\PYG{p}{,} \PYG{l+m+mi}{4}\PYG{p}{]}\PYG{p}{)}\PYG{p}{)}
\PYG{g+go}{array([  6,   5, 255])}
\PYG{g+gp}{\PYGZgt{}\PYGZgt{}\PYGZgt{} }\PYG{n}{np}\PYG{o}{.}\PYG{n}{array}\PYG{p}{(}\PYG{p}{[}\PYG{l+m+mi}{2}\PYG{p}{,} \PYG{l+m+mi}{5}\PYG{p}{,} \PYG{l+m+mi}{255}\PYG{p}{]}\PYG{p}{)} \PYG{o}{|} \PYG{n}{np}\PYG{o}{.}\PYG{n}{array}\PYG{p}{(}\PYG{p}{[}\PYG{l+m+mi}{4}\PYG{p}{,} \PYG{l+m+mi}{4}\PYG{p}{,} \PYG{l+m+mi}{4}\PYG{p}{]}\PYG{p}{)}
\PYG{g+go}{array([  6,   5, 255])}
\PYG{g+gp}{\PYGZgt{}\PYGZgt{}\PYGZgt{} }\PYG{n}{np}\PYG{o}{.}\PYG{n}{bitwise\PYGZus{}or}\PYG{p}{(}\PYG{n}{np}\PYG{o}{.}\PYG{n}{array}\PYG{p}{(}\PYG{p}{[}\PYG{l+m+mi}{2}\PYG{p}{,} \PYG{l+m+mi}{5}\PYG{p}{,} \PYG{l+m+mi}{255}\PYG{p}{,} \PYG{l+m+mi}{2147483647}\PYG{p}{]}\PYG{p}{,} \PYG{n}{dtype}\PYG{o}{=}\PYG{n}{np}\PYG{o}{.}\PYG{n}{int32}\PYG{p}{)}\PYG{p}{,}
\PYG{g+gp}{... }              \PYG{n}{np}\PYG{o}{.}\PYG{n}{array}\PYG{p}{(}\PYG{p}{[}\PYG{l+m+mi}{4}\PYG{p}{,} \PYG{l+m+mi}{4}\PYG{p}{,} \PYG{l+m+mi}{4}\PYG{p}{,} \PYG{l+m+mi}{2147483647}\PYG{p}{]}\PYG{p}{,} \PYG{n}{dtype}\PYG{o}{=}\PYG{n}{np}\PYG{o}{.}\PYG{n}{int32}\PYG{p}{)}\PYG{p}{)}
\PYG{g+go}{array([         6,          5,        255, 2147483647])}
\PYG{g+gp}{\PYGZgt{}\PYGZgt{}\PYGZgt{} }\PYG{n}{np}\PYG{o}{.}\PYG{n}{bitwise\PYGZus{}or}\PYG{p}{(}\PYG{p}{[}\PYG{k+kc}{True}\PYG{p}{,} \PYG{k+kc}{True}\PYG{p}{]}\PYG{p}{,} \PYG{p}{[}\PYG{k+kc}{False}\PYG{p}{,} \PYG{k+kc}{True}\PYG{p}{]}\PYG{p}{)}
\PYG{g+go}{array([ True,  True])}
\end{sphinxVerbatim}

\end{fulllineitems}

\index{bitwise\_xor() (in module symjax.tensor)@\spxentry{bitwise\_xor()}\spxextra{in module symjax.tensor}}

\begin{fulllineitems}
\phantomsection\label{\detokenize{modules/tensor:symjax.tensor.bitwise_xor}}\pysiglinewithargsret{\sphinxbfcode{\sphinxupquote{bitwise\_xor}}}{\emph{\DUrole{n}{x1}}, \emph{\DUrole{n}{x2}}}{}
Compute the bit\sphinxhyphen{}wise XOR of two arrays element\sphinxhyphen{}wise.

LAX\sphinxhyphen{}backend implementation of {\hyperref[\detokenize{modules/tensor:symjax.tensor.bitwise_xor}]{\sphinxcrossref{\sphinxcode{\sphinxupquote{bitwise\_xor()}}}}}.
ADDITIONOriginal docstring below.

LAX\sphinxhyphen{}backend implementation of {\hyperref[\detokenize{modules/tensor:symjax.tensor.bitwise_xor}]{\sphinxcrossref{\sphinxcode{\sphinxupquote{bitwise\_xor()}}}}}.
Original docstring below.

bitwise\_xor(x1, x2, /, out=None, {\color{red}\bfseries{}*}, where=True, casting=’same\_kind’, order=’K’, dtype=None, subok=True{[}, signature, extobj{]})

Computes the bit\sphinxhyphen{}wise XOR of the underlying binary representation of
the integers in the input arrays. This ufunc implements the C/Python
operator \sphinxcode{\sphinxupquote{\textasciicircum{}}}.
\begin{quote}\begin{description}
\item[{Returns}] \leavevmode
\sphinxstylestrong{out} \textendash{} Result.
This is a scalar if both \sphinxtitleref{x1} and \sphinxtitleref{x2} are scalars.

\item[{Return type}] \leavevmode
ndarray or scalar

\end{description}\end{quote}

\sphinxstrong{See also:}

{\hyperref[\detokenize{modules/tensor:symjax.tensor.logical_xor}]{\sphinxcrossref{\sphinxcode{\sphinxupquote{logical\_xor()}}}}}, {\hyperref[\detokenize{modules/tensor:symjax.tensor.bitwise_and}]{\sphinxcrossref{\sphinxcode{\sphinxupquote{bitwise\_and()}}}}}, {\hyperref[\detokenize{modules/tensor:symjax.tensor.bitwise_or}]{\sphinxcrossref{\sphinxcode{\sphinxupquote{bitwise\_or()}}}}}
\begin{description}
\item[{\sphinxcode{\sphinxupquote{binary\_repr()}}}] \leavevmode
Return the binary representation of the input number as a string.

\end{description}

\subsubsection*{Examples}

The number 13 is represented by \sphinxcode{\sphinxupquote{00001101}}. Likewise, 17 is
represented by \sphinxcode{\sphinxupquote{00010001}}.  The bit\sphinxhyphen{}wise XOR of 13 and 17 is
therefore \sphinxcode{\sphinxupquote{00011100}}, or 28:

\begin{sphinxVerbatim}[commandchars=\\\{\}]
\PYG{g+gp}{\PYGZgt{}\PYGZgt{}\PYGZgt{} }\PYG{n}{np}\PYG{o}{.}\PYG{n}{bitwise\PYGZus{}xor}\PYG{p}{(}\PYG{l+m+mi}{13}\PYG{p}{,} \PYG{l+m+mi}{17}\PYG{p}{)}
\PYG{g+go}{28}
\PYG{g+gp}{\PYGZgt{}\PYGZgt{}\PYGZgt{} }\PYG{n}{np}\PYG{o}{.}\PYG{n}{binary\PYGZus{}repr}\PYG{p}{(}\PYG{l+m+mi}{28}\PYG{p}{)}
\PYG{g+go}{\PYGZsq{}11100\PYGZsq{}}
\end{sphinxVerbatim}

\begin{sphinxVerbatim}[commandchars=\\\{\}]
\PYG{g+gp}{\PYGZgt{}\PYGZgt{}\PYGZgt{} }\PYG{n}{np}\PYG{o}{.}\PYG{n}{bitwise\PYGZus{}xor}\PYG{p}{(}\PYG{l+m+mi}{31}\PYG{p}{,} \PYG{l+m+mi}{5}\PYG{p}{)}
\PYG{g+go}{26}
\PYG{g+gp}{\PYGZgt{}\PYGZgt{}\PYGZgt{} }\PYG{n}{np}\PYG{o}{.}\PYG{n}{bitwise\PYGZus{}xor}\PYG{p}{(}\PYG{p}{[}\PYG{l+m+mi}{31}\PYG{p}{,}\PYG{l+m+mi}{3}\PYG{p}{]}\PYG{p}{,} \PYG{l+m+mi}{5}\PYG{p}{)}
\PYG{g+go}{array([26,  6])}
\end{sphinxVerbatim}

\begin{sphinxVerbatim}[commandchars=\\\{\}]
\PYG{g+gp}{\PYGZgt{}\PYGZgt{}\PYGZgt{} }\PYG{n}{np}\PYG{o}{.}\PYG{n}{bitwise\PYGZus{}xor}\PYG{p}{(}\PYG{p}{[}\PYG{l+m+mi}{31}\PYG{p}{,}\PYG{l+m+mi}{3}\PYG{p}{]}\PYG{p}{,} \PYG{p}{[}\PYG{l+m+mi}{5}\PYG{p}{,}\PYG{l+m+mi}{6}\PYG{p}{]}\PYG{p}{)}
\PYG{g+go}{array([26,  5])}
\PYG{g+gp}{\PYGZgt{}\PYGZgt{}\PYGZgt{} }\PYG{n}{np}\PYG{o}{.}\PYG{n}{bitwise\PYGZus{}xor}\PYG{p}{(}\PYG{p}{[}\PYG{k+kc}{True}\PYG{p}{,} \PYG{k+kc}{True}\PYG{p}{]}\PYG{p}{,} \PYG{p}{[}\PYG{k+kc}{False}\PYG{p}{,} \PYG{k+kc}{True}\PYG{p}{]}\PYG{p}{)}
\PYG{g+go}{array([ True, False])}
\end{sphinxVerbatim}

\end{fulllineitems}

\index{block() (in module symjax.tensor)@\spxentry{block()}\spxextra{in module symjax.tensor}}

\begin{fulllineitems}
\phantomsection\label{\detokenize{modules/tensor:symjax.tensor.block}}\pysiglinewithargsret{\sphinxbfcode{\sphinxupquote{block}}}{\emph{\DUrole{n}{arrays}}}{}
Assemble an nd\sphinxhyphen{}array from nested lists of blocks.

LAX\sphinxhyphen{}backend implementation of {\hyperref[\detokenize{modules/tensor:symjax.tensor.block}]{\sphinxcrossref{\sphinxcode{\sphinxupquote{block()}}}}}.
ADDITIONOriginal docstring below.

LAX\sphinxhyphen{}backend implementation of {\hyperref[\detokenize{modules/tensor:symjax.tensor.block}]{\sphinxcrossref{\sphinxcode{\sphinxupquote{block()}}}}}.
Original docstring below.

Blocks in the innermost lists are concatenated (see \sphinxtitleref{concatenate}) along
the last dimension (\sphinxhyphen{}1), then these are concatenated along the
second\sphinxhyphen{}last dimension (\sphinxhyphen{}2), and so on until the outermost list is reached.

Blocks can be of any dimension, but will not be broadcasted using the normal
rules. Instead, leading axes of size 1 are inserted, to make \sphinxcode{\sphinxupquote{block.ndim}}
the same for all blocks. This is primarily useful for working with scalars,
and means that code like \sphinxcode{\sphinxupquote{np.block({[}v, 1{]})}} is valid, where
\sphinxcode{\sphinxupquote{v.ndim == 1}}.

When the nested list is two levels deep, this allows block matrices to be
constructed from their components.

\DUrole{versionmodified,added}{New in version 1.13.0.}
\begin{quote}\begin{description}
\item[{Returns}] \leavevmode

\sphinxstylestrong{block\_array} \textendash{} The array assembled from the given blocks.

The dimensionality of the output is equal to the greatest of:
* the dimensionality of all the inputs
* the depth to which the input list is nested

\item[{Return type}] \leavevmode
ndarray

\item[{Raises}] \leavevmode
\sphinxstyleliteralstrong{\sphinxupquote{ValueError}} \textendash{} \begin{itemize}
\item {} 
If list depths are mismatched \sphinxhyphen{} for instance, \sphinxcode{\sphinxupquote{{[}{[}a, b{]}, c{]}}} is
      illegal, and should be spelt \sphinxcode{\sphinxupquote{{[}{[}a, b{]}, {[}c{]}{]}}}
    * If lists are empty \sphinxhyphen{} for instance, \sphinxcode{\sphinxupquote{{[}{[}a, b{]}, {[}{]}{]}}}

\end{itemize}

\end{description}\end{quote}

\sphinxstrong{See also:}

\begin{description}
\item[{{\hyperref[\detokenize{modules/tensor:symjax.tensor.concatenate}]{\sphinxcrossref{\sphinxcode{\sphinxupquote{concatenate()}}}}}}] \leavevmode
Join a sequence of arrays together.

\item[{{\hyperref[\detokenize{modules/tensor:symjax.tensor.stack}]{\sphinxcrossref{\sphinxcode{\sphinxupquote{stack()}}}}}}] \leavevmode
Stack arrays in sequence along a new dimension.

\item[{{\hyperref[\detokenize{modules/tensor:symjax.tensor.hstack}]{\sphinxcrossref{\sphinxcode{\sphinxupquote{hstack()}}}}}}] \leavevmode
Stack arrays in sequence horizontally (column wise).

\item[{{\hyperref[\detokenize{modules/tensor:symjax.tensor.vstack}]{\sphinxcrossref{\sphinxcode{\sphinxupquote{vstack()}}}}}}] \leavevmode
Stack arrays in sequence vertically (row wise).

\item[{{\hyperref[\detokenize{modules/tensor:symjax.tensor.dstack}]{\sphinxcrossref{\sphinxcode{\sphinxupquote{dstack()}}}}}}] \leavevmode
Stack arrays in sequence depth wise (along third dimension).

\item[{{\hyperref[\detokenize{modules/tensor:symjax.tensor.vsplit}]{\sphinxcrossref{\sphinxcode{\sphinxupquote{vsplit()}}}}}}] \leavevmode
Split array into a list of multiple sub\sphinxhyphen{}arrays vertically.

\end{description}

\subsubsection*{Notes}

When called with only scalars, \sphinxcode{\sphinxupquote{np.block}} is equivalent to an ndarray
call. So \sphinxcode{\sphinxupquote{np.block({[}{[}1, 2{]}, {[}3, 4{]}{]})}} is equivalent to
\sphinxcode{\sphinxupquote{np.array({[}{[}1, 2{]}, {[}3, 4{]}{]})}}.

This function does not enforce that the blocks lie on a fixed grid.
\sphinxcode{\sphinxupquote{np.block({[}{[}a, b{]}, {[}c, d{]}{]})}} is not restricted to arrays of the form:

\begin{sphinxVerbatim}[commandchars=\\\{\}]
\PYG{n}{AAAbb}
\PYG{n}{AAAbb}
\PYG{n}{cccDD}
\end{sphinxVerbatim}

But is also allowed to produce, for some \sphinxcode{\sphinxupquote{a, b, c, d}}:

\begin{sphinxVerbatim}[commandchars=\\\{\}]
\PYG{n}{AAAbb}
\PYG{n}{AAAbb}
\PYG{n}{cDDDD}
\end{sphinxVerbatim}

Since concatenation happens along the last axis first, \sphinxtitleref{block} is \_not\_
capable of producing the following directly:

\begin{sphinxVerbatim}[commandchars=\\\{\}]
\PYG{n}{AAAbb}
\PYG{n}{cccbb}
\PYG{n}{cccDD}
\end{sphinxVerbatim}

Matlab’s “square bracket stacking”, \sphinxcode{\sphinxupquote{{[}A, B, ...; p, q, ...{]}}}, is
equivalent to \sphinxcode{\sphinxupquote{np.block({[}{[}A, B, ...{]}, {[}p, q, ...{]}{]})}}.
\subsubsection*{Examples}

The most common use of this function is to build a block matrix

\begin{sphinxVerbatim}[commandchars=\\\{\}]
\PYG{g+gp}{\PYGZgt{}\PYGZgt{}\PYGZgt{} }\PYG{n}{A} \PYG{o}{=} \PYG{n}{np}\PYG{o}{.}\PYG{n}{eye}\PYG{p}{(}\PYG{l+m+mi}{2}\PYG{p}{)} \PYG{o}{*} \PYG{l+m+mi}{2}
\PYG{g+gp}{\PYGZgt{}\PYGZgt{}\PYGZgt{} }\PYG{n}{B} \PYG{o}{=} \PYG{n}{np}\PYG{o}{.}\PYG{n}{eye}\PYG{p}{(}\PYG{l+m+mi}{3}\PYG{p}{)} \PYG{o}{*} \PYG{l+m+mi}{3}
\PYG{g+gp}{\PYGZgt{}\PYGZgt{}\PYGZgt{} }\PYG{n}{np}\PYG{o}{.}\PYG{n}{block}\PYG{p}{(}\PYG{p}{[}
\PYG{g+gp}{... }    \PYG{p}{[}\PYG{n}{A}\PYG{p}{,}               \PYG{n}{np}\PYG{o}{.}\PYG{n}{zeros}\PYG{p}{(}\PYG{p}{(}\PYG{l+m+mi}{2}\PYG{p}{,} \PYG{l+m+mi}{3}\PYG{p}{)}\PYG{p}{)}\PYG{p}{]}\PYG{p}{,}
\PYG{g+gp}{... }    \PYG{p}{[}\PYG{n}{np}\PYG{o}{.}\PYG{n}{ones}\PYG{p}{(}\PYG{p}{(}\PYG{l+m+mi}{3}\PYG{p}{,} \PYG{l+m+mi}{2}\PYG{p}{)}\PYG{p}{)}\PYG{p}{,} \PYG{n}{B}               \PYG{p}{]}
\PYG{g+gp}{... }\PYG{p}{]}\PYG{p}{)}
\PYG{g+go}{array([[2., 0., 0., 0., 0.],}
\PYG{g+go}{       [0., 2., 0., 0., 0.],}
\PYG{g+go}{       [1., 1., 3., 0., 0.],}
\PYG{g+go}{       [1., 1., 0., 3., 0.],}
\PYG{g+go}{       [1., 1., 0., 0., 3.]])}
\end{sphinxVerbatim}

With a list of depth 1, \sphinxtitleref{block} can be used as \sphinxtitleref{hstack}

\begin{sphinxVerbatim}[commandchars=\\\{\}]
\PYG{g+gp}{\PYGZgt{}\PYGZgt{}\PYGZgt{} }\PYG{n}{np}\PYG{o}{.}\PYG{n}{block}\PYG{p}{(}\PYG{p}{[}\PYG{l+m+mi}{1}\PYG{p}{,} \PYG{l+m+mi}{2}\PYG{p}{,} \PYG{l+m+mi}{3}\PYG{p}{]}\PYG{p}{)}              \PYG{c+c1}{\PYGZsh{} hstack([1, 2, 3])}
\PYG{g+go}{array([1, 2, 3])}
\end{sphinxVerbatim}

\begin{sphinxVerbatim}[commandchars=\\\{\}]
\PYG{g+gp}{\PYGZgt{}\PYGZgt{}\PYGZgt{} }\PYG{n}{a} \PYG{o}{=} \PYG{n}{np}\PYG{o}{.}\PYG{n}{array}\PYG{p}{(}\PYG{p}{[}\PYG{l+m+mi}{1}\PYG{p}{,} \PYG{l+m+mi}{2}\PYG{p}{,} \PYG{l+m+mi}{3}\PYG{p}{]}\PYG{p}{)}
\PYG{g+gp}{\PYGZgt{}\PYGZgt{}\PYGZgt{} }\PYG{n}{b} \PYG{o}{=} \PYG{n}{np}\PYG{o}{.}\PYG{n}{array}\PYG{p}{(}\PYG{p}{[}\PYG{l+m+mi}{2}\PYG{p}{,} \PYG{l+m+mi}{3}\PYG{p}{,} \PYG{l+m+mi}{4}\PYG{p}{]}\PYG{p}{)}
\PYG{g+gp}{\PYGZgt{}\PYGZgt{}\PYGZgt{} }\PYG{n}{np}\PYG{o}{.}\PYG{n}{block}\PYG{p}{(}\PYG{p}{[}\PYG{n}{a}\PYG{p}{,} \PYG{n}{b}\PYG{p}{,} \PYG{l+m+mi}{10}\PYG{p}{]}\PYG{p}{)}             \PYG{c+c1}{\PYGZsh{} hstack([a, b, 10])}
\PYG{g+go}{array([ 1,  2,  3,  2,  3,  4, 10])}
\end{sphinxVerbatim}

\begin{sphinxVerbatim}[commandchars=\\\{\}]
\PYG{g+gp}{\PYGZgt{}\PYGZgt{}\PYGZgt{} }\PYG{n}{A} \PYG{o}{=} \PYG{n}{np}\PYG{o}{.}\PYG{n}{ones}\PYG{p}{(}\PYG{p}{(}\PYG{l+m+mi}{2}\PYG{p}{,} \PYG{l+m+mi}{2}\PYG{p}{)}\PYG{p}{,} \PYG{n+nb}{int}\PYG{p}{)}
\PYG{g+gp}{\PYGZgt{}\PYGZgt{}\PYGZgt{} }\PYG{n}{B} \PYG{o}{=} \PYG{l+m+mi}{2} \PYG{o}{*} \PYG{n}{A}
\PYG{g+gp}{\PYGZgt{}\PYGZgt{}\PYGZgt{} }\PYG{n}{np}\PYG{o}{.}\PYG{n}{block}\PYG{p}{(}\PYG{p}{[}\PYG{n}{A}\PYG{p}{,} \PYG{n}{B}\PYG{p}{]}\PYG{p}{)}                 \PYG{c+c1}{\PYGZsh{} hstack([A, B])}
\PYG{g+go}{array([[1, 1, 2, 2],}
\PYG{g+go}{       [1, 1, 2, 2]])}
\end{sphinxVerbatim}

With a list of depth 2, \sphinxtitleref{block} can be used in place of \sphinxtitleref{vstack}:

\begin{sphinxVerbatim}[commandchars=\\\{\}]
\PYG{g+gp}{\PYGZgt{}\PYGZgt{}\PYGZgt{} }\PYG{n}{a} \PYG{o}{=} \PYG{n}{np}\PYG{o}{.}\PYG{n}{array}\PYG{p}{(}\PYG{p}{[}\PYG{l+m+mi}{1}\PYG{p}{,} \PYG{l+m+mi}{2}\PYG{p}{,} \PYG{l+m+mi}{3}\PYG{p}{]}\PYG{p}{)}
\PYG{g+gp}{\PYGZgt{}\PYGZgt{}\PYGZgt{} }\PYG{n}{b} \PYG{o}{=} \PYG{n}{np}\PYG{o}{.}\PYG{n}{array}\PYG{p}{(}\PYG{p}{[}\PYG{l+m+mi}{2}\PYG{p}{,} \PYG{l+m+mi}{3}\PYG{p}{,} \PYG{l+m+mi}{4}\PYG{p}{]}\PYG{p}{)}
\PYG{g+gp}{\PYGZgt{}\PYGZgt{}\PYGZgt{} }\PYG{n}{np}\PYG{o}{.}\PYG{n}{block}\PYG{p}{(}\PYG{p}{[}\PYG{p}{[}\PYG{n}{a}\PYG{p}{]}\PYG{p}{,} \PYG{p}{[}\PYG{n}{b}\PYG{p}{]}\PYG{p}{]}\PYG{p}{)}             \PYG{c+c1}{\PYGZsh{} vstack([a, b])}
\PYG{g+go}{array([[1, 2, 3],}
\PYG{g+go}{       [2, 3, 4]])}
\end{sphinxVerbatim}

\begin{sphinxVerbatim}[commandchars=\\\{\}]
\PYG{g+gp}{\PYGZgt{}\PYGZgt{}\PYGZgt{} }\PYG{n}{A} \PYG{o}{=} \PYG{n}{np}\PYG{o}{.}\PYG{n}{ones}\PYG{p}{(}\PYG{p}{(}\PYG{l+m+mi}{2}\PYG{p}{,} \PYG{l+m+mi}{2}\PYG{p}{)}\PYG{p}{,} \PYG{n+nb}{int}\PYG{p}{)}
\PYG{g+gp}{\PYGZgt{}\PYGZgt{}\PYGZgt{} }\PYG{n}{B} \PYG{o}{=} \PYG{l+m+mi}{2} \PYG{o}{*} \PYG{n}{A}
\PYG{g+gp}{\PYGZgt{}\PYGZgt{}\PYGZgt{} }\PYG{n}{np}\PYG{o}{.}\PYG{n}{block}\PYG{p}{(}\PYG{p}{[}\PYG{p}{[}\PYG{n}{A}\PYG{p}{]}\PYG{p}{,} \PYG{p}{[}\PYG{n}{B}\PYG{p}{]}\PYG{p}{]}\PYG{p}{)}             \PYG{c+c1}{\PYGZsh{} vstack([A, B])}
\PYG{g+go}{array([[1, 1],}
\PYG{g+go}{       [1, 1],}
\PYG{g+go}{       [2, 2],}
\PYG{g+go}{       [2, 2]])}
\end{sphinxVerbatim}

It can also be used in places of \sphinxtitleref{atleast\_1d} and \sphinxtitleref{atleast\_2d}

\begin{sphinxVerbatim}[commandchars=\\\{\}]
\PYG{g+gp}{\PYGZgt{}\PYGZgt{}\PYGZgt{} }\PYG{n}{a} \PYG{o}{=} \PYG{n}{np}\PYG{o}{.}\PYG{n}{array}\PYG{p}{(}\PYG{l+m+mi}{0}\PYG{p}{)}
\PYG{g+gp}{\PYGZgt{}\PYGZgt{}\PYGZgt{} }\PYG{n}{b} \PYG{o}{=} \PYG{n}{np}\PYG{o}{.}\PYG{n}{array}\PYG{p}{(}\PYG{p}{[}\PYG{l+m+mi}{1}\PYG{p}{]}\PYG{p}{)}
\PYG{g+gp}{\PYGZgt{}\PYGZgt{}\PYGZgt{} }\PYG{n}{np}\PYG{o}{.}\PYG{n}{block}\PYG{p}{(}\PYG{p}{[}\PYG{n}{a}\PYG{p}{]}\PYG{p}{)}                    \PYG{c+c1}{\PYGZsh{} atleast\PYGZus{}1d(a)}
\PYG{g+go}{array([0])}
\PYG{g+gp}{\PYGZgt{}\PYGZgt{}\PYGZgt{} }\PYG{n}{np}\PYG{o}{.}\PYG{n}{block}\PYG{p}{(}\PYG{p}{[}\PYG{n}{b}\PYG{p}{]}\PYG{p}{)}                    \PYG{c+c1}{\PYGZsh{} atleast\PYGZus{}1d(b)}
\PYG{g+go}{array([1])}
\end{sphinxVerbatim}

\begin{sphinxVerbatim}[commandchars=\\\{\}]
\PYG{g+gp}{\PYGZgt{}\PYGZgt{}\PYGZgt{} }\PYG{n}{np}\PYG{o}{.}\PYG{n}{block}\PYG{p}{(}\PYG{p}{[}\PYG{p}{[}\PYG{n}{a}\PYG{p}{]}\PYG{p}{]}\PYG{p}{)}                  \PYG{c+c1}{\PYGZsh{} atleast\PYGZus{}2d(a)}
\PYG{g+go}{array([[0]])}
\PYG{g+gp}{\PYGZgt{}\PYGZgt{}\PYGZgt{} }\PYG{n}{np}\PYG{o}{.}\PYG{n}{block}\PYG{p}{(}\PYG{p}{[}\PYG{p}{[}\PYG{n}{b}\PYG{p}{]}\PYG{p}{]}\PYG{p}{)}                  \PYG{c+c1}{\PYGZsh{} atleast\PYGZus{}2d(b)}
\PYG{g+go}{array([[1]])}
\end{sphinxVerbatim}

\end{fulllineitems}

\index{can\_cast() (in module symjax.tensor)@\spxentry{can\_cast()}\spxextra{in module symjax.tensor}}

\begin{fulllineitems}
\phantomsection\label{\detokenize{modules/tensor:symjax.tensor.can_cast}}\pysiglinewithargsret{\sphinxbfcode{\sphinxupquote{can\_cast}}}{}{}~\begin{description}
\item[{Returns True if cast between data types can occur according to the}] \leavevmode
casting rule.  If from is a scalar or array scalar, also returns
True if the scalar value can be cast without overflow or truncation
to an integer.

\end{description}

LAX\sphinxhyphen{}backend implementation of {\hyperref[\detokenize{modules/tensor:symjax.tensor.can_cast}]{\sphinxcrossref{\sphinxcode{\sphinxupquote{can\_cast()}}}}}.
ADDITIONOriginal docstring below.
\begin{quote}
\begin{quote}

can\_cast({\color{red}\bfseries{}from\_}, to, casting=’safe’)
\end{quote}
\begin{description}
\item[{Returns}] \leavevmode\begin{description}
\item[{out}] \leavevmode{[}bool{]}
True if cast can occur according to the casting rule.

\end{description}

\DUrole{versionmodified,changed}{Changed in version 1.17.0: }Casting between a simple data type and a structured one is possible only
for “unsafe” casting.  Casting to multiple fields is allowed, but
casting from multiple fields is not.

\DUrole{versionmodified,changed}{Changed in version 1.9.0: }Casting from numeric to string types in ‘safe’ casting mode requires
that the string dtype length is long enough to store the maximum
integer/float value converted.

dtype, result\_type

Basic examples

\begin{sphinxVerbatim}[commandchars=\\\{\}]
\PYG{g+gp}{\PYGZgt{}\PYGZgt{}\PYGZgt{} }\PYG{n}{np}\PYG{o}{.}\PYG{n}{can\PYGZus{}cast}\PYG{p}{(}\PYG{n}{np}\PYG{o}{.}\PYG{n}{int32}\PYG{p}{,} \PYG{n}{np}\PYG{o}{.}\PYG{n}{int64}\PYG{p}{)}
\PYG{g+go}{True}
\PYG{g+gp}{\PYGZgt{}\PYGZgt{}\PYGZgt{} }\PYG{n}{np}\PYG{o}{.}\PYG{n}{can\PYGZus{}cast}\PYG{p}{(}\PYG{n}{np}\PYG{o}{.}\PYG{n}{float64}\PYG{p}{,} \PYG{n+nb}{complex}\PYG{p}{)}
\PYG{g+go}{True}
\PYG{g+gp}{\PYGZgt{}\PYGZgt{}\PYGZgt{} }\PYG{n}{np}\PYG{o}{.}\PYG{n}{can\PYGZus{}cast}\PYG{p}{(}\PYG{n+nb}{complex}\PYG{p}{,} \PYG{n+nb}{float}\PYG{p}{)}
\PYG{g+go}{False}
\end{sphinxVerbatim}

\begin{sphinxVerbatim}[commandchars=\\\{\}]
\PYG{g+gp}{\PYGZgt{}\PYGZgt{}\PYGZgt{} }\PYG{n}{np}\PYG{o}{.}\PYG{n}{can\PYGZus{}cast}\PYG{p}{(}\PYG{l+s+s1}{\PYGZsq{}}\PYG{l+s+s1}{i8}\PYG{l+s+s1}{\PYGZsq{}}\PYG{p}{,} \PYG{l+s+s1}{\PYGZsq{}}\PYG{l+s+s1}{f8}\PYG{l+s+s1}{\PYGZsq{}}\PYG{p}{)}
\PYG{g+go}{True}
\PYG{g+gp}{\PYGZgt{}\PYGZgt{}\PYGZgt{} }\PYG{n}{np}\PYG{o}{.}\PYG{n}{can\PYGZus{}cast}\PYG{p}{(}\PYG{l+s+s1}{\PYGZsq{}}\PYG{l+s+s1}{i8}\PYG{l+s+s1}{\PYGZsq{}}\PYG{p}{,} \PYG{l+s+s1}{\PYGZsq{}}\PYG{l+s+s1}{f4}\PYG{l+s+s1}{\PYGZsq{}}\PYG{p}{)}
\PYG{g+go}{False}
\PYG{g+gp}{\PYGZgt{}\PYGZgt{}\PYGZgt{} }\PYG{n}{np}\PYG{o}{.}\PYG{n}{can\PYGZus{}cast}\PYG{p}{(}\PYG{l+s+s1}{\PYGZsq{}}\PYG{l+s+s1}{i4}\PYG{l+s+s1}{\PYGZsq{}}\PYG{p}{,} \PYG{l+s+s1}{\PYGZsq{}}\PYG{l+s+s1}{S4}\PYG{l+s+s1}{\PYGZsq{}}\PYG{p}{)}
\PYG{g+go}{False}
\end{sphinxVerbatim}

Casting scalars

\begin{sphinxVerbatim}[commandchars=\\\{\}]
\PYG{g+gp}{\PYGZgt{}\PYGZgt{}\PYGZgt{} }\PYG{n}{np}\PYG{o}{.}\PYG{n}{can\PYGZus{}cast}\PYG{p}{(}\PYG{l+m+mi}{100}\PYG{p}{,} \PYG{l+s+s1}{\PYGZsq{}}\PYG{l+s+s1}{i1}\PYG{l+s+s1}{\PYGZsq{}}\PYG{p}{)}
\PYG{g+go}{True}
\PYG{g+gp}{\PYGZgt{}\PYGZgt{}\PYGZgt{} }\PYG{n}{np}\PYG{o}{.}\PYG{n}{can\PYGZus{}cast}\PYG{p}{(}\PYG{l+m+mi}{150}\PYG{p}{,} \PYG{l+s+s1}{\PYGZsq{}}\PYG{l+s+s1}{i1}\PYG{l+s+s1}{\PYGZsq{}}\PYG{p}{)}
\PYG{g+go}{False}
\PYG{g+gp}{\PYGZgt{}\PYGZgt{}\PYGZgt{} }\PYG{n}{np}\PYG{o}{.}\PYG{n}{can\PYGZus{}cast}\PYG{p}{(}\PYG{l+m+mi}{150}\PYG{p}{,} \PYG{l+s+s1}{\PYGZsq{}}\PYG{l+s+s1}{u1}\PYG{l+s+s1}{\PYGZsq{}}\PYG{p}{)}
\PYG{g+go}{True}
\end{sphinxVerbatim}

\begin{sphinxVerbatim}[commandchars=\\\{\}]
\PYG{g+gp}{\PYGZgt{}\PYGZgt{}\PYGZgt{} }\PYG{n}{np}\PYG{o}{.}\PYG{n}{can\PYGZus{}cast}\PYG{p}{(}\PYG{l+m+mf}{3.5e100}\PYG{p}{,} \PYG{n}{np}\PYG{o}{.}\PYG{n}{float32}\PYG{p}{)}
\PYG{g+go}{False}
\PYG{g+gp}{\PYGZgt{}\PYGZgt{}\PYGZgt{} }\PYG{n}{np}\PYG{o}{.}\PYG{n}{can\PYGZus{}cast}\PYG{p}{(}\PYG{l+m+mf}{1000.0}\PYG{p}{,} \PYG{n}{np}\PYG{o}{.}\PYG{n}{float32}\PYG{p}{)}
\PYG{g+go}{True}
\end{sphinxVerbatim}

Array scalar checks the value, array does not

\begin{sphinxVerbatim}[commandchars=\\\{\}]
\PYG{g+gp}{\PYGZgt{}\PYGZgt{}\PYGZgt{} }\PYG{n}{np}\PYG{o}{.}\PYG{n}{can\PYGZus{}cast}\PYG{p}{(}\PYG{n}{np}\PYG{o}{.}\PYG{n}{array}\PYG{p}{(}\PYG{l+m+mf}{1000.0}\PYG{p}{)}\PYG{p}{,} \PYG{n}{np}\PYG{o}{.}\PYG{n}{float32}\PYG{p}{)}
\PYG{g+go}{True}
\PYG{g+gp}{\PYGZgt{}\PYGZgt{}\PYGZgt{} }\PYG{n}{np}\PYG{o}{.}\PYG{n}{can\PYGZus{}cast}\PYG{p}{(}\PYG{n}{np}\PYG{o}{.}\PYG{n}{array}\PYG{p}{(}\PYG{p}{[}\PYG{l+m+mf}{1000.0}\PYG{p}{]}\PYG{p}{)}\PYG{p}{,} \PYG{n}{np}\PYG{o}{.}\PYG{n}{float32}\PYG{p}{)}
\PYG{g+go}{False}
\end{sphinxVerbatim}

Using the casting rules

\begin{sphinxVerbatim}[commandchars=\\\{\}]
\PYG{g+gp}{\PYGZgt{}\PYGZgt{}\PYGZgt{} }\PYG{n}{np}\PYG{o}{.}\PYG{n}{can\PYGZus{}cast}\PYG{p}{(}\PYG{l+s+s1}{\PYGZsq{}}\PYG{l+s+s1}{i8}\PYG{l+s+s1}{\PYGZsq{}}\PYG{p}{,} \PYG{l+s+s1}{\PYGZsq{}}\PYG{l+s+s1}{i8}\PYG{l+s+s1}{\PYGZsq{}}\PYG{p}{,} \PYG{l+s+s1}{\PYGZsq{}}\PYG{l+s+s1}{no}\PYG{l+s+s1}{\PYGZsq{}}\PYG{p}{)}
\PYG{g+go}{True}
\PYG{g+gp}{\PYGZgt{}\PYGZgt{}\PYGZgt{} }\PYG{n}{np}\PYG{o}{.}\PYG{n}{can\PYGZus{}cast}\PYG{p}{(}\PYG{l+s+s1}{\PYGZsq{}}\PYG{l+s+s1}{\PYGZlt{}i8}\PYG{l+s+s1}{\PYGZsq{}}\PYG{p}{,} \PYG{l+s+s1}{\PYGZsq{}}\PYG{l+s+s1}{\PYGZgt{}i8}\PYG{l+s+s1}{\PYGZsq{}}\PYG{p}{,} \PYG{l+s+s1}{\PYGZsq{}}\PYG{l+s+s1}{no}\PYG{l+s+s1}{\PYGZsq{}}\PYG{p}{)}
\PYG{g+go}{False}
\end{sphinxVerbatim}

\begin{sphinxVerbatim}[commandchars=\\\{\}]
\PYG{g+gp}{\PYGZgt{}\PYGZgt{}\PYGZgt{} }\PYG{n}{np}\PYG{o}{.}\PYG{n}{can\PYGZus{}cast}\PYG{p}{(}\PYG{l+s+s1}{\PYGZsq{}}\PYG{l+s+s1}{\PYGZlt{}i8}\PYG{l+s+s1}{\PYGZsq{}}\PYG{p}{,} \PYG{l+s+s1}{\PYGZsq{}}\PYG{l+s+s1}{\PYGZgt{}i8}\PYG{l+s+s1}{\PYGZsq{}}\PYG{p}{,} \PYG{l+s+s1}{\PYGZsq{}}\PYG{l+s+s1}{equiv}\PYG{l+s+s1}{\PYGZsq{}}\PYG{p}{)}
\PYG{g+go}{True}
\PYG{g+gp}{\PYGZgt{}\PYGZgt{}\PYGZgt{} }\PYG{n}{np}\PYG{o}{.}\PYG{n}{can\PYGZus{}cast}\PYG{p}{(}\PYG{l+s+s1}{\PYGZsq{}}\PYG{l+s+s1}{\PYGZlt{}i4}\PYG{l+s+s1}{\PYGZsq{}}\PYG{p}{,} \PYG{l+s+s1}{\PYGZsq{}}\PYG{l+s+s1}{\PYGZgt{}i8}\PYG{l+s+s1}{\PYGZsq{}}\PYG{p}{,} \PYG{l+s+s1}{\PYGZsq{}}\PYG{l+s+s1}{equiv}\PYG{l+s+s1}{\PYGZsq{}}\PYG{p}{)}
\PYG{g+go}{False}
\end{sphinxVerbatim}

\begin{sphinxVerbatim}[commandchars=\\\{\}]
\PYG{g+gp}{\PYGZgt{}\PYGZgt{}\PYGZgt{} }\PYG{n}{np}\PYG{o}{.}\PYG{n}{can\PYGZus{}cast}\PYG{p}{(}\PYG{l+s+s1}{\PYGZsq{}}\PYG{l+s+s1}{\PYGZlt{}i4}\PYG{l+s+s1}{\PYGZsq{}}\PYG{p}{,} \PYG{l+s+s1}{\PYGZsq{}}\PYG{l+s+s1}{\PYGZgt{}i8}\PYG{l+s+s1}{\PYGZsq{}}\PYG{p}{,} \PYG{l+s+s1}{\PYGZsq{}}\PYG{l+s+s1}{safe}\PYG{l+s+s1}{\PYGZsq{}}\PYG{p}{)}
\PYG{g+go}{True}
\PYG{g+gp}{\PYGZgt{}\PYGZgt{}\PYGZgt{} }\PYG{n}{np}\PYG{o}{.}\PYG{n}{can\PYGZus{}cast}\PYG{p}{(}\PYG{l+s+s1}{\PYGZsq{}}\PYG{l+s+s1}{\PYGZlt{}i8}\PYG{l+s+s1}{\PYGZsq{}}\PYG{p}{,} \PYG{l+s+s1}{\PYGZsq{}}\PYG{l+s+s1}{\PYGZgt{}i4}\PYG{l+s+s1}{\PYGZsq{}}\PYG{p}{,} \PYG{l+s+s1}{\PYGZsq{}}\PYG{l+s+s1}{safe}\PYG{l+s+s1}{\PYGZsq{}}\PYG{p}{)}
\PYG{g+go}{False}
\end{sphinxVerbatim}

\begin{sphinxVerbatim}[commandchars=\\\{\}]
\PYG{g+gp}{\PYGZgt{}\PYGZgt{}\PYGZgt{} }\PYG{n}{np}\PYG{o}{.}\PYG{n}{can\PYGZus{}cast}\PYG{p}{(}\PYG{l+s+s1}{\PYGZsq{}}\PYG{l+s+s1}{\PYGZlt{}i8}\PYG{l+s+s1}{\PYGZsq{}}\PYG{p}{,} \PYG{l+s+s1}{\PYGZsq{}}\PYG{l+s+s1}{\PYGZgt{}i4}\PYG{l+s+s1}{\PYGZsq{}}\PYG{p}{,} \PYG{l+s+s1}{\PYGZsq{}}\PYG{l+s+s1}{same\PYGZus{}kind}\PYG{l+s+s1}{\PYGZsq{}}\PYG{p}{)}
\PYG{g+go}{True}
\PYG{g+gp}{\PYGZgt{}\PYGZgt{}\PYGZgt{} }\PYG{n}{np}\PYG{o}{.}\PYG{n}{can\PYGZus{}cast}\PYG{p}{(}\PYG{l+s+s1}{\PYGZsq{}}\PYG{l+s+s1}{\PYGZlt{}i8}\PYG{l+s+s1}{\PYGZsq{}}\PYG{p}{,} \PYG{l+s+s1}{\PYGZsq{}}\PYG{l+s+s1}{\PYGZgt{}u4}\PYG{l+s+s1}{\PYGZsq{}}\PYG{p}{,} \PYG{l+s+s1}{\PYGZsq{}}\PYG{l+s+s1}{same\PYGZus{}kind}\PYG{l+s+s1}{\PYGZsq{}}\PYG{p}{)}
\PYG{g+go}{False}
\end{sphinxVerbatim}

\begin{sphinxVerbatim}[commandchars=\\\{\}]
\PYG{g+gp}{\PYGZgt{}\PYGZgt{}\PYGZgt{} }\PYG{n}{np}\PYG{o}{.}\PYG{n}{can\PYGZus{}cast}\PYG{p}{(}\PYG{l+s+s1}{\PYGZsq{}}\PYG{l+s+s1}{\PYGZlt{}i8}\PYG{l+s+s1}{\PYGZsq{}}\PYG{p}{,} \PYG{l+s+s1}{\PYGZsq{}}\PYG{l+s+s1}{\PYGZgt{}u4}\PYG{l+s+s1}{\PYGZsq{}}\PYG{p}{,} \PYG{l+s+s1}{\PYGZsq{}}\PYG{l+s+s1}{unsafe}\PYG{l+s+s1}{\PYGZsq{}}\PYG{p}{)}
\PYG{g+go}{True}
\end{sphinxVerbatim}

\end{description}
\end{quote}

\end{fulllineitems}

\index{ceil() (in module symjax.tensor)@\spxentry{ceil()}\spxextra{in module symjax.tensor}}

\begin{fulllineitems}
\phantomsection\label{\detokenize{modules/tensor:symjax.tensor.ceil}}\pysiglinewithargsret{\sphinxbfcode{\sphinxupquote{ceil}}}{\emph{\DUrole{n}{x}}}{}
Return the ceiling of the input, element\sphinxhyphen{}wise.

LAX\sphinxhyphen{}backend implementation of {\hyperref[\detokenize{modules/tensor:symjax.tensor.ceil}]{\sphinxcrossref{\sphinxcode{\sphinxupquote{ceil()}}}}}.
ADDITIONOriginal docstring below.

LAX\sphinxhyphen{}backend implementation of {\hyperref[\detokenize{modules/tensor:symjax.tensor.ceil}]{\sphinxcrossref{\sphinxcode{\sphinxupquote{ceil()}}}}}.
Original docstring below.

ceil(x, /, out=None, {\color{red}\bfseries{}*}, where=True, casting=’same\_kind’, order=’K’, dtype=None, subok=True{[}, signature, extobj{]})

The ceil of the scalar \sphinxtitleref{x} is the smallest integer \sphinxtitleref{i}, such that
\sphinxtitleref{i \textgreater{}= x}.  It is often denoted as \(\lceil x \rceil\).
\begin{quote}\begin{description}
\item[{Returns}] \leavevmode
\sphinxstylestrong{y} \textendash{} The ceiling of each element in \sphinxtitleref{x}, with \sphinxtitleref{float} dtype.
This is a scalar if \sphinxtitleref{x} is a scalar.

\item[{Return type}] \leavevmode
ndarray or scalar

\end{description}\end{quote}

\sphinxstrong{See also:}

{\hyperref[\detokenize{modules/tensor:symjax.tensor.floor}]{\sphinxcrossref{\sphinxcode{\sphinxupquote{floor()}}}}}, \sphinxcode{\sphinxupquote{trunc()}}, \sphinxcode{\sphinxupquote{rint()}}

\subsubsection*{Examples}

\begin{sphinxVerbatim}[commandchars=\\\{\}]
\PYG{g+gp}{\PYGZgt{}\PYGZgt{}\PYGZgt{} }\PYG{n}{a} \PYG{o}{=} \PYG{n}{np}\PYG{o}{.}\PYG{n}{array}\PYG{p}{(}\PYG{p}{[}\PYG{o}{\PYGZhy{}}\PYG{l+m+mf}{1.7}\PYG{p}{,} \PYG{o}{\PYGZhy{}}\PYG{l+m+mf}{1.5}\PYG{p}{,} \PYG{o}{\PYGZhy{}}\PYG{l+m+mf}{0.2}\PYG{p}{,} \PYG{l+m+mf}{0.2}\PYG{p}{,} \PYG{l+m+mf}{1.5}\PYG{p}{,} \PYG{l+m+mf}{1.7}\PYG{p}{,} \PYG{l+m+mf}{2.0}\PYG{p}{]}\PYG{p}{)}
\PYG{g+gp}{\PYGZgt{}\PYGZgt{}\PYGZgt{} }\PYG{n}{np}\PYG{o}{.}\PYG{n}{ceil}\PYG{p}{(}\PYG{n}{a}\PYG{p}{)}
\PYG{g+go}{array([\PYGZhy{}1., \PYGZhy{}1., \PYGZhy{}0.,  1.,  2.,  2.,  2.])}
\end{sphinxVerbatim}

\end{fulllineitems}

\index{clip() (in module symjax.tensor)@\spxentry{clip()}\spxextra{in module symjax.tensor}}

\begin{fulllineitems}
\phantomsection\label{\detokenize{modules/tensor:symjax.tensor.clip}}\pysiglinewithargsret{\sphinxbfcode{\sphinxupquote{clip}}}{\emph{\DUrole{n}{a}}, \emph{\DUrole{n}{a\_min}\DUrole{o}{=}\DUrole{default_value}{None}}, \emph{\DUrole{n}{a\_max}\DUrole{o}{=}\DUrole{default_value}{None}}}{}
Clip (limit) the values in an array.

LAX\sphinxhyphen{}backend implementation of {\hyperref[\detokenize{modules/tensor:symjax.tensor.clip}]{\sphinxcrossref{\sphinxcode{\sphinxupquote{clip()}}}}}.
ADDITIONOriginal docstring below.

LAX\sphinxhyphen{}backend implementation of {\hyperref[\detokenize{modules/tensor:symjax.tensor.clip}]{\sphinxcrossref{\sphinxcode{\sphinxupquote{clip()}}}}}.
Original docstring below.

Given an interval, values outside the interval are clipped to
the interval edges.  For example, if an interval of \sphinxcode{\sphinxupquote{{[}0, 1{]}}}
is specified, values smaller than 0 become 0, and values larger
than 1 become 1.

Equivalent to but faster than \sphinxcode{\sphinxupquote{np.maximum(a\_min, np.minimum(a, a\_max))}}.
No check is performed to ensure \sphinxcode{\sphinxupquote{a\_min \textless{} a\_max}}.
\begin{quote}\begin{description}
\item[{Returns}] \leavevmode
\sphinxstylestrong{clipped\_array} \textendash{} An array with the elements of \sphinxtitleref{a}, but where values
\textless{} \sphinxtitleref{a\_min} are replaced with \sphinxtitleref{a\_min}, and those \textgreater{} \sphinxtitleref{a\_max}
with \sphinxtitleref{a\_max}.

\item[{Return type}] \leavevmode
ndarray

\end{description}\end{quote}

\sphinxstrong{See also:}

\sphinxcode{\sphinxupquote{ufuncs\sphinxhyphen{}output\sphinxhyphen{}type()}}

\subsubsection*{Examples}

\begin{sphinxVerbatim}[commandchars=\\\{\}]
\PYG{g+gp}{\PYGZgt{}\PYGZgt{}\PYGZgt{} }\PYG{n}{a} \PYG{o}{=} \PYG{n}{np}\PYG{o}{.}\PYG{n}{arange}\PYG{p}{(}\PYG{l+m+mi}{10}\PYG{p}{)}
\PYG{g+gp}{\PYGZgt{}\PYGZgt{}\PYGZgt{} }\PYG{n}{np}\PYG{o}{.}\PYG{n}{clip}\PYG{p}{(}\PYG{n}{a}\PYG{p}{,} \PYG{l+m+mi}{1}\PYG{p}{,} \PYG{l+m+mi}{8}\PYG{p}{)}
\PYG{g+go}{array([1, 1, 2, 3, 4, 5, 6, 7, 8, 8])}
\PYG{g+gp}{\PYGZgt{}\PYGZgt{}\PYGZgt{} }\PYG{n}{a}
\PYG{g+go}{array([0, 1, 2, 3, 4, 5, 6, 7, 8, 9])}
\PYG{g+gp}{\PYGZgt{}\PYGZgt{}\PYGZgt{} }\PYG{n}{np}\PYG{o}{.}\PYG{n}{clip}\PYG{p}{(}\PYG{n}{a}\PYG{p}{,} \PYG{l+m+mi}{3}\PYG{p}{,} \PYG{l+m+mi}{6}\PYG{p}{,} \PYG{n}{out}\PYG{o}{=}\PYG{n}{a}\PYG{p}{)}
\PYG{g+go}{array([3, 3, 3, 3, 4, 5, 6, 6, 6, 6])}
\PYG{g+gp}{\PYGZgt{}\PYGZgt{}\PYGZgt{} }\PYG{n}{a} \PYG{o}{=} \PYG{n}{np}\PYG{o}{.}\PYG{n}{arange}\PYG{p}{(}\PYG{l+m+mi}{10}\PYG{p}{)}
\PYG{g+gp}{\PYGZgt{}\PYGZgt{}\PYGZgt{} }\PYG{n}{a}
\PYG{g+go}{array([0, 1, 2, 3, 4, 5, 6, 7, 8, 9])}
\PYG{g+gp}{\PYGZgt{}\PYGZgt{}\PYGZgt{} }\PYG{n}{np}\PYG{o}{.}\PYG{n}{clip}\PYG{p}{(}\PYG{n}{a}\PYG{p}{,} \PYG{p}{[}\PYG{l+m+mi}{3}\PYG{p}{,} \PYG{l+m+mi}{4}\PYG{p}{,} \PYG{l+m+mi}{1}\PYG{p}{,} \PYG{l+m+mi}{1}\PYG{p}{,} \PYG{l+m+mi}{1}\PYG{p}{,} \PYG{l+m+mi}{4}\PYG{p}{,} \PYG{l+m+mi}{4}\PYG{p}{,} \PYG{l+m+mi}{4}\PYG{p}{,} \PYG{l+m+mi}{4}\PYG{p}{,} \PYG{l+m+mi}{4}\PYG{p}{]}\PYG{p}{,} \PYG{l+m+mi}{8}\PYG{p}{)}
\PYG{g+go}{array([3, 4, 2, 3, 4, 5, 6, 7, 8, 8])}
\end{sphinxVerbatim}

\end{fulllineitems}

\index{column\_stack() (in module symjax.tensor)@\spxentry{column\_stack()}\spxextra{in module symjax.tensor}}

\begin{fulllineitems}
\phantomsection\label{\detokenize{modules/tensor:symjax.tensor.column_stack}}\pysiglinewithargsret{\sphinxbfcode{\sphinxupquote{column\_stack}}}{\emph{\DUrole{n}{tup}}}{}
Stack 1\sphinxhyphen{}D arrays as columns into a 2\sphinxhyphen{}D array.

LAX\sphinxhyphen{}backend implementation of {\hyperref[\detokenize{modules/tensor:symjax.tensor.column_stack}]{\sphinxcrossref{\sphinxcode{\sphinxupquote{column\_stack()}}}}}.
ADDITIONOriginal docstring below.

LAX\sphinxhyphen{}backend implementation of {\hyperref[\detokenize{modules/tensor:symjax.tensor.column_stack}]{\sphinxcrossref{\sphinxcode{\sphinxupquote{column\_stack()}}}}}.
Original docstring below.

Take a sequence of 1\sphinxhyphen{}D arrays and stack them as columns
to make a single 2\sphinxhyphen{}D array. 2\sphinxhyphen{}D arrays are stacked as\sphinxhyphen{}is,
just like with \sphinxtitleref{hstack}.  1\sphinxhyphen{}D arrays are turned into 2\sphinxhyphen{}D columns
first.
\begin{quote}\begin{description}
\item[{Returns}] \leavevmode
\sphinxstylestrong{stacked} \textendash{} The array formed by stacking the given arrays.

\item[{Return type}] \leavevmode
2\sphinxhyphen{}D array

\end{description}\end{quote}

\sphinxstrong{See also:}

{\hyperref[\detokenize{modules/tensor:symjax.tensor.stack}]{\sphinxcrossref{\sphinxcode{\sphinxupquote{stack()}}}}}, {\hyperref[\detokenize{modules/tensor:symjax.tensor.hstack}]{\sphinxcrossref{\sphinxcode{\sphinxupquote{hstack()}}}}}, {\hyperref[\detokenize{modules/tensor:symjax.tensor.vstack}]{\sphinxcrossref{\sphinxcode{\sphinxupquote{vstack()}}}}}, {\hyperref[\detokenize{modules/tensor:symjax.tensor.concatenate}]{\sphinxcrossref{\sphinxcode{\sphinxupquote{concatenate()}}}}}

\subsubsection*{Examples}

\begin{sphinxVerbatim}[commandchars=\\\{\}]
\PYG{g+gp}{\PYGZgt{}\PYGZgt{}\PYGZgt{} }\PYG{n}{a} \PYG{o}{=} \PYG{n}{np}\PYG{o}{.}\PYG{n}{array}\PYG{p}{(}\PYG{p}{(}\PYG{l+m+mi}{1}\PYG{p}{,}\PYG{l+m+mi}{2}\PYG{p}{,}\PYG{l+m+mi}{3}\PYG{p}{)}\PYG{p}{)}
\PYG{g+gp}{\PYGZgt{}\PYGZgt{}\PYGZgt{} }\PYG{n}{b} \PYG{o}{=} \PYG{n}{np}\PYG{o}{.}\PYG{n}{array}\PYG{p}{(}\PYG{p}{(}\PYG{l+m+mi}{2}\PYG{p}{,}\PYG{l+m+mi}{3}\PYG{p}{,}\PYG{l+m+mi}{4}\PYG{p}{)}\PYG{p}{)}
\PYG{g+gp}{\PYGZgt{}\PYGZgt{}\PYGZgt{} }\PYG{n}{np}\PYG{o}{.}\PYG{n}{column\PYGZus{}stack}\PYG{p}{(}\PYG{p}{(}\PYG{n}{a}\PYG{p}{,}\PYG{n}{b}\PYG{p}{)}\PYG{p}{)}
\PYG{g+go}{array([[1, 2],}
\PYG{g+go}{       [2, 3],}
\PYG{g+go}{       [3, 4]])}
\end{sphinxVerbatim}

\end{fulllineitems}

\index{concatenate() (in module symjax.tensor)@\spxentry{concatenate()}\spxextra{in module symjax.tensor}}

\begin{fulllineitems}
\phantomsection\label{\detokenize{modules/tensor:symjax.tensor.concatenate}}\pysiglinewithargsret{\sphinxbfcode{\sphinxupquote{concatenate}}}{\emph{\DUrole{n}{arrays}}, \emph{\DUrole{n}{axis}\DUrole{o}{=}\DUrole{default_value}{0}}}{}
Join a sequence of arrays along an existing axis.

LAX\sphinxhyphen{}backend implementation of {\hyperref[\detokenize{modules/tensor:symjax.tensor.concatenate}]{\sphinxcrossref{\sphinxcode{\sphinxupquote{concatenate()}}}}}.
ADDITIONOriginal docstring below.

LAX\sphinxhyphen{}backend implementation of {\hyperref[\detokenize{modules/tensor:symjax.tensor.concatenate}]{\sphinxcrossref{\sphinxcode{\sphinxupquote{concatenate()}}}}}.
Original docstring below.
\begin{quote}
\begin{quote}

concatenate((a1, a2, …), axis=0, out=None)
\end{quote}
\begin{description}
\item[{Returns}] \leavevmode\begin{description}
\item[{res}] \leavevmode{[}ndarray{]}
The concatenated array.

\end{description}

ma.concatenate : Concatenate function that preserves input masks.
array\_split : Split an array into multiple sub\sphinxhyphen{}arrays of equal or
\begin{quote}

near\sphinxhyphen{}equal size.
\end{quote}

split : Split array into a list of multiple sub\sphinxhyphen{}arrays of equal size.
hsplit : Split array into multiple sub\sphinxhyphen{}arrays horizontally (column wise)
vsplit : Split array into multiple sub\sphinxhyphen{}arrays vertically (row wise)
dsplit : Split array into multiple sub\sphinxhyphen{}arrays along the 3rd axis (depth).
stack : Stack a sequence of arrays along a new axis.
hstack : Stack arrays in sequence horizontally (column wise)
vstack : Stack arrays in sequence vertically (row wise)
dstack : Stack arrays in sequence depth wise (along third dimension)
block : Assemble arrays from blocks.

When one or more of the arrays to be concatenated is a MaskedArray,
this function will return a MaskedArray object instead of an ndarray,
but the input masks are \sphinxstyleemphasis{not} preserved. In cases where a MaskedArray
is expected as input, use the ma.concatenate function from the masked
array module instead.

\begin{sphinxVerbatim}[commandchars=\\\{\}]
\PYG{g+gp}{\PYGZgt{}\PYGZgt{}\PYGZgt{} }\PYG{n}{a} \PYG{o}{=} \PYG{n}{np}\PYG{o}{.}\PYG{n}{array}\PYG{p}{(}\PYG{p}{[}\PYG{p}{[}\PYG{l+m+mi}{1}\PYG{p}{,} \PYG{l+m+mi}{2}\PYG{p}{]}\PYG{p}{,} \PYG{p}{[}\PYG{l+m+mi}{3}\PYG{p}{,} \PYG{l+m+mi}{4}\PYG{p}{]}\PYG{p}{]}\PYG{p}{)}
\PYG{g+gp}{\PYGZgt{}\PYGZgt{}\PYGZgt{} }\PYG{n}{b} \PYG{o}{=} \PYG{n}{np}\PYG{o}{.}\PYG{n}{array}\PYG{p}{(}\PYG{p}{[}\PYG{p}{[}\PYG{l+m+mi}{5}\PYG{p}{,} \PYG{l+m+mi}{6}\PYG{p}{]}\PYG{p}{]}\PYG{p}{)}
\PYG{g+gp}{\PYGZgt{}\PYGZgt{}\PYGZgt{} }\PYG{n}{np}\PYG{o}{.}\PYG{n}{concatenate}\PYG{p}{(}\PYG{p}{(}\PYG{n}{a}\PYG{p}{,} \PYG{n}{b}\PYG{p}{)}\PYG{p}{,} \PYG{n}{axis}\PYG{o}{=}\PYG{l+m+mi}{0}\PYG{p}{)}
\PYG{g+go}{array([[1, 2],}
\PYG{g+go}{       [3, 4],}
\PYG{g+go}{       [5, 6]])}
\PYG{g+gp}{\PYGZgt{}\PYGZgt{}\PYGZgt{} }\PYG{n}{np}\PYG{o}{.}\PYG{n}{concatenate}\PYG{p}{(}\PYG{p}{(}\PYG{n}{a}\PYG{p}{,} \PYG{n}{b}\PYG{o}{.}\PYG{n}{T}\PYG{p}{)}\PYG{p}{,} \PYG{n}{axis}\PYG{o}{=}\PYG{l+m+mi}{1}\PYG{p}{)}
\PYG{g+go}{array([[1, 2, 5],}
\PYG{g+go}{       [3, 4, 6]])}
\PYG{g+gp}{\PYGZgt{}\PYGZgt{}\PYGZgt{} }\PYG{n}{np}\PYG{o}{.}\PYG{n}{concatenate}\PYG{p}{(}\PYG{p}{(}\PYG{n}{a}\PYG{p}{,} \PYG{n}{b}\PYG{p}{)}\PYG{p}{,} \PYG{n}{axis}\PYG{o}{=}\PYG{k+kc}{None}\PYG{p}{)}
\PYG{g+go}{array([1, 2, 3, 4, 5, 6])}
\end{sphinxVerbatim}

This function will not preserve masking of MaskedArray inputs.

\begin{sphinxVerbatim}[commandchars=\\\{\}]
\PYG{g+gp}{\PYGZgt{}\PYGZgt{}\PYGZgt{} }\PYG{n}{a} \PYG{o}{=} \PYG{n}{np}\PYG{o}{.}\PYG{n}{ma}\PYG{o}{.}\PYG{n}{arange}\PYG{p}{(}\PYG{l+m+mi}{3}\PYG{p}{)}
\PYG{g+gp}{\PYGZgt{}\PYGZgt{}\PYGZgt{} }\PYG{n}{a}\PYG{p}{[}\PYG{l+m+mi}{1}\PYG{p}{]} \PYG{o}{=} \PYG{n}{np}\PYG{o}{.}\PYG{n}{ma}\PYG{o}{.}\PYG{n}{masked}
\PYG{g+gp}{\PYGZgt{}\PYGZgt{}\PYGZgt{} }\PYG{n}{b} \PYG{o}{=} \PYG{n}{np}\PYG{o}{.}\PYG{n}{arange}\PYG{p}{(}\PYG{l+m+mi}{2}\PYG{p}{,} \PYG{l+m+mi}{5}\PYG{p}{)}
\PYG{g+gp}{\PYGZgt{}\PYGZgt{}\PYGZgt{} }\PYG{n}{a}
\PYG{g+go}{masked\PYGZus{}array(data=[0, \PYGZhy{}\PYGZhy{}, 2],}
\PYG{g+go}{             mask=[False,  True, False],}
\PYG{g+go}{       fill\PYGZus{}value=999999)}
\PYG{g+gp}{\PYGZgt{}\PYGZgt{}\PYGZgt{} }\PYG{n}{b}
\PYG{g+go}{array([2, 3, 4])}
\PYG{g+gp}{\PYGZgt{}\PYGZgt{}\PYGZgt{} }\PYG{n}{np}\PYG{o}{.}\PYG{n}{concatenate}\PYG{p}{(}\PYG{p}{[}\PYG{n}{a}\PYG{p}{,} \PYG{n}{b}\PYG{p}{]}\PYG{p}{)}
\PYG{g+go}{masked\PYGZus{}array(data=[0, 1, 2, 2, 3, 4],}
\PYG{g+go}{             mask=False,}
\PYG{g+go}{       fill\PYGZus{}value=999999)}
\PYG{g+gp}{\PYGZgt{}\PYGZgt{}\PYGZgt{} }\PYG{n}{np}\PYG{o}{.}\PYG{n}{ma}\PYG{o}{.}\PYG{n}{concatenate}\PYG{p}{(}\PYG{p}{[}\PYG{n}{a}\PYG{p}{,} \PYG{n}{b}\PYG{p}{]}\PYG{p}{)}
\PYG{g+go}{masked\PYGZus{}array(data=[0, \PYGZhy{}\PYGZhy{}, 2, 2, 3, 4],}
\PYG{g+go}{             mask=[False,  True, False, False, False, False],}
\PYG{g+go}{       fill\PYGZus{}value=999999)}
\end{sphinxVerbatim}

\end{description}
\end{quote}

\end{fulllineitems}

\index{conj() (in module symjax.tensor)@\spxentry{conj()}\spxextra{in module symjax.tensor}}

\begin{fulllineitems}
\phantomsection\label{\detokenize{modules/tensor:symjax.tensor.conj}}\pysiglinewithargsret{\sphinxbfcode{\sphinxupquote{conj}}}{\emph{\DUrole{n}{x}}}{}
Return the complex conjugate, element\sphinxhyphen{}wise.

LAX\sphinxhyphen{}backend implementation of {\hyperref[\detokenize{modules/tensor:symjax.tensor.conjugate}]{\sphinxcrossref{\sphinxcode{\sphinxupquote{conjugate()}}}}}.
ADDITIONOriginal docstring below.

LAX\sphinxhyphen{}backend implementation of {\hyperref[\detokenize{modules/tensor:symjax.tensor.conjugate}]{\sphinxcrossref{\sphinxcode{\sphinxupquote{conjugate()}}}}}.
Original docstring below.

conjugate(x, /, out=None, {\color{red}\bfseries{}*}, where=True, casting=’same\_kind’, order=’K’, dtype=None, subok=True{[}, signature, extobj{]})

The complex conjugate of a complex number is obtained by changing the
sign of its imaginary part.
\begin{quote}\begin{description}
\item[{Returns}] \leavevmode
\sphinxstylestrong{y} \textendash{} The complex conjugate of \sphinxtitleref{x}, with same dtype as \sphinxtitleref{y}.
This is a scalar if \sphinxtitleref{x} is a scalar.

\item[{Return type}] \leavevmode
ndarray

\end{description}\end{quote}
\subsubsection*{Notes}

\sphinxtitleref{conj} is an alias for \sphinxtitleref{conjugate}:

\begin{sphinxVerbatim}[commandchars=\\\{\}]
\PYG{g+gp}{\PYGZgt{}\PYGZgt{}\PYGZgt{} }\PYG{n}{np}\PYG{o}{.}\PYG{n}{conj} \PYG{o+ow}{is} \PYG{n}{np}\PYG{o}{.}\PYG{n}{conjugate}
\PYG{g+go}{True}
\end{sphinxVerbatim}
\subsubsection*{Examples}

\begin{sphinxVerbatim}[commandchars=\\\{\}]
\PYG{g+gp}{\PYGZgt{}\PYGZgt{}\PYGZgt{} }\PYG{n}{np}\PYG{o}{.}\PYG{n}{conjugate}\PYG{p}{(}\PYG{l+m+mi}{1}\PYG{o}{+}\PYG{l+m+mi}{2}\PYG{n}{j}\PYG{p}{)}
\PYG{g+go}{(1\PYGZhy{}2j)}
\end{sphinxVerbatim}

\begin{sphinxVerbatim}[commandchars=\\\{\}]
\PYG{g+gp}{\PYGZgt{}\PYGZgt{}\PYGZgt{} }\PYG{n}{x} \PYG{o}{=} \PYG{n}{np}\PYG{o}{.}\PYG{n}{eye}\PYG{p}{(}\PYG{l+m+mi}{2}\PYG{p}{)} \PYG{o}{+} \PYG{l+m+mi}{1}\PYG{n}{j} \PYG{o}{*} \PYG{n}{np}\PYG{o}{.}\PYG{n}{eye}\PYG{p}{(}\PYG{l+m+mi}{2}\PYG{p}{)}
\PYG{g+gp}{\PYGZgt{}\PYGZgt{}\PYGZgt{} }\PYG{n}{np}\PYG{o}{.}\PYG{n}{conjugate}\PYG{p}{(}\PYG{n}{x}\PYG{p}{)}
\PYG{g+go}{array([[ 1.\PYGZhy{}1.j,  0.\PYGZhy{}0.j],}
\PYG{g+go}{       [ 0.\PYGZhy{}0.j,  1.\PYGZhy{}1.j]])}
\end{sphinxVerbatim}

\end{fulllineitems}

\index{conjugate() (in module symjax.tensor)@\spxentry{conjugate()}\spxextra{in module symjax.tensor}}

\begin{fulllineitems}
\phantomsection\label{\detokenize{modules/tensor:symjax.tensor.conjugate}}\pysiglinewithargsret{\sphinxbfcode{\sphinxupquote{conjugate}}}{\emph{\DUrole{n}{x}}}{}
Return the complex conjugate, element\sphinxhyphen{}wise.

LAX\sphinxhyphen{}backend implementation of {\hyperref[\detokenize{modules/tensor:symjax.tensor.conjugate}]{\sphinxcrossref{\sphinxcode{\sphinxupquote{conjugate()}}}}}.
ADDITIONOriginal docstring below.

LAX\sphinxhyphen{}backend implementation of {\hyperref[\detokenize{modules/tensor:symjax.tensor.conjugate}]{\sphinxcrossref{\sphinxcode{\sphinxupquote{conjugate()}}}}}.
Original docstring below.

conjugate(x, /, out=None, {\color{red}\bfseries{}*}, where=True, casting=’same\_kind’, order=’K’, dtype=None, subok=True{[}, signature, extobj{]})

The complex conjugate of a complex number is obtained by changing the
sign of its imaginary part.
\begin{quote}\begin{description}
\item[{Returns}] \leavevmode
\sphinxstylestrong{y} \textendash{} The complex conjugate of \sphinxtitleref{x}, with same dtype as \sphinxtitleref{y}.
This is a scalar if \sphinxtitleref{x} is a scalar.

\item[{Return type}] \leavevmode
ndarray

\end{description}\end{quote}
\subsubsection*{Notes}

\sphinxtitleref{conj} is an alias for \sphinxtitleref{conjugate}:

\begin{sphinxVerbatim}[commandchars=\\\{\}]
\PYG{g+gp}{\PYGZgt{}\PYGZgt{}\PYGZgt{} }\PYG{n}{np}\PYG{o}{.}\PYG{n}{conj} \PYG{o+ow}{is} \PYG{n}{np}\PYG{o}{.}\PYG{n}{conjugate}
\PYG{g+go}{True}
\end{sphinxVerbatim}
\subsubsection*{Examples}

\begin{sphinxVerbatim}[commandchars=\\\{\}]
\PYG{g+gp}{\PYGZgt{}\PYGZgt{}\PYGZgt{} }\PYG{n}{np}\PYG{o}{.}\PYG{n}{conjugate}\PYG{p}{(}\PYG{l+m+mi}{1}\PYG{o}{+}\PYG{l+m+mi}{2}\PYG{n}{j}\PYG{p}{)}
\PYG{g+go}{(1\PYGZhy{}2j)}
\end{sphinxVerbatim}

\begin{sphinxVerbatim}[commandchars=\\\{\}]
\PYG{g+gp}{\PYGZgt{}\PYGZgt{}\PYGZgt{} }\PYG{n}{x} \PYG{o}{=} \PYG{n}{np}\PYG{o}{.}\PYG{n}{eye}\PYG{p}{(}\PYG{l+m+mi}{2}\PYG{p}{)} \PYG{o}{+} \PYG{l+m+mi}{1}\PYG{n}{j} \PYG{o}{*} \PYG{n}{np}\PYG{o}{.}\PYG{n}{eye}\PYG{p}{(}\PYG{l+m+mi}{2}\PYG{p}{)}
\PYG{g+gp}{\PYGZgt{}\PYGZgt{}\PYGZgt{} }\PYG{n}{np}\PYG{o}{.}\PYG{n}{conjugate}\PYG{p}{(}\PYG{n}{x}\PYG{p}{)}
\PYG{g+go}{array([[ 1.\PYGZhy{}1.j,  0.\PYGZhy{}0.j],}
\PYG{g+go}{       [ 0.\PYGZhy{}0.j,  1.\PYGZhy{}1.j]])}
\end{sphinxVerbatim}

\end{fulllineitems}

\index{corrcoef() (in module symjax.tensor)@\spxentry{corrcoef()}\spxextra{in module symjax.tensor}}

\begin{fulllineitems}
\phantomsection\label{\detokenize{modules/tensor:symjax.tensor.corrcoef}}\pysiglinewithargsret{\sphinxbfcode{\sphinxupquote{corrcoef}}}{\emph{\DUrole{n}{x}}, \emph{\DUrole{n}{y}\DUrole{o}{=}\DUrole{default_value}{None}}, \emph{\DUrole{n}{rowvar}\DUrole{o}{=}\DUrole{default_value}{True}}}{}
Return Pearson product\sphinxhyphen{}moment correlation coefficients.

LAX\sphinxhyphen{}backend implementation of {\hyperref[\detokenize{modules/tensor:symjax.tensor.corrcoef}]{\sphinxcrossref{\sphinxcode{\sphinxupquote{corrcoef()}}}}}.
ADDITIONOriginal docstring below.

LAX\sphinxhyphen{}backend implementation of {\hyperref[\detokenize{modules/tensor:symjax.tensor.corrcoef}]{\sphinxcrossref{\sphinxcode{\sphinxupquote{corrcoef()}}}}}.
Original docstring below.

Please refer to the documentation for \sphinxtitleref{cov} for more detail.  The
relationship between the correlation coefficient matrix, \sphinxtitleref{R}, and the
covariance matrix, \sphinxtitleref{C}, is
\begin{equation*}
\begin{split}R_{ij} = \frac{ C_{ij} } { \sqrt{ C_{ii} * C_{jj} } }\end{split}
\end{equation*}
The values of \sphinxtitleref{R} are between \sphinxhyphen{}1 and 1, inclusive.
\begin{quote}\begin{description}
\item[{Returns}] \leavevmode
\sphinxstylestrong{R} \textendash{} The correlation coefficient matrix of the variables.

\item[{Return type}] \leavevmode
ndarray

\end{description}\end{quote}

\sphinxstrong{See also:}

\begin{description}
\item[{{\hyperref[\detokenize{modules/tensor:symjax.tensor.cov}]{\sphinxcrossref{\sphinxcode{\sphinxupquote{cov()}}}}}}] \leavevmode
Covariance matrix

\end{description}

\subsubsection*{Notes}

Due to floating point rounding the resulting array may not be Hermitian,
the diagonal elements may not be 1, and the elements may not satisfy the
inequality abs(a) \textless{}= 1. The real and imaginary parts are clipped to the
interval {[}\sphinxhyphen{}1,  1{]} in an attempt to improve on that situation but is not
much help in the complex case.

This function accepts but discards arguments \sphinxtitleref{bias} and \sphinxtitleref{ddof}.  This is
for backwards compatibility with previous versions of this function.  These
arguments had no effect on the return values of the function and can be
safely ignored in this and previous versions of numpy.

\end{fulllineitems}

\index{cos() (in module symjax.tensor)@\spxentry{cos()}\spxextra{in module symjax.tensor}}

\begin{fulllineitems}
\phantomsection\label{\detokenize{modules/tensor:symjax.tensor.cos}}\pysiglinewithargsret{\sphinxbfcode{\sphinxupquote{cos}}}{\emph{\DUrole{n}{x}}}{}
Cosine element\sphinxhyphen{}wise.

LAX\sphinxhyphen{}backend implementation of {\hyperref[\detokenize{modules/tensor:symjax.tensor.cos}]{\sphinxcrossref{\sphinxcode{\sphinxupquote{cos()}}}}}.
ADDITIONOriginal docstring below.

LAX\sphinxhyphen{}backend implementation of {\hyperref[\detokenize{modules/tensor:symjax.tensor.cos}]{\sphinxcrossref{\sphinxcode{\sphinxupquote{cos()}}}}}.
Original docstring below.

cos(x, /, out=None, {\color{red}\bfseries{}*}, where=True, casting=’same\_kind’, order=’K’, dtype=None, subok=True{[}, signature, extobj{]})
\begin{quote}\begin{description}
\item[{Returns}] \leavevmode
\sphinxstylestrong{y} \textendash{} The corresponding cosine values.
This is a scalar if \sphinxtitleref{x} is a scalar.

\item[{Return type}] \leavevmode
ndarray

\end{description}\end{quote}
\subsubsection*{Notes}

If \sphinxtitleref{out} is provided, the function writes the result into it,
and returns a reference to \sphinxtitleref{out}.  (See Examples)
\subsubsection*{References}

M. Abramowitz and I. A. Stegun, Handbook of Mathematical Functions.
New York, NY: Dover, 1972.
\subsubsection*{Examples}

\begin{sphinxVerbatim}[commandchars=\\\{\}]
\PYG{g+gp}{\PYGZgt{}\PYGZgt{}\PYGZgt{} }\PYG{n}{np}\PYG{o}{.}\PYG{n}{cos}\PYG{p}{(}\PYG{n}{np}\PYG{o}{.}\PYG{n}{array}\PYG{p}{(}\PYG{p}{[}\PYG{l+m+mi}{0}\PYG{p}{,} \PYG{n}{np}\PYG{o}{.}\PYG{n}{pi}\PYG{o}{/}\PYG{l+m+mi}{2}\PYG{p}{,} \PYG{n}{np}\PYG{o}{.}\PYG{n}{pi}\PYG{p}{]}\PYG{p}{)}\PYG{p}{)}
\PYG{g+go}{array([  1.00000000e+00,   6.12303177e\PYGZhy{}17,  \PYGZhy{}1.00000000e+00])}
\PYG{g+go}{\PYGZgt{}\PYGZgt{}\PYGZgt{}}
\PYG{g+gp}{\PYGZgt{}\PYGZgt{}\PYGZgt{} }\PYG{c+c1}{\PYGZsh{} Example of providing the optional output parameter}
\PYG{g+gp}{\PYGZgt{}\PYGZgt{}\PYGZgt{} }\PYG{n}{out1} \PYG{o}{=} \PYG{n}{np}\PYG{o}{.}\PYG{n}{array}\PYG{p}{(}\PYG{p}{[}\PYG{l+m+mi}{0}\PYG{p}{]}\PYG{p}{,} \PYG{n}{dtype}\PYG{o}{=}\PYG{l+s+s1}{\PYGZsq{}}\PYG{l+s+s1}{d}\PYG{l+s+s1}{\PYGZsq{}}\PYG{p}{)}
\PYG{g+gp}{\PYGZgt{}\PYGZgt{}\PYGZgt{} }\PYG{n}{out2} \PYG{o}{=} \PYG{n}{np}\PYG{o}{.}\PYG{n}{cos}\PYG{p}{(}\PYG{p}{[}\PYG{l+m+mf}{0.1}\PYG{p}{]}\PYG{p}{,} \PYG{n}{out1}\PYG{p}{)}
\PYG{g+gp}{\PYGZgt{}\PYGZgt{}\PYGZgt{} }\PYG{n}{out2} \PYG{o+ow}{is} \PYG{n}{out1}
\PYG{g+go}{True}
\PYG{g+go}{\PYGZgt{}\PYGZgt{}\PYGZgt{}}
\PYG{g+gp}{\PYGZgt{}\PYGZgt{}\PYGZgt{} }\PYG{c+c1}{\PYGZsh{} Example of ValueError due to provision of shape mis\PYGZhy{}matched `out`}
\PYG{g+gp}{\PYGZgt{}\PYGZgt{}\PYGZgt{} }\PYG{n}{np}\PYG{o}{.}\PYG{n}{cos}\PYG{p}{(}\PYG{n}{np}\PYG{o}{.}\PYG{n}{zeros}\PYG{p}{(}\PYG{p}{(}\PYG{l+m+mi}{3}\PYG{p}{,}\PYG{l+m+mi}{3}\PYG{p}{)}\PYG{p}{)}\PYG{p}{,}\PYG{n}{np}\PYG{o}{.}\PYG{n}{zeros}\PYG{p}{(}\PYG{p}{(}\PYG{l+m+mi}{2}\PYG{p}{,}\PYG{l+m+mi}{2}\PYG{p}{)}\PYG{p}{)}\PYG{p}{)}
\PYG{g+gt}{Traceback (most recent call last):}
  File \PYG{n+nb}{\PYGZdq{}\PYGZlt{}stdin\PYGZgt{}\PYGZdq{}}, line \PYG{l+m}{1}, in \PYG{n}{\PYGZlt{}module\PYGZgt{}}
\PYG{g+gr}{ValueError}: \PYG{n}{operands could not be broadcast together with shapes (3,3) (2,2)}
\end{sphinxVerbatim}

\end{fulllineitems}

\index{cosh() (in module symjax.tensor)@\spxentry{cosh()}\spxextra{in module symjax.tensor}}

\begin{fulllineitems}
\phantomsection\label{\detokenize{modules/tensor:symjax.tensor.cosh}}\pysiglinewithargsret{\sphinxbfcode{\sphinxupquote{cosh}}}{\emph{\DUrole{n}{x}}}{}
Hyperbolic cosine, element\sphinxhyphen{}wise.

LAX\sphinxhyphen{}backend implementation of {\hyperref[\detokenize{modules/tensor:symjax.tensor.cosh}]{\sphinxcrossref{\sphinxcode{\sphinxupquote{cosh()}}}}}.
ADDITIONOriginal docstring below.

LAX\sphinxhyphen{}backend implementation of {\hyperref[\detokenize{modules/tensor:symjax.tensor.cosh}]{\sphinxcrossref{\sphinxcode{\sphinxupquote{cosh()}}}}}.
Original docstring below.

cosh(x, /, out=None, {\color{red}\bfseries{}*}, where=True, casting=’same\_kind’, order=’K’, dtype=None, subok=True{[}, signature, extobj{]})

Equivalent to \sphinxcode{\sphinxupquote{1/2 * (np.exp(x) + np.exp(\sphinxhyphen{}x))}} and \sphinxcode{\sphinxupquote{np.cos(1j*x)}}.
\begin{quote}\begin{description}
\item[{Returns}] \leavevmode
\sphinxstylestrong{out} \textendash{} Output array of same shape as \sphinxtitleref{x}.
This is a scalar if \sphinxtitleref{x} is a scalar.

\item[{Return type}] \leavevmode
ndarray or scalar

\end{description}\end{quote}
\subsubsection*{Examples}

\begin{sphinxVerbatim}[commandchars=\\\{\}]
\PYG{g+gp}{\PYGZgt{}\PYGZgt{}\PYGZgt{} }\PYG{n}{np}\PYG{o}{.}\PYG{n}{cosh}\PYG{p}{(}\PYG{l+m+mi}{0}\PYG{p}{)}
\PYG{g+go}{1.0}
\end{sphinxVerbatim}

The hyperbolic cosine describes the shape of a hanging cable:

\begin{sphinxVerbatim}[commandchars=\\\{\}]
\PYG{g+gp}{\PYGZgt{}\PYGZgt{}\PYGZgt{} }\PYG{k+kn}{import} \PYG{n+nn}{matplotlib}\PYG{n+nn}{.}\PYG{n+nn}{pyplot} \PYG{k}{as} \PYG{n+nn}{plt}
\PYG{g+gp}{\PYGZgt{}\PYGZgt{}\PYGZgt{} }\PYG{n}{x} \PYG{o}{=} \PYG{n}{np}\PYG{o}{.}\PYG{n}{linspace}\PYG{p}{(}\PYG{o}{\PYGZhy{}}\PYG{l+m+mi}{4}\PYG{p}{,} \PYG{l+m+mi}{4}\PYG{p}{,} \PYG{l+m+mi}{1000}\PYG{p}{)}
\PYG{g+gp}{\PYGZgt{}\PYGZgt{}\PYGZgt{} }\PYG{n}{plt}\PYG{o}{.}\PYG{n}{plot}\PYG{p}{(}\PYG{n}{x}\PYG{p}{,} \PYG{n}{np}\PYG{o}{.}\PYG{n}{cosh}\PYG{p}{(}\PYG{n}{x}\PYG{p}{)}\PYG{p}{)}
\PYG{g+gp}{\PYGZgt{}\PYGZgt{}\PYGZgt{} }\PYG{n}{plt}\PYG{o}{.}\PYG{n}{show}\PYG{p}{(}\PYG{p}{)}
\end{sphinxVerbatim}

\end{fulllineitems}

\index{count\_nonzero() (in module symjax.tensor)@\spxentry{count\_nonzero()}\spxextra{in module symjax.tensor}}

\begin{fulllineitems}
\phantomsection\label{\detokenize{modules/tensor:symjax.tensor.count_nonzero}}\pysiglinewithargsret{\sphinxbfcode{\sphinxupquote{count\_nonzero}}}{\emph{\DUrole{n}{a}}, \emph{\DUrole{n}{axis}\DUrole{o}{=}\DUrole{default_value}{None}}}{}
Counts the number of non\sphinxhyphen{}zero values in the array \sphinxcode{\sphinxupquote{a}}.

LAX\sphinxhyphen{}backend implementation of {\hyperref[\detokenize{modules/tensor:symjax.tensor.count_nonzero}]{\sphinxcrossref{\sphinxcode{\sphinxupquote{count\_nonzero()}}}}}.
ADDITIONOriginal docstring below.

LAX\sphinxhyphen{}backend implementation of {\hyperref[\detokenize{modules/tensor:symjax.tensor.count_nonzero}]{\sphinxcrossref{\sphinxcode{\sphinxupquote{count\_nonzero()}}}}}.
Original docstring below.

The word “non\sphinxhyphen{}zero” is in reference to the Python 2.x
built\sphinxhyphen{}in method \sphinxcode{\sphinxupquote{\_\_nonzero\_\_()}} (renamed \sphinxcode{\sphinxupquote{\_\_bool\_\_()}}
in Python 3.x) of Python objects that tests an object’s
“truthfulness”. For example, any number is considered
truthful if it is nonzero, whereas any string is considered
truthful if it is not the empty string. Thus, this function
(recursively) counts how many elements in \sphinxcode{\sphinxupquote{a}} (and in
sub\sphinxhyphen{}arrays thereof) have their \sphinxcode{\sphinxupquote{\_\_nonzero\_\_()}} or \sphinxcode{\sphinxupquote{\_\_bool\_\_()}}
method evaluated to \sphinxcode{\sphinxupquote{True}}.
\begin{quote}\begin{description}
\item[{Returns}] \leavevmode
\sphinxstylestrong{count} \textendash{} Number of non\sphinxhyphen{}zero values in the array along a given axis.
Otherwise, the total number of non\sphinxhyphen{}zero values in the array
is returned.

\item[{Return type}] \leavevmode
int or array of int

\end{description}\end{quote}

\sphinxstrong{See also:}

\begin{description}
\item[{{\hyperref[\detokenize{modules/tensor:symjax.tensor.nonzero}]{\sphinxcrossref{\sphinxcode{\sphinxupquote{nonzero()}}}}}}] \leavevmode
Return the coordinates of all the non\sphinxhyphen{}zero values.

\end{description}

\subsubsection*{Examples}

\begin{sphinxVerbatim}[commandchars=\\\{\}]
\PYG{g+gp}{\PYGZgt{}\PYGZgt{}\PYGZgt{} }\PYG{n}{np}\PYG{o}{.}\PYG{n}{count\PYGZus{}nonzero}\PYG{p}{(}\PYG{n}{np}\PYG{o}{.}\PYG{n}{eye}\PYG{p}{(}\PYG{l+m+mi}{4}\PYG{p}{)}\PYG{p}{)}
\PYG{g+go}{4}
\PYG{g+gp}{\PYGZgt{}\PYGZgt{}\PYGZgt{} }\PYG{n}{np}\PYG{o}{.}\PYG{n}{count\PYGZus{}nonzero}\PYG{p}{(}\PYG{p}{[}\PYG{p}{[}\PYG{l+m+mi}{0}\PYG{p}{,}\PYG{l+m+mi}{1}\PYG{p}{,}\PYG{l+m+mi}{7}\PYG{p}{,}\PYG{l+m+mi}{0}\PYG{p}{,}\PYG{l+m+mi}{0}\PYG{p}{]}\PYG{p}{,}\PYG{p}{[}\PYG{l+m+mi}{3}\PYG{p}{,}\PYG{l+m+mi}{0}\PYG{p}{,}\PYG{l+m+mi}{0}\PYG{p}{,}\PYG{l+m+mi}{2}\PYG{p}{,}\PYG{l+m+mi}{19}\PYG{p}{]}\PYG{p}{]}\PYG{p}{)}
\PYG{g+go}{5}
\PYG{g+gp}{\PYGZgt{}\PYGZgt{}\PYGZgt{} }\PYG{n}{np}\PYG{o}{.}\PYG{n}{count\PYGZus{}nonzero}\PYG{p}{(}\PYG{p}{[}\PYG{p}{[}\PYG{l+m+mi}{0}\PYG{p}{,}\PYG{l+m+mi}{1}\PYG{p}{,}\PYG{l+m+mi}{7}\PYG{p}{,}\PYG{l+m+mi}{0}\PYG{p}{,}\PYG{l+m+mi}{0}\PYG{p}{]}\PYG{p}{,}\PYG{p}{[}\PYG{l+m+mi}{3}\PYG{p}{,}\PYG{l+m+mi}{0}\PYG{p}{,}\PYG{l+m+mi}{0}\PYG{p}{,}\PYG{l+m+mi}{2}\PYG{p}{,}\PYG{l+m+mi}{19}\PYG{p}{]}\PYG{p}{]}\PYG{p}{,} \PYG{n}{axis}\PYG{o}{=}\PYG{l+m+mi}{0}\PYG{p}{)}
\PYG{g+go}{array([1, 1, 1, 1, 1])}
\PYG{g+gp}{\PYGZgt{}\PYGZgt{}\PYGZgt{} }\PYG{n}{np}\PYG{o}{.}\PYG{n}{count\PYGZus{}nonzero}\PYG{p}{(}\PYG{p}{[}\PYG{p}{[}\PYG{l+m+mi}{0}\PYG{p}{,}\PYG{l+m+mi}{1}\PYG{p}{,}\PYG{l+m+mi}{7}\PYG{p}{,}\PYG{l+m+mi}{0}\PYG{p}{,}\PYG{l+m+mi}{0}\PYG{p}{]}\PYG{p}{,}\PYG{p}{[}\PYG{l+m+mi}{3}\PYG{p}{,}\PYG{l+m+mi}{0}\PYG{p}{,}\PYG{l+m+mi}{0}\PYG{p}{,}\PYG{l+m+mi}{2}\PYG{p}{,}\PYG{l+m+mi}{19}\PYG{p}{]}\PYG{p}{]}\PYG{p}{,} \PYG{n}{axis}\PYG{o}{=}\PYG{l+m+mi}{1}\PYG{p}{)}
\PYG{g+go}{array([2, 3])}
\end{sphinxVerbatim}

\end{fulllineitems}

\index{cov() (in module symjax.tensor)@\spxentry{cov()}\spxextra{in module symjax.tensor}}

\begin{fulllineitems}
\phantomsection\label{\detokenize{modules/tensor:symjax.tensor.cov}}\pysiglinewithargsret{\sphinxbfcode{\sphinxupquote{cov}}}{\emph{\DUrole{n}{m}}, \emph{\DUrole{n}{y}\DUrole{o}{=}\DUrole{default_value}{None}}, \emph{\DUrole{n}{rowvar}\DUrole{o}{=}\DUrole{default_value}{True}}, \emph{\DUrole{n}{bias}\DUrole{o}{=}\DUrole{default_value}{False}}, \emph{\DUrole{n}{ddof}\DUrole{o}{=}\DUrole{default_value}{None}}, \emph{\DUrole{n}{fweights}\DUrole{o}{=}\DUrole{default_value}{None}}, \emph{\DUrole{n}{aweights}\DUrole{o}{=}\DUrole{default_value}{None}}}{}
Estimate a covariance matrix, given data and weights.

LAX\sphinxhyphen{}backend implementation of {\hyperref[\detokenize{modules/tensor:symjax.tensor.cov}]{\sphinxcrossref{\sphinxcode{\sphinxupquote{cov()}}}}}.
ADDITIONOriginal docstring below.

LAX\sphinxhyphen{}backend implementation of {\hyperref[\detokenize{modules/tensor:symjax.tensor.cov}]{\sphinxcrossref{\sphinxcode{\sphinxupquote{cov()}}}}}.
Original docstring below.

Covariance indicates the level to which two variables vary together.
If we examine N\sphinxhyphen{}dimensional samples, \(X = [x_1, x_2, ... x_N]^T\),
then the covariance matrix element \(C_{ij}\) is the covariance of
\(x_i\) and \(x_j\). The element \(C_{ii}\) is the variance
of \(x_i\).

See the notes for an outline of the algorithm.
\begin{quote}\begin{description}
\item[{Returns}] \leavevmode
\sphinxstylestrong{out} \textendash{} The covariance matrix of the variables.

\item[{Return type}] \leavevmode
ndarray

\end{description}\end{quote}

\sphinxstrong{See also:}

\begin{description}
\item[{{\hyperref[\detokenize{modules/tensor:symjax.tensor.corrcoef}]{\sphinxcrossref{\sphinxcode{\sphinxupquote{corrcoef()}}}}}}] \leavevmode
Normalized covariance matrix

\end{description}

\subsubsection*{Notes}

Assume that the observations are in the columns of the observation
array \sphinxtitleref{m} and let \sphinxcode{\sphinxupquote{f = fweights}} and \sphinxcode{\sphinxupquote{a = aweights}} for brevity. The
steps to compute the weighted covariance are as follows:

\begin{sphinxVerbatim}[commandchars=\\\{\}]
\PYG{g+gp}{\PYGZgt{}\PYGZgt{}\PYGZgt{} }\PYG{n}{m} \PYG{o}{=} \PYG{n}{np}\PYG{o}{.}\PYG{n}{arange}\PYG{p}{(}\PYG{l+m+mi}{10}\PYG{p}{,} \PYG{n}{dtype}\PYG{o}{=}\PYG{n}{np}\PYG{o}{.}\PYG{n}{float64}\PYG{p}{)}
\PYG{g+gp}{\PYGZgt{}\PYGZgt{}\PYGZgt{} }\PYG{n}{f} \PYG{o}{=} \PYG{n}{np}\PYG{o}{.}\PYG{n}{arange}\PYG{p}{(}\PYG{l+m+mi}{10}\PYG{p}{)} \PYG{o}{*} \PYG{l+m+mi}{2}
\PYG{g+gp}{\PYGZgt{}\PYGZgt{}\PYGZgt{} }\PYG{n}{a} \PYG{o}{=} \PYG{n}{np}\PYG{o}{.}\PYG{n}{arange}\PYG{p}{(}\PYG{l+m+mi}{10}\PYG{p}{)} \PYG{o}{*}\PYG{o}{*} \PYG{l+m+mf}{2.}
\PYG{g+gp}{\PYGZgt{}\PYGZgt{}\PYGZgt{} }\PYG{n}{ddof} \PYG{o}{=} \PYG{l+m+mi}{1}
\PYG{g+gp}{\PYGZgt{}\PYGZgt{}\PYGZgt{} }\PYG{n}{w} \PYG{o}{=} \PYG{n}{f} \PYG{o}{*} \PYG{n}{a}
\PYG{g+gp}{\PYGZgt{}\PYGZgt{}\PYGZgt{} }\PYG{n}{v1} \PYG{o}{=} \PYG{n}{np}\PYG{o}{.}\PYG{n}{sum}\PYG{p}{(}\PYG{n}{w}\PYG{p}{)}
\PYG{g+gp}{\PYGZgt{}\PYGZgt{}\PYGZgt{} }\PYG{n}{v2} \PYG{o}{=} \PYG{n}{np}\PYG{o}{.}\PYG{n}{sum}\PYG{p}{(}\PYG{n}{w} \PYG{o}{*} \PYG{n}{a}\PYG{p}{)}
\PYG{g+gp}{\PYGZgt{}\PYGZgt{}\PYGZgt{} }\PYG{n}{m} \PYG{o}{\PYGZhy{}}\PYG{o}{=} \PYG{n}{np}\PYG{o}{.}\PYG{n}{sum}\PYG{p}{(}\PYG{n}{m} \PYG{o}{*} \PYG{n}{w}\PYG{p}{,} \PYG{n}{axis}\PYG{o}{=}\PYG{k+kc}{None}\PYG{p}{,} \PYG{n}{keepdims}\PYG{o}{=}\PYG{k+kc}{True}\PYG{p}{)} \PYG{o}{/} \PYG{n}{v1}
\PYG{g+gp}{\PYGZgt{}\PYGZgt{}\PYGZgt{} }\PYG{n}{cov} \PYG{o}{=} \PYG{n}{np}\PYG{o}{.}\PYG{n}{dot}\PYG{p}{(}\PYG{n}{m} \PYG{o}{*} \PYG{n}{w}\PYG{p}{,} \PYG{n}{m}\PYG{o}{.}\PYG{n}{T}\PYG{p}{)} \PYG{o}{*} \PYG{n}{v1} \PYG{o}{/} \PYG{p}{(}\PYG{n}{v1}\PYG{o}{*}\PYG{o}{*}\PYG{l+m+mi}{2} \PYG{o}{\PYGZhy{}} \PYG{n}{ddof} \PYG{o}{*} \PYG{n}{v2}\PYG{p}{)}
\end{sphinxVerbatim}

Note that when \sphinxcode{\sphinxupquote{a == 1}}, the normalization factor
\sphinxcode{\sphinxupquote{v1 / (v1**2 \sphinxhyphen{} ddof * v2)}} goes over to \sphinxcode{\sphinxupquote{1 / (np.sum(f) \sphinxhyphen{} ddof)}}
as it should.
\subsubsection*{Examples}

Consider two variables, \(x_0\) and \(x_1\), which
correlate perfectly, but in opposite directions:

\begin{sphinxVerbatim}[commandchars=\\\{\}]
\PYG{g+gp}{\PYGZgt{}\PYGZgt{}\PYGZgt{} }\PYG{n}{x} \PYG{o}{=} \PYG{n}{np}\PYG{o}{.}\PYG{n}{array}\PYG{p}{(}\PYG{p}{[}\PYG{p}{[}\PYG{l+m+mi}{0}\PYG{p}{,} \PYG{l+m+mi}{2}\PYG{p}{]}\PYG{p}{,} \PYG{p}{[}\PYG{l+m+mi}{1}\PYG{p}{,} \PYG{l+m+mi}{1}\PYG{p}{]}\PYG{p}{,} \PYG{p}{[}\PYG{l+m+mi}{2}\PYG{p}{,} \PYG{l+m+mi}{0}\PYG{p}{]}\PYG{p}{]}\PYG{p}{)}\PYG{o}{.}\PYG{n}{T}
\PYG{g+gp}{\PYGZgt{}\PYGZgt{}\PYGZgt{} }\PYG{n}{x}
\PYG{g+go}{array([[0, 1, 2],}
\PYG{g+go}{       [2, 1, 0]])}
\end{sphinxVerbatim}

Note how \(x_0\) increases while \(x_1\) decreases. The covariance
matrix shows this clearly:

\begin{sphinxVerbatim}[commandchars=\\\{\}]
\PYG{g+gp}{\PYGZgt{}\PYGZgt{}\PYGZgt{} }\PYG{n}{np}\PYG{o}{.}\PYG{n}{cov}\PYG{p}{(}\PYG{n}{x}\PYG{p}{)}
\PYG{g+go}{array([[ 1., \PYGZhy{}1.],}
\PYG{g+go}{       [\PYGZhy{}1.,  1.]])}
\end{sphinxVerbatim}

Note that element \(C_{0,1}\), which shows the correlation between
\(x_0\) and \(x_1\), is negative.

Further, note how \sphinxtitleref{x} and \sphinxtitleref{y} are combined:

\begin{sphinxVerbatim}[commandchars=\\\{\}]
\PYG{g+gp}{\PYGZgt{}\PYGZgt{}\PYGZgt{} }\PYG{n}{x} \PYG{o}{=} \PYG{p}{[}\PYG{o}{\PYGZhy{}}\PYG{l+m+mf}{2.1}\PYG{p}{,} \PYG{o}{\PYGZhy{}}\PYG{l+m+mi}{1}\PYG{p}{,}  \PYG{l+m+mf}{4.3}\PYG{p}{]}
\PYG{g+gp}{\PYGZgt{}\PYGZgt{}\PYGZgt{} }\PYG{n}{y} \PYG{o}{=} \PYG{p}{[}\PYG{l+m+mi}{3}\PYG{p}{,}  \PYG{l+m+mf}{1.1}\PYG{p}{,}  \PYG{l+m+mf}{0.12}\PYG{p}{]}
\PYG{g+gp}{\PYGZgt{}\PYGZgt{}\PYGZgt{} }\PYG{n}{X} \PYG{o}{=} \PYG{n}{np}\PYG{o}{.}\PYG{n}{stack}\PYG{p}{(}\PYG{p}{(}\PYG{n}{x}\PYG{p}{,} \PYG{n}{y}\PYG{p}{)}\PYG{p}{,} \PYG{n}{axis}\PYG{o}{=}\PYG{l+m+mi}{0}\PYG{p}{)}
\PYG{g+gp}{\PYGZgt{}\PYGZgt{}\PYGZgt{} }\PYG{n}{np}\PYG{o}{.}\PYG{n}{cov}\PYG{p}{(}\PYG{n}{X}\PYG{p}{)}
\PYG{g+go}{array([[11.71      , \PYGZhy{}4.286     ], \PYGZsh{} may vary}
\PYG{g+go}{       [\PYGZhy{}4.286     ,  2.144133]])}
\PYG{g+gp}{\PYGZgt{}\PYGZgt{}\PYGZgt{} }\PYG{n}{np}\PYG{o}{.}\PYG{n}{cov}\PYG{p}{(}\PYG{n}{x}\PYG{p}{,} \PYG{n}{y}\PYG{p}{)}
\PYG{g+go}{array([[11.71      , \PYGZhy{}4.286     ], \PYGZsh{} may vary}
\PYG{g+go}{       [\PYGZhy{}4.286     ,  2.144133]])}
\PYG{g+gp}{\PYGZgt{}\PYGZgt{}\PYGZgt{} }\PYG{n}{np}\PYG{o}{.}\PYG{n}{cov}\PYG{p}{(}\PYG{n}{x}\PYG{p}{)}
\PYG{g+go}{array(11.71)}
\end{sphinxVerbatim}

\end{fulllineitems}

\index{cross() (in module symjax.tensor)@\spxentry{cross()}\spxextra{in module symjax.tensor}}

\begin{fulllineitems}
\phantomsection\label{\detokenize{modules/tensor:symjax.tensor.cross}}\pysiglinewithargsret{\sphinxbfcode{\sphinxupquote{cross}}}{\emph{\DUrole{n}{a}}, \emph{\DUrole{n}{b}}, \emph{\DUrole{n}{axisa}\DUrole{o}{=}\DUrole{default_value}{\sphinxhyphen{} 1}}, \emph{\DUrole{n}{axisb}\DUrole{o}{=}\DUrole{default_value}{\sphinxhyphen{} 1}}, \emph{\DUrole{n}{axisc}\DUrole{o}{=}\DUrole{default_value}{\sphinxhyphen{} 1}}, \emph{\DUrole{n}{axis}\DUrole{o}{=}\DUrole{default_value}{None}}}{}
Return the cross product of two (arrays of) vectors.

LAX\sphinxhyphen{}backend implementation of {\hyperref[\detokenize{modules/tensor:symjax.tensor.cross}]{\sphinxcrossref{\sphinxcode{\sphinxupquote{cross()}}}}}.
ADDITIONOriginal docstring below.

LAX\sphinxhyphen{}backend implementation of {\hyperref[\detokenize{modules/tensor:symjax.tensor.cross}]{\sphinxcrossref{\sphinxcode{\sphinxupquote{cross()}}}}}.
Original docstring below.

The cross product of \sphinxtitleref{a} and \sphinxtitleref{b} in \(R^3\) is a vector perpendicular
to both \sphinxtitleref{a} and \sphinxtitleref{b}.  If \sphinxtitleref{a} and \sphinxtitleref{b} are arrays of vectors, the vectors
are defined by the last axis of \sphinxtitleref{a} and \sphinxtitleref{b} by default, and these axes
can have dimensions 2 or 3.  Where the dimension of either \sphinxtitleref{a} or \sphinxtitleref{b} is
2, the third component of the input vector is assumed to be zero and the
cross product calculated accordingly.  In cases where both input vectors
have dimension 2, the z\sphinxhyphen{}component of the cross product is returned.
\begin{quote}\begin{description}
\item[{Returns}] \leavevmode
\sphinxstylestrong{c} \textendash{} Vector cross product(s).

\item[{Return type}] \leavevmode
ndarray

\item[{Raises}] \leavevmode
\sphinxstyleliteralstrong{\sphinxupquote{ValueError}} \textendash{} When the dimension of the vector(s) in \sphinxtitleref{a} and/or \sphinxtitleref{b} does not
    equal 2 or 3.

\end{description}\end{quote}

\sphinxstrong{See also:}

\begin{description}
\item[{{\hyperref[\detokenize{modules/tensor:symjax.tensor.inner}]{\sphinxcrossref{\sphinxcode{\sphinxupquote{inner()}}}}}}] \leavevmode
Inner product

\item[{{\hyperref[\detokenize{modules/tensor:symjax.tensor.outer}]{\sphinxcrossref{\sphinxcode{\sphinxupquote{outer()}}}}}}] \leavevmode
Outer product.

\item[{{\hyperref[\detokenize{modules/tensor:symjax.tensor.ix_}]{\sphinxcrossref{\sphinxcode{\sphinxupquote{ix\_()}}}}}}] \leavevmode
Construct index arrays.

\end{description}

\subsubsection*{Notes}

\DUrole{versionmodified,added}{New in version 1.9.0.}

Supports full broadcasting of the inputs.
\subsubsection*{Examples}

Vector cross\sphinxhyphen{}product.

\begin{sphinxVerbatim}[commandchars=\\\{\}]
\PYG{g+gp}{\PYGZgt{}\PYGZgt{}\PYGZgt{} }\PYG{n}{x} \PYG{o}{=} \PYG{p}{[}\PYG{l+m+mi}{1}\PYG{p}{,} \PYG{l+m+mi}{2}\PYG{p}{,} \PYG{l+m+mi}{3}\PYG{p}{]}
\PYG{g+gp}{\PYGZgt{}\PYGZgt{}\PYGZgt{} }\PYG{n}{y} \PYG{o}{=} \PYG{p}{[}\PYG{l+m+mi}{4}\PYG{p}{,} \PYG{l+m+mi}{5}\PYG{p}{,} \PYG{l+m+mi}{6}\PYG{p}{]}
\PYG{g+gp}{\PYGZgt{}\PYGZgt{}\PYGZgt{} }\PYG{n}{np}\PYG{o}{.}\PYG{n}{cross}\PYG{p}{(}\PYG{n}{x}\PYG{p}{,} \PYG{n}{y}\PYG{p}{)}
\PYG{g+go}{array([\PYGZhy{}3,  6, \PYGZhy{}3])}
\end{sphinxVerbatim}

One vector with dimension 2.

\begin{sphinxVerbatim}[commandchars=\\\{\}]
\PYG{g+gp}{\PYGZgt{}\PYGZgt{}\PYGZgt{} }\PYG{n}{x} \PYG{o}{=} \PYG{p}{[}\PYG{l+m+mi}{1}\PYG{p}{,} \PYG{l+m+mi}{2}\PYG{p}{]}
\PYG{g+gp}{\PYGZgt{}\PYGZgt{}\PYGZgt{} }\PYG{n}{y} \PYG{o}{=} \PYG{p}{[}\PYG{l+m+mi}{4}\PYG{p}{,} \PYG{l+m+mi}{5}\PYG{p}{,} \PYG{l+m+mi}{6}\PYG{p}{]}
\PYG{g+gp}{\PYGZgt{}\PYGZgt{}\PYGZgt{} }\PYG{n}{np}\PYG{o}{.}\PYG{n}{cross}\PYG{p}{(}\PYG{n}{x}\PYG{p}{,} \PYG{n}{y}\PYG{p}{)}
\PYG{g+go}{array([12, \PYGZhy{}6, \PYGZhy{}3])}
\end{sphinxVerbatim}

Equivalently:

\begin{sphinxVerbatim}[commandchars=\\\{\}]
\PYG{g+gp}{\PYGZgt{}\PYGZgt{}\PYGZgt{} }\PYG{n}{x} \PYG{o}{=} \PYG{p}{[}\PYG{l+m+mi}{1}\PYG{p}{,} \PYG{l+m+mi}{2}\PYG{p}{,} \PYG{l+m+mi}{0}\PYG{p}{]}
\PYG{g+gp}{\PYGZgt{}\PYGZgt{}\PYGZgt{} }\PYG{n}{y} \PYG{o}{=} \PYG{p}{[}\PYG{l+m+mi}{4}\PYG{p}{,} \PYG{l+m+mi}{5}\PYG{p}{,} \PYG{l+m+mi}{6}\PYG{p}{]}
\PYG{g+gp}{\PYGZgt{}\PYGZgt{}\PYGZgt{} }\PYG{n}{np}\PYG{o}{.}\PYG{n}{cross}\PYG{p}{(}\PYG{n}{x}\PYG{p}{,} \PYG{n}{y}\PYG{p}{)}
\PYG{g+go}{array([12, \PYGZhy{}6, \PYGZhy{}3])}
\end{sphinxVerbatim}

Both vectors with dimension 2.

\begin{sphinxVerbatim}[commandchars=\\\{\}]
\PYG{g+gp}{\PYGZgt{}\PYGZgt{}\PYGZgt{} }\PYG{n}{x} \PYG{o}{=} \PYG{p}{[}\PYG{l+m+mi}{1}\PYG{p}{,}\PYG{l+m+mi}{2}\PYG{p}{]}
\PYG{g+gp}{\PYGZgt{}\PYGZgt{}\PYGZgt{} }\PYG{n}{y} \PYG{o}{=} \PYG{p}{[}\PYG{l+m+mi}{4}\PYG{p}{,}\PYG{l+m+mi}{5}\PYG{p}{]}
\PYG{g+gp}{\PYGZgt{}\PYGZgt{}\PYGZgt{} }\PYG{n}{np}\PYG{o}{.}\PYG{n}{cross}\PYG{p}{(}\PYG{n}{x}\PYG{p}{,} \PYG{n}{y}\PYG{p}{)}
\PYG{g+go}{array(\PYGZhy{}3)}
\end{sphinxVerbatim}

Multiple vector cross\sphinxhyphen{}products. Note that the direction of the cross
product vector is defined by the \sphinxtitleref{right\sphinxhyphen{}hand rule}.

\begin{sphinxVerbatim}[commandchars=\\\{\}]
\PYG{g+gp}{\PYGZgt{}\PYGZgt{}\PYGZgt{} }\PYG{n}{x} \PYG{o}{=} \PYG{n}{np}\PYG{o}{.}\PYG{n}{array}\PYG{p}{(}\PYG{p}{[}\PYG{p}{[}\PYG{l+m+mi}{1}\PYG{p}{,}\PYG{l+m+mi}{2}\PYG{p}{,}\PYG{l+m+mi}{3}\PYG{p}{]}\PYG{p}{,} \PYG{p}{[}\PYG{l+m+mi}{4}\PYG{p}{,}\PYG{l+m+mi}{5}\PYG{p}{,}\PYG{l+m+mi}{6}\PYG{p}{]}\PYG{p}{]}\PYG{p}{)}
\PYG{g+gp}{\PYGZgt{}\PYGZgt{}\PYGZgt{} }\PYG{n}{y} \PYG{o}{=} \PYG{n}{np}\PYG{o}{.}\PYG{n}{array}\PYG{p}{(}\PYG{p}{[}\PYG{p}{[}\PYG{l+m+mi}{4}\PYG{p}{,}\PYG{l+m+mi}{5}\PYG{p}{,}\PYG{l+m+mi}{6}\PYG{p}{]}\PYG{p}{,} \PYG{p}{[}\PYG{l+m+mi}{1}\PYG{p}{,}\PYG{l+m+mi}{2}\PYG{p}{,}\PYG{l+m+mi}{3}\PYG{p}{]}\PYG{p}{]}\PYG{p}{)}
\PYG{g+gp}{\PYGZgt{}\PYGZgt{}\PYGZgt{} }\PYG{n}{np}\PYG{o}{.}\PYG{n}{cross}\PYG{p}{(}\PYG{n}{x}\PYG{p}{,} \PYG{n}{y}\PYG{p}{)}
\PYG{g+go}{array([[\PYGZhy{}3,  6, \PYGZhy{}3],}
\PYG{g+go}{       [ 3, \PYGZhy{}6,  3]])}
\end{sphinxVerbatim}

The orientation of \sphinxtitleref{c} can be changed using the \sphinxtitleref{axisc} keyword.

\begin{sphinxVerbatim}[commandchars=\\\{\}]
\PYG{g+gp}{\PYGZgt{}\PYGZgt{}\PYGZgt{} }\PYG{n}{np}\PYG{o}{.}\PYG{n}{cross}\PYG{p}{(}\PYG{n}{x}\PYG{p}{,} \PYG{n}{y}\PYG{p}{,} \PYG{n}{axisc}\PYG{o}{=}\PYG{l+m+mi}{0}\PYG{p}{)}
\PYG{g+go}{array([[\PYGZhy{}3,  3],}
\PYG{g+go}{       [ 6, \PYGZhy{}6],}
\PYG{g+go}{       [\PYGZhy{}3,  3]])}
\end{sphinxVerbatim}

Change the vector definition of \sphinxtitleref{x} and \sphinxtitleref{y} using \sphinxtitleref{axisa} and \sphinxtitleref{axisb}.

\begin{sphinxVerbatim}[commandchars=\\\{\}]
\PYG{g+gp}{\PYGZgt{}\PYGZgt{}\PYGZgt{} }\PYG{n}{x} \PYG{o}{=} \PYG{n}{np}\PYG{o}{.}\PYG{n}{array}\PYG{p}{(}\PYG{p}{[}\PYG{p}{[}\PYG{l+m+mi}{1}\PYG{p}{,}\PYG{l+m+mi}{2}\PYG{p}{,}\PYG{l+m+mi}{3}\PYG{p}{]}\PYG{p}{,} \PYG{p}{[}\PYG{l+m+mi}{4}\PYG{p}{,}\PYG{l+m+mi}{5}\PYG{p}{,}\PYG{l+m+mi}{6}\PYG{p}{]}\PYG{p}{,} \PYG{p}{[}\PYG{l+m+mi}{7}\PYG{p}{,} \PYG{l+m+mi}{8}\PYG{p}{,} \PYG{l+m+mi}{9}\PYG{p}{]}\PYG{p}{]}\PYG{p}{)}
\PYG{g+gp}{\PYGZgt{}\PYGZgt{}\PYGZgt{} }\PYG{n}{y} \PYG{o}{=} \PYG{n}{np}\PYG{o}{.}\PYG{n}{array}\PYG{p}{(}\PYG{p}{[}\PYG{p}{[}\PYG{l+m+mi}{7}\PYG{p}{,} \PYG{l+m+mi}{8}\PYG{p}{,} \PYG{l+m+mi}{9}\PYG{p}{]}\PYG{p}{,} \PYG{p}{[}\PYG{l+m+mi}{4}\PYG{p}{,}\PYG{l+m+mi}{5}\PYG{p}{,}\PYG{l+m+mi}{6}\PYG{p}{]}\PYG{p}{,} \PYG{p}{[}\PYG{l+m+mi}{1}\PYG{p}{,}\PYG{l+m+mi}{2}\PYG{p}{,}\PYG{l+m+mi}{3}\PYG{p}{]}\PYG{p}{]}\PYG{p}{)}
\PYG{g+gp}{\PYGZgt{}\PYGZgt{}\PYGZgt{} }\PYG{n}{np}\PYG{o}{.}\PYG{n}{cross}\PYG{p}{(}\PYG{n}{x}\PYG{p}{,} \PYG{n}{y}\PYG{p}{)}
\PYG{g+go}{array([[ \PYGZhy{}6,  12,  \PYGZhy{}6],}
\PYG{g+go}{       [  0,   0,   0],}
\PYG{g+go}{       [  6, \PYGZhy{}12,   6]])}
\PYG{g+gp}{\PYGZgt{}\PYGZgt{}\PYGZgt{} }\PYG{n}{np}\PYG{o}{.}\PYG{n}{cross}\PYG{p}{(}\PYG{n}{x}\PYG{p}{,} \PYG{n}{y}\PYG{p}{,} \PYG{n}{axisa}\PYG{o}{=}\PYG{l+m+mi}{0}\PYG{p}{,} \PYG{n}{axisb}\PYG{o}{=}\PYG{l+m+mi}{0}\PYG{p}{)}
\PYG{g+go}{array([[\PYGZhy{}24,  48, \PYGZhy{}24],}
\PYG{g+go}{       [\PYGZhy{}30,  60, \PYGZhy{}30],}
\PYG{g+go}{       [\PYGZhy{}36,  72, \PYGZhy{}36]])}
\end{sphinxVerbatim}

\end{fulllineitems}

\index{cumsum() (in module symjax.tensor)@\spxentry{cumsum()}\spxextra{in module symjax.tensor}}

\begin{fulllineitems}
\phantomsection\label{\detokenize{modules/tensor:symjax.tensor.cumsum}}\pysiglinewithargsret{\sphinxbfcode{\sphinxupquote{cumsum}}}{\emph{\DUrole{n}{a}}, \emph{\DUrole{n}{axis}\DUrole{o}{=}\DUrole{default_value}{None}}, \emph{\DUrole{n}{dtype}\DUrole{o}{=}\DUrole{default_value}{None}}}{}
Return the cumulative sum of the elements along a given axis.

LAX\sphinxhyphen{}backend implementation of {\hyperref[\detokenize{modules/tensor:symjax.tensor.cumsum}]{\sphinxcrossref{\sphinxcode{\sphinxupquote{cumsum()}}}}}.
ADDITIONOriginal docstring below.

LAX\sphinxhyphen{}backend implementation of {\hyperref[\detokenize{modules/tensor:symjax.tensor.cumsum}]{\sphinxcrossref{\sphinxcode{\sphinxupquote{cumsum()}}}}}.
Original docstring below.
\begin{quote}\begin{description}
\item[{Parameters}] \leavevmode
\sphinxstyleliteralstrong{\sphinxupquote{dtype}} (\sphinxstyleliteralemphasis{\sphinxupquote{dtype}}\sphinxstyleliteralemphasis{\sphinxupquote{, }}\sphinxstyleliteralemphasis{\sphinxupquote{optional}}) \textendash{} Type of the returned array and of the accumulator in which the
elements are summed.  If \sphinxtitleref{dtype} is not specified, it defaults
to the dtype of \sphinxtitleref{a}, unless \sphinxtitleref{a} has an integer dtype with a
precision less than that of the default platform integer.  In
that case, the default platform integer is used.

\item[{Returns}] \leavevmode
\sphinxstylestrong{cumsum\_along\_axis} \textendash{} A new array holding the result is returned unless \sphinxtitleref{out} is
specified, in which case a reference to \sphinxtitleref{out} is returned. The
result has the same size as \sphinxtitleref{a}, and the same shape as \sphinxtitleref{a} if
\sphinxtitleref{axis} is not None or \sphinxtitleref{a} is a 1\sphinxhyphen{}d array.

\item[{Return type}] \leavevmode
ndarray.

\end{description}\end{quote}

\sphinxstrong{See also:}

\begin{description}
\item[{{\hyperref[\detokenize{modules/tensor:symjax.tensor.sum}]{\sphinxcrossref{\sphinxcode{\sphinxupquote{sum()}}}}}}] \leavevmode
Sum array elements.

\item[{\sphinxcode{\sphinxupquote{trapz()}}}] \leavevmode
Integration of array values using the composite trapezoidal rule.

\item[{\sphinxcode{\sphinxupquote{diff()}}}] \leavevmode
Calculate the n\sphinxhyphen{}th discrete difference along given axis.

\end{description}

\subsubsection*{Notes}

Arithmetic is modular when using integer types, and no error is
raised on overflow.
\subsubsection*{Examples}

\begin{sphinxVerbatim}[commandchars=\\\{\}]
\PYG{g+gp}{\PYGZgt{}\PYGZgt{}\PYGZgt{} }\PYG{n}{a} \PYG{o}{=} \PYG{n}{np}\PYG{o}{.}\PYG{n}{array}\PYG{p}{(}\PYG{p}{[}\PYG{p}{[}\PYG{l+m+mi}{1}\PYG{p}{,}\PYG{l+m+mi}{2}\PYG{p}{,}\PYG{l+m+mi}{3}\PYG{p}{]}\PYG{p}{,} \PYG{p}{[}\PYG{l+m+mi}{4}\PYG{p}{,}\PYG{l+m+mi}{5}\PYG{p}{,}\PYG{l+m+mi}{6}\PYG{p}{]}\PYG{p}{]}\PYG{p}{)}
\PYG{g+gp}{\PYGZgt{}\PYGZgt{}\PYGZgt{} }\PYG{n}{a}
\PYG{g+go}{array([[1, 2, 3],}
\PYG{g+go}{       [4, 5, 6]])}
\PYG{g+gp}{\PYGZgt{}\PYGZgt{}\PYGZgt{} }\PYG{n}{np}\PYG{o}{.}\PYG{n}{cumsum}\PYG{p}{(}\PYG{n}{a}\PYG{p}{)}
\PYG{g+go}{array([ 1,  3,  6, 10, 15, 21])}
\PYG{g+gp}{\PYGZgt{}\PYGZgt{}\PYGZgt{} }\PYG{n}{np}\PYG{o}{.}\PYG{n}{cumsum}\PYG{p}{(}\PYG{n}{a}\PYG{p}{,} \PYG{n}{dtype}\PYG{o}{=}\PYG{n+nb}{float}\PYG{p}{)}     \PYG{c+c1}{\PYGZsh{} specifies type of output value(s)}
\PYG{g+go}{array([  1.,   3.,   6.,  10.,  15.,  21.])}
\end{sphinxVerbatim}

\begin{sphinxVerbatim}[commandchars=\\\{\}]
\PYG{g+gp}{\PYGZgt{}\PYGZgt{}\PYGZgt{} }\PYG{n}{np}\PYG{o}{.}\PYG{n}{cumsum}\PYG{p}{(}\PYG{n}{a}\PYG{p}{,}\PYG{n}{axis}\PYG{o}{=}\PYG{l+m+mi}{0}\PYG{p}{)}      \PYG{c+c1}{\PYGZsh{} sum over rows for each of the 3 columns}
\PYG{g+go}{array([[1, 2, 3],}
\PYG{g+go}{       [5, 7, 9]])}
\PYG{g+gp}{\PYGZgt{}\PYGZgt{}\PYGZgt{} }\PYG{n}{np}\PYG{o}{.}\PYG{n}{cumsum}\PYG{p}{(}\PYG{n}{a}\PYG{p}{,}\PYG{n}{axis}\PYG{o}{=}\PYG{l+m+mi}{1}\PYG{p}{)}      \PYG{c+c1}{\PYGZsh{} sum over columns for each of the 2 rows}
\PYG{g+go}{array([[ 1,  3,  6],}
\PYG{g+go}{       [ 4,  9, 15]])}
\end{sphinxVerbatim}

\end{fulllineitems}

\index{cumprod() (in module symjax.tensor)@\spxentry{cumprod()}\spxextra{in module symjax.tensor}}

\begin{fulllineitems}
\phantomsection\label{\detokenize{modules/tensor:symjax.tensor.cumprod}}\pysiglinewithargsret{\sphinxbfcode{\sphinxupquote{cumprod}}}{\emph{\DUrole{n}{a}}, \emph{\DUrole{n}{axis}\DUrole{o}{=}\DUrole{default_value}{None}}, \emph{\DUrole{n}{dtype}\DUrole{o}{=}\DUrole{default_value}{None}}}{}
Return the cumulative product of elements along a given axis.

LAX\sphinxhyphen{}backend implementation of {\hyperref[\detokenize{modules/tensor:symjax.tensor.cumprod}]{\sphinxcrossref{\sphinxcode{\sphinxupquote{cumprod()}}}}}.
ADDITIONOriginal docstring below.

LAX\sphinxhyphen{}backend implementation of {\hyperref[\detokenize{modules/tensor:symjax.tensor.cumprod}]{\sphinxcrossref{\sphinxcode{\sphinxupquote{cumprod()}}}}}.
Original docstring below.
\begin{quote}\begin{description}
\item[{Parameters}] \leavevmode
\sphinxstyleliteralstrong{\sphinxupquote{dtype}} (\sphinxstyleliteralemphasis{\sphinxupquote{dtype}}\sphinxstyleliteralemphasis{\sphinxupquote{, }}\sphinxstyleliteralemphasis{\sphinxupquote{optional}}) \textendash{} Type of the returned array, as well as of the accumulator in which
the elements are multiplied.  If \sphinxstyleemphasis{dtype} is not specified, it
defaults to the dtype of \sphinxtitleref{a}, unless \sphinxtitleref{a} has an integer dtype with
a precision less than that of the default platform integer.  In
that case, the default platform integer is used instead.

\item[{Returns}] \leavevmode
\sphinxstylestrong{cumprod} \textendash{} A new array holding the result is returned unless \sphinxtitleref{out} is
specified, in which case a reference to out is returned.

\item[{Return type}] \leavevmode
ndarray

\end{description}\end{quote}

\sphinxstrong{See also:}

\sphinxcode{\sphinxupquote{ufuncs\sphinxhyphen{}output\sphinxhyphen{}type()}}

\subsubsection*{Notes}

Arithmetic is modular when using integer types, and no error is
raised on overflow.
\subsubsection*{Examples}

\begin{sphinxVerbatim}[commandchars=\\\{\}]
\PYG{g+gp}{\PYGZgt{}\PYGZgt{}\PYGZgt{} }\PYG{n}{a} \PYG{o}{=} \PYG{n}{np}\PYG{o}{.}\PYG{n}{array}\PYG{p}{(}\PYG{p}{[}\PYG{l+m+mi}{1}\PYG{p}{,}\PYG{l+m+mi}{2}\PYG{p}{,}\PYG{l+m+mi}{3}\PYG{p}{]}\PYG{p}{)}
\PYG{g+gp}{\PYGZgt{}\PYGZgt{}\PYGZgt{} }\PYG{n}{np}\PYG{o}{.}\PYG{n}{cumprod}\PYG{p}{(}\PYG{n}{a}\PYG{p}{)} \PYG{c+c1}{\PYGZsh{} intermediate results 1, 1*2}
\PYG{g+gp}{... }              \PYG{c+c1}{\PYGZsh{} total product 1*2*3 = 6}
\PYG{g+go}{array([1, 2, 6])}
\PYG{g+gp}{\PYGZgt{}\PYGZgt{}\PYGZgt{} }\PYG{n}{a} \PYG{o}{=} \PYG{n}{np}\PYG{o}{.}\PYG{n}{array}\PYG{p}{(}\PYG{p}{[}\PYG{p}{[}\PYG{l+m+mi}{1}\PYG{p}{,} \PYG{l+m+mi}{2}\PYG{p}{,} \PYG{l+m+mi}{3}\PYG{p}{]}\PYG{p}{,} \PYG{p}{[}\PYG{l+m+mi}{4}\PYG{p}{,} \PYG{l+m+mi}{5}\PYG{p}{,} \PYG{l+m+mi}{6}\PYG{p}{]}\PYG{p}{]}\PYG{p}{)}
\PYG{g+gp}{\PYGZgt{}\PYGZgt{}\PYGZgt{} }\PYG{n}{np}\PYG{o}{.}\PYG{n}{cumprod}\PYG{p}{(}\PYG{n}{a}\PYG{p}{,} \PYG{n}{dtype}\PYG{o}{=}\PYG{n+nb}{float}\PYG{p}{)} \PYG{c+c1}{\PYGZsh{} specify type of output}
\PYG{g+go}{array([   1.,    2.,    6.,   24.,  120.,  720.])}
\end{sphinxVerbatim}

The cumulative product for each column (i.e., over the rows) of \sphinxtitleref{a}:

\begin{sphinxVerbatim}[commandchars=\\\{\}]
\PYG{g+gp}{\PYGZgt{}\PYGZgt{}\PYGZgt{} }\PYG{n}{np}\PYG{o}{.}\PYG{n}{cumprod}\PYG{p}{(}\PYG{n}{a}\PYG{p}{,} \PYG{n}{axis}\PYG{o}{=}\PYG{l+m+mi}{0}\PYG{p}{)}
\PYG{g+go}{array([[ 1,  2,  3],}
\PYG{g+go}{       [ 4, 10, 18]])}
\end{sphinxVerbatim}

The cumulative product for each row (i.e. over the columns) of \sphinxtitleref{a}:

\begin{sphinxVerbatim}[commandchars=\\\{\}]
\PYG{g+gp}{\PYGZgt{}\PYGZgt{}\PYGZgt{} }\PYG{n}{np}\PYG{o}{.}\PYG{n}{cumprod}\PYG{p}{(}\PYG{n}{a}\PYG{p}{,}\PYG{n}{axis}\PYG{o}{=}\PYG{l+m+mi}{1}\PYG{p}{)}
\PYG{g+go}{array([[  1,   2,   6],}
\PYG{g+go}{       [  4,  20, 120]])}
\end{sphinxVerbatim}

\end{fulllineitems}

\index{cumproduct() (in module symjax.tensor)@\spxentry{cumproduct()}\spxextra{in module symjax.tensor}}

\begin{fulllineitems}
\phantomsection\label{\detokenize{modules/tensor:symjax.tensor.cumproduct}}\pysiglinewithargsret{\sphinxbfcode{\sphinxupquote{cumproduct}}}{\emph{\DUrole{n}{a}}, \emph{\DUrole{n}{axis}\DUrole{o}{=}\DUrole{default_value}{None}}, \emph{\DUrole{n}{dtype}\DUrole{o}{=}\DUrole{default_value}{None}}}{}
Return the cumulative product of elements along a given axis.

LAX\sphinxhyphen{}backend implementation of {\hyperref[\detokenize{modules/tensor:symjax.tensor.cumprod}]{\sphinxcrossref{\sphinxcode{\sphinxupquote{cumprod()}}}}}.
ADDITIONOriginal docstring below.

LAX\sphinxhyphen{}backend implementation of {\hyperref[\detokenize{modules/tensor:symjax.tensor.cumprod}]{\sphinxcrossref{\sphinxcode{\sphinxupquote{cumprod()}}}}}.
Original docstring below.
\begin{quote}\begin{description}
\item[{Parameters}] \leavevmode
\sphinxstyleliteralstrong{\sphinxupquote{dtype}} (\sphinxstyleliteralemphasis{\sphinxupquote{dtype}}\sphinxstyleliteralemphasis{\sphinxupquote{, }}\sphinxstyleliteralemphasis{\sphinxupquote{optional}}) \textendash{} Type of the returned array, as well as of the accumulator in which
the elements are multiplied.  If \sphinxstyleemphasis{dtype} is not specified, it
defaults to the dtype of \sphinxtitleref{a}, unless \sphinxtitleref{a} has an integer dtype with
a precision less than that of the default platform integer.  In
that case, the default platform integer is used instead.

\item[{Returns}] \leavevmode
\sphinxstylestrong{cumprod} \textendash{} A new array holding the result is returned unless \sphinxtitleref{out} is
specified, in which case a reference to out is returned.

\item[{Return type}] \leavevmode
ndarray

\end{description}\end{quote}

\sphinxstrong{See also:}

\sphinxcode{\sphinxupquote{ufuncs\sphinxhyphen{}output\sphinxhyphen{}type()}}

\subsubsection*{Notes}

Arithmetic is modular when using integer types, and no error is
raised on overflow.
\subsubsection*{Examples}

\begin{sphinxVerbatim}[commandchars=\\\{\}]
\PYG{g+gp}{\PYGZgt{}\PYGZgt{}\PYGZgt{} }\PYG{n}{a} \PYG{o}{=} \PYG{n}{np}\PYG{o}{.}\PYG{n}{array}\PYG{p}{(}\PYG{p}{[}\PYG{l+m+mi}{1}\PYG{p}{,}\PYG{l+m+mi}{2}\PYG{p}{,}\PYG{l+m+mi}{3}\PYG{p}{]}\PYG{p}{)}
\PYG{g+gp}{\PYGZgt{}\PYGZgt{}\PYGZgt{} }\PYG{n}{np}\PYG{o}{.}\PYG{n}{cumprod}\PYG{p}{(}\PYG{n}{a}\PYG{p}{)} \PYG{c+c1}{\PYGZsh{} intermediate results 1, 1*2}
\PYG{g+gp}{... }              \PYG{c+c1}{\PYGZsh{} total product 1*2*3 = 6}
\PYG{g+go}{array([1, 2, 6])}
\PYG{g+gp}{\PYGZgt{}\PYGZgt{}\PYGZgt{} }\PYG{n}{a} \PYG{o}{=} \PYG{n}{np}\PYG{o}{.}\PYG{n}{array}\PYG{p}{(}\PYG{p}{[}\PYG{p}{[}\PYG{l+m+mi}{1}\PYG{p}{,} \PYG{l+m+mi}{2}\PYG{p}{,} \PYG{l+m+mi}{3}\PYG{p}{]}\PYG{p}{,} \PYG{p}{[}\PYG{l+m+mi}{4}\PYG{p}{,} \PYG{l+m+mi}{5}\PYG{p}{,} \PYG{l+m+mi}{6}\PYG{p}{]}\PYG{p}{]}\PYG{p}{)}
\PYG{g+gp}{\PYGZgt{}\PYGZgt{}\PYGZgt{} }\PYG{n}{np}\PYG{o}{.}\PYG{n}{cumprod}\PYG{p}{(}\PYG{n}{a}\PYG{p}{,} \PYG{n}{dtype}\PYG{o}{=}\PYG{n+nb}{float}\PYG{p}{)} \PYG{c+c1}{\PYGZsh{} specify type of output}
\PYG{g+go}{array([   1.,    2.,    6.,   24.,  120.,  720.])}
\end{sphinxVerbatim}

The cumulative product for each column (i.e., over the rows) of \sphinxtitleref{a}:

\begin{sphinxVerbatim}[commandchars=\\\{\}]
\PYG{g+gp}{\PYGZgt{}\PYGZgt{}\PYGZgt{} }\PYG{n}{np}\PYG{o}{.}\PYG{n}{cumprod}\PYG{p}{(}\PYG{n}{a}\PYG{p}{,} \PYG{n}{axis}\PYG{o}{=}\PYG{l+m+mi}{0}\PYG{p}{)}
\PYG{g+go}{array([[ 1,  2,  3],}
\PYG{g+go}{       [ 4, 10, 18]])}
\end{sphinxVerbatim}

The cumulative product for each row (i.e. over the columns) of \sphinxtitleref{a}:

\begin{sphinxVerbatim}[commandchars=\\\{\}]
\PYG{g+gp}{\PYGZgt{}\PYGZgt{}\PYGZgt{} }\PYG{n}{np}\PYG{o}{.}\PYG{n}{cumprod}\PYG{p}{(}\PYG{n}{a}\PYG{p}{,}\PYG{n}{axis}\PYG{o}{=}\PYG{l+m+mi}{1}\PYG{p}{)}
\PYG{g+go}{array([[  1,   2,   6],}
\PYG{g+go}{       [  4,  20, 120]])}
\end{sphinxVerbatim}

\end{fulllineitems}

\index{deg2rad() (in module symjax.tensor)@\spxentry{deg2rad()}\spxextra{in module symjax.tensor}}

\begin{fulllineitems}
\phantomsection\label{\detokenize{modules/tensor:symjax.tensor.deg2rad}}\pysiglinewithargsret{\sphinxbfcode{\sphinxupquote{deg2rad}}}{\emph{\DUrole{n}{x}}}{}
Convert angles from degrees to radians.

LAX\sphinxhyphen{}backend implementation of {\hyperref[\detokenize{modules/tensor:symjax.tensor.deg2rad}]{\sphinxcrossref{\sphinxcode{\sphinxupquote{deg2rad()}}}}}.
ADDITIONOriginal docstring below.

LAX\sphinxhyphen{}backend implementation of {\hyperref[\detokenize{modules/tensor:symjax.tensor.deg2rad}]{\sphinxcrossref{\sphinxcode{\sphinxupquote{deg2rad()}}}}}.
Original docstring below.

deg2rad(x, /, out=None, {\color{red}\bfseries{}*}, where=True, casting=’same\_kind’, order=’K’, dtype=None, subok=True{[}, signature, extobj{]})
\begin{quote}\begin{description}
\item[{Returns}] \leavevmode
\sphinxstylestrong{y} \textendash{} The corresponding angle in radians.
This is a scalar if \sphinxtitleref{x} is a scalar.

\item[{Return type}] \leavevmode
ndarray

\end{description}\end{quote}

\sphinxstrong{See also:}

\begin{description}
\item[{{\hyperref[\detokenize{modules/tensor:symjax.tensor.rad2deg}]{\sphinxcrossref{\sphinxcode{\sphinxupquote{rad2deg()}}}}}}] \leavevmode
Convert angles from radians to degrees.

\item[{\sphinxcode{\sphinxupquote{unwrap()}}}] \leavevmode
Remove large jumps in angle by wrapping.

\end{description}

\subsubsection*{Notes}

\DUrole{versionmodified,added}{New in version 1.3.0.}

\sphinxcode{\sphinxupquote{deg2rad(x)}} is \sphinxcode{\sphinxupquote{x * pi / 180}}.
\subsubsection*{Examples}

\begin{sphinxVerbatim}[commandchars=\\\{\}]
\PYG{g+gp}{\PYGZgt{}\PYGZgt{}\PYGZgt{} }\PYG{n}{np}\PYG{o}{.}\PYG{n}{deg2rad}\PYG{p}{(}\PYG{l+m+mi}{180}\PYG{p}{)}
\PYG{g+go}{3.1415926535897931}
\end{sphinxVerbatim}

\end{fulllineitems}

\index{degrees() (in module symjax.tensor)@\spxentry{degrees()}\spxextra{in module symjax.tensor}}

\begin{fulllineitems}
\phantomsection\label{\detokenize{modules/tensor:symjax.tensor.degrees}}\pysiglinewithargsret{\sphinxbfcode{\sphinxupquote{degrees}}}{\emph{\DUrole{n}{x}}}{}
Convert angles from radians to degrees.

LAX\sphinxhyphen{}backend implementation of {\hyperref[\detokenize{modules/tensor:symjax.tensor.rad2deg}]{\sphinxcrossref{\sphinxcode{\sphinxupquote{rad2deg()}}}}}.
ADDITIONOriginal docstring below.

LAX\sphinxhyphen{}backend implementation of {\hyperref[\detokenize{modules/tensor:symjax.tensor.rad2deg}]{\sphinxcrossref{\sphinxcode{\sphinxupquote{rad2deg()}}}}}.
Original docstring below.

rad2deg(x, /, out=None, {\color{red}\bfseries{}*}, where=True, casting=’same\_kind’, order=’K’, dtype=None, subok=True{[}, signature, extobj{]})
\begin{quote}\begin{description}
\item[{Returns}] \leavevmode
\sphinxstylestrong{y} \textendash{} The corresponding angle in degrees.
This is a scalar if \sphinxtitleref{x} is a scalar.

\item[{Return type}] \leavevmode
ndarray

\end{description}\end{quote}

\sphinxstrong{See also:}

\begin{description}
\item[{{\hyperref[\detokenize{modules/tensor:symjax.tensor.deg2rad}]{\sphinxcrossref{\sphinxcode{\sphinxupquote{deg2rad()}}}}}}] \leavevmode
Convert angles from degrees to radians.

\item[{\sphinxcode{\sphinxupquote{unwrap()}}}] \leavevmode
Remove large jumps in angle by wrapping.

\end{description}

\subsubsection*{Notes}

\DUrole{versionmodified,added}{New in version 1.3.0.}

rad2deg(x) is \sphinxcode{\sphinxupquote{180 * x / pi}}.
\subsubsection*{Examples}

\begin{sphinxVerbatim}[commandchars=\\\{\}]
\PYG{g+gp}{\PYGZgt{}\PYGZgt{}\PYGZgt{} }\PYG{n}{np}\PYG{o}{.}\PYG{n}{rad2deg}\PYG{p}{(}\PYG{n}{np}\PYG{o}{.}\PYG{n}{pi}\PYG{o}{/}\PYG{l+m+mi}{2}\PYG{p}{)}
\PYG{g+go}{90.0}
\end{sphinxVerbatim}

\end{fulllineitems}

\index{diag() (in module symjax.tensor)@\spxentry{diag()}\spxextra{in module symjax.tensor}}

\begin{fulllineitems}
\phantomsection\label{\detokenize{modules/tensor:symjax.tensor.diag}}\pysiglinewithargsret{\sphinxbfcode{\sphinxupquote{diag}}}{\emph{\DUrole{n}{v}}, \emph{\DUrole{n}{k}\DUrole{o}{=}\DUrole{default_value}{0}}}{}
Extract a diagonal or construct a diagonal array.

LAX\sphinxhyphen{}backend implementation of {\hyperref[\detokenize{modules/tensor:symjax.tensor.diag}]{\sphinxcrossref{\sphinxcode{\sphinxupquote{diag()}}}}}.
ADDITIONOriginal docstring below.

LAX\sphinxhyphen{}backend implementation of {\hyperref[\detokenize{modules/tensor:symjax.tensor.diag}]{\sphinxcrossref{\sphinxcode{\sphinxupquote{diag()}}}}}.
Original docstring below.

See the more detailed documentation for \sphinxcode{\sphinxupquote{numpy.diagonal}} if you use this
function to extract a diagonal and wish to write to the resulting array;
whether it returns a copy or a view depends on what version of numpy you
are using.
\begin{quote}\begin{description}
\item[{Returns}] \leavevmode
\sphinxstylestrong{out} \textendash{} The extracted diagonal or constructed diagonal array.

\item[{Return type}] \leavevmode
ndarray

\end{description}\end{quote}

\sphinxstrong{See also:}

\begin{description}
\item[{{\hyperref[\detokenize{modules/tensor:symjax.tensor.diagonal}]{\sphinxcrossref{\sphinxcode{\sphinxupquote{diagonal()}}}}}}] \leavevmode
Return specified diagonals.

\item[{\sphinxcode{\sphinxupquote{diagflat()}}}] \leavevmode
Create a 2\sphinxhyphen{}D array with the flattened input as a diagonal.

\item[{{\hyperref[\detokenize{modules/tensor:symjax.tensor.trace}]{\sphinxcrossref{\sphinxcode{\sphinxupquote{trace()}}}}}}] \leavevmode
Sum along diagonals.

\item[{{\hyperref[\detokenize{modules/tensor:symjax.tensor.triu}]{\sphinxcrossref{\sphinxcode{\sphinxupquote{triu()}}}}}}] \leavevmode
Upper triangle of an array.

\item[{{\hyperref[\detokenize{modules/tensor:symjax.tensor.tril}]{\sphinxcrossref{\sphinxcode{\sphinxupquote{tril()}}}}}}] \leavevmode
Lower triangle of an array.

\end{description}

\subsubsection*{Examples}

\begin{sphinxVerbatim}[commandchars=\\\{\}]
\PYG{g+gp}{\PYGZgt{}\PYGZgt{}\PYGZgt{} }\PYG{n}{x} \PYG{o}{=} \PYG{n}{np}\PYG{o}{.}\PYG{n}{arange}\PYG{p}{(}\PYG{l+m+mi}{9}\PYG{p}{)}\PYG{o}{.}\PYG{n}{reshape}\PYG{p}{(}\PYG{p}{(}\PYG{l+m+mi}{3}\PYG{p}{,}\PYG{l+m+mi}{3}\PYG{p}{)}\PYG{p}{)}
\PYG{g+gp}{\PYGZgt{}\PYGZgt{}\PYGZgt{} }\PYG{n}{x}
\PYG{g+go}{array([[0, 1, 2],}
\PYG{g+go}{       [3, 4, 5],}
\PYG{g+go}{       [6, 7, 8]])}
\end{sphinxVerbatim}

\begin{sphinxVerbatim}[commandchars=\\\{\}]
\PYG{g+gp}{\PYGZgt{}\PYGZgt{}\PYGZgt{} }\PYG{n}{np}\PYG{o}{.}\PYG{n}{diag}\PYG{p}{(}\PYG{n}{x}\PYG{p}{)}
\PYG{g+go}{array([0, 4, 8])}
\PYG{g+gp}{\PYGZgt{}\PYGZgt{}\PYGZgt{} }\PYG{n}{np}\PYG{o}{.}\PYG{n}{diag}\PYG{p}{(}\PYG{n}{x}\PYG{p}{,} \PYG{n}{k}\PYG{o}{=}\PYG{l+m+mi}{1}\PYG{p}{)}
\PYG{g+go}{array([1, 5])}
\PYG{g+gp}{\PYGZgt{}\PYGZgt{}\PYGZgt{} }\PYG{n}{np}\PYG{o}{.}\PYG{n}{diag}\PYG{p}{(}\PYG{n}{x}\PYG{p}{,} \PYG{n}{k}\PYG{o}{=}\PYG{o}{\PYGZhy{}}\PYG{l+m+mi}{1}\PYG{p}{)}
\PYG{g+go}{array([3, 7])}
\end{sphinxVerbatim}

\begin{sphinxVerbatim}[commandchars=\\\{\}]
\PYG{g+gp}{\PYGZgt{}\PYGZgt{}\PYGZgt{} }\PYG{n}{np}\PYG{o}{.}\PYG{n}{diag}\PYG{p}{(}\PYG{n}{np}\PYG{o}{.}\PYG{n}{diag}\PYG{p}{(}\PYG{n}{x}\PYG{p}{)}\PYG{p}{)}
\PYG{g+go}{array([[0, 0, 0],}
\PYG{g+go}{       [0, 4, 0],}
\PYG{g+go}{       [0, 0, 8]])}
\end{sphinxVerbatim}

\end{fulllineitems}

\index{diag\_indices() (in module symjax.tensor)@\spxentry{diag\_indices()}\spxextra{in module symjax.tensor}}

\begin{fulllineitems}
\phantomsection\label{\detokenize{modules/tensor:symjax.tensor.diag_indices}}\pysiglinewithargsret{\sphinxbfcode{\sphinxupquote{diag\_indices}}}{\emph{\DUrole{n}{n}}, \emph{\DUrole{n}{ndim}\DUrole{o}{=}\DUrole{default_value}{2}}}{}
Return the indices to access the main diagonal of an array.

LAX\sphinxhyphen{}backend implementation of {\hyperref[\detokenize{modules/tensor:symjax.tensor.diag_indices}]{\sphinxcrossref{\sphinxcode{\sphinxupquote{diag\_indices()}}}}}.
ADDITIONOriginal docstring below.

LAX\sphinxhyphen{}backend implementation of {\hyperref[\detokenize{modules/tensor:symjax.tensor.diag_indices}]{\sphinxcrossref{\sphinxcode{\sphinxupquote{diag\_indices()}}}}}.
Original docstring below.

This returns a tuple of indices that can be used to access the main
diagonal of an array \sphinxtitleref{a} with \sphinxcode{\sphinxupquote{a.ndim \textgreater{}= 2}} dimensions and shape
(n, n, …, n). For \sphinxcode{\sphinxupquote{a.ndim = 2}} this is the usual diagonal, for
\sphinxcode{\sphinxupquote{a.ndim \textgreater{} 2}} this is the set of indices to access \sphinxcode{\sphinxupquote{a{[}i, i, ..., i{]}}}
for \sphinxcode{\sphinxupquote{i = {[}0..n\sphinxhyphen{}1{]}}}.
\begin{quote}\begin{description}
\item[{Parameters}] \leavevmode
\sphinxstyleliteralstrong{\sphinxupquote{)}} \textendash{} 

\end{description}\end{quote}

\end{fulllineitems}

\index{diagonal() (in module symjax.tensor)@\spxentry{diagonal()}\spxextra{in module symjax.tensor}}

\begin{fulllineitems}
\phantomsection\label{\detokenize{modules/tensor:symjax.tensor.diagonal}}\pysiglinewithargsret{\sphinxbfcode{\sphinxupquote{diagonal}}}{\emph{\DUrole{n}{a}}, \emph{\DUrole{n}{offset}\DUrole{o}{=}\DUrole{default_value}{0}}, \emph{\DUrole{n}{axis1}\DUrole{o}{=}\DUrole{default_value}{0}}, \emph{\DUrole{n}{axis2}\DUrole{o}{=}\DUrole{default_value}{1}}}{}
Return specified diagonals.

LAX\sphinxhyphen{}backend implementation of {\hyperref[\detokenize{modules/tensor:symjax.tensor.diagonal}]{\sphinxcrossref{\sphinxcode{\sphinxupquote{diagonal()}}}}}.
ADDITIONOriginal docstring below.

LAX\sphinxhyphen{}backend implementation of {\hyperref[\detokenize{modules/tensor:symjax.tensor.diagonal}]{\sphinxcrossref{\sphinxcode{\sphinxupquote{diagonal()}}}}}.
Original docstring below.

If \sphinxtitleref{a} is 2\sphinxhyphen{}D, returns the diagonal of \sphinxtitleref{a} with the given offset,
i.e., the collection of elements of the form \sphinxcode{\sphinxupquote{a{[}i, i+offset{]}}}.  If
\sphinxtitleref{a} has more than two dimensions, then the axes specified by \sphinxtitleref{axis1}
and \sphinxtitleref{axis2} are used to determine the 2\sphinxhyphen{}D sub\sphinxhyphen{}array whose diagonal is
returned.  The shape of the resulting array can be determined by
removing \sphinxtitleref{axis1} and \sphinxtitleref{axis2} and appending an index to the right equal
to the size of the resulting diagonals.

In versions of NumPy prior to 1.7, this function always returned a new,
independent array containing a copy of the values in the diagonal.

In NumPy 1.7 and 1.8, it continues to return a copy of the diagonal,
but depending on this fact is deprecated. Writing to the resulting
array continues to work as it used to, but a FutureWarning is issued.

Starting in NumPy 1.9 it returns a read\sphinxhyphen{}only view on the original array.
Attempting to write to the resulting array will produce an error.

In some future release, it will return a read/write view and writing to
the returned array will alter your original array.  The returned array
will have the same type as the input array.

If you don’t write to the array returned by this function, then you can
just ignore all of the above.

If you depend on the current behavior, then we suggest copying the
returned array explicitly, i.e., use \sphinxcode{\sphinxupquote{np.diagonal(a).copy()}} instead
of just \sphinxcode{\sphinxupquote{np.diagonal(a)}}. This will work with both past and future
versions of NumPy.
\begin{quote}\begin{description}
\item[{Returns}] \leavevmode

\sphinxstylestrong{array\_of\_diagonals} \textendash{} If \sphinxtitleref{a} is 2\sphinxhyphen{}D, then a 1\sphinxhyphen{}D array containing the diagonal and of the
same type as \sphinxtitleref{a} is returned unless \sphinxtitleref{a} is a \sphinxtitleref{matrix}, in which case
a 1\sphinxhyphen{}D array rather than a (2\sphinxhyphen{}D) \sphinxtitleref{matrix} is returned in order to
maintain backward compatibility.

If \sphinxcode{\sphinxupquote{a.ndim \textgreater{} 2}}, then the dimensions specified by \sphinxtitleref{axis1} and \sphinxtitleref{axis2}
are removed, and a new axis inserted at the end corresponding to the
diagonal.

\item[{Return type}] \leavevmode
ndarray

\item[{Raises}] \leavevmode
\sphinxstyleliteralstrong{\sphinxupquote{ValueError}} \textendash{} If the dimension of \sphinxtitleref{a} is less than 2.

\end{description}\end{quote}

\sphinxstrong{See also:}

\begin{description}
\item[{{\hyperref[\detokenize{modules/tensor:symjax.tensor.diag}]{\sphinxcrossref{\sphinxcode{\sphinxupquote{diag()}}}}}}] \leavevmode
MATLAB work\sphinxhyphen{}a\sphinxhyphen{}like for 1\sphinxhyphen{}D and 2\sphinxhyphen{}D arrays.

\item[{\sphinxcode{\sphinxupquote{diagflat()}}}] \leavevmode
Create diagonal arrays.

\item[{{\hyperref[\detokenize{modules/tensor:symjax.tensor.trace}]{\sphinxcrossref{\sphinxcode{\sphinxupquote{trace()}}}}}}] \leavevmode
Sum along diagonals.

\end{description}

\subsubsection*{Examples}

\begin{sphinxVerbatim}[commandchars=\\\{\}]
\PYG{g+gp}{\PYGZgt{}\PYGZgt{}\PYGZgt{} }\PYG{n}{a} \PYG{o}{=} \PYG{n}{np}\PYG{o}{.}\PYG{n}{arange}\PYG{p}{(}\PYG{l+m+mi}{4}\PYG{p}{)}\PYG{o}{.}\PYG{n}{reshape}\PYG{p}{(}\PYG{l+m+mi}{2}\PYG{p}{,}\PYG{l+m+mi}{2}\PYG{p}{)}
\PYG{g+gp}{\PYGZgt{}\PYGZgt{}\PYGZgt{} }\PYG{n}{a}
\PYG{g+go}{array([[0, 1],}
\PYG{g+go}{       [2, 3]])}
\PYG{g+gp}{\PYGZgt{}\PYGZgt{}\PYGZgt{} }\PYG{n}{a}\PYG{o}{.}\PYG{n}{diagonal}\PYG{p}{(}\PYG{p}{)}
\PYG{g+go}{array([0, 3])}
\PYG{g+gp}{\PYGZgt{}\PYGZgt{}\PYGZgt{} }\PYG{n}{a}\PYG{o}{.}\PYG{n}{diagonal}\PYG{p}{(}\PYG{l+m+mi}{1}\PYG{p}{)}
\PYG{g+go}{array([1])}
\end{sphinxVerbatim}

A 3\sphinxhyphen{}D example:

\begin{sphinxVerbatim}[commandchars=\\\{\}]
\PYG{g+gp}{\PYGZgt{}\PYGZgt{}\PYGZgt{} }\PYG{n}{a} \PYG{o}{=} \PYG{n}{np}\PYG{o}{.}\PYG{n}{arange}\PYG{p}{(}\PYG{l+m+mi}{8}\PYG{p}{)}\PYG{o}{.}\PYG{n}{reshape}\PYG{p}{(}\PYG{l+m+mi}{2}\PYG{p}{,}\PYG{l+m+mi}{2}\PYG{p}{,}\PYG{l+m+mi}{2}\PYG{p}{)}\PYG{p}{;} \PYG{n}{a}
\PYG{g+go}{array([[[0, 1],}
\PYG{g+go}{        [2, 3]],}
\PYG{g+go}{       [[4, 5],}
\PYG{g+go}{        [6, 7]]])}
\PYG{g+gp}{\PYGZgt{}\PYGZgt{}\PYGZgt{} }\PYG{n}{a}\PYG{o}{.}\PYG{n}{diagonal}\PYG{p}{(}\PYG{l+m+mi}{0}\PYG{p}{,}  \PYG{c+c1}{\PYGZsh{} Main diagonals of two arrays created by skipping}
\PYG{g+gp}{... }           \PYG{l+m+mi}{0}\PYG{p}{,}  \PYG{c+c1}{\PYGZsh{} across the outer(left)\PYGZhy{}most axis last and}
\PYG{g+gp}{... }           \PYG{l+m+mi}{1}\PYG{p}{)}  \PYG{c+c1}{\PYGZsh{} the \PYGZdq{}middle\PYGZdq{} (row) axis first.}
\PYG{g+go}{array([[0, 6],}
\PYG{g+go}{       [1, 7]])}
\end{sphinxVerbatim}

The sub\sphinxhyphen{}arrays whose main diagonals we just obtained; note that each
corresponds to fixing the right\sphinxhyphen{}most (column) axis, and that the
diagonals are “packed” in rows.

\begin{sphinxVerbatim}[commandchars=\\\{\}]
\PYG{g+gp}{\PYGZgt{}\PYGZgt{}\PYGZgt{} }\PYG{n}{a}\PYG{p}{[}\PYG{p}{:}\PYG{p}{,}\PYG{p}{:}\PYG{p}{,}\PYG{l+m+mi}{0}\PYG{p}{]}  \PYG{c+c1}{\PYGZsh{} main diagonal is [0 6]}
\PYG{g+go}{array([[0, 2],}
\PYG{g+go}{       [4, 6]])}
\PYG{g+gp}{\PYGZgt{}\PYGZgt{}\PYGZgt{} }\PYG{n}{a}\PYG{p}{[}\PYG{p}{:}\PYG{p}{,}\PYG{p}{:}\PYG{p}{,}\PYG{l+m+mi}{1}\PYG{p}{]}  \PYG{c+c1}{\PYGZsh{} main diagonal is [1 7]}
\PYG{g+go}{array([[1, 3],}
\PYG{g+go}{       [5, 7]])}
\end{sphinxVerbatim}

The anti\sphinxhyphen{}diagonal can be obtained by reversing the order of elements
using either \sphinxtitleref{numpy.flipud} or \sphinxtitleref{numpy.fliplr}.

\begin{sphinxVerbatim}[commandchars=\\\{\}]
\PYG{g+gp}{\PYGZgt{}\PYGZgt{}\PYGZgt{} }\PYG{n}{a} \PYG{o}{=} \PYG{n}{np}\PYG{o}{.}\PYG{n}{arange}\PYG{p}{(}\PYG{l+m+mi}{9}\PYG{p}{)}\PYG{o}{.}\PYG{n}{reshape}\PYG{p}{(}\PYG{l+m+mi}{3}\PYG{p}{,} \PYG{l+m+mi}{3}\PYG{p}{)}
\PYG{g+gp}{\PYGZgt{}\PYGZgt{}\PYGZgt{} }\PYG{n}{a}
\PYG{g+go}{array([[0, 1, 2],}
\PYG{g+go}{       [3, 4, 5],}
\PYG{g+go}{       [6, 7, 8]])}
\PYG{g+gp}{\PYGZgt{}\PYGZgt{}\PYGZgt{} }\PYG{n}{np}\PYG{o}{.}\PYG{n}{fliplr}\PYG{p}{(}\PYG{n}{a}\PYG{p}{)}\PYG{o}{.}\PYG{n}{diagonal}\PYG{p}{(}\PYG{p}{)}  \PYG{c+c1}{\PYGZsh{} Horizontal flip}
\PYG{g+go}{array([2, 4, 6])}
\PYG{g+gp}{\PYGZgt{}\PYGZgt{}\PYGZgt{} }\PYG{n}{np}\PYG{o}{.}\PYG{n}{flipud}\PYG{p}{(}\PYG{n}{a}\PYG{p}{)}\PYG{o}{.}\PYG{n}{diagonal}\PYG{p}{(}\PYG{p}{)}  \PYG{c+c1}{\PYGZsh{} Vertical flip}
\PYG{g+go}{array([6, 4, 2])}
\end{sphinxVerbatim}

Note that the order in which the diagonal is retrieved varies depending
on the flip function.

\end{fulllineitems}

\index{divide() (in module symjax.tensor)@\spxentry{divide()}\spxextra{in module symjax.tensor}}

\begin{fulllineitems}
\phantomsection\label{\detokenize{modules/tensor:symjax.tensor.divide}}\pysiglinewithargsret{\sphinxbfcode{\sphinxupquote{divide}}}{\emph{\DUrole{n}{x1}}, \emph{\DUrole{n}{x2}}}{}
Returns a true division of the inputs, element\sphinxhyphen{}wise.

LAX\sphinxhyphen{}backend implementation of {\hyperref[\detokenize{modules/tensor:symjax.tensor.true_divide}]{\sphinxcrossref{\sphinxcode{\sphinxupquote{true\_divide()}}}}}.
ADDITIONOriginal docstring below.

LAX\sphinxhyphen{}backend implementation of {\hyperref[\detokenize{modules/tensor:symjax.tensor.true_divide}]{\sphinxcrossref{\sphinxcode{\sphinxupquote{true\_divide()}}}}}.
Original docstring below.

true\_divide(x1, x2, /, out=None, {\color{red}\bfseries{}*}, where=True, casting=’same\_kind’, order=’K’, dtype=None, subok=True{[}, signature, extobj{]})

Instead of the Python traditional ‘floor division’, this returns a true
division.  True division adjusts the output type to present the best
answer, regardless of input types.
\begin{quote}\begin{description}
\item[{Returns}] \leavevmode
\sphinxstylestrong{out} \textendash{} This is a scalar if both \sphinxtitleref{x1} and \sphinxtitleref{x2} are scalars.

\item[{Return type}] \leavevmode
ndarray or scalar

\end{description}\end{quote}
\subsubsection*{Notes}

The floor division operator \sphinxcode{\sphinxupquote{//}} was added in Python 2.2 making
\sphinxcode{\sphinxupquote{//}} and \sphinxcode{\sphinxupquote{/}} equivalent operators.  The default floor division
operation of \sphinxcode{\sphinxupquote{/}} can be replaced by true division with \sphinxcode{\sphinxupquote{from
\_\_future\_\_ import division}}.

In Python 3.0, \sphinxcode{\sphinxupquote{//}} is the floor division operator and \sphinxcode{\sphinxupquote{/}} the
true division operator.  The \sphinxcode{\sphinxupquote{true\_divide(x1, x2)}} function is
equivalent to true division in Python.
\subsubsection*{Examples}

\begin{sphinxVerbatim}[commandchars=\\\{\}]
\PYG{g+gp}{\PYGZgt{}\PYGZgt{}\PYGZgt{} }\PYG{n}{x} \PYG{o}{=} \PYG{n}{np}\PYG{o}{.}\PYG{n}{arange}\PYG{p}{(}\PYG{l+m+mi}{5}\PYG{p}{)}
\PYG{g+gp}{\PYGZgt{}\PYGZgt{}\PYGZgt{} }\PYG{n}{np}\PYG{o}{.}\PYG{n}{true\PYGZus{}divide}\PYG{p}{(}\PYG{n}{x}\PYG{p}{,} \PYG{l+m+mi}{4}\PYG{p}{)}
\PYG{g+go}{array([ 0.  ,  0.25,  0.5 ,  0.75,  1.  ])}
\end{sphinxVerbatim}

\begin{sphinxVerbatim}[commandchars=\\\{\}]
\PYG{g+gp}{\PYGZgt{}\PYGZgt{}\PYGZgt{} }\PYG{n}{x}\PYG{o}{/}\PYG{o}{/}\PYG{l+m+mi}{4}
\PYG{g+go}{array([0, 0, 0, 0, 1])}
\end{sphinxVerbatim}

\begin{sphinxVerbatim}[commandchars=\\\{\}]
\PYG{g+gp}{\PYGZgt{}\PYGZgt{}\PYGZgt{} }\PYG{k+kn}{from} \PYG{n+nn}{\PYGZus{}\PYGZus{}future\PYGZus{}\PYGZus{}} \PYG{k+kn}{import} \PYG{n}{division}
\PYG{g+gp}{\PYGZgt{}\PYGZgt{}\PYGZgt{} }\PYG{n}{x}\PYG{o}{/}\PYG{l+m+mi}{4}
\PYG{g+go}{array([ 0.  ,  0.25,  0.5 ,  0.75,  1.  ])}
\PYG{g+gp}{\PYGZgt{}\PYGZgt{}\PYGZgt{} }\PYG{n}{x}\PYG{o}{/}\PYG{o}{/}\PYG{l+m+mi}{4}
\PYG{g+go}{array([0, 0, 0, 0, 1])}
\end{sphinxVerbatim}

\end{fulllineitems}

\index{divmod() (in module symjax.tensor)@\spxentry{divmod()}\spxextra{in module symjax.tensor}}

\begin{fulllineitems}
\phantomsection\label{\detokenize{modules/tensor:symjax.tensor.divmod}}\pysiglinewithargsret{\sphinxbfcode{\sphinxupquote{divmod}}}{\emph{\DUrole{n}{x1}}, \emph{\DUrole{n}{x2}}}{}
Return element\sphinxhyphen{}wise quotient and remainder simultaneously.

LAX\sphinxhyphen{}backend implementation of {\hyperref[\detokenize{modules/tensor:symjax.tensor.divmod}]{\sphinxcrossref{\sphinxcode{\sphinxupquote{divmod()}}}}}.
ADDITIONOriginal docstring below.

LAX\sphinxhyphen{}backend implementation of {\hyperref[\detokenize{modules/tensor:symjax.tensor.divmod}]{\sphinxcrossref{\sphinxcode{\sphinxupquote{divmod()}}}}}.
Original docstring below.

divmod(x1, x2{[}, out1, out2{]}, / {[}, out=(None, None){]}, {\color{red}\bfseries{}*}, where=True, casting=’same\_kind’, order=’K’, dtype=None, subok=True{[}, signature, extobj{]})

\DUrole{versionmodified,added}{New in version 1.13.0.}

\sphinxcode{\sphinxupquote{np.divmod(x, y)}} is equivalent to \sphinxcode{\sphinxupquote{(x // y, x \% y)}}, but faster
because it avoids redundant work. It is used to implement the Python
built\sphinxhyphen{}in function \sphinxcode{\sphinxupquote{divmod}} on NumPy arrays.
\begin{quote}\begin{description}
\item[{Returns}] \leavevmode
\begin{itemize}
\item {} 
\sphinxstylestrong{out1} (\sphinxstyleemphasis{ndarray}) \textendash{} Element\sphinxhyphen{}wise quotient resulting from floor division.
This is a scalar if both \sphinxtitleref{x1} and \sphinxtitleref{x2} are scalars.

\item {} 
\sphinxstylestrong{out2} (\sphinxstyleemphasis{ndarray}) \textendash{} Element\sphinxhyphen{}wise remainder from floor division.
This is a scalar if both \sphinxtitleref{x1} and \sphinxtitleref{x2} are scalars.

\end{itemize}

\end{description}\end{quote}

\sphinxstrong{See also:}

\begin{description}
\item[{{\hyperref[\detokenize{modules/tensor:symjax.tensor.floor_divide}]{\sphinxcrossref{\sphinxcode{\sphinxupquote{floor\_divide()}}}}}}] \leavevmode
Equivalent to Python’s \sphinxcode{\sphinxupquote{//}} operator.

\item[{{\hyperref[\detokenize{modules/tensor:symjax.tensor.remainder}]{\sphinxcrossref{\sphinxcode{\sphinxupquote{remainder()}}}}}}] \leavevmode
Equivalent to Python’s \sphinxcode{\sphinxupquote{\%}} operator.

\item[{\sphinxcode{\sphinxupquote{modf()}}}] \leavevmode
Equivalent to \sphinxcode{\sphinxupquote{divmod(x, 1)}} for positive \sphinxcode{\sphinxupquote{x}} with the return values switched.

\end{description}

\subsubsection*{Examples}

\begin{sphinxVerbatim}[commandchars=\\\{\}]
\PYG{g+gp}{\PYGZgt{}\PYGZgt{}\PYGZgt{} }\PYG{n}{np}\PYG{o}{.}\PYG{n}{divmod}\PYG{p}{(}\PYG{n}{np}\PYG{o}{.}\PYG{n}{arange}\PYG{p}{(}\PYG{l+m+mi}{5}\PYG{p}{)}\PYG{p}{,} \PYG{l+m+mi}{3}\PYG{p}{)}
\PYG{g+go}{(array([0, 0, 0, 1, 1]), array([0, 1, 2, 0, 1]))}
\end{sphinxVerbatim}

\end{fulllineitems}

\index{dot() (in module symjax.tensor)@\spxentry{dot()}\spxextra{in module symjax.tensor}}

\begin{fulllineitems}
\phantomsection\label{\detokenize{modules/tensor:symjax.tensor.dot}}\pysiglinewithargsret{\sphinxbfcode{\sphinxupquote{dot}}}{\emph{\DUrole{n}{a}}, \emph{\DUrole{n}{b}}, \emph{\DUrole{n}{precision}\DUrole{o}{=}\DUrole{default_value}{None}}}{}
Dot product of two arrays. Specifically,

LAX\sphinxhyphen{}backend implementation of {\hyperref[\detokenize{modules/tensor:symjax.tensor.dot}]{\sphinxcrossref{\sphinxcode{\sphinxupquote{dot()}}}}}.
ADDITIONOriginal docstring below.

LAX\sphinxhyphen{}backend implementation of {\hyperref[\detokenize{modules/tensor:symjax.tensor.dot}]{\sphinxcrossref{\sphinxcode{\sphinxupquote{dot()}}}}}.
In addition to the original NumPy arguments listed below, also supports
\sphinxcode{\sphinxupquote{precision}} for extra control over matrix\sphinxhyphen{}multiplication precision
on supported devices. See \sphinxcode{\sphinxupquote{jax.lax.dot()}} for details.

Original docstring below.
\begin{quote}
\begin{quote}

dot(a, b, out=None)
\begin{itemize}
\item {} 
If both \sphinxtitleref{a} and \sphinxtitleref{b} are 1\sphinxhyphen{}D arrays, it is inner product of vectors
(without complex conjugation).

\item {} 
If both \sphinxtitleref{a} and \sphinxtitleref{b} are 2\sphinxhyphen{}D arrays, it is matrix multiplication,
but using {\hyperref[\detokenize{modules/tensor:symjax.tensor.matmul}]{\sphinxcrossref{\sphinxcode{\sphinxupquote{matmul()}}}}} or \sphinxcode{\sphinxupquote{a @ b}} is preferred.

\item {} 
If either \sphinxtitleref{a} or \sphinxtitleref{b} is 0\sphinxhyphen{}D (scalar), it is equivalent to {\hyperref[\detokenize{modules/tensor:symjax.tensor.multiply}]{\sphinxcrossref{\sphinxcode{\sphinxupquote{multiply()}}}}}
and using \sphinxcode{\sphinxupquote{numpy.multiply(a, b)}} or \sphinxcode{\sphinxupquote{a * b}} is preferred.

\item {} 
If \sphinxtitleref{a} is an N\sphinxhyphen{}D array and \sphinxtitleref{b} is a 1\sphinxhyphen{}D array, it is a sum product over
the last axis of \sphinxtitleref{a} and \sphinxtitleref{b}.

\item {} 
If \sphinxtitleref{a} is an N\sphinxhyphen{}D array and \sphinxtitleref{b} is an M\sphinxhyphen{}D array (where \sphinxcode{\sphinxupquote{M\textgreater{}=2}}), it is a
sum product over the last axis of \sphinxtitleref{a} and the second\sphinxhyphen{}to\sphinxhyphen{}last axis of \sphinxtitleref{b}:

\begin{sphinxVerbatim}[commandchars=\\\{\}]
\PYG{n}{dot}\PYG{p}{(}\PYG{n}{a}\PYG{p}{,} \PYG{n}{b}\PYG{p}{)}\PYG{p}{[}\PYG{n}{i}\PYG{p}{,}\PYG{n}{j}\PYG{p}{,}\PYG{n}{k}\PYG{p}{,}\PYG{n}{m}\PYG{p}{]} \PYG{o}{=} \PYG{n+nb}{sum}\PYG{p}{(}\PYG{n}{a}\PYG{p}{[}\PYG{n}{i}\PYG{p}{,}\PYG{n}{j}\PYG{p}{,}\PYG{p}{:}\PYG{p}{]} \PYG{o}{*} \PYG{n}{b}\PYG{p}{[}\PYG{n}{k}\PYG{p}{,}\PYG{p}{:}\PYG{p}{,}\PYG{n}{m}\PYG{p}{]}\PYG{p}{)}
\end{sphinxVerbatim}

\end{itemize}
\end{quote}
\begin{description}
\item[{Returns}] \leavevmode\begin{description}
\item[{output}] \leavevmode{[}ndarray{]}
Returns the dot product of \sphinxtitleref{a} and \sphinxtitleref{b}.  If \sphinxtitleref{a} and \sphinxtitleref{b} are both
scalars or both 1\sphinxhyphen{}D arrays then a scalar is returned; otherwise
an array is returned.
If \sphinxtitleref{out} is given, then it is returned.

\end{description}
\begin{description}
\item[{ValueError}] \leavevmode
If the last dimension of \sphinxtitleref{a} is not the same size as
the second\sphinxhyphen{}to\sphinxhyphen{}last dimension of \sphinxtitleref{b}.

\end{description}

vdot : Complex\sphinxhyphen{}conjugating dot product.
tensordot : Sum products over arbitrary axes.
einsum : Einstein summation convention.
matmul : ‘@’ operator as method with out parameter.

\begin{sphinxVerbatim}[commandchars=\\\{\}]
\PYG{g+gp}{\PYGZgt{}\PYGZgt{}\PYGZgt{} }\PYG{n}{np}\PYG{o}{.}\PYG{n}{dot}\PYG{p}{(}\PYG{l+m+mi}{3}\PYG{p}{,} \PYG{l+m+mi}{4}\PYG{p}{)}
\PYG{g+go}{12}
\end{sphinxVerbatim}

Neither argument is complex\sphinxhyphen{}conjugated:

\begin{sphinxVerbatim}[commandchars=\\\{\}]
\PYG{g+gp}{\PYGZgt{}\PYGZgt{}\PYGZgt{} }\PYG{n}{np}\PYG{o}{.}\PYG{n}{dot}\PYG{p}{(}\PYG{p}{[}\PYG{l+m+mi}{2}\PYG{n}{j}\PYG{p}{,} \PYG{l+m+mi}{3}\PYG{n}{j}\PYG{p}{]}\PYG{p}{,} \PYG{p}{[}\PYG{l+m+mi}{2}\PYG{n}{j}\PYG{p}{,} \PYG{l+m+mi}{3}\PYG{n}{j}\PYG{p}{]}\PYG{p}{)}
\PYG{g+go}{(\PYGZhy{}13+0j)}
\end{sphinxVerbatim}

For 2\sphinxhyphen{}D arrays it is the matrix product:

\begin{sphinxVerbatim}[commandchars=\\\{\}]
\PYG{g+gp}{\PYGZgt{}\PYGZgt{}\PYGZgt{} }\PYG{n}{a} \PYG{o}{=} \PYG{p}{[}\PYG{p}{[}\PYG{l+m+mi}{1}\PYG{p}{,} \PYG{l+m+mi}{0}\PYG{p}{]}\PYG{p}{,} \PYG{p}{[}\PYG{l+m+mi}{0}\PYG{p}{,} \PYG{l+m+mi}{1}\PYG{p}{]}\PYG{p}{]}
\PYG{g+gp}{\PYGZgt{}\PYGZgt{}\PYGZgt{} }\PYG{n}{b} \PYG{o}{=} \PYG{p}{[}\PYG{p}{[}\PYG{l+m+mi}{4}\PYG{p}{,} \PYG{l+m+mi}{1}\PYG{p}{]}\PYG{p}{,} \PYG{p}{[}\PYG{l+m+mi}{2}\PYG{p}{,} \PYG{l+m+mi}{2}\PYG{p}{]}\PYG{p}{]}
\PYG{g+gp}{\PYGZgt{}\PYGZgt{}\PYGZgt{} }\PYG{n}{np}\PYG{o}{.}\PYG{n}{dot}\PYG{p}{(}\PYG{n}{a}\PYG{p}{,} \PYG{n}{b}\PYG{p}{)}
\PYG{g+go}{array([[4, 1],}
\PYG{g+go}{       [2, 2]])}
\end{sphinxVerbatim}

\begin{sphinxVerbatim}[commandchars=\\\{\}]
\PYG{g+gp}{\PYGZgt{}\PYGZgt{}\PYGZgt{} }\PYG{n}{a} \PYG{o}{=} \PYG{n}{np}\PYG{o}{.}\PYG{n}{arange}\PYG{p}{(}\PYG{l+m+mi}{3}\PYG{o}{*}\PYG{l+m+mi}{4}\PYG{o}{*}\PYG{l+m+mi}{5}\PYG{o}{*}\PYG{l+m+mi}{6}\PYG{p}{)}\PYG{o}{.}\PYG{n}{reshape}\PYG{p}{(}\PYG{p}{(}\PYG{l+m+mi}{3}\PYG{p}{,}\PYG{l+m+mi}{4}\PYG{p}{,}\PYG{l+m+mi}{5}\PYG{p}{,}\PYG{l+m+mi}{6}\PYG{p}{)}\PYG{p}{)}
\PYG{g+gp}{\PYGZgt{}\PYGZgt{}\PYGZgt{} }\PYG{n}{b} \PYG{o}{=} \PYG{n}{np}\PYG{o}{.}\PYG{n}{arange}\PYG{p}{(}\PYG{l+m+mi}{3}\PYG{o}{*}\PYG{l+m+mi}{4}\PYG{o}{*}\PYG{l+m+mi}{5}\PYG{o}{*}\PYG{l+m+mi}{6}\PYG{p}{)}\PYG{p}{[}\PYG{p}{:}\PYG{p}{:}\PYG{o}{\PYGZhy{}}\PYG{l+m+mi}{1}\PYG{p}{]}\PYG{o}{.}\PYG{n}{reshape}\PYG{p}{(}\PYG{p}{(}\PYG{l+m+mi}{5}\PYG{p}{,}\PYG{l+m+mi}{4}\PYG{p}{,}\PYG{l+m+mi}{6}\PYG{p}{,}\PYG{l+m+mi}{3}\PYG{p}{)}\PYG{p}{)}
\PYG{g+gp}{\PYGZgt{}\PYGZgt{}\PYGZgt{} }\PYG{n}{np}\PYG{o}{.}\PYG{n}{dot}\PYG{p}{(}\PYG{n}{a}\PYG{p}{,} \PYG{n}{b}\PYG{p}{)}\PYG{p}{[}\PYG{l+m+mi}{2}\PYG{p}{,}\PYG{l+m+mi}{3}\PYG{p}{,}\PYG{l+m+mi}{2}\PYG{p}{,}\PYG{l+m+mi}{1}\PYG{p}{,}\PYG{l+m+mi}{2}\PYG{p}{,}\PYG{l+m+mi}{2}\PYG{p}{]}
\PYG{g+go}{499128}
\PYG{g+gp}{\PYGZgt{}\PYGZgt{}\PYGZgt{} }\PYG{n+nb}{sum}\PYG{p}{(}\PYG{n}{a}\PYG{p}{[}\PYG{l+m+mi}{2}\PYG{p}{,}\PYG{l+m+mi}{3}\PYG{p}{,}\PYG{l+m+mi}{2}\PYG{p}{,}\PYG{p}{:}\PYG{p}{]} \PYG{o}{*} \PYG{n}{b}\PYG{p}{[}\PYG{l+m+mi}{1}\PYG{p}{,}\PYG{l+m+mi}{2}\PYG{p}{,}\PYG{p}{:}\PYG{p}{,}\PYG{l+m+mi}{2}\PYG{p}{]}\PYG{p}{)}
\PYG{g+go}{499128}
\end{sphinxVerbatim}

\end{description}
\end{quote}

\end{fulllineitems}

\index{dsplit() (in module symjax.tensor)@\spxentry{dsplit()}\spxextra{in module symjax.tensor}}

\begin{fulllineitems}
\phantomsection\label{\detokenize{modules/tensor:symjax.tensor.dsplit}}\pysiglinewithargsret{\sphinxbfcode{\sphinxupquote{dsplit}}}{\emph{\DUrole{n}{ary}}, \emph{\DUrole{n}{indices\_or\_sections}}}{}
Split array into multiple sub\sphinxhyphen{}arrays along the 3rd axis (depth).

LAX\sphinxhyphen{}backend implementation of {\hyperref[\detokenize{modules/tensor:symjax.tensor.dsplit}]{\sphinxcrossref{\sphinxcode{\sphinxupquote{dsplit()}}}}}.
ADDITIONOriginal docstring below.

LA

\end{fulllineitems}

\index{dstack() (in module symjax.tensor)@\spxentry{dstack()}\spxextra{in module symjax.tensor}}

\begin{fulllineitems}
\phantomsection\label{\detokenize{modules/tensor:symjax.tensor.dstack}}\pysiglinewithargsret{\sphinxbfcode{\sphinxupquote{dstack}}}{\emph{\DUrole{n}{tup}}}{}
Stack arrays in sequence depth wise (along third axis).

LAX\sphinxhyphen{}backend implementation of {\hyperref[\detokenize{modules/tensor:symjax.tensor.dstack}]{\sphinxcrossref{\sphinxcode{\sphinxupquote{dstack()}}}}}.
ADDITIONOriginal docstring below.

LAX\sphinxhyphen{}backend implementation of {\hyperref[\detokenize{modules/tensor:symjax.tensor.dstack}]{\sphinxcrossref{\sphinxcode{\sphinxupquote{dstack()}}}}}.
Original docstring below.

This is equivalent to concatenation along the third axis after 2\sphinxhyphen{}D arrays
of shape \sphinxtitleref{(M,N)} have been reshaped to \sphinxtitleref{(M,N,1)} and 1\sphinxhyphen{}D arrays of shape
\sphinxtitleref{(N,)} have been reshaped to \sphinxtitleref{(1,N,1)}. Rebuilds arrays divided by
\sphinxtitleref{dsplit}.

This function makes most sense for arrays with up to 3 dimensions. For
instance, for pixel\sphinxhyphen{}data with a height (first axis), width (second axis),
and r/g/b channels (third axis). The functions \sphinxtitleref{concatenate}, \sphinxtitleref{stack} and
\sphinxtitleref{block} provide more general stacking and concatenation operations.
\begin{quote}\begin{description}
\item[{Returns}] \leavevmode
\sphinxstylestrong{stacked} \textendash{} The array formed by stacking the given arrays, will be at least 3\sphinxhyphen{}D.

\item[{Return type}] \leavevmode
ndarray

\end{description}\end{quote}

\sphinxstrong{See also:}

\begin{description}
\item[{{\hyperref[\detokenize{modules/tensor:symjax.tensor.stack}]{\sphinxcrossref{\sphinxcode{\sphinxupquote{stack()}}}}}}] \leavevmode
Join a sequence of arrays along a new axis.

\item[{{\hyperref[\detokenize{modules/tensor:symjax.tensor.vstack}]{\sphinxcrossref{\sphinxcode{\sphinxupquote{vstack()}}}}}}] \leavevmode
Stack along first axis.

\item[{{\hyperref[\detokenize{modules/tensor:symjax.tensor.hstack}]{\sphinxcrossref{\sphinxcode{\sphinxupquote{hstack()}}}}}}] \leavevmode
Stack along second axis.

\item[{{\hyperref[\detokenize{modules/tensor:symjax.tensor.concatenate}]{\sphinxcrossref{\sphinxcode{\sphinxupquote{concatenate()}}}}}}] \leavevmode
Join a sequence of arrays along an existing axis.

\item[{{\hyperref[\detokenize{modules/tensor:symjax.tensor.dsplit}]{\sphinxcrossref{\sphinxcode{\sphinxupquote{dsplit()}}}}}}] \leavevmode
Split array along third axis.

\end{description}

\subsubsection*{Examples}

\begin{sphinxVerbatim}[commandchars=\\\{\}]
\PYG{g+gp}{\PYGZgt{}\PYGZgt{}\PYGZgt{} }\PYG{n}{a} \PYG{o}{=} \PYG{n}{np}\PYG{o}{.}\PYG{n}{array}\PYG{p}{(}\PYG{p}{(}\PYG{l+m+mi}{1}\PYG{p}{,}\PYG{l+m+mi}{2}\PYG{p}{,}\PYG{l+m+mi}{3}\PYG{p}{)}\PYG{p}{)}
\PYG{g+gp}{\PYGZgt{}\PYGZgt{}\PYGZgt{} }\PYG{n}{b} \PYG{o}{=} \PYG{n}{np}\PYG{o}{.}\PYG{n}{array}\PYG{p}{(}\PYG{p}{(}\PYG{l+m+mi}{2}\PYG{p}{,}\PYG{l+m+mi}{3}\PYG{p}{,}\PYG{l+m+mi}{4}\PYG{p}{)}\PYG{p}{)}
\PYG{g+gp}{\PYGZgt{}\PYGZgt{}\PYGZgt{} }\PYG{n}{np}\PYG{o}{.}\PYG{n}{dstack}\PYG{p}{(}\PYG{p}{(}\PYG{n}{a}\PYG{p}{,}\PYG{n}{b}\PYG{p}{)}\PYG{p}{)}
\PYG{g+go}{array([[[1, 2],}
\PYG{g+go}{        [2, 3],}
\PYG{g+go}{        [3, 4]]])}
\end{sphinxVerbatim}

\begin{sphinxVerbatim}[commandchars=\\\{\}]
\PYG{g+gp}{\PYGZgt{}\PYGZgt{}\PYGZgt{} }\PYG{n}{a} \PYG{o}{=} \PYG{n}{np}\PYG{o}{.}\PYG{n}{array}\PYG{p}{(}\PYG{p}{[}\PYG{p}{[}\PYG{l+m+mi}{1}\PYG{p}{]}\PYG{p}{,}\PYG{p}{[}\PYG{l+m+mi}{2}\PYG{p}{]}\PYG{p}{,}\PYG{p}{[}\PYG{l+m+mi}{3}\PYG{p}{]}\PYG{p}{]}\PYG{p}{)}
\PYG{g+gp}{\PYGZgt{}\PYGZgt{}\PYGZgt{} }\PYG{n}{b} \PYG{o}{=} \PYG{n}{np}\PYG{o}{.}\PYG{n}{array}\PYG{p}{(}\PYG{p}{[}\PYG{p}{[}\PYG{l+m+mi}{2}\PYG{p}{]}\PYG{p}{,}\PYG{p}{[}\PYG{l+m+mi}{3}\PYG{p}{]}\PYG{p}{,}\PYG{p}{[}\PYG{l+m+mi}{4}\PYG{p}{]}\PYG{p}{]}\PYG{p}{)}
\PYG{g+gp}{\PYGZgt{}\PYGZgt{}\PYGZgt{} }\PYG{n}{np}\PYG{o}{.}\PYG{n}{dstack}\PYG{p}{(}\PYG{p}{(}\PYG{n}{a}\PYG{p}{,}\PYG{n}{b}\PYG{p}{)}\PYG{p}{)}
\PYG{g+go}{array([[[1, 2]],}
\PYG{g+go}{       [[2, 3]],}
\PYG{g+go}{       [[3, 4]]])}
\end{sphinxVerbatim}

\end{fulllineitems}

\index{einsum() (in module symjax.tensor)@\spxentry{einsum()}\spxextra{in module symjax.tensor}}

\begin{fulllineitems}
\phantomsection\label{\detokenize{modules/tensor:symjax.tensor.einsum}}\pysiglinewithargsret{\sphinxbfcode{\sphinxupquote{einsum}}}{\emph{\DUrole{o}{*}\DUrole{n}{operands}}, \emph{\DUrole{o}{**}\DUrole{n}{kwargs}}}{}
Evaluates the Einstein summation convention on the operands.

LAX\sphinxhyphen{}backend implementation of {\hyperref[\detokenize{modules/tensor:symjax.tensor.einsum}]{\sphinxcrossref{\sphinxcode{\sphinxupquote{einsum()}}}}}.
ADDITIONOriginal docstring below.

LAX\sphinxhyphen{}backend implementation of {\hyperref[\detokenize{modules/tensor:symjax.tensor.einsum}]{\sphinxcrossref{\sphinxcode{\sphinxupquote{einsum()}}}}}.
In addition to the original NumPy arguments listed below, also supports
\sphinxcode{\sphinxupquote{precision}} for extra control over matrix\sphinxhyphen{}multiplication precision
on supported devices. See \sphinxcode{\sphinxupquote{jax.lax.dot()}} for details.

Original docstring below.
\begin{quote}
\begin{quote}
\begin{description}
\item[{einsum(subscripts, {\color{red}\bfseries{}*}operands, out=None, dtype=None, order=’K’,}] \leavevmode
casting=’safe’, optimize=False)

\end{description}

Using the Einstein summation convention, many common multi\sphinxhyphen{}dimensional,
linear algebraic array operations can be represented in a simple fashion.
In \sphinxstyleemphasis{implicit} mode \sphinxtitleref{einsum} computes these values.

In \sphinxstyleemphasis{explicit} mode, \sphinxtitleref{einsum} provides further flexibility to compute
other array operations that might not be considered classical Einstein
summation operations, by disabling, or forcing summation over specified
subscript labels.

See the notes and examples for clarification.
\end{quote}
\begin{description}
\item[{Returns}] \leavevmode\begin{description}
\item[{output}] \leavevmode{[}ndarray{]}
The calculation based on the Einstein summation convention.

\end{description}

einsum\_path, dot, inner, outer, tensordot, linalg.multi\_dot

\DUrole{versionmodified,added}{New in version 1.6.0.}

The Einstein summation convention can be used to compute
many multi\sphinxhyphen{}dimensional, linear algebraic array operations. \sphinxtitleref{einsum}
provides a succinct way of representing these.

A non\sphinxhyphen{}exhaustive list of these operations,
which can be computed by \sphinxtitleref{einsum}, is shown below along with examples:
\begin{itemize}
\item {} 
Trace of an array, \sphinxcode{\sphinxupquote{numpy.trace()}}.

\item {} 
Return a diagonal, \sphinxcode{\sphinxupquote{numpy.diag()}}.

\item {} 
Array axis summations, \sphinxcode{\sphinxupquote{numpy.sum()}}.

\item {} 
Transpositions and permutations, \sphinxcode{\sphinxupquote{numpy.transpose()}}.

\item {} 
Matrix multiplication and dot product, \sphinxcode{\sphinxupquote{numpy.matmul()}} \sphinxcode{\sphinxupquote{numpy.dot()}}.

\item {} 
Vector inner and outer products, \sphinxcode{\sphinxupquote{numpy.inner()}} \sphinxcode{\sphinxupquote{numpy.outer()}}.

\item {} 
Broadcasting, element\sphinxhyphen{}wise and scalar multiplication, \sphinxcode{\sphinxupquote{numpy.multiply()}}.

\item {} 
Tensor contractions, \sphinxcode{\sphinxupquote{numpy.tensordot()}}.

\item {} 
Chained array operations, in efficient calculation order, \sphinxcode{\sphinxupquote{numpy.einsum\_path()}}.

\end{itemize}

The subscripts string is a comma\sphinxhyphen{}separated list of subscript labels,
where each label refers to a dimension of the corresponding operand.
Whenever a label is repeated it is summed, so \sphinxcode{\sphinxupquote{np.einsum(\textquotesingle{}i,i\textquotesingle{}, a, b)}}
is equivalent to \sphinxcode{\sphinxupquote{np.inner(a,b)}}. If a label
appears only once, it is not summed, so \sphinxcode{\sphinxupquote{np.einsum(\textquotesingle{}i\textquotesingle{}, a)}} produces a
view of \sphinxcode{\sphinxupquote{a}} with no changes. A further example \sphinxcode{\sphinxupquote{np.einsum(\textquotesingle{}ij,jk\textquotesingle{}, a, b)}}
describes traditional matrix multiplication and is equivalent to
\sphinxcode{\sphinxupquote{np.matmul(a,b)}}. Repeated subscript labels in one
operand take the diagonal. For example, \sphinxcode{\sphinxupquote{np.einsum(\textquotesingle{}ii\textquotesingle{}, a)}} is equivalent
to \sphinxcode{\sphinxupquote{np.trace(a)}}.

In \sphinxstyleemphasis{implicit mode}, the chosen subscripts are important
since the axes of the output are reordered alphabetically.  This
means that \sphinxcode{\sphinxupquote{np.einsum(\textquotesingle{}ij\textquotesingle{}, a)}} doesn’t affect a 2D array, while
\sphinxcode{\sphinxupquote{np.einsum(\textquotesingle{}ji\textquotesingle{}, a)}} takes its transpose. Additionally,
\sphinxcode{\sphinxupquote{np.einsum(\textquotesingle{}ij,jk\textquotesingle{}, a, b)}} returns a matrix multiplication, while,
\sphinxcode{\sphinxupquote{np.einsum(\textquotesingle{}ij,jh\textquotesingle{}, a, b)}} returns the transpose of the
multiplication since subscript ‘h’ precedes subscript ‘i’.

In \sphinxstyleemphasis{explicit mode} the output can be directly controlled by
specifying output subscript labels.  This requires the
identifier ‘\sphinxhyphen{}\textgreater{}’ as well as the list of output subscript labels.
This feature increases the flexibility of the function since
summing can be disabled or forced when required. The call
\sphinxcode{\sphinxupquote{np.einsum(\textquotesingle{}i\sphinxhyphen{}\textgreater{}\textquotesingle{}, a)}} is like \sphinxcode{\sphinxupquote{np.sum(a, axis=\sphinxhyphen{}1)}},
and \sphinxcode{\sphinxupquote{np.einsum(\textquotesingle{}ii\sphinxhyphen{}\textgreater{}i\textquotesingle{}, a)}} is like \sphinxcode{\sphinxupquote{np.diag(a)}}.
The difference is that \sphinxtitleref{einsum} does not allow broadcasting by default.
Additionally \sphinxcode{\sphinxupquote{np.einsum(\textquotesingle{}ij,jh\sphinxhyphen{}\textgreater{}ih\textquotesingle{}, a, b)}} directly specifies the
order of the output subscript labels and therefore returns matrix
multiplication, unlike the example above in implicit mode.

To enable and control broadcasting, use an ellipsis.  Default
NumPy\sphinxhyphen{}style broadcasting is done by adding an ellipsis
to the left of each term, like \sphinxcode{\sphinxupquote{np.einsum(\textquotesingle{}...ii\sphinxhyphen{}\textgreater{}...i\textquotesingle{}, a)}}.
To take the trace along the first and last axes,
you can do \sphinxcode{\sphinxupquote{np.einsum(\textquotesingle{}i...i\textquotesingle{}, a)}}, or to do a matrix\sphinxhyphen{}matrix
product with the left\sphinxhyphen{}most indices instead of rightmost, one can do
\sphinxcode{\sphinxupquote{np.einsum(\textquotesingle{}ij...,jk...\sphinxhyphen{}\textgreater{}ik...\textquotesingle{}, a, b)}}.

When there is only one operand, no axes are summed, and no output
parameter is provided, a view into the operand is returned instead
of a new array.  Thus, taking the diagonal as \sphinxcode{\sphinxupquote{np.einsum(\textquotesingle{}ii\sphinxhyphen{}\textgreater{}i\textquotesingle{}, a)}}
produces a view (changed in version 1.10.0).

\sphinxtitleref{einsum} also provides an alternative way to provide the subscripts
and operands as \sphinxcode{\sphinxupquote{einsum(op0, sublist0, op1, sublist1, ..., {[}sublistout{]})}}.
If the output shape is not provided in this format \sphinxtitleref{einsum} will be
calculated in implicit mode, otherwise it will be performed explicitly.
The examples below have corresponding \sphinxtitleref{einsum} calls with the two
parameter methods.

\DUrole{versionmodified,added}{New in version 1.10.0.}

Views returned from einsum are now writeable whenever the input array
is writeable. For example, \sphinxcode{\sphinxupquote{np.einsum(\textquotesingle{}ijk...\sphinxhyphen{}\textgreater{}kji...\textquotesingle{}, a)}} will now
have the same effect as \sphinxcode{\sphinxupquote{np.swapaxes(a, 0, 2)}}
and \sphinxcode{\sphinxupquote{np.einsum(\textquotesingle{}ii\sphinxhyphen{}\textgreater{}i\textquotesingle{}, a)}} will return a writeable view of the diagonal
of a 2D array.

\DUrole{versionmodified,added}{New in version 1.12.0.}

Added the \sphinxcode{\sphinxupquote{optimize}} argument which will optimize the contraction order
of an einsum expression. For a contraction with three or more operands this
can greatly increase the computational efficiency at the cost of a larger
memory footprint during computation.

Typically a ‘greedy’ algorithm is applied which empirical tests have shown
returns the optimal path in the majority of cases. In some cases ‘optimal’
will return the superlative path through a more expensive, exhaustive search.
For iterative calculations it may be advisable to calculate the optimal path
once and reuse that path by supplying it as an argument. An example is given
below.

See \sphinxcode{\sphinxupquote{numpy.einsum\_path()}} for more details.

\begin{sphinxVerbatim}[commandchars=\\\{\}]
\PYG{g+gp}{\PYGZgt{}\PYGZgt{}\PYGZgt{} }\PYG{n}{a} \PYG{o}{=} \PYG{n}{np}\PYG{o}{.}\PYG{n}{arange}\PYG{p}{(}\PYG{l+m+mi}{25}\PYG{p}{)}\PYG{o}{.}\PYG{n}{reshape}\PYG{p}{(}\PYG{l+m+mi}{5}\PYG{p}{,}\PYG{l+m+mi}{5}\PYG{p}{)}
\PYG{g+gp}{\PYGZgt{}\PYGZgt{}\PYGZgt{} }\PYG{n}{b} \PYG{o}{=} \PYG{n}{np}\PYG{o}{.}\PYG{n}{arange}\PYG{p}{(}\PYG{l+m+mi}{5}\PYG{p}{)}
\PYG{g+gp}{\PYGZgt{}\PYGZgt{}\PYGZgt{} }\PYG{n}{c} \PYG{o}{=} \PYG{n}{np}\PYG{o}{.}\PYG{n}{arange}\PYG{p}{(}\PYG{l+m+mi}{6}\PYG{p}{)}\PYG{o}{.}\PYG{n}{reshape}\PYG{p}{(}\PYG{l+m+mi}{2}\PYG{p}{,}\PYG{l+m+mi}{3}\PYG{p}{)}
\end{sphinxVerbatim}

Trace of a matrix:

\begin{sphinxVerbatim}[commandchars=\\\{\}]
\PYG{g+gp}{\PYGZgt{}\PYGZgt{}\PYGZgt{} }\PYG{n}{np}\PYG{o}{.}\PYG{n}{einsum}\PYG{p}{(}\PYG{l+s+s1}{\PYGZsq{}}\PYG{l+s+s1}{ii}\PYG{l+s+s1}{\PYGZsq{}}\PYG{p}{,} \PYG{n}{a}\PYG{p}{)}
\PYG{g+go}{60}
\PYG{g+gp}{\PYGZgt{}\PYGZgt{}\PYGZgt{} }\PYG{n}{np}\PYG{o}{.}\PYG{n}{einsum}\PYG{p}{(}\PYG{n}{a}\PYG{p}{,} \PYG{p}{[}\PYG{l+m+mi}{0}\PYG{p}{,}\PYG{l+m+mi}{0}\PYG{p}{]}\PYG{p}{)}
\PYG{g+go}{60}
\PYG{g+gp}{\PYGZgt{}\PYGZgt{}\PYGZgt{} }\PYG{n}{np}\PYG{o}{.}\PYG{n}{trace}\PYG{p}{(}\PYG{n}{a}\PYG{p}{)}
\PYG{g+go}{60}
\end{sphinxVerbatim}

Extract the diagonal (requires explicit form):

\begin{sphinxVerbatim}[commandchars=\\\{\}]
\PYG{g+gp}{\PYGZgt{}\PYGZgt{}\PYGZgt{} }\PYG{n}{np}\PYG{o}{.}\PYG{n}{einsum}\PYG{p}{(}\PYG{l+s+s1}{\PYGZsq{}}\PYG{l+s+s1}{ii\PYGZhy{}\PYGZgt{}i}\PYG{l+s+s1}{\PYGZsq{}}\PYG{p}{,} \PYG{n}{a}\PYG{p}{)}
\PYG{g+go}{array([ 0,  6, 12, 18, 24])}
\PYG{g+gp}{\PYGZgt{}\PYGZgt{}\PYGZgt{} }\PYG{n}{np}\PYG{o}{.}\PYG{n}{einsum}\PYG{p}{(}\PYG{n}{a}\PYG{p}{,} \PYG{p}{[}\PYG{l+m+mi}{0}\PYG{p}{,}\PYG{l+m+mi}{0}\PYG{p}{]}\PYG{p}{,} \PYG{p}{[}\PYG{l+m+mi}{0}\PYG{p}{]}\PYG{p}{)}
\PYG{g+go}{array([ 0,  6, 12, 18, 24])}
\PYG{g+gp}{\PYGZgt{}\PYGZgt{}\PYGZgt{} }\PYG{n}{np}\PYG{o}{.}\PYG{n}{diag}\PYG{p}{(}\PYG{n}{a}\PYG{p}{)}
\PYG{g+go}{array([ 0,  6, 12, 18, 24])}
\end{sphinxVerbatim}

Sum over an axis (requires explicit form):

\begin{sphinxVerbatim}[commandchars=\\\{\}]
\PYG{g+gp}{\PYGZgt{}\PYGZgt{}\PYGZgt{} }\PYG{n}{np}\PYG{o}{.}\PYG{n}{einsum}\PYG{p}{(}\PYG{l+s+s1}{\PYGZsq{}}\PYG{l+s+s1}{ij\PYGZhy{}\PYGZgt{}i}\PYG{l+s+s1}{\PYGZsq{}}\PYG{p}{,} \PYG{n}{a}\PYG{p}{)}
\PYG{g+go}{array([ 10,  35,  60,  85, 110])}
\PYG{g+gp}{\PYGZgt{}\PYGZgt{}\PYGZgt{} }\PYG{n}{np}\PYG{o}{.}\PYG{n}{einsum}\PYG{p}{(}\PYG{n}{a}\PYG{p}{,} \PYG{p}{[}\PYG{l+m+mi}{0}\PYG{p}{,}\PYG{l+m+mi}{1}\PYG{p}{]}\PYG{p}{,} \PYG{p}{[}\PYG{l+m+mi}{0}\PYG{p}{]}\PYG{p}{)}
\PYG{g+go}{array([ 10,  35,  60,  85, 110])}
\PYG{g+gp}{\PYGZgt{}\PYGZgt{}\PYGZgt{} }\PYG{n}{np}\PYG{o}{.}\PYG{n}{sum}\PYG{p}{(}\PYG{n}{a}\PYG{p}{,} \PYG{n}{axis}\PYG{o}{=}\PYG{l+m+mi}{1}\PYG{p}{)}
\PYG{g+go}{array([ 10,  35,  60,  85, 110])}
\end{sphinxVerbatim}

For higher dimensional arrays summing a single axis can be done with ellipsis:

\begin{sphinxVerbatim}[commandchars=\\\{\}]
\PYG{g+gp}{\PYGZgt{}\PYGZgt{}\PYGZgt{} }\PYG{n}{np}\PYG{o}{.}\PYG{n}{einsum}\PYG{p}{(}\PYG{l+s+s1}{\PYGZsq{}}\PYG{l+s+s1}{...j\PYGZhy{}\PYGZgt{}...}\PYG{l+s+s1}{\PYGZsq{}}\PYG{p}{,} \PYG{n}{a}\PYG{p}{)}
\PYG{g+go}{array([ 10,  35,  60,  85, 110])}
\PYG{g+gp}{\PYGZgt{}\PYGZgt{}\PYGZgt{} }\PYG{n}{np}\PYG{o}{.}\PYG{n}{einsum}\PYG{p}{(}\PYG{n}{a}\PYG{p}{,} \PYG{p}{[}\PYG{n+nb+bp}{Ellipsis}\PYG{p}{,}\PYG{l+m+mi}{1}\PYG{p}{]}\PYG{p}{,} \PYG{p}{[}\PYG{n+nb+bp}{Ellipsis}\PYG{p}{]}\PYG{p}{)}
\PYG{g+go}{array([ 10,  35,  60,  85, 110])}
\end{sphinxVerbatim}

Compute a matrix transpose, or reorder any number of axes:

\begin{sphinxVerbatim}[commandchars=\\\{\}]
\PYG{g+gp}{\PYGZgt{}\PYGZgt{}\PYGZgt{} }\PYG{n}{np}\PYG{o}{.}\PYG{n}{einsum}\PYG{p}{(}\PYG{l+s+s1}{\PYGZsq{}}\PYG{l+s+s1}{ji}\PYG{l+s+s1}{\PYGZsq{}}\PYG{p}{,} \PYG{n}{c}\PYG{p}{)}
\PYG{g+go}{array([[0, 3],}
\PYG{g+go}{       [1, 4],}
\PYG{g+go}{       [2, 5]])}
\PYG{g+gp}{\PYGZgt{}\PYGZgt{}\PYGZgt{} }\PYG{n}{np}\PYG{o}{.}\PYG{n}{einsum}\PYG{p}{(}\PYG{l+s+s1}{\PYGZsq{}}\PYG{l+s+s1}{ij\PYGZhy{}\PYGZgt{}ji}\PYG{l+s+s1}{\PYGZsq{}}\PYG{p}{,} \PYG{n}{c}\PYG{p}{)}
\PYG{g+go}{array([[0, 3],}
\PYG{g+go}{       [1, 4],}
\PYG{g+go}{       [2, 5]])}
\PYG{g+gp}{\PYGZgt{}\PYGZgt{}\PYGZgt{} }\PYG{n}{np}\PYG{o}{.}\PYG{n}{einsum}\PYG{p}{(}\PYG{n}{c}\PYG{p}{,} \PYG{p}{[}\PYG{l+m+mi}{1}\PYG{p}{,}\PYG{l+m+mi}{0}\PYG{p}{]}\PYG{p}{)}
\PYG{g+go}{array([[0, 3],}
\PYG{g+go}{       [1, 4],}
\PYG{g+go}{       [2, 5]])}
\PYG{g+gp}{\PYGZgt{}\PYGZgt{}\PYGZgt{} }\PYG{n}{np}\PYG{o}{.}\PYG{n}{transpose}\PYG{p}{(}\PYG{n}{c}\PYG{p}{)}
\PYG{g+go}{array([[0, 3],}
\PYG{g+go}{       [1, 4],}
\PYG{g+go}{       [2, 5]])}
\end{sphinxVerbatim}

Vector inner products:

\begin{sphinxVerbatim}[commandchars=\\\{\}]
\PYG{g+gp}{\PYGZgt{}\PYGZgt{}\PYGZgt{} }\PYG{n}{np}\PYG{o}{.}\PYG{n}{einsum}\PYG{p}{(}\PYG{l+s+s1}{\PYGZsq{}}\PYG{l+s+s1}{i,i}\PYG{l+s+s1}{\PYGZsq{}}\PYG{p}{,} \PYG{n}{b}\PYG{p}{,} \PYG{n}{b}\PYG{p}{)}
\PYG{g+go}{30}
\PYG{g+gp}{\PYGZgt{}\PYGZgt{}\PYGZgt{} }\PYG{n}{np}\PYG{o}{.}\PYG{n}{einsum}\PYG{p}{(}\PYG{n}{b}\PYG{p}{,} \PYG{p}{[}\PYG{l+m+mi}{0}\PYG{p}{]}\PYG{p}{,} \PYG{n}{b}\PYG{p}{,} \PYG{p}{[}\PYG{l+m+mi}{0}\PYG{p}{]}\PYG{p}{)}
\PYG{g+go}{30}
\PYG{g+gp}{\PYGZgt{}\PYGZgt{}\PYGZgt{} }\PYG{n}{np}\PYG{o}{.}\PYG{n}{inner}\PYG{p}{(}\PYG{n}{b}\PYG{p}{,}\PYG{n}{b}\PYG{p}{)}
\PYG{g+go}{30}
\end{sphinxVerbatim}

Matrix vector multiplication:

\begin{sphinxVerbatim}[commandchars=\\\{\}]
\PYG{g+gp}{\PYGZgt{}\PYGZgt{}\PYGZgt{} }\PYG{n}{np}\PYG{o}{.}\PYG{n}{einsum}\PYG{p}{(}\PYG{l+s+s1}{\PYGZsq{}}\PYG{l+s+s1}{ij,j}\PYG{l+s+s1}{\PYGZsq{}}\PYG{p}{,} \PYG{n}{a}\PYG{p}{,} \PYG{n}{b}\PYG{p}{)}
\PYG{g+go}{array([ 30,  80, 130, 180, 230])}
\PYG{g+gp}{\PYGZgt{}\PYGZgt{}\PYGZgt{} }\PYG{n}{np}\PYG{o}{.}\PYG{n}{einsum}\PYG{p}{(}\PYG{n}{a}\PYG{p}{,} \PYG{p}{[}\PYG{l+m+mi}{0}\PYG{p}{,}\PYG{l+m+mi}{1}\PYG{p}{]}\PYG{p}{,} \PYG{n}{b}\PYG{p}{,} \PYG{p}{[}\PYG{l+m+mi}{1}\PYG{p}{]}\PYG{p}{)}
\PYG{g+go}{array([ 30,  80, 130, 180, 230])}
\PYG{g+gp}{\PYGZgt{}\PYGZgt{}\PYGZgt{} }\PYG{n}{np}\PYG{o}{.}\PYG{n}{dot}\PYG{p}{(}\PYG{n}{a}\PYG{p}{,} \PYG{n}{b}\PYG{p}{)}
\PYG{g+go}{array([ 30,  80, 130, 180, 230])}
\PYG{g+gp}{\PYGZgt{}\PYGZgt{}\PYGZgt{} }\PYG{n}{np}\PYG{o}{.}\PYG{n}{einsum}\PYG{p}{(}\PYG{l+s+s1}{\PYGZsq{}}\PYG{l+s+s1}{...j,j}\PYG{l+s+s1}{\PYGZsq{}}\PYG{p}{,} \PYG{n}{a}\PYG{p}{,} \PYG{n}{b}\PYG{p}{)}
\PYG{g+go}{array([ 30,  80, 130, 180, 230])}
\end{sphinxVerbatim}

Broadcasting and scalar multiplication:

\begin{sphinxVerbatim}[commandchars=\\\{\}]
\PYG{g+gp}{\PYGZgt{}\PYGZgt{}\PYGZgt{} }\PYG{n}{np}\PYG{o}{.}\PYG{n}{einsum}\PYG{p}{(}\PYG{l+s+s1}{\PYGZsq{}}\PYG{l+s+s1}{..., ...}\PYG{l+s+s1}{\PYGZsq{}}\PYG{p}{,} \PYG{l+m+mi}{3}\PYG{p}{,} \PYG{n}{c}\PYG{p}{)}
\PYG{g+go}{array([[ 0,  3,  6],}
\PYG{g+go}{       [ 9, 12, 15]])}
\PYG{g+gp}{\PYGZgt{}\PYGZgt{}\PYGZgt{} }\PYG{n}{np}\PYG{o}{.}\PYG{n}{einsum}\PYG{p}{(}\PYG{l+s+s1}{\PYGZsq{}}\PYG{l+s+s1}{,ij}\PYG{l+s+s1}{\PYGZsq{}}\PYG{p}{,} \PYG{l+m+mi}{3}\PYG{p}{,} \PYG{n}{c}\PYG{p}{)}
\PYG{g+go}{array([[ 0,  3,  6],}
\PYG{g+go}{       [ 9, 12, 15]])}
\PYG{g+gp}{\PYGZgt{}\PYGZgt{}\PYGZgt{} }\PYG{n}{np}\PYG{o}{.}\PYG{n}{einsum}\PYG{p}{(}\PYG{l+m+mi}{3}\PYG{p}{,} \PYG{p}{[}\PYG{n+nb+bp}{Ellipsis}\PYG{p}{]}\PYG{p}{,} \PYG{n}{c}\PYG{p}{,} \PYG{p}{[}\PYG{n+nb+bp}{Ellipsis}\PYG{p}{]}\PYG{p}{)}
\PYG{g+go}{array([[ 0,  3,  6],}
\PYG{g+go}{       [ 9, 12, 15]])}
\PYG{g+gp}{\PYGZgt{}\PYGZgt{}\PYGZgt{} }\PYG{n}{np}\PYG{o}{.}\PYG{n}{multiply}\PYG{p}{(}\PYG{l+m+mi}{3}\PYG{p}{,} \PYG{n}{c}\PYG{p}{)}
\PYG{g+go}{array([[ 0,  3,  6],}
\PYG{g+go}{       [ 9, 12, 15]])}
\end{sphinxVerbatim}

Vector outer product:

\begin{sphinxVerbatim}[commandchars=\\\{\}]
\PYG{g+gp}{\PYGZgt{}\PYGZgt{}\PYGZgt{} }\PYG{n}{np}\PYG{o}{.}\PYG{n}{einsum}\PYG{p}{(}\PYG{l+s+s1}{\PYGZsq{}}\PYG{l+s+s1}{i,j}\PYG{l+s+s1}{\PYGZsq{}}\PYG{p}{,} \PYG{n}{np}\PYG{o}{.}\PYG{n}{arange}\PYG{p}{(}\PYG{l+m+mi}{2}\PYG{p}{)}\PYG{o}{+}\PYG{l+m+mi}{1}\PYG{p}{,} \PYG{n}{b}\PYG{p}{)}
\PYG{g+go}{array([[0, 1, 2, 3, 4],}
\PYG{g+go}{       [0, 2, 4, 6, 8]])}
\PYG{g+gp}{\PYGZgt{}\PYGZgt{}\PYGZgt{} }\PYG{n}{np}\PYG{o}{.}\PYG{n}{einsum}\PYG{p}{(}\PYG{n}{np}\PYG{o}{.}\PYG{n}{arange}\PYG{p}{(}\PYG{l+m+mi}{2}\PYG{p}{)}\PYG{o}{+}\PYG{l+m+mi}{1}\PYG{p}{,} \PYG{p}{[}\PYG{l+m+mi}{0}\PYG{p}{]}\PYG{p}{,} \PYG{n}{b}\PYG{p}{,} \PYG{p}{[}\PYG{l+m+mi}{1}\PYG{p}{]}\PYG{p}{)}
\PYG{g+go}{array([[0, 1, 2, 3, 4],}
\PYG{g+go}{       [0, 2, 4, 6, 8]])}
\PYG{g+gp}{\PYGZgt{}\PYGZgt{}\PYGZgt{} }\PYG{n}{np}\PYG{o}{.}\PYG{n}{outer}\PYG{p}{(}\PYG{n}{np}\PYG{o}{.}\PYG{n}{arange}\PYG{p}{(}\PYG{l+m+mi}{2}\PYG{p}{)}\PYG{o}{+}\PYG{l+m+mi}{1}\PYG{p}{,} \PYG{n}{b}\PYG{p}{)}
\PYG{g+go}{array([[0, 1, 2, 3, 4],}
\PYG{g+go}{       [0, 2, 4, 6, 8]])}
\end{sphinxVerbatim}

Tensor contraction:

\begin{sphinxVerbatim}[commandchars=\\\{\}]
\PYG{g+gp}{\PYGZgt{}\PYGZgt{}\PYGZgt{} }\PYG{n}{a} \PYG{o}{=} \PYG{n}{np}\PYG{o}{.}\PYG{n}{arange}\PYG{p}{(}\PYG{l+m+mf}{60.}\PYG{p}{)}\PYG{o}{.}\PYG{n}{reshape}\PYG{p}{(}\PYG{l+m+mi}{3}\PYG{p}{,}\PYG{l+m+mi}{4}\PYG{p}{,}\PYG{l+m+mi}{5}\PYG{p}{)}
\PYG{g+gp}{\PYGZgt{}\PYGZgt{}\PYGZgt{} }\PYG{n}{b} \PYG{o}{=} \PYG{n}{np}\PYG{o}{.}\PYG{n}{arange}\PYG{p}{(}\PYG{l+m+mf}{24.}\PYG{p}{)}\PYG{o}{.}\PYG{n}{reshape}\PYG{p}{(}\PYG{l+m+mi}{4}\PYG{p}{,}\PYG{l+m+mi}{3}\PYG{p}{,}\PYG{l+m+mi}{2}\PYG{p}{)}
\PYG{g+gp}{\PYGZgt{}\PYGZgt{}\PYGZgt{} }\PYG{n}{np}\PYG{o}{.}\PYG{n}{einsum}\PYG{p}{(}\PYG{l+s+s1}{\PYGZsq{}}\PYG{l+s+s1}{ijk,jil\PYGZhy{}\PYGZgt{}kl}\PYG{l+s+s1}{\PYGZsq{}}\PYG{p}{,} \PYG{n}{a}\PYG{p}{,} \PYG{n}{b}\PYG{p}{)}
\PYG{g+go}{array([[4400., 4730.],}
\PYG{g+go}{       [4532., 4874.],}
\PYG{g+go}{       [4664., 5018.],}
\PYG{g+go}{       [4796., 5162.],}
\PYG{g+go}{       [4928., 5306.]])}
\PYG{g+gp}{\PYGZgt{}\PYGZgt{}\PYGZgt{} }\PYG{n}{np}\PYG{o}{.}\PYG{n}{einsum}\PYG{p}{(}\PYG{n}{a}\PYG{p}{,} \PYG{p}{[}\PYG{l+m+mi}{0}\PYG{p}{,}\PYG{l+m+mi}{1}\PYG{p}{,}\PYG{l+m+mi}{2}\PYG{p}{]}\PYG{p}{,} \PYG{n}{b}\PYG{p}{,} \PYG{p}{[}\PYG{l+m+mi}{1}\PYG{p}{,}\PYG{l+m+mi}{0}\PYG{p}{,}\PYG{l+m+mi}{3}\PYG{p}{]}\PYG{p}{,} \PYG{p}{[}\PYG{l+m+mi}{2}\PYG{p}{,}\PYG{l+m+mi}{3}\PYG{p}{]}\PYG{p}{)}
\PYG{g+go}{array([[4400., 4730.],}
\PYG{g+go}{       [4532., 4874.],}
\PYG{g+go}{       [4664., 5018.],}
\PYG{g+go}{       [4796., 5162.],}
\PYG{g+go}{       [4928., 5306.]])}
\PYG{g+gp}{\PYGZgt{}\PYGZgt{}\PYGZgt{} }\PYG{n}{np}\PYG{o}{.}\PYG{n}{tensordot}\PYG{p}{(}\PYG{n}{a}\PYG{p}{,}\PYG{n}{b}\PYG{p}{,} \PYG{n}{axes}\PYG{o}{=}\PYG{p}{(}\PYG{p}{[}\PYG{l+m+mi}{1}\PYG{p}{,}\PYG{l+m+mi}{0}\PYG{p}{]}\PYG{p}{,}\PYG{p}{[}\PYG{l+m+mi}{0}\PYG{p}{,}\PYG{l+m+mi}{1}\PYG{p}{]}\PYG{p}{)}\PYG{p}{)}
\PYG{g+go}{array([[4400., 4730.],}
\PYG{g+go}{       [4532., 4874.],}
\PYG{g+go}{       [4664., 5018.],}
\PYG{g+go}{       [4796., 5162.],}
\PYG{g+go}{       [4928., 5306.]])}
\end{sphinxVerbatim}

Writeable returned arrays (since version 1.10.0):

\begin{sphinxVerbatim}[commandchars=\\\{\}]
\PYG{g+gp}{\PYGZgt{}\PYGZgt{}\PYGZgt{} }\PYG{n}{a} \PYG{o}{=} \PYG{n}{np}\PYG{o}{.}\PYG{n}{zeros}\PYG{p}{(}\PYG{p}{(}\PYG{l+m+mi}{3}\PYG{p}{,} \PYG{l+m+mi}{3}\PYG{p}{)}\PYG{p}{)}
\PYG{g+gp}{\PYGZgt{}\PYGZgt{}\PYGZgt{} }\PYG{n}{np}\PYG{o}{.}\PYG{n}{einsum}\PYG{p}{(}\PYG{l+s+s1}{\PYGZsq{}}\PYG{l+s+s1}{ii\PYGZhy{}\PYGZgt{}i}\PYG{l+s+s1}{\PYGZsq{}}\PYG{p}{,} \PYG{n}{a}\PYG{p}{)}\PYG{p}{[}\PYG{p}{:}\PYG{p}{]} \PYG{o}{=} \PYG{l+m+mi}{1}
\PYG{g+gp}{\PYGZgt{}\PYGZgt{}\PYGZgt{} }\PYG{n}{a}
\PYG{g+go}{array([[1., 0., 0.],}
\PYG{g+go}{       [0., 1., 0.],}
\PYG{g+go}{       [0., 0., 1.]])}
\end{sphinxVerbatim}

Example of ellipsis use:

\begin{sphinxVerbatim}[commandchars=\\\{\}]
\PYG{g+gp}{\PYGZgt{}\PYGZgt{}\PYGZgt{} }\PYG{n}{a} \PYG{o}{=} \PYG{n}{np}\PYG{o}{.}\PYG{n}{arange}\PYG{p}{(}\PYG{l+m+mi}{6}\PYG{p}{)}\PYG{o}{.}\PYG{n}{reshape}\PYG{p}{(}\PYG{p}{(}\PYG{l+m+mi}{3}\PYG{p}{,}\PYG{l+m+mi}{2}\PYG{p}{)}\PYG{p}{)}
\PYG{g+gp}{\PYGZgt{}\PYGZgt{}\PYGZgt{} }\PYG{n}{b} \PYG{o}{=} \PYG{n}{np}\PYG{o}{.}\PYG{n}{arange}\PYG{p}{(}\PYG{l+m+mi}{12}\PYG{p}{)}\PYG{o}{.}\PYG{n}{reshape}\PYG{p}{(}\PYG{p}{(}\PYG{l+m+mi}{4}\PYG{p}{,}\PYG{l+m+mi}{3}\PYG{p}{)}\PYG{p}{)}
\PYG{g+gp}{\PYGZgt{}\PYGZgt{}\PYGZgt{} }\PYG{n}{np}\PYG{o}{.}\PYG{n}{einsum}\PYG{p}{(}\PYG{l+s+s1}{\PYGZsq{}}\PYG{l+s+s1}{ki,jk\PYGZhy{}\PYGZgt{}ij}\PYG{l+s+s1}{\PYGZsq{}}\PYG{p}{,} \PYG{n}{a}\PYG{p}{,} \PYG{n}{b}\PYG{p}{)}
\PYG{g+go}{array([[10, 28, 46, 64],}
\PYG{g+go}{       [13, 40, 67, 94]])}
\PYG{g+gp}{\PYGZgt{}\PYGZgt{}\PYGZgt{} }\PYG{n}{np}\PYG{o}{.}\PYG{n}{einsum}\PYG{p}{(}\PYG{l+s+s1}{\PYGZsq{}}\PYG{l+s+s1}{ki,...k\PYGZhy{}\PYGZgt{}i...}\PYG{l+s+s1}{\PYGZsq{}}\PYG{p}{,} \PYG{n}{a}\PYG{p}{,} \PYG{n}{b}\PYG{p}{)}
\PYG{g+go}{array([[10, 28, 46, 64],}
\PYG{g+go}{       [13, 40, 67, 94]])}
\PYG{g+gp}{\PYGZgt{}\PYGZgt{}\PYGZgt{} }\PYG{n}{np}\PYG{o}{.}\PYG{n}{einsum}\PYG{p}{(}\PYG{l+s+s1}{\PYGZsq{}}\PYG{l+s+s1}{k...,jk}\PYG{l+s+s1}{\PYGZsq{}}\PYG{p}{,} \PYG{n}{a}\PYG{p}{,} \PYG{n}{b}\PYG{p}{)}
\PYG{g+go}{array([[10, 28, 46, 64],}
\PYG{g+go}{       [13, 40, 67, 94]])}
\end{sphinxVerbatim}

Chained array operations. For more complicated contractions, speed ups
might be achieved by repeatedly computing a ‘greedy’ path or pre\sphinxhyphen{}computing the
‘optimal’ path and repeatedly applying it, using an
\sphinxtitleref{einsum\_path} insertion (since version 1.12.0). Performance improvements can be
particularly significant with larger arrays:

\begin{sphinxVerbatim}[commandchars=\\\{\}]
\PYG{g+gp}{\PYGZgt{}\PYGZgt{}\PYGZgt{} }\PYG{n}{a} \PYG{o}{=} \PYG{n}{np}\PYG{o}{.}\PYG{n}{ones}\PYG{p}{(}\PYG{l+m+mi}{64}\PYG{p}{)}\PYG{o}{.}\PYG{n}{reshape}\PYG{p}{(}\PYG{l+m+mi}{2}\PYG{p}{,}\PYG{l+m+mi}{4}\PYG{p}{,}\PYG{l+m+mi}{8}\PYG{p}{)}
\end{sphinxVerbatim}

Basic \sphinxtitleref{einsum}: \textasciitilde{}1520ms  (benchmarked on 3.1GHz Intel i5.)

\begin{sphinxVerbatim}[commandchars=\\\{\}]
\PYG{g+gp}{\PYGZgt{}\PYGZgt{}\PYGZgt{} }\PYG{k}{for} \PYG{n}{iteration} \PYG{o+ow}{in} \PYG{n+nb}{range}\PYG{p}{(}\PYG{l+m+mi}{500}\PYG{p}{)}\PYG{p}{:}
\PYG{g+gp}{... }    \PYG{n}{\PYGZus{}} \PYG{o}{=} \PYG{n}{np}\PYG{o}{.}\PYG{n}{einsum}\PYG{p}{(}\PYG{l+s+s1}{\PYGZsq{}}\PYG{l+s+s1}{ijk,ilm,njm,nlk,abc\PYGZhy{}\PYGZgt{}}\PYG{l+s+s1}{\PYGZsq{}}\PYG{p}{,}\PYG{n}{a}\PYG{p}{,}\PYG{n}{a}\PYG{p}{,}\PYG{n}{a}\PYG{p}{,}\PYG{n}{a}\PYG{p}{,}\PYG{n}{a}\PYG{p}{)}
\end{sphinxVerbatim}

Sub\sphinxhyphen{}optimal \sphinxtitleref{einsum} (due to repeated path calculation time): \textasciitilde{}330ms

\begin{sphinxVerbatim}[commandchars=\\\{\}]
\PYG{g+gp}{\PYGZgt{}\PYGZgt{}\PYGZgt{} }\PYG{k}{for} \PYG{n}{iteration} \PYG{o+ow}{in} \PYG{n+nb}{range}\PYG{p}{(}\PYG{l+m+mi}{500}\PYG{p}{)}\PYG{p}{:}
\PYG{g+gp}{... }    \PYG{n}{\PYGZus{}} \PYG{o}{=} \PYG{n}{np}\PYG{o}{.}\PYG{n}{einsum}\PYG{p}{(}\PYG{l+s+s1}{\PYGZsq{}}\PYG{l+s+s1}{ijk,ilm,njm,nlk,abc\PYGZhy{}\PYGZgt{}}\PYG{l+s+s1}{\PYGZsq{}}\PYG{p}{,}\PYG{n}{a}\PYG{p}{,}\PYG{n}{a}\PYG{p}{,}\PYG{n}{a}\PYG{p}{,}\PYG{n}{a}\PYG{p}{,}\PYG{n}{a}\PYG{p}{,} \PYG{n}{optimize}\PYG{o}{=}\PYG{l+s+s1}{\PYGZsq{}}\PYG{l+s+s1}{optimal}\PYG{l+s+s1}{\PYGZsq{}}\PYG{p}{)}
\end{sphinxVerbatim}

Greedy \sphinxtitleref{einsum} (faster optimal path approximation): \textasciitilde{}160ms

\begin{sphinxVerbatim}[commandchars=\\\{\}]
\PYG{g+gp}{\PYGZgt{}\PYGZgt{}\PYGZgt{} }\PYG{k}{for} \PYG{n}{iteration} \PYG{o+ow}{in} \PYG{n+nb}{range}\PYG{p}{(}\PYG{l+m+mi}{500}\PYG{p}{)}\PYG{p}{:}
\PYG{g+gp}{... }    \PYG{n}{\PYGZus{}} \PYG{o}{=} \PYG{n}{np}\PYG{o}{.}\PYG{n}{einsum}\PYG{p}{(}\PYG{l+s+s1}{\PYGZsq{}}\PYG{l+s+s1}{ijk,ilm,njm,nlk,abc\PYGZhy{}\PYGZgt{}}\PYG{l+s+s1}{\PYGZsq{}}\PYG{p}{,}\PYG{n}{a}\PYG{p}{,}\PYG{n}{a}\PYG{p}{,}\PYG{n}{a}\PYG{p}{,}\PYG{n}{a}\PYG{p}{,}\PYG{n}{a}\PYG{p}{,} \PYG{n}{optimize}\PYG{o}{=}\PYG{l+s+s1}{\PYGZsq{}}\PYG{l+s+s1}{greedy}\PYG{l+s+s1}{\PYGZsq{}}\PYG{p}{)}
\end{sphinxVerbatim}

Optimal \sphinxtitleref{einsum} (best usage pattern in some use cases): \textasciitilde{}110ms

\begin{sphinxVerbatim}[commandchars=\\\{\}]
\PYG{g+gp}{\PYGZgt{}\PYGZgt{}\PYGZgt{} }\PYG{n}{path} \PYG{o}{=} \PYG{n}{np}\PYG{o}{.}\PYG{n}{einsum\PYGZus{}path}\PYG{p}{(}\PYG{l+s+s1}{\PYGZsq{}}\PYG{l+s+s1}{ijk,ilm,njm,nlk,abc\PYGZhy{}\PYGZgt{}}\PYG{l+s+s1}{\PYGZsq{}}\PYG{p}{,}\PYG{n}{a}\PYG{p}{,}\PYG{n}{a}\PYG{p}{,}\PYG{n}{a}\PYG{p}{,}\PYG{n}{a}\PYG{p}{,}\PYG{n}{a}\PYG{p}{,} \PYG{n}{optimize}\PYG{o}{=}\PYG{l+s+s1}{\PYGZsq{}}\PYG{l+s+s1}{optimal}\PYG{l+s+s1}{\PYGZsq{}}\PYG{p}{)}\PYG{p}{[}\PYG{l+m+mi}{0}\PYG{p}{]}
\PYG{g+gp}{\PYGZgt{}\PYGZgt{}\PYGZgt{} }\PYG{k}{for} \PYG{n}{iteration} \PYG{o+ow}{in} \PYG{n+nb}{range}\PYG{p}{(}\PYG{l+m+mi}{500}\PYG{p}{)}\PYG{p}{:}
\PYG{g+gp}{... }    \PYG{n}{\PYGZus{}} \PYG{o}{=} \PYG{n}{np}\PYG{o}{.}\PYG{n}{einsum}\PYG{p}{(}\PYG{l+s+s1}{\PYGZsq{}}\PYG{l+s+s1}{ijk,ilm,njm,nlk,abc\PYGZhy{}\PYGZgt{}}\PYG{l+s+s1}{\PYGZsq{}}\PYG{p}{,}\PYG{n}{a}\PYG{p}{,}\PYG{n}{a}\PYG{p}{,}\PYG{n}{a}\PYG{p}{,}\PYG{n}{a}\PYG{p}{,}\PYG{n}{a}\PYG{p}{,} \PYG{n}{optimize}\PYG{o}{=}\PYG{n}{path}\PYG{p}{)}
\end{sphinxVerbatim}

\end{description}
\end{quote}

\end{fulllineitems}

\index{equal() (in module symjax.tensor)@\spxentry{equal()}\spxextra{in module symjax.tensor}}

\begin{fulllineitems}
\phantomsection\label{\detokenize{modules/tensor:symjax.tensor.equal}}\pysiglinewithargsret{\sphinxbfcode{\sphinxupquote{equal}}}{\emph{\DUrole{n}{x1}}, \emph{\DUrole{n}{x2}}}{}
Return (x1 == x2) element\sphinxhyphen{}wise.

LAX\sphinxhyphen{}backend implementation of {\hyperref[\detokenize{modules/tensor:symjax.tensor.equal}]{\sphinxcrossref{\sphinxcode{\sphinxupquote{equal()}}}}}.
ADDITIONOriginal docstring below.

LAX\sphinxhyphen{}backend implementation of {\hyperref[\detokenize{modules/tensor:symjax.tensor.equal}]{\sphinxcrossref{\sphinxcode{\sphinxupquote{equal()}}}}}.
Original docstring below.

equal(x1, x2, /, out=None, {\color{red}\bfseries{}*}, where=True, casting=’same\_kind’, order=’K’, dtype=None, subok=True{[}, signature, extobj{]})
\begin{quote}\begin{description}
\item[{Returns}] \leavevmode
\sphinxstylestrong{out} \textendash{} Output array, element\sphinxhyphen{}wise comparison of \sphinxtitleref{x1} and \sphinxtitleref{x2}.
Typically of type bool, unless \sphinxcode{\sphinxupquote{dtype=object}} is passed.
This is a scalar if both \sphinxtitleref{x1} and \sphinxtitleref{x2} are scalars.

\item[{Return type}] \leavevmode
ndarray or scalar

\end{description}\end{quote}

\sphinxstrong{See also:}

{\hyperref[\detokenize{modules/tensor:symjax.tensor.not_equal}]{\sphinxcrossref{\sphinxcode{\sphinxupquote{not\_equal()}}}}}, {\hyperref[\detokenize{modules/tensor:symjax.tensor.greater_equal}]{\sphinxcrossref{\sphinxcode{\sphinxupquote{greater\_equal()}}}}}, {\hyperref[\detokenize{modules/tensor:symjax.tensor.less_equal}]{\sphinxcrossref{\sphinxcode{\sphinxupquote{less\_equal()}}}}}, {\hyperref[\detokenize{modules/tensor:symjax.tensor.greater}]{\sphinxcrossref{\sphinxcode{\sphinxupquote{greater()}}}}}, {\hyperref[\detokenize{modules/tensor:symjax.tensor.less}]{\sphinxcrossref{\sphinxcode{\sphinxupquote{less()}}}}}

\subsubsection*{Examples}

\begin{sphinxVerbatim}[commandchars=\\\{\}]
\PYG{g+gp}{\PYGZgt{}\PYGZgt{}\PYGZgt{} }\PYG{n}{np}\PYG{o}{.}\PYG{n}{equal}\PYG{p}{(}\PYG{p}{[}\PYG{l+m+mi}{0}\PYG{p}{,} \PYG{l+m+mi}{1}\PYG{p}{,} \PYG{l+m+mi}{3}\PYG{p}{]}\PYG{p}{,} \PYG{n}{np}\PYG{o}{.}\PYG{n}{arange}\PYG{p}{(}\PYG{l+m+mi}{3}\PYG{p}{)}\PYG{p}{)}
\PYG{g+go}{array([ True,  True, False])}
\end{sphinxVerbatim}

What is compared are values, not types. So an int (1) and an array of
length one can evaluate as True:

\begin{sphinxVerbatim}[commandchars=\\\{\}]
\PYG{g+gp}{\PYGZgt{}\PYGZgt{}\PYGZgt{} }\PYG{n}{np}\PYG{o}{.}\PYG{n}{equal}\PYG{p}{(}\PYG{l+m+mi}{1}\PYG{p}{,} \PYG{n}{np}\PYG{o}{.}\PYG{n}{ones}\PYG{p}{(}\PYG{l+m+mi}{1}\PYG{p}{)}\PYG{p}{)}
\PYG{g+go}{array([ True])}
\end{sphinxVerbatim}

\end{fulllineitems}

\index{empty() (in module symjax.tensor)@\spxentry{empty()}\spxextra{in module symjax.tensor}}

\begin{fulllineitems}
\phantomsection\label{\detokenize{modules/tensor:symjax.tensor.empty}}\pysiglinewithargsret{\sphinxbfcode{\sphinxupquote{empty}}}{\emph{\DUrole{n}{shape}}, \emph{\DUrole{n}{dtype}\DUrole{o}{=}\DUrole{default_value}{None}}}{}
Return a new array of given shape and type, filled with zeros.

LAX\sphinxhyphen{}backend implementation of {\hyperref[\detokenize{modules/tensor:symjax.tensor.zeros}]{\sphinxcrossref{\sphinxcode{\sphinxupquote{zeros()}}}}}.
ADDITIONOriginal docstring below.

LAX\sphinxhyphen{}backend implementation of {\hyperref[\detokenize{modules/tensor:symjax.tensor.zeros}]{\sphinxcrossref{\sphinxcode{\sphinxupquote{zeros()}}}}}.
Original docstring below.

zeros(shape, dtype=float, order=’C’)
\begin{quote}
\begin{quote}
\end{quote}
\begin{description}
\item[{Returns}] \leavevmode\begin{description}
\item[{out}] \leavevmode{[}ndarray{]}
Array of zeros with the given shape, dtype, and order.

\end{description}

zeros\_like : Return an array of zeros with shape and type of input.
empty : Return a new uninitialized array.
ones : Return a new array setting values to one.
full : Return a new array of given shape filled with value.

\begin{sphinxVerbatim}[commandchars=\\\{\}]
\PYG{g+gp}{\PYGZgt{}\PYGZgt{}\PYGZgt{} }\PYG{n}{np}\PYG{o}{.}\PYG{n}{zeros}\PYG{p}{(}\PYG{l+m+mi}{5}\PYG{p}{)}
\PYG{g+go}{array([ 0.,  0.,  0.,  0.,  0.])}
\end{sphinxVerbatim}

\begin{sphinxVerbatim}[commandchars=\\\{\}]
\PYG{g+gp}{\PYGZgt{}\PYGZgt{}\PYGZgt{} }\PYG{n}{np}\PYG{o}{.}\PYG{n}{zeros}\PYG{p}{(}\PYG{p}{(}\PYG{l+m+mi}{5}\PYG{p}{,}\PYG{p}{)}\PYG{p}{,} \PYG{n}{dtype}\PYG{o}{=}\PYG{n+nb}{int}\PYG{p}{)}
\PYG{g+go}{array([0, 0, 0, 0, 0])}
\end{sphinxVerbatim}

\begin{sphinxVerbatim}[commandchars=\\\{\}]
\PYG{g+gp}{\PYGZgt{}\PYGZgt{}\PYGZgt{} }\PYG{n}{np}\PYG{o}{.}\PYG{n}{zeros}\PYG{p}{(}\PYG{p}{(}\PYG{l+m+mi}{2}\PYG{p}{,} \PYG{l+m+mi}{1}\PYG{p}{)}\PYG{p}{)}
\PYG{g+go}{array([[ 0.],}
\PYG{g+go}{       [ 0.]])}
\end{sphinxVerbatim}

\begin{sphinxVerbatim}[commandchars=\\\{\}]
\PYG{g+gp}{\PYGZgt{}\PYGZgt{}\PYGZgt{} }\PYG{n}{s} \PYG{o}{=} \PYG{p}{(}\PYG{l+m+mi}{2}\PYG{p}{,}\PYG{l+m+mi}{2}\PYG{p}{)}
\PYG{g+gp}{\PYGZgt{}\PYGZgt{}\PYGZgt{} }\PYG{n}{np}\PYG{o}{.}\PYG{n}{zeros}\PYG{p}{(}\PYG{n}{s}\PYG{p}{)}
\PYG{g+go}{array([[ 0.,  0.],}
\PYG{g+go}{       [ 0.,  0.]])}
\end{sphinxVerbatim}

\begin{sphinxVerbatim}[commandchars=\\\{\}]
\PYG{g+gp}{\PYGZgt{}\PYGZgt{}\PYGZgt{} }\PYG{n}{np}\PYG{o}{.}\PYG{n}{zeros}\PYG{p}{(}\PYG{p}{(}\PYG{l+m+mi}{2}\PYG{p}{,}\PYG{p}{)}\PYG{p}{,} \PYG{n}{dtype}\PYG{o}{=}\PYG{p}{[}\PYG{p}{(}\PYG{l+s+s1}{\PYGZsq{}}\PYG{l+s+s1}{x}\PYG{l+s+s1}{\PYGZsq{}}\PYG{p}{,} \PYG{l+s+s1}{\PYGZsq{}}\PYG{l+s+s1}{i4}\PYG{l+s+s1}{\PYGZsq{}}\PYG{p}{)}\PYG{p}{,} \PYG{p}{(}\PYG{l+s+s1}{\PYGZsq{}}\PYG{l+s+s1}{y}\PYG{l+s+s1}{\PYGZsq{}}\PYG{p}{,} \PYG{l+s+s1}{\PYGZsq{}}\PYG{l+s+s1}{i4}\PYG{l+s+s1}{\PYGZsq{}}\PYG{p}{)}\PYG{p}{]}\PYG{p}{)} \PYG{c+c1}{\PYGZsh{} custom dtype}
\PYG{g+go}{array([(0, 0), (0, 0)],}
\PYG{g+go}{      dtype=[(\PYGZsq{}x\PYGZsq{}, \PYGZsq{}\PYGZlt{}i4\PYGZsq{}), (\PYGZsq{}y\PYGZsq{}, \PYGZsq{}\PYGZlt{}i4\PYGZsq{})])}
\end{sphinxVerbatim}

\end{description}
\end{quote}

\end{fulllineitems}

\index{empty\_like() (in module symjax.tensor)@\spxentry{empty\_like()}\spxextra{in module symjax.tensor}}

\begin{fulllineitems}
\phantomsection\label{\detokenize{modules/tensor:symjax.tensor.empty_like}}\pysiglinewithargsret{\sphinxbfcode{\sphinxupquote{empty\_like}}}{\emph{\DUrole{n}{x}}, \emph{\DUrole{n}{dtype}\DUrole{o}{=}\DUrole{default_value}{None}}}{}
Return an array of zeros with the same shape and type as a given array.

LAX\sphinxhyphen{}backend implementation of {\hyperref[\detokenize{modules/tensor:symjax.tensor.zeros_like}]{\sphinxcrossref{\sphinxcode{\sphinxupquote{zeros\_like()}}}}}.
ADDITIONOriginal docstring below.

LAX\sphinxhyphen{}backend implementation of {\hyperref[\detokenize{modules/tensor:symjax.tensor.zeros_like}]{\sphinxcrossref{\sphinxcode{\sphinxupquote{zeros\_like()}}}}}.
Original docstring below.
\begin{quote}\begin{description}
\item[{Parameters}] \leavevmode
\sphinxstyleliteralstrong{\sphinxupquote{dtype}} (\sphinxstyleliteralemphasis{\sphinxupquote{data\sphinxhyphen{}type}}\sphinxstyleliteralemphasis{\sphinxupquote{, }}\sphinxstyleliteralemphasis{\sphinxupquote{optional}}) \textendash{} Overrides the data type of the result.

\item[{Returns}] \leavevmode
\sphinxstylestrong{out} \textendash{} Array of zeros with the same shape and type as \sphinxtitleref{a}.

\item[{Return type}] \leavevmode
ndarray

\end{description}\end{quote}

\sphinxstrong{See also:}

\begin{description}
\item[{{\hyperref[\detokenize{modules/tensor:symjax.tensor.empty_like}]{\sphinxcrossref{\sphinxcode{\sphinxupquote{empty\_like()}}}}}}] \leavevmode
Return an empty array with shape and type of input.

\item[{{\hyperref[\detokenize{modules/tensor:symjax.tensor.ones_like}]{\sphinxcrossref{\sphinxcode{\sphinxupquote{ones\_like()}}}}}}] \leavevmode
Return an array of ones with shape and type of input.

\item[{{\hyperref[\detokenize{modules/tensor:symjax.tensor.full_like}]{\sphinxcrossref{\sphinxcode{\sphinxupquote{full\_like()}}}}}}] \leavevmode
Return a new array with shape of input filled with value.

\item[{{\hyperref[\detokenize{modules/tensor:symjax.tensor.zeros}]{\sphinxcrossref{\sphinxcode{\sphinxupquote{zeros()}}}}}}] \leavevmode
Return a new array setting values to zero.

\end{description}

\subsubsection*{Examples}

\begin{sphinxVerbatim}[commandchars=\\\{\}]
\PYG{g+gp}{\PYGZgt{}\PYGZgt{}\PYGZgt{} }\PYG{n}{x} \PYG{o}{=} \PYG{n}{np}\PYG{o}{.}\PYG{n}{arange}\PYG{p}{(}\PYG{l+m+mi}{6}\PYG{p}{)}
\PYG{g+gp}{\PYGZgt{}\PYGZgt{}\PYGZgt{} }\PYG{n}{x} \PYG{o}{=} \PYG{n}{x}\PYG{o}{.}\PYG{n}{reshape}\PYG{p}{(}\PYG{p}{(}\PYG{l+m+mi}{2}\PYG{p}{,} \PYG{l+m+mi}{3}\PYG{p}{)}\PYG{p}{)}
\PYG{g+gp}{\PYGZgt{}\PYGZgt{}\PYGZgt{} }\PYG{n}{x}
\PYG{g+go}{array([[0, 1, 2],}
\PYG{g+go}{       [3, 4, 5]])}
\PYG{g+gp}{\PYGZgt{}\PYGZgt{}\PYGZgt{} }\PYG{n}{np}\PYG{o}{.}\PYG{n}{zeros\PYGZus{}like}\PYG{p}{(}\PYG{n}{x}\PYG{p}{)}
\PYG{g+go}{array([[0, 0, 0],}
\PYG{g+go}{       [0, 0, 0]])}
\end{sphinxVerbatim}

\begin{sphinxVerbatim}[commandchars=\\\{\}]
\PYG{g+gp}{\PYGZgt{}\PYGZgt{}\PYGZgt{} }\PYG{n}{y} \PYG{o}{=} \PYG{n}{np}\PYG{o}{.}\PYG{n}{arange}\PYG{p}{(}\PYG{l+m+mi}{3}\PYG{p}{,} \PYG{n}{dtype}\PYG{o}{=}\PYG{n+nb}{float}\PYG{p}{)}
\PYG{g+gp}{\PYGZgt{}\PYGZgt{}\PYGZgt{} }\PYG{n}{y}
\PYG{g+go}{array([0., 1., 2.])}
\PYG{g+gp}{\PYGZgt{}\PYGZgt{}\PYGZgt{} }\PYG{n}{np}\PYG{o}{.}\PYG{n}{zeros\PYGZus{}like}\PYG{p}{(}\PYG{n}{y}\PYG{p}{)}
\PYG{g+go}{array([0.,  0.,  0.])}
\end{sphinxVerbatim}

\end{fulllineitems}

\index{exp() (in module symjax.tensor)@\spxentry{exp()}\spxextra{in module symjax.tensor}}

\begin{fulllineitems}
\phantomsection\label{\detokenize{modules/tensor:symjax.tensor.exp}}\pysiglinewithargsret{\sphinxbfcode{\sphinxupquote{exp}}}{\emph{\DUrole{n}{x}}}{}
Calculate the exponential of all elements in the input array.

LAX\sphinxhyphen{}backend implementation of {\hyperref[\detokenize{modules/tensor:symjax.tensor.exp}]{\sphinxcrossref{\sphinxcode{\sphinxupquote{exp()}}}}}.
ADDITIONOriginal docstring below.

LAX\sphinxhyphen{}backend implementation of {\hyperref[\detokenize{modules/tensor:symjax.tensor.exp}]{\sphinxcrossref{\sphinxcode{\sphinxupquote{exp()}}}}}.
Original docstring below.

exp(x, /, out=None, {\color{red}\bfseries{}*}, where=True, casting=’same\_kind’, order=’K’, dtype=None, subok=True{[}, signature, extobj{]})
\begin{quote}\begin{description}
\item[{Returns}] \leavevmode
\sphinxstylestrong{out} \textendash{} Output array, element\sphinxhyphen{}wise exponential of \sphinxtitleref{x}.
This is a scalar if \sphinxtitleref{x} is a scalar.

\item[{Return type}] \leavevmode
ndarray or scalar

\end{description}\end{quote}

\sphinxstrong{See also:}

\begin{description}
\item[{{\hyperref[\detokenize{modules/tensor:symjax.tensor.expm1}]{\sphinxcrossref{\sphinxcode{\sphinxupquote{expm1()}}}}}}] \leavevmode
Calculate \sphinxcode{\sphinxupquote{exp(x) \sphinxhyphen{} 1}} for all elements in the array.

\item[{{\hyperref[\detokenize{modules/tensor:symjax.tensor.exp2}]{\sphinxcrossref{\sphinxcode{\sphinxupquote{exp2()}}}}}}] \leavevmode
Calculate \sphinxcode{\sphinxupquote{2**x}} for all elements in the array.

\end{description}

\subsubsection*{Notes}

The irrational number \sphinxcode{\sphinxupquote{e}} is also known as Euler’s number.  It is
approximately 2.718281, and is the base of the natural logarithm,
\sphinxcode{\sphinxupquote{ln}} (this means that, if \(x = \ln y = \log_e y\),
then \(e^x = y\). For real input, \sphinxcode{\sphinxupquote{exp(x)}} is always positive.

For complex arguments, \sphinxcode{\sphinxupquote{x = a + ib}}, we can write
\(e^x = e^a e^{ib}\).  The first term, \(e^a\), is already
known (it is the real argument, described above).  The second term,
\(e^{ib}\), is \(\cos b + i \sin b\), a function with
magnitude 1 and a periodic phase.
\subsubsection*{References}
\subsubsection*{Examples}

Plot the magnitude and phase of \sphinxcode{\sphinxupquote{exp(x)}} in the complex plane:

\begin{sphinxVerbatim}[commandchars=\\\{\}]
\PYG{g+gp}{\PYGZgt{}\PYGZgt{}\PYGZgt{} }\PYG{k+kn}{import} \PYG{n+nn}{matplotlib}\PYG{n+nn}{.}\PYG{n+nn}{pyplot} \PYG{k}{as} \PYG{n+nn}{plt}
\end{sphinxVerbatim}

\begin{sphinxVerbatim}[commandchars=\\\{\}]
\PYG{g+gp}{\PYGZgt{}\PYGZgt{}\PYGZgt{} }\PYG{n}{x} \PYG{o}{=} \PYG{n}{np}\PYG{o}{.}\PYG{n}{linspace}\PYG{p}{(}\PYG{o}{\PYGZhy{}}\PYG{l+m+mi}{2}\PYG{o}{*}\PYG{n}{np}\PYG{o}{.}\PYG{n}{pi}\PYG{p}{,} \PYG{l+m+mi}{2}\PYG{o}{*}\PYG{n}{np}\PYG{o}{.}\PYG{n}{pi}\PYG{p}{,} \PYG{l+m+mi}{100}\PYG{p}{)}
\PYG{g+gp}{\PYGZgt{}\PYGZgt{}\PYGZgt{} }\PYG{n}{xx} \PYG{o}{=} \PYG{n}{x} \PYG{o}{+} \PYG{l+m+mi}{1}\PYG{n}{j} \PYG{o}{*} \PYG{n}{x}\PYG{p}{[}\PYG{p}{:}\PYG{p}{,} \PYG{n}{np}\PYG{o}{.}\PYG{n}{newaxis}\PYG{p}{]} \PYG{c+c1}{\PYGZsh{} a + ib over complex plane}
\PYG{g+gp}{\PYGZgt{}\PYGZgt{}\PYGZgt{} }\PYG{n}{out} \PYG{o}{=} \PYG{n}{np}\PYG{o}{.}\PYG{n}{exp}\PYG{p}{(}\PYG{n}{xx}\PYG{p}{)}
\end{sphinxVerbatim}

\begin{sphinxVerbatim}[commandchars=\\\{\}]
\PYG{g+gp}{\PYGZgt{}\PYGZgt{}\PYGZgt{} }\PYG{n}{plt}\PYG{o}{.}\PYG{n}{subplot}\PYG{p}{(}\PYG{l+m+mi}{121}\PYG{p}{)}
\PYG{g+gp}{\PYGZgt{}\PYGZgt{}\PYGZgt{} }\PYG{n}{plt}\PYG{o}{.}\PYG{n}{imshow}\PYG{p}{(}\PYG{n}{np}\PYG{o}{.}\PYG{n}{abs}\PYG{p}{(}\PYG{n}{out}\PYG{p}{)}\PYG{p}{,}
\PYG{g+gp}{... }           \PYG{n}{extent}\PYG{o}{=}\PYG{p}{[}\PYG{o}{\PYGZhy{}}\PYG{l+m+mi}{2}\PYG{o}{*}\PYG{n}{np}\PYG{o}{.}\PYG{n}{pi}\PYG{p}{,} \PYG{l+m+mi}{2}\PYG{o}{*}\PYG{n}{np}\PYG{o}{.}\PYG{n}{pi}\PYG{p}{,} \PYG{o}{\PYGZhy{}}\PYG{l+m+mi}{2}\PYG{o}{*}\PYG{n}{np}\PYG{o}{.}\PYG{n}{pi}\PYG{p}{,} \PYG{l+m+mi}{2}\PYG{o}{*}\PYG{n}{np}\PYG{o}{.}\PYG{n}{pi}\PYG{p}{]}\PYG{p}{,} \PYG{n}{cmap}\PYG{o}{=}\PYG{l+s+s1}{\PYGZsq{}}\PYG{l+s+s1}{gray}\PYG{l+s+s1}{\PYGZsq{}}\PYG{p}{)}
\PYG{g+gp}{\PYGZgt{}\PYGZgt{}\PYGZgt{} }\PYG{n}{plt}\PYG{o}{.}\PYG{n}{title}\PYG{p}{(}\PYG{l+s+s1}{\PYGZsq{}}\PYG{l+s+s1}{Magnitude of exp(x)}\PYG{l+s+s1}{\PYGZsq{}}\PYG{p}{)}
\end{sphinxVerbatim}

\begin{sphinxVerbatim}[commandchars=\\\{\}]
\PYG{g+gp}{\PYGZgt{}\PYGZgt{}\PYGZgt{} }\PYG{n}{plt}\PYG{o}{.}\PYG{n}{subplot}\PYG{p}{(}\PYG{l+m+mi}{122}\PYG{p}{)}
\PYG{g+gp}{\PYGZgt{}\PYGZgt{}\PYGZgt{} }\PYG{n}{plt}\PYG{o}{.}\PYG{n}{imshow}\PYG{p}{(}\PYG{n}{np}\PYG{o}{.}\PYG{n}{angle}\PYG{p}{(}\PYG{n}{out}\PYG{p}{)}\PYG{p}{,}
\PYG{g+gp}{... }           \PYG{n}{extent}\PYG{o}{=}\PYG{p}{[}\PYG{o}{\PYGZhy{}}\PYG{l+m+mi}{2}\PYG{o}{*}\PYG{n}{np}\PYG{o}{.}\PYG{n}{pi}\PYG{p}{,} \PYG{l+m+mi}{2}\PYG{o}{*}\PYG{n}{np}\PYG{o}{.}\PYG{n}{pi}\PYG{p}{,} \PYG{o}{\PYGZhy{}}\PYG{l+m+mi}{2}\PYG{o}{*}\PYG{n}{np}\PYG{o}{.}\PYG{n}{pi}\PYG{p}{,} \PYG{l+m+mi}{2}\PYG{o}{*}\PYG{n}{np}\PYG{o}{.}\PYG{n}{pi}\PYG{p}{]}\PYG{p}{,} \PYG{n}{cmap}\PYG{o}{=}\PYG{l+s+s1}{\PYGZsq{}}\PYG{l+s+s1}{hsv}\PYG{l+s+s1}{\PYGZsq{}}\PYG{p}{)}
\PYG{g+gp}{\PYGZgt{}\PYGZgt{}\PYGZgt{} }\PYG{n}{plt}\PYG{o}{.}\PYG{n}{title}\PYG{p}{(}\PYG{l+s+s1}{\PYGZsq{}}\PYG{l+s+s1}{Phase (angle) of exp(x)}\PYG{l+s+s1}{\PYGZsq{}}\PYG{p}{)}
\PYG{g+gp}{\PYGZgt{}\PYGZgt{}\PYGZgt{} }\PYG{n}{plt}\PYG{o}{.}\PYG{n}{show}\PYG{p}{(}\PYG{p}{)}
\end{sphinxVerbatim}

\end{fulllineitems}

\index{exp2() (in module symjax.tensor)@\spxentry{exp2()}\spxextra{in module symjax.tensor}}

\begin{fulllineitems}
\phantomsection\label{\detokenize{modules/tensor:symjax.tensor.exp2}}\pysiglinewithargsret{\sphinxbfcode{\sphinxupquote{exp2}}}{\emph{\DUrole{n}{x}}}{}
Calculate \sphinxtitleref{2**p} for all \sphinxtitleref{p} in the input array.

LAX\sphinxhyphen{}backend implementation of {\hyperref[\detokenize{modules/tensor:symjax.tensor.exp2}]{\sphinxcrossref{\sphinxcode{\sphinxupquote{exp2()}}}}}.
ADDITIONOriginal docstring below.

LAX\sphinxhyphen{}backend implementation of {\hyperref[\detokenize{modules/tensor:symjax.tensor.exp2}]{\sphinxcrossref{\sphinxcode{\sphinxupquote{exp2()}}}}}.
Original docstring below.

exp2(x, /, out=None, {\color{red}\bfseries{}*}, where=True, casting=’same\_kind’, order=’K’, dtype=None, subok=True{[}, signature, extobj{]})
\begin{quote}\begin{description}
\item[{Returns}] \leavevmode
\sphinxstylestrong{out} \textendash{} Element\sphinxhyphen{}wise 2 to the power \sphinxtitleref{x}.
This is a scalar if \sphinxtitleref{x} is a scalar.

\item[{Return type}] \leavevmode
ndarray or scalar

\end{description}\end{quote}

\sphinxstrong{See also:}

{\hyperref[\detokenize{modules/tensor:symjax.tensor.power}]{\sphinxcrossref{\sphinxcode{\sphinxupquote{power()}}}}}

\subsubsection*{Notes}

\DUrole{versionmodified,added}{New in version 1.3.0.}
\subsubsection*{Examples}

\begin{sphinxVerbatim}[commandchars=\\\{\}]
\PYG{g+gp}{\PYGZgt{}\PYGZgt{}\PYGZgt{} }\PYG{n}{np}\PYG{o}{.}\PYG{n}{exp2}\PYG{p}{(}\PYG{p}{[}\PYG{l+m+mi}{2}\PYG{p}{,} \PYG{l+m+mi}{3}\PYG{p}{]}\PYG{p}{)}
\PYG{g+go}{array([ 4.,  8.])}
\end{sphinxVerbatim}

\end{fulllineitems}

\index{expand\_dims() (in module symjax.tensor)@\spxentry{expand\_dims()}\spxextra{in module symjax.tensor}}

\begin{fulllineitems}
\phantomsection\label{\detokenize{modules/tensor:symjax.tensor.expand_dims}}\pysiglinewithargsret{\sphinxbfcode{\sphinxupquote{expand\_dims}}}{\emph{\DUrole{n}{a}}, \emph{\DUrole{n}{axis}}}{}
Expand the shape of an array.

LAX\sphinxhyphen{}backend implementation of {\hyperref[\detokenize{modules/tensor:symjax.tensor.expand_dims}]{\sphinxcrossref{\sphinxcode{\sphinxupquote{expand\_dims()}}}}}.
ADDITIONOriginal docstring below.

LAX\sphinxhyphen{}backend implementation of {\hyperref[\detokenize{modules/tensor:symjax.tensor.expand_dims}]{\sphinxcrossref{\sphinxcode{\sphinxupquote{expand\_dims()}}}}}.
Original docstring below.

Insert a new axis that will appear at the \sphinxtitleref{axis} position in the expanded
array shape.
\begin{quote}\begin{description}
\item[{Returns}] \leavevmode
\sphinxstylestrong{result} \textendash{} View of \sphinxtitleref{a} with the number of dimensions increased.

\item[{Return type}] \leavevmode
ndarray

\end{description}\end{quote}

\sphinxstrong{See also:}

\begin{description}
\item[{{\hyperref[\detokenize{modules/tensor:symjax.tensor.squeeze}]{\sphinxcrossref{\sphinxcode{\sphinxupquote{squeeze()}}}}}}] \leavevmode
The inverse operation, removing singleton dimensions

\item[{{\hyperref[\detokenize{modules/tensor:symjax.tensor.reshape}]{\sphinxcrossref{\sphinxcode{\sphinxupquote{reshape()}}}}}}] \leavevmode
Insert, remove, and combine dimensions, and resize existing ones

\end{description}

\sphinxcode{\sphinxupquote{doc.indexing()}}, {\hyperref[\detokenize{modules/tensor:symjax.tensor.atleast_1d}]{\sphinxcrossref{\sphinxcode{\sphinxupquote{atleast\_1d()}}}}}, {\hyperref[\detokenize{modules/tensor:symjax.tensor.atleast_2d}]{\sphinxcrossref{\sphinxcode{\sphinxupquote{atleast\_2d()}}}}}, {\hyperref[\detokenize{modules/tensor:symjax.tensor.atleast_3d}]{\sphinxcrossref{\sphinxcode{\sphinxupquote{atleast\_3d()}}}}}

\subsubsection*{Examples}

\begin{sphinxVerbatim}[commandchars=\\\{\}]
\PYG{g+gp}{\PYGZgt{}\PYGZgt{}\PYGZgt{} }\PYG{n}{x} \PYG{o}{=} \PYG{n}{np}\PYG{o}{.}\PYG{n}{array}\PYG{p}{(}\PYG{p}{[}\PYG{l+m+mi}{1}\PYG{p}{,} \PYG{l+m+mi}{2}\PYG{p}{]}\PYG{p}{)}
\PYG{g+gp}{\PYGZgt{}\PYGZgt{}\PYGZgt{} }\PYG{n}{x}\PYG{o}{.}\PYG{n}{shape}
\PYG{g+go}{(2,)}
\end{sphinxVerbatim}

The following is equivalent to \sphinxcode{\sphinxupquote{x{[}np.newaxis, :{]}}} or \sphinxcode{\sphinxupquote{x{[}np.newaxis{]}}}:

\begin{sphinxVerbatim}[commandchars=\\\{\}]
\PYG{g+gp}{\PYGZgt{}\PYGZgt{}\PYGZgt{} }\PYG{n}{y} \PYG{o}{=} \PYG{n}{np}\PYG{o}{.}\PYG{n}{expand\PYGZus{}dims}\PYG{p}{(}\PYG{n}{x}\PYG{p}{,} \PYG{n}{axis}\PYG{o}{=}\PYG{l+m+mi}{0}\PYG{p}{)}
\PYG{g+gp}{\PYGZgt{}\PYGZgt{}\PYGZgt{} }\PYG{n}{y}
\PYG{g+go}{array([[1, 2]])}
\PYG{g+gp}{\PYGZgt{}\PYGZgt{}\PYGZgt{} }\PYG{n}{y}\PYG{o}{.}\PYG{n}{shape}
\PYG{g+go}{(1, 2)}
\end{sphinxVerbatim}

The following is equivalent to \sphinxcode{\sphinxupquote{x{[}:, np.newaxis{]}}}:

\begin{sphinxVerbatim}[commandchars=\\\{\}]
\PYG{g+gp}{\PYGZgt{}\PYGZgt{}\PYGZgt{} }\PYG{n}{y} \PYG{o}{=} \PYG{n}{np}\PYG{o}{.}\PYG{n}{expand\PYGZus{}dims}\PYG{p}{(}\PYG{n}{x}\PYG{p}{,} \PYG{n}{axis}\PYG{o}{=}\PYG{l+m+mi}{1}\PYG{p}{)}
\PYG{g+gp}{\PYGZgt{}\PYGZgt{}\PYGZgt{} }\PYG{n}{y}
\PYG{g+go}{array([[1],}
\PYG{g+go}{       [2]])}
\PYG{g+gp}{\PYGZgt{}\PYGZgt{}\PYGZgt{} }\PYG{n}{y}\PYG{o}{.}\PYG{n}{shape}
\PYG{g+go}{(2, 1)}
\end{sphinxVerbatim}

\sphinxcode{\sphinxupquote{axis}} may also be a tuple:

\begin{sphinxVerbatim}[commandchars=\\\{\}]
\PYG{g+gp}{\PYGZgt{}\PYGZgt{}\PYGZgt{} }\PYG{n}{y} \PYG{o}{=} \PYG{n}{np}\PYG{o}{.}\PYG{n}{expand\PYGZus{}dims}\PYG{p}{(}\PYG{n}{x}\PYG{p}{,} \PYG{n}{axis}\PYG{o}{=}\PYG{p}{(}\PYG{l+m+mi}{0}\PYG{p}{,} \PYG{l+m+mi}{1}\PYG{p}{)}\PYG{p}{)}
\PYG{g+gp}{\PYGZgt{}\PYGZgt{}\PYGZgt{} }\PYG{n}{y}
\PYG{g+go}{array([[[1, 2]]])}
\end{sphinxVerbatim}

\begin{sphinxVerbatim}[commandchars=\\\{\}]
\PYG{g+gp}{\PYGZgt{}\PYGZgt{}\PYGZgt{} }\PYG{n}{y} \PYG{o}{=} \PYG{n}{np}\PYG{o}{.}\PYG{n}{expand\PYGZus{}dims}\PYG{p}{(}\PYG{n}{x}\PYG{p}{,} \PYG{n}{axis}\PYG{o}{=}\PYG{p}{(}\PYG{l+m+mi}{2}\PYG{p}{,} \PYG{l+m+mi}{0}\PYG{p}{)}\PYG{p}{)}
\PYG{g+gp}{\PYGZgt{}\PYGZgt{}\PYGZgt{} }\PYG{n}{y}
\PYG{g+go}{array([[[1],}
\PYG{g+go}{        [2]]])}
\end{sphinxVerbatim}

Note that some examples may use \sphinxcode{\sphinxupquote{None}} instead of \sphinxcode{\sphinxupquote{np.newaxis}}.  These
are the same objects:

\begin{sphinxVerbatim}[commandchars=\\\{\}]
\PYG{g+gp}{\PYGZgt{}\PYGZgt{}\PYGZgt{} }\PYG{n}{np}\PYG{o}{.}\PYG{n}{newaxis} \PYG{o+ow}{is} \PYG{k+kc}{None}
\PYG{g+go}{True}
\end{sphinxVerbatim}

\end{fulllineitems}

\index{expm1() (in module symjax.tensor)@\spxentry{expm1()}\spxextra{in module symjax.tensor}}

\begin{fulllineitems}
\phantomsection\label{\detokenize{modules/tensor:symjax.tensor.expm1}}\pysiglinewithargsret{\sphinxbfcode{\sphinxupquote{expm1}}}{\emph{\DUrole{n}{x}}}{}
Calculate \sphinxcode{\sphinxupquote{exp(x) \sphinxhyphen{} 1}} for all elements in the array.

LAX\sphinxhyphen{}backend implementation of {\hyperref[\detokenize{modules/tensor:symjax.tensor.expm1}]{\sphinxcrossref{\sphinxcode{\sphinxupquote{expm1()}}}}}.
ADDITIONOriginal docstring below.

LAX\sphinxhyphen{}backend implementation of {\hyperref[\detokenize{modules/tensor:symjax.tensor.expm1}]{\sphinxcrossref{\sphinxcode{\sphinxupquote{expm1()}}}}}.
Original docstring below.

expm1(x, /, out=None, {\color{red}\bfseries{}*}, where=True, casting=’same\_kind’, order=’K’, dtype=None, subok=True{[}, signature, extobj{]})
\begin{quote}\begin{description}
\item[{Returns}] \leavevmode
\sphinxstylestrong{out} \textendash{} Element\sphinxhyphen{}wise exponential minus one: \sphinxcode{\sphinxupquote{out = exp(x) \sphinxhyphen{} 1}}.
This is a scalar if \sphinxtitleref{x} is a scalar.

\item[{Return type}] \leavevmode
ndarray or scalar

\end{description}\end{quote}

\sphinxstrong{See also:}

\begin{description}
\item[{{\hyperref[\detokenize{modules/tensor:symjax.tensor.log1p}]{\sphinxcrossref{\sphinxcode{\sphinxupquote{log1p()}}}}}}] \leavevmode
\sphinxcode{\sphinxupquote{log(1 + x)}}, the inverse of expm1.

\end{description}

\subsubsection*{Notes}

This function provides greater precision than \sphinxcode{\sphinxupquote{exp(x) \sphinxhyphen{} 1}}
for small values of \sphinxcode{\sphinxupquote{x}}.
\subsubsection*{Examples}

The true value of \sphinxcode{\sphinxupquote{exp(1e\sphinxhyphen{}10) \sphinxhyphen{} 1}} is \sphinxcode{\sphinxupquote{1.00000000005e\sphinxhyphen{}10}} to
about 32 significant digits. This example shows the superiority of
expm1 in this case.

\begin{sphinxVerbatim}[commandchars=\\\{\}]
\PYG{g+gp}{\PYGZgt{}\PYGZgt{}\PYGZgt{} }\PYG{n}{np}\PYG{o}{.}\PYG{n}{expm1}\PYG{p}{(}\PYG{l+m+mf}{1e\PYGZhy{}10}\PYG{p}{)}
\PYG{g+go}{1.00000000005e\PYGZhy{}10}
\PYG{g+gp}{\PYGZgt{}\PYGZgt{}\PYGZgt{} }\PYG{n}{np}\PYG{o}{.}\PYG{n}{exp}\PYG{p}{(}\PYG{l+m+mf}{1e\PYGZhy{}10}\PYG{p}{)} \PYG{o}{\PYGZhy{}} \PYG{l+m+mi}{1}
\PYG{g+go}{1.000000082740371e\PYGZhy{}10}
\end{sphinxVerbatim}

\end{fulllineitems}

\index{eye() (in module symjax.tensor)@\spxentry{eye()}\spxextra{in module symjax.tensor}}

\begin{fulllineitems}
\phantomsection\label{\detokenize{modules/tensor:symjax.tensor.eye}}\pysiglinewithargsret{\sphinxbfcode{\sphinxupquote{eye}}}{\emph{\DUrole{n}{N}}, \emph{\DUrole{n}{M}\DUrole{o}{=}\DUrole{default_value}{None}}, \emph{\DUrole{n}{k}\DUrole{o}{=}\DUrole{default_value}{0}}, \emph{\DUrole{n}{dtype}\DUrole{o}{=}\DUrole{default_value}{None}}}{}
Return a 2\sphinxhyphen{}D array with ones on the diagonal and zeros elsewhere.

LAX\sphinxhyphen{}backend implementation of {\hyperref[\detokenize{modules/tensor:symjax.tensor.eye}]{\sphinxcrossref{\sphinxcode{\sphinxupquote{eye()}}}}}.
ADDITIONOriginal docstring below.

LAX\sphinxhyphen{}backend implementation of {\hyperref[\detokenize{modules/tensor:symjax.tensor.eye}]{\sphinxcrossref{\sphinxcode{\sphinxupquote{eye()}}}}}.
Original docstring below.
\begin{quote}\begin{description}
\item[{Parameters}] \leavevmode
\sphinxstyleliteralstrong{\sphinxupquote{dtype}} (\sphinxstyleliteralemphasis{\sphinxupquote{data\sphinxhyphen{}type}}\sphinxstyleliteralemphasis{\sphinxupquote{, }}\sphinxstyleliteralemphasis{\sphinxupquote{optional}}) \textendash{} 

\item[{Returns}] \leavevmode
\sphinxstylestrong{I} \textendash{} An array where all elements are equal to zero, except for the \sphinxtitleref{k}\sphinxhyphen{}th
diagonal, whose values are equal to one.

\item[{Return type}] \leavevmode
ndarray of shape (N,M)

\end{description}\end{quote}

\sphinxstrong{See also:}

\begin{description}
\item[{{\hyperref[\detokenize{modules/tensor:symjax.tensor.identity}]{\sphinxcrossref{\sphinxcode{\sphinxupquote{identity()}}}}}}] \leavevmode
(almost) equivalent function

\item[{{\hyperref[\detokenize{modules/tensor:symjax.tensor.diag}]{\sphinxcrossref{\sphinxcode{\sphinxupquote{diag()}}}}}}] \leavevmode
diagonal 2\sphinxhyphen{}D array from a 1\sphinxhyphen{}D array specified by the user.

\end{description}

\subsubsection*{Examples}

\begin{sphinxVerbatim}[commandchars=\\\{\}]
\PYG{g+gp}{\PYGZgt{}\PYGZgt{}\PYGZgt{} }\PYG{n}{np}\PYG{o}{.}\PYG{n}{eye}\PYG{p}{(}\PYG{l+m+mi}{2}\PYG{p}{,} \PYG{n}{dtype}\PYG{o}{=}\PYG{n+nb}{int}\PYG{p}{)}
\PYG{g+go}{array([[1, 0],}
\PYG{g+go}{       [0, 1]])}
\PYG{g+gp}{\PYGZgt{}\PYGZgt{}\PYGZgt{} }\PYG{n}{np}\PYG{o}{.}\PYG{n}{eye}\PYG{p}{(}\PYG{l+m+mi}{3}\PYG{p}{,} \PYG{n}{k}\PYG{o}{=}\PYG{l+m+mi}{1}\PYG{p}{)}
\PYG{g+go}{array([[0.,  1.,  0.],}
\PYG{g+go}{       [0.,  0.,  1.],}
\PYG{g+go}{       [0.,  0.,  0.]])}
\end{sphinxVerbatim}

\end{fulllineitems}

\index{fabs() (in module symjax.tensor)@\spxentry{fabs()}\spxextra{in module symjax.tensor}}

\begin{fulllineitems}
\phantomsection\label{\detokenize{modules/tensor:symjax.tensor.fabs}}\pysiglinewithargsret{\sphinxbfcode{\sphinxupquote{fabs}}}{\emph{\DUrole{n}{x}}}{}
Compute the absolute values element\sphinxhyphen{}wise.

LAX\sphinxhyphen{}backend implementation of {\hyperref[\detokenize{modules/tensor:symjax.tensor.fabs}]{\sphinxcrossref{\sphinxcode{\sphinxupquote{fabs()}}}}}.
ADDITIONOriginal docstring below.

LAX\sphinxhyphen{}backend implementation of {\hyperref[\detokenize{modules/tensor:symjax.tensor.fabs}]{\sphinxcrossref{\sphinxcode{\sphinxupquote{fabs()}}}}}.
Original docstring below.

fabs(x, /, out=None, {\color{red}\bfseries{}*}, where=True, casting=’same\_kind’, order=’K’, dtype=None, subok=True{[}, signature, extobj{]})

This function returns the absolute values (positive magnitude) of the
data in \sphinxtitleref{x}. Complex values are not handled, use \sphinxtitleref{absolute} to find the
absolute values of complex data.
\begin{quote}\begin{description}
\item[{Returns}] \leavevmode
\sphinxstylestrong{y} \textendash{} The absolute values of \sphinxtitleref{x}, the returned values are always floats.
This is a scalar if \sphinxtitleref{x} is a scalar.

\item[{Return type}] \leavevmode
ndarray or scalar

\end{description}\end{quote}

\sphinxstrong{See also:}

\begin{description}
\item[{{\hyperref[\detokenize{modules/tensor:symjax.tensor.absolute}]{\sphinxcrossref{\sphinxcode{\sphinxupquote{absolute()}}}}}}] \leavevmode
Absolute values including \sphinxtitleref{complex} types.

\end{description}

\subsubsection*{Examples}

\begin{sphinxVerbatim}[commandchars=\\\{\}]
\PYG{g+gp}{\PYGZgt{}\PYGZgt{}\PYGZgt{} }\PYG{n}{np}\PYG{o}{.}\PYG{n}{fabs}\PYG{p}{(}\PYG{o}{\PYGZhy{}}\PYG{l+m+mi}{1}\PYG{p}{)}
\PYG{g+go}{1.0}
\PYG{g+gp}{\PYGZgt{}\PYGZgt{}\PYGZgt{} }\PYG{n}{np}\PYG{o}{.}\PYG{n}{fabs}\PYG{p}{(}\PYG{p}{[}\PYG{o}{\PYGZhy{}}\PYG{l+m+mf}{1.2}\PYG{p}{,} \PYG{l+m+mf}{1.2}\PYG{p}{]}\PYG{p}{)}
\PYG{g+go}{array([ 1.2,  1.2])}
\end{sphinxVerbatim}

\end{fulllineitems}

\index{flip() (in module symjax.tensor)@\spxentry{flip()}\spxextra{in module symjax.tensor}}

\begin{fulllineitems}
\phantomsection\label{\detokenize{modules/tensor:symjax.tensor.flip}}\pysiglinewithargsret{\sphinxbfcode{\sphinxupquote{flip}}}{\emph{\DUrole{n}{m}}, \emph{\DUrole{n}{axis}\DUrole{o}{=}\DUrole{default_value}{None}}}{}
Reverse the order of elements in an array along the given axis.

LAX\sphinxhyphen{}backend implementation of {\hyperref[\detokenize{modules/tensor:symjax.tensor.flip}]{\sphinxcrossref{\sphinxcode{\sphinxupquote{flip()}}}}}.
ADDITIONOriginal docstring below.

LAX\sphinxhyphen{}backend implementation of {\hyperref[\detokenize{modules/tensor:symjax.tensor.flip}]{\sphinxcrossref{\sphinxcode{\sphinxupquote{flip()}}}}}.
Original docstring below.

The shape of the array is preserved, but the elements are reordered.

\DUrole{versionmodified,added}{New in version 1.12.0.}
\begin{quote}\begin{description}
\item[{Returns}] \leavevmode
\sphinxstylestrong{out} \textendash{} A view of \sphinxtitleref{m} with the entries of axis reversed.  Since a view is
returned, this operation is done in constant time.

\item[{Return type}] \leavevmode
array\_like

\end{description}\end{quote}

\sphinxstrong{See also:}

\begin{description}
\item[{{\hyperref[\detokenize{modules/tensor:symjax.tensor.flipud}]{\sphinxcrossref{\sphinxcode{\sphinxupquote{flipud()}}}}}}] \leavevmode
Flip an array vertically (axis=0).

\item[{{\hyperref[\detokenize{modules/tensor:symjax.tensor.fliplr}]{\sphinxcrossref{\sphinxcode{\sphinxupquote{fliplr()}}}}}}] \leavevmode
Flip an array horizontally (axis=1).

\end{description}

\subsubsection*{Notes}

flip(m, 0) is equivalent to flipud(m).

flip(m, 1) is equivalent to fliplr(m).

flip(m, n) corresponds to \sphinxcode{\sphinxupquote{m{[}...,::\sphinxhyphen{}1,...{]}}} with \sphinxcode{\sphinxupquote{::\sphinxhyphen{}1}} at position n.

flip(m) corresponds to \sphinxcode{\sphinxupquote{m{[}::\sphinxhyphen{}1,::\sphinxhyphen{}1,...,::\sphinxhyphen{}1{]}}} with \sphinxcode{\sphinxupquote{::\sphinxhyphen{}1}} at all
positions.

flip(m, (0, 1)) corresponds to \sphinxcode{\sphinxupquote{m{[}::\sphinxhyphen{}1,::\sphinxhyphen{}1,...{]}}} with \sphinxcode{\sphinxupquote{::\sphinxhyphen{}1}} at
position 0 and position 1.
\subsubsection*{Examples}

\begin{sphinxVerbatim}[commandchars=\\\{\}]
\PYG{g+gp}{\PYGZgt{}\PYGZgt{}\PYGZgt{} }\PYG{n}{A} \PYG{o}{=} \PYG{n}{np}\PYG{o}{.}\PYG{n}{arange}\PYG{p}{(}\PYG{l+m+mi}{8}\PYG{p}{)}\PYG{o}{.}\PYG{n}{reshape}\PYG{p}{(}\PYG{p}{(}\PYG{l+m+mi}{2}\PYG{p}{,}\PYG{l+m+mi}{2}\PYG{p}{,}\PYG{l+m+mi}{2}\PYG{p}{)}\PYG{p}{)}
\PYG{g+gp}{\PYGZgt{}\PYGZgt{}\PYGZgt{} }\PYG{n}{A}
\PYG{g+go}{array([[[0, 1],}
\PYG{g+go}{        [2, 3]],}
\PYG{g+go}{       [[4, 5],}
\PYG{g+go}{        [6, 7]]])}
\PYG{g+gp}{\PYGZgt{}\PYGZgt{}\PYGZgt{} }\PYG{n}{np}\PYG{o}{.}\PYG{n}{flip}\PYG{p}{(}\PYG{n}{A}\PYG{p}{,} \PYG{l+m+mi}{0}\PYG{p}{)}
\PYG{g+go}{array([[[4, 5],}
\PYG{g+go}{        [6, 7]],}
\PYG{g+go}{       [[0, 1],}
\PYG{g+go}{        [2, 3]]])}
\PYG{g+gp}{\PYGZgt{}\PYGZgt{}\PYGZgt{} }\PYG{n}{np}\PYG{o}{.}\PYG{n}{flip}\PYG{p}{(}\PYG{n}{A}\PYG{p}{,} \PYG{l+m+mi}{1}\PYG{p}{)}
\PYG{g+go}{array([[[2, 3],}
\PYG{g+go}{        [0, 1]],}
\PYG{g+go}{       [[6, 7],}
\PYG{g+go}{        [4, 5]]])}
\PYG{g+gp}{\PYGZgt{}\PYGZgt{}\PYGZgt{} }\PYG{n}{np}\PYG{o}{.}\PYG{n}{flip}\PYG{p}{(}\PYG{n}{A}\PYG{p}{)}
\PYG{g+go}{array([[[7, 6],}
\PYG{g+go}{        [5, 4]],}
\PYG{g+go}{       [[3, 2],}
\PYG{g+go}{        [1, 0]]])}
\PYG{g+gp}{\PYGZgt{}\PYGZgt{}\PYGZgt{} }\PYG{n}{np}\PYG{o}{.}\PYG{n}{flip}\PYG{p}{(}\PYG{n}{A}\PYG{p}{,} \PYG{p}{(}\PYG{l+m+mi}{0}\PYG{p}{,} \PYG{l+m+mi}{2}\PYG{p}{)}\PYG{p}{)}
\PYG{g+go}{array([[[5, 4],}
\PYG{g+go}{        [7, 6]],}
\PYG{g+go}{       [[1, 0],}
\PYG{g+go}{        [3, 2]]])}
\PYG{g+gp}{\PYGZgt{}\PYGZgt{}\PYGZgt{} }\PYG{n}{A} \PYG{o}{=} \PYG{n}{np}\PYG{o}{.}\PYG{n}{random}\PYG{o}{.}\PYG{n}{randn}\PYG{p}{(}\PYG{l+m+mi}{3}\PYG{p}{,}\PYG{l+m+mi}{4}\PYG{p}{,}\PYG{l+m+mi}{5}\PYG{p}{)}
\PYG{g+gp}{\PYGZgt{}\PYGZgt{}\PYGZgt{} }\PYG{n}{np}\PYG{o}{.}\PYG{n}{all}\PYG{p}{(}\PYG{n}{np}\PYG{o}{.}\PYG{n}{flip}\PYG{p}{(}\PYG{n}{A}\PYG{p}{,}\PYG{l+m+mi}{2}\PYG{p}{)} \PYG{o}{==} \PYG{n}{A}\PYG{p}{[}\PYG{p}{:}\PYG{p}{,}\PYG{p}{:}\PYG{p}{,}\PYG{p}{:}\PYG{p}{:}\PYG{o}{\PYGZhy{}}\PYG{l+m+mi}{1}\PYG{p}{,}\PYG{o}{.}\PYG{o}{.}\PYG{o}{.}\PYG{p}{]}\PYG{p}{)}
\PYG{g+go}{True}
\end{sphinxVerbatim}

\end{fulllineitems}

\index{fliplr() (in module symjax.tensor)@\spxentry{fliplr()}\spxextra{in module symjax.tensor}}

\begin{fulllineitems}
\phantomsection\label{\detokenize{modules/tensor:symjax.tensor.fliplr}}\pysiglinewithargsret{\sphinxbfcode{\sphinxupquote{fliplr}}}{\emph{\DUrole{n}{m}}}{}
Flip array in the left/right direction.

LAX\sphinxhyphen{}backend implementation of {\hyperref[\detokenize{modules/tensor:symjax.tensor.fliplr}]{\sphinxcrossref{\sphinxcode{\sphinxupquote{fliplr()}}}}}.
ADDITIONOriginal docstring below.

LAX\sphinxhyphen{}backend implementation of {\hyperref[\detokenize{modules/tensor:symjax.tensor.fliplr}]{\sphinxcrossref{\sphinxcode{\sphinxupquote{fliplr()}}}}}.
Original docstring below.

Flip the entries in each row in the left/right direction.
Columns are preserved, but appear in a different order than before.
\begin{quote}\begin{description}
\item[{Returns}] \leavevmode
\sphinxstylestrong{f} \textendash{} A view of \sphinxtitleref{m} with the columns reversed.  Since a view
is returned, this operation is \(\mathcal O(1)\).

\item[{Return type}] \leavevmode
ndarray

\end{description}\end{quote}

\sphinxstrong{See also:}

\begin{description}
\item[{{\hyperref[\detokenize{modules/tensor:symjax.tensor.flipud}]{\sphinxcrossref{\sphinxcode{\sphinxupquote{flipud()}}}}}}] \leavevmode
Flip array in the up/down direction.

\item[{{\hyperref[\detokenize{modules/tensor:symjax.tensor.rot90}]{\sphinxcrossref{\sphinxcode{\sphinxupquote{rot90()}}}}}}] \leavevmode
Rotate array counterclockwise.

\end{description}

\subsubsection*{Notes}

Equivalent to m{[}:,::\sphinxhyphen{}1{]}. Requires the array to be at least 2\sphinxhyphen{}D.
\subsubsection*{Examples}

\begin{sphinxVerbatim}[commandchars=\\\{\}]
\PYG{g+gp}{\PYGZgt{}\PYGZgt{}\PYGZgt{} }\PYG{n}{A} \PYG{o}{=} \PYG{n}{np}\PYG{o}{.}\PYG{n}{diag}\PYG{p}{(}\PYG{p}{[}\PYG{l+m+mf}{1.}\PYG{p}{,}\PYG{l+m+mf}{2.}\PYG{p}{,}\PYG{l+m+mf}{3.}\PYG{p}{]}\PYG{p}{)}
\PYG{g+gp}{\PYGZgt{}\PYGZgt{}\PYGZgt{} }\PYG{n}{A}
\PYG{g+go}{array([[1.,  0.,  0.],}
\PYG{g+go}{       [0.,  2.,  0.],}
\PYG{g+go}{       [0.,  0.,  3.]])}
\PYG{g+gp}{\PYGZgt{}\PYGZgt{}\PYGZgt{} }\PYG{n}{np}\PYG{o}{.}\PYG{n}{fliplr}\PYG{p}{(}\PYG{n}{A}\PYG{p}{)}
\PYG{g+go}{array([[0.,  0.,  1.],}
\PYG{g+go}{       [0.,  2.,  0.],}
\PYG{g+go}{       [3.,  0.,  0.]])}
\end{sphinxVerbatim}

\begin{sphinxVerbatim}[commandchars=\\\{\}]
\PYG{g+gp}{\PYGZgt{}\PYGZgt{}\PYGZgt{} }\PYG{n}{A} \PYG{o}{=} \PYG{n}{np}\PYG{o}{.}\PYG{n}{random}\PYG{o}{.}\PYG{n}{randn}\PYG{p}{(}\PYG{l+m+mi}{2}\PYG{p}{,}\PYG{l+m+mi}{3}\PYG{p}{,}\PYG{l+m+mi}{5}\PYG{p}{)}
\PYG{g+gp}{\PYGZgt{}\PYGZgt{}\PYGZgt{} }\PYG{n}{np}\PYG{o}{.}\PYG{n}{all}\PYG{p}{(}\PYG{n}{np}\PYG{o}{.}\PYG{n}{fliplr}\PYG{p}{(}\PYG{n}{A}\PYG{p}{)} \PYG{o}{==} \PYG{n}{A}\PYG{p}{[}\PYG{p}{:}\PYG{p}{,}\PYG{p}{:}\PYG{p}{:}\PYG{o}{\PYGZhy{}}\PYG{l+m+mi}{1}\PYG{p}{,}\PYG{o}{.}\PYG{o}{.}\PYG{o}{.}\PYG{p}{]}\PYG{p}{)}
\PYG{g+go}{True}
\end{sphinxVerbatim}

\end{fulllineitems}

\index{flipud() (in module symjax.tensor)@\spxentry{flipud()}\spxextra{in module symjax.tensor}}

\begin{fulllineitems}
\phantomsection\label{\detokenize{modules/tensor:symjax.tensor.flipud}}\pysiglinewithargsret{\sphinxbfcode{\sphinxupquote{flipud}}}{\emph{\DUrole{n}{m}}}{}
Flip array in the up/down direction.

LAX\sphinxhyphen{}backend implementation of {\hyperref[\detokenize{modules/tensor:symjax.tensor.flipud}]{\sphinxcrossref{\sphinxcode{\sphinxupquote{flipud()}}}}}.
ADDITIONOriginal docstring below.

LAX\sphinxhyphen{}backend implementation of {\hyperref[\detokenize{modules/tensor:symjax.tensor.flipud}]{\sphinxcrossref{\sphinxcode{\sphinxupquote{flipud()}}}}}.
Original docstring below.

Flip the entries in each column in the up/down direction.
Rows are preserved, but appear in a different order than before.
\begin{quote}\begin{description}
\item[{Returns}] \leavevmode
\sphinxstylestrong{out} \textendash{} A view of \sphinxtitleref{m} with the rows reversed.  Since a view is
returned, this operation is \(\mathcal O(1)\).

\item[{Return type}] \leavevmode
array\_like

\end{description}\end{quote}

\sphinxstrong{See also:}

\begin{description}
\item[{{\hyperref[\detokenize{modules/tensor:symjax.tensor.fliplr}]{\sphinxcrossref{\sphinxcode{\sphinxupquote{fliplr()}}}}}}] \leavevmode
Flip array in the left/right direction.

\item[{{\hyperref[\detokenize{modules/tensor:symjax.tensor.rot90}]{\sphinxcrossref{\sphinxcode{\sphinxupquote{rot90()}}}}}}] \leavevmode
Rotate array counterclockwise.

\end{description}

\subsubsection*{Notes}

Equivalent to \sphinxcode{\sphinxupquote{m{[}::\sphinxhyphen{}1,...{]}}}.
Does not require the array to be two\sphinxhyphen{}dimensional.
\subsubsection*{Examples}

\begin{sphinxVerbatim}[commandchars=\\\{\}]
\PYG{g+gp}{\PYGZgt{}\PYGZgt{}\PYGZgt{} }\PYG{n}{A} \PYG{o}{=} \PYG{n}{np}\PYG{o}{.}\PYG{n}{diag}\PYG{p}{(}\PYG{p}{[}\PYG{l+m+mf}{1.0}\PYG{p}{,} \PYG{l+m+mi}{2}\PYG{p}{,} \PYG{l+m+mi}{3}\PYG{p}{]}\PYG{p}{)}
\PYG{g+gp}{\PYGZgt{}\PYGZgt{}\PYGZgt{} }\PYG{n}{A}
\PYG{g+go}{array([[1.,  0.,  0.],}
\PYG{g+go}{       [0.,  2.,  0.],}
\PYG{g+go}{       [0.,  0.,  3.]])}
\PYG{g+gp}{\PYGZgt{}\PYGZgt{}\PYGZgt{} }\PYG{n}{np}\PYG{o}{.}\PYG{n}{flipud}\PYG{p}{(}\PYG{n}{A}\PYG{p}{)}
\PYG{g+go}{array([[0.,  0.,  3.],}
\PYG{g+go}{       [0.,  2.,  0.],}
\PYG{g+go}{       [1.,  0.,  0.]])}
\end{sphinxVerbatim}

\begin{sphinxVerbatim}[commandchars=\\\{\}]
\PYG{g+gp}{\PYGZgt{}\PYGZgt{}\PYGZgt{} }\PYG{n}{A} \PYG{o}{=} \PYG{n}{np}\PYG{o}{.}\PYG{n}{random}\PYG{o}{.}\PYG{n}{randn}\PYG{p}{(}\PYG{l+m+mi}{2}\PYG{p}{,}\PYG{l+m+mi}{3}\PYG{p}{,}\PYG{l+m+mi}{5}\PYG{p}{)}
\PYG{g+gp}{\PYGZgt{}\PYGZgt{}\PYGZgt{} }\PYG{n}{np}\PYG{o}{.}\PYG{n}{all}\PYG{p}{(}\PYG{n}{np}\PYG{o}{.}\PYG{n}{flipud}\PYG{p}{(}\PYG{n}{A}\PYG{p}{)} \PYG{o}{==} \PYG{n}{A}\PYG{p}{[}\PYG{p}{:}\PYG{p}{:}\PYG{o}{\PYGZhy{}}\PYG{l+m+mi}{1}\PYG{p}{,}\PYG{o}{.}\PYG{o}{.}\PYG{o}{.}\PYG{p}{]}\PYG{p}{)}
\PYG{g+go}{True}
\end{sphinxVerbatim}

\begin{sphinxVerbatim}[commandchars=\\\{\}]
\PYG{g+gp}{\PYGZgt{}\PYGZgt{}\PYGZgt{} }\PYG{n}{np}\PYG{o}{.}\PYG{n}{flipud}\PYG{p}{(}\PYG{p}{[}\PYG{l+m+mi}{1}\PYG{p}{,}\PYG{l+m+mi}{2}\PYG{p}{]}\PYG{p}{)}
\PYG{g+go}{array([2, 1])}
\end{sphinxVerbatim}

\end{fulllineitems}

\index{float\_power() (in module symjax.tensor)@\spxentry{float\_power()}\spxextra{in module symjax.tensor}}

\begin{fulllineitems}
\phantomsection\label{\detokenize{modules/tensor:symjax.tensor.float_power}}\pysiglinewithargsret{\sphinxbfcode{\sphinxupquote{float\_power}}}{\emph{\DUrole{n}{x1}}, \emph{\DUrole{n}{x2}}}{}
First array elements raised to powers from second array, element\sphinxhyphen{}wise.

LAX\sphinxhyphen{}backend implementation of {\hyperref[\detokenize{modules/tensor:symjax.tensor.float_power}]{\sphinxcrossref{\sphinxcode{\sphinxupquote{float\_power()}}}}}.
ADDITIONOriginal docstring below.

LAX\sphinxhyphen{}backend implementation of {\hyperref[\detokenize{modules/tensor:symjax.tensor.float_power}]{\sphinxcrossref{\sphinxcode{\sphinxupquote{float\_power()}}}}}.
Original docstring below.

float\_power(x1, x2, /, out=None, {\color{red}\bfseries{}*}, where=True, casting=’same\_kind’, order=’K’, dtype=None, subok=True{[}, signature, extobj{]})

Raise each base in \sphinxtitleref{x1} to the positionally\sphinxhyphen{}corresponding power in \sphinxtitleref{x2}.
\sphinxtitleref{x1} and \sphinxtitleref{x2} must be broadcastable to the same shape. This differs from
the power function in that integers, float16, and float32  are promoted to
floats with a minimum precision of float64 so that the result is always
inexact.  The intent is that the function will return a usable result for
negative powers and seldom overflow for positive powers.

\DUrole{versionmodified,added}{New in version 1.12.0.}
\begin{quote}\begin{description}
\item[{Returns}] \leavevmode
\sphinxstylestrong{y} \textendash{} The bases in \sphinxtitleref{x1} raised to the exponents in \sphinxtitleref{x2}.
This is a scalar if both \sphinxtitleref{x1} and \sphinxtitleref{x2} are scalars.

\item[{Return type}] \leavevmode
ndarray

\end{description}\end{quote}

\sphinxstrong{See also:}

\begin{description}
\item[{{\hyperref[\detokenize{modules/tensor:symjax.tensor.power}]{\sphinxcrossref{\sphinxcode{\sphinxupquote{power()}}}}}}] \leavevmode
power function that preserves type

\end{description}

\subsubsection*{Examples}

Cube each element in a list.

\begin{sphinxVerbatim}[commandchars=\\\{\}]
\PYG{g+gp}{\PYGZgt{}\PYGZgt{}\PYGZgt{} }\PYG{n}{x1} \PYG{o}{=} \PYG{n+nb}{range}\PYG{p}{(}\PYG{l+m+mi}{6}\PYG{p}{)}
\PYG{g+gp}{\PYGZgt{}\PYGZgt{}\PYGZgt{} }\PYG{n}{x1}
\PYG{g+go}{[0, 1, 2, 3, 4, 5]}
\PYG{g+gp}{\PYGZgt{}\PYGZgt{}\PYGZgt{} }\PYG{n}{np}\PYG{o}{.}\PYG{n}{float\PYGZus{}power}\PYG{p}{(}\PYG{n}{x1}\PYG{p}{,} \PYG{l+m+mi}{3}\PYG{p}{)}
\PYG{g+go}{array([   0.,    1.,    8.,   27.,   64.,  125.])}
\end{sphinxVerbatim}

Raise the bases to different exponents.

\begin{sphinxVerbatim}[commandchars=\\\{\}]
\PYG{g+gp}{\PYGZgt{}\PYGZgt{}\PYGZgt{} }\PYG{n}{x2} \PYG{o}{=} \PYG{p}{[}\PYG{l+m+mf}{1.0}\PYG{p}{,} \PYG{l+m+mf}{2.0}\PYG{p}{,} \PYG{l+m+mf}{3.0}\PYG{p}{,} \PYG{l+m+mf}{3.0}\PYG{p}{,} \PYG{l+m+mf}{2.0}\PYG{p}{,} \PYG{l+m+mf}{1.0}\PYG{p}{]}
\PYG{g+gp}{\PYGZgt{}\PYGZgt{}\PYGZgt{} }\PYG{n}{np}\PYG{o}{.}\PYG{n}{float\PYGZus{}power}\PYG{p}{(}\PYG{n}{x1}\PYG{p}{,} \PYG{n}{x2}\PYG{p}{)}
\PYG{g+go}{array([  0.,   1.,   8.,  27.,  16.,   5.])}
\end{sphinxVerbatim}

The effect of broadcasting.

\begin{sphinxVerbatim}[commandchars=\\\{\}]
\PYG{g+gp}{\PYGZgt{}\PYGZgt{}\PYGZgt{} }\PYG{n}{x2} \PYG{o}{=} \PYG{n}{np}\PYG{o}{.}\PYG{n}{array}\PYG{p}{(}\PYG{p}{[}\PYG{p}{[}\PYG{l+m+mi}{1}\PYG{p}{,} \PYG{l+m+mi}{2}\PYG{p}{,} \PYG{l+m+mi}{3}\PYG{p}{,} \PYG{l+m+mi}{3}\PYG{p}{,} \PYG{l+m+mi}{2}\PYG{p}{,} \PYG{l+m+mi}{1}\PYG{p}{]}\PYG{p}{,} \PYG{p}{[}\PYG{l+m+mi}{1}\PYG{p}{,} \PYG{l+m+mi}{2}\PYG{p}{,} \PYG{l+m+mi}{3}\PYG{p}{,} \PYG{l+m+mi}{3}\PYG{p}{,} \PYG{l+m+mi}{2}\PYG{p}{,} \PYG{l+m+mi}{1}\PYG{p}{]}\PYG{p}{]}\PYG{p}{)}
\PYG{g+gp}{\PYGZgt{}\PYGZgt{}\PYGZgt{} }\PYG{n}{x2}
\PYG{g+go}{array([[1, 2, 3, 3, 2, 1],}
\PYG{g+go}{       [1, 2, 3, 3, 2, 1]])}
\PYG{g+gp}{\PYGZgt{}\PYGZgt{}\PYGZgt{} }\PYG{n}{np}\PYG{o}{.}\PYG{n}{float\PYGZus{}power}\PYG{p}{(}\PYG{n}{x1}\PYG{p}{,} \PYG{n}{x2}\PYG{p}{)}
\PYG{g+go}{array([[  0.,   1.,   8.,  27.,  16.,   5.],}
\PYG{g+go}{       [  0.,   1.,   8.,  27.,  16.,   5.]])}
\end{sphinxVerbatim}

\end{fulllineitems}

\index{floor() (in module symjax.tensor)@\spxentry{floor()}\spxextra{in module symjax.tensor}}

\begin{fulllineitems}
\phantomsection\label{\detokenize{modules/tensor:symjax.tensor.floor}}\pysiglinewithargsret{\sphinxbfcode{\sphinxupquote{floor}}}{\emph{\DUrole{n}{x}}}{}
Return the floor of the input, element\sphinxhyphen{}wise.

LAX\sphinxhyphen{}backend implementation of {\hyperref[\detokenize{modules/tensor:symjax.tensor.floor}]{\sphinxcrossref{\sphinxcode{\sphinxupquote{floor()}}}}}.
ADDITIONOriginal docstring below.

LAX\sphinxhyphen{}backend implementation of {\hyperref[\detokenize{modules/tensor:symjax.tensor.floor}]{\sphinxcrossref{\sphinxcode{\sphinxupquote{floor()}}}}}.
Original docstring below.

floor(x, /, out=None, {\color{red}\bfseries{}*}, where=True, casting=’same\_kind’, order=’K’, dtype=None, subok=True{[}, signature, extobj{]})

The floor of the scalar \sphinxtitleref{x} is the largest integer \sphinxtitleref{i}, such that
\sphinxtitleref{i \textless{}= x}.  It is often denoted as \(\lfloor x \rfloor\).
\begin{quote}\begin{description}
\item[{Returns}] \leavevmode
\sphinxstylestrong{y} \textendash{} The floor of each element in \sphinxtitleref{x}.
This is a scalar if \sphinxtitleref{x} is a scalar.

\item[{Return type}] \leavevmode
ndarray or scalar

\end{description}\end{quote}

\sphinxstrong{See also:}

{\hyperref[\detokenize{modules/tensor:symjax.tensor.ceil}]{\sphinxcrossref{\sphinxcode{\sphinxupquote{ceil()}}}}}, \sphinxcode{\sphinxupquote{trunc()}}, \sphinxcode{\sphinxupquote{rint()}}

\subsubsection*{Notes}

Some spreadsheet programs calculate the “floor\sphinxhyphen{}towards\sphinxhyphen{}zero”, in other
words \sphinxcode{\sphinxupquote{floor(\sphinxhyphen{}2.5) == \sphinxhyphen{}2}}.  NumPy instead uses the definition of
\sphinxtitleref{floor} where \sphinxtitleref{floor(\sphinxhyphen{}2.5) == \sphinxhyphen{}3}.
\subsubsection*{Examples}

\begin{sphinxVerbatim}[commandchars=\\\{\}]
\PYG{g+gp}{\PYGZgt{}\PYGZgt{}\PYGZgt{} }\PYG{n}{a} \PYG{o}{=} \PYG{n}{np}\PYG{o}{.}\PYG{n}{array}\PYG{p}{(}\PYG{p}{[}\PYG{o}{\PYGZhy{}}\PYG{l+m+mf}{1.7}\PYG{p}{,} \PYG{o}{\PYGZhy{}}\PYG{l+m+mf}{1.5}\PYG{p}{,} \PYG{o}{\PYGZhy{}}\PYG{l+m+mf}{0.2}\PYG{p}{,} \PYG{l+m+mf}{0.2}\PYG{p}{,} \PYG{l+m+mf}{1.5}\PYG{p}{,} \PYG{l+m+mf}{1.7}\PYG{p}{,} \PYG{l+m+mf}{2.0}\PYG{p}{]}\PYG{p}{)}
\PYG{g+gp}{\PYGZgt{}\PYGZgt{}\PYGZgt{} }\PYG{n}{np}\PYG{o}{.}\PYG{n}{floor}\PYG{p}{(}\PYG{n}{a}\PYG{p}{)}
\PYG{g+go}{array([\PYGZhy{}2., \PYGZhy{}2., \PYGZhy{}1.,  0.,  1.,  1.,  2.])}
\end{sphinxVerbatim}

\end{fulllineitems}

\index{floor\_divide() (in module symjax.tensor)@\spxentry{floor\_divide()}\spxextra{in module symjax.tensor}}

\begin{fulllineitems}
\phantomsection\label{\detokenize{modules/tensor:symjax.tensor.floor_divide}}\pysiglinewithargsret{\sphinxbfcode{\sphinxupquote{floor\_divide}}}{\emph{\DUrole{n}{x1}}, \emph{\DUrole{n}{x2}}}{}
Return the largest integer smaller or equal to the division of the inputs.
It is equivalent to the Python \sphinxcode{\sphinxupquote{//}} operator and pairs with the
Python \sphinxcode{\sphinxupquote{\%}} (\sphinxtitleref{remainder}), function so that \sphinxcode{\sphinxupquote{a = a \% b + b * (a // b)}}
up to roundoff.

LAX\sphinxhyphen{}backend implementation of {\hyperref[\detokenize{modules/tensor:symjax.tensor.floor_divide}]{\sphinxcrossref{\sphinxcode{\sphinxupquote{floor\_divide()}}}}}.
ADDITIONOriginal docstring below.

LAX\sphinxhyphen{}backend implementation of {\hyperref[\detokenize{modules/tensor:symjax.tensor.floor_divide}]{\sphinxcrossref{\sphinxcode{\sphinxupquote{floor\_divide()}}}}}.
Original docstring below.

floor\_divide(x1, x2, /, out=None, {\color{red}\bfseries{}*}, where=True, casting=’same\_kind’, order=’K’, dtype=None, subok=True{[}, signature, extobj{]})
\begin{quote}\begin{description}
\item[{Returns}] \leavevmode
\sphinxstylestrong{y} \textendash{} y = floor(\sphinxtitleref{x1}/\sphinxtitleref{x2})
This is a scalar if both \sphinxtitleref{x1} and \sphinxtitleref{x2} are scalars.

\item[{Return type}] \leavevmode
ndarray

\end{description}\end{quote}

\sphinxstrong{See also:}

\begin{description}
\item[{{\hyperref[\detokenize{modules/tensor:symjax.tensor.remainder}]{\sphinxcrossref{\sphinxcode{\sphinxupquote{remainder()}}}}}}] \leavevmode
Remainder complementary to floor\_divide.

\item[{{\hyperref[\detokenize{modules/tensor:symjax.tensor.divmod}]{\sphinxcrossref{\sphinxcode{\sphinxupquote{divmod()}}}}}}] \leavevmode
Simultaneous floor division and remainder.

\item[{{\hyperref[\detokenize{modules/tensor:symjax.tensor.divide}]{\sphinxcrossref{\sphinxcode{\sphinxupquote{divide()}}}}}}] \leavevmode
Standard division.

\item[{{\hyperref[\detokenize{modules/tensor:symjax.tensor.floor}]{\sphinxcrossref{\sphinxcode{\sphinxupquote{floor()}}}}}}] \leavevmode
Round a number to the nearest integer toward minus infinity.

\item[{{\hyperref[\detokenize{modules/tensor:symjax.tensor.ceil}]{\sphinxcrossref{\sphinxcode{\sphinxupquote{ceil()}}}}}}] \leavevmode
Round a number to the nearest integer toward infinity.

\end{description}

\subsubsection*{Examples}

\begin{sphinxVerbatim}[commandchars=\\\{\}]
\PYG{g+gp}{\PYGZgt{}\PYGZgt{}\PYGZgt{} }\PYG{n}{np}\PYG{o}{.}\PYG{n}{floor\PYGZus{}divide}\PYG{p}{(}\PYG{l+m+mi}{7}\PYG{p}{,}\PYG{l+m+mi}{3}\PYG{p}{)}
\PYG{g+go}{2}
\PYG{g+gp}{\PYGZgt{}\PYGZgt{}\PYGZgt{} }\PYG{n}{np}\PYG{o}{.}\PYG{n}{floor\PYGZus{}divide}\PYG{p}{(}\PYG{p}{[}\PYG{l+m+mf}{1.}\PYG{p}{,} \PYG{l+m+mf}{2.}\PYG{p}{,} \PYG{l+m+mf}{3.}\PYG{p}{,} \PYG{l+m+mf}{4.}\PYG{p}{]}\PYG{p}{,} \PYG{l+m+mf}{2.5}\PYG{p}{)}
\PYG{g+go}{array([ 0.,  0.,  1.,  1.])}
\end{sphinxVerbatim}

\end{fulllineitems}

\index{fmod() (in module symjax.tensor)@\spxentry{fmod()}\spxextra{in module symjax.tensor}}

\begin{fulllineitems}
\phantomsection\label{\detokenize{modules/tensor:symjax.tensor.fmod}}\pysiglinewithargsret{\sphinxbfcode{\sphinxupquote{fmod}}}{\emph{\DUrole{n}{x1}}, \emph{\DUrole{n}{x2}}}{}
Return the element\sphinxhyphen{}wise remainder of division.

LAX\sphinxhyphen{}backend implementation of {\hyperref[\detokenize{modules/tensor:symjax.tensor.fmod}]{\sphinxcrossref{\sphinxcode{\sphinxupquote{fmod()}}}}}.
ADDITIONOriginal docstring below.

LAX\sphinxhyphen{}backend implementation of {\hyperref[\detokenize{modules/tensor:symjax.tensor.fmod}]{\sphinxcrossref{\sphinxcode{\sphinxupquote{fmod()}}}}}.
Original docstring below.

fmod(x1, x2, /, out=None, {\color{red}\bfseries{}*}, where=True, casting=’same\_kind’, order=’K’, dtype=None, subok=True{[}, signature, extobj{]})

This is the NumPy implementation of the C library function fmod, the
remainder has the same sign as the dividend \sphinxtitleref{x1}. It is equivalent to
the Matlab(TM) \sphinxcode{\sphinxupquote{rem}} function and should not be confused with the
Python modulus operator \sphinxcode{\sphinxupquote{x1 \% x2}}.
\begin{quote}\begin{description}
\item[{Returns}] \leavevmode
\sphinxstylestrong{y} \textendash{} The remainder of the division of \sphinxtitleref{x1} by \sphinxtitleref{x2}.
This is a scalar if both \sphinxtitleref{x1} and \sphinxtitleref{x2} are scalars.

\item[{Return type}] \leavevmode
array\_like

\end{description}\end{quote}

\sphinxstrong{See also:}

\begin{description}
\item[{{\hyperref[\detokenize{modules/tensor:symjax.tensor.remainder}]{\sphinxcrossref{\sphinxcode{\sphinxupquote{remainder()}}}}}}] \leavevmode
Equivalent to the Python \sphinxcode{\sphinxupquote{\%}} operator.

\end{description}

{\hyperref[\detokenize{modules/tensor:symjax.tensor.divide}]{\sphinxcrossref{\sphinxcode{\sphinxupquote{divide()}}}}}

\subsubsection*{Notes}

The result of the modulo operation for negative dividend and divisors
is bound by conventions. For \sphinxtitleref{fmod}, the sign of result is the sign of
the dividend, while for \sphinxtitleref{remainder} the sign of the result is the sign
of the divisor. The \sphinxtitleref{fmod} function is equivalent to the Matlab(TM)
\sphinxcode{\sphinxupquote{rem}} function.
\subsubsection*{Examples}

\begin{sphinxVerbatim}[commandchars=\\\{\}]
\PYG{g+gp}{\PYGZgt{}\PYGZgt{}\PYGZgt{} }\PYG{n}{np}\PYG{o}{.}\PYG{n}{fmod}\PYG{p}{(}\PYG{p}{[}\PYG{o}{\PYGZhy{}}\PYG{l+m+mi}{3}\PYG{p}{,} \PYG{o}{\PYGZhy{}}\PYG{l+m+mi}{2}\PYG{p}{,} \PYG{o}{\PYGZhy{}}\PYG{l+m+mi}{1}\PYG{p}{,} \PYG{l+m+mi}{1}\PYG{p}{,} \PYG{l+m+mi}{2}\PYG{p}{,} \PYG{l+m+mi}{3}\PYG{p}{]}\PYG{p}{,} \PYG{l+m+mi}{2}\PYG{p}{)}
\PYG{g+go}{array([\PYGZhy{}1,  0, \PYGZhy{}1,  1,  0,  1])}
\PYG{g+gp}{\PYGZgt{}\PYGZgt{}\PYGZgt{} }\PYG{n}{np}\PYG{o}{.}\PYG{n}{remainder}\PYG{p}{(}\PYG{p}{[}\PYG{o}{\PYGZhy{}}\PYG{l+m+mi}{3}\PYG{p}{,} \PYG{o}{\PYGZhy{}}\PYG{l+m+mi}{2}\PYG{p}{,} \PYG{o}{\PYGZhy{}}\PYG{l+m+mi}{1}\PYG{p}{,} \PYG{l+m+mi}{1}\PYG{p}{,} \PYG{l+m+mi}{2}\PYG{p}{,} \PYG{l+m+mi}{3}\PYG{p}{]}\PYG{p}{,} \PYG{l+m+mi}{2}\PYG{p}{)}
\PYG{g+go}{array([1, 0, 1, 1, 0, 1])}
\end{sphinxVerbatim}

\begin{sphinxVerbatim}[commandchars=\\\{\}]
\PYG{g+gp}{\PYGZgt{}\PYGZgt{}\PYGZgt{} }\PYG{n}{np}\PYG{o}{.}\PYG{n}{fmod}\PYG{p}{(}\PYG{p}{[}\PYG{l+m+mi}{5}\PYG{p}{,} \PYG{l+m+mi}{3}\PYG{p}{]}\PYG{p}{,} \PYG{p}{[}\PYG{l+m+mi}{2}\PYG{p}{,} \PYG{l+m+mf}{2.}\PYG{p}{]}\PYG{p}{)}
\PYG{g+go}{array([ 1.,  1.])}
\PYG{g+gp}{\PYGZgt{}\PYGZgt{}\PYGZgt{} }\PYG{n}{a} \PYG{o}{=} \PYG{n}{np}\PYG{o}{.}\PYG{n}{arange}\PYG{p}{(}\PYG{o}{\PYGZhy{}}\PYG{l+m+mi}{3}\PYG{p}{,} \PYG{l+m+mi}{3}\PYG{p}{)}\PYG{o}{.}\PYG{n}{reshape}\PYG{p}{(}\PYG{l+m+mi}{3}\PYG{p}{,} \PYG{l+m+mi}{2}\PYG{p}{)}
\PYG{g+gp}{\PYGZgt{}\PYGZgt{}\PYGZgt{} }\PYG{n}{a}
\PYG{g+go}{array([[\PYGZhy{}3, \PYGZhy{}2],}
\PYG{g+go}{       [\PYGZhy{}1,  0],}
\PYG{g+go}{       [ 1,  2]])}
\PYG{g+gp}{\PYGZgt{}\PYGZgt{}\PYGZgt{} }\PYG{n}{np}\PYG{o}{.}\PYG{n}{fmod}\PYG{p}{(}\PYG{n}{a}\PYG{p}{,} \PYG{p}{[}\PYG{l+m+mi}{2}\PYG{p}{,}\PYG{l+m+mi}{2}\PYG{p}{]}\PYG{p}{)}
\PYG{g+go}{array([[\PYGZhy{}1,  0],}
\PYG{g+go}{       [\PYGZhy{}1,  0],}
\PYG{g+go}{       [ 1,  0]])}
\end{sphinxVerbatim}

\end{fulllineitems}

\index{full() (in module symjax.tensor)@\spxentry{full()}\spxextra{in module symjax.tensor}}

\begin{fulllineitems}
\phantomsection\label{\detokenize{modules/tensor:symjax.tensor.full}}\pysiglinewithargsret{\sphinxbfcode{\sphinxupquote{full}}}{\emph{\DUrole{n}{shape}}, \emph{\DUrole{n}{fill\_value}}, \emph{\DUrole{n}{dtype}\DUrole{o}{=}\DUrole{default_value}{None}}}{}
Return a new array of given shape and type, filled with \sphinxtitleref{fill\_value}.

LAX\sphinxhyphen{}backend implementation of {\hyperref[\detokenize{modules/tensor:symjax.tensor.full}]{\sphinxcrossref{\sphinxcode{\sphinxupquote{full()}}}}}.
ADDITIONOriginal docstring below.

LAX\sphinxhyphen{}backend implementation of {\hyperref[\detokenize{modules/tensor:symjax.tensor.full}]{\sphinxcrossref{\sphinxcode{\sphinxupquote{full()}}}}}.
Original docstring below.
\begin{quote}\begin{description}
\item[{Parameters}] \leavevmode\begin{itemize}
\item {} 
\sphinxstyleliteralstrong{\sphinxupquote{shape}} (\sphinxstyleliteralemphasis{\sphinxupquote{int}}\sphinxstyleliteralemphasis{\sphinxupquote{ or }}\sphinxstyleliteralemphasis{\sphinxupquote{sequence of ints}}) \textendash{} Shape of the new array, e.g., \sphinxcode{\sphinxupquote{(2, 3)}} or \sphinxcode{\sphinxupquote{2}}.

\item {} 
\sphinxstyleliteralstrong{\sphinxupquote{dtype}} (\sphinxstyleliteralemphasis{\sphinxupquote{data\sphinxhyphen{}type}}\sphinxstyleliteralemphasis{\sphinxupquote{, }}\sphinxstyleliteralemphasis{\sphinxupquote{optional}}) \textendash{} \begin{description}
\item[{The desired data\sphinxhyphen{}type for the array  The default, None, means}] \leavevmode
\sphinxtitleref{np.array(fill\_value).dtype}.

\end{description}

\end{itemize}

\item[{Returns}] \leavevmode
\sphinxstylestrong{out} \textendash{} Array of \sphinxtitleref{fill\_value} with the given shape, dtype, and order.

\item[{Return type}] \leavevmode
ndarray

\end{description}\end{quote}

\sphinxstrong{See also:}

\begin{description}
\item[{{\hyperref[\detokenize{modules/tensor:symjax.tensor.full_like}]{\sphinxcrossref{\sphinxcode{\sphinxupquote{full\_like()}}}}}}] \leavevmode
Return a new array with shape of input filled with value.

\item[{{\hyperref[\detokenize{modules/tensor:symjax.tensor.empty}]{\sphinxcrossref{\sphinxcode{\sphinxupquote{empty()}}}}}}] \leavevmode
Return a new uninitialized array.

\item[{{\hyperref[\detokenize{modules/tensor:symjax.tensor.ones}]{\sphinxcrossref{\sphinxcode{\sphinxupquote{ones()}}}}}}] \leavevmode
Return a new array setting values to one.

\item[{{\hyperref[\detokenize{modules/tensor:symjax.tensor.zeros}]{\sphinxcrossref{\sphinxcode{\sphinxupquote{zeros()}}}}}}] \leavevmode
Return a new array setting values to zero.

\end{description}

\subsubsection*{Examples}

\begin{sphinxVerbatim}[commandchars=\\\{\}]
\PYG{g+gp}{\PYGZgt{}\PYGZgt{}\PYGZgt{} }\PYG{n}{np}\PYG{o}{.}\PYG{n}{full}\PYG{p}{(}\PYG{p}{(}\PYG{l+m+mi}{2}\PYG{p}{,} \PYG{l+m+mi}{2}\PYG{p}{)}\PYG{p}{,} \PYG{n}{np}\PYG{o}{.}\PYG{n}{inf}\PYG{p}{)}
\PYG{g+go}{array([[inf, inf],}
\PYG{g+go}{       [inf, inf]])}
\PYG{g+gp}{\PYGZgt{}\PYGZgt{}\PYGZgt{} }\PYG{n}{np}\PYG{o}{.}\PYG{n}{full}\PYG{p}{(}\PYG{p}{(}\PYG{l+m+mi}{2}\PYG{p}{,} \PYG{l+m+mi}{2}\PYG{p}{)}\PYG{p}{,} \PYG{l+m+mi}{10}\PYG{p}{)}
\PYG{g+go}{array([[10, 10],}
\PYG{g+go}{       [10, 10]])}
\end{sphinxVerbatim}

\end{fulllineitems}

\index{full\_like() (in module symjax.tensor)@\spxentry{full\_like()}\spxextra{in module symjax.tensor}}

\begin{fulllineitems}
\phantomsection\label{\detokenize{modules/tensor:symjax.tensor.full_like}}\pysiglinewithargsret{\sphinxbfcode{\sphinxupquote{full\_like}}}{\emph{\DUrole{n}{a}}, \emph{\DUrole{n}{fill\_value}}, \emph{\DUrole{n}{dtype}\DUrole{o}{=}\DUrole{default_value}{None}}}{}
Return a full array with the same shape and type as a given array.

LAX\sphinxhyphen{}backend implementation of {\hyperref[\detokenize{modules/tensor:symjax.tensor.full_like}]{\sphinxcrossref{\sphinxcode{\sphinxupquote{full\_like()}}}}}.
ADDITIONOriginal docstring below.

LAX\sphinxhyphen{}backend implementation of {\hyperref[\detokenize{modules/tensor:symjax.tensor.full_like}]{\sphinxcrossref{\sphinxcode{\sphinxupquote{full\_like()}}}}}.
Original docstring below.
\begin{quote}\begin{description}
\item[{Parameters}] \leavevmode
\sphinxstyleliteralstrong{\sphinxupquote{dtype}} (\sphinxstyleliteralemphasis{\sphinxupquote{data\sphinxhyphen{}type}}\sphinxstyleliteralemphasis{\sphinxupquote{, }}\sphinxstyleliteralemphasis{\sphinxupquote{optional}}) \textendash{} Overrides the data type of the result.

\item[{Returns}] \leavevmode
\sphinxstylestrong{out} \textendash{} Array of \sphinxtitleref{fill\_value} with the same shape and type as \sphinxtitleref{a}.

\item[{Return type}] \leavevmode
ndarray

\end{description}\end{quote}

\sphinxstrong{See also:}

\begin{description}
\item[{{\hyperref[\detokenize{modules/tensor:symjax.tensor.empty_like}]{\sphinxcrossref{\sphinxcode{\sphinxupquote{empty\_like()}}}}}}] \leavevmode
Return an empty array with shape and type of input.

\item[{{\hyperref[\detokenize{modules/tensor:symjax.tensor.ones_like}]{\sphinxcrossref{\sphinxcode{\sphinxupquote{ones\_like()}}}}}}] \leavevmode
Return an array of ones with shape and type of input.

\item[{{\hyperref[\detokenize{modules/tensor:symjax.tensor.zeros_like}]{\sphinxcrossref{\sphinxcode{\sphinxupquote{zeros\_like()}}}}}}] \leavevmode
Return an array of zeros with shape and type of input.

\item[{{\hyperref[\detokenize{modules/tensor:symjax.tensor.full}]{\sphinxcrossref{\sphinxcode{\sphinxupquote{full()}}}}}}] \leavevmode
Return a new array of given shape filled with value.

\end{description}

\subsubsection*{Examples}

\begin{sphinxVerbatim}[commandchars=\\\{\}]
\PYG{g+gp}{\PYGZgt{}\PYGZgt{}\PYGZgt{} }\PYG{n}{x} \PYG{o}{=} \PYG{n}{np}\PYG{o}{.}\PYG{n}{arange}\PYG{p}{(}\PYG{l+m+mi}{6}\PYG{p}{,} \PYG{n}{dtype}\PYG{o}{=}\PYG{n+nb}{int}\PYG{p}{)}
\PYG{g+gp}{\PYGZgt{}\PYGZgt{}\PYGZgt{} }\PYG{n}{np}\PYG{o}{.}\PYG{n}{full\PYGZus{}like}\PYG{p}{(}\PYG{n}{x}\PYG{p}{,} \PYG{l+m+mi}{1}\PYG{p}{)}
\PYG{g+go}{array([1, 1, 1, 1, 1, 1])}
\PYG{g+gp}{\PYGZgt{}\PYGZgt{}\PYGZgt{} }\PYG{n}{np}\PYG{o}{.}\PYG{n}{full\PYGZus{}like}\PYG{p}{(}\PYG{n}{x}\PYG{p}{,} \PYG{l+m+mf}{0.1}\PYG{p}{)}
\PYG{g+go}{array([0, 0, 0, 0, 0, 0])}
\PYG{g+gp}{\PYGZgt{}\PYGZgt{}\PYGZgt{} }\PYG{n}{np}\PYG{o}{.}\PYG{n}{full\PYGZus{}like}\PYG{p}{(}\PYG{n}{x}\PYG{p}{,} \PYG{l+m+mf}{0.1}\PYG{p}{,} \PYG{n}{dtype}\PYG{o}{=}\PYG{n}{np}\PYG{o}{.}\PYG{n}{double}\PYG{p}{)}
\PYG{g+go}{array([0.1, 0.1, 0.1, 0.1, 0.1, 0.1])}
\PYG{g+gp}{\PYGZgt{}\PYGZgt{}\PYGZgt{} }\PYG{n}{np}\PYG{o}{.}\PYG{n}{full\PYGZus{}like}\PYG{p}{(}\PYG{n}{x}\PYG{p}{,} \PYG{n}{np}\PYG{o}{.}\PYG{n}{nan}\PYG{p}{,} \PYG{n}{dtype}\PYG{o}{=}\PYG{n}{np}\PYG{o}{.}\PYG{n}{double}\PYG{p}{)}
\PYG{g+go}{array([nan, nan, nan, nan, nan, nan])}
\end{sphinxVerbatim}

\begin{sphinxVerbatim}[commandchars=\\\{\}]
\PYG{g+gp}{\PYGZgt{}\PYGZgt{}\PYGZgt{} }\PYG{n}{y} \PYG{o}{=} \PYG{n}{np}\PYG{o}{.}\PYG{n}{arange}\PYG{p}{(}\PYG{l+m+mi}{6}\PYG{p}{,} \PYG{n}{dtype}\PYG{o}{=}\PYG{n}{np}\PYG{o}{.}\PYG{n}{double}\PYG{p}{)}
\PYG{g+gp}{\PYGZgt{}\PYGZgt{}\PYGZgt{} }\PYG{n}{np}\PYG{o}{.}\PYG{n}{full\PYGZus{}like}\PYG{p}{(}\PYG{n}{y}\PYG{p}{,} \PYG{l+m+mf}{0.1}\PYG{p}{)}
\PYG{g+go}{array([0.1,  0.1,  0.1,  0.1,  0.1,  0.1])}
\end{sphinxVerbatim}

\end{fulllineitems}

\index{gcd() (in module symjax.tensor)@\spxentry{gcd()}\spxextra{in module symjax.tensor}}

\begin{fulllineitems}
\phantomsection\label{\detokenize{modules/tensor:symjax.tensor.gcd}}\pysiglinewithargsret{\sphinxbfcode{\sphinxupquote{gcd}}}{\emph{\DUrole{n}{x1}}, \emph{\DUrole{n}{x2}}}{}
Returns the greatest common divisor of \sphinxcode{\sphinxupquote{|x1|}} and \sphinxcode{\sphinxupquote{|x2|}}

LAX\sphinxhyphen{}backend implementation of {\hyperref[\detokenize{modules/tensor:symjax.tensor.gcd}]{\sphinxcrossref{\sphinxcode{\sphinxupquote{gcd()}}}}}.
ADDITIONOriginal docstring below.

LAX\sphinxhyphen{}backend implementation of {\hyperref[\detokenize{modules/tensor:symjax.tensor.gcd}]{\sphinxcrossref{\sphinxcode{\sphinxupquote{gcd()}}}}}.
Original docstring below.

gcd(x1, x2, /, out=None, {\color{red}\bfseries{}*}, where=True, casting=’same\_kind’, order=’K’, dtype=None, subok=True{[}, signature, extobj{]})
\begin{quote}\begin{description}
\item[{Returns}] \leavevmode
\sphinxstylestrong{y} \textendash{} The greatest common divisor of the absolute value of the inputs
This is a scalar if both \sphinxtitleref{x1} and \sphinxtitleref{x2} are scalars.

\item[{Return type}] \leavevmode
ndarray or scalar

\end{description}\end{quote}

\sphinxstrong{See also:}

\begin{description}
\item[{{\hyperref[\detokenize{modules/tensor:symjax.tensor.lcm}]{\sphinxcrossref{\sphinxcode{\sphinxupquote{lcm()}}}}}}] \leavevmode
The lowest common multiple

\end{description}

\subsubsection*{Examples}

\begin{sphinxVerbatim}[commandchars=\\\{\}]
\PYG{g+gp}{\PYGZgt{}\PYGZgt{}\PYGZgt{} }\PYG{n}{np}\PYG{o}{.}\PYG{n}{gcd}\PYG{p}{(}\PYG{l+m+mi}{12}\PYG{p}{,} \PYG{l+m+mi}{20}\PYG{p}{)}
\PYG{g+go}{4}
\PYG{g+gp}{\PYGZgt{}\PYGZgt{}\PYGZgt{} }\PYG{n}{np}\PYG{o}{.}\PYG{n}{gcd}\PYG{o}{.}\PYG{n}{reduce}\PYG{p}{(}\PYG{p}{[}\PYG{l+m+mi}{15}\PYG{p}{,} \PYG{l+m+mi}{25}\PYG{p}{,} \PYG{l+m+mi}{35}\PYG{p}{]}\PYG{p}{)}
\PYG{g+go}{5}
\PYG{g+gp}{\PYGZgt{}\PYGZgt{}\PYGZgt{} }\PYG{n}{np}\PYG{o}{.}\PYG{n}{gcd}\PYG{p}{(}\PYG{n}{np}\PYG{o}{.}\PYG{n}{arange}\PYG{p}{(}\PYG{l+m+mi}{6}\PYG{p}{)}\PYG{p}{,} \PYG{l+m+mi}{20}\PYG{p}{)}
\PYG{g+go}{array([20,  1,  2,  1,  4,  5])}
\end{sphinxVerbatim}

\end{fulllineitems}

\index{geomspace() (in module symjax.tensor)@\spxentry{geomspace()}\spxextra{in module symjax.tensor}}

\begin{fulllineitems}
\phantomsection\label{\detokenize{modules/tensor:symjax.tensor.geomspace}}\pysiglinewithargsret{\sphinxbfcode{\sphinxupquote{geomspace}}}{\emph{\DUrole{n}{start}}, \emph{\DUrole{n}{stop}}, \emph{\DUrole{n}{num}\DUrole{o}{=}\DUrole{default_value}{50}}, \emph{\DUrole{n}{endpoint}\DUrole{o}{=}\DUrole{default_value}{True}}, \emph{\DUrole{n}{dtype}\DUrole{o}{=}\DUrole{default_value}{None}}, \emph{\DUrole{n}{axis}\DUrole{o}{=}\DUrole{default_value}{0}}}{}
Return numbers spaced evenly on a log scale (a geometric progression).

LAX\sphinxhyphen{}backend implementation of {\hyperref[\detokenize{modules/tensor:symjax.tensor.geomspace}]{\sphinxcrossref{\sphinxcode{\sphinxupquote{geomspace()}}}}}.
ADDITIONOriginal docstring below.

LAX\sphinxhyphen{}backend implementation of {\hyperref[\detokenize{modules/tensor:symjax.tensor.geomspace}]{\sphinxcrossref{\sphinxcode{\sphinxupquote{geomspace()}}}}}.
Original docstring below.

This is similar to \sphinxtitleref{logspace}, but with endpoints specified directly.
Each output sample is a constant multiple of the previous.

\DUrole{versionmodified,changed}{Changed in version 1.16.0: }Non\sphinxhyphen{}scalar \sphinxtitleref{start} and \sphinxtitleref{stop} are now supported.
\begin{quote}\begin{description}
\item[{Parameters}] \leavevmode
\sphinxstyleliteralstrong{\sphinxupquote{dtype}} (\sphinxstyleliteralemphasis{\sphinxupquote{dtype}}) \textendash{} The type of the output array.  If \sphinxtitleref{dtype} is not given, infer the data
type from the other input arguments.

\item[{Returns}] \leavevmode
\sphinxstylestrong{samples} \textendash{} \sphinxtitleref{num} samples, equally spaced on a log scale.

\item[{Return type}] \leavevmode
ndarray

\end{description}\end{quote}

\sphinxstrong{See also:}

\begin{description}
\item[{{\hyperref[\detokenize{modules/tensor:symjax.tensor.logspace}]{\sphinxcrossref{\sphinxcode{\sphinxupquote{logspace()}}}}}}] \leavevmode
Similar to geomspace, but with endpoints specified using log and base.

\item[{{\hyperref[\detokenize{modules/tensor:symjax.tensor.linspace}]{\sphinxcrossref{\sphinxcode{\sphinxupquote{linspace()}}}}}}] \leavevmode
Similar to geomspace, but with arithmetic instead of geometric progression.

\item[{{\hyperref[\detokenize{modules/tensor:symjax.tensor.arange}]{\sphinxcrossref{\sphinxcode{\sphinxupquote{arange()}}}}}}] \leavevmode
Similar to linspace, with the step size specified instead of the number of samples.

\end{description}

\subsubsection*{Notes}

If the inputs or dtype are complex, the output will follow a logarithmic
spiral in the complex plane.  (There are an infinite number of spirals
passing through two points; the output will follow the shortest such path.)
\subsubsection*{Examples}

\begin{sphinxVerbatim}[commandchars=\\\{\}]
\PYG{g+gp}{\PYGZgt{}\PYGZgt{}\PYGZgt{} }\PYG{n}{np}\PYG{o}{.}\PYG{n}{geomspace}\PYG{p}{(}\PYG{l+m+mi}{1}\PYG{p}{,} \PYG{l+m+mi}{1000}\PYG{p}{,} \PYG{n}{num}\PYG{o}{=}\PYG{l+m+mi}{4}\PYG{p}{)}
\PYG{g+go}{array([    1.,    10.,   100.,  1000.])}
\PYG{g+gp}{\PYGZgt{}\PYGZgt{}\PYGZgt{} }\PYG{n}{np}\PYG{o}{.}\PYG{n}{geomspace}\PYG{p}{(}\PYG{l+m+mi}{1}\PYG{p}{,} \PYG{l+m+mi}{1000}\PYG{p}{,} \PYG{n}{num}\PYG{o}{=}\PYG{l+m+mi}{3}\PYG{p}{,} \PYG{n}{endpoint}\PYG{o}{=}\PYG{k+kc}{False}\PYG{p}{)}
\PYG{g+go}{array([   1.,   10.,  100.])}
\PYG{g+gp}{\PYGZgt{}\PYGZgt{}\PYGZgt{} }\PYG{n}{np}\PYG{o}{.}\PYG{n}{geomspace}\PYG{p}{(}\PYG{l+m+mi}{1}\PYG{p}{,} \PYG{l+m+mi}{1000}\PYG{p}{,} \PYG{n}{num}\PYG{o}{=}\PYG{l+m+mi}{4}\PYG{p}{,} \PYG{n}{endpoint}\PYG{o}{=}\PYG{k+kc}{False}\PYG{p}{)}
\PYG{g+go}{array([   1.        ,    5.62341325,   31.6227766 ,  177.827941  ])}
\PYG{g+gp}{\PYGZgt{}\PYGZgt{}\PYGZgt{} }\PYG{n}{np}\PYG{o}{.}\PYG{n}{geomspace}\PYG{p}{(}\PYG{l+m+mi}{1}\PYG{p}{,} \PYG{l+m+mi}{256}\PYG{p}{,} \PYG{n}{num}\PYG{o}{=}\PYG{l+m+mi}{9}\PYG{p}{)}
\PYG{g+go}{array([   1.,    2.,    4.,    8.,   16.,   32.,   64.,  128.,  256.])}
\end{sphinxVerbatim}

Note that the above may not produce exact integers:

\begin{sphinxVerbatim}[commandchars=\\\{\}]
\PYG{g+gp}{\PYGZgt{}\PYGZgt{}\PYGZgt{} }\PYG{n}{np}\PYG{o}{.}\PYG{n}{geomspace}\PYG{p}{(}\PYG{l+m+mi}{1}\PYG{p}{,} \PYG{l+m+mi}{256}\PYG{p}{,} \PYG{n}{num}\PYG{o}{=}\PYG{l+m+mi}{9}\PYG{p}{,} \PYG{n}{dtype}\PYG{o}{=}\PYG{n+nb}{int}\PYG{p}{)}
\PYG{g+go}{array([  1,   2,   4,   7,  16,  32,  63, 127, 256])}
\PYG{g+gp}{\PYGZgt{}\PYGZgt{}\PYGZgt{} }\PYG{n}{np}\PYG{o}{.}\PYG{n}{around}\PYG{p}{(}\PYG{n}{np}\PYG{o}{.}\PYG{n}{geomspace}\PYG{p}{(}\PYG{l+m+mi}{1}\PYG{p}{,} \PYG{l+m+mi}{256}\PYG{p}{,} \PYG{n}{num}\PYG{o}{=}\PYG{l+m+mi}{9}\PYG{p}{)}\PYG{p}{)}\PYG{o}{.}\PYG{n}{astype}\PYG{p}{(}\PYG{n+nb}{int}\PYG{p}{)}
\PYG{g+go}{array([  1,   2,   4,   8,  16,  32,  64, 128, 256])}
\end{sphinxVerbatim}

Negative, decreasing, and complex inputs are allowed:

\begin{sphinxVerbatim}[commandchars=\\\{\}]
\PYG{g+gp}{\PYGZgt{}\PYGZgt{}\PYGZgt{} }\PYG{n}{np}\PYG{o}{.}\PYG{n}{geomspace}\PYG{p}{(}\PYG{l+m+mi}{1000}\PYG{p}{,} \PYG{l+m+mi}{1}\PYG{p}{,} \PYG{n}{num}\PYG{o}{=}\PYG{l+m+mi}{4}\PYG{p}{)}
\PYG{g+go}{array([1000.,  100.,   10.,    1.])}
\PYG{g+gp}{\PYGZgt{}\PYGZgt{}\PYGZgt{} }\PYG{n}{np}\PYG{o}{.}\PYG{n}{geomspace}\PYG{p}{(}\PYG{o}{\PYGZhy{}}\PYG{l+m+mi}{1000}\PYG{p}{,} \PYG{o}{\PYGZhy{}}\PYG{l+m+mi}{1}\PYG{p}{,} \PYG{n}{num}\PYG{o}{=}\PYG{l+m+mi}{4}\PYG{p}{)}
\PYG{g+go}{array([\PYGZhy{}1000.,  \PYGZhy{}100.,   \PYGZhy{}10.,    \PYGZhy{}1.])}
\PYG{g+gp}{\PYGZgt{}\PYGZgt{}\PYGZgt{} }\PYG{n}{np}\PYG{o}{.}\PYG{n}{geomspace}\PYG{p}{(}\PYG{l+m+mi}{1}\PYG{n}{j}\PYG{p}{,} \PYG{l+m+mi}{1000}\PYG{n}{j}\PYG{p}{,} \PYG{n}{num}\PYG{o}{=}\PYG{l+m+mi}{4}\PYG{p}{)}  \PYG{c+c1}{\PYGZsh{} Straight line}
\PYG{g+go}{array([0.   +1.j, 0.  +10.j, 0. +100.j, 0.+1000.j])}
\PYG{g+gp}{\PYGZgt{}\PYGZgt{}\PYGZgt{} }\PYG{n}{np}\PYG{o}{.}\PYG{n}{geomspace}\PYG{p}{(}\PYG{o}{\PYGZhy{}}\PYG{l+m+mi}{1}\PYG{o}{+}\PYG{l+m+mi}{0}\PYG{n}{j}\PYG{p}{,} \PYG{l+m+mi}{1}\PYG{o}{+}\PYG{l+m+mi}{0}\PYG{n}{j}\PYG{p}{,} \PYG{n}{num}\PYG{o}{=}\PYG{l+m+mi}{5}\PYG{p}{)}  \PYG{c+c1}{\PYGZsh{} Circle}
\PYG{g+go}{array([\PYGZhy{}1.00000000e+00+1.22464680e\PYGZhy{}16j, \PYGZhy{}7.07106781e\PYGZhy{}01+7.07106781e\PYGZhy{}01j,}
\PYG{g+go}{        6.12323400e\PYGZhy{}17+1.00000000e+00j,  7.07106781e\PYGZhy{}01+7.07106781e\PYGZhy{}01j,}
\PYG{g+go}{        1.00000000e+00+0.00000000e+00j])}
\end{sphinxVerbatim}

Graphical illustration of \sphinxcode{\sphinxupquote{endpoint}} parameter:

\begin{sphinxVerbatim}[commandchars=\\\{\}]
\PYG{g+gp}{\PYGZgt{}\PYGZgt{}\PYGZgt{} }\PYG{k+kn}{import} \PYG{n+nn}{matplotlib}\PYG{n+nn}{.}\PYG{n+nn}{pyplot} \PYG{k}{as} \PYG{n+nn}{plt}
\PYG{g+gp}{\PYGZgt{}\PYGZgt{}\PYGZgt{} }\PYG{n}{N} \PYG{o}{=} \PYG{l+m+mi}{10}
\PYG{g+gp}{\PYGZgt{}\PYGZgt{}\PYGZgt{} }\PYG{n}{y} \PYG{o}{=} \PYG{n}{np}\PYG{o}{.}\PYG{n}{zeros}\PYG{p}{(}\PYG{n}{N}\PYG{p}{)}
\PYG{g+gp}{\PYGZgt{}\PYGZgt{}\PYGZgt{} }\PYG{n}{plt}\PYG{o}{.}\PYG{n}{semilogx}\PYG{p}{(}\PYG{n}{np}\PYG{o}{.}\PYG{n}{geomspace}\PYG{p}{(}\PYG{l+m+mi}{1}\PYG{p}{,} \PYG{l+m+mi}{1000}\PYG{p}{,} \PYG{n}{N}\PYG{p}{,} \PYG{n}{endpoint}\PYG{o}{=}\PYG{k+kc}{True}\PYG{p}{)}\PYG{p}{,} \PYG{n}{y} \PYG{o}{+} \PYG{l+m+mi}{1}\PYG{p}{,} \PYG{l+s+s1}{\PYGZsq{}}\PYG{l+s+s1}{o}\PYG{l+s+s1}{\PYGZsq{}}\PYG{p}{)}
\PYG{g+go}{[\PYGZlt{}matplotlib.lines.Line2D object at 0x...\PYGZgt{}]}
\PYG{g+gp}{\PYGZgt{}\PYGZgt{}\PYGZgt{} }\PYG{n}{plt}\PYG{o}{.}\PYG{n}{semilogx}\PYG{p}{(}\PYG{n}{np}\PYG{o}{.}\PYG{n}{geomspace}\PYG{p}{(}\PYG{l+m+mi}{1}\PYG{p}{,} \PYG{l+m+mi}{1000}\PYG{p}{,} \PYG{n}{N}\PYG{p}{,} \PYG{n}{endpoint}\PYG{o}{=}\PYG{k+kc}{False}\PYG{p}{)}\PYG{p}{,} \PYG{n}{y} \PYG{o}{+} \PYG{l+m+mi}{2}\PYG{p}{,} \PYG{l+s+s1}{\PYGZsq{}}\PYG{l+s+s1}{o}\PYG{l+s+s1}{\PYGZsq{}}\PYG{p}{)}
\PYG{g+go}{[\PYGZlt{}matplotlib.lines.Line2D object at 0x...\PYGZgt{}]}
\PYG{g+gp}{\PYGZgt{}\PYGZgt{}\PYGZgt{} }\PYG{n}{plt}\PYG{o}{.}\PYG{n}{axis}\PYG{p}{(}\PYG{p}{[}\PYG{l+m+mf}{0.5}\PYG{p}{,} \PYG{l+m+mi}{2000}\PYG{p}{,} \PYG{l+m+mi}{0}\PYG{p}{,} \PYG{l+m+mi}{3}\PYG{p}{]}\PYG{p}{)}
\PYG{g+go}{[0.5, 2000, 0, 3]}
\PYG{g+gp}{\PYGZgt{}\PYGZgt{}\PYGZgt{} }\PYG{n}{plt}\PYG{o}{.}\PYG{n}{grid}\PYG{p}{(}\PYG{k+kc}{True}\PYG{p}{,} \PYG{n}{color}\PYG{o}{=}\PYG{l+s+s1}{\PYGZsq{}}\PYG{l+s+s1}{0.7}\PYG{l+s+s1}{\PYGZsq{}}\PYG{p}{,} \PYG{n}{linestyle}\PYG{o}{=}\PYG{l+s+s1}{\PYGZsq{}}\PYG{l+s+s1}{\PYGZhy{}}\PYG{l+s+s1}{\PYGZsq{}}\PYG{p}{,} \PYG{n}{which}\PYG{o}{=}\PYG{l+s+s1}{\PYGZsq{}}\PYG{l+s+s1}{both}\PYG{l+s+s1}{\PYGZsq{}}\PYG{p}{,} \PYG{n}{axis}\PYG{o}{=}\PYG{l+s+s1}{\PYGZsq{}}\PYG{l+s+s1}{both}\PYG{l+s+s1}{\PYGZsq{}}\PYG{p}{)}
\PYG{g+gp}{\PYGZgt{}\PYGZgt{}\PYGZgt{} }\PYG{n}{plt}\PYG{o}{.}\PYG{n}{show}\PYG{p}{(}\PYG{p}{)}
\end{sphinxVerbatim}

\end{fulllineitems}

\index{greater() (in module symjax.tensor)@\spxentry{greater()}\spxextra{in module symjax.tensor}}

\begin{fulllineitems}
\phantomsection\label{\detokenize{modules/tensor:symjax.tensor.greater}}\pysiglinewithargsret{\sphinxbfcode{\sphinxupquote{greater}}}{\emph{\DUrole{n}{x1}}, \emph{\DUrole{n}{x2}}}{}
Return the truth value of (x1 \textgreater{} x2) element\sphinxhyphen{}wise.

LAX\sphinxhyphen{}backend implementation of {\hyperref[\detokenize{modules/tensor:symjax.tensor.greater}]{\sphinxcrossref{\sphinxcode{\sphinxupquote{greater()}}}}}.
ADDITIONOriginal docstring below.

LAX\sphinxhyphen{}backend implementation of {\hyperref[\detokenize{modules/tensor:symjax.tensor.greater}]{\sphinxcrossref{\sphinxcode{\sphinxupquote{greater()}}}}}.
Original docstring below.

greater(x1, x2, /, out=None, {\color{red}\bfseries{}*}, where=True, casting=’same\_kind’, order=’K’, dtype=None, subok=True{[}, signature, extobj{]})
\begin{quote}\begin{description}
\item[{Returns}] \leavevmode
\sphinxstylestrong{out} \textendash{} Output array, element\sphinxhyphen{}wise comparison of \sphinxtitleref{x1} and \sphinxtitleref{x2}.
Typically of type bool, unless \sphinxcode{\sphinxupquote{dtype=object}} is passed.
This is a scalar if both \sphinxtitleref{x1} and \sphinxtitleref{x2} are scalars.

\item[{Return type}] \leavevmode
ndarray or scalar

\end{description}\end{quote}

\sphinxstrong{See also:}

{\hyperref[\detokenize{modules/tensor:symjax.tensor.greater_equal}]{\sphinxcrossref{\sphinxcode{\sphinxupquote{greater\_equal()}}}}}, {\hyperref[\detokenize{modules/tensor:symjax.tensor.less}]{\sphinxcrossref{\sphinxcode{\sphinxupquote{less()}}}}}, {\hyperref[\detokenize{modules/tensor:symjax.tensor.less_equal}]{\sphinxcrossref{\sphinxcode{\sphinxupquote{less\_equal()}}}}}, {\hyperref[\detokenize{modules/tensor:symjax.tensor.equal}]{\sphinxcrossref{\sphinxcode{\sphinxupquote{equal()}}}}}, {\hyperref[\detokenize{modules/tensor:symjax.tensor.not_equal}]{\sphinxcrossref{\sphinxcode{\sphinxupquote{not\_equal()}}}}}

\subsubsection*{Examples}

\begin{sphinxVerbatim}[commandchars=\\\{\}]
\PYG{g+gp}{\PYGZgt{}\PYGZgt{}\PYGZgt{} }\PYG{n}{np}\PYG{o}{.}\PYG{n}{greater}\PYG{p}{(}\PYG{p}{[}\PYG{l+m+mi}{4}\PYG{p}{,}\PYG{l+m+mi}{2}\PYG{p}{]}\PYG{p}{,}\PYG{p}{[}\PYG{l+m+mi}{2}\PYG{p}{,}\PYG{l+m+mi}{2}\PYG{p}{]}\PYG{p}{)}
\PYG{g+go}{array([ True, False])}
\end{sphinxVerbatim}

If the inputs are ndarrays, then np.greater is equivalent to ‘\textgreater{}’.

\begin{sphinxVerbatim}[commandchars=\\\{\}]
\PYG{g+gp}{\PYGZgt{}\PYGZgt{}\PYGZgt{} }\PYG{n}{a} \PYG{o}{=} \PYG{n}{np}\PYG{o}{.}\PYG{n}{array}\PYG{p}{(}\PYG{p}{[}\PYG{l+m+mi}{4}\PYG{p}{,}\PYG{l+m+mi}{2}\PYG{p}{]}\PYG{p}{)}
\PYG{g+gp}{\PYGZgt{}\PYGZgt{}\PYGZgt{} }\PYG{n}{b} \PYG{o}{=} \PYG{n}{np}\PYG{o}{.}\PYG{n}{array}\PYG{p}{(}\PYG{p}{[}\PYG{l+m+mi}{2}\PYG{p}{,}\PYG{l+m+mi}{2}\PYG{p}{]}\PYG{p}{)}
\PYG{g+gp}{\PYGZgt{}\PYGZgt{}\PYGZgt{} }\PYG{n}{a} \PYG{o}{\PYGZgt{}} \PYG{n}{b}
\PYG{g+go}{array([ True, False])}
\end{sphinxVerbatim}

\end{fulllineitems}

\index{greater\_equal() (in module symjax.tensor)@\spxentry{greater\_equal()}\spxextra{in module symjax.tensor}}

\begin{fulllineitems}
\phantomsection\label{\detokenize{modules/tensor:symjax.tensor.greater_equal}}\pysiglinewithargsret{\sphinxbfcode{\sphinxupquote{greater\_equal}}}{\emph{\DUrole{n}{x1}}, \emph{\DUrole{n}{x2}}}{}
Return the truth value of (x1 \textgreater{}= x2) element\sphinxhyphen{}wise.

LAX\sphinxhyphen{}backend implementation of {\hyperref[\detokenize{modules/tensor:symjax.tensor.greater_equal}]{\sphinxcrossref{\sphinxcode{\sphinxupquote{greater\_equal()}}}}}.
ADDITIONOriginal docstring below.

LAX\sphinxhyphen{}backend implementation of {\hyperref[\detokenize{modules/tensor:symjax.tensor.greater_equal}]{\sphinxcrossref{\sphinxcode{\sphinxupquote{greater\_equal()}}}}}.
Original docstring below.

greater\_equal(x1, x2, /, out=None, {\color{red}\bfseries{}*}, where=True, casting=’same\_kind’, order=’K’, dtype=None, subok=True{[}, signature, extobj{]})
\begin{quote}\begin{description}
\item[{Returns}] \leavevmode
\sphinxstylestrong{out} \textendash{} Output array, element\sphinxhyphen{}wise comparison of \sphinxtitleref{x1} and \sphinxtitleref{x2}.
Typically of type bool, unless \sphinxcode{\sphinxupquote{dtype=object}} is passed.
This is a scalar if both \sphinxtitleref{x1} and \sphinxtitleref{x2} are scalars.

\item[{Return type}] \leavevmode
bool or ndarray of bool

\end{description}\end{quote}

\sphinxstrong{See also:}

{\hyperref[\detokenize{modules/tensor:symjax.tensor.greater}]{\sphinxcrossref{\sphinxcode{\sphinxupquote{greater()}}}}}, {\hyperref[\detokenize{modules/tensor:symjax.tensor.less}]{\sphinxcrossref{\sphinxcode{\sphinxupquote{less()}}}}}, {\hyperref[\detokenize{modules/tensor:symjax.tensor.less_equal}]{\sphinxcrossref{\sphinxcode{\sphinxupquote{less\_equal()}}}}}, {\hyperref[\detokenize{modules/tensor:symjax.tensor.equal}]{\sphinxcrossref{\sphinxcode{\sphinxupquote{equal()}}}}}, {\hyperref[\detokenize{modules/tensor:symjax.tensor.not_equal}]{\sphinxcrossref{\sphinxcode{\sphinxupquote{not\_equal()}}}}}

\subsubsection*{Examples}

\begin{sphinxVerbatim}[commandchars=\\\{\}]
\PYG{g+gp}{\PYGZgt{}\PYGZgt{}\PYGZgt{} }\PYG{n}{np}\PYG{o}{.}\PYG{n}{greater\PYGZus{}equal}\PYG{p}{(}\PYG{p}{[}\PYG{l+m+mi}{4}\PYG{p}{,} \PYG{l+m+mi}{2}\PYG{p}{,} \PYG{l+m+mi}{1}\PYG{p}{]}\PYG{p}{,} \PYG{p}{[}\PYG{l+m+mi}{2}\PYG{p}{,} \PYG{l+m+mi}{2}\PYG{p}{,} \PYG{l+m+mi}{2}\PYG{p}{]}\PYG{p}{)}
\PYG{g+go}{array([ True, True, False])}
\end{sphinxVerbatim}

\end{fulllineitems}

\index{heaviside() (in module symjax.tensor)@\spxentry{heaviside()}\spxextra{in module symjax.tensor}}

\begin{fulllineitems}
\phantomsection\label{\detokenize{modules/tensor:symjax.tensor.heaviside}}\pysiglinewithargsret{\sphinxbfcode{\sphinxupquote{heaviside}}}{\emph{\DUrole{n}{x1}}, \emph{\DUrole{n}{x2}}}{}
Compute the Heaviside step function.

LAX\sphinxhyphen{}backend implementation of {\hyperref[\detokenize{modules/tensor:symjax.tensor.heaviside}]{\sphinxcrossref{\sphinxcode{\sphinxupquote{heaviside()}}}}}.
ADDITIONOriginal docstring below.

LAX\sphinxhyphen{}backend implementation of {\hyperref[\detokenize{modules/tensor:symjax.tensor.heaviside}]{\sphinxcrossref{\sphinxcode{\sphinxupquote{heaviside()}}}}}.
Original docstring below.

heaviside(x1, x2, /, out=None, {\color{red}\bfseries{}*}, where=True, casting=’same\_kind’, order=’K’, dtype=None, subok=True{[}, signature, extobj{]})

The Heaviside step function is defined as:

\begin{sphinxVerbatim}[commandchars=\\\{\}]
                      \PYG{l+m+mi}{0}   \PYG{k}{if} \PYG{n}{x1} \PYG{o}{\PYGZlt{}} \PYG{l+m+mi}{0}
\PYG{n}{heaviside}\PYG{p}{(}\PYG{n}{x1}\PYG{p}{,} \PYG{n}{x2}\PYG{p}{)} \PYG{o}{=}  \PYG{n}{x2}   \PYG{k}{if} \PYG{n}{x1} \PYG{o}{==} \PYG{l+m+mi}{0}
                      \PYG{l+m+mi}{1}   \PYG{k}{if} \PYG{n}{x1} \PYG{o}{\PYGZgt{}} \PYG{l+m+mi}{0}
\end{sphinxVerbatim}

where \sphinxtitleref{x2} is often taken to be 0.5, but 0 and 1 are also sometimes used.
\begin{quote}\begin{description}
\item[{Returns}] \leavevmode
\sphinxstylestrong{out} \textendash{} The output array, element\sphinxhyphen{}wise Heaviside step function of \sphinxtitleref{x1}.
This is a scalar if both \sphinxtitleref{x1} and \sphinxtitleref{x2} are scalars.

\item[{Return type}] \leavevmode
ndarray or scalar

\end{description}\end{quote}
\subsubsection*{Notes}

\DUrole{versionmodified,added}{New in version 1.13.0.}
\subsubsection*{References}
\subsubsection*{Examples}

\begin{sphinxVerbatim}[commandchars=\\\{\}]
\PYG{g+gp}{\PYGZgt{}\PYGZgt{}\PYGZgt{} }\PYG{n}{np}\PYG{o}{.}\PYG{n}{heaviside}\PYG{p}{(}\PYG{p}{[}\PYG{o}{\PYGZhy{}}\PYG{l+m+mf}{1.5}\PYG{p}{,} \PYG{l+m+mi}{0}\PYG{p}{,} \PYG{l+m+mf}{2.0}\PYG{p}{]}\PYG{p}{,} \PYG{l+m+mf}{0.5}\PYG{p}{)}
\PYG{g+go}{array([ 0. ,  0.5,  1. ])}
\PYG{g+gp}{\PYGZgt{}\PYGZgt{}\PYGZgt{} }\PYG{n}{np}\PYG{o}{.}\PYG{n}{heaviside}\PYG{p}{(}\PYG{p}{[}\PYG{o}{\PYGZhy{}}\PYG{l+m+mf}{1.5}\PYG{p}{,} \PYG{l+m+mi}{0}\PYG{p}{,} \PYG{l+m+mf}{2.0}\PYG{p}{]}\PYG{p}{,} \PYG{l+m+mi}{1}\PYG{p}{)}
\PYG{g+go}{array([ 0.,  1.,  1.])}
\end{sphinxVerbatim}

\end{fulllineitems}

\index{hsplit() (in module symjax.tensor)@\spxentry{hsplit()}\spxextra{in module symjax.tensor}}

\begin{fulllineitems}
\phantomsection\label{\detokenize{modules/tensor:symjax.tensor.hsplit}}\pysiglinewithargsret{\sphinxbfcode{\sphinxupquote{hsplit}}}{\emph{\DUrole{n}{ary}}, \emph{\DUrole{n}{indices\_or\_sections}}}{}
Split an array into multiple sub\sphinxhyphen{}arrays horizontally (column\sphinxhyphen{}wise).

LAX\sphinxhyphen{}backend implementation of {\hyperref[\detokenize{modules/tensor:symjax.tensor.hsplit}]{\sphinxcrossref{\sphinxcode{\sphinxupquote{hsplit()}}}}}.
ADDITIONOriginal docstring below.

LA

\end{fulllineitems}

\index{hstack() (in module symjax.tensor)@\spxentry{hstack()}\spxextra{in module symjax.tensor}}

\begin{fulllineitems}
\phantomsection\label{\detokenize{modules/tensor:symjax.tensor.hstack}}\pysiglinewithargsret{\sphinxbfcode{\sphinxupquote{hstack}}}{\emph{\DUrole{n}{tup}}}{}
Stack arrays in sequence horizontally (column wise).

LAX\sphinxhyphen{}backend implementation of {\hyperref[\detokenize{modules/tensor:symjax.tensor.hstack}]{\sphinxcrossref{\sphinxcode{\sphinxupquote{hstack()}}}}}.
ADDITIONOriginal docstring below.

LAX\sphinxhyphen{}backend implementation of {\hyperref[\detokenize{modules/tensor:symjax.tensor.hstack}]{\sphinxcrossref{\sphinxcode{\sphinxupquote{hstack()}}}}}.
Original docstring below.

This is equivalent to concatenation along the second axis, except for 1\sphinxhyphen{}D
arrays where it concatenates along the first axis. Rebuilds arrays divided
by \sphinxtitleref{hsplit}.

This function makes most sense for arrays with up to 3 dimensions. For
instance, for pixel\sphinxhyphen{}data with a height (first axis), width (second axis),
and r/g/b channels (third axis). The functions \sphinxtitleref{concatenate}, \sphinxtitleref{stack} and
\sphinxtitleref{block} provide more general stacking and concatenation operations.
\begin{quote}\begin{description}
\item[{Returns}] \leavevmode
\sphinxstylestrong{stacked} \textendash{} The array formed by stacking the given arrays.

\item[{Return type}] \leavevmode
ndarray

\end{description}\end{quote}

\sphinxstrong{See also:}

\begin{description}
\item[{{\hyperref[\detokenize{modules/tensor:symjax.tensor.stack}]{\sphinxcrossref{\sphinxcode{\sphinxupquote{stack()}}}}}}] \leavevmode
Join a sequence of arrays along a new axis.

\item[{{\hyperref[\detokenize{modules/tensor:symjax.tensor.vstack}]{\sphinxcrossref{\sphinxcode{\sphinxupquote{vstack()}}}}}}] \leavevmode
Stack arrays in sequence vertically (row wise).

\item[{{\hyperref[\detokenize{modules/tensor:symjax.tensor.dstack}]{\sphinxcrossref{\sphinxcode{\sphinxupquote{dstack()}}}}}}] \leavevmode
Stack arrays in sequence depth wise (along third axis).

\item[{{\hyperref[\detokenize{modules/tensor:symjax.tensor.concatenate}]{\sphinxcrossref{\sphinxcode{\sphinxupquote{concatenate()}}}}}}] \leavevmode
Join a sequence of arrays along an existing axis.

\item[{{\hyperref[\detokenize{modules/tensor:symjax.tensor.hsplit}]{\sphinxcrossref{\sphinxcode{\sphinxupquote{hsplit()}}}}}}] \leavevmode
Split array along second axis.

\item[{{\hyperref[\detokenize{modules/tensor:symjax.tensor.block}]{\sphinxcrossref{\sphinxcode{\sphinxupquote{block()}}}}}}] \leavevmode
Assemble arrays from blocks.

\end{description}

\subsubsection*{Examples}

\begin{sphinxVerbatim}[commandchars=\\\{\}]
\PYG{g+gp}{\PYGZgt{}\PYGZgt{}\PYGZgt{} }\PYG{n}{a} \PYG{o}{=} \PYG{n}{np}\PYG{o}{.}\PYG{n}{array}\PYG{p}{(}\PYG{p}{(}\PYG{l+m+mi}{1}\PYG{p}{,}\PYG{l+m+mi}{2}\PYG{p}{,}\PYG{l+m+mi}{3}\PYG{p}{)}\PYG{p}{)}
\PYG{g+gp}{\PYGZgt{}\PYGZgt{}\PYGZgt{} }\PYG{n}{b} \PYG{o}{=} \PYG{n}{np}\PYG{o}{.}\PYG{n}{array}\PYG{p}{(}\PYG{p}{(}\PYG{l+m+mi}{2}\PYG{p}{,}\PYG{l+m+mi}{3}\PYG{p}{,}\PYG{l+m+mi}{4}\PYG{p}{)}\PYG{p}{)}
\PYG{g+gp}{\PYGZgt{}\PYGZgt{}\PYGZgt{} }\PYG{n}{np}\PYG{o}{.}\PYG{n}{hstack}\PYG{p}{(}\PYG{p}{(}\PYG{n}{a}\PYG{p}{,}\PYG{n}{b}\PYG{p}{)}\PYG{p}{)}
\PYG{g+go}{array([1, 2, 3, 2, 3, 4])}
\PYG{g+gp}{\PYGZgt{}\PYGZgt{}\PYGZgt{} }\PYG{n}{a} \PYG{o}{=} \PYG{n}{np}\PYG{o}{.}\PYG{n}{array}\PYG{p}{(}\PYG{p}{[}\PYG{p}{[}\PYG{l+m+mi}{1}\PYG{p}{]}\PYG{p}{,}\PYG{p}{[}\PYG{l+m+mi}{2}\PYG{p}{]}\PYG{p}{,}\PYG{p}{[}\PYG{l+m+mi}{3}\PYG{p}{]}\PYG{p}{]}\PYG{p}{)}
\PYG{g+gp}{\PYGZgt{}\PYGZgt{}\PYGZgt{} }\PYG{n}{b} \PYG{o}{=} \PYG{n}{np}\PYG{o}{.}\PYG{n}{array}\PYG{p}{(}\PYG{p}{[}\PYG{p}{[}\PYG{l+m+mi}{2}\PYG{p}{]}\PYG{p}{,}\PYG{p}{[}\PYG{l+m+mi}{3}\PYG{p}{]}\PYG{p}{,}\PYG{p}{[}\PYG{l+m+mi}{4}\PYG{p}{]}\PYG{p}{]}\PYG{p}{)}
\PYG{g+gp}{\PYGZgt{}\PYGZgt{}\PYGZgt{} }\PYG{n}{np}\PYG{o}{.}\PYG{n}{hstack}\PYG{p}{(}\PYG{p}{(}\PYG{n}{a}\PYG{p}{,}\PYG{n}{b}\PYG{p}{)}\PYG{p}{)}
\PYG{g+go}{array([[1, 2],}
\PYG{g+go}{       [2, 3],}
\PYG{g+go}{       [3, 4]])}
\end{sphinxVerbatim}

\end{fulllineitems}

\index{identity() (in module symjax.tensor)@\spxentry{identity()}\spxextra{in module symjax.tensor}}

\begin{fulllineitems}
\phantomsection\label{\detokenize{modules/tensor:symjax.tensor.identity}}\pysiglinewithargsret{\sphinxbfcode{\sphinxupquote{identity}}}{\emph{\DUrole{n}{n}}, \emph{\DUrole{n}{dtype}\DUrole{o}{=}\DUrole{default_value}{None}}}{}
Return the identity array.

LAX\sphinxhyphen{}backend implementation of {\hyperref[\detokenize{modules/tensor:symjax.tensor.identity}]{\sphinxcrossref{\sphinxcode{\sphinxupquote{identity()}}}}}.
ADDITIONOriginal docstring below.

LAX\sphinxhyphen{}backend implementation of {\hyperref[\detokenize{modules/tensor:symjax.tensor.identity}]{\sphinxcrossref{\sphinxcode{\sphinxupquote{identity()}}}}}.
Original docstring below.

The identity array is a square array with ones on
the main diagonal.
\begin{quote}\begin{description}
\item[{Parameters}] \leavevmode
\sphinxstyleliteralstrong{\sphinxupquote{dtype}} (\sphinxstyleliteralemphasis{\sphinxupquote{data\sphinxhyphen{}type}}\sphinxstyleliteralemphasis{\sphinxupquote{, }}\sphinxstyleliteralemphasis{\sphinxupquote{optional}}) \textendash{} Data\sphinxhyphen{}type of the output.  Defaults to \sphinxcode{\sphinxupquote{float}}.

\item[{Returns}] \leavevmode
\sphinxstylestrong{out} \textendash{} \sphinxtitleref{n} x \sphinxtitleref{n} array with its main diagonal set to one,
and all other elements 0.

\item[{Return type}] \leavevmode
ndarray

\end{description}\end{quote}
\subsubsection*{Examples}

\begin{sphinxVerbatim}[commandchars=\\\{\}]
\PYG{g+gp}{\PYGZgt{}\PYGZgt{}\PYGZgt{} }\PYG{n}{np}\PYG{o}{.}\PYG{n}{identity}\PYG{p}{(}\PYG{l+m+mi}{3}\PYG{p}{)}
\PYG{g+go}{array([[1.,  0.,  0.],}
\PYG{g+go}{       [0.,  1.,  0.],}
\PYG{g+go}{       [0.,  0.,  1.]])}
\end{sphinxVerbatim}

\end{fulllineitems}

\index{imag() (in module symjax.tensor)@\spxentry{imag()}\spxextra{in module symjax.tensor}}

\begin{fulllineitems}
\phantomsection\label{\detokenize{modules/tensor:symjax.tensor.imag}}\pysiglinewithargsret{\sphinxbfcode{\sphinxupquote{imag}}}{\emph{\DUrole{n}{val}}}{}
Return the imaginary part of the complex argument.

LAX\sphinxhyphen{}backend implementation of {\hyperref[\detokenize{modules/tensor:symjax.tensor.imag}]{\sphinxcrossref{\sphinxcode{\sphinxupquote{imag()}}}}}.
ADDITIONOriginal docstring below.

LAX\sphinxhyphen{}backend implementation of {\hyperref[\detokenize{modules/tensor:symjax.tensor.imag}]{\sphinxcrossref{\sphinxcode{\sphinxupquote{imag()}}}}}.
Original docstring below.
\begin{quote}\begin{description}
\item[{Returns}] \leavevmode
\sphinxstylestrong{out} \textendash{} The imaginary component of the complex argument. If \sphinxtitleref{val} is real,
the type of \sphinxtitleref{val} is used for the output.  If \sphinxtitleref{val} has complex
elements, the returned type is float.

\item[{Return type}] \leavevmode
ndarray or scalar

\end{description}\end{quote}

\sphinxstrong{See also:}

{\hyperref[\detokenize{modules/tensor:symjax.tensor.real}]{\sphinxcrossref{\sphinxcode{\sphinxupquote{real()}}}}}, {\hyperref[\detokenize{modules/tensor:symjax.tensor.angle}]{\sphinxcrossref{\sphinxcode{\sphinxupquote{angle()}}}}}, \sphinxcode{\sphinxupquote{real\_if\_close()}}

\subsubsection*{Examples}

\begin{sphinxVerbatim}[commandchars=\\\{\}]
\PYG{g+gp}{\PYGZgt{}\PYGZgt{}\PYGZgt{} }\PYG{n}{a} \PYG{o}{=} \PYG{n}{np}\PYG{o}{.}\PYG{n}{array}\PYG{p}{(}\PYG{p}{[}\PYG{l+m+mi}{1}\PYG{o}{+}\PYG{l+m+mi}{2}\PYG{n}{j}\PYG{p}{,} \PYG{l+m+mi}{3}\PYG{o}{+}\PYG{l+m+mi}{4}\PYG{n}{j}\PYG{p}{,} \PYG{l+m+mi}{5}\PYG{o}{+}\PYG{l+m+mi}{6}\PYG{n}{j}\PYG{p}{]}\PYG{p}{)}
\PYG{g+gp}{\PYGZgt{}\PYGZgt{}\PYGZgt{} }\PYG{n}{a}\PYG{o}{.}\PYG{n}{imag}
\PYG{g+go}{array([2.,  4.,  6.])}
\PYG{g+gp}{\PYGZgt{}\PYGZgt{}\PYGZgt{} }\PYG{n}{a}\PYG{o}{.}\PYG{n}{imag} \PYG{o}{=} \PYG{n}{np}\PYG{o}{.}\PYG{n}{array}\PYG{p}{(}\PYG{p}{[}\PYG{l+m+mi}{8}\PYG{p}{,} \PYG{l+m+mi}{10}\PYG{p}{,} \PYG{l+m+mi}{12}\PYG{p}{]}\PYG{p}{)}
\PYG{g+gp}{\PYGZgt{}\PYGZgt{}\PYGZgt{} }\PYG{n}{a}
\PYG{g+go}{array([1. +8.j,  3.+10.j,  5.+12.j])}
\PYG{g+gp}{\PYGZgt{}\PYGZgt{}\PYGZgt{} }\PYG{n}{np}\PYG{o}{.}\PYG{n}{imag}\PYG{p}{(}\PYG{l+m+mi}{1} \PYG{o}{+} \PYG{l+m+mi}{1}\PYG{n}{j}\PYG{p}{)}
\PYG{g+go}{1.0}
\end{sphinxVerbatim}

\end{fulllineitems}

\index{inner() (in module symjax.tensor)@\spxentry{inner()}\spxextra{in module symjax.tensor}}

\begin{fulllineitems}
\phantomsection\label{\detokenize{modules/tensor:symjax.tensor.inner}}\pysiglinewithargsret{\sphinxbfcode{\sphinxupquote{inner}}}{\emph{\DUrole{n}{a}}, \emph{\DUrole{n}{b}}, \emph{\DUrole{n}{precision}\DUrole{o}{=}\DUrole{default_value}{None}}}{}
Inner product of two arrays.

LAX\sphinxhyphen{}backend implementation of {\hyperref[\detokenize{modules/tensor:symjax.tensor.inner}]{\sphinxcrossref{\sphinxcode{\sphinxupquote{inner()}}}}}.
ADDITIONOriginal docstring below.

LAX\sphinxhyphen{}backend implementation of {\hyperref[\detokenize{modules/tensor:symjax.tensor.inner}]{\sphinxcrossref{\sphinxcode{\sphinxupquote{inner()}}}}}.
In addition to the original NumPy arguments listed below, also supports
\sphinxcode{\sphinxupquote{precision}} for extra control over matrix\sphinxhyphen{}multiplication precision
on supported devices. See \sphinxcode{\sphinxupquote{jax.lax.dot()}} for details.

Original docstring below.
\begin{quote}
\begin{quote}

inner(a, b)

Ordinary inner product of vectors for 1\sphinxhyphen{}D arrays (without complex
conjugation), in higher dimensions a sum product over the last axes.
\end{quote}
\begin{description}
\item[{Returns}] \leavevmode\begin{description}
\item[{out}] \leavevmode{[}ndarray{]}
\sphinxtitleref{out.shape = a.shape{[}:\sphinxhyphen{}1{]} + b.shape{[}:\sphinxhyphen{}1{]}}

\end{description}
\begin{description}
\item[{ValueError}] \leavevmode
If the last dimension of \sphinxtitleref{a} and \sphinxtitleref{b} has different size.

\end{description}

tensordot : Sum products over arbitrary axes.
dot : Generalised matrix product, using second last dimension of \sphinxtitleref{b}.
einsum : Einstein summation convention.

For vectors (1\sphinxhyphen{}D arrays) it computes the ordinary inner\sphinxhyphen{}product:

\begin{sphinxVerbatim}[commandchars=\\\{\}]
\PYG{n}{np}\PYG{o}{.}\PYG{n}{inner}\PYG{p}{(}\PYG{n}{a}\PYG{p}{,} \PYG{n}{b}\PYG{p}{)} \PYG{o}{=} \PYG{n+nb}{sum}\PYG{p}{(}\PYG{n}{a}\PYG{p}{[}\PYG{p}{:}\PYG{p}{]}\PYG{o}{*}\PYG{n}{b}\PYG{p}{[}\PYG{p}{:}\PYG{p}{]}\PYG{p}{)}
\end{sphinxVerbatim}

More generally, if \sphinxtitleref{ndim(a) = r \textgreater{} 0} and \sphinxtitleref{ndim(b) = s \textgreater{} 0}:

\begin{sphinxVerbatim}[commandchars=\\\{\}]
\PYG{n}{np}\PYG{o}{.}\PYG{n}{inner}\PYG{p}{(}\PYG{n}{a}\PYG{p}{,} \PYG{n}{b}\PYG{p}{)} \PYG{o}{=} \PYG{n}{np}\PYG{o}{.}\PYG{n}{tensordot}\PYG{p}{(}\PYG{n}{a}\PYG{p}{,} \PYG{n}{b}\PYG{p}{,} \PYG{n}{axes}\PYG{o}{=}\PYG{p}{(}\PYG{o}{\PYGZhy{}}\PYG{l+m+mi}{1}\PYG{p}{,}\PYG{o}{\PYGZhy{}}\PYG{l+m+mi}{1}\PYG{p}{)}\PYG{p}{)}
\end{sphinxVerbatim}

or explicitly:

\begin{sphinxVerbatim}[commandchars=\\\{\}]
\PYG{n}{np}\PYG{o}{.}\PYG{n}{inner}\PYG{p}{(}\PYG{n}{a}\PYG{p}{,} \PYG{n}{b}\PYG{p}{)}\PYG{p}{[}\PYG{n}{i0}\PYG{p}{,}\PYG{o}{.}\PYG{o}{.}\PYG{o}{.}\PYG{p}{,}\PYG{n}{ir}\PYG{o}{\PYGZhy{}}\PYG{l+m+mi}{1}\PYG{p}{,}\PYG{n}{j0}\PYG{p}{,}\PYG{o}{.}\PYG{o}{.}\PYG{o}{.}\PYG{p}{,}\PYG{n}{js}\PYG{o}{\PYGZhy{}}\PYG{l+m+mi}{1}\PYG{p}{]}
     \PYG{o}{=} \PYG{n+nb}{sum}\PYG{p}{(}\PYG{n}{a}\PYG{p}{[}\PYG{n}{i0}\PYG{p}{,}\PYG{o}{.}\PYG{o}{.}\PYG{o}{.}\PYG{p}{,}\PYG{n}{ir}\PYG{o}{\PYGZhy{}}\PYG{l+m+mi}{1}\PYG{p}{,}\PYG{p}{:}\PYG{p}{]}\PYG{o}{*}\PYG{n}{b}\PYG{p}{[}\PYG{n}{j0}\PYG{p}{,}\PYG{o}{.}\PYG{o}{.}\PYG{o}{.}\PYG{p}{,}\PYG{n}{js}\PYG{o}{\PYGZhy{}}\PYG{l+m+mi}{1}\PYG{p}{,}\PYG{p}{:}\PYG{p}{]}\PYG{p}{)}
\end{sphinxVerbatim}

In addition \sphinxtitleref{a} or \sphinxtitleref{b} may be scalars, in which case:

\begin{sphinxVerbatim}[commandchars=\\\{\}]
\PYG{n}{np}\PYG{o}{.}\PYG{n}{inner}\PYG{p}{(}\PYG{n}{a}\PYG{p}{,}\PYG{n}{b}\PYG{p}{)} \PYG{o}{=} \PYG{n}{a}\PYG{o}{*}\PYG{n}{b}
\end{sphinxVerbatim}

Ordinary inner product for vectors:

\begin{sphinxVerbatim}[commandchars=\\\{\}]
\PYG{g+gp}{\PYGZgt{}\PYGZgt{}\PYGZgt{} }\PYG{n}{a} \PYG{o}{=} \PYG{n}{np}\PYG{o}{.}\PYG{n}{array}\PYG{p}{(}\PYG{p}{[}\PYG{l+m+mi}{1}\PYG{p}{,}\PYG{l+m+mi}{2}\PYG{p}{,}\PYG{l+m+mi}{3}\PYG{p}{]}\PYG{p}{)}
\PYG{g+gp}{\PYGZgt{}\PYGZgt{}\PYGZgt{} }\PYG{n}{b} \PYG{o}{=} \PYG{n}{np}\PYG{o}{.}\PYG{n}{array}\PYG{p}{(}\PYG{p}{[}\PYG{l+m+mi}{0}\PYG{p}{,}\PYG{l+m+mi}{1}\PYG{p}{,}\PYG{l+m+mi}{0}\PYG{p}{]}\PYG{p}{)}
\PYG{g+gp}{\PYGZgt{}\PYGZgt{}\PYGZgt{} }\PYG{n}{np}\PYG{o}{.}\PYG{n}{inner}\PYG{p}{(}\PYG{n}{a}\PYG{p}{,} \PYG{n}{b}\PYG{p}{)}
\PYG{g+go}{2}
\end{sphinxVerbatim}

A multidimensional example:

\begin{sphinxVerbatim}[commandchars=\\\{\}]
\PYG{g+gp}{\PYGZgt{}\PYGZgt{}\PYGZgt{} }\PYG{n}{a} \PYG{o}{=} \PYG{n}{np}\PYG{o}{.}\PYG{n}{arange}\PYG{p}{(}\PYG{l+m+mi}{24}\PYG{p}{)}\PYG{o}{.}\PYG{n}{reshape}\PYG{p}{(}\PYG{p}{(}\PYG{l+m+mi}{2}\PYG{p}{,}\PYG{l+m+mi}{3}\PYG{p}{,}\PYG{l+m+mi}{4}\PYG{p}{)}\PYG{p}{)}
\PYG{g+gp}{\PYGZgt{}\PYGZgt{}\PYGZgt{} }\PYG{n}{b} \PYG{o}{=} \PYG{n}{np}\PYG{o}{.}\PYG{n}{arange}\PYG{p}{(}\PYG{l+m+mi}{4}\PYG{p}{)}
\PYG{g+gp}{\PYGZgt{}\PYGZgt{}\PYGZgt{} }\PYG{n}{np}\PYG{o}{.}\PYG{n}{inner}\PYG{p}{(}\PYG{n}{a}\PYG{p}{,} \PYG{n}{b}\PYG{p}{)}
\PYG{g+go}{array([[ 14,  38,  62],}
\PYG{g+go}{       [ 86, 110, 134]])}
\end{sphinxVerbatim}

An example where \sphinxtitleref{b} is a scalar:

\begin{sphinxVerbatim}[commandchars=\\\{\}]
\PYG{g+gp}{\PYGZgt{}\PYGZgt{}\PYGZgt{} }\PYG{n}{np}\PYG{o}{.}\PYG{n}{inner}\PYG{p}{(}\PYG{n}{np}\PYG{o}{.}\PYG{n}{eye}\PYG{p}{(}\PYG{l+m+mi}{2}\PYG{p}{)}\PYG{p}{,} \PYG{l+m+mi}{7}\PYG{p}{)}
\PYG{g+go}{array([[7., 0.],}
\PYG{g+go}{       [0., 7.]])}
\end{sphinxVerbatim}

\end{description}
\end{quote}

\end{fulllineitems}

\index{isclose() (in module symjax.tensor)@\spxentry{isclose()}\spxextra{in module symjax.tensor}}

\begin{fulllineitems}
\phantomsection\label{\detokenize{modules/tensor:symjax.tensor.isclose}}\pysiglinewithargsret{\sphinxbfcode{\sphinxupquote{isclose}}}{\emph{\DUrole{n}{a}}, \emph{\DUrole{n}{b}}, \emph{\DUrole{n}{rtol}\DUrole{o}{=}\DUrole{default_value}{1e\sphinxhyphen{}05}}, \emph{\DUrole{n}{atol}\DUrole{o}{=}\DUrole{default_value}{1e\sphinxhyphen{}08}}, \emph{\DUrole{n}{equal\_nan}\DUrole{o}{=}\DUrole{default_value}{False}}}{}~\begin{description}
\item[{Returns a boolean array where two arrays are element\sphinxhyphen{}wise equal within a}] \leavevmode
tolerance.

\end{description}

LAX\sphinxhyphen{}backend implementation of {\hyperref[\detokenize{modules/tensor:symjax.tensor.isclose}]{\sphinxcrossref{\sphinxcode{\sphinxupquote{isclose()}}}}}.
ADDITIONOriginal docstring below.

LAX\sphinxhyphen{}backend implementation of {\hyperref[\detokenize{modules/tensor:symjax.tensor.isclose}]{\sphinxcrossref{\sphinxcode{\sphinxupquote{isclose()}}}}}.
Original docstring below.

The tolerance values are positive, typically very small numbers.  The
relative difference (\sphinxtitleref{rtol} * abs(\sphinxtitleref{b})) and the absolute difference
\sphinxtitleref{atol} are added together to compare against the absolute difference
between \sphinxtitleref{a} and \sphinxtitleref{b}.

\begin{sphinxadmonition}{warning}{Warning:}
The default \sphinxtitleref{atol} is not appropriate for comparing numbers
that are much smaller than one (see Notes).
\end{sphinxadmonition}
\begin{quote}\begin{description}
\item[{Returns}] \leavevmode
\sphinxstylestrong{y} \textendash{} Returns a boolean array of where \sphinxtitleref{a} and \sphinxtitleref{b} are equal within the
given tolerance. If both \sphinxtitleref{a} and \sphinxtitleref{b} are scalars, returns a single
boolean value.

\item[{Return type}] \leavevmode
array\_like

\end{description}\end{quote}

\sphinxstrong{See also:}

{\hyperref[\detokenize{modules/tensor:symjax.tensor.allclose}]{\sphinxcrossref{\sphinxcode{\sphinxupquote{allclose()}}}}}

\subsubsection*{Notes}

\DUrole{versionmodified,added}{New in version 1.7.0.}

For finite values, isclose uses the following equation to test whether
two floating point values are equivalent.
\begin{quote}

absolute(\sphinxtitleref{a} \sphinxhyphen{} \sphinxtitleref{b}) \textless{}= (\sphinxtitleref{atol} + \sphinxtitleref{rtol} * absolute(\sphinxtitleref{b}))
\end{quote}

Unlike the built\sphinxhyphen{}in \sphinxtitleref{math.isclose}, the above equation is not symmetric
in \sphinxtitleref{a} and \sphinxtitleref{b} \textendash{} it assumes \sphinxtitleref{b} is the reference value \textendash{} so that
\sphinxtitleref{isclose(a, b)} might be different from \sphinxtitleref{isclose(b, a)}. Furthermore,
the default value of atol is not zero, and is used to determine what
small values should be considered close to zero. The default value is
appropriate for expected values of order unity: if the expected values
are significantly smaller than one, it can result in false positives.
\sphinxtitleref{atol} should be carefully selected for the use case at hand. A zero value
for \sphinxtitleref{atol} will result in \sphinxtitleref{False} if either \sphinxtitleref{a} or \sphinxtitleref{b} is zero.
\subsubsection*{Examples}

\begin{sphinxVerbatim}[commandchars=\\\{\}]
\PYG{g+gp}{\PYGZgt{}\PYGZgt{}\PYGZgt{} }\PYG{n}{np}\PYG{o}{.}\PYG{n}{isclose}\PYG{p}{(}\PYG{p}{[}\PYG{l+m+mf}{1e10}\PYG{p}{,}\PYG{l+m+mf}{1e\PYGZhy{}7}\PYG{p}{]}\PYG{p}{,} \PYG{p}{[}\PYG{l+m+mf}{1.00001e10}\PYG{p}{,}\PYG{l+m+mf}{1e\PYGZhy{}8}\PYG{p}{]}\PYG{p}{)}
\PYG{g+go}{array([ True, False])}
\PYG{g+gp}{\PYGZgt{}\PYGZgt{}\PYGZgt{} }\PYG{n}{np}\PYG{o}{.}\PYG{n}{isclose}\PYG{p}{(}\PYG{p}{[}\PYG{l+m+mf}{1e10}\PYG{p}{,}\PYG{l+m+mf}{1e\PYGZhy{}8}\PYG{p}{]}\PYG{p}{,} \PYG{p}{[}\PYG{l+m+mf}{1.00001e10}\PYG{p}{,}\PYG{l+m+mf}{1e\PYGZhy{}9}\PYG{p}{]}\PYG{p}{)}
\PYG{g+go}{array([ True, True])}
\PYG{g+gp}{\PYGZgt{}\PYGZgt{}\PYGZgt{} }\PYG{n}{np}\PYG{o}{.}\PYG{n}{isclose}\PYG{p}{(}\PYG{p}{[}\PYG{l+m+mf}{1e10}\PYG{p}{,}\PYG{l+m+mf}{1e\PYGZhy{}8}\PYG{p}{]}\PYG{p}{,} \PYG{p}{[}\PYG{l+m+mf}{1.0001e10}\PYG{p}{,}\PYG{l+m+mf}{1e\PYGZhy{}9}\PYG{p}{]}\PYG{p}{)}
\PYG{g+go}{array([False,  True])}
\PYG{g+gp}{\PYGZgt{}\PYGZgt{}\PYGZgt{} }\PYG{n}{np}\PYG{o}{.}\PYG{n}{isclose}\PYG{p}{(}\PYG{p}{[}\PYG{l+m+mf}{1.0}\PYG{p}{,} \PYG{n}{np}\PYG{o}{.}\PYG{n}{nan}\PYG{p}{]}\PYG{p}{,} \PYG{p}{[}\PYG{l+m+mf}{1.0}\PYG{p}{,} \PYG{n}{np}\PYG{o}{.}\PYG{n}{nan}\PYG{p}{]}\PYG{p}{)}
\PYG{g+go}{array([ True, False])}
\PYG{g+gp}{\PYGZgt{}\PYGZgt{}\PYGZgt{} }\PYG{n}{np}\PYG{o}{.}\PYG{n}{isclose}\PYG{p}{(}\PYG{p}{[}\PYG{l+m+mf}{1.0}\PYG{p}{,} \PYG{n}{np}\PYG{o}{.}\PYG{n}{nan}\PYG{p}{]}\PYG{p}{,} \PYG{p}{[}\PYG{l+m+mf}{1.0}\PYG{p}{,} \PYG{n}{np}\PYG{o}{.}\PYG{n}{nan}\PYG{p}{]}\PYG{p}{,} \PYG{n}{equal\PYGZus{}nan}\PYG{o}{=}\PYG{k+kc}{True}\PYG{p}{)}
\PYG{g+go}{array([ True, True])}
\PYG{g+gp}{\PYGZgt{}\PYGZgt{}\PYGZgt{} }\PYG{n}{np}\PYG{o}{.}\PYG{n}{isclose}\PYG{p}{(}\PYG{p}{[}\PYG{l+m+mf}{1e\PYGZhy{}8}\PYG{p}{,} \PYG{l+m+mf}{1e\PYGZhy{}7}\PYG{p}{]}\PYG{p}{,} \PYG{p}{[}\PYG{l+m+mf}{0.0}\PYG{p}{,} \PYG{l+m+mf}{0.0}\PYG{p}{]}\PYG{p}{)}
\PYG{g+go}{array([ True, False])}
\PYG{g+gp}{\PYGZgt{}\PYGZgt{}\PYGZgt{} }\PYG{n}{np}\PYG{o}{.}\PYG{n}{isclose}\PYG{p}{(}\PYG{p}{[}\PYG{l+m+mf}{1e\PYGZhy{}100}\PYG{p}{,} \PYG{l+m+mf}{1e\PYGZhy{}7}\PYG{p}{]}\PYG{p}{,} \PYG{p}{[}\PYG{l+m+mf}{0.0}\PYG{p}{,} \PYG{l+m+mf}{0.0}\PYG{p}{]}\PYG{p}{,} \PYG{n}{atol}\PYG{o}{=}\PYG{l+m+mf}{0.0}\PYG{p}{)}
\PYG{g+go}{array([False, False])}
\PYG{g+gp}{\PYGZgt{}\PYGZgt{}\PYGZgt{} }\PYG{n}{np}\PYG{o}{.}\PYG{n}{isclose}\PYG{p}{(}\PYG{p}{[}\PYG{l+m+mf}{1e\PYGZhy{}10}\PYG{p}{,} \PYG{l+m+mf}{1e\PYGZhy{}10}\PYG{p}{]}\PYG{p}{,} \PYG{p}{[}\PYG{l+m+mf}{1e\PYGZhy{}20}\PYG{p}{,} \PYG{l+m+mf}{0.0}\PYG{p}{]}\PYG{p}{)}
\PYG{g+go}{array([ True,  True])}
\PYG{g+gp}{\PYGZgt{}\PYGZgt{}\PYGZgt{} }\PYG{n}{np}\PYG{o}{.}\PYG{n}{isclose}\PYG{p}{(}\PYG{p}{[}\PYG{l+m+mf}{1e\PYGZhy{}10}\PYG{p}{,} \PYG{l+m+mf}{1e\PYGZhy{}10}\PYG{p}{]}\PYG{p}{,} \PYG{p}{[}\PYG{l+m+mf}{1e\PYGZhy{}20}\PYG{p}{,} \PYG{l+m+mf}{0.999999e\PYGZhy{}10}\PYG{p}{]}\PYG{p}{,} \PYG{n}{atol}\PYG{o}{=}\PYG{l+m+mf}{0.0}\PYG{p}{)}
\PYG{g+go}{array([False,  True])}
\end{sphinxVerbatim}

\end{fulllineitems}

\index{iscomplex() (in module symjax.tensor)@\spxentry{iscomplex()}\spxextra{in module symjax.tensor}}

\begin{fulllineitems}
\phantomsection\label{\detokenize{modules/tensor:symjax.tensor.iscomplex}}\pysiglinewithargsret{\sphinxbfcode{\sphinxupquote{iscomplex}}}{\emph{\DUrole{n}{x}}}{}
Returns a bool array, where True if input element is complex.

LAX\sphinxhyphen{}backend implementation of {\hyperref[\detokenize{modules/tensor:symjax.tensor.iscomplex}]{\sphinxcrossref{\sphinxcode{\sphinxupquote{iscomplex()}}}}}.
ADDITIONOriginal docstring below.

LAX\sphinxhyphen{}backend implementation of {\hyperref[\detokenize{modules/tensor:symjax.tensor.iscomplex}]{\sphinxcrossref{\sphinxcode{\sphinxupquote{iscomplex()}}}}}.
Original docstring below.

What is tested is whether the input has a non\sphinxhyphen{}zero imaginary part, not if
the input type is complex.
\begin{quote}\begin{description}
\item[{Returns}] \leavevmode
\sphinxstylestrong{out} \textendash{} Output array.

\item[{Return type}] \leavevmode
ndarray of bools

\end{description}\end{quote}

\sphinxstrong{See also:}

{\hyperref[\detokenize{modules/tensor:symjax.tensor.isreal}]{\sphinxcrossref{\sphinxcode{\sphinxupquote{isreal()}}}}}
\begin{description}
\item[{\sphinxcode{\sphinxupquote{iscomplexobj()}}}] \leavevmode
Return True if x is a complex type or an array of complex numbers.

\end{description}

\subsubsection*{Examples}

\begin{sphinxVerbatim}[commandchars=\\\{\}]
\PYG{g+gp}{\PYGZgt{}\PYGZgt{}\PYGZgt{} }\PYG{n}{np}\PYG{o}{.}\PYG{n}{iscomplex}\PYG{p}{(}\PYG{p}{[}\PYG{l+m+mi}{1}\PYG{o}{+}\PYG{l+m+mi}{1}\PYG{n}{j}\PYG{p}{,} \PYG{l+m+mi}{1}\PYG{o}{+}\PYG{l+m+mi}{0}\PYG{n}{j}\PYG{p}{,} \PYG{l+m+mf}{4.5}\PYG{p}{,} \PYG{l+m+mi}{3}\PYG{p}{,} \PYG{l+m+mi}{2}\PYG{p}{,} \PYG{l+m+mi}{2}\PYG{n}{j}\PYG{p}{]}\PYG{p}{)}
\PYG{g+go}{array([ True, False, False, False, False,  True])}
\end{sphinxVerbatim}

\end{fulllineitems}

\index{isfinite() (in module symjax.tensor)@\spxentry{isfinite()}\spxextra{in module symjax.tensor}}

\begin{fulllineitems}
\phantomsection\label{\detokenize{modules/tensor:symjax.tensor.isfinite}}\pysiglinewithargsret{\sphinxbfcode{\sphinxupquote{isfinite}}}{\emph{\DUrole{n}{x}}}{}
Test element\sphinxhyphen{}wise for finiteness (not infinity or not Not a Number).

LAX\sphinxhyphen{}backend implementation of {\hyperref[\detokenize{modules/tensor:symjax.tensor.isfinite}]{\sphinxcrossref{\sphinxcode{\sphinxupquote{isfinite()}}}}}.
ADDITIONOriginal docstring below.

LAX\sphinxhyphen{}backend implementation of {\hyperref[\detokenize{modules/tensor:symjax.tensor.isfinite}]{\sphinxcrossref{\sphinxcode{\sphinxupquote{isfinite()}}}}}.
Original docstring below.

isfinite(x, /, out=None, {\color{red}\bfseries{}*}, where=True, casting=’same\_kind’, order=’K’, dtype=None, subok=True{[}, signature, extobj{]})

The result is returned as a boolean array.
\begin{quote}\begin{description}
\item[{Returns}] \leavevmode
\sphinxstylestrong{y} \textendash{} True where \sphinxcode{\sphinxupquote{x}} is not positive infinity, negative infinity,
or NaN; false otherwise.
This is a scalar if \sphinxtitleref{x} is a scalar.

\item[{Return type}] \leavevmode
ndarray, bool

\end{description}\end{quote}

\sphinxstrong{See also:}

{\hyperref[\detokenize{modules/tensor:symjax.tensor.isinf}]{\sphinxcrossref{\sphinxcode{\sphinxupquote{isinf()}}}}}, \sphinxcode{\sphinxupquote{isneginf()}}, \sphinxcode{\sphinxupquote{isposinf()}}, {\hyperref[\detokenize{modules/tensor:symjax.tensor.isnan}]{\sphinxcrossref{\sphinxcode{\sphinxupquote{isnan()}}}}}

\subsubsection*{Notes}

Not a Number, positive infinity and negative infinity are considered
to be non\sphinxhyphen{}finite.

NumPy uses the IEEE Standard for Binary Floating\sphinxhyphen{}Point for Arithmetic
(IEEE 754). This means that Not a Number is not equivalent to infinity.
Also that positive infinity is not equivalent to negative infinity. But
infinity is equivalent to positive infinity.  Errors result if the
second argument is also supplied when \sphinxtitleref{x} is a scalar input, or if
first and second arguments have different shapes.
\subsubsection*{Examples}

\begin{sphinxVerbatim}[commandchars=\\\{\}]
\PYG{g+gp}{\PYGZgt{}\PYGZgt{}\PYGZgt{} }\PYG{n}{np}\PYG{o}{.}\PYG{n}{isfinite}\PYG{p}{(}\PYG{l+m+mi}{1}\PYG{p}{)}
\PYG{g+go}{True}
\PYG{g+gp}{\PYGZgt{}\PYGZgt{}\PYGZgt{} }\PYG{n}{np}\PYG{o}{.}\PYG{n}{isfinite}\PYG{p}{(}\PYG{l+m+mi}{0}\PYG{p}{)}
\PYG{g+go}{True}
\PYG{g+gp}{\PYGZgt{}\PYGZgt{}\PYGZgt{} }\PYG{n}{np}\PYG{o}{.}\PYG{n}{isfinite}\PYG{p}{(}\PYG{n}{np}\PYG{o}{.}\PYG{n}{nan}\PYG{p}{)}
\PYG{g+go}{False}
\PYG{g+gp}{\PYGZgt{}\PYGZgt{}\PYGZgt{} }\PYG{n}{np}\PYG{o}{.}\PYG{n}{isfinite}\PYG{p}{(}\PYG{n}{np}\PYG{o}{.}\PYG{n}{inf}\PYG{p}{)}
\PYG{g+go}{False}
\PYG{g+gp}{\PYGZgt{}\PYGZgt{}\PYGZgt{} }\PYG{n}{np}\PYG{o}{.}\PYG{n}{isfinite}\PYG{p}{(}\PYG{n}{np}\PYG{o}{.}\PYG{n}{NINF}\PYG{p}{)}
\PYG{g+go}{False}
\PYG{g+gp}{\PYGZgt{}\PYGZgt{}\PYGZgt{} }\PYG{n}{np}\PYG{o}{.}\PYG{n}{isfinite}\PYG{p}{(}\PYG{p}{[}\PYG{n}{np}\PYG{o}{.}\PYG{n}{log}\PYG{p}{(}\PYG{o}{\PYGZhy{}}\PYG{l+m+mf}{1.}\PYG{p}{)}\PYG{p}{,}\PYG{l+m+mf}{1.}\PYG{p}{,}\PYG{n}{np}\PYG{o}{.}\PYG{n}{log}\PYG{p}{(}\PYG{l+m+mi}{0}\PYG{p}{)}\PYG{p}{]}\PYG{p}{)}
\PYG{g+go}{array([False,  True, False])}
\end{sphinxVerbatim}

\begin{sphinxVerbatim}[commandchars=\\\{\}]
\PYG{g+gp}{\PYGZgt{}\PYGZgt{}\PYGZgt{} }\PYG{n}{x} \PYG{o}{=} \PYG{n}{np}\PYG{o}{.}\PYG{n}{array}\PYG{p}{(}\PYG{p}{[}\PYG{o}{\PYGZhy{}}\PYG{n}{np}\PYG{o}{.}\PYG{n}{inf}\PYG{p}{,} \PYG{l+m+mf}{0.}\PYG{p}{,} \PYG{n}{np}\PYG{o}{.}\PYG{n}{inf}\PYG{p}{]}\PYG{p}{)}
\PYG{g+gp}{\PYGZgt{}\PYGZgt{}\PYGZgt{} }\PYG{n}{y} \PYG{o}{=} \PYG{n}{np}\PYG{o}{.}\PYG{n}{array}\PYG{p}{(}\PYG{p}{[}\PYG{l+m+mi}{2}\PYG{p}{,} \PYG{l+m+mi}{2}\PYG{p}{,} \PYG{l+m+mi}{2}\PYG{p}{]}\PYG{p}{)}
\PYG{g+gp}{\PYGZgt{}\PYGZgt{}\PYGZgt{} }\PYG{n}{np}\PYG{o}{.}\PYG{n}{isfinite}\PYG{p}{(}\PYG{n}{x}\PYG{p}{,} \PYG{n}{y}\PYG{p}{)}
\PYG{g+go}{array([0, 1, 0])}
\PYG{g+gp}{\PYGZgt{}\PYGZgt{}\PYGZgt{} }\PYG{n}{y}
\PYG{g+go}{array([0, 1, 0])}
\end{sphinxVerbatim}

\end{fulllineitems}

\index{isinf() (in module symjax.tensor)@\spxentry{isinf()}\spxextra{in module symjax.tensor}}

\begin{fulllineitems}
\phantomsection\label{\detokenize{modules/tensor:symjax.tensor.isinf}}\pysiglinewithargsret{\sphinxbfcode{\sphinxupquote{isinf}}}{\emph{\DUrole{n}{x}}}{}
Test element\sphinxhyphen{}wise for positive or negative infinity.

LAX\sphinxhyphen{}backend implementation of {\hyperref[\detokenize{modules/tensor:symjax.tensor.isinf}]{\sphinxcrossref{\sphinxcode{\sphinxupquote{isinf()}}}}}.
ADDITIONOriginal docstring below.

LAX\sphinxhyphen{}backend implementation of {\hyperref[\detokenize{modules/tensor:symjax.tensor.isinf}]{\sphinxcrossref{\sphinxcode{\sphinxupquote{isinf()}}}}}.
Original docstring below.

isinf(x, /, out=None, {\color{red}\bfseries{}*}, where=True, casting=’same\_kind’, order=’K’, dtype=None, subok=True{[}, signature, extobj{]})

Returns a boolean array of the same shape as \sphinxtitleref{x}, True where \sphinxcode{\sphinxupquote{x ==
+/\sphinxhyphen{}inf}}, otherwise False.
\begin{quote}\begin{description}
\item[{Returns}] \leavevmode
\sphinxstylestrong{y} \textendash{} True where \sphinxcode{\sphinxupquote{x}} is positive or negative infinity, false otherwise.
This is a scalar if \sphinxtitleref{x} is a scalar.

\item[{Return type}] \leavevmode
bool (scalar) or boolean ndarray

\end{description}\end{quote}

\sphinxstrong{See also:}

\sphinxcode{\sphinxupquote{isneginf()}}, \sphinxcode{\sphinxupquote{isposinf()}}, {\hyperref[\detokenize{modules/tensor:symjax.tensor.isnan}]{\sphinxcrossref{\sphinxcode{\sphinxupquote{isnan()}}}}}, {\hyperref[\detokenize{modules/tensor:symjax.tensor.isfinite}]{\sphinxcrossref{\sphinxcode{\sphinxupquote{isfinite()}}}}}

\subsubsection*{Notes}

NumPy uses the IEEE Standard for Binary Floating\sphinxhyphen{}Point for Arithmetic
(IEEE 754).

Errors result if the second argument is supplied when the first
argument is a scalar, or if the first and second arguments have
different shapes.
\subsubsection*{Examples}

\begin{sphinxVerbatim}[commandchars=\\\{\}]
\PYG{g+gp}{\PYGZgt{}\PYGZgt{}\PYGZgt{} }\PYG{n}{np}\PYG{o}{.}\PYG{n}{isinf}\PYG{p}{(}\PYG{n}{np}\PYG{o}{.}\PYG{n}{inf}\PYG{p}{)}
\PYG{g+go}{True}
\PYG{g+gp}{\PYGZgt{}\PYGZgt{}\PYGZgt{} }\PYG{n}{np}\PYG{o}{.}\PYG{n}{isinf}\PYG{p}{(}\PYG{n}{np}\PYG{o}{.}\PYG{n}{nan}\PYG{p}{)}
\PYG{g+go}{False}
\PYG{g+gp}{\PYGZgt{}\PYGZgt{}\PYGZgt{} }\PYG{n}{np}\PYG{o}{.}\PYG{n}{isinf}\PYG{p}{(}\PYG{n}{np}\PYG{o}{.}\PYG{n}{NINF}\PYG{p}{)}
\PYG{g+go}{True}
\PYG{g+gp}{\PYGZgt{}\PYGZgt{}\PYGZgt{} }\PYG{n}{np}\PYG{o}{.}\PYG{n}{isinf}\PYG{p}{(}\PYG{p}{[}\PYG{n}{np}\PYG{o}{.}\PYG{n}{inf}\PYG{p}{,} \PYG{o}{\PYGZhy{}}\PYG{n}{np}\PYG{o}{.}\PYG{n}{inf}\PYG{p}{,} \PYG{l+m+mf}{1.0}\PYG{p}{,} \PYG{n}{np}\PYG{o}{.}\PYG{n}{nan}\PYG{p}{]}\PYG{p}{)}
\PYG{g+go}{array([ True,  True, False, False])}
\end{sphinxVerbatim}

\begin{sphinxVerbatim}[commandchars=\\\{\}]
\PYG{g+gp}{\PYGZgt{}\PYGZgt{}\PYGZgt{} }\PYG{n}{x} \PYG{o}{=} \PYG{n}{np}\PYG{o}{.}\PYG{n}{array}\PYG{p}{(}\PYG{p}{[}\PYG{o}{\PYGZhy{}}\PYG{n}{np}\PYG{o}{.}\PYG{n}{inf}\PYG{p}{,} \PYG{l+m+mf}{0.}\PYG{p}{,} \PYG{n}{np}\PYG{o}{.}\PYG{n}{inf}\PYG{p}{]}\PYG{p}{)}
\PYG{g+gp}{\PYGZgt{}\PYGZgt{}\PYGZgt{} }\PYG{n}{y} \PYG{o}{=} \PYG{n}{np}\PYG{o}{.}\PYG{n}{array}\PYG{p}{(}\PYG{p}{[}\PYG{l+m+mi}{2}\PYG{p}{,} \PYG{l+m+mi}{2}\PYG{p}{,} \PYG{l+m+mi}{2}\PYG{p}{]}\PYG{p}{)}
\PYG{g+gp}{\PYGZgt{}\PYGZgt{}\PYGZgt{} }\PYG{n}{np}\PYG{o}{.}\PYG{n}{isinf}\PYG{p}{(}\PYG{n}{x}\PYG{p}{,} \PYG{n}{y}\PYG{p}{)}
\PYG{g+go}{array([1, 0, 1])}
\PYG{g+gp}{\PYGZgt{}\PYGZgt{}\PYGZgt{} }\PYG{n}{y}
\PYG{g+go}{array([1, 0, 1])}
\end{sphinxVerbatim}

\end{fulllineitems}

\index{isnan() (in module symjax.tensor)@\spxentry{isnan()}\spxextra{in module symjax.tensor}}

\begin{fulllineitems}
\phantomsection\label{\detokenize{modules/tensor:symjax.tensor.isnan}}\pysiglinewithargsret{\sphinxbfcode{\sphinxupquote{isnan}}}{\emph{\DUrole{n}{x}}}{}
Test element\sphinxhyphen{}wise for NaN and return result as a boolean array.

LAX\sphinxhyphen{}backend implementation of {\hyperref[\detokenize{modules/tensor:symjax.tensor.isnan}]{\sphinxcrossref{\sphinxcode{\sphinxupquote{isnan()}}}}}.
ADDITIONOriginal docstring below.

LAX\sphinxhyphen{}backend implementation of {\hyperref[\detokenize{modules/tensor:symjax.tensor.isnan}]{\sphinxcrossref{\sphinxcode{\sphinxupquote{isnan()}}}}}.
Original docstring below.

isnan(x, /, out=None, {\color{red}\bfseries{}*}, where=True, casting=’same\_kind’, order=’K’, dtype=None, subok=True{[}, signature, extobj{]})
\begin{quote}\begin{description}
\item[{Returns}] \leavevmode
\sphinxstylestrong{y} \textendash{} True where \sphinxcode{\sphinxupquote{x}} is NaN, false otherwise.
This is a scalar if \sphinxtitleref{x} is a scalar.

\item[{Return type}] \leavevmode
ndarray or bool

\end{description}\end{quote}

\sphinxstrong{See also:}

{\hyperref[\detokenize{modules/tensor:symjax.tensor.isinf}]{\sphinxcrossref{\sphinxcode{\sphinxupquote{isinf()}}}}}, \sphinxcode{\sphinxupquote{isneginf()}}, \sphinxcode{\sphinxupquote{isposinf()}}, {\hyperref[\detokenize{modules/tensor:symjax.tensor.isfinite}]{\sphinxcrossref{\sphinxcode{\sphinxupquote{isfinite()}}}}}, \sphinxcode{\sphinxupquote{isnat()}}

\subsubsection*{Notes}

NumPy uses the IEEE Standard for Binary Floating\sphinxhyphen{}Point for Arithmetic
(IEEE 754). This means that Not a Number is not equivalent to infinity.
\subsubsection*{Examples}

\begin{sphinxVerbatim}[commandchars=\\\{\}]
\PYG{g+gp}{\PYGZgt{}\PYGZgt{}\PYGZgt{} }\PYG{n}{np}\PYG{o}{.}\PYG{n}{isnan}\PYG{p}{(}\PYG{n}{np}\PYG{o}{.}\PYG{n}{nan}\PYG{p}{)}
\PYG{g+go}{True}
\PYG{g+gp}{\PYGZgt{}\PYGZgt{}\PYGZgt{} }\PYG{n}{np}\PYG{o}{.}\PYG{n}{isnan}\PYG{p}{(}\PYG{n}{np}\PYG{o}{.}\PYG{n}{inf}\PYG{p}{)}
\PYG{g+go}{False}
\PYG{g+gp}{\PYGZgt{}\PYGZgt{}\PYGZgt{} }\PYG{n}{np}\PYG{o}{.}\PYG{n}{isnan}\PYG{p}{(}\PYG{p}{[}\PYG{n}{np}\PYG{o}{.}\PYG{n}{log}\PYG{p}{(}\PYG{o}{\PYGZhy{}}\PYG{l+m+mf}{1.}\PYG{p}{)}\PYG{p}{,}\PYG{l+m+mf}{1.}\PYG{p}{,}\PYG{n}{np}\PYG{o}{.}\PYG{n}{log}\PYG{p}{(}\PYG{l+m+mi}{0}\PYG{p}{)}\PYG{p}{]}\PYG{p}{)}
\PYG{g+go}{array([ True, False, False])}
\end{sphinxVerbatim}

\end{fulllineitems}

\index{isreal() (in module symjax.tensor)@\spxentry{isreal()}\spxextra{in module symjax.tensor}}

\begin{fulllineitems}
\phantomsection\label{\detokenize{modules/tensor:symjax.tensor.isreal}}\pysiglinewithargsret{\sphinxbfcode{\sphinxupquote{isreal}}}{\emph{\DUrole{n}{x}}}{}
Returns a bool array, where True if input element is real.

LAX\sphinxhyphen{}backend implementation of {\hyperref[\detokenize{modules/tensor:symjax.tensor.isreal}]{\sphinxcrossref{\sphinxcode{\sphinxupquote{isreal()}}}}}.
ADDITIONOriginal docstring below.

LAX\sphinxhyphen{}backend implementation of {\hyperref[\detokenize{modules/tensor:symjax.tensor.isreal}]{\sphinxcrossref{\sphinxcode{\sphinxupquote{isreal()}}}}}.
Original docstring below.

If element has complex type with zero complex part, the return value
for that element is True.
\begin{quote}\begin{description}
\item[{Returns}] \leavevmode
\sphinxstylestrong{out} \textendash{} Boolean array of same shape as \sphinxtitleref{x}.

\item[{Return type}] \leavevmode
ndarray, bool

\end{description}\end{quote}

\sphinxstrong{See also:}

{\hyperref[\detokenize{modules/tensor:symjax.tensor.iscomplex}]{\sphinxcrossref{\sphinxcode{\sphinxupquote{iscomplex()}}}}}
\begin{description}
\item[{\sphinxcode{\sphinxupquote{isrealobj()}}}] \leavevmode
Return True if x is not a complex type.

\end{description}

\subsubsection*{Examples}

\begin{sphinxVerbatim}[commandchars=\\\{\}]
\PYG{g+gp}{\PYGZgt{}\PYGZgt{}\PYGZgt{} }\PYG{n}{np}\PYG{o}{.}\PYG{n}{isreal}\PYG{p}{(}\PYG{p}{[}\PYG{l+m+mi}{1}\PYG{o}{+}\PYG{l+m+mi}{1}\PYG{n}{j}\PYG{p}{,} \PYG{l+m+mi}{1}\PYG{o}{+}\PYG{l+m+mi}{0}\PYG{n}{j}\PYG{p}{,} \PYG{l+m+mf}{4.5}\PYG{p}{,} \PYG{l+m+mi}{3}\PYG{p}{,} \PYG{l+m+mi}{2}\PYG{p}{,} \PYG{l+m+mi}{2}\PYG{n}{j}\PYG{p}{]}\PYG{p}{)}
\PYG{g+go}{array([False,  True,  True,  True,  True, False])}
\end{sphinxVerbatim}

\end{fulllineitems}

\index{isscalar() (in module symjax.tensor)@\spxentry{isscalar()}\spxextra{in module symjax.tensor}}

\begin{fulllineitems}
\phantomsection\label{\detokenize{modules/tensor:symjax.tensor.isscalar}}\pysiglinewithargsret{\sphinxbfcode{\sphinxupquote{isscalar}}}{\emph{\DUrole{n}{num}}}{}
Returns True if the type of \sphinxtitleref{element} is a scalar type.

LAX\sphinxhyphen{}backend implementation of {\hyperref[\detokenize{modules/tensor:symjax.tensor.isscalar}]{\sphinxcrossref{\sphinxcode{\sphinxupquote{isscalar()}}}}}.
ADDITIONOriginal docstring below.

LAX\sphinxhyphen{}backend implementation of {\hyperref[\detokenize{modules/tensor:symjax.tensor.isscalar}]{\sphinxcrossref{\sphinxcode{\sphinxupquote{isscalar()}}}}}.
Original docstring below.
\begin{quote}\begin{description}
\item[{Returns}] \leavevmode
\sphinxstylestrong{val} \textendash{} True if \sphinxtitleref{element} is a scalar type, False if it is not.

\item[{Return type}] \leavevmode
bool

\end{description}\end{quote}

\sphinxstrong{See also:}

\begin{description}
\item[{\sphinxcode{\sphinxupquote{ndim()}}}] \leavevmode
Get the number of dimensions of an array

\end{description}

\subsubsection*{Notes}

If you need a stricter way to identify a \sphinxstyleemphasis{numerical} scalar, use
\sphinxcode{\sphinxupquote{isinstance(x, numbers.Number)}}, as that returns \sphinxcode{\sphinxupquote{False}} for most
non\sphinxhyphen{}numerical elements such as strings.

In most cases \sphinxcode{\sphinxupquote{np.ndim(x) == 0}} should be used instead of this function,
as that will also return true for 0d arrays. This is how numpy overloads
functions in the style of the \sphinxcode{\sphinxupquote{dx}} arguments to \sphinxtitleref{gradient} and the \sphinxcode{\sphinxupquote{bins}}
argument to \sphinxtitleref{histogram}. Some key differences:

\begin{savenotes}\sphinxattablestart
\centering
\begin{tabulary}{\linewidth}[t]{|T|T|T|}
\hline
\sphinxstyletheadfamily 
x
&\sphinxstyletheadfamily 
\sphinxcode{\sphinxupquote{isscalar(x)}}
&\sphinxstyletheadfamily 
\sphinxcode{\sphinxupquote{np.ndim(x) == 0}}
\\
\hline
PEP 3141 numeric objects (including
builtins)
&
\sphinxcode{\sphinxupquote{True}}
&
\sphinxcode{\sphinxupquote{True}}
\\
\hline
builtin string and buffer objects
&
\sphinxcode{\sphinxupquote{True}}
&
\sphinxcode{\sphinxupquote{True}}
\\
\hline
other builtin objects, like
\sphinxtitleref{pathlib.Path}, \sphinxtitleref{Exception},
the result of \sphinxtitleref{re.compile}
&
\sphinxcode{\sphinxupquote{False}}
&
\sphinxcode{\sphinxupquote{True}}
\\
\hline
third\sphinxhyphen{}party objects like
\sphinxtitleref{matplotlib.figure.Figure}
&
\sphinxcode{\sphinxupquote{False}}
&
\sphinxcode{\sphinxupquote{True}}
\\
\hline
zero\sphinxhyphen{}dimensional numpy arrays
&
\sphinxcode{\sphinxupquote{False}}
&
\sphinxcode{\sphinxupquote{True}}
\\
\hline
other numpy arrays
&
\sphinxcode{\sphinxupquote{False}}
&
\sphinxcode{\sphinxupquote{False}}
\\
\hline
\sphinxtitleref{list}, \sphinxtitleref{tuple}, and other sequence
objects
&
\sphinxcode{\sphinxupquote{False}}
&
\sphinxcode{\sphinxupquote{False}}
\\
\hline
\end{tabulary}
\par
\sphinxattableend\end{savenotes}
\subsubsection*{Examples}

\begin{sphinxVerbatim}[commandchars=\\\{\}]
\PYG{g+gp}{\PYGZgt{}\PYGZgt{}\PYGZgt{} }\PYG{n}{np}\PYG{o}{.}\PYG{n}{isscalar}\PYG{p}{(}\PYG{l+m+mf}{3.1}\PYG{p}{)}
\PYG{g+go}{True}
\PYG{g+gp}{\PYGZgt{}\PYGZgt{}\PYGZgt{} }\PYG{n}{np}\PYG{o}{.}\PYG{n}{isscalar}\PYG{p}{(}\PYG{n}{np}\PYG{o}{.}\PYG{n}{array}\PYG{p}{(}\PYG{l+m+mf}{3.1}\PYG{p}{)}\PYG{p}{)}
\PYG{g+go}{False}
\PYG{g+gp}{\PYGZgt{}\PYGZgt{}\PYGZgt{} }\PYG{n}{np}\PYG{o}{.}\PYG{n}{isscalar}\PYG{p}{(}\PYG{p}{[}\PYG{l+m+mf}{3.1}\PYG{p}{]}\PYG{p}{)}
\PYG{g+go}{False}
\PYG{g+gp}{\PYGZgt{}\PYGZgt{}\PYGZgt{} }\PYG{n}{np}\PYG{o}{.}\PYG{n}{isscalar}\PYG{p}{(}\PYG{k+kc}{False}\PYG{p}{)}
\PYG{g+go}{True}
\PYG{g+gp}{\PYGZgt{}\PYGZgt{}\PYGZgt{} }\PYG{n}{np}\PYG{o}{.}\PYG{n}{isscalar}\PYG{p}{(}\PYG{l+s+s1}{\PYGZsq{}}\PYG{l+s+s1}{numpy}\PYG{l+s+s1}{\PYGZsq{}}\PYG{p}{)}
\PYG{g+go}{True}
\end{sphinxVerbatim}

NumPy supports PEP 3141 numbers:

\begin{sphinxVerbatim}[commandchars=\\\{\}]
\PYG{g+gp}{\PYGZgt{}\PYGZgt{}\PYGZgt{} }\PYG{k+kn}{from} \PYG{n+nn}{fractions} \PYG{k+kn}{import} \PYG{n}{Fraction}
\PYG{g+gp}{\PYGZgt{}\PYGZgt{}\PYGZgt{} }\PYG{n}{np}\PYG{o}{.}\PYG{n}{isscalar}\PYG{p}{(}\PYG{n}{Fraction}\PYG{p}{(}\PYG{l+m+mi}{5}\PYG{p}{,} \PYG{l+m+mi}{17}\PYG{p}{)}\PYG{p}{)}
\PYG{g+go}{True}
\PYG{g+gp}{\PYGZgt{}\PYGZgt{}\PYGZgt{} }\PYG{k+kn}{from} \PYG{n+nn}{numbers} \PYG{k+kn}{import} \PYG{n}{Number}
\PYG{g+gp}{\PYGZgt{}\PYGZgt{}\PYGZgt{} }\PYG{n}{np}\PYG{o}{.}\PYG{n}{isscalar}\PYG{p}{(}\PYG{n}{Number}\PYG{p}{(}\PYG{p}{)}\PYG{p}{)}
\PYG{g+go}{True}
\end{sphinxVerbatim}

\end{fulllineitems}

\index{ix\_() (in module symjax.tensor)@\spxentry{ix\_()}\spxextra{in module symjax.tensor}}

\begin{fulllineitems}
\phantomsection\label{\detokenize{modules/tensor:symjax.tensor.ix_}}\pysiglinewithargsret{\sphinxbfcode{\sphinxupquote{ix\_}}}{\emph{\DUrole{o}{*}\DUrole{n}{args}}}{}
Construct an open mesh from multiple sequences.

LAX\sphinxhyphen{}backend implementation of {\hyperref[\detokenize{modules/tensor:symjax.tensor.ix_}]{\sphinxcrossref{\sphinxcode{\sphinxupquote{ix\_()}}}}}.
ADDITIONOriginal docstring below.

LAX\sphinxhyphen{}backend implementation of {\hyperref[\detokenize{modules/tensor:symjax.tensor.ix_}]{\sphinxcrossref{\sphinxcode{\sphinxupquote{ix\_()}}}}}.
Original docstring below.

This function takes N 1\sphinxhyphen{}D sequences and returns N outputs with N
dimensions each, such that the shape is 1 in all but one dimension
and the dimension with the non\sphinxhyphen{}unit shape value cycles through all
N dimensions.

Using \sphinxtitleref{ix\_} one can quickly construct index arrays that will index
the cross product. \sphinxcode{\sphinxupquote{a{[}np.ix\_({[}1,3{]},{[}2,5{]}){]}}} returns the array
\sphinxcode{\sphinxupquote{{[}{[}a{[}1,2{]} a{[}1,5{]}{]}, {[}a{[}3,2{]} a{[}3,5{]}{]}{]}}}.
\begin{quote}\begin{description}
\item[{Parameters}] \leavevmode
\sphinxstyleliteralstrong{\sphinxupquote{args}} (\sphinxstyleliteralemphasis{\sphinxupquote{1\sphinxhyphen{}D sequences}}) \textendash{} Each sequence should be of integer or boolean type.
Boolean sequences will be interpreted as boolean masks for the
corresponding dimension (equivalent to passing in
\sphinxcode{\sphinxupquote{np.nonzero(boolean\_sequence)}}).

\item[{Returns}] \leavevmode
\sphinxstylestrong{out} \textendash{} N arrays with N dimensions each, with N the number of input
sequences. Together these arrays form an open mesh.

\item[{Return type}] \leavevmode
tuple of ndarrays

\end{description}\end{quote}

\sphinxstrong{See also:}

\sphinxcode{\sphinxupquote{ogrid()}}, \sphinxcode{\sphinxupquote{mgrid()}}, {\hyperref[\detokenize{modules/tensor:symjax.tensor.meshgrid}]{\sphinxcrossref{\sphinxcode{\sphinxupquote{meshgrid()}}}}}

\subsubsection*{Examples}

\begin{sphinxVerbatim}[commandchars=\\\{\}]
\PYG{g+gp}{\PYGZgt{}\PYGZgt{}\PYGZgt{} }\PYG{n}{a} \PYG{o}{=} \PYG{n}{np}\PYG{o}{.}\PYG{n}{arange}\PYG{p}{(}\PYG{l+m+mi}{10}\PYG{p}{)}\PYG{o}{.}\PYG{n}{reshape}\PYG{p}{(}\PYG{l+m+mi}{2}\PYG{p}{,} \PYG{l+m+mi}{5}\PYG{p}{)}
\PYG{g+gp}{\PYGZgt{}\PYGZgt{}\PYGZgt{} }\PYG{n}{a}
\PYG{g+go}{array([[0, 1, 2, 3, 4],}
\PYG{g+go}{       [5, 6, 7, 8, 9]])}
\PYG{g+gp}{\PYGZgt{}\PYGZgt{}\PYGZgt{} }\PYG{n}{ixgrid} \PYG{o}{=} \PYG{n}{np}\PYG{o}{.}\PYG{n}{ix\PYGZus{}}\PYG{p}{(}\PYG{p}{[}\PYG{l+m+mi}{0}\PYG{p}{,} \PYG{l+m+mi}{1}\PYG{p}{]}\PYG{p}{,} \PYG{p}{[}\PYG{l+m+mi}{2}\PYG{p}{,} \PYG{l+m+mi}{4}\PYG{p}{]}\PYG{p}{)}
\PYG{g+gp}{\PYGZgt{}\PYGZgt{}\PYGZgt{} }\PYG{n}{ixgrid}
\PYG{g+go}{(array([[0],}
\PYG{g+go}{       [1]]), array([[2, 4]]))}
\PYG{g+gp}{\PYGZgt{}\PYGZgt{}\PYGZgt{} }\PYG{n}{ixgrid}\PYG{p}{[}\PYG{l+m+mi}{0}\PYG{p}{]}\PYG{o}{.}\PYG{n}{shape}\PYG{p}{,} \PYG{n}{ixgrid}\PYG{p}{[}\PYG{l+m+mi}{1}\PYG{p}{]}\PYG{o}{.}\PYG{n}{shape}
\PYG{g+go}{((2, 1), (1, 2))}
\PYG{g+gp}{\PYGZgt{}\PYGZgt{}\PYGZgt{} }\PYG{n}{a}\PYG{p}{[}\PYG{n}{ixgrid}\PYG{p}{]}
\PYG{g+go}{array([[2, 4],}
\PYG{g+go}{       [7, 9]])}
\end{sphinxVerbatim}

\begin{sphinxVerbatim}[commandchars=\\\{\}]
\PYG{g+gp}{\PYGZgt{}\PYGZgt{}\PYGZgt{} }\PYG{n}{ixgrid} \PYG{o}{=} \PYG{n}{np}\PYG{o}{.}\PYG{n}{ix\PYGZus{}}\PYG{p}{(}\PYG{p}{[}\PYG{k+kc}{True}\PYG{p}{,} \PYG{k+kc}{True}\PYG{p}{]}\PYG{p}{,} \PYG{p}{[}\PYG{l+m+mi}{2}\PYG{p}{,} \PYG{l+m+mi}{4}\PYG{p}{]}\PYG{p}{)}
\PYG{g+gp}{\PYGZgt{}\PYGZgt{}\PYGZgt{} }\PYG{n}{a}\PYG{p}{[}\PYG{n}{ixgrid}\PYG{p}{]}
\PYG{g+go}{array([[2, 4],}
\PYG{g+go}{       [7, 9]])}
\PYG{g+gp}{\PYGZgt{}\PYGZgt{}\PYGZgt{} }\PYG{n}{ixgrid} \PYG{o}{=} \PYG{n}{np}\PYG{o}{.}\PYG{n}{ix\PYGZus{}}\PYG{p}{(}\PYG{p}{[}\PYG{k+kc}{True}\PYG{p}{,} \PYG{k+kc}{True}\PYG{p}{]}\PYG{p}{,} \PYG{p}{[}\PYG{k+kc}{False}\PYG{p}{,} \PYG{k+kc}{False}\PYG{p}{,} \PYG{k+kc}{True}\PYG{p}{,} \PYG{k+kc}{False}\PYG{p}{,} \PYG{k+kc}{True}\PYG{p}{]}\PYG{p}{)}
\PYG{g+gp}{\PYGZgt{}\PYGZgt{}\PYGZgt{} }\PYG{n}{a}\PYG{p}{[}\PYG{n}{ixgrid}\PYG{p}{]}
\PYG{g+go}{array([[2, 4],}
\PYG{g+go}{       [7, 9]])}
\end{sphinxVerbatim}

\end{fulllineitems}

\index{kron() (in module symjax.tensor)@\spxentry{kron()}\spxextra{in module symjax.tensor}}

\begin{fulllineitems}
\phantomsection\label{\detokenize{modules/tensor:symjax.tensor.kron}}\pysiglinewithargsret{\sphinxbfcode{\sphinxupquote{kron}}}{\emph{\DUrole{n}{a}}, \emph{\DUrole{n}{b}}}{}
Kronecker product of two arrays.

LAX\sphinxhyphen{}backend implementation of {\hyperref[\detokenize{modules/tensor:symjax.tensor.kron}]{\sphinxcrossref{\sphinxcode{\sphinxupquote{kron()}}}}}.
ADDITIONOriginal docstring below.

LAX\sphinxhyphen{}backend implementation of {\hyperref[\detokenize{modules/tensor:symjax.tensor.kron}]{\sphinxcrossref{\sphinxcode{\sphinxupquote{kron()}}}}}.
Original docstring below.

Computes the Kronecker product, a composite array made of blocks of the
second array scaled by the first.
\begin{quote}\begin{description}
\item[{Returns}] \leavevmode
\sphinxstylestrong{out}

\item[{Return type}] \leavevmode
ndarray

\end{description}\end{quote}

\sphinxstrong{See also:}

\begin{description}
\item[{{\hyperref[\detokenize{modules/tensor:symjax.tensor.outer}]{\sphinxcrossref{\sphinxcode{\sphinxupquote{outer()}}}}}}] \leavevmode
The outer product

\end{description}

\subsubsection*{Notes}

The function assumes that the number of dimensions of \sphinxtitleref{a} and \sphinxtitleref{b}
are the same, if necessary prepending the smallest with ones.
If \sphinxtitleref{a.shape = (r0,r1,..,rN)} and \sphinxtitleref{b.shape = (s0,s1,…,sN)},
the Kronecker product has shape \sphinxtitleref{(r0*s0, r1*s1, …, rN*SN)}.
The elements are products of elements from \sphinxtitleref{a} and \sphinxtitleref{b}, organized
explicitly by:

\begin{sphinxVerbatim}[commandchars=\\\{\}]
\PYG{n}{kron}\PYG{p}{(}\PYG{n}{a}\PYG{p}{,}\PYG{n}{b}\PYG{p}{)}\PYG{p}{[}\PYG{n}{k0}\PYG{p}{,}\PYG{n}{k1}\PYG{p}{,}\PYG{o}{.}\PYG{o}{.}\PYG{o}{.}\PYG{p}{,}\PYG{n}{kN}\PYG{p}{]} \PYG{o}{=} \PYG{n}{a}\PYG{p}{[}\PYG{n}{i0}\PYG{p}{,}\PYG{n}{i1}\PYG{p}{,}\PYG{o}{.}\PYG{o}{.}\PYG{o}{.}\PYG{p}{,}\PYG{n}{iN}\PYG{p}{]} \PYG{o}{*} \PYG{n}{b}\PYG{p}{[}\PYG{n}{j0}\PYG{p}{,}\PYG{n}{j1}\PYG{p}{,}\PYG{o}{.}\PYG{o}{.}\PYG{o}{.}\PYG{p}{,}\PYG{n}{jN}\PYG{p}{]}
\end{sphinxVerbatim}

where:

\begin{sphinxVerbatim}[commandchars=\\\{\}]
\PYG{n}{kt} \PYG{o}{=} \PYG{n}{it} \PYG{o}{*} \PYG{n}{st} \PYG{o}{+} \PYG{n}{jt}\PYG{p}{,}  \PYG{n}{t} \PYG{o}{=} \PYG{l+m+mi}{0}\PYG{p}{,}\PYG{o}{.}\PYG{o}{.}\PYG{o}{.}\PYG{p}{,}\PYG{n}{N}
\end{sphinxVerbatim}

In the common 2\sphinxhyphen{}D case (N=1), the block structure can be visualized:

\begin{sphinxVerbatim}[commandchars=\\\{\}]
\PYG{p}{[}\PYG{p}{[} \PYG{n}{a}\PYG{p}{[}\PYG{l+m+mi}{0}\PYG{p}{,}\PYG{l+m+mi}{0}\PYG{p}{]}\PYG{o}{*}\PYG{n}{b}\PYG{p}{,}   \PYG{n}{a}\PYG{p}{[}\PYG{l+m+mi}{0}\PYG{p}{,}\PYG{l+m+mi}{1}\PYG{p}{]}\PYG{o}{*}\PYG{n}{b}\PYG{p}{,}  \PYG{o}{.}\PYG{o}{.}\PYG{o}{.} \PYG{p}{,} \PYG{n}{a}\PYG{p}{[}\PYG{l+m+mi}{0}\PYG{p}{,}\PYG{o}{\PYGZhy{}}\PYG{l+m+mi}{1}\PYG{p}{]}\PYG{o}{*}\PYG{n}{b}  \PYG{p}{]}\PYG{p}{,}
 \PYG{p}{[}  \PYG{o}{.}\PYG{o}{.}\PYG{o}{.}                              \PYG{o}{.}\PYG{o}{.}\PYG{o}{.}   \PYG{p}{]}\PYG{p}{,}
 \PYG{p}{[} \PYG{n}{a}\PYG{p}{[}\PYG{o}{\PYGZhy{}}\PYG{l+m+mi}{1}\PYG{p}{,}\PYG{l+m+mi}{0}\PYG{p}{]}\PYG{o}{*}\PYG{n}{b}\PYG{p}{,}  \PYG{n}{a}\PYG{p}{[}\PYG{o}{\PYGZhy{}}\PYG{l+m+mi}{1}\PYG{p}{,}\PYG{l+m+mi}{1}\PYG{p}{]}\PYG{o}{*}\PYG{n}{b}\PYG{p}{,} \PYG{o}{.}\PYG{o}{.}\PYG{o}{.} \PYG{p}{,} \PYG{n}{a}\PYG{p}{[}\PYG{o}{\PYGZhy{}}\PYG{l+m+mi}{1}\PYG{p}{,}\PYG{o}{\PYGZhy{}}\PYG{l+m+mi}{1}\PYG{p}{]}\PYG{o}{*}\PYG{n}{b} \PYG{p}{]}\PYG{p}{]}
\end{sphinxVerbatim}
\subsubsection*{Examples}

\begin{sphinxVerbatim}[commandchars=\\\{\}]
\PYG{g+gp}{\PYGZgt{}\PYGZgt{}\PYGZgt{} }\PYG{n}{np}\PYG{o}{.}\PYG{n}{kron}\PYG{p}{(}\PYG{p}{[}\PYG{l+m+mi}{1}\PYG{p}{,}\PYG{l+m+mi}{10}\PYG{p}{,}\PYG{l+m+mi}{100}\PYG{p}{]}\PYG{p}{,} \PYG{p}{[}\PYG{l+m+mi}{5}\PYG{p}{,}\PYG{l+m+mi}{6}\PYG{p}{,}\PYG{l+m+mi}{7}\PYG{p}{]}\PYG{p}{)}
\PYG{g+go}{array([  5,   6,   7, ..., 500, 600, 700])}
\PYG{g+gp}{\PYGZgt{}\PYGZgt{}\PYGZgt{} }\PYG{n}{np}\PYG{o}{.}\PYG{n}{kron}\PYG{p}{(}\PYG{p}{[}\PYG{l+m+mi}{5}\PYG{p}{,}\PYG{l+m+mi}{6}\PYG{p}{,}\PYG{l+m+mi}{7}\PYG{p}{]}\PYG{p}{,} \PYG{p}{[}\PYG{l+m+mi}{1}\PYG{p}{,}\PYG{l+m+mi}{10}\PYG{p}{,}\PYG{l+m+mi}{100}\PYG{p}{]}\PYG{p}{)}
\PYG{g+go}{array([  5,  50, 500, ...,   7,  70, 700])}
\end{sphinxVerbatim}

\begin{sphinxVerbatim}[commandchars=\\\{\}]
\PYG{g+gp}{\PYGZgt{}\PYGZgt{}\PYGZgt{} }\PYG{n}{np}\PYG{o}{.}\PYG{n}{kron}\PYG{p}{(}\PYG{n}{np}\PYG{o}{.}\PYG{n}{eye}\PYG{p}{(}\PYG{l+m+mi}{2}\PYG{p}{)}\PYG{p}{,} \PYG{n}{np}\PYG{o}{.}\PYG{n}{ones}\PYG{p}{(}\PYG{p}{(}\PYG{l+m+mi}{2}\PYG{p}{,}\PYG{l+m+mi}{2}\PYG{p}{)}\PYG{p}{)}\PYG{p}{)}
\PYG{g+go}{array([[1.,  1.,  0.,  0.],}
\PYG{g+go}{       [1.,  1.,  0.,  0.],}
\PYG{g+go}{       [0.,  0.,  1.,  1.],}
\PYG{g+go}{       [0.,  0.,  1.,  1.]])}
\end{sphinxVerbatim}

\begin{sphinxVerbatim}[commandchars=\\\{\}]
\PYG{g+gp}{\PYGZgt{}\PYGZgt{}\PYGZgt{} }\PYG{n}{a} \PYG{o}{=} \PYG{n}{np}\PYG{o}{.}\PYG{n}{arange}\PYG{p}{(}\PYG{l+m+mi}{100}\PYG{p}{)}\PYG{o}{.}\PYG{n}{reshape}\PYG{p}{(}\PYG{p}{(}\PYG{l+m+mi}{2}\PYG{p}{,}\PYG{l+m+mi}{5}\PYG{p}{,}\PYG{l+m+mi}{2}\PYG{p}{,}\PYG{l+m+mi}{5}\PYG{p}{)}\PYG{p}{)}
\PYG{g+gp}{\PYGZgt{}\PYGZgt{}\PYGZgt{} }\PYG{n}{b} \PYG{o}{=} \PYG{n}{np}\PYG{o}{.}\PYG{n}{arange}\PYG{p}{(}\PYG{l+m+mi}{24}\PYG{p}{)}\PYG{o}{.}\PYG{n}{reshape}\PYG{p}{(}\PYG{p}{(}\PYG{l+m+mi}{2}\PYG{p}{,}\PYG{l+m+mi}{3}\PYG{p}{,}\PYG{l+m+mi}{4}\PYG{p}{)}\PYG{p}{)}
\PYG{g+gp}{\PYGZgt{}\PYGZgt{}\PYGZgt{} }\PYG{n}{c} \PYG{o}{=} \PYG{n}{np}\PYG{o}{.}\PYG{n}{kron}\PYG{p}{(}\PYG{n}{a}\PYG{p}{,}\PYG{n}{b}\PYG{p}{)}
\PYG{g+gp}{\PYGZgt{}\PYGZgt{}\PYGZgt{} }\PYG{n}{c}\PYG{o}{.}\PYG{n}{shape}
\PYG{g+go}{(2, 10, 6, 20)}
\PYG{g+gp}{\PYGZgt{}\PYGZgt{}\PYGZgt{} }\PYG{n}{I} \PYG{o}{=} \PYG{p}{(}\PYG{l+m+mi}{1}\PYG{p}{,}\PYG{l+m+mi}{3}\PYG{p}{,}\PYG{l+m+mi}{0}\PYG{p}{,}\PYG{l+m+mi}{2}\PYG{p}{)}
\PYG{g+gp}{\PYGZgt{}\PYGZgt{}\PYGZgt{} }\PYG{n}{J} \PYG{o}{=} \PYG{p}{(}\PYG{l+m+mi}{0}\PYG{p}{,}\PYG{l+m+mi}{2}\PYG{p}{,}\PYG{l+m+mi}{1}\PYG{p}{)}
\PYG{g+gp}{\PYGZgt{}\PYGZgt{}\PYGZgt{} }\PYG{n}{J1} \PYG{o}{=} \PYG{p}{(}\PYG{l+m+mi}{0}\PYG{p}{,}\PYG{p}{)} \PYG{o}{+} \PYG{n}{J}             \PYG{c+c1}{\PYGZsh{} extend to ndim=4}
\PYG{g+gp}{\PYGZgt{}\PYGZgt{}\PYGZgt{} }\PYG{n}{S1} \PYG{o}{=} \PYG{p}{(}\PYG{l+m+mi}{1}\PYG{p}{,}\PYG{p}{)} \PYG{o}{+} \PYG{n}{b}\PYG{o}{.}\PYG{n}{shape}
\PYG{g+gp}{\PYGZgt{}\PYGZgt{}\PYGZgt{} }\PYG{n}{K} \PYG{o}{=} \PYG{n+nb}{tuple}\PYG{p}{(}\PYG{n}{np}\PYG{o}{.}\PYG{n}{array}\PYG{p}{(}\PYG{n}{I}\PYG{p}{)} \PYG{o}{*} \PYG{n}{np}\PYG{o}{.}\PYG{n}{array}\PYG{p}{(}\PYG{n}{S1}\PYG{p}{)} \PYG{o}{+} \PYG{n}{np}\PYG{o}{.}\PYG{n}{array}\PYG{p}{(}\PYG{n}{J1}\PYG{p}{)}\PYG{p}{)}
\PYG{g+gp}{\PYGZgt{}\PYGZgt{}\PYGZgt{} }\PYG{n}{c}\PYG{p}{[}\PYG{n}{K}\PYG{p}{]} \PYG{o}{==} \PYG{n}{a}\PYG{p}{[}\PYG{n}{I}\PYG{p}{]}\PYG{o}{*}\PYG{n}{b}\PYG{p}{[}\PYG{n}{J}\PYG{p}{]}
\PYG{g+go}{True}
\end{sphinxVerbatim}

\end{fulllineitems}

\index{lcm() (in module symjax.tensor)@\spxentry{lcm()}\spxextra{in module symjax.tensor}}

\begin{fulllineitems}
\phantomsection\label{\detokenize{modules/tensor:symjax.tensor.lcm}}\pysiglinewithargsret{\sphinxbfcode{\sphinxupquote{lcm}}}{\emph{\DUrole{n}{x1}}, \emph{\DUrole{n}{x2}}}{}
Returns the lowest common multiple of \sphinxcode{\sphinxupquote{|x1|}} and \sphinxcode{\sphinxupquote{|x2|}}

LAX\sphinxhyphen{}backend implementation of {\hyperref[\detokenize{modules/tensor:symjax.tensor.lcm}]{\sphinxcrossref{\sphinxcode{\sphinxupquote{lcm()}}}}}.
ADDITIONOriginal docstring below.

LAX\sphinxhyphen{}backend implementation of {\hyperref[\detokenize{modules/tensor:symjax.tensor.lcm}]{\sphinxcrossref{\sphinxcode{\sphinxupquote{lcm()}}}}}.
Original docstring below.

lcm(x1, x2, /, out=None, {\color{red}\bfseries{}*}, where=True, casting=’same\_kind’, order=’K’, dtype=None, subok=True{[}, signature, extobj{]})
\begin{quote}\begin{description}
\item[{Returns}] \leavevmode
\sphinxstylestrong{y} \textendash{} The lowest common multiple of the absolute value of the inputs
This is a scalar if both \sphinxtitleref{x1} and \sphinxtitleref{x2} are scalars.

\item[{Return type}] \leavevmode
ndarray or scalar

\end{description}\end{quote}

\sphinxstrong{See also:}

\begin{description}
\item[{{\hyperref[\detokenize{modules/tensor:symjax.tensor.gcd}]{\sphinxcrossref{\sphinxcode{\sphinxupquote{gcd()}}}}}}] \leavevmode
The greatest common divisor

\end{description}

\subsubsection*{Examples}

\begin{sphinxVerbatim}[commandchars=\\\{\}]
\PYG{g+gp}{\PYGZgt{}\PYGZgt{}\PYGZgt{} }\PYG{n}{np}\PYG{o}{.}\PYG{n}{lcm}\PYG{p}{(}\PYG{l+m+mi}{12}\PYG{p}{,} \PYG{l+m+mi}{20}\PYG{p}{)}
\PYG{g+go}{60}
\PYG{g+gp}{\PYGZgt{}\PYGZgt{}\PYGZgt{} }\PYG{n}{np}\PYG{o}{.}\PYG{n}{lcm}\PYG{o}{.}\PYG{n}{reduce}\PYG{p}{(}\PYG{p}{[}\PYG{l+m+mi}{3}\PYG{p}{,} \PYG{l+m+mi}{12}\PYG{p}{,} \PYG{l+m+mi}{20}\PYG{p}{]}\PYG{p}{)}
\PYG{g+go}{60}
\PYG{g+gp}{\PYGZgt{}\PYGZgt{}\PYGZgt{} }\PYG{n}{np}\PYG{o}{.}\PYG{n}{lcm}\PYG{o}{.}\PYG{n}{reduce}\PYG{p}{(}\PYG{p}{[}\PYG{l+m+mi}{40}\PYG{p}{,} \PYG{l+m+mi}{12}\PYG{p}{,} \PYG{l+m+mi}{20}\PYG{p}{]}\PYG{p}{)}
\PYG{g+go}{120}
\PYG{g+gp}{\PYGZgt{}\PYGZgt{}\PYGZgt{} }\PYG{n}{np}\PYG{o}{.}\PYG{n}{lcm}\PYG{p}{(}\PYG{n}{np}\PYG{o}{.}\PYG{n}{arange}\PYG{p}{(}\PYG{l+m+mi}{6}\PYG{p}{)}\PYG{p}{,} \PYG{l+m+mi}{20}\PYG{p}{)}
\PYG{g+go}{array([ 0, 20, 20, 60, 20, 20])}
\end{sphinxVerbatim}

\end{fulllineitems}

\index{left\_shift() (in module symjax.tensor)@\spxentry{left\_shift()}\spxextra{in module symjax.tensor}}

\begin{fulllineitems}
\phantomsection\label{\detokenize{modules/tensor:symjax.tensor.left_shift}}\pysiglinewithargsret{\sphinxbfcode{\sphinxupquote{left\_shift}}}{\emph{\DUrole{n}{x1}}, \emph{\DUrole{n}{x2}}}{}
Shift the bits of an integer to the left.

LAX\sphinxhyphen{}backend implementation of {\hyperref[\detokenize{modules/tensor:symjax.tensor.left_shift}]{\sphinxcrossref{\sphinxcode{\sphinxupquote{left\_shift()}}}}}.
ADDITIONOriginal docstring below.

LAX\sphinxhyphen{}backend implementation of {\hyperref[\detokenize{modules/tensor:symjax.tensor.left_shift}]{\sphinxcrossref{\sphinxcode{\sphinxupquote{left\_shift()}}}}}.
Original docstring below.

left\_shift(x1, x2, /, out=None, {\color{red}\bfseries{}*}, where=True, casting=’same\_kind’, order=’K’, dtype=None, subok=True{[}, signature, extobj{]})

Bits are shifted to the left by appending \sphinxtitleref{x2} 0s at the right of \sphinxtitleref{x1}.
Since the internal representation of numbers is in binary format, this
operation is equivalent to multiplying \sphinxtitleref{x1} by \sphinxcode{\sphinxupquote{2**x2}}.
\begin{quote}\begin{description}
\item[{Returns}] \leavevmode
\sphinxstylestrong{out} \textendash{} Return \sphinxtitleref{x1} with bits shifted \sphinxtitleref{x2} times to the left.
This is a scalar if both \sphinxtitleref{x1} and \sphinxtitleref{x2} are scalars.

\item[{Return type}] \leavevmode
array of integer type

\end{description}\end{quote}

\sphinxstrong{See also:}

\begin{description}
\item[{\sphinxcode{\sphinxupquote{right\_shift()}}}] \leavevmode
Shift the bits of an integer to the right.

\item[{\sphinxcode{\sphinxupquote{binary\_repr()}}}] \leavevmode
Return the binary representation of the input number as a string.

\end{description}

\subsubsection*{Examples}

\begin{sphinxVerbatim}[commandchars=\\\{\}]
\PYG{g+gp}{\PYGZgt{}\PYGZgt{}\PYGZgt{} }\PYG{n}{np}\PYG{o}{.}\PYG{n}{binary\PYGZus{}repr}\PYG{p}{(}\PYG{l+m+mi}{5}\PYG{p}{)}
\PYG{g+go}{\PYGZsq{}101\PYGZsq{}}
\PYG{g+gp}{\PYGZgt{}\PYGZgt{}\PYGZgt{} }\PYG{n}{np}\PYG{o}{.}\PYG{n}{left\PYGZus{}shift}\PYG{p}{(}\PYG{l+m+mi}{5}\PYG{p}{,} \PYG{l+m+mi}{2}\PYG{p}{)}
\PYG{g+go}{20}
\PYG{g+gp}{\PYGZgt{}\PYGZgt{}\PYGZgt{} }\PYG{n}{np}\PYG{o}{.}\PYG{n}{binary\PYGZus{}repr}\PYG{p}{(}\PYG{l+m+mi}{20}\PYG{p}{)}
\PYG{g+go}{\PYGZsq{}10100\PYGZsq{}}
\end{sphinxVerbatim}

\begin{sphinxVerbatim}[commandchars=\\\{\}]
\PYG{g+gp}{\PYGZgt{}\PYGZgt{}\PYGZgt{} }\PYG{n}{np}\PYG{o}{.}\PYG{n}{left\PYGZus{}shift}\PYG{p}{(}\PYG{l+m+mi}{5}\PYG{p}{,} \PYG{p}{[}\PYG{l+m+mi}{1}\PYG{p}{,}\PYG{l+m+mi}{2}\PYG{p}{,}\PYG{l+m+mi}{3}\PYG{p}{]}\PYG{p}{)}
\PYG{g+go}{array([10, 20, 40])}
\end{sphinxVerbatim}

\end{fulllineitems}

\index{less() (in module symjax.tensor)@\spxentry{less()}\spxextra{in module symjax.tensor}}

\begin{fulllineitems}
\phantomsection\label{\detokenize{modules/tensor:symjax.tensor.less}}\pysiglinewithargsret{\sphinxbfcode{\sphinxupquote{less}}}{\emph{\DUrole{n}{x1}}, \emph{\DUrole{n}{x2}}}{}
Return the truth value of (x1 \textless{} x2) element\sphinxhyphen{}wise.

LAX\sphinxhyphen{}backend implementation of {\hyperref[\detokenize{modules/tensor:symjax.tensor.less}]{\sphinxcrossref{\sphinxcode{\sphinxupquote{less()}}}}}.
ADDITIONOriginal docstring below.

LAX\sphinxhyphen{}backend implementation of {\hyperref[\detokenize{modules/tensor:symjax.tensor.less}]{\sphinxcrossref{\sphinxcode{\sphinxupquote{less()}}}}}.
Original docstring below.

less(x1, x2, /, out=None, {\color{red}\bfseries{}*}, where=True, casting=’same\_kind’, order=’K’, dtype=None, subok=True{[}, signature, extobj{]})
\begin{quote}\begin{description}
\item[{Returns}] \leavevmode
\sphinxstylestrong{out} \textendash{} Output array, element\sphinxhyphen{}wise comparison of \sphinxtitleref{x1} and \sphinxtitleref{x2}.
Typically of type bool, unless \sphinxcode{\sphinxupquote{dtype=object}} is passed.
This is a scalar if both \sphinxtitleref{x1} and \sphinxtitleref{x2} are scalars.

\item[{Return type}] \leavevmode
ndarray or scalar

\end{description}\end{quote}

\sphinxstrong{See also:}

{\hyperref[\detokenize{modules/tensor:symjax.tensor.greater}]{\sphinxcrossref{\sphinxcode{\sphinxupquote{greater()}}}}}, {\hyperref[\detokenize{modules/tensor:symjax.tensor.less_equal}]{\sphinxcrossref{\sphinxcode{\sphinxupquote{less\_equal()}}}}}, {\hyperref[\detokenize{modules/tensor:symjax.tensor.greater_equal}]{\sphinxcrossref{\sphinxcode{\sphinxupquote{greater\_equal()}}}}}, {\hyperref[\detokenize{modules/tensor:symjax.tensor.equal}]{\sphinxcrossref{\sphinxcode{\sphinxupquote{equal()}}}}}, {\hyperref[\detokenize{modules/tensor:symjax.tensor.not_equal}]{\sphinxcrossref{\sphinxcode{\sphinxupquote{not\_equal()}}}}}

\subsubsection*{Examples}

\begin{sphinxVerbatim}[commandchars=\\\{\}]
\PYG{g+gp}{\PYGZgt{}\PYGZgt{}\PYGZgt{} }\PYG{n}{np}\PYG{o}{.}\PYG{n}{less}\PYG{p}{(}\PYG{p}{[}\PYG{l+m+mi}{1}\PYG{p}{,} \PYG{l+m+mi}{2}\PYG{p}{]}\PYG{p}{,} \PYG{p}{[}\PYG{l+m+mi}{2}\PYG{p}{,} \PYG{l+m+mi}{2}\PYG{p}{]}\PYG{p}{)}
\PYG{g+go}{array([ True, False])}
\end{sphinxVerbatim}

\end{fulllineitems}

\index{less\_equal() (in module symjax.tensor)@\spxentry{less\_equal()}\spxextra{in module symjax.tensor}}

\begin{fulllineitems}
\phantomsection\label{\detokenize{modules/tensor:symjax.tensor.less_equal}}\pysiglinewithargsret{\sphinxbfcode{\sphinxupquote{less\_equal}}}{\emph{\DUrole{n}{x1}}, \emph{\DUrole{n}{x2}}}{}
Return the truth value of (x1 =\textless{} x2) element\sphinxhyphen{}wise.

LAX\sphinxhyphen{}backend implementation of {\hyperref[\detokenize{modules/tensor:symjax.tensor.less_equal}]{\sphinxcrossref{\sphinxcode{\sphinxupquote{less\_equal()}}}}}.
ADDITIONOriginal docstring below.

LAX\sphinxhyphen{}backend implementation of {\hyperref[\detokenize{modules/tensor:symjax.tensor.less_equal}]{\sphinxcrossref{\sphinxcode{\sphinxupquote{less\_equal()}}}}}.
Original docstring below.

less\_equal(x1, x2, /, out=None, {\color{red}\bfseries{}*}, where=True, casting=’same\_kind’, order=’K’, dtype=None, subok=True{[}, signature, extobj{]})
\begin{quote}\begin{description}
\item[{Returns}] \leavevmode
\sphinxstylestrong{out} \textendash{} Output array, element\sphinxhyphen{}wise comparison of \sphinxtitleref{x1} and \sphinxtitleref{x2}.
Typically of type bool, unless \sphinxcode{\sphinxupquote{dtype=object}} is passed.
This is a scalar if both \sphinxtitleref{x1} and \sphinxtitleref{x2} are scalars.

\item[{Return type}] \leavevmode
ndarray or scalar

\end{description}\end{quote}

\sphinxstrong{See also:}

{\hyperref[\detokenize{modules/tensor:symjax.tensor.greater}]{\sphinxcrossref{\sphinxcode{\sphinxupquote{greater()}}}}}, {\hyperref[\detokenize{modules/tensor:symjax.tensor.less}]{\sphinxcrossref{\sphinxcode{\sphinxupquote{less()}}}}}, {\hyperref[\detokenize{modules/tensor:symjax.tensor.greater_equal}]{\sphinxcrossref{\sphinxcode{\sphinxupquote{greater\_equal()}}}}}, {\hyperref[\detokenize{modules/tensor:symjax.tensor.equal}]{\sphinxcrossref{\sphinxcode{\sphinxupquote{equal()}}}}}, {\hyperref[\detokenize{modules/tensor:symjax.tensor.not_equal}]{\sphinxcrossref{\sphinxcode{\sphinxupquote{not\_equal()}}}}}

\subsubsection*{Examples}

\begin{sphinxVerbatim}[commandchars=\\\{\}]
\PYG{g+gp}{\PYGZgt{}\PYGZgt{}\PYGZgt{} }\PYG{n}{np}\PYG{o}{.}\PYG{n}{less\PYGZus{}equal}\PYG{p}{(}\PYG{p}{[}\PYG{l+m+mi}{4}\PYG{p}{,} \PYG{l+m+mi}{2}\PYG{p}{,} \PYG{l+m+mi}{1}\PYG{p}{]}\PYG{p}{,} \PYG{p}{[}\PYG{l+m+mi}{2}\PYG{p}{,} \PYG{l+m+mi}{2}\PYG{p}{,} \PYG{l+m+mi}{2}\PYG{p}{]}\PYG{p}{)}
\PYG{g+go}{array([False,  True,  True])}
\end{sphinxVerbatim}

\end{fulllineitems}

\index{linspace() (in module symjax.tensor)@\spxentry{linspace()}\spxextra{in module symjax.tensor}}

\begin{fulllineitems}
\phantomsection\label{\detokenize{modules/tensor:symjax.tensor.linspace}}\pysiglinewithargsret{\sphinxbfcode{\sphinxupquote{linspace}}}{\emph{\DUrole{n}{start}}, \emph{\DUrole{n}{stop}}, \emph{\DUrole{n}{num}\DUrole{o}{=}\DUrole{default_value}{50}}, \emph{\DUrole{n}{endpoint}\DUrole{o}{=}\DUrole{default_value}{True}}, \emph{\DUrole{n}{retstep}\DUrole{o}{=}\DUrole{default_value}{False}}, \emph{\DUrole{n}{dtype}\DUrole{o}{=}\DUrole{default_value}{None}}, \emph{\DUrole{n}{axis}\DUrole{o}{=}\DUrole{default_value}{0}}}{}
Return evenly spaced numbers over a specified interval.

LAX\sphinxhyphen{}backend implementation of {\hyperref[\detokenize{modules/tensor:symjax.tensor.linspace}]{\sphinxcrossref{\sphinxcode{\sphinxupquote{linspace()}}}}}.
ADDITIONOriginal docstring below.

LAX\sphinxhyphen{}backend implementation of {\hyperref[\detokenize{modules/tensor:symjax.tensor.linspace}]{\sphinxcrossref{\sphinxcode{\sphinxupquote{linspace()}}}}}.
Original docstring below.

Returns \sphinxtitleref{num} evenly spaced samples, calculated over the
interval {[}\sphinxtitleref{start}, \sphinxtitleref{stop}{]}.

The endpoint of the interval can optionally be excluded.

\DUrole{versionmodified,changed}{Changed in version 1.16.0: }Non\sphinxhyphen{}scalar \sphinxtitleref{start} and \sphinxtitleref{stop} are now supported.
\begin{quote}\begin{description}
\item[{Parameters}] \leavevmode
\sphinxstyleliteralstrong{\sphinxupquote{dtype}} (\sphinxstyleliteralemphasis{\sphinxupquote{dtype}}\sphinxstyleliteralemphasis{\sphinxupquote{, }}\sphinxstyleliteralemphasis{\sphinxupquote{optional}}) \textendash{} The type of the output array.  If \sphinxtitleref{dtype} is not given, infer the data
type from the other input arguments.

\item[{Returns}] \leavevmode
\begin{itemize}
\item {} 
\sphinxstylestrong{samples} (\sphinxstyleemphasis{ndarray}) \textendash{} There are \sphinxtitleref{num} equally spaced samples in the closed interval
\sphinxcode{\sphinxupquote{{[}start, stop{]}}} or the half\sphinxhyphen{}open interval \sphinxcode{\sphinxupquote{{[}start, stop)}}
(depending on whether \sphinxtitleref{endpoint} is True or False).

\item {} 
\sphinxstylestrong{step} (\sphinxstyleemphasis{float, optional}) \textendash{} Only returned if \sphinxtitleref{retstep} is True

Size of spacing between samples.

\end{itemize}

\end{description}\end{quote}

\sphinxstrong{See also:}

\begin{description}
\item[{{\hyperref[\detokenize{modules/tensor:symjax.tensor.arange}]{\sphinxcrossref{\sphinxcode{\sphinxupquote{arange()}}}}}}] \leavevmode
Similar to \sphinxtitleref{linspace}, but uses a step size (instead of the number of samples).

\item[{{\hyperref[\detokenize{modules/tensor:symjax.tensor.geomspace}]{\sphinxcrossref{\sphinxcode{\sphinxupquote{geomspace()}}}}}}] \leavevmode
Similar to \sphinxtitleref{linspace}, but with numbers spaced evenly on a log scale (a geometric progression).

\item[{{\hyperref[\detokenize{modules/tensor:symjax.tensor.logspace}]{\sphinxcrossref{\sphinxcode{\sphinxupquote{logspace()}}}}}}] \leavevmode
Similar to \sphinxtitleref{geomspace}, but with the end points specified as logarithms.

\end{description}

\subsubsection*{Examples}

\begin{sphinxVerbatim}[commandchars=\\\{\}]
\PYG{g+gp}{\PYGZgt{}\PYGZgt{}\PYGZgt{} }\PYG{n}{np}\PYG{o}{.}\PYG{n}{linspace}\PYG{p}{(}\PYG{l+m+mf}{2.0}\PYG{p}{,} \PYG{l+m+mf}{3.0}\PYG{p}{,} \PYG{n}{num}\PYG{o}{=}\PYG{l+m+mi}{5}\PYG{p}{)}
\PYG{g+go}{array([2.  , 2.25, 2.5 , 2.75, 3.  ])}
\PYG{g+gp}{\PYGZgt{}\PYGZgt{}\PYGZgt{} }\PYG{n}{np}\PYG{o}{.}\PYG{n}{linspace}\PYG{p}{(}\PYG{l+m+mf}{2.0}\PYG{p}{,} \PYG{l+m+mf}{3.0}\PYG{p}{,} \PYG{n}{num}\PYG{o}{=}\PYG{l+m+mi}{5}\PYG{p}{,} \PYG{n}{endpoint}\PYG{o}{=}\PYG{k+kc}{False}\PYG{p}{)}
\PYG{g+go}{array([2. ,  2.2,  2.4,  2.6,  2.8])}
\PYG{g+gp}{\PYGZgt{}\PYGZgt{}\PYGZgt{} }\PYG{n}{np}\PYG{o}{.}\PYG{n}{linspace}\PYG{p}{(}\PYG{l+m+mf}{2.0}\PYG{p}{,} \PYG{l+m+mf}{3.0}\PYG{p}{,} \PYG{n}{num}\PYG{o}{=}\PYG{l+m+mi}{5}\PYG{p}{,} \PYG{n}{retstep}\PYG{o}{=}\PYG{k+kc}{True}\PYG{p}{)}
\PYG{g+go}{(array([2.  ,  2.25,  2.5 ,  2.75,  3.  ]), 0.25)}
\end{sphinxVerbatim}

Graphical illustration:

\begin{sphinxVerbatim}[commandchars=\\\{\}]
\PYG{g+gp}{\PYGZgt{}\PYGZgt{}\PYGZgt{} }\PYG{k+kn}{import} \PYG{n+nn}{matplotlib}\PYG{n+nn}{.}\PYG{n+nn}{pyplot} \PYG{k}{as} \PYG{n+nn}{plt}
\PYG{g+gp}{\PYGZgt{}\PYGZgt{}\PYGZgt{} }\PYG{n}{N} \PYG{o}{=} \PYG{l+m+mi}{8}
\PYG{g+gp}{\PYGZgt{}\PYGZgt{}\PYGZgt{} }\PYG{n}{y} \PYG{o}{=} \PYG{n}{np}\PYG{o}{.}\PYG{n}{zeros}\PYG{p}{(}\PYG{n}{N}\PYG{p}{)}
\PYG{g+gp}{\PYGZgt{}\PYGZgt{}\PYGZgt{} }\PYG{n}{x1} \PYG{o}{=} \PYG{n}{np}\PYG{o}{.}\PYG{n}{linspace}\PYG{p}{(}\PYG{l+m+mi}{0}\PYG{p}{,} \PYG{l+m+mi}{10}\PYG{p}{,} \PYG{n}{N}\PYG{p}{,} \PYG{n}{endpoint}\PYG{o}{=}\PYG{k+kc}{True}\PYG{p}{)}
\PYG{g+gp}{\PYGZgt{}\PYGZgt{}\PYGZgt{} }\PYG{n}{x2} \PYG{o}{=} \PYG{n}{np}\PYG{o}{.}\PYG{n}{linspace}\PYG{p}{(}\PYG{l+m+mi}{0}\PYG{p}{,} \PYG{l+m+mi}{10}\PYG{p}{,} \PYG{n}{N}\PYG{p}{,} \PYG{n}{endpoint}\PYG{o}{=}\PYG{k+kc}{False}\PYG{p}{)}
\PYG{g+gp}{\PYGZgt{}\PYGZgt{}\PYGZgt{} }\PYG{n}{plt}\PYG{o}{.}\PYG{n}{plot}\PYG{p}{(}\PYG{n}{x1}\PYG{p}{,} \PYG{n}{y}\PYG{p}{,} \PYG{l+s+s1}{\PYGZsq{}}\PYG{l+s+s1}{o}\PYG{l+s+s1}{\PYGZsq{}}\PYG{p}{)}
\PYG{g+go}{[\PYGZlt{}matplotlib.lines.Line2D object at 0x...\PYGZgt{}]}
\PYG{g+gp}{\PYGZgt{}\PYGZgt{}\PYGZgt{} }\PYG{n}{plt}\PYG{o}{.}\PYG{n}{plot}\PYG{p}{(}\PYG{n}{x2}\PYG{p}{,} \PYG{n}{y} \PYG{o}{+} \PYG{l+m+mf}{0.5}\PYG{p}{,} \PYG{l+s+s1}{\PYGZsq{}}\PYG{l+s+s1}{o}\PYG{l+s+s1}{\PYGZsq{}}\PYG{p}{)}
\PYG{g+go}{[\PYGZlt{}matplotlib.lines.Line2D object at 0x...\PYGZgt{}]}
\PYG{g+gp}{\PYGZgt{}\PYGZgt{}\PYGZgt{} }\PYG{n}{plt}\PYG{o}{.}\PYG{n}{ylim}\PYG{p}{(}\PYG{p}{[}\PYG{o}{\PYGZhy{}}\PYG{l+m+mf}{0.5}\PYG{p}{,} \PYG{l+m+mi}{1}\PYG{p}{]}\PYG{p}{)}
\PYG{g+go}{(\PYGZhy{}0.5, 1)}
\PYG{g+gp}{\PYGZgt{}\PYGZgt{}\PYGZgt{} }\PYG{n}{plt}\PYG{o}{.}\PYG{n}{show}\PYG{p}{(}\PYG{p}{)}
\end{sphinxVerbatim}

\end{fulllineitems}

\index{log() (in module symjax.tensor)@\spxentry{log()}\spxextra{in module symjax.tensor}}

\begin{fulllineitems}
\phantomsection\label{\detokenize{modules/tensor:symjax.tensor.log}}\pysiglinewithargsret{\sphinxbfcode{\sphinxupquote{log}}}{\emph{\DUrole{n}{x}}}{}
Natural logarithm, element\sphinxhyphen{}wise.

LAX\sphinxhyphen{}backend implementation of {\hyperref[\detokenize{modules/tensor:symjax.tensor.log}]{\sphinxcrossref{\sphinxcode{\sphinxupquote{log()}}}}}.
ADDITIONOriginal docstring below.

LAX\sphinxhyphen{}backend implementation of {\hyperref[\detokenize{modules/tensor:symjax.tensor.log}]{\sphinxcrossref{\sphinxcode{\sphinxupquote{log()}}}}}.
Original docstring below.

log(x, /, out=None, {\color{red}\bfseries{}*}, where=True, casting=’same\_kind’, order=’K’, dtype=None, subok=True{[}, signature, extobj{]})

The natural logarithm \sphinxtitleref{log} is the inverse of the exponential function,
so that \sphinxtitleref{log(exp(x)) = x}. The natural logarithm is logarithm in base
\sphinxtitleref{e}.
\begin{quote}\begin{description}
\item[{Returns}] \leavevmode
\sphinxstylestrong{y} \textendash{} The natural logarithm of \sphinxtitleref{x}, element\sphinxhyphen{}wise.
This is a scalar if \sphinxtitleref{x} is a scalar.

\item[{Return type}] \leavevmode
ndarray

\end{description}\end{quote}

\sphinxstrong{See also:}

{\hyperref[\detokenize{modules/tensor:symjax.tensor.log10}]{\sphinxcrossref{\sphinxcode{\sphinxupquote{log10()}}}}}, {\hyperref[\detokenize{modules/tensor:symjax.tensor.log2}]{\sphinxcrossref{\sphinxcode{\sphinxupquote{log2()}}}}}, {\hyperref[\detokenize{modules/tensor:symjax.tensor.log1p}]{\sphinxcrossref{\sphinxcode{\sphinxupquote{log1p()}}}}}, \sphinxcode{\sphinxupquote{emath.log()}}

\subsubsection*{Notes}

Logarithm is a multivalued function: for each \sphinxtitleref{x} there is an infinite
number of \sphinxtitleref{z} such that \sphinxtitleref{exp(z) = x}. The convention is to return the
\sphinxtitleref{z} whose imaginary part lies in \sphinxtitleref{{[}\sphinxhyphen{}pi, pi{]}}.

For real\sphinxhyphen{}valued input data types, \sphinxtitleref{log} always returns real output. For
each value that cannot be expressed as a real number or infinity, it
yields \sphinxcode{\sphinxupquote{nan}} and sets the \sphinxtitleref{invalid} floating point error flag.

For complex\sphinxhyphen{}valued input, \sphinxtitleref{log} is a complex analytical function that
has a branch cut \sphinxtitleref{{[}\sphinxhyphen{}inf, 0{]}} and is continuous from above on it. \sphinxtitleref{log}
handles the floating\sphinxhyphen{}point negative zero as an infinitesimal negative
number, conforming to the C99 standard.
\subsubsection*{References}
\subsubsection*{Examples}

\begin{sphinxVerbatim}[commandchars=\\\{\}]
\PYG{g+gp}{\PYGZgt{}\PYGZgt{}\PYGZgt{} }\PYG{n}{np}\PYG{o}{.}\PYG{n}{log}\PYG{p}{(}\PYG{p}{[}\PYG{l+m+mi}{1}\PYG{p}{,} \PYG{n}{np}\PYG{o}{.}\PYG{n}{e}\PYG{p}{,} \PYG{n}{np}\PYG{o}{.}\PYG{n}{e}\PYG{o}{*}\PYG{o}{*}\PYG{l+m+mi}{2}\PYG{p}{,} \PYG{l+m+mi}{0}\PYG{p}{]}\PYG{p}{)}
\PYG{g+go}{array([  0.,   1.,   2., \PYGZhy{}Inf])}
\end{sphinxVerbatim}

\end{fulllineitems}

\index{log10() (in module symjax.tensor)@\spxentry{log10()}\spxextra{in module symjax.tensor}}

\begin{fulllineitems}
\phantomsection\label{\detokenize{modules/tensor:symjax.tensor.log10}}\pysiglinewithargsret{\sphinxbfcode{\sphinxupquote{log10}}}{\emph{\DUrole{n}{x}}}{}
Return the base 10 logarithm of the input array, element\sphinxhyphen{}wise.

LAX\sphinxhyphen{}backend implementation of {\hyperref[\detokenize{modules/tensor:symjax.tensor.log10}]{\sphinxcrossref{\sphinxcode{\sphinxupquote{log10()}}}}}.
ADDITIONOriginal docstring below.

LAX\sphinxhyphen{}backend implementation of {\hyperref[\detokenize{modules/tensor:symjax.tensor.log10}]{\sphinxcrossref{\sphinxcode{\sphinxupquote{log10()}}}}}.
Original docstring below.

log10(x, /, out=None, {\color{red}\bfseries{}*}, where=True, casting=’same\_kind’, order=’K’, dtype=None, subok=True{[}, signature, extobj{]})
\begin{quote}\begin{description}
\item[{Returns}] \leavevmode
\sphinxstylestrong{y} \textendash{} The logarithm to the base 10 of \sphinxtitleref{x}, element\sphinxhyphen{}wise. NaNs are
returned where x is negative.
This is a scalar if \sphinxtitleref{x} is a scalar.

\item[{Return type}] \leavevmode
ndarray

\end{description}\end{quote}

\sphinxstrong{See also:}

\sphinxcode{\sphinxupquote{emath.log10()}}

\subsubsection*{Notes}

Logarithm is a multivalued function: for each \sphinxtitleref{x} there is an infinite
number of \sphinxtitleref{z} such that \sphinxtitleref{10**z = x}. The convention is to return the
\sphinxtitleref{z} whose imaginary part lies in \sphinxtitleref{{[}\sphinxhyphen{}pi, pi{]}}.

For real\sphinxhyphen{}valued input data types, \sphinxtitleref{log10} always returns real output.
For each value that cannot be expressed as a real number or infinity,
it yields \sphinxcode{\sphinxupquote{nan}} and sets the \sphinxtitleref{invalid} floating point error flag.

For complex\sphinxhyphen{}valued input, \sphinxtitleref{log10} is a complex analytical function that
has a branch cut \sphinxtitleref{{[}\sphinxhyphen{}inf, 0{]}} and is continuous from above on it.
\sphinxtitleref{log10} handles the floating\sphinxhyphen{}point negative zero as an infinitesimal
negative number, conforming to the C99 standard.
\subsubsection*{References}
\subsubsection*{Examples}

\begin{sphinxVerbatim}[commandchars=\\\{\}]
\PYG{g+gp}{\PYGZgt{}\PYGZgt{}\PYGZgt{} }\PYG{n}{np}\PYG{o}{.}\PYG{n}{log10}\PYG{p}{(}\PYG{p}{[}\PYG{l+m+mf}{1e\PYGZhy{}15}\PYG{p}{,} \PYG{o}{\PYGZhy{}}\PYG{l+m+mf}{3.}\PYG{p}{]}\PYG{p}{)}
\PYG{g+go}{array([\PYGZhy{}15.,  nan])}
\end{sphinxVerbatim}

\end{fulllineitems}

\index{log1p() (in module symjax.tensor)@\spxentry{log1p()}\spxextra{in module symjax.tensor}}

\begin{fulllineitems}
\phantomsection\label{\detokenize{modules/tensor:symjax.tensor.log1p}}\pysiglinewithargsret{\sphinxbfcode{\sphinxupquote{log1p}}}{\emph{\DUrole{n}{x}}}{}
Return the natural logarithm of one plus the input array, element\sphinxhyphen{}wise.

LAX\sphinxhyphen{}backend implementation of {\hyperref[\detokenize{modules/tensor:symjax.tensor.log1p}]{\sphinxcrossref{\sphinxcode{\sphinxupquote{log1p()}}}}}.
ADDITIONOriginal docstring below.

LAX\sphinxhyphen{}backend implementation of {\hyperref[\detokenize{modules/tensor:symjax.tensor.log1p}]{\sphinxcrossref{\sphinxcode{\sphinxupquote{log1p()}}}}}.
Original docstring below.

log1p(x, /, out=None, {\color{red}\bfseries{}*}, where=True, casting=’same\_kind’, order=’K’, dtype=None, subok=True{[}, signature, extobj{]})

Calculates \sphinxcode{\sphinxupquote{log(1 + x)}}.
\begin{quote}\begin{description}
\item[{Returns}] \leavevmode
\sphinxstylestrong{y} \textendash{} Natural logarithm of \sphinxtitleref{1 + x}, element\sphinxhyphen{}wise.
This is a scalar if \sphinxtitleref{x} is a scalar.

\item[{Return type}] \leavevmode
ndarray

\end{description}\end{quote}

\sphinxstrong{See also:}

\begin{description}
\item[{{\hyperref[\detokenize{modules/tensor:symjax.tensor.expm1}]{\sphinxcrossref{\sphinxcode{\sphinxupquote{expm1()}}}}}}] \leavevmode
\sphinxcode{\sphinxupquote{exp(x) \sphinxhyphen{} 1}}, the inverse of \sphinxtitleref{log1p}.

\end{description}

\subsubsection*{Notes}

For real\sphinxhyphen{}valued input, \sphinxtitleref{log1p} is accurate also for \sphinxtitleref{x} so small
that \sphinxtitleref{1 + x == 1} in floating\sphinxhyphen{}point accuracy.

Logarithm is a multivalued function: for each \sphinxtitleref{x} there is an infinite
number of \sphinxtitleref{z} such that \sphinxtitleref{exp(z) = 1 + x}. The convention is to return
the \sphinxtitleref{z} whose imaginary part lies in \sphinxtitleref{{[}\sphinxhyphen{}pi, pi{]}}.

For real\sphinxhyphen{}valued input data types, \sphinxtitleref{log1p} always returns real output.
For each value that cannot be expressed as a real number or infinity,
it yields \sphinxcode{\sphinxupquote{nan}} and sets the \sphinxtitleref{invalid} floating point error flag.

For complex\sphinxhyphen{}valued input, \sphinxtitleref{log1p} is a complex analytical function that
has a branch cut \sphinxtitleref{{[}\sphinxhyphen{}inf, \sphinxhyphen{}1{]}} and is continuous from above on it.
\sphinxtitleref{log1p} handles the floating\sphinxhyphen{}point negative zero as an infinitesimal
negative number, conforming to the C99 standard.
\subsubsection*{References}
\subsubsection*{Examples}

\begin{sphinxVerbatim}[commandchars=\\\{\}]
\PYG{g+gp}{\PYGZgt{}\PYGZgt{}\PYGZgt{} }\PYG{n}{np}\PYG{o}{.}\PYG{n}{log1p}\PYG{p}{(}\PYG{l+m+mf}{1e\PYGZhy{}99}\PYG{p}{)}
\PYG{g+go}{1e\PYGZhy{}99}
\PYG{g+gp}{\PYGZgt{}\PYGZgt{}\PYGZgt{} }\PYG{n}{np}\PYG{o}{.}\PYG{n}{log}\PYG{p}{(}\PYG{l+m+mi}{1} \PYG{o}{+} \PYG{l+m+mf}{1e\PYGZhy{}99}\PYG{p}{)}
\PYG{g+go}{0.0}
\end{sphinxVerbatim}

\end{fulllineitems}

\index{log2() (in module symjax.tensor)@\spxentry{log2()}\spxextra{in module symjax.tensor}}

\begin{fulllineitems}
\phantomsection\label{\detokenize{modules/tensor:symjax.tensor.log2}}\pysiglinewithargsret{\sphinxbfcode{\sphinxupquote{log2}}}{\emph{\DUrole{n}{x}}}{}
Base\sphinxhyphen{}2 logarithm of \sphinxtitleref{x}.

LAX\sphinxhyphen{}backend implementation of {\hyperref[\detokenize{modules/tensor:symjax.tensor.log2}]{\sphinxcrossref{\sphinxcode{\sphinxupquote{log2()}}}}}.
ADDITIONOriginal docstring below.

LAX\sphinxhyphen{}backend implementation of {\hyperref[\detokenize{modules/tensor:symjax.tensor.log2}]{\sphinxcrossref{\sphinxcode{\sphinxupquote{log2()}}}}}.
Original docstring below.

log2(x, /, out=None, {\color{red}\bfseries{}*}, where=True, casting=’same\_kind’, order=’K’, dtype=None, subok=True{[}, signature, extobj{]})
\begin{quote}\begin{description}
\item[{Returns}] \leavevmode
\sphinxstylestrong{y} \textendash{} Base\sphinxhyphen{}2 logarithm of \sphinxtitleref{x}.
This is a scalar if \sphinxtitleref{x} is a scalar.

\item[{Return type}] \leavevmode
ndarray

\end{description}\end{quote}

\sphinxstrong{See also:}

{\hyperref[\detokenize{modules/tensor:symjax.tensor.log}]{\sphinxcrossref{\sphinxcode{\sphinxupquote{log()}}}}}, {\hyperref[\detokenize{modules/tensor:symjax.tensor.log10}]{\sphinxcrossref{\sphinxcode{\sphinxupquote{log10()}}}}}, {\hyperref[\detokenize{modules/tensor:symjax.tensor.log1p}]{\sphinxcrossref{\sphinxcode{\sphinxupquote{log1p()}}}}}, \sphinxcode{\sphinxupquote{emath.log2()}}

\subsubsection*{Notes}

\DUrole{versionmodified,added}{New in version 1.3.0.}

Logarithm is a multivalued function: for each \sphinxtitleref{x} there is an infinite
number of \sphinxtitleref{z} such that \sphinxtitleref{2**z = x}. The convention is to return the \sphinxtitleref{z}
whose imaginary part lies in \sphinxtitleref{{[}\sphinxhyphen{}pi, pi{]}}.

For real\sphinxhyphen{}valued input data types, \sphinxtitleref{log2} always returns real output.
For each value that cannot be expressed as a real number or infinity,
it yields \sphinxcode{\sphinxupquote{nan}} and sets the \sphinxtitleref{invalid} floating point error flag.

For complex\sphinxhyphen{}valued input, \sphinxtitleref{log2} is a complex analytical function that
has a branch cut \sphinxtitleref{{[}\sphinxhyphen{}inf, 0{]}} and is continuous from above on it. \sphinxtitleref{log2}
handles the floating\sphinxhyphen{}point negative zero as an infinitesimal negative
number, conforming to the C99 standard.
\subsubsection*{Examples}

\begin{sphinxVerbatim}[commandchars=\\\{\}]
\PYG{g+gp}{\PYGZgt{}\PYGZgt{}\PYGZgt{} }\PYG{n}{x} \PYG{o}{=} \PYG{n}{np}\PYG{o}{.}\PYG{n}{array}\PYG{p}{(}\PYG{p}{[}\PYG{l+m+mi}{0}\PYG{p}{,} \PYG{l+m+mi}{1}\PYG{p}{,} \PYG{l+m+mi}{2}\PYG{p}{,} \PYG{l+m+mi}{2}\PYG{o}{*}\PYG{o}{*}\PYG{l+m+mi}{4}\PYG{p}{]}\PYG{p}{)}
\PYG{g+gp}{\PYGZgt{}\PYGZgt{}\PYGZgt{} }\PYG{n}{np}\PYG{o}{.}\PYG{n}{log2}\PYG{p}{(}\PYG{n}{x}\PYG{p}{)}
\PYG{g+go}{array([\PYGZhy{}Inf,   0.,   1.,   4.])}
\end{sphinxVerbatim}

\begin{sphinxVerbatim}[commandchars=\\\{\}]
\PYG{g+gp}{\PYGZgt{}\PYGZgt{}\PYGZgt{} }\PYG{n}{xi} \PYG{o}{=} \PYG{n}{np}\PYG{o}{.}\PYG{n}{array}\PYG{p}{(}\PYG{p}{[}\PYG{l+m+mi}{0}\PYG{o}{+}\PYG{l+m+mf}{1.}\PYG{n}{j}\PYG{p}{,} \PYG{l+m+mi}{1}\PYG{p}{,} \PYG{l+m+mi}{2}\PYG{o}{+}\PYG{l+m+mf}{0.}\PYG{n}{j}\PYG{p}{,} \PYG{l+m+mf}{4.}\PYG{n}{j}\PYG{p}{]}\PYG{p}{)}
\PYG{g+gp}{\PYGZgt{}\PYGZgt{}\PYGZgt{} }\PYG{n}{np}\PYG{o}{.}\PYG{n}{log2}\PYG{p}{(}\PYG{n}{xi}\PYG{p}{)}
\PYG{g+go}{array([ 0.+2.26618007j,  0.+0.j        ,  1.+0.j        ,  2.+2.26618007j])}
\end{sphinxVerbatim}

\end{fulllineitems}

\index{logical\_and() (in module symjax.tensor)@\spxentry{logical\_and()}\spxextra{in module symjax.tensor}}

\begin{fulllineitems}
\phantomsection\label{\detokenize{modules/tensor:symjax.tensor.logical_and}}\pysiglinewithargsret{\sphinxbfcode{\sphinxupquote{logical\_and}}}{\emph{\DUrole{o}{*}\DUrole{n}{args}}}{}
Compute the truth value of x1 AND x2 element\sphinxhyphen{}wise.

LAX\sphinxhyphen{}backend implementation of {\hyperref[\detokenize{modules/tensor:symjax.tensor.logical_and}]{\sphinxcrossref{\sphinxcode{\sphinxupquote{logical\_and()}}}}}.
ADDITIONOriginal docstring below.

LAX\sphinxhyphen{}backend implementation of {\hyperref[\detokenize{modules/tensor:symjax.tensor.logical_and}]{\sphinxcrossref{\sphinxcode{\sphinxupquote{logical\_and()}}}}}.
Original docstring below.

logical\_and(x1, x2, /, out=None, {\color{red}\bfseries{}*}, where=True, casting=’same\_kind’, order=’K’, dtype=None, subok=True{[}, signature, extobj{]})
\begin{quote}\begin{description}
\item[{Returns}] \leavevmode
\sphinxstylestrong{y} \textendash{} Boolean result of the logical AND operation applied to the elements
of \sphinxtitleref{x1} and \sphinxtitleref{x2}; the shape is determined by broadcasting.
This is a scalar if both \sphinxtitleref{x1} and \sphinxtitleref{x2} are scalars.

\item[{Return type}] \leavevmode
ndarray or bool

\end{description}\end{quote}

\sphinxstrong{See also:}

{\hyperref[\detokenize{modules/tensor:symjax.tensor.logical_or}]{\sphinxcrossref{\sphinxcode{\sphinxupquote{logical\_or()}}}}}, {\hyperref[\detokenize{modules/tensor:symjax.tensor.logical_not}]{\sphinxcrossref{\sphinxcode{\sphinxupquote{logical\_not()}}}}}, {\hyperref[\detokenize{modules/tensor:symjax.tensor.logical_xor}]{\sphinxcrossref{\sphinxcode{\sphinxupquote{logical\_xor()}}}}}, {\hyperref[\detokenize{modules/tensor:symjax.tensor.bitwise_and}]{\sphinxcrossref{\sphinxcode{\sphinxupquote{bitwise\_and()}}}}}

\subsubsection*{Examples}

\begin{sphinxVerbatim}[commandchars=\\\{\}]
\PYG{g+gp}{\PYGZgt{}\PYGZgt{}\PYGZgt{} }\PYG{n}{np}\PYG{o}{.}\PYG{n}{logical\PYGZus{}and}\PYG{p}{(}\PYG{k+kc}{True}\PYG{p}{,} \PYG{k+kc}{False}\PYG{p}{)}
\PYG{g+go}{False}
\PYG{g+gp}{\PYGZgt{}\PYGZgt{}\PYGZgt{} }\PYG{n}{np}\PYG{o}{.}\PYG{n}{logical\PYGZus{}and}\PYG{p}{(}\PYG{p}{[}\PYG{k+kc}{True}\PYG{p}{,} \PYG{k+kc}{False}\PYG{p}{]}\PYG{p}{,} \PYG{p}{[}\PYG{k+kc}{False}\PYG{p}{,} \PYG{k+kc}{False}\PYG{p}{]}\PYG{p}{)}
\PYG{g+go}{array([False, False])}
\end{sphinxVerbatim}

\begin{sphinxVerbatim}[commandchars=\\\{\}]
\PYG{g+gp}{\PYGZgt{}\PYGZgt{}\PYGZgt{} }\PYG{n}{x} \PYG{o}{=} \PYG{n}{np}\PYG{o}{.}\PYG{n}{arange}\PYG{p}{(}\PYG{l+m+mi}{5}\PYG{p}{)}
\PYG{g+gp}{\PYGZgt{}\PYGZgt{}\PYGZgt{} }\PYG{n}{np}\PYG{o}{.}\PYG{n}{logical\PYGZus{}and}\PYG{p}{(}\PYG{n}{x}\PYG{o}{\PYGZgt{}}\PYG{l+m+mi}{1}\PYG{p}{,} \PYG{n}{x}\PYG{o}{\PYGZlt{}}\PYG{l+m+mi}{4}\PYG{p}{)}
\PYG{g+go}{array([False, False,  True,  True, False])}
\end{sphinxVerbatim}

\end{fulllineitems}

\index{logical\_not() (in module symjax.tensor)@\spxentry{logical\_not()}\spxextra{in module symjax.tensor}}

\begin{fulllineitems}
\phantomsection\label{\detokenize{modules/tensor:symjax.tensor.logical_not}}\pysiglinewithargsret{\sphinxbfcode{\sphinxupquote{logical\_not}}}{\emph{\DUrole{o}{*}\DUrole{n}{args}}}{}
Compute the truth value of NOT x element\sphinxhyphen{}wise.

LAX\sphinxhyphen{}backend implementation of {\hyperref[\detokenize{modules/tensor:symjax.tensor.logical_not}]{\sphinxcrossref{\sphinxcode{\sphinxupquote{logical\_not()}}}}}.
ADDITIONOriginal docstring below.

LAX\sphinxhyphen{}backend implementation of {\hyperref[\detokenize{modules/tensor:symjax.tensor.logical_not}]{\sphinxcrossref{\sphinxcode{\sphinxupquote{logical\_not()}}}}}.
Original docstring below.

logical\_not(x, /, out=None, {\color{red}\bfseries{}*}, where=True, casting=’same\_kind’, order=’K’, dtype=None, subok=True{[}, signature, extobj{]})
\begin{quote}\begin{description}
\item[{Returns}] \leavevmode
\sphinxstylestrong{y} \textendash{} Boolean result with the same shape as \sphinxtitleref{x} of the NOT operation
on elements of \sphinxtitleref{x}.
This is a scalar if \sphinxtitleref{x} is a scalar.

\item[{Return type}] \leavevmode
bool or ndarray of bool

\end{description}\end{quote}

\sphinxstrong{See also:}

{\hyperref[\detokenize{modules/tensor:symjax.tensor.logical_and}]{\sphinxcrossref{\sphinxcode{\sphinxupquote{logical\_and()}}}}}, {\hyperref[\detokenize{modules/tensor:symjax.tensor.logical_or}]{\sphinxcrossref{\sphinxcode{\sphinxupquote{logical\_or()}}}}}, {\hyperref[\detokenize{modules/tensor:symjax.tensor.logical_xor}]{\sphinxcrossref{\sphinxcode{\sphinxupquote{logical\_xor()}}}}}

\subsubsection*{Examples}

\begin{sphinxVerbatim}[commandchars=\\\{\}]
\PYG{g+gp}{\PYGZgt{}\PYGZgt{}\PYGZgt{} }\PYG{n}{np}\PYG{o}{.}\PYG{n}{logical\PYGZus{}not}\PYG{p}{(}\PYG{l+m+mi}{3}\PYG{p}{)}
\PYG{g+go}{False}
\PYG{g+gp}{\PYGZgt{}\PYGZgt{}\PYGZgt{} }\PYG{n}{np}\PYG{o}{.}\PYG{n}{logical\PYGZus{}not}\PYG{p}{(}\PYG{p}{[}\PYG{k+kc}{True}\PYG{p}{,} \PYG{k+kc}{False}\PYG{p}{,} \PYG{l+m+mi}{0}\PYG{p}{,} \PYG{l+m+mi}{1}\PYG{p}{]}\PYG{p}{)}
\PYG{g+go}{array([False,  True,  True, False])}
\end{sphinxVerbatim}

\begin{sphinxVerbatim}[commandchars=\\\{\}]
\PYG{g+gp}{\PYGZgt{}\PYGZgt{}\PYGZgt{} }\PYG{n}{x} \PYG{o}{=} \PYG{n}{np}\PYG{o}{.}\PYG{n}{arange}\PYG{p}{(}\PYG{l+m+mi}{5}\PYG{p}{)}
\PYG{g+gp}{\PYGZgt{}\PYGZgt{}\PYGZgt{} }\PYG{n}{np}\PYG{o}{.}\PYG{n}{logical\PYGZus{}not}\PYG{p}{(}\PYG{n}{x}\PYG{o}{\PYGZlt{}}\PYG{l+m+mi}{3}\PYG{p}{)}
\PYG{g+go}{array([False, False, False,  True,  True])}
\end{sphinxVerbatim}

\end{fulllineitems}

\index{logical\_or() (in module symjax.tensor)@\spxentry{logical\_or()}\spxextra{in module symjax.tensor}}

\begin{fulllineitems}
\phantomsection\label{\detokenize{modules/tensor:symjax.tensor.logical_or}}\pysiglinewithargsret{\sphinxbfcode{\sphinxupquote{logical\_or}}}{\emph{\DUrole{o}{*}\DUrole{n}{args}}}{}
Compute the truth value of x1 OR x2 element\sphinxhyphen{}wise.

LAX\sphinxhyphen{}backend implementation of {\hyperref[\detokenize{modules/tensor:symjax.tensor.logical_or}]{\sphinxcrossref{\sphinxcode{\sphinxupquote{logical\_or()}}}}}.
ADDITIONOriginal docstring below.

LAX\sphinxhyphen{}backend implementation of {\hyperref[\detokenize{modules/tensor:symjax.tensor.logical_or}]{\sphinxcrossref{\sphinxcode{\sphinxupquote{logical\_or()}}}}}.
Original docstring below.

logical\_or(x1, x2, /, out=None, {\color{red}\bfseries{}*}, where=True, casting=’same\_kind’, order=’K’, dtype=None, subok=True{[}, signature, extobj{]})
\begin{quote}\begin{description}
\item[{Returns}] \leavevmode
\sphinxstylestrong{y} \textendash{} Boolean result of the logical OR operation applied to the elements
of \sphinxtitleref{x1} and \sphinxtitleref{x2}; the shape is determined by broadcasting.
This is a scalar if both \sphinxtitleref{x1} and \sphinxtitleref{x2} are scalars.

\item[{Return type}] \leavevmode
ndarray or bool

\end{description}\end{quote}

\sphinxstrong{See also:}

{\hyperref[\detokenize{modules/tensor:symjax.tensor.logical_and}]{\sphinxcrossref{\sphinxcode{\sphinxupquote{logical\_and()}}}}}, {\hyperref[\detokenize{modules/tensor:symjax.tensor.logical_not}]{\sphinxcrossref{\sphinxcode{\sphinxupquote{logical\_not()}}}}}, {\hyperref[\detokenize{modules/tensor:symjax.tensor.logical_xor}]{\sphinxcrossref{\sphinxcode{\sphinxupquote{logical\_xor()}}}}}, {\hyperref[\detokenize{modules/tensor:symjax.tensor.bitwise_or}]{\sphinxcrossref{\sphinxcode{\sphinxupquote{bitwise\_or()}}}}}

\subsubsection*{Examples}

\begin{sphinxVerbatim}[commandchars=\\\{\}]
\PYG{g+gp}{\PYGZgt{}\PYGZgt{}\PYGZgt{} }\PYG{n}{np}\PYG{o}{.}\PYG{n}{logical\PYGZus{}or}\PYG{p}{(}\PYG{k+kc}{True}\PYG{p}{,} \PYG{k+kc}{False}\PYG{p}{)}
\PYG{g+go}{True}
\PYG{g+gp}{\PYGZgt{}\PYGZgt{}\PYGZgt{} }\PYG{n}{np}\PYG{o}{.}\PYG{n}{logical\PYGZus{}or}\PYG{p}{(}\PYG{p}{[}\PYG{k+kc}{True}\PYG{p}{,} \PYG{k+kc}{False}\PYG{p}{]}\PYG{p}{,} \PYG{p}{[}\PYG{k+kc}{False}\PYG{p}{,} \PYG{k+kc}{False}\PYG{p}{]}\PYG{p}{)}
\PYG{g+go}{array([ True, False])}
\end{sphinxVerbatim}

\begin{sphinxVerbatim}[commandchars=\\\{\}]
\PYG{g+gp}{\PYGZgt{}\PYGZgt{}\PYGZgt{} }\PYG{n}{x} \PYG{o}{=} \PYG{n}{np}\PYG{o}{.}\PYG{n}{arange}\PYG{p}{(}\PYG{l+m+mi}{5}\PYG{p}{)}
\PYG{g+gp}{\PYGZgt{}\PYGZgt{}\PYGZgt{} }\PYG{n}{np}\PYG{o}{.}\PYG{n}{logical\PYGZus{}or}\PYG{p}{(}\PYG{n}{x} \PYG{o}{\PYGZlt{}} \PYG{l+m+mi}{1}\PYG{p}{,} \PYG{n}{x} \PYG{o}{\PYGZgt{}} \PYG{l+m+mi}{3}\PYG{p}{)}
\PYG{g+go}{array([ True, False, False, False,  True])}
\end{sphinxVerbatim}

\end{fulllineitems}

\index{logical\_xor() (in module symjax.tensor)@\spxentry{logical\_xor()}\spxextra{in module symjax.tensor}}

\begin{fulllineitems}
\phantomsection\label{\detokenize{modules/tensor:symjax.tensor.logical_xor}}\pysiglinewithargsret{\sphinxbfcode{\sphinxupquote{logical\_xor}}}{\emph{\DUrole{o}{*}\DUrole{n}{args}}}{}
Compute the truth value of x1 XOR x2, element\sphinxhyphen{}wise.

LAX\sphinxhyphen{}backend implementation of {\hyperref[\detokenize{modules/tensor:symjax.tensor.logical_xor}]{\sphinxcrossref{\sphinxcode{\sphinxupquote{logical\_xor()}}}}}.
ADDITIONOriginal docstring below.

LAX\sphinxhyphen{}backend implementation of {\hyperref[\detokenize{modules/tensor:symjax.tensor.logical_xor}]{\sphinxcrossref{\sphinxcode{\sphinxupquote{logical\_xor()}}}}}.
Original docstring below.

logical\_xor(x1, x2, /, out=None, {\color{red}\bfseries{}*}, where=True, casting=’same\_kind’, order=’K’, dtype=None, subok=True{[}, signature, extobj{]})
\begin{quote}\begin{description}
\item[{Returns}] \leavevmode
\sphinxstylestrong{y} \textendash{} Boolean result of the logical XOR operation applied to the elements
of \sphinxtitleref{x1} and \sphinxtitleref{x2}; the shape is determined by broadcasting.
This is a scalar if both \sphinxtitleref{x1} and \sphinxtitleref{x2} are scalars.

\item[{Return type}] \leavevmode
bool or ndarray of bool

\end{description}\end{quote}

\sphinxstrong{See also:}

{\hyperref[\detokenize{modules/tensor:symjax.tensor.logical_and}]{\sphinxcrossref{\sphinxcode{\sphinxupquote{logical\_and()}}}}}, {\hyperref[\detokenize{modules/tensor:symjax.tensor.logical_or}]{\sphinxcrossref{\sphinxcode{\sphinxupquote{logical\_or()}}}}}, {\hyperref[\detokenize{modules/tensor:symjax.tensor.logical_not}]{\sphinxcrossref{\sphinxcode{\sphinxupquote{logical\_not()}}}}}, {\hyperref[\detokenize{modules/tensor:symjax.tensor.bitwise_xor}]{\sphinxcrossref{\sphinxcode{\sphinxupquote{bitwise\_xor()}}}}}

\subsubsection*{Examples}

\begin{sphinxVerbatim}[commandchars=\\\{\}]
\PYG{g+gp}{\PYGZgt{}\PYGZgt{}\PYGZgt{} }\PYG{n}{np}\PYG{o}{.}\PYG{n}{logical\PYGZus{}xor}\PYG{p}{(}\PYG{k+kc}{True}\PYG{p}{,} \PYG{k+kc}{False}\PYG{p}{)}
\PYG{g+go}{True}
\PYG{g+gp}{\PYGZgt{}\PYGZgt{}\PYGZgt{} }\PYG{n}{np}\PYG{o}{.}\PYG{n}{logical\PYGZus{}xor}\PYG{p}{(}\PYG{p}{[}\PYG{k+kc}{True}\PYG{p}{,} \PYG{k+kc}{True}\PYG{p}{,} \PYG{k+kc}{False}\PYG{p}{,} \PYG{k+kc}{False}\PYG{p}{]}\PYG{p}{,} \PYG{p}{[}\PYG{k+kc}{True}\PYG{p}{,} \PYG{k+kc}{False}\PYG{p}{,} \PYG{k+kc}{True}\PYG{p}{,} \PYG{k+kc}{False}\PYG{p}{]}\PYG{p}{)}
\PYG{g+go}{array([False,  True,  True, False])}
\end{sphinxVerbatim}

\begin{sphinxVerbatim}[commandchars=\\\{\}]
\PYG{g+gp}{\PYGZgt{}\PYGZgt{}\PYGZgt{} }\PYG{n}{x} \PYG{o}{=} \PYG{n}{np}\PYG{o}{.}\PYG{n}{arange}\PYG{p}{(}\PYG{l+m+mi}{5}\PYG{p}{)}
\PYG{g+gp}{\PYGZgt{}\PYGZgt{}\PYGZgt{} }\PYG{n}{np}\PYG{o}{.}\PYG{n}{logical\PYGZus{}xor}\PYG{p}{(}\PYG{n}{x} \PYG{o}{\PYGZlt{}} \PYG{l+m+mi}{1}\PYG{p}{,} \PYG{n}{x} \PYG{o}{\PYGZgt{}} \PYG{l+m+mi}{3}\PYG{p}{)}
\PYG{g+go}{array([ True, False, False, False,  True])}
\end{sphinxVerbatim}

Simple example showing support of broadcasting

\begin{sphinxVerbatim}[commandchars=\\\{\}]
\PYG{g+gp}{\PYGZgt{}\PYGZgt{}\PYGZgt{} }\PYG{n}{np}\PYG{o}{.}\PYG{n}{logical\PYGZus{}xor}\PYG{p}{(}\PYG{l+m+mi}{0}\PYG{p}{,} \PYG{n}{np}\PYG{o}{.}\PYG{n}{eye}\PYG{p}{(}\PYG{l+m+mi}{2}\PYG{p}{)}\PYG{p}{)}
\PYG{g+go}{array([[ True, False],}
\PYG{g+go}{       [False,  True]])}
\end{sphinxVerbatim}

\end{fulllineitems}

\index{logspace() (in module symjax.tensor)@\spxentry{logspace()}\spxextra{in module symjax.tensor}}

\begin{fulllineitems}
\phantomsection\label{\detokenize{modules/tensor:symjax.tensor.logspace}}\pysiglinewithargsret{\sphinxbfcode{\sphinxupquote{logspace}}}{\emph{\DUrole{n}{start}}, \emph{\DUrole{n}{stop}}, \emph{\DUrole{n}{num}\DUrole{o}{=}\DUrole{default_value}{50}}, \emph{\DUrole{n}{endpoint}\DUrole{o}{=}\DUrole{default_value}{True}}, \emph{\DUrole{n}{base}\DUrole{o}{=}\DUrole{default_value}{10.0}}, \emph{\DUrole{n}{dtype}\DUrole{o}{=}\DUrole{default_value}{None}}, \emph{\DUrole{n}{axis}\DUrole{o}{=}\DUrole{default_value}{0}}}{}
Return numbers spaced evenly on a log scale.

LAX\sphinxhyphen{}backend implementation of {\hyperref[\detokenize{modules/tensor:symjax.tensor.logspace}]{\sphinxcrossref{\sphinxcode{\sphinxupquote{logspace()}}}}}.
ADDITIONOriginal docstring below.

LAX\sphinxhyphen{}backend implementation of {\hyperref[\detokenize{modules/tensor:symjax.tensor.logspace}]{\sphinxcrossref{\sphinxcode{\sphinxupquote{logspace()}}}}}.
Original docstring below.

In linear space, the sequence starts at \sphinxcode{\sphinxupquote{base ** start}}
(\sphinxtitleref{base} to the power of \sphinxtitleref{start}) and ends with \sphinxcode{\sphinxupquote{base ** stop}}
(see \sphinxtitleref{endpoint} below).

\DUrole{versionmodified,changed}{Changed in version 1.16.0: }Non\sphinxhyphen{}scalar \sphinxtitleref{start} and \sphinxtitleref{stop} are now supported.
\begin{quote}\begin{description}
\item[{Parameters}] \leavevmode
\sphinxstyleliteralstrong{\sphinxupquote{dtype}} (\sphinxstyleliteralemphasis{\sphinxupquote{dtype}}) \textendash{} The type of the output array.  If \sphinxtitleref{dtype} is not given, infer the data
type from the other input arguments.

\item[{Returns}] \leavevmode
\sphinxstylestrong{samples} \textendash{} \sphinxtitleref{num} samples, equally spaced on a log scale.

\item[{Return type}] \leavevmode
ndarray

\end{description}\end{quote}

\sphinxstrong{See also:}

\begin{description}
\item[{{\hyperref[\detokenize{modules/tensor:symjax.tensor.arange}]{\sphinxcrossref{\sphinxcode{\sphinxupquote{arange()}}}}}}] \leavevmode
Similar to linspace, with the step size specified instead of the number of samples. Note that, when used with a float endpoint, the endpoint may or may not be included.

\item[{{\hyperref[\detokenize{modules/tensor:symjax.tensor.linspace}]{\sphinxcrossref{\sphinxcode{\sphinxupquote{linspace()}}}}}}] \leavevmode
Similar to logspace, but with the samples uniformly distributed in linear space, instead of log space.

\item[{{\hyperref[\detokenize{modules/tensor:symjax.tensor.geomspace}]{\sphinxcrossref{\sphinxcode{\sphinxupquote{geomspace()}}}}}}] \leavevmode
Similar to logspace, but with endpoints specified directly.

\end{description}

\subsubsection*{Notes}

Logspace is equivalent to the code

\begin{sphinxVerbatim}[commandchars=\\\{\}]
\PYG{g+gp}{\PYGZgt{}\PYGZgt{}\PYGZgt{} }\PYG{n}{y} \PYG{o}{=} \PYG{n}{np}\PYG{o}{.}\PYG{n}{linspace}\PYG{p}{(}\PYG{n}{start}\PYG{p}{,} \PYG{n}{stop}\PYG{p}{,} \PYG{n}{num}\PYG{o}{=}\PYG{n}{num}\PYG{p}{,} \PYG{n}{endpoint}\PYG{o}{=}\PYG{n}{endpoint}\PYG{p}{)}
\PYG{g+gp}{... }
\PYG{g+gp}{\PYGZgt{}\PYGZgt{}\PYGZgt{} }\PYG{n}{power}\PYG{p}{(}\PYG{n}{base}\PYG{p}{,} \PYG{n}{y}\PYG{p}{)}\PYG{o}{.}\PYG{n}{astype}\PYG{p}{(}\PYG{n}{dtype}\PYG{p}{)}
\PYG{g+gp}{... }
\end{sphinxVerbatim}
\subsubsection*{Examples}

\begin{sphinxVerbatim}[commandchars=\\\{\}]
\PYG{g+gp}{\PYGZgt{}\PYGZgt{}\PYGZgt{} }\PYG{n}{np}\PYG{o}{.}\PYG{n}{logspace}\PYG{p}{(}\PYG{l+m+mf}{2.0}\PYG{p}{,} \PYG{l+m+mf}{3.0}\PYG{p}{,} \PYG{n}{num}\PYG{o}{=}\PYG{l+m+mi}{4}\PYG{p}{)}
\PYG{g+go}{array([ 100.        ,  215.443469  ,  464.15888336, 1000.        ])}
\PYG{g+gp}{\PYGZgt{}\PYGZgt{}\PYGZgt{} }\PYG{n}{np}\PYG{o}{.}\PYG{n}{logspace}\PYG{p}{(}\PYG{l+m+mf}{2.0}\PYG{p}{,} \PYG{l+m+mf}{3.0}\PYG{p}{,} \PYG{n}{num}\PYG{o}{=}\PYG{l+m+mi}{4}\PYG{p}{,} \PYG{n}{endpoint}\PYG{o}{=}\PYG{k+kc}{False}\PYG{p}{)}
\PYG{g+go}{array([100.        ,  177.827941  ,  316.22776602,  562.34132519])}
\PYG{g+gp}{\PYGZgt{}\PYGZgt{}\PYGZgt{} }\PYG{n}{np}\PYG{o}{.}\PYG{n}{logspace}\PYG{p}{(}\PYG{l+m+mf}{2.0}\PYG{p}{,} \PYG{l+m+mf}{3.0}\PYG{p}{,} \PYG{n}{num}\PYG{o}{=}\PYG{l+m+mi}{4}\PYG{p}{,} \PYG{n}{base}\PYG{o}{=}\PYG{l+m+mf}{2.0}\PYG{p}{)}
\PYG{g+go}{array([4.        ,  5.0396842 ,  6.34960421,  8.        ])}
\end{sphinxVerbatim}

Graphical illustration:

\begin{sphinxVerbatim}[commandchars=\\\{\}]
\PYG{g+gp}{\PYGZgt{}\PYGZgt{}\PYGZgt{} }\PYG{k+kn}{import} \PYG{n+nn}{matplotlib}\PYG{n+nn}{.}\PYG{n+nn}{pyplot} \PYG{k}{as} \PYG{n+nn}{plt}
\PYG{g+gp}{\PYGZgt{}\PYGZgt{}\PYGZgt{} }\PYG{n}{N} \PYG{o}{=} \PYG{l+m+mi}{10}
\PYG{g+gp}{\PYGZgt{}\PYGZgt{}\PYGZgt{} }\PYG{n}{x1} \PYG{o}{=} \PYG{n}{np}\PYG{o}{.}\PYG{n}{logspace}\PYG{p}{(}\PYG{l+m+mf}{0.1}\PYG{p}{,} \PYG{l+m+mi}{1}\PYG{p}{,} \PYG{n}{N}\PYG{p}{,} \PYG{n}{endpoint}\PYG{o}{=}\PYG{k+kc}{True}\PYG{p}{)}
\PYG{g+gp}{\PYGZgt{}\PYGZgt{}\PYGZgt{} }\PYG{n}{x2} \PYG{o}{=} \PYG{n}{np}\PYG{o}{.}\PYG{n}{logspace}\PYG{p}{(}\PYG{l+m+mf}{0.1}\PYG{p}{,} \PYG{l+m+mi}{1}\PYG{p}{,} \PYG{n}{N}\PYG{p}{,} \PYG{n}{endpoint}\PYG{o}{=}\PYG{k+kc}{False}\PYG{p}{)}
\PYG{g+gp}{\PYGZgt{}\PYGZgt{}\PYGZgt{} }\PYG{n}{y} \PYG{o}{=} \PYG{n}{np}\PYG{o}{.}\PYG{n}{zeros}\PYG{p}{(}\PYG{n}{N}\PYG{p}{)}
\PYG{g+gp}{\PYGZgt{}\PYGZgt{}\PYGZgt{} }\PYG{n}{plt}\PYG{o}{.}\PYG{n}{plot}\PYG{p}{(}\PYG{n}{x1}\PYG{p}{,} \PYG{n}{y}\PYG{p}{,} \PYG{l+s+s1}{\PYGZsq{}}\PYG{l+s+s1}{o}\PYG{l+s+s1}{\PYGZsq{}}\PYG{p}{)}
\PYG{g+go}{[\PYGZlt{}matplotlib.lines.Line2D object at 0x...\PYGZgt{}]}
\PYG{g+gp}{\PYGZgt{}\PYGZgt{}\PYGZgt{} }\PYG{n}{plt}\PYG{o}{.}\PYG{n}{plot}\PYG{p}{(}\PYG{n}{x2}\PYG{p}{,} \PYG{n}{y} \PYG{o}{+} \PYG{l+m+mf}{0.5}\PYG{p}{,} \PYG{l+s+s1}{\PYGZsq{}}\PYG{l+s+s1}{o}\PYG{l+s+s1}{\PYGZsq{}}\PYG{p}{)}
\PYG{g+go}{[\PYGZlt{}matplotlib.lines.Line2D object at 0x...\PYGZgt{}]}
\PYG{g+gp}{\PYGZgt{}\PYGZgt{}\PYGZgt{} }\PYG{n}{plt}\PYG{o}{.}\PYG{n}{ylim}\PYG{p}{(}\PYG{p}{[}\PYG{o}{\PYGZhy{}}\PYG{l+m+mf}{0.5}\PYG{p}{,} \PYG{l+m+mi}{1}\PYG{p}{]}\PYG{p}{)}
\PYG{g+go}{(\PYGZhy{}0.5, 1)}
\PYG{g+gp}{\PYGZgt{}\PYGZgt{}\PYGZgt{} }\PYG{n}{plt}\PYG{o}{.}\PYG{n}{show}\PYG{p}{(}\PYG{p}{)}
\end{sphinxVerbatim}

\end{fulllineitems}

\index{matmul() (in module symjax.tensor)@\spxentry{matmul()}\spxextra{in module symjax.tensor}}

\begin{fulllineitems}
\phantomsection\label{\detokenize{modules/tensor:symjax.tensor.matmul}}\pysiglinewithargsret{\sphinxbfcode{\sphinxupquote{matmul}}}{\emph{\DUrole{n}{a}}, \emph{\DUrole{n}{b}}, \emph{\DUrole{n}{precision}\DUrole{o}{=}\DUrole{default_value}{None}}}{}
Matrix product of two arrays.

LAX\sphinxhyphen{}backend implementation of {\hyperref[\detokenize{modules/tensor:symjax.tensor.matmul}]{\sphinxcrossref{\sphinxcode{\sphinxupquote{matmul()}}}}}.
ADDITIONOriginal docstring below.

LAX\sphinxhyphen{}backend implementation of {\hyperref[\detokenize{modules/tensor:symjax.tensor.matmul}]{\sphinxcrossref{\sphinxcode{\sphinxupquote{matmul()}}}}}.
In addition to the original NumPy arguments listed below, also supports
\sphinxcode{\sphinxupquote{precision}} for extra control over matrix\sphinxhyphen{}multiplication precision
on supported devices. See \sphinxcode{\sphinxupquote{jax.lax.dot()}} for details.

Original docstring below.

matmul(x1, x2, /, out=None, {\color{red}\bfseries{}*}, casting=’same\_kind’, order=’K’, dtype=None, subok=True{[}, signature, extobj{]})
\begin{quote}\begin{description}
\item[{Returns}] \leavevmode
\sphinxstylestrong{y} \textendash{} The matrix product of the inputs.
This is a scalar only when both x1, x2 are 1\sphinxhyphen{}d vectors.

\item[{Return type}] \leavevmode
ndarray

\item[{Raises}] \leavevmode
\sphinxstyleliteralstrong{\sphinxupquote{ValueError}} \textendash{} If the last dimension of \sphinxtitleref{a} is not the same size as
    the second\sphinxhyphen{}to\sphinxhyphen{}last dimension of \sphinxtitleref{b}.
    
    If a scalar value is passed in.

\end{description}\end{quote}

\sphinxstrong{See also:}

\begin{description}
\item[{{\hyperref[\detokenize{modules/tensor:symjax.tensor.vdot}]{\sphinxcrossref{\sphinxcode{\sphinxupquote{vdot()}}}}}}] \leavevmode
Complex\sphinxhyphen{}conjugating dot product.

\item[{{\hyperref[\detokenize{modules/tensor:symjax.tensor.tensordot}]{\sphinxcrossref{\sphinxcode{\sphinxupquote{tensordot()}}}}}}] \leavevmode
Sum products over arbitrary axes.

\item[{{\hyperref[\detokenize{modules/tensor:symjax.tensor.einsum}]{\sphinxcrossref{\sphinxcode{\sphinxupquote{einsum()}}}}}}] \leavevmode
Einstein summation convention.

\item[{{\hyperref[\detokenize{modules/tensor:symjax.tensor.dot}]{\sphinxcrossref{\sphinxcode{\sphinxupquote{dot()}}}}}}] \leavevmode
alternative matrix product with different broadcasting rules.

\end{description}

\subsubsection*{Notes}

The behavior depends on the arguments in the following way.
\begin{itemize}
\item {} 
If both arguments are 2\sphinxhyphen{}D they are multiplied like conventional
matrices.

\item {} 
If either argument is N\sphinxhyphen{}D, N \textgreater{} 2, it is treated as a stack of
matrices residing in the last two indexes and broadcast accordingly.

\item {} 
If the first argument is 1\sphinxhyphen{}D, it is promoted to a matrix by
prepending a 1 to its dimensions. After matrix multiplication
the prepended 1 is removed.

\item {} 
If the second argument is 1\sphinxhyphen{}D, it is promoted to a matrix by
appending a 1 to its dimensions. After matrix multiplication
the appended 1 is removed.

\end{itemize}

\sphinxcode{\sphinxupquote{matmul}} differs from \sphinxcode{\sphinxupquote{dot}} in two important ways:
\begin{itemize}
\item {} 
Multiplication by scalars is not allowed, use \sphinxcode{\sphinxupquote{*}} instead.

\item {} 
Stacks of matrices are broadcast together as if the matrices
were elements, respecting the signature \sphinxcode{\sphinxupquote{(n,k),(k,m)\sphinxhyphen{}\textgreater{}(n,m)}}:

\begin{sphinxVerbatim}[commandchars=\\\{\}]
\PYG{g+gp}{\PYGZgt{}\PYGZgt{}\PYGZgt{} }\PYG{n}{a} \PYG{o}{=} \PYG{n}{np}\PYG{o}{.}\PYG{n}{ones}\PYG{p}{(}\PYG{p}{[}\PYG{l+m+mi}{9}\PYG{p}{,} \PYG{l+m+mi}{5}\PYG{p}{,} \PYG{l+m+mi}{7}\PYG{p}{,} \PYG{l+m+mi}{4}\PYG{p}{]}\PYG{p}{)}
\PYG{g+gp}{\PYGZgt{}\PYGZgt{}\PYGZgt{} }\PYG{n}{c} \PYG{o}{=} \PYG{n}{np}\PYG{o}{.}\PYG{n}{ones}\PYG{p}{(}\PYG{p}{[}\PYG{l+m+mi}{9}\PYG{p}{,} \PYG{l+m+mi}{5}\PYG{p}{,} \PYG{l+m+mi}{4}\PYG{p}{,} \PYG{l+m+mi}{3}\PYG{p}{]}\PYG{p}{)}
\PYG{g+gp}{\PYGZgt{}\PYGZgt{}\PYGZgt{} }\PYG{n}{np}\PYG{o}{.}\PYG{n}{dot}\PYG{p}{(}\PYG{n}{a}\PYG{p}{,} \PYG{n}{c}\PYG{p}{)}\PYG{o}{.}\PYG{n}{shape}
\PYG{g+go}{(9, 5, 7, 9, 5, 3)}
\PYG{g+gp}{\PYGZgt{}\PYGZgt{}\PYGZgt{} }\PYG{n}{np}\PYG{o}{.}\PYG{n}{matmul}\PYG{p}{(}\PYG{n}{a}\PYG{p}{,} \PYG{n}{c}\PYG{p}{)}\PYG{o}{.}\PYG{n}{shape}
\PYG{g+go}{(9, 5, 7, 3)}
\PYG{g+gp}{\PYGZgt{}\PYGZgt{}\PYGZgt{} }\PYG{c+c1}{\PYGZsh{} n is 7, k is 4, m is 3}
\end{sphinxVerbatim}

\end{itemize}

The matmul function implements the semantics of the \sphinxtitleref{@} operator introduced
in Python 3.5 following PEP465.
\subsubsection*{Examples}

For 2\sphinxhyphen{}D arrays it is the matrix product:

\begin{sphinxVerbatim}[commandchars=\\\{\}]
\PYG{g+gp}{\PYGZgt{}\PYGZgt{}\PYGZgt{} }\PYG{n}{a} \PYG{o}{=} \PYG{n}{np}\PYG{o}{.}\PYG{n}{array}\PYG{p}{(}\PYG{p}{[}\PYG{p}{[}\PYG{l+m+mi}{1}\PYG{p}{,} \PYG{l+m+mi}{0}\PYG{p}{]}\PYG{p}{,}
\PYG{g+gp}{... }              \PYG{p}{[}\PYG{l+m+mi}{0}\PYG{p}{,} \PYG{l+m+mi}{1}\PYG{p}{]}\PYG{p}{]}\PYG{p}{)}
\PYG{g+gp}{\PYGZgt{}\PYGZgt{}\PYGZgt{} }\PYG{n}{b} \PYG{o}{=} \PYG{n}{np}\PYG{o}{.}\PYG{n}{array}\PYG{p}{(}\PYG{p}{[}\PYG{p}{[}\PYG{l+m+mi}{4}\PYG{p}{,} \PYG{l+m+mi}{1}\PYG{p}{]}\PYG{p}{,}
\PYG{g+gp}{... }              \PYG{p}{[}\PYG{l+m+mi}{2}\PYG{p}{,} \PYG{l+m+mi}{2}\PYG{p}{]}\PYG{p}{]}\PYG{p}{)}
\PYG{g+gp}{\PYGZgt{}\PYGZgt{}\PYGZgt{} }\PYG{n}{np}\PYG{o}{.}\PYG{n}{matmul}\PYG{p}{(}\PYG{n}{a}\PYG{p}{,} \PYG{n}{b}\PYG{p}{)}
\PYG{g+go}{array([[4, 1],}
\PYG{g+go}{       [2, 2]])}
\end{sphinxVerbatim}

For 2\sphinxhyphen{}D mixed with 1\sphinxhyphen{}D, the result is the usual.

\begin{sphinxVerbatim}[commandchars=\\\{\}]
\PYG{g+gp}{\PYGZgt{}\PYGZgt{}\PYGZgt{} }\PYG{n}{a} \PYG{o}{=} \PYG{n}{np}\PYG{o}{.}\PYG{n}{array}\PYG{p}{(}\PYG{p}{[}\PYG{p}{[}\PYG{l+m+mi}{1}\PYG{p}{,} \PYG{l+m+mi}{0}\PYG{p}{]}\PYG{p}{,}
\PYG{g+gp}{... }              \PYG{p}{[}\PYG{l+m+mi}{0}\PYG{p}{,} \PYG{l+m+mi}{1}\PYG{p}{]}\PYG{p}{]}\PYG{p}{)}
\PYG{g+gp}{\PYGZgt{}\PYGZgt{}\PYGZgt{} }\PYG{n}{b} \PYG{o}{=} \PYG{n}{np}\PYG{o}{.}\PYG{n}{array}\PYG{p}{(}\PYG{p}{[}\PYG{l+m+mi}{1}\PYG{p}{,} \PYG{l+m+mi}{2}\PYG{p}{]}\PYG{p}{)}
\PYG{g+gp}{\PYGZgt{}\PYGZgt{}\PYGZgt{} }\PYG{n}{np}\PYG{o}{.}\PYG{n}{matmul}\PYG{p}{(}\PYG{n}{a}\PYG{p}{,} \PYG{n}{b}\PYG{p}{)}
\PYG{g+go}{array([1, 2])}
\PYG{g+gp}{\PYGZgt{}\PYGZgt{}\PYGZgt{} }\PYG{n}{np}\PYG{o}{.}\PYG{n}{matmul}\PYG{p}{(}\PYG{n}{b}\PYG{p}{,} \PYG{n}{a}\PYG{p}{)}
\PYG{g+go}{array([1, 2])}
\end{sphinxVerbatim}

Broadcasting is conventional for stacks of arrays

\begin{sphinxVerbatim}[commandchars=\\\{\}]
\PYG{g+gp}{\PYGZgt{}\PYGZgt{}\PYGZgt{} }\PYG{n}{a} \PYG{o}{=} \PYG{n}{np}\PYG{o}{.}\PYG{n}{arange}\PYG{p}{(}\PYG{l+m+mi}{2} \PYG{o}{*} \PYG{l+m+mi}{2} \PYG{o}{*} \PYG{l+m+mi}{4}\PYG{p}{)}\PYG{o}{.}\PYG{n}{reshape}\PYG{p}{(}\PYG{p}{(}\PYG{l+m+mi}{2}\PYG{p}{,} \PYG{l+m+mi}{2}\PYG{p}{,} \PYG{l+m+mi}{4}\PYG{p}{)}\PYG{p}{)}
\PYG{g+gp}{\PYGZgt{}\PYGZgt{}\PYGZgt{} }\PYG{n}{b} \PYG{o}{=} \PYG{n}{np}\PYG{o}{.}\PYG{n}{arange}\PYG{p}{(}\PYG{l+m+mi}{2} \PYG{o}{*} \PYG{l+m+mi}{2} \PYG{o}{*} \PYG{l+m+mi}{4}\PYG{p}{)}\PYG{o}{.}\PYG{n}{reshape}\PYG{p}{(}\PYG{p}{(}\PYG{l+m+mi}{2}\PYG{p}{,} \PYG{l+m+mi}{4}\PYG{p}{,} \PYG{l+m+mi}{2}\PYG{p}{)}\PYG{p}{)}
\PYG{g+gp}{\PYGZgt{}\PYGZgt{}\PYGZgt{} }\PYG{n}{np}\PYG{o}{.}\PYG{n}{matmul}\PYG{p}{(}\PYG{n}{a}\PYG{p}{,}\PYG{n}{b}\PYG{p}{)}\PYG{o}{.}\PYG{n}{shape}
\PYG{g+go}{(2, 2, 2)}
\PYG{g+gp}{\PYGZgt{}\PYGZgt{}\PYGZgt{} }\PYG{n}{np}\PYG{o}{.}\PYG{n}{matmul}\PYG{p}{(}\PYG{n}{a}\PYG{p}{,} \PYG{n}{b}\PYG{p}{)}\PYG{p}{[}\PYG{l+m+mi}{0}\PYG{p}{,} \PYG{l+m+mi}{1}\PYG{p}{,} \PYG{l+m+mi}{1}\PYG{p}{]}
\PYG{g+go}{98}
\PYG{g+gp}{\PYGZgt{}\PYGZgt{}\PYGZgt{} }\PYG{n+nb}{sum}\PYG{p}{(}\PYG{n}{a}\PYG{p}{[}\PYG{l+m+mi}{0}\PYG{p}{,} \PYG{l+m+mi}{1}\PYG{p}{,} \PYG{p}{:}\PYG{p}{]} \PYG{o}{*} \PYG{n}{b}\PYG{p}{[}\PYG{l+m+mi}{0} \PYG{p}{,} \PYG{p}{:}\PYG{p}{,} \PYG{l+m+mi}{1}\PYG{p}{]}\PYG{p}{)}
\PYG{g+go}{98}
\end{sphinxVerbatim}

Vector, vector returns the scalar inner product, but neither argument
is complex\sphinxhyphen{}conjugated:

\begin{sphinxVerbatim}[commandchars=\\\{\}]
\PYG{g+gp}{\PYGZgt{}\PYGZgt{}\PYGZgt{} }\PYG{n}{np}\PYG{o}{.}\PYG{n}{matmul}\PYG{p}{(}\PYG{p}{[}\PYG{l+m+mi}{2}\PYG{n}{j}\PYG{p}{,} \PYG{l+m+mi}{3}\PYG{n}{j}\PYG{p}{]}\PYG{p}{,} \PYG{p}{[}\PYG{l+m+mi}{2}\PYG{n}{j}\PYG{p}{,} \PYG{l+m+mi}{3}\PYG{n}{j}\PYG{p}{]}\PYG{p}{)}
\PYG{g+go}{(\PYGZhy{}13+0j)}
\end{sphinxVerbatim}

Scalar multiplication raises an error.

\begin{sphinxVerbatim}[commandchars=\\\{\}]
\PYG{g+gp}{\PYGZgt{}\PYGZgt{}\PYGZgt{} }\PYG{n}{np}\PYG{o}{.}\PYG{n}{matmul}\PYG{p}{(}\PYG{p}{[}\PYG{l+m+mi}{1}\PYG{p}{,}\PYG{l+m+mi}{2}\PYG{p}{]}\PYG{p}{,} \PYG{l+m+mi}{3}\PYG{p}{)}
\PYG{g+gt}{Traceback (most recent call last):}
\PYG{c}{...}
\PYG{g+gr}{ValueError}: \PYG{n}{matmul: Input operand 1 does not have enough dimensions ...}
\end{sphinxVerbatim}

\DUrole{versionmodified,added}{New in version 1.10.0.}

\end{fulllineitems}

\index{max() (in module symjax.tensor)@\spxentry{max()}\spxextra{in module symjax.tensor}}

\begin{fulllineitems}
\phantomsection\label{\detokenize{modules/tensor:symjax.tensor.max}}\pysiglinewithargsret{\sphinxbfcode{\sphinxupquote{max}}}{\emph{\DUrole{n}{a}}, \emph{\DUrole{n}{axis}\DUrole{o}{=}\DUrole{default_value}{None}}, \emph{\DUrole{n}{dtype}\DUrole{o}{=}\DUrole{default_value}{None}}, \emph{\DUrole{n}{out}\DUrole{o}{=}\DUrole{default_value}{None}}, \emph{\DUrole{n}{keepdims}\DUrole{o}{=}\DUrole{default_value}{False}}}{}
Return the maximum of an array or maximum along an axis.

LAX\sphinxhyphen{}backend implementation of {\hyperref[\detokenize{modules/tensor:symjax.tensor.amax}]{\sphinxcrossref{\sphinxcode{\sphinxupquote{amax()}}}}}.
ADDITIONOriginal docstring below.

LAX\sphinxhyphen{}backend implementation of {\hyperref[\detokenize{modules/tensor:symjax.tensor.amax}]{\sphinxcrossref{\sphinxcode{\sphinxupquote{amax()}}}}}.
Original docstring below.
\begin{quote}\begin{description}
\item[{Returns}] \leavevmode
\sphinxstylestrong{amax} \textendash{} Maximum of \sphinxtitleref{a}. If \sphinxtitleref{axis} is None, the result is a scalar value.
If \sphinxtitleref{axis} is given, the result is an array of dimension
\sphinxcode{\sphinxupquote{a.ndim \sphinxhyphen{} 1}}.

\item[{Return type}] \leavevmode
ndarray or scalar

\end{description}\end{quote}

\sphinxstrong{See also:}

\begin{description}
\item[{{\hyperref[\detokenize{modules/tensor:symjax.tensor.amin}]{\sphinxcrossref{\sphinxcode{\sphinxupquote{amin()}}}}}}] \leavevmode
The minimum value of an array along a given axis, propagating any NaNs.

\item[{{\hyperref[\detokenize{modules/tensor:symjax.tensor.nanmax}]{\sphinxcrossref{\sphinxcode{\sphinxupquote{nanmax()}}}}}}] \leavevmode
The maximum value of an array along a given axis, ignoring any NaNs.

\item[{{\hyperref[\detokenize{modules/tensor:symjax.tensor.maximum}]{\sphinxcrossref{\sphinxcode{\sphinxupquote{maximum()}}}}}}] \leavevmode
Element\sphinxhyphen{}wise maximum of two arrays, propagating any NaNs.

\item[{\sphinxcode{\sphinxupquote{fmax()}}}] \leavevmode
Element\sphinxhyphen{}wise maximum of two arrays, ignoring any NaNs.

\item[{{\hyperref[\detokenize{modules/tensor:symjax.tensor.argmax}]{\sphinxcrossref{\sphinxcode{\sphinxupquote{argmax()}}}}}}] \leavevmode
Return the indices of the maximum values.

\end{description}

{\hyperref[\detokenize{modules/tensor:symjax.tensor.nanmin}]{\sphinxcrossref{\sphinxcode{\sphinxupquote{nanmin()}}}}}, {\hyperref[\detokenize{modules/tensor:symjax.tensor.minimum}]{\sphinxcrossref{\sphinxcode{\sphinxupquote{minimum()}}}}}, \sphinxcode{\sphinxupquote{fmin()}}

\subsubsection*{Notes}

NaN values are propagated, that is if at least one item is NaN, the
corresponding max value will be NaN as well. To ignore NaN values
(MATLAB behavior), please use nanmax.

Don’t use \sphinxtitleref{amax} for element\sphinxhyphen{}wise comparison of 2 arrays; when
\sphinxcode{\sphinxupquote{a.shape{[}0{]}}} is 2, \sphinxcode{\sphinxupquote{maximum(a{[}0{]}, a{[}1{]})}} is faster than
\sphinxcode{\sphinxupquote{amax(a, axis=0)}}.
\subsubsection*{Examples}

\begin{sphinxVerbatim}[commandchars=\\\{\}]
\PYG{g+gp}{\PYGZgt{}\PYGZgt{}\PYGZgt{} }\PYG{n}{a} \PYG{o}{=} \PYG{n}{np}\PYG{o}{.}\PYG{n}{arange}\PYG{p}{(}\PYG{l+m+mi}{4}\PYG{p}{)}\PYG{o}{.}\PYG{n}{reshape}\PYG{p}{(}\PYG{p}{(}\PYG{l+m+mi}{2}\PYG{p}{,}\PYG{l+m+mi}{2}\PYG{p}{)}\PYG{p}{)}
\PYG{g+gp}{\PYGZgt{}\PYGZgt{}\PYGZgt{} }\PYG{n}{a}
\PYG{g+go}{array([[0, 1],}
\PYG{g+go}{       [2, 3]])}
\PYG{g+gp}{\PYGZgt{}\PYGZgt{}\PYGZgt{} }\PYG{n}{np}\PYG{o}{.}\PYG{n}{amax}\PYG{p}{(}\PYG{n}{a}\PYG{p}{)}           \PYG{c+c1}{\PYGZsh{} Maximum of the flattened array}
\PYG{g+go}{3}
\PYG{g+gp}{\PYGZgt{}\PYGZgt{}\PYGZgt{} }\PYG{n}{np}\PYG{o}{.}\PYG{n}{amax}\PYG{p}{(}\PYG{n}{a}\PYG{p}{,} \PYG{n}{axis}\PYG{o}{=}\PYG{l+m+mi}{0}\PYG{p}{)}   \PYG{c+c1}{\PYGZsh{} Maxima along the first axis}
\PYG{g+go}{array([2, 3])}
\PYG{g+gp}{\PYGZgt{}\PYGZgt{}\PYGZgt{} }\PYG{n}{np}\PYG{o}{.}\PYG{n}{amax}\PYG{p}{(}\PYG{n}{a}\PYG{p}{,} \PYG{n}{axis}\PYG{o}{=}\PYG{l+m+mi}{1}\PYG{p}{)}   \PYG{c+c1}{\PYGZsh{} Maxima along the second axis}
\PYG{g+go}{array([1, 3])}
\PYG{g+gp}{\PYGZgt{}\PYGZgt{}\PYGZgt{} }\PYG{n}{np}\PYG{o}{.}\PYG{n}{amax}\PYG{p}{(}\PYG{n}{a}\PYG{p}{,} \PYG{n}{where}\PYG{o}{=}\PYG{p}{[}\PYG{k+kc}{False}\PYG{p}{,} \PYG{k+kc}{True}\PYG{p}{]}\PYG{p}{,} \PYG{n}{initial}\PYG{o}{=}\PYG{o}{\PYGZhy{}}\PYG{l+m+mi}{1}\PYG{p}{,} \PYG{n}{axis}\PYG{o}{=}\PYG{l+m+mi}{0}\PYG{p}{)}
\PYG{g+go}{array([\PYGZhy{}1,  3])}
\PYG{g+gp}{\PYGZgt{}\PYGZgt{}\PYGZgt{} }\PYG{n}{b} \PYG{o}{=} \PYG{n}{np}\PYG{o}{.}\PYG{n}{arange}\PYG{p}{(}\PYG{l+m+mi}{5}\PYG{p}{,} \PYG{n}{dtype}\PYG{o}{=}\PYG{n+nb}{float}\PYG{p}{)}
\PYG{g+gp}{\PYGZgt{}\PYGZgt{}\PYGZgt{} }\PYG{n}{b}\PYG{p}{[}\PYG{l+m+mi}{2}\PYG{p}{]} \PYG{o}{=} \PYG{n}{np}\PYG{o}{.}\PYG{n}{NaN}
\PYG{g+gp}{\PYGZgt{}\PYGZgt{}\PYGZgt{} }\PYG{n}{np}\PYG{o}{.}\PYG{n}{amax}\PYG{p}{(}\PYG{n}{b}\PYG{p}{)}
\PYG{g+go}{nan}
\PYG{g+gp}{\PYGZgt{}\PYGZgt{}\PYGZgt{} }\PYG{n}{np}\PYG{o}{.}\PYG{n}{amax}\PYG{p}{(}\PYG{n}{b}\PYG{p}{,} \PYG{n}{where}\PYG{o}{=}\PYG{o}{\PYGZti{}}\PYG{n}{np}\PYG{o}{.}\PYG{n}{isnan}\PYG{p}{(}\PYG{n}{b}\PYG{p}{)}\PYG{p}{,} \PYG{n}{initial}\PYG{o}{=}\PYG{o}{\PYGZhy{}}\PYG{l+m+mi}{1}\PYG{p}{)}
\PYG{g+go}{4.0}
\PYG{g+gp}{\PYGZgt{}\PYGZgt{}\PYGZgt{} }\PYG{n}{np}\PYG{o}{.}\PYG{n}{nanmax}\PYG{p}{(}\PYG{n}{b}\PYG{p}{)}
\PYG{g+go}{4.0}
\end{sphinxVerbatim}

You can use an initial value to compute the maximum of an empty slice, or
to initialize it to a different value:

\begin{sphinxVerbatim}[commandchars=\\\{\}]
\PYG{g+gp}{\PYGZgt{}\PYGZgt{}\PYGZgt{} }\PYG{n}{np}\PYG{o}{.}\PYG{n}{max}\PYG{p}{(}\PYG{p}{[}\PYG{p}{[}\PYG{o}{\PYGZhy{}}\PYG{l+m+mi}{50}\PYG{p}{]}\PYG{p}{,} \PYG{p}{[}\PYG{l+m+mi}{10}\PYG{p}{]}\PYG{p}{]}\PYG{p}{,} \PYG{n}{axis}\PYG{o}{=}\PYG{o}{\PYGZhy{}}\PYG{l+m+mi}{1}\PYG{p}{,} \PYG{n}{initial}\PYG{o}{=}\PYG{l+m+mi}{0}\PYG{p}{)}
\PYG{g+go}{array([ 0, 10])}
\end{sphinxVerbatim}

Notice that the initial value is used as one of the elements for which the
maximum is determined, unlike for the default argument Python’s max
function, which is only used for empty iterables.

\begin{sphinxVerbatim}[commandchars=\\\{\}]
\PYG{g+gp}{\PYGZgt{}\PYGZgt{}\PYGZgt{} }\PYG{n}{np}\PYG{o}{.}\PYG{n}{max}\PYG{p}{(}\PYG{p}{[}\PYG{l+m+mi}{5}\PYG{p}{]}\PYG{p}{,} \PYG{n}{initial}\PYG{o}{=}\PYG{l+m+mi}{6}\PYG{p}{)}
\PYG{g+go}{6}
\PYG{g+gp}{\PYGZgt{}\PYGZgt{}\PYGZgt{} }\PYG{n+nb}{max}\PYG{p}{(}\PYG{p}{[}\PYG{l+m+mi}{5}\PYG{p}{]}\PYG{p}{,} \PYG{n}{default}\PYG{o}{=}\PYG{l+m+mi}{6}\PYG{p}{)}
\PYG{g+go}{5}
\end{sphinxVerbatim}

\end{fulllineitems}

\index{maximum() (in module symjax.tensor)@\spxentry{maximum()}\spxextra{in module symjax.tensor}}

\begin{fulllineitems}
\phantomsection\label{\detokenize{modules/tensor:symjax.tensor.maximum}}\pysiglinewithargsret{\sphinxbfcode{\sphinxupquote{maximum}}}{\emph{\DUrole{n}{x1}}, \emph{\DUrole{n}{x2}}}{}
Element\sphinxhyphen{}wise maximum of array elements.

LAX\sphinxhyphen{}backend implementation of {\hyperref[\detokenize{modules/tensor:symjax.tensor.maximum}]{\sphinxcrossref{\sphinxcode{\sphinxupquote{maximum()}}}}}.
ADDITIONOriginal docstring below.

LAX\sphinxhyphen{}backend implementation of {\hyperref[\detokenize{modules/tensor:symjax.tensor.maximum}]{\sphinxcrossref{\sphinxcode{\sphinxupquote{maximum()}}}}}.
Original docstring below.

maximum(x1, x2, /, out=None, {\color{red}\bfseries{}*}, where=True, casting=’same\_kind’, order=’K’, dtype=None, subok=True{[}, signature, extobj{]})

Compare two arrays and returns a new array containing the element\sphinxhyphen{}wise
maxima. If one of the elements being compared is a NaN, then that
element is returned. If both elements are NaNs then the first is
returned. The latter distinction is important for complex NaNs, which
are defined as at least one of the real or imaginary parts being a NaN.
The net effect is that NaNs are propagated.
\begin{quote}\begin{description}
\item[{Returns}] \leavevmode
\sphinxstylestrong{y} \textendash{} The maximum of \sphinxtitleref{x1} and \sphinxtitleref{x2}, element\sphinxhyphen{}wise.
This is a scalar if both \sphinxtitleref{x1} and \sphinxtitleref{x2} are scalars.

\item[{Return type}] \leavevmode
ndarray or scalar

\end{description}\end{quote}

\sphinxstrong{See also:}

\begin{description}
\item[{{\hyperref[\detokenize{modules/tensor:symjax.tensor.minimum}]{\sphinxcrossref{\sphinxcode{\sphinxupquote{minimum()}}}}}}] \leavevmode
Element\sphinxhyphen{}wise minimum of two arrays, propagates NaNs.

\item[{\sphinxcode{\sphinxupquote{fmax()}}}] \leavevmode
Element\sphinxhyphen{}wise maximum of two arrays, ignores NaNs.

\item[{{\hyperref[\detokenize{modules/tensor:symjax.tensor.amax}]{\sphinxcrossref{\sphinxcode{\sphinxupquote{amax()}}}}}}] \leavevmode
The maximum value of an array along a given axis, propagates NaNs.

\item[{{\hyperref[\detokenize{modules/tensor:symjax.tensor.nanmax}]{\sphinxcrossref{\sphinxcode{\sphinxupquote{nanmax()}}}}}}] \leavevmode
The maximum value of an array along a given axis, ignores NaNs.

\end{description}

\sphinxcode{\sphinxupquote{fmin()}}, {\hyperref[\detokenize{modules/tensor:symjax.tensor.amin}]{\sphinxcrossref{\sphinxcode{\sphinxupquote{amin()}}}}}, {\hyperref[\detokenize{modules/tensor:symjax.tensor.nanmin}]{\sphinxcrossref{\sphinxcode{\sphinxupquote{nanmin()}}}}}

\subsubsection*{Notes}

The maximum is equivalent to \sphinxcode{\sphinxupquote{np.where(x1 \textgreater{}= x2, x1, x2)}} when
neither x1 nor x2 are nans, but it is faster and does proper
broadcasting.
\subsubsection*{Examples}

\begin{sphinxVerbatim}[commandchars=\\\{\}]
\PYG{g+gp}{\PYGZgt{}\PYGZgt{}\PYGZgt{} }\PYG{n}{np}\PYG{o}{.}\PYG{n}{maximum}\PYG{p}{(}\PYG{p}{[}\PYG{l+m+mi}{2}\PYG{p}{,} \PYG{l+m+mi}{3}\PYG{p}{,} \PYG{l+m+mi}{4}\PYG{p}{]}\PYG{p}{,} \PYG{p}{[}\PYG{l+m+mi}{1}\PYG{p}{,} \PYG{l+m+mi}{5}\PYG{p}{,} \PYG{l+m+mi}{2}\PYG{p}{]}\PYG{p}{)}
\PYG{g+go}{array([2, 5, 4])}
\end{sphinxVerbatim}

\begin{sphinxVerbatim}[commandchars=\\\{\}]
\PYG{g+gp}{\PYGZgt{}\PYGZgt{}\PYGZgt{} }\PYG{n}{np}\PYG{o}{.}\PYG{n}{maximum}\PYG{p}{(}\PYG{n}{np}\PYG{o}{.}\PYG{n}{eye}\PYG{p}{(}\PYG{l+m+mi}{2}\PYG{p}{)}\PYG{p}{,} \PYG{p}{[}\PYG{l+m+mf}{0.5}\PYG{p}{,} \PYG{l+m+mi}{2}\PYG{p}{]}\PYG{p}{)} \PYG{c+c1}{\PYGZsh{} broadcasting}
\PYG{g+go}{array([[ 1. ,  2. ],}
\PYG{g+go}{       [ 0.5,  2. ]])}
\end{sphinxVerbatim}

\begin{sphinxVerbatim}[commandchars=\\\{\}]
\PYG{g+gp}{\PYGZgt{}\PYGZgt{}\PYGZgt{} }\PYG{n}{np}\PYG{o}{.}\PYG{n}{maximum}\PYG{p}{(}\PYG{p}{[}\PYG{n}{np}\PYG{o}{.}\PYG{n}{nan}\PYG{p}{,} \PYG{l+m+mi}{0}\PYG{p}{,} \PYG{n}{np}\PYG{o}{.}\PYG{n}{nan}\PYG{p}{]}\PYG{p}{,} \PYG{p}{[}\PYG{l+m+mi}{0}\PYG{p}{,} \PYG{n}{np}\PYG{o}{.}\PYG{n}{nan}\PYG{p}{,} \PYG{n}{np}\PYG{o}{.}\PYG{n}{nan}\PYG{p}{]}\PYG{p}{)}
\PYG{g+go}{array([nan, nan, nan])}
\PYG{g+gp}{\PYGZgt{}\PYGZgt{}\PYGZgt{} }\PYG{n}{np}\PYG{o}{.}\PYG{n}{maximum}\PYG{p}{(}\PYG{n}{np}\PYG{o}{.}\PYG{n}{Inf}\PYG{p}{,} \PYG{l+m+mi}{1}\PYG{p}{)}
\PYG{g+go}{inf}
\end{sphinxVerbatim}

\end{fulllineitems}

\index{mean() (in module symjax.tensor)@\spxentry{mean()}\spxextra{in module symjax.tensor}}

\begin{fulllineitems}
\phantomsection\label{\detokenize{modules/tensor:symjax.tensor.mean}}\pysiglinewithargsret{\sphinxbfcode{\sphinxupquote{mean}}}{\emph{\DUrole{n}{a}}, \emph{\DUrole{n}{axis}\DUrole{o}{=}\DUrole{default_value}{None}}, \emph{\DUrole{n}{dtype}\DUrole{o}{=}\DUrole{default_value}{None}}, \emph{\DUrole{n}{out}\DUrole{o}{=}\DUrole{default_value}{None}}, \emph{\DUrole{n}{keepdims}\DUrole{o}{=}\DUrole{default_value}{False}}}{}
Compute the arithmetic mean along the specified axis.

LAX\sphinxhyphen{}backend implementation of {\hyperref[\detokenize{modules/tensor:symjax.tensor.mean}]{\sphinxcrossref{\sphinxcode{\sphinxupquote{mean()}}}}}.
ADDITIONOriginal docstring below.

LAX\sphinxhyphen{}backend implementation of {\hyperref[\detokenize{modules/tensor:symjax.tensor.mean}]{\sphinxcrossref{\sphinxcode{\sphinxupquote{mean()}}}}}.
Original docstring below.

Returns the average of the array elements.  The average is taken over
the flattened array by default, otherwise over the specified axis.
\sphinxtitleref{float64} intermediate and return values are used for integer inputs.
\begin{quote}\begin{description}
\item[{Parameters}] \leavevmode
\sphinxstyleliteralstrong{\sphinxupquote{dtype}} (\sphinxstyleliteralemphasis{\sphinxupquote{data\sphinxhyphen{}type}}\sphinxstyleliteralemphasis{\sphinxupquote{, }}\sphinxstyleliteralemphasis{\sphinxupquote{optional}}) \textendash{} Type to use in computing the mean.  For integer inputs, the default
is \sphinxtitleref{float64}; for floating point inputs, it is the same as the
input dtype.

\item[{Returns}] \leavevmode
\sphinxstylestrong{m} \textendash{} If \sphinxtitleref{out=None}, returns a new array containing the mean values,
otherwise a reference to the output array is returned.

\item[{Return type}] \leavevmode
ndarray, see dtype parameter above

\end{description}\end{quote}

\sphinxstrong{See also:}

\begin{description}
\item[{\sphinxcode{\sphinxupquote{average()}}}] \leavevmode
Weighted average

\end{description}

{\hyperref[\detokenize{modules/tensor:symjax.tensor.std}]{\sphinxcrossref{\sphinxcode{\sphinxupquote{std()}}}}}, {\hyperref[\detokenize{modules/tensor:symjax.tensor.var}]{\sphinxcrossref{\sphinxcode{\sphinxupquote{var()}}}}}, \sphinxcode{\sphinxupquote{nanmean()}}, \sphinxcode{\sphinxupquote{nanstd()}}, \sphinxcode{\sphinxupquote{nanvar()}}

\subsubsection*{Notes}

The arithmetic mean is the sum of the elements along the axis divided
by the number of elements.

Note that for floating\sphinxhyphen{}point input, the mean is computed using the
same precision the input has.  Depending on the input data, this can
cause the results to be inaccurate, especially for \sphinxtitleref{float32} (see
example below).  Specifying a higher\sphinxhyphen{}precision accumulator using the
\sphinxtitleref{dtype} keyword can alleviate this issue.

By default, \sphinxtitleref{float16} results are computed using \sphinxtitleref{float32} intermediates
for extra precision.
\subsubsection*{Examples}

\begin{sphinxVerbatim}[commandchars=\\\{\}]
\PYG{g+gp}{\PYGZgt{}\PYGZgt{}\PYGZgt{} }\PYG{n}{a} \PYG{o}{=} \PYG{n}{np}\PYG{o}{.}\PYG{n}{array}\PYG{p}{(}\PYG{p}{[}\PYG{p}{[}\PYG{l+m+mi}{1}\PYG{p}{,} \PYG{l+m+mi}{2}\PYG{p}{]}\PYG{p}{,} \PYG{p}{[}\PYG{l+m+mi}{3}\PYG{p}{,} \PYG{l+m+mi}{4}\PYG{p}{]}\PYG{p}{]}\PYG{p}{)}
\PYG{g+gp}{\PYGZgt{}\PYGZgt{}\PYGZgt{} }\PYG{n}{np}\PYG{o}{.}\PYG{n}{mean}\PYG{p}{(}\PYG{n}{a}\PYG{p}{)}
\PYG{g+go}{2.5}
\PYG{g+gp}{\PYGZgt{}\PYGZgt{}\PYGZgt{} }\PYG{n}{np}\PYG{o}{.}\PYG{n}{mean}\PYG{p}{(}\PYG{n}{a}\PYG{p}{,} \PYG{n}{axis}\PYG{o}{=}\PYG{l+m+mi}{0}\PYG{p}{)}
\PYG{g+go}{array([2., 3.])}
\PYG{g+gp}{\PYGZgt{}\PYGZgt{}\PYGZgt{} }\PYG{n}{np}\PYG{o}{.}\PYG{n}{mean}\PYG{p}{(}\PYG{n}{a}\PYG{p}{,} \PYG{n}{axis}\PYG{o}{=}\PYG{l+m+mi}{1}\PYG{p}{)}
\PYG{g+go}{array([1.5, 3.5])}
\end{sphinxVerbatim}

In single precision, \sphinxtitleref{mean} can be inaccurate:

\begin{sphinxVerbatim}[commandchars=\\\{\}]
\PYG{g+gp}{\PYGZgt{}\PYGZgt{}\PYGZgt{} }\PYG{n}{a} \PYG{o}{=} \PYG{n}{np}\PYG{o}{.}\PYG{n}{zeros}\PYG{p}{(}\PYG{p}{(}\PYG{l+m+mi}{2}\PYG{p}{,} \PYG{l+m+mi}{512}\PYG{o}{*}\PYG{l+m+mi}{512}\PYG{p}{)}\PYG{p}{,} \PYG{n}{dtype}\PYG{o}{=}\PYG{n}{np}\PYG{o}{.}\PYG{n}{float32}\PYG{p}{)}
\PYG{g+gp}{\PYGZgt{}\PYGZgt{}\PYGZgt{} }\PYG{n}{a}\PYG{p}{[}\PYG{l+m+mi}{0}\PYG{p}{,} \PYG{p}{:}\PYG{p}{]} \PYG{o}{=} \PYG{l+m+mf}{1.0}
\PYG{g+gp}{\PYGZgt{}\PYGZgt{}\PYGZgt{} }\PYG{n}{a}\PYG{p}{[}\PYG{l+m+mi}{1}\PYG{p}{,} \PYG{p}{:}\PYG{p}{]} \PYG{o}{=} \PYG{l+m+mf}{0.1}
\PYG{g+gp}{\PYGZgt{}\PYGZgt{}\PYGZgt{} }\PYG{n}{np}\PYG{o}{.}\PYG{n}{mean}\PYG{p}{(}\PYG{n}{a}\PYG{p}{)}
\PYG{g+go}{0.54999924}
\end{sphinxVerbatim}

Computing the mean in float64 is more accurate:

\begin{sphinxVerbatim}[commandchars=\\\{\}]
\PYG{g+gp}{\PYGZgt{}\PYGZgt{}\PYGZgt{} }\PYG{n}{np}\PYG{o}{.}\PYG{n}{mean}\PYG{p}{(}\PYG{n}{a}\PYG{p}{,} \PYG{n}{dtype}\PYG{o}{=}\PYG{n}{np}\PYG{o}{.}\PYG{n}{float64}\PYG{p}{)}
\PYG{g+go}{0.55000000074505806 \PYGZsh{} may vary}
\end{sphinxVerbatim}

\end{fulllineitems}

\index{median() (in module symjax.tensor)@\spxentry{median()}\spxextra{in module symjax.tensor}}

\begin{fulllineitems}
\phantomsection\label{\detokenize{modules/tensor:symjax.tensor.median}}\pysiglinewithargsret{\sphinxbfcode{\sphinxupquote{median}}}{\emph{\DUrole{n}{a}}, \emph{\DUrole{n}{axis}\DUrole{o}{=}\DUrole{default_value}{None}}, \emph{\DUrole{n}{out}\DUrole{o}{=}\DUrole{default_value}{None}}, \emph{\DUrole{n}{overwrite\_input}\DUrole{o}{=}\DUrole{default_value}{False}}, \emph{\DUrole{n}{keepdims}\DUrole{o}{=}\DUrole{default_value}{False}}}{}
Compute the median along the specified axis.

LAX\sphinxhyphen{}backend implementation of {\hyperref[\detokenize{modules/tensor:symjax.tensor.median}]{\sphinxcrossref{\sphinxcode{\sphinxupquote{median()}}}}}.
ADDITIONOriginal docstring below.

LAX\sphinxhyphen{}backend implementation of {\hyperref[\detokenize{modules/tensor:symjax.tensor.median}]{\sphinxcrossref{\sphinxcode{\sphinxupquote{median()}}}}}.
Original docstring below.

Returns the median of the array elements.
\begin{quote}\begin{description}
\item[{Returns}] \leavevmode
\sphinxstylestrong{median} \textendash{} A new array holding the result. If the input contains integers
or floats smaller than \sphinxcode{\sphinxupquote{float64}}, then the output data\sphinxhyphen{}type is
\sphinxcode{\sphinxupquote{np.float64}}.  Otherwise, the data\sphinxhyphen{}type of the output is the
same as that of the input. If \sphinxtitleref{out} is specified, that array is
returned instead.

\item[{Return type}] \leavevmode
ndarray

\end{description}\end{quote}

\sphinxstrong{See also:}

{\hyperref[\detokenize{modules/tensor:symjax.tensor.mean}]{\sphinxcrossref{\sphinxcode{\sphinxupquote{mean()}}}}}, {\hyperref[\detokenize{modules/tensor:symjax.tensor.percentile}]{\sphinxcrossref{\sphinxcode{\sphinxupquote{percentile()}}}}}

\subsubsection*{Notes}

Given a vector \sphinxcode{\sphinxupquote{V}} of length \sphinxcode{\sphinxupquote{N}}, the median of \sphinxcode{\sphinxupquote{V}} is the
middle value of a sorted copy of \sphinxcode{\sphinxupquote{V}}, \sphinxcode{\sphinxupquote{V\_sorted}} \sphinxhyphen{} i
e., \sphinxcode{\sphinxupquote{V\_sorted{[}(N\sphinxhyphen{}1)/2{]}}}, when \sphinxcode{\sphinxupquote{N}} is odd, and the average of the
two middle values of \sphinxcode{\sphinxupquote{V\_sorted}} when \sphinxcode{\sphinxupquote{N}} is even.
\subsubsection*{Examples}

\begin{sphinxVerbatim}[commandchars=\\\{\}]
\PYG{g+gp}{\PYGZgt{}\PYGZgt{}\PYGZgt{} }\PYG{n}{a} \PYG{o}{=} \PYG{n}{np}\PYG{o}{.}\PYG{n}{array}\PYG{p}{(}\PYG{p}{[}\PYG{p}{[}\PYG{l+m+mi}{10}\PYG{p}{,} \PYG{l+m+mi}{7}\PYG{p}{,} \PYG{l+m+mi}{4}\PYG{p}{]}\PYG{p}{,} \PYG{p}{[}\PYG{l+m+mi}{3}\PYG{p}{,} \PYG{l+m+mi}{2}\PYG{p}{,} \PYG{l+m+mi}{1}\PYG{p}{]}\PYG{p}{]}\PYG{p}{)}
\PYG{g+gp}{\PYGZgt{}\PYGZgt{}\PYGZgt{} }\PYG{n}{a}
\PYG{g+go}{array([[10,  7,  4],}
\PYG{g+go}{       [ 3,  2,  1]])}
\PYG{g+gp}{\PYGZgt{}\PYGZgt{}\PYGZgt{} }\PYG{n}{np}\PYG{o}{.}\PYG{n}{median}\PYG{p}{(}\PYG{n}{a}\PYG{p}{)}
\PYG{g+go}{3.5}
\PYG{g+gp}{\PYGZgt{}\PYGZgt{}\PYGZgt{} }\PYG{n}{np}\PYG{o}{.}\PYG{n}{median}\PYG{p}{(}\PYG{n}{a}\PYG{p}{,} \PYG{n}{axis}\PYG{o}{=}\PYG{l+m+mi}{0}\PYG{p}{)}
\PYG{g+go}{array([6.5, 4.5, 2.5])}
\PYG{g+gp}{\PYGZgt{}\PYGZgt{}\PYGZgt{} }\PYG{n}{np}\PYG{o}{.}\PYG{n}{median}\PYG{p}{(}\PYG{n}{a}\PYG{p}{,} \PYG{n}{axis}\PYG{o}{=}\PYG{l+m+mi}{1}\PYG{p}{)}
\PYG{g+go}{array([7.,  2.])}
\PYG{g+gp}{\PYGZgt{}\PYGZgt{}\PYGZgt{} }\PYG{n}{m} \PYG{o}{=} \PYG{n}{np}\PYG{o}{.}\PYG{n}{median}\PYG{p}{(}\PYG{n}{a}\PYG{p}{,} \PYG{n}{axis}\PYG{o}{=}\PYG{l+m+mi}{0}\PYG{p}{)}
\PYG{g+gp}{\PYGZgt{}\PYGZgt{}\PYGZgt{} }\PYG{n}{out} \PYG{o}{=} \PYG{n}{np}\PYG{o}{.}\PYG{n}{zeros\PYGZus{}like}\PYG{p}{(}\PYG{n}{m}\PYG{p}{)}
\PYG{g+gp}{\PYGZgt{}\PYGZgt{}\PYGZgt{} }\PYG{n}{np}\PYG{o}{.}\PYG{n}{median}\PYG{p}{(}\PYG{n}{a}\PYG{p}{,} \PYG{n}{axis}\PYG{o}{=}\PYG{l+m+mi}{0}\PYG{p}{,} \PYG{n}{out}\PYG{o}{=}\PYG{n}{m}\PYG{p}{)}
\PYG{g+go}{array([6.5,  4.5,  2.5])}
\PYG{g+gp}{\PYGZgt{}\PYGZgt{}\PYGZgt{} }\PYG{n}{m}
\PYG{g+go}{array([6.5,  4.5,  2.5])}
\PYG{g+gp}{\PYGZgt{}\PYGZgt{}\PYGZgt{} }\PYG{n}{b} \PYG{o}{=} \PYG{n}{a}\PYG{o}{.}\PYG{n}{copy}\PYG{p}{(}\PYG{p}{)}
\PYG{g+gp}{\PYGZgt{}\PYGZgt{}\PYGZgt{} }\PYG{n}{np}\PYG{o}{.}\PYG{n}{median}\PYG{p}{(}\PYG{n}{b}\PYG{p}{,} \PYG{n}{axis}\PYG{o}{=}\PYG{l+m+mi}{1}\PYG{p}{,} \PYG{n}{overwrite\PYGZus{}input}\PYG{o}{=}\PYG{k+kc}{True}\PYG{p}{)}
\PYG{g+go}{array([7.,  2.])}
\PYG{g+gp}{\PYGZgt{}\PYGZgt{}\PYGZgt{} }\PYG{k}{assert} \PYG{o+ow}{not} \PYG{n}{np}\PYG{o}{.}\PYG{n}{all}\PYG{p}{(}\PYG{n}{a}\PYG{o}{==}\PYG{n}{b}\PYG{p}{)}
\PYG{g+gp}{\PYGZgt{}\PYGZgt{}\PYGZgt{} }\PYG{n}{b} \PYG{o}{=} \PYG{n}{a}\PYG{o}{.}\PYG{n}{copy}\PYG{p}{(}\PYG{p}{)}
\PYG{g+gp}{\PYGZgt{}\PYGZgt{}\PYGZgt{} }\PYG{n}{np}\PYG{o}{.}\PYG{n}{median}\PYG{p}{(}\PYG{n}{b}\PYG{p}{,} \PYG{n}{axis}\PYG{o}{=}\PYG{k+kc}{None}\PYG{p}{,} \PYG{n}{overwrite\PYGZus{}input}\PYG{o}{=}\PYG{k+kc}{True}\PYG{p}{)}
\PYG{g+go}{3.5}
\PYG{g+gp}{\PYGZgt{}\PYGZgt{}\PYGZgt{} }\PYG{k}{assert} \PYG{o+ow}{not} \PYG{n}{np}\PYG{o}{.}\PYG{n}{all}\PYG{p}{(}\PYG{n}{a}\PYG{o}{==}\PYG{n}{b}\PYG{p}{)}
\end{sphinxVerbatim}

\end{fulllineitems}

\index{meshgrid() (in module symjax.tensor)@\spxentry{meshgrid()}\spxextra{in module symjax.tensor}}

\begin{fulllineitems}
\phantomsection\label{\detokenize{modules/tensor:symjax.tensor.meshgrid}}\pysiglinewithargsret{\sphinxbfcode{\sphinxupquote{meshgrid}}}{\emph{\DUrole{o}{*}\DUrole{n}{args}}, \emph{\DUrole{o}{**}\DUrole{n}{kwargs}}}{}
Return coordinate matrices from coordinate vectors.

LAX\sphinxhyphen{}backend implementation of {\hyperref[\detokenize{modules/tensor:symjax.tensor.meshgrid}]{\sphinxcrossref{\sphinxcode{\sphinxupquote{meshgrid()}}}}}.
ADDITIONOriginal docstring below.

LAX\sphinxhyphen{}backend implementation of {\hyperref[\detokenize{modules/tensor:symjax.tensor.meshgrid}]{\sphinxcrossref{\sphinxcode{\sphinxupquote{meshgrid()}}}}}.
Original docstring below.

Make N\sphinxhyphen{}D coordinate arrays for vectorized evaluations of
N\sphinxhyphen{}D scalar/vector fields over N\sphinxhyphen{}D grids, given
one\sphinxhyphen{}dimensional coordinate arrays x1, x2,…, xn.

\DUrole{versionmodified,changed}{Changed in version 1.9: }1\sphinxhyphen{}D and 0\sphinxhyphen{}D cases are allowed.
\begin{quote}\begin{description}
\item[{Returns}] \leavevmode
\sphinxstylestrong{X1, X2,…, XN} \textendash{} For vectors \sphinxtitleref{x1}, \sphinxtitleref{x2},…, ‘xn’ with lengths \sphinxcode{\sphinxupquote{Ni=len(xi)}} ,
return \sphinxcode{\sphinxupquote{(N1, N2, N3,...Nn)}} shaped arrays if indexing=’ij’
or \sphinxcode{\sphinxupquote{(N2, N1, N3,...Nn)}} shaped arrays if indexing=’xy’
with the elements of \sphinxtitleref{xi} repeated to fill the matrix along
the first dimension for \sphinxtitleref{x1}, the second for \sphinxtitleref{x2} and so on.

\item[{Return type}] \leavevmode
ndarray

\end{description}\end{quote}
\subsubsection*{Notes}

This function supports both indexing conventions through the indexing
keyword argument.  Giving the string ‘ij’ returns a meshgrid with
matrix indexing, while ‘xy’ returns a meshgrid with Cartesian indexing.
In the 2\sphinxhyphen{}D case with inputs of length M and N, the outputs are of shape
(N, M) for ‘xy’ indexing and (M, N) for ‘ij’ indexing.  In the 3\sphinxhyphen{}D case
with inputs of length M, N and P, outputs are of shape (N, M, P) for
‘xy’ indexing and (M, N, P) for ‘ij’ indexing.  The difference is
illustrated by the following code snippet:

\begin{sphinxVerbatim}[commandchars=\\\{\}]
\PYG{n}{xv}\PYG{p}{,} \PYG{n}{yv} \PYG{o}{=} \PYG{n}{np}\PYG{o}{.}\PYG{n}{meshgrid}\PYG{p}{(}\PYG{n}{x}\PYG{p}{,} \PYG{n}{y}\PYG{p}{,} \PYG{n}{sparse}\PYG{o}{=}\PYG{k+kc}{False}\PYG{p}{,} \PYG{n}{indexing}\PYG{o}{=}\PYG{l+s+s1}{\PYGZsq{}}\PYG{l+s+s1}{ij}\PYG{l+s+s1}{\PYGZsq{}}\PYG{p}{)}
\PYG{k}{for} \PYG{n}{i} \PYG{o+ow}{in} \PYG{n+nb}{range}\PYG{p}{(}\PYG{n}{nx}\PYG{p}{)}\PYG{p}{:}
    \PYG{k}{for} \PYG{n}{j} \PYG{o+ow}{in} \PYG{n+nb}{range}\PYG{p}{(}\PYG{n}{ny}\PYG{p}{)}\PYG{p}{:}
        \PYG{c+c1}{\PYGZsh{} treat xv[i,j], yv[i,j]}

\PYG{n}{xv}\PYG{p}{,} \PYG{n}{yv} \PYG{o}{=} \PYG{n}{np}\PYG{o}{.}\PYG{n}{meshgrid}\PYG{p}{(}\PYG{n}{x}\PYG{p}{,} \PYG{n}{y}\PYG{p}{,} \PYG{n}{sparse}\PYG{o}{=}\PYG{k+kc}{False}\PYG{p}{,} \PYG{n}{indexing}\PYG{o}{=}\PYG{l+s+s1}{\PYGZsq{}}\PYG{l+s+s1}{xy}\PYG{l+s+s1}{\PYGZsq{}}\PYG{p}{)}
\PYG{k}{for} \PYG{n}{i} \PYG{o+ow}{in} \PYG{n+nb}{range}\PYG{p}{(}\PYG{n}{nx}\PYG{p}{)}\PYG{p}{:}
    \PYG{k}{for} \PYG{n}{j} \PYG{o+ow}{in} \PYG{n+nb}{range}\PYG{p}{(}\PYG{n}{ny}\PYG{p}{)}\PYG{p}{:}
        \PYG{c+c1}{\PYGZsh{} treat xv[j,i], yv[j,i]}
\end{sphinxVerbatim}

In the 1\sphinxhyphen{}D and 0\sphinxhyphen{}D case, the indexing and sparse keywords have no effect.

\sphinxstrong{See also:}

\begin{description}
\item[{\sphinxcode{\sphinxupquote{index\_tricks.mgrid()}}}] \leavevmode
Construct a multi\sphinxhyphen{}dimensional “meshgrid” using indexing notation.

\item[{\sphinxcode{\sphinxupquote{index\_tricks.ogrid()}}}] \leavevmode
Construct an open multi\sphinxhyphen{}dimensional “meshgrid” using indexing notation.

\end{description}

\subsubsection*{Examples}

\begin{sphinxVerbatim}[commandchars=\\\{\}]
\PYG{g+gp}{\PYGZgt{}\PYGZgt{}\PYGZgt{} }\PYG{n}{nx}\PYG{p}{,} \PYG{n}{ny} \PYG{o}{=} \PYG{p}{(}\PYG{l+m+mi}{3}\PYG{p}{,} \PYG{l+m+mi}{2}\PYG{p}{)}
\PYG{g+gp}{\PYGZgt{}\PYGZgt{}\PYGZgt{} }\PYG{n}{x} \PYG{o}{=} \PYG{n}{np}\PYG{o}{.}\PYG{n}{linspace}\PYG{p}{(}\PYG{l+m+mi}{0}\PYG{p}{,} \PYG{l+m+mi}{1}\PYG{p}{,} \PYG{n}{nx}\PYG{p}{)}
\PYG{g+gp}{\PYGZgt{}\PYGZgt{}\PYGZgt{} }\PYG{n}{y} \PYG{o}{=} \PYG{n}{np}\PYG{o}{.}\PYG{n}{linspace}\PYG{p}{(}\PYG{l+m+mi}{0}\PYG{p}{,} \PYG{l+m+mi}{1}\PYG{p}{,} \PYG{n}{ny}\PYG{p}{)}
\PYG{g+gp}{\PYGZgt{}\PYGZgt{}\PYGZgt{} }\PYG{n}{xv}\PYG{p}{,} \PYG{n}{yv} \PYG{o}{=} \PYG{n}{np}\PYG{o}{.}\PYG{n}{meshgrid}\PYG{p}{(}\PYG{n}{x}\PYG{p}{,} \PYG{n}{y}\PYG{p}{)}
\PYG{g+gp}{\PYGZgt{}\PYGZgt{}\PYGZgt{} }\PYG{n}{xv}
\PYG{g+go}{array([[0. , 0.5, 1. ],}
\PYG{g+go}{       [0. , 0.5, 1. ]])}
\PYG{g+gp}{\PYGZgt{}\PYGZgt{}\PYGZgt{} }\PYG{n}{yv}
\PYG{g+go}{array([[0.,  0.,  0.],}
\PYG{g+go}{       [1.,  1.,  1.]])}
\PYG{g+gp}{\PYGZgt{}\PYGZgt{}\PYGZgt{} }\PYG{n}{xv}\PYG{p}{,} \PYG{n}{yv} \PYG{o}{=} \PYG{n}{np}\PYG{o}{.}\PYG{n}{meshgrid}\PYG{p}{(}\PYG{n}{x}\PYG{p}{,} \PYG{n}{y}\PYG{p}{,} \PYG{n}{sparse}\PYG{o}{=}\PYG{k+kc}{True}\PYG{p}{)}  \PYG{c+c1}{\PYGZsh{} make sparse output arrays}
\PYG{g+gp}{\PYGZgt{}\PYGZgt{}\PYGZgt{} }\PYG{n}{xv}
\PYG{g+go}{array([[0. ,  0.5,  1. ]])}
\PYG{g+gp}{\PYGZgt{}\PYGZgt{}\PYGZgt{} }\PYG{n}{yv}
\PYG{g+go}{array([[0.],}
\PYG{g+go}{       [1.]])}
\end{sphinxVerbatim}

\sphinxtitleref{meshgrid} is very useful to evaluate functions on a grid.

\begin{sphinxVerbatim}[commandchars=\\\{\}]
\PYG{g+gp}{\PYGZgt{}\PYGZgt{}\PYGZgt{} }\PYG{k+kn}{import} \PYG{n+nn}{matplotlib}\PYG{n+nn}{.}\PYG{n+nn}{pyplot} \PYG{k}{as} \PYG{n+nn}{plt}
\PYG{g+gp}{\PYGZgt{}\PYGZgt{}\PYGZgt{} }\PYG{n}{x} \PYG{o}{=} \PYG{n}{np}\PYG{o}{.}\PYG{n}{arange}\PYG{p}{(}\PYG{o}{\PYGZhy{}}\PYG{l+m+mi}{5}\PYG{p}{,} \PYG{l+m+mi}{5}\PYG{p}{,} \PYG{l+m+mf}{0.1}\PYG{p}{)}
\PYG{g+gp}{\PYGZgt{}\PYGZgt{}\PYGZgt{} }\PYG{n}{y} \PYG{o}{=} \PYG{n}{np}\PYG{o}{.}\PYG{n}{arange}\PYG{p}{(}\PYG{o}{\PYGZhy{}}\PYG{l+m+mi}{5}\PYG{p}{,} \PYG{l+m+mi}{5}\PYG{p}{,} \PYG{l+m+mf}{0.1}\PYG{p}{)}
\PYG{g+gp}{\PYGZgt{}\PYGZgt{}\PYGZgt{} }\PYG{n}{xx}\PYG{p}{,} \PYG{n}{yy} \PYG{o}{=} \PYG{n}{np}\PYG{o}{.}\PYG{n}{meshgrid}\PYG{p}{(}\PYG{n}{x}\PYG{p}{,} \PYG{n}{y}\PYG{p}{,} \PYG{n}{sparse}\PYG{o}{=}\PYG{k+kc}{True}\PYG{p}{)}
\PYG{g+gp}{\PYGZgt{}\PYGZgt{}\PYGZgt{} }\PYG{n}{z} \PYG{o}{=} \PYG{n}{np}\PYG{o}{.}\PYG{n}{sin}\PYG{p}{(}\PYG{n}{xx}\PYG{o}{*}\PYG{o}{*}\PYG{l+m+mi}{2} \PYG{o}{+} \PYG{n}{yy}\PYG{o}{*}\PYG{o}{*}\PYG{l+m+mi}{2}\PYG{p}{)} \PYG{o}{/} \PYG{p}{(}\PYG{n}{xx}\PYG{o}{*}\PYG{o}{*}\PYG{l+m+mi}{2} \PYG{o}{+} \PYG{n}{yy}\PYG{o}{*}\PYG{o}{*}\PYG{l+m+mi}{2}\PYG{p}{)}
\PYG{g+gp}{\PYGZgt{}\PYGZgt{}\PYGZgt{} }\PYG{n}{h} \PYG{o}{=} \PYG{n}{plt}\PYG{o}{.}\PYG{n}{contourf}\PYG{p}{(}\PYG{n}{x}\PYG{p}{,}\PYG{n}{y}\PYG{p}{,}\PYG{n}{z}\PYG{p}{)}
\PYG{g+gp}{\PYGZgt{}\PYGZgt{}\PYGZgt{} }\PYG{n}{plt}\PYG{o}{.}\PYG{n}{show}\PYG{p}{(}\PYG{p}{)}
\end{sphinxVerbatim}

\end{fulllineitems}

\index{min() (in module symjax.tensor)@\spxentry{min()}\spxextra{in module symjax.tensor}}

\begin{fulllineitems}
\phantomsection\label{\detokenize{modules/tensor:symjax.tensor.min}}\pysiglinewithargsret{\sphinxbfcode{\sphinxupquote{min}}}{\emph{\DUrole{n}{a}}, \emph{\DUrole{n}{axis}\DUrole{o}{=}\DUrole{default_value}{None}}, \emph{\DUrole{n}{dtype}\DUrole{o}{=}\DUrole{default_value}{None}}, \emph{\DUrole{n}{out}\DUrole{o}{=}\DUrole{default_value}{None}}, \emph{\DUrole{n}{keepdims}\DUrole{o}{=}\DUrole{default_value}{False}}}{}
Return the minimum of an array or minimum along an axis.

LAX\sphinxhyphen{}backend implementation of {\hyperref[\detokenize{modules/tensor:symjax.tensor.amin}]{\sphinxcrossref{\sphinxcode{\sphinxupquote{amin()}}}}}.
ADDITIONOriginal docstring below.

LAX\sphinxhyphen{}backend implementation of {\hyperref[\detokenize{modules/tensor:symjax.tensor.amin}]{\sphinxcrossref{\sphinxcode{\sphinxupquote{amin()}}}}}.
Original docstring below.
\begin{quote}\begin{description}
\item[{Returns}] \leavevmode
\sphinxstylestrong{amin} \textendash{} Minimum of \sphinxtitleref{a}. If \sphinxtitleref{axis} is None, the result is a scalar value.
If \sphinxtitleref{axis} is given, the result is an array of dimension
\sphinxcode{\sphinxupquote{a.ndim \sphinxhyphen{} 1}}.

\item[{Return type}] \leavevmode
ndarray or scalar

\end{description}\end{quote}

\sphinxstrong{See also:}

\begin{description}
\item[{{\hyperref[\detokenize{modules/tensor:symjax.tensor.amax}]{\sphinxcrossref{\sphinxcode{\sphinxupquote{amax()}}}}}}] \leavevmode
The maximum value of an array along a given axis, propagating any NaNs.

\item[{{\hyperref[\detokenize{modules/tensor:symjax.tensor.nanmin}]{\sphinxcrossref{\sphinxcode{\sphinxupquote{nanmin()}}}}}}] \leavevmode
The minimum value of an array along a given axis, ignoring any NaNs.

\item[{{\hyperref[\detokenize{modules/tensor:symjax.tensor.minimum}]{\sphinxcrossref{\sphinxcode{\sphinxupquote{minimum()}}}}}}] \leavevmode
Element\sphinxhyphen{}wise minimum of two arrays, propagating any NaNs.

\item[{\sphinxcode{\sphinxupquote{fmin()}}}] \leavevmode
Element\sphinxhyphen{}wise minimum of two arrays, ignoring any NaNs.

\item[{{\hyperref[\detokenize{modules/tensor:symjax.tensor.argmin}]{\sphinxcrossref{\sphinxcode{\sphinxupquote{argmin()}}}}}}] \leavevmode
Return the indices of the minimum values.

\end{description}

{\hyperref[\detokenize{modules/tensor:symjax.tensor.nanmax}]{\sphinxcrossref{\sphinxcode{\sphinxupquote{nanmax()}}}}}, {\hyperref[\detokenize{modules/tensor:symjax.tensor.maximum}]{\sphinxcrossref{\sphinxcode{\sphinxupquote{maximum()}}}}}, \sphinxcode{\sphinxupquote{fmax()}}

\subsubsection*{Notes}

NaN values are propagated, that is if at least one item is NaN, the
corresponding min value will be NaN as well. To ignore NaN values
(MATLAB behavior), please use nanmin.

Don’t use \sphinxtitleref{amin} for element\sphinxhyphen{}wise comparison of 2 arrays; when
\sphinxcode{\sphinxupquote{a.shape{[}0{]}}} is 2, \sphinxcode{\sphinxupquote{minimum(a{[}0{]}, a{[}1{]})}} is faster than
\sphinxcode{\sphinxupquote{amin(a, axis=0)}}.
\subsubsection*{Examples}

\begin{sphinxVerbatim}[commandchars=\\\{\}]
\PYG{g+gp}{\PYGZgt{}\PYGZgt{}\PYGZgt{} }\PYG{n}{a} \PYG{o}{=} \PYG{n}{np}\PYG{o}{.}\PYG{n}{arange}\PYG{p}{(}\PYG{l+m+mi}{4}\PYG{p}{)}\PYG{o}{.}\PYG{n}{reshape}\PYG{p}{(}\PYG{p}{(}\PYG{l+m+mi}{2}\PYG{p}{,}\PYG{l+m+mi}{2}\PYG{p}{)}\PYG{p}{)}
\PYG{g+gp}{\PYGZgt{}\PYGZgt{}\PYGZgt{} }\PYG{n}{a}
\PYG{g+go}{array([[0, 1],}
\PYG{g+go}{       [2, 3]])}
\PYG{g+gp}{\PYGZgt{}\PYGZgt{}\PYGZgt{} }\PYG{n}{np}\PYG{o}{.}\PYG{n}{amin}\PYG{p}{(}\PYG{n}{a}\PYG{p}{)}           \PYG{c+c1}{\PYGZsh{} Minimum of the flattened array}
\PYG{g+go}{0}
\PYG{g+gp}{\PYGZgt{}\PYGZgt{}\PYGZgt{} }\PYG{n}{np}\PYG{o}{.}\PYG{n}{amin}\PYG{p}{(}\PYG{n}{a}\PYG{p}{,} \PYG{n}{axis}\PYG{o}{=}\PYG{l+m+mi}{0}\PYG{p}{)}   \PYG{c+c1}{\PYGZsh{} Minima along the first axis}
\PYG{g+go}{array([0, 1])}
\PYG{g+gp}{\PYGZgt{}\PYGZgt{}\PYGZgt{} }\PYG{n}{np}\PYG{o}{.}\PYG{n}{amin}\PYG{p}{(}\PYG{n}{a}\PYG{p}{,} \PYG{n}{axis}\PYG{o}{=}\PYG{l+m+mi}{1}\PYG{p}{)}   \PYG{c+c1}{\PYGZsh{} Minima along the second axis}
\PYG{g+go}{array([0, 2])}
\PYG{g+gp}{\PYGZgt{}\PYGZgt{}\PYGZgt{} }\PYG{n}{np}\PYG{o}{.}\PYG{n}{amin}\PYG{p}{(}\PYG{n}{a}\PYG{p}{,} \PYG{n}{where}\PYG{o}{=}\PYG{p}{[}\PYG{k+kc}{False}\PYG{p}{,} \PYG{k+kc}{True}\PYG{p}{]}\PYG{p}{,} \PYG{n}{initial}\PYG{o}{=}\PYG{l+m+mi}{10}\PYG{p}{,} \PYG{n}{axis}\PYG{o}{=}\PYG{l+m+mi}{0}\PYG{p}{)}
\PYG{g+go}{array([10,  1])}
\end{sphinxVerbatim}

\begin{sphinxVerbatim}[commandchars=\\\{\}]
\PYG{g+gp}{\PYGZgt{}\PYGZgt{}\PYGZgt{} }\PYG{n}{b} \PYG{o}{=} \PYG{n}{np}\PYG{o}{.}\PYG{n}{arange}\PYG{p}{(}\PYG{l+m+mi}{5}\PYG{p}{,} \PYG{n}{dtype}\PYG{o}{=}\PYG{n+nb}{float}\PYG{p}{)}
\PYG{g+gp}{\PYGZgt{}\PYGZgt{}\PYGZgt{} }\PYG{n}{b}\PYG{p}{[}\PYG{l+m+mi}{2}\PYG{p}{]} \PYG{o}{=} \PYG{n}{np}\PYG{o}{.}\PYG{n}{NaN}
\PYG{g+gp}{\PYGZgt{}\PYGZgt{}\PYGZgt{} }\PYG{n}{np}\PYG{o}{.}\PYG{n}{amin}\PYG{p}{(}\PYG{n}{b}\PYG{p}{)}
\PYG{g+go}{nan}
\PYG{g+gp}{\PYGZgt{}\PYGZgt{}\PYGZgt{} }\PYG{n}{np}\PYG{o}{.}\PYG{n}{amin}\PYG{p}{(}\PYG{n}{b}\PYG{p}{,} \PYG{n}{where}\PYG{o}{=}\PYG{o}{\PYGZti{}}\PYG{n}{np}\PYG{o}{.}\PYG{n}{isnan}\PYG{p}{(}\PYG{n}{b}\PYG{p}{)}\PYG{p}{,} \PYG{n}{initial}\PYG{o}{=}\PYG{l+m+mi}{10}\PYG{p}{)}
\PYG{g+go}{0.0}
\PYG{g+gp}{\PYGZgt{}\PYGZgt{}\PYGZgt{} }\PYG{n}{np}\PYG{o}{.}\PYG{n}{nanmin}\PYG{p}{(}\PYG{n}{b}\PYG{p}{)}
\PYG{g+go}{0.0}
\end{sphinxVerbatim}

\begin{sphinxVerbatim}[commandchars=\\\{\}]
\PYG{g+gp}{\PYGZgt{}\PYGZgt{}\PYGZgt{} }\PYG{n}{np}\PYG{o}{.}\PYG{n}{min}\PYG{p}{(}\PYG{p}{[}\PYG{p}{[}\PYG{o}{\PYGZhy{}}\PYG{l+m+mi}{50}\PYG{p}{]}\PYG{p}{,} \PYG{p}{[}\PYG{l+m+mi}{10}\PYG{p}{]}\PYG{p}{]}\PYG{p}{,} \PYG{n}{axis}\PYG{o}{=}\PYG{o}{\PYGZhy{}}\PYG{l+m+mi}{1}\PYG{p}{,} \PYG{n}{initial}\PYG{o}{=}\PYG{l+m+mi}{0}\PYG{p}{)}
\PYG{g+go}{array([\PYGZhy{}50,   0])}
\end{sphinxVerbatim}

Notice that the initial value is used as one of the elements for which the
minimum is determined, unlike for the default argument Python’s max
function, which is only used for empty iterables.

Notice that this isn’t the same as Python’s \sphinxcode{\sphinxupquote{default}} argument.

\begin{sphinxVerbatim}[commandchars=\\\{\}]
\PYG{g+gp}{\PYGZgt{}\PYGZgt{}\PYGZgt{} }\PYG{n}{np}\PYG{o}{.}\PYG{n}{min}\PYG{p}{(}\PYG{p}{[}\PYG{l+m+mi}{6}\PYG{p}{]}\PYG{p}{,} \PYG{n}{initial}\PYG{o}{=}\PYG{l+m+mi}{5}\PYG{p}{)}
\PYG{g+go}{5}
\PYG{g+gp}{\PYGZgt{}\PYGZgt{}\PYGZgt{} }\PYG{n+nb}{min}\PYG{p}{(}\PYG{p}{[}\PYG{l+m+mi}{6}\PYG{p}{]}\PYG{p}{,} \PYG{n}{default}\PYG{o}{=}\PYG{l+m+mi}{5}\PYG{p}{)}
\PYG{g+go}{6}
\end{sphinxVerbatim}

\end{fulllineitems}

\index{minimum() (in module symjax.tensor)@\spxentry{minimum()}\spxextra{in module symjax.tensor}}

\begin{fulllineitems}
\phantomsection\label{\detokenize{modules/tensor:symjax.tensor.minimum}}\pysiglinewithargsret{\sphinxbfcode{\sphinxupquote{minimum}}}{\emph{\DUrole{n}{x1}}, \emph{\DUrole{n}{x2}}}{}
Element\sphinxhyphen{}wise minimum of array elements.

LAX\sphinxhyphen{}backend implementation of {\hyperref[\detokenize{modules/tensor:symjax.tensor.minimum}]{\sphinxcrossref{\sphinxcode{\sphinxupquote{minimum()}}}}}.
ADDITIONOriginal docstring below.

LAX\sphinxhyphen{}backend implementation of {\hyperref[\detokenize{modules/tensor:symjax.tensor.minimum}]{\sphinxcrossref{\sphinxcode{\sphinxupquote{minimum()}}}}}.
Original docstring below.

minimum(x1, x2, /, out=None, {\color{red}\bfseries{}*}, where=True, casting=’same\_kind’, order=’K’, dtype=None, subok=True{[}, signature, extobj{]})

Compare two arrays and returns a new array containing the element\sphinxhyphen{}wise
minima. If one of the elements being compared is a NaN, then that
element is returned. If both elements are NaNs then the first is
returned. The latter distinction is important for complex NaNs, which
are defined as at least one of the real or imaginary parts being a NaN.
The net effect is that NaNs are propagated.
\begin{quote}\begin{description}
\item[{Returns}] \leavevmode
\sphinxstylestrong{y} \textendash{} The minimum of \sphinxtitleref{x1} and \sphinxtitleref{x2}, element\sphinxhyphen{}wise.
This is a scalar if both \sphinxtitleref{x1} and \sphinxtitleref{x2} are scalars.

\item[{Return type}] \leavevmode
ndarray or scalar

\end{description}\end{quote}

\sphinxstrong{See also:}

\begin{description}
\item[{{\hyperref[\detokenize{modules/tensor:symjax.tensor.maximum}]{\sphinxcrossref{\sphinxcode{\sphinxupquote{maximum()}}}}}}] \leavevmode
Element\sphinxhyphen{}wise maximum of two arrays, propagates NaNs.

\item[{\sphinxcode{\sphinxupquote{fmin()}}}] \leavevmode
Element\sphinxhyphen{}wise minimum of two arrays, ignores NaNs.

\item[{{\hyperref[\detokenize{modules/tensor:symjax.tensor.amin}]{\sphinxcrossref{\sphinxcode{\sphinxupquote{amin()}}}}}}] \leavevmode
The minimum value of an array along a given axis, propagates NaNs.

\item[{{\hyperref[\detokenize{modules/tensor:symjax.tensor.nanmin}]{\sphinxcrossref{\sphinxcode{\sphinxupquote{nanmin()}}}}}}] \leavevmode
The minimum value of an array along a given axis, ignores NaNs.

\end{description}

\sphinxcode{\sphinxupquote{fmax()}}, {\hyperref[\detokenize{modules/tensor:symjax.tensor.amax}]{\sphinxcrossref{\sphinxcode{\sphinxupquote{amax()}}}}}, {\hyperref[\detokenize{modules/tensor:symjax.tensor.nanmax}]{\sphinxcrossref{\sphinxcode{\sphinxupquote{nanmax()}}}}}

\subsubsection*{Notes}

The minimum is equivalent to \sphinxcode{\sphinxupquote{np.where(x1 \textless{}= x2, x1, x2)}} when
neither x1 nor x2 are NaNs, but it is faster and does proper
broadcasting.
\subsubsection*{Examples}

\begin{sphinxVerbatim}[commandchars=\\\{\}]
\PYG{g+gp}{\PYGZgt{}\PYGZgt{}\PYGZgt{} }\PYG{n}{np}\PYG{o}{.}\PYG{n}{minimum}\PYG{p}{(}\PYG{p}{[}\PYG{l+m+mi}{2}\PYG{p}{,} \PYG{l+m+mi}{3}\PYG{p}{,} \PYG{l+m+mi}{4}\PYG{p}{]}\PYG{p}{,} \PYG{p}{[}\PYG{l+m+mi}{1}\PYG{p}{,} \PYG{l+m+mi}{5}\PYG{p}{,} \PYG{l+m+mi}{2}\PYG{p}{]}\PYG{p}{)}
\PYG{g+go}{array([1, 3, 2])}
\end{sphinxVerbatim}

\begin{sphinxVerbatim}[commandchars=\\\{\}]
\PYG{g+gp}{\PYGZgt{}\PYGZgt{}\PYGZgt{} }\PYG{n}{np}\PYG{o}{.}\PYG{n}{minimum}\PYG{p}{(}\PYG{n}{np}\PYG{o}{.}\PYG{n}{eye}\PYG{p}{(}\PYG{l+m+mi}{2}\PYG{p}{)}\PYG{p}{,} \PYG{p}{[}\PYG{l+m+mf}{0.5}\PYG{p}{,} \PYG{l+m+mi}{2}\PYG{p}{]}\PYG{p}{)} \PYG{c+c1}{\PYGZsh{} broadcasting}
\PYG{g+go}{array([[ 0.5,  0. ],}
\PYG{g+go}{       [ 0. ,  1. ]])}
\end{sphinxVerbatim}

\begin{sphinxVerbatim}[commandchars=\\\{\}]
\PYG{g+gp}{\PYGZgt{}\PYGZgt{}\PYGZgt{} }\PYG{n}{np}\PYG{o}{.}\PYG{n}{minimum}\PYG{p}{(}\PYG{p}{[}\PYG{n}{np}\PYG{o}{.}\PYG{n}{nan}\PYG{p}{,} \PYG{l+m+mi}{0}\PYG{p}{,} \PYG{n}{np}\PYG{o}{.}\PYG{n}{nan}\PYG{p}{]}\PYG{p}{,}\PYG{p}{[}\PYG{l+m+mi}{0}\PYG{p}{,} \PYG{n}{np}\PYG{o}{.}\PYG{n}{nan}\PYG{p}{,} \PYG{n}{np}\PYG{o}{.}\PYG{n}{nan}\PYG{p}{]}\PYG{p}{)}
\PYG{g+go}{array([nan, nan, nan])}
\PYG{g+gp}{\PYGZgt{}\PYGZgt{}\PYGZgt{} }\PYG{n}{np}\PYG{o}{.}\PYG{n}{minimum}\PYG{p}{(}\PYG{o}{\PYGZhy{}}\PYG{n}{np}\PYG{o}{.}\PYG{n}{Inf}\PYG{p}{,} \PYG{l+m+mi}{1}\PYG{p}{)}
\PYG{g+go}{\PYGZhy{}inf}
\end{sphinxVerbatim}

\end{fulllineitems}

\index{mod() (in module symjax.tensor)@\spxentry{mod()}\spxextra{in module symjax.tensor}}

\begin{fulllineitems}
\phantomsection\label{\detokenize{modules/tensor:symjax.tensor.mod}}\pysiglinewithargsret{\sphinxbfcode{\sphinxupquote{mod}}}{\emph{\DUrole{n}{x1}}, \emph{\DUrole{n}{x2}}}{}
Return element\sphinxhyphen{}wise remainder of division.

LAX\sphinxhyphen{}backend implementation of {\hyperref[\detokenize{modules/tensor:symjax.tensor.remainder}]{\sphinxcrossref{\sphinxcode{\sphinxupquote{remainder()}}}}}.
ADDITIONOriginal docstring below.

LAX\sphinxhyphen{}backend implementation of {\hyperref[\detokenize{modules/tensor:symjax.tensor.remainder}]{\sphinxcrossref{\sphinxcode{\sphinxupquote{remainder()}}}}}.
Original docstring below.

remainder(x1, x2, /, out=None, {\color{red}\bfseries{}*}, where=True, casting=’same\_kind’, order=’K’, dtype=None, subok=True{[}, signature, extobj{]})

Computes the remainder complementary to the \sphinxtitleref{floor\_divide} function.  It is
equivalent to the Python modulus operator\textasciigrave{}\textasciigrave{}x1 \% x2\textasciigrave{}\textasciigrave{} and has the same sign
as the divisor \sphinxtitleref{x2}. The MATLAB function equivalent to \sphinxcode{\sphinxupquote{np.remainder}}
is \sphinxcode{\sphinxupquote{mod}}.

\begin{sphinxadmonition}{warning}{Warning:}
This should not be confused with:
\begin{itemize}
\item {} 
Python 3.7’s \sphinxtitleref{math.remainder} and C’s \sphinxcode{\sphinxupquote{remainder}}, which
computes the IEEE remainder, which are the complement to
\sphinxcode{\sphinxupquote{round(x1 / x2)}}.

\item {} 
The MATLAB \sphinxcode{\sphinxupquote{rem}} function and or the C \sphinxcode{\sphinxupquote{\%}} operator which is the
complement to \sphinxcode{\sphinxupquote{int(x1 / x2)}}.

\end{itemize}
\end{sphinxadmonition}
\begin{quote}\begin{description}
\item[{Returns}] \leavevmode
\sphinxstylestrong{y} \textendash{} The element\sphinxhyphen{}wise remainder of the quotient \sphinxcode{\sphinxupquote{floor\_divide(x1, x2)}}.
This is a scalar if both \sphinxtitleref{x1} and \sphinxtitleref{x2} are scalars.

\item[{Return type}] \leavevmode
ndarray

\end{description}\end{quote}

\sphinxstrong{See also:}

\begin{description}
\item[{{\hyperref[\detokenize{modules/tensor:symjax.tensor.floor_divide}]{\sphinxcrossref{\sphinxcode{\sphinxupquote{floor\_divide()}}}}}}] \leavevmode
Equivalent of Python \sphinxcode{\sphinxupquote{//}} operator.

\item[{{\hyperref[\detokenize{modules/tensor:symjax.tensor.divmod}]{\sphinxcrossref{\sphinxcode{\sphinxupquote{divmod()}}}}}}] \leavevmode
Simultaneous floor division and remainder.

\item[{{\hyperref[\detokenize{modules/tensor:symjax.tensor.fmod}]{\sphinxcrossref{\sphinxcode{\sphinxupquote{fmod()}}}}}}] \leavevmode
Equivalent of the MATLAB \sphinxcode{\sphinxupquote{rem}} function.

\end{description}

{\hyperref[\detokenize{modules/tensor:symjax.tensor.divide}]{\sphinxcrossref{\sphinxcode{\sphinxupquote{divide()}}}}}, {\hyperref[\detokenize{modules/tensor:symjax.tensor.floor}]{\sphinxcrossref{\sphinxcode{\sphinxupquote{floor()}}}}}

\subsubsection*{Notes}

Returns 0 when \sphinxtitleref{x2} is 0 and both \sphinxtitleref{x1} and \sphinxtitleref{x2} are (arrays of)
integers.
\sphinxcode{\sphinxupquote{mod}} is an alias of \sphinxcode{\sphinxupquote{remainder}}.
\subsubsection*{Examples}

\begin{sphinxVerbatim}[commandchars=\\\{\}]
\PYG{g+gp}{\PYGZgt{}\PYGZgt{}\PYGZgt{} }\PYG{n}{np}\PYG{o}{.}\PYG{n}{remainder}\PYG{p}{(}\PYG{p}{[}\PYG{l+m+mi}{4}\PYG{p}{,} \PYG{l+m+mi}{7}\PYG{p}{]}\PYG{p}{,} \PYG{p}{[}\PYG{l+m+mi}{2}\PYG{p}{,} \PYG{l+m+mi}{3}\PYG{p}{]}\PYG{p}{)}
\PYG{g+go}{array([0, 1])}
\PYG{g+gp}{\PYGZgt{}\PYGZgt{}\PYGZgt{} }\PYG{n}{np}\PYG{o}{.}\PYG{n}{remainder}\PYG{p}{(}\PYG{n}{np}\PYG{o}{.}\PYG{n}{arange}\PYG{p}{(}\PYG{l+m+mi}{7}\PYG{p}{)}\PYG{p}{,} \PYG{l+m+mi}{5}\PYG{p}{)}
\PYG{g+go}{array([0, 1, 2, 3, 4, 0, 1])}
\end{sphinxVerbatim}

\end{fulllineitems}

\index{moveaxis() (in module symjax.tensor)@\spxentry{moveaxis()}\spxextra{in module symjax.tensor}}

\begin{fulllineitems}
\phantomsection\label{\detokenize{modules/tensor:symjax.tensor.moveaxis}}\pysiglinewithargsret{\sphinxbfcode{\sphinxupquote{moveaxis}}}{\emph{\DUrole{n}{a}}, \emph{\DUrole{n}{source}}, \emph{\DUrole{n}{destination}}}{}
Move axes of an array to new positions.

LAX\sphinxhyphen{}backend implementation of {\hyperref[\detokenize{modules/tensor:symjax.tensor.moveaxis}]{\sphinxcrossref{\sphinxcode{\sphinxupquote{moveaxis()}}}}}.
ADDITIONOriginal docstring below.

LAX\sphinxhyphen{}backend implementation of {\hyperref[\detokenize{modules/tensor:symjax.tensor.moveaxis}]{\sphinxcrossref{\sphinxcode{\sphinxupquote{moveaxis()}}}}}.
Original docstring below.

Other axes remain in their original order.

\DUrole{versionmodified,added}{New in version 1.11.0.}
\begin{quote}\begin{description}
\item[{Returns}] \leavevmode
\sphinxstylestrong{result} \textendash{} Array with moved axes. This array is a view of the input array.

\item[{Return type}] \leavevmode
np.ndarray

\end{description}\end{quote}

\sphinxstrong{See also:}

\begin{description}
\item[{{\hyperref[\detokenize{modules/tensor:symjax.tensor.transpose}]{\sphinxcrossref{\sphinxcode{\sphinxupquote{transpose()}}}}}}] \leavevmode
Permute the dimensions of an array.

\item[{{\hyperref[\detokenize{modules/tensor:symjax.tensor.swapaxes}]{\sphinxcrossref{\sphinxcode{\sphinxupquote{swapaxes()}}}}}}] \leavevmode
Interchange two axes of an array.

\end{description}

\subsubsection*{Examples}

\begin{sphinxVerbatim}[commandchars=\\\{\}]
\PYG{g+gp}{\PYGZgt{}\PYGZgt{}\PYGZgt{} }\PYG{n}{x} \PYG{o}{=} \PYG{n}{np}\PYG{o}{.}\PYG{n}{zeros}\PYG{p}{(}\PYG{p}{(}\PYG{l+m+mi}{3}\PYG{p}{,} \PYG{l+m+mi}{4}\PYG{p}{,} \PYG{l+m+mi}{5}\PYG{p}{)}\PYG{p}{)}
\PYG{g+gp}{\PYGZgt{}\PYGZgt{}\PYGZgt{} }\PYG{n}{np}\PYG{o}{.}\PYG{n}{moveaxis}\PYG{p}{(}\PYG{n}{x}\PYG{p}{,} \PYG{l+m+mi}{0}\PYG{p}{,} \PYG{o}{\PYGZhy{}}\PYG{l+m+mi}{1}\PYG{p}{)}\PYG{o}{.}\PYG{n}{shape}
\PYG{g+go}{(4, 5, 3)}
\PYG{g+gp}{\PYGZgt{}\PYGZgt{}\PYGZgt{} }\PYG{n}{np}\PYG{o}{.}\PYG{n}{moveaxis}\PYG{p}{(}\PYG{n}{x}\PYG{p}{,} \PYG{o}{\PYGZhy{}}\PYG{l+m+mi}{1}\PYG{p}{,} \PYG{l+m+mi}{0}\PYG{p}{)}\PYG{o}{.}\PYG{n}{shape}
\PYG{g+go}{(5, 3, 4)}
\end{sphinxVerbatim}

These all achieve the same result:

\begin{sphinxVerbatim}[commandchars=\\\{\}]
\PYG{g+gp}{\PYGZgt{}\PYGZgt{}\PYGZgt{} }\PYG{n}{np}\PYG{o}{.}\PYG{n}{transpose}\PYG{p}{(}\PYG{n}{x}\PYG{p}{)}\PYG{o}{.}\PYG{n}{shape}
\PYG{g+go}{(5, 4, 3)}
\PYG{g+gp}{\PYGZgt{}\PYGZgt{}\PYGZgt{} }\PYG{n}{np}\PYG{o}{.}\PYG{n}{swapaxes}\PYG{p}{(}\PYG{n}{x}\PYG{p}{,} \PYG{l+m+mi}{0}\PYG{p}{,} \PYG{o}{\PYGZhy{}}\PYG{l+m+mi}{1}\PYG{p}{)}\PYG{o}{.}\PYG{n}{shape}
\PYG{g+go}{(5, 4, 3)}
\PYG{g+gp}{\PYGZgt{}\PYGZgt{}\PYGZgt{} }\PYG{n}{np}\PYG{o}{.}\PYG{n}{moveaxis}\PYG{p}{(}\PYG{n}{x}\PYG{p}{,} \PYG{p}{[}\PYG{l+m+mi}{0}\PYG{p}{,} \PYG{l+m+mi}{1}\PYG{p}{]}\PYG{p}{,} \PYG{p}{[}\PYG{o}{\PYGZhy{}}\PYG{l+m+mi}{1}\PYG{p}{,} \PYG{o}{\PYGZhy{}}\PYG{l+m+mi}{2}\PYG{p}{]}\PYG{p}{)}\PYG{o}{.}\PYG{n}{shape}
\PYG{g+go}{(5, 4, 3)}
\PYG{g+gp}{\PYGZgt{}\PYGZgt{}\PYGZgt{} }\PYG{n}{np}\PYG{o}{.}\PYG{n}{moveaxis}\PYG{p}{(}\PYG{n}{x}\PYG{p}{,} \PYG{p}{[}\PYG{l+m+mi}{0}\PYG{p}{,} \PYG{l+m+mi}{1}\PYG{p}{,} \PYG{l+m+mi}{2}\PYG{p}{]}\PYG{p}{,} \PYG{p}{[}\PYG{o}{\PYGZhy{}}\PYG{l+m+mi}{1}\PYG{p}{,} \PYG{o}{\PYGZhy{}}\PYG{l+m+mi}{2}\PYG{p}{,} \PYG{o}{\PYGZhy{}}\PYG{l+m+mi}{3}\PYG{p}{]}\PYG{p}{)}\PYG{o}{.}\PYG{n}{shape}
\PYG{g+go}{(5, 4, 3)}
\end{sphinxVerbatim}

\end{fulllineitems}

\index{multiply() (in module symjax.tensor)@\spxentry{multiply()}\spxextra{in module symjax.tensor}}

\begin{fulllineitems}
\phantomsection\label{\detokenize{modules/tensor:symjax.tensor.multiply}}\pysiglinewithargsret{\sphinxbfcode{\sphinxupquote{multiply}}}{\emph{\DUrole{n}{x1}}, \emph{\DUrole{n}{x2}}}{}
Multiply arguments element\sphinxhyphen{}wise.

LAX\sphinxhyphen{}backend implementation of {\hyperref[\detokenize{modules/tensor:symjax.tensor.multiply}]{\sphinxcrossref{\sphinxcode{\sphinxupquote{multiply()}}}}}.
ADDITIONOriginal docstring below.

LAX\sphinxhyphen{}backend implementation of {\hyperref[\detokenize{modules/tensor:symjax.tensor.multiply}]{\sphinxcrossref{\sphinxcode{\sphinxupquote{multiply()}}}}}.
Original docstring below.

multiply(x1, x2, /, out=None, {\color{red}\bfseries{}*}, where=True, casting=’same\_kind’, order=’K’, dtype=None, subok=True{[}, signature, extobj{]})
\begin{quote}\begin{description}
\item[{Returns}] \leavevmode
\sphinxstylestrong{y} \textendash{} The product of \sphinxtitleref{x1} and \sphinxtitleref{x2}, element\sphinxhyphen{}wise.
This is a scalar if both \sphinxtitleref{x1} and \sphinxtitleref{x2} are scalars.

\item[{Return type}] \leavevmode
ndarray

\end{description}\end{quote}
\subsubsection*{Notes}

Equivalent to \sphinxtitleref{x1} * \sphinxtitleref{x2} in terms of array broadcasting.
\subsubsection*{Examples}

\begin{sphinxVerbatim}[commandchars=\\\{\}]
\PYG{g+gp}{\PYGZgt{}\PYGZgt{}\PYGZgt{} }\PYG{n}{np}\PYG{o}{.}\PYG{n}{multiply}\PYG{p}{(}\PYG{l+m+mf}{2.0}\PYG{p}{,} \PYG{l+m+mf}{4.0}\PYG{p}{)}
\PYG{g+go}{8.0}
\end{sphinxVerbatim}

\begin{sphinxVerbatim}[commandchars=\\\{\}]
\PYG{g+gp}{\PYGZgt{}\PYGZgt{}\PYGZgt{} }\PYG{n}{x1} \PYG{o}{=} \PYG{n}{np}\PYG{o}{.}\PYG{n}{arange}\PYG{p}{(}\PYG{l+m+mf}{9.0}\PYG{p}{)}\PYG{o}{.}\PYG{n}{reshape}\PYG{p}{(}\PYG{p}{(}\PYG{l+m+mi}{3}\PYG{p}{,} \PYG{l+m+mi}{3}\PYG{p}{)}\PYG{p}{)}
\PYG{g+gp}{\PYGZgt{}\PYGZgt{}\PYGZgt{} }\PYG{n}{x2} \PYG{o}{=} \PYG{n}{np}\PYG{o}{.}\PYG{n}{arange}\PYG{p}{(}\PYG{l+m+mf}{3.0}\PYG{p}{)}
\PYG{g+gp}{\PYGZgt{}\PYGZgt{}\PYGZgt{} }\PYG{n}{np}\PYG{o}{.}\PYG{n}{multiply}\PYG{p}{(}\PYG{n}{x1}\PYG{p}{,} \PYG{n}{x2}\PYG{p}{)}
\PYG{g+go}{array([[  0.,   1.,   4.],}
\PYG{g+go}{       [  0.,   4.,  10.],}
\PYG{g+go}{       [  0.,   7.,  16.]])}
\end{sphinxVerbatim}

\end{fulllineitems}

\index{nan\_to\_num() (in module symjax.tensor)@\spxentry{nan\_to\_num()}\spxextra{in module symjax.tensor}}

\begin{fulllineitems}
\phantomsection\label{\detokenize{modules/tensor:symjax.tensor.nan_to_num}}\pysiglinewithargsret{\sphinxbfcode{\sphinxupquote{nan\_to\_num}}}{\emph{\DUrole{n}{x}}, \emph{\DUrole{n}{copy}\DUrole{o}{=}\DUrole{default_value}{True}}}{}~\begin{description}
\item[{Replace NaN with zero and infinity with large finite numbers (default}] \leavevmode
behaviour) or with the numbers defined by the user using the \sphinxtitleref{nan},
\sphinxtitleref{posinf} and/or \sphinxtitleref{neginf} keywords.

\end{description}

LAX\sphinxhyphen{}backend implementation of {\hyperref[\detokenize{modules/tensor:symjax.tensor.nan_to_num}]{\sphinxcrossref{\sphinxcode{\sphinxupquote{nan\_to\_num()}}}}}.
ADDITIONOriginal docstring below.

LAX\sphinxhyphen{}backend implementation of {\hyperref[\detokenize{modules/tensor:symjax.tensor.nan_to_num}]{\sphinxcrossref{\sphinxcode{\sphinxupquote{nan\_to\_num()}}}}}.
Original docstring below.

If \sphinxtitleref{x} is inexact, NaN is replaced by zero or by the user defined value in
\sphinxtitleref{nan} keyword, infinity is replaced by the largest finite floating point
values representable by \sphinxcode{\sphinxupquote{x.dtype}} or by the user defined value in
\sphinxtitleref{posinf} keyword and \sphinxhyphen{}infinity is replaced by the most negative finite
floating point values representable by \sphinxcode{\sphinxupquote{x.dtype}} or by the user defined
value in \sphinxtitleref{neginf} keyword.

For complex dtypes, the above is applied to each of the real and
imaginary components of \sphinxtitleref{x} separately.

If \sphinxtitleref{x} is not inexact, then no replacements are made.
\begin{quote}\begin{description}
\item[{Returns}] \leavevmode
\sphinxstylestrong{out} \textendash{} \sphinxtitleref{x}, with the non\sphinxhyphen{}finite values replaced. If \sphinxtitleref{copy} is False, this may
be \sphinxtitleref{x} itself.

\item[{Return type}] \leavevmode
ndarray

\end{description}\end{quote}

\sphinxstrong{See also:}

\begin{description}
\item[{{\hyperref[\detokenize{modules/tensor:symjax.tensor.isinf}]{\sphinxcrossref{\sphinxcode{\sphinxupquote{isinf()}}}}}}] \leavevmode
Shows which elements are positive or negative infinity.

\item[{\sphinxcode{\sphinxupquote{isneginf()}}}] \leavevmode
Shows which elements are negative infinity.

\item[{\sphinxcode{\sphinxupquote{isposinf()}}}] \leavevmode
Shows which elements are positive infinity.

\item[{{\hyperref[\detokenize{modules/tensor:symjax.tensor.isnan}]{\sphinxcrossref{\sphinxcode{\sphinxupquote{isnan()}}}}}}] \leavevmode
Shows which elements are Not a Number (NaN).

\item[{{\hyperref[\detokenize{modules/tensor:symjax.tensor.isfinite}]{\sphinxcrossref{\sphinxcode{\sphinxupquote{isfinite()}}}}}}] \leavevmode
Shows which elements are finite (not NaN, not infinity)

\end{description}

\subsubsection*{Notes}

NumPy uses the IEEE Standard for Binary Floating\sphinxhyphen{}Point for Arithmetic
(IEEE 754). This means that Not a Number is not equivalent to infinity.
\subsubsection*{Examples}

\begin{sphinxVerbatim}[commandchars=\\\{\}]
\PYG{g+gp}{\PYGZgt{}\PYGZgt{}\PYGZgt{} }\PYG{n}{np}\PYG{o}{.}\PYG{n}{nan\PYGZus{}to\PYGZus{}num}\PYG{p}{(}\PYG{n}{np}\PYG{o}{.}\PYG{n}{inf}\PYG{p}{)}
\PYG{g+go}{1.7976931348623157e+308}
\PYG{g+gp}{\PYGZgt{}\PYGZgt{}\PYGZgt{} }\PYG{n}{np}\PYG{o}{.}\PYG{n}{nan\PYGZus{}to\PYGZus{}num}\PYG{p}{(}\PYG{o}{\PYGZhy{}}\PYG{n}{np}\PYG{o}{.}\PYG{n}{inf}\PYG{p}{)}
\PYG{g+go}{\PYGZhy{}1.7976931348623157e+308}
\PYG{g+gp}{\PYGZgt{}\PYGZgt{}\PYGZgt{} }\PYG{n}{np}\PYG{o}{.}\PYG{n}{nan\PYGZus{}to\PYGZus{}num}\PYG{p}{(}\PYG{n}{np}\PYG{o}{.}\PYG{n}{nan}\PYG{p}{)}
\PYG{g+go}{0.0}
\PYG{g+gp}{\PYGZgt{}\PYGZgt{}\PYGZgt{} }\PYG{n}{x} \PYG{o}{=} \PYG{n}{np}\PYG{o}{.}\PYG{n}{array}\PYG{p}{(}\PYG{p}{[}\PYG{n}{np}\PYG{o}{.}\PYG{n}{inf}\PYG{p}{,} \PYG{o}{\PYGZhy{}}\PYG{n}{np}\PYG{o}{.}\PYG{n}{inf}\PYG{p}{,} \PYG{n}{np}\PYG{o}{.}\PYG{n}{nan}\PYG{p}{,} \PYG{o}{\PYGZhy{}}\PYG{l+m+mi}{128}\PYG{p}{,} \PYG{l+m+mi}{128}\PYG{p}{]}\PYG{p}{)}
\PYG{g+gp}{\PYGZgt{}\PYGZgt{}\PYGZgt{} }\PYG{n}{np}\PYG{o}{.}\PYG{n}{nan\PYGZus{}to\PYGZus{}num}\PYG{p}{(}\PYG{n}{x}\PYG{p}{)}
\PYG{g+go}{array([ 1.79769313e+308, \PYGZhy{}1.79769313e+308,  0.00000000e+000, \PYGZsh{} may vary}
\PYG{g+go}{       \PYGZhy{}1.28000000e+002,  1.28000000e+002])}
\PYG{g+gp}{\PYGZgt{}\PYGZgt{}\PYGZgt{} }\PYG{n}{np}\PYG{o}{.}\PYG{n}{nan\PYGZus{}to\PYGZus{}num}\PYG{p}{(}\PYG{n}{x}\PYG{p}{,} \PYG{n}{nan}\PYG{o}{=}\PYG{o}{\PYGZhy{}}\PYG{l+m+mi}{9999}\PYG{p}{,} \PYG{n}{posinf}\PYG{o}{=}\PYG{l+m+mi}{33333333}\PYG{p}{,} \PYG{n}{neginf}\PYG{o}{=}\PYG{l+m+mi}{33333333}\PYG{p}{)}
\PYG{g+go}{array([ 3.3333333e+07,  3.3333333e+07, \PYGZhy{}9.9990000e+03,}
\PYG{g+go}{       \PYGZhy{}1.2800000e+02,  1.2800000e+02])}
\PYG{g+gp}{\PYGZgt{}\PYGZgt{}\PYGZgt{} }\PYG{n}{y} \PYG{o}{=} \PYG{n}{np}\PYG{o}{.}\PYG{n}{array}\PYG{p}{(}\PYG{p}{[}\PYG{n+nb}{complex}\PYG{p}{(}\PYG{n}{np}\PYG{o}{.}\PYG{n}{inf}\PYG{p}{,} \PYG{n}{np}\PYG{o}{.}\PYG{n}{nan}\PYG{p}{)}\PYG{p}{,} \PYG{n}{np}\PYG{o}{.}\PYG{n}{nan}\PYG{p}{,} \PYG{n+nb}{complex}\PYG{p}{(}\PYG{n}{np}\PYG{o}{.}\PYG{n}{nan}\PYG{p}{,} \PYG{n}{np}\PYG{o}{.}\PYG{n}{inf}\PYG{p}{)}\PYG{p}{]}\PYG{p}{)}
\PYG{g+go}{array([  1.79769313e+308,  \PYGZhy{}1.79769313e+308,   0.00000000e+000, \PYGZsh{} may vary}
\PYG{g+go}{     \PYGZhy{}1.28000000e+002,   1.28000000e+002])}
\PYG{g+gp}{\PYGZgt{}\PYGZgt{}\PYGZgt{} }\PYG{n}{np}\PYG{o}{.}\PYG{n}{nan\PYGZus{}to\PYGZus{}num}\PYG{p}{(}\PYG{n}{y}\PYG{p}{)}
\PYG{g+go}{array([  1.79769313e+308 +0.00000000e+000j, \PYGZsh{} may vary}
\PYG{g+go}{         0.00000000e+000 +0.00000000e+000j,}
\PYG{g+go}{         0.00000000e+000 +1.79769313e+308j])}
\PYG{g+gp}{\PYGZgt{}\PYGZgt{}\PYGZgt{} }\PYG{n}{np}\PYG{o}{.}\PYG{n}{nan\PYGZus{}to\PYGZus{}num}\PYG{p}{(}\PYG{n}{y}\PYG{p}{,} \PYG{n}{nan}\PYG{o}{=}\PYG{l+m+mi}{111111}\PYG{p}{,} \PYG{n}{posinf}\PYG{o}{=}\PYG{l+m+mi}{222222}\PYG{p}{)}
\PYG{g+go}{array([222222.+111111.j, 111111.     +0.j, 111111.+222222.j])}
\end{sphinxVerbatim}

\end{fulllineitems}

\index{nancumprod() (in module symjax.tensor)@\spxentry{nancumprod()}\spxextra{in module symjax.tensor}}

\begin{fulllineitems}
\phantomsection\label{\detokenize{modules/tensor:symjax.tensor.nancumprod}}\pysiglinewithargsret{\sphinxbfcode{\sphinxupquote{nancumprod}}}{\emph{\DUrole{n}{a}}, \emph{\DUrole{n}{axis}\DUrole{o}{=}\DUrole{default_value}{None}}, \emph{\DUrole{n}{dtype}\DUrole{o}{=}\DUrole{default_value}{None}}}{}~\begin{description}
\item[{Return the cumulative product of array elements over a given axis treating Not a}] \leavevmode
Numbers (NaNs) as one.  The cumulative product does not change when NaNs are
encountered and leading NaNs are replaced by ones.

\end{description}

LAX\sphinxhyphen{}backend implementation of {\hyperref[\detokenize{modules/tensor:symjax.tensor.nancumprod}]{\sphinxcrossref{\sphinxcode{\sphinxupquote{nancumprod()}}}}}.
ADDITIONOriginal docstring below.

LAX\sphinxhyphen{}backend implementation of {\hyperref[\detokenize{modules/tensor:symjax.tensor.nancumprod}]{\sphinxcrossref{\sphinxcode{\sphinxupquote{nancumprod()}}}}}.
Original docstring below.

Ones are returned for slices that are all\sphinxhyphen{}NaN or empty.

\DUrole{versionmodified,added}{New in version 1.12.0.}
\begin{quote}\begin{description}
\item[{Parameters}] \leavevmode
\sphinxstyleliteralstrong{\sphinxupquote{dtype}} (\sphinxstyleliteralemphasis{\sphinxupquote{dtype}}\sphinxstyleliteralemphasis{\sphinxupquote{, }}\sphinxstyleliteralemphasis{\sphinxupquote{optional}}) \textendash{} Type of the returned array, as well as of the accumulator in which
the elements are multiplied.  If \sphinxstyleemphasis{dtype} is not specified, it
defaults to the dtype of \sphinxtitleref{a}, unless \sphinxtitleref{a} has an integer dtype with
a precision less than that of the default platform integer.  In
that case, the default platform integer is used instead.

\item[{Returns}] \leavevmode
\sphinxstylestrong{nancumprod} \textendash{} A new array holding the result is returned unless \sphinxtitleref{out} is
specified, in which case it is returned.

\item[{Return type}] \leavevmode
ndarray

\end{description}\end{quote}

\sphinxstrong{See also:}

\begin{description}
\item[{\sphinxcode{\sphinxupquote{numpy.cumprod()}}}] \leavevmode
Cumulative product across array propagating NaNs.

\item[{{\hyperref[\detokenize{modules/tensor:symjax.tensor.isnan}]{\sphinxcrossref{\sphinxcode{\sphinxupquote{isnan()}}}}}}] \leavevmode
Show which elements are NaN.

\end{description}

\subsubsection*{Examples}

\begin{sphinxVerbatim}[commandchars=\\\{\}]
\PYG{g+gp}{\PYGZgt{}\PYGZgt{}\PYGZgt{} }\PYG{n}{np}\PYG{o}{.}\PYG{n}{nancumprod}\PYG{p}{(}\PYG{l+m+mi}{1}\PYG{p}{)}
\PYG{g+go}{array([1])}
\PYG{g+gp}{\PYGZgt{}\PYGZgt{}\PYGZgt{} }\PYG{n}{np}\PYG{o}{.}\PYG{n}{nancumprod}\PYG{p}{(}\PYG{p}{[}\PYG{l+m+mi}{1}\PYG{p}{]}\PYG{p}{)}
\PYG{g+go}{array([1])}
\PYG{g+gp}{\PYGZgt{}\PYGZgt{}\PYGZgt{} }\PYG{n}{np}\PYG{o}{.}\PYG{n}{nancumprod}\PYG{p}{(}\PYG{p}{[}\PYG{l+m+mi}{1}\PYG{p}{,} \PYG{n}{np}\PYG{o}{.}\PYG{n}{nan}\PYG{p}{]}\PYG{p}{)}
\PYG{g+go}{array([1.,  1.])}
\PYG{g+gp}{\PYGZgt{}\PYGZgt{}\PYGZgt{} }\PYG{n}{a} \PYG{o}{=} \PYG{n}{np}\PYG{o}{.}\PYG{n}{array}\PYG{p}{(}\PYG{p}{[}\PYG{p}{[}\PYG{l+m+mi}{1}\PYG{p}{,} \PYG{l+m+mi}{2}\PYG{p}{]}\PYG{p}{,} \PYG{p}{[}\PYG{l+m+mi}{3}\PYG{p}{,} \PYG{n}{np}\PYG{o}{.}\PYG{n}{nan}\PYG{p}{]}\PYG{p}{]}\PYG{p}{)}
\PYG{g+gp}{\PYGZgt{}\PYGZgt{}\PYGZgt{} }\PYG{n}{np}\PYG{o}{.}\PYG{n}{nancumprod}\PYG{p}{(}\PYG{n}{a}\PYG{p}{)}
\PYG{g+go}{array([1.,  2.,  6.,  6.])}
\PYG{g+gp}{\PYGZgt{}\PYGZgt{}\PYGZgt{} }\PYG{n}{np}\PYG{o}{.}\PYG{n}{nancumprod}\PYG{p}{(}\PYG{n}{a}\PYG{p}{,} \PYG{n}{axis}\PYG{o}{=}\PYG{l+m+mi}{0}\PYG{p}{)}
\PYG{g+go}{array([[1.,  2.],}
\PYG{g+go}{       [3.,  2.]])}
\PYG{g+gp}{\PYGZgt{}\PYGZgt{}\PYGZgt{} }\PYG{n}{np}\PYG{o}{.}\PYG{n}{nancumprod}\PYG{p}{(}\PYG{n}{a}\PYG{p}{,} \PYG{n}{axis}\PYG{o}{=}\PYG{l+m+mi}{1}\PYG{p}{)}
\PYG{g+go}{array([[1.,  2.],}
\PYG{g+go}{       [3.,  3.]])}
\end{sphinxVerbatim}

\end{fulllineitems}

\index{nancumsum() (in module symjax.tensor)@\spxentry{nancumsum()}\spxextra{in module symjax.tensor}}

\begin{fulllineitems}
\phantomsection\label{\detokenize{modules/tensor:symjax.tensor.nancumsum}}\pysiglinewithargsret{\sphinxbfcode{\sphinxupquote{nancumsum}}}{\emph{\DUrole{n}{a}}, \emph{\DUrole{n}{axis}\DUrole{o}{=}\DUrole{default_value}{None}}, \emph{\DUrole{n}{dtype}\DUrole{o}{=}\DUrole{default_value}{None}}}{}~\begin{description}
\item[{Return the cumulative sum of array elements over a given axis treating Not a}] \leavevmode
Numbers (NaNs) as zero.  The cumulative sum does not change when NaNs are
encountered and leading NaNs are replaced by zeros.

\end{description}

LAX\sphinxhyphen{}backend implementation of {\hyperref[\detokenize{modules/tensor:symjax.tensor.nancumsum}]{\sphinxcrossref{\sphinxcode{\sphinxupquote{nancumsum()}}}}}.
ADDITIONOriginal docstring below.

LAX\sphinxhyphen{}backend implementation of {\hyperref[\detokenize{modules/tensor:symjax.tensor.nancumsum}]{\sphinxcrossref{\sphinxcode{\sphinxupquote{nancumsum()}}}}}.
Original docstring below.

Zeros are returned for slices that are all\sphinxhyphen{}NaN or empty.

\DUrole{versionmodified,added}{New in version 1.12.0.}
\begin{quote}\begin{description}
\item[{Parameters}] \leavevmode
\sphinxstyleliteralstrong{\sphinxupquote{dtype}} (\sphinxstyleliteralemphasis{\sphinxupquote{dtype}}\sphinxstyleliteralemphasis{\sphinxupquote{, }}\sphinxstyleliteralemphasis{\sphinxupquote{optional}}) \textendash{} Type of the returned array and of the accumulator in which the
elements are summed.  If \sphinxtitleref{dtype} is not specified, it defaults
to the dtype of \sphinxtitleref{a}, unless \sphinxtitleref{a} has an integer dtype with a
precision less than that of the default platform integer.  In
that case, the default platform integer is used.

\item[{Returns}] \leavevmode
\sphinxstylestrong{nancumsum} \textendash{} A new array holding the result is returned unless \sphinxtitleref{out} is
specified, in which it is returned. The result has the same
size as \sphinxtitleref{a}, and the same shape as \sphinxtitleref{a} if \sphinxtitleref{axis} is not None
or \sphinxtitleref{a} is a 1\sphinxhyphen{}d array.

\item[{Return type}] \leavevmode
ndarray.

\end{description}\end{quote}

\sphinxstrong{See also:}

\begin{description}
\item[{\sphinxcode{\sphinxupquote{numpy.cumsum()}}}] \leavevmode
Cumulative sum across array propagating NaNs.

\item[{{\hyperref[\detokenize{modules/tensor:symjax.tensor.isnan}]{\sphinxcrossref{\sphinxcode{\sphinxupquote{isnan()}}}}}}] \leavevmode
Show which elements are NaN.

\end{description}

\subsubsection*{Examples}

\begin{sphinxVerbatim}[commandchars=\\\{\}]
\PYG{g+gp}{\PYGZgt{}\PYGZgt{}\PYGZgt{} }\PYG{n}{np}\PYG{o}{.}\PYG{n}{nancumsum}\PYG{p}{(}\PYG{l+m+mi}{1}\PYG{p}{)}
\PYG{g+go}{array([1])}
\PYG{g+gp}{\PYGZgt{}\PYGZgt{}\PYGZgt{} }\PYG{n}{np}\PYG{o}{.}\PYG{n}{nancumsum}\PYG{p}{(}\PYG{p}{[}\PYG{l+m+mi}{1}\PYG{p}{]}\PYG{p}{)}
\PYG{g+go}{array([1])}
\PYG{g+gp}{\PYGZgt{}\PYGZgt{}\PYGZgt{} }\PYG{n}{np}\PYG{o}{.}\PYG{n}{nancumsum}\PYG{p}{(}\PYG{p}{[}\PYG{l+m+mi}{1}\PYG{p}{,} \PYG{n}{np}\PYG{o}{.}\PYG{n}{nan}\PYG{p}{]}\PYG{p}{)}
\PYG{g+go}{array([1.,  1.])}
\PYG{g+gp}{\PYGZgt{}\PYGZgt{}\PYGZgt{} }\PYG{n}{a} \PYG{o}{=} \PYG{n}{np}\PYG{o}{.}\PYG{n}{array}\PYG{p}{(}\PYG{p}{[}\PYG{p}{[}\PYG{l+m+mi}{1}\PYG{p}{,} \PYG{l+m+mi}{2}\PYG{p}{]}\PYG{p}{,} \PYG{p}{[}\PYG{l+m+mi}{3}\PYG{p}{,} \PYG{n}{np}\PYG{o}{.}\PYG{n}{nan}\PYG{p}{]}\PYG{p}{]}\PYG{p}{)}
\PYG{g+gp}{\PYGZgt{}\PYGZgt{}\PYGZgt{} }\PYG{n}{np}\PYG{o}{.}\PYG{n}{nancumsum}\PYG{p}{(}\PYG{n}{a}\PYG{p}{)}
\PYG{g+go}{array([1.,  3.,  6.,  6.])}
\PYG{g+gp}{\PYGZgt{}\PYGZgt{}\PYGZgt{} }\PYG{n}{np}\PYG{o}{.}\PYG{n}{nancumsum}\PYG{p}{(}\PYG{n}{a}\PYG{p}{,} \PYG{n}{axis}\PYG{o}{=}\PYG{l+m+mi}{0}\PYG{p}{)}
\PYG{g+go}{array([[1.,  2.],}
\PYG{g+go}{       [4.,  2.]])}
\PYG{g+gp}{\PYGZgt{}\PYGZgt{}\PYGZgt{} }\PYG{n}{np}\PYG{o}{.}\PYG{n}{nancumsum}\PYG{p}{(}\PYG{n}{a}\PYG{p}{,} \PYG{n}{axis}\PYG{o}{=}\PYG{l+m+mi}{1}\PYG{p}{)}
\PYG{g+go}{array([[1.,  3.],}
\PYG{g+go}{       [3.,  3.]])}
\end{sphinxVerbatim}

\end{fulllineitems}

\index{nanmax() (in module symjax.tensor)@\spxentry{nanmax()}\spxextra{in module symjax.tensor}}

\begin{fulllineitems}
\phantomsection\label{\detokenize{modules/tensor:symjax.tensor.nanmax}}\pysiglinewithargsret{\sphinxbfcode{\sphinxupquote{nanmax}}}{\emph{\DUrole{n}{a}}, \emph{\DUrole{n}{axis}\DUrole{o}{=}\DUrole{default_value}{None}}, \emph{\DUrole{n}{out}\DUrole{o}{=}\DUrole{default_value}{None}}, \emph{\DUrole{n}{keepdims}\DUrole{o}{=}\DUrole{default_value}{False}}, \emph{\DUrole{o}{**}\DUrole{n}{kwargs}}}{}~\begin{description}
\item[{Return the maximum of an array or maximum along an axis, ignoring any}] \leavevmode
NaNs.  When all\sphinxhyphen{}NaN slices are encountered a \sphinxcode{\sphinxupquote{RuntimeWarning}} is
raised and NaN is returned for that slice.

\end{description}

LAX\sphinxhyphen{}backend implementation of {\hyperref[\detokenize{modules/tensor:symjax.tensor.nanmax}]{\sphinxcrossref{\sphinxcode{\sphinxupquote{nanmax()}}}}}.
ADDITIONOriginal docstring below.

LAX\sphinxhyphen{}backend implementation of {\hyperref[\detokenize{modules/tensor:symjax.tensor.nanmax}]{\sphinxcrossref{\sphinxcode{\sphinxupquote{nanmax()}}}}}.
Original docstring below.
\begin{quote}\begin{description}
\item[{Returns}] \leavevmode
\sphinxstylestrong{nanmax} \textendash{} An array with the same shape as \sphinxtitleref{a}, with the specified axis removed.
If \sphinxtitleref{a} is a 0\sphinxhyphen{}d array, or if axis is None, an ndarray scalar is
returned.  The same dtype as \sphinxtitleref{a} is returned.

\item[{Return type}] \leavevmode
ndarray

\end{description}\end{quote}

\sphinxstrong{See also:}

\begin{description}
\item[{{\hyperref[\detokenize{modules/tensor:symjax.tensor.nanmin}]{\sphinxcrossref{\sphinxcode{\sphinxupquote{nanmin()}}}}}}] \leavevmode
The minimum value of an array along a given axis, ignoring any NaNs.

\item[{{\hyperref[\detokenize{modules/tensor:symjax.tensor.amax}]{\sphinxcrossref{\sphinxcode{\sphinxupquote{amax()}}}}}}] \leavevmode
The maximum value of an array along a given axis, propagating any NaNs.

\item[{\sphinxcode{\sphinxupquote{fmax()}}}] \leavevmode
Element\sphinxhyphen{}wise maximum of two arrays, ignoring any NaNs.

\item[{{\hyperref[\detokenize{modules/tensor:symjax.tensor.maximum}]{\sphinxcrossref{\sphinxcode{\sphinxupquote{maximum()}}}}}}] \leavevmode
Element\sphinxhyphen{}wise maximum of two arrays, propagating any NaNs.

\item[{{\hyperref[\detokenize{modules/tensor:symjax.tensor.isnan}]{\sphinxcrossref{\sphinxcode{\sphinxupquote{isnan()}}}}}}] \leavevmode
Shows which elements are Not a Number (NaN).

\item[{{\hyperref[\detokenize{modules/tensor:symjax.tensor.isfinite}]{\sphinxcrossref{\sphinxcode{\sphinxupquote{isfinite()}}}}}}] \leavevmode
Shows which elements are neither NaN nor infinity.

\end{description}

{\hyperref[\detokenize{modules/tensor:symjax.tensor.amin}]{\sphinxcrossref{\sphinxcode{\sphinxupquote{amin()}}}}}, \sphinxcode{\sphinxupquote{fmin()}}, {\hyperref[\detokenize{modules/tensor:symjax.tensor.minimum}]{\sphinxcrossref{\sphinxcode{\sphinxupquote{minimum()}}}}}

\subsubsection*{Notes}

NumPy uses the IEEE Standard for Binary Floating\sphinxhyphen{}Point for Arithmetic
(IEEE 754). This means that Not a Number is not equivalent to infinity.
Positive infinity is treated as a very large number and negative
infinity is treated as a very small (i.e. negative) number.

If the input has a integer type the function is equivalent to np.max.
\subsubsection*{Examples}

\begin{sphinxVerbatim}[commandchars=\\\{\}]
\PYG{g+gp}{\PYGZgt{}\PYGZgt{}\PYGZgt{} }\PYG{n}{a} \PYG{o}{=} \PYG{n}{np}\PYG{o}{.}\PYG{n}{array}\PYG{p}{(}\PYG{p}{[}\PYG{p}{[}\PYG{l+m+mi}{1}\PYG{p}{,} \PYG{l+m+mi}{2}\PYG{p}{]}\PYG{p}{,} \PYG{p}{[}\PYG{l+m+mi}{3}\PYG{p}{,} \PYG{n}{np}\PYG{o}{.}\PYG{n}{nan}\PYG{p}{]}\PYG{p}{]}\PYG{p}{)}
\PYG{g+gp}{\PYGZgt{}\PYGZgt{}\PYGZgt{} }\PYG{n}{np}\PYG{o}{.}\PYG{n}{nanmax}\PYG{p}{(}\PYG{n}{a}\PYG{p}{)}
\PYG{g+go}{3.0}
\PYG{g+gp}{\PYGZgt{}\PYGZgt{}\PYGZgt{} }\PYG{n}{np}\PYG{o}{.}\PYG{n}{nanmax}\PYG{p}{(}\PYG{n}{a}\PYG{p}{,} \PYG{n}{axis}\PYG{o}{=}\PYG{l+m+mi}{0}\PYG{p}{)}
\PYG{g+go}{array([3.,  2.])}
\PYG{g+gp}{\PYGZgt{}\PYGZgt{}\PYGZgt{} }\PYG{n}{np}\PYG{o}{.}\PYG{n}{nanmax}\PYG{p}{(}\PYG{n}{a}\PYG{p}{,} \PYG{n}{axis}\PYG{o}{=}\PYG{l+m+mi}{1}\PYG{p}{)}
\PYG{g+go}{array([2.,  3.])}
\end{sphinxVerbatim}

When positive infinity and negative infinity are present:

\begin{sphinxVerbatim}[commandchars=\\\{\}]
\PYG{g+gp}{\PYGZgt{}\PYGZgt{}\PYGZgt{} }\PYG{n}{np}\PYG{o}{.}\PYG{n}{nanmax}\PYG{p}{(}\PYG{p}{[}\PYG{l+m+mi}{1}\PYG{p}{,} \PYG{l+m+mi}{2}\PYG{p}{,} \PYG{n}{np}\PYG{o}{.}\PYG{n}{nan}\PYG{p}{,} \PYG{n}{np}\PYG{o}{.}\PYG{n}{NINF}\PYG{p}{]}\PYG{p}{)}
\PYG{g+go}{2.0}
\PYG{g+gp}{\PYGZgt{}\PYGZgt{}\PYGZgt{} }\PYG{n}{np}\PYG{o}{.}\PYG{n}{nanmax}\PYG{p}{(}\PYG{p}{[}\PYG{l+m+mi}{1}\PYG{p}{,} \PYG{l+m+mi}{2}\PYG{p}{,} \PYG{n}{np}\PYG{o}{.}\PYG{n}{nan}\PYG{p}{,} \PYG{n}{np}\PYG{o}{.}\PYG{n}{inf}\PYG{p}{]}\PYG{p}{)}
\PYG{g+go}{inf}
\end{sphinxVerbatim}

\end{fulllineitems}

\index{nanmin() (in module symjax.tensor)@\spxentry{nanmin()}\spxextra{in module symjax.tensor}}

\begin{fulllineitems}
\phantomsection\label{\detokenize{modules/tensor:symjax.tensor.nanmin}}\pysiglinewithargsret{\sphinxbfcode{\sphinxupquote{nanmin}}}{\emph{\DUrole{n}{a}}, \emph{\DUrole{n}{axis}\DUrole{o}{=}\DUrole{default_value}{None}}, \emph{\DUrole{n}{out}\DUrole{o}{=}\DUrole{default_value}{None}}, \emph{\DUrole{n}{keepdims}\DUrole{o}{=}\DUrole{default_value}{False}}, \emph{\DUrole{o}{**}\DUrole{n}{kwargs}}}{}~\begin{description}
\item[{Return minimum of an array or minimum along an axis, ignoring any NaNs.}] \leavevmode
When all\sphinxhyphen{}NaN slices are encountered a \sphinxcode{\sphinxupquote{RuntimeWarning}} is raised and
Nan is returned for that slice.

\end{description}

LAX\sphinxhyphen{}backend implementation of {\hyperref[\detokenize{modules/tensor:symjax.tensor.nanmin}]{\sphinxcrossref{\sphinxcode{\sphinxupquote{nanmin()}}}}}.
ADDITIONOriginal docstring below.

LAX\sphinxhyphen{}backend implementation of {\hyperref[\detokenize{modules/tensor:symjax.tensor.nanmin}]{\sphinxcrossref{\sphinxcode{\sphinxupquote{nanmin()}}}}}.
Original docstring below.
\begin{quote}\begin{description}
\item[{Returns}] \leavevmode
\sphinxstylestrong{nanmin} \textendash{} An array with the same shape as \sphinxtitleref{a}, with the specified axis
removed.  If \sphinxtitleref{a} is a 0\sphinxhyphen{}d array, or if axis is None, an ndarray
scalar is returned.  The same dtype as \sphinxtitleref{a} is returned.

\item[{Return type}] \leavevmode
ndarray

\end{description}\end{quote}

\sphinxstrong{See also:}

\begin{description}
\item[{{\hyperref[\detokenize{modules/tensor:symjax.tensor.nanmax}]{\sphinxcrossref{\sphinxcode{\sphinxupquote{nanmax()}}}}}}] \leavevmode
The maximum value of an array along a given axis, ignoring any NaNs.

\item[{{\hyperref[\detokenize{modules/tensor:symjax.tensor.amin}]{\sphinxcrossref{\sphinxcode{\sphinxupquote{amin()}}}}}}] \leavevmode
The minimum value of an array along a given axis, propagating any NaNs.

\item[{\sphinxcode{\sphinxupquote{fmin()}}}] \leavevmode
Element\sphinxhyphen{}wise minimum of two arrays, ignoring any NaNs.

\item[{{\hyperref[\detokenize{modules/tensor:symjax.tensor.minimum}]{\sphinxcrossref{\sphinxcode{\sphinxupquote{minimum()}}}}}}] \leavevmode
Element\sphinxhyphen{}wise minimum of two arrays, propagating any NaNs.

\item[{{\hyperref[\detokenize{modules/tensor:symjax.tensor.isnan}]{\sphinxcrossref{\sphinxcode{\sphinxupquote{isnan()}}}}}}] \leavevmode
Shows which elements are Not a Number (NaN).

\item[{{\hyperref[\detokenize{modules/tensor:symjax.tensor.isfinite}]{\sphinxcrossref{\sphinxcode{\sphinxupquote{isfinite()}}}}}}] \leavevmode
Shows which elements are neither NaN nor infinity.

\end{description}

{\hyperref[\detokenize{modules/tensor:symjax.tensor.amax}]{\sphinxcrossref{\sphinxcode{\sphinxupquote{amax()}}}}}, \sphinxcode{\sphinxupquote{fmax()}}, {\hyperref[\detokenize{modules/tensor:symjax.tensor.maximum}]{\sphinxcrossref{\sphinxcode{\sphinxupquote{maximum()}}}}}

\subsubsection*{Notes}

NumPy uses the IEEE Standard for Binary Floating\sphinxhyphen{}Point for Arithmetic
(IEEE 754). This means that Not a Number is not equivalent to infinity.
Positive infinity is treated as a very large number and negative
infinity is treated as a very small (i.e. negative) number.

If the input has a integer type the function is equivalent to np.min.
\subsubsection*{Examples}

\begin{sphinxVerbatim}[commandchars=\\\{\}]
\PYG{g+gp}{\PYGZgt{}\PYGZgt{}\PYGZgt{} }\PYG{n}{a} \PYG{o}{=} \PYG{n}{np}\PYG{o}{.}\PYG{n}{array}\PYG{p}{(}\PYG{p}{[}\PYG{p}{[}\PYG{l+m+mi}{1}\PYG{p}{,} \PYG{l+m+mi}{2}\PYG{p}{]}\PYG{p}{,} \PYG{p}{[}\PYG{l+m+mi}{3}\PYG{p}{,} \PYG{n}{np}\PYG{o}{.}\PYG{n}{nan}\PYG{p}{]}\PYG{p}{]}\PYG{p}{)}
\PYG{g+gp}{\PYGZgt{}\PYGZgt{}\PYGZgt{} }\PYG{n}{np}\PYG{o}{.}\PYG{n}{nanmin}\PYG{p}{(}\PYG{n}{a}\PYG{p}{)}
\PYG{g+go}{1.0}
\PYG{g+gp}{\PYGZgt{}\PYGZgt{}\PYGZgt{} }\PYG{n}{np}\PYG{o}{.}\PYG{n}{nanmin}\PYG{p}{(}\PYG{n}{a}\PYG{p}{,} \PYG{n}{axis}\PYG{o}{=}\PYG{l+m+mi}{0}\PYG{p}{)}
\PYG{g+go}{array([1.,  2.])}
\PYG{g+gp}{\PYGZgt{}\PYGZgt{}\PYGZgt{} }\PYG{n}{np}\PYG{o}{.}\PYG{n}{nanmin}\PYG{p}{(}\PYG{n}{a}\PYG{p}{,} \PYG{n}{axis}\PYG{o}{=}\PYG{l+m+mi}{1}\PYG{p}{)}
\PYG{g+go}{array([1.,  3.])}
\end{sphinxVerbatim}

When positive infinity and negative infinity are present:

\begin{sphinxVerbatim}[commandchars=\\\{\}]
\PYG{g+gp}{\PYGZgt{}\PYGZgt{}\PYGZgt{} }\PYG{n}{np}\PYG{o}{.}\PYG{n}{nanmin}\PYG{p}{(}\PYG{p}{[}\PYG{l+m+mi}{1}\PYG{p}{,} \PYG{l+m+mi}{2}\PYG{p}{,} \PYG{n}{np}\PYG{o}{.}\PYG{n}{nan}\PYG{p}{,} \PYG{n}{np}\PYG{o}{.}\PYG{n}{inf}\PYG{p}{]}\PYG{p}{)}
\PYG{g+go}{1.0}
\PYG{g+gp}{\PYGZgt{}\PYGZgt{}\PYGZgt{} }\PYG{n}{np}\PYG{o}{.}\PYG{n}{nanmin}\PYG{p}{(}\PYG{p}{[}\PYG{l+m+mi}{1}\PYG{p}{,} \PYG{l+m+mi}{2}\PYG{p}{,} \PYG{n}{np}\PYG{o}{.}\PYG{n}{nan}\PYG{p}{,} \PYG{n}{np}\PYG{o}{.}\PYG{n}{NINF}\PYG{p}{]}\PYG{p}{)}
\PYG{g+go}{\PYGZhy{}inf}
\end{sphinxVerbatim}

\end{fulllineitems}

\index{nanprod() (in module symjax.tensor)@\spxentry{nanprod()}\spxextra{in module symjax.tensor}}

\begin{fulllineitems}
\phantomsection\label{\detokenize{modules/tensor:symjax.tensor.nanprod}}\pysiglinewithargsret{\sphinxbfcode{\sphinxupquote{nanprod}}}{\emph{\DUrole{n}{a}}, \emph{\DUrole{n}{axis}\DUrole{o}{=}\DUrole{default_value}{None}}, \emph{\DUrole{n}{out}\DUrole{o}{=}\DUrole{default_value}{None}}, \emph{\DUrole{n}{keepdims}\DUrole{o}{=}\DUrole{default_value}{False}}, \emph{\DUrole{o}{**}\DUrole{n}{kwargs}}}{}~\begin{description}
\item[{Return the product of array elements over a given axis treating Not a}] \leavevmode
Numbers (NaNs) as ones.

\end{description}

LAX\sphinxhyphen{}backend implementation of {\hyperref[\detokenize{modules/tensor:symjax.tensor.nanprod}]{\sphinxcrossref{\sphinxcode{\sphinxupquote{nanprod()}}}}}.
ADDITIONOriginal docstring below.

LAX\sphinxhyphen{}backend implementation of {\hyperref[\detokenize{modules/tensor:symjax.tensor.nanprod}]{\sphinxcrossref{\sphinxcode{\sphinxupquote{nanprod()}}}}}.
Original docstring below.

One is returned for slices that are all\sphinxhyphen{}NaN or empty.

\DUrole{versionmodified,added}{New in version 1.10.0.}
\begin{quote}\begin{description}
\item[{Returns}] \leavevmode
\sphinxstylestrong{nanprod} \textendash{} A new array holding the result is returned unless \sphinxtitleref{out} is
specified, in which case it is returned.

\item[{Return type}] \leavevmode
ndarray

\end{description}\end{quote}

\sphinxstrong{See also:}

\begin{description}
\item[{\sphinxcode{\sphinxupquote{numpy.prod()}}}] \leavevmode
Product across array propagating NaNs.

\item[{{\hyperref[\detokenize{modules/tensor:symjax.tensor.isnan}]{\sphinxcrossref{\sphinxcode{\sphinxupquote{isnan()}}}}}}] \leavevmode
Show which elements are NaN.

\end{description}

\subsubsection*{Examples}

\begin{sphinxVerbatim}[commandchars=\\\{\}]
\PYG{g+gp}{\PYGZgt{}\PYGZgt{}\PYGZgt{} }\PYG{n}{np}\PYG{o}{.}\PYG{n}{nanprod}\PYG{p}{(}\PYG{l+m+mi}{1}\PYG{p}{)}
\PYG{g+go}{1}
\PYG{g+gp}{\PYGZgt{}\PYGZgt{}\PYGZgt{} }\PYG{n}{np}\PYG{o}{.}\PYG{n}{nanprod}\PYG{p}{(}\PYG{p}{[}\PYG{l+m+mi}{1}\PYG{p}{]}\PYG{p}{)}
\PYG{g+go}{1}
\PYG{g+gp}{\PYGZgt{}\PYGZgt{}\PYGZgt{} }\PYG{n}{np}\PYG{o}{.}\PYG{n}{nanprod}\PYG{p}{(}\PYG{p}{[}\PYG{l+m+mi}{1}\PYG{p}{,} \PYG{n}{np}\PYG{o}{.}\PYG{n}{nan}\PYG{p}{]}\PYG{p}{)}
\PYG{g+go}{1.0}
\PYG{g+gp}{\PYGZgt{}\PYGZgt{}\PYGZgt{} }\PYG{n}{a} \PYG{o}{=} \PYG{n}{np}\PYG{o}{.}\PYG{n}{array}\PYG{p}{(}\PYG{p}{[}\PYG{p}{[}\PYG{l+m+mi}{1}\PYG{p}{,} \PYG{l+m+mi}{2}\PYG{p}{]}\PYG{p}{,} \PYG{p}{[}\PYG{l+m+mi}{3}\PYG{p}{,} \PYG{n}{np}\PYG{o}{.}\PYG{n}{nan}\PYG{p}{]}\PYG{p}{]}\PYG{p}{)}
\PYG{g+gp}{\PYGZgt{}\PYGZgt{}\PYGZgt{} }\PYG{n}{np}\PYG{o}{.}\PYG{n}{nanprod}\PYG{p}{(}\PYG{n}{a}\PYG{p}{)}
\PYG{g+go}{6.0}
\PYG{g+gp}{\PYGZgt{}\PYGZgt{}\PYGZgt{} }\PYG{n}{np}\PYG{o}{.}\PYG{n}{nanprod}\PYG{p}{(}\PYG{n}{a}\PYG{p}{,} \PYG{n}{axis}\PYG{o}{=}\PYG{l+m+mi}{0}\PYG{p}{)}
\PYG{g+go}{array([3., 2.])}
\end{sphinxVerbatim}

\end{fulllineitems}

\index{nansum() (in module symjax.tensor)@\spxentry{nansum()}\spxextra{in module symjax.tensor}}

\begin{fulllineitems}
\phantomsection\label{\detokenize{modules/tensor:symjax.tensor.nansum}}\pysiglinewithargsret{\sphinxbfcode{\sphinxupquote{nansum}}}{\emph{\DUrole{n}{a}}, \emph{\DUrole{n}{axis}\DUrole{o}{=}\DUrole{default_value}{None}}, \emph{\DUrole{n}{out}\DUrole{o}{=}\DUrole{default_value}{None}}, \emph{\DUrole{n}{keepdims}\DUrole{o}{=}\DUrole{default_value}{False}}, \emph{\DUrole{o}{**}\DUrole{n}{kwargs}}}{}~\begin{description}
\item[{Return the sum of array elements over a given axis treating Not a}] \leavevmode
Numbers (NaNs) as zero.

\end{description}

LAX\sphinxhyphen{}backend implementation of {\hyperref[\detokenize{modules/tensor:symjax.tensor.nansum}]{\sphinxcrossref{\sphinxcode{\sphinxupquote{nansum()}}}}}.
ADDITIONOriginal docstring below.

LAX\sphinxhyphen{}backend implementation of {\hyperref[\detokenize{modules/tensor:symjax.tensor.nansum}]{\sphinxcrossref{\sphinxcode{\sphinxupquote{nansum()}}}}}.
Original docstring below.

In NumPy versions \textless{}= 1.9.0 Nan is returned for slices that are all\sphinxhyphen{}NaN or
empty. In later versions zero is returned.
\begin{quote}\begin{description}
\item[{Returns}] \leavevmode
\sphinxstylestrong{nansum} \textendash{} A new array holding the result is returned unless \sphinxtitleref{out} is
specified, in which it is returned. The result has the same
size as \sphinxtitleref{a}, and the same shape as \sphinxtitleref{a} if \sphinxtitleref{axis} is not None
or \sphinxtitleref{a} is a 1\sphinxhyphen{}d array.

\item[{Return type}] \leavevmode
ndarray.

\end{description}\end{quote}

\sphinxstrong{See also:}

\begin{description}
\item[{\sphinxcode{\sphinxupquote{numpy.sum()}}}] \leavevmode
Sum across array propagating NaNs.

\item[{{\hyperref[\detokenize{modules/tensor:symjax.tensor.isnan}]{\sphinxcrossref{\sphinxcode{\sphinxupquote{isnan()}}}}}}] \leavevmode
Show which elements are NaN.

\item[{{\hyperref[\detokenize{modules/tensor:symjax.tensor.isfinite}]{\sphinxcrossref{\sphinxcode{\sphinxupquote{isfinite()}}}}}}] \leavevmode
Show which elements are not NaN or +/\sphinxhyphen{}inf.

\end{description}

\subsubsection*{Notes}

If both positive and negative infinity are present, the sum will be Not
A Number (NaN).
\subsubsection*{Examples}

\begin{sphinxVerbatim}[commandchars=\\\{\}]
\PYG{g+gp}{\PYGZgt{}\PYGZgt{}\PYGZgt{} }\PYG{n}{np}\PYG{o}{.}\PYG{n}{nansum}\PYG{p}{(}\PYG{l+m+mi}{1}\PYG{p}{)}
\PYG{g+go}{1}
\PYG{g+gp}{\PYGZgt{}\PYGZgt{}\PYGZgt{} }\PYG{n}{np}\PYG{o}{.}\PYG{n}{nansum}\PYG{p}{(}\PYG{p}{[}\PYG{l+m+mi}{1}\PYG{p}{]}\PYG{p}{)}
\PYG{g+go}{1}
\PYG{g+gp}{\PYGZgt{}\PYGZgt{}\PYGZgt{} }\PYG{n}{np}\PYG{o}{.}\PYG{n}{nansum}\PYG{p}{(}\PYG{p}{[}\PYG{l+m+mi}{1}\PYG{p}{,} \PYG{n}{np}\PYG{o}{.}\PYG{n}{nan}\PYG{p}{]}\PYG{p}{)}
\PYG{g+go}{1.0}
\PYG{g+gp}{\PYGZgt{}\PYGZgt{}\PYGZgt{} }\PYG{n}{a} \PYG{o}{=} \PYG{n}{np}\PYG{o}{.}\PYG{n}{array}\PYG{p}{(}\PYG{p}{[}\PYG{p}{[}\PYG{l+m+mi}{1}\PYG{p}{,} \PYG{l+m+mi}{1}\PYG{p}{]}\PYG{p}{,} \PYG{p}{[}\PYG{l+m+mi}{1}\PYG{p}{,} \PYG{n}{np}\PYG{o}{.}\PYG{n}{nan}\PYG{p}{]}\PYG{p}{]}\PYG{p}{)}
\PYG{g+gp}{\PYGZgt{}\PYGZgt{}\PYGZgt{} }\PYG{n}{np}\PYG{o}{.}\PYG{n}{nansum}\PYG{p}{(}\PYG{n}{a}\PYG{p}{)}
\PYG{g+go}{3.0}
\PYG{g+gp}{\PYGZgt{}\PYGZgt{}\PYGZgt{} }\PYG{n}{np}\PYG{o}{.}\PYG{n}{nansum}\PYG{p}{(}\PYG{n}{a}\PYG{p}{,} \PYG{n}{axis}\PYG{o}{=}\PYG{l+m+mi}{0}\PYG{p}{)}
\PYG{g+go}{array([2.,  1.])}
\PYG{g+gp}{\PYGZgt{}\PYGZgt{}\PYGZgt{} }\PYG{n}{np}\PYG{o}{.}\PYG{n}{nansum}\PYG{p}{(}\PYG{p}{[}\PYG{l+m+mi}{1}\PYG{p}{,} \PYG{n}{np}\PYG{o}{.}\PYG{n}{nan}\PYG{p}{,} \PYG{n}{np}\PYG{o}{.}\PYG{n}{inf}\PYG{p}{]}\PYG{p}{)}
\PYG{g+go}{inf}
\PYG{g+gp}{\PYGZgt{}\PYGZgt{}\PYGZgt{} }\PYG{n}{np}\PYG{o}{.}\PYG{n}{nansum}\PYG{p}{(}\PYG{p}{[}\PYG{l+m+mi}{1}\PYG{p}{,} \PYG{n}{np}\PYG{o}{.}\PYG{n}{nan}\PYG{p}{,} \PYG{n}{np}\PYG{o}{.}\PYG{n}{NINF}\PYG{p}{]}\PYG{p}{)}
\PYG{g+go}{\PYGZhy{}inf}
\PYG{g+gp}{\PYGZgt{}\PYGZgt{}\PYGZgt{} }\PYG{k+kn}{from} \PYG{n+nn}{numpy}\PYG{n+nn}{.}\PYG{n+nn}{testing} \PYG{k+kn}{import} \PYG{n}{suppress\PYGZus{}warnings}
\PYG{g+gp}{\PYGZgt{}\PYGZgt{}\PYGZgt{} }\PYG{k}{with} \PYG{n}{suppress\PYGZus{}warnings}\PYG{p}{(}\PYG{p}{)} \PYG{k}{as} \PYG{n}{sup}\PYG{p}{:}
\PYG{g+gp}{... }    \PYG{n}{sup}\PYG{o}{.}\PYG{n}{filter}\PYG{p}{(}\PYG{n+ne}{RuntimeWarning}\PYG{p}{)}
\PYG{g+gp}{... }    \PYG{n}{np}\PYG{o}{.}\PYG{n}{nansum}\PYG{p}{(}\PYG{p}{[}\PYG{l+m+mi}{1}\PYG{p}{,} \PYG{n}{np}\PYG{o}{.}\PYG{n}{nan}\PYG{p}{,} \PYG{n}{np}\PYG{o}{.}\PYG{n}{inf}\PYG{p}{,} \PYG{o}{\PYGZhy{}}\PYG{n}{np}\PYG{o}{.}\PYG{n}{inf}\PYG{p}{]}\PYG{p}{)} \PYG{c+c1}{\PYGZsh{} both +/\PYGZhy{} infinity present}
\PYG{g+go}{nan}
\end{sphinxVerbatim}

\end{fulllineitems}

\index{nonzero() (in module symjax.tensor)@\spxentry{nonzero()}\spxextra{in module symjax.tensor}}

\begin{fulllineitems}
\phantomsection\label{\detokenize{modules/tensor:symjax.tensor.nonzero}}\pysiglinewithargsret{\sphinxbfcode{\sphinxupquote{nonzero}}}{\emph{\DUrole{n}{a}}}{}
Return the indices of the elements that are non\sphinxhyphen{}zero.

LAX\sphinxhyphen{}backend implementation of {\hyperref[\detokenize{modules/tensor:symjax.tensor.nonzero}]{\sphinxcrossref{\sphinxcode{\sphinxupquote{nonzero()}}}}}.
ADDITIONOriginal docstring below.

LAX\sphinxhyphen{}backend implementation of {\hyperref[\detokenize{modules/tensor:symjax.tensor.nonzero}]{\sphinxcrossref{\sphinxcode{\sphinxupquote{nonzero()}}}}}.
At present, JAX does not support JIT\sphinxhyphen{}compilation of \sphinxcode{\sphinxupquote{jax.numpy.nonzero()}}
because its output shape is data\sphinxhyphen{}dependent.

Original docstring below.

Returns a tuple of arrays, one for each dimension of \sphinxtitleref{a},
containing the indices of the non\sphinxhyphen{}zero elements in that
dimension. The values in \sphinxtitleref{a} are always tested and returned in
row\sphinxhyphen{}major, C\sphinxhyphen{}style order.

To group the indices by element, rather than dimension, use \sphinxtitleref{argwhere},
which returns a row for each non\sphinxhyphen{}zero element.

\begin{sphinxadmonition}{note}{Note:}
When called on a zero\sphinxhyphen{}d array or scalar, \sphinxcode{\sphinxupquote{nonzero(a)}} is treated
as \sphinxcode{\sphinxupquote{nonzero(atleast1d(a))}}.

\DUrole{versionmodified,deprecated}{Deprecated since version 1.17.0: }Use \sphinxtitleref{atleast1d} explicitly if this behavior is deliberate.
\end{sphinxadmonition}
\begin{quote}\begin{description}
\item[{Returns}] \leavevmode
\sphinxstylestrong{tuple\_of\_arrays} \textendash{} Indices of elements that are non\sphinxhyphen{}zero.

\item[{Return type}] \leavevmode
tuple

\end{description}\end{quote}

\sphinxstrong{See also:}

\begin{description}
\item[{\sphinxcode{\sphinxupquote{flatnonzero()}}}] \leavevmode
Return indices that are non\sphinxhyphen{}zero in the flattened version of the input array.

\item[{\sphinxcode{\sphinxupquote{ndarray.nonzero()}}}] \leavevmode
Equivalent ndarray method.

\item[{{\hyperref[\detokenize{modules/tensor:symjax.tensor.count_nonzero}]{\sphinxcrossref{\sphinxcode{\sphinxupquote{count\_nonzero()}}}}}}] \leavevmode
Counts the number of non\sphinxhyphen{}zero elements in the input array.

\end{description}

\subsubsection*{Notes}

While the nonzero values can be obtained with \sphinxcode{\sphinxupquote{a{[}nonzero(a){]}}}, it is
recommended to use \sphinxcode{\sphinxupquote{x{[}x.astype(bool){]}}} or \sphinxcode{\sphinxupquote{x{[}x != 0{]}}} instead, which
will correctly handle 0\sphinxhyphen{}d arrays.
\subsubsection*{Examples}

\begin{sphinxVerbatim}[commandchars=\\\{\}]
\PYG{g+gp}{\PYGZgt{}\PYGZgt{}\PYGZgt{} }\PYG{n}{x} \PYG{o}{=} \PYG{n}{np}\PYG{o}{.}\PYG{n}{array}\PYG{p}{(}\PYG{p}{[}\PYG{p}{[}\PYG{l+m+mi}{3}\PYG{p}{,} \PYG{l+m+mi}{0}\PYG{p}{,} \PYG{l+m+mi}{0}\PYG{p}{]}\PYG{p}{,} \PYG{p}{[}\PYG{l+m+mi}{0}\PYG{p}{,} \PYG{l+m+mi}{4}\PYG{p}{,} \PYG{l+m+mi}{0}\PYG{p}{]}\PYG{p}{,} \PYG{p}{[}\PYG{l+m+mi}{5}\PYG{p}{,} \PYG{l+m+mi}{6}\PYG{p}{,} \PYG{l+m+mi}{0}\PYG{p}{]}\PYG{p}{]}\PYG{p}{)}
\PYG{g+gp}{\PYGZgt{}\PYGZgt{}\PYGZgt{} }\PYG{n}{x}
\PYG{g+go}{array([[3, 0, 0],}
\PYG{g+go}{       [0, 4, 0],}
\PYG{g+go}{       [5, 6, 0]])}
\PYG{g+gp}{\PYGZgt{}\PYGZgt{}\PYGZgt{} }\PYG{n}{np}\PYG{o}{.}\PYG{n}{nonzero}\PYG{p}{(}\PYG{n}{x}\PYG{p}{)}
\PYG{g+go}{(array([0, 1, 2, 2]), array([0, 1, 0, 1]))}
\end{sphinxVerbatim}

\begin{sphinxVerbatim}[commandchars=\\\{\}]
\PYG{g+gp}{\PYGZgt{}\PYGZgt{}\PYGZgt{} }\PYG{n}{x}\PYG{p}{[}\PYG{n}{np}\PYG{o}{.}\PYG{n}{nonzero}\PYG{p}{(}\PYG{n}{x}\PYG{p}{)}\PYG{p}{]}
\PYG{g+go}{array([3, 4, 5, 6])}
\PYG{g+gp}{\PYGZgt{}\PYGZgt{}\PYGZgt{} }\PYG{n}{np}\PYG{o}{.}\PYG{n}{transpose}\PYG{p}{(}\PYG{n}{np}\PYG{o}{.}\PYG{n}{nonzero}\PYG{p}{(}\PYG{n}{x}\PYG{p}{)}\PYG{p}{)}
\PYG{g+go}{array([[0, 0],}
\PYG{g+go}{       [1, 1],}
\PYG{g+go}{       [2, 0],}
\PYG{g+go}{       [2, 1]])}
\end{sphinxVerbatim}

A common use for \sphinxcode{\sphinxupquote{nonzero}} is to find the indices of an array, where
a condition is True.  Given an array \sphinxtitleref{a}, the condition \sphinxtitleref{a} \textgreater{} 3 is a
boolean array and since False is interpreted as 0, np.nonzero(a \textgreater{} 3)
yields the indices of the \sphinxtitleref{a} where the condition is true.

\begin{sphinxVerbatim}[commandchars=\\\{\}]
\PYG{g+gp}{\PYGZgt{}\PYGZgt{}\PYGZgt{} }\PYG{n}{a} \PYG{o}{=} \PYG{n}{np}\PYG{o}{.}\PYG{n}{array}\PYG{p}{(}\PYG{p}{[}\PYG{p}{[}\PYG{l+m+mi}{1}\PYG{p}{,} \PYG{l+m+mi}{2}\PYG{p}{,} \PYG{l+m+mi}{3}\PYG{p}{]}\PYG{p}{,} \PYG{p}{[}\PYG{l+m+mi}{4}\PYG{p}{,} \PYG{l+m+mi}{5}\PYG{p}{,} \PYG{l+m+mi}{6}\PYG{p}{]}\PYG{p}{,} \PYG{p}{[}\PYG{l+m+mi}{7}\PYG{p}{,} \PYG{l+m+mi}{8}\PYG{p}{,} \PYG{l+m+mi}{9}\PYG{p}{]}\PYG{p}{]}\PYG{p}{)}
\PYG{g+gp}{\PYGZgt{}\PYGZgt{}\PYGZgt{} }\PYG{n}{a} \PYG{o}{\PYGZgt{}} \PYG{l+m+mi}{3}
\PYG{g+go}{array([[False, False, False],}
\PYG{g+go}{       [ True,  True,  True],}
\PYG{g+go}{       [ True,  True,  True]])}
\PYG{g+gp}{\PYGZgt{}\PYGZgt{}\PYGZgt{} }\PYG{n}{np}\PYG{o}{.}\PYG{n}{nonzero}\PYG{p}{(}\PYG{n}{a} \PYG{o}{\PYGZgt{}} \PYG{l+m+mi}{3}\PYG{p}{)}
\PYG{g+go}{(array([1, 1, 1, 2, 2, 2]), array([0, 1, 2, 0, 1, 2]))}
\end{sphinxVerbatim}

Using this result to index \sphinxtitleref{a} is equivalent to using the mask directly:

\begin{sphinxVerbatim}[commandchars=\\\{\}]
\PYG{g+gp}{\PYGZgt{}\PYGZgt{}\PYGZgt{} }\PYG{n}{a}\PYG{p}{[}\PYG{n}{np}\PYG{o}{.}\PYG{n}{nonzero}\PYG{p}{(}\PYG{n}{a} \PYG{o}{\PYGZgt{}} \PYG{l+m+mi}{3}\PYG{p}{)}\PYG{p}{]}
\PYG{g+go}{array([4, 5, 6, 7, 8, 9])}
\PYG{g+gp}{\PYGZgt{}\PYGZgt{}\PYGZgt{} }\PYG{n}{a}\PYG{p}{[}\PYG{n}{a} \PYG{o}{\PYGZgt{}} \PYG{l+m+mi}{3}\PYG{p}{]}  \PYG{c+c1}{\PYGZsh{} prefer this spelling}
\PYG{g+go}{array([4, 5, 6, 7, 8, 9])}
\end{sphinxVerbatim}

\sphinxcode{\sphinxupquote{nonzero}} can also be called as a method of the array.

\begin{sphinxVerbatim}[commandchars=\\\{\}]
\PYG{g+gp}{\PYGZgt{}\PYGZgt{}\PYGZgt{} }\PYG{p}{(}\PYG{n}{a} \PYG{o}{\PYGZgt{}} \PYG{l+m+mi}{3}\PYG{p}{)}\PYG{o}{.}\PYG{n}{nonzero}\PYG{p}{(}\PYG{p}{)}
\PYG{g+go}{(array([1, 1, 1, 2, 2, 2]), array([0, 1, 2, 0, 1, 2]))}
\end{sphinxVerbatim}

\end{fulllineitems}

\index{not\_equal() (in module symjax.tensor)@\spxentry{not\_equal()}\spxextra{in module symjax.tensor}}

\begin{fulllineitems}
\phantomsection\label{\detokenize{modules/tensor:symjax.tensor.not_equal}}\pysiglinewithargsret{\sphinxbfcode{\sphinxupquote{not\_equal}}}{\emph{\DUrole{n}{x1}}, \emph{\DUrole{n}{x2}}}{}
Return (x1 != x2) element\sphinxhyphen{}wise.

LAX\sphinxhyphen{}backend implementation of {\hyperref[\detokenize{modules/tensor:symjax.tensor.not_equal}]{\sphinxcrossref{\sphinxcode{\sphinxupquote{not\_equal()}}}}}.
ADDITIONOriginal docstring below.

LAX\sphinxhyphen{}backend implementation of {\hyperref[\detokenize{modules/tensor:symjax.tensor.not_equal}]{\sphinxcrossref{\sphinxcode{\sphinxupquote{not\_equal()}}}}}.
Original docstring below.

not\_equal(x1, x2, /, out=None, {\color{red}\bfseries{}*}, where=True, casting=’same\_kind’, order=’K’, dtype=None, subok=True{[}, signature, extobj{]})
\begin{quote}\begin{description}
\item[{Returns}] \leavevmode
\sphinxstylestrong{out} \textendash{} Output array, element\sphinxhyphen{}wise comparison of \sphinxtitleref{x1} and \sphinxtitleref{x2}.
Typically of type bool, unless \sphinxcode{\sphinxupquote{dtype=object}} is passed.
This is a scalar if both \sphinxtitleref{x1} and \sphinxtitleref{x2} are scalars.

\item[{Return type}] \leavevmode
ndarray or scalar

\end{description}\end{quote}

\sphinxstrong{See also:}

{\hyperref[\detokenize{modules/tensor:symjax.tensor.equal}]{\sphinxcrossref{\sphinxcode{\sphinxupquote{equal()}}}}}, {\hyperref[\detokenize{modules/tensor:symjax.tensor.greater}]{\sphinxcrossref{\sphinxcode{\sphinxupquote{greater()}}}}}, {\hyperref[\detokenize{modules/tensor:symjax.tensor.greater_equal}]{\sphinxcrossref{\sphinxcode{\sphinxupquote{greater\_equal()}}}}}, {\hyperref[\detokenize{modules/tensor:symjax.tensor.less}]{\sphinxcrossref{\sphinxcode{\sphinxupquote{less()}}}}}, {\hyperref[\detokenize{modules/tensor:symjax.tensor.less_equal}]{\sphinxcrossref{\sphinxcode{\sphinxupquote{less\_equal()}}}}}

\subsubsection*{Examples}

\begin{sphinxVerbatim}[commandchars=\\\{\}]
\PYG{g+gp}{\PYGZgt{}\PYGZgt{}\PYGZgt{} }\PYG{n}{np}\PYG{o}{.}\PYG{n}{not\PYGZus{}equal}\PYG{p}{(}\PYG{p}{[}\PYG{l+m+mf}{1.}\PYG{p}{,}\PYG{l+m+mf}{2.}\PYG{p}{]}\PYG{p}{,} \PYG{p}{[}\PYG{l+m+mf}{1.}\PYG{p}{,} \PYG{l+m+mf}{3.}\PYG{p}{]}\PYG{p}{)}
\PYG{g+go}{array([False,  True])}
\PYG{g+gp}{\PYGZgt{}\PYGZgt{}\PYGZgt{} }\PYG{n}{np}\PYG{o}{.}\PYG{n}{not\PYGZus{}equal}\PYG{p}{(}\PYG{p}{[}\PYG{l+m+mi}{1}\PYG{p}{,} \PYG{l+m+mi}{2}\PYG{p}{]}\PYG{p}{,} \PYG{p}{[}\PYG{p}{[}\PYG{l+m+mi}{1}\PYG{p}{,} \PYG{l+m+mi}{3}\PYG{p}{]}\PYG{p}{,}\PYG{p}{[}\PYG{l+m+mi}{1}\PYG{p}{,} \PYG{l+m+mi}{4}\PYG{p}{]}\PYG{p}{]}\PYG{p}{)}
\PYG{g+go}{array([[False,  True],}
\PYG{g+go}{       [False,  True]])}
\end{sphinxVerbatim}

\end{fulllineitems}

\index{ones() (in module symjax.tensor)@\spxentry{ones()}\spxextra{in module symjax.tensor}}

\begin{fulllineitems}
\phantomsection\label{\detokenize{modules/tensor:symjax.tensor.ones}}\pysiglinewithargsret{\sphinxbfcode{\sphinxupquote{ones}}}{\emph{\DUrole{n}{shape}}, \emph{\DUrole{n}{dtype}\DUrole{o}{=}\DUrole{default_value}{None}}}{}
Return a new array of given shape and type, filled with ones.

LAX\sphinxhyphen{}backend implementation of {\hyperref[\detokenize{modules/tensor:symjax.tensor.ones}]{\sphinxcrossref{\sphinxcode{\sphinxupquote{ones()}}}}}.
ADDITIONOriginal docstring below.

LAX\sphinxhyphen{}backend implementation of {\hyperref[\detokenize{modules/tensor:symjax.tensor.ones}]{\sphinxcrossref{\sphinxcode{\sphinxupquote{ones()}}}}}.
Original docstring below.
\begin{quote}\begin{description}
\item[{Parameters}] \leavevmode\begin{itemize}
\item {} 
\sphinxstyleliteralstrong{\sphinxupquote{shape}} (\sphinxstyleliteralemphasis{\sphinxupquote{int}}\sphinxstyleliteralemphasis{\sphinxupquote{ or }}\sphinxstyleliteralemphasis{\sphinxupquote{sequence of ints}}) \textendash{} Shape of the new array, e.g., \sphinxcode{\sphinxupquote{(2, 3)}} or \sphinxcode{\sphinxupquote{2}}.

\item {} 
\sphinxstyleliteralstrong{\sphinxupquote{dtype}} (\sphinxstyleliteralemphasis{\sphinxupquote{data\sphinxhyphen{}type}}\sphinxstyleliteralemphasis{\sphinxupquote{, }}\sphinxstyleliteralemphasis{\sphinxupquote{optional}}) \textendash{} The desired data\sphinxhyphen{}type for the array, e.g., \sphinxtitleref{numpy.int8}.  Default is
\sphinxtitleref{numpy.float64}.

\end{itemize}

\item[{Returns}] \leavevmode
\sphinxstylestrong{out} \textendash{} Array of ones with the given shape, dtype, and order.

\item[{Return type}] \leavevmode
ndarray

\end{description}\end{quote}

\sphinxstrong{See also:}

\begin{description}
\item[{{\hyperref[\detokenize{modules/tensor:symjax.tensor.ones_like}]{\sphinxcrossref{\sphinxcode{\sphinxupquote{ones\_like()}}}}}}] \leavevmode
Return an array of ones with shape and type of input.

\item[{{\hyperref[\detokenize{modules/tensor:symjax.tensor.empty}]{\sphinxcrossref{\sphinxcode{\sphinxupquote{empty()}}}}}}] \leavevmode
Return a new uninitialized array.

\item[{{\hyperref[\detokenize{modules/tensor:symjax.tensor.zeros}]{\sphinxcrossref{\sphinxcode{\sphinxupquote{zeros()}}}}}}] \leavevmode
Return a new array setting values to zero.

\item[{{\hyperref[\detokenize{modules/tensor:symjax.tensor.full}]{\sphinxcrossref{\sphinxcode{\sphinxupquote{full()}}}}}}] \leavevmode
Return a new array of given shape filled with value.

\end{description}

\subsubsection*{Examples}

\begin{sphinxVerbatim}[commandchars=\\\{\}]
\PYG{g+gp}{\PYGZgt{}\PYGZgt{}\PYGZgt{} }\PYG{n}{np}\PYG{o}{.}\PYG{n}{ones}\PYG{p}{(}\PYG{l+m+mi}{5}\PYG{p}{)}
\PYG{g+go}{array([1., 1., 1., 1., 1.])}
\end{sphinxVerbatim}

\begin{sphinxVerbatim}[commandchars=\\\{\}]
\PYG{g+gp}{\PYGZgt{}\PYGZgt{}\PYGZgt{} }\PYG{n}{np}\PYG{o}{.}\PYG{n}{ones}\PYG{p}{(}\PYG{p}{(}\PYG{l+m+mi}{5}\PYG{p}{,}\PYG{p}{)}\PYG{p}{,} \PYG{n}{dtype}\PYG{o}{=}\PYG{n+nb}{int}\PYG{p}{)}
\PYG{g+go}{array([1, 1, 1, 1, 1])}
\end{sphinxVerbatim}

\begin{sphinxVerbatim}[commandchars=\\\{\}]
\PYG{g+gp}{\PYGZgt{}\PYGZgt{}\PYGZgt{} }\PYG{n}{np}\PYG{o}{.}\PYG{n}{ones}\PYG{p}{(}\PYG{p}{(}\PYG{l+m+mi}{2}\PYG{p}{,} \PYG{l+m+mi}{1}\PYG{p}{)}\PYG{p}{)}
\PYG{g+go}{array([[1.],}
\PYG{g+go}{       [1.]])}
\end{sphinxVerbatim}

\begin{sphinxVerbatim}[commandchars=\\\{\}]
\PYG{g+gp}{\PYGZgt{}\PYGZgt{}\PYGZgt{} }\PYG{n}{s} \PYG{o}{=} \PYG{p}{(}\PYG{l+m+mi}{2}\PYG{p}{,}\PYG{l+m+mi}{2}\PYG{p}{)}
\PYG{g+gp}{\PYGZgt{}\PYGZgt{}\PYGZgt{} }\PYG{n}{np}\PYG{o}{.}\PYG{n}{ones}\PYG{p}{(}\PYG{n}{s}\PYG{p}{)}
\PYG{g+go}{array([[1.,  1.],}
\PYG{g+go}{       [1.,  1.]])}
\end{sphinxVerbatim}

\end{fulllineitems}

\index{ones\_like() (in module symjax.tensor)@\spxentry{ones\_like()}\spxextra{in module symjax.tensor}}

\begin{fulllineitems}
\phantomsection\label{\detokenize{modules/tensor:symjax.tensor.ones_like}}\pysiglinewithargsret{\sphinxbfcode{\sphinxupquote{ones\_like}}}{\emph{\DUrole{n}{x}}, \emph{\DUrole{n}{dtype}\DUrole{o}{=}\DUrole{default_value}{None}}}{}
Return an array of ones with the same shape and type as a given array.

LAX\sphinxhyphen{}backend implementation of {\hyperref[\detokenize{modules/tensor:symjax.tensor.ones_like}]{\sphinxcrossref{\sphinxcode{\sphinxupquote{ones\_like()}}}}}.
ADDITIONOriginal docstring below.

LAX\sphinxhyphen{}backend implementation of {\hyperref[\detokenize{modules/tensor:symjax.tensor.ones_like}]{\sphinxcrossref{\sphinxcode{\sphinxupquote{ones\_like()}}}}}.
Original docstring below.
\begin{quote}\begin{description}
\item[{Parameters}] \leavevmode
\sphinxstyleliteralstrong{\sphinxupquote{dtype}} (\sphinxstyleliteralemphasis{\sphinxupquote{data\sphinxhyphen{}type}}\sphinxstyleliteralemphasis{\sphinxupquote{, }}\sphinxstyleliteralemphasis{\sphinxupquote{optional}}) \textendash{} Overrides the data type of the result.

\item[{Returns}] \leavevmode
\sphinxstylestrong{out} \textendash{} Array of ones with the same shape and type as \sphinxtitleref{a}.

\item[{Return type}] \leavevmode
ndarray

\end{description}\end{quote}

\sphinxstrong{See also:}

\begin{description}
\item[{{\hyperref[\detokenize{modules/tensor:symjax.tensor.empty_like}]{\sphinxcrossref{\sphinxcode{\sphinxupquote{empty\_like()}}}}}}] \leavevmode
Return an empty array with shape and type of input.

\item[{{\hyperref[\detokenize{modules/tensor:symjax.tensor.zeros_like}]{\sphinxcrossref{\sphinxcode{\sphinxupquote{zeros\_like()}}}}}}] \leavevmode
Return an array of zeros with shape and type of input.

\item[{{\hyperref[\detokenize{modules/tensor:symjax.tensor.full_like}]{\sphinxcrossref{\sphinxcode{\sphinxupquote{full\_like()}}}}}}] \leavevmode
Return a new array with shape of input filled with value.

\item[{{\hyperref[\detokenize{modules/tensor:symjax.tensor.ones}]{\sphinxcrossref{\sphinxcode{\sphinxupquote{ones()}}}}}}] \leavevmode
Return a new array setting values to one.

\end{description}

\subsubsection*{Examples}

\begin{sphinxVerbatim}[commandchars=\\\{\}]
\PYG{g+gp}{\PYGZgt{}\PYGZgt{}\PYGZgt{} }\PYG{n}{x} \PYG{o}{=} \PYG{n}{np}\PYG{o}{.}\PYG{n}{arange}\PYG{p}{(}\PYG{l+m+mi}{6}\PYG{p}{)}
\PYG{g+gp}{\PYGZgt{}\PYGZgt{}\PYGZgt{} }\PYG{n}{x} \PYG{o}{=} \PYG{n}{x}\PYG{o}{.}\PYG{n}{reshape}\PYG{p}{(}\PYG{p}{(}\PYG{l+m+mi}{2}\PYG{p}{,} \PYG{l+m+mi}{3}\PYG{p}{)}\PYG{p}{)}
\PYG{g+gp}{\PYGZgt{}\PYGZgt{}\PYGZgt{} }\PYG{n}{x}
\PYG{g+go}{array([[0, 1, 2],}
\PYG{g+go}{       [3, 4, 5]])}
\PYG{g+gp}{\PYGZgt{}\PYGZgt{}\PYGZgt{} }\PYG{n}{np}\PYG{o}{.}\PYG{n}{ones\PYGZus{}like}\PYG{p}{(}\PYG{n}{x}\PYG{p}{)}
\PYG{g+go}{array([[1, 1, 1],}
\PYG{g+go}{       [1, 1, 1]])}
\end{sphinxVerbatim}

\begin{sphinxVerbatim}[commandchars=\\\{\}]
\PYG{g+gp}{\PYGZgt{}\PYGZgt{}\PYGZgt{} }\PYG{n}{y} \PYG{o}{=} \PYG{n}{np}\PYG{o}{.}\PYG{n}{arange}\PYG{p}{(}\PYG{l+m+mi}{3}\PYG{p}{,} \PYG{n}{dtype}\PYG{o}{=}\PYG{n+nb}{float}\PYG{p}{)}
\PYG{g+gp}{\PYGZgt{}\PYGZgt{}\PYGZgt{} }\PYG{n}{y}
\PYG{g+go}{array([0., 1., 2.])}
\PYG{g+gp}{\PYGZgt{}\PYGZgt{}\PYGZgt{} }\PYG{n}{np}\PYG{o}{.}\PYG{n}{ones\PYGZus{}like}\PYG{p}{(}\PYG{n}{y}\PYG{p}{)}
\PYG{g+go}{array([1.,  1.,  1.])}
\end{sphinxVerbatim}

\end{fulllineitems}

\index{outer() (in module symjax.tensor)@\spxentry{outer()}\spxextra{in module symjax.tensor}}

\begin{fulllineitems}
\phantomsection\label{\detokenize{modules/tensor:symjax.tensor.outer}}\pysiglinewithargsret{\sphinxbfcode{\sphinxupquote{outer}}}{\emph{\DUrole{n}{a}}, \emph{\DUrole{n}{b}}, \emph{\DUrole{n}{out}\DUrole{o}{=}\DUrole{default_value}{None}}}{}
Compute the outer product of two vectors.

LAX\sphinxhyphen{}backend implementation of {\hyperref[\detokenize{modules/tensor:symjax.tensor.outer}]{\sphinxcrossref{\sphinxcode{\sphinxupquote{outer()}}}}}.
ADDITIONOriginal docstring below.

LAX\sphinxhyphen{}backend implementation of {\hyperref[\detokenize{modules/tensor:symjax.tensor.outer}]{\sphinxcrossref{\sphinxcode{\sphinxupquote{outer()}}}}}.
Original docstring below.

Given two vectors, \sphinxcode{\sphinxupquote{a = {[}a0, a1, ..., aM{]}}} and
\sphinxcode{\sphinxupquote{b = {[}b0, b1, ..., bN{]}}},
the outer product {\color{red}\bfseries{}{[}1{]}\_} is:

\begin{sphinxVerbatim}[commandchars=\\\{\}]
\PYG{p}{[}\PYG{p}{[}\PYG{n}{a0}\PYG{o}{*}\PYG{n}{b0}  \PYG{n}{a0}\PYG{o}{*}\PYG{n}{b1} \PYG{o}{.}\PYG{o}{.}\PYG{o}{.} \PYG{n}{a0}\PYG{o}{*}\PYG{n}{bN} \PYG{p}{]}
 \PYG{p}{[}\PYG{n}{a1}\PYG{o}{*}\PYG{n}{b0}    \PYG{o}{.}
 \PYG{p}{[} \PYG{o}{.}\PYG{o}{.}\PYG{o}{.}          \PYG{o}{.}
 \PYG{p}{[}\PYG{n}{aM}\PYG{o}{*}\PYG{n}{b0}            \PYG{n}{aM}\PYG{o}{*}\PYG{n}{bN} \PYG{p}{]}\PYG{p}{]}
\end{sphinxVerbatim}
\begin{quote}\begin{description}
\item[{Returns}] \leavevmode
\sphinxstylestrong{out} \textendash{} \sphinxcode{\sphinxupquote{out{[}i, j{]} = a{[}i{]} * b{[}j{]}}}

\item[{Return type}] \leavevmode
(M, N) ndarray

\end{description}\end{quote}

\sphinxstrong{See also:}

{\hyperref[\detokenize{modules/tensor:symjax.tensor.inner}]{\sphinxcrossref{\sphinxcode{\sphinxupquote{inner()}}}}}
\begin{description}
\item[{{\hyperref[\detokenize{modules/tensor:symjax.tensor.einsum}]{\sphinxcrossref{\sphinxcode{\sphinxupquote{einsum()}}}}}}] \leavevmode
\sphinxcode{\sphinxupquote{einsum(\textquotesingle{}i,j\sphinxhyphen{}\textgreater{}ij\textquotesingle{}, a.ravel(), b.ravel())}} is the equivalent.

\item[{\sphinxcode{\sphinxupquote{ufunc.outer()}}}] \leavevmode
A generalization to N dimensions and other operations. \sphinxcode{\sphinxupquote{np.multiply.outer(a.ravel(), b.ravel())}} is the equivalent.

\end{description}

\subsubsection*{References}
\subsubsection*{Examples}

Make a (\sphinxstyleemphasis{very} coarse) grid for computing a Mandelbrot set:

\begin{sphinxVerbatim}[commandchars=\\\{\}]
\PYG{g+gp}{\PYGZgt{}\PYGZgt{}\PYGZgt{} }\PYG{n}{rl} \PYG{o}{=} \PYG{n}{np}\PYG{o}{.}\PYG{n}{outer}\PYG{p}{(}\PYG{n}{np}\PYG{o}{.}\PYG{n}{ones}\PYG{p}{(}\PYG{p}{(}\PYG{l+m+mi}{5}\PYG{p}{,}\PYG{p}{)}\PYG{p}{)}\PYG{p}{,} \PYG{n}{np}\PYG{o}{.}\PYG{n}{linspace}\PYG{p}{(}\PYG{o}{\PYGZhy{}}\PYG{l+m+mi}{2}\PYG{p}{,} \PYG{l+m+mi}{2}\PYG{p}{,} \PYG{l+m+mi}{5}\PYG{p}{)}\PYG{p}{)}
\PYG{g+gp}{\PYGZgt{}\PYGZgt{}\PYGZgt{} }\PYG{n}{rl}
\PYG{g+go}{array([[\PYGZhy{}2., \PYGZhy{}1.,  0.,  1.,  2.],}
\PYG{g+go}{       [\PYGZhy{}2., \PYGZhy{}1.,  0.,  1.,  2.],}
\PYG{g+go}{       [\PYGZhy{}2., \PYGZhy{}1.,  0.,  1.,  2.],}
\PYG{g+go}{       [\PYGZhy{}2., \PYGZhy{}1.,  0.,  1.,  2.],}
\PYG{g+go}{       [\PYGZhy{}2., \PYGZhy{}1.,  0.,  1.,  2.]])}
\PYG{g+gp}{\PYGZgt{}\PYGZgt{}\PYGZgt{} }\PYG{n}{im} \PYG{o}{=} \PYG{n}{np}\PYG{o}{.}\PYG{n}{outer}\PYG{p}{(}\PYG{l+m+mi}{1}\PYG{n}{j}\PYG{o}{*}\PYG{n}{np}\PYG{o}{.}\PYG{n}{linspace}\PYG{p}{(}\PYG{l+m+mi}{2}\PYG{p}{,} \PYG{o}{\PYGZhy{}}\PYG{l+m+mi}{2}\PYG{p}{,} \PYG{l+m+mi}{5}\PYG{p}{)}\PYG{p}{,} \PYG{n}{np}\PYG{o}{.}\PYG{n}{ones}\PYG{p}{(}\PYG{p}{(}\PYG{l+m+mi}{5}\PYG{p}{,}\PYG{p}{)}\PYG{p}{)}\PYG{p}{)}
\PYG{g+gp}{\PYGZgt{}\PYGZgt{}\PYGZgt{} }\PYG{n}{im}
\PYG{g+go}{array([[0.+2.j, 0.+2.j, 0.+2.j, 0.+2.j, 0.+2.j],}
\PYG{g+go}{       [0.+1.j, 0.+1.j, 0.+1.j, 0.+1.j, 0.+1.j],}
\PYG{g+go}{       [0.+0.j, 0.+0.j, 0.+0.j, 0.+0.j, 0.+0.j],}
\PYG{g+go}{       [0.\PYGZhy{}1.j, 0.\PYGZhy{}1.j, 0.\PYGZhy{}1.j, 0.\PYGZhy{}1.j, 0.\PYGZhy{}1.j],}
\PYG{g+go}{       [0.\PYGZhy{}2.j, 0.\PYGZhy{}2.j, 0.\PYGZhy{}2.j, 0.\PYGZhy{}2.j, 0.\PYGZhy{}2.j]])}
\PYG{g+gp}{\PYGZgt{}\PYGZgt{}\PYGZgt{} }\PYG{n}{grid} \PYG{o}{=} \PYG{n}{rl} \PYG{o}{+} \PYG{n}{im}
\PYG{g+gp}{\PYGZgt{}\PYGZgt{}\PYGZgt{} }\PYG{n}{grid}
\PYG{g+go}{array([[\PYGZhy{}2.+2.j, \PYGZhy{}1.+2.j,  0.+2.j,  1.+2.j,  2.+2.j],}
\PYG{g+go}{       [\PYGZhy{}2.+1.j, \PYGZhy{}1.+1.j,  0.+1.j,  1.+1.j,  2.+1.j],}
\PYG{g+go}{       [\PYGZhy{}2.+0.j, \PYGZhy{}1.+0.j,  0.+0.j,  1.+0.j,  2.+0.j],}
\PYG{g+go}{       [\PYGZhy{}2.\PYGZhy{}1.j, \PYGZhy{}1.\PYGZhy{}1.j,  0.\PYGZhy{}1.j,  1.\PYGZhy{}1.j,  2.\PYGZhy{}1.j],}
\PYG{g+go}{       [\PYGZhy{}2.\PYGZhy{}2.j, \PYGZhy{}1.\PYGZhy{}2.j,  0.\PYGZhy{}2.j,  1.\PYGZhy{}2.j,  2.\PYGZhy{}2.j]])}
\end{sphinxVerbatim}

An example using a “vector” of letters:

\begin{sphinxVerbatim}[commandchars=\\\{\}]
\PYG{g+gp}{\PYGZgt{}\PYGZgt{}\PYGZgt{} }\PYG{n}{x} \PYG{o}{=} \PYG{n}{np}\PYG{o}{.}\PYG{n}{array}\PYG{p}{(}\PYG{p}{[}\PYG{l+s+s1}{\PYGZsq{}}\PYG{l+s+s1}{a}\PYG{l+s+s1}{\PYGZsq{}}\PYG{p}{,} \PYG{l+s+s1}{\PYGZsq{}}\PYG{l+s+s1}{b}\PYG{l+s+s1}{\PYGZsq{}}\PYG{p}{,} \PYG{l+s+s1}{\PYGZsq{}}\PYG{l+s+s1}{c}\PYG{l+s+s1}{\PYGZsq{}}\PYG{p}{]}\PYG{p}{,} \PYG{n}{dtype}\PYG{o}{=}\PYG{n+nb}{object}\PYG{p}{)}
\PYG{g+gp}{\PYGZgt{}\PYGZgt{}\PYGZgt{} }\PYG{n}{np}\PYG{o}{.}\PYG{n}{outer}\PYG{p}{(}\PYG{n}{x}\PYG{p}{,} \PYG{p}{[}\PYG{l+m+mi}{1}\PYG{p}{,} \PYG{l+m+mi}{2}\PYG{p}{,} \PYG{l+m+mi}{3}\PYG{p}{]}\PYG{p}{)}
\PYG{g+go}{array([[\PYGZsq{}a\PYGZsq{}, \PYGZsq{}aa\PYGZsq{}, \PYGZsq{}aaa\PYGZsq{}],}
\PYG{g+go}{       [\PYGZsq{}b\PYGZsq{}, \PYGZsq{}bb\PYGZsq{}, \PYGZsq{}bbb\PYGZsq{}],}
\PYG{g+go}{       [\PYGZsq{}c\PYGZsq{}, \PYGZsq{}cc\PYGZsq{}, \PYGZsq{}ccc\PYGZsq{}]], dtype=object)}
\end{sphinxVerbatim}

\end{fulllineitems}

\index{pad() (in module symjax.tensor)@\spxentry{pad()}\spxextra{in module symjax.tensor}}

\begin{fulllineitems}
\phantomsection\label{\detokenize{modules/tensor:symjax.tensor.pad}}\pysiglinewithargsret{\sphinxbfcode{\sphinxupquote{pad}}}{\emph{\DUrole{n}{array}}, \emph{\DUrole{n}{pad\_width}}, \emph{\DUrole{n}{mode}\DUrole{o}{=}\DUrole{default_value}{\textquotesingle{}constant\textquotesingle{}}}, \emph{\DUrole{n}{constant\_values}\DUrole{o}{=}\DUrole{default_value}{0}}}{}
Pad an array.

LAX\sphinxhyphen{}backend implementation of {\hyperref[\detokenize{modules/tensor:symjax.tensor.pad}]{\sphinxcrossref{\sphinxcode{\sphinxupquote{pad()}}}}}.
ADDITIONOriginal docstring below.

LAX\sphinxhyphen{}backend implementation of {\hyperref[\detokenize{modules/tensor:symjax.tensor.pad}]{\sphinxcrossref{\sphinxcode{\sphinxupquote{pad()}}}}}.
Original docstring below.
\begin{quote}\begin{description}
\item[{Returns}] \leavevmode
\sphinxstylestrong{pad} \textendash{} Padded array of rank equal to \sphinxtitleref{array} with shape increased
according to \sphinxtitleref{pad\_width}.

\item[{Return type}] \leavevmode
ndarray

\end{description}\end{quote}
\subsubsection*{Notes}

\DUrole{versionmodified,added}{New in version 1.7.0.}

For an array with rank greater than 1, some of the padding of later
axes is calculated from padding of previous axes.  This is easiest to
think about with a rank 2 array where the corners of the padded array
are calculated by using padded values from the first axis.

The padding function, if used, should modify a rank 1 array in\sphinxhyphen{}place. It
has the following signature:

\begin{sphinxVerbatim}[commandchars=\\\{\}]
\PYG{n}{padding\PYGZus{}func}\PYG{p}{(}\PYG{n}{vector}\PYG{p}{,} \PYG{n}{iaxis\PYGZus{}pad\PYGZus{}width}\PYG{p}{,} \PYG{n}{iaxis}\PYG{p}{,} \PYG{n}{kwargs}\PYG{p}{)}
\end{sphinxVerbatim}

where
\begin{quote}
\begin{description}
\item[{vector}] \leavevmode{[}ndarray{]}
A rank 1 array already padded with zeros.  Padded values are
vector{[}:iaxis\_pad\_width{[}0{]}{]} and vector{[}\sphinxhyphen{}iaxis\_pad\_width{[}1{]}:{]}.

\item[{iaxis\_pad\_width}] \leavevmode{[}tuple{]}
A 2\sphinxhyphen{}tuple of ints, iaxis\_pad\_width{[}0{]} represents the number of
values padded at the beginning of vector where
iaxis\_pad\_width{[}1{]} represents the number of values padded at
the end of vector.

\item[{iaxis}] \leavevmode{[}int{]}
The axis currently being calculated.

\item[{kwargs}] \leavevmode{[}dict{]}
Any keyword arguments the function requires.

\end{description}
\end{quote}
\subsubsection*{Examples}

\begin{sphinxVerbatim}[commandchars=\\\{\}]
\PYG{g+gp}{\PYGZgt{}\PYGZgt{}\PYGZgt{} }\PYG{n}{a} \PYG{o}{=} \PYG{p}{[}\PYG{l+m+mi}{1}\PYG{p}{,} \PYG{l+m+mi}{2}\PYG{p}{,} \PYG{l+m+mi}{3}\PYG{p}{,} \PYG{l+m+mi}{4}\PYG{p}{,} \PYG{l+m+mi}{5}\PYG{p}{]}
\PYG{g+gp}{\PYGZgt{}\PYGZgt{}\PYGZgt{} }\PYG{n}{np}\PYG{o}{.}\PYG{n}{pad}\PYG{p}{(}\PYG{n}{a}\PYG{p}{,} \PYG{p}{(}\PYG{l+m+mi}{2}\PYG{p}{,} \PYG{l+m+mi}{3}\PYG{p}{)}\PYG{p}{,} \PYG{l+s+s1}{\PYGZsq{}}\PYG{l+s+s1}{constant}\PYG{l+s+s1}{\PYGZsq{}}\PYG{p}{,} \PYG{n}{constant\PYGZus{}values}\PYG{o}{=}\PYG{p}{(}\PYG{l+m+mi}{4}\PYG{p}{,} \PYG{l+m+mi}{6}\PYG{p}{)}\PYG{p}{)}
\PYG{g+go}{array([4, 4, 1, ..., 6, 6, 6])}
\end{sphinxVerbatim}

\begin{sphinxVerbatim}[commandchars=\\\{\}]
\PYG{g+gp}{\PYGZgt{}\PYGZgt{}\PYGZgt{} }\PYG{n}{np}\PYG{o}{.}\PYG{n}{pad}\PYG{p}{(}\PYG{n}{a}\PYG{p}{,} \PYG{p}{(}\PYG{l+m+mi}{2}\PYG{p}{,} \PYG{l+m+mi}{3}\PYG{p}{)}\PYG{p}{,} \PYG{l+s+s1}{\PYGZsq{}}\PYG{l+s+s1}{edge}\PYG{l+s+s1}{\PYGZsq{}}\PYG{p}{)}
\PYG{g+go}{array([1, 1, 1, ..., 5, 5, 5])}
\end{sphinxVerbatim}

\begin{sphinxVerbatim}[commandchars=\\\{\}]
\PYG{g+gp}{\PYGZgt{}\PYGZgt{}\PYGZgt{} }\PYG{n}{np}\PYG{o}{.}\PYG{n}{pad}\PYG{p}{(}\PYG{n}{a}\PYG{p}{,} \PYG{p}{(}\PYG{l+m+mi}{2}\PYG{p}{,} \PYG{l+m+mi}{3}\PYG{p}{)}\PYG{p}{,} \PYG{l+s+s1}{\PYGZsq{}}\PYG{l+s+s1}{linear\PYGZus{}ramp}\PYG{l+s+s1}{\PYGZsq{}}\PYG{p}{,} \PYG{n}{end\PYGZus{}values}\PYG{o}{=}\PYG{p}{(}\PYG{l+m+mi}{5}\PYG{p}{,} \PYG{o}{\PYGZhy{}}\PYG{l+m+mi}{4}\PYG{p}{)}\PYG{p}{)}
\PYG{g+go}{array([ 5,  3,  1,  2,  3,  4,  5,  2, \PYGZhy{}1, \PYGZhy{}4])}
\end{sphinxVerbatim}

\begin{sphinxVerbatim}[commandchars=\\\{\}]
\PYG{g+gp}{\PYGZgt{}\PYGZgt{}\PYGZgt{} }\PYG{n}{np}\PYG{o}{.}\PYG{n}{pad}\PYG{p}{(}\PYG{n}{a}\PYG{p}{,} \PYG{p}{(}\PYG{l+m+mi}{2}\PYG{p}{,}\PYG{p}{)}\PYG{p}{,} \PYG{l+s+s1}{\PYGZsq{}}\PYG{l+s+s1}{maximum}\PYG{l+s+s1}{\PYGZsq{}}\PYG{p}{)}
\PYG{g+go}{array([5, 5, 1, 2, 3, 4, 5, 5, 5])}
\end{sphinxVerbatim}

\begin{sphinxVerbatim}[commandchars=\\\{\}]
\PYG{g+gp}{\PYGZgt{}\PYGZgt{}\PYGZgt{} }\PYG{n}{np}\PYG{o}{.}\PYG{n}{pad}\PYG{p}{(}\PYG{n}{a}\PYG{p}{,} \PYG{p}{(}\PYG{l+m+mi}{2}\PYG{p}{,}\PYG{p}{)}\PYG{p}{,} \PYG{l+s+s1}{\PYGZsq{}}\PYG{l+s+s1}{mean}\PYG{l+s+s1}{\PYGZsq{}}\PYG{p}{)}
\PYG{g+go}{array([3, 3, 1, 2, 3, 4, 5, 3, 3])}
\end{sphinxVerbatim}

\begin{sphinxVerbatim}[commandchars=\\\{\}]
\PYG{g+gp}{\PYGZgt{}\PYGZgt{}\PYGZgt{} }\PYG{n}{np}\PYG{o}{.}\PYG{n}{pad}\PYG{p}{(}\PYG{n}{a}\PYG{p}{,} \PYG{p}{(}\PYG{l+m+mi}{2}\PYG{p}{,}\PYG{p}{)}\PYG{p}{,} \PYG{l+s+s1}{\PYGZsq{}}\PYG{l+s+s1}{median}\PYG{l+s+s1}{\PYGZsq{}}\PYG{p}{)}
\PYG{g+go}{array([3, 3, 1, 2, 3, 4, 5, 3, 3])}
\end{sphinxVerbatim}

\begin{sphinxVerbatim}[commandchars=\\\{\}]
\PYG{g+gp}{\PYGZgt{}\PYGZgt{}\PYGZgt{} }\PYG{n}{a} \PYG{o}{=} \PYG{p}{[}\PYG{p}{[}\PYG{l+m+mi}{1}\PYG{p}{,} \PYG{l+m+mi}{2}\PYG{p}{]}\PYG{p}{,} \PYG{p}{[}\PYG{l+m+mi}{3}\PYG{p}{,} \PYG{l+m+mi}{4}\PYG{p}{]}\PYG{p}{]}
\PYG{g+gp}{\PYGZgt{}\PYGZgt{}\PYGZgt{} }\PYG{n}{np}\PYG{o}{.}\PYG{n}{pad}\PYG{p}{(}\PYG{n}{a}\PYG{p}{,} \PYG{p}{(}\PYG{p}{(}\PYG{l+m+mi}{3}\PYG{p}{,} \PYG{l+m+mi}{2}\PYG{p}{)}\PYG{p}{,} \PYG{p}{(}\PYG{l+m+mi}{2}\PYG{p}{,} \PYG{l+m+mi}{3}\PYG{p}{)}\PYG{p}{)}\PYG{p}{,} \PYG{l+s+s1}{\PYGZsq{}}\PYG{l+s+s1}{minimum}\PYG{l+s+s1}{\PYGZsq{}}\PYG{p}{)}
\PYG{g+go}{array([[1, 1, 1, 2, 1, 1, 1],}
\PYG{g+go}{       [1, 1, 1, 2, 1, 1, 1],}
\PYG{g+go}{       [1, 1, 1, 2, 1, 1, 1],}
\PYG{g+go}{       [1, 1, 1, 2, 1, 1, 1],}
\PYG{g+go}{       [3, 3, 3, 4, 3, 3, 3],}
\PYG{g+go}{       [1, 1, 1, 2, 1, 1, 1],}
\PYG{g+go}{       [1, 1, 1, 2, 1, 1, 1]])}
\end{sphinxVerbatim}

\begin{sphinxVerbatim}[commandchars=\\\{\}]
\PYG{g+gp}{\PYGZgt{}\PYGZgt{}\PYGZgt{} }\PYG{n}{a} \PYG{o}{=} \PYG{p}{[}\PYG{l+m+mi}{1}\PYG{p}{,} \PYG{l+m+mi}{2}\PYG{p}{,} \PYG{l+m+mi}{3}\PYG{p}{,} \PYG{l+m+mi}{4}\PYG{p}{,} \PYG{l+m+mi}{5}\PYG{p}{]}
\PYG{g+gp}{\PYGZgt{}\PYGZgt{}\PYGZgt{} }\PYG{n}{np}\PYG{o}{.}\PYG{n}{pad}\PYG{p}{(}\PYG{n}{a}\PYG{p}{,} \PYG{p}{(}\PYG{l+m+mi}{2}\PYG{p}{,} \PYG{l+m+mi}{3}\PYG{p}{)}\PYG{p}{,} \PYG{l+s+s1}{\PYGZsq{}}\PYG{l+s+s1}{reflect}\PYG{l+s+s1}{\PYGZsq{}}\PYG{p}{)}
\PYG{g+go}{array([3, 2, 1, 2, 3, 4, 5, 4, 3, 2])}
\end{sphinxVerbatim}

\begin{sphinxVerbatim}[commandchars=\\\{\}]
\PYG{g+gp}{\PYGZgt{}\PYGZgt{}\PYGZgt{} }\PYG{n}{np}\PYG{o}{.}\PYG{n}{pad}\PYG{p}{(}\PYG{n}{a}\PYG{p}{,} \PYG{p}{(}\PYG{l+m+mi}{2}\PYG{p}{,} \PYG{l+m+mi}{3}\PYG{p}{)}\PYG{p}{,} \PYG{l+s+s1}{\PYGZsq{}}\PYG{l+s+s1}{reflect}\PYG{l+s+s1}{\PYGZsq{}}\PYG{p}{,} \PYG{n}{reflect\PYGZus{}type}\PYG{o}{=}\PYG{l+s+s1}{\PYGZsq{}}\PYG{l+s+s1}{odd}\PYG{l+s+s1}{\PYGZsq{}}\PYG{p}{)}
\PYG{g+go}{array([\PYGZhy{}1,  0,  1,  2,  3,  4,  5,  6,  7,  8])}
\end{sphinxVerbatim}

\begin{sphinxVerbatim}[commandchars=\\\{\}]
\PYG{g+gp}{\PYGZgt{}\PYGZgt{}\PYGZgt{} }\PYG{n}{np}\PYG{o}{.}\PYG{n}{pad}\PYG{p}{(}\PYG{n}{a}\PYG{p}{,} \PYG{p}{(}\PYG{l+m+mi}{2}\PYG{p}{,} \PYG{l+m+mi}{3}\PYG{p}{)}\PYG{p}{,} \PYG{l+s+s1}{\PYGZsq{}}\PYG{l+s+s1}{symmetric}\PYG{l+s+s1}{\PYGZsq{}}\PYG{p}{)}
\PYG{g+go}{array([2, 1, 1, 2, 3, 4, 5, 5, 4, 3])}
\end{sphinxVerbatim}

\begin{sphinxVerbatim}[commandchars=\\\{\}]
\PYG{g+gp}{\PYGZgt{}\PYGZgt{}\PYGZgt{} }\PYG{n}{np}\PYG{o}{.}\PYG{n}{pad}\PYG{p}{(}\PYG{n}{a}\PYG{p}{,} \PYG{p}{(}\PYG{l+m+mi}{2}\PYG{p}{,} \PYG{l+m+mi}{3}\PYG{p}{)}\PYG{p}{,} \PYG{l+s+s1}{\PYGZsq{}}\PYG{l+s+s1}{symmetric}\PYG{l+s+s1}{\PYGZsq{}}\PYG{p}{,} \PYG{n}{reflect\PYGZus{}type}\PYG{o}{=}\PYG{l+s+s1}{\PYGZsq{}}\PYG{l+s+s1}{odd}\PYG{l+s+s1}{\PYGZsq{}}\PYG{p}{)}
\PYG{g+go}{array([0, 1, 1, 2, 3, 4, 5, 5, 6, 7])}
\end{sphinxVerbatim}

\begin{sphinxVerbatim}[commandchars=\\\{\}]
\PYG{g+gp}{\PYGZgt{}\PYGZgt{}\PYGZgt{} }\PYG{n}{np}\PYG{o}{.}\PYG{n}{pad}\PYG{p}{(}\PYG{n}{a}\PYG{p}{,} \PYG{p}{(}\PYG{l+m+mi}{2}\PYG{p}{,} \PYG{l+m+mi}{3}\PYG{p}{)}\PYG{p}{,} \PYG{l+s+s1}{\PYGZsq{}}\PYG{l+s+s1}{wrap}\PYG{l+s+s1}{\PYGZsq{}}\PYG{p}{)}
\PYG{g+go}{array([4, 5, 1, 2, 3, 4, 5, 1, 2, 3])}
\end{sphinxVerbatim}

\begin{sphinxVerbatim}[commandchars=\\\{\}]
\PYG{g+gp}{\PYGZgt{}\PYGZgt{}\PYGZgt{} }\PYG{k}{def} \PYG{n+nf}{pad\PYGZus{}with}\PYG{p}{(}\PYG{n}{vector}\PYG{p}{,} \PYG{n}{pad\PYGZus{}width}\PYG{p}{,} \PYG{n}{iaxis}\PYG{p}{,} \PYG{n}{kwargs}\PYG{p}{)}\PYG{p}{:}
\PYG{g+gp}{... }    \PYG{n}{pad\PYGZus{}value} \PYG{o}{=} \PYG{n}{kwargs}\PYG{o}{.}\PYG{n}{get}\PYG{p}{(}\PYG{l+s+s1}{\PYGZsq{}}\PYG{l+s+s1}{padder}\PYG{l+s+s1}{\PYGZsq{}}\PYG{p}{,} \PYG{l+m+mi}{10}\PYG{p}{)}
\PYG{g+gp}{... }    \PYG{n}{vector}\PYG{p}{[}\PYG{p}{:}\PYG{n}{pad\PYGZus{}width}\PYG{p}{[}\PYG{l+m+mi}{0}\PYG{p}{]}\PYG{p}{]} \PYG{o}{=} \PYG{n}{pad\PYGZus{}value}
\PYG{g+gp}{... }    \PYG{n}{vector}\PYG{p}{[}\PYG{o}{\PYGZhy{}}\PYG{n}{pad\PYGZus{}width}\PYG{p}{[}\PYG{l+m+mi}{1}\PYG{p}{]}\PYG{p}{:}\PYG{p}{]} \PYG{o}{=} \PYG{n}{pad\PYGZus{}value}
\PYG{g+gp}{\PYGZgt{}\PYGZgt{}\PYGZgt{} }\PYG{n}{a} \PYG{o}{=} \PYG{n}{np}\PYG{o}{.}\PYG{n}{arange}\PYG{p}{(}\PYG{l+m+mi}{6}\PYG{p}{)}
\PYG{g+gp}{\PYGZgt{}\PYGZgt{}\PYGZgt{} }\PYG{n}{a} \PYG{o}{=} \PYG{n}{a}\PYG{o}{.}\PYG{n}{reshape}\PYG{p}{(}\PYG{p}{(}\PYG{l+m+mi}{2}\PYG{p}{,} \PYG{l+m+mi}{3}\PYG{p}{)}\PYG{p}{)}
\PYG{g+gp}{\PYGZgt{}\PYGZgt{}\PYGZgt{} }\PYG{n}{np}\PYG{o}{.}\PYG{n}{pad}\PYG{p}{(}\PYG{n}{a}\PYG{p}{,} \PYG{l+m+mi}{2}\PYG{p}{,} \PYG{n}{pad\PYGZus{}with}\PYG{p}{)}
\PYG{g+go}{array([[10, 10, 10, 10, 10, 10, 10],}
\PYG{g+go}{       [10, 10, 10, 10, 10, 10, 10],}
\PYG{g+go}{       [10, 10,  0,  1,  2, 10, 10],}
\PYG{g+go}{       [10, 10,  3,  4,  5, 10, 10],}
\PYG{g+go}{       [10, 10, 10, 10, 10, 10, 10],}
\PYG{g+go}{       [10, 10, 10, 10, 10, 10, 10]])}
\PYG{g+gp}{\PYGZgt{}\PYGZgt{}\PYGZgt{} }\PYG{n}{np}\PYG{o}{.}\PYG{n}{pad}\PYG{p}{(}\PYG{n}{a}\PYG{p}{,} \PYG{l+m+mi}{2}\PYG{p}{,} \PYG{n}{pad\PYGZus{}with}\PYG{p}{,} \PYG{n}{padder}\PYG{o}{=}\PYG{l+m+mi}{100}\PYG{p}{)}
\PYG{g+go}{array([[100, 100, 100, 100, 100, 100, 100],}
\PYG{g+go}{       [100, 100, 100, 100, 100, 100, 100],}
\PYG{g+go}{       [100, 100,   0,   1,   2, 100, 100],}
\PYG{g+go}{       [100, 100,   3,   4,   5, 100, 100],}
\PYG{g+go}{       [100, 100, 100, 100, 100, 100, 100],}
\PYG{g+go}{       [100, 100, 100, 100, 100, 100, 100]])}
\end{sphinxVerbatim}

\end{fulllineitems}

\index{percentile() (in module symjax.tensor)@\spxentry{percentile()}\spxextra{in module symjax.tensor}}

\begin{fulllineitems}
\phantomsection\label{\detokenize{modules/tensor:symjax.tensor.percentile}}\pysiglinewithargsret{\sphinxbfcode{\sphinxupquote{percentile}}}{\emph{\DUrole{n}{a}}, \emph{\DUrole{n}{q}}, \emph{\DUrole{n}{axis}\DUrole{o}{=}\DUrole{default_value}{None}}, \emph{\DUrole{n}{out}\DUrole{o}{=}\DUrole{default_value}{None}}, \emph{\DUrole{n}{overwrite\_input}\DUrole{o}{=}\DUrole{default_value}{False}}, \emph{\DUrole{n}{interpolation}\DUrole{o}{=}\DUrole{default_value}{\textquotesingle{}linear\textquotesingle{}}}, \emph{\DUrole{n}{keepdims}\DUrole{o}{=}\DUrole{default_value}{False}}}{}
Compute the q\sphinxhyphen{}th percentile of the data along the specified axis.

LAX\sphinxhyphen{}backend implementation of {\hyperref[\detokenize{modules/tensor:symjax.tensor.percentile}]{\sphinxcrossref{\sphinxcode{\sphinxupquote{percentile()}}}}}.
ADDITIONOriginal docstring below.

LAX\sphinxhyphen{}backend implementation of {\hyperref[\detokenize{modules/tensor:symjax.tensor.percentile}]{\sphinxcrossref{\sphinxcode{\sphinxupquote{percentile()}}}}}.
Original docstring below.

Returns the q\sphinxhyphen{}th percentile(s) of the array elements.
\begin{quote}\begin{description}
\item[{Returns}] \leavevmode
\sphinxstylestrong{percentile} \textendash{} If \sphinxtitleref{q} is a single percentile and \sphinxtitleref{axis=None}, then the result
is a scalar. If multiple percentiles are given, first axis of
the result corresponds to the percentiles. The other axes are
the axes that remain after the reduction of \sphinxtitleref{a}. If the input
contains integers or floats smaller than \sphinxcode{\sphinxupquote{float64}}, the output
data\sphinxhyphen{}type is \sphinxcode{\sphinxupquote{float64}}. Otherwise, the output data\sphinxhyphen{}type is the
same as that of the input. If \sphinxtitleref{out} is specified, that array is
returned instead.

\item[{Return type}] \leavevmode
scalar or ndarray

\end{description}\end{quote}

\sphinxstrong{See also:}

{\hyperref[\detokenize{modules/tensor:symjax.tensor.mean}]{\sphinxcrossref{\sphinxcode{\sphinxupquote{mean()}}}}}
\begin{description}
\item[{{\hyperref[\detokenize{modules/tensor:symjax.tensor.median}]{\sphinxcrossref{\sphinxcode{\sphinxupquote{median()}}}}}}] \leavevmode
equivalent to \sphinxcode{\sphinxupquote{percentile(..., 50)}}

\end{description}

\sphinxcode{\sphinxupquote{nanpercentile()}}
\begin{description}
\item[{{\hyperref[\detokenize{modules/tensor:symjax.tensor.quantile}]{\sphinxcrossref{\sphinxcode{\sphinxupquote{quantile()}}}}}}] \leavevmode
equivalent to percentile, except with q in the range {[}0, 1{]}.

\end{description}

\subsubsection*{Notes}

Given a vector \sphinxcode{\sphinxupquote{V}} of length \sphinxcode{\sphinxupquote{N}}, the q\sphinxhyphen{}th percentile of
\sphinxcode{\sphinxupquote{V}} is the value \sphinxcode{\sphinxupquote{q/100}} of the way from the minimum to the
maximum in a sorted copy of \sphinxcode{\sphinxupquote{V}}. The values and distances of
the two nearest neighbors as well as the \sphinxtitleref{interpolation} parameter
will determine the percentile if the normalized ranking does not
match the location of \sphinxcode{\sphinxupquote{q}} exactly. This function is the same as
the median if \sphinxcode{\sphinxupquote{q=50}}, the same as the minimum if \sphinxcode{\sphinxupquote{q=0}} and the
same as the maximum if \sphinxcode{\sphinxupquote{q=100}}.
\subsubsection*{Examples}

\begin{sphinxVerbatim}[commandchars=\\\{\}]
\PYG{g+gp}{\PYGZgt{}\PYGZgt{}\PYGZgt{} }\PYG{n}{a} \PYG{o}{=} \PYG{n}{np}\PYG{o}{.}\PYG{n}{array}\PYG{p}{(}\PYG{p}{[}\PYG{p}{[}\PYG{l+m+mi}{10}\PYG{p}{,} \PYG{l+m+mi}{7}\PYG{p}{,} \PYG{l+m+mi}{4}\PYG{p}{]}\PYG{p}{,} \PYG{p}{[}\PYG{l+m+mi}{3}\PYG{p}{,} \PYG{l+m+mi}{2}\PYG{p}{,} \PYG{l+m+mi}{1}\PYG{p}{]}\PYG{p}{]}\PYG{p}{)}
\PYG{g+gp}{\PYGZgt{}\PYGZgt{}\PYGZgt{} }\PYG{n}{a}
\PYG{g+go}{array([[10,  7,  4],}
\PYG{g+go}{       [ 3,  2,  1]])}
\PYG{g+gp}{\PYGZgt{}\PYGZgt{}\PYGZgt{} }\PYG{n}{np}\PYG{o}{.}\PYG{n}{percentile}\PYG{p}{(}\PYG{n}{a}\PYG{p}{,} \PYG{l+m+mi}{50}\PYG{p}{)}
\PYG{g+go}{3.5}
\PYG{g+gp}{\PYGZgt{}\PYGZgt{}\PYGZgt{} }\PYG{n}{np}\PYG{o}{.}\PYG{n}{percentile}\PYG{p}{(}\PYG{n}{a}\PYG{p}{,} \PYG{l+m+mi}{50}\PYG{p}{,} \PYG{n}{axis}\PYG{o}{=}\PYG{l+m+mi}{0}\PYG{p}{)}
\PYG{g+go}{array([6.5, 4.5, 2.5])}
\PYG{g+gp}{\PYGZgt{}\PYGZgt{}\PYGZgt{} }\PYG{n}{np}\PYG{o}{.}\PYG{n}{percentile}\PYG{p}{(}\PYG{n}{a}\PYG{p}{,} \PYG{l+m+mi}{50}\PYG{p}{,} \PYG{n}{axis}\PYG{o}{=}\PYG{l+m+mi}{1}\PYG{p}{)}
\PYG{g+go}{array([7.,  2.])}
\PYG{g+gp}{\PYGZgt{}\PYGZgt{}\PYGZgt{} }\PYG{n}{np}\PYG{o}{.}\PYG{n}{percentile}\PYG{p}{(}\PYG{n}{a}\PYG{p}{,} \PYG{l+m+mi}{50}\PYG{p}{,} \PYG{n}{axis}\PYG{o}{=}\PYG{l+m+mi}{1}\PYG{p}{,} \PYG{n}{keepdims}\PYG{o}{=}\PYG{k+kc}{True}\PYG{p}{)}
\PYG{g+go}{array([[7.],}
\PYG{g+go}{       [2.]])}
\end{sphinxVerbatim}

\begin{sphinxVerbatim}[commandchars=\\\{\}]
\PYG{g+gp}{\PYGZgt{}\PYGZgt{}\PYGZgt{} }\PYG{n}{m} \PYG{o}{=} \PYG{n}{np}\PYG{o}{.}\PYG{n}{percentile}\PYG{p}{(}\PYG{n}{a}\PYG{p}{,} \PYG{l+m+mi}{50}\PYG{p}{,} \PYG{n}{axis}\PYG{o}{=}\PYG{l+m+mi}{0}\PYG{p}{)}
\PYG{g+gp}{\PYGZgt{}\PYGZgt{}\PYGZgt{} }\PYG{n}{out} \PYG{o}{=} \PYG{n}{np}\PYG{o}{.}\PYG{n}{zeros\PYGZus{}like}\PYG{p}{(}\PYG{n}{m}\PYG{p}{)}
\PYG{g+gp}{\PYGZgt{}\PYGZgt{}\PYGZgt{} }\PYG{n}{np}\PYG{o}{.}\PYG{n}{percentile}\PYG{p}{(}\PYG{n}{a}\PYG{p}{,} \PYG{l+m+mi}{50}\PYG{p}{,} \PYG{n}{axis}\PYG{o}{=}\PYG{l+m+mi}{0}\PYG{p}{,} \PYG{n}{out}\PYG{o}{=}\PYG{n}{out}\PYG{p}{)}
\PYG{g+go}{array([6.5, 4.5, 2.5])}
\PYG{g+gp}{\PYGZgt{}\PYGZgt{}\PYGZgt{} }\PYG{n}{m}
\PYG{g+go}{array([6.5, 4.5, 2.5])}
\end{sphinxVerbatim}

\begin{sphinxVerbatim}[commandchars=\\\{\}]
\PYG{g+gp}{\PYGZgt{}\PYGZgt{}\PYGZgt{} }\PYG{n}{b} \PYG{o}{=} \PYG{n}{a}\PYG{o}{.}\PYG{n}{copy}\PYG{p}{(}\PYG{p}{)}
\PYG{g+gp}{\PYGZgt{}\PYGZgt{}\PYGZgt{} }\PYG{n}{np}\PYG{o}{.}\PYG{n}{percentile}\PYG{p}{(}\PYG{n}{b}\PYG{p}{,} \PYG{l+m+mi}{50}\PYG{p}{,} \PYG{n}{axis}\PYG{o}{=}\PYG{l+m+mi}{1}\PYG{p}{,} \PYG{n}{overwrite\PYGZus{}input}\PYG{o}{=}\PYG{k+kc}{True}\PYG{p}{)}
\PYG{g+go}{array([7.,  2.])}
\PYG{g+gp}{\PYGZgt{}\PYGZgt{}\PYGZgt{} }\PYG{k}{assert} \PYG{o+ow}{not} \PYG{n}{np}\PYG{o}{.}\PYG{n}{all}\PYG{p}{(}\PYG{n}{a} \PYG{o}{==} \PYG{n}{b}\PYG{p}{)}
\end{sphinxVerbatim}

The different types of interpolation can be visualized graphically:

\begin{figure}[htbp]
\centering

\noindent\sphinxincludegraphics{{tensor-1}.pdf}
\end{figure}

\end{fulllineitems}

\index{polyval() (in module symjax.tensor)@\spxentry{polyval()}\spxextra{in module symjax.tensor}}

\begin{fulllineitems}
\phantomsection\label{\detokenize{modules/tensor:symjax.tensor.polyval}}\pysiglinewithargsret{\sphinxbfcode{\sphinxupquote{polyval}}}{\emph{\DUrole{n}{p}}, \emph{\DUrole{n}{x}}}{}
Evaluate a polynomial at specific values.

LAX\sphinxhyphen{}backend implementation of {\hyperref[\detokenize{modules/tensor:symjax.tensor.polyval}]{\sphinxcrossref{\sphinxcode{\sphinxupquote{polyval()}}}}}.
ADDITIONOriginal docstring below.

LAX\sphinxhyphen{}backend implementation of {\hyperref[\detokenize{modules/tensor:symjax.tensor.polyval}]{\sphinxcrossref{\sphinxcode{\sphinxupquote{polyval()}}}}}.
Original docstring below.

If \sphinxtitleref{p} is of length N, this function returns the value:
\begin{quote}

\sphinxcode{\sphinxupquote{p{[}0{]}*x**(N\sphinxhyphen{}1) + p{[}1{]}*x**(N\sphinxhyphen{}2) + ... + p{[}N\sphinxhyphen{}2{]}*x + p{[}N\sphinxhyphen{}1{]}}}
\end{quote}

If \sphinxtitleref{x} is a sequence, then \sphinxtitleref{p(x)} is returned for each element of \sphinxtitleref{x}.
If \sphinxtitleref{x} is another polynomial then the composite polynomial \sphinxtitleref{p(x(t))}
is returned.
\begin{quote}\begin{description}
\item[{Returns}] \leavevmode
\sphinxstylestrong{values} \textendash{} If \sphinxtitleref{x} is a poly1d instance, the result is the composition of the two
polynomials, i.e., \sphinxtitleref{x} is “substituted” in \sphinxtitleref{p} and the simplified
result is returned. In addition, the type of \sphinxtitleref{x} \sphinxhyphen{} array\_like or
poly1d \sphinxhyphen{} governs the type of the output: \sphinxtitleref{x} array\_like =\textgreater{} \sphinxtitleref{values}
array\_like, \sphinxtitleref{x} a poly1d object =\textgreater{} \sphinxtitleref{values} is also.

\item[{Return type}] \leavevmode
ndarray or poly1d

\end{description}\end{quote}

\sphinxstrong{See also:}

\begin{description}
\item[{\sphinxcode{\sphinxupquote{poly1d()}}}] \leavevmode
A polynomial class.

\end{description}

\subsubsection*{Notes}

Horner’s scheme {\color{red}\bfseries{}{[}1{]}\_} is used to evaluate the polynomial. Even so,
for polynomials of high degree the values may be inaccurate due to
rounding errors. Use carefully.

If \sphinxtitleref{x} is a subtype of \sphinxtitleref{ndarray} the return value will be of the same type.
\subsubsection*{References}
\subsubsection*{Examples}

\begin{sphinxVerbatim}[commandchars=\\\{\}]
\PYG{g+gp}{\PYGZgt{}\PYGZgt{}\PYGZgt{} }\PYG{n}{np}\PYG{o}{.}\PYG{n}{polyval}\PYG{p}{(}\PYG{p}{[}\PYG{l+m+mi}{3}\PYG{p}{,}\PYG{l+m+mi}{0}\PYG{p}{,}\PYG{l+m+mi}{1}\PYG{p}{]}\PYG{p}{,} \PYG{l+m+mi}{5}\PYG{p}{)}  \PYG{c+c1}{\PYGZsh{} 3 * 5**2 + 0 * 5**1 + 1}
\PYG{g+go}{76}
\PYG{g+gp}{\PYGZgt{}\PYGZgt{}\PYGZgt{} }\PYG{n}{np}\PYG{o}{.}\PYG{n}{polyval}\PYG{p}{(}\PYG{p}{[}\PYG{l+m+mi}{3}\PYG{p}{,}\PYG{l+m+mi}{0}\PYG{p}{,}\PYG{l+m+mi}{1}\PYG{p}{]}\PYG{p}{,} \PYG{n}{np}\PYG{o}{.}\PYG{n}{poly1d}\PYG{p}{(}\PYG{l+m+mi}{5}\PYG{p}{)}\PYG{p}{)}
\PYG{g+go}{poly1d([76.])}
\PYG{g+gp}{\PYGZgt{}\PYGZgt{}\PYGZgt{} }\PYG{n}{np}\PYG{o}{.}\PYG{n}{polyval}\PYG{p}{(}\PYG{n}{np}\PYG{o}{.}\PYG{n}{poly1d}\PYG{p}{(}\PYG{p}{[}\PYG{l+m+mi}{3}\PYG{p}{,}\PYG{l+m+mi}{0}\PYG{p}{,}\PYG{l+m+mi}{1}\PYG{p}{]}\PYG{p}{)}\PYG{p}{,} \PYG{l+m+mi}{5}\PYG{p}{)}
\PYG{g+go}{76}
\PYG{g+gp}{\PYGZgt{}\PYGZgt{}\PYGZgt{} }\PYG{n}{np}\PYG{o}{.}\PYG{n}{polyval}\PYG{p}{(}\PYG{n}{np}\PYG{o}{.}\PYG{n}{poly1d}\PYG{p}{(}\PYG{p}{[}\PYG{l+m+mi}{3}\PYG{p}{,}\PYG{l+m+mi}{0}\PYG{p}{,}\PYG{l+m+mi}{1}\PYG{p}{]}\PYG{p}{)}\PYG{p}{,} \PYG{n}{np}\PYG{o}{.}\PYG{n}{poly1d}\PYG{p}{(}\PYG{l+m+mi}{5}\PYG{p}{)}\PYG{p}{)}
\PYG{g+go}{poly1d([76.])}
\end{sphinxVerbatim}

\end{fulllineitems}

\index{power() (in module symjax.tensor)@\spxentry{power()}\spxextra{in module symjax.tensor}}

\begin{fulllineitems}
\phantomsection\label{\detokenize{modules/tensor:symjax.tensor.power}}\pysiglinewithargsret{\sphinxbfcode{\sphinxupquote{power}}}{\emph{\DUrole{n}{x1}}, \emph{\DUrole{n}{x2}}}{}
First array elements raised to powers from second array, element\sphinxhyphen{}wise.

LAX\sphinxhyphen{}backend implementation of {\hyperref[\detokenize{modules/tensor:symjax.tensor.power}]{\sphinxcrossref{\sphinxcode{\sphinxupquote{power()}}}}}.
ADDITIONOriginal docstring below.

LAX\sphinxhyphen{}backend implementation of {\hyperref[\detokenize{modules/tensor:symjax.tensor.power}]{\sphinxcrossref{\sphinxcode{\sphinxupquote{power()}}}}}.
Original docstring below.

power(x1, x2, /, out=None, {\color{red}\bfseries{}*}, where=True, casting=’same\_kind’, order=’K’, dtype=None, subok=True{[}, signature, extobj{]})

Raise each base in \sphinxtitleref{x1} to the positionally\sphinxhyphen{}corresponding power in
\sphinxtitleref{x2}.  \sphinxtitleref{x1} and \sphinxtitleref{x2} must be broadcastable to the same shape. Note that an
integer type raised to a negative integer power will raise a ValueError.
\begin{quote}\begin{description}
\item[{Returns}] \leavevmode
\sphinxstylestrong{y} \textendash{} The bases in \sphinxtitleref{x1} raised to the exponents in \sphinxtitleref{x2}.
This is a scalar if both \sphinxtitleref{x1} and \sphinxtitleref{x2} are scalars.

\item[{Return type}] \leavevmode
ndarray

\end{description}\end{quote}

\sphinxstrong{See also:}

\begin{description}
\item[{{\hyperref[\detokenize{modules/tensor:symjax.tensor.float_power}]{\sphinxcrossref{\sphinxcode{\sphinxupquote{float\_power()}}}}}}] \leavevmode
power function that promotes integers to float

\end{description}

\subsubsection*{Examples}

Cube each element in a list.

\begin{sphinxVerbatim}[commandchars=\\\{\}]
\PYG{g+gp}{\PYGZgt{}\PYGZgt{}\PYGZgt{} }\PYG{n}{x1} \PYG{o}{=} \PYG{n+nb}{range}\PYG{p}{(}\PYG{l+m+mi}{6}\PYG{p}{)}
\PYG{g+gp}{\PYGZgt{}\PYGZgt{}\PYGZgt{} }\PYG{n}{x1}
\PYG{g+go}{[0, 1, 2, 3, 4, 5]}
\PYG{g+gp}{\PYGZgt{}\PYGZgt{}\PYGZgt{} }\PYG{n}{np}\PYG{o}{.}\PYG{n}{power}\PYG{p}{(}\PYG{n}{x1}\PYG{p}{,} \PYG{l+m+mi}{3}\PYG{p}{)}
\PYG{g+go}{array([  0,   1,   8,  27,  64, 125])}
\end{sphinxVerbatim}

Raise the bases to different exponents.

\begin{sphinxVerbatim}[commandchars=\\\{\}]
\PYG{g+gp}{\PYGZgt{}\PYGZgt{}\PYGZgt{} }\PYG{n}{x2} \PYG{o}{=} \PYG{p}{[}\PYG{l+m+mf}{1.0}\PYG{p}{,} \PYG{l+m+mf}{2.0}\PYG{p}{,} \PYG{l+m+mf}{3.0}\PYG{p}{,} \PYG{l+m+mf}{3.0}\PYG{p}{,} \PYG{l+m+mf}{2.0}\PYG{p}{,} \PYG{l+m+mf}{1.0}\PYG{p}{]}
\PYG{g+gp}{\PYGZgt{}\PYGZgt{}\PYGZgt{} }\PYG{n}{np}\PYG{o}{.}\PYG{n}{power}\PYG{p}{(}\PYG{n}{x1}\PYG{p}{,} \PYG{n}{x2}\PYG{p}{)}
\PYG{g+go}{array([  0.,   1.,   8.,  27.,  16.,   5.])}
\end{sphinxVerbatim}

The effect of broadcasting.

\begin{sphinxVerbatim}[commandchars=\\\{\}]
\PYG{g+gp}{\PYGZgt{}\PYGZgt{}\PYGZgt{} }\PYG{n}{x2} \PYG{o}{=} \PYG{n}{np}\PYG{o}{.}\PYG{n}{array}\PYG{p}{(}\PYG{p}{[}\PYG{p}{[}\PYG{l+m+mi}{1}\PYG{p}{,} \PYG{l+m+mi}{2}\PYG{p}{,} \PYG{l+m+mi}{3}\PYG{p}{,} \PYG{l+m+mi}{3}\PYG{p}{,} \PYG{l+m+mi}{2}\PYG{p}{,} \PYG{l+m+mi}{1}\PYG{p}{]}\PYG{p}{,} \PYG{p}{[}\PYG{l+m+mi}{1}\PYG{p}{,} \PYG{l+m+mi}{2}\PYG{p}{,} \PYG{l+m+mi}{3}\PYG{p}{,} \PYG{l+m+mi}{3}\PYG{p}{,} \PYG{l+m+mi}{2}\PYG{p}{,} \PYG{l+m+mi}{1}\PYG{p}{]}\PYG{p}{]}\PYG{p}{)}
\PYG{g+gp}{\PYGZgt{}\PYGZgt{}\PYGZgt{} }\PYG{n}{x2}
\PYG{g+go}{array([[1, 2, 3, 3, 2, 1],}
\PYG{g+go}{       [1, 2, 3, 3, 2, 1]])}
\PYG{g+gp}{\PYGZgt{}\PYGZgt{}\PYGZgt{} }\PYG{n}{np}\PYG{o}{.}\PYG{n}{power}\PYG{p}{(}\PYG{n}{x1}\PYG{p}{,} \PYG{n}{x2}\PYG{p}{)}
\PYG{g+go}{array([[ 0,  1,  8, 27, 16,  5],}
\PYG{g+go}{       [ 0,  1,  8, 27, 16,  5]])}
\end{sphinxVerbatim}

\end{fulllineitems}

\index{positive() (in module symjax.tensor)@\spxentry{positive()}\spxextra{in module symjax.tensor}}

\begin{fulllineitems}
\phantomsection\label{\detokenize{modules/tensor:symjax.tensor.positive}}\pysiglinewithargsret{\sphinxbfcode{\sphinxupquote{positive}}}{\emph{\DUrole{n}{x}}}{}
Numerical positive, element\sphinxhyphen{}wise.

LAX\sphinxhyphen{}backend implementation of {\hyperref[\detokenize{modules/tensor:symjax.tensor.positive}]{\sphinxcrossref{\sphinxcode{\sphinxupquote{positive()}}}}}.
ADDITIONOriginal docstring below.

LAX\sphinxhyphen{}backend implementation of {\hyperref[\detokenize{modules/tensor:symjax.tensor.positive}]{\sphinxcrossref{\sphinxcode{\sphinxupquote{positive()}}}}}.
Original docstring below.

positive(x, /, out=None, {\color{red}\bfseries{}*}, where=True, casting=’same\_kind’, order=’K’, dtype=None, subok=True{[}, signature, extobj{]})

\DUrole{versionmodified,added}{New in version 1.13.0.}
\begin{quote}\begin{description}
\item[{Returns}] \leavevmode
\sphinxstylestrong{y} \textendash{} Returned array or scalar: \sphinxtitleref{y = +x}.
This is a scalar if \sphinxtitleref{x} is a scalar.

\item[{Return type}] \leavevmode
ndarray or scalar

\end{description}\end{quote}
\subsubsection*{Notes}

Equivalent to \sphinxtitleref{x.copy()}, but only defined for types that support
arithmetic.

\end{fulllineitems}

\index{prod() (in module symjax.tensor)@\spxentry{prod()}\spxextra{in module symjax.tensor}}

\begin{fulllineitems}
\phantomsection\label{\detokenize{modules/tensor:symjax.tensor.prod}}\pysiglinewithargsret{\sphinxbfcode{\sphinxupquote{prod}}}{\emph{\DUrole{n}{a}}, \emph{\DUrole{n}{axis}\DUrole{o}{=}\DUrole{default_value}{None}}, \emph{\DUrole{n}{dtype}\DUrole{o}{=}\DUrole{default_value}{None}}, \emph{\DUrole{n}{out}\DUrole{o}{=}\DUrole{default_value}{None}}, \emph{\DUrole{n}{keepdims}\DUrole{o}{=}\DUrole{default_value}{False}}}{}
Return the product of array elements over a given axis.

LAX\sphinxhyphen{}backend implementation of {\hyperref[\detokenize{modules/tensor:symjax.tensor.prod}]{\sphinxcrossref{\sphinxcode{\sphinxupquote{prod()}}}}}.
ADDITIONOriginal docstring below.

LAX\sphinxhyphen{}backend implementation of {\hyperref[\detokenize{modules/tensor:symjax.tensor.prod}]{\sphinxcrossref{\sphinxcode{\sphinxupquote{prod()}}}}}.
Original docstring below.
\begin{quote}\begin{description}
\item[{Parameters}] \leavevmode
\sphinxstyleliteralstrong{\sphinxupquote{dtype}} (\sphinxstyleliteralemphasis{\sphinxupquote{dtype}}\sphinxstyleliteralemphasis{\sphinxupquote{, }}\sphinxstyleliteralemphasis{\sphinxupquote{optional}}) \textendash{} The type of the returned array, as well as of the accumulator in
which the elements are multiplied.  The dtype of \sphinxtitleref{a} is used by
default unless \sphinxtitleref{a} has an integer dtype of less precision than the
default platform integer.  In that case, if \sphinxtitleref{a} is signed then the
platform integer is used while if \sphinxtitleref{a} is unsigned then an unsigned
integer of the same precision as the platform integer is used.

\item[{Returns}] \leavevmode
\sphinxstylestrong{product\_along\_axis} \textendash{} An array shaped as \sphinxtitleref{a} but with the specified axis removed.
Returns a reference to \sphinxtitleref{out} if specified.

\item[{Return type}] \leavevmode
ndarray, see \sphinxtitleref{dtype} parameter above.

\end{description}\end{quote}

\sphinxstrong{See also:}

\begin{description}
\item[{\sphinxcode{\sphinxupquote{ndarray.prod()}}}] \leavevmode
equivalent method

\end{description}

\sphinxcode{\sphinxupquote{ufuncs\sphinxhyphen{}output\sphinxhyphen{}type()}}

\subsubsection*{Notes}

Arithmetic is modular when using integer types, and no error is
raised on overflow.  That means that, on a 32\sphinxhyphen{}bit platform:

\begin{sphinxVerbatim}[commandchars=\\\{\}]
\PYG{g+gp}{\PYGZgt{}\PYGZgt{}\PYGZgt{} }\PYG{n}{x} \PYG{o}{=} \PYG{n}{np}\PYG{o}{.}\PYG{n}{array}\PYG{p}{(}\PYG{p}{[}\PYG{l+m+mi}{536870910}\PYG{p}{,} \PYG{l+m+mi}{536870910}\PYG{p}{,} \PYG{l+m+mi}{536870910}\PYG{p}{,} \PYG{l+m+mi}{536870910}\PYG{p}{]}\PYG{p}{)}
\PYG{g+gp}{\PYGZgt{}\PYGZgt{}\PYGZgt{} }\PYG{n}{np}\PYG{o}{.}\PYG{n}{prod}\PYG{p}{(}\PYG{n}{x}\PYG{p}{)}
\PYG{g+go}{16 \PYGZsh{} may vary}
\end{sphinxVerbatim}

The product of an empty array is the neutral element 1:

\begin{sphinxVerbatim}[commandchars=\\\{\}]
\PYG{g+gp}{\PYGZgt{}\PYGZgt{}\PYGZgt{} }\PYG{n}{np}\PYG{o}{.}\PYG{n}{prod}\PYG{p}{(}\PYG{p}{[}\PYG{p}{]}\PYG{p}{)}
\PYG{g+go}{1.0}
\end{sphinxVerbatim}
\subsubsection*{Examples}

By default, calculate the product of all elements:

\begin{sphinxVerbatim}[commandchars=\\\{\}]
\PYG{g+gp}{\PYGZgt{}\PYGZgt{}\PYGZgt{} }\PYG{n}{np}\PYG{o}{.}\PYG{n}{prod}\PYG{p}{(}\PYG{p}{[}\PYG{l+m+mf}{1.}\PYG{p}{,}\PYG{l+m+mf}{2.}\PYG{p}{]}\PYG{p}{)}
\PYG{g+go}{2.0}
\end{sphinxVerbatim}

Even when the input array is two\sphinxhyphen{}dimensional:

\begin{sphinxVerbatim}[commandchars=\\\{\}]
\PYG{g+gp}{\PYGZgt{}\PYGZgt{}\PYGZgt{} }\PYG{n}{np}\PYG{o}{.}\PYG{n}{prod}\PYG{p}{(}\PYG{p}{[}\PYG{p}{[}\PYG{l+m+mf}{1.}\PYG{p}{,}\PYG{l+m+mf}{2.}\PYG{p}{]}\PYG{p}{,}\PYG{p}{[}\PYG{l+m+mf}{3.}\PYG{p}{,}\PYG{l+m+mf}{4.}\PYG{p}{]}\PYG{p}{]}\PYG{p}{)}
\PYG{g+go}{24.0}
\end{sphinxVerbatim}

But we can also specify the axis over which to multiply:

\begin{sphinxVerbatim}[commandchars=\\\{\}]
\PYG{g+gp}{\PYGZgt{}\PYGZgt{}\PYGZgt{} }\PYG{n}{np}\PYG{o}{.}\PYG{n}{prod}\PYG{p}{(}\PYG{p}{[}\PYG{p}{[}\PYG{l+m+mf}{1.}\PYG{p}{,}\PYG{l+m+mf}{2.}\PYG{p}{]}\PYG{p}{,}\PYG{p}{[}\PYG{l+m+mf}{3.}\PYG{p}{,}\PYG{l+m+mf}{4.}\PYG{p}{]}\PYG{p}{]}\PYG{p}{,} \PYG{n}{axis}\PYG{o}{=}\PYG{l+m+mi}{1}\PYG{p}{)}
\PYG{g+go}{array([  2.,  12.])}
\end{sphinxVerbatim}

Or select specific elements to include:

\begin{sphinxVerbatim}[commandchars=\\\{\}]
\PYG{g+gp}{\PYGZgt{}\PYGZgt{}\PYGZgt{} }\PYG{n}{np}\PYG{o}{.}\PYG{n}{prod}\PYG{p}{(}\PYG{p}{[}\PYG{l+m+mf}{1.}\PYG{p}{,} \PYG{n}{np}\PYG{o}{.}\PYG{n}{nan}\PYG{p}{,} \PYG{l+m+mf}{3.}\PYG{p}{]}\PYG{p}{,} \PYG{n}{where}\PYG{o}{=}\PYG{p}{[}\PYG{k+kc}{True}\PYG{p}{,} \PYG{k+kc}{False}\PYG{p}{,} \PYG{k+kc}{True}\PYG{p}{]}\PYG{p}{)}
\PYG{g+go}{3.0}
\end{sphinxVerbatim}

If the type of \sphinxtitleref{x} is unsigned, then the output type is
the unsigned platform integer:

\begin{sphinxVerbatim}[commandchars=\\\{\}]
\PYG{g+gp}{\PYGZgt{}\PYGZgt{}\PYGZgt{} }\PYG{n}{x} \PYG{o}{=} \PYG{n}{np}\PYG{o}{.}\PYG{n}{array}\PYG{p}{(}\PYG{p}{[}\PYG{l+m+mi}{1}\PYG{p}{,} \PYG{l+m+mi}{2}\PYG{p}{,} \PYG{l+m+mi}{3}\PYG{p}{]}\PYG{p}{,} \PYG{n}{dtype}\PYG{o}{=}\PYG{n}{np}\PYG{o}{.}\PYG{n}{uint8}\PYG{p}{)}
\PYG{g+gp}{\PYGZgt{}\PYGZgt{}\PYGZgt{} }\PYG{n}{np}\PYG{o}{.}\PYG{n}{prod}\PYG{p}{(}\PYG{n}{x}\PYG{p}{)}\PYG{o}{.}\PYG{n}{dtype} \PYG{o}{==} \PYG{n}{np}\PYG{o}{.}\PYG{n}{uint}
\PYG{g+go}{True}
\end{sphinxVerbatim}

If \sphinxtitleref{x} is of a signed integer type, then the output type
is the default platform integer:

\begin{sphinxVerbatim}[commandchars=\\\{\}]
\PYG{g+gp}{\PYGZgt{}\PYGZgt{}\PYGZgt{} }\PYG{n}{x} \PYG{o}{=} \PYG{n}{np}\PYG{o}{.}\PYG{n}{array}\PYG{p}{(}\PYG{p}{[}\PYG{l+m+mi}{1}\PYG{p}{,} \PYG{l+m+mi}{2}\PYG{p}{,} \PYG{l+m+mi}{3}\PYG{p}{]}\PYG{p}{,} \PYG{n}{dtype}\PYG{o}{=}\PYG{n}{np}\PYG{o}{.}\PYG{n}{int8}\PYG{p}{)}
\PYG{g+gp}{\PYGZgt{}\PYGZgt{}\PYGZgt{} }\PYG{n}{np}\PYG{o}{.}\PYG{n}{prod}\PYG{p}{(}\PYG{n}{x}\PYG{p}{)}\PYG{o}{.}\PYG{n}{dtype} \PYG{o}{==} \PYG{n+nb}{int}
\PYG{g+go}{True}
\end{sphinxVerbatim}

You can also start the product with a value other than one:

\begin{sphinxVerbatim}[commandchars=\\\{\}]
\PYG{g+gp}{\PYGZgt{}\PYGZgt{}\PYGZgt{} }\PYG{n}{np}\PYG{o}{.}\PYG{n}{prod}\PYG{p}{(}\PYG{p}{[}\PYG{l+m+mi}{1}\PYG{p}{,} \PYG{l+m+mi}{2}\PYG{p}{]}\PYG{p}{,} \PYG{n}{initial}\PYG{o}{=}\PYG{l+m+mi}{5}\PYG{p}{)}
\PYG{g+go}{10}
\end{sphinxVerbatim}

\end{fulllineitems}

\index{product() (in module symjax.tensor)@\spxentry{product()}\spxextra{in module symjax.tensor}}

\begin{fulllineitems}
\phantomsection\label{\detokenize{modules/tensor:symjax.tensor.product}}\pysiglinewithargsret{\sphinxbfcode{\sphinxupquote{product}}}{\emph{\DUrole{n}{a}}, \emph{\DUrole{n}{axis}\DUrole{o}{=}\DUrole{default_value}{None}}, \emph{\DUrole{n}{dtype}\DUrole{o}{=}\DUrole{default_value}{None}}, \emph{\DUrole{n}{out}\DUrole{o}{=}\DUrole{default_value}{None}}, \emph{\DUrole{n}{keepdims}\DUrole{o}{=}\DUrole{default_value}{False}}}{}
Return the product of array elements over a given axis.

LAX\sphinxhyphen{}backend implementation of {\hyperref[\detokenize{modules/tensor:symjax.tensor.prod}]{\sphinxcrossref{\sphinxcode{\sphinxupquote{prod()}}}}}.
ADDITIONOriginal docstring below.

LAX\sphinxhyphen{}backend implementation of {\hyperref[\detokenize{modules/tensor:symjax.tensor.prod}]{\sphinxcrossref{\sphinxcode{\sphinxupquote{prod()}}}}}.
Original docstring below.
\begin{quote}\begin{description}
\item[{Parameters}] \leavevmode
\sphinxstyleliteralstrong{\sphinxupquote{dtype}} (\sphinxstyleliteralemphasis{\sphinxupquote{dtype}}\sphinxstyleliteralemphasis{\sphinxupquote{, }}\sphinxstyleliteralemphasis{\sphinxupquote{optional}}) \textendash{} The type of the returned array, as well as of the accumulator in
which the elements are multiplied.  The dtype of \sphinxtitleref{a} is used by
default unless \sphinxtitleref{a} has an integer dtype of less precision than the
default platform integer.  In that case, if \sphinxtitleref{a} is signed then the
platform integer is used while if \sphinxtitleref{a} is unsigned then an unsigned
integer of the same precision as the platform integer is used.

\item[{Returns}] \leavevmode
\sphinxstylestrong{product\_along\_axis} \textendash{} An array shaped as \sphinxtitleref{a} but with the specified axis removed.
Returns a reference to \sphinxtitleref{out} if specified.

\item[{Return type}] \leavevmode
ndarray, see \sphinxtitleref{dtype} parameter above.

\end{description}\end{quote}

\sphinxstrong{See also:}

\begin{description}
\item[{\sphinxcode{\sphinxupquote{ndarray.prod()}}}] \leavevmode
equivalent method

\end{description}

\sphinxcode{\sphinxupquote{ufuncs\sphinxhyphen{}output\sphinxhyphen{}type()}}

\subsubsection*{Notes}

Arithmetic is modular when using integer types, and no error is
raised on overflow.  That means that, on a 32\sphinxhyphen{}bit platform:

\begin{sphinxVerbatim}[commandchars=\\\{\}]
\PYG{g+gp}{\PYGZgt{}\PYGZgt{}\PYGZgt{} }\PYG{n}{x} \PYG{o}{=} \PYG{n}{np}\PYG{o}{.}\PYG{n}{array}\PYG{p}{(}\PYG{p}{[}\PYG{l+m+mi}{536870910}\PYG{p}{,} \PYG{l+m+mi}{536870910}\PYG{p}{,} \PYG{l+m+mi}{536870910}\PYG{p}{,} \PYG{l+m+mi}{536870910}\PYG{p}{]}\PYG{p}{)}
\PYG{g+gp}{\PYGZgt{}\PYGZgt{}\PYGZgt{} }\PYG{n}{np}\PYG{o}{.}\PYG{n}{prod}\PYG{p}{(}\PYG{n}{x}\PYG{p}{)}
\PYG{g+go}{16 \PYGZsh{} may vary}
\end{sphinxVerbatim}

The product of an empty array is the neutral element 1:

\begin{sphinxVerbatim}[commandchars=\\\{\}]
\PYG{g+gp}{\PYGZgt{}\PYGZgt{}\PYGZgt{} }\PYG{n}{np}\PYG{o}{.}\PYG{n}{prod}\PYG{p}{(}\PYG{p}{[}\PYG{p}{]}\PYG{p}{)}
\PYG{g+go}{1.0}
\end{sphinxVerbatim}
\subsubsection*{Examples}

By default, calculate the product of all elements:

\begin{sphinxVerbatim}[commandchars=\\\{\}]
\PYG{g+gp}{\PYGZgt{}\PYGZgt{}\PYGZgt{} }\PYG{n}{np}\PYG{o}{.}\PYG{n}{prod}\PYG{p}{(}\PYG{p}{[}\PYG{l+m+mf}{1.}\PYG{p}{,}\PYG{l+m+mf}{2.}\PYG{p}{]}\PYG{p}{)}
\PYG{g+go}{2.0}
\end{sphinxVerbatim}

Even when the input array is two\sphinxhyphen{}dimensional:

\begin{sphinxVerbatim}[commandchars=\\\{\}]
\PYG{g+gp}{\PYGZgt{}\PYGZgt{}\PYGZgt{} }\PYG{n}{np}\PYG{o}{.}\PYG{n}{prod}\PYG{p}{(}\PYG{p}{[}\PYG{p}{[}\PYG{l+m+mf}{1.}\PYG{p}{,}\PYG{l+m+mf}{2.}\PYG{p}{]}\PYG{p}{,}\PYG{p}{[}\PYG{l+m+mf}{3.}\PYG{p}{,}\PYG{l+m+mf}{4.}\PYG{p}{]}\PYG{p}{]}\PYG{p}{)}
\PYG{g+go}{24.0}
\end{sphinxVerbatim}

But we can also specify the axis over which to multiply:

\begin{sphinxVerbatim}[commandchars=\\\{\}]
\PYG{g+gp}{\PYGZgt{}\PYGZgt{}\PYGZgt{} }\PYG{n}{np}\PYG{o}{.}\PYG{n}{prod}\PYG{p}{(}\PYG{p}{[}\PYG{p}{[}\PYG{l+m+mf}{1.}\PYG{p}{,}\PYG{l+m+mf}{2.}\PYG{p}{]}\PYG{p}{,}\PYG{p}{[}\PYG{l+m+mf}{3.}\PYG{p}{,}\PYG{l+m+mf}{4.}\PYG{p}{]}\PYG{p}{]}\PYG{p}{,} \PYG{n}{axis}\PYG{o}{=}\PYG{l+m+mi}{1}\PYG{p}{)}
\PYG{g+go}{array([  2.,  12.])}
\end{sphinxVerbatim}

Or select specific elements to include:

\begin{sphinxVerbatim}[commandchars=\\\{\}]
\PYG{g+gp}{\PYGZgt{}\PYGZgt{}\PYGZgt{} }\PYG{n}{np}\PYG{o}{.}\PYG{n}{prod}\PYG{p}{(}\PYG{p}{[}\PYG{l+m+mf}{1.}\PYG{p}{,} \PYG{n}{np}\PYG{o}{.}\PYG{n}{nan}\PYG{p}{,} \PYG{l+m+mf}{3.}\PYG{p}{]}\PYG{p}{,} \PYG{n}{where}\PYG{o}{=}\PYG{p}{[}\PYG{k+kc}{True}\PYG{p}{,} \PYG{k+kc}{False}\PYG{p}{,} \PYG{k+kc}{True}\PYG{p}{]}\PYG{p}{)}
\PYG{g+go}{3.0}
\end{sphinxVerbatim}

If the type of \sphinxtitleref{x} is unsigned, then the output type is
the unsigned platform integer:

\begin{sphinxVerbatim}[commandchars=\\\{\}]
\PYG{g+gp}{\PYGZgt{}\PYGZgt{}\PYGZgt{} }\PYG{n}{x} \PYG{o}{=} \PYG{n}{np}\PYG{o}{.}\PYG{n}{array}\PYG{p}{(}\PYG{p}{[}\PYG{l+m+mi}{1}\PYG{p}{,} \PYG{l+m+mi}{2}\PYG{p}{,} \PYG{l+m+mi}{3}\PYG{p}{]}\PYG{p}{,} \PYG{n}{dtype}\PYG{o}{=}\PYG{n}{np}\PYG{o}{.}\PYG{n}{uint8}\PYG{p}{)}
\PYG{g+gp}{\PYGZgt{}\PYGZgt{}\PYGZgt{} }\PYG{n}{np}\PYG{o}{.}\PYG{n}{prod}\PYG{p}{(}\PYG{n}{x}\PYG{p}{)}\PYG{o}{.}\PYG{n}{dtype} \PYG{o}{==} \PYG{n}{np}\PYG{o}{.}\PYG{n}{uint}
\PYG{g+go}{True}
\end{sphinxVerbatim}

If \sphinxtitleref{x} is of a signed integer type, then the output type
is the default platform integer:

\begin{sphinxVerbatim}[commandchars=\\\{\}]
\PYG{g+gp}{\PYGZgt{}\PYGZgt{}\PYGZgt{} }\PYG{n}{x} \PYG{o}{=} \PYG{n}{np}\PYG{o}{.}\PYG{n}{array}\PYG{p}{(}\PYG{p}{[}\PYG{l+m+mi}{1}\PYG{p}{,} \PYG{l+m+mi}{2}\PYG{p}{,} \PYG{l+m+mi}{3}\PYG{p}{]}\PYG{p}{,} \PYG{n}{dtype}\PYG{o}{=}\PYG{n}{np}\PYG{o}{.}\PYG{n}{int8}\PYG{p}{)}
\PYG{g+gp}{\PYGZgt{}\PYGZgt{}\PYGZgt{} }\PYG{n}{np}\PYG{o}{.}\PYG{n}{prod}\PYG{p}{(}\PYG{n}{x}\PYG{p}{)}\PYG{o}{.}\PYG{n}{dtype} \PYG{o}{==} \PYG{n+nb}{int}
\PYG{g+go}{True}
\end{sphinxVerbatim}

You can also start the product with a value other than one:

\begin{sphinxVerbatim}[commandchars=\\\{\}]
\PYG{g+gp}{\PYGZgt{}\PYGZgt{}\PYGZgt{} }\PYG{n}{np}\PYG{o}{.}\PYG{n}{prod}\PYG{p}{(}\PYG{p}{[}\PYG{l+m+mi}{1}\PYG{p}{,} \PYG{l+m+mi}{2}\PYG{p}{]}\PYG{p}{,} \PYG{n}{initial}\PYG{o}{=}\PYG{l+m+mi}{5}\PYG{p}{)}
\PYG{g+go}{10}
\end{sphinxVerbatim}

\end{fulllineitems}

\index{promote\_types() (in module symjax.tensor)@\spxentry{promote\_types()}\spxextra{in module symjax.tensor}}

\begin{fulllineitems}
\phantomsection\label{\detokenize{modules/tensor:symjax.tensor.promote_types}}\pysiglinewithargsret{\sphinxbfcode{\sphinxupquote{promote\_types}}}{\emph{\DUrole{n}{a}}, \emph{\DUrole{n}{b}}}{}
Returns the type to which a binary operation should cast its arguments.

LAX\sphinxhyphen{}backend implementation of {\hyperref[\detokenize{modules/tensor:symjax.tensor.promote_types}]{\sphinxcrossref{\sphinxcode{\sphinxupquote{promote\_types()}}}}}.
ADDITIONOriginal docstring below.
\begin{quote}

Returns:
A \sphinxcode{\sphinxupquote{numpy.dtype}} object.
\end{quote}

\end{fulllineitems}

\index{quantile() (in module symjax.tensor)@\spxentry{quantile()}\spxextra{in module symjax.tensor}}

\begin{fulllineitems}
\phantomsection\label{\detokenize{modules/tensor:symjax.tensor.quantile}}\pysiglinewithargsret{\sphinxbfcode{\sphinxupquote{quantile}}}{\emph{\DUrole{n}{a}}, \emph{\DUrole{n}{q}}, \emph{\DUrole{n}{axis}\DUrole{o}{=}\DUrole{default_value}{None}}, \emph{\DUrole{n}{out}\DUrole{o}{=}\DUrole{default_value}{None}}, \emph{\DUrole{n}{overwrite\_input}\DUrole{o}{=}\DUrole{default_value}{False}}, \emph{\DUrole{n}{interpolation}\DUrole{o}{=}\DUrole{default_value}{\textquotesingle{}linear\textquotesingle{}}}, \emph{\DUrole{n}{keepdims}\DUrole{o}{=}\DUrole{default_value}{False}}}{}
Compute the q\sphinxhyphen{}th quantile of the data along the specified axis.
\begin{quote}

\DUrole{versionmodified,added}{New in version 1.15.0.}
\end{quote}

LAX\sphinxhyphen{}backend implementation of {\hyperref[\detokenize{modules/tensor:symjax.tensor.quantile}]{\sphinxcrossref{\sphinxcode{\sphinxupquote{quantile()}}}}}.
ADDITIONOriginal docstring below.

LAX\sphinxhyphen{}backend implementation of {\hyperref[\detokenize{modules/tensor:symjax.tensor.quantile}]{\sphinxcrossref{\sphinxcode{\sphinxupquote{quantile()}}}}}.
Original docstring below.
\begin{quote}\begin{description}
\item[{Returns}] \leavevmode
\sphinxstylestrong{quantile} \textendash{} If \sphinxtitleref{q} is a single quantile and \sphinxtitleref{axis=None}, then the result
is a scalar. If multiple quantiles are given, first axis of
the result corresponds to the quantiles. The other axes are
the axes that remain after the reduction of \sphinxtitleref{a}. If the input
contains integers or floats smaller than \sphinxcode{\sphinxupquote{float64}}, the output
data\sphinxhyphen{}type is \sphinxcode{\sphinxupquote{float64}}. Otherwise, the output data\sphinxhyphen{}type is the
same as that of the input. If \sphinxtitleref{out} is specified, that array is
returned instead.

\item[{Return type}] \leavevmode
scalar or ndarray

\end{description}\end{quote}

\sphinxstrong{See also:}

{\hyperref[\detokenize{modules/tensor:symjax.tensor.mean}]{\sphinxcrossref{\sphinxcode{\sphinxupquote{mean()}}}}}
\begin{description}
\item[{{\hyperref[\detokenize{modules/tensor:symjax.tensor.percentile}]{\sphinxcrossref{\sphinxcode{\sphinxupquote{percentile()}}}}}}] \leavevmode
equivalent to quantile, but with q in the range {[}0, 100{]}.

\item[{{\hyperref[\detokenize{modules/tensor:symjax.tensor.median}]{\sphinxcrossref{\sphinxcode{\sphinxupquote{median()}}}}}}] \leavevmode
equivalent to \sphinxcode{\sphinxupquote{quantile(..., 0.5)}}

\end{description}

\sphinxcode{\sphinxupquote{nanquantile()}}

\subsubsection*{Notes}

Given a vector \sphinxcode{\sphinxupquote{V}} of length \sphinxcode{\sphinxupquote{N}}, the q\sphinxhyphen{}th quantile of
\sphinxcode{\sphinxupquote{V}} is the value \sphinxcode{\sphinxupquote{q}} of the way from the minimum to the
maximum in a sorted copy of \sphinxcode{\sphinxupquote{V}}. The values and distances of
the two nearest neighbors as well as the \sphinxtitleref{interpolation} parameter
will determine the quantile if the normalized ranking does not
match the location of \sphinxcode{\sphinxupquote{q}} exactly. This function is the same as
the median if \sphinxcode{\sphinxupquote{q=0.5}}, the same as the minimum if \sphinxcode{\sphinxupquote{q=0.0}} and the
same as the maximum if \sphinxcode{\sphinxupquote{q=1.0}}.
\subsubsection*{Examples}

\begin{sphinxVerbatim}[commandchars=\\\{\}]
\PYG{g+gp}{\PYGZgt{}\PYGZgt{}\PYGZgt{} }\PYG{n}{a} \PYG{o}{=} \PYG{n}{np}\PYG{o}{.}\PYG{n}{array}\PYG{p}{(}\PYG{p}{[}\PYG{p}{[}\PYG{l+m+mi}{10}\PYG{p}{,} \PYG{l+m+mi}{7}\PYG{p}{,} \PYG{l+m+mi}{4}\PYG{p}{]}\PYG{p}{,} \PYG{p}{[}\PYG{l+m+mi}{3}\PYG{p}{,} \PYG{l+m+mi}{2}\PYG{p}{,} \PYG{l+m+mi}{1}\PYG{p}{]}\PYG{p}{]}\PYG{p}{)}
\PYG{g+gp}{\PYGZgt{}\PYGZgt{}\PYGZgt{} }\PYG{n}{a}
\PYG{g+go}{array([[10,  7,  4],}
\PYG{g+go}{       [ 3,  2,  1]])}
\PYG{g+gp}{\PYGZgt{}\PYGZgt{}\PYGZgt{} }\PYG{n}{np}\PYG{o}{.}\PYG{n}{quantile}\PYG{p}{(}\PYG{n}{a}\PYG{p}{,} \PYG{l+m+mf}{0.5}\PYG{p}{)}
\PYG{g+go}{3.5}
\PYG{g+gp}{\PYGZgt{}\PYGZgt{}\PYGZgt{} }\PYG{n}{np}\PYG{o}{.}\PYG{n}{quantile}\PYG{p}{(}\PYG{n}{a}\PYG{p}{,} \PYG{l+m+mf}{0.5}\PYG{p}{,} \PYG{n}{axis}\PYG{o}{=}\PYG{l+m+mi}{0}\PYG{p}{)}
\PYG{g+go}{array([6.5, 4.5, 2.5])}
\PYG{g+gp}{\PYGZgt{}\PYGZgt{}\PYGZgt{} }\PYG{n}{np}\PYG{o}{.}\PYG{n}{quantile}\PYG{p}{(}\PYG{n}{a}\PYG{p}{,} \PYG{l+m+mf}{0.5}\PYG{p}{,} \PYG{n}{axis}\PYG{o}{=}\PYG{l+m+mi}{1}\PYG{p}{)}
\PYG{g+go}{array([7.,  2.])}
\PYG{g+gp}{\PYGZgt{}\PYGZgt{}\PYGZgt{} }\PYG{n}{np}\PYG{o}{.}\PYG{n}{quantile}\PYG{p}{(}\PYG{n}{a}\PYG{p}{,} \PYG{l+m+mf}{0.5}\PYG{p}{,} \PYG{n}{axis}\PYG{o}{=}\PYG{l+m+mi}{1}\PYG{p}{,} \PYG{n}{keepdims}\PYG{o}{=}\PYG{k+kc}{True}\PYG{p}{)}
\PYG{g+go}{array([[7.],}
\PYG{g+go}{       [2.]])}
\PYG{g+gp}{\PYGZgt{}\PYGZgt{}\PYGZgt{} }\PYG{n}{m} \PYG{o}{=} \PYG{n}{np}\PYG{o}{.}\PYG{n}{quantile}\PYG{p}{(}\PYG{n}{a}\PYG{p}{,} \PYG{l+m+mf}{0.5}\PYG{p}{,} \PYG{n}{axis}\PYG{o}{=}\PYG{l+m+mi}{0}\PYG{p}{)}
\PYG{g+gp}{\PYGZgt{}\PYGZgt{}\PYGZgt{} }\PYG{n}{out} \PYG{o}{=} \PYG{n}{np}\PYG{o}{.}\PYG{n}{zeros\PYGZus{}like}\PYG{p}{(}\PYG{n}{m}\PYG{p}{)}
\PYG{g+gp}{\PYGZgt{}\PYGZgt{}\PYGZgt{} }\PYG{n}{np}\PYG{o}{.}\PYG{n}{quantile}\PYG{p}{(}\PYG{n}{a}\PYG{p}{,} \PYG{l+m+mf}{0.5}\PYG{p}{,} \PYG{n}{axis}\PYG{o}{=}\PYG{l+m+mi}{0}\PYG{p}{,} \PYG{n}{out}\PYG{o}{=}\PYG{n}{out}\PYG{p}{)}
\PYG{g+go}{array([6.5, 4.5, 2.5])}
\PYG{g+gp}{\PYGZgt{}\PYGZgt{}\PYGZgt{} }\PYG{n}{m}
\PYG{g+go}{array([6.5, 4.5, 2.5])}
\PYG{g+gp}{\PYGZgt{}\PYGZgt{}\PYGZgt{} }\PYG{n}{b} \PYG{o}{=} \PYG{n}{a}\PYG{o}{.}\PYG{n}{copy}\PYG{p}{(}\PYG{p}{)}
\PYG{g+gp}{\PYGZgt{}\PYGZgt{}\PYGZgt{} }\PYG{n}{np}\PYG{o}{.}\PYG{n}{quantile}\PYG{p}{(}\PYG{n}{b}\PYG{p}{,} \PYG{l+m+mf}{0.5}\PYG{p}{,} \PYG{n}{axis}\PYG{o}{=}\PYG{l+m+mi}{1}\PYG{p}{,} \PYG{n}{overwrite\PYGZus{}input}\PYG{o}{=}\PYG{k+kc}{True}\PYG{p}{)}
\PYG{g+go}{array([7.,  2.])}
\PYG{g+gp}{\PYGZgt{}\PYGZgt{}\PYGZgt{} }\PYG{k}{assert} \PYG{o+ow}{not} \PYG{n}{np}\PYG{o}{.}\PYG{n}{all}\PYG{p}{(}\PYG{n}{a} \PYG{o}{==} \PYG{n}{b}\PYG{p}{)}
\end{sphinxVerbatim}

\end{fulllineitems}

\index{rad2deg() (in module symjax.tensor)@\spxentry{rad2deg()}\spxextra{in module symjax.tensor}}

\begin{fulllineitems}
\phantomsection\label{\detokenize{modules/tensor:symjax.tensor.rad2deg}}\pysiglinewithargsret{\sphinxbfcode{\sphinxupquote{rad2deg}}}{\emph{\DUrole{n}{x}}}{}
Convert angles from radians to degrees.

LAX\sphinxhyphen{}backend implementation of {\hyperref[\detokenize{modules/tensor:symjax.tensor.rad2deg}]{\sphinxcrossref{\sphinxcode{\sphinxupquote{rad2deg()}}}}}.
ADDITIONOriginal docstring below.

LAX\sphinxhyphen{}backend implementation of {\hyperref[\detokenize{modules/tensor:symjax.tensor.rad2deg}]{\sphinxcrossref{\sphinxcode{\sphinxupquote{rad2deg()}}}}}.
Original docstring below.

rad2deg(x, /, out=None, {\color{red}\bfseries{}*}, where=True, casting=’same\_kind’, order=’K’, dtype=None, subok=True{[}, signature, extobj{]})
\begin{quote}\begin{description}
\item[{Returns}] \leavevmode
\sphinxstylestrong{y} \textendash{} The corresponding angle in degrees.
This is a scalar if \sphinxtitleref{x} is a scalar.

\item[{Return type}] \leavevmode
ndarray

\end{description}\end{quote}

\sphinxstrong{See also:}

\begin{description}
\item[{{\hyperref[\detokenize{modules/tensor:symjax.tensor.deg2rad}]{\sphinxcrossref{\sphinxcode{\sphinxupquote{deg2rad()}}}}}}] \leavevmode
Convert angles from degrees to radians.

\item[{\sphinxcode{\sphinxupquote{unwrap()}}}] \leavevmode
Remove large jumps in angle by wrapping.

\end{description}

\subsubsection*{Notes}

\DUrole{versionmodified,added}{New in version 1.3.0.}

rad2deg(x) is \sphinxcode{\sphinxupquote{180 * x / pi}}.
\subsubsection*{Examples}

\begin{sphinxVerbatim}[commandchars=\\\{\}]
\PYG{g+gp}{\PYGZgt{}\PYGZgt{}\PYGZgt{} }\PYG{n}{np}\PYG{o}{.}\PYG{n}{rad2deg}\PYG{p}{(}\PYG{n}{np}\PYG{o}{.}\PYG{n}{pi}\PYG{o}{/}\PYG{l+m+mi}{2}\PYG{p}{)}
\PYG{g+go}{90.0}
\end{sphinxVerbatim}

\end{fulllineitems}

\index{radians() (in module symjax.tensor)@\spxentry{radians()}\spxextra{in module symjax.tensor}}

\begin{fulllineitems}
\phantomsection\label{\detokenize{modules/tensor:symjax.tensor.radians}}\pysiglinewithargsret{\sphinxbfcode{\sphinxupquote{radians}}}{\emph{\DUrole{n}{x}}}{}
Convert angles from degrees to radians.

LAX\sphinxhyphen{}backend implementation of {\hyperref[\detokenize{modules/tensor:symjax.tensor.deg2rad}]{\sphinxcrossref{\sphinxcode{\sphinxupquote{deg2rad()}}}}}.
ADDITIONOriginal docstring below.

LAX\sphinxhyphen{}backend implementation of {\hyperref[\detokenize{modules/tensor:symjax.tensor.deg2rad}]{\sphinxcrossref{\sphinxcode{\sphinxupquote{deg2rad()}}}}}.
Original docstring below.

deg2rad(x, /, out=None, {\color{red}\bfseries{}*}, where=True, casting=’same\_kind’, order=’K’, dtype=None, subok=True{[}, signature, extobj{]})
\begin{quote}\begin{description}
\item[{Returns}] \leavevmode
\sphinxstylestrong{y} \textendash{} The corresponding angle in radians.
This is a scalar if \sphinxtitleref{x} is a scalar.

\item[{Return type}] \leavevmode
ndarray

\end{description}\end{quote}

\sphinxstrong{See also:}

\begin{description}
\item[{{\hyperref[\detokenize{modules/tensor:symjax.tensor.rad2deg}]{\sphinxcrossref{\sphinxcode{\sphinxupquote{rad2deg()}}}}}}] \leavevmode
Convert angles from radians to degrees.

\item[{\sphinxcode{\sphinxupquote{unwrap()}}}] \leavevmode
Remove large jumps in angle by wrapping.

\end{description}

\subsubsection*{Notes}

\DUrole{versionmodified,added}{New in version 1.3.0.}

\sphinxcode{\sphinxupquote{deg2rad(x)}} is \sphinxcode{\sphinxupquote{x * pi / 180}}.
\subsubsection*{Examples}

\begin{sphinxVerbatim}[commandchars=\\\{\}]
\PYG{g+gp}{\PYGZgt{}\PYGZgt{}\PYGZgt{} }\PYG{n}{np}\PYG{o}{.}\PYG{n}{deg2rad}\PYG{p}{(}\PYG{l+m+mi}{180}\PYG{p}{)}
\PYG{g+go}{3.1415926535897931}
\end{sphinxVerbatim}

\end{fulllineitems}

\index{ravel() (in module symjax.tensor)@\spxentry{ravel()}\spxextra{in module symjax.tensor}}

\begin{fulllineitems}
\phantomsection\label{\detokenize{modules/tensor:symjax.tensor.ravel}}\pysiglinewithargsret{\sphinxbfcode{\sphinxupquote{ravel}}}{\emph{\DUrole{n}{a}}, \emph{\DUrole{n}{order}\DUrole{o}{=}\DUrole{default_value}{\textquotesingle{}C\textquotesingle{}}}}{}
Return a contiguous flattened array.

LAX\sphinxhyphen{}backend implementation of {\hyperref[\detokenize{modules/tensor:symjax.tensor.ravel}]{\sphinxcrossref{\sphinxcode{\sphinxupquote{ravel()}}}}}.
ADDITIONOriginal docstring below.

LAX\sphinxhyphen{}backend implementation of {\hyperref[\detokenize{modules/tensor:symjax.tensor.ravel}]{\sphinxcrossref{\sphinxcode{\sphinxupquote{ravel()}}}}}.
Original docstring below.

A 1\sphinxhyphen{}D array, containing the elements of the input, is returned.  A copy is
made only if needed.

As of NumPy 1.10, the returned array will have the same type as the input
array. (for example, a masked array will be returned for a masked array
input)
\begin{quote}\begin{description}
\item[{Returns}] \leavevmode
\sphinxstylestrong{y} \textendash{} y is an array of the same subtype as \sphinxtitleref{a}, with shape \sphinxcode{\sphinxupquote{(a.size,)}}.
Note that matrices are special cased for backward compatibility, if \sphinxtitleref{a}
is a matrix, then y is a 1\sphinxhyphen{}D ndarray.

\item[{Return type}] \leavevmode
array\_like

\end{description}\end{quote}

\sphinxstrong{See also:}

\begin{description}
\item[{\sphinxcode{\sphinxupquote{ndarray.flat()}}}] \leavevmode
1\sphinxhyphen{}D iterator over an array.

\item[{\sphinxcode{\sphinxupquote{ndarray.flatten()}}}] \leavevmode
1\sphinxhyphen{}D array copy of the elements of an array in row\sphinxhyphen{}major order.

\item[{\sphinxcode{\sphinxupquote{ndarray.reshape()}}}] \leavevmode
Change the shape of an array without changing its data.

\end{description}

\subsubsection*{Notes}

In row\sphinxhyphen{}major, C\sphinxhyphen{}style order, in two dimensions, the row index
varies the slowest, and the column index the quickest.  This can
be generalized to multiple dimensions, where row\sphinxhyphen{}major order
implies that the index along the first axis varies slowest, and
the index along the last quickest.  The opposite holds for
column\sphinxhyphen{}major, Fortran\sphinxhyphen{}style index ordering.

When a view is desired in as many cases as possible, \sphinxcode{\sphinxupquote{arr.reshape(\sphinxhyphen{}1)}}
may be preferable.
\subsubsection*{Examples}

It is equivalent to \sphinxcode{\sphinxupquote{reshape(\sphinxhyphen{}1, order=order)}}.

\begin{sphinxVerbatim}[commandchars=\\\{\}]
\PYG{g+gp}{\PYGZgt{}\PYGZgt{}\PYGZgt{} }\PYG{n}{x} \PYG{o}{=} \PYG{n}{np}\PYG{o}{.}\PYG{n}{array}\PYG{p}{(}\PYG{p}{[}\PYG{p}{[}\PYG{l+m+mi}{1}\PYG{p}{,} \PYG{l+m+mi}{2}\PYG{p}{,} \PYG{l+m+mi}{3}\PYG{p}{]}\PYG{p}{,} \PYG{p}{[}\PYG{l+m+mi}{4}\PYG{p}{,} \PYG{l+m+mi}{5}\PYG{p}{,} \PYG{l+m+mi}{6}\PYG{p}{]}\PYG{p}{]}\PYG{p}{)}
\PYG{g+gp}{\PYGZgt{}\PYGZgt{}\PYGZgt{} }\PYG{n}{np}\PYG{o}{.}\PYG{n}{ravel}\PYG{p}{(}\PYG{n}{x}\PYG{p}{)}
\PYG{g+go}{array([1, 2, 3, 4, 5, 6])}
\end{sphinxVerbatim}

\begin{sphinxVerbatim}[commandchars=\\\{\}]
\PYG{g+gp}{\PYGZgt{}\PYGZgt{}\PYGZgt{} }\PYG{n}{x}\PYG{o}{.}\PYG{n}{reshape}\PYG{p}{(}\PYG{o}{\PYGZhy{}}\PYG{l+m+mi}{1}\PYG{p}{)}
\PYG{g+go}{array([1, 2, 3, 4, 5, 6])}
\end{sphinxVerbatim}

\begin{sphinxVerbatim}[commandchars=\\\{\}]
\PYG{g+gp}{\PYGZgt{}\PYGZgt{}\PYGZgt{} }\PYG{n}{np}\PYG{o}{.}\PYG{n}{ravel}\PYG{p}{(}\PYG{n}{x}\PYG{p}{,} \PYG{n}{order}\PYG{o}{=}\PYG{l+s+s1}{\PYGZsq{}}\PYG{l+s+s1}{F}\PYG{l+s+s1}{\PYGZsq{}}\PYG{p}{)}
\PYG{g+go}{array([1, 4, 2, 5, 3, 6])}
\end{sphinxVerbatim}

When \sphinxcode{\sphinxupquote{order}} is ‘A’, it will preserve the array’s ‘C’ or ‘F’ ordering:

\begin{sphinxVerbatim}[commandchars=\\\{\}]
\PYG{g+gp}{\PYGZgt{}\PYGZgt{}\PYGZgt{} }\PYG{n}{np}\PYG{o}{.}\PYG{n}{ravel}\PYG{p}{(}\PYG{n}{x}\PYG{o}{.}\PYG{n}{T}\PYG{p}{)}
\PYG{g+go}{array([1, 4, 2, 5, 3, 6])}
\PYG{g+gp}{\PYGZgt{}\PYGZgt{}\PYGZgt{} }\PYG{n}{np}\PYG{o}{.}\PYG{n}{ravel}\PYG{p}{(}\PYG{n}{x}\PYG{o}{.}\PYG{n}{T}\PYG{p}{,} \PYG{n}{order}\PYG{o}{=}\PYG{l+s+s1}{\PYGZsq{}}\PYG{l+s+s1}{A}\PYG{l+s+s1}{\PYGZsq{}}\PYG{p}{)}
\PYG{g+go}{array([1, 2, 3, 4, 5, 6])}
\end{sphinxVerbatim}

When \sphinxcode{\sphinxupquote{order}} is ‘K’, it will preserve orderings that are neither ‘C’
nor ‘F’, but won’t reverse axes:

\begin{sphinxVerbatim}[commandchars=\\\{\}]
\PYG{g+gp}{\PYGZgt{}\PYGZgt{}\PYGZgt{} }\PYG{n}{a} \PYG{o}{=} \PYG{n}{np}\PYG{o}{.}\PYG{n}{arange}\PYG{p}{(}\PYG{l+m+mi}{3}\PYG{p}{)}\PYG{p}{[}\PYG{p}{:}\PYG{p}{:}\PYG{o}{\PYGZhy{}}\PYG{l+m+mi}{1}\PYG{p}{]}\PYG{p}{;} \PYG{n}{a}
\PYG{g+go}{array([2, 1, 0])}
\PYG{g+gp}{\PYGZgt{}\PYGZgt{}\PYGZgt{} }\PYG{n}{a}\PYG{o}{.}\PYG{n}{ravel}\PYG{p}{(}\PYG{n}{order}\PYG{o}{=}\PYG{l+s+s1}{\PYGZsq{}}\PYG{l+s+s1}{C}\PYG{l+s+s1}{\PYGZsq{}}\PYG{p}{)}
\PYG{g+go}{array([2, 1, 0])}
\PYG{g+gp}{\PYGZgt{}\PYGZgt{}\PYGZgt{} }\PYG{n}{a}\PYG{o}{.}\PYG{n}{ravel}\PYG{p}{(}\PYG{n}{order}\PYG{o}{=}\PYG{l+s+s1}{\PYGZsq{}}\PYG{l+s+s1}{K}\PYG{l+s+s1}{\PYGZsq{}}\PYG{p}{)}
\PYG{g+go}{array([2, 1, 0])}
\end{sphinxVerbatim}

\begin{sphinxVerbatim}[commandchars=\\\{\}]
\PYG{g+gp}{\PYGZgt{}\PYGZgt{}\PYGZgt{} }\PYG{n}{a} \PYG{o}{=} \PYG{n}{np}\PYG{o}{.}\PYG{n}{arange}\PYG{p}{(}\PYG{l+m+mi}{12}\PYG{p}{)}\PYG{o}{.}\PYG{n}{reshape}\PYG{p}{(}\PYG{l+m+mi}{2}\PYG{p}{,}\PYG{l+m+mi}{3}\PYG{p}{,}\PYG{l+m+mi}{2}\PYG{p}{)}\PYG{o}{.}\PYG{n}{swapaxes}\PYG{p}{(}\PYG{l+m+mi}{1}\PYG{p}{,}\PYG{l+m+mi}{2}\PYG{p}{)}\PYG{p}{;} \PYG{n}{a}
\PYG{g+go}{array([[[ 0,  2,  4],}
\PYG{g+go}{        [ 1,  3,  5]],}
\PYG{g+go}{       [[ 6,  8, 10],}
\PYG{g+go}{        [ 7,  9, 11]]])}
\PYG{g+gp}{\PYGZgt{}\PYGZgt{}\PYGZgt{} }\PYG{n}{a}\PYG{o}{.}\PYG{n}{ravel}\PYG{p}{(}\PYG{n}{order}\PYG{o}{=}\PYG{l+s+s1}{\PYGZsq{}}\PYG{l+s+s1}{C}\PYG{l+s+s1}{\PYGZsq{}}\PYG{p}{)}
\PYG{g+go}{array([ 0,  2,  4,  1,  3,  5,  6,  8, 10,  7,  9, 11])}
\PYG{g+gp}{\PYGZgt{}\PYGZgt{}\PYGZgt{} }\PYG{n}{a}\PYG{o}{.}\PYG{n}{ravel}\PYG{p}{(}\PYG{n}{order}\PYG{o}{=}\PYG{l+s+s1}{\PYGZsq{}}\PYG{l+s+s1}{K}\PYG{l+s+s1}{\PYGZsq{}}\PYG{p}{)}
\PYG{g+go}{array([ 0,  1,  2,  3,  4,  5,  6,  7,  8,  9, 10, 11])}
\end{sphinxVerbatim}

\end{fulllineitems}

\index{real() (in module symjax.tensor)@\spxentry{real()}\spxextra{in module symjax.tensor}}

\begin{fulllineitems}
\phantomsection\label{\detokenize{modules/tensor:symjax.tensor.real}}\pysiglinewithargsret{\sphinxbfcode{\sphinxupquote{real}}}{\emph{\DUrole{n}{val}}}{}
Return the real part of the complex argument.

LAX\sphinxhyphen{}backend implementation of {\hyperref[\detokenize{modules/tensor:symjax.tensor.real}]{\sphinxcrossref{\sphinxcode{\sphinxupquote{real()}}}}}.
ADDITIONOriginal docstring below.

LAX\sphinxhyphen{}backend implementation of {\hyperref[\detokenize{modules/tensor:symjax.tensor.real}]{\sphinxcrossref{\sphinxcode{\sphinxupquote{real()}}}}}.
Original docstring below.
\begin{quote}\begin{description}
\item[{Returns}] \leavevmode
\sphinxstylestrong{out} \textendash{} The real component of the complex argument. If \sphinxtitleref{val} is real, the type
of \sphinxtitleref{val} is used for the output.  If \sphinxtitleref{val} has complex elements, the
returned type is float.

\item[{Return type}] \leavevmode
ndarray or scalar

\end{description}\end{quote}

\sphinxstrong{See also:}

\sphinxcode{\sphinxupquote{real\_if\_close()}}, {\hyperref[\detokenize{modules/tensor:symjax.tensor.imag}]{\sphinxcrossref{\sphinxcode{\sphinxupquote{imag()}}}}}, {\hyperref[\detokenize{modules/tensor:symjax.tensor.angle}]{\sphinxcrossref{\sphinxcode{\sphinxupquote{angle()}}}}}

\subsubsection*{Examples}

\begin{sphinxVerbatim}[commandchars=\\\{\}]
\PYG{g+gp}{\PYGZgt{}\PYGZgt{}\PYGZgt{} }\PYG{n}{a} \PYG{o}{=} \PYG{n}{np}\PYG{o}{.}\PYG{n}{array}\PYG{p}{(}\PYG{p}{[}\PYG{l+m+mi}{1}\PYG{o}{+}\PYG{l+m+mi}{2}\PYG{n}{j}\PYG{p}{,} \PYG{l+m+mi}{3}\PYG{o}{+}\PYG{l+m+mi}{4}\PYG{n}{j}\PYG{p}{,} \PYG{l+m+mi}{5}\PYG{o}{+}\PYG{l+m+mi}{6}\PYG{n}{j}\PYG{p}{]}\PYG{p}{)}
\PYG{g+gp}{\PYGZgt{}\PYGZgt{}\PYGZgt{} }\PYG{n}{a}\PYG{o}{.}\PYG{n}{real}
\PYG{g+go}{array([1.,  3.,  5.])}
\PYG{g+gp}{\PYGZgt{}\PYGZgt{}\PYGZgt{} }\PYG{n}{a}\PYG{o}{.}\PYG{n}{real} \PYG{o}{=} \PYG{l+m+mi}{9}
\PYG{g+gp}{\PYGZgt{}\PYGZgt{}\PYGZgt{} }\PYG{n}{a}
\PYG{g+go}{array([9.+2.j,  9.+4.j,  9.+6.j])}
\PYG{g+gp}{\PYGZgt{}\PYGZgt{}\PYGZgt{} }\PYG{n}{a}\PYG{o}{.}\PYG{n}{real} \PYG{o}{=} \PYG{n}{np}\PYG{o}{.}\PYG{n}{array}\PYG{p}{(}\PYG{p}{[}\PYG{l+m+mi}{9}\PYG{p}{,} \PYG{l+m+mi}{8}\PYG{p}{,} \PYG{l+m+mi}{7}\PYG{p}{]}\PYG{p}{)}
\PYG{g+gp}{\PYGZgt{}\PYGZgt{}\PYGZgt{} }\PYG{n}{a}
\PYG{g+go}{array([9.+2.j,  8.+4.j,  7.+6.j])}
\PYG{g+gp}{\PYGZgt{}\PYGZgt{}\PYGZgt{} }\PYG{n}{np}\PYG{o}{.}\PYG{n}{real}\PYG{p}{(}\PYG{l+m+mi}{1} \PYG{o}{+} \PYG{l+m+mi}{1}\PYG{n}{j}\PYG{p}{)}
\PYG{g+go}{1.0}
\end{sphinxVerbatim}

\end{fulllineitems}

\index{remainder() (in module symjax.tensor)@\spxentry{remainder()}\spxextra{in module symjax.tensor}}

\begin{fulllineitems}
\phantomsection\label{\detokenize{modules/tensor:symjax.tensor.remainder}}\pysiglinewithargsret{\sphinxbfcode{\sphinxupquote{remainder}}}{\emph{\DUrole{n}{x1}}, \emph{\DUrole{n}{x2}}}{}
Return element\sphinxhyphen{}wise remainder of division.

LAX\sphinxhyphen{}backend implementation of {\hyperref[\detokenize{modules/tensor:symjax.tensor.remainder}]{\sphinxcrossref{\sphinxcode{\sphinxupquote{remainder()}}}}}.
ADDITIONOriginal docstring below.

LAX\sphinxhyphen{}backend implementation of {\hyperref[\detokenize{modules/tensor:symjax.tensor.remainder}]{\sphinxcrossref{\sphinxcode{\sphinxupquote{remainder()}}}}}.
Original docstring below.

remainder(x1, x2, /, out=None, {\color{red}\bfseries{}*}, where=True, casting=’same\_kind’, order=’K’, dtype=None, subok=True{[}, signature, extobj{]})

Computes the remainder complementary to the \sphinxtitleref{floor\_divide} function.  It is
equivalent to the Python modulus operator\textasciigrave{}\textasciigrave{}x1 \% x2\textasciigrave{}\textasciigrave{} and has the same sign
as the divisor \sphinxtitleref{x2}. The MATLAB function equivalent to \sphinxcode{\sphinxupquote{np.remainder}}
is \sphinxcode{\sphinxupquote{mod}}.

\begin{sphinxadmonition}{warning}{Warning:}
This should not be confused with:
\begin{itemize}
\item {} 
Python 3.7’s \sphinxtitleref{math.remainder} and C’s \sphinxcode{\sphinxupquote{remainder}}, which
computes the IEEE remainder, which are the complement to
\sphinxcode{\sphinxupquote{round(x1 / x2)}}.

\item {} 
The MATLAB \sphinxcode{\sphinxupquote{rem}} function and or the C \sphinxcode{\sphinxupquote{\%}} operator which is the
complement to \sphinxcode{\sphinxupquote{int(x1 / x2)}}.

\end{itemize}
\end{sphinxadmonition}
\begin{quote}\begin{description}
\item[{Returns}] \leavevmode
\sphinxstylestrong{y} \textendash{} The element\sphinxhyphen{}wise remainder of the quotient \sphinxcode{\sphinxupquote{floor\_divide(x1, x2)}}.
This is a scalar if both \sphinxtitleref{x1} and \sphinxtitleref{x2} are scalars.

\item[{Return type}] \leavevmode
ndarray

\end{description}\end{quote}

\sphinxstrong{See also:}

\begin{description}
\item[{{\hyperref[\detokenize{modules/tensor:symjax.tensor.floor_divide}]{\sphinxcrossref{\sphinxcode{\sphinxupquote{floor\_divide()}}}}}}] \leavevmode
Equivalent of Python \sphinxcode{\sphinxupquote{//}} operator.

\item[{{\hyperref[\detokenize{modules/tensor:symjax.tensor.divmod}]{\sphinxcrossref{\sphinxcode{\sphinxupquote{divmod()}}}}}}] \leavevmode
Simultaneous floor division and remainder.

\item[{{\hyperref[\detokenize{modules/tensor:symjax.tensor.fmod}]{\sphinxcrossref{\sphinxcode{\sphinxupquote{fmod()}}}}}}] \leavevmode
Equivalent of the MATLAB \sphinxcode{\sphinxupquote{rem}} function.

\end{description}

{\hyperref[\detokenize{modules/tensor:symjax.tensor.divide}]{\sphinxcrossref{\sphinxcode{\sphinxupquote{divide()}}}}}, {\hyperref[\detokenize{modules/tensor:symjax.tensor.floor}]{\sphinxcrossref{\sphinxcode{\sphinxupquote{floor()}}}}}

\subsubsection*{Notes}

Returns 0 when \sphinxtitleref{x2} is 0 and both \sphinxtitleref{x1} and \sphinxtitleref{x2} are (arrays of)
integers.
\sphinxcode{\sphinxupquote{mod}} is an alias of \sphinxcode{\sphinxupquote{remainder}}.
\subsubsection*{Examples}

\begin{sphinxVerbatim}[commandchars=\\\{\}]
\PYG{g+gp}{\PYGZgt{}\PYGZgt{}\PYGZgt{} }\PYG{n}{np}\PYG{o}{.}\PYG{n}{remainder}\PYG{p}{(}\PYG{p}{[}\PYG{l+m+mi}{4}\PYG{p}{,} \PYG{l+m+mi}{7}\PYG{p}{]}\PYG{p}{,} \PYG{p}{[}\PYG{l+m+mi}{2}\PYG{p}{,} \PYG{l+m+mi}{3}\PYG{p}{]}\PYG{p}{)}
\PYG{g+go}{array([0, 1])}
\PYG{g+gp}{\PYGZgt{}\PYGZgt{}\PYGZgt{} }\PYG{n}{np}\PYG{o}{.}\PYG{n}{remainder}\PYG{p}{(}\PYG{n}{np}\PYG{o}{.}\PYG{n}{arange}\PYG{p}{(}\PYG{l+m+mi}{7}\PYG{p}{)}\PYG{p}{,} \PYG{l+m+mi}{5}\PYG{p}{)}
\PYG{g+go}{array([0, 1, 2, 3, 4, 0, 1])}
\end{sphinxVerbatim}

\end{fulllineitems}

\index{repeat() (in module symjax.tensor)@\spxentry{repeat()}\spxextra{in module symjax.tensor}}

\begin{fulllineitems}
\phantomsection\label{\detokenize{modules/tensor:symjax.tensor.repeat}}\pysiglinewithargsret{\sphinxbfcode{\sphinxupquote{repeat}}}{\emph{\DUrole{n}{a}}, \emph{\DUrole{n}{repeats}}, \emph{\DUrole{n}{axis}\DUrole{o}{=}\DUrole{default_value}{None}}}{}
Repeat elements of an array.

LAX\sphinxhyphen{}backend implementation of {\hyperref[\detokenize{modules/tensor:symjax.tensor.repeat}]{\sphinxcrossref{\sphinxcode{\sphinxupquote{repeat()}}}}}.
ADDITIONOriginal docstring below.

LAX\sphinxhyphen{}backend implementation of {\hyperref[\detokenize{modules/tensor:symjax.tensor.repeat}]{\sphinxcrossref{\sphinxcode{\sphinxupquote{repeat()}}}}}.
Original docstring below.
\begin{quote}\begin{description}
\item[{Returns}] \leavevmode
\sphinxstylestrong{repeated\_array} \textendash{} Output array which has the same shape as \sphinxtitleref{a}, except along
the given axis.

\item[{Return type}] \leavevmode
ndarray

\end{description}\end{quote}

\sphinxstrong{See also:}

\begin{description}
\item[{{\hyperref[\detokenize{modules/tensor:symjax.tensor.tile}]{\sphinxcrossref{\sphinxcode{\sphinxupquote{tile()}}}}}}] \leavevmode
Tile an array.

\end{description}

\subsubsection*{Examples}

\begin{sphinxVerbatim}[commandchars=\\\{\}]
\PYG{g+gp}{\PYGZgt{}\PYGZgt{}\PYGZgt{} }\PYG{n}{np}\PYG{o}{.}\PYG{n}{repeat}\PYG{p}{(}\PYG{l+m+mi}{3}\PYG{p}{,} \PYG{l+m+mi}{4}\PYG{p}{)}
\PYG{g+go}{array([3, 3, 3, 3])}
\PYG{g+gp}{\PYGZgt{}\PYGZgt{}\PYGZgt{} }\PYG{n}{x} \PYG{o}{=} \PYG{n}{np}\PYG{o}{.}\PYG{n}{array}\PYG{p}{(}\PYG{p}{[}\PYG{p}{[}\PYG{l+m+mi}{1}\PYG{p}{,}\PYG{l+m+mi}{2}\PYG{p}{]}\PYG{p}{,}\PYG{p}{[}\PYG{l+m+mi}{3}\PYG{p}{,}\PYG{l+m+mi}{4}\PYG{p}{]}\PYG{p}{]}\PYG{p}{)}
\PYG{g+gp}{\PYGZgt{}\PYGZgt{}\PYGZgt{} }\PYG{n}{np}\PYG{o}{.}\PYG{n}{repeat}\PYG{p}{(}\PYG{n}{x}\PYG{p}{,} \PYG{l+m+mi}{2}\PYG{p}{)}
\PYG{g+go}{array([1, 1, 2, 2, 3, 3, 4, 4])}
\PYG{g+gp}{\PYGZgt{}\PYGZgt{}\PYGZgt{} }\PYG{n}{np}\PYG{o}{.}\PYG{n}{repeat}\PYG{p}{(}\PYG{n}{x}\PYG{p}{,} \PYG{l+m+mi}{3}\PYG{p}{,} \PYG{n}{axis}\PYG{o}{=}\PYG{l+m+mi}{1}\PYG{p}{)}
\PYG{g+go}{array([[1, 1, 1, 2, 2, 2],}
\PYG{g+go}{       [3, 3, 3, 4, 4, 4]])}
\PYG{g+gp}{\PYGZgt{}\PYGZgt{}\PYGZgt{} }\PYG{n}{np}\PYG{o}{.}\PYG{n}{repeat}\PYG{p}{(}\PYG{n}{x}\PYG{p}{,} \PYG{p}{[}\PYG{l+m+mi}{1}\PYG{p}{,} \PYG{l+m+mi}{2}\PYG{p}{]}\PYG{p}{,} \PYG{n}{axis}\PYG{o}{=}\PYG{l+m+mi}{0}\PYG{p}{)}
\PYG{g+go}{array([[1, 2],}
\PYG{g+go}{       [3, 4],}
\PYG{g+go}{       [3, 4]])}
\end{sphinxVerbatim}

\end{fulllineitems}

\index{reshape() (in module symjax.tensor)@\spxentry{reshape()}\spxextra{in module symjax.tensor}}

\begin{fulllineitems}
\phantomsection\label{\detokenize{modules/tensor:symjax.tensor.reshape}}\pysiglinewithargsret{\sphinxbfcode{\sphinxupquote{reshape}}}{\emph{\DUrole{n}{a}}, \emph{\DUrole{n}{newshape}}, \emph{\DUrole{n}{order}\DUrole{o}{=}\DUrole{default_value}{\textquotesingle{}C\textquotesingle{}}}}{}
Gives a new shape to an array without changing its data.

LAX\sphinxhyphen{}backend implementation of {\hyperref[\detokenize{modules/tensor:symjax.tensor.reshape}]{\sphinxcrossref{\sphinxcode{\sphinxupquote{reshape()}}}}}.
ADDITIONOriginal docstring below.

LAX\sphinxhyphen{}backend implementation of {\hyperref[\detokenize{modules/tensor:symjax.tensor.reshape}]{\sphinxcrossref{\sphinxcode{\sphinxupquote{reshape()}}}}}.
Original docstring below.
\begin{quote}\begin{description}
\item[{Returns}] \leavevmode
\sphinxstylestrong{reshaped\_array} \textendash{} This will be a new view object if possible; otherwise, it will
be a copy.  Note there is no guarantee of the \sphinxstyleemphasis{memory layout} (C\sphinxhyphen{} or
Fortran\sphinxhyphen{} contiguous) of the returned array.

\item[{Return type}] \leavevmode
ndarray

\end{description}\end{quote}

\sphinxstrong{See also:}

\begin{description}
\item[{\sphinxcode{\sphinxupquote{ndarray.reshape()}}}] \leavevmode
Equivalent method.

\end{description}

\subsubsection*{Notes}

It is not always possible to change the shape of an array without
copying the data. If you want an error to be raised when the data is copied,
you should assign the new shape to the shape attribute of the array:

\begin{sphinxVerbatim}[commandchars=\\\{\}]
\PYG{g+gp}{\PYGZgt{}\PYGZgt{}\PYGZgt{} }\PYG{n}{a} \PYG{o}{=} \PYG{n}{np}\PYG{o}{.}\PYG{n}{zeros}\PYG{p}{(}\PYG{p}{(}\PYG{l+m+mi}{10}\PYG{p}{,} \PYG{l+m+mi}{2}\PYG{p}{)}\PYG{p}{)}

\PYG{g+go}{\PYGZsh{} A transpose makes the array non\PYGZhy{}contiguous}
\PYG{g+gp}{\PYGZgt{}\PYGZgt{}\PYGZgt{} }\PYG{n}{b} \PYG{o}{=} \PYG{n}{a}\PYG{o}{.}\PYG{n}{T}

\PYG{g+go}{\PYGZsh{} Taking a view makes it possible to modify the shape without modifying}
\PYG{g+go}{\PYGZsh{} the initial object.}
\PYG{g+gp}{\PYGZgt{}\PYGZgt{}\PYGZgt{} }\PYG{n}{c} \PYG{o}{=} \PYG{n}{b}\PYG{o}{.}\PYG{n}{view}\PYG{p}{(}\PYG{p}{)}
\PYG{g+gp}{\PYGZgt{}\PYGZgt{}\PYGZgt{} }\PYG{n}{c}\PYG{o}{.}\PYG{n}{shape} \PYG{o}{=} \PYG{p}{(}\PYG{l+m+mi}{20}\PYG{p}{)}
\PYG{g+gt}{Traceback (most recent call last):}
   \PYG{c}{...}
\PYG{g+gr}{AttributeError}: \PYG{n}{incompatible shape for a non\PYGZhy{}contiguous array}
\end{sphinxVerbatim}

The \sphinxtitleref{order} keyword gives the index ordering both for \sphinxstyleemphasis{fetching} the values
from \sphinxtitleref{a}, and then \sphinxstyleemphasis{placing} the values into the output array.
For example, let’s say you have an array:

\begin{sphinxVerbatim}[commandchars=\\\{\}]
\PYG{g+gp}{\PYGZgt{}\PYGZgt{}\PYGZgt{} }\PYG{n}{a} \PYG{o}{=} \PYG{n}{np}\PYG{o}{.}\PYG{n}{arange}\PYG{p}{(}\PYG{l+m+mi}{6}\PYG{p}{)}\PYG{o}{.}\PYG{n}{reshape}\PYG{p}{(}\PYG{p}{(}\PYG{l+m+mi}{3}\PYG{p}{,} \PYG{l+m+mi}{2}\PYG{p}{)}\PYG{p}{)}
\PYG{g+gp}{\PYGZgt{}\PYGZgt{}\PYGZgt{} }\PYG{n}{a}
\PYG{g+go}{array([[0, 1],}
\PYG{g+go}{       [2, 3],}
\PYG{g+go}{       [4, 5]])}
\end{sphinxVerbatim}

You can think of reshaping as first raveling the array (using the given
index order), then inserting the elements from the raveled array into the
new array using the same kind of index ordering as was used for the
raveling.

\begin{sphinxVerbatim}[commandchars=\\\{\}]
\PYG{g+gp}{\PYGZgt{}\PYGZgt{}\PYGZgt{} }\PYG{n}{np}\PYG{o}{.}\PYG{n}{reshape}\PYG{p}{(}\PYG{n}{a}\PYG{p}{,} \PYG{p}{(}\PYG{l+m+mi}{2}\PYG{p}{,} \PYG{l+m+mi}{3}\PYG{p}{)}\PYG{p}{)} \PYG{c+c1}{\PYGZsh{} C\PYGZhy{}like index ordering}
\PYG{g+go}{array([[0, 1, 2],}
\PYG{g+go}{       [3, 4, 5]])}
\PYG{g+gp}{\PYGZgt{}\PYGZgt{}\PYGZgt{} }\PYG{n}{np}\PYG{o}{.}\PYG{n}{reshape}\PYG{p}{(}\PYG{n}{np}\PYG{o}{.}\PYG{n}{ravel}\PYG{p}{(}\PYG{n}{a}\PYG{p}{)}\PYG{p}{,} \PYG{p}{(}\PYG{l+m+mi}{2}\PYG{p}{,} \PYG{l+m+mi}{3}\PYG{p}{)}\PYG{p}{)} \PYG{c+c1}{\PYGZsh{} equivalent to C ravel then C reshape}
\PYG{g+go}{array([[0, 1, 2],}
\PYG{g+go}{       [3, 4, 5]])}
\PYG{g+gp}{\PYGZgt{}\PYGZgt{}\PYGZgt{} }\PYG{n}{np}\PYG{o}{.}\PYG{n}{reshape}\PYG{p}{(}\PYG{n}{a}\PYG{p}{,} \PYG{p}{(}\PYG{l+m+mi}{2}\PYG{p}{,} \PYG{l+m+mi}{3}\PYG{p}{)}\PYG{p}{,} \PYG{n}{order}\PYG{o}{=}\PYG{l+s+s1}{\PYGZsq{}}\PYG{l+s+s1}{F}\PYG{l+s+s1}{\PYGZsq{}}\PYG{p}{)} \PYG{c+c1}{\PYGZsh{} Fortran\PYGZhy{}like index ordering}
\PYG{g+go}{array([[0, 4, 3],}
\PYG{g+go}{       [2, 1, 5]])}
\PYG{g+gp}{\PYGZgt{}\PYGZgt{}\PYGZgt{} }\PYG{n}{np}\PYG{o}{.}\PYG{n}{reshape}\PYG{p}{(}\PYG{n}{np}\PYG{o}{.}\PYG{n}{ravel}\PYG{p}{(}\PYG{n}{a}\PYG{p}{,} \PYG{n}{order}\PYG{o}{=}\PYG{l+s+s1}{\PYGZsq{}}\PYG{l+s+s1}{F}\PYG{l+s+s1}{\PYGZsq{}}\PYG{p}{)}\PYG{p}{,} \PYG{p}{(}\PYG{l+m+mi}{2}\PYG{p}{,} \PYG{l+m+mi}{3}\PYG{p}{)}\PYG{p}{,} \PYG{n}{order}\PYG{o}{=}\PYG{l+s+s1}{\PYGZsq{}}\PYG{l+s+s1}{F}\PYG{l+s+s1}{\PYGZsq{}}\PYG{p}{)}
\PYG{g+go}{array([[0, 4, 3],}
\PYG{g+go}{       [2, 1, 5]])}
\end{sphinxVerbatim}
\subsubsection*{Examples}

\begin{sphinxVerbatim}[commandchars=\\\{\}]
\PYG{g+gp}{\PYGZgt{}\PYGZgt{}\PYGZgt{} }\PYG{n}{a} \PYG{o}{=} \PYG{n}{np}\PYG{o}{.}\PYG{n}{array}\PYG{p}{(}\PYG{p}{[}\PYG{p}{[}\PYG{l+m+mi}{1}\PYG{p}{,}\PYG{l+m+mi}{2}\PYG{p}{,}\PYG{l+m+mi}{3}\PYG{p}{]}\PYG{p}{,} \PYG{p}{[}\PYG{l+m+mi}{4}\PYG{p}{,}\PYG{l+m+mi}{5}\PYG{p}{,}\PYG{l+m+mi}{6}\PYG{p}{]}\PYG{p}{]}\PYG{p}{)}
\PYG{g+gp}{\PYGZgt{}\PYGZgt{}\PYGZgt{} }\PYG{n}{np}\PYG{o}{.}\PYG{n}{reshape}\PYG{p}{(}\PYG{n}{a}\PYG{p}{,} \PYG{l+m+mi}{6}\PYG{p}{)}
\PYG{g+go}{array([1, 2, 3, 4, 5, 6])}
\PYG{g+gp}{\PYGZgt{}\PYGZgt{}\PYGZgt{} }\PYG{n}{np}\PYG{o}{.}\PYG{n}{reshape}\PYG{p}{(}\PYG{n}{a}\PYG{p}{,} \PYG{l+m+mi}{6}\PYG{p}{,} \PYG{n}{order}\PYG{o}{=}\PYG{l+s+s1}{\PYGZsq{}}\PYG{l+s+s1}{F}\PYG{l+s+s1}{\PYGZsq{}}\PYG{p}{)}
\PYG{g+go}{array([1, 4, 2, 5, 3, 6])}
\end{sphinxVerbatim}

\begin{sphinxVerbatim}[commandchars=\\\{\}]
\PYG{g+gp}{\PYGZgt{}\PYGZgt{}\PYGZgt{} }\PYG{n}{np}\PYG{o}{.}\PYG{n}{reshape}\PYG{p}{(}\PYG{n}{a}\PYG{p}{,} \PYG{p}{(}\PYG{l+m+mi}{3}\PYG{p}{,}\PYG{o}{\PYGZhy{}}\PYG{l+m+mi}{1}\PYG{p}{)}\PYG{p}{)}       \PYG{c+c1}{\PYGZsh{} the unspecified value is inferred to be 2}
\PYG{g+go}{array([[1, 2],}
\PYG{g+go}{       [3, 4],}
\PYG{g+go}{       [5, 6]])}
\end{sphinxVerbatim}

\end{fulllineitems}

\index{roll() (in module symjax.tensor)@\spxentry{roll()}\spxextra{in module symjax.tensor}}

\begin{fulllineitems}
\phantomsection\label{\detokenize{modules/tensor:symjax.tensor.roll}}\pysiglinewithargsret{\sphinxbfcode{\sphinxupquote{roll}}}{\emph{\DUrole{n}{a}}, \emph{\DUrole{n}{shift}}, \emph{\DUrole{n}{axis}\DUrole{o}{=}\DUrole{default_value}{None}}}{}
Roll array elements along a given axis.

LAX\sphinxhyphen{}backend implementation of {\hyperref[\detokenize{modules/tensor:symjax.tensor.roll}]{\sphinxcrossref{\sphinxcode{\sphinxupquote{roll()}}}}}.
ADDITIONOriginal docstring below.

LAX\sphinxhyphen{}backend implementation of {\hyperref[\detokenize{modules/tensor:symjax.tensor.roll}]{\sphinxcrossref{\sphinxcode{\sphinxupquote{roll()}}}}}.
Original docstring below.

Elements that roll beyond the last position are re\sphinxhyphen{}introduced at
the first.
\begin{quote}\begin{description}
\item[{Returns}] \leavevmode
\sphinxstylestrong{res} \textendash{} Output array, with the same shape as \sphinxtitleref{a}.

\item[{Return type}] \leavevmode
ndarray

\end{description}\end{quote}

\sphinxstrong{See also:}

\begin{description}
\item[{\sphinxcode{\sphinxupquote{rollaxis()}}}] \leavevmode
Roll the specified axis backwards, until it lies in a given position.

\end{description}

\subsubsection*{Notes}

\DUrole{versionmodified,added}{New in version 1.12.0.}

Supports rolling over multiple dimensions simultaneously.
\subsubsection*{Examples}

\begin{sphinxVerbatim}[commandchars=\\\{\}]
\PYG{g+gp}{\PYGZgt{}\PYGZgt{}\PYGZgt{} }\PYG{n}{x} \PYG{o}{=} \PYG{n}{np}\PYG{o}{.}\PYG{n}{arange}\PYG{p}{(}\PYG{l+m+mi}{10}\PYG{p}{)}
\PYG{g+gp}{\PYGZgt{}\PYGZgt{}\PYGZgt{} }\PYG{n}{np}\PYG{o}{.}\PYG{n}{roll}\PYG{p}{(}\PYG{n}{x}\PYG{p}{,} \PYG{l+m+mi}{2}\PYG{p}{)}
\PYG{g+go}{array([8, 9, 0, 1, 2, 3, 4, 5, 6, 7])}
\PYG{g+gp}{\PYGZgt{}\PYGZgt{}\PYGZgt{} }\PYG{n}{np}\PYG{o}{.}\PYG{n}{roll}\PYG{p}{(}\PYG{n}{x}\PYG{p}{,} \PYG{o}{\PYGZhy{}}\PYG{l+m+mi}{2}\PYG{p}{)}
\PYG{g+go}{array([2, 3, 4, 5, 6, 7, 8, 9, 0, 1])}
\end{sphinxVerbatim}

\begin{sphinxVerbatim}[commandchars=\\\{\}]
\PYG{g+gp}{\PYGZgt{}\PYGZgt{}\PYGZgt{} }\PYG{n}{x2} \PYG{o}{=} \PYG{n}{np}\PYG{o}{.}\PYG{n}{reshape}\PYG{p}{(}\PYG{n}{x}\PYG{p}{,} \PYG{p}{(}\PYG{l+m+mi}{2}\PYG{p}{,}\PYG{l+m+mi}{5}\PYG{p}{)}\PYG{p}{)}
\PYG{g+gp}{\PYGZgt{}\PYGZgt{}\PYGZgt{} }\PYG{n}{x2}
\PYG{g+go}{array([[0, 1, 2, 3, 4],}
\PYG{g+go}{       [5, 6, 7, 8, 9]])}
\PYG{g+gp}{\PYGZgt{}\PYGZgt{}\PYGZgt{} }\PYG{n}{np}\PYG{o}{.}\PYG{n}{roll}\PYG{p}{(}\PYG{n}{x2}\PYG{p}{,} \PYG{l+m+mi}{1}\PYG{p}{)}
\PYG{g+go}{array([[9, 0, 1, 2, 3],}
\PYG{g+go}{       [4, 5, 6, 7, 8]])}
\PYG{g+gp}{\PYGZgt{}\PYGZgt{}\PYGZgt{} }\PYG{n}{np}\PYG{o}{.}\PYG{n}{roll}\PYG{p}{(}\PYG{n}{x2}\PYG{p}{,} \PYG{o}{\PYGZhy{}}\PYG{l+m+mi}{1}\PYG{p}{)}
\PYG{g+go}{array([[1, 2, 3, 4, 5],}
\PYG{g+go}{       [6, 7, 8, 9, 0]])}
\PYG{g+gp}{\PYGZgt{}\PYGZgt{}\PYGZgt{} }\PYG{n}{np}\PYG{o}{.}\PYG{n}{roll}\PYG{p}{(}\PYG{n}{x2}\PYG{p}{,} \PYG{l+m+mi}{1}\PYG{p}{,} \PYG{n}{axis}\PYG{o}{=}\PYG{l+m+mi}{0}\PYG{p}{)}
\PYG{g+go}{array([[5, 6, 7, 8, 9],}
\PYG{g+go}{       [0, 1, 2, 3, 4]])}
\PYG{g+gp}{\PYGZgt{}\PYGZgt{}\PYGZgt{} }\PYG{n}{np}\PYG{o}{.}\PYG{n}{roll}\PYG{p}{(}\PYG{n}{x2}\PYG{p}{,} \PYG{o}{\PYGZhy{}}\PYG{l+m+mi}{1}\PYG{p}{,} \PYG{n}{axis}\PYG{o}{=}\PYG{l+m+mi}{0}\PYG{p}{)}
\PYG{g+go}{array([[5, 6, 7, 8, 9],}
\PYG{g+go}{       [0, 1, 2, 3, 4]])}
\PYG{g+gp}{\PYGZgt{}\PYGZgt{}\PYGZgt{} }\PYG{n}{np}\PYG{o}{.}\PYG{n}{roll}\PYG{p}{(}\PYG{n}{x2}\PYG{p}{,} \PYG{l+m+mi}{1}\PYG{p}{,} \PYG{n}{axis}\PYG{o}{=}\PYG{l+m+mi}{1}\PYG{p}{)}
\PYG{g+go}{array([[4, 0, 1, 2, 3],}
\PYG{g+go}{       [9, 5, 6, 7, 8]])}
\PYG{g+gp}{\PYGZgt{}\PYGZgt{}\PYGZgt{} }\PYG{n}{np}\PYG{o}{.}\PYG{n}{roll}\PYG{p}{(}\PYG{n}{x2}\PYG{p}{,} \PYG{o}{\PYGZhy{}}\PYG{l+m+mi}{1}\PYG{p}{,} \PYG{n}{axis}\PYG{o}{=}\PYG{l+m+mi}{1}\PYG{p}{)}
\PYG{g+go}{array([[1, 2, 3, 4, 0],}
\PYG{g+go}{       [6, 7, 8, 9, 5]])}
\end{sphinxVerbatim}

\end{fulllineitems}

\index{rot90() (in module symjax.tensor)@\spxentry{rot90()}\spxextra{in module symjax.tensor}}

\begin{fulllineitems}
\phantomsection\label{\detokenize{modules/tensor:symjax.tensor.rot90}}\pysiglinewithargsret{\sphinxbfcode{\sphinxupquote{rot90}}}{\emph{\DUrole{n}{m}}, \emph{\DUrole{n}{k}\DUrole{o}{=}\DUrole{default_value}{1}}, \emph{\DUrole{n}{axes}\DUrole{o}{=}\DUrole{default_value}{0, 1}}}{}
Rotate an array by 90 degrees in the plane specified by axes.

LAX\sphinxhyphen{}backend implementation of {\hyperref[\detokenize{modules/tensor:symjax.tensor.rot90}]{\sphinxcrossref{\sphinxcode{\sphinxupquote{rot90()}}}}}.
ADDITIONOriginal docstring below.

LAX\sphinxhyphen{}backend implementation of {\hyperref[\detokenize{modules/tensor:symjax.tensor.rot90}]{\sphinxcrossref{\sphinxcode{\sphinxupquote{rot90()}}}}}.
Original docstring below.

Rotation direction is from the first towards the second axis.
\begin{quote}\begin{description}
\item[{Returns}] \leavevmode
\sphinxstylestrong{y} \textendash{} A rotated view of \sphinxtitleref{m}.

\item[{Return type}] \leavevmode
ndarray

\end{description}\end{quote}

\sphinxstrong{See also:}

\begin{description}
\item[{{\hyperref[\detokenize{modules/tensor:symjax.tensor.flip}]{\sphinxcrossref{\sphinxcode{\sphinxupquote{flip()}}}}}}] \leavevmode
Reverse the order of elements in an array along the given axis.

\item[{{\hyperref[\detokenize{modules/tensor:symjax.tensor.fliplr}]{\sphinxcrossref{\sphinxcode{\sphinxupquote{fliplr()}}}}}}] \leavevmode
Flip an array horizontally.

\item[{{\hyperref[\detokenize{modules/tensor:symjax.tensor.flipud}]{\sphinxcrossref{\sphinxcode{\sphinxupquote{flipud()}}}}}}] \leavevmode
Flip an array vertically.

\end{description}

\subsubsection*{Notes}

rot90(m, k=1, axes=(1,0)) is the reverse of rot90(m, k=1, axes=(0,1))
rot90(m, k=1, axes=(1,0)) is equivalent to rot90(m, k=\sphinxhyphen{}1, axes=(0,1))
\subsubsection*{Examples}

\begin{sphinxVerbatim}[commandchars=\\\{\}]
\PYG{g+gp}{\PYGZgt{}\PYGZgt{}\PYGZgt{} }\PYG{n}{m} \PYG{o}{=} \PYG{n}{np}\PYG{o}{.}\PYG{n}{array}\PYG{p}{(}\PYG{p}{[}\PYG{p}{[}\PYG{l+m+mi}{1}\PYG{p}{,}\PYG{l+m+mi}{2}\PYG{p}{]}\PYG{p}{,}\PYG{p}{[}\PYG{l+m+mi}{3}\PYG{p}{,}\PYG{l+m+mi}{4}\PYG{p}{]}\PYG{p}{]}\PYG{p}{,} \PYG{n+nb}{int}\PYG{p}{)}
\PYG{g+gp}{\PYGZgt{}\PYGZgt{}\PYGZgt{} }\PYG{n}{m}
\PYG{g+go}{array([[1, 2],}
\PYG{g+go}{       [3, 4]])}
\PYG{g+gp}{\PYGZgt{}\PYGZgt{}\PYGZgt{} }\PYG{n}{np}\PYG{o}{.}\PYG{n}{rot90}\PYG{p}{(}\PYG{n}{m}\PYG{p}{)}
\PYG{g+go}{array([[2, 4],}
\PYG{g+go}{       [1, 3]])}
\PYG{g+gp}{\PYGZgt{}\PYGZgt{}\PYGZgt{} }\PYG{n}{np}\PYG{o}{.}\PYG{n}{rot90}\PYG{p}{(}\PYG{n}{m}\PYG{p}{,} \PYG{l+m+mi}{2}\PYG{p}{)}
\PYG{g+go}{array([[4, 3],}
\PYG{g+go}{       [2, 1]])}
\PYG{g+gp}{\PYGZgt{}\PYGZgt{}\PYGZgt{} }\PYG{n}{m} \PYG{o}{=} \PYG{n}{np}\PYG{o}{.}\PYG{n}{arange}\PYG{p}{(}\PYG{l+m+mi}{8}\PYG{p}{)}\PYG{o}{.}\PYG{n}{reshape}\PYG{p}{(}\PYG{p}{(}\PYG{l+m+mi}{2}\PYG{p}{,}\PYG{l+m+mi}{2}\PYG{p}{,}\PYG{l+m+mi}{2}\PYG{p}{)}\PYG{p}{)}
\PYG{g+gp}{\PYGZgt{}\PYGZgt{}\PYGZgt{} }\PYG{n}{np}\PYG{o}{.}\PYG{n}{rot90}\PYG{p}{(}\PYG{n}{m}\PYG{p}{,} \PYG{l+m+mi}{1}\PYG{p}{,} \PYG{p}{(}\PYG{l+m+mi}{1}\PYG{p}{,}\PYG{l+m+mi}{2}\PYG{p}{)}\PYG{p}{)}
\PYG{g+go}{array([[[1, 3],}
\PYG{g+go}{        [0, 2]],}
\PYG{g+go}{       [[5, 7],}
\PYG{g+go}{        [4, 6]]])}
\end{sphinxVerbatim}

\end{fulllineitems}

\index{round() (in module symjax.tensor)@\spxentry{round()}\spxextra{in module symjax.tensor}}

\begin{fulllineitems}
\phantomsection\label{\detokenize{modules/tensor:symjax.tensor.round}}\pysiglinewithargsret{\sphinxbfcode{\sphinxupquote{round}}}{\emph{\DUrole{n}{a}}, \emph{\DUrole{n}{decimals}\DUrole{o}{=}\DUrole{default_value}{0}}}{}
Round an array to the given number of decimals.

LAX\sphinxhyphen{}backend implementation of \sphinxcode{\sphinxupquote{round\_()}}.
ADDITIONOriginal docstring below.

LA

\end{fulllineitems}

\index{row\_stack() (in module symjax.tensor)@\spxentry{row\_stack()}\spxextra{in module symjax.tensor}}

\begin{fulllineitems}
\phantomsection\label{\detokenize{modules/tensor:symjax.tensor.row_stack}}\pysiglinewithargsret{\sphinxbfcode{\sphinxupquote{row\_stack}}}{\emph{\DUrole{n}{tup}}}{}
Stack arrays in sequence vertically (row wise).

LAX\sphinxhyphen{}backend implementation of {\hyperref[\detokenize{modules/tensor:symjax.tensor.vstack}]{\sphinxcrossref{\sphinxcode{\sphinxupquote{vstack()}}}}}.
ADDITIONOriginal docstring below.

LAX\sphinxhyphen{}backend implementation of {\hyperref[\detokenize{modules/tensor:symjax.tensor.vstack}]{\sphinxcrossref{\sphinxcode{\sphinxupquote{vstack()}}}}}.
Original docstring below.

This is equivalent to concatenation along the first axis after 1\sphinxhyphen{}D arrays
of shape \sphinxtitleref{(N,)} have been reshaped to \sphinxtitleref{(1,N)}. Rebuilds arrays divided by
\sphinxtitleref{vsplit}.

This function makes most sense for arrays with up to 3 dimensions. For
instance, for pixel\sphinxhyphen{}data with a height (first axis), width (second axis),
and r/g/b channels (third axis). The functions \sphinxtitleref{concatenate}, \sphinxtitleref{stack} and
\sphinxtitleref{block} provide more general stacking and concatenation operations.
\begin{quote}\begin{description}
\item[{Returns}] \leavevmode
\sphinxstylestrong{stacked} \textendash{} The array formed by stacking the given arrays, will be at least 2\sphinxhyphen{}D.

\item[{Return type}] \leavevmode
ndarray

\end{description}\end{quote}

\sphinxstrong{See also:}

\begin{description}
\item[{{\hyperref[\detokenize{modules/tensor:symjax.tensor.stack}]{\sphinxcrossref{\sphinxcode{\sphinxupquote{stack()}}}}}}] \leavevmode
Join a sequence of arrays along a new axis.

\item[{{\hyperref[\detokenize{modules/tensor:symjax.tensor.hstack}]{\sphinxcrossref{\sphinxcode{\sphinxupquote{hstack()}}}}}}] \leavevmode
Stack arrays in sequence horizontally (column wise).

\item[{{\hyperref[\detokenize{modules/tensor:symjax.tensor.dstack}]{\sphinxcrossref{\sphinxcode{\sphinxupquote{dstack()}}}}}}] \leavevmode
Stack arrays in sequence depth wise (along third dimension).

\item[{{\hyperref[\detokenize{modules/tensor:symjax.tensor.concatenate}]{\sphinxcrossref{\sphinxcode{\sphinxupquote{concatenate()}}}}}}] \leavevmode
Join a sequence of arrays along an existing axis.

\item[{{\hyperref[\detokenize{modules/tensor:symjax.tensor.vsplit}]{\sphinxcrossref{\sphinxcode{\sphinxupquote{vsplit()}}}}}}] \leavevmode
Split array into a list of multiple sub\sphinxhyphen{}arrays vertically.

\item[{{\hyperref[\detokenize{modules/tensor:symjax.tensor.block}]{\sphinxcrossref{\sphinxcode{\sphinxupquote{block()}}}}}}] \leavevmode
Assemble arrays from blocks.

\end{description}

\subsubsection*{Examples}

\begin{sphinxVerbatim}[commandchars=\\\{\}]
\PYG{g+gp}{\PYGZgt{}\PYGZgt{}\PYGZgt{} }\PYG{n}{a} \PYG{o}{=} \PYG{n}{np}\PYG{o}{.}\PYG{n}{array}\PYG{p}{(}\PYG{p}{[}\PYG{l+m+mi}{1}\PYG{p}{,} \PYG{l+m+mi}{2}\PYG{p}{,} \PYG{l+m+mi}{3}\PYG{p}{]}\PYG{p}{)}
\PYG{g+gp}{\PYGZgt{}\PYGZgt{}\PYGZgt{} }\PYG{n}{b} \PYG{o}{=} \PYG{n}{np}\PYG{o}{.}\PYG{n}{array}\PYG{p}{(}\PYG{p}{[}\PYG{l+m+mi}{2}\PYG{p}{,} \PYG{l+m+mi}{3}\PYG{p}{,} \PYG{l+m+mi}{4}\PYG{p}{]}\PYG{p}{)}
\PYG{g+gp}{\PYGZgt{}\PYGZgt{}\PYGZgt{} }\PYG{n}{np}\PYG{o}{.}\PYG{n}{vstack}\PYG{p}{(}\PYG{p}{(}\PYG{n}{a}\PYG{p}{,}\PYG{n}{b}\PYG{p}{)}\PYG{p}{)}
\PYG{g+go}{array([[1, 2, 3],}
\PYG{g+go}{       [2, 3, 4]])}
\end{sphinxVerbatim}

\begin{sphinxVerbatim}[commandchars=\\\{\}]
\PYG{g+gp}{\PYGZgt{}\PYGZgt{}\PYGZgt{} }\PYG{n}{a} \PYG{o}{=} \PYG{n}{np}\PYG{o}{.}\PYG{n}{array}\PYG{p}{(}\PYG{p}{[}\PYG{p}{[}\PYG{l+m+mi}{1}\PYG{p}{]}\PYG{p}{,} \PYG{p}{[}\PYG{l+m+mi}{2}\PYG{p}{]}\PYG{p}{,} \PYG{p}{[}\PYG{l+m+mi}{3}\PYG{p}{]}\PYG{p}{]}\PYG{p}{)}
\PYG{g+gp}{\PYGZgt{}\PYGZgt{}\PYGZgt{} }\PYG{n}{b} \PYG{o}{=} \PYG{n}{np}\PYG{o}{.}\PYG{n}{array}\PYG{p}{(}\PYG{p}{[}\PYG{p}{[}\PYG{l+m+mi}{2}\PYG{p}{]}\PYG{p}{,} \PYG{p}{[}\PYG{l+m+mi}{3}\PYG{p}{]}\PYG{p}{,} \PYG{p}{[}\PYG{l+m+mi}{4}\PYG{p}{]}\PYG{p}{]}\PYG{p}{)}
\PYG{g+gp}{\PYGZgt{}\PYGZgt{}\PYGZgt{} }\PYG{n}{np}\PYG{o}{.}\PYG{n}{vstack}\PYG{p}{(}\PYG{p}{(}\PYG{n}{a}\PYG{p}{,}\PYG{n}{b}\PYG{p}{)}\PYG{p}{)}
\PYG{g+go}{array([[1],}
\PYG{g+go}{       [2],}
\PYG{g+go}{       [3],}
\PYG{g+go}{       [2],}
\PYG{g+go}{       [3],}
\PYG{g+go}{       [4]])}
\end{sphinxVerbatim}

\end{fulllineitems}

\index{sign() (in module symjax.tensor)@\spxentry{sign()}\spxextra{in module symjax.tensor}}

\begin{fulllineitems}
\phantomsection\label{\detokenize{modules/tensor:symjax.tensor.sign}}\pysiglinewithargsret{\sphinxbfcode{\sphinxupquote{sign}}}{\emph{\DUrole{n}{x}}}{}
Returns an element\sphinxhyphen{}wise indication of the sign of a number.

LAX\sphinxhyphen{}backend implementation of {\hyperref[\detokenize{modules/tensor:symjax.tensor.sign}]{\sphinxcrossref{\sphinxcode{\sphinxupquote{sign()}}}}}.
ADDITIONOriginal docstring below.

LAX\sphinxhyphen{}backend implementation of {\hyperref[\detokenize{modules/tensor:symjax.tensor.sign}]{\sphinxcrossref{\sphinxcode{\sphinxupquote{sign()}}}}}.
Original docstring below.

sign(x, /, out=None, {\color{red}\bfseries{}*}, where=True, casting=’same\_kind’, order=’K’, dtype=None, subok=True{[}, signature, extobj{]})

The \sphinxtitleref{sign} function returns \sphinxcode{\sphinxupquote{\sphinxhyphen{}1 if x \textless{} 0, 0 if x==0, 1 if x \textgreater{} 0}}.  nan
is returned for nan inputs.

For complex inputs, the \sphinxtitleref{sign} function returns
\sphinxcode{\sphinxupquote{sign(x.real) + 0j if x.real != 0 else sign(x.imag) + 0j}}.

complex(nan, 0) is returned for complex nan inputs.
\begin{quote}\begin{description}
\item[{Returns}] \leavevmode
\sphinxstylestrong{y} \textendash{} The sign of \sphinxtitleref{x}.
This is a scalar if \sphinxtitleref{x} is a scalar.

\item[{Return type}] \leavevmode
ndarray

\end{description}\end{quote}
\subsubsection*{Notes}

There is more than one definition of sign in common use for complex
numbers.  The definition used here is equivalent to \(x/\sqrt{x*x}\)
which is different from a common alternative, \(x/|x|\).
\subsubsection*{Examples}

\begin{sphinxVerbatim}[commandchars=\\\{\}]
\PYG{g+gp}{\PYGZgt{}\PYGZgt{}\PYGZgt{} }\PYG{n}{np}\PYG{o}{.}\PYG{n}{sign}\PYG{p}{(}\PYG{p}{[}\PYG{o}{\PYGZhy{}}\PYG{l+m+mf}{5.}\PYG{p}{,} \PYG{l+m+mf}{4.5}\PYG{p}{]}\PYG{p}{)}
\PYG{g+go}{array([\PYGZhy{}1.,  1.])}
\PYG{g+gp}{\PYGZgt{}\PYGZgt{}\PYGZgt{} }\PYG{n}{np}\PYG{o}{.}\PYG{n}{sign}\PYG{p}{(}\PYG{l+m+mi}{0}\PYG{p}{)}
\PYG{g+go}{0}
\PYG{g+gp}{\PYGZgt{}\PYGZgt{}\PYGZgt{} }\PYG{n}{np}\PYG{o}{.}\PYG{n}{sign}\PYG{p}{(}\PYG{l+m+mi}{5}\PYG{o}{\PYGZhy{}}\PYG{l+m+mi}{2}\PYG{n}{j}\PYG{p}{)}
\PYG{g+go}{(1+0j)}
\end{sphinxVerbatim}

\end{fulllineitems}

\index{signbit() (in module symjax.tensor)@\spxentry{signbit()}\spxextra{in module symjax.tensor}}

\begin{fulllineitems}
\phantomsection\label{\detokenize{modules/tensor:symjax.tensor.signbit}}\pysiglinewithargsret{\sphinxbfcode{\sphinxupquote{signbit}}}{\emph{\DUrole{n}{x}}}{}
Returns element\sphinxhyphen{}wise True where signbit is set (less than zero).

LAX\sphinxhyphen{}backend implementation of {\hyperref[\detokenize{modules/tensor:symjax.tensor.signbit}]{\sphinxcrossref{\sphinxcode{\sphinxupquote{signbit()}}}}}.
ADDITIONOriginal docstring below.

LAX\sphinxhyphen{}backend implementation of {\hyperref[\detokenize{modules/tensor:symjax.tensor.signbit}]{\sphinxcrossref{\sphinxcode{\sphinxupquote{signbit()}}}}}.
Original docstring below.

signbit(x, /, out=None, {\color{red}\bfseries{}*}, where=True, casting=’same\_kind’, order=’K’, dtype=None, subok=True{[}, signature, extobj{]})
\begin{quote}\begin{description}
\item[{Returns}] \leavevmode
\sphinxstylestrong{result} \textendash{} Output array, or reference to \sphinxtitleref{out} if that was supplied.
This is a scalar if \sphinxtitleref{x} is a scalar.

\item[{Return type}] \leavevmode
ndarray of bool

\end{description}\end{quote}
\subsubsection*{Examples}

\begin{sphinxVerbatim}[commandchars=\\\{\}]
\PYG{g+gp}{\PYGZgt{}\PYGZgt{}\PYGZgt{} }\PYG{n}{np}\PYG{o}{.}\PYG{n}{signbit}\PYG{p}{(}\PYG{o}{\PYGZhy{}}\PYG{l+m+mf}{1.2}\PYG{p}{)}
\PYG{g+go}{True}
\PYG{g+gp}{\PYGZgt{}\PYGZgt{}\PYGZgt{} }\PYG{n}{np}\PYG{o}{.}\PYG{n}{signbit}\PYG{p}{(}\PYG{n}{np}\PYG{o}{.}\PYG{n}{array}\PYG{p}{(}\PYG{p}{[}\PYG{l+m+mi}{1}\PYG{p}{,} \PYG{o}{\PYGZhy{}}\PYG{l+m+mf}{2.3}\PYG{p}{,} \PYG{l+m+mf}{2.1}\PYG{p}{]}\PYG{p}{)}\PYG{p}{)}
\PYG{g+go}{array([False,  True, False])}
\end{sphinxVerbatim}

\end{fulllineitems}

\index{sin() (in module symjax.tensor)@\spxentry{sin()}\spxextra{in module symjax.tensor}}

\begin{fulllineitems}
\phantomsection\label{\detokenize{modules/tensor:symjax.tensor.sin}}\pysiglinewithargsret{\sphinxbfcode{\sphinxupquote{sin}}}{\emph{\DUrole{n}{x}}}{}
Trigonometric sine, element\sphinxhyphen{}wise.

LAX\sphinxhyphen{}backend implementation of {\hyperref[\detokenize{modules/tensor:symjax.tensor.sin}]{\sphinxcrossref{\sphinxcode{\sphinxupquote{sin()}}}}}.
ADDITIONOriginal docstring below.

LAX\sphinxhyphen{}backend implementation of {\hyperref[\detokenize{modules/tensor:symjax.tensor.sin}]{\sphinxcrossref{\sphinxcode{\sphinxupquote{sin()}}}}}.
Original docstring below.

sin(x, /, out=None, {\color{red}\bfseries{}*}, where=True, casting=’same\_kind’, order=’K’, dtype=None, subok=True{[}, signature, extobj{]})
\begin{quote}\begin{description}
\item[{Returns}] \leavevmode
\sphinxstylestrong{y} \textendash{} The sine of each element of x.
This is a scalar if \sphinxtitleref{x} is a scalar.

\item[{Return type}] \leavevmode
array\_like

\end{description}\end{quote}

\sphinxstrong{See also:}

{\hyperref[\detokenize{modules/tensor:symjax.tensor.arcsin}]{\sphinxcrossref{\sphinxcode{\sphinxupquote{arcsin()}}}}}, {\hyperref[\detokenize{modules/tensor:symjax.tensor.sinh}]{\sphinxcrossref{\sphinxcode{\sphinxupquote{sinh()}}}}}, {\hyperref[\detokenize{modules/tensor:symjax.tensor.cos}]{\sphinxcrossref{\sphinxcode{\sphinxupquote{cos()}}}}}

\subsubsection*{Notes}

The sine is one of the fundamental functions of trigonometry (the
mathematical study of triangles).  Consider a circle of radius 1
centered on the origin.  A ray comes in from the \(+x\) axis, makes
an angle at the origin (measured counter\sphinxhyphen{}clockwise from that axis), and
departs from the origin.  The \(y\) coordinate of the outgoing
ray’s intersection with the unit circle is the sine of that angle.  It
ranges from \sphinxhyphen{}1 for \(x=3\pi / 2\) to +1 for \(\pi / 2.\)  The
function has zeroes where the angle is a multiple of \(\pi\).
Sines of angles between \(\pi\) and \(2\pi\) are negative.
The numerous properties of the sine and related functions are included
in any standard trigonometry text.
\subsubsection*{Examples}

Print sine of one angle:

\begin{sphinxVerbatim}[commandchars=\\\{\}]
\PYG{g+gp}{\PYGZgt{}\PYGZgt{}\PYGZgt{} }\PYG{n}{np}\PYG{o}{.}\PYG{n}{sin}\PYG{p}{(}\PYG{n}{np}\PYG{o}{.}\PYG{n}{pi}\PYG{o}{/}\PYG{l+m+mf}{2.}\PYG{p}{)}
\PYG{g+go}{1.0}
\end{sphinxVerbatim}

Print sines of an array of angles given in degrees:

\begin{sphinxVerbatim}[commandchars=\\\{\}]
\PYG{g+gp}{\PYGZgt{}\PYGZgt{}\PYGZgt{} }\PYG{n}{np}\PYG{o}{.}\PYG{n}{sin}\PYG{p}{(}\PYG{n}{np}\PYG{o}{.}\PYG{n}{array}\PYG{p}{(}\PYG{p}{(}\PYG{l+m+mf}{0.}\PYG{p}{,} \PYG{l+m+mf}{30.}\PYG{p}{,} \PYG{l+m+mf}{45.}\PYG{p}{,} \PYG{l+m+mf}{60.}\PYG{p}{,} \PYG{l+m+mf}{90.}\PYG{p}{)}\PYG{p}{)} \PYG{o}{*} \PYG{n}{np}\PYG{o}{.}\PYG{n}{pi} \PYG{o}{/} \PYG{l+m+mf}{180.} \PYG{p}{)}
\PYG{g+go}{array([ 0.        ,  0.5       ,  0.70710678,  0.8660254 ,  1.        ])}
\end{sphinxVerbatim}

Plot the sine function:

\begin{sphinxVerbatim}[commandchars=\\\{\}]
\PYG{g+gp}{\PYGZgt{}\PYGZgt{}\PYGZgt{} }\PYG{k+kn}{import} \PYG{n+nn}{matplotlib}\PYG{n+nn}{.}\PYG{n+nn}{pylab} \PYG{k}{as} \PYG{n+nn}{plt}
\PYG{g+gp}{\PYGZgt{}\PYGZgt{}\PYGZgt{} }\PYG{n}{x} \PYG{o}{=} \PYG{n}{np}\PYG{o}{.}\PYG{n}{linspace}\PYG{p}{(}\PYG{o}{\PYGZhy{}}\PYG{n}{np}\PYG{o}{.}\PYG{n}{pi}\PYG{p}{,} \PYG{n}{np}\PYG{o}{.}\PYG{n}{pi}\PYG{p}{,} \PYG{l+m+mi}{201}\PYG{p}{)}
\PYG{g+gp}{\PYGZgt{}\PYGZgt{}\PYGZgt{} }\PYG{n}{plt}\PYG{o}{.}\PYG{n}{plot}\PYG{p}{(}\PYG{n}{x}\PYG{p}{,} \PYG{n}{np}\PYG{o}{.}\PYG{n}{sin}\PYG{p}{(}\PYG{n}{x}\PYG{p}{)}\PYG{p}{)}
\PYG{g+gp}{\PYGZgt{}\PYGZgt{}\PYGZgt{} }\PYG{n}{plt}\PYG{o}{.}\PYG{n}{xlabel}\PYG{p}{(}\PYG{l+s+s1}{\PYGZsq{}}\PYG{l+s+s1}{Angle [rad]}\PYG{l+s+s1}{\PYGZsq{}}\PYG{p}{)}
\PYG{g+gp}{\PYGZgt{}\PYGZgt{}\PYGZgt{} }\PYG{n}{plt}\PYG{o}{.}\PYG{n}{ylabel}\PYG{p}{(}\PYG{l+s+s1}{\PYGZsq{}}\PYG{l+s+s1}{sin(x)}\PYG{l+s+s1}{\PYGZsq{}}\PYG{p}{)}
\PYG{g+gp}{\PYGZgt{}\PYGZgt{}\PYGZgt{} }\PYG{n}{plt}\PYG{o}{.}\PYG{n}{axis}\PYG{p}{(}\PYG{l+s+s1}{\PYGZsq{}}\PYG{l+s+s1}{tight}\PYG{l+s+s1}{\PYGZsq{}}\PYG{p}{)}
\PYG{g+gp}{\PYGZgt{}\PYGZgt{}\PYGZgt{} }\PYG{n}{plt}\PYG{o}{.}\PYG{n}{show}\PYG{p}{(}\PYG{p}{)}
\end{sphinxVerbatim}

\end{fulllineitems}

\index{sinc() (in module symjax.tensor)@\spxentry{sinc()}\spxextra{in module symjax.tensor}}

\begin{fulllineitems}
\phantomsection\label{\detokenize{modules/tensor:symjax.tensor.sinc}}\pysiglinewithargsret{\sphinxbfcode{\sphinxupquote{sinc}}}{\emph{\DUrole{n}{x}}}{}
Return the sinc function.

LAX\sphinxhyphen{}backend implementation of {\hyperref[\detokenize{modules/tensor:symjax.tensor.sinc}]{\sphinxcrossref{\sphinxcode{\sphinxupquote{sinc()}}}}}.
ADDITIONOriginal docstring below.

LAX\sphinxhyphen{}backend implementation of {\hyperref[\detokenize{modules/tensor:symjax.tensor.sinc}]{\sphinxcrossref{\sphinxcode{\sphinxupquote{sinc()}}}}}.
Original docstring below.
\begin{quote}
\begin{quote}

The sinc function is \(\sin(\pi x)/(\pi x)\).
\end{quote}
\begin{description}
\item[{Returns}] \leavevmode\begin{description}
\item[{out}] \leavevmode{[}ndarray{]}
\sphinxcode{\sphinxupquote{sinc(x)}}, which has the same shape as the input.

\end{description}

\sphinxcode{\sphinxupquote{sinc(0)}} is the limit value 1.

The name sinc is short for “sine cardinal” or “sinus cardinalis”.

The sinc function is used in various signal processing applications,
including in anti\sphinxhyphen{}aliasing, in the construction of a Lanczos resampling
filter, and in interpolation.

For bandlimited interpolation of discrete\sphinxhyphen{}time signals, the ideal
interpolation kernel is proportional to the sinc function.

\begin{sphinxVerbatim}[commandchars=\\\{\}]
\PYG{g+gp}{\PYGZgt{}\PYGZgt{}\PYGZgt{} }\PYG{k+kn}{import} \PYG{n+nn}{matplotlib}\PYG{n+nn}{.}\PYG{n+nn}{pyplot} \PYG{k}{as} \PYG{n+nn}{plt}
\PYG{g+gp}{\PYGZgt{}\PYGZgt{}\PYGZgt{} }\PYG{n}{x} \PYG{o}{=} \PYG{n}{np}\PYG{o}{.}\PYG{n}{linspace}\PYG{p}{(}\PYG{o}{\PYGZhy{}}\PYG{l+m+mi}{4}\PYG{p}{,} \PYG{l+m+mi}{4}\PYG{p}{,} \PYG{l+m+mi}{41}\PYG{p}{)}
\PYG{g+gp}{\PYGZgt{}\PYGZgt{}\PYGZgt{} }\PYG{n}{np}\PYG{o}{.}\PYG{n}{sinc}\PYG{p}{(}\PYG{n}{x}\PYG{p}{)}
\PYG{g+go}{ array([\PYGZhy{}3.89804309e\PYGZhy{}17,  \PYGZhy{}4.92362781e\PYGZhy{}02,  \PYGZhy{}8.40918587e\PYGZhy{}02, \PYGZsh{} may vary}
\PYG{g+go}{        \PYGZhy{}8.90384387e\PYGZhy{}02,  \PYGZhy{}5.84680802e\PYGZhy{}02,   3.89804309e\PYGZhy{}17,}
\PYG{g+go}{        6.68206631e\PYGZhy{}02,   1.16434881e\PYGZhy{}01,   1.26137788e\PYGZhy{}01,}
\PYG{g+go}{        8.50444803e\PYGZhy{}02,  \PYGZhy{}3.89804309e\PYGZhy{}17,  \PYGZhy{}1.03943254e\PYGZhy{}01,}
\PYG{g+go}{        \PYGZhy{}1.89206682e\PYGZhy{}01,  \PYGZhy{}2.16236208e\PYGZhy{}01,  \PYGZhy{}1.55914881e\PYGZhy{}01,}
\PYG{g+go}{        3.89804309e\PYGZhy{}17,   2.33872321e\PYGZhy{}01,   5.04551152e\PYGZhy{}01,}
\PYG{g+go}{        7.56826729e\PYGZhy{}01,   9.35489284e\PYGZhy{}01,   1.00000000e+00,}
\PYG{g+go}{        9.35489284e\PYGZhy{}01,   7.56826729e\PYGZhy{}01,   5.04551152e\PYGZhy{}01,}
\PYG{g+go}{        2.33872321e\PYGZhy{}01,   3.89804309e\PYGZhy{}17,  \PYGZhy{}1.55914881e\PYGZhy{}01,}
\PYG{g+go}{       \PYGZhy{}2.16236208e\PYGZhy{}01,  \PYGZhy{}1.89206682e\PYGZhy{}01,  \PYGZhy{}1.03943254e\PYGZhy{}01,}
\PYG{g+go}{       \PYGZhy{}3.89804309e\PYGZhy{}17,   8.50444803e\PYGZhy{}02,   1.26137788e\PYGZhy{}01,}
\PYG{g+go}{        1.16434881e\PYGZhy{}01,   6.68206631e\PYGZhy{}02,   3.89804309e\PYGZhy{}17,}
\PYG{g+go}{        \PYGZhy{}5.84680802e\PYGZhy{}02,  \PYGZhy{}8.90384387e\PYGZhy{}02,  \PYGZhy{}8.40918587e\PYGZhy{}02,}
\PYG{g+go}{        \PYGZhy{}4.92362781e\PYGZhy{}02,  \PYGZhy{}3.89804309e\PYGZhy{}17])}
\end{sphinxVerbatim}

\begin{sphinxVerbatim}[commandchars=\\\{\}]
\PYG{g+gp}{\PYGZgt{}\PYGZgt{}\PYGZgt{} }\PYG{n}{plt}\PYG{o}{.}\PYG{n}{plot}\PYG{p}{(}\PYG{n}{x}\PYG{p}{,} \PYG{n}{np}\PYG{o}{.}\PYG{n}{sinc}\PYG{p}{(}\PYG{n}{x}\PYG{p}{)}\PYG{p}{)}
\PYG{g+go}{[\PYGZlt{}matplotlib.lines.Line2D object at 0x...\PYGZgt{}]}
\PYG{g+gp}{\PYGZgt{}\PYGZgt{}\PYGZgt{} }\PYG{n}{plt}\PYG{o}{.}\PYG{n}{title}\PYG{p}{(}\PYG{l+s+s2}{\PYGZdq{}}\PYG{l+s+s2}{Sinc Function}\PYG{l+s+s2}{\PYGZdq{}}\PYG{p}{)}
\PYG{g+go}{Text(0.5, 1.0, \PYGZsq{}Sinc Function\PYGZsq{})}
\PYG{g+gp}{\PYGZgt{}\PYGZgt{}\PYGZgt{} }\PYG{n}{plt}\PYG{o}{.}\PYG{n}{ylabel}\PYG{p}{(}\PYG{l+s+s2}{\PYGZdq{}}\PYG{l+s+s2}{Amplitude}\PYG{l+s+s2}{\PYGZdq{}}\PYG{p}{)}
\PYG{g+go}{Text(0, 0.5, \PYGZsq{}Amplitude\PYGZsq{})}
\PYG{g+gp}{\PYGZgt{}\PYGZgt{}\PYGZgt{} }\PYG{n}{plt}\PYG{o}{.}\PYG{n}{xlabel}\PYG{p}{(}\PYG{l+s+s2}{\PYGZdq{}}\PYG{l+s+s2}{X}\PYG{l+s+s2}{\PYGZdq{}}\PYG{p}{)}
\PYG{g+go}{Text(0.5, 0, \PYGZsq{}X\PYGZsq{})}
\PYG{g+gp}{\PYGZgt{}\PYGZgt{}\PYGZgt{} }\PYG{n}{plt}\PYG{o}{.}\PYG{n}{show}\PYG{p}{(}\PYG{p}{)}
\end{sphinxVerbatim}

\end{description}
\end{quote}

\end{fulllineitems}

\index{sinh() (in module symjax.tensor)@\spxentry{sinh()}\spxextra{in module symjax.tensor}}

\begin{fulllineitems}
\phantomsection\label{\detokenize{modules/tensor:symjax.tensor.sinh}}\pysiglinewithargsret{\sphinxbfcode{\sphinxupquote{sinh}}}{\emph{\DUrole{n}{x}}}{}
Hyperbolic sine, element\sphinxhyphen{}wise.

LAX\sphinxhyphen{}backend implementation of {\hyperref[\detokenize{modules/tensor:symjax.tensor.sinh}]{\sphinxcrossref{\sphinxcode{\sphinxupquote{sinh()}}}}}.
ADDITIONOriginal docstring below.

LAX\sphinxhyphen{}backend implementation of {\hyperref[\detokenize{modules/tensor:symjax.tensor.sinh}]{\sphinxcrossref{\sphinxcode{\sphinxupquote{sinh()}}}}}.
Original docstring below.

sinh(x, /, out=None, {\color{red}\bfseries{}*}, where=True, casting=’same\_kind’, order=’K’, dtype=None, subok=True{[}, signature, extobj{]})

Equivalent to \sphinxcode{\sphinxupquote{1/2 * (np.exp(x) \sphinxhyphen{} np.exp(\sphinxhyphen{}x))}} or
\sphinxcode{\sphinxupquote{\sphinxhyphen{}1j * np.sin(1j*x)}}.
\begin{quote}\begin{description}
\item[{Returns}] \leavevmode
\sphinxstylestrong{y} \textendash{} The corresponding hyperbolic sine values.
This is a scalar if \sphinxtitleref{x} is a scalar.

\item[{Return type}] \leavevmode
ndarray

\end{description}\end{quote}
\subsubsection*{Notes}

If \sphinxtitleref{out} is provided, the function writes the result into it,
and returns a reference to \sphinxtitleref{out}.  (See Examples)
\subsubsection*{References}

M. Abramowitz and I. A. Stegun, Handbook of Mathematical Functions.
New York, NY: Dover, 1972, pg. 83.
\subsubsection*{Examples}

\begin{sphinxVerbatim}[commandchars=\\\{\}]
\PYG{g+gp}{\PYGZgt{}\PYGZgt{}\PYGZgt{} }\PYG{n}{np}\PYG{o}{.}\PYG{n}{sinh}\PYG{p}{(}\PYG{l+m+mi}{0}\PYG{p}{)}
\PYG{g+go}{0.0}
\PYG{g+gp}{\PYGZgt{}\PYGZgt{}\PYGZgt{} }\PYG{n}{np}\PYG{o}{.}\PYG{n}{sinh}\PYG{p}{(}\PYG{n}{np}\PYG{o}{.}\PYG{n}{pi}\PYG{o}{*}\PYG{l+m+mi}{1}\PYG{n}{j}\PYG{o}{/}\PYG{l+m+mi}{2}\PYG{p}{)}
\PYG{g+go}{1j}
\PYG{g+gp}{\PYGZgt{}\PYGZgt{}\PYGZgt{} }\PYG{n}{np}\PYG{o}{.}\PYG{n}{sinh}\PYG{p}{(}\PYG{n}{np}\PYG{o}{.}\PYG{n}{pi}\PYG{o}{*}\PYG{l+m+mi}{1}\PYG{n}{j}\PYG{p}{)} \PYG{c+c1}{\PYGZsh{} (exact value is 0)}
\PYG{g+go}{1.2246063538223773e\PYGZhy{}016j}
\PYG{g+gp}{\PYGZgt{}\PYGZgt{}\PYGZgt{} }\PYG{c+c1}{\PYGZsh{} Discrepancy due to vagaries of floating point arithmetic.}
\end{sphinxVerbatim}

\begin{sphinxVerbatim}[commandchars=\\\{\}]
\PYG{g+gp}{\PYGZgt{}\PYGZgt{}\PYGZgt{} }\PYG{c+c1}{\PYGZsh{} Example of providing the optional output parameter}
\PYG{g+gp}{\PYGZgt{}\PYGZgt{}\PYGZgt{} }\PYG{n}{out1} \PYG{o}{=} \PYG{n}{np}\PYG{o}{.}\PYG{n}{array}\PYG{p}{(}\PYG{p}{[}\PYG{l+m+mi}{0}\PYG{p}{]}\PYG{p}{,} \PYG{n}{dtype}\PYG{o}{=}\PYG{l+s+s1}{\PYGZsq{}}\PYG{l+s+s1}{d}\PYG{l+s+s1}{\PYGZsq{}}\PYG{p}{)}
\PYG{g+gp}{\PYGZgt{}\PYGZgt{}\PYGZgt{} }\PYG{n}{out2} \PYG{o}{=} \PYG{n}{np}\PYG{o}{.}\PYG{n}{sinh}\PYG{p}{(}\PYG{p}{[}\PYG{l+m+mf}{0.1}\PYG{p}{]}\PYG{p}{,} \PYG{n}{out1}\PYG{p}{)}
\PYG{g+gp}{\PYGZgt{}\PYGZgt{}\PYGZgt{} }\PYG{n}{out2} \PYG{o+ow}{is} \PYG{n}{out1}
\PYG{g+go}{True}
\end{sphinxVerbatim}

\begin{sphinxVerbatim}[commandchars=\\\{\}]
\PYG{g+gp}{\PYGZgt{}\PYGZgt{}\PYGZgt{} }\PYG{c+c1}{\PYGZsh{} Example of ValueError due to provision of shape mis\PYGZhy{}matched `out`}
\PYG{g+gp}{\PYGZgt{}\PYGZgt{}\PYGZgt{} }\PYG{n}{np}\PYG{o}{.}\PYG{n}{sinh}\PYG{p}{(}\PYG{n}{np}\PYG{o}{.}\PYG{n}{zeros}\PYG{p}{(}\PYG{p}{(}\PYG{l+m+mi}{3}\PYG{p}{,}\PYG{l+m+mi}{3}\PYG{p}{)}\PYG{p}{)}\PYG{p}{,}\PYG{n}{np}\PYG{o}{.}\PYG{n}{zeros}\PYG{p}{(}\PYG{p}{(}\PYG{l+m+mi}{2}\PYG{p}{,}\PYG{l+m+mi}{2}\PYG{p}{)}\PYG{p}{)}\PYG{p}{)}
\PYG{g+gt}{Traceback (most recent call last):}
  File \PYG{n+nb}{\PYGZdq{}\PYGZlt{}stdin\PYGZgt{}\PYGZdq{}}, line \PYG{l+m}{1}, in \PYG{n}{\PYGZlt{}module\PYGZgt{}}
\PYG{g+gr}{ValueError}: \PYG{n}{operands could not be broadcast together with shapes (3,3) (2,2)}
\end{sphinxVerbatim}

\end{fulllineitems}

\index{sort() (in module symjax.tensor)@\spxentry{sort()}\spxextra{in module symjax.tensor}}

\begin{fulllineitems}
\phantomsection\label{\detokenize{modules/tensor:symjax.tensor.sort}}\pysiglinewithargsret{\sphinxbfcode{\sphinxupquote{sort}}}{\emph{\DUrole{n}{a}}, \emph{\DUrole{n}{axis}\DUrole{o}{=}\DUrole{default_value}{\sphinxhyphen{} 1}}, \emph{\DUrole{n}{kind}\DUrole{o}{=}\DUrole{default_value}{\textquotesingle{}quicksort\textquotesingle{}}}, \emph{\DUrole{n}{order}\DUrole{o}{=}\DUrole{default_value}{None}}}{}
Return a sorted copy of an array.

LAX\sphinxhyphen{}backend implementation of {\hyperref[\detokenize{modules/tensor:symjax.tensor.sort}]{\sphinxcrossref{\sphinxcode{\sphinxupquote{sort()}}}}}.
ADDITIONOriginal docstring below.

LAX\sphinxhyphen{}backend implementation of {\hyperref[\detokenize{modules/tensor:symjax.tensor.sort}]{\sphinxcrossref{\sphinxcode{\sphinxupquote{sort()}}}}}.
Original docstring below.
\begin{quote}\begin{description}
\item[{Returns}] \leavevmode
\sphinxstylestrong{sorted\_array} \textendash{} Array of the same type and shape as \sphinxtitleref{a}.

\item[{Return type}] \leavevmode
ndarray

\end{description}\end{quote}

\sphinxstrong{See also:}

\begin{description}
\item[{\sphinxcode{\sphinxupquote{ndarray.sort()}}}] \leavevmode
Method to sort an array in\sphinxhyphen{}place.

\item[{{\hyperref[\detokenize{modules/tensor:symjax.tensor.argsort}]{\sphinxcrossref{\sphinxcode{\sphinxupquote{argsort()}}}}}}] \leavevmode
Indirect sort.

\item[{\sphinxcode{\sphinxupquote{lexsort()}}}] \leavevmode
Indirect stable sort on multiple keys.

\item[{\sphinxcode{\sphinxupquote{searchsorted()}}}] \leavevmode
Find elements in a sorted array.

\item[{\sphinxcode{\sphinxupquote{partition()}}}] \leavevmode
Partial sort.

\end{description}

\subsubsection*{Notes}

The various sorting algorithms are characterized by their average speed,
worst case performance, work space size, and whether they are stable. A
stable sort keeps items with the same key in the same relative
order. The four algorithms implemented in NumPy have the following
properties:

\begin{savenotes}\sphinxattablestart
\centering
\begin{tabulary}{\linewidth}[t]{|T|T|T|T|T|}
\hline
\sphinxstyletheadfamily 
kind
&\sphinxstyletheadfamily 
speed
&\sphinxstyletheadfamily 
worst case
&\sphinxstyletheadfamily 
work space
&\sphinxstyletheadfamily 
stable
\\
\hline
‘quicksort’
&
1
&
O(n\textasciicircum{}2)
&
0
&
no
\\
\hline
‘heapsort’
&
3
&
O(n*log(n))
&
0
&
no
\\
\hline
‘mergesort’
&
2
&
O(n*log(n))
&
\textasciitilde{}n/2
&
yes
\\
\hline
‘timsort’
&
2
&
O(n*log(n))
&
\textasciitilde{}n/2
&
yes
\\
\hline
\end{tabulary}
\par
\sphinxattableend\end{savenotes}

\begin{sphinxadmonition}{note}{Note:}
The datatype determines which of ‘mergesort’ or ‘timsort’
is actually used, even if ‘mergesort’ is specified. User selection
at a finer scale is not currently available.
\end{sphinxadmonition}

All the sort algorithms make temporary copies of the data when
sorting along any but the last axis.  Consequently, sorting along
the last axis is faster and uses less space than sorting along
any other axis.

The sort order for complex numbers is lexicographic. If both the real
and imaginary parts are non\sphinxhyphen{}nan then the order is determined by the
real parts except when they are equal, in which case the order is
determined by the imaginary parts.

Previous to numpy 1.4.0 sorting real and complex arrays containing nan
values led to undefined behaviour. In numpy versions \textgreater{}= 1.4.0 nan
values are sorted to the end. The extended sort order is:
\begin{itemize}
\item {} 
Real: {[}R, nan{]}

\item {} 
Complex: {[}R + Rj, R + nanj, nan + Rj, nan + nanj{]}

\end{itemize}

where R is a non\sphinxhyphen{}nan real value. Complex values with the same nan
placements are sorted according to the non\sphinxhyphen{}nan part if it exists.
Non\sphinxhyphen{}nan values are sorted as before.

\DUrole{versionmodified,added}{New in version 1.12.0.}

quicksort has been changed to \sphinxhref{https://en.wikipedia.org/wiki/Introsort}{introsort}.
When sorting does not make enough progress it switches to
\sphinxhref{https://en.wikipedia.org/wiki/Heapsort}{heapsort}.
This implementation makes quicksort O(n*log(n)) in the worst case.

‘stable’ automatically chooses the best stable sorting algorithm
for the data type being sorted.
It, along with ‘mergesort’ is currently mapped to
\sphinxhref{https://en.wikipedia.org/wiki/Timsort}{timsort}
or \sphinxhref{https://en.wikipedia.org/wiki/Radix\_sort}{radix sort}
depending on the data type.
API forward compatibility currently limits the
ability to select the implementation and it is hardwired for the different
data types.

\DUrole{versionmodified,added}{New in version 1.17.0.}

Timsort is added for better performance on already or nearly
sorted data. On random data timsort is almost identical to
mergesort. It is now used for stable sort while quicksort is still the
default sort if none is chosen. For timsort details, refer to
\sphinxhref{https://github.com/python/cpython/blob/3.7/Objects/listsort.txt}{CPython listsort.txt}.
‘mergesort’ and ‘stable’ are mapped to radix sort for integer data types. Radix sort is an
O(n) sort instead of O(n log n).

\DUrole{versionmodified,changed}{Changed in version 1.17.0.}

NaT now sorts to the end of arrays for consistency with NaN.
\subsubsection*{Examples}

\begin{sphinxVerbatim}[commandchars=\\\{\}]
\PYG{g+gp}{\PYGZgt{}\PYGZgt{}\PYGZgt{} }\PYG{n}{a} \PYG{o}{=} \PYG{n}{np}\PYG{o}{.}\PYG{n}{array}\PYG{p}{(}\PYG{p}{[}\PYG{p}{[}\PYG{l+m+mi}{1}\PYG{p}{,}\PYG{l+m+mi}{4}\PYG{p}{]}\PYG{p}{,}\PYG{p}{[}\PYG{l+m+mi}{3}\PYG{p}{,}\PYG{l+m+mi}{1}\PYG{p}{]}\PYG{p}{]}\PYG{p}{)}
\PYG{g+gp}{\PYGZgt{}\PYGZgt{}\PYGZgt{} }\PYG{n}{np}\PYG{o}{.}\PYG{n}{sort}\PYG{p}{(}\PYG{n}{a}\PYG{p}{)}                \PYG{c+c1}{\PYGZsh{} sort along the last axis}
\PYG{g+go}{array([[1, 4],}
\PYG{g+go}{       [1, 3]])}
\PYG{g+gp}{\PYGZgt{}\PYGZgt{}\PYGZgt{} }\PYG{n}{np}\PYG{o}{.}\PYG{n}{sort}\PYG{p}{(}\PYG{n}{a}\PYG{p}{,} \PYG{n}{axis}\PYG{o}{=}\PYG{k+kc}{None}\PYG{p}{)}     \PYG{c+c1}{\PYGZsh{} sort the flattened array}
\PYG{g+go}{array([1, 1, 3, 4])}
\PYG{g+gp}{\PYGZgt{}\PYGZgt{}\PYGZgt{} }\PYG{n}{np}\PYG{o}{.}\PYG{n}{sort}\PYG{p}{(}\PYG{n}{a}\PYG{p}{,} \PYG{n}{axis}\PYG{o}{=}\PYG{l+m+mi}{0}\PYG{p}{)}        \PYG{c+c1}{\PYGZsh{} sort along the first axis}
\PYG{g+go}{array([[1, 1],}
\PYG{g+go}{       [3, 4]])}
\end{sphinxVerbatim}

Use the \sphinxtitleref{order} keyword to specify a field to use when sorting a
structured array:

\begin{sphinxVerbatim}[commandchars=\\\{\}]
\PYG{g+gp}{\PYGZgt{}\PYGZgt{}\PYGZgt{} }\PYG{n}{dtype} \PYG{o}{=} \PYG{p}{[}\PYG{p}{(}\PYG{l+s+s1}{\PYGZsq{}}\PYG{l+s+s1}{name}\PYG{l+s+s1}{\PYGZsq{}}\PYG{p}{,} \PYG{l+s+s1}{\PYGZsq{}}\PYG{l+s+s1}{S10}\PYG{l+s+s1}{\PYGZsq{}}\PYG{p}{)}\PYG{p}{,} \PYG{p}{(}\PYG{l+s+s1}{\PYGZsq{}}\PYG{l+s+s1}{height}\PYG{l+s+s1}{\PYGZsq{}}\PYG{p}{,} \PYG{n+nb}{float}\PYG{p}{)}\PYG{p}{,} \PYG{p}{(}\PYG{l+s+s1}{\PYGZsq{}}\PYG{l+s+s1}{age}\PYG{l+s+s1}{\PYGZsq{}}\PYG{p}{,} \PYG{n+nb}{int}\PYG{p}{)}\PYG{p}{]}
\PYG{g+gp}{\PYGZgt{}\PYGZgt{}\PYGZgt{} }\PYG{n}{values} \PYG{o}{=} \PYG{p}{[}\PYG{p}{(}\PYG{l+s+s1}{\PYGZsq{}}\PYG{l+s+s1}{Arthur}\PYG{l+s+s1}{\PYGZsq{}}\PYG{p}{,} \PYG{l+m+mf}{1.8}\PYG{p}{,} \PYG{l+m+mi}{41}\PYG{p}{)}\PYG{p}{,} \PYG{p}{(}\PYG{l+s+s1}{\PYGZsq{}}\PYG{l+s+s1}{Lancelot}\PYG{l+s+s1}{\PYGZsq{}}\PYG{p}{,} \PYG{l+m+mf}{1.9}\PYG{p}{,} \PYG{l+m+mi}{38}\PYG{p}{)}\PYG{p}{,}
\PYG{g+gp}{... }          \PYG{p}{(}\PYG{l+s+s1}{\PYGZsq{}}\PYG{l+s+s1}{Galahad}\PYG{l+s+s1}{\PYGZsq{}}\PYG{p}{,} \PYG{l+m+mf}{1.7}\PYG{p}{,} \PYG{l+m+mi}{38}\PYG{p}{)}\PYG{p}{]}
\PYG{g+gp}{\PYGZgt{}\PYGZgt{}\PYGZgt{} }\PYG{n}{a} \PYG{o}{=} \PYG{n}{np}\PYG{o}{.}\PYG{n}{array}\PYG{p}{(}\PYG{n}{values}\PYG{p}{,} \PYG{n}{dtype}\PYG{o}{=}\PYG{n}{dtype}\PYG{p}{)}       \PYG{c+c1}{\PYGZsh{} create a structured array}
\PYG{g+gp}{\PYGZgt{}\PYGZgt{}\PYGZgt{} }\PYG{n}{np}\PYG{o}{.}\PYG{n}{sort}\PYG{p}{(}\PYG{n}{a}\PYG{p}{,} \PYG{n}{order}\PYG{o}{=}\PYG{l+s+s1}{\PYGZsq{}}\PYG{l+s+s1}{height}\PYG{l+s+s1}{\PYGZsq{}}\PYG{p}{)}                        
\PYG{g+go}{array([(\PYGZsq{}Galahad\PYGZsq{}, 1.7, 38), (\PYGZsq{}Arthur\PYGZsq{}, 1.8, 41),}
\PYG{g+go}{       (\PYGZsq{}Lancelot\PYGZsq{}, 1.8999999999999999, 38)],}
\PYG{g+go}{      dtype=[(\PYGZsq{}name\PYGZsq{}, \PYGZsq{}|S10\PYGZsq{}), (\PYGZsq{}height\PYGZsq{}, \PYGZsq{}\PYGZlt{}f8\PYGZsq{}), (\PYGZsq{}age\PYGZsq{}, \PYGZsq{}\PYGZlt{}i4\PYGZsq{})])}
\end{sphinxVerbatim}

Sort by age, then height if ages are equal:

\begin{sphinxVerbatim}[commandchars=\\\{\}]
\PYG{g+gp}{\PYGZgt{}\PYGZgt{}\PYGZgt{} }\PYG{n}{np}\PYG{o}{.}\PYG{n}{sort}\PYG{p}{(}\PYG{n}{a}\PYG{p}{,} \PYG{n}{order}\PYG{o}{=}\PYG{p}{[}\PYG{l+s+s1}{\PYGZsq{}}\PYG{l+s+s1}{age}\PYG{l+s+s1}{\PYGZsq{}}\PYG{p}{,} \PYG{l+s+s1}{\PYGZsq{}}\PYG{l+s+s1}{height}\PYG{l+s+s1}{\PYGZsq{}}\PYG{p}{]}\PYG{p}{)}               
\PYG{g+go}{array([(\PYGZsq{}Galahad\PYGZsq{}, 1.7, 38), (\PYGZsq{}Lancelot\PYGZsq{}, 1.8999999999999999, 38),}
\PYG{g+go}{       (\PYGZsq{}Arthur\PYGZsq{}, 1.8, 41)],}
\PYG{g+go}{      dtype=[(\PYGZsq{}name\PYGZsq{}, \PYGZsq{}|S10\PYGZsq{}), (\PYGZsq{}height\PYGZsq{}, \PYGZsq{}\PYGZlt{}f8\PYGZsq{}), (\PYGZsq{}age\PYGZsq{}, \PYGZsq{}\PYGZlt{}i4\PYGZsq{})])}
\end{sphinxVerbatim}

\end{fulllineitems}

\index{split() (in module symjax.tensor)@\spxentry{split()}\spxextra{in module symjax.tensor}}

\begin{fulllineitems}
\phantomsection\label{\detokenize{modules/tensor:symjax.tensor.split}}\pysiglinewithargsret{\sphinxbfcode{\sphinxupquote{split}}}{\emph{\DUrole{n}{ary}}, \emph{\DUrole{n}{indices\_or\_sections}}, \emph{\DUrole{n}{axis}\DUrole{o}{=}\DUrole{default_value}{0}}}{}
Split an array into multiple sub\sphinxhyphen{}arrays as views into \sphinxtitleref{ary}.

LAX\sphinxhyphen{}backend implementation of {\hyperref[\detokenize{modules/tensor:symjax.tensor.split}]{\sphinxcrossref{\sphinxcode{\sphinxupquote{split()}}}}}.
ADDITIONOriginal docstring below.

LAX\sphinxhyphen{}backend implementation of {\hyperref[\detokenize{modules/tensor:symjax.tensor.split}]{\sphinxcrossref{\sphinxcode{\sphinxupquote{split()}}}}}.
Original docstring below.
\begin{quote}\begin{description}
\item[{Returns}] \leavevmode
\sphinxstylestrong{sub\sphinxhyphen{}arrays} \textendash{} A list of sub\sphinxhyphen{}arrays as views into \sphinxtitleref{ary}.

\item[{Return type}] \leavevmode
list of ndarrays

\item[{Raises}] \leavevmode
\sphinxstyleliteralstrong{\sphinxupquote{ValueError}} \textendash{} If \sphinxtitleref{indices\_or\_sections} is given as an integer, but
    a split does not result in equal division.

\end{description}\end{quote}

\sphinxstrong{See also:}

\begin{description}
\item[{\sphinxcode{\sphinxupquote{array\_split()}}}] \leavevmode
Split an array into multiple sub\sphinxhyphen{}arrays of equal or near\sphinxhyphen{}equal size.  Does not raise an exception if an equal division cannot be made.

\item[{{\hyperref[\detokenize{modules/tensor:symjax.tensor.hsplit}]{\sphinxcrossref{\sphinxcode{\sphinxupquote{hsplit()}}}}}}] \leavevmode
Split array into multiple sub\sphinxhyphen{}arrays horizontally (column\sphinxhyphen{}wise).

\item[{{\hyperref[\detokenize{modules/tensor:symjax.tensor.vsplit}]{\sphinxcrossref{\sphinxcode{\sphinxupquote{vsplit()}}}}}}] \leavevmode
Split array into multiple sub\sphinxhyphen{}arrays vertically (row wise).

\item[{{\hyperref[\detokenize{modules/tensor:symjax.tensor.dsplit}]{\sphinxcrossref{\sphinxcode{\sphinxupquote{dsplit()}}}}}}] \leavevmode
Split array into multiple sub\sphinxhyphen{}arrays along the 3rd axis (depth).

\item[{{\hyperref[\detokenize{modules/tensor:symjax.tensor.concatenate}]{\sphinxcrossref{\sphinxcode{\sphinxupquote{concatenate()}}}}}}] \leavevmode
Join a sequence of arrays along an existing axis.

\item[{{\hyperref[\detokenize{modules/tensor:symjax.tensor.stack}]{\sphinxcrossref{\sphinxcode{\sphinxupquote{stack()}}}}}}] \leavevmode
Join a sequence of arrays along a new axis.

\item[{{\hyperref[\detokenize{modules/tensor:symjax.tensor.hstack}]{\sphinxcrossref{\sphinxcode{\sphinxupquote{hstack()}}}}}}] \leavevmode
Stack arrays in sequence horizontally (column wise).

\item[{{\hyperref[\detokenize{modules/tensor:symjax.tensor.vstack}]{\sphinxcrossref{\sphinxcode{\sphinxupquote{vstack()}}}}}}] \leavevmode
Stack arrays in sequence vertically (row wise).

\item[{{\hyperref[\detokenize{modules/tensor:symjax.tensor.dstack}]{\sphinxcrossref{\sphinxcode{\sphinxupquote{dstack()}}}}}}] \leavevmode
Stack arrays in sequence depth wise (along third dimension).

\end{description}

\subsubsection*{Examples}

\begin{sphinxVerbatim}[commandchars=\\\{\}]
\PYG{g+gp}{\PYGZgt{}\PYGZgt{}\PYGZgt{} }\PYG{n}{x} \PYG{o}{=} \PYG{n}{np}\PYG{o}{.}\PYG{n}{arange}\PYG{p}{(}\PYG{l+m+mf}{9.0}\PYG{p}{)}
\PYG{g+gp}{\PYGZgt{}\PYGZgt{}\PYGZgt{} }\PYG{n}{np}\PYG{o}{.}\PYG{n}{split}\PYG{p}{(}\PYG{n}{x}\PYG{p}{,} \PYG{l+m+mi}{3}\PYG{p}{)}
\PYG{g+go}{[array([0.,  1.,  2.]), array([3.,  4.,  5.]), array([6.,  7.,  8.])]}
\end{sphinxVerbatim}

\begin{sphinxVerbatim}[commandchars=\\\{\}]
\PYG{g+gp}{\PYGZgt{}\PYGZgt{}\PYGZgt{} }\PYG{n}{x} \PYG{o}{=} \PYG{n}{np}\PYG{o}{.}\PYG{n}{arange}\PYG{p}{(}\PYG{l+m+mf}{8.0}\PYG{p}{)}
\PYG{g+gp}{\PYGZgt{}\PYGZgt{}\PYGZgt{} }\PYG{n}{np}\PYG{o}{.}\PYG{n}{split}\PYG{p}{(}\PYG{n}{x}\PYG{p}{,} \PYG{p}{[}\PYG{l+m+mi}{3}\PYG{p}{,} \PYG{l+m+mi}{5}\PYG{p}{,} \PYG{l+m+mi}{6}\PYG{p}{,} \PYG{l+m+mi}{10}\PYG{p}{]}\PYG{p}{)}
\PYG{g+go}{[array([0.,  1.,  2.]),}
\PYG{g+go}{ array([3.,  4.]),}
\PYG{g+go}{ array([5.]),}
\PYG{g+go}{ array([6.,  7.]),}
\PYG{g+go}{ array([], dtype=float64)]}
\end{sphinxVerbatim}

\end{fulllineitems}

\index{sqrt() (in module symjax.tensor)@\spxentry{sqrt()}\spxextra{in module symjax.tensor}}

\begin{fulllineitems}
\phantomsection\label{\detokenize{modules/tensor:symjax.tensor.sqrt}}\pysiglinewithargsret{\sphinxbfcode{\sphinxupquote{sqrt}}}{\emph{\DUrole{n}{x}}}{}
Return the non\sphinxhyphen{}negative square\sphinxhyphen{}root of an array, element\sphinxhyphen{}wise.

LAX\sphinxhyphen{}backend implementation of {\hyperref[\detokenize{modules/tensor:symjax.tensor.sqrt}]{\sphinxcrossref{\sphinxcode{\sphinxupquote{sqrt()}}}}}.
ADDITIONOriginal docstring below.

LAX\sphinxhyphen{}backend implementation of {\hyperref[\detokenize{modules/tensor:symjax.tensor.sqrt}]{\sphinxcrossref{\sphinxcode{\sphinxupquote{sqrt()}}}}}.
Original docstring below.

sqrt(x, /, out=None, {\color{red}\bfseries{}*}, where=True, casting=’same\_kind’, order=’K’, dtype=None, subok=True{[}, signature, extobj{]})
\begin{quote}\begin{description}
\item[{Returns}] \leavevmode
\sphinxstylestrong{y} \textendash{} An array of the same shape as \sphinxtitleref{x}, containing the positive
square\sphinxhyphen{}root of each element in \sphinxtitleref{x}.  If any element in \sphinxtitleref{x} is
complex, a complex array is returned (and the square\sphinxhyphen{}roots of
negative reals are calculated).  If all of the elements in \sphinxtitleref{x}
are real, so is \sphinxtitleref{y}, with negative elements returning \sphinxcode{\sphinxupquote{nan}}.
If \sphinxtitleref{out} was provided, \sphinxtitleref{y} is a reference to it.
This is a scalar if \sphinxtitleref{x} is a scalar.

\item[{Return type}] \leavevmode
ndarray

\end{description}\end{quote}

\sphinxstrong{See also:}

\begin{description}
\item[{\sphinxcode{\sphinxupquote{lib.scimath.sqrt()}}}] \leavevmode
A version which returns complex numbers when given negative reals.

\end{description}

\subsubsection*{Notes}

\sphinxstyleemphasis{sqrt} has\textendash{}consistent with common convention\textendash{}as its branch cut the
real “interval” {[}\sphinxtitleref{\sphinxhyphen{}inf}, 0), and is continuous from above on it.
A branch cut is a curve in the complex plane across which a given
complex function fails to be continuous.
\subsubsection*{Examples}

\begin{sphinxVerbatim}[commandchars=\\\{\}]
\PYG{g+gp}{\PYGZgt{}\PYGZgt{}\PYGZgt{} }\PYG{n}{np}\PYG{o}{.}\PYG{n}{sqrt}\PYG{p}{(}\PYG{p}{[}\PYG{l+m+mi}{1}\PYG{p}{,}\PYG{l+m+mi}{4}\PYG{p}{,}\PYG{l+m+mi}{9}\PYG{p}{]}\PYG{p}{)}
\PYG{g+go}{array([ 1.,  2.,  3.])}
\end{sphinxVerbatim}

\begin{sphinxVerbatim}[commandchars=\\\{\}]
\PYG{g+gp}{\PYGZgt{}\PYGZgt{}\PYGZgt{} }\PYG{n}{np}\PYG{o}{.}\PYG{n}{sqrt}\PYG{p}{(}\PYG{p}{[}\PYG{l+m+mi}{4}\PYG{p}{,} \PYG{o}{\PYGZhy{}}\PYG{l+m+mi}{1}\PYG{p}{,} \PYG{o}{\PYGZhy{}}\PYG{l+m+mi}{3}\PYG{o}{+}\PYG{l+m+mi}{4}\PYG{n}{J}\PYG{p}{]}\PYG{p}{)}
\PYG{g+go}{array([ 2.+0.j,  0.+1.j,  1.+2.j])}
\end{sphinxVerbatim}

\begin{sphinxVerbatim}[commandchars=\\\{\}]
\PYG{g+gp}{\PYGZgt{}\PYGZgt{}\PYGZgt{} }\PYG{n}{np}\PYG{o}{.}\PYG{n}{sqrt}\PYG{p}{(}\PYG{p}{[}\PYG{l+m+mi}{4}\PYG{p}{,} \PYG{o}{\PYGZhy{}}\PYG{l+m+mi}{1}\PYG{p}{,} \PYG{n}{np}\PYG{o}{.}\PYG{n}{inf}\PYG{p}{]}\PYG{p}{)}
\PYG{g+go}{array([ 2., nan, inf])}
\end{sphinxVerbatim}

\end{fulllineitems}

\index{square() (in module symjax.tensor)@\spxentry{square()}\spxextra{in module symjax.tensor}}

\begin{fulllineitems}
\phantomsection\label{\detokenize{modules/tensor:symjax.tensor.square}}\pysiglinewithargsret{\sphinxbfcode{\sphinxupquote{square}}}{\emph{\DUrole{n}{x}}}{}
Return the element\sphinxhyphen{}wise square of the input.

LAX\sphinxhyphen{}backend implementation of {\hyperref[\detokenize{modules/tensor:symjax.tensor.square}]{\sphinxcrossref{\sphinxcode{\sphinxupquote{square()}}}}}.
ADDITIONOriginal docstring below.

LAX\sphinxhyphen{}backend implementation of {\hyperref[\detokenize{modules/tensor:symjax.tensor.square}]{\sphinxcrossref{\sphinxcode{\sphinxupquote{square()}}}}}.
Original docstring below.

square(x, /, out=None, {\color{red}\bfseries{}*}, where=True, casting=’same\_kind’, order=’K’, dtype=None, subok=True{[}, signature, extobj{]})
\begin{quote}\begin{description}
\item[{Returns}] \leavevmode
\sphinxstylestrong{out} \textendash{} Element\sphinxhyphen{}wise \sphinxtitleref{x*x}, of the same shape and dtype as \sphinxtitleref{x}.
This is a scalar if \sphinxtitleref{x} is a scalar.

\item[{Return type}] \leavevmode
ndarray or scalar

\end{description}\end{quote}

\sphinxstrong{See also:}

\sphinxcode{\sphinxupquote{numpy.linalg.matrix\_power()}}, {\hyperref[\detokenize{modules/tensor:symjax.tensor.sqrt}]{\sphinxcrossref{\sphinxcode{\sphinxupquote{sqrt()}}}}}, {\hyperref[\detokenize{modules/tensor:symjax.tensor.power}]{\sphinxcrossref{\sphinxcode{\sphinxupquote{power()}}}}}

\subsubsection*{Examples}

\begin{sphinxVerbatim}[commandchars=\\\{\}]
\PYG{g+gp}{\PYGZgt{}\PYGZgt{}\PYGZgt{} }\PYG{n}{np}\PYG{o}{.}\PYG{n}{square}\PYG{p}{(}\PYG{p}{[}\PYG{o}{\PYGZhy{}}\PYG{l+m+mi}{1}\PYG{n}{j}\PYG{p}{,} \PYG{l+m+mi}{1}\PYG{p}{]}\PYG{p}{)}
\PYG{g+go}{array([\PYGZhy{}1.\PYGZhy{}0.j,  1.+0.j])}
\end{sphinxVerbatim}

\end{fulllineitems}

\index{squeeze() (in module symjax.tensor)@\spxentry{squeeze()}\spxextra{in module symjax.tensor}}

\begin{fulllineitems}
\phantomsection\label{\detokenize{modules/tensor:symjax.tensor.squeeze}}\pysiglinewithargsret{\sphinxbfcode{\sphinxupquote{squeeze}}}{\emph{\DUrole{n}{a}}, \emph{\DUrole{n}{axis}\DUrole{o}{=}\DUrole{default_value}{None}}}{}
Remove single\sphinxhyphen{}dimensional entries from the shape of an array.

LAX\sphinxhyphen{}backend implementation of {\hyperref[\detokenize{modules/tensor:symjax.tensor.squeeze}]{\sphinxcrossref{\sphinxcode{\sphinxupquote{squeeze()}}}}}.
ADDITIONOriginal docstring below.

LAX\sphinxhyphen{}backend implementation of {\hyperref[\detokenize{modules/tensor:symjax.tensor.squeeze}]{\sphinxcrossref{\sphinxcode{\sphinxupquote{squeeze()}}}}}.
Original docstring below.
\begin{quote}\begin{description}
\item[{Returns}] \leavevmode
\sphinxstylestrong{squeezed} \textendash{} The input array, but with all or a subset of the
dimensions of length 1 removed. This is always \sphinxtitleref{a} itself
or a view into \sphinxtitleref{a}.

\item[{Return type}] \leavevmode
ndarray

\item[{Raises}] \leavevmode
\sphinxstyleliteralstrong{\sphinxupquote{ValueError}} \textendash{} If \sphinxtitleref{axis} is not None, and an axis being squeezed is not of length 1

\end{description}\end{quote}

\sphinxstrong{See also:}

\begin{description}
\item[{{\hyperref[\detokenize{modules/tensor:symjax.tensor.expand_dims}]{\sphinxcrossref{\sphinxcode{\sphinxupquote{expand\_dims()}}}}}}] \leavevmode
The inverse operation, adding singleton dimensions

\item[{{\hyperref[\detokenize{modules/tensor:symjax.tensor.reshape}]{\sphinxcrossref{\sphinxcode{\sphinxupquote{reshape()}}}}}}] \leavevmode
Insert, remove, and combine dimensions, and resize existing ones

\end{description}

\subsubsection*{Examples}

\begin{sphinxVerbatim}[commandchars=\\\{\}]
\PYG{g+gp}{\PYGZgt{}\PYGZgt{}\PYGZgt{} }\PYG{n}{x} \PYG{o}{=} \PYG{n}{np}\PYG{o}{.}\PYG{n}{array}\PYG{p}{(}\PYG{p}{[}\PYG{p}{[}\PYG{p}{[}\PYG{l+m+mi}{0}\PYG{p}{]}\PYG{p}{,} \PYG{p}{[}\PYG{l+m+mi}{1}\PYG{p}{]}\PYG{p}{,} \PYG{p}{[}\PYG{l+m+mi}{2}\PYG{p}{]}\PYG{p}{]}\PYG{p}{]}\PYG{p}{)}
\PYG{g+gp}{\PYGZgt{}\PYGZgt{}\PYGZgt{} }\PYG{n}{x}\PYG{o}{.}\PYG{n}{shape}
\PYG{g+go}{(1, 3, 1)}
\PYG{g+gp}{\PYGZgt{}\PYGZgt{}\PYGZgt{} }\PYG{n}{np}\PYG{o}{.}\PYG{n}{squeeze}\PYG{p}{(}\PYG{n}{x}\PYG{p}{)}\PYG{o}{.}\PYG{n}{shape}
\PYG{g+go}{(3,)}
\PYG{g+gp}{\PYGZgt{}\PYGZgt{}\PYGZgt{} }\PYG{n}{np}\PYG{o}{.}\PYG{n}{squeeze}\PYG{p}{(}\PYG{n}{x}\PYG{p}{,} \PYG{n}{axis}\PYG{o}{=}\PYG{l+m+mi}{0}\PYG{p}{)}\PYG{o}{.}\PYG{n}{shape}
\PYG{g+go}{(3, 1)}
\PYG{g+gp}{\PYGZgt{}\PYGZgt{}\PYGZgt{} }\PYG{n}{np}\PYG{o}{.}\PYG{n}{squeeze}\PYG{p}{(}\PYG{n}{x}\PYG{p}{,} \PYG{n}{axis}\PYG{o}{=}\PYG{l+m+mi}{1}\PYG{p}{)}\PYG{o}{.}\PYG{n}{shape}
\PYG{g+gt}{Traceback (most recent call last):}
\PYG{c}{...}
\PYG{g+gr}{ValueError}: \PYG{n}{cannot select an axis to squeeze out which has size not equal to one}
\PYG{g+gp}{\PYGZgt{}\PYGZgt{}\PYGZgt{} }\PYG{n}{np}\PYG{o}{.}\PYG{n}{squeeze}\PYG{p}{(}\PYG{n}{x}\PYG{p}{,} \PYG{n}{axis}\PYG{o}{=}\PYG{l+m+mi}{2}\PYG{p}{)}\PYG{o}{.}\PYG{n}{shape}
\PYG{g+go}{(1, 3)}
\end{sphinxVerbatim}

\end{fulllineitems}

\index{stack() (in module symjax.tensor)@\spxentry{stack()}\spxextra{in module symjax.tensor}}

\begin{fulllineitems}
\phantomsection\label{\detokenize{modules/tensor:symjax.tensor.stack}}\pysiglinewithargsret{\sphinxbfcode{\sphinxupquote{stack}}}{\emph{\DUrole{n}{arrays}}, \emph{\DUrole{n}{axis}\DUrole{o}{=}\DUrole{default_value}{0}}}{}
Join a sequence of arrays along a new axis.

LAX\sphinxhyphen{}backend implementation of {\hyperref[\detokenize{modules/tensor:symjax.tensor.stack}]{\sphinxcrossref{\sphinxcode{\sphinxupquote{stack()}}}}}.
ADDITIONOriginal docstring below.

LAX\sphinxhyphen{}backend implementation of {\hyperref[\detokenize{modules/tensor:symjax.tensor.stack}]{\sphinxcrossref{\sphinxcode{\sphinxupquote{stack()}}}}}.
Original docstring below.

The \sphinxcode{\sphinxupquote{axis}} parameter specifies the index of the new axis in the
dimensions of the result. For example, if \sphinxcode{\sphinxupquote{axis=0}} it will be the first
dimension and if \sphinxcode{\sphinxupquote{axis=\sphinxhyphen{}1}} it will be the last dimension.

\DUrole{versionmodified,added}{New in version 1.10.0.}
\begin{quote}\begin{description}
\item[{Returns}] \leavevmode
\sphinxstylestrong{stacked} \textendash{} The stacked array has one more dimension than the input arrays.

\item[{Return type}] \leavevmode
ndarray

\end{description}\end{quote}

\sphinxstrong{See also:}

\begin{description}
\item[{{\hyperref[\detokenize{modules/tensor:symjax.tensor.concatenate}]{\sphinxcrossref{\sphinxcode{\sphinxupquote{concatenate()}}}}}}] \leavevmode
Join a sequence of arrays along an existing axis.

\item[{{\hyperref[\detokenize{modules/tensor:symjax.tensor.split}]{\sphinxcrossref{\sphinxcode{\sphinxupquote{split()}}}}}}] \leavevmode
Split array into a list of multiple sub\sphinxhyphen{}arrays of equal size.

\item[{{\hyperref[\detokenize{modules/tensor:symjax.tensor.block}]{\sphinxcrossref{\sphinxcode{\sphinxupquote{block()}}}}}}] \leavevmode
Assemble arrays from blocks.

\end{description}

\subsubsection*{Examples}

\begin{sphinxVerbatim}[commandchars=\\\{\}]
\PYG{g+gp}{\PYGZgt{}\PYGZgt{}\PYGZgt{} }\PYG{n}{arrays} \PYG{o}{=} \PYG{p}{[}\PYG{n}{np}\PYG{o}{.}\PYG{n}{random}\PYG{o}{.}\PYG{n}{randn}\PYG{p}{(}\PYG{l+m+mi}{3}\PYG{p}{,} \PYG{l+m+mi}{4}\PYG{p}{)} \PYG{k}{for} \PYG{n}{\PYGZus{}} \PYG{o+ow}{in} \PYG{n+nb}{range}\PYG{p}{(}\PYG{l+m+mi}{10}\PYG{p}{)}\PYG{p}{]}
\PYG{g+gp}{\PYGZgt{}\PYGZgt{}\PYGZgt{} }\PYG{n}{np}\PYG{o}{.}\PYG{n}{stack}\PYG{p}{(}\PYG{n}{arrays}\PYG{p}{,} \PYG{n}{axis}\PYG{o}{=}\PYG{l+m+mi}{0}\PYG{p}{)}\PYG{o}{.}\PYG{n}{shape}
\PYG{g+go}{(10, 3, 4)}
\end{sphinxVerbatim}

\begin{sphinxVerbatim}[commandchars=\\\{\}]
\PYG{g+gp}{\PYGZgt{}\PYGZgt{}\PYGZgt{} }\PYG{n}{np}\PYG{o}{.}\PYG{n}{stack}\PYG{p}{(}\PYG{n}{arrays}\PYG{p}{,} \PYG{n}{axis}\PYG{o}{=}\PYG{l+m+mi}{1}\PYG{p}{)}\PYG{o}{.}\PYG{n}{shape}
\PYG{g+go}{(3, 10, 4)}
\end{sphinxVerbatim}

\begin{sphinxVerbatim}[commandchars=\\\{\}]
\PYG{g+gp}{\PYGZgt{}\PYGZgt{}\PYGZgt{} }\PYG{n}{np}\PYG{o}{.}\PYG{n}{stack}\PYG{p}{(}\PYG{n}{arrays}\PYG{p}{,} \PYG{n}{axis}\PYG{o}{=}\PYG{l+m+mi}{2}\PYG{p}{)}\PYG{o}{.}\PYG{n}{shape}
\PYG{g+go}{(3, 4, 10)}
\end{sphinxVerbatim}

\begin{sphinxVerbatim}[commandchars=\\\{\}]
\PYG{g+gp}{\PYGZgt{}\PYGZgt{}\PYGZgt{} }\PYG{n}{a} \PYG{o}{=} \PYG{n}{np}\PYG{o}{.}\PYG{n}{array}\PYG{p}{(}\PYG{p}{[}\PYG{l+m+mi}{1}\PYG{p}{,} \PYG{l+m+mi}{2}\PYG{p}{,} \PYG{l+m+mi}{3}\PYG{p}{]}\PYG{p}{)}
\PYG{g+gp}{\PYGZgt{}\PYGZgt{}\PYGZgt{} }\PYG{n}{b} \PYG{o}{=} \PYG{n}{np}\PYG{o}{.}\PYG{n}{array}\PYG{p}{(}\PYG{p}{[}\PYG{l+m+mi}{2}\PYG{p}{,} \PYG{l+m+mi}{3}\PYG{p}{,} \PYG{l+m+mi}{4}\PYG{p}{]}\PYG{p}{)}
\PYG{g+gp}{\PYGZgt{}\PYGZgt{}\PYGZgt{} }\PYG{n}{np}\PYG{o}{.}\PYG{n}{stack}\PYG{p}{(}\PYG{p}{(}\PYG{n}{a}\PYG{p}{,} \PYG{n}{b}\PYG{p}{)}\PYG{p}{)}
\PYG{g+go}{array([[1, 2, 3],}
\PYG{g+go}{       [2, 3, 4]])}
\end{sphinxVerbatim}

\begin{sphinxVerbatim}[commandchars=\\\{\}]
\PYG{g+gp}{\PYGZgt{}\PYGZgt{}\PYGZgt{} }\PYG{n}{np}\PYG{o}{.}\PYG{n}{stack}\PYG{p}{(}\PYG{p}{(}\PYG{n}{a}\PYG{p}{,} \PYG{n}{b}\PYG{p}{)}\PYG{p}{,} \PYG{n}{axis}\PYG{o}{=}\PYG{o}{\PYGZhy{}}\PYG{l+m+mi}{1}\PYG{p}{)}
\PYG{g+go}{array([[1, 2],}
\PYG{g+go}{       [2, 3],}
\PYG{g+go}{       [3, 4]])}
\end{sphinxVerbatim}

\end{fulllineitems}

\index{std() (in module symjax.tensor)@\spxentry{std()}\spxextra{in module symjax.tensor}}

\begin{fulllineitems}
\phantomsection\label{\detokenize{modules/tensor:symjax.tensor.std}}\pysiglinewithargsret{\sphinxbfcode{\sphinxupquote{std}}}{\emph{\DUrole{n}{a}}, \emph{\DUrole{n}{axis}\DUrole{o}{=}\DUrole{default_value}{None}}, \emph{\DUrole{n}{dtype}\DUrole{o}{=}\DUrole{default_value}{None}}, \emph{\DUrole{n}{out}\DUrole{o}{=}\DUrole{default_value}{None}}, \emph{\DUrole{n}{ddof}\DUrole{o}{=}\DUrole{default_value}{0}}, \emph{\DUrole{n}{keepdims}\DUrole{o}{=}\DUrole{default_value}{False}}}{}
Compute the standard deviation along the specified axis.

LAX\sphinxhyphen{}backend implementation of {\hyperref[\detokenize{modules/tensor:symjax.tensor.std}]{\sphinxcrossref{\sphinxcode{\sphinxupquote{std()}}}}}.
ADDITIONOriginal docstring below.

LAX\sphinxhyphen{}backend implementation of {\hyperref[\detokenize{modules/tensor:symjax.tensor.std}]{\sphinxcrossref{\sphinxcode{\sphinxupquote{std()}}}}}.
Original docstring below.

Returns the standard deviation, a measure of the spread of a distribution,
of the array elements. The standard deviation is computed for the
flattened array by default, otherwise over the specified axis.
\begin{quote}\begin{description}
\item[{Parameters}] \leavevmode
\sphinxstyleliteralstrong{\sphinxupquote{dtype}} (\sphinxstyleliteralemphasis{\sphinxupquote{dtype}}\sphinxstyleliteralemphasis{\sphinxupquote{, }}\sphinxstyleliteralemphasis{\sphinxupquote{optional}}) \textendash{} Type to use in computing the standard deviation. For arrays of
integer type the default is float64, for arrays of float types it is
the same as the array type.

\item[{Returns}] \leavevmode
\sphinxstylestrong{standard\_deviation} \textendash{} If \sphinxtitleref{out} is None, return a new array containing the standard deviation,
otherwise return a reference to the output array.

\item[{Return type}] \leavevmode
ndarray, see dtype parameter above.

\end{description}\end{quote}

\sphinxstrong{See also:}

{\hyperref[\detokenize{modules/tensor:symjax.tensor.var}]{\sphinxcrossref{\sphinxcode{\sphinxupquote{var()}}}}}, {\hyperref[\detokenize{modules/tensor:symjax.tensor.mean}]{\sphinxcrossref{\sphinxcode{\sphinxupquote{mean()}}}}}, \sphinxcode{\sphinxupquote{nanmean()}}, \sphinxcode{\sphinxupquote{nanstd()}}, \sphinxcode{\sphinxupquote{nanvar()}}, \sphinxcode{\sphinxupquote{ufuncs\sphinxhyphen{}output\sphinxhyphen{}type()}}

\subsubsection*{Notes}

The standard deviation is the square root of the average of the squared
deviations from the mean, i.e., \sphinxcode{\sphinxupquote{std = sqrt(mean(abs(x \sphinxhyphen{} x.mean())**2))}}.

The average squared deviation is normally calculated as
\sphinxcode{\sphinxupquote{x.sum() / N}}, where \sphinxcode{\sphinxupquote{N = len(x)}}.  If, however, \sphinxtitleref{ddof} is specified,
the divisor \sphinxcode{\sphinxupquote{N \sphinxhyphen{} ddof}} is used instead. In standard statistical
practice, \sphinxcode{\sphinxupquote{ddof=1}} provides an unbiased estimator of the variance
of the infinite population. \sphinxcode{\sphinxupquote{ddof=0}} provides a maximum likelihood
estimate of the variance for normally distributed variables. The
standard deviation computed in this function is the square root of
the estimated variance, so even with \sphinxcode{\sphinxupquote{ddof=1}}, it will not be an
unbiased estimate of the standard deviation per se.

Note that, for complex numbers, \sphinxtitleref{std} takes the absolute
value before squaring, so that the result is always real and nonnegative.

For floating\sphinxhyphen{}point input, the \sphinxstyleemphasis{std} is computed using the same
precision the input has. Depending on the input data, this can cause
the results to be inaccurate, especially for float32 (see example below).
Specifying a higher\sphinxhyphen{}accuracy accumulator using the \sphinxtitleref{dtype} keyword can
alleviate this issue.
\subsubsection*{Examples}

\begin{sphinxVerbatim}[commandchars=\\\{\}]
\PYG{g+gp}{\PYGZgt{}\PYGZgt{}\PYGZgt{} }\PYG{n}{a} \PYG{o}{=} \PYG{n}{np}\PYG{o}{.}\PYG{n}{array}\PYG{p}{(}\PYG{p}{[}\PYG{p}{[}\PYG{l+m+mi}{1}\PYG{p}{,} \PYG{l+m+mi}{2}\PYG{p}{]}\PYG{p}{,} \PYG{p}{[}\PYG{l+m+mi}{3}\PYG{p}{,} \PYG{l+m+mi}{4}\PYG{p}{]}\PYG{p}{]}\PYG{p}{)}
\PYG{g+gp}{\PYGZgt{}\PYGZgt{}\PYGZgt{} }\PYG{n}{np}\PYG{o}{.}\PYG{n}{std}\PYG{p}{(}\PYG{n}{a}\PYG{p}{)}
\PYG{g+go}{1.1180339887498949 \PYGZsh{} may vary}
\PYG{g+gp}{\PYGZgt{}\PYGZgt{}\PYGZgt{} }\PYG{n}{np}\PYG{o}{.}\PYG{n}{std}\PYG{p}{(}\PYG{n}{a}\PYG{p}{,} \PYG{n}{axis}\PYG{o}{=}\PYG{l+m+mi}{0}\PYG{p}{)}
\PYG{g+go}{array([1.,  1.])}
\PYG{g+gp}{\PYGZgt{}\PYGZgt{}\PYGZgt{} }\PYG{n}{np}\PYG{o}{.}\PYG{n}{std}\PYG{p}{(}\PYG{n}{a}\PYG{p}{,} \PYG{n}{axis}\PYG{o}{=}\PYG{l+m+mi}{1}\PYG{p}{)}
\PYG{g+go}{array([0.5,  0.5])}
\end{sphinxVerbatim}

In single precision, std() can be inaccurate:

\begin{sphinxVerbatim}[commandchars=\\\{\}]
\PYG{g+gp}{\PYGZgt{}\PYGZgt{}\PYGZgt{} }\PYG{n}{a} \PYG{o}{=} \PYG{n}{np}\PYG{o}{.}\PYG{n}{zeros}\PYG{p}{(}\PYG{p}{(}\PYG{l+m+mi}{2}\PYG{p}{,} \PYG{l+m+mi}{512}\PYG{o}{*}\PYG{l+m+mi}{512}\PYG{p}{)}\PYG{p}{,} \PYG{n}{dtype}\PYG{o}{=}\PYG{n}{np}\PYG{o}{.}\PYG{n}{float32}\PYG{p}{)}
\PYG{g+gp}{\PYGZgt{}\PYGZgt{}\PYGZgt{} }\PYG{n}{a}\PYG{p}{[}\PYG{l+m+mi}{0}\PYG{p}{,} \PYG{p}{:}\PYG{p}{]} \PYG{o}{=} \PYG{l+m+mf}{1.0}
\PYG{g+gp}{\PYGZgt{}\PYGZgt{}\PYGZgt{} }\PYG{n}{a}\PYG{p}{[}\PYG{l+m+mi}{1}\PYG{p}{,} \PYG{p}{:}\PYG{p}{]} \PYG{o}{=} \PYG{l+m+mf}{0.1}
\PYG{g+gp}{\PYGZgt{}\PYGZgt{}\PYGZgt{} }\PYG{n}{np}\PYG{o}{.}\PYG{n}{std}\PYG{p}{(}\PYG{n}{a}\PYG{p}{)}
\PYG{g+go}{0.45000005}
\end{sphinxVerbatim}

Computing the standard deviation in float64 is more accurate:

\begin{sphinxVerbatim}[commandchars=\\\{\}]
\PYG{g+gp}{\PYGZgt{}\PYGZgt{}\PYGZgt{} }\PYG{n}{np}\PYG{o}{.}\PYG{n}{std}\PYG{p}{(}\PYG{n}{a}\PYG{p}{,} \PYG{n}{dtype}\PYG{o}{=}\PYG{n}{np}\PYG{o}{.}\PYG{n}{float64}\PYG{p}{)}
\PYG{g+go}{0.44999999925494177 \PYGZsh{} may vary}
\end{sphinxVerbatim}

\end{fulllineitems}

\index{subtract() (in module symjax.tensor)@\spxentry{subtract()}\spxextra{in module symjax.tensor}}

\begin{fulllineitems}
\phantomsection\label{\detokenize{modules/tensor:symjax.tensor.subtract}}\pysiglinewithargsret{\sphinxbfcode{\sphinxupquote{subtract}}}{\emph{\DUrole{n}{x1}}, \emph{\DUrole{n}{x2}}}{}
Subtract arguments, element\sphinxhyphen{}wise.

LAX\sphinxhyphen{}backend implementation of {\hyperref[\detokenize{modules/tensor:symjax.tensor.subtract}]{\sphinxcrossref{\sphinxcode{\sphinxupquote{subtract()}}}}}.
ADDITIONOriginal docstring below.

LAX\sphinxhyphen{}backend implementation of {\hyperref[\detokenize{modules/tensor:symjax.tensor.subtract}]{\sphinxcrossref{\sphinxcode{\sphinxupquote{subtract()}}}}}.
Original docstring below.

subtract(x1, x2, /, out=None, {\color{red}\bfseries{}*}, where=True, casting=’same\_kind’, order=’K’, dtype=None, subok=True{[}, signature, extobj{]})
\begin{quote}\begin{description}
\item[{Returns}] \leavevmode
\sphinxstylestrong{y} \textendash{} The difference of \sphinxtitleref{x1} and \sphinxtitleref{x2}, element\sphinxhyphen{}wise.
This is a scalar if both \sphinxtitleref{x1} and \sphinxtitleref{x2} are scalars.

\item[{Return type}] \leavevmode
ndarray

\end{description}\end{quote}
\subsubsection*{Notes}

Equivalent to \sphinxcode{\sphinxupquote{x1 \sphinxhyphen{} x2}} in terms of array broadcasting.
\subsubsection*{Examples}

\begin{sphinxVerbatim}[commandchars=\\\{\}]
\PYG{g+gp}{\PYGZgt{}\PYGZgt{}\PYGZgt{} }\PYG{n}{np}\PYG{o}{.}\PYG{n}{subtract}\PYG{p}{(}\PYG{l+m+mf}{1.0}\PYG{p}{,} \PYG{l+m+mf}{4.0}\PYG{p}{)}
\PYG{g+go}{\PYGZhy{}3.0}
\end{sphinxVerbatim}

\begin{sphinxVerbatim}[commandchars=\\\{\}]
\PYG{g+gp}{\PYGZgt{}\PYGZgt{}\PYGZgt{} }\PYG{n}{x1} \PYG{o}{=} \PYG{n}{np}\PYG{o}{.}\PYG{n}{arange}\PYG{p}{(}\PYG{l+m+mf}{9.0}\PYG{p}{)}\PYG{o}{.}\PYG{n}{reshape}\PYG{p}{(}\PYG{p}{(}\PYG{l+m+mi}{3}\PYG{p}{,} \PYG{l+m+mi}{3}\PYG{p}{)}\PYG{p}{)}
\PYG{g+gp}{\PYGZgt{}\PYGZgt{}\PYGZgt{} }\PYG{n}{x2} \PYG{o}{=} \PYG{n}{np}\PYG{o}{.}\PYG{n}{arange}\PYG{p}{(}\PYG{l+m+mf}{3.0}\PYG{p}{)}
\PYG{g+gp}{\PYGZgt{}\PYGZgt{}\PYGZgt{} }\PYG{n}{np}\PYG{o}{.}\PYG{n}{subtract}\PYG{p}{(}\PYG{n}{x1}\PYG{p}{,} \PYG{n}{x2}\PYG{p}{)}
\PYG{g+go}{array([[ 0.,  0.,  0.],}
\PYG{g+go}{       [ 3.,  3.,  3.],}
\PYG{g+go}{       [ 6.,  6.,  6.]])}
\end{sphinxVerbatim}

\end{fulllineitems}

\index{sum() (in module symjax.tensor)@\spxentry{sum()}\spxextra{in module symjax.tensor}}

\begin{fulllineitems}
\phantomsection\label{\detokenize{modules/tensor:symjax.tensor.sum}}\pysiglinewithargsret{\sphinxbfcode{\sphinxupquote{sum}}}{\emph{\DUrole{n}{a}}, \emph{\DUrole{n}{axis}\DUrole{o}{=}\DUrole{default_value}{None}}, \emph{\DUrole{n}{dtype}\DUrole{o}{=}\DUrole{default_value}{None}}, \emph{\DUrole{n}{out}\DUrole{o}{=}\DUrole{default_value}{None}}, \emph{\DUrole{n}{keepdims}\DUrole{o}{=}\DUrole{default_value}{False}}}{}
Sum of array elements over a given axis.

LAX\sphinxhyphen{}backend implementation of {\hyperref[\detokenize{modules/tensor:symjax.tensor.sum}]{\sphinxcrossref{\sphinxcode{\sphinxupquote{sum()}}}}}.
ADDITIONOriginal docstring below.

LAX\sphinxhyphen{}backend implementation of {\hyperref[\detokenize{modules/tensor:symjax.tensor.sum}]{\sphinxcrossref{\sphinxcode{\sphinxupquote{sum()}}}}}.
Original docstring below.
\begin{quote}\begin{description}
\item[{Parameters}] \leavevmode
\sphinxstyleliteralstrong{\sphinxupquote{dtype}} (\sphinxstyleliteralemphasis{\sphinxupquote{dtype}}\sphinxstyleliteralemphasis{\sphinxupquote{, }}\sphinxstyleliteralemphasis{\sphinxupquote{optional}}) \textendash{} The type of the returned array and of the accumulator in which the
elements are summed.  The dtype of \sphinxtitleref{a} is used by default unless \sphinxtitleref{a}
has an integer dtype of less precision than the default platform
integer.  In that case, if \sphinxtitleref{a} is signed then the platform integer
is used while if \sphinxtitleref{a} is unsigned then an unsigned integer of the
same precision as the platform integer is used.

\item[{Returns}] \leavevmode
\sphinxstylestrong{sum\_along\_axis} \textendash{} An array with the same shape as \sphinxtitleref{a}, with the specified
axis removed.   If \sphinxtitleref{a} is a 0\sphinxhyphen{}d array, or if \sphinxtitleref{axis} is None, a scalar
is returned.  If an output array is specified, a reference to
\sphinxtitleref{out} is returned.

\item[{Return type}] \leavevmode
ndarray

\end{description}\end{quote}

\sphinxstrong{See also:}

\begin{description}
\item[{\sphinxcode{\sphinxupquote{ndarray.sum()}}}] \leavevmode
Equivalent method.

\item[{\sphinxcode{\sphinxupquote{add.reduce()}}}] \leavevmode
Equivalent functionality of \sphinxtitleref{add}.

\item[{{\hyperref[\detokenize{modules/tensor:symjax.tensor.cumsum}]{\sphinxcrossref{\sphinxcode{\sphinxupquote{cumsum()}}}}}}] \leavevmode
Cumulative sum of array elements.

\item[{\sphinxcode{\sphinxupquote{trapz()}}}] \leavevmode
Integration of array values using the composite trapezoidal rule.

\end{description}

{\hyperref[\detokenize{modules/tensor:symjax.tensor.mean}]{\sphinxcrossref{\sphinxcode{\sphinxupquote{mean()}}}}}, \sphinxcode{\sphinxupquote{average()}}

\subsubsection*{Notes}

Arithmetic is modular when using integer types, and no error is
raised on overflow.

The sum of an empty array is the neutral element 0:

\begin{sphinxVerbatim}[commandchars=\\\{\}]
\PYG{g+gp}{\PYGZgt{}\PYGZgt{}\PYGZgt{} }\PYG{n}{np}\PYG{o}{.}\PYG{n}{sum}\PYG{p}{(}\PYG{p}{[}\PYG{p}{]}\PYG{p}{)}
\PYG{g+go}{0.0}
\end{sphinxVerbatim}

For floating point numbers the numerical precision of sum (and
\sphinxcode{\sphinxupquote{np.add.reduce}}) is in general limited by directly adding each number
individually to the result causing rounding errors in every step.
However, often numpy will use a  numerically better approach (partial
pairwise summation) leading to improved precision in many use\sphinxhyphen{}cases.
This improved precision is always provided when no \sphinxcode{\sphinxupquote{axis}} is given.
When \sphinxcode{\sphinxupquote{axis}} is given, it will depend on which axis is summed.
Technically, to provide the best speed possible, the improved precision
is only used when the summation is along the fast axis in memory.
Note that the exact precision may vary depending on other parameters.
In contrast to NumPy, Python’s \sphinxcode{\sphinxupquote{math.fsum}} function uses a slower but
more precise approach to summation.
Especially when summing a large number of lower precision floating point
numbers, such as \sphinxcode{\sphinxupquote{float32}}, numerical errors can become significant.
In such cases it can be advisable to use \sphinxtitleref{dtype=”float64”} to use a higher
precision for the output.
\subsubsection*{Examples}

\begin{sphinxVerbatim}[commandchars=\\\{\}]
\PYG{g+gp}{\PYGZgt{}\PYGZgt{}\PYGZgt{} }\PYG{n}{np}\PYG{o}{.}\PYG{n}{sum}\PYG{p}{(}\PYG{p}{[}\PYG{l+m+mf}{0.5}\PYG{p}{,} \PYG{l+m+mf}{1.5}\PYG{p}{]}\PYG{p}{)}
\PYG{g+go}{2.0}
\PYG{g+gp}{\PYGZgt{}\PYGZgt{}\PYGZgt{} }\PYG{n}{np}\PYG{o}{.}\PYG{n}{sum}\PYG{p}{(}\PYG{p}{[}\PYG{l+m+mf}{0.5}\PYG{p}{,} \PYG{l+m+mf}{0.7}\PYG{p}{,} \PYG{l+m+mf}{0.2}\PYG{p}{,} \PYG{l+m+mf}{1.5}\PYG{p}{]}\PYG{p}{,} \PYG{n}{dtype}\PYG{o}{=}\PYG{n}{np}\PYG{o}{.}\PYG{n}{int32}\PYG{p}{)}
\PYG{g+go}{1}
\PYG{g+gp}{\PYGZgt{}\PYGZgt{}\PYGZgt{} }\PYG{n}{np}\PYG{o}{.}\PYG{n}{sum}\PYG{p}{(}\PYG{p}{[}\PYG{p}{[}\PYG{l+m+mi}{0}\PYG{p}{,} \PYG{l+m+mi}{1}\PYG{p}{]}\PYG{p}{,} \PYG{p}{[}\PYG{l+m+mi}{0}\PYG{p}{,} \PYG{l+m+mi}{5}\PYG{p}{]}\PYG{p}{]}\PYG{p}{)}
\PYG{g+go}{6}
\PYG{g+gp}{\PYGZgt{}\PYGZgt{}\PYGZgt{} }\PYG{n}{np}\PYG{o}{.}\PYG{n}{sum}\PYG{p}{(}\PYG{p}{[}\PYG{p}{[}\PYG{l+m+mi}{0}\PYG{p}{,} \PYG{l+m+mi}{1}\PYG{p}{]}\PYG{p}{,} \PYG{p}{[}\PYG{l+m+mi}{0}\PYG{p}{,} \PYG{l+m+mi}{5}\PYG{p}{]}\PYG{p}{]}\PYG{p}{,} \PYG{n}{axis}\PYG{o}{=}\PYG{l+m+mi}{0}\PYG{p}{)}
\PYG{g+go}{array([0, 6])}
\PYG{g+gp}{\PYGZgt{}\PYGZgt{}\PYGZgt{} }\PYG{n}{np}\PYG{o}{.}\PYG{n}{sum}\PYG{p}{(}\PYG{p}{[}\PYG{p}{[}\PYG{l+m+mi}{0}\PYG{p}{,} \PYG{l+m+mi}{1}\PYG{p}{]}\PYG{p}{,} \PYG{p}{[}\PYG{l+m+mi}{0}\PYG{p}{,} \PYG{l+m+mi}{5}\PYG{p}{]}\PYG{p}{]}\PYG{p}{,} \PYG{n}{axis}\PYG{o}{=}\PYG{l+m+mi}{1}\PYG{p}{)}
\PYG{g+go}{array([1, 5])}
\PYG{g+gp}{\PYGZgt{}\PYGZgt{}\PYGZgt{} }\PYG{n}{np}\PYG{o}{.}\PYG{n}{sum}\PYG{p}{(}\PYG{p}{[}\PYG{p}{[}\PYG{l+m+mi}{0}\PYG{p}{,} \PYG{l+m+mi}{1}\PYG{p}{]}\PYG{p}{,} \PYG{p}{[}\PYG{n}{np}\PYG{o}{.}\PYG{n}{nan}\PYG{p}{,} \PYG{l+m+mi}{5}\PYG{p}{]}\PYG{p}{]}\PYG{p}{,} \PYG{n}{where}\PYG{o}{=}\PYG{p}{[}\PYG{k+kc}{False}\PYG{p}{,} \PYG{k+kc}{True}\PYG{p}{]}\PYG{p}{,} \PYG{n}{axis}\PYG{o}{=}\PYG{l+m+mi}{1}\PYG{p}{)}
\PYG{g+go}{array([1., 5.])}
\end{sphinxVerbatim}

If the accumulator is too small, overflow occurs:

\begin{sphinxVerbatim}[commandchars=\\\{\}]
\PYG{g+gp}{\PYGZgt{}\PYGZgt{}\PYGZgt{} }\PYG{n}{np}\PYG{o}{.}\PYG{n}{ones}\PYG{p}{(}\PYG{l+m+mi}{128}\PYG{p}{,} \PYG{n}{dtype}\PYG{o}{=}\PYG{n}{np}\PYG{o}{.}\PYG{n}{int8}\PYG{p}{)}\PYG{o}{.}\PYG{n}{sum}\PYG{p}{(}\PYG{n}{dtype}\PYG{o}{=}\PYG{n}{np}\PYG{o}{.}\PYG{n}{int8}\PYG{p}{)}
\PYG{g+go}{\PYGZhy{}128}
\end{sphinxVerbatim}

You can also start the sum with a value other than zero:

\begin{sphinxVerbatim}[commandchars=\\\{\}]
\PYG{g+gp}{\PYGZgt{}\PYGZgt{}\PYGZgt{} }\PYG{n}{np}\PYG{o}{.}\PYG{n}{sum}\PYG{p}{(}\PYG{p}{[}\PYG{l+m+mi}{10}\PYG{p}{]}\PYG{p}{,} \PYG{n}{initial}\PYG{o}{=}\PYG{l+m+mi}{5}\PYG{p}{)}
\PYG{g+go}{15}
\end{sphinxVerbatim}

\end{fulllineitems}

\index{swapaxes() (in module symjax.tensor)@\spxentry{swapaxes()}\spxextra{in module symjax.tensor}}

\begin{fulllineitems}
\phantomsection\label{\detokenize{modules/tensor:symjax.tensor.swapaxes}}\pysiglinewithargsret{\sphinxbfcode{\sphinxupquote{swapaxes}}}{\emph{\DUrole{n}{a}}, \emph{\DUrole{n}{axis1}}, \emph{\DUrole{n}{axis2}}}{}
Interchange two axes of an array.

LAX\sphinxhyphen{}backend implementation of {\hyperref[\detokenize{modules/tensor:symjax.tensor.swapaxes}]{\sphinxcrossref{\sphinxcode{\sphinxupquote{swapaxes()}}}}}.
ADDITIONOriginal docstring below.

LAX\sphinxhyphen{}backend implementation of {\hyperref[\detokenize{modules/tensor:symjax.tensor.swapaxes}]{\sphinxcrossref{\sphinxcode{\sphinxupquote{swapaxes()}}}}}.
Original docstring below.
\begin{quote}\begin{description}
\item[{Returns}] \leavevmode
\sphinxstylestrong{a\_swapped} \textendash{} For NumPy \textgreater{}= 1.10.0, if \sphinxtitleref{a} is an ndarray, then a view of \sphinxtitleref{a} is
returned; otherwise a new array is created. For earlier NumPy
versions a view of \sphinxtitleref{a} is returned only if the order of the
axes is changed, otherwise the input array is returned.

\item[{Return type}] \leavevmode
ndarray

\end{description}\end{quote}
\subsubsection*{Examples}

\begin{sphinxVerbatim}[commandchars=\\\{\}]
\PYG{g+gp}{\PYGZgt{}\PYGZgt{}\PYGZgt{} }\PYG{n}{x} \PYG{o}{=} \PYG{n}{np}\PYG{o}{.}\PYG{n}{array}\PYG{p}{(}\PYG{p}{[}\PYG{p}{[}\PYG{l+m+mi}{1}\PYG{p}{,}\PYG{l+m+mi}{2}\PYG{p}{,}\PYG{l+m+mi}{3}\PYG{p}{]}\PYG{p}{]}\PYG{p}{)}
\PYG{g+gp}{\PYGZgt{}\PYGZgt{}\PYGZgt{} }\PYG{n}{np}\PYG{o}{.}\PYG{n}{swapaxes}\PYG{p}{(}\PYG{n}{x}\PYG{p}{,}\PYG{l+m+mi}{0}\PYG{p}{,}\PYG{l+m+mi}{1}\PYG{p}{)}
\PYG{g+go}{array([[1],}
\PYG{g+go}{       [2],}
\PYG{g+go}{       [3]])}
\end{sphinxVerbatim}

\begin{sphinxVerbatim}[commandchars=\\\{\}]
\PYG{g+gp}{\PYGZgt{}\PYGZgt{}\PYGZgt{} }\PYG{n}{x} \PYG{o}{=} \PYG{n}{np}\PYG{o}{.}\PYG{n}{array}\PYG{p}{(}\PYG{p}{[}\PYG{p}{[}\PYG{p}{[}\PYG{l+m+mi}{0}\PYG{p}{,}\PYG{l+m+mi}{1}\PYG{p}{]}\PYG{p}{,}\PYG{p}{[}\PYG{l+m+mi}{2}\PYG{p}{,}\PYG{l+m+mi}{3}\PYG{p}{]}\PYG{p}{]}\PYG{p}{,}\PYG{p}{[}\PYG{p}{[}\PYG{l+m+mi}{4}\PYG{p}{,}\PYG{l+m+mi}{5}\PYG{p}{]}\PYG{p}{,}\PYG{p}{[}\PYG{l+m+mi}{6}\PYG{p}{,}\PYG{l+m+mi}{7}\PYG{p}{]}\PYG{p}{]}\PYG{p}{]}\PYG{p}{)}
\PYG{g+gp}{\PYGZgt{}\PYGZgt{}\PYGZgt{} }\PYG{n}{x}
\PYG{g+go}{array([[[0, 1],}
\PYG{g+go}{        [2, 3]],}
\PYG{g+go}{       [[4, 5],}
\PYG{g+go}{        [6, 7]]])}
\end{sphinxVerbatim}

\begin{sphinxVerbatim}[commandchars=\\\{\}]
\PYG{g+gp}{\PYGZgt{}\PYGZgt{}\PYGZgt{} }\PYG{n}{np}\PYG{o}{.}\PYG{n}{swapaxes}\PYG{p}{(}\PYG{n}{x}\PYG{p}{,}\PYG{l+m+mi}{0}\PYG{p}{,}\PYG{l+m+mi}{2}\PYG{p}{)}
\PYG{g+go}{array([[[0, 4],}
\PYG{g+go}{        [2, 6]],}
\PYG{g+go}{       [[1, 5],}
\PYG{g+go}{        [3, 7]]])}
\end{sphinxVerbatim}

\end{fulllineitems}

\index{take() (in module symjax.tensor)@\spxentry{take()}\spxextra{in module symjax.tensor}}

\begin{fulllineitems}
\phantomsection\label{\detokenize{modules/tensor:symjax.tensor.take}}\pysiglinewithargsret{\sphinxbfcode{\sphinxupquote{take}}}{\emph{\DUrole{n}{a}}, \emph{\DUrole{n}{indices}}, \emph{\DUrole{n}{axis}\DUrole{o}{=}\DUrole{default_value}{None}}, \emph{\DUrole{n}{out}\DUrole{o}{=}\DUrole{default_value}{None}}, \emph{\DUrole{n}{mode}\DUrole{o}{=}\DUrole{default_value}{None}}}{}
Take elements from an array along an axis.

LAX\sphinxhyphen{}backend implementation of {\hyperref[\detokenize{modules/tensor:symjax.tensor.take}]{\sphinxcrossref{\sphinxcode{\sphinxupquote{take()}}}}}.
ADDITIONOriginal docstring below.

LAX\sphinxhyphen{}backend implementation of {\hyperref[\detokenize{modules/tensor:symjax.tensor.take}]{\sphinxcrossref{\sphinxcode{\sphinxupquote{take()}}}}}.
Original docstring below.

When axis is not None, this function does the same thing as “fancy”
indexing (indexing arrays using arrays); however, it can be easier to use
if you need elements along a given axis. A call such as
\sphinxcode{\sphinxupquote{np.take(arr, indices, axis=3)}} is equivalent to
\sphinxcode{\sphinxupquote{arr{[}:,:,:,indices,...{]}}}.

Explained without fancy indexing, this is equivalent to the following use
of \sphinxtitleref{ndindex}, which sets each of \sphinxcode{\sphinxupquote{ii}}, \sphinxcode{\sphinxupquote{jj}}, and \sphinxcode{\sphinxupquote{kk}} to a tuple of
indices:

\begin{sphinxVerbatim}[commandchars=\\\{\}]
\PYG{n}{Ni}\PYG{p}{,} \PYG{n}{Nk} \PYG{o}{=} \PYG{n}{a}\PYG{o}{.}\PYG{n}{shape}\PYG{p}{[}\PYG{p}{:}\PYG{n}{axis}\PYG{p}{]}\PYG{p}{,} \PYG{n}{a}\PYG{o}{.}\PYG{n}{shape}\PYG{p}{[}\PYG{n}{axis}\PYG{o}{+}\PYG{l+m+mi}{1}\PYG{p}{:}\PYG{p}{]}
\PYG{n}{Nj} \PYG{o}{=} \PYG{n}{indices}\PYG{o}{.}\PYG{n}{shape}
\PYG{k}{for} \PYG{n}{ii} \PYG{o+ow}{in} \PYG{n}{ndindex}\PYG{p}{(}\PYG{n}{Ni}\PYG{p}{)}\PYG{p}{:}
    \PYG{k}{for} \PYG{n}{jj} \PYG{o+ow}{in} \PYG{n}{ndindex}\PYG{p}{(}\PYG{n}{Nj}\PYG{p}{)}\PYG{p}{:}
        \PYG{k}{for} \PYG{n}{kk} \PYG{o+ow}{in} \PYG{n}{ndindex}\PYG{p}{(}\PYG{n}{Nk}\PYG{p}{)}\PYG{p}{:}
            \PYG{n}{out}\PYG{p}{[}\PYG{n}{ii} \PYG{o}{+} \PYG{n}{jj} \PYG{o}{+} \PYG{n}{kk}\PYG{p}{]} \PYG{o}{=} \PYG{n}{a}\PYG{p}{[}\PYG{n}{ii} \PYG{o}{+} \PYG{p}{(}\PYG{n}{indices}\PYG{p}{[}\PYG{n}{jj}\PYG{p}{]}\PYG{p}{,}\PYG{p}{)} \PYG{o}{+} \PYG{n}{kk}\PYG{p}{]}
\end{sphinxVerbatim}
\begin{quote}\begin{description}
\item[{Returns}] \leavevmode
\sphinxstylestrong{out} \textendash{} The returned array has the same type as \sphinxtitleref{a}.

\item[{Return type}] \leavevmode
ndarray (Ni…, Nj…, Nk…)

\end{description}\end{quote}

\sphinxstrong{See also:}

\begin{description}
\item[{\sphinxcode{\sphinxupquote{compress()}}}] \leavevmode
Take elements using a boolean mask

\item[{\sphinxcode{\sphinxupquote{ndarray.take()}}}] \leavevmode
equivalent method

\item[{{\hyperref[\detokenize{modules/tensor:symjax.tensor.take_along_axis}]{\sphinxcrossref{\sphinxcode{\sphinxupquote{take\_along\_axis()}}}}}}] \leavevmode
Take elements by matching the array and the index arrays

\end{description}

\subsubsection*{Notes}

By eliminating the inner loop in the description above, and using \sphinxtitleref{s\_} to
build simple slice objects, \sphinxtitleref{take} can be expressed  in terms of applying
fancy indexing to each 1\sphinxhyphen{}d slice:

\begin{sphinxVerbatim}[commandchars=\\\{\}]
\PYG{n}{Ni}\PYG{p}{,} \PYG{n}{Nk} \PYG{o}{=} \PYG{n}{a}\PYG{o}{.}\PYG{n}{shape}\PYG{p}{[}\PYG{p}{:}\PYG{n}{axis}\PYG{p}{]}\PYG{p}{,} \PYG{n}{a}\PYG{o}{.}\PYG{n}{shape}\PYG{p}{[}\PYG{n}{axis}\PYG{o}{+}\PYG{l+m+mi}{1}\PYG{p}{:}\PYG{p}{]}
\PYG{k}{for} \PYG{n}{ii} \PYG{o+ow}{in} \PYG{n}{ndindex}\PYG{p}{(}\PYG{n}{Ni}\PYG{p}{)}\PYG{p}{:}
    \PYG{k}{for} \PYG{n}{kk} \PYG{o+ow}{in} \PYG{n}{ndindex}\PYG{p}{(}\PYG{n}{Nj}\PYG{p}{)}\PYG{p}{:}
        \PYG{n}{out}\PYG{p}{[}\PYG{n}{ii} \PYG{o}{+} \PYG{n}{s\PYGZus{}}\PYG{p}{[}\PYG{o}{.}\PYG{o}{.}\PYG{o}{.}\PYG{p}{,}\PYG{p}{]} \PYG{o}{+} \PYG{n}{kk}\PYG{p}{]} \PYG{o}{=} \PYG{n}{a}\PYG{p}{[}\PYG{n}{ii} \PYG{o}{+} \PYG{n}{s\PYGZus{}}\PYG{p}{[}\PYG{p}{:}\PYG{p}{,}\PYG{p}{]} \PYG{o}{+} \PYG{n}{kk}\PYG{p}{]}\PYG{p}{[}\PYG{n}{indices}\PYG{p}{]}
\end{sphinxVerbatim}

For this reason, it is equivalent to (but faster than) the following use
of \sphinxtitleref{apply\_along\_axis}:

\begin{sphinxVerbatim}[commandchars=\\\{\}]
\PYG{n}{out} \PYG{o}{=} \PYG{n}{np}\PYG{o}{.}\PYG{n}{apply\PYGZus{}along\PYGZus{}axis}\PYG{p}{(}\PYG{k}{lambda} \PYG{n}{a\PYGZus{}1d}\PYG{p}{:} \PYG{n}{a\PYGZus{}1d}\PYG{p}{[}\PYG{n}{indices}\PYG{p}{]}\PYG{p}{,} \PYG{n}{axis}\PYG{p}{,} \PYG{n}{a}\PYG{p}{)}
\end{sphinxVerbatim}
\subsubsection*{Examples}

\begin{sphinxVerbatim}[commandchars=\\\{\}]
\PYG{g+gp}{\PYGZgt{}\PYGZgt{}\PYGZgt{} }\PYG{n}{a} \PYG{o}{=} \PYG{p}{[}\PYG{l+m+mi}{4}\PYG{p}{,} \PYG{l+m+mi}{3}\PYG{p}{,} \PYG{l+m+mi}{5}\PYG{p}{,} \PYG{l+m+mi}{7}\PYG{p}{,} \PYG{l+m+mi}{6}\PYG{p}{,} \PYG{l+m+mi}{8}\PYG{p}{]}
\PYG{g+gp}{\PYGZgt{}\PYGZgt{}\PYGZgt{} }\PYG{n}{indices} \PYG{o}{=} \PYG{p}{[}\PYG{l+m+mi}{0}\PYG{p}{,} \PYG{l+m+mi}{1}\PYG{p}{,} \PYG{l+m+mi}{4}\PYG{p}{]}
\PYG{g+gp}{\PYGZgt{}\PYGZgt{}\PYGZgt{} }\PYG{n}{np}\PYG{o}{.}\PYG{n}{take}\PYG{p}{(}\PYG{n}{a}\PYG{p}{,} \PYG{n}{indices}\PYG{p}{)}
\PYG{g+go}{array([4, 3, 6])}
\end{sphinxVerbatim}

In this example if \sphinxtitleref{a} is an ndarray, “fancy” indexing can be used.

\begin{sphinxVerbatim}[commandchars=\\\{\}]
\PYG{g+gp}{\PYGZgt{}\PYGZgt{}\PYGZgt{} }\PYG{n}{a} \PYG{o}{=} \PYG{n}{np}\PYG{o}{.}\PYG{n}{array}\PYG{p}{(}\PYG{n}{a}\PYG{p}{)}
\PYG{g+gp}{\PYGZgt{}\PYGZgt{}\PYGZgt{} }\PYG{n}{a}\PYG{p}{[}\PYG{n}{indices}\PYG{p}{]}
\PYG{g+go}{array([4, 3, 6])}
\end{sphinxVerbatim}

If \sphinxtitleref{indices} is not one dimensional, the output also has these dimensions.

\begin{sphinxVerbatim}[commandchars=\\\{\}]
\PYG{g+gp}{\PYGZgt{}\PYGZgt{}\PYGZgt{} }\PYG{n}{np}\PYG{o}{.}\PYG{n}{take}\PYG{p}{(}\PYG{n}{a}\PYG{p}{,} \PYG{p}{[}\PYG{p}{[}\PYG{l+m+mi}{0}\PYG{p}{,} \PYG{l+m+mi}{1}\PYG{p}{]}\PYG{p}{,} \PYG{p}{[}\PYG{l+m+mi}{2}\PYG{p}{,} \PYG{l+m+mi}{3}\PYG{p}{]}\PYG{p}{]}\PYG{p}{)}
\PYG{g+go}{array([[4, 3],}
\PYG{g+go}{       [5, 7]])}
\end{sphinxVerbatim}

\end{fulllineitems}

\index{take\_along\_axis() (in module symjax.tensor)@\spxentry{take\_along\_axis()}\spxextra{in module symjax.tensor}}

\begin{fulllineitems}
\phantomsection\label{\detokenize{modules/tensor:symjax.tensor.take_along_axis}}\pysiglinewithargsret{\sphinxbfcode{\sphinxupquote{take\_along\_axis}}}{\emph{\DUrole{n}{arr}}, \emph{\DUrole{n}{indices}}, \emph{\DUrole{n}{axis}}}{}
Take values from the input array by matching 1d index and data slices.

LAX\sphinxhyphen{}backend implementation of {\hyperref[\detokenize{modules/tensor:symjax.tensor.take_along_axis}]{\sphinxcrossref{\sphinxcode{\sphinxupquote{take\_along\_axis()}}}}}.
ADDITIONOriginal docstring below.

LAX\sphinxhyphen{}backend implementation of {\hyperref[\detokenize{modules/tensor:symjax.tensor.take_along_axis}]{\sphinxcrossref{\sphinxcode{\sphinxupquote{take\_along\_axis()}}}}}.
Original docstring below.
\begin{quote}
\begin{quote}

This iterates over matching 1d slices oriented along the specified axis in
the index and data arrays, and uses the former to look up values in the
latter. These slices can be different lengths.

Functions returning an index along an axis, like \sphinxtitleref{argsort} and
\sphinxtitleref{argpartition}, produce suitable indices for this function.

\DUrole{versionmodified,added}{New in version 1.15.0.}
\end{quote}
\begin{description}
\item[{Returns}] \leavevmode\begin{description}
\item[{out: ndarray (Ni…, J, Nk…)}] \leavevmode
The indexed result.

\end{description}

This is equivalent to (but faster than) the following use of \sphinxtitleref{ndindex} and
\sphinxtitleref{s\_}, which sets each of \sphinxcode{\sphinxupquote{ii}} and \sphinxcode{\sphinxupquote{kk}} to a tuple of indices:

\begin{sphinxVerbatim}[commandchars=\\\{\}]
\PYG{n}{Ni}\PYG{p}{,} \PYG{n}{M}\PYG{p}{,} \PYG{n}{Nk} \PYG{o}{=} \PYG{n}{a}\PYG{o}{.}\PYG{n}{shape}\PYG{p}{[}\PYG{p}{:}\PYG{n}{axis}\PYG{p}{]}\PYG{p}{,} \PYG{n}{a}\PYG{o}{.}\PYG{n}{shape}\PYG{p}{[}\PYG{n}{axis}\PYG{p}{]}\PYG{p}{,} \PYG{n}{a}\PYG{o}{.}\PYG{n}{shape}\PYG{p}{[}\PYG{n}{axis}\PYG{o}{+}\PYG{l+m+mi}{1}\PYG{p}{:}\PYG{p}{]}
\PYG{n}{J} \PYG{o}{=} \PYG{n}{indices}\PYG{o}{.}\PYG{n}{shape}\PYG{p}{[}\PYG{n}{axis}\PYG{p}{]}  \PYG{c+c1}{\PYGZsh{} Need not equal M}
\PYG{n}{out} \PYG{o}{=} \PYG{n}{np}\PYG{o}{.}\PYG{n}{empty}\PYG{p}{(}\PYG{n}{Ni} \PYG{o}{+} \PYG{p}{(}\PYG{n}{J}\PYG{p}{,}\PYG{p}{)} \PYG{o}{+} \PYG{n}{Nk}\PYG{p}{)}

\PYG{k}{for} \PYG{n}{ii} \PYG{o+ow}{in} \PYG{n}{ndindex}\PYG{p}{(}\PYG{n}{Ni}\PYG{p}{)}\PYG{p}{:}
    \PYG{k}{for} \PYG{n}{kk} \PYG{o+ow}{in} \PYG{n}{ndindex}\PYG{p}{(}\PYG{n}{Nk}\PYG{p}{)}\PYG{p}{:}
        \PYG{n}{a\PYGZus{}1d}       \PYG{o}{=} \PYG{n}{a}      \PYG{p}{[}\PYG{n}{ii} \PYG{o}{+} \PYG{n}{s\PYGZus{}}\PYG{p}{[}\PYG{p}{:}\PYG{p}{,}\PYG{p}{]} \PYG{o}{+} \PYG{n}{kk}\PYG{p}{]}
        \PYG{n}{indices\PYGZus{}1d} \PYG{o}{=} \PYG{n}{indices}\PYG{p}{[}\PYG{n}{ii} \PYG{o}{+} \PYG{n}{s\PYGZus{}}\PYG{p}{[}\PYG{p}{:}\PYG{p}{,}\PYG{p}{]} \PYG{o}{+} \PYG{n}{kk}\PYG{p}{]}
        \PYG{n}{out\PYGZus{}1d}     \PYG{o}{=} \PYG{n}{out}    \PYG{p}{[}\PYG{n}{ii} \PYG{o}{+} \PYG{n}{s\PYGZus{}}\PYG{p}{[}\PYG{p}{:}\PYG{p}{,}\PYG{p}{]} \PYG{o}{+} \PYG{n}{kk}\PYG{p}{]}
        \PYG{k}{for} \PYG{n}{j} \PYG{o+ow}{in} \PYG{n+nb}{range}\PYG{p}{(}\PYG{n}{J}\PYG{p}{)}\PYG{p}{:}
            \PYG{n}{out\PYGZus{}1d}\PYG{p}{[}\PYG{n}{j}\PYG{p}{]} \PYG{o}{=} \PYG{n}{a\PYGZus{}1d}\PYG{p}{[}\PYG{n}{indices\PYGZus{}1d}\PYG{p}{[}\PYG{n}{j}\PYG{p}{]}\PYG{p}{]}
\end{sphinxVerbatim}

Equivalently, eliminating the inner loop, the last two lines would be:

\begin{sphinxVerbatim}[commandchars=\\\{\}]
\PYG{n}{out\PYGZus{}1d}\PYG{p}{[}\PYG{p}{:}\PYG{p}{]} \PYG{o}{=} \PYG{n}{a\PYGZus{}1d}\PYG{p}{[}\PYG{n}{indices\PYGZus{}1d}\PYG{p}{]}
\end{sphinxVerbatim}

take : Take along an axis, using the same indices for every 1d slice
put\_along\_axis :
\begin{quote}

Put values into the destination array by matching 1d index and data slices
\end{quote}

For this sample array

\begin{sphinxVerbatim}[commandchars=\\\{\}]
\PYG{g+gp}{\PYGZgt{}\PYGZgt{}\PYGZgt{} }\PYG{n}{a} \PYG{o}{=} \PYG{n}{np}\PYG{o}{.}\PYG{n}{array}\PYG{p}{(}\PYG{p}{[}\PYG{p}{[}\PYG{l+m+mi}{10}\PYG{p}{,} \PYG{l+m+mi}{30}\PYG{p}{,} \PYG{l+m+mi}{20}\PYG{p}{]}\PYG{p}{,} \PYG{p}{[}\PYG{l+m+mi}{60}\PYG{p}{,} \PYG{l+m+mi}{40}\PYG{p}{,} \PYG{l+m+mi}{50}\PYG{p}{]}\PYG{p}{]}\PYG{p}{)}
\end{sphinxVerbatim}

We can sort either by using sort directly, or argsort and this function

\begin{sphinxVerbatim}[commandchars=\\\{\}]
\PYG{g+gp}{\PYGZgt{}\PYGZgt{}\PYGZgt{} }\PYG{n}{np}\PYG{o}{.}\PYG{n}{sort}\PYG{p}{(}\PYG{n}{a}\PYG{p}{,} \PYG{n}{axis}\PYG{o}{=}\PYG{l+m+mi}{1}\PYG{p}{)}
\PYG{g+go}{array([[10, 20, 30],}
\PYG{g+go}{       [40, 50, 60]])}
\PYG{g+gp}{\PYGZgt{}\PYGZgt{}\PYGZgt{} }\PYG{n}{ai} \PYG{o}{=} \PYG{n}{np}\PYG{o}{.}\PYG{n}{argsort}\PYG{p}{(}\PYG{n}{a}\PYG{p}{,} \PYG{n}{axis}\PYG{o}{=}\PYG{l+m+mi}{1}\PYG{p}{)}\PYG{p}{;} \PYG{n}{ai}
\PYG{g+go}{array([[0, 2, 1],}
\PYG{g+go}{       [1, 2, 0]])}
\PYG{g+gp}{\PYGZgt{}\PYGZgt{}\PYGZgt{} }\PYG{n}{np}\PYG{o}{.}\PYG{n}{take\PYGZus{}along\PYGZus{}axis}\PYG{p}{(}\PYG{n}{a}\PYG{p}{,} \PYG{n}{ai}\PYG{p}{,} \PYG{n}{axis}\PYG{o}{=}\PYG{l+m+mi}{1}\PYG{p}{)}
\PYG{g+go}{array([[10, 20, 30],}
\PYG{g+go}{       [40, 50, 60]])}
\end{sphinxVerbatim}

The same works for max and min, if you expand the dimensions:

\begin{sphinxVerbatim}[commandchars=\\\{\}]
\PYG{g+gp}{\PYGZgt{}\PYGZgt{}\PYGZgt{} }\PYG{n}{np}\PYG{o}{.}\PYG{n}{expand\PYGZus{}dims}\PYG{p}{(}\PYG{n}{np}\PYG{o}{.}\PYG{n}{max}\PYG{p}{(}\PYG{n}{a}\PYG{p}{,} \PYG{n}{axis}\PYG{o}{=}\PYG{l+m+mi}{1}\PYG{p}{)}\PYG{p}{,} \PYG{n}{axis}\PYG{o}{=}\PYG{l+m+mi}{1}\PYG{p}{)}
\PYG{g+go}{array([[30],}
\PYG{g+go}{       [60]])}
\PYG{g+gp}{\PYGZgt{}\PYGZgt{}\PYGZgt{} }\PYG{n}{ai} \PYG{o}{=} \PYG{n}{np}\PYG{o}{.}\PYG{n}{expand\PYGZus{}dims}\PYG{p}{(}\PYG{n}{np}\PYG{o}{.}\PYG{n}{argmax}\PYG{p}{(}\PYG{n}{a}\PYG{p}{,} \PYG{n}{axis}\PYG{o}{=}\PYG{l+m+mi}{1}\PYG{p}{)}\PYG{p}{,} \PYG{n}{axis}\PYG{o}{=}\PYG{l+m+mi}{1}\PYG{p}{)}
\PYG{g+gp}{\PYGZgt{}\PYGZgt{}\PYGZgt{} }\PYG{n}{ai}
\PYG{g+go}{array([[1],}
\PYG{g+go}{       [0]])}
\PYG{g+gp}{\PYGZgt{}\PYGZgt{}\PYGZgt{} }\PYG{n}{np}\PYG{o}{.}\PYG{n}{take\PYGZus{}along\PYGZus{}axis}\PYG{p}{(}\PYG{n}{a}\PYG{p}{,} \PYG{n}{ai}\PYG{p}{,} \PYG{n}{axis}\PYG{o}{=}\PYG{l+m+mi}{1}\PYG{p}{)}
\PYG{g+go}{array([[30],}
\PYG{g+go}{       [60]])}
\end{sphinxVerbatim}

If we want to get the max and min at the same time, we can stack the
indices first

\begin{sphinxVerbatim}[commandchars=\\\{\}]
\PYG{g+gp}{\PYGZgt{}\PYGZgt{}\PYGZgt{} }\PYG{n}{ai\PYGZus{}min} \PYG{o}{=} \PYG{n}{np}\PYG{o}{.}\PYG{n}{expand\PYGZus{}dims}\PYG{p}{(}\PYG{n}{np}\PYG{o}{.}\PYG{n}{argmin}\PYG{p}{(}\PYG{n}{a}\PYG{p}{,} \PYG{n}{axis}\PYG{o}{=}\PYG{l+m+mi}{1}\PYG{p}{)}\PYG{p}{,} \PYG{n}{axis}\PYG{o}{=}\PYG{l+m+mi}{1}\PYG{p}{)}
\PYG{g+gp}{\PYGZgt{}\PYGZgt{}\PYGZgt{} }\PYG{n}{ai\PYGZus{}max} \PYG{o}{=} \PYG{n}{np}\PYG{o}{.}\PYG{n}{expand\PYGZus{}dims}\PYG{p}{(}\PYG{n}{np}\PYG{o}{.}\PYG{n}{argmax}\PYG{p}{(}\PYG{n}{a}\PYG{p}{,} \PYG{n}{axis}\PYG{o}{=}\PYG{l+m+mi}{1}\PYG{p}{)}\PYG{p}{,} \PYG{n}{axis}\PYG{o}{=}\PYG{l+m+mi}{1}\PYG{p}{)}
\PYG{g+gp}{\PYGZgt{}\PYGZgt{}\PYGZgt{} }\PYG{n}{ai} \PYG{o}{=} \PYG{n}{np}\PYG{o}{.}\PYG{n}{concatenate}\PYG{p}{(}\PYG{p}{[}\PYG{n}{ai\PYGZus{}min}\PYG{p}{,} \PYG{n}{ai\PYGZus{}max}\PYG{p}{]}\PYG{p}{,} \PYG{n}{axis}\PYG{o}{=}\PYG{l+m+mi}{1}\PYG{p}{)}
\PYG{g+gp}{\PYGZgt{}\PYGZgt{}\PYGZgt{} }\PYG{n}{ai}
\PYG{g+go}{array([[0, 1],}
\PYG{g+go}{       [1, 0]])}
\PYG{g+gp}{\PYGZgt{}\PYGZgt{}\PYGZgt{} }\PYG{n}{np}\PYG{o}{.}\PYG{n}{take\PYGZus{}along\PYGZus{}axis}\PYG{p}{(}\PYG{n}{a}\PYG{p}{,} \PYG{n}{ai}\PYG{p}{,} \PYG{n}{axis}\PYG{o}{=}\PYG{l+m+mi}{1}\PYG{p}{)}
\PYG{g+go}{array([[10, 30],}
\PYG{g+go}{       [40, 60]])}
\end{sphinxVerbatim}

\end{description}
\end{quote}

\end{fulllineitems}

\index{tan() (in module symjax.tensor)@\spxentry{tan()}\spxextra{in module symjax.tensor}}

\begin{fulllineitems}
\phantomsection\label{\detokenize{modules/tensor:symjax.tensor.tan}}\pysiglinewithargsret{\sphinxbfcode{\sphinxupquote{tan}}}{\emph{\DUrole{n}{x}}}{}
Compute tangent element\sphinxhyphen{}wise.

LAX\sphinxhyphen{}backend implementation of {\hyperref[\detokenize{modules/tensor:symjax.tensor.tan}]{\sphinxcrossref{\sphinxcode{\sphinxupquote{tan()}}}}}.
ADDITIONOriginal docstring below.

LAX\sphinxhyphen{}backend implementation of {\hyperref[\detokenize{modules/tensor:symjax.tensor.tan}]{\sphinxcrossref{\sphinxcode{\sphinxupquote{tan()}}}}}.
Original docstring below.

tan(x, /, out=None, {\color{red}\bfseries{}*}, where=True, casting=’same\_kind’, order=’K’, dtype=None, subok=True{[}, signature, extobj{]})

Equivalent to \sphinxcode{\sphinxupquote{np.sin(x)/np.cos(x)}} element\sphinxhyphen{}wise.
\begin{quote}\begin{description}
\item[{Returns}] \leavevmode
\sphinxstylestrong{y} \textendash{} The corresponding tangent values.
This is a scalar if \sphinxtitleref{x} is a scalar.

\item[{Return type}] \leavevmode
ndarray

\end{description}\end{quote}
\subsubsection*{Notes}

If \sphinxtitleref{out} is provided, the function writes the result into it,
and returns a reference to \sphinxtitleref{out}.  (See Examples)
\subsubsection*{References}

M. Abramowitz and I. A. Stegun, Handbook of Mathematical Functions.
New York, NY: Dover, 1972.
\subsubsection*{Examples}

\begin{sphinxVerbatim}[commandchars=\\\{\}]
\PYG{g+gp}{\PYGZgt{}\PYGZgt{}\PYGZgt{} }\PYG{k+kn}{from} \PYG{n+nn}{math} \PYG{k+kn}{import} \PYG{n}{pi}
\PYG{g+gp}{\PYGZgt{}\PYGZgt{}\PYGZgt{} }\PYG{n}{np}\PYG{o}{.}\PYG{n}{tan}\PYG{p}{(}\PYG{n}{np}\PYG{o}{.}\PYG{n}{array}\PYG{p}{(}\PYG{p}{[}\PYG{o}{\PYGZhy{}}\PYG{n}{pi}\PYG{p}{,}\PYG{n}{pi}\PYG{o}{/}\PYG{l+m+mi}{2}\PYG{p}{,}\PYG{n}{pi}\PYG{p}{]}\PYG{p}{)}\PYG{p}{)}
\PYG{g+go}{array([  1.22460635e\PYGZhy{}16,   1.63317787e+16,  \PYGZhy{}1.22460635e\PYGZhy{}16])}
\PYG{g+go}{\PYGZgt{}\PYGZgt{}\PYGZgt{}}
\PYG{g+gp}{\PYGZgt{}\PYGZgt{}\PYGZgt{} }\PYG{c+c1}{\PYGZsh{} Example of providing the optional output parameter illustrating}
\PYG{g+gp}{\PYGZgt{}\PYGZgt{}\PYGZgt{} }\PYG{c+c1}{\PYGZsh{} that what is returned is a reference to said parameter}
\PYG{g+gp}{\PYGZgt{}\PYGZgt{}\PYGZgt{} }\PYG{n}{out1} \PYG{o}{=} \PYG{n}{np}\PYG{o}{.}\PYG{n}{array}\PYG{p}{(}\PYG{p}{[}\PYG{l+m+mi}{0}\PYG{p}{]}\PYG{p}{,} \PYG{n}{dtype}\PYG{o}{=}\PYG{l+s+s1}{\PYGZsq{}}\PYG{l+s+s1}{d}\PYG{l+s+s1}{\PYGZsq{}}\PYG{p}{)}
\PYG{g+gp}{\PYGZgt{}\PYGZgt{}\PYGZgt{} }\PYG{n}{out2} \PYG{o}{=} \PYG{n}{np}\PYG{o}{.}\PYG{n}{cos}\PYG{p}{(}\PYG{p}{[}\PYG{l+m+mf}{0.1}\PYG{p}{]}\PYG{p}{,} \PYG{n}{out1}\PYG{p}{)}
\PYG{g+gp}{\PYGZgt{}\PYGZgt{}\PYGZgt{} }\PYG{n}{out2} \PYG{o+ow}{is} \PYG{n}{out1}
\PYG{g+go}{True}
\PYG{g+go}{\PYGZgt{}\PYGZgt{}\PYGZgt{}}
\PYG{g+gp}{\PYGZgt{}\PYGZgt{}\PYGZgt{} }\PYG{c+c1}{\PYGZsh{} Example of ValueError due to provision of shape mis\PYGZhy{}matched `out`}
\PYG{g+gp}{\PYGZgt{}\PYGZgt{}\PYGZgt{} }\PYG{n}{np}\PYG{o}{.}\PYG{n}{cos}\PYG{p}{(}\PYG{n}{np}\PYG{o}{.}\PYG{n}{zeros}\PYG{p}{(}\PYG{p}{(}\PYG{l+m+mi}{3}\PYG{p}{,}\PYG{l+m+mi}{3}\PYG{p}{)}\PYG{p}{)}\PYG{p}{,}\PYG{n}{np}\PYG{o}{.}\PYG{n}{zeros}\PYG{p}{(}\PYG{p}{(}\PYG{l+m+mi}{2}\PYG{p}{,}\PYG{l+m+mi}{2}\PYG{p}{)}\PYG{p}{)}\PYG{p}{)}
\PYG{g+gt}{Traceback (most recent call last):}
  File \PYG{n+nb}{\PYGZdq{}\PYGZlt{}stdin\PYGZgt{}\PYGZdq{}}, line \PYG{l+m}{1}, in \PYG{n}{\PYGZlt{}module\PYGZgt{}}
\PYG{g+gr}{ValueError}: \PYG{n}{operands could not be broadcast together with shapes (3,3) (2,2)}
\end{sphinxVerbatim}

\end{fulllineitems}

\index{tanh() (in module symjax.tensor)@\spxentry{tanh()}\spxextra{in module symjax.tensor}}

\begin{fulllineitems}
\phantomsection\label{\detokenize{modules/tensor:symjax.tensor.tanh}}\pysiglinewithargsret{\sphinxbfcode{\sphinxupquote{tanh}}}{\emph{\DUrole{n}{x}}}{}
Compute hyperbolic tangent element\sphinxhyphen{}wise.

LAX\sphinxhyphen{}backend implementation of {\hyperref[\detokenize{modules/tensor:symjax.tensor.tanh}]{\sphinxcrossref{\sphinxcode{\sphinxupquote{tanh()}}}}}.
ADDITIONOriginal docstring below.

LAX\sphinxhyphen{}backend implementation of {\hyperref[\detokenize{modules/tensor:symjax.tensor.tanh}]{\sphinxcrossref{\sphinxcode{\sphinxupquote{tanh()}}}}}.
Original docstring below.

tanh(x, /, out=None, {\color{red}\bfseries{}*}, where=True, casting=’same\_kind’, order=’K’, dtype=None, subok=True{[}, signature, extobj{]})

Equivalent to \sphinxcode{\sphinxupquote{np.sinh(x)/np.cosh(x)}} or \sphinxcode{\sphinxupquote{\sphinxhyphen{}1j * np.tan(1j*x)}}.
\begin{quote}\begin{description}
\item[{Returns}] \leavevmode
\sphinxstylestrong{y} \textendash{} The corresponding hyperbolic tangent values.
This is a scalar if \sphinxtitleref{x} is a scalar.

\item[{Return type}] \leavevmode
ndarray

\end{description}\end{quote}
\subsubsection*{Notes}

If \sphinxtitleref{out} is provided, the function writes the result into it,
and returns a reference to \sphinxtitleref{out}.  (See Examples)
\subsubsection*{References}
\subsubsection*{Examples}

\begin{sphinxVerbatim}[commandchars=\\\{\}]
\PYG{g+gp}{\PYGZgt{}\PYGZgt{}\PYGZgt{} }\PYG{n}{np}\PYG{o}{.}\PYG{n}{tanh}\PYG{p}{(}\PYG{p}{(}\PYG{l+m+mi}{0}\PYG{p}{,} \PYG{n}{np}\PYG{o}{.}\PYG{n}{pi}\PYG{o}{*}\PYG{l+m+mi}{1}\PYG{n}{j}\PYG{p}{,} \PYG{n}{np}\PYG{o}{.}\PYG{n}{pi}\PYG{o}{*}\PYG{l+m+mi}{1}\PYG{n}{j}\PYG{o}{/}\PYG{l+m+mi}{2}\PYG{p}{)}\PYG{p}{)}
\PYG{g+go}{array([ 0. +0.00000000e+00j,  0. \PYGZhy{}1.22460635e\PYGZhy{}16j,  0. +1.63317787e+16j])}
\end{sphinxVerbatim}

\begin{sphinxVerbatim}[commandchars=\\\{\}]
\PYG{g+gp}{\PYGZgt{}\PYGZgt{}\PYGZgt{} }\PYG{c+c1}{\PYGZsh{} Example of providing the optional output parameter illustrating}
\PYG{g+gp}{\PYGZgt{}\PYGZgt{}\PYGZgt{} }\PYG{c+c1}{\PYGZsh{} that what is returned is a reference to said parameter}
\PYG{g+gp}{\PYGZgt{}\PYGZgt{}\PYGZgt{} }\PYG{n}{out1} \PYG{o}{=} \PYG{n}{np}\PYG{o}{.}\PYG{n}{array}\PYG{p}{(}\PYG{p}{[}\PYG{l+m+mi}{0}\PYG{p}{]}\PYG{p}{,} \PYG{n}{dtype}\PYG{o}{=}\PYG{l+s+s1}{\PYGZsq{}}\PYG{l+s+s1}{d}\PYG{l+s+s1}{\PYGZsq{}}\PYG{p}{)}
\PYG{g+gp}{\PYGZgt{}\PYGZgt{}\PYGZgt{} }\PYG{n}{out2} \PYG{o}{=} \PYG{n}{np}\PYG{o}{.}\PYG{n}{tanh}\PYG{p}{(}\PYG{p}{[}\PYG{l+m+mf}{0.1}\PYG{p}{]}\PYG{p}{,} \PYG{n}{out1}\PYG{p}{)}
\PYG{g+gp}{\PYGZgt{}\PYGZgt{}\PYGZgt{} }\PYG{n}{out2} \PYG{o+ow}{is} \PYG{n}{out1}
\PYG{g+go}{True}
\end{sphinxVerbatim}

\begin{sphinxVerbatim}[commandchars=\\\{\}]
\PYG{g+gp}{\PYGZgt{}\PYGZgt{}\PYGZgt{} }\PYG{c+c1}{\PYGZsh{} Example of ValueError due to provision of shape mis\PYGZhy{}matched `out`}
\PYG{g+gp}{\PYGZgt{}\PYGZgt{}\PYGZgt{} }\PYG{n}{np}\PYG{o}{.}\PYG{n}{tanh}\PYG{p}{(}\PYG{n}{np}\PYG{o}{.}\PYG{n}{zeros}\PYG{p}{(}\PYG{p}{(}\PYG{l+m+mi}{3}\PYG{p}{,}\PYG{l+m+mi}{3}\PYG{p}{)}\PYG{p}{)}\PYG{p}{,}\PYG{n}{np}\PYG{o}{.}\PYG{n}{zeros}\PYG{p}{(}\PYG{p}{(}\PYG{l+m+mi}{2}\PYG{p}{,}\PYG{l+m+mi}{2}\PYG{p}{)}\PYG{p}{)}\PYG{p}{)}
\PYG{g+gt}{Traceback (most recent call last):}
  File \PYG{n+nb}{\PYGZdq{}\PYGZlt{}stdin\PYGZgt{}\PYGZdq{}}, line \PYG{l+m}{1}, in \PYG{n}{\PYGZlt{}module\PYGZgt{}}
\PYG{g+gr}{ValueError}: \PYG{n}{operands could not be broadcast together with shapes (3,3) (2,2)}
\end{sphinxVerbatim}

\end{fulllineitems}

\index{tensordot() (in module symjax.tensor)@\spxentry{tensordot()}\spxextra{in module symjax.tensor}}

\begin{fulllineitems}
\phantomsection\label{\detokenize{modules/tensor:symjax.tensor.tensordot}}\pysiglinewithargsret{\sphinxbfcode{\sphinxupquote{tensordot}}}{\emph{\DUrole{n}{a}}, \emph{\DUrole{n}{b}}, \emph{\DUrole{n}{axes}\DUrole{o}{=}\DUrole{default_value}{2}}, \emph{\DUrole{n}{precision}\DUrole{o}{=}\DUrole{default_value}{None}}}{}
Compute tensor dot product along specified axes.

LAX\sphinxhyphen{}backend implementation of {\hyperref[\detokenize{modules/tensor:symjax.tensor.tensordot}]{\sphinxcrossref{\sphinxcode{\sphinxupquote{tensordot()}}}}}.
ADDITIONOriginal docstring below.

LAX\sphinxhyphen{}backend implementation of {\hyperref[\detokenize{modules/tensor:symjax.tensor.tensordot}]{\sphinxcrossref{\sphinxcode{\sphinxupquote{tensordot()}}}}}.
In addition to the original NumPy arguments listed below, also supports
\sphinxcode{\sphinxupquote{precision}} for extra control over matrix\sphinxhyphen{}multiplication precision
on supported devices. See \sphinxcode{\sphinxupquote{jax.lax.dot()}} for details.

Original docstring below.

Given two tensors, \sphinxtitleref{a} and \sphinxtitleref{b}, and an array\_like object containing
two array\_like objects, \sphinxcode{\sphinxupquote{(a\_axes, b\_axes)}}, sum the products of
\sphinxtitleref{a}’s and \sphinxtitleref{b}’s elements (components) over the axes specified by
\sphinxcode{\sphinxupquote{a\_axes}} and \sphinxcode{\sphinxupquote{b\_axes}}. The third argument can be a single non\sphinxhyphen{}negative
integer\_like scalar, \sphinxcode{\sphinxupquote{N}}; if it is such, then the last \sphinxcode{\sphinxupquote{N}} dimensions
of \sphinxtitleref{a} and the first \sphinxcode{\sphinxupquote{N}} dimensions of \sphinxtitleref{b} are summed over.
\begin{quote}\begin{description}
\item[{Returns}] \leavevmode
\sphinxstylestrong{output} \textendash{} The tensor dot product of the input.

\item[{Return type}] \leavevmode
ndarray

\end{description}\end{quote}

\sphinxstrong{See also:}

{\hyperref[\detokenize{modules/tensor:symjax.tensor.dot}]{\sphinxcrossref{\sphinxcode{\sphinxupquote{dot()}}}}}, {\hyperref[\detokenize{modules/tensor:symjax.tensor.einsum}]{\sphinxcrossref{\sphinxcode{\sphinxupquote{einsum()}}}}}

\subsubsection*{Notes}
\begin{description}
\item[{Three common use cases are:}] \leavevmode\begin{itemize}
\item {} 
\sphinxcode{\sphinxupquote{axes = 0}} : tensor product \(a\otimes b\)

\item {} 
\sphinxcode{\sphinxupquote{axes = 1}} : tensor dot product \(a\cdot b\)

\item {} 
\sphinxcode{\sphinxupquote{axes = 2}} : (default) tensor double contraction \(a:b\)

\end{itemize}

\end{description}

When \sphinxtitleref{axes} is integer\_like, the sequence for evaluation will be: first
the \sphinxhyphen{}Nth axis in \sphinxtitleref{a} and 0th axis in \sphinxtitleref{b}, and the \sphinxhyphen{}1th axis in \sphinxtitleref{a} and
Nth axis in \sphinxtitleref{b} last.

When there is more than one axis to sum over \sphinxhyphen{} and they are not the last
(first) axes of \sphinxtitleref{a} (\sphinxtitleref{b}) \sphinxhyphen{} the argument \sphinxtitleref{axes} should consist of
two sequences of the same length, with the first axis to sum over given
first in both sequences, the second axis second, and so forth.

The shape of the result consists of the non\sphinxhyphen{}contracted axes of the
first tensor, followed by the non\sphinxhyphen{}contracted axes of the second.
\subsubsection*{Examples}

A “traditional” example:

\begin{sphinxVerbatim}[commandchars=\\\{\}]
\PYG{g+gp}{\PYGZgt{}\PYGZgt{}\PYGZgt{} }\PYG{n}{a} \PYG{o}{=} \PYG{n}{np}\PYG{o}{.}\PYG{n}{arange}\PYG{p}{(}\PYG{l+m+mf}{60.}\PYG{p}{)}\PYG{o}{.}\PYG{n}{reshape}\PYG{p}{(}\PYG{l+m+mi}{3}\PYG{p}{,}\PYG{l+m+mi}{4}\PYG{p}{,}\PYG{l+m+mi}{5}\PYG{p}{)}
\PYG{g+gp}{\PYGZgt{}\PYGZgt{}\PYGZgt{} }\PYG{n}{b} \PYG{o}{=} \PYG{n}{np}\PYG{o}{.}\PYG{n}{arange}\PYG{p}{(}\PYG{l+m+mf}{24.}\PYG{p}{)}\PYG{o}{.}\PYG{n}{reshape}\PYG{p}{(}\PYG{l+m+mi}{4}\PYG{p}{,}\PYG{l+m+mi}{3}\PYG{p}{,}\PYG{l+m+mi}{2}\PYG{p}{)}
\PYG{g+gp}{\PYGZgt{}\PYGZgt{}\PYGZgt{} }\PYG{n}{c} \PYG{o}{=} \PYG{n}{np}\PYG{o}{.}\PYG{n}{tensordot}\PYG{p}{(}\PYG{n}{a}\PYG{p}{,}\PYG{n}{b}\PYG{p}{,} \PYG{n}{axes}\PYG{o}{=}\PYG{p}{(}\PYG{p}{[}\PYG{l+m+mi}{1}\PYG{p}{,}\PYG{l+m+mi}{0}\PYG{p}{]}\PYG{p}{,}\PYG{p}{[}\PYG{l+m+mi}{0}\PYG{p}{,}\PYG{l+m+mi}{1}\PYG{p}{]}\PYG{p}{)}\PYG{p}{)}
\PYG{g+gp}{\PYGZgt{}\PYGZgt{}\PYGZgt{} }\PYG{n}{c}\PYG{o}{.}\PYG{n}{shape}
\PYG{g+go}{(5, 2)}
\PYG{g+gp}{\PYGZgt{}\PYGZgt{}\PYGZgt{} }\PYG{n}{c}
\PYG{g+go}{array([[4400., 4730.],}
\PYG{g+go}{       [4532., 4874.],}
\PYG{g+go}{       [4664., 5018.],}
\PYG{g+go}{       [4796., 5162.],}
\PYG{g+go}{       [4928., 5306.]])}
\PYG{g+gp}{\PYGZgt{}\PYGZgt{}\PYGZgt{} }\PYG{c+c1}{\PYGZsh{} A slower but equivalent way of computing the same...}
\PYG{g+gp}{\PYGZgt{}\PYGZgt{}\PYGZgt{} }\PYG{n}{d} \PYG{o}{=} \PYG{n}{np}\PYG{o}{.}\PYG{n}{zeros}\PYG{p}{(}\PYG{p}{(}\PYG{l+m+mi}{5}\PYG{p}{,}\PYG{l+m+mi}{2}\PYG{p}{)}\PYG{p}{)}
\PYG{g+gp}{\PYGZgt{}\PYGZgt{}\PYGZgt{} }\PYG{k}{for} \PYG{n}{i} \PYG{o+ow}{in} \PYG{n+nb}{range}\PYG{p}{(}\PYG{l+m+mi}{5}\PYG{p}{)}\PYG{p}{:}
\PYG{g+gp}{... }  \PYG{k}{for} \PYG{n}{j} \PYG{o+ow}{in} \PYG{n+nb}{range}\PYG{p}{(}\PYG{l+m+mi}{2}\PYG{p}{)}\PYG{p}{:}
\PYG{g+gp}{... }    \PYG{k}{for} \PYG{n}{k} \PYG{o+ow}{in} \PYG{n+nb}{range}\PYG{p}{(}\PYG{l+m+mi}{3}\PYG{p}{)}\PYG{p}{:}
\PYG{g+gp}{... }      \PYG{k}{for} \PYG{n}{n} \PYG{o+ow}{in} \PYG{n+nb}{range}\PYG{p}{(}\PYG{l+m+mi}{4}\PYG{p}{)}\PYG{p}{:}
\PYG{g+gp}{... }        \PYG{n}{d}\PYG{p}{[}\PYG{n}{i}\PYG{p}{,}\PYG{n}{j}\PYG{p}{]} \PYG{o}{+}\PYG{o}{=} \PYG{n}{a}\PYG{p}{[}\PYG{n}{k}\PYG{p}{,}\PYG{n}{n}\PYG{p}{,}\PYG{n}{i}\PYG{p}{]} \PYG{o}{*} \PYG{n}{b}\PYG{p}{[}\PYG{n}{n}\PYG{p}{,}\PYG{n}{k}\PYG{p}{,}\PYG{n}{j}\PYG{p}{]}
\PYG{g+gp}{\PYGZgt{}\PYGZgt{}\PYGZgt{} }\PYG{n}{c} \PYG{o}{==} \PYG{n}{d}
\PYG{g+go}{array([[ True,  True],}
\PYG{g+go}{       [ True,  True],}
\PYG{g+go}{       [ True,  True],}
\PYG{g+go}{       [ True,  True],}
\PYG{g+go}{       [ True,  True]])}
\end{sphinxVerbatim}

An extended example taking advantage of the overloading of + and *:

\begin{sphinxVerbatim}[commandchars=\\\{\}]
\PYG{g+gp}{\PYGZgt{}\PYGZgt{}\PYGZgt{} }\PYG{n}{a} \PYG{o}{=} \PYG{n}{np}\PYG{o}{.}\PYG{n}{array}\PYG{p}{(}\PYG{n+nb}{range}\PYG{p}{(}\PYG{l+m+mi}{1}\PYG{p}{,} \PYG{l+m+mi}{9}\PYG{p}{)}\PYG{p}{)}
\PYG{g+gp}{\PYGZgt{}\PYGZgt{}\PYGZgt{} }\PYG{n}{a}\PYG{o}{.}\PYG{n}{shape} \PYG{o}{=} \PYG{p}{(}\PYG{l+m+mi}{2}\PYG{p}{,} \PYG{l+m+mi}{2}\PYG{p}{,} \PYG{l+m+mi}{2}\PYG{p}{)}
\PYG{g+gp}{\PYGZgt{}\PYGZgt{}\PYGZgt{} }\PYG{n}{A} \PYG{o}{=} \PYG{n}{np}\PYG{o}{.}\PYG{n}{array}\PYG{p}{(}\PYG{p}{(}\PYG{l+s+s1}{\PYGZsq{}}\PYG{l+s+s1}{a}\PYG{l+s+s1}{\PYGZsq{}}\PYG{p}{,} \PYG{l+s+s1}{\PYGZsq{}}\PYG{l+s+s1}{b}\PYG{l+s+s1}{\PYGZsq{}}\PYG{p}{,} \PYG{l+s+s1}{\PYGZsq{}}\PYG{l+s+s1}{c}\PYG{l+s+s1}{\PYGZsq{}}\PYG{p}{,} \PYG{l+s+s1}{\PYGZsq{}}\PYG{l+s+s1}{d}\PYG{l+s+s1}{\PYGZsq{}}\PYG{p}{)}\PYG{p}{,} \PYG{n}{dtype}\PYG{o}{=}\PYG{n+nb}{object}\PYG{p}{)}
\PYG{g+gp}{\PYGZgt{}\PYGZgt{}\PYGZgt{} }\PYG{n}{A}\PYG{o}{.}\PYG{n}{shape} \PYG{o}{=} \PYG{p}{(}\PYG{l+m+mi}{2}\PYG{p}{,} \PYG{l+m+mi}{2}\PYG{p}{)}
\PYG{g+gp}{\PYGZgt{}\PYGZgt{}\PYGZgt{} }\PYG{n}{a}\PYG{p}{;} \PYG{n}{A}
\PYG{g+go}{array([[[1, 2],}
\PYG{g+go}{        [3, 4]],}
\PYG{g+go}{       [[5, 6],}
\PYG{g+go}{        [7, 8]]])}
\PYG{g+go}{array([[\PYGZsq{}a\PYGZsq{}, \PYGZsq{}b\PYGZsq{}],}
\PYG{g+go}{       [\PYGZsq{}c\PYGZsq{}, \PYGZsq{}d\PYGZsq{}]], dtype=object)}
\end{sphinxVerbatim}

\begin{sphinxVerbatim}[commandchars=\\\{\}]
\PYG{g+gp}{\PYGZgt{}\PYGZgt{}\PYGZgt{} }\PYG{n}{np}\PYG{o}{.}\PYG{n}{tensordot}\PYG{p}{(}\PYG{n}{a}\PYG{p}{,} \PYG{n}{A}\PYG{p}{)} \PYG{c+c1}{\PYGZsh{} third argument default is 2 for double\PYGZhy{}contraction}
\PYG{g+go}{array([\PYGZsq{}abbcccdddd\PYGZsq{}, \PYGZsq{}aaaaabbbbbbcccccccdddddddd\PYGZsq{}], dtype=object)}
\end{sphinxVerbatim}

\begin{sphinxVerbatim}[commandchars=\\\{\}]
\PYG{g+gp}{\PYGZgt{}\PYGZgt{}\PYGZgt{} }\PYG{n}{np}\PYG{o}{.}\PYG{n}{tensordot}\PYG{p}{(}\PYG{n}{a}\PYG{p}{,} \PYG{n}{A}\PYG{p}{,} \PYG{l+m+mi}{1}\PYG{p}{)}
\PYG{g+go}{array([[[\PYGZsq{}acc\PYGZsq{}, \PYGZsq{}bdd\PYGZsq{}],}
\PYG{g+go}{        [\PYGZsq{}aaacccc\PYGZsq{}, \PYGZsq{}bbbdddd\PYGZsq{}]],}
\PYG{g+go}{       [[\PYGZsq{}aaaaacccccc\PYGZsq{}, \PYGZsq{}bbbbbdddddd\PYGZsq{}],}
\PYG{g+go}{        [\PYGZsq{}aaaaaaacccccccc\PYGZsq{}, \PYGZsq{}bbbbbbbdddddddd\PYGZsq{}]]], dtype=object)}
\end{sphinxVerbatim}

\begin{sphinxVerbatim}[commandchars=\\\{\}]
\PYG{g+gp}{\PYGZgt{}\PYGZgt{}\PYGZgt{} }\PYG{n}{np}\PYG{o}{.}\PYG{n}{tensordot}\PYG{p}{(}\PYG{n}{a}\PYG{p}{,} \PYG{n}{A}\PYG{p}{,} \PYG{l+m+mi}{0}\PYG{p}{)} \PYG{c+c1}{\PYGZsh{} tensor product (result too long to incl.)}
\PYG{g+go}{array([[[[[\PYGZsq{}a\PYGZsq{}, \PYGZsq{}b\PYGZsq{}],}
\PYG{g+go}{          [\PYGZsq{}c\PYGZsq{}, \PYGZsq{}d\PYGZsq{}]],}
\PYG{g+go}{          ...}
\end{sphinxVerbatim}

\begin{sphinxVerbatim}[commandchars=\\\{\}]
\PYG{g+gp}{\PYGZgt{}\PYGZgt{}\PYGZgt{} }\PYG{n}{np}\PYG{o}{.}\PYG{n}{tensordot}\PYG{p}{(}\PYG{n}{a}\PYG{p}{,} \PYG{n}{A}\PYG{p}{,} \PYG{p}{(}\PYG{l+m+mi}{0}\PYG{p}{,} \PYG{l+m+mi}{1}\PYG{p}{)}\PYG{p}{)}
\PYG{g+go}{array([[[\PYGZsq{}abbbbb\PYGZsq{}, \PYGZsq{}cddddd\PYGZsq{}],}
\PYG{g+go}{        [\PYGZsq{}aabbbbbb\PYGZsq{}, \PYGZsq{}ccdddddd\PYGZsq{}]],}
\PYG{g+go}{       [[\PYGZsq{}aaabbbbbbb\PYGZsq{}, \PYGZsq{}cccddddddd\PYGZsq{}],}
\PYG{g+go}{        [\PYGZsq{}aaaabbbbbbbb\PYGZsq{}, \PYGZsq{}ccccdddddddd\PYGZsq{}]]], dtype=object)}
\end{sphinxVerbatim}

\begin{sphinxVerbatim}[commandchars=\\\{\}]
\PYG{g+gp}{\PYGZgt{}\PYGZgt{}\PYGZgt{} }\PYG{n}{np}\PYG{o}{.}\PYG{n}{tensordot}\PYG{p}{(}\PYG{n}{a}\PYG{p}{,} \PYG{n}{A}\PYG{p}{,} \PYG{p}{(}\PYG{l+m+mi}{2}\PYG{p}{,} \PYG{l+m+mi}{1}\PYG{p}{)}\PYG{p}{)}
\PYG{g+go}{array([[[\PYGZsq{}abb\PYGZsq{}, \PYGZsq{}cdd\PYGZsq{}],}
\PYG{g+go}{        [\PYGZsq{}aaabbbb\PYGZsq{}, \PYGZsq{}cccdddd\PYGZsq{}]],}
\PYG{g+go}{       [[\PYGZsq{}aaaaabbbbbb\PYGZsq{}, \PYGZsq{}cccccdddddd\PYGZsq{}],}
\PYG{g+go}{        [\PYGZsq{}aaaaaaabbbbbbbb\PYGZsq{}, \PYGZsq{}cccccccdddddddd\PYGZsq{}]]], dtype=object)}
\end{sphinxVerbatim}

\begin{sphinxVerbatim}[commandchars=\\\{\}]
\PYG{g+gp}{\PYGZgt{}\PYGZgt{}\PYGZgt{} }\PYG{n}{np}\PYG{o}{.}\PYG{n}{tensordot}\PYG{p}{(}\PYG{n}{a}\PYG{p}{,} \PYG{n}{A}\PYG{p}{,} \PYG{p}{(}\PYG{p}{(}\PYG{l+m+mi}{0}\PYG{p}{,} \PYG{l+m+mi}{1}\PYG{p}{)}\PYG{p}{,} \PYG{p}{(}\PYG{l+m+mi}{0}\PYG{p}{,} \PYG{l+m+mi}{1}\PYG{p}{)}\PYG{p}{)}\PYG{p}{)}
\PYG{g+go}{array([\PYGZsq{}abbbcccccddddddd\PYGZsq{}, \PYGZsq{}aabbbbccccccdddddddd\PYGZsq{}], dtype=object)}
\end{sphinxVerbatim}

\begin{sphinxVerbatim}[commandchars=\\\{\}]
\PYG{g+gp}{\PYGZgt{}\PYGZgt{}\PYGZgt{} }\PYG{n}{np}\PYG{o}{.}\PYG{n}{tensordot}\PYG{p}{(}\PYG{n}{a}\PYG{p}{,} \PYG{n}{A}\PYG{p}{,} \PYG{p}{(}\PYG{p}{(}\PYG{l+m+mi}{2}\PYG{p}{,} \PYG{l+m+mi}{1}\PYG{p}{)}\PYG{p}{,} \PYG{p}{(}\PYG{l+m+mi}{1}\PYG{p}{,} \PYG{l+m+mi}{0}\PYG{p}{)}\PYG{p}{)}\PYG{p}{)}
\PYG{g+go}{array([\PYGZsq{}acccbbdddd\PYGZsq{}, \PYGZsq{}aaaaacccccccbbbbbbdddddddd\PYGZsq{}], dtype=object)}
\end{sphinxVerbatim}

\end{fulllineitems}

\index{tile() (in module symjax.tensor)@\spxentry{tile()}\spxextra{in module symjax.tensor}}

\begin{fulllineitems}
\phantomsection\label{\detokenize{modules/tensor:symjax.tensor.tile}}\pysiglinewithargsret{\sphinxbfcode{\sphinxupquote{tile}}}{\emph{\DUrole{n}{a}}, \emph{\DUrole{n}{reps}}}{}
Construct an array by repeating A the number of times given by reps.

LAX\sphinxhyphen{}backend implementation of {\hyperref[\detokenize{modules/tensor:symjax.tensor.tile}]{\sphinxcrossref{\sphinxcode{\sphinxupquote{tile()}}}}}.
ADDITIONOriginal docstring below.

LAX\sphinxhyphen{}backend implementation of {\hyperref[\detokenize{modules/tensor:symjax.tensor.tile}]{\sphinxcrossref{\sphinxcode{\sphinxupquote{tile()}}}}}.
Original docstring below.

If \sphinxtitleref{reps} has length \sphinxcode{\sphinxupquote{d}}, the result will have dimension of
\sphinxcode{\sphinxupquote{max(d, A.ndim)}}.

If \sphinxcode{\sphinxupquote{A.ndim \textless{} d}}, \sphinxtitleref{A} is promoted to be d\sphinxhyphen{}dimensional by prepending new
axes. So a shape (3,) array is promoted to (1, 3) for 2\sphinxhyphen{}D replication,
or shape (1, 1, 3) for 3\sphinxhyphen{}D replication. If this is not the desired
behavior, promote \sphinxtitleref{A} to d\sphinxhyphen{}dimensions manually before calling this
function.

If \sphinxcode{\sphinxupquote{A.ndim \textgreater{} d}}, \sphinxtitleref{reps} is promoted to \sphinxtitleref{A}.ndim by pre\sphinxhyphen{}pending 1’s to it.
Thus for an \sphinxtitleref{A} of shape (2, 3, 4, 5), a \sphinxtitleref{reps} of (2, 2) is treated as
(1, 1, 2, 2).

Note : Although tile may be used for broadcasting, it is strongly
recommended to use numpy’s broadcasting operations and functions.
\begin{quote}\begin{description}
\item[{Returns}] \leavevmode
\sphinxstylestrong{c} \textendash{} The tiled output array.

\item[{Return type}] \leavevmode
ndarray

\end{description}\end{quote}

\sphinxstrong{See also:}

\begin{description}
\item[{{\hyperref[\detokenize{modules/tensor:symjax.tensor.repeat}]{\sphinxcrossref{\sphinxcode{\sphinxupquote{repeat()}}}}}}] \leavevmode
Repeat elements of an array.

\item[{\sphinxcode{\sphinxupquote{broadcast\_to()}}}] \leavevmode
Broadcast an array to a new shape

\end{description}

\subsubsection*{Examples}

\begin{sphinxVerbatim}[commandchars=\\\{\}]
\PYG{g+gp}{\PYGZgt{}\PYGZgt{}\PYGZgt{} }\PYG{n}{a} \PYG{o}{=} \PYG{n}{np}\PYG{o}{.}\PYG{n}{array}\PYG{p}{(}\PYG{p}{[}\PYG{l+m+mi}{0}\PYG{p}{,} \PYG{l+m+mi}{1}\PYG{p}{,} \PYG{l+m+mi}{2}\PYG{p}{]}\PYG{p}{)}
\PYG{g+gp}{\PYGZgt{}\PYGZgt{}\PYGZgt{} }\PYG{n}{np}\PYG{o}{.}\PYG{n}{tile}\PYG{p}{(}\PYG{n}{a}\PYG{p}{,} \PYG{l+m+mi}{2}\PYG{p}{)}
\PYG{g+go}{array([0, 1, 2, 0, 1, 2])}
\PYG{g+gp}{\PYGZgt{}\PYGZgt{}\PYGZgt{} }\PYG{n}{np}\PYG{o}{.}\PYG{n}{tile}\PYG{p}{(}\PYG{n}{a}\PYG{p}{,} \PYG{p}{(}\PYG{l+m+mi}{2}\PYG{p}{,} \PYG{l+m+mi}{2}\PYG{p}{)}\PYG{p}{)}
\PYG{g+go}{array([[0, 1, 2, 0, 1, 2],}
\PYG{g+go}{       [0, 1, 2, 0, 1, 2]])}
\PYG{g+gp}{\PYGZgt{}\PYGZgt{}\PYGZgt{} }\PYG{n}{np}\PYG{o}{.}\PYG{n}{tile}\PYG{p}{(}\PYG{n}{a}\PYG{p}{,} \PYG{p}{(}\PYG{l+m+mi}{2}\PYG{p}{,} \PYG{l+m+mi}{1}\PYG{p}{,} \PYG{l+m+mi}{2}\PYG{p}{)}\PYG{p}{)}
\PYG{g+go}{array([[[0, 1, 2, 0, 1, 2]],}
\PYG{g+go}{       [[0, 1, 2, 0, 1, 2]]])}
\end{sphinxVerbatim}

\begin{sphinxVerbatim}[commandchars=\\\{\}]
\PYG{g+gp}{\PYGZgt{}\PYGZgt{}\PYGZgt{} }\PYG{n}{b} \PYG{o}{=} \PYG{n}{np}\PYG{o}{.}\PYG{n}{array}\PYG{p}{(}\PYG{p}{[}\PYG{p}{[}\PYG{l+m+mi}{1}\PYG{p}{,} \PYG{l+m+mi}{2}\PYG{p}{]}\PYG{p}{,} \PYG{p}{[}\PYG{l+m+mi}{3}\PYG{p}{,} \PYG{l+m+mi}{4}\PYG{p}{]}\PYG{p}{]}\PYG{p}{)}
\PYG{g+gp}{\PYGZgt{}\PYGZgt{}\PYGZgt{} }\PYG{n}{np}\PYG{o}{.}\PYG{n}{tile}\PYG{p}{(}\PYG{n}{b}\PYG{p}{,} \PYG{l+m+mi}{2}\PYG{p}{)}
\PYG{g+go}{array([[1, 2, 1, 2],}
\PYG{g+go}{       [3, 4, 3, 4]])}
\PYG{g+gp}{\PYGZgt{}\PYGZgt{}\PYGZgt{} }\PYG{n}{np}\PYG{o}{.}\PYG{n}{tile}\PYG{p}{(}\PYG{n}{b}\PYG{p}{,} \PYG{p}{(}\PYG{l+m+mi}{2}\PYG{p}{,} \PYG{l+m+mi}{1}\PYG{p}{)}\PYG{p}{)}
\PYG{g+go}{array([[1, 2],}
\PYG{g+go}{       [3, 4],}
\PYG{g+go}{       [1, 2],}
\PYG{g+go}{       [3, 4]])}
\end{sphinxVerbatim}

\begin{sphinxVerbatim}[commandchars=\\\{\}]
\PYG{g+gp}{\PYGZgt{}\PYGZgt{}\PYGZgt{} }\PYG{n}{c} \PYG{o}{=} \PYG{n}{np}\PYG{o}{.}\PYG{n}{array}\PYG{p}{(}\PYG{p}{[}\PYG{l+m+mi}{1}\PYG{p}{,}\PYG{l+m+mi}{2}\PYG{p}{,}\PYG{l+m+mi}{3}\PYG{p}{,}\PYG{l+m+mi}{4}\PYG{p}{]}\PYG{p}{)}
\PYG{g+gp}{\PYGZgt{}\PYGZgt{}\PYGZgt{} }\PYG{n}{np}\PYG{o}{.}\PYG{n}{tile}\PYG{p}{(}\PYG{n}{c}\PYG{p}{,}\PYG{p}{(}\PYG{l+m+mi}{4}\PYG{p}{,}\PYG{l+m+mi}{1}\PYG{p}{)}\PYG{p}{)}
\PYG{g+go}{array([[1, 2, 3, 4],}
\PYG{g+go}{       [1, 2, 3, 4],}
\PYG{g+go}{       [1, 2, 3, 4],}
\PYG{g+go}{       [1, 2, 3, 4]])}
\end{sphinxVerbatim}

\end{fulllineitems}

\index{trace() (in module symjax.tensor)@\spxentry{trace()}\spxextra{in module symjax.tensor}}

\begin{fulllineitems}
\phantomsection\label{\detokenize{modules/tensor:symjax.tensor.trace}}\pysiglinewithargsret{\sphinxbfcode{\sphinxupquote{trace}}}{\emph{\DUrole{n}{a}}, \emph{\DUrole{n}{offset}\DUrole{o}{=}\DUrole{default_value}{0}}, \emph{\DUrole{n}{axis1}\DUrole{o}{=}\DUrole{default_value}{0}}, \emph{\DUrole{n}{axis2}\DUrole{o}{=}\DUrole{default_value}{1}}, \emph{\DUrole{n}{dtype}\DUrole{o}{=}\DUrole{default_value}{None}}, \emph{\DUrole{n}{out}\DUrole{o}{=}\DUrole{default_value}{None}}}{}
Return the sum along diagonals of the array.

LAX\sphinxhyphen{}backend implementation of {\hyperref[\detokenize{modules/tensor:symjax.tensor.trace}]{\sphinxcrossref{\sphinxcode{\sphinxupquote{trace()}}}}}.
ADDITIONOriginal docstring below.

LAX\sphinxhyphen{}backend implementation of {\hyperref[\detokenize{modules/tensor:symjax.tensor.trace}]{\sphinxcrossref{\sphinxcode{\sphinxupquote{trace()}}}}}.
Original docstring below.

If \sphinxtitleref{a} is 2\sphinxhyphen{}D, the sum along its diagonal with the given offset
is returned, i.e., the sum of elements \sphinxcode{\sphinxupquote{a{[}i,i+offset{]}}} for all i.

If \sphinxtitleref{a} has more than two dimensions, then the axes specified by axis1 and
axis2 are used to determine the 2\sphinxhyphen{}D sub\sphinxhyphen{}arrays whose traces are returned.
The shape of the resulting array is the same as that of \sphinxtitleref{a} with \sphinxtitleref{axis1}
and \sphinxtitleref{axis2} removed.
\begin{quote}\begin{description}
\item[{Parameters}] \leavevmode
\sphinxstyleliteralstrong{\sphinxupquote{dtype}} (\sphinxstyleliteralemphasis{\sphinxupquote{dtype}}\sphinxstyleliteralemphasis{\sphinxupquote{, }}\sphinxstyleliteralemphasis{\sphinxupquote{optional}}) \textendash{} Determines the data\sphinxhyphen{}type of the returned array and of the accumulator
where the elements are summed. If dtype has the value None and \sphinxtitleref{a} is
of integer type of precision less than the default integer
precision, then the default integer precision is used. Otherwise,
the precision is the same as that of \sphinxtitleref{a}.

\item[{Returns}] \leavevmode
\sphinxstylestrong{sum\_along\_diagonals} \textendash{} If \sphinxtitleref{a} is 2\sphinxhyphen{}D, the sum along the diagonal is returned.  If \sphinxtitleref{a} has
larger dimensions, then an array of sums along diagonals is returned.

\item[{Return type}] \leavevmode
ndarray

\end{description}\end{quote}

\sphinxstrong{See also:}

{\hyperref[\detokenize{modules/tensor:symjax.tensor.diag}]{\sphinxcrossref{\sphinxcode{\sphinxupquote{diag()}}}}}, {\hyperref[\detokenize{modules/tensor:symjax.tensor.diagonal}]{\sphinxcrossref{\sphinxcode{\sphinxupquote{diagonal()}}}}}, \sphinxcode{\sphinxupquote{diagflat()}}

\subsubsection*{Examples}

\begin{sphinxVerbatim}[commandchars=\\\{\}]
\PYG{g+gp}{\PYGZgt{}\PYGZgt{}\PYGZgt{} }\PYG{n}{np}\PYG{o}{.}\PYG{n}{trace}\PYG{p}{(}\PYG{n}{np}\PYG{o}{.}\PYG{n}{eye}\PYG{p}{(}\PYG{l+m+mi}{3}\PYG{p}{)}\PYG{p}{)}
\PYG{g+go}{3.0}
\PYG{g+gp}{\PYGZgt{}\PYGZgt{}\PYGZgt{} }\PYG{n}{a} \PYG{o}{=} \PYG{n}{np}\PYG{o}{.}\PYG{n}{arange}\PYG{p}{(}\PYG{l+m+mi}{8}\PYG{p}{)}\PYG{o}{.}\PYG{n}{reshape}\PYG{p}{(}\PYG{p}{(}\PYG{l+m+mi}{2}\PYG{p}{,}\PYG{l+m+mi}{2}\PYG{p}{,}\PYG{l+m+mi}{2}\PYG{p}{)}\PYG{p}{)}
\PYG{g+gp}{\PYGZgt{}\PYGZgt{}\PYGZgt{} }\PYG{n}{np}\PYG{o}{.}\PYG{n}{trace}\PYG{p}{(}\PYG{n}{a}\PYG{p}{)}
\PYG{g+go}{array([6, 8])}
\end{sphinxVerbatim}

\begin{sphinxVerbatim}[commandchars=\\\{\}]
\PYG{g+gp}{\PYGZgt{}\PYGZgt{}\PYGZgt{} }\PYG{n}{a} \PYG{o}{=} \PYG{n}{np}\PYG{o}{.}\PYG{n}{arange}\PYG{p}{(}\PYG{l+m+mi}{24}\PYG{p}{)}\PYG{o}{.}\PYG{n}{reshape}\PYG{p}{(}\PYG{p}{(}\PYG{l+m+mi}{2}\PYG{p}{,}\PYG{l+m+mi}{2}\PYG{p}{,}\PYG{l+m+mi}{2}\PYG{p}{,}\PYG{l+m+mi}{3}\PYG{p}{)}\PYG{p}{)}
\PYG{g+gp}{\PYGZgt{}\PYGZgt{}\PYGZgt{} }\PYG{n}{np}\PYG{o}{.}\PYG{n}{trace}\PYG{p}{(}\PYG{n}{a}\PYG{p}{)}\PYG{o}{.}\PYG{n}{shape}
\PYG{g+go}{(2, 3)}
\end{sphinxVerbatim}

\end{fulllineitems}

\index{transpose() (in module symjax.tensor)@\spxentry{transpose()}\spxextra{in module symjax.tensor}}

\begin{fulllineitems}
\phantomsection\label{\detokenize{modules/tensor:symjax.tensor.transpose}}\pysiglinewithargsret{\sphinxbfcode{\sphinxupquote{transpose}}}{\emph{\DUrole{n}{a}}, \emph{\DUrole{n}{axes}\DUrole{o}{=}\DUrole{default_value}{None}}}{}
Permute the dimensions of an array.

LAX\sphinxhyphen{}backend implementation of {\hyperref[\detokenize{modules/tensor:symjax.tensor.transpose}]{\sphinxcrossref{\sphinxcode{\sphinxupquote{transpose()}}}}}.
ADDITIONOriginal docstring below.

LAX\sphinxhyphen{}backend implementation of {\hyperref[\detokenize{modules/tensor:symjax.tensor.transpose}]{\sphinxcrossref{\sphinxcode{\sphinxupquote{transpose()}}}}}.
Original docstring below.
\begin{quote}\begin{description}
\item[{Returns}] \leavevmode
\sphinxstylestrong{p} \textendash{} \sphinxtitleref{a} with its axes permuted.  A view is returned whenever
possible.

\item[{Return type}] \leavevmode
ndarray

\end{description}\end{quote}

\sphinxstrong{See also:}

{\hyperref[\detokenize{modules/tensor:symjax.tensor.moveaxis}]{\sphinxcrossref{\sphinxcode{\sphinxupquote{moveaxis()}}}}}, {\hyperref[\detokenize{modules/tensor:symjax.tensor.argsort}]{\sphinxcrossref{\sphinxcode{\sphinxupquote{argsort()}}}}}

\subsubsection*{Notes}

Use \sphinxtitleref{transpose(a, argsort(axes))} to invert the transposition of tensors
when using the \sphinxtitleref{axes} keyword argument.

Transposing a 1\sphinxhyphen{}D array returns an unchanged view of the original array.
\subsubsection*{Examples}

\begin{sphinxVerbatim}[commandchars=\\\{\}]
\PYG{g+gp}{\PYGZgt{}\PYGZgt{}\PYGZgt{} }\PYG{n}{x} \PYG{o}{=} \PYG{n}{np}\PYG{o}{.}\PYG{n}{arange}\PYG{p}{(}\PYG{l+m+mi}{4}\PYG{p}{)}\PYG{o}{.}\PYG{n}{reshape}\PYG{p}{(}\PYG{p}{(}\PYG{l+m+mi}{2}\PYG{p}{,}\PYG{l+m+mi}{2}\PYG{p}{)}\PYG{p}{)}
\PYG{g+gp}{\PYGZgt{}\PYGZgt{}\PYGZgt{} }\PYG{n}{x}
\PYG{g+go}{array([[0, 1],}
\PYG{g+go}{       [2, 3]])}
\end{sphinxVerbatim}

\begin{sphinxVerbatim}[commandchars=\\\{\}]
\PYG{g+gp}{\PYGZgt{}\PYGZgt{}\PYGZgt{} }\PYG{n}{np}\PYG{o}{.}\PYG{n}{transpose}\PYG{p}{(}\PYG{n}{x}\PYG{p}{)}
\PYG{g+go}{array([[0, 2],}
\PYG{g+go}{       [1, 3]])}
\end{sphinxVerbatim}

\begin{sphinxVerbatim}[commandchars=\\\{\}]
\PYG{g+gp}{\PYGZgt{}\PYGZgt{}\PYGZgt{} }\PYG{n}{x} \PYG{o}{=} \PYG{n}{np}\PYG{o}{.}\PYG{n}{ones}\PYG{p}{(}\PYG{p}{(}\PYG{l+m+mi}{1}\PYG{p}{,} \PYG{l+m+mi}{2}\PYG{p}{,} \PYG{l+m+mi}{3}\PYG{p}{)}\PYG{p}{)}
\PYG{g+gp}{\PYGZgt{}\PYGZgt{}\PYGZgt{} }\PYG{n}{np}\PYG{o}{.}\PYG{n}{transpose}\PYG{p}{(}\PYG{n}{x}\PYG{p}{,} \PYG{p}{(}\PYG{l+m+mi}{1}\PYG{p}{,} \PYG{l+m+mi}{0}\PYG{p}{,} \PYG{l+m+mi}{2}\PYG{p}{)}\PYG{p}{)}\PYG{o}{.}\PYG{n}{shape}
\PYG{g+go}{(2, 1, 3)}
\end{sphinxVerbatim}

\end{fulllineitems}

\index{tri() (in module symjax.tensor)@\spxentry{tri()}\spxextra{in module symjax.tensor}}

\begin{fulllineitems}
\phantomsection\label{\detokenize{modules/tensor:symjax.tensor.tri}}\pysiglinewithargsret{\sphinxbfcode{\sphinxupquote{tri}}}{\emph{\DUrole{n}{N}}, \emph{\DUrole{n}{M}\DUrole{o}{=}\DUrole{default_value}{None}}, \emph{\DUrole{n}{k}\DUrole{o}{=}\DUrole{default_value}{0}}, \emph{\DUrole{n}{dtype}\DUrole{o}{=}\DUrole{default_value}{None}}}{}
An array with ones at and below the given diagonal and zeros elsewhere.

LAX\sphinxhyphen{}backend implementation of {\hyperref[\detokenize{modules/tensor:symjax.tensor.tri}]{\sphinxcrossref{\sphinxcode{\sphinxupquote{tri()}}}}}.
ADDITIONOriginal docstring below.

LAX\sphinxhyphen{}backend implementation of {\hyperref[\detokenize{modules/tensor:symjax.tensor.tri}]{\sphinxcrossref{\sphinxcode{\sphinxupquote{tri()}}}}}.
Original docstring below.
\begin{quote}\begin{description}
\item[{Parameters}] \leavevmode
\sphinxstyleliteralstrong{\sphinxupquote{dtype}} (\sphinxstyleliteralemphasis{\sphinxupquote{dtype}}\sphinxstyleliteralemphasis{\sphinxupquote{, }}\sphinxstyleliteralemphasis{\sphinxupquote{optional}}) \textendash{} Data type of the returned array.  The default is float.

\item[{Returns}] \leavevmode
\sphinxstylestrong{tri} \textendash{} Array with its lower triangle filled with ones and zero elsewhere;
in other words \sphinxcode{\sphinxupquote{T{[}i,j{]} == 1}} for \sphinxcode{\sphinxupquote{j \textless{}= i + k}}, 0 otherwise.

\item[{Return type}] \leavevmode
ndarray of shape (N, M)

\end{description}\end{quote}
\subsubsection*{Examples}

\begin{sphinxVerbatim}[commandchars=\\\{\}]
\PYG{g+gp}{\PYGZgt{}\PYGZgt{}\PYGZgt{} }\PYG{n}{np}\PYG{o}{.}\PYG{n}{tri}\PYG{p}{(}\PYG{l+m+mi}{3}\PYG{p}{,} \PYG{l+m+mi}{5}\PYG{p}{,} \PYG{l+m+mi}{2}\PYG{p}{,} \PYG{n}{dtype}\PYG{o}{=}\PYG{n+nb}{int}\PYG{p}{)}
\PYG{g+go}{array([[1, 1, 1, 0, 0],}
\PYG{g+go}{       [1, 1, 1, 1, 0],}
\PYG{g+go}{       [1, 1, 1, 1, 1]])}
\end{sphinxVerbatim}

\begin{sphinxVerbatim}[commandchars=\\\{\}]
\PYG{g+gp}{\PYGZgt{}\PYGZgt{}\PYGZgt{} }\PYG{n}{np}\PYG{o}{.}\PYG{n}{tri}\PYG{p}{(}\PYG{l+m+mi}{3}\PYG{p}{,} \PYG{l+m+mi}{5}\PYG{p}{,} \PYG{o}{\PYGZhy{}}\PYG{l+m+mi}{1}\PYG{p}{)}
\PYG{g+go}{array([[0.,  0.,  0.,  0.,  0.],}
\PYG{g+go}{       [1.,  0.,  0.,  0.,  0.],}
\PYG{g+go}{       [1.,  1.,  0.,  0.,  0.]])}
\end{sphinxVerbatim}

\end{fulllineitems}

\index{tril() (in module symjax.tensor)@\spxentry{tril()}\spxextra{in module symjax.tensor}}

\begin{fulllineitems}
\phantomsection\label{\detokenize{modules/tensor:symjax.tensor.tril}}\pysiglinewithargsret{\sphinxbfcode{\sphinxupquote{tril}}}{\emph{\DUrole{n}{m}}, \emph{\DUrole{n}{k}\DUrole{o}{=}\DUrole{default_value}{0}}}{}
Lower triangle of an array.

LAX\sphinxhyphen{}backend implementation of {\hyperref[\detokenize{modules/tensor:symjax.tensor.tril}]{\sphinxcrossref{\sphinxcode{\sphinxupquote{tril()}}}}}.
ADDITIONOriginal docstring below.

LAX\sphinxhyphen{}backend implementation of {\hyperref[\detokenize{modules/tensor:symjax.tensor.tril}]{\sphinxcrossref{\sphinxcode{\sphinxupquote{tril()}}}}}.
Original docstring below.

Return a copy of an array with elements above the \sphinxtitleref{k}\sphinxhyphen{}th diagonal zeroed.
\begin{quote}\begin{description}
\item[{Returns}] \leavevmode
\sphinxstylestrong{tril} \textendash{} Lower triangle of \sphinxtitleref{m}, of same shape and data\sphinxhyphen{}type as \sphinxtitleref{m}.

\item[{Return type}] \leavevmode
ndarray, shape (M, N)

\end{description}\end{quote}

\sphinxstrong{See also:}

\begin{description}
\item[{{\hyperref[\detokenize{modules/tensor:symjax.tensor.triu}]{\sphinxcrossref{\sphinxcode{\sphinxupquote{triu()}}}}}}] \leavevmode
same thing, only for the upper triangle

\end{description}

\subsubsection*{Examples}

\begin{sphinxVerbatim}[commandchars=\\\{\}]
\PYG{g+gp}{\PYGZgt{}\PYGZgt{}\PYGZgt{} }\PYG{n}{np}\PYG{o}{.}\PYG{n}{tril}\PYG{p}{(}\PYG{p}{[}\PYG{p}{[}\PYG{l+m+mi}{1}\PYG{p}{,}\PYG{l+m+mi}{2}\PYG{p}{,}\PYG{l+m+mi}{3}\PYG{p}{]}\PYG{p}{,}\PYG{p}{[}\PYG{l+m+mi}{4}\PYG{p}{,}\PYG{l+m+mi}{5}\PYG{p}{,}\PYG{l+m+mi}{6}\PYG{p}{]}\PYG{p}{,}\PYG{p}{[}\PYG{l+m+mi}{7}\PYG{p}{,}\PYG{l+m+mi}{8}\PYG{p}{,}\PYG{l+m+mi}{9}\PYG{p}{]}\PYG{p}{,}\PYG{p}{[}\PYG{l+m+mi}{10}\PYG{p}{,}\PYG{l+m+mi}{11}\PYG{p}{,}\PYG{l+m+mi}{12}\PYG{p}{]}\PYG{p}{]}\PYG{p}{,} \PYG{o}{\PYGZhy{}}\PYG{l+m+mi}{1}\PYG{p}{)}
\PYG{g+go}{array([[ 0,  0,  0],}
\PYG{g+go}{       [ 4,  0,  0],}
\PYG{g+go}{       [ 7,  8,  0],}
\PYG{g+go}{       [10, 11, 12]])}
\end{sphinxVerbatim}

\end{fulllineitems}

\index{tril\_indices() (in module symjax.tensor)@\spxentry{tril\_indices()}\spxextra{in module symjax.tensor}}

\begin{fulllineitems}
\phantomsection\label{\detokenize{modules/tensor:symjax.tensor.tril_indices}}\pysiglinewithargsret{\sphinxbfcode{\sphinxupquote{tril\_indices}}}{\emph{\DUrole{o}{*}\DUrole{n}{args}}, \emph{\DUrole{o}{**}\DUrole{n}{kwargs}}}{}
Return the indices for the lower\sphinxhyphen{}triangle of an (n, m) array.

LAX\sphinxhyphen{}backend implementation of {\hyperref[\detokenize{modules/tensor:symjax.tensor.tril_indices}]{\sphinxcrossref{\sphinxcode{\sphinxupquote{tril\_indices()}}}}}.
ADDITIONOriginal docstring below.

LAX\sphinxhyphen{}backend implementation of {\hyperref[\detokenize{modules/tensor:symjax.tensor.tril_indices}]{\sphinxcrossref{\sphinxcode{\sphinxupquote{tril\_indices()}}}}}.
Original docstring below.
\begin{quote}
\begin{quote}
\end{quote}
\begin{description}
\item[{Returns}] \leavevmode\begin{description}
\item[{inds}] \leavevmode{[}tuple of arrays{]}
The indices for the triangle. The returned tuple contains two arrays,
each with the indices along one dimension of the array.

\end{description}

triu\_indices : similar function, for upper\sphinxhyphen{}triangular.
mask\_indices : generic function accepting an arbitrary mask function.
tril, triu

\DUrole{versionmodified,added}{New in version 1.4.0.}

Compute two different sets of indices to access 4x4 arrays, one for the
lower triangular part starting at the main diagonal, and one starting two
diagonals further right:

\begin{sphinxVerbatim}[commandchars=\\\{\}]
\PYG{g+gp}{\PYGZgt{}\PYGZgt{}\PYGZgt{} }\PYG{n}{il1} \PYG{o}{=} \PYG{n}{np}\PYG{o}{.}\PYG{n}{tril\PYGZus{}indices}\PYG{p}{(}\PYG{l+m+mi}{4}\PYG{p}{)}
\PYG{g+gp}{\PYGZgt{}\PYGZgt{}\PYGZgt{} }\PYG{n}{il2} \PYG{o}{=} \PYG{n}{np}\PYG{o}{.}\PYG{n}{tril\PYGZus{}indices}\PYG{p}{(}\PYG{l+m+mi}{4}\PYG{p}{,} \PYG{l+m+mi}{2}\PYG{p}{)}
\end{sphinxVerbatim}

Here is how they can be used with a sample array:

\begin{sphinxVerbatim}[commandchars=\\\{\}]
\PYG{g+gp}{\PYGZgt{}\PYGZgt{}\PYGZgt{} }\PYG{n}{a} \PYG{o}{=} \PYG{n}{np}\PYG{o}{.}\PYG{n}{arange}\PYG{p}{(}\PYG{l+m+mi}{16}\PYG{p}{)}\PYG{o}{.}\PYG{n}{reshape}\PYG{p}{(}\PYG{l+m+mi}{4}\PYG{p}{,} \PYG{l+m+mi}{4}\PYG{p}{)}
\PYG{g+gp}{\PYGZgt{}\PYGZgt{}\PYGZgt{} }\PYG{n}{a}
\PYG{g+go}{array([[ 0,  1,  2,  3],}
\PYG{g+go}{       [ 4,  5,  6,  7],}
\PYG{g+go}{       [ 8,  9, 10, 11],}
\PYG{g+go}{       [12, 13, 14, 15]])}
\end{sphinxVerbatim}

Both for indexing:

\begin{sphinxVerbatim}[commandchars=\\\{\}]
\PYG{g+gp}{\PYGZgt{}\PYGZgt{}\PYGZgt{} }\PYG{n}{a}\PYG{p}{[}\PYG{n}{il1}\PYG{p}{]}
\PYG{g+go}{array([ 0,  4,  5, ..., 13, 14, 15])}
\end{sphinxVerbatim}

And for assigning values:

\begin{sphinxVerbatim}[commandchars=\\\{\}]
\PYG{g+gp}{\PYGZgt{}\PYGZgt{}\PYGZgt{} }\PYG{n}{a}\PYG{p}{[}\PYG{n}{il1}\PYG{p}{]} \PYG{o}{=} \PYG{o}{\PYGZhy{}}\PYG{l+m+mi}{1}
\PYG{g+gp}{\PYGZgt{}\PYGZgt{}\PYGZgt{} }\PYG{n}{a}
\PYG{g+go}{array([[\PYGZhy{}1,  1,  2,  3],}
\PYG{g+go}{       [\PYGZhy{}1, \PYGZhy{}1,  6,  7],}
\PYG{g+go}{       [\PYGZhy{}1, \PYGZhy{}1, \PYGZhy{}1, 11],}
\PYG{g+go}{       [\PYGZhy{}1, \PYGZhy{}1, \PYGZhy{}1, \PYGZhy{}1]])}
\end{sphinxVerbatim}

These cover almost the whole array (two diagonals right of the main one):

\begin{sphinxVerbatim}[commandchars=\\\{\}]
\PYG{g+gp}{\PYGZgt{}\PYGZgt{}\PYGZgt{} }\PYG{n}{a}\PYG{p}{[}\PYG{n}{il2}\PYG{p}{]} \PYG{o}{=} \PYG{o}{\PYGZhy{}}\PYG{l+m+mi}{10}
\PYG{g+gp}{\PYGZgt{}\PYGZgt{}\PYGZgt{} }\PYG{n}{a}
\PYG{g+go}{array([[\PYGZhy{}10, \PYGZhy{}10, \PYGZhy{}10,   3],}
\PYG{g+go}{       [\PYGZhy{}10, \PYGZhy{}10, \PYGZhy{}10, \PYGZhy{}10],}
\PYG{g+go}{       [\PYGZhy{}10, \PYGZhy{}10, \PYGZhy{}10, \PYGZhy{}10],}
\PYG{g+go}{       [\PYGZhy{}10, \PYGZhy{}10, \PYGZhy{}10, \PYGZhy{}10]])}
\end{sphinxVerbatim}

\end{description}
\end{quote}

\end{fulllineitems}

\index{triu() (in module symjax.tensor)@\spxentry{triu()}\spxextra{in module symjax.tensor}}

\begin{fulllineitems}
\phantomsection\label{\detokenize{modules/tensor:symjax.tensor.triu}}\pysiglinewithargsret{\sphinxbfcode{\sphinxupquote{triu}}}{\emph{\DUrole{n}{m}}, \emph{\DUrole{n}{k}\DUrole{o}{=}\DUrole{default_value}{0}}}{}
Upper triangle of an array.

LAX\sphinxhyphen{}backend implementation of {\hyperref[\detokenize{modules/tensor:symjax.tensor.triu}]{\sphinxcrossref{\sphinxcode{\sphinxupquote{triu()}}}}}.
ADDITIONOriginal docstring below.

LA

\end{fulllineitems}

\index{triu\_indices() (in module symjax.tensor)@\spxentry{triu\_indices()}\spxextra{in module symjax.tensor}}

\begin{fulllineitems}
\phantomsection\label{\detokenize{modules/tensor:symjax.tensor.triu_indices}}\pysiglinewithargsret{\sphinxbfcode{\sphinxupquote{triu\_indices}}}{\emph{\DUrole{o}{*}\DUrole{n}{args}}, \emph{\DUrole{o}{**}\DUrole{n}{kwargs}}}{}
Return the indices for the upper\sphinxhyphen{}triangle of an (n, m) array.

LAX\sphinxhyphen{}backend implementation of {\hyperref[\detokenize{modules/tensor:symjax.tensor.triu_indices}]{\sphinxcrossref{\sphinxcode{\sphinxupquote{triu\_indices()}}}}}.
ADDITIONOriginal docstring below.

LAX\sphinxhyphen{}backend implementation of {\hyperref[\detokenize{modules/tensor:symjax.tensor.triu_indices}]{\sphinxcrossref{\sphinxcode{\sphinxupquote{triu\_indices()}}}}}.
Original docstring below.
\begin{quote}
\begin{quote}
\end{quote}
\begin{description}
\item[{Returns}] \leavevmode\begin{description}
\item[{inds}] \leavevmode{[}tuple, shape(2) of ndarrays, shape(\sphinxtitleref{n}){]}
The indices for the triangle. The returned tuple contains two arrays,
each with the indices along one dimension of the array.  Can be used
to slice a ndarray of shape(\sphinxtitleref{n}, \sphinxtitleref{n}).

\end{description}

tril\_indices : similar function, for lower\sphinxhyphen{}triangular.
mask\_indices : generic function accepting an arbitrary mask function.
triu, tril

\DUrole{versionmodified,added}{New in version 1.4.0.}

Compute two different sets of indices to access 4x4 arrays, one for the
upper triangular part starting at the main diagonal, and one starting two
diagonals further right:

\begin{sphinxVerbatim}[commandchars=\\\{\}]
\PYG{g+gp}{\PYGZgt{}\PYGZgt{}\PYGZgt{} }\PYG{n}{iu1} \PYG{o}{=} \PYG{n}{np}\PYG{o}{.}\PYG{n}{triu\PYGZus{}indices}\PYG{p}{(}\PYG{l+m+mi}{4}\PYG{p}{)}
\PYG{g+gp}{\PYGZgt{}\PYGZgt{}\PYGZgt{} }\PYG{n}{iu2} \PYG{o}{=} \PYG{n}{np}\PYG{o}{.}\PYG{n}{triu\PYGZus{}indices}\PYG{p}{(}\PYG{l+m+mi}{4}\PYG{p}{,} \PYG{l+m+mi}{2}\PYG{p}{)}
\end{sphinxVerbatim}

Here is how they can be used with a sample array:

\begin{sphinxVerbatim}[commandchars=\\\{\}]
\PYG{g+gp}{\PYGZgt{}\PYGZgt{}\PYGZgt{} }\PYG{n}{a} \PYG{o}{=} \PYG{n}{np}\PYG{o}{.}\PYG{n}{arange}\PYG{p}{(}\PYG{l+m+mi}{16}\PYG{p}{)}\PYG{o}{.}\PYG{n}{reshape}\PYG{p}{(}\PYG{l+m+mi}{4}\PYG{p}{,} \PYG{l+m+mi}{4}\PYG{p}{)}
\PYG{g+gp}{\PYGZgt{}\PYGZgt{}\PYGZgt{} }\PYG{n}{a}
\PYG{g+go}{array([[ 0,  1,  2,  3],}
\PYG{g+go}{       [ 4,  5,  6,  7],}
\PYG{g+go}{       [ 8,  9, 10, 11],}
\PYG{g+go}{       [12, 13, 14, 15]])}
\end{sphinxVerbatim}

Both for indexing:

\begin{sphinxVerbatim}[commandchars=\\\{\}]
\PYG{g+gp}{\PYGZgt{}\PYGZgt{}\PYGZgt{} }\PYG{n}{a}\PYG{p}{[}\PYG{n}{iu1}\PYG{p}{]}
\PYG{g+go}{array([ 0,  1,  2, ..., 10, 11, 15])}
\end{sphinxVerbatim}

And for assigning values:

\begin{sphinxVerbatim}[commandchars=\\\{\}]
\PYG{g+gp}{\PYGZgt{}\PYGZgt{}\PYGZgt{} }\PYG{n}{a}\PYG{p}{[}\PYG{n}{iu1}\PYG{p}{]} \PYG{o}{=} \PYG{o}{\PYGZhy{}}\PYG{l+m+mi}{1}
\PYG{g+gp}{\PYGZgt{}\PYGZgt{}\PYGZgt{} }\PYG{n}{a}
\PYG{g+go}{array([[\PYGZhy{}1, \PYGZhy{}1, \PYGZhy{}1, \PYGZhy{}1],}
\PYG{g+go}{       [ 4, \PYGZhy{}1, \PYGZhy{}1, \PYGZhy{}1],}
\PYG{g+go}{       [ 8,  9, \PYGZhy{}1, \PYGZhy{}1],}
\PYG{g+go}{       [12, 13, 14, \PYGZhy{}1]])}
\end{sphinxVerbatim}

These cover only a small part of the whole array (two diagonals right
of the main one):

\begin{sphinxVerbatim}[commandchars=\\\{\}]
\PYG{g+gp}{\PYGZgt{}\PYGZgt{}\PYGZgt{} }\PYG{n}{a}\PYG{p}{[}\PYG{n}{iu2}\PYG{p}{]} \PYG{o}{=} \PYG{o}{\PYGZhy{}}\PYG{l+m+mi}{10}
\PYG{g+gp}{\PYGZgt{}\PYGZgt{}\PYGZgt{} }\PYG{n}{a}
\PYG{g+go}{array([[ \PYGZhy{}1,  \PYGZhy{}1, \PYGZhy{}10, \PYGZhy{}10],}
\PYG{g+go}{       [  4,  \PYGZhy{}1,  \PYGZhy{}1, \PYGZhy{}10],}
\PYG{g+go}{       [  8,   9,  \PYGZhy{}1,  \PYGZhy{}1],}
\PYG{g+go}{       [ 12,  13,  14,  \PYGZhy{}1]])}
\end{sphinxVerbatim}

\end{description}
\end{quote}

\end{fulllineitems}

\index{true\_divide() (in module symjax.tensor)@\spxentry{true\_divide()}\spxextra{in module symjax.tensor}}

\begin{fulllineitems}
\phantomsection\label{\detokenize{modules/tensor:symjax.tensor.true_divide}}\pysiglinewithargsret{\sphinxbfcode{\sphinxupquote{true\_divide}}}{\emph{\DUrole{n}{x1}}, \emph{\DUrole{n}{x2}}}{}
Returns a true division of the inputs, element\sphinxhyphen{}wise.

LAX\sphinxhyphen{}backend implementation of {\hyperref[\detokenize{modules/tensor:symjax.tensor.true_divide}]{\sphinxcrossref{\sphinxcode{\sphinxupquote{true\_divide()}}}}}.
ADDITIONOriginal docstring below.

LAX\sphinxhyphen{}backend implementation of {\hyperref[\detokenize{modules/tensor:symjax.tensor.true_divide}]{\sphinxcrossref{\sphinxcode{\sphinxupquote{true\_divide()}}}}}.
Original docstring below.

true\_divide(x1, x2, /, out=None, {\color{red}\bfseries{}*}, where=True, casting=’same\_kind’, order=’K’, dtype=None, subok=True{[}, signature, extobj{]})

Instead of the Python traditional ‘floor division’, this returns a true
division.  True division adjusts the output type to present the best
answer, regardless of input types.
\begin{quote}\begin{description}
\item[{Returns}] \leavevmode
\sphinxstylestrong{out} \textendash{} This is a scalar if both \sphinxtitleref{x1} and \sphinxtitleref{x2} are scalars.

\item[{Return type}] \leavevmode
ndarray or scalar

\end{description}\end{quote}
\subsubsection*{Notes}

The floor division operator \sphinxcode{\sphinxupquote{//}} was added in Python 2.2 making
\sphinxcode{\sphinxupquote{//}} and \sphinxcode{\sphinxupquote{/}} equivalent operators.  The default floor division
operation of \sphinxcode{\sphinxupquote{/}} can be replaced by true division with \sphinxcode{\sphinxupquote{from
\_\_future\_\_ import division}}.

In Python 3.0, \sphinxcode{\sphinxupquote{//}} is the floor division operator and \sphinxcode{\sphinxupquote{/}} the
true division operator.  The \sphinxcode{\sphinxupquote{true\_divide(x1, x2)}} function is
equivalent to true division in Python.
\subsubsection*{Examples}

\begin{sphinxVerbatim}[commandchars=\\\{\}]
\PYG{g+gp}{\PYGZgt{}\PYGZgt{}\PYGZgt{} }\PYG{n}{x} \PYG{o}{=} \PYG{n}{np}\PYG{o}{.}\PYG{n}{arange}\PYG{p}{(}\PYG{l+m+mi}{5}\PYG{p}{)}
\PYG{g+gp}{\PYGZgt{}\PYGZgt{}\PYGZgt{} }\PYG{n}{np}\PYG{o}{.}\PYG{n}{true\PYGZus{}divide}\PYG{p}{(}\PYG{n}{x}\PYG{p}{,} \PYG{l+m+mi}{4}\PYG{p}{)}
\PYG{g+go}{array([ 0.  ,  0.25,  0.5 ,  0.75,  1.  ])}
\end{sphinxVerbatim}

\begin{sphinxVerbatim}[commandchars=\\\{\}]
\PYG{g+gp}{\PYGZgt{}\PYGZgt{}\PYGZgt{} }\PYG{n}{x}\PYG{o}{/}\PYG{o}{/}\PYG{l+m+mi}{4}
\PYG{g+go}{array([0, 0, 0, 0, 1])}
\end{sphinxVerbatim}

\begin{sphinxVerbatim}[commandchars=\\\{\}]
\PYG{g+gp}{\PYGZgt{}\PYGZgt{}\PYGZgt{} }\PYG{k+kn}{from} \PYG{n+nn}{\PYGZus{}\PYGZus{}future\PYGZus{}\PYGZus{}} \PYG{k+kn}{import} \PYG{n}{division}
\PYG{g+gp}{\PYGZgt{}\PYGZgt{}\PYGZgt{} }\PYG{n}{x}\PYG{o}{/}\PYG{l+m+mi}{4}
\PYG{g+go}{array([ 0.  ,  0.25,  0.5 ,  0.75,  1.  ])}
\PYG{g+gp}{\PYGZgt{}\PYGZgt{}\PYGZgt{} }\PYG{n}{x}\PYG{o}{/}\PYG{o}{/}\PYG{l+m+mi}{4}
\PYG{g+go}{array([0, 0, 0, 0, 1])}
\end{sphinxVerbatim}

\end{fulllineitems}

\index{var() (in module symjax.tensor)@\spxentry{var()}\spxextra{in module symjax.tensor}}

\begin{fulllineitems}
\phantomsection\label{\detokenize{modules/tensor:symjax.tensor.var}}\pysiglinewithargsret{\sphinxbfcode{\sphinxupquote{var}}}{\emph{\DUrole{n}{a}}, \emph{\DUrole{n}{axis}\DUrole{o}{=}\DUrole{default_value}{None}}, \emph{\DUrole{n}{dtype}\DUrole{o}{=}\DUrole{default_value}{None}}, \emph{\DUrole{n}{out}\DUrole{o}{=}\DUrole{default_value}{None}}, \emph{\DUrole{n}{ddof}\DUrole{o}{=}\DUrole{default_value}{0}}, \emph{\DUrole{n}{keepdims}\DUrole{o}{=}\DUrole{default_value}{False}}}{}
Compute the variance along the specified axis.

LAX\sphinxhyphen{}backend implementation of {\hyperref[\detokenize{modules/tensor:symjax.tensor.var}]{\sphinxcrossref{\sphinxcode{\sphinxupquote{var()}}}}}.
ADDITIONOriginal docstring below.

LAX\sphinxhyphen{}backend implementation of {\hyperref[\detokenize{modules/tensor:symjax.tensor.var}]{\sphinxcrossref{\sphinxcode{\sphinxupquote{var()}}}}}.
Original docstring below.

Returns the variance of the array elements, a measure of the spread of a
distribution.  The variance is computed for the flattened array by
default, otherwise over the specified axis.
\begin{quote}\begin{description}
\item[{Parameters}] \leavevmode
\sphinxstyleliteralstrong{\sphinxupquote{dtype}} (\sphinxstyleliteralemphasis{\sphinxupquote{data\sphinxhyphen{}type}}\sphinxstyleliteralemphasis{\sphinxupquote{, }}\sphinxstyleliteralemphasis{\sphinxupquote{optional}}) \textendash{} Type to use in computing the variance.  For arrays of integer type
the default is \sphinxtitleref{float64}; for arrays of float types it is the same as
the array type.

\item[{Returns}] \leavevmode
\sphinxstylestrong{variance} \textendash{} If \sphinxcode{\sphinxupquote{out=None}}, returns a new array containing the variance;
otherwise, a reference to the output array is returned.

\item[{Return type}] \leavevmode
ndarray, see dtype parameter above

\end{description}\end{quote}

\sphinxstrong{See also:}

{\hyperref[\detokenize{modules/tensor:symjax.tensor.std}]{\sphinxcrossref{\sphinxcode{\sphinxupquote{std()}}}}}, {\hyperref[\detokenize{modules/tensor:symjax.tensor.mean}]{\sphinxcrossref{\sphinxcode{\sphinxupquote{mean()}}}}}, \sphinxcode{\sphinxupquote{nanmean()}}, \sphinxcode{\sphinxupquote{nanstd()}}, \sphinxcode{\sphinxupquote{nanvar()}}, \sphinxcode{\sphinxupquote{ufuncs\sphinxhyphen{}output\sphinxhyphen{}type()}}

\subsubsection*{Notes}

The variance is the average of the squared deviations from the mean,
i.e.,  \sphinxcode{\sphinxupquote{var = mean(abs(x \sphinxhyphen{} x.mean())**2)}}.

The mean is normally calculated as \sphinxcode{\sphinxupquote{x.sum() / N}}, where \sphinxcode{\sphinxupquote{N = len(x)}}.
If, however, \sphinxtitleref{ddof} is specified, the divisor \sphinxcode{\sphinxupquote{N \sphinxhyphen{} ddof}} is used
instead.  In standard statistical practice, \sphinxcode{\sphinxupquote{ddof=1}} provides an
unbiased estimator of the variance of a hypothetical infinite population.
\sphinxcode{\sphinxupquote{ddof=0}} provides a maximum likelihood estimate of the variance for
normally distributed variables.

Note that for complex numbers, the absolute value is taken before
squaring, so that the result is always real and nonnegative.

For floating\sphinxhyphen{}point input, the variance is computed using the same
precision the input has.  Depending on the input data, this can cause
the results to be inaccurate, especially for \sphinxtitleref{float32} (see example
below).  Specifying a higher\sphinxhyphen{}accuracy accumulator using the \sphinxcode{\sphinxupquote{dtype}}
keyword can alleviate this issue.
\subsubsection*{Examples}

\begin{sphinxVerbatim}[commandchars=\\\{\}]
\PYG{g+gp}{\PYGZgt{}\PYGZgt{}\PYGZgt{} }\PYG{n}{a} \PYG{o}{=} \PYG{n}{np}\PYG{o}{.}\PYG{n}{array}\PYG{p}{(}\PYG{p}{[}\PYG{p}{[}\PYG{l+m+mi}{1}\PYG{p}{,} \PYG{l+m+mi}{2}\PYG{p}{]}\PYG{p}{,} \PYG{p}{[}\PYG{l+m+mi}{3}\PYG{p}{,} \PYG{l+m+mi}{4}\PYG{p}{]}\PYG{p}{]}\PYG{p}{)}
\PYG{g+gp}{\PYGZgt{}\PYGZgt{}\PYGZgt{} }\PYG{n}{np}\PYG{o}{.}\PYG{n}{var}\PYG{p}{(}\PYG{n}{a}\PYG{p}{)}
\PYG{g+go}{1.25}
\PYG{g+gp}{\PYGZgt{}\PYGZgt{}\PYGZgt{} }\PYG{n}{np}\PYG{o}{.}\PYG{n}{var}\PYG{p}{(}\PYG{n}{a}\PYG{p}{,} \PYG{n}{axis}\PYG{o}{=}\PYG{l+m+mi}{0}\PYG{p}{)}
\PYG{g+go}{array([1.,  1.])}
\PYG{g+gp}{\PYGZgt{}\PYGZgt{}\PYGZgt{} }\PYG{n}{np}\PYG{o}{.}\PYG{n}{var}\PYG{p}{(}\PYG{n}{a}\PYG{p}{,} \PYG{n}{axis}\PYG{o}{=}\PYG{l+m+mi}{1}\PYG{p}{)}
\PYG{g+go}{array([0.25,  0.25])}
\end{sphinxVerbatim}

In single precision, var() can be inaccurate:

\begin{sphinxVerbatim}[commandchars=\\\{\}]
\PYG{g+gp}{\PYGZgt{}\PYGZgt{}\PYGZgt{} }\PYG{n}{a} \PYG{o}{=} \PYG{n}{np}\PYG{o}{.}\PYG{n}{zeros}\PYG{p}{(}\PYG{p}{(}\PYG{l+m+mi}{2}\PYG{p}{,} \PYG{l+m+mi}{512}\PYG{o}{*}\PYG{l+m+mi}{512}\PYG{p}{)}\PYG{p}{,} \PYG{n}{dtype}\PYG{o}{=}\PYG{n}{np}\PYG{o}{.}\PYG{n}{float32}\PYG{p}{)}
\PYG{g+gp}{\PYGZgt{}\PYGZgt{}\PYGZgt{} }\PYG{n}{a}\PYG{p}{[}\PYG{l+m+mi}{0}\PYG{p}{,} \PYG{p}{:}\PYG{p}{]} \PYG{o}{=} \PYG{l+m+mf}{1.0}
\PYG{g+gp}{\PYGZgt{}\PYGZgt{}\PYGZgt{} }\PYG{n}{a}\PYG{p}{[}\PYG{l+m+mi}{1}\PYG{p}{,} \PYG{p}{:}\PYG{p}{]} \PYG{o}{=} \PYG{l+m+mf}{0.1}
\PYG{g+gp}{\PYGZgt{}\PYGZgt{}\PYGZgt{} }\PYG{n}{np}\PYG{o}{.}\PYG{n}{var}\PYG{p}{(}\PYG{n}{a}\PYG{p}{)}
\PYG{g+go}{0.20250003}
\end{sphinxVerbatim}

Computing the variance in float64 is more accurate:

\begin{sphinxVerbatim}[commandchars=\\\{\}]
\PYG{g+gp}{\PYGZgt{}\PYGZgt{}\PYGZgt{} }\PYG{n}{np}\PYG{o}{.}\PYG{n}{var}\PYG{p}{(}\PYG{n}{a}\PYG{p}{,} \PYG{n}{dtype}\PYG{o}{=}\PYG{n}{np}\PYG{o}{.}\PYG{n}{float64}\PYG{p}{)}
\PYG{g+go}{0.20249999932944759 \PYGZsh{} may vary}
\PYG{g+gp}{\PYGZgt{}\PYGZgt{}\PYGZgt{} }\PYG{p}{(}\PYG{p}{(}\PYG{l+m+mi}{1}\PYG{o}{\PYGZhy{}}\PYG{l+m+mf}{0.55}\PYG{p}{)}\PYG{o}{*}\PYG{o}{*}\PYG{l+m+mi}{2} \PYG{o}{+} \PYG{p}{(}\PYG{l+m+mf}{0.1}\PYG{o}{\PYGZhy{}}\PYG{l+m+mf}{0.55}\PYG{p}{)}\PYG{o}{*}\PYG{o}{*}\PYG{l+m+mi}{2}\PYG{p}{)}\PYG{o}{/}\PYG{l+m+mi}{2}
\PYG{g+go}{0.2025}
\end{sphinxVerbatim}

\end{fulllineitems}

\index{vdot() (in module symjax.tensor)@\spxentry{vdot()}\spxextra{in module symjax.tensor}}

\begin{fulllineitems}
\phantomsection\label{\detokenize{modules/tensor:symjax.tensor.vdot}}\pysiglinewithargsret{\sphinxbfcode{\sphinxupquote{vdot}}}{\emph{\DUrole{n}{a}}, \emph{\DUrole{n}{b}}, \emph{\DUrole{n}{precision}\DUrole{o}{=}\DUrole{default_value}{None}}}{}
Return the dot product of two vectors.

LAX\sphinxhyphen{}backend implementation of {\hyperref[\detokenize{modules/tensor:symjax.tensor.vdot}]{\sphinxcrossref{\sphinxcode{\sphinxupquote{vdot()}}}}}.
ADDITIONOriginal docstring below.

LAX\sphinxhyphen{}backend implementation of {\hyperref[\detokenize{modules/tensor:symjax.tensor.vdot}]{\sphinxcrossref{\sphinxcode{\sphinxupquote{vdot()}}}}}.
In addition to the original NumPy arguments listed below, also supports
\sphinxcode{\sphinxupquote{precision}} for extra control over matrix\sphinxhyphen{}multiplication precision
on supported devices. See \sphinxcode{\sphinxupquote{jax.lax.dot()}} for details.

Original docstring below.
\begin{quote}
\begin{quote}

vdot(a, b)

The vdot(\sphinxtitleref{a}, \sphinxtitleref{b}) function handles complex numbers differently than
dot(\sphinxtitleref{a}, \sphinxtitleref{b}).  If the first argument is complex the complex conjugate
of the first argument is used for the calculation of the dot product.

Note that \sphinxtitleref{vdot} handles multidimensional arrays differently than \sphinxtitleref{dot}:
it does \sphinxstyleemphasis{not} perform a matrix product, but flattens input arguments
to 1\sphinxhyphen{}D vectors first. Consequently, it should only be used for vectors.
\end{quote}
\begin{description}
\item[{Returns}] \leavevmode\begin{description}
\item[{output}] \leavevmode{[}ndarray{]}
Dot product of \sphinxtitleref{a} and \sphinxtitleref{b}.  Can be an int, float, or
complex depending on the types of \sphinxtitleref{a} and \sphinxtitleref{b}.

\end{description}
\begin{description}
\item[{dot}] \leavevmode{[}Return the dot product without using the complex conjugate of the{]}
first argument.

\end{description}

\begin{sphinxVerbatim}[commandchars=\\\{\}]
\PYG{g+gp}{\PYGZgt{}\PYGZgt{}\PYGZgt{} }\PYG{n}{a} \PYG{o}{=} \PYG{n}{np}\PYG{o}{.}\PYG{n}{array}\PYG{p}{(}\PYG{p}{[}\PYG{l+m+mi}{1}\PYG{o}{+}\PYG{l+m+mi}{2}\PYG{n}{j}\PYG{p}{,}\PYG{l+m+mi}{3}\PYG{o}{+}\PYG{l+m+mi}{4}\PYG{n}{j}\PYG{p}{]}\PYG{p}{)}
\PYG{g+gp}{\PYGZgt{}\PYGZgt{}\PYGZgt{} }\PYG{n}{b} \PYG{o}{=} \PYG{n}{np}\PYG{o}{.}\PYG{n}{array}\PYG{p}{(}\PYG{p}{[}\PYG{l+m+mi}{5}\PYG{o}{+}\PYG{l+m+mi}{6}\PYG{n}{j}\PYG{p}{,}\PYG{l+m+mi}{7}\PYG{o}{+}\PYG{l+m+mi}{8}\PYG{n}{j}\PYG{p}{]}\PYG{p}{)}
\PYG{g+gp}{\PYGZgt{}\PYGZgt{}\PYGZgt{} }\PYG{n}{np}\PYG{o}{.}\PYG{n}{vdot}\PYG{p}{(}\PYG{n}{a}\PYG{p}{,} \PYG{n}{b}\PYG{p}{)}
\PYG{g+go}{(70\PYGZhy{}8j)}
\PYG{g+gp}{\PYGZgt{}\PYGZgt{}\PYGZgt{} }\PYG{n}{np}\PYG{o}{.}\PYG{n}{vdot}\PYG{p}{(}\PYG{n}{b}\PYG{p}{,} \PYG{n}{a}\PYG{p}{)}
\PYG{g+go}{(70+8j)}
\end{sphinxVerbatim}

Note that higher\sphinxhyphen{}dimensional arrays are flattened!

\begin{sphinxVerbatim}[commandchars=\\\{\}]
\PYG{g+gp}{\PYGZgt{}\PYGZgt{}\PYGZgt{} }\PYG{n}{a} \PYG{o}{=} \PYG{n}{np}\PYG{o}{.}\PYG{n}{array}\PYG{p}{(}\PYG{p}{[}\PYG{p}{[}\PYG{l+m+mi}{1}\PYG{p}{,} \PYG{l+m+mi}{4}\PYG{p}{]}\PYG{p}{,} \PYG{p}{[}\PYG{l+m+mi}{5}\PYG{p}{,} \PYG{l+m+mi}{6}\PYG{p}{]}\PYG{p}{]}\PYG{p}{)}
\PYG{g+gp}{\PYGZgt{}\PYGZgt{}\PYGZgt{} }\PYG{n}{b} \PYG{o}{=} \PYG{n}{np}\PYG{o}{.}\PYG{n}{array}\PYG{p}{(}\PYG{p}{[}\PYG{p}{[}\PYG{l+m+mi}{4}\PYG{p}{,} \PYG{l+m+mi}{1}\PYG{p}{]}\PYG{p}{,} \PYG{p}{[}\PYG{l+m+mi}{2}\PYG{p}{,} \PYG{l+m+mi}{2}\PYG{p}{]}\PYG{p}{]}\PYG{p}{)}
\PYG{g+gp}{\PYGZgt{}\PYGZgt{}\PYGZgt{} }\PYG{n}{np}\PYG{o}{.}\PYG{n}{vdot}\PYG{p}{(}\PYG{n}{a}\PYG{p}{,} \PYG{n}{b}\PYG{p}{)}
\PYG{g+go}{30}
\PYG{g+gp}{\PYGZgt{}\PYGZgt{}\PYGZgt{} }\PYG{n}{np}\PYG{o}{.}\PYG{n}{vdot}\PYG{p}{(}\PYG{n}{b}\PYG{p}{,} \PYG{n}{a}\PYG{p}{)}
\PYG{g+go}{30}
\PYG{g+gp}{\PYGZgt{}\PYGZgt{}\PYGZgt{} }\PYG{l+m+mi}{1}\PYG{o}{*}\PYG{l+m+mi}{4} \PYG{o}{+} \PYG{l+m+mi}{4}\PYG{o}{*}\PYG{l+m+mi}{1} \PYG{o}{+} \PYG{l+m+mi}{5}\PYG{o}{*}\PYG{l+m+mi}{2} \PYG{o}{+} \PYG{l+m+mi}{6}\PYG{o}{*}\PYG{l+m+mi}{2}
\PYG{g+go}{30}
\end{sphinxVerbatim}

\end{description}
\end{quote}

\end{fulllineitems}

\index{vsplit() (in module symjax.tensor)@\spxentry{vsplit()}\spxextra{in module symjax.tensor}}

\begin{fulllineitems}
\phantomsection\label{\detokenize{modules/tensor:symjax.tensor.vsplit}}\pysiglinewithargsret{\sphinxbfcode{\sphinxupquote{vsplit}}}{\emph{\DUrole{n}{ary}}, \emph{\DUrole{n}{indices\_or\_sections}}}{}
Split an array into multiple sub\sphinxhyphen{}arrays vertically (row\sphinxhyphen{}wise).

LAX\sphinxhyphen{}backend implementation of {\hyperref[\detokenize{modules/tensor:symjax.tensor.vsplit}]{\sphinxcrossref{\sphinxcode{\sphinxupquote{vsplit()}}}}}.
ADDITIONOriginal docstring below.

LA

\end{fulllineitems}

\index{vstack() (in module symjax.tensor)@\spxentry{vstack()}\spxextra{in module symjax.tensor}}

\begin{fulllineitems}
\phantomsection\label{\detokenize{modules/tensor:symjax.tensor.vstack}}\pysiglinewithargsret{\sphinxbfcode{\sphinxupquote{vstack}}}{\emph{\DUrole{n}{tup}}}{}
Stack arrays in sequence vertically (row wise).

LAX\sphinxhyphen{}backend implementation of {\hyperref[\detokenize{modules/tensor:symjax.tensor.vstack}]{\sphinxcrossref{\sphinxcode{\sphinxupquote{vstack()}}}}}.
ADDITIONOriginal docstring below.

LAX\sphinxhyphen{}backend implementation of {\hyperref[\detokenize{modules/tensor:symjax.tensor.vstack}]{\sphinxcrossref{\sphinxcode{\sphinxupquote{vstack()}}}}}.
Original docstring below.

This is equivalent to concatenation along the first axis after 1\sphinxhyphen{}D arrays
of shape \sphinxtitleref{(N,)} have been reshaped to \sphinxtitleref{(1,N)}. Rebuilds arrays divided by
\sphinxtitleref{vsplit}.

This function makes most sense for arrays with up to 3 dimensions. For
instance, for pixel\sphinxhyphen{}data with a height (first axis), width (second axis),
and r/g/b channels (third axis). The functions \sphinxtitleref{concatenate}, \sphinxtitleref{stack} and
\sphinxtitleref{block} provide more general stacking and concatenation operations.
\begin{quote}\begin{description}
\item[{Returns}] \leavevmode
\sphinxstylestrong{stacked} \textendash{} The array formed by stacking the given arrays, will be at least 2\sphinxhyphen{}D.

\item[{Return type}] \leavevmode
ndarray

\end{description}\end{quote}

\sphinxstrong{See also:}

\begin{description}
\item[{{\hyperref[\detokenize{modules/tensor:symjax.tensor.stack}]{\sphinxcrossref{\sphinxcode{\sphinxupquote{stack()}}}}}}] \leavevmode
Join a sequence of arrays along a new axis.

\item[{{\hyperref[\detokenize{modules/tensor:symjax.tensor.hstack}]{\sphinxcrossref{\sphinxcode{\sphinxupquote{hstack()}}}}}}] \leavevmode
Stack arrays in sequence horizontally (column wise).

\item[{{\hyperref[\detokenize{modules/tensor:symjax.tensor.dstack}]{\sphinxcrossref{\sphinxcode{\sphinxupquote{dstack()}}}}}}] \leavevmode
Stack arrays in sequence depth wise (along third dimension).

\item[{{\hyperref[\detokenize{modules/tensor:symjax.tensor.concatenate}]{\sphinxcrossref{\sphinxcode{\sphinxupquote{concatenate()}}}}}}] \leavevmode
Join a sequence of arrays along an existing axis.

\item[{{\hyperref[\detokenize{modules/tensor:symjax.tensor.vsplit}]{\sphinxcrossref{\sphinxcode{\sphinxupquote{vsplit()}}}}}}] \leavevmode
Split array into a list of multiple sub\sphinxhyphen{}arrays vertically.

\item[{{\hyperref[\detokenize{modules/tensor:symjax.tensor.block}]{\sphinxcrossref{\sphinxcode{\sphinxupquote{block()}}}}}}] \leavevmode
Assemble arrays from blocks.

\end{description}

\subsubsection*{Examples}

\begin{sphinxVerbatim}[commandchars=\\\{\}]
\PYG{g+gp}{\PYGZgt{}\PYGZgt{}\PYGZgt{} }\PYG{n}{a} \PYG{o}{=} \PYG{n}{np}\PYG{o}{.}\PYG{n}{array}\PYG{p}{(}\PYG{p}{[}\PYG{l+m+mi}{1}\PYG{p}{,} \PYG{l+m+mi}{2}\PYG{p}{,} \PYG{l+m+mi}{3}\PYG{p}{]}\PYG{p}{)}
\PYG{g+gp}{\PYGZgt{}\PYGZgt{}\PYGZgt{} }\PYG{n}{b} \PYG{o}{=} \PYG{n}{np}\PYG{o}{.}\PYG{n}{array}\PYG{p}{(}\PYG{p}{[}\PYG{l+m+mi}{2}\PYG{p}{,} \PYG{l+m+mi}{3}\PYG{p}{,} \PYG{l+m+mi}{4}\PYG{p}{]}\PYG{p}{)}
\PYG{g+gp}{\PYGZgt{}\PYGZgt{}\PYGZgt{} }\PYG{n}{np}\PYG{o}{.}\PYG{n}{vstack}\PYG{p}{(}\PYG{p}{(}\PYG{n}{a}\PYG{p}{,}\PYG{n}{b}\PYG{p}{)}\PYG{p}{)}
\PYG{g+go}{array([[1, 2, 3],}
\PYG{g+go}{       [2, 3, 4]])}
\end{sphinxVerbatim}

\begin{sphinxVerbatim}[commandchars=\\\{\}]
\PYG{g+gp}{\PYGZgt{}\PYGZgt{}\PYGZgt{} }\PYG{n}{a} \PYG{o}{=} \PYG{n}{np}\PYG{o}{.}\PYG{n}{array}\PYG{p}{(}\PYG{p}{[}\PYG{p}{[}\PYG{l+m+mi}{1}\PYG{p}{]}\PYG{p}{,} \PYG{p}{[}\PYG{l+m+mi}{2}\PYG{p}{]}\PYG{p}{,} \PYG{p}{[}\PYG{l+m+mi}{3}\PYG{p}{]}\PYG{p}{]}\PYG{p}{)}
\PYG{g+gp}{\PYGZgt{}\PYGZgt{}\PYGZgt{} }\PYG{n}{b} \PYG{o}{=} \PYG{n}{np}\PYG{o}{.}\PYG{n}{array}\PYG{p}{(}\PYG{p}{[}\PYG{p}{[}\PYG{l+m+mi}{2}\PYG{p}{]}\PYG{p}{,} \PYG{p}{[}\PYG{l+m+mi}{3}\PYG{p}{]}\PYG{p}{,} \PYG{p}{[}\PYG{l+m+mi}{4}\PYG{p}{]}\PYG{p}{]}\PYG{p}{)}
\PYG{g+gp}{\PYGZgt{}\PYGZgt{}\PYGZgt{} }\PYG{n}{np}\PYG{o}{.}\PYG{n}{vstack}\PYG{p}{(}\PYG{p}{(}\PYG{n}{a}\PYG{p}{,}\PYG{n}{b}\PYG{p}{)}\PYG{p}{)}
\PYG{g+go}{array([[1],}
\PYG{g+go}{       [2],}
\PYG{g+go}{       [3],}
\PYG{g+go}{       [2],}
\PYG{g+go}{       [3],}
\PYG{g+go}{       [4]])}
\end{sphinxVerbatim}

\end{fulllineitems}

\index{zeros() (in module symjax.tensor)@\spxentry{zeros()}\spxextra{in module symjax.tensor}}

\begin{fulllineitems}
\phantomsection\label{\detokenize{modules/tensor:symjax.tensor.zeros}}\pysiglinewithargsret{\sphinxbfcode{\sphinxupquote{zeros}}}{\emph{\DUrole{n}{shape}}, \emph{\DUrole{n}{dtype}\DUrole{o}{=}\DUrole{default_value}{None}}}{}
Return a new array of given shape and type, filled with zeros.

LAX\sphinxhyphen{}backend implementation of {\hyperref[\detokenize{modules/tensor:symjax.tensor.zeros}]{\sphinxcrossref{\sphinxcode{\sphinxupquote{zeros()}}}}}.
ADDITIONOriginal docstring below.

LAX\sphinxhyphen{}backend implementation of {\hyperref[\detokenize{modules/tensor:symjax.tensor.zeros}]{\sphinxcrossref{\sphinxcode{\sphinxupquote{zeros()}}}}}.
Original docstring below.

zeros(shape, dtype=float, order=’C’)
\begin{quote}
\begin{quote}
\end{quote}
\begin{description}
\item[{Returns}] \leavevmode\begin{description}
\item[{out}] \leavevmode{[}ndarray{]}
Array of zeros with the given shape, dtype, and order.

\end{description}

zeros\_like : Return an array of zeros with shape and type of input.
empty : Return a new uninitialized array.
ones : Return a new array setting values to one.
full : Return a new array of given shape filled with value.

\begin{sphinxVerbatim}[commandchars=\\\{\}]
\PYG{g+gp}{\PYGZgt{}\PYGZgt{}\PYGZgt{} }\PYG{n}{np}\PYG{o}{.}\PYG{n}{zeros}\PYG{p}{(}\PYG{l+m+mi}{5}\PYG{p}{)}
\PYG{g+go}{array([ 0.,  0.,  0.,  0.,  0.])}
\end{sphinxVerbatim}

\begin{sphinxVerbatim}[commandchars=\\\{\}]
\PYG{g+gp}{\PYGZgt{}\PYGZgt{}\PYGZgt{} }\PYG{n}{np}\PYG{o}{.}\PYG{n}{zeros}\PYG{p}{(}\PYG{p}{(}\PYG{l+m+mi}{5}\PYG{p}{,}\PYG{p}{)}\PYG{p}{,} \PYG{n}{dtype}\PYG{o}{=}\PYG{n+nb}{int}\PYG{p}{)}
\PYG{g+go}{array([0, 0, 0, 0, 0])}
\end{sphinxVerbatim}

\begin{sphinxVerbatim}[commandchars=\\\{\}]
\PYG{g+gp}{\PYGZgt{}\PYGZgt{}\PYGZgt{} }\PYG{n}{np}\PYG{o}{.}\PYG{n}{zeros}\PYG{p}{(}\PYG{p}{(}\PYG{l+m+mi}{2}\PYG{p}{,} \PYG{l+m+mi}{1}\PYG{p}{)}\PYG{p}{)}
\PYG{g+go}{array([[ 0.],}
\PYG{g+go}{       [ 0.]])}
\end{sphinxVerbatim}

\begin{sphinxVerbatim}[commandchars=\\\{\}]
\PYG{g+gp}{\PYGZgt{}\PYGZgt{}\PYGZgt{} }\PYG{n}{s} \PYG{o}{=} \PYG{p}{(}\PYG{l+m+mi}{2}\PYG{p}{,}\PYG{l+m+mi}{2}\PYG{p}{)}
\PYG{g+gp}{\PYGZgt{}\PYGZgt{}\PYGZgt{} }\PYG{n}{np}\PYG{o}{.}\PYG{n}{zeros}\PYG{p}{(}\PYG{n}{s}\PYG{p}{)}
\PYG{g+go}{array([[ 0.,  0.],}
\PYG{g+go}{       [ 0.,  0.]])}
\end{sphinxVerbatim}

\begin{sphinxVerbatim}[commandchars=\\\{\}]
\PYG{g+gp}{\PYGZgt{}\PYGZgt{}\PYGZgt{} }\PYG{n}{np}\PYG{o}{.}\PYG{n}{zeros}\PYG{p}{(}\PYG{p}{(}\PYG{l+m+mi}{2}\PYG{p}{,}\PYG{p}{)}\PYG{p}{,} \PYG{n}{dtype}\PYG{o}{=}\PYG{p}{[}\PYG{p}{(}\PYG{l+s+s1}{\PYGZsq{}}\PYG{l+s+s1}{x}\PYG{l+s+s1}{\PYGZsq{}}\PYG{p}{,} \PYG{l+s+s1}{\PYGZsq{}}\PYG{l+s+s1}{i4}\PYG{l+s+s1}{\PYGZsq{}}\PYG{p}{)}\PYG{p}{,} \PYG{p}{(}\PYG{l+s+s1}{\PYGZsq{}}\PYG{l+s+s1}{y}\PYG{l+s+s1}{\PYGZsq{}}\PYG{p}{,} \PYG{l+s+s1}{\PYGZsq{}}\PYG{l+s+s1}{i4}\PYG{l+s+s1}{\PYGZsq{}}\PYG{p}{)}\PYG{p}{]}\PYG{p}{)} \PYG{c+c1}{\PYGZsh{} custom dtype}
\PYG{g+go}{array([(0, 0), (0, 0)],}
\PYG{g+go}{      dtype=[(\PYGZsq{}x\PYGZsq{}, \PYGZsq{}\PYGZlt{}i4\PYGZsq{}), (\PYGZsq{}y\PYGZsq{}, \PYGZsq{}\PYGZlt{}i4\PYGZsq{})])}
\end{sphinxVerbatim}

\end{description}
\end{quote}

\end{fulllineitems}

\index{zeros\_like() (in module symjax.tensor)@\spxentry{zeros\_like()}\spxextra{in module symjax.tensor}}

\begin{fulllineitems}
\phantomsection\label{\detokenize{modules/tensor:symjax.tensor.zeros_like}}\pysiglinewithargsret{\sphinxbfcode{\sphinxupquote{zeros\_like}}}{\emph{\DUrole{n}{x}}, \emph{\DUrole{n}{dtype}\DUrole{o}{=}\DUrole{default_value}{None}}}{}
Return an array of zeros with the same shape and type as a given array.

LAX\sphinxhyphen{}backend implementation of {\hyperref[\detokenize{modules/tensor:symjax.tensor.zeros_like}]{\sphinxcrossref{\sphinxcode{\sphinxupquote{zeros\_like()}}}}}.
ADDITIONOriginal docstring below.

LAX\sphinxhyphen{}backend implementation of {\hyperref[\detokenize{modules/tensor:symjax.tensor.zeros_like}]{\sphinxcrossref{\sphinxcode{\sphinxupquote{zeros\_like()}}}}}.
Original docstring below.
\begin{quote}\begin{description}
\item[{Parameters}] \leavevmode
\sphinxstyleliteralstrong{\sphinxupquote{dtype}} (\sphinxstyleliteralemphasis{\sphinxupquote{data\sphinxhyphen{}type}}\sphinxstyleliteralemphasis{\sphinxupquote{, }}\sphinxstyleliteralemphasis{\sphinxupquote{optional}}) \textendash{} Overrides the data type of the result.

\item[{Returns}] \leavevmode
\sphinxstylestrong{out} \textendash{} Array of zeros with the same shape and type as \sphinxtitleref{a}.

\item[{Return type}] \leavevmode
ndarray

\end{description}\end{quote}

\sphinxstrong{See also:}

\begin{description}
\item[{{\hyperref[\detokenize{modules/tensor:symjax.tensor.empty_like}]{\sphinxcrossref{\sphinxcode{\sphinxupquote{empty\_like()}}}}}}] \leavevmode
Return an empty array with shape and type of input.

\item[{{\hyperref[\detokenize{modules/tensor:symjax.tensor.ones_like}]{\sphinxcrossref{\sphinxcode{\sphinxupquote{ones\_like()}}}}}}] \leavevmode
Return an array of ones with shape and type of input.

\item[{{\hyperref[\detokenize{modules/tensor:symjax.tensor.full_like}]{\sphinxcrossref{\sphinxcode{\sphinxupquote{full\_like()}}}}}}] \leavevmode
Return a new array with shape of input filled with value.

\item[{{\hyperref[\detokenize{modules/tensor:symjax.tensor.zeros}]{\sphinxcrossref{\sphinxcode{\sphinxupquote{zeros()}}}}}}] \leavevmode
Return a new array setting values to zero.

\end{description}

\subsubsection*{Examples}

\begin{sphinxVerbatim}[commandchars=\\\{\}]
\PYG{g+gp}{\PYGZgt{}\PYGZgt{}\PYGZgt{} }\PYG{n}{x} \PYG{o}{=} \PYG{n}{np}\PYG{o}{.}\PYG{n}{arange}\PYG{p}{(}\PYG{l+m+mi}{6}\PYG{p}{)}
\PYG{g+gp}{\PYGZgt{}\PYGZgt{}\PYGZgt{} }\PYG{n}{x} \PYG{o}{=} \PYG{n}{x}\PYG{o}{.}\PYG{n}{reshape}\PYG{p}{(}\PYG{p}{(}\PYG{l+m+mi}{2}\PYG{p}{,} \PYG{l+m+mi}{3}\PYG{p}{)}\PYG{p}{)}
\PYG{g+gp}{\PYGZgt{}\PYGZgt{}\PYGZgt{} }\PYG{n}{x}
\PYG{g+go}{array([[0, 1, 2],}
\PYG{g+go}{       [3, 4, 5]])}
\PYG{g+gp}{\PYGZgt{}\PYGZgt{}\PYGZgt{} }\PYG{n}{np}\PYG{o}{.}\PYG{n}{zeros\PYGZus{}like}\PYG{p}{(}\PYG{n}{x}\PYG{p}{)}
\PYG{g+go}{array([[0, 0, 0],}
\PYG{g+go}{       [0, 0, 0]])}
\end{sphinxVerbatim}

\begin{sphinxVerbatim}[commandchars=\\\{\}]
\PYG{g+gp}{\PYGZgt{}\PYGZgt{}\PYGZgt{} }\PYG{n}{y} \PYG{o}{=} \PYG{n}{np}\PYG{o}{.}\PYG{n}{arange}\PYG{p}{(}\PYG{l+m+mi}{3}\PYG{p}{,} \PYG{n}{dtype}\PYG{o}{=}\PYG{n+nb}{float}\PYG{p}{)}
\PYG{g+gp}{\PYGZgt{}\PYGZgt{}\PYGZgt{} }\PYG{n}{y}
\PYG{g+go}{array([0., 1., 2.])}
\PYG{g+gp}{\PYGZgt{}\PYGZgt{}\PYGZgt{} }\PYG{n}{np}\PYG{o}{.}\PYG{n}{zeros\PYGZus{}like}\PYG{p}{(}\PYG{n}{y}\PYG{p}{)}
\PYG{g+go}{array([0.,  0.,  0.])}
\end{sphinxVerbatim}

\end{fulllineitems}

\subsection{\sphinxstyleliteralintitle{\sphinxupquote{symjax.tensor.pdfs}}}
\label{\detokenize{modules/pdfs:module-symjax.tensor.pdfs}}\label{\detokenize{modules/pdfs:symjax-tensor-pdfs}}\label{\detokenize{modules/pdfs::doc}}\index{module@\spxentry{module}!symjax.tensor.pdfs@\spxentry{symjax.tensor.pdfs}}\index{symjax.tensor.pdfs@\spxentry{symjax.tensor.pdfs}!module@\spxentry{module}}

\subsubsection{Multivariate Normal}
\label{\detokenize{modules/pdfs:multivariate-normal}}\index{multivariate\_normal (class in symjax.tensor.pdfs)@\spxentry{multivariate\_normal}\spxextra{class in symjax.tensor.pdfs}}

\begin{fulllineitems}
\phantomsection\label{\detokenize{modules/pdfs:symjax.tensor.pdfs.multivariate_normal}}\pysigline{\sphinxbfcode{\sphinxupquote{class }}\sphinxbfcode{\sphinxupquote{multivariate\_normal}}}~\index{logpdf() (multivariate\_normal method)@\spxentry{logpdf()}\spxextra{multivariate\_normal method}}

\begin{fulllineitems}
\phantomsection\label{\detokenize{modules/pdfs:symjax.tensor.pdfs.multivariate_normal.logpdf}}\pysiglinewithargsret{\sphinxbfcode{\sphinxupquote{logpdf}}}{\emph{\DUrole{n}{mean}}, \emph{\DUrole{n}{cov}}}{}
Log of the multivariate normal probability density function.
\begin{quote}\begin{description}
\item[{Parameters}] \leavevmode
\sphinxstyleliteralstrong{\sphinxupquote{x}} (\sphinxstyleliteralemphasis{\sphinxupquote{array\_like}}) \textendash{} 

\item[{Returns}] \leavevmode
\sphinxstylestrong{pdf} \textendash{} Log of the probability density function evaluated at \sphinxtitleref{x}

\item[{Return type}] \leavevmode
ndarray

\end{description}\end{quote}

\end{fulllineitems}

\index{pdf() (multivariate\_normal method)@\spxentry{pdf()}\spxextra{multivariate\_normal method}}

\begin{fulllineitems}
\phantomsection\label{\detokenize{modules/pdfs:symjax.tensor.pdfs.multivariate_normal.pdf}}\pysiglinewithargsret{\sphinxbfcode{\sphinxupquote{pdf}}}{\emph{\DUrole{n}{mean}}, \emph{\DUrole{n}{cov}}}{}
Multivariate normal probability density function.
:param x: Quantiles, with the last axis of \sphinxtitleref{x} denoting the components.
:type x: array\_like
\begin{quote}\begin{description}
\item[{Returns}] \leavevmode
\sphinxstylestrong{pdf} \textendash{} Probability density function evaluated at \sphinxtitleref{x}

\item[{Return type}] \leavevmode
ndarray

\end{description}\end{quote}

\end{fulllineitems}

\end{fulllineitems}

\subsection{\sphinxstyleliteralintitle{\sphinxupquote{symjax.tensor.signal}}}
\label{\detokenize{modules/signal:module-symjax.tensor.signal}}\label{\detokenize{modules/signal:symjax-tensor-signal}}\label{\detokenize{modules/signal::doc}}\index{module@\spxentry{module}!symjax.tensor.signal@\spxentry{symjax.tensor.signal}}\index{symjax.tensor.signal@\spxentry{symjax.tensor.signal}!module@\spxentry{module}}

\subsubsection{Apodization Windows}
\label{\detokenize{modules/signal:apodization-windows}}

\begin{savenotes}\sphinxatlongtablestart\begin{longtable}[c]{\X{1}{2}\X{1}{2}}
\hline

\endfirsthead

\multicolumn{2}{c}%
{\makebox[0pt]{\sphinxtablecontinued{\tablename\ \thetable{} \textendash{} continued from previous page}}}\\
\hline

\endhead

\hline
\multicolumn{2}{r}{\makebox[0pt][r]{\sphinxtablecontinued{continues on next page}}}\\
\endfoot

\endlastfoot

{\hyperref[\detokenize{modules/signal:symjax.tensor.signal.blackman}]{\sphinxcrossref{\sphinxcode{\sphinxupquote{blackman}}}}}(M)
&
Return the Blackman window.
\\
\hline
{\hyperref[\detokenize{modules/signal:symjax.tensor.signal.bartlett}]{\sphinxcrossref{\sphinxcode{\sphinxupquote{bartlett}}}}}(M)
&
Return the Bartlett window.
\\
\hline
{\hyperref[\detokenize{modules/signal:symjax.tensor.signal.hamming}]{\sphinxcrossref{\sphinxcode{\sphinxupquote{hamming}}}}}(M)
&
Return the Hamming window.
\\
\hline
{\hyperref[\detokenize{modules/signal:symjax.tensor.signal.hanning}]{\sphinxcrossref{\sphinxcode{\sphinxupquote{hanning}}}}}(M)
&
Return the Hanning window.
\\
\hline
{\hyperref[\detokenize{modules/signal:symjax.tensor.signal.kaiser}]{\sphinxcrossref{\sphinxcode{\sphinxupquote{kaiser}}}}}(M, beta)
&
Return the Kaiser window.
\\
\hline
\end{longtable}\sphinxatlongtableend\end{savenotes}

\subsubsection{Fourier Transforms}
\label{\detokenize{modules/signal:fourier-transforms}}

\begin{savenotes}\sphinxatlongtablestart\begin{longtable}[c]{\X{1}{2}\X{1}{2}}
\hline

\endfirsthead

\multicolumn{2}{c}%
{\makebox[0pt]{\sphinxtablecontinued{\tablename\ \thetable{} \textendash{} continued from previous page}}}\\
\hline

\endhead

\hline
\multicolumn{2}{r}{\makebox[0pt][r]{\sphinxtablecontinued{continues on next page}}}\\
\endfoot

\endlastfoot

{\hyperref[\detokenize{modules/signal:symjax.tensor.signal.fft}]{\sphinxcrossref{\sphinxcode{\sphinxupquote{fft}}}}}(a{[}, n, axis, norm{]})
&
Compute the one\sphinxhyphen{}dimensional discrete Fourier Transform.
\\
\hline
{\hyperref[\detokenize{modules/signal:symjax.tensor.signal.ifft}]{\sphinxcrossref{\sphinxcode{\sphinxupquote{ifft}}}}}(a{[}, n, axis, norm{]})
&
Compute the one\sphinxhyphen{}dimensional inverse discrete Fourier Transform.
\\
\hline
{\hyperref[\detokenize{modules/signal:symjax.tensor.signal.fft2}]{\sphinxcrossref{\sphinxcode{\sphinxupquote{fft2}}}}}(a{[}, s, axes, norm{]})
&
Compute the 2\sphinxhyphen{}dimensional discrete Fourier Transform
\\
\hline
{\hyperref[\detokenize{modules/signal:symjax.tensor.signal.ifft2}]{\sphinxcrossref{\sphinxcode{\sphinxupquote{ifft2}}}}}(a{[}, s, axes, norm{]})
&
Compute the 2\sphinxhyphen{}dimensional inverse discrete Fourier Transform.
\\
\hline
{\hyperref[\detokenize{modules/signal:symjax.tensor.signal.fftn}]{\sphinxcrossref{\sphinxcode{\sphinxupquote{fftn}}}}}(a{[}, s, axes, norm{]})
&
Compute the N\sphinxhyphen{}dimensional discrete Fourier Transform.
\\
\hline
{\hyperref[\detokenize{modules/signal:symjax.tensor.signal.ifftn}]{\sphinxcrossref{\sphinxcode{\sphinxupquote{ifftn}}}}}(a{[}, s, axes, norm{]})
&
Compute the N\sphinxhyphen{}dimensional inverse discrete Fourier Transform.
\\
\hline
{\hyperref[\detokenize{modules/signal:symjax.tensor.signal.rfft}]{\sphinxcrossref{\sphinxcode{\sphinxupquote{rfft}}}}}(a{[}, n, axis, norm{]})
&
Compute the one\sphinxhyphen{}dimensional discrete Fourier Transform for real input.
\\
\hline
{\hyperref[\detokenize{modules/signal:symjax.tensor.signal.irfft}]{\sphinxcrossref{\sphinxcode{\sphinxupquote{irfft}}}}}(a{[}, n, axis, norm{]})
&
Compute the inverse of the n\sphinxhyphen{}point DFT for real input.
\\
\hline
{\hyperref[\detokenize{modules/signal:symjax.tensor.signal.rfft2}]{\sphinxcrossref{\sphinxcode{\sphinxupquote{rfft2}}}}}(a{[}, s, axes, norm{]})
&
Compute the 2\sphinxhyphen{}dimensional FFT of a real array.
\\
\hline
{\hyperref[\detokenize{modules/signal:symjax.tensor.signal.irfft2}]{\sphinxcrossref{\sphinxcode{\sphinxupquote{irfft2}}}}}(a{[}, s, axes, norm{]})
&
Compute the 2\sphinxhyphen{}dimensional inverse FFT of a real array.
\\
\hline
{\hyperref[\detokenize{modules/signal:symjax.tensor.signal.rfftn}]{\sphinxcrossref{\sphinxcode{\sphinxupquote{rfftn}}}}}(a{[}, s, axes, norm{]})
&
Compute the N\sphinxhyphen{}dimensional discrete Fourier Transform for real input.
\\
\hline
{\hyperref[\detokenize{modules/signal:symjax.tensor.signal.irfftn}]{\sphinxcrossref{\sphinxcode{\sphinxupquote{irfftn}}}}}(a{[}, s, axes, norm{]})
&
Compute the inverse of the N\sphinxhyphen{}dimensional FFT of real input.
\\
\hline
{\hyperref[\detokenize{modules/signal:symjax.tensor.signal.fftfreq}]{\sphinxcrossref{\sphinxcode{\sphinxupquote{fftfreq}}}}}(n{[}, d{]})
&
Return the Discrete Fourier Transform sample frequencies.
\\
\hline
{\hyperref[\detokenize{modules/signal:symjax.tensor.signal.rfftfreq}]{\sphinxcrossref{\sphinxcode{\sphinxupquote{rfftfreq}}}}}(n{[}, d{]})
&
Return the Discrete Fourier Transform sample frequencies
\\
\hline
\end{longtable}\sphinxatlongtableend\end{savenotes}

\subsubsection{Additional Time\sphinxhyphen{}Frequency Representations}
\label{\detokenize{modules/signal:additional-time-frequency-representations}}

\begin{savenotes}\sphinxatlongtablestart\begin{longtable}[c]{\X{1}{2}\X{1}{2}}
\hline

\endfirsthead

\multicolumn{2}{c}%
{\makebox[0pt]{\sphinxtablecontinued{\tablename\ \thetable{} \textendash{} continued from previous page}}}\\
\hline

\endhead

\hline
\multicolumn{2}{r}{\makebox[0pt][r]{\sphinxtablecontinued{continues on next page}}}\\
\endfoot

\endlastfoot

{\hyperref[\detokenize{modules/signal:symjax.tensor.signal.mfcc}]{\sphinxcrossref{\sphinxcode{\sphinxupquote{mfcc}}}}}(signal, window, hop, n\_filter, …{[}, …{]})
&
\sphinxurl{https://librosa.github.io/librosa/\_modules/librosa/feature/spectral.html\#mfcc}
\\
\hline
{\hyperref[\detokenize{modules/signal:symjax.tensor.signal.dct}]{\sphinxcrossref{\sphinxcode{\sphinxupquote{dct}}}}}(signal{[}, axes{]})
&
\sphinxurl{https://dsp.stackexchange.com/questions/2807/fast-cosine-transform-via-fft}
\\
\hline
{\hyperref[\detokenize{modules/signal:symjax.tensor.signal.wvd}]{\sphinxcrossref{\sphinxcode{\sphinxupquote{wvd}}}}}(signal, window, hop, L{[}, apod, mode{]})
&

\\
\hline
{\hyperref[\detokenize{modules/signal:symjax.tensor.signal.hilbert_transform}]{\sphinxcrossref{\sphinxcode{\sphinxupquote{hilbert\_transform}}}}}(signal)
&
the time should be the last dimension return the analytical signal
\\
\hline
\end{longtable}\sphinxatlongtableend\end{savenotes}

\subsubsection{Detailed Descritpions}
\label{\detokenize{modules/signal:detailed-descritpions}}\index{blackman() (in module symjax.tensor.signal)@\spxentry{blackman()}\spxextra{in module symjax.tensor.signal}}

\begin{fulllineitems}
\phantomsection\label{\detokenize{modules/signal:symjax.tensor.signal.blackman}}\pysiglinewithargsret{\sphinxbfcode{\sphinxupquote{blackman}}}{\emph{\DUrole{n}{M}}}{}
Return the Blackman window.

LAX\sphinxhyphen{}backend implementation of {\hyperref[\detokenize{modules/signal:symjax.tensor.signal.blackman}]{\sphinxcrossref{\sphinxcode{\sphinxupquote{blackman()}}}}}.
ADDITIONOriginal docstring below.

The Blackman window is a taper formed by using the first three
terms of a summation of cosines. It was designed to have close to the
minimal leakage possible.  It is close to optimal, only slightly worse
than a Kaiser window.
\begin{quote}\begin{description}
\item[{Returns}] \leavevmode
\sphinxstylestrong{out} \textendash{} The window, with the maximum value normalized to one (the value one
appears only if the number of samples is odd).

\item[{Return type}] \leavevmode
ndarray

\end{description}\end{quote}

\sphinxstrong{See also:}

{\hyperref[\detokenize{modules/signal:symjax.tensor.signal.bartlett}]{\sphinxcrossref{\sphinxcode{\sphinxupquote{bartlett()}}}}}, {\hyperref[\detokenize{modules/signal:symjax.tensor.signal.hamming}]{\sphinxcrossref{\sphinxcode{\sphinxupquote{hamming()}}}}}, {\hyperref[\detokenize{modules/signal:symjax.tensor.signal.hanning}]{\sphinxcrossref{\sphinxcode{\sphinxupquote{hanning()}}}}}, {\hyperref[\detokenize{modules/signal:symjax.tensor.signal.kaiser}]{\sphinxcrossref{\sphinxcode{\sphinxupquote{kaiser()}}}}}

\subsubsection*{Notes}

The Blackman window is defined as
\begin{equation*}
\begin{split}w(n) = 0.42 - 0.5 \cos(2\pi n/M) + 0.08 \cos(4\pi n/M)\end{split}
\end{equation*}
Most references to the Blackman window come from the signal processing
literature, where it is used as one of many windowing functions for
smoothing values.  It is also known as an apodization (which means
“removing the foot”, i.e. smoothing discontinuities at the beginning
and end of the sampled signal) or tapering function. It is known as a
“near optimal” tapering function, almost as good (by some measures)
as the kaiser window.
\subsubsection*{References}

Blackman, R.B. and Tukey, J.W., (1958) The measurement of power spectra,
Dover Publications, New York.

Oppenheim, A.V., and R.W. Schafer. Discrete\sphinxhyphen{}Time Signal Processing.
Upper Saddle River, NJ: Prentice\sphinxhyphen{}Hall, 1999, pp. 468\sphinxhyphen{}471.
\subsubsection*{Examples}

\begin{sphinxVerbatim}[commandchars=\\\{\}]
\PYG{g+gp}{\PYGZgt{}\PYGZgt{}\PYGZgt{} }\PYG{k+kn}{import} \PYG{n+nn}{matplotlib}\PYG{n+nn}{.}\PYG{n+nn}{pyplot} \PYG{k}{as} \PYG{n+nn}{plt}
\PYG{g+gp}{\PYGZgt{}\PYGZgt{}\PYGZgt{} }\PYG{n}{np}\PYG{o}{.}\PYG{n}{blackman}\PYG{p}{(}\PYG{l+m+mi}{12}\PYG{p}{)}
\PYG{g+go}{array([\PYGZhy{}1.38777878e\PYGZhy{}17,   3.26064346e\PYGZhy{}02,   1.59903635e\PYGZhy{}01, \PYGZsh{} may vary}
\PYG{g+go}{        4.14397981e\PYGZhy{}01,   7.36045180e\PYGZhy{}01,   9.67046769e\PYGZhy{}01,}
\PYG{g+go}{        9.67046769e\PYGZhy{}01,   7.36045180e\PYGZhy{}01,   4.14397981e\PYGZhy{}01,}
\PYG{g+go}{        1.59903635e\PYGZhy{}01,   3.26064346e\PYGZhy{}02,  \PYGZhy{}1.38777878e\PYGZhy{}17])}
\end{sphinxVerbatim}

Plot the window and the frequency response:

\begin{sphinxVerbatim}[commandchars=\\\{\}]
\PYG{g+gp}{\PYGZgt{}\PYGZgt{}\PYGZgt{} }\PYG{k+kn}{from} \PYG{n+nn}{numpy}\PYG{n+nn}{.}\PYG{n+nn}{fft} \PYG{k+kn}{import} \PYG{n}{fft}\PYG{p}{,} \PYG{n}{fftshift}
\PYG{g+gp}{\PYGZgt{}\PYGZgt{}\PYGZgt{} }\PYG{n}{window} \PYG{o}{=} \PYG{n}{np}\PYG{o}{.}\PYG{n}{blackman}\PYG{p}{(}\PYG{l+m+mi}{51}\PYG{p}{)}
\PYG{g+gp}{\PYGZgt{}\PYGZgt{}\PYGZgt{} }\PYG{n}{plt}\PYG{o}{.}\PYG{n}{plot}\PYG{p}{(}\PYG{n}{window}\PYG{p}{)}
\PYG{g+go}{[\PYGZlt{}matplotlib.lines.Line2D object at 0x...\PYGZgt{}]}
\PYG{g+gp}{\PYGZgt{}\PYGZgt{}\PYGZgt{} }\PYG{n}{plt}\PYG{o}{.}\PYG{n}{title}\PYG{p}{(}\PYG{l+s+s2}{\PYGZdq{}}\PYG{l+s+s2}{Blackman window}\PYG{l+s+s2}{\PYGZdq{}}\PYG{p}{)}
\PYG{g+go}{Text(0.5, 1.0, \PYGZsq{}Blackman window\PYGZsq{})}
\PYG{g+gp}{\PYGZgt{}\PYGZgt{}\PYGZgt{} }\PYG{n}{plt}\PYG{o}{.}\PYG{n}{ylabel}\PYG{p}{(}\PYG{l+s+s2}{\PYGZdq{}}\PYG{l+s+s2}{Amplitude}\PYG{l+s+s2}{\PYGZdq{}}\PYG{p}{)}
\PYG{g+go}{Text(0, 0.5, \PYGZsq{}Amplitude\PYGZsq{})}
\PYG{g+gp}{\PYGZgt{}\PYGZgt{}\PYGZgt{} }\PYG{n}{plt}\PYG{o}{.}\PYG{n}{xlabel}\PYG{p}{(}\PYG{l+s+s2}{\PYGZdq{}}\PYG{l+s+s2}{Sample}\PYG{l+s+s2}{\PYGZdq{}}\PYG{p}{)}
\PYG{g+go}{Text(0.5, 0, \PYGZsq{}Sample\PYGZsq{})}
\PYG{g+gp}{\PYGZgt{}\PYGZgt{}\PYGZgt{} }\PYG{n}{plt}\PYG{o}{.}\PYG{n}{show}\PYG{p}{(}\PYG{p}{)}
\end{sphinxVerbatim}

\begin{sphinxVerbatim}[commandchars=\\\{\}]
\PYG{g+gp}{\PYGZgt{}\PYGZgt{}\PYGZgt{} }\PYG{n}{plt}\PYG{o}{.}\PYG{n}{figure}\PYG{p}{(}\PYG{p}{)}
\PYG{g+go}{\PYGZlt{}Figure size 640x480 with 0 Axes\PYGZgt{}}
\PYG{g+gp}{\PYGZgt{}\PYGZgt{}\PYGZgt{} }\PYG{n}{A} \PYG{o}{=} \PYG{n}{fft}\PYG{p}{(}\PYG{n}{window}\PYG{p}{,} \PYG{l+m+mi}{2048}\PYG{p}{)} \PYG{o}{/} \PYG{l+m+mf}{25.5}
\PYG{g+gp}{\PYGZgt{}\PYGZgt{}\PYGZgt{} }\PYG{n}{mag} \PYG{o}{=} \PYG{n}{np}\PYG{o}{.}\PYG{n}{abs}\PYG{p}{(}\PYG{n}{fftshift}\PYG{p}{(}\PYG{n}{A}\PYG{p}{)}\PYG{p}{)}
\PYG{g+gp}{\PYGZgt{}\PYGZgt{}\PYGZgt{} }\PYG{n}{freq} \PYG{o}{=} \PYG{n}{np}\PYG{o}{.}\PYG{n}{linspace}\PYG{p}{(}\PYG{o}{\PYGZhy{}}\PYG{l+m+mf}{0.5}\PYG{p}{,} \PYG{l+m+mf}{0.5}\PYG{p}{,} \PYG{n+nb}{len}\PYG{p}{(}\PYG{n}{A}\PYG{p}{)}\PYG{p}{)}
\PYG{g+gp}{\PYGZgt{}\PYGZgt{}\PYGZgt{} }\PYG{k}{with} \PYG{n}{np}\PYG{o}{.}\PYG{n}{errstate}\PYG{p}{(}\PYG{n}{divide}\PYG{o}{=}\PYG{l+s+s1}{\PYGZsq{}}\PYG{l+s+s1}{ignore}\PYG{l+s+s1}{\PYGZsq{}}\PYG{p}{,} \PYG{n}{invalid}\PYG{o}{=}\PYG{l+s+s1}{\PYGZsq{}}\PYG{l+s+s1}{ignore}\PYG{l+s+s1}{\PYGZsq{}}\PYG{p}{)}\PYG{p}{:}
\PYG{g+gp}{... }    \PYG{n}{response} \PYG{o}{=} \PYG{l+m+mi}{20} \PYG{o}{*} \PYG{n}{np}\PYG{o}{.}\PYG{n}{log10}\PYG{p}{(}\PYG{n}{mag}\PYG{p}{)}
\PYG{g+gp}{...}
\PYG{g+gp}{\PYGZgt{}\PYGZgt{}\PYGZgt{} }\PYG{n}{response} \PYG{o}{=} \PYG{n}{np}\PYG{o}{.}\PYG{n}{clip}\PYG{p}{(}\PYG{n}{response}\PYG{p}{,} \PYG{o}{\PYGZhy{}}\PYG{l+m+mi}{100}\PYG{p}{,} \PYG{l+m+mi}{100}\PYG{p}{)}
\PYG{g+gp}{\PYGZgt{}\PYGZgt{}\PYGZgt{} }\PYG{n}{plt}\PYG{o}{.}\PYG{n}{plot}\PYG{p}{(}\PYG{n}{freq}\PYG{p}{,} \PYG{n}{response}\PYG{p}{)}
\PYG{g+go}{[\PYGZlt{}matplotlib.lines.Line2D object at 0x...\PYGZgt{}]}
\PYG{g+gp}{\PYGZgt{}\PYGZgt{}\PYGZgt{} }\PYG{n}{plt}\PYG{o}{.}\PYG{n}{title}\PYG{p}{(}\PYG{l+s+s2}{\PYGZdq{}}\PYG{l+s+s2}{Frequency response of Blackman window}\PYG{l+s+s2}{\PYGZdq{}}\PYG{p}{)}
\PYG{g+go}{Text(0.5, 1.0, \PYGZsq{}Frequency response of Blackman window\PYGZsq{})}
\PYG{g+gp}{\PYGZgt{}\PYGZgt{}\PYGZgt{} }\PYG{n}{plt}\PYG{o}{.}\PYG{n}{ylabel}\PYG{p}{(}\PYG{l+s+s2}{\PYGZdq{}}\PYG{l+s+s2}{Magnitude [dB]}\PYG{l+s+s2}{\PYGZdq{}}\PYG{p}{)}
\PYG{g+go}{Text(0, 0.5, \PYGZsq{}Magnitude [dB]\PYGZsq{})}
\PYG{g+gp}{\PYGZgt{}\PYGZgt{}\PYGZgt{} }\PYG{n}{plt}\PYG{o}{.}\PYG{n}{xlabel}\PYG{p}{(}\PYG{l+s+s2}{\PYGZdq{}}\PYG{l+s+s2}{Normalized frequency [cycles per sample]}\PYG{l+s+s2}{\PYGZdq{}}\PYG{p}{)}
\PYG{g+go}{Text(0.5, 0, \PYGZsq{}Normalized frequency [cycles per sample]\PYGZsq{})}
\PYG{g+gp}{\PYGZgt{}\PYGZgt{}\PYGZgt{} }\PYG{n}{\PYGZus{}} \PYG{o}{=} \PYG{n}{plt}\PYG{o}{.}\PYG{n}{axis}\PYG{p}{(}\PYG{l+s+s1}{\PYGZsq{}}\PYG{l+s+s1}{tight}\PYG{l+s+s1}{\PYGZsq{}}\PYG{p}{)}
\PYG{g+gp}{\PYGZgt{}\PYGZgt{}\PYGZgt{} }\PYG{n}{plt}\PYG{o}{.}\PYG{n}{show}\PYG{p}{(}\PYG{p}{)}
\end{sphinxVerbatim}

\end{fulllineitems}

\index{bartlett() (in module symjax.tensor.signal)@\spxentry{bartlett()}\spxextra{in module symjax.tensor.signal}}

\begin{fulllineitems}
\phantomsection\label{\detokenize{modules/signal:symjax.tensor.signal.bartlett}}\pysiglinewithargsret{\sphinxbfcode{\sphinxupquote{bartlett}}}{\emph{\DUrole{n}{M}}}{}
Return the Bartlett window.

LAX\sphinxhyphen{}backend implementation of {\hyperref[\detokenize{modules/signal:symjax.tensor.signal.bartlett}]{\sphinxcrossref{\sphinxcode{\sphinxupquote{bartlett()}}}}}.
ADDITIONOriginal docstring below.

The Bartlett window is very similar to a triangular window, except
that the end points are at zero.  It is often used in signal
processing for tapering a signal, without generating too much
ripple in the frequency domain.
\begin{quote}\begin{description}
\item[{Returns}] \leavevmode
\sphinxstylestrong{out} \textendash{} The triangular window, with the maximum value normalized to one
(the value one appears only if the number of samples is odd), with
the first and last samples equal to zero.

\item[{Return type}] \leavevmode
array

\end{description}\end{quote}

\sphinxstrong{See also:}

{\hyperref[\detokenize{modules/signal:symjax.tensor.signal.blackman}]{\sphinxcrossref{\sphinxcode{\sphinxupquote{blackman()}}}}}, {\hyperref[\detokenize{modules/signal:symjax.tensor.signal.hamming}]{\sphinxcrossref{\sphinxcode{\sphinxupquote{hamming()}}}}}, {\hyperref[\detokenize{modules/signal:symjax.tensor.signal.hanning}]{\sphinxcrossref{\sphinxcode{\sphinxupquote{hanning()}}}}}, {\hyperref[\detokenize{modules/signal:symjax.tensor.signal.kaiser}]{\sphinxcrossref{\sphinxcode{\sphinxupquote{kaiser()}}}}}

\subsubsection*{Notes}

The Bartlett window is defined as
\begin{equation*}
\begin{split}w(n) = \frac{2}{M-1} \left(
\frac{M-1}{2} - \left|n - \frac{M-1}{2}\right|
\right)\end{split}
\end{equation*}
Most references to the Bartlett window come from the signal
processing literature, where it is used as one of many windowing
functions for smoothing values.  Note that convolution with this
window produces linear interpolation.  It is also known as an
apodization (which means”removing the foot”, i.e. smoothing
discontinuities at the beginning and end of the sampled signal) or
tapering function. The fourier transform of the Bartlett is the product
of two sinc functions.
Note the excellent discussion in Kanasewich.
\subsubsection*{References}
\subsubsection*{Examples}

\begin{sphinxVerbatim}[commandchars=\\\{\}]
\PYG{g+gp}{\PYGZgt{}\PYGZgt{}\PYGZgt{} }\PYG{k+kn}{import} \PYG{n+nn}{matplotlib}\PYG{n+nn}{.}\PYG{n+nn}{pyplot} \PYG{k}{as} \PYG{n+nn}{plt}
\PYG{g+gp}{\PYGZgt{}\PYGZgt{}\PYGZgt{} }\PYG{n}{np}\PYG{o}{.}\PYG{n}{bartlett}\PYG{p}{(}\PYG{l+m+mi}{12}\PYG{p}{)}
\PYG{g+go}{array([ 0.        ,  0.18181818,  0.36363636,  0.54545455,  0.72727273, \PYGZsh{} may vary}
\PYG{g+go}{        0.90909091,  0.90909091,  0.72727273,  0.54545455,  0.36363636,}
\PYG{g+go}{        0.18181818,  0.        ])}
\end{sphinxVerbatim}

Plot the window and its frequency response (requires SciPy and matplotlib):

\begin{sphinxVerbatim}[commandchars=\\\{\}]
\PYG{g+gp}{\PYGZgt{}\PYGZgt{}\PYGZgt{} }\PYG{k+kn}{from} \PYG{n+nn}{numpy}\PYG{n+nn}{.}\PYG{n+nn}{fft} \PYG{k+kn}{import} \PYG{n}{fft}\PYG{p}{,} \PYG{n}{fftshift}
\PYG{g+gp}{\PYGZgt{}\PYGZgt{}\PYGZgt{} }\PYG{n}{window} \PYG{o}{=} \PYG{n}{np}\PYG{o}{.}\PYG{n}{bartlett}\PYG{p}{(}\PYG{l+m+mi}{51}\PYG{p}{)}
\PYG{g+gp}{\PYGZgt{}\PYGZgt{}\PYGZgt{} }\PYG{n}{plt}\PYG{o}{.}\PYG{n}{plot}\PYG{p}{(}\PYG{n}{window}\PYG{p}{)}
\PYG{g+go}{[\PYGZlt{}matplotlib.lines.Line2D object at 0x...\PYGZgt{}]}
\PYG{g+gp}{\PYGZgt{}\PYGZgt{}\PYGZgt{} }\PYG{n}{plt}\PYG{o}{.}\PYG{n}{title}\PYG{p}{(}\PYG{l+s+s2}{\PYGZdq{}}\PYG{l+s+s2}{Bartlett window}\PYG{l+s+s2}{\PYGZdq{}}\PYG{p}{)}
\PYG{g+go}{Text(0.5, 1.0, \PYGZsq{}Bartlett window\PYGZsq{})}
\PYG{g+gp}{\PYGZgt{}\PYGZgt{}\PYGZgt{} }\PYG{n}{plt}\PYG{o}{.}\PYG{n}{ylabel}\PYG{p}{(}\PYG{l+s+s2}{\PYGZdq{}}\PYG{l+s+s2}{Amplitude}\PYG{l+s+s2}{\PYGZdq{}}\PYG{p}{)}
\PYG{g+go}{Text(0, 0.5, \PYGZsq{}Amplitude\PYGZsq{})}
\PYG{g+gp}{\PYGZgt{}\PYGZgt{}\PYGZgt{} }\PYG{n}{plt}\PYG{o}{.}\PYG{n}{xlabel}\PYG{p}{(}\PYG{l+s+s2}{\PYGZdq{}}\PYG{l+s+s2}{Sample}\PYG{l+s+s2}{\PYGZdq{}}\PYG{p}{)}
\PYG{g+go}{Text(0.5, 0, \PYGZsq{}Sample\PYGZsq{})}
\PYG{g+gp}{\PYGZgt{}\PYGZgt{}\PYGZgt{} }\PYG{n}{plt}\PYG{o}{.}\PYG{n}{show}\PYG{p}{(}\PYG{p}{)}
\end{sphinxVerbatim}

\begin{sphinxVerbatim}[commandchars=\\\{\}]
\PYG{g+gp}{\PYGZgt{}\PYGZgt{}\PYGZgt{} }\PYG{n}{plt}\PYG{o}{.}\PYG{n}{figure}\PYG{p}{(}\PYG{p}{)}
\PYG{g+go}{\PYGZlt{}Figure size 640x480 with 0 Axes\PYGZgt{}}
\PYG{g+gp}{\PYGZgt{}\PYGZgt{}\PYGZgt{} }\PYG{n}{A} \PYG{o}{=} \PYG{n}{fft}\PYG{p}{(}\PYG{n}{window}\PYG{p}{,} \PYG{l+m+mi}{2048}\PYG{p}{)} \PYG{o}{/} \PYG{l+m+mf}{25.5}
\PYG{g+gp}{\PYGZgt{}\PYGZgt{}\PYGZgt{} }\PYG{n}{mag} \PYG{o}{=} \PYG{n}{np}\PYG{o}{.}\PYG{n}{abs}\PYG{p}{(}\PYG{n}{fftshift}\PYG{p}{(}\PYG{n}{A}\PYG{p}{)}\PYG{p}{)}
\PYG{g+gp}{\PYGZgt{}\PYGZgt{}\PYGZgt{} }\PYG{n}{freq} \PYG{o}{=} \PYG{n}{np}\PYG{o}{.}\PYG{n}{linspace}\PYG{p}{(}\PYG{o}{\PYGZhy{}}\PYG{l+m+mf}{0.5}\PYG{p}{,} \PYG{l+m+mf}{0.5}\PYG{p}{,} \PYG{n+nb}{len}\PYG{p}{(}\PYG{n}{A}\PYG{p}{)}\PYG{p}{)}
\PYG{g+gp}{\PYGZgt{}\PYGZgt{}\PYGZgt{} }\PYG{k}{with} \PYG{n}{np}\PYG{o}{.}\PYG{n}{errstate}\PYG{p}{(}\PYG{n}{divide}\PYG{o}{=}\PYG{l+s+s1}{\PYGZsq{}}\PYG{l+s+s1}{ignore}\PYG{l+s+s1}{\PYGZsq{}}\PYG{p}{,} \PYG{n}{invalid}\PYG{o}{=}\PYG{l+s+s1}{\PYGZsq{}}\PYG{l+s+s1}{ignore}\PYG{l+s+s1}{\PYGZsq{}}\PYG{p}{)}\PYG{p}{:}
\PYG{g+gp}{... }    \PYG{n}{response} \PYG{o}{=} \PYG{l+m+mi}{20} \PYG{o}{*} \PYG{n}{np}\PYG{o}{.}\PYG{n}{log10}\PYG{p}{(}\PYG{n}{mag}\PYG{p}{)}
\PYG{g+gp}{...}
\PYG{g+gp}{\PYGZgt{}\PYGZgt{}\PYGZgt{} }\PYG{n}{response} \PYG{o}{=} \PYG{n}{np}\PYG{o}{.}\PYG{n}{clip}\PYG{p}{(}\PYG{n}{response}\PYG{p}{,} \PYG{o}{\PYGZhy{}}\PYG{l+m+mi}{100}\PYG{p}{,} \PYG{l+m+mi}{100}\PYG{p}{)}
\PYG{g+gp}{\PYGZgt{}\PYGZgt{}\PYGZgt{} }\PYG{n}{plt}\PYG{o}{.}\PYG{n}{plot}\PYG{p}{(}\PYG{n}{freq}\PYG{p}{,} \PYG{n}{response}\PYG{p}{)}
\PYG{g+go}{[\PYGZlt{}matplotlib.lines.Line2D object at 0x...\PYGZgt{}]}
\PYG{g+gp}{\PYGZgt{}\PYGZgt{}\PYGZgt{} }\PYG{n}{plt}\PYG{o}{.}\PYG{n}{title}\PYG{p}{(}\PYG{l+s+s2}{\PYGZdq{}}\PYG{l+s+s2}{Frequency response of Bartlett window}\PYG{l+s+s2}{\PYGZdq{}}\PYG{p}{)}
\PYG{g+go}{Text(0.5, 1.0, \PYGZsq{}Frequency response of Bartlett window\PYGZsq{})}
\PYG{g+gp}{\PYGZgt{}\PYGZgt{}\PYGZgt{} }\PYG{n}{plt}\PYG{o}{.}\PYG{n}{ylabel}\PYG{p}{(}\PYG{l+s+s2}{\PYGZdq{}}\PYG{l+s+s2}{Magnitude [dB]}\PYG{l+s+s2}{\PYGZdq{}}\PYG{p}{)}
\PYG{g+go}{Text(0, 0.5, \PYGZsq{}Magnitude [dB]\PYGZsq{})}
\PYG{g+gp}{\PYGZgt{}\PYGZgt{}\PYGZgt{} }\PYG{n}{plt}\PYG{o}{.}\PYG{n}{xlabel}\PYG{p}{(}\PYG{l+s+s2}{\PYGZdq{}}\PYG{l+s+s2}{Normalized frequency [cycles per sample]}\PYG{l+s+s2}{\PYGZdq{}}\PYG{p}{)}
\PYG{g+go}{Text(0.5, 0, \PYGZsq{}Normalized frequency [cycles per sample]\PYGZsq{})}
\PYG{g+gp}{\PYGZgt{}\PYGZgt{}\PYGZgt{} }\PYG{n}{\PYGZus{}} \PYG{o}{=} \PYG{n}{plt}\PYG{o}{.}\PYG{n}{axis}\PYG{p}{(}\PYG{l+s+s1}{\PYGZsq{}}\PYG{l+s+s1}{tight}\PYG{l+s+s1}{\PYGZsq{}}\PYG{p}{)}
\PYG{g+gp}{\PYGZgt{}\PYGZgt{}\PYGZgt{} }\PYG{n}{plt}\PYG{o}{.}\PYG{n}{show}\PYG{p}{(}\PYG{p}{)}
\end{sphinxVerbatim}

\end{fulllineitems}

\index{hamming() (in module symjax.tensor.signal)@\spxentry{hamming()}\spxextra{in module symjax.tensor.signal}}

\begin{fulllineitems}
\phantomsection\label{\detokenize{modules/signal:symjax.tensor.signal.hamming}}\pysiglinewithargsret{\sphinxbfcode{\sphinxupquote{hamming}}}{\emph{\DUrole{n}{M}}}{}
Return the Hamming window.

LAX\sphinxhyphen{}backend implementation of {\hyperref[\detokenize{modules/signal:symjax.tensor.signal.hamming}]{\sphinxcrossref{\sphinxcode{\sphinxupquote{hamming()}}}}}.
ADDITIONOriginal docstring below.

The Hamming window is a taper formed by using a weighted cosine.
\begin{quote}\begin{description}
\item[{Returns}] \leavevmode
\sphinxstylestrong{out} \textendash{} The window, with the maximum value normalized to one (the value
one appears only if the number of samples is odd).

\item[{Return type}] \leavevmode
ndarray

\end{description}\end{quote}

\sphinxstrong{See also:}

{\hyperref[\detokenize{modules/signal:symjax.tensor.signal.bartlett}]{\sphinxcrossref{\sphinxcode{\sphinxupquote{bartlett()}}}}}, {\hyperref[\detokenize{modules/signal:symjax.tensor.signal.blackman}]{\sphinxcrossref{\sphinxcode{\sphinxupquote{blackman()}}}}}, {\hyperref[\detokenize{modules/signal:symjax.tensor.signal.hanning}]{\sphinxcrossref{\sphinxcode{\sphinxupquote{hanning()}}}}}, {\hyperref[\detokenize{modules/signal:symjax.tensor.signal.kaiser}]{\sphinxcrossref{\sphinxcode{\sphinxupquote{kaiser()}}}}}

\subsubsection*{Notes}

The Hamming window is defined as
\begin{equation*}
\begin{split}w(n) = 0.54 - 0.46cos\left(\frac{2\pi{n}}{M-1}\right)
\qquad 0 \leq n \leq M-1\end{split}
\end{equation*}
The Hamming was named for R. W. Hamming, an associate of J. W. Tukey
and is described in Blackman and Tukey. It was recommended for
smoothing the truncated autocovariance function in the time domain.
Most references to the Hamming window come from the signal processing
literature, where it is used as one of many windowing functions for
smoothing values.  It is also known as an apodization (which means
“removing the foot”, i.e. smoothing discontinuities at the beginning
and end of the sampled signal) or tapering function.
\subsubsection*{References}
\subsubsection*{Examples}

\begin{sphinxVerbatim}[commandchars=\\\{\}]
\PYG{g+gp}{\PYGZgt{}\PYGZgt{}\PYGZgt{} }\PYG{n}{np}\PYG{o}{.}\PYG{n}{hamming}\PYG{p}{(}\PYG{l+m+mi}{12}\PYG{p}{)}
\PYG{g+go}{array([ 0.08      ,  0.15302337,  0.34890909,  0.60546483,  0.84123594, \PYGZsh{} may vary}
\PYG{g+go}{        0.98136677,  0.98136677,  0.84123594,  0.60546483,  0.34890909,}
\PYG{g+go}{        0.15302337,  0.08      ])}
\end{sphinxVerbatim}

Plot the window and the frequency response:

\begin{sphinxVerbatim}[commandchars=\\\{\}]
\PYG{g+gp}{\PYGZgt{}\PYGZgt{}\PYGZgt{} }\PYG{k+kn}{import} \PYG{n+nn}{matplotlib}\PYG{n+nn}{.}\PYG{n+nn}{pyplot} \PYG{k}{as} \PYG{n+nn}{plt}
\PYG{g+gp}{\PYGZgt{}\PYGZgt{}\PYGZgt{} }\PYG{k+kn}{from} \PYG{n+nn}{numpy}\PYG{n+nn}{.}\PYG{n+nn}{fft} \PYG{k+kn}{import} \PYG{n}{fft}\PYG{p}{,} \PYG{n}{fftshift}
\PYG{g+gp}{\PYGZgt{}\PYGZgt{}\PYGZgt{} }\PYG{n}{window} \PYG{o}{=} \PYG{n}{np}\PYG{o}{.}\PYG{n}{hamming}\PYG{p}{(}\PYG{l+m+mi}{51}\PYG{p}{)}
\PYG{g+gp}{\PYGZgt{}\PYGZgt{}\PYGZgt{} }\PYG{n}{plt}\PYG{o}{.}\PYG{n}{plot}\PYG{p}{(}\PYG{n}{window}\PYG{p}{)}
\PYG{g+go}{[\PYGZlt{}matplotlib.lines.Line2D object at 0x...\PYGZgt{}]}
\PYG{g+gp}{\PYGZgt{}\PYGZgt{}\PYGZgt{} }\PYG{n}{plt}\PYG{o}{.}\PYG{n}{title}\PYG{p}{(}\PYG{l+s+s2}{\PYGZdq{}}\PYG{l+s+s2}{Hamming window}\PYG{l+s+s2}{\PYGZdq{}}\PYG{p}{)}
\PYG{g+go}{Text(0.5, 1.0, \PYGZsq{}Hamming window\PYGZsq{})}
\PYG{g+gp}{\PYGZgt{}\PYGZgt{}\PYGZgt{} }\PYG{n}{plt}\PYG{o}{.}\PYG{n}{ylabel}\PYG{p}{(}\PYG{l+s+s2}{\PYGZdq{}}\PYG{l+s+s2}{Amplitude}\PYG{l+s+s2}{\PYGZdq{}}\PYG{p}{)}
\PYG{g+go}{Text(0, 0.5, \PYGZsq{}Amplitude\PYGZsq{})}
\PYG{g+gp}{\PYGZgt{}\PYGZgt{}\PYGZgt{} }\PYG{n}{plt}\PYG{o}{.}\PYG{n}{xlabel}\PYG{p}{(}\PYG{l+s+s2}{\PYGZdq{}}\PYG{l+s+s2}{Sample}\PYG{l+s+s2}{\PYGZdq{}}\PYG{p}{)}
\PYG{g+go}{Text(0.5, 0, \PYGZsq{}Sample\PYGZsq{})}
\PYG{g+gp}{\PYGZgt{}\PYGZgt{}\PYGZgt{} }\PYG{n}{plt}\PYG{o}{.}\PYG{n}{show}\PYG{p}{(}\PYG{p}{)}
\end{sphinxVerbatim}

\begin{sphinxVerbatim}[commandchars=\\\{\}]
\PYG{g+gp}{\PYGZgt{}\PYGZgt{}\PYGZgt{} }\PYG{n}{plt}\PYG{o}{.}\PYG{n}{figure}\PYG{p}{(}\PYG{p}{)}
\PYG{g+go}{\PYGZlt{}Figure size 640x480 with 0 Axes\PYGZgt{}}
\PYG{g+gp}{\PYGZgt{}\PYGZgt{}\PYGZgt{} }\PYG{n}{A} \PYG{o}{=} \PYG{n}{fft}\PYG{p}{(}\PYG{n}{window}\PYG{p}{,} \PYG{l+m+mi}{2048}\PYG{p}{)} \PYG{o}{/} \PYG{l+m+mf}{25.5}
\PYG{g+gp}{\PYGZgt{}\PYGZgt{}\PYGZgt{} }\PYG{n}{mag} \PYG{o}{=} \PYG{n}{np}\PYG{o}{.}\PYG{n}{abs}\PYG{p}{(}\PYG{n}{fftshift}\PYG{p}{(}\PYG{n}{A}\PYG{p}{)}\PYG{p}{)}
\PYG{g+gp}{\PYGZgt{}\PYGZgt{}\PYGZgt{} }\PYG{n}{freq} \PYG{o}{=} \PYG{n}{np}\PYG{o}{.}\PYG{n}{linspace}\PYG{p}{(}\PYG{o}{\PYGZhy{}}\PYG{l+m+mf}{0.5}\PYG{p}{,} \PYG{l+m+mf}{0.5}\PYG{p}{,} \PYG{n+nb}{len}\PYG{p}{(}\PYG{n}{A}\PYG{p}{)}\PYG{p}{)}
\PYG{g+gp}{\PYGZgt{}\PYGZgt{}\PYGZgt{} }\PYG{n}{response} \PYG{o}{=} \PYG{l+m+mi}{20} \PYG{o}{*} \PYG{n}{np}\PYG{o}{.}\PYG{n}{log10}\PYG{p}{(}\PYG{n}{mag}\PYG{p}{)}
\PYG{g+gp}{\PYGZgt{}\PYGZgt{}\PYGZgt{} }\PYG{n}{response} \PYG{o}{=} \PYG{n}{np}\PYG{o}{.}\PYG{n}{clip}\PYG{p}{(}\PYG{n}{response}\PYG{p}{,} \PYG{o}{\PYGZhy{}}\PYG{l+m+mi}{100}\PYG{p}{,} \PYG{l+m+mi}{100}\PYG{p}{)}
\PYG{g+gp}{\PYGZgt{}\PYGZgt{}\PYGZgt{} }\PYG{n}{plt}\PYG{o}{.}\PYG{n}{plot}\PYG{p}{(}\PYG{n}{freq}\PYG{p}{,} \PYG{n}{response}\PYG{p}{)}
\PYG{g+go}{[\PYGZlt{}matplotlib.lines.Line2D object at 0x...\PYGZgt{}]}
\PYG{g+gp}{\PYGZgt{}\PYGZgt{}\PYGZgt{} }\PYG{n}{plt}\PYG{o}{.}\PYG{n}{title}\PYG{p}{(}\PYG{l+s+s2}{\PYGZdq{}}\PYG{l+s+s2}{Frequency response of Hamming window}\PYG{l+s+s2}{\PYGZdq{}}\PYG{p}{)}
\PYG{g+go}{Text(0.5, 1.0, \PYGZsq{}Frequency response of Hamming window\PYGZsq{})}
\PYG{g+gp}{\PYGZgt{}\PYGZgt{}\PYGZgt{} }\PYG{n}{plt}\PYG{o}{.}\PYG{n}{ylabel}\PYG{p}{(}\PYG{l+s+s2}{\PYGZdq{}}\PYG{l+s+s2}{Magnitude [dB]}\PYG{l+s+s2}{\PYGZdq{}}\PYG{p}{)}
\PYG{g+go}{Text(0, 0.5, \PYGZsq{}Magnitude [dB]\PYGZsq{})}
\PYG{g+gp}{\PYGZgt{}\PYGZgt{}\PYGZgt{} }\PYG{n}{plt}\PYG{o}{.}\PYG{n}{xlabel}\PYG{p}{(}\PYG{l+s+s2}{\PYGZdq{}}\PYG{l+s+s2}{Normalized frequency [cycles per sample]}\PYG{l+s+s2}{\PYGZdq{}}\PYG{p}{)}
\PYG{g+go}{Text(0.5, 0, \PYGZsq{}Normalized frequency [cycles per sample]\PYGZsq{})}
\PYG{g+gp}{\PYGZgt{}\PYGZgt{}\PYGZgt{} }\PYG{n}{plt}\PYG{o}{.}\PYG{n}{axis}\PYG{p}{(}\PYG{l+s+s1}{\PYGZsq{}}\PYG{l+s+s1}{tight}\PYG{l+s+s1}{\PYGZsq{}}\PYG{p}{)}
\PYG{g+gp}{...}
\PYG{g+gp}{\PYGZgt{}\PYGZgt{}\PYGZgt{} }\PYG{n}{plt}\PYG{o}{.}\PYG{n}{show}\PYG{p}{(}\PYG{p}{)}
\end{sphinxVerbatim}

\end{fulllineitems}

\index{hanning() (in module symjax.tensor.signal)@\spxentry{hanning()}\spxextra{in module symjax.tensor.signal}}

\begin{fulllineitems}
\phantomsection\label{\detokenize{modules/signal:symjax.tensor.signal.hanning}}\pysiglinewithargsret{\sphinxbfcode{\sphinxupquote{hanning}}}{\emph{\DUrole{n}{M}}}{}
Return the Hanning window.

LAX\sphinxhyphen{}backend implementation of {\hyperref[\detokenize{modules/signal:symjax.tensor.signal.hanning}]{\sphinxcrossref{\sphinxcode{\sphinxupquote{hanning()}}}}}.
ADDITIONOriginal docstring below.

The Hanning window is a taper formed by using a weighted cosine.
\begin{quote}\begin{description}
\item[{Returns}] \leavevmode
\sphinxstylestrong{out} \textendash{} The window, with the maximum value normalized to one (the value
one appears only if \sphinxtitleref{M} is odd).

\item[{Return type}] \leavevmode
ndarray, shape(M,)

\end{description}\end{quote}

\sphinxstrong{See also:}

{\hyperref[\detokenize{modules/signal:symjax.tensor.signal.bartlett}]{\sphinxcrossref{\sphinxcode{\sphinxupquote{bartlett()}}}}}, {\hyperref[\detokenize{modules/signal:symjax.tensor.signal.blackman}]{\sphinxcrossref{\sphinxcode{\sphinxupquote{blackman()}}}}}, {\hyperref[\detokenize{modules/signal:symjax.tensor.signal.hamming}]{\sphinxcrossref{\sphinxcode{\sphinxupquote{hamming()}}}}}, {\hyperref[\detokenize{modules/signal:symjax.tensor.signal.kaiser}]{\sphinxcrossref{\sphinxcode{\sphinxupquote{kaiser()}}}}}

\subsubsection*{Notes}

The Hanning window is defined as
\begin{equation*}
\begin{split}w(n) = 0.5 - 0.5cos\left(\frac{2\pi{n}}{M-1}\right)
\qquad 0 \leq n \leq M-1\end{split}
\end{equation*}
The Hanning was named for Julius von Hann, an Austrian meteorologist.
It is also known as the Cosine Bell. Some authors prefer that it be
called a Hann window, to help avoid confusion with the very similar
Hamming window.

Most references to the Hanning window come from the signal processing
literature, where it is used as one of many windowing functions for
smoothing values.  It is also known as an apodization (which means
“removing the foot”, i.e. smoothing discontinuities at the beginning
and end of the sampled signal) or tapering function.
\subsubsection*{References}
\subsubsection*{Examples}

\begin{sphinxVerbatim}[commandchars=\\\{\}]
\PYG{g+gp}{\PYGZgt{}\PYGZgt{}\PYGZgt{} }\PYG{n}{np}\PYG{o}{.}\PYG{n}{hanning}\PYG{p}{(}\PYG{l+m+mi}{12}\PYG{p}{)}
\PYG{g+go}{array([0.        , 0.07937323, 0.29229249, 0.57115742, 0.82743037,}
\PYG{g+go}{       0.97974649, 0.97974649, 0.82743037, 0.57115742, 0.29229249,}
\PYG{g+go}{       0.07937323, 0.        ])}
\end{sphinxVerbatim}

Plot the window and its frequency response:

\begin{sphinxVerbatim}[commandchars=\\\{\}]
\PYG{g+gp}{\PYGZgt{}\PYGZgt{}\PYGZgt{} }\PYG{k+kn}{import} \PYG{n+nn}{matplotlib}\PYG{n+nn}{.}\PYG{n+nn}{pyplot} \PYG{k}{as} \PYG{n+nn}{plt}
\PYG{g+gp}{\PYGZgt{}\PYGZgt{}\PYGZgt{} }\PYG{k+kn}{from} \PYG{n+nn}{numpy}\PYG{n+nn}{.}\PYG{n+nn}{fft} \PYG{k+kn}{import} \PYG{n}{fft}\PYG{p}{,} \PYG{n}{fftshift}
\PYG{g+gp}{\PYGZgt{}\PYGZgt{}\PYGZgt{} }\PYG{n}{window} \PYG{o}{=} \PYG{n}{np}\PYG{o}{.}\PYG{n}{hanning}\PYG{p}{(}\PYG{l+m+mi}{51}\PYG{p}{)}
\PYG{g+gp}{\PYGZgt{}\PYGZgt{}\PYGZgt{} }\PYG{n}{plt}\PYG{o}{.}\PYG{n}{plot}\PYG{p}{(}\PYG{n}{window}\PYG{p}{)}
\PYG{g+go}{[\PYGZlt{}matplotlib.lines.Line2D object at 0x...\PYGZgt{}]}
\PYG{g+gp}{\PYGZgt{}\PYGZgt{}\PYGZgt{} }\PYG{n}{plt}\PYG{o}{.}\PYG{n}{title}\PYG{p}{(}\PYG{l+s+s2}{\PYGZdq{}}\PYG{l+s+s2}{Hann window}\PYG{l+s+s2}{\PYGZdq{}}\PYG{p}{)}
\PYG{g+go}{Text(0.5, 1.0, \PYGZsq{}Hann window\PYGZsq{})}
\PYG{g+gp}{\PYGZgt{}\PYGZgt{}\PYGZgt{} }\PYG{n}{plt}\PYG{o}{.}\PYG{n}{ylabel}\PYG{p}{(}\PYG{l+s+s2}{\PYGZdq{}}\PYG{l+s+s2}{Amplitude}\PYG{l+s+s2}{\PYGZdq{}}\PYG{p}{)}
\PYG{g+go}{Text(0, 0.5, \PYGZsq{}Amplitude\PYGZsq{})}
\PYG{g+gp}{\PYGZgt{}\PYGZgt{}\PYGZgt{} }\PYG{n}{plt}\PYG{o}{.}\PYG{n}{xlabel}\PYG{p}{(}\PYG{l+s+s2}{\PYGZdq{}}\PYG{l+s+s2}{Sample}\PYG{l+s+s2}{\PYGZdq{}}\PYG{p}{)}
\PYG{g+go}{Text(0.5, 0, \PYGZsq{}Sample\PYGZsq{})}
\PYG{g+gp}{\PYGZgt{}\PYGZgt{}\PYGZgt{} }\PYG{n}{plt}\PYG{o}{.}\PYG{n}{show}\PYG{p}{(}\PYG{p}{)}
\end{sphinxVerbatim}

\begin{sphinxVerbatim}[commandchars=\\\{\}]
\PYG{g+gp}{\PYGZgt{}\PYGZgt{}\PYGZgt{} }\PYG{n}{plt}\PYG{o}{.}\PYG{n}{figure}\PYG{p}{(}\PYG{p}{)}
\PYG{g+go}{\PYGZlt{}Figure size 640x480 with 0 Axes\PYGZgt{}}
\PYG{g+gp}{\PYGZgt{}\PYGZgt{}\PYGZgt{} }\PYG{n}{A} \PYG{o}{=} \PYG{n}{fft}\PYG{p}{(}\PYG{n}{window}\PYG{p}{,} \PYG{l+m+mi}{2048}\PYG{p}{)} \PYG{o}{/} \PYG{l+m+mf}{25.5}
\PYG{g+gp}{\PYGZgt{}\PYGZgt{}\PYGZgt{} }\PYG{n}{mag} \PYG{o}{=} \PYG{n}{np}\PYG{o}{.}\PYG{n}{abs}\PYG{p}{(}\PYG{n}{fftshift}\PYG{p}{(}\PYG{n}{A}\PYG{p}{)}\PYG{p}{)}
\PYG{g+gp}{\PYGZgt{}\PYGZgt{}\PYGZgt{} }\PYG{n}{freq} \PYG{o}{=} \PYG{n}{np}\PYG{o}{.}\PYG{n}{linspace}\PYG{p}{(}\PYG{o}{\PYGZhy{}}\PYG{l+m+mf}{0.5}\PYG{p}{,} \PYG{l+m+mf}{0.5}\PYG{p}{,} \PYG{n+nb}{len}\PYG{p}{(}\PYG{n}{A}\PYG{p}{)}\PYG{p}{)}
\PYG{g+gp}{\PYGZgt{}\PYGZgt{}\PYGZgt{} }\PYG{k}{with} \PYG{n}{np}\PYG{o}{.}\PYG{n}{errstate}\PYG{p}{(}\PYG{n}{divide}\PYG{o}{=}\PYG{l+s+s1}{\PYGZsq{}}\PYG{l+s+s1}{ignore}\PYG{l+s+s1}{\PYGZsq{}}\PYG{p}{,} \PYG{n}{invalid}\PYG{o}{=}\PYG{l+s+s1}{\PYGZsq{}}\PYG{l+s+s1}{ignore}\PYG{l+s+s1}{\PYGZsq{}}\PYG{p}{)}\PYG{p}{:}
\PYG{g+gp}{... }    \PYG{n}{response} \PYG{o}{=} \PYG{l+m+mi}{20} \PYG{o}{*} \PYG{n}{np}\PYG{o}{.}\PYG{n}{log10}\PYG{p}{(}\PYG{n}{mag}\PYG{p}{)}
\PYG{g+gp}{...}
\PYG{g+gp}{\PYGZgt{}\PYGZgt{}\PYGZgt{} }\PYG{n}{response} \PYG{o}{=} \PYG{n}{np}\PYG{o}{.}\PYG{n}{clip}\PYG{p}{(}\PYG{n}{response}\PYG{p}{,} \PYG{o}{\PYGZhy{}}\PYG{l+m+mi}{100}\PYG{p}{,} \PYG{l+m+mi}{100}\PYG{p}{)}
\PYG{g+gp}{\PYGZgt{}\PYGZgt{}\PYGZgt{} }\PYG{n}{plt}\PYG{o}{.}\PYG{n}{plot}\PYG{p}{(}\PYG{n}{freq}\PYG{p}{,} \PYG{n}{response}\PYG{p}{)}
\PYG{g+go}{[\PYGZlt{}matplotlib.lines.Line2D object at 0x...\PYGZgt{}]}
\PYG{g+gp}{\PYGZgt{}\PYGZgt{}\PYGZgt{} }\PYG{n}{plt}\PYG{o}{.}\PYG{n}{title}\PYG{p}{(}\PYG{l+s+s2}{\PYGZdq{}}\PYG{l+s+s2}{Frequency response of the Hann window}\PYG{l+s+s2}{\PYGZdq{}}\PYG{p}{)}
\PYG{g+go}{Text(0.5, 1.0, \PYGZsq{}Frequency response of the Hann window\PYGZsq{})}
\PYG{g+gp}{\PYGZgt{}\PYGZgt{}\PYGZgt{} }\PYG{n}{plt}\PYG{o}{.}\PYG{n}{ylabel}\PYG{p}{(}\PYG{l+s+s2}{\PYGZdq{}}\PYG{l+s+s2}{Magnitude [dB]}\PYG{l+s+s2}{\PYGZdq{}}\PYG{p}{)}
\PYG{g+go}{Text(0, 0.5, \PYGZsq{}Magnitude [dB]\PYGZsq{})}
\PYG{g+gp}{\PYGZgt{}\PYGZgt{}\PYGZgt{} }\PYG{n}{plt}\PYG{o}{.}\PYG{n}{xlabel}\PYG{p}{(}\PYG{l+s+s2}{\PYGZdq{}}\PYG{l+s+s2}{Normalized frequency [cycles per sample]}\PYG{l+s+s2}{\PYGZdq{}}\PYG{p}{)}
\PYG{g+go}{Text(0.5, 0, \PYGZsq{}Normalized frequency [cycles per sample]\PYGZsq{})}
\PYG{g+gp}{\PYGZgt{}\PYGZgt{}\PYGZgt{} }\PYG{n}{plt}\PYG{o}{.}\PYG{n}{axis}\PYG{p}{(}\PYG{l+s+s1}{\PYGZsq{}}\PYG{l+s+s1}{tight}\PYG{l+s+s1}{\PYGZsq{}}\PYG{p}{)}
\PYG{g+gp}{...}
\PYG{g+gp}{\PYGZgt{}\PYGZgt{}\PYGZgt{} }\PYG{n}{plt}\PYG{o}{.}\PYG{n}{show}\PYG{p}{(}\PYG{p}{)}
\end{sphinxVerbatim}

\end{fulllineitems}

\index{kaiser() (in module symjax.tensor.signal)@\spxentry{kaiser()}\spxextra{in module symjax.tensor.signal}}

\begin{fulllineitems}
\phantomsection\label{\detokenize{modules/signal:symjax.tensor.signal.kaiser}}\pysiglinewithargsret{\sphinxbfcode{\sphinxupquote{kaiser}}}{\emph{\DUrole{n}{M}}, \emph{\DUrole{n}{beta}}}{}
Return the Kaiser window.

LAX\sphinxhyphen{}backend implementation of {\hyperref[\detokenize{modules/signal:symjax.tensor.signal.kaiser}]{\sphinxcrossref{\sphinxcode{\sphinxupquote{kaiser()}}}}}.
ADDITIONOriginal docstring below.

The Kaiser window is a taper formed by using a Bessel function.
\begin{quote}\begin{description}
\item[{Returns}] \leavevmode
\sphinxstylestrong{out} \textendash{} The window, with the maximum value normalized to one (the value
one appears only if the number of samples is odd).

\item[{Return type}] \leavevmode
array

\end{description}\end{quote}

\sphinxstrong{See also:}

{\hyperref[\detokenize{modules/signal:symjax.tensor.signal.bartlett}]{\sphinxcrossref{\sphinxcode{\sphinxupquote{bartlett()}}}}}, {\hyperref[\detokenize{modules/signal:symjax.tensor.signal.blackman}]{\sphinxcrossref{\sphinxcode{\sphinxupquote{blackman()}}}}}, {\hyperref[\detokenize{modules/signal:symjax.tensor.signal.hamming}]{\sphinxcrossref{\sphinxcode{\sphinxupquote{hamming()}}}}}, {\hyperref[\detokenize{modules/signal:symjax.tensor.signal.hanning}]{\sphinxcrossref{\sphinxcode{\sphinxupquote{hanning()}}}}}

\subsubsection*{Notes}

The Kaiser window is defined as
\begin{equation*}
\begin{split}w(n) = I_0\left( \beta \sqrt{1-\frac{4n^2}{(M-1)^2}}
\right)/I_0(\beta)\end{split}
\end{equation*}
with
\begin{equation*}
\begin{split}\quad -\frac{M-1}{2} \leq n \leq \frac{M-1}{2},\end{split}
\end{equation*}
where \(I_0\) is the modified zeroth\sphinxhyphen{}order Bessel function.

The Kaiser was named for Jim Kaiser, who discovered a simple
approximation to the DPSS window based on Bessel functions.  The Kaiser
window is a very good approximation to the Digital Prolate Spheroidal
Sequence, or Slepian window, which is the transform which maximizes the
energy in the main lobe of the window relative to total energy.

The Kaiser can approximate many other windows by varying the beta
parameter.

\begin{savenotes}\sphinxattablestart
\centering
\begin{tabulary}{\linewidth}[t]{|T|T|}
\hline
\sphinxstyletheadfamily 
beta
&\sphinxstyletheadfamily 
Window shape
\\
\hline
0
&
Rectangular
\\
\hline
5
&
Similar to a Hamming
\\
\hline
6
&
Similar to a Hanning
\\
\hline
8.6
&
Similar to a Blackman
\\
\hline
\end{tabulary}
\par
\sphinxattableend\end{savenotes}

A beta value of 14 is probably a good starting point. Note that as beta
gets large, the window narrows, and so the number of samples needs to be
large enough to sample the increasingly narrow spike, otherwise NaNs will
get returned.

Most references to the Kaiser window come from the signal processing
literature, where it is used as one of many windowing functions for
smoothing values.  It is also known as an apodization (which means
“removing the foot”, i.e. smoothing discontinuities at the beginning
and end of the sampled signal) or tapering function.
\subsubsection*{References}
\subsubsection*{Examples}

\begin{sphinxVerbatim}[commandchars=\\\{\}]
\PYG{g+gp}{\PYGZgt{}\PYGZgt{}\PYGZgt{} }\PYG{k+kn}{import} \PYG{n+nn}{matplotlib}\PYG{n+nn}{.}\PYG{n+nn}{pyplot} \PYG{k}{as} \PYG{n+nn}{plt}
\PYG{g+gp}{\PYGZgt{}\PYGZgt{}\PYGZgt{} }\PYG{n}{np}\PYG{o}{.}\PYG{n}{kaiser}\PYG{p}{(}\PYG{l+m+mi}{12}\PYG{p}{,} \PYG{l+m+mi}{14}\PYG{p}{)}
\PYG{g+go}{ array([7.72686684e\PYGZhy{}06, 3.46009194e\PYGZhy{}03, 4.65200189e\PYGZhy{}02, \PYGZsh{} may vary}
\PYG{g+go}{        2.29737120e\PYGZhy{}01, 5.99885316e\PYGZhy{}01, 9.45674898e\PYGZhy{}01,}
\PYG{g+go}{        9.45674898e\PYGZhy{}01, 5.99885316e\PYGZhy{}01, 2.29737120e\PYGZhy{}01,}
\PYG{g+go}{        4.65200189e\PYGZhy{}02, 3.46009194e\PYGZhy{}03, 7.72686684e\PYGZhy{}06])}
\end{sphinxVerbatim}

Plot the window and the frequency response:

\begin{sphinxVerbatim}[commandchars=\\\{\}]
\PYG{g+gp}{\PYGZgt{}\PYGZgt{}\PYGZgt{} }\PYG{k+kn}{from} \PYG{n+nn}{numpy}\PYG{n+nn}{.}\PYG{n+nn}{fft} \PYG{k+kn}{import} \PYG{n}{fft}\PYG{p}{,} \PYG{n}{fftshift}
\PYG{g+gp}{\PYGZgt{}\PYGZgt{}\PYGZgt{} }\PYG{n}{window} \PYG{o}{=} \PYG{n}{np}\PYG{o}{.}\PYG{n}{kaiser}\PYG{p}{(}\PYG{l+m+mi}{51}\PYG{p}{,} \PYG{l+m+mi}{14}\PYG{p}{)}
\PYG{g+gp}{\PYGZgt{}\PYGZgt{}\PYGZgt{} }\PYG{n}{plt}\PYG{o}{.}\PYG{n}{plot}\PYG{p}{(}\PYG{n}{window}\PYG{p}{)}
\PYG{g+go}{[\PYGZlt{}matplotlib.lines.Line2D object at 0x...\PYGZgt{}]}
\PYG{g+gp}{\PYGZgt{}\PYGZgt{}\PYGZgt{} }\PYG{n}{plt}\PYG{o}{.}\PYG{n}{title}\PYG{p}{(}\PYG{l+s+s2}{\PYGZdq{}}\PYG{l+s+s2}{Kaiser window}\PYG{l+s+s2}{\PYGZdq{}}\PYG{p}{)}
\PYG{g+go}{Text(0.5, 1.0, \PYGZsq{}Kaiser window\PYGZsq{})}
\PYG{g+gp}{\PYGZgt{}\PYGZgt{}\PYGZgt{} }\PYG{n}{plt}\PYG{o}{.}\PYG{n}{ylabel}\PYG{p}{(}\PYG{l+s+s2}{\PYGZdq{}}\PYG{l+s+s2}{Amplitude}\PYG{l+s+s2}{\PYGZdq{}}\PYG{p}{)}
\PYG{g+go}{Text(0, 0.5, \PYGZsq{}Amplitude\PYGZsq{})}
\PYG{g+gp}{\PYGZgt{}\PYGZgt{}\PYGZgt{} }\PYG{n}{plt}\PYG{o}{.}\PYG{n}{xlabel}\PYG{p}{(}\PYG{l+s+s2}{\PYGZdq{}}\PYG{l+s+s2}{Sample}\PYG{l+s+s2}{\PYGZdq{}}\PYG{p}{)}
\PYG{g+go}{Text(0.5, 0, \PYGZsq{}Sample\PYGZsq{})}
\PYG{g+gp}{\PYGZgt{}\PYGZgt{}\PYGZgt{} }\PYG{n}{plt}\PYG{o}{.}\PYG{n}{show}\PYG{p}{(}\PYG{p}{)}
\end{sphinxVerbatim}

\begin{sphinxVerbatim}[commandchars=\\\{\}]
\PYG{g+gp}{\PYGZgt{}\PYGZgt{}\PYGZgt{} }\PYG{n}{plt}\PYG{o}{.}\PYG{n}{figure}\PYG{p}{(}\PYG{p}{)}
\PYG{g+go}{\PYGZlt{}Figure size 640x480 with 0 Axes\PYGZgt{}}
\PYG{g+gp}{\PYGZgt{}\PYGZgt{}\PYGZgt{} }\PYG{n}{A} \PYG{o}{=} \PYG{n}{fft}\PYG{p}{(}\PYG{n}{window}\PYG{p}{,} \PYG{l+m+mi}{2048}\PYG{p}{)} \PYG{o}{/} \PYG{l+m+mf}{25.5}
\PYG{g+gp}{\PYGZgt{}\PYGZgt{}\PYGZgt{} }\PYG{n}{mag} \PYG{o}{=} \PYG{n}{np}\PYG{o}{.}\PYG{n}{abs}\PYG{p}{(}\PYG{n}{fftshift}\PYG{p}{(}\PYG{n}{A}\PYG{p}{)}\PYG{p}{)}
\PYG{g+gp}{\PYGZgt{}\PYGZgt{}\PYGZgt{} }\PYG{n}{freq} \PYG{o}{=} \PYG{n}{np}\PYG{o}{.}\PYG{n}{linspace}\PYG{p}{(}\PYG{o}{\PYGZhy{}}\PYG{l+m+mf}{0.5}\PYG{p}{,} \PYG{l+m+mf}{0.5}\PYG{p}{,} \PYG{n+nb}{len}\PYG{p}{(}\PYG{n}{A}\PYG{p}{)}\PYG{p}{)}
\PYG{g+gp}{\PYGZgt{}\PYGZgt{}\PYGZgt{} }\PYG{n}{response} \PYG{o}{=} \PYG{l+m+mi}{20} \PYG{o}{*} \PYG{n}{np}\PYG{o}{.}\PYG{n}{log10}\PYG{p}{(}\PYG{n}{mag}\PYG{p}{)}
\PYG{g+gp}{\PYGZgt{}\PYGZgt{}\PYGZgt{} }\PYG{n}{response} \PYG{o}{=} \PYG{n}{np}\PYG{o}{.}\PYG{n}{clip}\PYG{p}{(}\PYG{n}{response}\PYG{p}{,} \PYG{o}{\PYGZhy{}}\PYG{l+m+mi}{100}\PYG{p}{,} \PYG{l+m+mi}{100}\PYG{p}{)}
\PYG{g+gp}{\PYGZgt{}\PYGZgt{}\PYGZgt{} }\PYG{n}{plt}\PYG{o}{.}\PYG{n}{plot}\PYG{p}{(}\PYG{n}{freq}\PYG{p}{,} \PYG{n}{response}\PYG{p}{)}
\PYG{g+go}{[\PYGZlt{}matplotlib.lines.Line2D object at 0x...\PYGZgt{}]}
\PYG{g+gp}{\PYGZgt{}\PYGZgt{}\PYGZgt{} }\PYG{n}{plt}\PYG{o}{.}\PYG{n}{title}\PYG{p}{(}\PYG{l+s+s2}{\PYGZdq{}}\PYG{l+s+s2}{Frequency response of Kaiser window}\PYG{l+s+s2}{\PYGZdq{}}\PYG{p}{)}
\PYG{g+go}{Text(0.5, 1.0, \PYGZsq{}Frequency response of Kaiser window\PYGZsq{})}
\PYG{g+gp}{\PYGZgt{}\PYGZgt{}\PYGZgt{} }\PYG{n}{plt}\PYG{o}{.}\PYG{n}{ylabel}\PYG{p}{(}\PYG{l+s+s2}{\PYGZdq{}}\PYG{l+s+s2}{Magnitude [dB]}\PYG{l+s+s2}{\PYGZdq{}}\PYG{p}{)}
\PYG{g+go}{Text(0, 0.5, \PYGZsq{}Magnitude [dB]\PYGZsq{})}
\PYG{g+gp}{\PYGZgt{}\PYGZgt{}\PYGZgt{} }\PYG{n}{plt}\PYG{o}{.}\PYG{n}{xlabel}\PYG{p}{(}\PYG{l+s+s2}{\PYGZdq{}}\PYG{l+s+s2}{Normalized frequency [cycles per sample]}\PYG{l+s+s2}{\PYGZdq{}}\PYG{p}{)}
\PYG{g+go}{Text(0.5, 0, \PYGZsq{}Normalized frequency [cycles per sample]\PYGZsq{})}
\PYG{g+gp}{\PYGZgt{}\PYGZgt{}\PYGZgt{} }\PYG{n}{plt}\PYG{o}{.}\PYG{n}{axis}\PYG{p}{(}\PYG{l+s+s1}{\PYGZsq{}}\PYG{l+s+s1}{tight}\PYG{l+s+s1}{\PYGZsq{}}\PYG{p}{)}
\PYG{g+go}{(\PYGZhy{}0.5, 0.5, \PYGZhy{}100.0, ...) \PYGZsh{} may vary}
\PYG{g+gp}{\PYGZgt{}\PYGZgt{}\PYGZgt{} }\PYG{n}{plt}\PYG{o}{.}\PYG{n}{show}\PYG{p}{(}\PYG{p}{)}
\end{sphinxVerbatim}

\end{fulllineitems}

\index{fft() (in module symjax.tensor.signal)@\spxentry{fft()}\spxextra{in module symjax.tensor.signal}}

\begin{fulllineitems}
\phantomsection\label{\detokenize{modules/signal:symjax.tensor.signal.fft}}\pysiglinewithargsret{\sphinxbfcode{\sphinxupquote{fft}}}{\emph{\DUrole{n}{a}}, \emph{\DUrole{n}{n}\DUrole{o}{=}\DUrole{default_value}{None}}, \emph{\DUrole{n}{axis}\DUrole{o}{=}\DUrole{default_value}{\sphinxhyphen{} 1}}, \emph{\DUrole{n}{norm}\DUrole{o}{=}\DUrole{default_value}{None}}}{}
Compute the one\sphinxhyphen{}dimensional discrete Fourier Transform.

LAX\sphinxhyphen{}backend implementation of {\hyperref[\detokenize{modules/signal:symjax.tensor.signal.fft}]{\sphinxcrossref{\sphinxcode{\sphinxupquote{fft()}}}}}.
ADDITIONOriginal docstring below.

LAX\sphinxhyphen{}backend implementation of {\hyperref[\detokenize{modules/signal:symjax.tensor.signal.fft}]{\sphinxcrossref{\sphinxcode{\sphinxupquote{fft()}}}}}.
Original docstring below.

This function computes the one\sphinxhyphen{}dimensional \sphinxstyleemphasis{n}\sphinxhyphen{}point discrete Fourier
Transform (DFT) with the efficient Fast Fourier Transform (FFT)
algorithm {[}CT{]}.
\begin{quote}\begin{description}
\item[{Returns}] \leavevmode
\sphinxstylestrong{out} \textendash{} The truncated or zero\sphinxhyphen{}padded input, transformed along the axis
indicated by \sphinxtitleref{axis}, or the last one if \sphinxtitleref{axis} is not specified.

\item[{Return type}] \leavevmode
complex ndarray

\item[{Raises}] \leavevmode
\sphinxstyleliteralstrong{\sphinxupquote{IndexError}} \textendash{} if \sphinxtitleref{axes} is larger than the last axis of \sphinxtitleref{a}.

\end{description}\end{quote}

\sphinxstrong{See also:}

\begin{description}
\item[{\sphinxcode{\sphinxupquote{numpy.fft()}}}] \leavevmode
for definition of the DFT and conventions used.

\item[{{\hyperref[\detokenize{modules/signal:symjax.tensor.signal.ifft}]{\sphinxcrossref{\sphinxcode{\sphinxupquote{ifft()}}}}}}] \leavevmode
The inverse of \sphinxtitleref{fft}.

\item[{{\hyperref[\detokenize{modules/signal:symjax.tensor.signal.fft2}]{\sphinxcrossref{\sphinxcode{\sphinxupquote{fft2()}}}}}}] \leavevmode
The two\sphinxhyphen{}dimensional FFT.

\item[{{\hyperref[\detokenize{modules/signal:symjax.tensor.signal.fftn}]{\sphinxcrossref{\sphinxcode{\sphinxupquote{fftn()}}}}}}] \leavevmode
The \sphinxstyleemphasis{n}\sphinxhyphen{}dimensional FFT.

\item[{{\hyperref[\detokenize{modules/signal:symjax.tensor.signal.rfftn}]{\sphinxcrossref{\sphinxcode{\sphinxupquote{rfftn()}}}}}}] \leavevmode
The \sphinxstyleemphasis{n}\sphinxhyphen{}dimensional FFT of real input.

\item[{{\hyperref[\detokenize{modules/signal:symjax.tensor.signal.fftfreq}]{\sphinxcrossref{\sphinxcode{\sphinxupquote{fftfreq()}}}}}}] \leavevmode
Frequency bins for given FFT parameters.

\end{description}

\subsubsection*{Notes}

FFT (Fast Fourier Transform) refers to a way the discrete Fourier
Transform (DFT) can be calculated efficiently, by using symmetries in the
calculated terms.  The symmetry is highest when \sphinxtitleref{n} is a power of 2, and
the transform is therefore most efficient for these sizes.

The DFT is defined, with the conventions used in this implementation, in
the documentation for the \sphinxtitleref{numpy.fft} module.
\subsubsection*{References}
\subsubsection*{Examples}

\begin{sphinxVerbatim}[commandchars=\\\{\}]
\PYG{g+gp}{\PYGZgt{}\PYGZgt{}\PYGZgt{} }\PYG{n}{np}\PYG{o}{.}\PYG{n}{fft}\PYG{o}{.}\PYG{n}{fft}\PYG{p}{(}\PYG{n}{np}\PYG{o}{.}\PYG{n}{exp}\PYG{p}{(}\PYG{l+m+mi}{2}\PYG{n}{j} \PYG{o}{*} \PYG{n}{np}\PYG{o}{.}\PYG{n}{pi} \PYG{o}{*} \PYG{n}{np}\PYG{o}{.}\PYG{n}{arange}\PYG{p}{(}\PYG{l+m+mi}{8}\PYG{p}{)} \PYG{o}{/} \PYG{l+m+mi}{8}\PYG{p}{)}\PYG{p}{)}
\PYG{g+go}{array([\PYGZhy{}2.33486982e\PYGZhy{}16+1.14423775e\PYGZhy{}17j,  8.00000000e+00\PYGZhy{}1.25557246e\PYGZhy{}15j,}
\PYG{g+go}{        2.33486982e\PYGZhy{}16+2.33486982e\PYGZhy{}16j,  0.00000000e+00+1.22464680e\PYGZhy{}16j,}
\PYG{g+go}{       \PYGZhy{}1.14423775e\PYGZhy{}17+2.33486982e\PYGZhy{}16j,  0.00000000e+00+5.20784380e\PYGZhy{}16j,}
\PYG{g+go}{        1.14423775e\PYGZhy{}17+1.14423775e\PYGZhy{}17j,  0.00000000e+00+1.22464680e\PYGZhy{}16j])}
\end{sphinxVerbatim}

In this example, real input has an FFT which is Hermitian, i.e., symmetric
in the real part and anti\sphinxhyphen{}symmetric in the imaginary part, as described in
the \sphinxtitleref{numpy.fft} documentation:

\begin{sphinxVerbatim}[commandchars=\\\{\}]
\PYG{g+gp}{\PYGZgt{}\PYGZgt{}\PYGZgt{} }\PYG{k+kn}{import} \PYG{n+nn}{matplotlib}\PYG{n+nn}{.}\PYG{n+nn}{pyplot} \PYG{k}{as} \PYG{n+nn}{plt}
\PYG{g+gp}{\PYGZgt{}\PYGZgt{}\PYGZgt{} }\PYG{n}{t} \PYG{o}{=} \PYG{n}{np}\PYG{o}{.}\PYG{n}{arange}\PYG{p}{(}\PYG{l+m+mi}{256}\PYG{p}{)}
\PYG{g+gp}{\PYGZgt{}\PYGZgt{}\PYGZgt{} }\PYG{n}{sp} \PYG{o}{=} \PYG{n}{np}\PYG{o}{.}\PYG{n}{fft}\PYG{o}{.}\PYG{n}{fft}\PYG{p}{(}\PYG{n}{np}\PYG{o}{.}\PYG{n}{sin}\PYG{p}{(}\PYG{n}{t}\PYG{p}{)}\PYG{p}{)}
\PYG{g+gp}{\PYGZgt{}\PYGZgt{}\PYGZgt{} }\PYG{n}{freq} \PYG{o}{=} \PYG{n}{np}\PYG{o}{.}\PYG{n}{fft}\PYG{o}{.}\PYG{n}{fftfreq}\PYG{p}{(}\PYG{n}{t}\PYG{o}{.}\PYG{n}{shape}\PYG{p}{[}\PYG{o}{\PYGZhy{}}\PYG{l+m+mi}{1}\PYG{p}{]}\PYG{p}{)}
\PYG{g+gp}{\PYGZgt{}\PYGZgt{}\PYGZgt{} }\PYG{n}{plt}\PYG{o}{.}\PYG{n}{plot}\PYG{p}{(}\PYG{n}{freq}\PYG{p}{,} \PYG{n}{sp}\PYG{o}{.}\PYG{n}{real}\PYG{p}{,} \PYG{n}{freq}\PYG{p}{,} \PYG{n}{sp}\PYG{o}{.}\PYG{n}{imag}\PYG{p}{)}
\PYG{g+go}{[\PYGZlt{}matplotlib.lines.Line2D object at 0x...\PYGZgt{}, \PYGZlt{}matplotlib.lines.Line2D object at 0x...\PYGZgt{}]}
\PYG{g+gp}{\PYGZgt{}\PYGZgt{}\PYGZgt{} }\PYG{n}{plt}\PYG{o}{.}\PYG{n}{show}\PYG{p}{(}\PYG{p}{)}
\end{sphinxVerbatim}

\end{fulllineitems}

\index{ifft() (in module symjax.tensor.signal)@\spxentry{ifft()}\spxextra{in module symjax.tensor.signal}}

\begin{fulllineitems}
\phantomsection\label{\detokenize{modules/signal:symjax.tensor.signal.ifft}}\pysiglinewithargsret{\sphinxbfcode{\sphinxupquote{ifft}}}{\emph{\DUrole{n}{a}}, \emph{\DUrole{n}{n}\DUrole{o}{=}\DUrole{default_value}{None}}, \emph{\DUrole{n}{axis}\DUrole{o}{=}\DUrole{default_value}{\sphinxhyphen{} 1}}, \emph{\DUrole{n}{norm}\DUrole{o}{=}\DUrole{default_value}{None}}}{}
Compute the one\sphinxhyphen{}dimensional inverse discrete Fourier Transform.

LAX\sphinxhyphen{}backend implementation of {\hyperref[\detokenize{modules/signal:symjax.tensor.signal.ifft}]{\sphinxcrossref{\sphinxcode{\sphinxupquote{ifft()}}}}}.
ADDITIONOriginal docstring below.

LAX\sphinxhyphen{}backend implementation of {\hyperref[\detokenize{modules/signal:symjax.tensor.signal.ifft}]{\sphinxcrossref{\sphinxcode{\sphinxupquote{ifft()}}}}}.
Original docstring below.

This function computes the inverse of the one\sphinxhyphen{}dimensional \sphinxstyleemphasis{n}\sphinxhyphen{}point
discrete Fourier transform computed by \sphinxtitleref{fft}.  In other words,
\sphinxcode{\sphinxupquote{ifft(fft(a)) == a}} to within numerical accuracy.
For a general description of the algorithm and definitions,
see \sphinxtitleref{numpy.fft}.

The input should be ordered in the same way as is returned by \sphinxtitleref{fft},
i.e.,
\begin{itemize}
\item {} 
\sphinxcode{\sphinxupquote{a{[}0{]}}} should contain the zero frequency term,

\item {} 
\sphinxcode{\sphinxupquote{a{[}1:n//2{]}}} should contain the positive\sphinxhyphen{}frequency terms,

\item {} 
\sphinxcode{\sphinxupquote{a{[}n//2 + 1:{]}}} should contain the negative\sphinxhyphen{}frequency terms, in
increasing order starting from the most negative frequency.

\end{itemize}

For an even number of input points, \sphinxcode{\sphinxupquote{A{[}n//2{]}}} represents the sum of
the values at the positive and negative Nyquist frequencies, as the two
are aliased together. See \sphinxtitleref{numpy.fft} for details.
\begin{quote}\begin{description}
\item[{Returns}] \leavevmode
\sphinxstylestrong{out} \textendash{} The truncated or zero\sphinxhyphen{}padded input, transformed along the axis
indicated by \sphinxtitleref{axis}, or the last one if \sphinxtitleref{axis} is not specified.

\item[{Return type}] \leavevmode
complex ndarray

\item[{Raises}] \leavevmode
\sphinxstyleliteralstrong{\sphinxupquote{IndexError}} \textendash{} If \sphinxtitleref{axes} is larger than the last axis of \sphinxtitleref{a}.

\end{description}\end{quote}

\sphinxstrong{See also:}

\begin{description}
\item[{\sphinxcode{\sphinxupquote{numpy.fft()}}}] \leavevmode
An introduction, with definitions and general explanations.

\item[{{\hyperref[\detokenize{modules/signal:symjax.tensor.signal.fft}]{\sphinxcrossref{\sphinxcode{\sphinxupquote{fft()}}}}}}] \leavevmode
The one\sphinxhyphen{}dimensional (forward) FFT, of which \sphinxtitleref{ifft} is the inverse

\item[{{\hyperref[\detokenize{modules/signal:symjax.tensor.signal.ifft2}]{\sphinxcrossref{\sphinxcode{\sphinxupquote{ifft2()}}}}}}] \leavevmode
The two\sphinxhyphen{}dimensional inverse FFT.

\item[{{\hyperref[\detokenize{modules/signal:symjax.tensor.signal.ifftn}]{\sphinxcrossref{\sphinxcode{\sphinxupquote{ifftn()}}}}}}] \leavevmode
The n\sphinxhyphen{}dimensional inverse FFT.

\end{description}

\subsubsection*{Notes}

If the input parameter \sphinxtitleref{n} is larger than the size of the input, the input
is padded by appending zeros at the end.  Even though this is the common
approach, it might lead to surprising results.  If a different padding is
desired, it must be performed before calling \sphinxtitleref{ifft}.
\subsubsection*{Examples}

\begin{sphinxVerbatim}[commandchars=\\\{\}]
\PYG{g+gp}{\PYGZgt{}\PYGZgt{}\PYGZgt{} }\PYG{n}{np}\PYG{o}{.}\PYG{n}{fft}\PYG{o}{.}\PYG{n}{ifft}\PYG{p}{(}\PYG{p}{[}\PYG{l+m+mi}{0}\PYG{p}{,} \PYG{l+m+mi}{4}\PYG{p}{,} \PYG{l+m+mi}{0}\PYG{p}{,} \PYG{l+m+mi}{0}\PYG{p}{]}\PYG{p}{)}
\PYG{g+go}{array([ 1.+0.j,  0.+1.j, \PYGZhy{}1.+0.j,  0.\PYGZhy{}1.j]) \PYGZsh{} may vary}
\end{sphinxVerbatim}

Create and plot a band\sphinxhyphen{}limited signal with random phases:

\begin{sphinxVerbatim}[commandchars=\\\{\}]
\PYG{g+gp}{\PYGZgt{}\PYGZgt{}\PYGZgt{} }\PYG{k+kn}{import} \PYG{n+nn}{matplotlib}\PYG{n+nn}{.}\PYG{n+nn}{pyplot} \PYG{k}{as} \PYG{n+nn}{plt}
\PYG{g+gp}{\PYGZgt{}\PYGZgt{}\PYGZgt{} }\PYG{n}{t} \PYG{o}{=} \PYG{n}{np}\PYG{o}{.}\PYG{n}{arange}\PYG{p}{(}\PYG{l+m+mi}{400}\PYG{p}{)}
\PYG{g+gp}{\PYGZgt{}\PYGZgt{}\PYGZgt{} }\PYG{n}{n} \PYG{o}{=} \PYG{n}{np}\PYG{o}{.}\PYG{n}{zeros}\PYG{p}{(}\PYG{p}{(}\PYG{l+m+mi}{400}\PYG{p}{,}\PYG{p}{)}\PYG{p}{,} \PYG{n}{dtype}\PYG{o}{=}\PYG{n+nb}{complex}\PYG{p}{)}
\PYG{g+gp}{\PYGZgt{}\PYGZgt{}\PYGZgt{} }\PYG{n}{n}\PYG{p}{[}\PYG{l+m+mi}{40}\PYG{p}{:}\PYG{l+m+mi}{60}\PYG{p}{]} \PYG{o}{=} \PYG{n}{np}\PYG{o}{.}\PYG{n}{exp}\PYG{p}{(}\PYG{l+m+mi}{1}\PYG{n}{j}\PYG{o}{*}\PYG{n}{np}\PYG{o}{.}\PYG{n}{random}\PYG{o}{.}\PYG{n}{uniform}\PYG{p}{(}\PYG{l+m+mi}{0}\PYG{p}{,} \PYG{l+m+mi}{2}\PYG{o}{*}\PYG{n}{np}\PYG{o}{.}\PYG{n}{pi}\PYG{p}{,} \PYG{p}{(}\PYG{l+m+mi}{20}\PYG{p}{,}\PYG{p}{)}\PYG{p}{)}\PYG{p}{)}
\PYG{g+gp}{\PYGZgt{}\PYGZgt{}\PYGZgt{} }\PYG{n}{s} \PYG{o}{=} \PYG{n}{np}\PYG{o}{.}\PYG{n}{fft}\PYG{o}{.}\PYG{n}{ifft}\PYG{p}{(}\PYG{n}{n}\PYG{p}{)}
\PYG{g+gp}{\PYGZgt{}\PYGZgt{}\PYGZgt{} }\PYG{n}{plt}\PYG{o}{.}\PYG{n}{plot}\PYG{p}{(}\PYG{n}{t}\PYG{p}{,} \PYG{n}{s}\PYG{o}{.}\PYG{n}{real}\PYG{p}{,} \PYG{l+s+s1}{\PYGZsq{}}\PYG{l+s+s1}{b\PYGZhy{}}\PYG{l+s+s1}{\PYGZsq{}}\PYG{p}{,} \PYG{n}{t}\PYG{p}{,} \PYG{n}{s}\PYG{o}{.}\PYG{n}{imag}\PYG{p}{,} \PYG{l+s+s1}{\PYGZsq{}}\PYG{l+s+s1}{r\PYGZhy{}\PYGZhy{}}\PYG{l+s+s1}{\PYGZsq{}}\PYG{p}{)}
\PYG{g+go}{[\PYGZlt{}matplotlib.lines.Line2D object at ...\PYGZgt{}, \PYGZlt{}matplotlib.lines.Line2D object at ...\PYGZgt{}]}
\PYG{g+gp}{\PYGZgt{}\PYGZgt{}\PYGZgt{} }\PYG{n}{plt}\PYG{o}{.}\PYG{n}{legend}\PYG{p}{(}\PYG{p}{(}\PYG{l+s+s1}{\PYGZsq{}}\PYG{l+s+s1}{real}\PYG{l+s+s1}{\PYGZsq{}}\PYG{p}{,} \PYG{l+s+s1}{\PYGZsq{}}\PYG{l+s+s1}{imaginary}\PYG{l+s+s1}{\PYGZsq{}}\PYG{p}{)}\PYG{p}{)}
\PYG{g+go}{\PYGZlt{}matplotlib.legend.Legend object at ...\PYGZgt{}}
\PYG{g+gp}{\PYGZgt{}\PYGZgt{}\PYGZgt{} }\PYG{n}{plt}\PYG{o}{.}\PYG{n}{show}\PYG{p}{(}\PYG{p}{)}
\end{sphinxVerbatim}

\end{fulllineitems}

\index{fft2() (in module symjax.tensor.signal)@\spxentry{fft2()}\spxextra{in module symjax.tensor.signal}}

\begin{fulllineitems}
\phantomsection\label{\detokenize{modules/signal:symjax.tensor.signal.fft2}}\pysiglinewithargsret{\sphinxbfcode{\sphinxupquote{fft2}}}{\emph{\DUrole{n}{a}}, \emph{\DUrole{n}{s}\DUrole{o}{=}\DUrole{default_value}{None}}, \emph{\DUrole{n}{axes}\DUrole{o}{=}\DUrole{default_value}{\sphinxhyphen{} 2, \sphinxhyphen{} 1}}, \emph{\DUrole{n}{norm}\DUrole{o}{=}\DUrole{default_value}{None}}}{}
Compute the 2\sphinxhyphen{}dimensional discrete Fourier Transform

LAX\sphinxhyphen{}backend implementation of {\hyperref[\detokenize{modules/signal:symjax.tensor.signal.fft2}]{\sphinxcrossref{\sphinxcode{\sphinxupquote{fft2()}}}}}.
ADDITIONOriginal docstring below.

LAX\sphinxhyphen{}backend implementation of {\hyperref[\detokenize{modules/signal:symjax.tensor.signal.fft2}]{\sphinxcrossref{\sphinxcode{\sphinxupquote{fft2()}}}}}.
Original docstring below.

This function computes the \sphinxstyleemphasis{n}\sphinxhyphen{}dimensional discrete Fourier Transform
over any axes in an \sphinxstyleemphasis{M}\sphinxhyphen{}dimensional array by means of the
Fast Fourier Transform (FFT).  By default, the transform is computed over
the last two axes of the input array, i.e., a 2\sphinxhyphen{}dimensional FFT.
\begin{quote}\begin{description}
\item[{Returns}] \leavevmode
\sphinxstylestrong{out} \textendash{} The truncated or zero\sphinxhyphen{}padded input, transformed along the axes
indicated by \sphinxtitleref{axes}, or the last two axes if \sphinxtitleref{axes} is not given.

\item[{Return type}] \leavevmode
complex ndarray

\item[{Raises}] \leavevmode\begin{itemize}
\item {} 
\sphinxstyleliteralstrong{\sphinxupquote{ValueError}} \textendash{} If \sphinxtitleref{s} and \sphinxtitleref{axes} have different length, or \sphinxtitleref{axes} not given and
    \sphinxcode{\sphinxupquote{len(s) != 2}}.

\item {} 
\sphinxstyleliteralstrong{\sphinxupquote{IndexError}} \textendash{} If an element of \sphinxtitleref{axes} is larger than than the number of axes of \sphinxtitleref{a}.

\end{itemize}

\end{description}\end{quote}

\sphinxstrong{See also:}

\begin{description}
\item[{\sphinxcode{\sphinxupquote{numpy.fft()}}}] \leavevmode
Overall view of discrete Fourier transforms, with definitions and conventions used.

\item[{{\hyperref[\detokenize{modules/signal:symjax.tensor.signal.ifft2}]{\sphinxcrossref{\sphinxcode{\sphinxupquote{ifft2()}}}}}}] \leavevmode
The inverse two\sphinxhyphen{}dimensional FFT.

\item[{{\hyperref[\detokenize{modules/signal:symjax.tensor.signal.fft}]{\sphinxcrossref{\sphinxcode{\sphinxupquote{fft()}}}}}}] \leavevmode
The one\sphinxhyphen{}dimensional FFT.

\item[{{\hyperref[\detokenize{modules/signal:symjax.tensor.signal.fftn}]{\sphinxcrossref{\sphinxcode{\sphinxupquote{fftn()}}}}}}] \leavevmode
The \sphinxstyleemphasis{n}\sphinxhyphen{}dimensional FFT.

\item[{\sphinxcode{\sphinxupquote{fftshift()}}}] \leavevmode
Shifts zero\sphinxhyphen{}frequency terms to the center of the array. For two\sphinxhyphen{}dimensional input, swaps first and third quadrants, and second and fourth quadrants.

\end{description}

\subsubsection*{Notes}

\sphinxtitleref{fft2} is just \sphinxtitleref{fftn} with a different default for \sphinxtitleref{axes}.

The output, analogously to \sphinxtitleref{fft}, contains the term for zero frequency in
the low\sphinxhyphen{}order corner of the transformed axes, the positive frequency terms
in the first half of these axes, the term for the Nyquist frequency in the
middle of the axes and the negative frequency terms in the second half of
the axes, in order of decreasingly negative frequency.

See \sphinxtitleref{fftn} for details and a plotting example, and \sphinxtitleref{numpy.fft} for
definitions and conventions used.
\subsubsection*{Examples}

\begin{sphinxVerbatim}[commandchars=\\\{\}]
\PYG{g+gp}{\PYGZgt{}\PYGZgt{}\PYGZgt{} }\PYG{n}{a} \PYG{o}{=} \PYG{n}{np}\PYG{o}{.}\PYG{n}{mgrid}\PYG{p}{[}\PYG{p}{:}\PYG{l+m+mi}{5}\PYG{p}{,} \PYG{p}{:}\PYG{l+m+mi}{5}\PYG{p}{]}\PYG{p}{[}\PYG{l+m+mi}{0}\PYG{p}{]}
\PYG{g+gp}{\PYGZgt{}\PYGZgt{}\PYGZgt{} }\PYG{n}{np}\PYG{o}{.}\PYG{n}{fft}\PYG{o}{.}\PYG{n}{fft2}\PYG{p}{(}\PYG{n}{a}\PYG{p}{)}
\PYG{g+go}{array([[ 50.  +0.j        ,   0.  +0.j        ,   0.  +0.j        , \PYGZsh{} may vary}
\PYG{g+go}{          0.  +0.j        ,   0.  +0.j        ],}
\PYG{g+go}{       [\PYGZhy{}12.5+17.20477401j,   0.  +0.j        ,   0.  +0.j        ,}
\PYG{g+go}{          0.  +0.j        ,   0.  +0.j        ],}
\PYG{g+go}{       [\PYGZhy{}12.5 +4.0614962j ,   0.  +0.j        ,   0.  +0.j        ,}
\PYG{g+go}{          0.  +0.j        ,   0.  +0.j        ],}
\PYG{g+go}{       [\PYGZhy{}12.5 \PYGZhy{}4.0614962j ,   0.  +0.j        ,   0.  +0.j        ,}
\PYG{g+go}{          0.  +0.j        ,   0.  +0.j        ],}
\PYG{g+go}{       [\PYGZhy{}12.5\PYGZhy{}17.20477401j,   0.  +0.j        ,   0.  +0.j        ,}
\PYG{g+go}{          0.  +0.j        ,   0.  +0.j        ]])}
\end{sphinxVerbatim}

\end{fulllineitems}

\index{ifft2() (in module symjax.tensor.signal)@\spxentry{ifft2()}\spxextra{in module symjax.tensor.signal}}

\begin{fulllineitems}
\phantomsection\label{\detokenize{modules/signal:symjax.tensor.signal.ifft2}}\pysiglinewithargsret{\sphinxbfcode{\sphinxupquote{ifft2}}}{\emph{\DUrole{n}{a}}, \emph{\DUrole{n}{s}\DUrole{o}{=}\DUrole{default_value}{None}}, \emph{\DUrole{n}{axes}\DUrole{o}{=}\DUrole{default_value}{\sphinxhyphen{} 2, \sphinxhyphen{} 1}}, \emph{\DUrole{n}{norm}\DUrole{o}{=}\DUrole{default_value}{None}}}{}
Compute the 2\sphinxhyphen{}dimensional inverse discrete Fourier Transform.

LAX\sphinxhyphen{}backend implementation of {\hyperref[\detokenize{modules/signal:symjax.tensor.signal.ifft2}]{\sphinxcrossref{\sphinxcode{\sphinxupquote{ifft2()}}}}}.
ADDITIONOriginal docstring below.

LAX\sphinxhyphen{}backend implementation of {\hyperref[\detokenize{modules/signal:symjax.tensor.signal.ifft2}]{\sphinxcrossref{\sphinxcode{\sphinxupquote{ifft2()}}}}}.
Original docstring below.

This function computes the inverse of the 2\sphinxhyphen{}dimensional discrete Fourier
Transform over any number of axes in an M\sphinxhyphen{}dimensional array by means of
the Fast Fourier Transform (FFT).  In other words, \sphinxcode{\sphinxupquote{ifft2(fft2(a)) == a}}
to within numerical accuracy.  By default, the inverse transform is
computed over the last two axes of the input array.

The input, analogously to \sphinxtitleref{ifft}, should be ordered in the same way as is
returned by \sphinxtitleref{fft2}, i.e. it should have the term for zero frequency
in the low\sphinxhyphen{}order corner of the two axes, the positive frequency terms in
the first half of these axes, the term for the Nyquist frequency in the
middle of the axes and the negative frequency terms in the second half of
both axes, in order of decreasingly negative frequency.
\begin{quote}\begin{description}
\item[{Returns}] \leavevmode
\sphinxstylestrong{out} \textendash{} The truncated or zero\sphinxhyphen{}padded input, transformed along the axes
indicated by \sphinxtitleref{axes}, or the last two axes if \sphinxtitleref{axes} is not given.

\item[{Return type}] \leavevmode
complex ndarray

\item[{Raises}] \leavevmode\begin{itemize}
\item {} 
\sphinxstyleliteralstrong{\sphinxupquote{ValueError}} \textendash{} If \sphinxtitleref{s} and \sphinxtitleref{axes} have different length, or \sphinxtitleref{axes} not given and
    \sphinxcode{\sphinxupquote{len(s) != 2}}.

\item {} 
\sphinxstyleliteralstrong{\sphinxupquote{IndexError}} \textendash{} If an element of \sphinxtitleref{axes} is larger than than the number of axes of \sphinxtitleref{a}.

\end{itemize}

\end{description}\end{quote}

\sphinxstrong{See also:}

\begin{description}
\item[{\sphinxcode{\sphinxupquote{numpy.fft()}}}] \leavevmode
Overall view of discrete Fourier transforms, with definitions and conventions used.

\item[{{\hyperref[\detokenize{modules/signal:symjax.tensor.signal.fft2}]{\sphinxcrossref{\sphinxcode{\sphinxupquote{fft2()}}}}}}] \leavevmode
The forward 2\sphinxhyphen{}dimensional FFT, of which \sphinxtitleref{ifft2} is the inverse.

\item[{{\hyperref[\detokenize{modules/signal:symjax.tensor.signal.ifftn}]{\sphinxcrossref{\sphinxcode{\sphinxupquote{ifftn()}}}}}}] \leavevmode
The inverse of the \sphinxstyleemphasis{n}\sphinxhyphen{}dimensional FFT.

\item[{{\hyperref[\detokenize{modules/signal:symjax.tensor.signal.fft}]{\sphinxcrossref{\sphinxcode{\sphinxupquote{fft()}}}}}}] \leavevmode
The one\sphinxhyphen{}dimensional FFT.

\item[{{\hyperref[\detokenize{modules/signal:symjax.tensor.signal.ifft}]{\sphinxcrossref{\sphinxcode{\sphinxupquote{ifft()}}}}}}] \leavevmode
The one\sphinxhyphen{}dimensional inverse FFT.

\end{description}

\subsubsection*{Notes}

\sphinxtitleref{ifft2} is just \sphinxtitleref{ifftn} with a different default for \sphinxtitleref{axes}.

See \sphinxtitleref{ifftn} for details and a plotting example, and \sphinxtitleref{numpy.fft} for
definition and conventions used.

Zero\sphinxhyphen{}padding, analogously with \sphinxtitleref{ifft}, is performed by appending zeros to
the input along the specified dimension.  Although this is the common
approach, it might lead to surprising results.  If another form of zero
padding is desired, it must be performed before \sphinxtitleref{ifft2} is called.
\subsubsection*{Examples}

\begin{sphinxVerbatim}[commandchars=\\\{\}]
\PYG{g+gp}{\PYGZgt{}\PYGZgt{}\PYGZgt{} }\PYG{n}{a} \PYG{o}{=} \PYG{l+m+mi}{4} \PYG{o}{*} \PYG{n}{np}\PYG{o}{.}\PYG{n}{eye}\PYG{p}{(}\PYG{l+m+mi}{4}\PYG{p}{)}
\PYG{g+gp}{\PYGZgt{}\PYGZgt{}\PYGZgt{} }\PYG{n}{np}\PYG{o}{.}\PYG{n}{fft}\PYG{o}{.}\PYG{n}{ifft2}\PYG{p}{(}\PYG{n}{a}\PYG{p}{)}
\PYG{g+go}{array([[1.+0.j,  0.+0.j,  0.+0.j,  0.+0.j], \PYGZsh{} may vary}
\PYG{g+go}{       [0.+0.j,  0.+0.j,  0.+0.j,  1.+0.j],}
\PYG{g+go}{       [0.+0.j,  0.+0.j,  1.+0.j,  0.+0.j],}
\PYG{g+go}{       [0.+0.j,  1.+0.j,  0.+0.j,  0.+0.j]])}
\end{sphinxVerbatim}

\end{fulllineitems}

\index{fftn() (in module symjax.tensor.signal)@\spxentry{fftn()}\spxextra{in module symjax.tensor.signal}}

\begin{fulllineitems}
\phantomsection\label{\detokenize{modules/signal:symjax.tensor.signal.fftn}}\pysiglinewithargsret{\sphinxbfcode{\sphinxupquote{fftn}}}{\emph{\DUrole{n}{a}}, \emph{\DUrole{n}{s}\DUrole{o}{=}\DUrole{default_value}{None}}, \emph{\DUrole{n}{axes}\DUrole{o}{=}\DUrole{default_value}{None}}, \emph{\DUrole{n}{norm}\DUrole{o}{=}\DUrole{default_value}{None}}}{}
Compute the N\sphinxhyphen{}dimensional discrete Fourier Transform.

LAX\sphinxhyphen{}backend implementation of {\hyperref[\detokenize{modules/signal:symjax.tensor.signal.fftn}]{\sphinxcrossref{\sphinxcode{\sphinxupquote{fftn()}}}}}.
ADDITIONOriginal docstring below.

LAX\sphinxhyphen{}backend implementation of {\hyperref[\detokenize{modules/signal:symjax.tensor.signal.fftn}]{\sphinxcrossref{\sphinxcode{\sphinxupquote{fftn()}}}}}.
Original docstring below.

This function computes the \sphinxstyleemphasis{N}\sphinxhyphen{}dimensional discrete Fourier Transform over
any number of axes in an \sphinxstyleemphasis{M}\sphinxhyphen{}dimensional array by means of the Fast Fourier
Transform (FFT).
\begin{quote}\begin{description}
\item[{Returns}] \leavevmode
\sphinxstylestrong{out} \textendash{} The truncated or zero\sphinxhyphen{}padded input, transformed along the axes
indicated by \sphinxtitleref{axes}, or by a combination of \sphinxtitleref{s} and \sphinxtitleref{a},
as explained in the parameters section above.

\item[{Return type}] \leavevmode
complex ndarray

\item[{Raises}] \leavevmode\begin{itemize}
\item {} 
\sphinxstyleliteralstrong{\sphinxupquote{ValueError}} \textendash{} If \sphinxtitleref{s} and \sphinxtitleref{axes} have different length.

\item {} 
\sphinxstyleliteralstrong{\sphinxupquote{IndexError}} \textendash{} If an element of \sphinxtitleref{axes} is larger than than the number of axes of \sphinxtitleref{a}.

\end{itemize}

\end{description}\end{quote}

\sphinxstrong{See also:}

\begin{description}
\item[{\sphinxcode{\sphinxupquote{numpy.fft()}}}] \leavevmode
Overall view of discrete Fourier transforms, with definitions and conventions used.

\item[{{\hyperref[\detokenize{modules/signal:symjax.tensor.signal.ifftn}]{\sphinxcrossref{\sphinxcode{\sphinxupquote{ifftn()}}}}}}] \leavevmode
The inverse of \sphinxtitleref{fftn}, the inverse \sphinxstyleemphasis{n}\sphinxhyphen{}dimensional FFT.

\item[{{\hyperref[\detokenize{modules/signal:symjax.tensor.signal.fft}]{\sphinxcrossref{\sphinxcode{\sphinxupquote{fft()}}}}}}] \leavevmode
The one\sphinxhyphen{}dimensional FFT, with definitions and conventions used.

\item[{{\hyperref[\detokenize{modules/signal:symjax.tensor.signal.rfftn}]{\sphinxcrossref{\sphinxcode{\sphinxupquote{rfftn()}}}}}}] \leavevmode
The \sphinxstyleemphasis{n}\sphinxhyphen{}dimensional FFT of real input.

\item[{{\hyperref[\detokenize{modules/signal:symjax.tensor.signal.fft2}]{\sphinxcrossref{\sphinxcode{\sphinxupquote{fft2()}}}}}}] \leavevmode
The two\sphinxhyphen{}dimensional FFT.

\item[{\sphinxcode{\sphinxupquote{fftshift()}}}] \leavevmode
Shifts zero\sphinxhyphen{}frequency terms to centre of array

\end{description}

\subsubsection*{Notes}

The output, analogously to \sphinxtitleref{fft}, contains the term for zero frequency in
the low\sphinxhyphen{}order corner of all axes, the positive frequency terms in the
first half of all axes, the term for the Nyquist frequency in the middle
of all axes and the negative frequency terms in the second half of all
axes, in order of decreasingly negative frequency.

See \sphinxtitleref{numpy.fft} for details, definitions and conventions used.
\subsubsection*{Examples}

\begin{sphinxVerbatim}[commandchars=\\\{\}]
\PYG{g+gp}{\PYGZgt{}\PYGZgt{}\PYGZgt{} }\PYG{n}{a} \PYG{o}{=} \PYG{n}{np}\PYG{o}{.}\PYG{n}{mgrid}\PYG{p}{[}\PYG{p}{:}\PYG{l+m+mi}{3}\PYG{p}{,} \PYG{p}{:}\PYG{l+m+mi}{3}\PYG{p}{,} \PYG{p}{:}\PYG{l+m+mi}{3}\PYG{p}{]}\PYG{p}{[}\PYG{l+m+mi}{0}\PYG{p}{]}
\PYG{g+gp}{\PYGZgt{}\PYGZgt{}\PYGZgt{} }\PYG{n}{np}\PYG{o}{.}\PYG{n}{fft}\PYG{o}{.}\PYG{n}{fftn}\PYG{p}{(}\PYG{n}{a}\PYG{p}{,} \PYG{n}{axes}\PYG{o}{=}\PYG{p}{(}\PYG{l+m+mi}{1}\PYG{p}{,} \PYG{l+m+mi}{2}\PYG{p}{)}\PYG{p}{)}
\PYG{g+go}{array([[[ 0.+0.j,   0.+0.j,   0.+0.j], \PYGZsh{} may vary}
\PYG{g+go}{        [ 0.+0.j,   0.+0.j,   0.+0.j],}
\PYG{g+go}{        [ 0.+0.j,   0.+0.j,   0.+0.j]],}
\PYG{g+go}{       [[ 9.+0.j,   0.+0.j,   0.+0.j],}
\PYG{g+go}{        [ 0.+0.j,   0.+0.j,   0.+0.j],}
\PYG{g+go}{        [ 0.+0.j,   0.+0.j,   0.+0.j]],}
\PYG{g+go}{       [[18.+0.j,   0.+0.j,   0.+0.j],}
\PYG{g+go}{        [ 0.+0.j,   0.+0.j,   0.+0.j],}
\PYG{g+go}{        [ 0.+0.j,   0.+0.j,   0.+0.j]]])}
\PYG{g+gp}{\PYGZgt{}\PYGZgt{}\PYGZgt{} }\PYG{n}{np}\PYG{o}{.}\PYG{n}{fft}\PYG{o}{.}\PYG{n}{fftn}\PYG{p}{(}\PYG{n}{a}\PYG{p}{,} \PYG{p}{(}\PYG{l+m+mi}{2}\PYG{p}{,} \PYG{l+m+mi}{2}\PYG{p}{)}\PYG{p}{,} \PYG{n}{axes}\PYG{o}{=}\PYG{p}{(}\PYG{l+m+mi}{0}\PYG{p}{,} \PYG{l+m+mi}{1}\PYG{p}{)}\PYG{p}{)}
\PYG{g+go}{array([[[ 2.+0.j,  2.+0.j,  2.+0.j], \PYGZsh{} may vary}
\PYG{g+go}{        [ 0.+0.j,  0.+0.j,  0.+0.j]],}
\PYG{g+go}{       [[\PYGZhy{}2.+0.j, \PYGZhy{}2.+0.j, \PYGZhy{}2.+0.j],}
\PYG{g+go}{        [ 0.+0.j,  0.+0.j,  0.+0.j]]])}
\end{sphinxVerbatim}

\begin{sphinxVerbatim}[commandchars=\\\{\}]
\PYG{g+gp}{\PYGZgt{}\PYGZgt{}\PYGZgt{} }\PYG{k+kn}{import} \PYG{n+nn}{matplotlib}\PYG{n+nn}{.}\PYG{n+nn}{pyplot} \PYG{k}{as} \PYG{n+nn}{plt}
\PYG{g+gp}{\PYGZgt{}\PYGZgt{}\PYGZgt{} }\PYG{p}{[}\PYG{n}{X}\PYG{p}{,} \PYG{n}{Y}\PYG{p}{]} \PYG{o}{=} \PYG{n}{np}\PYG{o}{.}\PYG{n}{meshgrid}\PYG{p}{(}\PYG{l+m+mi}{2} \PYG{o}{*} \PYG{n}{np}\PYG{o}{.}\PYG{n}{pi} \PYG{o}{*} \PYG{n}{np}\PYG{o}{.}\PYG{n}{arange}\PYG{p}{(}\PYG{l+m+mi}{200}\PYG{p}{)} \PYG{o}{/} \PYG{l+m+mi}{12}\PYG{p}{,}
\PYG{g+gp}{... }                     \PYG{l+m+mi}{2} \PYG{o}{*} \PYG{n}{np}\PYG{o}{.}\PYG{n}{pi} \PYG{o}{*} \PYG{n}{np}\PYG{o}{.}\PYG{n}{arange}\PYG{p}{(}\PYG{l+m+mi}{200}\PYG{p}{)} \PYG{o}{/} \PYG{l+m+mi}{34}\PYG{p}{)}
\PYG{g+gp}{\PYGZgt{}\PYGZgt{}\PYGZgt{} }\PYG{n}{S} \PYG{o}{=} \PYG{n}{np}\PYG{o}{.}\PYG{n}{sin}\PYG{p}{(}\PYG{n}{X}\PYG{p}{)} \PYG{o}{+} \PYG{n}{np}\PYG{o}{.}\PYG{n}{cos}\PYG{p}{(}\PYG{n}{Y}\PYG{p}{)} \PYG{o}{+} \PYG{n}{np}\PYG{o}{.}\PYG{n}{random}\PYG{o}{.}\PYG{n}{uniform}\PYG{p}{(}\PYG{l+m+mi}{0}\PYG{p}{,} \PYG{l+m+mi}{1}\PYG{p}{,} \PYG{n}{X}\PYG{o}{.}\PYG{n}{shape}\PYG{p}{)}
\PYG{g+gp}{\PYGZgt{}\PYGZgt{}\PYGZgt{} }\PYG{n}{FS} \PYG{o}{=} \PYG{n}{np}\PYG{o}{.}\PYG{n}{fft}\PYG{o}{.}\PYG{n}{fftn}\PYG{p}{(}\PYG{n}{S}\PYG{p}{)}
\PYG{g+gp}{\PYGZgt{}\PYGZgt{}\PYGZgt{} }\PYG{n}{plt}\PYG{o}{.}\PYG{n}{imshow}\PYG{p}{(}\PYG{n}{np}\PYG{o}{.}\PYG{n}{log}\PYG{p}{(}\PYG{n}{np}\PYG{o}{.}\PYG{n}{abs}\PYG{p}{(}\PYG{n}{np}\PYG{o}{.}\PYG{n}{fft}\PYG{o}{.}\PYG{n}{fftshift}\PYG{p}{(}\PYG{n}{FS}\PYG{p}{)}\PYG{p}{)}\PYG{o}{*}\PYG{o}{*}\PYG{l+m+mi}{2}\PYG{p}{)}\PYG{p}{)}
\PYG{g+go}{\PYGZlt{}matplotlib.image.AxesImage object at 0x...\PYGZgt{}}
\PYG{g+gp}{\PYGZgt{}\PYGZgt{}\PYGZgt{} }\PYG{n}{plt}\PYG{o}{.}\PYG{n}{show}\PYG{p}{(}\PYG{p}{)}
\end{sphinxVerbatim}

\end{fulllineitems}

\index{ifftn() (in module symjax.tensor.signal)@\spxentry{ifftn()}\spxextra{in module symjax.tensor.signal}}

\begin{fulllineitems}
\phantomsection\label{\detokenize{modules/signal:symjax.tensor.signal.ifftn}}\pysiglinewithargsret{\sphinxbfcode{\sphinxupquote{ifftn}}}{\emph{\DUrole{n}{a}}, \emph{\DUrole{n}{s}\DUrole{o}{=}\DUrole{default_value}{None}}, \emph{\DUrole{n}{axes}\DUrole{o}{=}\DUrole{default_value}{None}}, \emph{\DUrole{n}{norm}\DUrole{o}{=}\DUrole{default_value}{None}}}{}
Compute the N\sphinxhyphen{}dimensional inverse discrete Fourier Transform.

LAX\sphinxhyphen{}backend implementation of {\hyperref[\detokenize{modules/signal:symjax.tensor.signal.ifftn}]{\sphinxcrossref{\sphinxcode{\sphinxupquote{ifftn()}}}}}.
ADDITIONOriginal docstring below.

LAX\sphinxhyphen{}backend implementation of {\hyperref[\detokenize{modules/signal:symjax.tensor.signal.ifftn}]{\sphinxcrossref{\sphinxcode{\sphinxupquote{ifftn()}}}}}.
Original docstring below.

This function computes the inverse of the N\sphinxhyphen{}dimensional discrete
Fourier Transform over any number of axes in an M\sphinxhyphen{}dimensional array by
means of the Fast Fourier Transform (FFT).  In other words,
\sphinxcode{\sphinxupquote{ifftn(fftn(a)) == a}} to within numerical accuracy.
For a description of the definitions and conventions used, see \sphinxtitleref{numpy.fft}.

The input, analogously to \sphinxtitleref{ifft}, should be ordered in the same way as is
returned by \sphinxtitleref{fftn}, i.e. it should have the term for zero frequency
in all axes in the low\sphinxhyphen{}order corner, the positive frequency terms in the
first half of all axes, the term for the Nyquist frequency in the middle
of all axes and the negative frequency terms in the second half of all
axes, in order of decreasingly negative frequency.
\begin{quote}\begin{description}
\item[{Returns}] \leavevmode
\sphinxstylestrong{out} \textendash{} The truncated or zero\sphinxhyphen{}padded input, transformed along the axes
indicated by \sphinxtitleref{axes}, or by a combination of \sphinxtitleref{s} or \sphinxtitleref{a},
as explained in the parameters section above.

\item[{Return type}] \leavevmode
complex ndarray

\item[{Raises}] \leavevmode\begin{itemize}
\item {} 
\sphinxstyleliteralstrong{\sphinxupquote{ValueError}} \textendash{} If \sphinxtitleref{s} and \sphinxtitleref{axes} have different length.

\item {} 
\sphinxstyleliteralstrong{\sphinxupquote{IndexError}} \textendash{} If an element of \sphinxtitleref{axes} is larger than than the number of axes of \sphinxtitleref{a}.

\end{itemize}

\end{description}\end{quote}

\sphinxstrong{See also:}

\begin{description}
\item[{\sphinxcode{\sphinxupquote{numpy.fft()}}}] \leavevmode
Overall view of discrete Fourier transforms, with definitions and conventions used.

\item[{{\hyperref[\detokenize{modules/signal:symjax.tensor.signal.fftn}]{\sphinxcrossref{\sphinxcode{\sphinxupquote{fftn()}}}}}}] \leavevmode
The forward \sphinxstyleemphasis{n}\sphinxhyphen{}dimensional FFT, of which \sphinxtitleref{ifftn} is the inverse.

\item[{{\hyperref[\detokenize{modules/signal:symjax.tensor.signal.ifft}]{\sphinxcrossref{\sphinxcode{\sphinxupquote{ifft()}}}}}}] \leavevmode
The one\sphinxhyphen{}dimensional inverse FFT.

\item[{{\hyperref[\detokenize{modules/signal:symjax.tensor.signal.ifft2}]{\sphinxcrossref{\sphinxcode{\sphinxupquote{ifft2()}}}}}}] \leavevmode
The two\sphinxhyphen{}dimensional inverse FFT.

\item[{\sphinxcode{\sphinxupquote{ifftshift()}}}] \leavevmode
Undoes \sphinxtitleref{fftshift}, shifts zero\sphinxhyphen{}frequency terms to beginning of array.

\end{description}

\subsubsection*{Notes}

See \sphinxtitleref{numpy.fft} for definitions and conventions used.

Zero\sphinxhyphen{}padding, analogously with \sphinxtitleref{ifft}, is performed by appending zeros to
the input along the specified dimension.  Although this is the common
approach, it might lead to surprising results.  If another form of zero
padding is desired, it must be performed before \sphinxtitleref{ifftn} is called.
\subsubsection*{Examples}

\begin{sphinxVerbatim}[commandchars=\\\{\}]
\PYG{g+gp}{\PYGZgt{}\PYGZgt{}\PYGZgt{} }\PYG{n}{a} \PYG{o}{=} \PYG{n}{np}\PYG{o}{.}\PYG{n}{eye}\PYG{p}{(}\PYG{l+m+mi}{4}\PYG{p}{)}
\PYG{g+gp}{\PYGZgt{}\PYGZgt{}\PYGZgt{} }\PYG{n}{np}\PYG{o}{.}\PYG{n}{fft}\PYG{o}{.}\PYG{n}{ifftn}\PYG{p}{(}\PYG{n}{np}\PYG{o}{.}\PYG{n}{fft}\PYG{o}{.}\PYG{n}{fftn}\PYG{p}{(}\PYG{n}{a}\PYG{p}{,} \PYG{n}{axes}\PYG{o}{=}\PYG{p}{(}\PYG{l+m+mi}{0}\PYG{p}{,}\PYG{p}{)}\PYG{p}{)}\PYG{p}{,} \PYG{n}{axes}\PYG{o}{=}\PYG{p}{(}\PYG{l+m+mi}{1}\PYG{p}{,}\PYG{p}{)}\PYG{p}{)}
\PYG{g+go}{array([[1.+0.j,  0.+0.j,  0.+0.j,  0.+0.j], \PYGZsh{} may vary}
\PYG{g+go}{       [0.+0.j,  1.+0.j,  0.+0.j,  0.+0.j],}
\PYG{g+go}{       [0.+0.j,  0.+0.j,  1.+0.j,  0.+0.j],}
\PYG{g+go}{       [0.+0.j,  0.+0.j,  0.+0.j,  1.+0.j]])}
\end{sphinxVerbatim}

Create and plot an image with band\sphinxhyphen{}limited frequency content:

\begin{sphinxVerbatim}[commandchars=\\\{\}]
\PYG{g+gp}{\PYGZgt{}\PYGZgt{}\PYGZgt{} }\PYG{k+kn}{import} \PYG{n+nn}{matplotlib}\PYG{n+nn}{.}\PYG{n+nn}{pyplot} \PYG{k}{as} \PYG{n+nn}{plt}
\PYG{g+gp}{\PYGZgt{}\PYGZgt{}\PYGZgt{} }\PYG{n}{n} \PYG{o}{=} \PYG{n}{np}\PYG{o}{.}\PYG{n}{zeros}\PYG{p}{(}\PYG{p}{(}\PYG{l+m+mi}{200}\PYG{p}{,}\PYG{l+m+mi}{200}\PYG{p}{)}\PYG{p}{,} \PYG{n}{dtype}\PYG{o}{=}\PYG{n+nb}{complex}\PYG{p}{)}
\PYG{g+gp}{\PYGZgt{}\PYGZgt{}\PYGZgt{} }\PYG{n}{n}\PYG{p}{[}\PYG{l+m+mi}{60}\PYG{p}{:}\PYG{l+m+mi}{80}\PYG{p}{,} \PYG{l+m+mi}{20}\PYG{p}{:}\PYG{l+m+mi}{40}\PYG{p}{]} \PYG{o}{=} \PYG{n}{np}\PYG{o}{.}\PYG{n}{exp}\PYG{p}{(}\PYG{l+m+mi}{1}\PYG{n}{j}\PYG{o}{*}\PYG{n}{np}\PYG{o}{.}\PYG{n}{random}\PYG{o}{.}\PYG{n}{uniform}\PYG{p}{(}\PYG{l+m+mi}{0}\PYG{p}{,} \PYG{l+m+mi}{2}\PYG{o}{*}\PYG{n}{np}\PYG{o}{.}\PYG{n}{pi}\PYG{p}{,} \PYG{p}{(}\PYG{l+m+mi}{20}\PYG{p}{,} \PYG{l+m+mi}{20}\PYG{p}{)}\PYG{p}{)}\PYG{p}{)}
\PYG{g+gp}{\PYGZgt{}\PYGZgt{}\PYGZgt{} }\PYG{n}{im} \PYG{o}{=} \PYG{n}{np}\PYG{o}{.}\PYG{n}{fft}\PYG{o}{.}\PYG{n}{ifftn}\PYG{p}{(}\PYG{n}{n}\PYG{p}{)}\PYG{o}{.}\PYG{n}{real}
\PYG{g+gp}{\PYGZgt{}\PYGZgt{}\PYGZgt{} }\PYG{n}{plt}\PYG{o}{.}\PYG{n}{imshow}\PYG{p}{(}\PYG{n}{im}\PYG{p}{)}
\PYG{g+go}{\PYGZlt{}matplotlib.image.AxesImage object at 0x...\PYGZgt{}}
\PYG{g+gp}{\PYGZgt{}\PYGZgt{}\PYGZgt{} }\PYG{n}{plt}\PYG{o}{.}\PYG{n}{show}\PYG{p}{(}\PYG{p}{)}
\end{sphinxVerbatim}

\end{fulllineitems}

\index{rfft() (in module symjax.tensor.signal)@\spxentry{rfft()}\spxextra{in module symjax.tensor.signal}}

\begin{fulllineitems}
\phantomsection\label{\detokenize{modules/signal:symjax.tensor.signal.rfft}}\pysiglinewithargsret{\sphinxbfcode{\sphinxupquote{rfft}}}{\emph{\DUrole{n}{a}}, \emph{\DUrole{n}{n}\DUrole{o}{=}\DUrole{default_value}{None}}, \emph{\DUrole{n}{axis}\DUrole{o}{=}\DUrole{default_value}{\sphinxhyphen{} 1}}, \emph{\DUrole{n}{norm}\DUrole{o}{=}\DUrole{default_value}{None}}}{}
Compute the one\sphinxhyphen{}dimensional discrete Fourier Transform for real input.

LAX\sphinxhyphen{}backend implementation of {\hyperref[\detokenize{modules/signal:symjax.tensor.signal.rfft}]{\sphinxcrossref{\sphinxcode{\sphinxupquote{rfft()}}}}}.
ADDITIONOriginal docstring below.

LAX\sphinxhyphen{}backend implementation of {\hyperref[\detokenize{modules/signal:symjax.tensor.signal.rfft}]{\sphinxcrossref{\sphinxcode{\sphinxupquote{rfft()}}}}}.
Original docstring below.

This function computes the one\sphinxhyphen{}dimensional \sphinxstyleemphasis{n}\sphinxhyphen{}point discrete Fourier
Transform (DFT) of a real\sphinxhyphen{}valued array by means of an efficient algorithm
called the Fast Fourier Transform (FFT).
\begin{quote}\begin{description}
\item[{Returns}] \leavevmode
\sphinxstylestrong{out} \textendash{} The truncated or zero\sphinxhyphen{}padded input, transformed along the axis
indicated by \sphinxtitleref{axis}, or the last one if \sphinxtitleref{axis} is not specified.
If \sphinxtitleref{n} is even, the length of the transformed axis is \sphinxcode{\sphinxupquote{(n/2)+1}}.
If \sphinxtitleref{n} is odd, the length is \sphinxcode{\sphinxupquote{(n+1)/2}}.

\item[{Return type}] \leavevmode
complex ndarray

\item[{Raises}] \leavevmode
\sphinxstyleliteralstrong{\sphinxupquote{IndexError}} \textendash{} If \sphinxtitleref{axis} is larger than the last axis of \sphinxtitleref{a}.

\end{description}\end{quote}

\sphinxstrong{See also:}

\begin{description}
\item[{\sphinxcode{\sphinxupquote{numpy.fft()}}}] \leavevmode
For definition of the DFT and conventions used.

\item[{{\hyperref[\detokenize{modules/signal:symjax.tensor.signal.irfft}]{\sphinxcrossref{\sphinxcode{\sphinxupquote{irfft()}}}}}}] \leavevmode
The inverse of \sphinxtitleref{rfft}.

\item[{{\hyperref[\detokenize{modules/signal:symjax.tensor.signal.fft}]{\sphinxcrossref{\sphinxcode{\sphinxupquote{fft()}}}}}}] \leavevmode
The one\sphinxhyphen{}dimensional FFT of general (complex) input.

\item[{{\hyperref[\detokenize{modules/signal:symjax.tensor.signal.fftn}]{\sphinxcrossref{\sphinxcode{\sphinxupquote{fftn()}}}}}}] \leavevmode
The \sphinxstyleemphasis{n}\sphinxhyphen{}dimensional FFT.

\item[{{\hyperref[\detokenize{modules/signal:symjax.tensor.signal.rfftn}]{\sphinxcrossref{\sphinxcode{\sphinxupquote{rfftn()}}}}}}] \leavevmode
The \sphinxstyleemphasis{n}\sphinxhyphen{}dimensional FFT of real input.

\end{description}

\subsubsection*{Notes}

When the DFT is computed for purely real input, the output is
Hermitian\sphinxhyphen{}symmetric, i.e. the negative frequency terms are just the complex
conjugates of the corresponding positive\sphinxhyphen{}frequency terms, and the
negative\sphinxhyphen{}frequency terms are therefore redundant.  This function does not
compute the negative frequency terms, and the length of the transformed
axis of the output is therefore \sphinxcode{\sphinxupquote{n//2 + 1}}.

When \sphinxcode{\sphinxupquote{A = rfft(a)}} and fs is the sampling frequency, \sphinxcode{\sphinxupquote{A{[}0{]}}} contains
the zero\sphinxhyphen{}frequency term 0*fs, which is real due to Hermitian symmetry.

If \sphinxtitleref{n} is even, \sphinxcode{\sphinxupquote{A{[}\sphinxhyphen{}1{]}}} contains the term representing both positive
and negative Nyquist frequency (+fs/2 and \sphinxhyphen{}fs/2), and must also be purely
real. If \sphinxtitleref{n} is odd, there is no term at fs/2; \sphinxcode{\sphinxupquote{A{[}\sphinxhyphen{}1{]}}} contains
the largest positive frequency (fs/2*(n\sphinxhyphen{}1)/n), and is complex in the
general case.

If the input \sphinxtitleref{a} contains an imaginary part, it is silently discarded.
\subsubsection*{Examples}

\begin{sphinxVerbatim}[commandchars=\\\{\}]
\PYG{g+gp}{\PYGZgt{}\PYGZgt{}\PYGZgt{} }\PYG{n}{np}\PYG{o}{.}\PYG{n}{fft}\PYG{o}{.}\PYG{n}{fft}\PYG{p}{(}\PYG{p}{[}\PYG{l+m+mi}{0}\PYG{p}{,} \PYG{l+m+mi}{1}\PYG{p}{,} \PYG{l+m+mi}{0}\PYG{p}{,} \PYG{l+m+mi}{0}\PYG{p}{]}\PYG{p}{)}
\PYG{g+go}{array([ 1.+0.j,  0.\PYGZhy{}1.j, \PYGZhy{}1.+0.j,  0.+1.j]) \PYGZsh{} may vary}
\PYG{g+gp}{\PYGZgt{}\PYGZgt{}\PYGZgt{} }\PYG{n}{np}\PYG{o}{.}\PYG{n}{fft}\PYG{o}{.}\PYG{n}{rfft}\PYG{p}{(}\PYG{p}{[}\PYG{l+m+mi}{0}\PYG{p}{,} \PYG{l+m+mi}{1}\PYG{p}{,} \PYG{l+m+mi}{0}\PYG{p}{,} \PYG{l+m+mi}{0}\PYG{p}{]}\PYG{p}{)}
\PYG{g+go}{array([ 1.+0.j,  0.\PYGZhy{}1.j, \PYGZhy{}1.+0.j]) \PYGZsh{} may vary}
\end{sphinxVerbatim}

Notice how the final element of the \sphinxtitleref{fft} output is the complex conjugate
of the second element, for real input. For \sphinxtitleref{rfft}, this symmetry is
exploited to compute only the non\sphinxhyphen{}negative frequency terms.

\end{fulllineitems}

\index{irfft() (in module symjax.tensor.signal)@\spxentry{irfft()}\spxextra{in module symjax.tensor.signal}}

\begin{fulllineitems}
\phantomsection\label{\detokenize{modules/signal:symjax.tensor.signal.irfft}}\pysiglinewithargsret{\sphinxbfcode{\sphinxupquote{irfft}}}{\emph{\DUrole{n}{a}}, \emph{\DUrole{n}{n}\DUrole{o}{=}\DUrole{default_value}{None}}, \emph{\DUrole{n}{axis}\DUrole{o}{=}\DUrole{default_value}{\sphinxhyphen{} 1}}, \emph{\DUrole{n}{norm}\DUrole{o}{=}\DUrole{default_value}{None}}}{}
Compute the inverse of the n\sphinxhyphen{}point DFT for real input.

LAX\sphinxhyphen{}backend implementation of {\hyperref[\detokenize{modules/signal:symjax.tensor.signal.irfft}]{\sphinxcrossref{\sphinxcode{\sphinxupquote{irfft()}}}}}.
ADDITIONOriginal docstring below.

LAX\sphinxhyphen{}backend implementation of {\hyperref[\detokenize{modules/signal:symjax.tensor.signal.irfft}]{\sphinxcrossref{\sphinxcode{\sphinxupquote{irfft()}}}}}.
Original docstring below.

This function computes the inverse of the one\sphinxhyphen{}dimensional \sphinxstyleemphasis{n}\sphinxhyphen{}point
discrete Fourier Transform of real input computed by \sphinxtitleref{rfft}.
In other words, \sphinxcode{\sphinxupquote{irfft(rfft(a), len(a)) == a}} to within numerical
accuracy. (See Notes below for why \sphinxcode{\sphinxupquote{len(a)}} is necessary here.)

The input is expected to be in the form returned by \sphinxtitleref{rfft}, i.e. the
real zero\sphinxhyphen{}frequency term followed by the complex positive frequency terms
in order of increasing frequency.  Since the discrete Fourier Transform of
real input is Hermitian\sphinxhyphen{}symmetric, the negative frequency terms are taken
to be the complex conjugates of the corresponding positive frequency terms.
\begin{quote}\begin{description}
\item[{Returns}] \leavevmode
\sphinxstylestrong{out} \textendash{} The truncated or zero\sphinxhyphen{}padded input, transformed along the axis
indicated by \sphinxtitleref{axis}, or the last one if \sphinxtitleref{axis} is not specified.
The length of the transformed axis is \sphinxtitleref{n}, or, if \sphinxtitleref{n} is not given,
\sphinxcode{\sphinxupquote{2*(m\sphinxhyphen{}1)}} where \sphinxcode{\sphinxupquote{m}} is the length of the transformed axis of the
input. To get an odd number of output points, \sphinxtitleref{n} must be specified.

\item[{Return type}] \leavevmode
ndarray

\item[{Raises}] \leavevmode
\sphinxstyleliteralstrong{\sphinxupquote{IndexError}} \textendash{} If \sphinxtitleref{axis} is larger than the last axis of \sphinxtitleref{a}.

\end{description}\end{quote}

\sphinxstrong{See also:}

\begin{description}
\item[{\sphinxcode{\sphinxupquote{numpy.fft()}}}] \leavevmode
For definition of the DFT and conventions used.

\item[{{\hyperref[\detokenize{modules/signal:symjax.tensor.signal.rfft}]{\sphinxcrossref{\sphinxcode{\sphinxupquote{rfft()}}}}}}] \leavevmode
The one\sphinxhyphen{}dimensional FFT of real input, of which \sphinxtitleref{irfft} is inverse.

\item[{{\hyperref[\detokenize{modules/signal:symjax.tensor.signal.fft}]{\sphinxcrossref{\sphinxcode{\sphinxupquote{fft()}}}}}}] \leavevmode
The one\sphinxhyphen{}dimensional FFT.

\item[{{\hyperref[\detokenize{modules/signal:symjax.tensor.signal.irfft2}]{\sphinxcrossref{\sphinxcode{\sphinxupquote{irfft2()}}}}}}] \leavevmode
The inverse of the two\sphinxhyphen{}dimensional FFT of real input.

\item[{{\hyperref[\detokenize{modules/signal:symjax.tensor.signal.irfftn}]{\sphinxcrossref{\sphinxcode{\sphinxupquote{irfftn()}}}}}}] \leavevmode
The inverse of the \sphinxstyleemphasis{n}\sphinxhyphen{}dimensional FFT of real input.

\end{description}

\subsubsection*{Notes}

Returns the real valued \sphinxtitleref{n}\sphinxhyphen{}point inverse discrete Fourier transform
of \sphinxtitleref{a}, where \sphinxtitleref{a} contains the non\sphinxhyphen{}negative frequency terms of a
Hermitian\sphinxhyphen{}symmetric sequence. \sphinxtitleref{n} is the length of the result, not the
input.

If you specify an \sphinxtitleref{n} such that \sphinxtitleref{a} must be zero\sphinxhyphen{}padded or truncated, the
extra/removed values will be added/removed at high frequencies. One can
thus resample a series to \sphinxtitleref{m} points via Fourier interpolation by:
\sphinxcode{\sphinxupquote{a\_resamp = irfft(rfft(a), m)}}.

The correct interpretation of the hermitian input depends on the length of
the original data, as given by \sphinxtitleref{n}. This is because each input shape could
correspond to either an odd or even length signal. By default, \sphinxtitleref{irfft}
assumes an even output length which puts the last entry at the Nyquist
frequency; aliasing with its symmetric counterpart. By Hermitian symmetry,
the value is thus treated as purely real. To avoid losing information, the
correct length of the real input \sphinxstylestrong{must} be given.
\subsubsection*{Examples}

\begin{sphinxVerbatim}[commandchars=\\\{\}]
\PYG{g+gp}{\PYGZgt{}\PYGZgt{}\PYGZgt{} }\PYG{n}{np}\PYG{o}{.}\PYG{n}{fft}\PYG{o}{.}\PYG{n}{ifft}\PYG{p}{(}\PYG{p}{[}\PYG{l+m+mi}{1}\PYG{p}{,} \PYG{o}{\PYGZhy{}}\PYG{l+m+mi}{1}\PYG{n}{j}\PYG{p}{,} \PYG{o}{\PYGZhy{}}\PYG{l+m+mi}{1}\PYG{p}{,} \PYG{l+m+mi}{1}\PYG{n}{j}\PYG{p}{]}\PYG{p}{)}
\PYG{g+go}{array([0.+0.j,  1.+0.j,  0.+0.j,  0.+0.j]) \PYGZsh{} may vary}
\PYG{g+gp}{\PYGZgt{}\PYGZgt{}\PYGZgt{} }\PYG{n}{np}\PYG{o}{.}\PYG{n}{fft}\PYG{o}{.}\PYG{n}{irfft}\PYG{p}{(}\PYG{p}{[}\PYG{l+m+mi}{1}\PYG{p}{,} \PYG{o}{\PYGZhy{}}\PYG{l+m+mi}{1}\PYG{n}{j}\PYG{p}{,} \PYG{o}{\PYGZhy{}}\PYG{l+m+mi}{1}\PYG{p}{]}\PYG{p}{)}
\PYG{g+go}{array([0.,  1.,  0.,  0.])}
\end{sphinxVerbatim}

Notice how the last term in the input to the ordinary \sphinxtitleref{ifft} is the
complex conjugate of the second term, and the output has zero imaginary
part everywhere.  When calling \sphinxtitleref{irfft}, the negative frequencies are not
specified, and the output array is purely real.

\end{fulllineitems}

\index{rfft2() (in module symjax.tensor.signal)@\spxentry{rfft2()}\spxextra{in module symjax.tensor.signal}}

\begin{fulllineitems}
\phantomsection\label{\detokenize{modules/signal:symjax.tensor.signal.rfft2}}\pysiglinewithargsret{\sphinxbfcode{\sphinxupquote{rfft2}}}{\emph{\DUrole{n}{a}}, \emph{\DUrole{n}{s}\DUrole{o}{=}\DUrole{default_value}{None}}, \emph{\DUrole{n}{axes}\DUrole{o}{=}\DUrole{default_value}{\sphinxhyphen{} 2, \sphinxhyphen{} 1}}, \emph{\DUrole{n}{norm}\DUrole{o}{=}\DUrole{default_value}{None}}}{}
Compute the 2\sphinxhyphen{}dimensional FFT of a real array.

LAX\sphinxhyphen{}backend implementation of {\hyperref[\detokenize{modules/signal:symjax.tensor.signal.rfft2}]{\sphinxcrossref{\sphinxcode{\sphinxupquote{rfft2()}}}}}.
ADDITIONOriginal docstring below.

LAX\sphinxhyphen{}backend implementation of {\hyperref[\detokenize{modules/signal:symjax.tensor.signal.rfft2}]{\sphinxcrossref{\sphinxcode{\sphinxupquote{rfft2()}}}}}.
Original docstring below.
\begin{quote}\begin{description}
\item[{Returns}] \leavevmode
\sphinxstylestrong{out} \textendash{} The result of the real 2\sphinxhyphen{}D FFT.

\item[{Return type}] \leavevmode
ndarray

\end{description}\end{quote}

\sphinxstrong{See also:}

\begin{description}
\item[{{\hyperref[\detokenize{modules/signal:symjax.tensor.signal.rfftn}]{\sphinxcrossref{\sphinxcode{\sphinxupquote{rfftn()}}}}}}] \leavevmode
Compute the N\sphinxhyphen{}dimensional discrete Fourier Transform for real input.

\end{description}

\subsubsection*{Notes}

This is really just \sphinxtitleref{rfftn} with different default behavior.
For more details see \sphinxtitleref{rfftn}.

\end{fulllineitems}

\index{irfft2() (in module symjax.tensor.signal)@\spxentry{irfft2()}\spxextra{in module symjax.tensor.signal}}

\begin{fulllineitems}
\phantomsection\label{\detokenize{modules/signal:symjax.tensor.signal.irfft2}}\pysiglinewithargsret{\sphinxbfcode{\sphinxupquote{irfft2}}}{\emph{\DUrole{n}{a}}, \emph{\DUrole{n}{s}\DUrole{o}{=}\DUrole{default_value}{None}}, \emph{\DUrole{n}{axes}\DUrole{o}{=}\DUrole{default_value}{\sphinxhyphen{} 2, \sphinxhyphen{} 1}}, \emph{\DUrole{n}{norm}\DUrole{o}{=}\DUrole{default_value}{None}}}{}
Compute the 2\sphinxhyphen{}dimensional inverse FFT of a real array.

LAX\sphinxhyphen{}backend implementation of {\hyperref[\detokenize{modules/signal:symjax.tensor.signal.irfft2}]{\sphinxcrossref{\sphinxcode{\sphinxupquote{irfft2()}}}}}.
ADDITIONOriginal docstring below.

LAX\sphinxhyphen{}backend implementation of {\hyperref[\detokenize{modules/signal:symjax.tensor.signal.irfft2}]{\sphinxcrossref{\sphinxcode{\sphinxupquote{irfft2()}}}}}.
Original docstring below.
\begin{quote}\begin{description}
\item[{Returns}] \leavevmode
\sphinxstylestrong{out} \textendash{} The result of the inverse real 2\sphinxhyphen{}D FFT.

\item[{Return type}] \leavevmode
ndarray

\end{description}\end{quote}

\sphinxstrong{See also:}

\begin{description}
\item[{{\hyperref[\detokenize{modules/signal:symjax.tensor.signal.irfftn}]{\sphinxcrossref{\sphinxcode{\sphinxupquote{irfftn()}}}}}}] \leavevmode
Compute the inverse of the N\sphinxhyphen{}dimensional FFT of real input.

\end{description}

\subsubsection*{Notes}

This is really \sphinxtitleref{irfftn} with different defaults.
For more details see \sphinxtitleref{irfftn}.

\end{fulllineitems}

\index{rfftn() (in module symjax.tensor.signal)@\spxentry{rfftn()}\spxextra{in module symjax.tensor.signal}}

\begin{fulllineitems}
\phantomsection\label{\detokenize{modules/signal:symjax.tensor.signal.rfftn}}\pysiglinewithargsret{\sphinxbfcode{\sphinxupquote{rfftn}}}{\emph{\DUrole{n}{a}}, \emph{\DUrole{n}{s}\DUrole{o}{=}\DUrole{default_value}{None}}, \emph{\DUrole{n}{axes}\DUrole{o}{=}\DUrole{default_value}{None}}, \emph{\DUrole{n}{norm}\DUrole{o}{=}\DUrole{default_value}{None}}}{}
Compute the N\sphinxhyphen{}dimensional discrete Fourier Transform for real input.

LAX\sphinxhyphen{}backend implementation of {\hyperref[\detokenize{modules/signal:symjax.tensor.signal.rfftn}]{\sphinxcrossref{\sphinxcode{\sphinxupquote{rfftn()}}}}}.
ADDITIONOriginal docstring below.

LAX\sphinxhyphen{}backend implementation of {\hyperref[\detokenize{modules/signal:symjax.tensor.signal.rfftn}]{\sphinxcrossref{\sphinxcode{\sphinxupquote{rfftn()}}}}}.
Original docstring below.

This function computes the N\sphinxhyphen{}dimensional discrete Fourier Transform over
any number of axes in an M\sphinxhyphen{}dimensional real array by means of the Fast
Fourier Transform (FFT).  By default, all axes are transformed, with the
real transform performed over the last axis, while the remaining
transforms are complex.
\begin{quote}\begin{description}
\item[{Returns}] \leavevmode
\sphinxstylestrong{out} \textendash{} The truncated or zero\sphinxhyphen{}padded input, transformed along the axes
indicated by \sphinxtitleref{axes}, or by a combination of \sphinxtitleref{s} and \sphinxtitleref{a},
as explained in the parameters section above.
The length of the last axis transformed will be \sphinxcode{\sphinxupquote{s{[}\sphinxhyphen{}1{]}//2+1}},
while the remaining transformed axes will have lengths according to
\sphinxtitleref{s}, or unchanged from the input.

\item[{Return type}] \leavevmode
complex ndarray

\item[{Raises}] \leavevmode\begin{itemize}
\item {} 
\sphinxstyleliteralstrong{\sphinxupquote{ValueError}} \textendash{} If \sphinxtitleref{s} and \sphinxtitleref{axes} have different length.

\item {} 
\sphinxstyleliteralstrong{\sphinxupquote{IndexError}} \textendash{} If an element of \sphinxtitleref{axes} is larger than than the number of axes of \sphinxtitleref{a}.

\end{itemize}

\end{description}\end{quote}

\sphinxstrong{See also:}

\begin{description}
\item[{{\hyperref[\detokenize{modules/signal:symjax.tensor.signal.irfftn}]{\sphinxcrossref{\sphinxcode{\sphinxupquote{irfftn()}}}}}}] \leavevmode
The inverse of \sphinxtitleref{rfftn}, i.e. the inverse of the n\sphinxhyphen{}dimensional FFT of real input.

\item[{{\hyperref[\detokenize{modules/signal:symjax.tensor.signal.fft}]{\sphinxcrossref{\sphinxcode{\sphinxupquote{fft()}}}}}}] \leavevmode
The one\sphinxhyphen{}dimensional FFT, with definitions and conventions used.

\item[{{\hyperref[\detokenize{modules/signal:symjax.tensor.signal.rfft}]{\sphinxcrossref{\sphinxcode{\sphinxupquote{rfft()}}}}}}] \leavevmode
The one\sphinxhyphen{}dimensional FFT of real input.

\item[{{\hyperref[\detokenize{modules/signal:symjax.tensor.signal.fftn}]{\sphinxcrossref{\sphinxcode{\sphinxupquote{fftn()}}}}}}] \leavevmode
The n\sphinxhyphen{}dimensional FFT.

\item[{{\hyperref[\detokenize{modules/signal:symjax.tensor.signal.rfft2}]{\sphinxcrossref{\sphinxcode{\sphinxupquote{rfft2()}}}}}}] \leavevmode
The two\sphinxhyphen{}dimensional FFT of real input.

\end{description}

\subsubsection*{Notes}

The transform for real input is performed over the last transformation
axis, as by \sphinxtitleref{rfft}, then the transform over the remaining axes is
performed as by \sphinxtitleref{fftn}.  The order of the output is as for \sphinxtitleref{rfft} for the
final transformation axis, and as for \sphinxtitleref{fftn} for the remaining
transformation axes.

See \sphinxtitleref{fft} for details, definitions and conventions used.
\subsubsection*{Examples}

\begin{sphinxVerbatim}[commandchars=\\\{\}]
\PYG{g+gp}{\PYGZgt{}\PYGZgt{}\PYGZgt{} }\PYG{n}{a} \PYG{o}{=} \PYG{n}{np}\PYG{o}{.}\PYG{n}{ones}\PYG{p}{(}\PYG{p}{(}\PYG{l+m+mi}{2}\PYG{p}{,} \PYG{l+m+mi}{2}\PYG{p}{,} \PYG{l+m+mi}{2}\PYG{p}{)}\PYG{p}{)}
\PYG{g+gp}{\PYGZgt{}\PYGZgt{}\PYGZgt{} }\PYG{n}{np}\PYG{o}{.}\PYG{n}{fft}\PYG{o}{.}\PYG{n}{rfftn}\PYG{p}{(}\PYG{n}{a}\PYG{p}{)}
\PYG{g+go}{array([[[8.+0.j,  0.+0.j], \PYGZsh{} may vary}
\PYG{g+go}{        [0.+0.j,  0.+0.j]],}
\PYG{g+go}{       [[0.+0.j,  0.+0.j],}
\PYG{g+go}{        [0.+0.j,  0.+0.j]]])}
\end{sphinxVerbatim}

\begin{sphinxVerbatim}[commandchars=\\\{\}]
\PYG{g+gp}{\PYGZgt{}\PYGZgt{}\PYGZgt{} }\PYG{n}{np}\PYG{o}{.}\PYG{n}{fft}\PYG{o}{.}\PYG{n}{rfftn}\PYG{p}{(}\PYG{n}{a}\PYG{p}{,} \PYG{n}{axes}\PYG{o}{=}\PYG{p}{(}\PYG{l+m+mi}{2}\PYG{p}{,} \PYG{l+m+mi}{0}\PYG{p}{)}\PYG{p}{)}
\PYG{g+go}{array([[[4.+0.j,  0.+0.j], \PYGZsh{} may vary}
\PYG{g+go}{        [4.+0.j,  0.+0.j]],}
\PYG{g+go}{       [[0.+0.j,  0.+0.j],}
\PYG{g+go}{        [0.+0.j,  0.+0.j]]])}
\end{sphinxVerbatim}

\end{fulllineitems}

\index{irfftn() (in module symjax.tensor.signal)@\spxentry{irfftn()}\spxextra{in module symjax.tensor.signal}}

\begin{fulllineitems}
\phantomsection\label{\detokenize{modules/signal:symjax.tensor.signal.irfftn}}\pysiglinewithargsret{\sphinxbfcode{\sphinxupquote{irfftn}}}{\emph{\DUrole{n}{a}}, \emph{\DUrole{n}{s}\DUrole{o}{=}\DUrole{default_value}{None}}, \emph{\DUrole{n}{axes}\DUrole{o}{=}\DUrole{default_value}{None}}, \emph{\DUrole{n}{norm}\DUrole{o}{=}\DUrole{default_value}{None}}}{}
Compute the inverse of the N\sphinxhyphen{}dimensional FFT of real input.

LAX\sphinxhyphen{}backend implementation of {\hyperref[\detokenize{modules/signal:symjax.tensor.signal.irfftn}]{\sphinxcrossref{\sphinxcode{\sphinxupquote{irfftn()}}}}}.
ADDITIONOriginal docstring below.

LAX\sphinxhyphen{}backend implementation of {\hyperref[\detokenize{modules/signal:symjax.tensor.signal.irfftn}]{\sphinxcrossref{\sphinxcode{\sphinxupquote{irfftn()}}}}}.
Original docstring below.

This function computes the inverse of the N\sphinxhyphen{}dimensional discrete
Fourier Transform for real input over any number of axes in an
M\sphinxhyphen{}dimensional array by means of the Fast Fourier Transform (FFT).  In
other words, \sphinxcode{\sphinxupquote{irfftn(rfftn(a), a.shape) == a}} to within numerical
accuracy. (The \sphinxcode{\sphinxupquote{a.shape}} is necessary like \sphinxcode{\sphinxupquote{len(a)}} is for \sphinxtitleref{irfft},
and for the same reason.)

The input should be ordered in the same way as is returned by \sphinxtitleref{rfftn},
i.e. as for \sphinxtitleref{irfft} for the final transformation axis, and as for \sphinxtitleref{ifftn}
along all the other axes.
\begin{quote}\begin{description}
\item[{Returns}] \leavevmode
\sphinxstylestrong{out} \textendash{} The truncated or zero\sphinxhyphen{}padded input, transformed along the axes
indicated by \sphinxtitleref{axes}, or by a combination of \sphinxtitleref{s} or \sphinxtitleref{a},
as explained in the parameters section above.
The length of each transformed axis is as given by the corresponding
element of \sphinxtitleref{s}, or the length of the input in every axis except for the
last one if \sphinxtitleref{s} is not given.  In the final transformed axis the length
of the output when \sphinxtitleref{s} is not given is \sphinxcode{\sphinxupquote{2*(m\sphinxhyphen{}1)}} where \sphinxcode{\sphinxupquote{m}} is the
length of the final transformed axis of the input.  To get an odd
number of output points in the final axis, \sphinxtitleref{s} must be specified.

\item[{Return type}] \leavevmode
ndarray

\item[{Raises}] \leavevmode\begin{itemize}
\item {} 
\sphinxstyleliteralstrong{\sphinxupquote{ValueError}} \textendash{} If \sphinxtitleref{s} and \sphinxtitleref{axes} have different length.

\item {} 
\sphinxstyleliteralstrong{\sphinxupquote{IndexError}} \textendash{} If an element of \sphinxtitleref{axes} is larger than than the number of axes of \sphinxtitleref{a}.

\end{itemize}

\end{description}\end{quote}

\sphinxstrong{See also:}

\begin{description}
\item[{{\hyperref[\detokenize{modules/signal:symjax.tensor.signal.rfftn}]{\sphinxcrossref{\sphinxcode{\sphinxupquote{rfftn()}}}}}}] \leavevmode
The forward n\sphinxhyphen{}dimensional FFT of real input, of which \sphinxtitleref{ifftn} is the inverse.

\item[{{\hyperref[\detokenize{modules/signal:symjax.tensor.signal.fft}]{\sphinxcrossref{\sphinxcode{\sphinxupquote{fft()}}}}}}] \leavevmode
The one\sphinxhyphen{}dimensional FFT, with definitions and conventions used.

\item[{{\hyperref[\detokenize{modules/signal:symjax.tensor.signal.irfft}]{\sphinxcrossref{\sphinxcode{\sphinxupquote{irfft()}}}}}}] \leavevmode
The inverse of the one\sphinxhyphen{}dimensional FFT of real input.

\item[{{\hyperref[\detokenize{modules/signal:symjax.tensor.signal.irfft2}]{\sphinxcrossref{\sphinxcode{\sphinxupquote{irfft2()}}}}}}] \leavevmode
The inverse of the two\sphinxhyphen{}dimensional FFT of real input.

\end{description}

\subsubsection*{Notes}

See \sphinxtitleref{fft} for definitions and conventions used.

See \sphinxtitleref{rfft} for definitions and conventions used for real input.

The correct interpretation of the hermitian input depends on the shape of
the original data, as given by \sphinxtitleref{s}. This is because each input shape could
correspond to either an odd or even length signal. By default, \sphinxtitleref{irfftn}
assumes an even output length which puts the last entry at the Nyquist
frequency; aliasing with its symmetric counterpart. When performing the
final complex to real transform, the last value is thus treated as purely
real. To avoid losing information, the correct shape of the real input
\sphinxstylestrong{must} be given.
\subsubsection*{Examples}

\begin{sphinxVerbatim}[commandchars=\\\{\}]
\PYG{g+gp}{\PYGZgt{}\PYGZgt{}\PYGZgt{} }\PYG{n}{a} \PYG{o}{=} \PYG{n}{np}\PYG{o}{.}\PYG{n}{zeros}\PYG{p}{(}\PYG{p}{(}\PYG{l+m+mi}{3}\PYG{p}{,} \PYG{l+m+mi}{2}\PYG{p}{,} \PYG{l+m+mi}{2}\PYG{p}{)}\PYG{p}{)}
\PYG{g+gp}{\PYGZgt{}\PYGZgt{}\PYGZgt{} }\PYG{n}{a}\PYG{p}{[}\PYG{l+m+mi}{0}\PYG{p}{,} \PYG{l+m+mi}{0}\PYG{p}{,} \PYG{l+m+mi}{0}\PYG{p}{]} \PYG{o}{=} \PYG{l+m+mi}{3} \PYG{o}{*} \PYG{l+m+mi}{2} \PYG{o}{*} \PYG{l+m+mi}{2}
\PYG{g+gp}{\PYGZgt{}\PYGZgt{}\PYGZgt{} }\PYG{n}{np}\PYG{o}{.}\PYG{n}{fft}\PYG{o}{.}\PYG{n}{irfftn}\PYG{p}{(}\PYG{n}{a}\PYG{p}{)}
\PYG{g+go}{array([[[1.,  1.],}
\PYG{g+go}{        [1.,  1.]],}
\PYG{g+go}{       [[1.,  1.],}
\PYG{g+go}{        [1.,  1.]],}
\PYG{g+go}{       [[1.,  1.],}
\PYG{g+go}{        [1.,  1.]]])}
\end{sphinxVerbatim}

\end{fulllineitems}

\index{fftfreq() (in module symjax.tensor.signal)@\spxentry{fftfreq()}\spxextra{in module symjax.tensor.signal}}

\begin{fulllineitems}
\phantomsection\label{\detokenize{modules/signal:symjax.tensor.signal.fftfreq}}\pysiglinewithargsret{\sphinxbfcode{\sphinxupquote{fftfreq}}}{\emph{\DUrole{n}{n}}, \emph{\DUrole{n}{d}\DUrole{o}{=}\DUrole{default_value}{1.0}}}{}
Return the Discrete Fourier Transform sample frequencies.

LAX\sphinxhyphen{}backend implementation of {\hyperref[\detokenize{modules/signal:symjax.tensor.signal.fftfreq}]{\sphinxcrossref{\sphinxcode{\sphinxupquote{fftfreq()}}}}}.
ADDITIONOriginal docstring below.

LAX\sphinxhyphen{}backend implementation of {\hyperref[\detokenize{modules/signal:symjax.tensor.signal.fftfreq}]{\sphinxcrossref{\sphinxcode{\sphinxupquote{fftfreq()}}}}}.
Original docstring below.

The returned float array \sphinxtitleref{f} contains the frequency bin centers in cycles
per unit of the sample spacing (with zero at the start).  For instance, if
the sample spacing is in seconds, then the frequency unit is cycles/second.

Given a window length \sphinxtitleref{n} and a sample spacing \sphinxtitleref{d}:

\begin{sphinxVerbatim}[commandchars=\\\{\}]
\PYG{n}{f} \PYG{o}{=} \PYG{p}{[}\PYG{l+m+mi}{0}\PYG{p}{,} \PYG{l+m+mi}{1}\PYG{p}{,} \PYG{o}{.}\PYG{o}{.}\PYG{o}{.}\PYG{p}{,}   \PYG{n}{n}\PYG{o}{/}\PYG{l+m+mi}{2}\PYG{o}{\PYGZhy{}}\PYG{l+m+mi}{1}\PYG{p}{,}     \PYG{o}{\PYGZhy{}}\PYG{n}{n}\PYG{o}{/}\PYG{l+m+mi}{2}\PYG{p}{,} \PYG{o}{.}\PYG{o}{.}\PYG{o}{.}\PYG{p}{,} \PYG{o}{\PYGZhy{}}\PYG{l+m+mi}{1}\PYG{p}{]} \PYG{o}{/} \PYG{p}{(}\PYG{n}{d}\PYG{o}{*}\PYG{n}{n}\PYG{p}{)}   \PYG{k}{if} \PYG{n}{n} \PYG{o+ow}{is} \PYG{n}{even}
\PYG{n}{f} \PYG{o}{=} \PYG{p}{[}\PYG{l+m+mi}{0}\PYG{p}{,} \PYG{l+m+mi}{1}\PYG{p}{,} \PYG{o}{.}\PYG{o}{.}\PYG{o}{.}\PYG{p}{,} \PYG{p}{(}\PYG{n}{n}\PYG{o}{\PYGZhy{}}\PYG{l+m+mi}{1}\PYG{p}{)}\PYG{o}{/}\PYG{l+m+mi}{2}\PYG{p}{,} \PYG{o}{\PYGZhy{}}\PYG{p}{(}\PYG{n}{n}\PYG{o}{\PYGZhy{}}\PYG{l+m+mi}{1}\PYG{p}{)}\PYG{o}{/}\PYG{l+m+mi}{2}\PYG{p}{,} \PYG{o}{.}\PYG{o}{.}\PYG{o}{.}\PYG{p}{,} \PYG{o}{\PYGZhy{}}\PYG{l+m+mi}{1}\PYG{p}{]} \PYG{o}{/} \PYG{p}{(}\PYG{n}{d}\PYG{o}{*}\PYG{n}{n}\PYG{p}{)}   \PYG{k}{if} \PYG{n}{n} \PYG{o+ow}{is} \PYG{n}{odd}
\end{sphinxVerbatim}
\begin{quote}\begin{description}
\item[{Returns}] \leavevmode
\sphinxstylestrong{f} \textendash{} Array of length \sphinxtitleref{n} containing the sample frequencies.

\item[{Return type}] \leavevmode
ndarray

\end{description}\end{quote}
\subsubsection*{Examples}

\begin{sphinxVerbatim}[commandchars=\\\{\}]
\PYG{g+gp}{\PYGZgt{}\PYGZgt{}\PYGZgt{} }\PYG{n}{signal} \PYG{o}{=} \PYG{n}{np}\PYG{o}{.}\PYG{n}{array}\PYG{p}{(}\PYG{p}{[}\PYG{o}{\PYGZhy{}}\PYG{l+m+mi}{2}\PYG{p}{,} \PYG{l+m+mi}{8}\PYG{p}{,} \PYG{l+m+mi}{6}\PYG{p}{,} \PYG{l+m+mi}{4}\PYG{p}{,} \PYG{l+m+mi}{1}\PYG{p}{,} \PYG{l+m+mi}{0}\PYG{p}{,} \PYG{l+m+mi}{3}\PYG{p}{,} \PYG{l+m+mi}{5}\PYG{p}{]}\PYG{p}{,} \PYG{n}{dtype}\PYG{o}{=}\PYG{n+nb}{float}\PYG{p}{)}
\PYG{g+gp}{\PYGZgt{}\PYGZgt{}\PYGZgt{} }\PYG{n}{fourier} \PYG{o}{=} \PYG{n}{np}\PYG{o}{.}\PYG{n}{fft}\PYG{o}{.}\PYG{n}{fft}\PYG{p}{(}\PYG{n}{signal}\PYG{p}{)}
\PYG{g+gp}{\PYGZgt{}\PYGZgt{}\PYGZgt{} }\PYG{n}{n} \PYG{o}{=} \PYG{n}{signal}\PYG{o}{.}\PYG{n}{size}
\PYG{g+gp}{\PYGZgt{}\PYGZgt{}\PYGZgt{} }\PYG{n}{timestep} \PYG{o}{=} \PYG{l+m+mf}{0.1}
\PYG{g+gp}{\PYGZgt{}\PYGZgt{}\PYGZgt{} }\PYG{n}{freq} \PYG{o}{=} \PYG{n}{np}\PYG{o}{.}\PYG{n}{fft}\PYG{o}{.}\PYG{n}{fftfreq}\PYG{p}{(}\PYG{n}{n}\PYG{p}{,} \PYG{n}{d}\PYG{o}{=}\PYG{n}{timestep}\PYG{p}{)}
\PYG{g+gp}{\PYGZgt{}\PYGZgt{}\PYGZgt{} }\PYG{n}{freq}
\PYG{g+go}{array([ 0.  ,  1.25,  2.5 , ..., \PYGZhy{}3.75, \PYGZhy{}2.5 , \PYGZhy{}1.25])}
\end{sphinxVerbatim}

\end{fulllineitems}

\index{rfftfreq() (in module symjax.tensor.signal)@\spxentry{rfftfreq()}\spxextra{in module symjax.tensor.signal}}

\begin{fulllineitems}
\phantomsection\label{\detokenize{modules/signal:symjax.tensor.signal.rfftfreq}}\pysiglinewithargsret{\sphinxbfcode{\sphinxupquote{rfftfreq}}}{\emph{\DUrole{n}{n}}, \emph{\DUrole{n}{d}\DUrole{o}{=}\DUrole{default_value}{1.0}}}{}~\begin{description}
\item[{Return the Discrete Fourier Transform sample frequencies}] \leavevmode
(for usage with rfft, irfft).

\end{description}

LAX\sphinxhyphen{}backend implementation of {\hyperref[\detokenize{modules/signal:symjax.tensor.signal.rfftfreq}]{\sphinxcrossref{\sphinxcode{\sphinxupquote{rfftfreq()}}}}}.
ADDITIONOriginal docstring below.

LAX\sphinxhyphen{}backend implementation of {\hyperref[\detokenize{modules/signal:symjax.tensor.signal.rfftfreq}]{\sphinxcrossref{\sphinxcode{\sphinxupquote{rfftfreq()}}}}}.
Original docstring below.

The returned float array \sphinxtitleref{f} contains the frequency bin centers in cycles
per unit of the sample spacing (with zero at the start).  For instance, if
the sample spacing is in seconds, then the frequency unit is cycles/second.

Given a window length \sphinxtitleref{n} and a sample spacing \sphinxtitleref{d}:

\begin{sphinxVerbatim}[commandchars=\\\{\}]
\PYG{n}{f} \PYG{o}{=} \PYG{p}{[}\PYG{l+m+mi}{0}\PYG{p}{,} \PYG{l+m+mi}{1}\PYG{p}{,} \PYG{o}{.}\PYG{o}{.}\PYG{o}{.}\PYG{p}{,}     \PYG{n}{n}\PYG{o}{/}\PYG{l+m+mi}{2}\PYG{o}{\PYGZhy{}}\PYG{l+m+mi}{1}\PYG{p}{,}     \PYG{n}{n}\PYG{o}{/}\PYG{l+m+mi}{2}\PYG{p}{]} \PYG{o}{/} \PYG{p}{(}\PYG{n}{d}\PYG{o}{*}\PYG{n}{n}\PYG{p}{)}   \PYG{k}{if} \PYG{n}{n} \PYG{o+ow}{is} \PYG{n}{even}
\PYG{n}{f} \PYG{o}{=} \PYG{p}{[}\PYG{l+m+mi}{0}\PYG{p}{,} \PYG{l+m+mi}{1}\PYG{p}{,} \PYG{o}{.}\PYG{o}{.}\PYG{o}{.}\PYG{p}{,} \PYG{p}{(}\PYG{n}{n}\PYG{o}{\PYGZhy{}}\PYG{l+m+mi}{1}\PYG{p}{)}\PYG{o}{/}\PYG{l+m+mi}{2}\PYG{o}{\PYGZhy{}}\PYG{l+m+mi}{1}\PYG{p}{,} \PYG{p}{(}\PYG{n}{n}\PYG{o}{\PYGZhy{}}\PYG{l+m+mi}{1}\PYG{p}{)}\PYG{o}{/}\PYG{l+m+mi}{2}\PYG{p}{]} \PYG{o}{/} \PYG{p}{(}\PYG{n}{d}\PYG{o}{*}\PYG{n}{n}\PYG{p}{)}   \PYG{k}{if} \PYG{n}{n} \PYG{o+ow}{is} \PYG{n}{odd}
\end{sphinxVerbatim}

Unlike \sphinxtitleref{fftfreq} (but like \sphinxtitleref{scipy.fftpack.rfftfreq})
the Nyquist frequency component is considered to be positive.
\begin{quote}\begin{description}
\item[{Returns}] \leavevmode
\sphinxstylestrong{f} \textendash{} Array of length \sphinxcode{\sphinxupquote{n//2 + 1}} containing the sample frequencies.

\item[{Return type}] \leavevmode
ndarray

\end{description}\end{quote}
\subsubsection*{Examples}

\begin{sphinxVerbatim}[commandchars=\\\{\}]
\PYG{g+gp}{\PYGZgt{}\PYGZgt{}\PYGZgt{} }\PYG{n}{signal} \PYG{o}{=} \PYG{n}{np}\PYG{o}{.}\PYG{n}{array}\PYG{p}{(}\PYG{p}{[}\PYG{o}{\PYGZhy{}}\PYG{l+m+mi}{2}\PYG{p}{,} \PYG{l+m+mi}{8}\PYG{p}{,} \PYG{l+m+mi}{6}\PYG{p}{,} \PYG{l+m+mi}{4}\PYG{p}{,} \PYG{l+m+mi}{1}\PYG{p}{,} \PYG{l+m+mi}{0}\PYG{p}{,} \PYG{l+m+mi}{3}\PYG{p}{,} \PYG{l+m+mi}{5}\PYG{p}{,} \PYG{o}{\PYGZhy{}}\PYG{l+m+mi}{3}\PYG{p}{,} \PYG{l+m+mi}{4}\PYG{p}{]}\PYG{p}{,} \PYG{n}{dtype}\PYG{o}{=}\PYG{n+nb}{float}\PYG{p}{)}
\PYG{g+gp}{\PYGZgt{}\PYGZgt{}\PYGZgt{} }\PYG{n}{fourier} \PYG{o}{=} \PYG{n}{np}\PYG{o}{.}\PYG{n}{fft}\PYG{o}{.}\PYG{n}{rfft}\PYG{p}{(}\PYG{n}{signal}\PYG{p}{)}
\PYG{g+gp}{\PYGZgt{}\PYGZgt{}\PYGZgt{} }\PYG{n}{n} \PYG{o}{=} \PYG{n}{signal}\PYG{o}{.}\PYG{n}{size}
\PYG{g+gp}{\PYGZgt{}\PYGZgt{}\PYGZgt{} }\PYG{n}{sample\PYGZus{}rate} \PYG{o}{=} \PYG{l+m+mi}{100}
\PYG{g+gp}{\PYGZgt{}\PYGZgt{}\PYGZgt{} }\PYG{n}{freq} \PYG{o}{=} \PYG{n}{np}\PYG{o}{.}\PYG{n}{fft}\PYG{o}{.}\PYG{n}{fftfreq}\PYG{p}{(}\PYG{n}{n}\PYG{p}{,} \PYG{n}{d}\PYG{o}{=}\PYG{l+m+mf}{1.}\PYG{o}{/}\PYG{n}{sample\PYGZus{}rate}\PYG{p}{)}
\PYG{g+gp}{\PYGZgt{}\PYGZgt{}\PYGZgt{} }\PYG{n}{freq}
\PYG{g+go}{array([  0.,  10.,  20., ..., \PYGZhy{}30., \PYGZhy{}20., \PYGZhy{}10.])}
\PYG{g+gp}{\PYGZgt{}\PYGZgt{}\PYGZgt{} }\PYG{n}{freq} \PYG{o}{=} \PYG{n}{np}\PYG{o}{.}\PYG{n}{fft}\PYG{o}{.}\PYG{n}{rfftfreq}\PYG{p}{(}\PYG{n}{n}\PYG{p}{,} \PYG{n}{d}\PYG{o}{=}\PYG{l+m+mf}{1.}\PYG{o}{/}\PYG{n}{sample\PYGZus{}rate}\PYG{p}{)}
\PYG{g+gp}{\PYGZgt{}\PYGZgt{}\PYGZgt{} }\PYG{n}{freq}
\PYG{g+go}{array([  0.,  10.,  20.,  30.,  40.,  50.])}
\end{sphinxVerbatim}

\end{fulllineitems}

\index{mfcc() (in module symjax.tensor.signal)@\spxentry{mfcc()}\spxextra{in module symjax.tensor.signal}}

\begin{fulllineitems}
\phantomsection\label{\detokenize{modules/signal:symjax.tensor.signal.mfcc}}\pysiglinewithargsret{\sphinxbfcode{\sphinxupquote{mfcc}}}{\emph{signal}, \emph{window}, \emph{hop}, \emph{n\_filter}, \emph{low\_freq}, \emph{high\_freq}, \emph{nyquist}, \emph{n\_mfcc}, \emph{nfft=None}, \emph{mode=\textquotesingle{}valid\textquotesingle{}}, \emph{apod=\textless{}function hanning\textgreater{}}}{}
\sphinxurl{https://librosa.github.io/librosa/\_modules/librosa/feature/spectral.html\#mfcc}

\end{fulllineitems}

\index{dct() (in module symjax.tensor.signal)@\spxentry{dct()}\spxextra{in module symjax.tensor.signal}}

\begin{fulllineitems}
\phantomsection\label{\detokenize{modules/signal:symjax.tensor.signal.dct}}\pysiglinewithargsret{\sphinxbfcode{\sphinxupquote{dct}}}{\emph{\DUrole{n}{signal}}, \emph{\DUrole{n}{axes}\DUrole{o}{=}\DUrole{default_value}{\sphinxhyphen{} 1}}}{}
\sphinxurl{https://dsp.stackexchange.com/questions/2807/fast-cosine-transform-via-fft}

\end{fulllineitems}

\index{wvd() (in module symjax.tensor.signal)@\spxentry{wvd()}\spxextra{in module symjax.tensor.signal}}

\begin{fulllineitems}
\phantomsection\label{\detokenize{modules/signal:symjax.tensor.signal.wvd}}\pysiglinewithargsret{\sphinxbfcode{\sphinxupquote{wvd}}}{\emph{signal}, \emph{window}, \emph{hop}, \emph{L}, \emph{apod=\textless{}function hanning\textgreater{}}, \emph{mode=\textquotesingle{}valid\textquotesingle{}}}{}
\end{fulllineitems}

\index{hilbert\_transform() (in module symjax.tensor.signal)@\spxentry{hilbert\_transform()}\spxextra{in module symjax.tensor.signal}}

\begin{fulllineitems}
\phantomsection\label{\detokenize{modules/signal:symjax.tensor.signal.hilbert_transform}}\pysiglinewithargsret{\sphinxbfcode{\sphinxupquote{hilbert\_transform}}}{\emph{\DUrole{n}{signal}}}{}
the time should be the last dimension
return the analytical signal

\end{fulllineitems}

\subsection{\sphinxstyleliteralintitle{\sphinxupquote{symjax.tensor.random}}}
\label{\detokenize{modules/random:module-symjax.tensor.random}}\label{\detokenize{modules/random:symjax-tensor-random}}\label{\detokenize{modules/random::doc}}\index{module@\spxentry{module}!symjax.tensor.random@\spxentry{symjax.tensor.random}}\index{symjax.tensor.random@\spxentry{symjax.tensor.random}!module@\spxentry{module}}

\begin{savenotes}\sphinxatlongtablestart\begin{longtable}[c]{\X{1}{2}\X{1}{2}}
\hline

\endfirsthead

\multicolumn{2}{c}%
{\makebox[0pt]{\sphinxtablecontinued{\tablename\ \thetable{} \textendash{} continued from previous page}}}\\
\hline

\endhead

\hline
\multicolumn{2}{r}{\makebox[0pt][r]{\sphinxtablecontinued{continues on next page}}}\\
\endfoot

\endlastfoot

{\hyperref[\detokenize{modules/random:symjax.tensor.random.bernoulli}]{\sphinxcrossref{\sphinxcode{\sphinxupquote{bernoulli}}}}}(key{[}, p, shape{]})
&
Sample Bernoulli random values with given shape and mean.
\\
\hline
{\hyperref[\detokenize{modules/random:symjax.tensor.random.beta}]{\sphinxcrossref{\sphinxcode{\sphinxupquote{beta}}}}}(key, a, b{[}, shape, dtype{]})
&
Sample Beta random values with given shape and float dtype.
\\
\hline
{\hyperref[\detokenize{modules/random:symjax.tensor.random.cauchy}]{\sphinxcrossref{\sphinxcode{\sphinxupquote{cauchy}}}}}(key{[}, shape, dtype{]})
&
Sample Cauchy random values with given shape and float dtype.
\\
\hline
{\hyperref[\detokenize{modules/random:symjax.tensor.random.dirichlet}]{\sphinxcrossref{\sphinxcode{\sphinxupquote{dirichlet}}}}}(key, alpha{[}, shape, dtype{]})
&
Sample Dirichlet random values with given shape and float dtype.
\\
\hline
{\hyperref[\detokenize{modules/random:symjax.tensor.random.gamma}]{\sphinxcrossref{\sphinxcode{\sphinxupquote{gamma}}}}}(key, a{[}, shape, dtype{]})
&
Sample Gamma random values with given shape and float dtype.
\\
\hline
{\hyperref[\detokenize{modules/random:symjax.tensor.random.gumbel}]{\sphinxcrossref{\sphinxcode{\sphinxupquote{gumbel}}}}}(key{[}, shape, dtype{]})
&
Sample Gumbel random values with given shape and float dtype.
\\
\hline
{\hyperref[\detokenize{modules/random:symjax.tensor.random.laplace}]{\sphinxcrossref{\sphinxcode{\sphinxupquote{laplace}}}}}(key{[}, shape, dtype{]})
&
Sample Laplace random values with given shape and float dtype.
\\
\hline
{\hyperref[\detokenize{modules/random:symjax.tensor.random.logit}]{\sphinxcrossref{\sphinxcode{\sphinxupquote{logit}}}}}
&
Logit ufunc for ndarrays.
\\
\hline
{\hyperref[\detokenize{modules/random:symjax.tensor.random.multivariate_normal}]{\sphinxcrossref{\sphinxcode{\sphinxupquote{multivariate\_normal}}}}}(key, mean, cov{[}, shape, …{]})
&
Sample multivariate normal random values with given mean and covariance.
\\
\hline
{\hyperref[\detokenize{modules/random:symjax.tensor.random.normal}]{\sphinxcrossref{\sphinxcode{\sphinxupquote{normal}}}}}(key{[}, shape, dtype{]})
&
Sample standard normal random values with given shape and float dtype.
\\
\hline
{\hyperref[\detokenize{modules/random:symjax.tensor.random.pareto}]{\sphinxcrossref{\sphinxcode{\sphinxupquote{pareto}}}}}(key, b{[}, shape, dtype{]})
&
Sample Pareto random values with given shape and float dtype.
\\
\hline
{\hyperref[\detokenize{modules/random:symjax.tensor.random.randint}]{\sphinxcrossref{\sphinxcode{\sphinxupquote{randint}}}}}(key, shape, minval, maxval{[}, dtype{]})
&
Sample uniform random values in {[}minval, maxval) with given shape/dtype.
\\
\hline
{\hyperref[\detokenize{modules/random:symjax.tensor.random.shuffle}]{\sphinxcrossref{\sphinxcode{\sphinxupquote{shuffle}}}}}(key, x{[}, axis{]})
&
Shuffle the elements of an array uniformly at random along an axis.
\\
\hline
{\hyperref[\detokenize{modules/random:symjax.tensor.random.truncated_normal}]{\sphinxcrossref{\sphinxcode{\sphinxupquote{truncated\_normal}}}}}(key, lower, upper{[}, shape, …{]})
&
Sample truncated standard normal random values with given shape and dtype.
\\
\hline
{\hyperref[\detokenize{modules/random:symjax.tensor.random.uniform}]{\sphinxcrossref{\sphinxcode{\sphinxupquote{uniform}}}}}(key{[}, shape, dtype, minval, maxval{]})
&
Sample uniform random values in {[}minval, maxval) with given shape/dtype.
\\
\hline
\end{longtable}\sphinxatlongtableend\end{savenotes}

\subsubsection{Detailed Description}
\label{\detokenize{modules/random:detailed-description}}\index{bernoulli() (in module symjax.tensor.random)@\spxentry{bernoulli()}\spxextra{in module symjax.tensor.random}}

\begin{fulllineitems}
\phantomsection\label{\detokenize{modules/random:symjax.tensor.random.bernoulli}}\pysiglinewithargsret{\sphinxbfcode{\sphinxupquote{bernoulli}}}{\emph{\DUrole{n}{key}\DUrole{p}{:} \DUrole{n}{jax.numpy.lax\_numpy.ndarray}}, \emph{\DUrole{n}{p}\DUrole{p}{:} \DUrole{n}{jax.numpy.lax\_numpy.ndarray} \DUrole{o}{=} \DUrole{default_value}{0.5}}, \emph{\DUrole{n}{shape}\DUrole{p}{:} \DUrole{n}{Optional\DUrole{p}{{[}}Sequence\DUrole{p}{{[}}int\DUrole{p}{{]}}\DUrole{p}{{]}}} \DUrole{o}{=} \DUrole{default_value}{None}}}{{ $\rightarrow$ jax.numpy.lax\_numpy.ndarray}}
Sample Bernoulli random values with given shape and mean.

LAX\sphinxhyphen{}backend implementation of {\hyperref[\detokenize{modules/random:symjax.tensor.random.bernoulli}]{\sphinxcrossref{\sphinxcode{\sphinxupquote{bernoulli()}}}}}.
ADDITIONOriginal docstring below.
\begin{quote}

Returns:
A random array with boolean dtype and shape given by \sphinxcode{\sphinxupquote{shape}} if \sphinxcode{\sphinxupquote{shape}}
is not None, or else \sphinxcode{\sphinxupquote{p.shape}}.
\end{quote}

\end{fulllineitems}

\index{beta() (in module symjax.tensor.random)@\spxentry{beta()}\spxextra{in module symjax.tensor.random}}

\begin{fulllineitems}
\phantomsection\label{\detokenize{modules/random:symjax.tensor.random.beta}}\pysiglinewithargsret{\sphinxbfcode{\sphinxupquote{beta}}}{\emph{key: jax.numpy.lax\_numpy.ndarray, a: Union{[}float, jax.numpy.lax\_numpy.ndarray{]}, b: Union{[}float, jax.numpy.lax\_numpy.ndarray{]}, shape: Optional{[}Sequence{[}int{]}{]} = None, dtype: numpy.dtype = \textless{}class \textquotesingle{}numpy.float64\textquotesingle{}\textgreater{}}}{{ $\rightarrow$ jax.numpy.lax\_numpy.ndarray}}
Sample Beta random values with given shape and float dtype.

LAX\sphinxhyphen{}backend implementation of {\hyperref[\detokenize{modules/random:symjax.tensor.random.beta}]{\sphinxcrossref{\sphinxcode{\sphinxupquote{beta()}}}}}.
ADDITIONOriginal docstring below.
\begin{quote}

Returns:
A random array with the specified dtype and shape given by \sphinxcode{\sphinxupquote{shape}} if
\sphinxcode{\sphinxupquote{shape}} is not None, or else by broadcasting \sphinxcode{\sphinxupquote{a}} and \sphinxcode{\sphinxupquote{b}}.
\end{quote}

\end{fulllineitems}

\index{cauchy() (in module symjax.tensor.random)@\spxentry{cauchy()}\spxextra{in module symjax.tensor.random}}

\begin{fulllineitems}
\phantomsection\label{\detokenize{modules/random:symjax.tensor.random.cauchy}}\pysiglinewithargsret{\sphinxbfcode{\sphinxupquote{cauchy}}}{\emph{key}, \emph{shape=()}, \emph{dtype=\textless{}class \textquotesingle{}numpy.float64\textquotesingle{}\textgreater{}}}{}
Sample Cauchy random values with given shape and float dtype.

LAX\sphinxhyphen{}backend implementation of {\hyperref[\detokenize{modules/random:symjax.tensor.random.cauchy}]{\sphinxcrossref{\sphinxcode{\sphinxupquote{cauchy()}}}}}.
ADDITIONOriginal docstring below.
\begin{quote}

Returns:
A random array with the specified shape and dtype.
\end{quote}

\end{fulllineitems}

\index{dirichlet() (in module symjax.tensor.random)@\spxentry{dirichlet()}\spxextra{in module symjax.tensor.random}}

\begin{fulllineitems}
\phantomsection\label{\detokenize{modules/random:symjax.tensor.random.dirichlet}}\pysiglinewithargsret{\sphinxbfcode{\sphinxupquote{dirichlet}}}{\emph{key}, \emph{alpha}, \emph{shape=None}, \emph{dtype=\textless{}class \textquotesingle{}numpy.float64\textquotesingle{}\textgreater{}}}{}
Sample Dirichlet random values with given shape and float dtype.

LAX\sphinxhyphen{}backend implementation of {\hyperref[\detokenize{modules/random:symjax.tensor.random.dirichlet}]{\sphinxcrossref{\sphinxcode{\sphinxupquote{dirichlet()}}}}}.
ADDITIONOriginal docstring below.
\begin{quote}

Returns:
A random array with the specified dtype and shape given by
\sphinxcode{\sphinxupquote{shape + (alpha.shape{[}\sphinxhyphen{}1{]},)}} if \sphinxcode{\sphinxupquote{shape}} is not None, or else
\sphinxcode{\sphinxupquote{alpha.shape}}.
\end{quote}

\end{fulllineitems}

\index{gamma() (in module symjax.tensor.random)@\spxentry{gamma()}\spxextra{in module symjax.tensor.random}}

\begin{fulllineitems}
\phantomsection\label{\detokenize{modules/random:symjax.tensor.random.gamma}}\pysiglinewithargsret{\sphinxbfcode{\sphinxupquote{gamma}}}{\emph{key}, \emph{a}, \emph{shape=None}, \emph{dtype=\textless{}class \textquotesingle{}numpy.float64\textquotesingle{}\textgreater{}}}{}
Sample Gamma random values with given shape and float dtype.

LAX\sphinxhyphen{}backend implementation of {\hyperref[\detokenize{modules/random:symjax.tensor.random.gamma}]{\sphinxcrossref{\sphinxcode{\sphinxupquote{gamma()}}}}}.
ADDITIONOriginal docstring below.
\begin{quote}

Returns:
A random array with the specified dtype and with shape given by \sphinxcode{\sphinxupquote{shape}} if
\sphinxcode{\sphinxupquote{shape}} is not None, or else by \sphinxcode{\sphinxupquote{a.shape}}.
\end{quote}

\end{fulllineitems}

\index{gumbel() (in module symjax.tensor.random)@\spxentry{gumbel()}\spxextra{in module symjax.tensor.random}}

\begin{fulllineitems}
\phantomsection\label{\detokenize{modules/random:symjax.tensor.random.gumbel}}\pysiglinewithargsret{\sphinxbfcode{\sphinxupquote{gumbel}}}{\emph{key}, \emph{shape=()}, \emph{dtype=\textless{}class \textquotesingle{}numpy.float64\textquotesingle{}\textgreater{}}}{}
Sample Gumbel random values with given shape and float dtype.

LAX\sphinxhyphen{}backend implementation of {\hyperref[\detokenize{modules/random:symjax.tensor.random.gumbel}]{\sphinxcrossref{\sphinxcode{\sphinxupquote{gumbel()}}}}}.
ADDITIONOriginal docstring below.
\begin{quote}

Returns:
A random array with the specified shape and dtype.
\end{quote}

\end{fulllineitems}

\index{laplace() (in module symjax.tensor.random)@\spxentry{laplace()}\spxextra{in module symjax.tensor.random}}

\begin{fulllineitems}
\phantomsection\label{\detokenize{modules/random:symjax.tensor.random.laplace}}\pysiglinewithargsret{\sphinxbfcode{\sphinxupquote{laplace}}}{\emph{key}, \emph{shape=()}, \emph{dtype=\textless{}class \textquotesingle{}numpy.float64\textquotesingle{}\textgreater{}}}{}
Sample Laplace random values with given shape and float dtype.

LAX\sphinxhyphen{}backend implementation of {\hyperref[\detokenize{modules/random:symjax.tensor.random.laplace}]{\sphinxcrossref{\sphinxcode{\sphinxupquote{laplace()}}}}}.
ADDITIONOriginal docstring below.
\begin{quote}

Returns:
A random array with the specified shape and dtype.
\end{quote}

\end{fulllineitems}

\index{logit() (in module symjax.tensor.random)@\spxentry{logit()}\spxextra{in module symjax.tensor.random}}

\begin{fulllineitems}
\phantomsection\label{\detokenize{modules/random:symjax.tensor.random.logit}}\pysiglinewithargsret{\sphinxbfcode{\sphinxupquote{logit}}}{\emph{\DUrole{n}{x}}}{}
Logit ufunc for ndarrays.

LAX\sphinxhyphen{}backend implementation of {\hyperref[\detokenize{modules/random:symjax.tensor.random.logit}]{\sphinxcrossref{\sphinxcode{\sphinxupquote{logit()}}}}}.
ADDITIONOriginal docstring below.

LAX\sphinxhyphen{}backend implementation of {\hyperref[\detokenize{modules/random:symjax.tensor.random.logit}]{\sphinxcrossref{\sphinxcode{\sphinxupquote{logit()}}}}}.
Original docstring below.

logit(x, /, out=None, {\color{red}\bfseries{}*}, where=True, casting=’same\_kind’, order=’K’, dtype=None, subok=True{[}, signature, extobj{]})

logit(x)

The logit function is defined as logit(p) = log(p/(1\sphinxhyphen{}p)).
Note that logit(0) = \sphinxhyphen{}inf, logit(1) = inf, and logit(p)
for p\textless{}0 or p\textgreater{}1 yields nan.
\begin{quote}\begin{description}
\item[{Returns}] \leavevmode
\sphinxstylestrong{out} \textendash{} An ndarray of the same shape as x. Its entries
are logit of the corresponding entry of x.

\item[{Return type}] \leavevmode
ndarray

\end{description}\end{quote}

\sphinxstrong{See also:}

\sphinxcode{\sphinxupquote{expit()}}

\subsubsection*{Notes}

As a ufunc logit takes a number of optional
keyword arguments. For more information
see \sphinxhref{https://docs.scipy.org/doc/numpy/reference/ufuncs.html}{ufuncs}

\DUrole{versionmodified,added}{New in version 0.10.0.}
\subsubsection*{Examples}

\begin{sphinxVerbatim}[commandchars=\\\{\}]
\PYG{g+gp}{\PYGZgt{}\PYGZgt{}\PYGZgt{} }\PYG{k+kn}{from} \PYG{n+nn}{scipy}\PYG{n+nn}{.}\PYG{n+nn}{special} \PYG{k+kn}{import} \PYG{n}{logit}\PYG{p}{,} \PYG{n}{expit}
\end{sphinxVerbatim}

\begin{sphinxVerbatim}[commandchars=\\\{\}]
\PYG{g+gp}{\PYGZgt{}\PYGZgt{}\PYGZgt{} }\PYG{n}{logit}\PYG{p}{(}\PYG{p}{[}\PYG{l+m+mi}{0}\PYG{p}{,} \PYG{l+m+mf}{0.25}\PYG{p}{,} \PYG{l+m+mf}{0.5}\PYG{p}{,} \PYG{l+m+mf}{0.75}\PYG{p}{,} \PYG{l+m+mi}{1}\PYG{p}{]}\PYG{p}{)}
\PYG{g+go}{array([       \PYGZhy{}inf, \PYGZhy{}1.09861229,  0.        ,  1.09861229,         inf])}
\end{sphinxVerbatim}

\sphinxtitleref{expit} is the inverse of \sphinxtitleref{logit}:

\begin{sphinxVerbatim}[commandchars=\\\{\}]
\PYG{g+gp}{\PYGZgt{}\PYGZgt{}\PYGZgt{} }\PYG{n}{expit}\PYG{p}{(}\PYG{n}{logit}\PYG{p}{(}\PYG{p}{[}\PYG{l+m+mf}{0.1}\PYG{p}{,} \PYG{l+m+mf}{0.75}\PYG{p}{,} \PYG{l+m+mf}{0.999}\PYG{p}{]}\PYG{p}{)}\PYG{p}{)}
\PYG{g+go}{array([ 0.1  ,  0.75 ,  0.999])}
\end{sphinxVerbatim}

Plot logit(x) for x in {[}0, 1{]}:

\begin{sphinxVerbatim}[commandchars=\\\{\}]
\PYG{g+gp}{\PYGZgt{}\PYGZgt{}\PYGZgt{} }\PYG{k+kn}{import} \PYG{n+nn}{matplotlib}\PYG{n+nn}{.}\PYG{n+nn}{pyplot} \PYG{k}{as} \PYG{n+nn}{plt}
\PYG{g+gp}{\PYGZgt{}\PYGZgt{}\PYGZgt{} }\PYG{n}{x} \PYG{o}{=} \PYG{n}{np}\PYG{o}{.}\PYG{n}{linspace}\PYG{p}{(}\PYG{l+m+mi}{0}\PYG{p}{,} \PYG{l+m+mi}{1}\PYG{p}{,} \PYG{l+m+mi}{501}\PYG{p}{)}
\PYG{g+gp}{\PYGZgt{}\PYGZgt{}\PYGZgt{} }\PYG{n}{y} \PYG{o}{=} \PYG{n}{logit}\PYG{p}{(}\PYG{n}{x}\PYG{p}{)}
\PYG{g+gp}{\PYGZgt{}\PYGZgt{}\PYGZgt{} }\PYG{n}{plt}\PYG{o}{.}\PYG{n}{plot}\PYG{p}{(}\PYG{n}{x}\PYG{p}{,} \PYG{n}{y}\PYG{p}{)}
\PYG{g+gp}{\PYGZgt{}\PYGZgt{}\PYGZgt{} }\PYG{n}{plt}\PYG{o}{.}\PYG{n}{grid}\PYG{p}{(}\PYG{p}{)}
\PYG{g+gp}{\PYGZgt{}\PYGZgt{}\PYGZgt{} }\PYG{n}{plt}\PYG{o}{.}\PYG{n}{ylim}\PYG{p}{(}\PYG{o}{\PYGZhy{}}\PYG{l+m+mi}{6}\PYG{p}{,} \PYG{l+m+mi}{6}\PYG{p}{)}
\PYG{g+gp}{\PYGZgt{}\PYGZgt{}\PYGZgt{} }\PYG{n}{plt}\PYG{o}{.}\PYG{n}{xlabel}\PYG{p}{(}\PYG{l+s+s1}{\PYGZsq{}}\PYG{l+s+s1}{x}\PYG{l+s+s1}{\PYGZsq{}}\PYG{p}{)}
\PYG{g+gp}{\PYGZgt{}\PYGZgt{}\PYGZgt{} }\PYG{n}{plt}\PYG{o}{.}\PYG{n}{title}\PYG{p}{(}\PYG{l+s+s1}{\PYGZsq{}}\PYG{l+s+s1}{logit(x)}\PYG{l+s+s1}{\PYGZsq{}}\PYG{p}{)}
\PYG{g+gp}{\PYGZgt{}\PYGZgt{}\PYGZgt{} }\PYG{n}{plt}\PYG{o}{.}\PYG{n}{show}\PYG{p}{(}\PYG{p}{)}
\end{sphinxVerbatim}

\end{fulllineitems}

\index{multivariate\_normal() (in module symjax.tensor.random)@\spxentry{multivariate\_normal()}\spxextra{in module symjax.tensor.random}}

\begin{fulllineitems}
\phantomsection\label{\detokenize{modules/random:symjax.tensor.random.multivariate_normal}}\pysiglinewithargsret{\sphinxbfcode{\sphinxupquote{multivariate\_normal}}}{\emph{key: jax.numpy.lax\_numpy.ndarray}, \emph{mean: jax.numpy.lax\_numpy.ndarray}, \emph{cov: jax.numpy.lax\_numpy.ndarray}, \emph{shape: Optional{[}Sequence{[}int{]}{]} = None}, \emph{dtype: numpy.dtype = \textless{}class \textquotesingle{}numpy.float64\textquotesingle{}\textgreater{}}}{{ $\rightarrow$ jax.numpy.lax\_numpy.ndarray}}
Sample multivariate normal random values with given mean and covariance.

LAX\sphinxhyphen{}backend implementation of {\hyperref[\detokenize{modules/random:symjax.tensor.random.multivariate_normal}]{\sphinxcrossref{\sphinxcode{\sphinxupquote{multivariate\_normal()}}}}}.
ADDITIONOriginal docstring below.
\begin{quote}

Returns:
A random array with the specified dtype and shape given by
\sphinxcode{\sphinxupquote{shape + mean.shape{[}\sphinxhyphen{}1:{]}}} if \sphinxcode{\sphinxupquote{shape}} is not None, or else
\sphinxcode{\sphinxupquote{broadcast\_shapes(mean.shape{[}:\sphinxhyphen{}1{]}, cov.shape{[}:\sphinxhyphen{}2{]}) + mean.shape{[}\sphinxhyphen{}1:{]}}}.
\end{quote}

\end{fulllineitems}

\index{normal() (in module symjax.tensor.random)@\spxentry{normal()}\spxextra{in module symjax.tensor.random}}

\begin{fulllineitems}
\phantomsection\label{\detokenize{modules/random:symjax.tensor.random.normal}}\pysiglinewithargsret{\sphinxbfcode{\sphinxupquote{normal}}}{\emph{key: jax.numpy.lax\_numpy.ndarray}, \emph{shape: Sequence{[}int{]} = ()}, \emph{dtype: numpy.dtype = \textless{}class \textquotesingle{}numpy.float64\textquotesingle{}\textgreater{}}}{{ $\rightarrow$ jax.numpy.lax\_numpy.ndarray}}
Sample standard normal random values with given shape and float dtype.

LAX\sphinxhyphen{}backend implementation of {\hyperref[\detokenize{modules/random:symjax.tensor.random.normal}]{\sphinxcrossref{\sphinxcode{\sphinxupquote{normal()}}}}}.
ADDITIONOriginal docstring below.
\begin{quote}

Returns:
A random array with the specified shape and dtype.
\end{quote}

\end{fulllineitems}

\index{pareto() (in module symjax.tensor.random)@\spxentry{pareto()}\spxextra{in module symjax.tensor.random}}

\begin{fulllineitems}
\phantomsection\label{\detokenize{modules/random:symjax.tensor.random.pareto}}\pysiglinewithargsret{\sphinxbfcode{\sphinxupquote{pareto}}}{\emph{key}, \emph{b}, \emph{shape=None}, \emph{dtype=\textless{}class \textquotesingle{}numpy.float64\textquotesingle{}\textgreater{}}}{}
Sample Pareto random values with given shape and float dtype.

LAX\sphinxhyphen{}backend implementation of {\hyperref[\detokenize{modules/random:symjax.tensor.random.pareto}]{\sphinxcrossref{\sphinxcode{\sphinxupquote{pareto()}}}}}.
ADDITIONOriginal docstring below.
\begin{quote}

Returns:
A random array with the specified dtype and with shape given by \sphinxcode{\sphinxupquote{shape}} if
\sphinxcode{\sphinxupquote{shape}} is not None, or else by \sphinxcode{\sphinxupquote{b.shape}}.
\end{quote}

\end{fulllineitems}

\index{randint() (in module symjax.tensor.random)@\spxentry{randint()}\spxextra{in module symjax.tensor.random}}

\begin{fulllineitems}
\phantomsection\label{\detokenize{modules/random:symjax.tensor.random.randint}}\pysiglinewithargsret{\sphinxbfcode{\sphinxupquote{randint}}}{\emph{key: jax.numpy.lax\_numpy.ndarray, shape: Sequence{[}int{]}, minval: Union{[}int, jax.numpy.lax\_numpy.ndarray{]}, maxval: Union{[}int, jax.numpy.lax\_numpy.ndarray{]}, dtype: numpy.dtype = \textless{}class \textquotesingle{}numpy.int64\textquotesingle{}\textgreater{}}}{}
Sample uniform random values in {[}minval, maxval) with given shape/dtype.

LAX\sphinxhyphen{}backend implementation of {\hyperref[\detokenize{modules/random:symjax.tensor.random.randint}]{\sphinxcrossref{\sphinxcode{\sphinxupquote{randint()}}}}}.
ADDITIONOriginal docstring below.
\begin{quote}

Returns:
A random array with the specified shape and dtype.
\end{quote}

\end{fulllineitems}

\index{shuffle() (in module symjax.tensor.random)@\spxentry{shuffle()}\spxextra{in module symjax.tensor.random}}

\begin{fulllineitems}
\phantomsection\label{\detokenize{modules/random:symjax.tensor.random.shuffle}}\pysiglinewithargsret{\sphinxbfcode{\sphinxupquote{shuffle}}}{\emph{\DUrole{n}{key}\DUrole{p}{:} \DUrole{n}{jax.numpy.lax\_numpy.ndarray}}, \emph{\DUrole{n}{x}\DUrole{p}{:} \DUrole{n}{jax.numpy.lax\_numpy.ndarray}}, \emph{\DUrole{n}{axis}\DUrole{p}{:} \DUrole{n}{int} \DUrole{o}{=} \DUrole{default_value}{0}}}{{ $\rightarrow$ jax.numpy.lax\_numpy.ndarray}}
Shuffle the elements of an array uniformly at random along an axis.

LAX\sphinxhyphen{}backend implementation of {\hyperref[\detokenize{modules/random:symjax.tensor.random.shuffle}]{\sphinxcrossref{\sphinxcode{\sphinxupquote{shuffle()}}}}}.
ADDITIONOriginal docstring below.
\begin{quote}

Returns:
A shuffled version of x.
\end{quote}

\end{fulllineitems}

\index{truncated\_normal() (in module symjax.tensor.random)@\spxentry{truncated\_normal()}\spxextra{in module symjax.tensor.random}}

\begin{fulllineitems}
\phantomsection\label{\detokenize{modules/random:symjax.tensor.random.truncated_normal}}\pysiglinewithargsret{\sphinxbfcode{\sphinxupquote{truncated\_normal}}}{\emph{key: jax.numpy.lax\_numpy.ndarray, lower: Union{[}float, jax.numpy.lax\_numpy.ndarray{]}, upper: Union{[}float, jax.numpy.lax\_numpy.ndarray{]}, shape: Optional{[}Sequence{[}int{]}{]} = None, dtype: numpy.dtype = \textless{}class \textquotesingle{}numpy.float64\textquotesingle{}\textgreater{}}}{{ $\rightarrow$ jax.numpy.lax\_numpy.ndarray}}
Sample truncated standard normal random values with given shape and dtype.

LAX\sphinxhyphen{}backend implementation of {\hyperref[\detokenize{modules/random:symjax.tensor.random.truncated_normal}]{\sphinxcrossref{\sphinxcode{\sphinxupquote{truncated\_normal()}}}}}.
ADDITIONOriginal docstring below.
\begin{quote}

Returns:
A random array with the specified dtype and shape given by \sphinxcode{\sphinxupquote{shape}} if
\sphinxcode{\sphinxupquote{shape}} is not None, or else by broadcasting \sphinxcode{\sphinxupquote{lower}} and \sphinxcode{\sphinxupquote{upper}}.
\end{quote}

\end{fulllineitems}

\index{uniform() (in module symjax.tensor.random)@\spxentry{uniform()}\spxextra{in module symjax.tensor.random}}

\begin{fulllineitems}
\phantomsection\label{\detokenize{modules/random:symjax.tensor.random.uniform}}\pysiglinewithargsret{\sphinxbfcode{\sphinxupquote{uniform}}}{\emph{key: jax.numpy.lax\_numpy.ndarray}, \emph{shape: Sequence{[}int{]} = ()}, \emph{dtype: numpy.dtype = \textless{}class \textquotesingle{}numpy.float64\textquotesingle{}\textgreater{}}, \emph{minval: Union{[}float}, \emph{jax.numpy.lax\_numpy.ndarray{]} = 0.0}, \emph{maxval: Union{[}float}, \emph{jax.numpy.lax\_numpy.ndarray{]} = 1.0}}{{ $\rightarrow$ jax.numpy.lax\_numpy.ndarray}}
Sample uniform random values in {[}minval, maxval) with given shape/dtype.

LAX\sphinxhyphen{}backend implementation of {\hyperref[\detokenize{modules/random:symjax.tensor.random.uniform}]{\sphinxcrossref{\sphinxcode{\sphinxupquote{uniform()}}}}}.
ADDITIONOriginal docstring below.
\begin{quote}

Returns:
A random array with the specified shape and dtype.
\end{quote}

\end{fulllineitems}

\subsection{\sphinxstyleliteralintitle{\sphinxupquote{symjax.utils}}}
\label{\detokenize{modules/utils:symjax-utils}}\label{\detokenize{modules/utils::doc}}\index{utils (in module symjax)@\spxentry{utils}\spxextra{in module symjax}}

\begin{fulllineitems}
\phantomsection\label{\detokenize{modules/utils:symjax.utils}}\pysigline{\sphinxbfcode{\sphinxupquote{utils}}}
alias of {\hyperref[\detokenize{modules/utils:module-symjax.utils}]{\sphinxcrossref{\sphinxcode{\sphinxupquote{symjax.utils}}}}}

\end{fulllineitems}

\phantomsection\label{\detokenize{modules/utils:module-symjax.utils}}\index{module@\spxentry{module}!symjax.utils@\spxentry{symjax.utils}}\index{symjax.utils@\spxentry{symjax.utils}!module@\spxentry{module}}

\section{Deep Learning}
\label{\detokenize{index:deep-learning}}

\subsection{\sphinxstyleliteralintitle{\sphinxupquote{symjax.datasets}}}
\label{\detokenize{modules/datasets:symjax-datasets}}\label{\detokenize{modules/datasets::doc}}

\subsubsection{Images}
\label{\detokenize{modules/datasets:images}}

\begin{savenotes}\sphinxatlongtablestart\begin{longtable}[c]{\X{1}{2}\X{1}{2}}
\hline

\endfirsthead

\multicolumn{2}{c}%
{\makebox[0pt]{\sphinxtablecontinued{\tablename\ \thetable{} \textendash{} continued from previous page}}}\\
\hline

\endhead

\hline
\multicolumn{2}{r}{\makebox[0pt][r]{\sphinxtablecontinued{continues on next page}}}\\
\endfoot

\endlastfoot

{\hyperref[\detokenize{modules/datasets:symjax.datasets.mnist}]{\sphinxcrossref{\sphinxcode{\sphinxupquote{symjax.datasets.mnist}}}}}
&
Grayscale digit classification.
\\
\hline
{\hyperref[\detokenize{modules/datasets:symjax.datasets.fashionmnist}]{\sphinxcrossref{\sphinxcode{\sphinxupquote{symjax.datasets.fashionmnist}}}}}
&
Grayscale image classification
\\
\hline
{\hyperref[\detokenize{modules/datasets:symjax.datasets.dsprites}]{\sphinxcrossref{\sphinxcode{\sphinxupquote{symjax.datasets.dsprites}}}}}
&
greyscale image classification and disentanglement
\\
\hline
{\hyperref[\detokenize{modules/datasets:symjax.datasets.svhn}]{\sphinxcrossref{\sphinxcode{\sphinxupquote{symjax.datasets.svhn}}}}}
&
Street number classification.
\\
\hline
{\hyperref[\detokenize{modules/datasets:symjax.datasets.cifar10}]{\sphinxcrossref{\sphinxcode{\sphinxupquote{symjax.datasets.cifar10}}}}}
&
Image classification.
\\
\hline
{\hyperref[\detokenize{modules/datasets:symjax.datasets.cifar100}]{\sphinxcrossref{\sphinxcode{\sphinxupquote{symjax.datasets.cifar100}}}}}
&
Image classification.
\\
\hline
{\hyperref[\detokenize{modules/datasets:symjax.datasets.ibeans}]{\sphinxcrossref{\sphinxcode{\sphinxupquote{symjax.datasets.ibeans}}}}}
&
Plant images classification.
\\
\hline
{\hyperref[\detokenize{modules/datasets:symjax.datasets.cassava}]{\sphinxcrossref{\sphinxcode{\sphinxupquote{symjax.datasets.cassava}}}}}
&
Plant images classification.
\\
\hline
{\hyperref[\detokenize{modules/datasets:symjax.datasets.stl10}]{\sphinxcrossref{\sphinxcode{\sphinxupquote{symjax.datasets.stl10}}}}}
&
Image classification with extra unlabeled images.
\\
\hline
{\hyperref[\detokenize{modules/datasets:symjax.datasets.tinyimagenet}]{\sphinxcrossref{\sphinxcode{\sphinxupquote{symjax.datasets.tinyimagenet}}}}}
&
Tiny Imagenet has 200 classes.
\\
\hline
\end{longtable}\sphinxatlongtableend\end{savenotes}

\subsubsection{Audio}
\label{\detokenize{modules/datasets:audio}}

\begin{savenotes}\sphinxatlongtablestart\begin{longtable}[c]{\X{1}{2}\X{1}{2}}
\hline

\endfirsthead

\multicolumn{2}{c}%
{\makebox[0pt]{\sphinxtablecontinued{\tablename\ \thetable{} \textendash{} continued from previous page}}}\\
\hline

\endhead

\hline
\multicolumn{2}{r}{\makebox[0pt][r]{\sphinxtablecontinued{continues on next page}}}\\
\endfoot

\endlastfoot

{\hyperref[\detokenize{modules/datasets:symjax.datasets.audiomnist}]{\sphinxcrossref{\sphinxcode{\sphinxupquote{symjax.datasets.audiomnist}}}}}
&
digit recognition
\\
\hline
{\hyperref[\detokenize{modules/datasets:symjax.datasets.esc}]{\sphinxcrossref{\sphinxcode{\sphinxupquote{symjax.datasets.esc}}}}}
&
ESC\sphinxhyphen{}10/50: Environmental Sound Classification
\\
\hline
{\hyperref[\detokenize{modules/datasets:symjax.datasets.warblr}]{\sphinxcrossref{\sphinxcode{\sphinxupquote{symjax.datasets.warblr}}}}}
&
Binary audio classification, presence or absence of a bird.
\\
\hline
{\hyperref[\detokenize{modules/datasets:symjax.datasets.gtzan}]{\sphinxcrossref{\sphinxcode{\sphinxupquote{symjax.datasets.gtzan}}}}}
&
music genre classification
\\
\hline
{\hyperref[\detokenize{modules/datasets:symjax.datasets.dclde}]{\sphinxcrossref{\sphinxcode{\sphinxupquote{symjax.datasets.dclde}}}}}
&

\\
\hline
{\hyperref[\detokenize{modules/datasets:symjax.datasets.irmas}]{\sphinxcrossref{\sphinxcode{\sphinxupquote{symjax.datasets.irmas}}}}}
&
music instrument classification
\\
\hline
{\hyperref[\detokenize{modules/datasets:symjax.datasets.vocalset}]{\sphinxcrossref{\sphinxcode{\sphinxupquote{symjax.datasets.vocalset}}}}}
&
singer/technique/vowel of singing voices
\\
\hline
{\hyperref[\detokenize{modules/datasets:symjax.datasets.freefield1010}]{\sphinxcrossref{\sphinxcode{\sphinxupquote{symjax.datasets.freefield1010}}}}}
&
Audio binary classification, presence or absence of bird songs.
\\
\hline
{\hyperref[\detokenize{modules/datasets:symjax.datasets.birdvox_70k}]{\sphinxcrossref{\sphinxcode{\sphinxupquote{symjax.datasets.birdvox\_70k}}}}}
&
a dataset for avian flight call detection in half\sphinxhyphen{}second clips
\\
\hline
{\hyperref[\detokenize{modules/datasets:symjax.datasets.birdvox_dcase_20k}]{\sphinxcrossref{\sphinxcode{\sphinxupquote{symjax.datasets.birdvox\_dcase\_20k}}}}}
&
Binary bird detection classification
\\
\hline
\end{longtable}\sphinxatlongtableend\end{savenotes}

\subsubsection{Detailed description (Image)}
\label{\detokenize{modules/datasets:detailed-description-image}}\index{mnist (class in symjax.datasets)@\spxentry{mnist}\spxextra{class in symjax.datasets}}

\begin{fulllineitems}
\phantomsection\label{\detokenize{modules/datasets:symjax.datasets.mnist}}\pysigline{\sphinxbfcode{\sphinxupquote{class }}\sphinxbfcode{\sphinxupquote{mnist}}}
Grayscale digit classification.

The \sphinxhref{http://yann.lecun.com/exdb/mnist/}{MNIST} database of handwritten
digits, available from this page, has a training set of 60,000 examples,
and a test set of 10,000 examples. It is a subset of a larger set available
from NIST. The digits have been size\sphinxhyphen{}normalized and centered in a
fixed\sphinxhyphen{}size image. It is a good database for people who want to try learning
techniques and pattern recognition methods on real\sphinxhyphen{}world data while
spending minimal efforts on preprocessing and formatting.
\index{download() (mnist static method)@\spxentry{download()}\spxextra{mnist static method}}

\begin{fulllineitems}
\phantomsection\label{\detokenize{modules/datasets:symjax.datasets.mnist.download}}\pysiglinewithargsret{\sphinxbfcode{\sphinxupquote{static }}\sphinxbfcode{\sphinxupquote{download}}}{\emph{\DUrole{n}{path}}}{}
Download the MNIST dataset and store the result into the given
path
\begin{quote}\begin{description}
\item[{Parameters}] \leavevmode
\sphinxstyleliteralstrong{\sphinxupquote{path}} (\sphinxstyleliteralemphasis{\sphinxupquote{str}}) \textendash{} the path where the downloaded files will be stored. If the
directory does not exist, it is created.

\end{description}\end{quote}

\end{fulllineitems}

\index{load() (mnist static method)@\spxentry{load()}\spxextra{mnist static method}}

\begin{fulllineitems}
\phantomsection\label{\detokenize{modules/datasets:symjax.datasets.mnist.load}}\pysiglinewithargsret{\sphinxbfcode{\sphinxupquote{static }}\sphinxbfcode{\sphinxupquote{load}}}{\emph{\DUrole{n}{path}\DUrole{o}{=}\DUrole{default_value}{None}}}{}~\begin{quote}\begin{description}
\item[{Parameters}] \leavevmode
\sphinxstyleliteralstrong{\sphinxupquote{path}} (\sphinxstyleliteralemphasis{\sphinxupquote{str}}\sphinxstyleliteralemphasis{\sphinxupquote{ (}}\sphinxstyleliteralemphasis{\sphinxupquote{optional}}\sphinxstyleliteralemphasis{\sphinxupquote{)}}) \textendash{} default (\$DATASET\_PATH), the path to look for the data and
where the data will be downloaded if not present

\item[{Returns}] \leavevmode
\begin{itemize}
\item {} 
\sphinxstylestrong{train\_images} (\sphinxstyleemphasis{array})

\item {} 
\sphinxstylestrong{train\_labels} (\sphinxstyleemphasis{array})

\item {} 
\sphinxstylestrong{valid\_images} (\sphinxstyleemphasis{array})

\item {} 
\sphinxstylestrong{valid\_labels} (\sphinxstyleemphasis{array})

\item {} 
\sphinxstylestrong{test\_images} (\sphinxstyleemphasis{array})

\item {} 
\sphinxstylestrong{test\_labels} (\sphinxstyleemphasis{array})

\end{itemize}

\end{description}\end{quote}

\end{fulllineitems}

\end{fulllineitems}

\index{fashionmnist (class in symjax.datasets)@\spxentry{fashionmnist}\spxextra{class in symjax.datasets}}

\begin{fulllineitems}
\phantomsection\label{\detokenize{modules/datasets:symjax.datasets.fashionmnist}}\pysigline{\sphinxbfcode{\sphinxupquote{class }}\sphinxbfcode{\sphinxupquote{fashionmnist}}}
Grayscale image classification

\sphinxhref{https://jobs.zalando.com/tech/}{Zalando} ‘s article image classification.
\sphinxhref{https://github.com/zalandoresearch/fashion-mnist}{Fashion\sphinxhyphen{}MNIST} is
a dataset of \sphinxhref{https://jobs.zalando.com/tech/}{Zalando} ‘s article
images consisting of a training set of 60,000 examples and a test set
of 10,000 examples. Each example is a 28x28 grayscale image, associated
with a label from 10 classes. We intend Fashion\sphinxhyphen{}MNIST to serve as a direct
drop\sphinxhyphen{}in replacement for the original MNIST dataset for benchmarking
machine learning algorithms. It shares the same image size and structure
of training and testing splits.
\index{download() (fashionmnist method)@\spxentry{download()}\spxextra{fashionmnist method}}

\begin{fulllineitems}
\phantomsection\label{\detokenize{modules/datasets:symjax.datasets.fashionmnist.download}}\pysiglinewithargsret{\sphinxbfcode{\sphinxupquote{download}}}{}{}
Download the fashion\sphinxhyphen{}MNIST dataset and store the result into the given
path
\begin{quote}\begin{description}
\item[{Parameters}] \leavevmode
\sphinxstyleliteralstrong{\sphinxupquote{path}} (\sphinxstyleliteralemphasis{\sphinxupquote{str}}) \textendash{} the path where the downloaded files will be stored. If the
directory does not exist, it is created.

\end{description}\end{quote}

\end{fulllineitems}

\index{load() (fashionmnist method)@\spxentry{load()}\spxextra{fashionmnist method}}

\begin{fulllineitems}
\phantomsection\label{\detokenize{modules/datasets:symjax.datasets.fashionmnist.load}}\pysiglinewithargsret{\sphinxbfcode{\sphinxupquote{load}}}{}{}~\begin{quote}\begin{description}
\item[{Parameters}] \leavevmode
\sphinxstyleliteralstrong{\sphinxupquote{path}} (\sphinxstyleliteralemphasis{\sphinxupquote{str}}\sphinxstyleliteralemphasis{\sphinxupquote{ (}}\sphinxstyleliteralemphasis{\sphinxupquote{optional}}\sphinxstyleliteralemphasis{\sphinxupquote{)}}) \textendash{} default (\$DATASET\_PATH), the path to look for the data and
where the data will be downloaded if not present

\item[{Returns}] \leavevmode
\begin{itemize}
\item {} 
\sphinxstylestrong{train\_images} (\sphinxstyleemphasis{array})

\item {} 
\sphinxstylestrong{train\_labels} (\sphinxstyleemphasis{array})

\item {} 
\sphinxstylestrong{test\_images} (\sphinxstyleemphasis{array})

\item {} 
\sphinxstylestrong{test\_labels} (\sphinxstyleemphasis{array})

\end{itemize}

\end{description}\end{quote}

\end{fulllineitems}

\end{fulllineitems}

\index{dsprites (class in symjax.datasets)@\spxentry{dsprites}\spxextra{class in symjax.datasets}}

\begin{fulllineitems}
\phantomsection\label{\detokenize{modules/datasets:symjax.datasets.dsprites}}\pysigline{\sphinxbfcode{\sphinxupquote{class }}\sphinxbfcode{\sphinxupquote{dsprites}}}
greyscale image classification and disentanglement

This dataset consists of 737,280 images of 2D shapes, procedurally generated
from 5 ground truth independent latent factors, controlling the shape, scale,
rotation and position of a sprite. This data can be used to assess the
disentanglement properties of unsupervised learning methods.

dSprites is a dataset of 2D shapes procedurally generated from 6 ground
truth independent latent factors. These factors are color, shape, scale,
rotation, x and y positions of a sprite.

All possible combinations of these latents are present exactly once,
generating N = 737280 total images.
\index{download() (dsprites method)@\spxentry{download()}\spxextra{dsprites method}}

\begin{fulllineitems}
\phantomsection\label{\detokenize{modules/datasets:symjax.datasets.dsprites.download}}\pysiglinewithargsret{\sphinxbfcode{\sphinxupquote{download}}}{}{}
Download the MNIST dataset and store the result into the given
path
\begin{quote}\begin{description}
\item[{Parameters}] \leavevmode
\sphinxstyleliteralstrong{\sphinxupquote{path}} (\sphinxstyleliteralemphasis{\sphinxupquote{str}}) \textendash{} the path where the downloaded files will be stored. If the
directory does not exist, it is created.

\end{description}\end{quote}

\end{fulllineitems}

\index{load() (dsprites method)@\spxentry{load()}\spxextra{dsprites method}}

\begin{fulllineitems}
\phantomsection\label{\detokenize{modules/datasets:symjax.datasets.dsprites.load}}\pysiglinewithargsret{\sphinxbfcode{\sphinxupquote{load}}}{}{}~\begin{quote}\begin{description}
\item[{Parameters}] \leavevmode
\sphinxstyleliteralstrong{\sphinxupquote{path}} (\sphinxstyleliteralemphasis{\sphinxupquote{str}}\sphinxstyleliteralemphasis{\sphinxupquote{ (}}\sphinxstyleliteralemphasis{\sphinxupquote{optional}}\sphinxstyleliteralemphasis{\sphinxupquote{)}}) \textendash{} default (\$DATASET\_PATH), the path to look for the data and
where the data will be downloaded if not present

\item[{Returns}] \leavevmode
\begin{itemize}
\item {} 
\sphinxstylestrong{images} (\sphinxstyleemphasis{array})

\item {} 
\sphinxstylestrong{latent} (\sphinxstyleemphasis{array})

\item {} 
\sphinxstylestrong{classes} (\sphinxstyleemphasis{array})

\end{itemize}

\end{description}\end{quote}

\end{fulllineitems}

\end{fulllineitems}

\index{svhn (class in symjax.datasets)@\spxentry{svhn}\spxextra{class in symjax.datasets}}

\begin{fulllineitems}
\phantomsection\label{\detokenize{modules/datasets:symjax.datasets.svhn}}\pysigline{\sphinxbfcode{\sphinxupquote{class }}\sphinxbfcode{\sphinxupquote{svhn}}}
Street number classification.

The \sphinxhref{http://ufldl.stanford.edu/housenumbers/}{SVHN}
dataset is a real\sphinxhyphen{}world
image dataset for developing machine learning and object
recognition algorithms with minimal requirement on data
preprocessing and formatting. It can be seen as similar in flavor
to MNIST (e.g., the images are of small cropped digits), but
incorporates an order of magnitude more labeled data (over 600,000
digit images) and comes from a significantly harder, unsolved,
real world problem (recognizing digits and numbers in natural
scene images). SVHN is obtained from house numbers in Google
Street View images.
\index{download() (svhn method)@\spxentry{download()}\spxextra{svhn method}}

\begin{fulllineitems}
\phantomsection\label{\detokenize{modules/datasets:symjax.datasets.svhn.download}}\pysiglinewithargsret{\sphinxbfcode{\sphinxupquote{download}}}{}{}
Download the SVHN dataset and store the result into the given
path
\begin{quote}\begin{description}
\item[{Parameters}] \leavevmode
\sphinxstyleliteralstrong{\sphinxupquote{path}} (\sphinxstyleliteralemphasis{\sphinxupquote{str}}) \textendash{} the path where the downloaded files will be stored. If the
directory does not exist, it is created.

\end{description}\end{quote}

\end{fulllineitems}

\index{load() (svhn method)@\spxentry{load()}\spxextra{svhn method}}

\begin{fulllineitems}
\phantomsection\label{\detokenize{modules/datasets:symjax.datasets.svhn.load}}\pysiglinewithargsret{\sphinxbfcode{\sphinxupquote{load}}}{}{}~\begin{quote}\begin{description}
\item[{Parameters}] \leavevmode
\sphinxstyleliteralstrong{\sphinxupquote{path}} (\sphinxstyleliteralemphasis{\sphinxupquote{str}}\sphinxstyleliteralemphasis{\sphinxupquote{ (}}\sphinxstyleliteralemphasis{\sphinxupquote{optional}}\sphinxstyleliteralemphasis{\sphinxupquote{)}}) \textendash{} default \$DATASET\_PATH, the path to look for the data and
where the data will be downloaded if not present

\item[{Returns}] \leavevmode
\begin{itemize}
\item {} 
\sphinxstylestrong{train\_images} (\sphinxstyleemphasis{array})

\item {} 
\sphinxstylestrong{train\_labels} (\sphinxstyleemphasis{array})

\item {} 
\sphinxstylestrong{test\_images} (\sphinxstyleemphasis{array})

\item {} 
\sphinxstylestrong{test\_labels} (\sphinxstyleemphasis{array})

\end{itemize}

\end{description}\end{quote}

\end{fulllineitems}

\end{fulllineitems}

\index{cifar10 (class in symjax.datasets)@\spxentry{cifar10}\spxextra{class in symjax.datasets}}

\begin{fulllineitems}
\phantomsection\label{\detokenize{modules/datasets:symjax.datasets.cifar10}}\pysigline{\sphinxbfcode{\sphinxupquote{class }}\sphinxbfcode{\sphinxupquote{cifar10}}}
Image classification.

The \sphinxhref{https://www.cs.toronto.edu/~kriz/cifar.html}{CIFAR\sphinxhyphen{}10} dataset
was collected by Alex Krizhevsky, Vinod Nair, and Geoffrey
Hinton. It consists of 60000 32x32 colour images in 10 classes, with
6000 images per class. There are 50000 training images and 10000 test images.
The dataset is divided into five training batches and one test batch,
each with 10000 images. The test batch contains exactly 1000 randomly
selected images from each class. The training batches contain the
remaining images in random order, but some training batches may
contain more images from one class than another. Between them, the
training batches contain exactly 5000 images from each class.
\index{download() (cifar10 method)@\spxentry{download()}\spxextra{cifar10 method}}

\begin{fulllineitems}
\phantomsection\label{\detokenize{modules/datasets:symjax.datasets.cifar10.download}}\pysiglinewithargsret{\sphinxbfcode{\sphinxupquote{download}}}{}{}
Download the MNIST dataset and store the result into the given
path
\begin{quote}\begin{description}
\item[{Parameters}] \leavevmode
\sphinxstyleliteralstrong{\sphinxupquote{path}} (\sphinxstyleliteralemphasis{\sphinxupquote{str}}) \textendash{} the path where the downloaded files will be stored. If the
directory does not exist, it is created.

\end{description}\end{quote}

\end{fulllineitems}

\index{load() (cifar10 method)@\spxentry{load()}\spxextra{cifar10 method}}

\begin{fulllineitems}
\phantomsection\label{\detokenize{modules/datasets:symjax.datasets.cifar10.load}}\pysiglinewithargsret{\sphinxbfcode{\sphinxupquote{load}}}{}{}~\begin{quote}\begin{description}
\item[{Parameters}] \leavevmode
\sphinxstyleliteralstrong{\sphinxupquote{path}} (\sphinxstyleliteralemphasis{\sphinxupquote{str}}\sphinxstyleliteralemphasis{\sphinxupquote{ (}}\sphinxstyleliteralemphasis{\sphinxupquote{optional}}\sphinxstyleliteralemphasis{\sphinxupquote{)}}) \textendash{} default (\$DATASET\_PATH), the path to look for the data and
where the data will be downloaded if not present

\item[{Returns}] \leavevmode
\begin{itemize}
\item {} 
\sphinxstylestrong{train\_images} (\sphinxstyleemphasis{array})

\item {} 
\sphinxstylestrong{train\_labels} (\sphinxstyleemphasis{array})

\item {} 
\sphinxstylestrong{test\_images} (\sphinxstyleemphasis{array})

\item {} 
\sphinxstylestrong{test\_labels} (\sphinxstyleemphasis{array})

\end{itemize}

\end{description}\end{quote}

\end{fulllineitems}

\end{fulllineitems}

\index{cifar100 (class in symjax.datasets)@\spxentry{cifar100}\spxextra{class in symjax.datasets}}

\begin{fulllineitems}
\phantomsection\label{\detokenize{modules/datasets:symjax.datasets.cifar100}}\pysigline{\sphinxbfcode{\sphinxupquote{class }}\sphinxbfcode{\sphinxupquote{cifar100}}}
Image classification.

The \sphinxhref{https://www.cs.toronto.edu/~kriz/cifar.html}{CIFAR\sphinxhyphen{}100} dataset is
just like the CIFAR\sphinxhyphen{}10, except it has 100 classes containing 600 images
each. There are 500 training images and 100 testing images per class.
The 100 classes in the CIFAR\sphinxhyphen{}100 are grouped into 20 superclasses. Each
image comes with a “fine” label (the class to which it belongs) and a
“coarse” label (the superclass to which it belongs).
\index{download() (cifar100 method)@\spxentry{download()}\spxextra{cifar100 method}}

\begin{fulllineitems}
\phantomsection\label{\detokenize{modules/datasets:symjax.datasets.cifar100.download}}\pysiglinewithargsret{\sphinxbfcode{\sphinxupquote{download}}}{}{}
Download the CIFAR100 dataset and store the result into the given
path
\begin{quote}\begin{description}
\item[{Parameters}] \leavevmode
\sphinxstyleliteralstrong{\sphinxupquote{path}} (\sphinxstyleliteralemphasis{\sphinxupquote{str}}) \textendash{} the path where the downloaded files will be stored. If the
directory does not exist, it is created.

\end{description}\end{quote}

\end{fulllineitems}

\end{fulllineitems}

\index{ibeans (class in symjax.datasets)@\spxentry{ibeans}\spxextra{class in symjax.datasets}}

\begin{fulllineitems}
\phantomsection\label{\detokenize{modules/datasets:symjax.datasets.ibeans}}\pysigline{\sphinxbfcode{\sphinxupquote{class }}\sphinxbfcode{\sphinxupquote{ibeans}}}
Plant images classification.

This dataset is of leaf images taken in the field in different
districts in Uganda by the Makerere AI lab in collaboration with the
National Crops Resources Research Institute (NaCRRI), the national
body in charge of research in agriculture in Uganda.

The goal is to build a robust machine learning model that is able
to distinguish between diseases in the Bean plants. Beans are an
important cereal food crop for Africa grown by many small\sphinxhyphen{}holder
farmers \sphinxhyphen{} they are a significant source of proteins for school\sphinxhyphen{}age
going children in East Africa.

The data is of leaf images representing 3 classes: the healthy class of
images, and two disease classes including Angular Leaf Spot and Bean
Rust diseases. The model should be able to distinguish between these 3
classes with high accuracy. The end goal is to build a robust, model
that can be deployed on a mobile device and used in the field by a
farmer.

The data includes leaf images taken in the field. The figure above
depicts examples of the types of images per class. Images were taken
from the field/garden a basic smartphone.

The images were then annotated by experts from NaCRRI who determined
for each image which disease was manifested. The experts were part of
the data collection team and images were annotated directly during the
data collection process in the field.

Class   Examples
Healthy class   428
Angular Leaf Spot   432
Bean Rust   436
Total:  1,296

Data Released   20\sphinxhyphen{}January\sphinxhyphen{}2020
License     MIT
Credits     Makerere AI Lab
\index{download() (ibeans static method)@\spxentry{download()}\spxextra{ibeans static method}}

\begin{fulllineitems}
\phantomsection\label{\detokenize{modules/datasets:symjax.datasets.ibeans.download}}\pysiglinewithargsret{\sphinxbfcode{\sphinxupquote{static }}\sphinxbfcode{\sphinxupquote{download}}}{\emph{\DUrole{n}{path}}}{}
Download the ibeans dataset and store the result into the given
path
\begin{quote}\begin{description}
\item[{Parameters}] \leavevmode
\sphinxstyleliteralstrong{\sphinxupquote{path}} (\sphinxstyleliteralemphasis{\sphinxupquote{str}}) \textendash{} the path where the downloaded files will be stored. If the
directory does not exist, it is created.

\end{description}\end{quote}

\end{fulllineitems}

\index{load() (ibeans static method)@\spxentry{load()}\spxextra{ibeans static method}}

\begin{fulllineitems}
\phantomsection\label{\detokenize{modules/datasets:symjax.datasets.ibeans.load}}\pysiglinewithargsret{\sphinxbfcode{\sphinxupquote{static }}\sphinxbfcode{\sphinxupquote{load}}}{\emph{\DUrole{n}{path}\DUrole{o}{=}\DUrole{default_value}{None}}}{}~\begin{quote}\begin{description}
\item[{Parameters}] \leavevmode
\sphinxstyleliteralstrong{\sphinxupquote{path}} (\sphinxstyleliteralemphasis{\sphinxupquote{str}}\sphinxstyleliteralemphasis{\sphinxupquote{ (}}\sphinxstyleliteralemphasis{\sphinxupquote{optional}}\sphinxstyleliteralemphasis{\sphinxupquote{)}}) \textendash{} default (\$DATASET\_PATH), the path to look for the data and
where the data will be downloaded if not present

\item[{Returns}] \leavevmode
\begin{itemize}
\item {} 
\sphinxstylestrong{train\_images} (\sphinxstyleemphasis{array})

\item {} 
\sphinxstylestrong{train\_labels} (\sphinxstyleemphasis{array})

\item {} 
\sphinxstylestrong{valid\_images} (\sphinxstyleemphasis{array})

\item {} 
\sphinxstylestrong{valid\_labels} (\sphinxstyleemphasis{array})

\item {} 
\sphinxstylestrong{test\_images} (\sphinxstyleemphasis{array})

\item {} 
\sphinxstylestrong{test\_labels} (\sphinxstyleemphasis{array})

\end{itemize}

\end{description}\end{quote}

\end{fulllineitems}

\end{fulllineitems}

\index{cassava (class in symjax.datasets)@\spxentry{cassava}\spxextra{class in symjax.datasets}}

\begin{fulllineitems}
\phantomsection\label{\detokenize{modules/datasets:symjax.datasets.cassava}}\pysigline{\sphinxbfcode{\sphinxupquote{class }}\sphinxbfcode{\sphinxupquote{cassava}}}
Plant images classification.

The data consists of two folders, a training folder that contains 5
subfolders that contain the respective images for the different 5 classes
and a test folder containing test images.

Participants are to train their models using the images in the training
folder and provide a submission file like the sample provided which contains
the image name exactly matching the image name in the test folder and the
corresponding class prediction with labels corresponding to the disease
categories, cmd, healthy, cgm, cbsd, cbb.

Please cite this paper if you use the dataset for your project:
\sphinxurl{https://arxiv.org/pdf/1908.02900.pdf}
\index{download() (cassava static method)@\spxentry{download()}\spxextra{cassava static method}}

\begin{fulllineitems}
\phantomsection\label{\detokenize{modules/datasets:symjax.datasets.cassava.download}}\pysiglinewithargsret{\sphinxbfcode{\sphinxupquote{static }}\sphinxbfcode{\sphinxupquote{download}}}{\emph{\DUrole{n}{path}}}{}
Download the cassava dataset and store the result into the given
path
\begin{quote}\begin{description}
\item[{Parameters}] \leavevmode
\sphinxstyleliteralstrong{\sphinxupquote{path}} (\sphinxstyleliteralemphasis{\sphinxupquote{str}}) \textendash{} the path where the downloaded files will be stored. If the
directory does not exist, it is created.

\end{description}\end{quote}

\end{fulllineitems}

\index{load() (cassava static method)@\spxentry{load()}\spxextra{cassava static method}}

\begin{fulllineitems}
\phantomsection\label{\detokenize{modules/datasets:symjax.datasets.cassava.load}}\pysiglinewithargsret{\sphinxbfcode{\sphinxupquote{static }}\sphinxbfcode{\sphinxupquote{load}}}{\emph{\DUrole{n}{path}\DUrole{o}{=}\DUrole{default_value}{None}}}{}~\begin{quote}\begin{description}
\item[{Parameters}] \leavevmode
\sphinxstyleliteralstrong{\sphinxupquote{path}} (\sphinxstyleliteralemphasis{\sphinxupquote{str}}\sphinxstyleliteralemphasis{\sphinxupquote{ (}}\sphinxstyleliteralemphasis{\sphinxupquote{optional}}\sphinxstyleliteralemphasis{\sphinxupquote{)}}) \textendash{} default (\$DATASET\_PATH), the path to look for the data and
where the data will be downloaded if not present

\item[{Returns}] \leavevmode
\begin{itemize}
\item {} 
\sphinxstylestrong{train\_images} (\sphinxstyleemphasis{array})

\item {} 
\sphinxstylestrong{train\_labels} (\sphinxstyleemphasis{array})

\item {} 
\sphinxstylestrong{valid\_images} (\sphinxstyleemphasis{array})

\item {} 
\sphinxstylestrong{valid\_labels} (\sphinxstyleemphasis{array})

\item {} 
\sphinxstylestrong{test\_images} (\sphinxstyleemphasis{array})

\item {} 
\sphinxstylestrong{test\_labels} (\sphinxstyleemphasis{array})

\end{itemize}

\end{description}\end{quote}

\end{fulllineitems}

\end{fulllineitems}

\index{stl10 (class in symjax.datasets)@\spxentry{stl10}\spxextra{class in symjax.datasets}}

\begin{fulllineitems}
\phantomsection\label{\detokenize{modules/datasets:symjax.datasets.stl10}}\pysigline{\sphinxbfcode{\sphinxupquote{class }}\sphinxbfcode{\sphinxupquote{stl10}}}
Image classification with extra unlabeled images.

The \sphinxhref{https://cs.stanford.edu/~acoates/stl10/}{STL\sphinxhyphen{}10} dataset is an image
recognition dataset for developing unsupervised feature learning,
deep learning, self\sphinxhyphen{}taught learning algorithms. It is inspired by the
CIFAR\sphinxhyphen{}10 dataset but with
some modifications. In particular, each class has fewer labeled
training examples than in CIFAR\sphinxhyphen{}10, but a very
large set of unlabeled examples is provided to learn image models prior
to supervised training. The primary challenge is to make use of the
unlabeled data (which comes from a similar but different distribution from
the labeled data) to build a useful prior. We also expect that the higher
resolution of this dataset (96x96) will make it a challenging benchmark
for developing more scalable unsupervised learning methods.
\index{load() (stl10 static method)@\spxentry{load()}\spxextra{stl10 static method}}

\begin{fulllineitems}
\phantomsection\label{\detokenize{modules/datasets:symjax.datasets.stl10.load}}\pysiglinewithargsret{\sphinxbfcode{\sphinxupquote{static }}\sphinxbfcode{\sphinxupquote{load}}}{\emph{\DUrole{n}{path}\DUrole{o}{=}\DUrole{default_value}{None}}}{}
load the data
\begin{quote}\begin{description}
\item[{Parameters}] \leavevmode
\sphinxstyleliteralstrong{\sphinxupquote{path}} (\sphinxstyleliteralemphasis{\sphinxupquote{str}}\sphinxstyleliteralemphasis{\sphinxupquote{ (}}\sphinxstyleliteralemphasis{\sphinxupquote{optional}}\sphinxstyleliteralemphasis{\sphinxupquote{)}}) \textendash{} the path to look for the data and where it will be downloaded if
not present

\item[{Returns}] \leavevmode
\begin{itemize}
\item {} 
\sphinxstylestrong{train\_images} (\sphinxstyleemphasis{array}) \textendash{} the training images

\item {} 
\sphinxstylestrong{train\_labels} (\sphinxstyleemphasis{array}) \textendash{} the training labels

\item {} 
\sphinxstylestrong{test\_images} (\sphinxstyleemphasis{array}) \textendash{} the test images

\item {} 
\sphinxstylestrong{test\_labels} (\sphinxstyleemphasis{array}) \textendash{} the test labels

\item {} 
\sphinxstylestrong{extra\_images} (\sphinxstyleemphasis{array}) \textendash{} the unlabeled additional images

\end{itemize}

\end{description}\end{quote}

\end{fulllineitems}

\end{fulllineitems}

\index{tinyimagenet (class in symjax.datasets)@\spxentry{tinyimagenet}\spxextra{class in symjax.datasets}}

\begin{fulllineitems}
\phantomsection\label{\detokenize{modules/datasets:symjax.datasets.tinyimagenet}}\pysigline{\sphinxbfcode{\sphinxupquote{class }}\sphinxbfcode{\sphinxupquote{tinyimagenet}}}
Tiny Imagenet has 200 classes. Each class has 500 training images, 50
validation images, and 50 test images. We have released the training and
validation sets with images and annotations. We provide both class labels an
bounding boxes as annotations; however, you are asked only to predict the
class label of each image without localizing the objects. The test set is
released without labels. You can download the whole tiny ImageNet dataset
here.

\end{fulllineitems}

\subsubsection{Detailed description (Audio)}
\label{\detokenize{modules/datasets:detailed-description-audio}}\index{audiomnist (class in symjax.datasets)@\spxentry{audiomnist}\spxextra{class in symjax.datasets}}

\begin{fulllineitems}
\phantomsection\label{\detokenize{modules/datasets:symjax.datasets.audiomnist}}\pysigline{\sphinxbfcode{\sphinxupquote{class }}\sphinxbfcode{\sphinxupquote{audiomnist}}}~\begin{description}
\item[{digit recognition}] \leavevmode
\sphinxurl{https://github.com/soerenab/AudioMNIST}

\end{description}

A simple audio/speech dataset consisting of recordings of spoken digits in
wav files at 8kHz. The recordings are trimmed so that they have near
minimal silence at the beginnings and ends.

FSDD is an open dataset, which means it will grow over time as data is
contributed. In order to enable reproducibility and accurate citation the
dataset is versioned using Zenodo DOI as well as git tags.

Current status
\begin{quote}

4 speakers
2,000 recordings (50 of each digit per speaker)
English pronunciations
\end{quote}

\end{fulllineitems}

\index{esc (class in symjax.datasets)@\spxentry{esc}\spxextra{class in symjax.datasets}}

\begin{fulllineitems}
\phantomsection\label{\detokenize{modules/datasets:symjax.datasets.esc}}\pysigline{\sphinxbfcode{\sphinxupquote{class }}\sphinxbfcode{\sphinxupquote{esc}}}
ESC\sphinxhyphen{}10/50: Environmental Sound Classification

\sphinxurl{https://github.com/karolpiczak/ESC-50\#download}

The ESC\sphinxhyphen{}50 dataset is a labeled collection of 2000 environmental audio
recordings suitable for benchmarking methods of environmental sound
classification.

The dataset consists of 5\sphinxhyphen{}second\sphinxhyphen{}long recordings organized into 50
semantical classes (with 40 examples per class) loosely arranged into 5
major categories:
\begin{quote}

Animals
Natural soundscapes \& water sounds
Human, non\sphinxhyphen{}speech sounds
Interior/domestic sounds
Exterior/urban noises
\end{quote}

Clips in this dataset have been manually extracted from public field
recordings gathered by the Freesound.org project. The dataset has been
prearranged into 5 folds for comparable cross\sphinxhyphen{}validation, making sure
that fragments from the same original source file are contained in a
single fold.
\index{load() (esc method)@\spxentry{load()}\spxextra{esc method}}

\begin{fulllineitems}
\phantomsection\label{\detokenize{modules/datasets:symjax.datasets.esc.load}}\pysiglinewithargsret{\sphinxbfcode{\sphinxupquote{load}}}{}{}
ESC 50.

\sphinxurl{https://github.com/karolpiczak/ESC-50\#download}
\begin{quote}\begin{description}
\item[{Parameters}] \leavevmode
\sphinxstyleliteralstrong{\sphinxupquote{path}} (\sphinxstyleliteralemphasis{\sphinxupquote{str}}\sphinxstyleliteralemphasis{\sphinxupquote{ (}}\sphinxstyleliteralemphasis{\sphinxupquote{optional}}\sphinxstyleliteralemphasis{\sphinxupquote{)}}) \textendash{} default \$DATASET\_path), the path to look for the data and
where the data will be downloaded if not present

\item[{Returns}] \leavevmode
\begin{itemize}
\item {} 
\sphinxstylestrong{wavs} (\sphinxstyleemphasis{array}) \textendash{} the wavs as a numpy array (matrix) with first dimension the data
and second dimension time

\item {} 
\sphinxstylestrong{fine\_labels} (\sphinxstyleemphasis{array}) \textendash{} the labels of the final classes (50 different ones) as a integer
vector

\item {} 
\sphinxstylestrong{coarse\_labels} (\sphinxstyleemphasis{array}) \textendash{} the labels of the classes big cateogry (5 of them)

\item {} 
\sphinxstylestrong{folds} (\sphinxstyleemphasis{array}) \textendash{} the fold as an integer from 1 to 5 specifying how to split the data
one should not split a fold into train and set as it would
make the same recording (but different subparts) be present in train
and test, biasing optimistically the results.

\item {} 
\sphinxstylestrong{esc10} (\sphinxstyleemphasis{array}) \textendash{} the boolean vector specifying if the corresponding datum (wav, label,
…) is in the ESC\sphinxhyphen{}10 dataset or not. That is, to load the ESC\sphinxhyphen{}10
dataset simply load ESC\sphinxhyphen{}50 and use this boolean vector to extract
only the ESC\sphinxhyphen{}10 data.

\end{itemize}

\end{description}\end{quote}

\end{fulllineitems}

\end{fulllineitems}

\index{warblr (class in symjax.datasets)@\spxentry{warblr}\spxextra{class in symjax.datasets}}

\begin{fulllineitems}
\phantomsection\label{\detokenize{modules/datasets:symjax.datasets.warblr}}\pysigline{\sphinxbfcode{\sphinxupquote{class }}\sphinxbfcode{\sphinxupquote{warblr}}}
Binary audio classification, presence or absence of a bird.

\sphinxhref{http://machine-listening.eecs.qmul.ac.uk/bird-audio-detection-challenge/\#downloads}{Warblr}
comes from a UK bird\sphinxhyphen{}sound crowdsourcing
research spinout called Warblr. From this initiative we have
10,000 ten\sphinxhyphen{}second smartphone audio recordings from around the UK.
The audio totals around 44 hours duration. The audio will be
published by Warblr under a Creative Commons licence. The audio
covers a wide distribution of UK locations and environments, and
includes weather noise, traffic noise, human speech and even human
bird imitations. It is directly representative of the data that is
collected from a mobile crowdsourcing initiative.
\index{download() (warblr method)@\spxentry{download()}\spxextra{warblr method}}

\begin{fulllineitems}
\phantomsection\label{\detokenize{modules/datasets:symjax.datasets.warblr.download}}\pysiglinewithargsret{\sphinxbfcode{\sphinxupquote{download}}}{}{}
Download the data

\end{fulllineitems}

\index{load() (warblr method)@\spxentry{load()}\spxextra{warblr method}}

\begin{fulllineitems}
\phantomsection\label{\detokenize{modules/datasets:symjax.datasets.warblr.load}}\pysiglinewithargsret{\sphinxbfcode{\sphinxupquote{load}}}{}{}
Load the data given a path

\end{fulllineitems}

\end{fulllineitems}

\index{gtzan (class in symjax.datasets)@\spxentry{gtzan}\spxextra{class in symjax.datasets}}

\begin{fulllineitems}
\phantomsection\label{\detokenize{modules/datasets:symjax.datasets.gtzan}}\pysigline{\sphinxbfcode{\sphinxupquote{class }}\sphinxbfcode{\sphinxupquote{gtzan}}}
music genre classification

This dataset was used for the well known paper in genre classification
“Musical genre classification of audio signals” by G. Tzanetakis
and P. Cook in IEEE Transactions on Audio and Speech Processing 2002.

Unfortunately the database was collected gradually and very early on in my
research so I have no titles (and obviously no copyright permission etc).
The files were collected in 2000\sphinxhyphen{}2001 from a variety of sources including
personal CDs, radio, microphone recordings, in order to represent a variety
of recording conditions. Nevetheless I have been providing it to researchers
upon request mainly for comparison purposes etc. Please contact George
Tzanetakis (\sphinxhref{mailto:gtzan@cs.uvic.ca}{gtzan@cs.uvic.ca}) if you intend to publish experimental results
using this dataset.

There are some practical and conceptual issues with this dataset, described
in “The GTZAN dataset: Its contents, its faults, their effects on
evaluation, and its future use” by B. Sturm on arXiv 2013.

\end{fulllineitems}

\index{dclde (in module symjax.datasets)@\spxentry{dclde}\spxextra{in module symjax.datasets}}

\begin{fulllineitems}
\phantomsection\label{\detokenize{modules/datasets:symjax.datasets.dclde}}\pysigline{\sphinxbfcode{\sphinxupquote{dclde}}}
alias of {\hyperref[\detokenize{modules/datasets:symjax.datasets.dclde}]{\sphinxcrossref{\sphinxcode{\sphinxupquote{symjax.datasets.dclde}}}}}

\end{fulllineitems}

\index{irmas (class in symjax.datasets)@\spxentry{irmas}\spxextra{class in symjax.datasets}}

\begin{fulllineitems}
\phantomsection\label{\detokenize{modules/datasets:symjax.datasets.irmas}}\pysigline{\sphinxbfcode{\sphinxupquote{class }}\sphinxbfcode{\sphinxupquote{irmas}}}
music instrument classification

ref \sphinxurl{https://zenodo.org/record/1290750\#.WzCwSRyxXMU}

This dataset includes musical audio excerpts with annotations of the
predominant instrument(s) present. It was used for the evaluation in the
following article:

Bosch, J. J., Janer, J., Fuhrmann, F., \& Herrera, P. “A Comparison of Sound
Segregation Techniques for Predominant Instrument Recognition in Musical
Audio Signals”, in Proc. ISMIR (pp. 559\sphinxhyphen{}564), 2012

Please Acknowledge IRMAS in Academic Research

IRMAS is intended to be used for training and testing methods for the
automatic recognition of predominant instruments in musical audio. The
instruments considered are: cello, clarinet, flute, acoustic guitar,
electric guitar, organ, piano, saxophone, trumpet, violin, and human singing
voice. This dataset is derived from the one compiled by Ferdinand Fuhrmann
in his PhD thesis, with the difference that we provide audio data in stereo
format, the annotations in the testing dataset are limited to specific
pitched instruments, and there is a different amount and lenght of excerpts.

\end{fulllineitems}

\index{vocalset (class in symjax.datasets)@\spxentry{vocalset}\spxextra{class in symjax.datasets}}

\begin{fulllineitems}
\phantomsection\label{\detokenize{modules/datasets:symjax.datasets.vocalset}}\pysigline{\sphinxbfcode{\sphinxupquote{class }}\sphinxbfcode{\sphinxupquote{vocalset}}}
singer/technique/vowel of singing voices

source: \sphinxurl{https://zenodo.org/record/1442513\#.W7OaFBNKjx4}

We present VocalSet, a singing voice dataset consisting of 10.1 hours
of monophonic recorded audio of professional singers demonstrating both
standard and extended vocal techniques on all 5 vowels. Existing
singing voice datasets aim to capture a focused subset of singing
voice characteristics, and generally consist of just a few singers.
VocalSet contains recordings from 20 different singers (9 male, 11
female) and a range of voice types.  VocalSet aims to improve the
state of existing singing voice datasets and singing voice research by
capturing not only a range of vowels, but also a diverse set of voices
on many different vocal techniques, sung in contexts of scales,
arpeggios, long tones, and excerpts.
\index{load() (vocalset method)@\spxentry{load()}\spxextra{vocalset method}}

\begin{fulllineitems}
\phantomsection\label{\detokenize{modules/datasets:symjax.datasets.vocalset.load}}\pysiglinewithargsret{\sphinxbfcode{\sphinxupquote{load}}}{}{}~\begin{quote}\begin{description}
\item[{Parameters}] \leavevmode
\sphinxstyleliteralstrong{\sphinxupquote{path}} (\sphinxstyleliteralemphasis{\sphinxupquote{str}}\sphinxstyleliteralemphasis{\sphinxupquote{ (}}\sphinxstyleliteralemphasis{\sphinxupquote{optional}}\sphinxstyleliteralemphasis{\sphinxupquote{)}}) \textendash{} a string where to load the data and download if not present

\item[{Returns}] \leavevmode
\begin{itemize}
\item {} 
\sphinxstylestrong{singers} (\sphinxstyleemphasis{list}) \textendash{} the list of singers as strings, 11 males and 9 females as in male1,
male2, …

\item {} 
\sphinxstylestrong{genders} (\sphinxstyleemphasis{list}) \textendash{} the list of genders of the singers as in male, male, female, …

\item {} 
\sphinxstylestrong{vowels} (\sphinxstyleemphasis{list}) \textendash{} the vowels being pronunced

\item {} 
\sphinxstylestrong{data} (\sphinxstyleemphasis{list}) \textendash{} the list of waveforms, not all equal length

\end{itemize}

\end{description}\end{quote}

\end{fulllineitems}

\end{fulllineitems}

\index{freefield1010 (class in symjax.datasets)@\spxentry{freefield1010}\spxextra{class in symjax.datasets}}

\begin{fulllineitems}
\phantomsection\label{\detokenize{modules/datasets:symjax.datasets.freefield1010}}\pysigline{\sphinxbfcode{\sphinxupquote{class }}\sphinxbfcode{\sphinxupquote{freefield1010}}}
Audio binary classification, presence or absence of bird songs.
\sphinxhref{http://machine-listening.eecs.qmul.ac.uk/bird-audio-detection-challenge/\#downloads}{freefield1010}.
is a collection of over 7,000 excerpts from field recordings
around the world, gathered by the FreeSound project, and then standardised
for research. This collection is very diverse in location and environment,
and for the BAD Challenge we have newly annotated it for the
presence/absence of birds.

\end{fulllineitems}

\index{birdvox\_70k (class in symjax.datasets)@\spxentry{birdvox\_70k}\spxextra{class in symjax.datasets}}

\begin{fulllineitems}
\phantomsection\label{\detokenize{modules/datasets:symjax.datasets.birdvox_70k}}\pysigline{\sphinxbfcode{\sphinxupquote{class }}\sphinxbfcode{\sphinxupquote{birdvox\_70k}}}
a dataset for avian flight call detection in half\sphinxhyphen{}second clips

Version 1.0, April 2018.

Created By

Vincent Lostanlen (1, 2, 3), Justin Salamon (2, 3), Andrew Farnsworth (1),
Steve Kelling (1), and Juan Pablo Bello (2, 3).

(1): Cornell Lab of Ornithology (CLO)
(2): Center for Urban Science and Progress, New York University
(3): Music and Audio Research Lab, New York University

\sphinxurl{https://wp.nyu.edu/birdvox}

Description

The BirdVox\sphinxhyphen{}70k dataset contains 70k half\sphinxhyphen{}second clips from 6 audio
recordings in the BirdVox\sphinxhyphen{}full\sphinxhyphen{}night dataset, each about ten hours in
duration. These recordings come from ROBIN autonomous recording units,
placed near Ithaca, NY, USA during the fall 2015. They were captured on the
night of September 23rd, 2015, by six different sensors, originally
numbered 1, 2, 3, 5, 7, and 10.

Andrew Farnsworth used the Raven software to pinpoint every avian flight
call in time and frequency. He found 35402 flight calls in total.
He estimates that about 25 different species of passerines (thrushes,
warblers, and sparrows) are present in this recording. Species are not
labeled in BirdVox\sphinxhyphen{}70k, but it is possible to tell apart thrushes from
warblers and sparrows by looking at the center frequencies of their calls.
The annotation process took 102 hours.

The dataset can be used, among other things, for the research,development
and testing of bioacoustic classification models, including the
reproduction of the results reported in {[}1{]}.

For details on the hardware of ROBIN recording units, we refer the reader
to {[}2{]}.

{[}1{]} V. Lostanlen, J. Salamon, A. Farnsworth, S. Kelling, J. Bello. BirdVox\sphinxhyphen{}full\sphinxhyphen{}night: a dataset and benchmark for avian flight call detection. Proc. IEEE ICASSP, 2018.

{[}2{]} J. Salamon, J. P. Bello, A. Farnsworth, M. Robbins, S. Keen, H. Klinck, and S. Kelling. Towards the Automatic Classification of Avian Flight Calls for Bioacoustic Monitoring. PLoS One, 2016.

@inproceedings\{lostanlen2018icassp,
title = \{BirdVox\sphinxhyphen{}full\sphinxhyphen{}night: a dataset and benchmark for avian flight call detection\},
author = \{Lostanlen, Vincent and Salamon, Justin and Farnsworth, Andrew and Kelling, Steve and Bello, Juan Pablo\},
booktitle = \{Proc. IEEE ICASSP\},
year = \{2018\},
published = \{IEEE\},
venue = \{Calgary, Canada\},
month = \{April\},
\}
\index{download() (birdvox\_70k static method)@\spxentry{download()}\spxextra{birdvox\_70k static method}}

\begin{fulllineitems}
\phantomsection\label{\detokenize{modules/datasets:symjax.datasets.birdvox_70k.download}}\pysiglinewithargsret{\sphinxbfcode{\sphinxupquote{static }}\sphinxbfcode{\sphinxupquote{download}}}{\emph{\DUrole{n}{path}}}{}
Download the Birdvox dataset and store the result into the given
path
\begin{quote}\begin{description}
\item[{Parameters}] \leavevmode
\sphinxstyleliteralstrong{\sphinxupquote{path}} (\sphinxstyleliteralemphasis{\sphinxupquote{str}}) \textendash{} the path where the downloaded files will be stored. If the
directory does not exist, it is created.

\end{description}\end{quote}

\end{fulllineitems}

\index{load() (birdvox\_70k static method)@\spxentry{load()}\spxextra{birdvox\_70k static method}}

\begin{fulllineitems}
\phantomsection\label{\detokenize{modules/datasets:symjax.datasets.birdvox_70k.load}}\pysiglinewithargsret{\sphinxbfcode{\sphinxupquote{static }}\sphinxbfcode{\sphinxupquote{load}}}{\emph{\DUrole{n}{path}\DUrole{o}{=}\DUrole{default_value}{None}}}{}~\begin{quote}\begin{description}
\item[{Parameters}] \leavevmode
\sphinxstyleliteralstrong{\sphinxupquote{path}} (\sphinxstyleliteralemphasis{\sphinxupquote{str}}\sphinxstyleliteralemphasis{\sphinxupquote{ (}}\sphinxstyleliteralemphasis{\sphinxupquote{optional}}\sphinxstyleliteralemphasis{\sphinxupquote{)}}) \textendash{} default (\$DATASET\_PATH), the path to look for the data and
where the data will be downloaded if not present

\item[{Returns}] \leavevmode
\begin{itemize}
\item {} 
\sphinxstylestrong{wavs} (\sphinxstyleemphasis{array(70804, 12000)}) \textendash{} the waveforms in the time amplitude domain

\item {} 
\sphinxstylestrong{labels} (\sphinxstyleemphasis{array(70804,)}) \textendash{} binary values representing the presence or not of an avian

\item {} 
\sphinxstylestrong{recording} (\sphinxstyleemphasis{array(70804,)}) \textendash{} the file number from which the sample has been extracted

\end{itemize}

\end{description}\end{quote}

\end{fulllineitems}

\end{fulllineitems}

\index{birdvox\_dcase\_20k (class in symjax.datasets)@\spxentry{birdvox\_dcase\_20k}\spxextra{class in symjax.datasets}}

\begin{fulllineitems}
\phantomsection\label{\detokenize{modules/datasets:symjax.datasets.birdvox_dcase_20k}}\pysigline{\sphinxbfcode{\sphinxupquote{class }}\sphinxbfcode{\sphinxupquote{birdvox\_dcase\_20k}}}
Binary bird detection classification

Dataset is 16.5Go compressed.

BirdVox\sphinxhyphen{}DCASE\sphinxhyphen{}20k: a dataset for bird audio detection in 10\sphinxhyphen{}second
clips

Version 2.0, March 2018.

Created By

Vincent Lostanlen (1, 2, 3), Justin Salamon (2, 3), Andrew Farnsworth
(1), Steve Kelling (1), and Juan Pablo Bello (2, 3).

(1): Cornell Lab of Ornithology (CLO)
(2): Center for Urban Science and Progress, New York University
(3): Music and Audio Research Lab, New York University

\sphinxurl{https://wp.nyu.edu/birdvox}

Description

The BirdVox\sphinxhyphen{}DCASE\sphinxhyphen{}20k dataset contains 20,000 ten\sphinxhyphen{}second audio
recordings. These recordings come from ROBIN autonomous recording
units, placed near Ithaca, NY, USA during the fall 2015. They were
captured on the night of September 23rd, 2015, by six different
sensors, originally numbered 1, 2, 3, 5, 7, and 10.

Out of these 20,000 recording, 10,017 (50.09\%) contain at least one
bird vocalization (either song, call, or chatter).

The dataset is a derivative work of the BirdVox\sphinxhyphen{}full\sphinxhyphen{}night dataset
{[}1{]}, containing almost as much data but formatted into ten\sphinxhyphen{}second
excerpts rather than ten\sphinxhyphen{}hour full night recordings.

In addition, the BirdVox\sphinxhyphen{}DCASE\sphinxhyphen{}20k dataset is provided as a
development set in the context of the “Bird Audio Detection”
challenge, organized by DCASE (Detection and Classification of
Acoustic Scenes and Events) and the IEEE Signal Processing Society.

The dataset can be used, among other things, for the development and
evaluation of bioacoustic classification models.

We refer the reader to {[}1{]} for details on the distribution of the
data and {[}2{]} for details on the hardware of ROBIN recording units.

{[}1{]} V. Lostanlen, J. Salamon, A. Farnsworth, S. Kelling, J.P. Bello.
“BirdVox\sphinxhyphen{}full\sphinxhyphen{}night: a dataset and benchmark for avian flight call
detection”, Proc. IEEE ICASSP, 2018.

{[}2{]} J. Salamon, J. P. Bello, A. Farnsworth, M. Robbins, S. Keen,
H. Klinck, and S. Kelling. Towards the Automatic Classification of
Avian Flight Calls for Bioacoustic Monitoring. PLoS One, 2016.

Data Files

The wav folder contains the recordings as WAV files, sampled at
44,1 kHz, with a single channel (mono). The original sample rate
was 24 kHz.

The name of each wav file is a random 128\sphinxhyphen{}bit UUID (Universal
Unique IDentifier) string, which is randomized with respect to the
origin of the recording in BirdVox\sphinxhyphen{}full\sphinxhyphen{}night, both in terms of
time (UTC hour at the start of the excerpt) and space (location of
the sensor).

The origin of each 10\sphinxhyphen{}second excerpt is known by the challenge
organizers, but not disclosed to the participants.

Please Acknowledge BirdVox\sphinxhyphen{}DCASE\sphinxhyphen{}20k in Academic Research

When BirdVox\sphinxhyphen{}70k is used for academic research, we would highly
appreciate it if  scientific publications of works partly based on
this dataset cite the following publication:

V. Lostanlen, J. Salamon, A. Farnsworth, S. Kelling, J. Bello.
“BirdVox\sphinxhyphen{}full\sphinxhyphen{}night: a dataset and benchmark for avian flight call
detection”, Proc. IEEE ICASSP, 2018.

@inproceedings\{lostanlen2018icassp,
title = \{BirdVox\sphinxhyphen{}full\sphinxhyphen{}night: a dataset and benchmark for avian
flight call detection\},
author = \{Lostanlen, Vincent and Salamon, Justin and Farnsworth,
Andrew and Kelling, Steve and Bello, Juan Pablo\},
booktitle = \{Proc. IEEE ICASSP\},
year = \{2018\},
published = \{IEEE\},
venue = \{Calgary, Canada\},
month = \{April\},
\}

The creation of this dataset was supported by NSF grants 1125098
(BIRDCAST) and 1633259 (BIRDVOX), a Google Faculty Award, the Leon
Levy Foundation, and two anonymous donors.
\index{download() (birdvox\_dcase\_20k static method)@\spxentry{download()}\spxextra{birdvox\_dcase\_20k static method}}

\begin{fulllineitems}
\phantomsection\label{\detokenize{modules/datasets:symjax.datasets.birdvox_dcase_20k.download}}\pysiglinewithargsret{\sphinxbfcode{\sphinxupquote{static }}\sphinxbfcode{\sphinxupquote{download}}}{\emph{\DUrole{n}{path}}}{}
Download the Birdvox dataset and store the result into the given
path
\begin{quote}\begin{description}
\item[{Parameters}] \leavevmode
\sphinxstyleliteralstrong{\sphinxupquote{path}} (\sphinxstyleliteralemphasis{\sphinxupquote{str}}) \textendash{} the path where the downloaded files will be stored. If the
directory does not exist, it is created.

\end{description}\end{quote}

\end{fulllineitems}

\index{load() (birdvox\_dcase\_20k static method)@\spxentry{load()}\spxextra{birdvox\_dcase\_20k static method}}

\begin{fulllineitems}
\phantomsection\label{\detokenize{modules/datasets:symjax.datasets.birdvox_dcase_20k.load}}\pysiglinewithargsret{\sphinxbfcode{\sphinxupquote{static }}\sphinxbfcode{\sphinxupquote{load}}}{\emph{\DUrole{n}{path}\DUrole{o}{=}\DUrole{default_value}{None}}}{}~\begin{quote}\begin{description}
\item[{Parameters}] \leavevmode
\sphinxstyleliteralstrong{\sphinxupquote{path}} (\sphinxstyleliteralemphasis{\sphinxupquote{str}}\sphinxstyleliteralemphasis{\sphinxupquote{ (}}\sphinxstyleliteralemphasis{\sphinxupquote{optional}}\sphinxstyleliteralemphasis{\sphinxupquote{)}}) \textendash{} default (\$DATASET\_PATH), the path to look for the data and
where the data will be downloaded if not present

\item[{Returns}] \leavevmode
\begin{itemize}
\item {} 
\sphinxstylestrong{wavs} (\sphinxstyleemphasis{array}) \textendash{} the waveforms in the time amplitude domain

\item {} 
\sphinxstylestrong{labels} (\sphinxstyleemphasis{array}) \textendash{} binary values representing the presence or not of an avian

\item {} 
\sphinxstylestrong{recording} (\sphinxstyleemphasis{array}) \textendash{} the file number from which the sample has been extracted

\end{itemize}

\end{description}\end{quote}

\end{fulllineitems}

\end{fulllineitems}

\subsection{\sphinxstyleliteralintitle{\sphinxupquote{symjax.initializers}}}
\label{\detokenize{modules/initializers:symjax-initializers}}\label{\detokenize{modules/initializers::doc}}\index{initializers (in module symjax)@\spxentry{initializers}\spxextra{in module symjax}}

\begin{fulllineitems}
\phantomsection\label{\detokenize{modules/initializers:symjax.initializers}}\pysigline{\sphinxbfcode{\sphinxupquote{initializers}}}
alias of {\hyperref[\detokenize{modules/initializers:module-symjax.initializers}]{\sphinxcrossref{\sphinxcode{\sphinxupquote{symjax.initializers}}}}}

\end{fulllineitems}

\phantomsection\label{\detokenize{modules/initializers:module-symjax.initializers}}\index{module@\spxentry{module}!symjax.initializers@\spxentry{symjax.initializers}}\index{symjax.initializers@\spxentry{symjax.initializers}!module@\spxentry{module}}\index{glorot() (in module symjax.initializers)@\spxentry{glorot()}\spxextra{in module symjax.initializers}}

\begin{fulllineitems}
\phantomsection\label{\detokenize{modules/initializers:symjax.initializers.glorot}}\pysiglinewithargsret{\sphinxbfcode{\sphinxupquote{glorot}}}{\emph{shape}, \emph{gain=1}, \emph{distribution=\textless{}function normal\textgreater{}}}{}
Glorot weight initialization.
This is also known as Xavier initialization {\color{red}\bfseries{}{[}1{]}\_}.
\begin{quote}\begin{description}
\item[{Parameters}] \leavevmode\begin{itemize}
\item {} 
\sphinxstyleliteralstrong{\sphinxupquote{initializer}} (\sphinxstyleliteralemphasis{\sphinxupquote{lasagne.init.Initializer}}) \textendash{} Initializer used to sample the weights, must accept \sphinxtitleref{std} in its
constructor to sample from a distribution with a given standard
deviation.

\item {} 
\sphinxstyleliteralstrong{\sphinxupquote{gain}} (\sphinxstyleliteralemphasis{\sphinxupquote{float}}\sphinxstyleliteralemphasis{\sphinxupquote{ or }}\sphinxstyleliteralemphasis{\sphinxupquote{\textquotesingle{}relu\textquotesingle{}}}) \textendash{} Scaling factor for the weights. Set this to \sphinxcode{\sphinxupquote{1.0}} for linear and
sigmoid units, to ‘relu’ or \sphinxcode{\sphinxupquote{sqrt(2)}} for rectified linear units, and
to \sphinxcode{\sphinxupquote{sqrt(2/(1+alpha**2))}} for leaky rectified linear units with
leakiness \sphinxcode{\sphinxupquote{alpha}}. Other transfer functions may need different
factors.

\item {} 
\sphinxstyleliteralstrong{\sphinxupquote{c01b}} (\sphinxstyleliteralemphasis{\sphinxupquote{bool}}) \textendash{} For a \sphinxcode{\sphinxupquote{lasagne.layers.cuda\_convnet.Conv2DCCLayer}} constructed
with \sphinxcode{\sphinxupquote{dimshuffle=False}}, \sphinxtitleref{c01b} must be set to \sphinxcode{\sphinxupquote{True}} to compute
the correct fan\sphinxhyphen{}in and fan\sphinxhyphen{}out.

\end{itemize}

\end{description}\end{quote}
\subsubsection*{References}
\subsubsection*{Notes}

For a \sphinxcode{\sphinxupquote{DenseLayer}}, if \sphinxcode{\sphinxupquote{gain=\textquotesingle{}relu\textquotesingle{}}}
and \sphinxcode{\sphinxupquote{initializer=Uniform}}, the weights are initialized as
.. math:

\begin{sphinxVerbatim}[commandchars=\\\{\}]
\PYG{n}{a} \PYG{o}{\PYGZam{}}\PYG{o}{=} \PYGZbs{}\PYG{n}{sqrt}\PYG{p}{\PYGZob{}}\PYGZbs{}\PYG{n}{frac}\PYG{p}{\PYGZob{}}\PYG{l+m+mi}{12}\PYG{p}{\PYGZcb{}}\PYG{p}{\PYGZob{}}\PYG{n}{fan\PYGZus{}}\PYG{p}{\PYGZob{}}\PYG{o+ow}{in}\PYG{p}{\PYGZcb{}}\PYG{o}{+}\PYG{n}{fan\PYGZus{}}\PYG{p}{\PYGZob{}}\PYG{n}{out}\PYG{p}{\PYGZcb{}}\PYG{p}{\PYGZcb{}}\PYG{p}{\PYGZcb{}}\PYGZbs{}\PYGZbs{}
\PYG{n}{W} \PYG{o}{\PYGZam{}}\PYGZbs{}\PYG{n}{sim} \PYG{n}{U}\PYG{p}{[}\PYG{o}{\PYGZhy{}}\PYG{n}{a}\PYG{p}{,} \PYG{n}{a}\PYG{p}{]}
\end{sphinxVerbatim}

If \sphinxcode{\sphinxupquote{gain=1}} and \sphinxcode{\sphinxupquote{initializer=Normal}}, the weights are initialized as
.. math:

\begin{sphinxVerbatim}[commandchars=\\\{\}]
\PYGZbs{}\PYG{n}{sigma} \PYG{o}{\PYGZam{}}\PYG{o}{=} \PYGZbs{}\PYG{n}{sqrt}\PYG{p}{\PYGZob{}}\PYGZbs{}\PYG{n}{frac}\PYG{p}{\PYGZob{}}\PYG{l+m+mi}{2}\PYG{p}{\PYGZcb{}}\PYG{p}{\PYGZob{}}\PYG{n}{fan\PYGZus{}}\PYG{p}{\PYGZob{}}\PYG{o+ow}{in}\PYG{p}{\PYGZcb{}}\PYG{o}{+}\PYG{n}{fan\PYGZus{}}\PYG{p}{\PYGZob{}}\PYG{n}{out}\PYG{p}{\PYGZcb{}}\PYG{p}{\PYGZcb{}}\PYG{p}{\PYGZcb{}}\PYGZbs{}\PYGZbs{}
\PYG{n}{W} \PYG{o}{\PYGZam{}}\PYGZbs{}\PYG{n}{sim} \PYG{n}{N}\PYG{p}{(}\PYG{l+m+mi}{0}\PYG{p}{,} \PYGZbs{}\PYG{n}{sigma}\PYG{p}{)}
\end{sphinxVerbatim}

\end{fulllineitems}

\index{he() (in module symjax.initializers)@\spxentry{he()}\spxextra{in module symjax.initializers}}

\begin{fulllineitems}
\phantomsection\label{\detokenize{modules/initializers:symjax.initializers.he}}\pysiglinewithargsret{\sphinxbfcode{\sphinxupquote{he}}}{\emph{shape}, \emph{gain=1.4142135623730951}, \emph{distribution=\textless{}function normal\textgreater{}}}{}
He weight initialization.
Weights are initialized with a standard deviation of
\(\sigma = gain \sqrt{\frac{1}{fan_{in}}}\) {\color{red}\bfseries{}{[}1{]}\_}.
\begin{quote}\begin{description}
\item[{Parameters}] \leavevmode\begin{itemize}
\item {} 
\sphinxstyleliteralstrong{\sphinxupquote{initializer}} (\sphinxstyleliteralemphasis{\sphinxupquote{lasagne.init.Initializer}}) \textendash{} Initializer used to sample the weights, must accept \sphinxtitleref{std} in its
constructor to sample from a distribution with a given standard
deviation.

\item {} 
\sphinxstyleliteralstrong{\sphinxupquote{gain}} (\sphinxstyleliteralemphasis{\sphinxupquote{float}}\sphinxstyleliteralemphasis{\sphinxupquote{ or }}\sphinxstyleliteralemphasis{\sphinxupquote{\textquotesingle{}relu\textquotesingle{}}}) \textendash{} Scaling factor for the weights. Set this to \sphinxcode{\sphinxupquote{1.0}} for linear and
sigmoid units, to ‘relu’ or \sphinxcode{\sphinxupquote{sqrt(2)}} for rectified linear units, and
to \sphinxcode{\sphinxupquote{sqrt(2/(1+alpha**2))}} for leaky rectified linear units with
leakiness \sphinxcode{\sphinxupquote{alpha}}. Other transfer functions may need different
factors.

\item {} 
\sphinxstyleliteralstrong{\sphinxupquote{c01b}} (\sphinxstyleliteralemphasis{\sphinxupquote{bool}}) \textendash{} For a \sphinxcode{\sphinxupquote{lasagne.layers.cuda\_convnet.Conv2DCCLayer}} constructed
with \sphinxcode{\sphinxupquote{dimshuffle=False}}, \sphinxtitleref{c01b} must be set to \sphinxcode{\sphinxupquote{True}} to compute
the correct fan\sphinxhyphen{}in and fan\sphinxhyphen{}out.

\end{itemize}

\end{description}\end{quote}
\subsubsection*{References}

\sphinxstrong{See also:}

\begin{description}
\item[{\sphinxcode{\sphinxupquote{HeNormal()}}}] \leavevmode
Shortcut with Gaussian initializer.

\item[{\sphinxcode{\sphinxupquote{HeUniform()}}}] \leavevmode
Shortcut with uniform initializer.

\end{description}

\end{fulllineitems}

\index{normal() (in module symjax.initializers)@\spxentry{normal()}\spxextra{in module symjax.initializers}}

\begin{fulllineitems}
\phantomsection\label{\detokenize{modules/initializers:symjax.initializers.normal}}\pysiglinewithargsret{\sphinxbfcode{\sphinxupquote{normal}}}{\emph{\DUrole{n}{shape}}, \emph{\DUrole{n}{mean}\DUrole{o}{=}\DUrole{default_value}{0.0}}, \emph{\DUrole{n}{std}\DUrole{o}{=}\DUrole{default_value}{1.0}}}{}
Sample initial weights from the Gaussian distribution.
Initial weight parameters are sampled from N(mean, std).
\begin{quote}\begin{description}
\item[{Parameters}] \leavevmode\begin{itemize}
\item {} 
\sphinxstyleliteralstrong{\sphinxupquote{std}} (\sphinxstyleliteralemphasis{\sphinxupquote{float}}) \textendash{} Std of initial parameters.

\item {} 
\sphinxstyleliteralstrong{\sphinxupquote{mean}} (\sphinxstyleliteralemphasis{\sphinxupquote{float}}) \textendash{} Mean of initial parameters.

\end{itemize}

\end{description}\end{quote}

\end{fulllineitems}

\index{uniform() (in module symjax.initializers)@\spxentry{uniform()}\spxextra{in module symjax.initializers}}

\begin{fulllineitems}
\phantomsection\label{\detokenize{modules/initializers:symjax.initializers.uniform}}\pysiglinewithargsret{\sphinxbfcode{\sphinxupquote{uniform}}}{\emph{\DUrole{n}{shape}}, \emph{\DUrole{n}{range}\DUrole{o}{=}\DUrole{default_value}{0.01}}, \emph{\DUrole{n}{std}\DUrole{o}{=}\DUrole{default_value}{None}}, \emph{\DUrole{n}{mean}\DUrole{o}{=}\DUrole{default_value}{0.0}}}{}
Sample initial weights from the uniform distribution.
Parameters are sampled from U(a, b).
\begin{quote}\begin{description}
\item[{Parameters}] \leavevmode\begin{itemize}
\item {} 
\sphinxstyleliteralstrong{\sphinxupquote{range}} (\sphinxstyleliteralemphasis{\sphinxupquote{float}}\sphinxstyleliteralemphasis{\sphinxupquote{ or }}\sphinxstyleliteralemphasis{\sphinxupquote{tuple}}) \textendash{} When std is None then range determines a, b. If range is a float the
weights are sampled from U(\sphinxhyphen{}range, range). If range is a tuple the
weights are sampled from U(range{[}0{]}, range{[}1{]}).

\item {} 
\sphinxstyleliteralstrong{\sphinxupquote{std}} (\sphinxstyleliteralemphasis{\sphinxupquote{float}}\sphinxstyleliteralemphasis{\sphinxupquote{ or }}\sphinxstyleliteralemphasis{\sphinxupquote{None}}) \textendash{} If std is a float then the weights are sampled from
U(mean \sphinxhyphen{} numpy.sqrt(3) * std, mean + numpy.sqrt(3) * std).

\item {} 
\sphinxstyleliteralstrong{\sphinxupquote{mean}} (\sphinxstyleliteralemphasis{\sphinxupquote{float}}) \textendash{} see std for description.

\end{itemize}

\end{description}\end{quote}

\end{fulllineitems}

\subsection{\sphinxstyleliteralintitle{\sphinxupquote{symjax.layers}}}
\label{\detokenize{modules/layers:module-symjax.layers}}\label{\detokenize{modules/layers:symjax-layers}}\label{\detokenize{modules/layers::doc}}\index{module@\spxentry{module}!symjax.layers@\spxentry{symjax.layers}}\index{symjax.layers@\spxentry{symjax.layers}!module@\spxentry{module}}

\subsubsection{Renormalization}
\label{\detokenize{modules/layers:renormalization}}

\begin{savenotes}\sphinxatlongtablestart\begin{longtable}[c]{\X{1}{2}\X{1}{2}}
\hline

\endfirsthead

\multicolumn{2}{c}%
{\makebox[0pt]{\sphinxtablecontinued{\tablename\ \thetable{} \textendash{} continued from previous page}}}\\
\hline

\endhead

\hline
\multicolumn{2}{r}{\makebox[0pt][r]{\sphinxtablecontinued{continues on next page}}}\\
\endfoot

\endlastfoot

\sphinxcode{\sphinxupquote{BatchNormalization}}(input\_or\_shape, axis, …)
&
batch\sphinxhyphen{}normalization layer
\\
\hline
\end{longtable}\sphinxatlongtableend\end{savenotes}

\subsubsection{Data Augmentation}
\label{\detokenize{modules/layers:data-augmentation}}

\begin{savenotes}\sphinxatlongtablestart\begin{longtable}[c]{\X{1}{2}\X{1}{2}}
\hline

\endfirsthead

\multicolumn{2}{c}%
{\makebox[0pt]{\sphinxtablecontinued{\tablename\ \thetable{} \textendash{} continued from previous page}}}\\
\hline

\endhead

\hline
\multicolumn{2}{r}{\makebox[0pt][r]{\sphinxtablecontinued{continues on next page}}}\\
\endfoot

\endlastfoot

\sphinxcode{\sphinxupquote{RandomCrop}}(input\_or\_shape, crop\_shape, …)
&
random crop selection form the input
\\
\hline
\sphinxcode{\sphinxupquote{RandomFlip}}(input\_or\_shape, p, axis, …{[}, seed{]})
&
random axis flip on the input
\\
\hline
\sphinxcode{\sphinxupquote{Dropout}}(input\_or\_shape, p, deterministic{[}, seed{]})
&
binary mask onto the input
\\
\hline
\end{longtable}\sphinxatlongtableend\end{savenotes}

\subsubsection{Convolution}
\label{\detokenize{modules/layers:convolution}}

\begin{savenotes}\sphinxatlongtablestart\begin{longtable}[c]{\X{1}{2}\X{1}{2}}
\hline

\endfirsthead

\multicolumn{2}{c}%
{\makebox[0pt]{\sphinxtablecontinued{\tablename\ \thetable{} \textendash{} continued from previous page}}}\\
\hline

\endhead

\hline
\multicolumn{2}{r}{\makebox[0pt][r]{\sphinxtablecontinued{continues on next page}}}\\
\endfoot

\endlastfoot

\sphinxcode{\sphinxupquote{Conv1D}}(input\_or\_shape, n\_filters, filter\_length)
&
1\sphinxhyphen{}D (time) convolution
\\
\hline
\sphinxcode{\sphinxupquote{Conv2D}}(input\_or\_shape, n\_filters, filter\_shape)
&
2\sphinxhyphen{}D (spatial) convolution
\\
\hline
\end{longtable}\sphinxatlongtableend\end{savenotes}

\subsubsection{Pooling}
\label{\detokenize{modules/layers:pooling}}

\begin{savenotes}\sphinxatlongtablestart\begin{longtable}[c]{\X{1}{2}\X{1}{2}}
\hline

\endfirsthead

\multicolumn{2}{c}%
{\makebox[0pt]{\sphinxtablecontinued{\tablename\ \thetable{} \textendash{} continued from previous page}}}\\
\hline

\endhead

\hline
\multicolumn{2}{r}{\makebox[0pt][r]{\sphinxtablecontinued{continues on next page}}}\\
\endfoot

\endlastfoot

\sphinxcode{\sphinxupquote{Pool1D}}(input\_or\_shape, pool\_shape{[}, …{]})
&
2\sphinxhyphen{}D (spatial) pooling
\\
\hline
\sphinxcode{\sphinxupquote{Pool2D}}(input\_or\_shape, pool\_shape{[}, …{]})
&
2\sphinxhyphen{}D (spatial) pooling
\\
\hline
\end{longtable}\sphinxatlongtableend\end{savenotes}

\subsubsection{Utilities}
\label{\detokenize{modules/layers:utilities}}

\begin{savenotes}\sphinxatlongtablestart\begin{longtable}[c]{\X{1}{2}\X{1}{2}}
\hline

\endfirsthead

\multicolumn{2}{c}%
{\makebox[0pt]{\sphinxtablecontinued{\tablename\ \thetable{} \textendash{} continued from previous page}}}\\
\hline

\endhead

\hline
\multicolumn{2}{r}{\makebox[0pt][r]{\sphinxtablecontinued{continues on next page}}}\\
\endfoot

\endlastfoot

\sphinxcode{\sphinxupquote{forward}}(input, layers)
&
perform a forward path in the layers given an input
\\
\hline
\end{longtable}\sphinxatlongtableend\end{savenotes}

\subsection{\sphinxstyleliteralintitle{\sphinxupquote{symjax.optimizers}}}
\label{\detokenize{modules/optimizers:symjax-optimizers}}\label{\detokenize{modules/optimizers::doc}}\index{optimizers (in module symjax)@\spxentry{optimizers}\spxextra{in module symjax}}

\begin{fulllineitems}
\phantomsection\label{\detokenize{modules/optimizers:symjax.optimizers}}\pysigline{\sphinxbfcode{\sphinxupquote{optimizers}}}
alias of {\hyperref[\detokenize{modules/optimizers:module-symjax.optimizers}]{\sphinxcrossref{\sphinxcode{\sphinxupquote{symjax.optimizers}}}}}

\end{fulllineitems}

\phantomsection\label{\detokenize{modules/optimizers:module-symjax.optimizers}}\index{module@\spxentry{module}!symjax.optimizers@\spxentry{symjax.optimizers}}\index{symjax.optimizers@\spxentry{symjax.optimizers}!module@\spxentry{module}}\index{Adam (class in symjax.optimizers)@\spxentry{Adam}\spxextra{class in symjax.optimizers}}

\begin{fulllineitems}
\phantomsection\label{\detokenize{modules/optimizers:symjax.optimizers.Adam}}\pysiglinewithargsret{\sphinxbfcode{\sphinxupquote{class }}\sphinxbfcode{\sphinxupquote{Adam}}}{\emph{\DUrole{n}{grads\_or\_loss}}, \emph{\DUrole{n}{learning\_rate}}, \emph{\DUrole{n}{beta1}\DUrole{o}{=}\DUrole{default_value}{0.9}}, \emph{\DUrole{n}{beta2}\DUrole{o}{=}\DUrole{default_value}{0.999}}, \emph{\DUrole{n}{epsilon}\DUrole{o}{=}\DUrole{default_value}{1e\sphinxhyphen{}06}}, \emph{\DUrole{n}{params}\DUrole{o}{=}\DUrole{default_value}{None}}}{}
Adaptive Gradient Based Optimization with renormalization
\begin{quote}\begin{description}
\item[{Parameters}] \leavevmode\begin{itemize}
\item {} 
\sphinxstyleliteralstrong{\sphinxupquote{grads\_or\_loss}} (\sphinxstyleliteralemphasis{\sphinxupquote{scalar tensor}}\sphinxstyleliteralemphasis{\sphinxupquote{ or }}\sphinxstyleliteralemphasis{\sphinxupquote{list of gradients}}) \textendash{} either the loss (scalar of Tensor type) to be differentied
or the list of gradients already computed and possibly altered
manually (such as clipping)

\item {} 
\sphinxstyleliteralstrong{\sphinxupquote{params}} (\sphinxstyleliteralemphasis{\sphinxupquote{list of parameters to update}}) \textendash{} if grads\_or\_loss is al list then it should be ordered w.r.t. the
given parameters

\item {} 
\sphinxstyleliteralstrong{\sphinxupquote{learning\_rate}} (\sphinxstyleliteralemphasis{\sphinxupquote{constant}}\sphinxstyleliteralemphasis{\sphinxupquote{ or }}\sphinxstyleliteralemphasis{\sphinxupquote{Tensor}}) \textendash{} the learning rate use to update the parameters

\item {} 
\sphinxstyleliteralstrong{\sphinxupquote{beta1}} (\sphinxstyleliteralemphasis{\sphinxupquote{constant}}\sphinxstyleliteralemphasis{\sphinxupquote{ or }}\sphinxstyleliteralemphasis{\sphinxupquote{Tensor}}) \textendash{} the value of the exponential moving average of the average of the
gradients through time (updates)

\item {} 
\sphinxstyleliteralstrong{\sphinxupquote{beta2}} (\sphinxstyleliteralemphasis{\sphinxupquote{constant}}\sphinxstyleliteralemphasis{\sphinxupquote{ or }}\sphinxstyleliteralemphasis{\sphinxupquote{Tensor}}) \textendash{} the value of the exponential moving average of the variance of the
gradients through time

\end{itemize}

\end{description}\end{quote}
\index{updates (Adam attribute)@\spxentry{updates}\spxextra{Adam attribute}}

\begin{fulllineitems}
\phantomsection\label{\detokenize{modules/optimizers:symjax.optimizers.Adam.updates}}\pysigline{\sphinxbfcode{\sphinxupquote{updates}}}~\begin{quote}\begin{description}
\item[{Type}] \leavevmode
list of updates

\end{description}\end{quote}

\end{fulllineitems}

\index{variables (Adam attribute)@\spxentry{variables}\spxextra{Adam attribute}}

\begin{fulllineitems}
\phantomsection\label{\detokenize{modules/optimizers:symjax.optimizers.Adam.variables}}\pysigline{\sphinxbfcode{\sphinxupquote{variables}}}~\begin{quote}\begin{description}
\item[{Type}] \leavevmode
list of variables

\end{description}\end{quote}

\end{fulllineitems}

\end{fulllineitems}

\index{NesterovMomentum (class in symjax.optimizers)@\spxentry{NesterovMomentum}\spxextra{class in symjax.optimizers}}

\begin{fulllineitems}
\phantomsection\label{\detokenize{modules/optimizers:symjax.optimizers.NesterovMomentum}}\pysiglinewithargsret{\sphinxbfcode{\sphinxupquote{class }}\sphinxbfcode{\sphinxupquote{NesterovMomentum}}}{\emph{\DUrole{n}{grads\_or\_loss}}, \emph{\DUrole{n}{learning\_rate}}, \emph{\DUrole{n}{momentum}}, \emph{\DUrole{n}{params}\DUrole{o}{=}\DUrole{default_value}{None}}}{}
Nesterov momentum Optimization
\begin{quote}\begin{description}
\item[{Parameters}] \leavevmode\begin{itemize}
\item {} 
\sphinxstyleliteralstrong{\sphinxupquote{grads\_or\_loss}} (\sphinxstyleliteralemphasis{\sphinxupquote{scalar tensor}}\sphinxstyleliteralemphasis{\sphinxupquote{ or }}\sphinxstyleliteralemphasis{\sphinxupquote{list of gradients}}) \textendash{} either the loss (scalar of Tensor type) to be differentied
or the list of gradients already computed and possibly altered
manually (such as clipping)

\item {} 
\sphinxstyleliteralstrong{\sphinxupquote{params}} (\sphinxstyleliteralemphasis{\sphinxupquote{list of parameters to update}}) \textendash{} if grads\_or\_loss is al list then it should be ordered w.r.t. the
given parameters

\item {} 
\sphinxstyleliteralstrong{\sphinxupquote{learning\_rate}} (\sphinxstyleliteralemphasis{\sphinxupquote{constant}}\sphinxstyleliteralemphasis{\sphinxupquote{ or }}\sphinxstyleliteralemphasis{\sphinxupquote{Tensor}}) \textendash{} the learning rate use to update the parameters

\end{itemize}

\end{description}\end{quote}
\index{updates (NesterovMomentum attribute)@\spxentry{updates}\spxextra{NesterovMomentum attribute}}

\begin{fulllineitems}
\phantomsection\label{\detokenize{modules/optimizers:symjax.optimizers.NesterovMomentum.updates}}\pysigline{\sphinxbfcode{\sphinxupquote{updates}}}~\begin{quote}\begin{description}
\item[{Type}] \leavevmode
list of updates

\end{description}\end{quote}

\end{fulllineitems}

\index{variables (NesterovMomentum attribute)@\spxentry{variables}\spxextra{NesterovMomentum attribute}}

\begin{fulllineitems}
\phantomsection\label{\detokenize{modules/optimizers:symjax.optimizers.NesterovMomentum.variables}}\pysigline{\sphinxbfcode{\sphinxupquote{variables}}}~\begin{quote}\begin{description}
\item[{Type}] \leavevmode
list of variables

\end{description}\end{quote}

\end{fulllineitems}

\end{fulllineitems}

\index{SGD (class in symjax.optimizers)@\spxentry{SGD}\spxextra{class in symjax.optimizers}}

\begin{fulllineitems}
\phantomsection\label{\detokenize{modules/optimizers:symjax.optimizers.SGD}}\pysiglinewithargsret{\sphinxbfcode{\sphinxupquote{class }}\sphinxbfcode{\sphinxupquote{SGD}}}{\emph{\DUrole{n}{grads\_or\_loss}}, \emph{\DUrole{n}{learning\_rate}}, \emph{\DUrole{n}{params}\DUrole{o}{=}\DUrole{default_value}{None}}}{}
Gradient Descent Optimization
\begin{quote}\begin{description}
\item[{Parameters}] \leavevmode\begin{itemize}
\item {} 
\sphinxstyleliteralstrong{\sphinxupquote{grads\_or\_loss}} (\sphinxstyleliteralemphasis{\sphinxupquote{scalar tensor}}\sphinxstyleliteralemphasis{\sphinxupquote{ or }}\sphinxstyleliteralemphasis{\sphinxupquote{list of gradients}}) \textendash{} either the loss (scalar of Tensor type) to be differentied
or the list of gradients already computed and possibly altered
manually (such as clipping)

\item {} 
\sphinxstyleliteralstrong{\sphinxupquote{params}} (\sphinxstyleliteralemphasis{\sphinxupquote{list of parameters to update}}) \textendash{} if grads\_or\_loss is al list then it should be ordered w.r.t. the
given parameters

\item {} 
\sphinxstyleliteralstrong{\sphinxupquote{learning\_rate}} (\sphinxstyleliteralemphasis{\sphinxupquote{constant}}\sphinxstyleliteralemphasis{\sphinxupquote{ or }}\sphinxstyleliteralemphasis{\sphinxupquote{Tensor}}) \textendash{} the learning rate use to update the parameters

\end{itemize}

\end{description}\end{quote}
\index{updates (SGD attribute)@\spxentry{updates}\spxextra{SGD attribute}}

\begin{fulllineitems}
\phantomsection\label{\detokenize{modules/optimizers:symjax.optimizers.SGD.updates}}\pysigline{\sphinxbfcode{\sphinxupquote{updates}}}~\begin{quote}\begin{description}
\item[{Type}] \leavevmode
list of updates

\end{description}\end{quote}

\end{fulllineitems}

\index{variables (SGD attribute)@\spxentry{variables}\spxextra{SGD attribute}}

\begin{fulllineitems}
\phantomsection\label{\detokenize{modules/optimizers:symjax.optimizers.SGD.variables}}\pysigline{\sphinxbfcode{\sphinxupquote{variables}}}~\begin{quote}\begin{description}
\item[{Type}] \leavevmode
list of variables

\end{description}\end{quote}

\end{fulllineitems}

\end{fulllineitems}

\chapter{Indices and tables}
\label{\detokenize{index:indices-and-tables}}\begin{itemize}
\item {} 
\DUrole{xref,std,std-ref}{genindex}

\item {} 
\DUrole{xref,std,std-ref}{modindex}

\item {} 
\DUrole{xref,std,std-ref}{search}

\end{itemize}

\begin{sphinxthebibliography}{CT}
\bibitem[CT]{modules/signal:ct}
Cooley, James W., and John W. Tukey, 1965, “An algorithm for the
machine calculation of complex Fourier series,” \sphinxstyleemphasis{Math. Comput.}
19: 297\sphinxhyphen{}301.
\end{sphinxthebibliography}

\renewcommand{\indexname}{Python Module Index}
\begin{sphinxtheindex}
\let\bigletter\sphinxstyleindexlettergroup
\bigletter{s}
\item\relax\sphinxstyleindexentry{symjax}\sphinxstyleindexpageref{modules/symjax:\detokenize{module-symjax}}
\item\relax\sphinxstyleindexentry{symjax.initializers}\sphinxstyleindexpageref{modules/initializers:\detokenize{module-symjax.initializers}}
\item\relax\sphinxstyleindexentry{symjax.layers}\sphinxstyleindexpageref{modules/layers:\detokenize{module-symjax.layers}}
\item\relax\sphinxstyleindexentry{symjax.optimizers}\sphinxstyleindexpageref{modules/optimizers:\detokenize{module-symjax.optimizers}}
\item\relax\sphinxstyleindexentry{symjax.tensor}\sphinxstyleindexpageref{modules/tensor:\detokenize{module-symjax.tensor}}
\item\relax\sphinxstyleindexentry{symjax.tensor.control\_flow}\sphinxstyleindexpageref{modules/tensor:\detokenize{module-symjax.tensor.control_flow}}
\item\relax\sphinxstyleindexentry{symjax.tensor.index\_ops}\sphinxstyleindexpageref{modules/tensor:\detokenize{module-symjax.tensor.index_ops}}
\item\relax\sphinxstyleindexentry{symjax.tensor.ops\_math}\sphinxstyleindexpageref{modules/tensor:\detokenize{module-symjax.tensor.ops_math}}
\item\relax\sphinxstyleindexentry{symjax.tensor.pdfs}\sphinxstyleindexpageref{modules/pdfs:\detokenize{module-symjax.tensor.pdfs}}
\item\relax\sphinxstyleindexentry{symjax.tensor.random}\sphinxstyleindexpageref{modules/random:\detokenize{module-symjax.tensor.random}}
\item\relax\sphinxstyleindexentry{symjax.tensor.signal}\sphinxstyleindexpageref{modules/signal:\detokenize{module-symjax.tensor.signal}}
\item\relax\sphinxstyleindexentry{symjax.utils}\sphinxstyleindexpageref{modules/utils:\detokenize{module-symjax.utils}}
\end{sphinxtheindex}

\renewcommand{\indexname}{Index}
\printindex
\end{document}